\documentclass{article}%
\usepackage{amsmath}
\usepackage{amsfonts}
\usepackage{amssymb}
\usepackage{graphicx}%
\newtheorem{theorem}{Theorem}

\newtheorem{axiom}[theorem]{Axiom}

\newtheorem{corollary}[theorem]{Corollary}

\newtheorem{definition}[theorem]{Definition}

\newtheorem{lemma}[theorem]{Lemma}
\newtheorem{notation}[theorem]{Notation}
\newtheorem{problem}[theorem]{Problem}
\newtheorem{proposition}[theorem]{Proposition}

\newenvironment{proof}[1][Proof]{\noindent\textbf{#1.} }{\ \rule{0.5em}{0.5em}}
\begin{document}

\title{Mathematics for theoretical physics}
\author{Jean Claude Dutailly\\Paris}
\maketitle

\begin{abstract}
This book intends to give the main definitions and theorems in mathematics
which could be useful for workers in theoretical physics. It gives an
extensive and precise coverage of the subjects which are addressed, in a
consistent and intelligible manner.The first part addresses the Foundations
(mathematical logic, set theory, categories), the second Algebra (algebraic
strucutes, groups, vector spaces tensors, matrices, Clifford algebra).\ The
third Analysis (general topology, measure theory, Banach Spaces, Spectral
theory).\ The fourth Differential Geometry (derivatives, manifolds, tensorial
bundle, pseudo-riemannian manifolds, symplectic manifolds).\ The fifth Lie
Algebras, Lie Groups.and representation theory. The sixth Fiber bundles and
jets. The last one Functional Analysis (differential operators, distributions,
ODE, PDE, variational calculus).\ Several signficant new results are presented
(distributions over vector bundles, functional derivative, spin bundle and
manifolds with boundary).

\end{abstract}

The purpose of this book is to give a comprehensive collection of precise
definitions and results in advanced mathematics, which can be useful to
workers in mathematics or physics.

The specificities of this book are :

- it is self contained : any definition or notation used can be found within

- it is precise : any theorem lists the precise conditions which must be met
for its use

- it is easy to use : the book proceeds from the simple to the most advanced
topics, but in any part the necessary definitions are reminded so that the
reader can enter quickly into the subject

- it is comprehensive : it addresses the basic concepts but reaches most of
the advanced topics which are required nowadays

- it is pedagogical : the key points and usual misunderstandings are
underlined so that the reader can get a strong grasp of the tools which are presented.

The first option is unusual for a book of this kind. Usually a book starts
with the assumption that the reader has already some background
knowledge.\ The problem is that nobody has the same background.\ So a great
deal is dedicated to remind some basic stuff, in an abbreviated way, which
does not left much scope to their understanding, and is limited to specific
cases. In fact, starting from the very beginning, it has been easy, step by
step, to expose each concept in the most general settings. And, by proceeding
this way, to extend the scope of many results so that they can be made
available to the - unavoidable - special case that the reader may face.
Overall it gives a fresh, unified view of the mathematics, but still
affordable because it avoids as far as possible the sophisticated language
which is fashionable. The goal is that the reader understands clearly and
effortlessly, not to prove the extent of the author's knowledge.

The definitions chosen here meet the "generally accepted definitions" in
mathematics. However, as they come in many flavors according to the authors
and their field of interest, we have striven to take definitions which are
both the most general and the most easy to use.

Of course this cannot be achieved with some drawbacks. So many demonstrations
are omitted. More precisely the chosen option is the following :

- whenever a demonstration is short, it is given entirely, at least as an
example of "how it works"

- when a demonstration is too long and involves either technical or specific
conditions, a precise reference to where the demonstration can be found is
given. Anyway the theorem is written in accordance with the notations and
definitions of this book, and a special attention has been given that they
match the reference.

- exceptionally, when this is a well known theorem, whose demonstration can be
found easily in any book on the subject, there is no reference.

The bibliography is short. Indeed due to the scope which is covered it could
be enormous.\ So it is strictly limited to the works which are referenced in
the text, with a priority to the most easily available sources.

This is not mainly a research paper, even if the unification of the concepts
is, in many ways, new, but some significant results appear here for the first
time, to my knowledge.

- distributions over vector bundles

- a rigorous definition of functional derivatives

- a manifold with boundary can be defined by a unique function

and several other results about Clifford algebras, spin bundles and
differential geometry.

This second edition adds complements about Fock spaces, and corrects minor errors.

\bigskip\footnote{j.c.dutailly@free.fr}

\begin{center}
\newpage
\end{center}

\part{FOUNDATIONS}

\bigskip

In this first part we start with what makes the real foundations of today
mathematics : logic, set theory and categories. The two last subsections are
natural in this book, and they will be mainly dedicated to a long list of
definitions, mandatory to fix the language that is used in the rest of the
book. A section about logic seems appropriate, even if it gives just an
overview of the topic, because this is a subject that is rarely addressed,
except in specialized publications.

\section{LOGIC}

\label{Logic}

For a mathematician logic can be addressed from two points of view :

- the conventions and rules that any mathematical text should follow in order
to be deemed "right"

- the consistency and limitations of any formal theory using these logical rules.

It is the scope of a branch of mathematics of its own : "mathematical logic"

Indeed logic is not limited to a bylaw for mathematicians : there are also
theorems in logic.\ To produce these theorems one distinguishes the object of
the investigation ("language-object" or "theory") and the language used to
proceed to the demonstrations in mathematical logic, which is informal (plain
English).\ It seems strange to use a weak form of "logic" to prove results
about the more formal theories but it is related to one of the most important
feature of any scientific discourse : that it must be perceived and accepted
by other workers in the field as "sensible" and "convincing". And in fact
there are several schools in logic : some do not accept any nonnumerable
construct, or the principle of non contradiction, which makes logic a
confusing branch of mathematics. But whatever the interest of exotic lines of
reasoning in specific fields, for the vast majority of mathematicians, in
their daily work, there is a set of "generally accepted logical principles".

On this topic we follow mainly Kleene where definitions and theorems can be found.

\subsection{Propositional logic}

\label{Propositional logic}

Logic can be considered from two points of view : the first ("models") which
is focused on telling what are true or false statements, and the second
("demonstration") which strives to build demonstrations from premises. This
distinction is at the heart of many issues in mathematical logic.

\subsubsection{Models}

\paragraph{Formulas\newline}

\begin{definition}
An \textbf{atom\footnote{The name of an object is in boldface the first time
it appears (in its definition)}} is any given sentence accepted in the theory.
\end{definition}

The atoms are denoted as Latin letters A,B,..

\begin{definition}
The \textbf{logical operators} are :

$\sim:$ equivalent

$\Rightarrow:$ imply

$\wedge:$ and (both)

$\vee:$ or (possibly both)

$\urcorner:$ negation
\end{definition}

(notation and list depending on the authors)

\begin{definition}
A \textbf{formula} is any finite sequence of atoms linked by logical operators.
\end{definition}

One can build formulas from other formulas using these operators. A formula is
"well-built" (it is deemed acceptable in the theory) if it is constructed
according to the previous rules.

Examples : if $"3+2=x","\sqrt{5}-3>2","x%
{{}^2}%
+2x-1=0"$ are atoms then $\left(  \left(  3+2=x\right)  \wedge\left(  x%
{{}^2}%
+2x-1=0\right)  \right)  \Rightarrow\left(  \sqrt{5}-3>2\right)  $ is a well
built formula.

In building a formula we do not question the meaning or the validity of the
atoms (this the job of the theory which is investigated) : we only follow
rules to build formulas from given atoms. When building formulas with the
operators it is always good to use brackets to delimit the scope of the
operators.\ However there is a rule of precedence (by decreasing order):
$\sim\ >\hspace{0in}\Rightarrow\hspace{0in}>\wedge>\vee>\hspace{0in}\urcorner$

\paragraph{Truth-tables\newline}

The previous rules give only the "grammar" : how to build accepted formulas. A
formula can be well built but meaningless, or can have a meaning only if
certain conditions are met. Logic is the way to tell if something is true or false.

\begin{definition}
To each atom of a theory is attached a "\textbf{truth-table}", with only two
values : true (T) or false (F) exclusively.
\end{definition}

\begin{definition}
A \textbf{model} for a theory is the list of its atoms and their truth-table.
\end{definition}

\begin{definition}
A \textbf{proposition} is any formula issued from a model
\end{definition}

The rules telling how the operators work to deduce the truth table of a
formula from the tables of its atoms are the following (A,B are any formula) :

\bigskip

$%
\begin{bmatrix}
A & B & \left(  A\sim B\right)  & \left(  A\Rightarrow B\right)  & \left(
A\wedge B\right)  & \left(  A\vee B\right) \\
T & T & T & T & T & T\\
T & F & F & F & F & T\\
F & T & F & T & F & T\\
F & F & T & T & F & F
\end{bmatrix}%
\begin{bmatrix}
A & \left(  \urcorner A\right) \\
T & F\\
F & T
\end{bmatrix}
$

\bigskip

The only non obvious rule is for $\Rightarrow.$ It is the only one which
provides a full and practical set of rules, but other possibilities are
mentioned in quantum physics.

\paragraph{Valid formulas\newline}

With these rules the truth-table of any formula can be computed (formulas have
only a finite number of atoms). The formulas which are always true (their
truth-table presents only T) are of particular interest.

\begin{definition}
A formula A of a model is said to be \textbf{valid} if it is always true.\ It
is then denoted $\vDash A$.
\end{definition}

\begin{definition}
A formula B is a \textbf{valid consequence} of A if $\vDash\left(
A\Rightarrow B\right)  $.\ This is denoted : $A\vDash B.$
\end{definition}

More generally one writes : $A_{1},..A_{m}\vDash B$

Valid formulas are crucial in logic. There are two different categories of
valid formulas:

- formulas which are always valid, whatever the model : they provide the
"model" of propositional calculus in mathematical logic, as they tell how to
produce "true" statements without any assumption about the meaning of the formulas.

- formulas which are valid in some model only : they describe the properties
assigned to some atoms in the theory which is modelled. So, from the logical
point of view, they define the theory itself.

The following formula are always valid in any model (and most of them are of
constant use in mathematics). Indeed they are just the traduction of the
previous tables.

\bigskip

1. first set (they play a specific role in logic):

$\left(  A\wedge B\right)  \Rightarrow A;\left(  A\wedge B\right)  \Rightarrow
B$

$A\Rightarrow\left(  A\vee B\right)  ;B\Rightarrow\left(  A\vee B\right)  $

$\urcorner\urcorner A\Rightarrow A$

$A\Rightarrow\left(  B\Rightarrow A\right)  $

$\left(  A\sim B\right)  \Rightarrow\left(  A\Rightarrow B\right)  ;\left(
A\sim B\right)  \Rightarrow\left(  B\Rightarrow A\right)  $

$\left(  A\Rightarrow B\right)  \Rightarrow\left(  \left(  A\Rightarrow\left(
B\Rightarrow C\right)  \right)  \Rightarrow\left(  A\Rightarrow C\right)
\right)  $

$A\Rightarrow\left(  B\Rightarrow\left(  A\wedge B\right)  \right)  $

$\left(  A\Rightarrow B\right)  \Rightarrow\left(  \left(  A\Rightarrow
\urcorner B\right)  \Rightarrow\urcorner A\right)  $

$\left(  A\Rightarrow B\right)  \Rightarrow\left(  \left(  B\Rightarrow
A\right)  \Rightarrow\left(  A\sim B\right)  \right)  $

2. Others (there are infinitely many others formulas which are always valid)

$A\Rightarrow A;$

$A\sim A;\left(  A\sim B\right)  \sim\left(  B\sim A\right)  ;\left(  \left(
A\sim B\right)  \wedge\left(  B\sim C\right)  \right)  \Rightarrow\left(
A\sim C\right)  $

$\left(  A\Rightarrow B\right)  \sim\left(  \left(  \urcorner A\right)
\Rightarrow\left(  \urcorner B\right)  \right)  $

$\urcorner A\Rightarrow\left(  A\Rightarrow B\right)  $

$\urcorner\urcorner A\sim A;\urcorner\left(  A\wedge\left(  \urcorner
A\right)  \right)  ;A\vee\left(  \urcorner A\right)  $

$\urcorner\left(  A\vee B\right)  \sim\left(  \left(  \urcorner A\right)
\wedge\left(  \urcorner B\right)  \right)  ;\urcorner\left(  A\wedge B\right)
\sim\left(  \left(  \urcorner A\right)  \vee\left(  \urcorner B\right)
\right)  ;\urcorner\left(  A\Rightarrow B\right)  \sim\left(  A\wedge\left(
\urcorner B\right)  \right)  $

Notice that $\vDash A\vee\left(  \urcorner A\right)  $ meaning that a formula
is either true or false is an obvious consequence of the rules which have been
set up here.

\bigskip

An example of formula which is valid in a specific model : in a set theory the
expressions "$a\in A","A\subset B"$ are atoms, they are true or false (but
their value is beyond pure logic). And $"\left(  \left(  a\in A\right)
\wedge\left(  A\subset B\right)  \right)  \Rightarrow\left(  a\in B\right)  "$
is a formula.\ To say that it is always true expresses a fundamental property
of set theory (but we could also postulate that it is not always true, and we
would have another set theory).

\begin{theorem}
If $\vDash A$ and $\vDash\left(  A\Rightarrow B\right)  $ then : $\vDash B$
\end{theorem}

\begin{theorem}
$\vDash A\sim B$ iff\footnote{We will use often the usual abbreviation "iff"
for "if and only if"} A and B have same tables.
\end{theorem}

\begin{theorem}
Duality: Let be E a formula built only with atoms $A_{1},..A_{m}$ , their
negation $\urcorner A_{1},..\urcorner A_{m}$\ , the operators $\vee,\wedge, $
and E' the formula deduced from E by substituting $\vee$ with $\wedge,\wedge$
with $\vee,A_{i}$ with $\urcorner A_{i},\urcorner A_{i}$ with $A_{i}$ then :

If $\vDash E$ then $\vDash\urcorner E^{\prime}$

If $\vDash\urcorner E$ then $\vDash E^{\prime}$

With the same procedure for another similar formula F:

If $\vDash E\Rightarrow F$ then $\vDash F^{\prime}\Rightarrow E^{\prime}$

If $\vDash E\sim F$ then $\vDash E^{\prime}\sim F^{\prime}$
\end{theorem}

\subsubsection{Demonstration}

Usually one does not proceed by truth tables but by demonstrations. In a
formal theory, axioms, hypotheses and theorems can be written as formulas. A
demonstration is a sequence of formulas using logical rules and rules of
inference, starting from axioms or hypotheses and ending by the proven result.

In deductive logic a \textbf{formula} is always true. They are built according
to the following rules by linking formulas with the logical operators above :

i) There is a given set of formulas $\left(  A_{1},A_{2},...A_{m},..\right)
$\ (possibly infinite) called the \textbf{axioms} of the theory

ii) There is an inference rule : if A is a formula, and $\left(  A\Rightarrow
B\right)  $ is a formula, then $\left(  B\right)  $ is a formula.

iii) Any formula built from other formulas with logical operators and using
the "first set" of rules above is a formula

For instance if A,B are formulas, then $\left(  \left(  A\wedge B\right)
\Rightarrow A\right)  $ is a formula.

The formulas are listed, line by line. The last line gives a "true" formula
which is said to be proven.

\begin{definition}
A \textbf{demonstration} is a \textit{finite} sequence of formulas where the
last one B is the proven formula, and this is denoted : $\Vdash B.$ B is
\textbf{provable}.
\end{definition}

Similarly B is deduced from $A_{1},A_{2},..$ is denoted : $A_{1},A_{2}%
,..A_{m},..\Vdash B:$ . In this picture there are logical rules (the "first
set" of formulas and the inference rule) and "non logical" formulas (the
axioms). The set of logical rules can vary according to the authors, but is
roughly always the same.\ The critical part is the set of axioms which is
specific to the theory which is under review.

\begin{theorem}
$A_{1},A_{2},...A_{m}\Vdash A_{p}$ with 1%
$<$%
p$\leq m$
\end{theorem}

\begin{theorem}
If $A_{1},A_{2},...A_{m}\Vdash B_{1},A_{1},A_{2},...A_{m}\Vdash B_{2}%
,...A_{1},A_{2},...A_{m}\Vdash B_{p}$ and $B_{1},B_{2},...B_{p}\Vdash C$ then
$A_{1},A_{2},...A_{m}\Vdash C$
\end{theorem}

\begin{theorem}
If $\Vdash\left(  A\Rightarrow B\right)  $ then $A\Vdash B$ and conversely :
if $A\Vdash B$ then $\Vdash\left(  A\Rightarrow B\right)  $
\end{theorem}

\bigskip

\subsection{Predicates}

\label{Predicates}

In propositional logic there can be an infinite number of atoms (models) or
axioms (demonstration) but, in principle, they should be listed prior to any
computation.\ This is clearly a strong limitation.\ So the previous picture is
extended to \textbf{predicates}, meaning formulas including variables and functions.

\subsubsection{Models with predicates}

\paragraph{Predicate\newline}

\begin{definition}
A \textbf{variable} is a symbol which takes its value in a given collection D
(the \textbf{domain}).
\end{definition}

They are denoted x,y,z,...It is assumed that the domain D is always the same
for all the variables and it is not empty. A variable can appears in different
places, with the usual meaning that in this case the same value must be
assigned to these variables.

\begin{definition}
A \textbf{propositional function} is a symbol, with definite places for one or
more variables, such that when each variable is replaced by one of its value
in the domain, the function becomes a proposition.
\end{definition}

They are denoted : $P(x,y),Q(r),...$There is a truth-table assigned to the
function for all the combinations of variables.

\begin{definition}
A \textbf{quantizer} is a logical operator acting \textit{on the variables}.\ 
\end{definition}

They are :

$\forall:$ for any value of the variable (in the domain D)

$\exists:$ there exists a value of the variable (in the domain D)

A quantizer acts, on one variable only, each time it appears : $\forall
x,\exists y,..$\ . This variable is then \textbf{bound}. A variable which is
not bound is \textbf{free}. A \ quantizer cannot act on a previously bound
variable (one cannot have \ $\forall x,\exists x$ in the same formula). As
previously it is always good to use different symbols for the variables and
brackets to precise the scope of the operators.

\begin{definition}
A \textbf{predicate} is a sentence comprised of propositions, quantizers
preceding variables, and propositional functions linked by logical operators.
\end{definition}

Examples of predicates :

$\left(  \left(  \forall x,\left(  x+3>z\right)  \right)  \wedge A\right)
\Rightarrow\urcorner\left(  \exists y,\left(  \sqrt{y^{2}-1}=a\right)
\right)  \vee\left(  z=0\right)  $

$\forall n\left(  \left(  n>N\right)  \wedge\left(  \exists p,\left(
p+a>n\right)  \right)  \right)  \Rightarrow B$

To evaluate a predicate one needs a truth-rule for the quantizers
$\forall,\exists:$

- a formula $\left(  \forall x,A\left(  x\right)  \right)  $ is T if A(x) is T
for all values of x

- a formula $\left(  \exists x,A(x))\right)  $ is T if A(x) has at least one
value equal to T

With these rules whenever all the variables in a predicate are bound, this
predicate, for the truth table purpose, becomes a proposition.

Notice that the quantizers act only on variables, not formulas.\ This is
specific to \textbf{first order predicates}.\ In higher orders predicates
calculus there are expressions like $"\forall A",$ and the theory has
significantly different outcomes.

\paragraph{Valid consequence\newline}

With these rules it is possible, in principle, to compute the truth table of
any predicate.

\begin{definition}
A predicate A is \textbf{D-valid,} denoted\textbf{\ }$^{D}\vDash A$ if it is
valid whatever the value of the free variables in D. It is \textbf{valid} if
is D-valid whatever the domain D.
\end{definition}

The propositions listed previously in the "first set" are valid for any D.

$\vDash A\sim B$ iff for any domain D \ A and B have the same truth-table.

\subsubsection{Demonstration with predicates}

The same new elements are added : variables, quantizers, propositional
functions. Variables and quantizers are defined as above (in the model
framework) with the same conditions of use.

A formula is built according to the following rules by linking formulas with
the logical operators and quantizers :

i) There is a given set of formulas $\left(  A_{1},A_{2},...A_{m},..\right)
$\ (possibly infinite) called the \textbf{axioms} of the theory

ii) There are three inference rules :

- if A is a formula, and $\left(  A\Rightarrow B\right)  $ is a formula, then
$\left(  B\right)  $ is a formula

- If C is a formula where x is not present and A(x) a formula, then :

if $C\Rightarrow A(x)$ is a formula, then $C\Rightarrow\forall xA(x)$ is a formula

if $A\left(  x\right)  \Rightarrow C$ is a formula, then $\exists
xA(x)\Rightarrow C$ is a formula

iii) Any formula built from other formulas with logical operators and using
the "first set" of rules above plus :

$\forall xA\left(  x\right)  \Rightarrow A\left(  r\right)  $

$A\left(  r\right)  \Rightarrow\exists xA(x)$

where r is free, is a formula

\begin{definition}
B is \textbf{provable} if there is a finite sequence of formulas where the
last one is B, which is denoted : $\Vdash B.$
\end{definition}

B can be deduced from $A_{1},A_{2},...A_{m}$ if B is provable starting with
the formulas $A_{1},A_{2},...A_{m}$ ,and is denoted : $A_{1},A_{2}%
,...A_{m}\Vdash B$

\bigskip

\subsection{Formal theories}

\label{Formal theories}

\subsubsection{Definitions}

The previous definitions and theorems give a framework to review the logic of
formal theories. A formal theory uses a symbolic language in which terms are
defined, relations between some of these terms are deemed "true" to express
their characteristics, and logical rules are used to evaluate formulas or
deduce theorems. There are many refinements and complications but, roughly,
the logical rules always come back to some kind of predicates logic as exposed
in the previous section. But there are two different points of view : the
models on one hand and the demonstration on the other : the same theory can be
described using a model (model type theory) or axioms and deductions
(deductive type).

Models are related to the \textbf{semantic} of the theory. Indeed they are
based on the assumption that for every atom there is some truth-table that
could be exhibited, meaning that there is some "extra-logic" to compute the
result. And the non purely logical formulas which are valid (always true in
the model) characterize the properties of the objects modelled by the theory.

Demonstrations are related to the \textbf{syntactic} part of the theory.\ They
deal only with formulas without any concern about their meaning : either they
are logical formulas (the first set) or they are axioms, and in both cases
they are assumed to be "true", in the meaning that they are worth to be used
in a demonstration. The axioms sum up the non logical part of the system.\ The
axioms on one hand and the logical rules on the other hand are all that is
necessary to work.

Both model theories and deductive theories use logical rules (either to
compute truth-tables or to list formulas), so they have a common ground.\ And
the non-logical formulas which are valid in a model are the equivalent of the
axioms of a deductive theory. So the two points of view are not opposed, but
proceed from the two meanings of logic.

In reviewing the logic of a formal theory the main questions that arise are :

- which are the axioms needed to account for the theory (as usual one wants to
have as few of them as possible) ?

- can we assert that there is no formula A such that both A and its negation
$\urcorner A$ can be proven ?

- can we prove any valid formula ?

- is it possible to list all the valid formulas of the theory ?

A formal theory of the model type is said to be \textbf{sound} (or
\textbf{consistent}) if only valid formulas can be proven. Conversely a formal
theory of the deductive type is said to be \textbf{complete} if any valid
formula can be proven.

\subsubsection{Completeness of the predicate calculus}

Predicate logic (\textbf{first order logic}) can be seen as a theory by
itself. From a set of atoms, variables and propositional functions \ one can
build formulas by using the logical operators for predicates. There are
formulas which are always valid in the propositional calculus, and there are
similar formulas in the predicates calculus, whatever the domain D. Starting
with these formulas, and using the set of logical rules and the inference
rules as above one can build a deductive theory.

The \textbf{G\"{o}del's completeness theorem} says that any valid formula can
be proven, and conversely that only valid formulas can be proven. So one can
write in the first order logic : $\vDash A$ iff $\Vdash A.$

It must be clear that this result, which justifies the apparatus of first
order logic, stands only for the formulas (such as those listed above) which
are valid in any model : indeed they are the pure logical relations, and do
not involve any "non logical" axioms.

The \textbf{G\"{o}del's} \textbf{compactness} \textbf{theorem} says in
addition that if a formula can be proven from a set of formulas, it can also
be proven by a finite set of formulas : there is always a demonstration using
a finite number of steps and formulas.

These results are specific to first order logic, and does not hold for higher
order of logic (when the quantizers act on formulas and not only on
variables). Thus one can say that mathematical logic (at least under the form
of first order propositional calculus) has a strong foundation.

\subsubsection{Incompleteness theorems}

At the beginning of the XX$%
{{}^\circ}%
$ century mathematicians were looking forward to a set of axioms and logical
rules which could give solid foundations to mathematics (the "Hilbert's
program"). Two theories are crucial for this purpose : set theory and natural
number (arithmetic). Indeed set theory is the language of modern mathematics,
and natural numbers are a prerequisite for the rule of inference, and even to
define infinity. Such formal theories use the rules of first order logic, but
require also additional "non logical" axioms. The axioms required in a formal
set theory (such as Zermelo-Frankel's) or in arithmetic (such as Peano's) are
well known. There are several systems, more or less equivalent.

A formal theory is said to be \textbf{effectively generated} if its set of
axioms is a recursively enumerable set. This means that there is a computer
program that, in principle, could enumerate all the axioms of the theory.
\textbf{G\"{o}del's first incompleteness theorem} states that any effectively
generated theory capable of expressing elementary arithmetic cannot be both
consistent and complete. In particular, for any consistent, effectively
generated formal theory that proves certain basic arithmetic truths, there is
an arithmetical statement that is true but not provable in the theory (Kleene
p. 250). In fact the "truth" of the statement must be understood as : neither
the statement or its negation can be proven.\ As the statement is true or
false the statement itself or its converse is true. All usual theories of
arithmetic fall under the scope of this theorem. So one can say that in
mathematics the previous result ( $\vDash A$ iff $\Vdash A)$ does not stand.

This result is not really a surprise : in any formal theory one can build
infinitely many predicates, which are grammatically correct. To say that there
is always a way to prove any such predicate (or its converse) is certainly a
crude assumption. It is linked to the possibility to write computer programs
to automatically check demonstrations.

\subsubsection{Decidable and computable theories}

The incompleteness theorems are closely related to the concepts of
\textbf{decidable} and \textbf{computable}.

In a formal deductive theory computer programs can be written to formalize
demonstrations (an example is "Isabelle" see the Internet), so that they can
be made safer. One can go further and ask if it is possible to design a
program such that it could, for any statement of the theory, check if it is
valid (model side) or provable (deducible side). If so the theory is said
\textbf{decidable}.

The answer is yes for the propositional calculus (without predicate), because
it is always possible to compute the truth table, but it is no in general for
predicates calculus. And it is no for theories modelling arithmetic.

Decidability is an aspect of computability : one looks for a program which
could, starting from a large class of inputs, compute an answer which is yes
or no.

Computability is studied through \textbf{T\"{u}ring machines} which are
schematic computers. A T\"{u}ring machine is comprised of an input system (a
flow of binary data read bit by bit), a program (the computer has p states,
including an end, and it goes from one state to another according to its
present state and the bit that has been read), and an output system (the
computer writes a bit). A T\"{u}ring machine can compute integer functions
(the input, output and parameters are integers).\ One demonstration of the
G\"{o}del incompleteness theorem shows that there are functions that cannot be
computed : notably the function telling, for any given input, in how many
steps the computer would stop. If we look for a program that can give more
than a "Yes/No" answer one has the so-called function problems, which study
not only the possibility but the efficiency (in terms of resources used) of
algorithms. The \textbf{complexity} of a given problem is measured by the
ratio of the number of steps required by a T\"{u}ring machine to compute the
function, to the size in bits of the input (the problem).

\newpage

\section{SET\ THEORY}

\subsection{Axiomatic}

\label{Axiomatic}

Set theory was founded by Cantor and Dedekind in early XX$%
{{}^\circ}%
$ century. The initial set theory was impaired by paradoxes, which are usually
the consequences of an inadequate definition of a "set of sets". Several
improved versions were proposed, and its most common , formalized by
Zermello-Fraenkel, is denoted ZFC when it includes the axiom of choice. For
the details see Wikipedia "Zermelo--Fraenkel set theory".

\subsubsection{Axioms of ZFC}

Some of the axioms listed below are redundant, as they can be deduced from
others, depending of the presentation.

\begin{axiom}
Axiom of extensionality : Two sets are equal (are the same set) if they have
the same elements.
\end{axiom}

$\left(  A=B\right)  \sim\left(  \left(  \forall x\left(  x\in A\sim x\in
B\right)  \right)  \wedge\left(  \forall x\left(  A\in x\sim B\in x\right)
\right)  \right)  $

\begin{axiom}
Axiom of regularity (also called the Axiom of foundation) : Every non-empty
set A contains a member B such that A and B are disjoint sets.
\end{axiom}

\begin{axiom}
Axiom schema of specification (also called the axiom schema of separation or
of restricted comprehension) : If A is a set, and P(x) is any property which
may characterize the elements x of A, then there is a subset B of A containing
those x in A which satisfy the property.
\end{axiom}

The axiom of specification can be used to prove the existence of one unique
empty set, denoted $\varnothing$, once the existence of at least one set is established.

\begin{axiom}
Axiom of pairing : If A and B are sets, then there exists a set which contains
A and B as elements.
\end{axiom}

\begin{axiom}
Axiom of union : For any set S there is a set A containing every set that is a
member of some member of S.
\end{axiom}

\begin{axiom}
Axiom schema of replacement : If the domain of a definable function f is a
set, and f(x) is a set for any x in that domain, then the range of f is a
subclass of a set, subject to a restriction needed to avoid paradoxes.
\end{axiom}

\begin{axiom}
Axiom of infinity : Let S(x) abbreviate $x\cup\left\{  x\right\}  $, where x
is some set. Then there exists a set X such that the empty set is a member of
X and, whenever a set y is a member of X, then S(y) is also a member of X.
\end{axiom}

More colloquially, there exists a set X having infinitely many members.

\begin{axiom}
Axiom of power set : For any set A there is a set, called the power set of A
whose elements are all the subsets of A.
\end{axiom}

\begin{axiom}
Well-ordering theorem : For any set X, there is a binary relation R which
well-orders X.
\end{axiom}

This means R is an order\ relation on X such that every non empty subset of X
has a member which is minimal under R (see below the definition of order relation).

\begin{axiom}
The axiom of choice (AC) : Let X be a set whose members are all non-empty.
Then there exists a function f from X to the union of the members of X, called
a "choice function", such that for all Y $\in$ X one has f(Y) $\in$ Y.
\end{axiom}

To tell it plainly : if we have a collection (possibly infinite) of sets, it
is always possible to choose an element in each set. The axiom of choice is
equivalent to the Well-ordering theorem, given the other 8 axioms. AC is
characterized as non constructive because it asserts the existence of a set of
chosen elements, but says nothing about how to choose them.

\subsubsection{Extensions}

There are several axiomatic extensions of ZFC, which strive to incorporate
larger structures without the hindrance of "too large sets". Usually they
introduce a distinction between "sets" (ordinary sets) and "classes" or
"universes". A universe is comprised of sets, but is not a set itself and does
not meet the axioms of sets. This precaution precludes the possibility of
defining sets by recursion : any set must be defined before it can be used.
von Neumann organizes sets according to a hierarchy based on ordinal numbers,
"New foundation" (Jensen, Holmes) is another system based on a different hierarchy.

\bigskip

We give below the extension used by Kashiwara and Schapira which is typical of
these extensions, and will be used later\ in categories theory.

A \textbf{universe} U is an object satisfying the following properties :

1. $\varnothing\in U$

2. $u\in U\Rightarrow u\subset U$

3. $u\in U\Rightarrow\left\{  u\right\}  \in U$ (the set with the unique
element u)

4. $u\in U\Rightarrow2^{u}\in U$ (the set of all subsets of u)

5. if for each member of the family (see below) $\left(  u_{i}\right)  _{i\in
I}$\ of sets $u_{i}\in U$ then $\cup_{i\in I}u_{i}\in U$

6. $%
\mathbb{N}
\in U$

A universe is a "collection of sets" , with the implicit restriction that all
its elements are known (there is no recursive definition) so that the ususal
paradoxes are avoided. As a consequence :

7. $u\in U\Rightarrow\cup_{x\in u}x\in U$

8. $u,v\in U\Rightarrow u\times v\in U$

9. $u\subset v\in U\Rightarrow u\in U$

10.if for each member of the family (see below) of sets $\left(  u_{i}\right)
_{i\in I}$ $u_{i}\in U$ then $%
{\displaystyle\prod_{i\in I}}
u_{i}\in U$

An axiom is added to the ZFC system : for any set x there exists an universe U
such that $x\in U$

A set X is \textbf{U-small} if there is a bijection between X and a set of U.

\subsubsection{Operations on sets}

In formal set theories :

"x belongs to X" $:x\in X$ is an atom (it is always true or false).\ In "fuzzy
logic" it can be neither.

"A is included in B" : $A\subset B$ where A,B are any sets, is an atom. It is
true if every element of A belongs to B

We have also the notation (that we will use, rarely, indifferently) :
$A\sqsubseteq B$ meaning $A\subset B$ and possibly A = B

From the previous axioms and these atoms are defined the following operators
on sets:

\begin{definition}
The \textbf{Union} of the sets A and B, denoted A $\cup$ B, is the set of all
objects that are a member of A, or B, or both.
\end{definition}

\begin{definition}
The \textbf{Intersection} of the sets A and B, denoted A $\cap$ B, is the set
of all objects that are members of both A and B.
\end{definition}

\begin{definition}
The \textbf{Set difference} of U and A, denoted U
$\backslash$
A is the set of all members of U that are not members of A.
\end{definition}

Example : The set difference \{1,2,3\}
$\backslash$
\{2,3,4\} is \{1\} , while, conversely, the set difference \{2,3,4\}
$\backslash$
\{1,2,3\} is \{4\} .

\begin{definition}
A \textbf{subset} of a set A is a set B such that all its elements belong to A
\end{definition}

\begin{definition}
The \textbf{complement} of a subset A with respect to a set U is the set
difference U
$\backslash$
A
\end{definition}

If the choice of U is clear from the context, the notation $A^{c}$ will be
used. Another notation is $\complement_{U}^{A}=A^{c}$

\begin{definition}
The \textbf{Symmetric difference} of the sets A and B, denoted $A\triangle
B=\left(  A\cup B\right)  \backslash\left(  A\cap B\right)  $ is the set of
all objects that are a member of exactly one of A and B (elements which are in
one of the sets, but not in both).
\end{definition}

\begin{definition}
The \textbf{Cartesian product} of A and B, denoted A $\times$\ B, is the set
whose members are all possible \textit{ordered} pairs (a,b) where a is a
member of A and b is a member of B.
\end{definition}

The Cartesian product of sets can be extended to an infinite number of sets
(see below)

\begin{definition}
The \textbf{Power set} of a set A is the set whose members are all possible
subsets of A. It is denoted $2^{A}.$
\end{definition}

\begin{theorem}
Union and intersection are associative and distributive
\end{theorem}

$A\cup\left(  B\cup C\right)  =\left(  A\cup B\right)  \cup C$

$A\cap\left(  B\cap C\right)  =\left(  A\cap B\right)  \cap C$

$A\cap\left(  B\cup C\right)  =\left(  A\cap B\right)  \cup\left(  A\cap
C\right)  $

$A\cup\left(  B\cap C\right)  =\left(  A\cup B\right)  \cap\left(  A\cup
C\right)  $

\begin{theorem}
Symmetric difference is commutative, associative and distributive with respect
to intersection.
\end{theorem}

$\complement_{B}^{A\cup B}=\left(  A\cup B\right)  ^{c}=A^{c}\cap
B^{c},\left(  A\cap B\right)  ^{c}=A^{c}\cup B^{c}$

Remark : there are more sophisticated operators involving an infinite number
of sets (see Measure).

\bigskip

\subsection{Maps}

\label{Maps}

\subsubsection{Definitions}

\begin{definition}
A \textbf{map} f from a set E to a set F, denoted $f:E\rightarrow
F::y=f\left(  x\right)  $ is a relation which associates to each element x of
E one element y=f(x) of F.

x in f(x) is called the \textbf{argument}, f(x) is the \textbf{value} of f for
the argument x.

E is the \textbf{domain} of f, F is the \textbf{codomain} of f.

The set $f(E)=\left\{  y=f(x),x\in E\right\}  $ is the \textbf{range} (or
\textbf{image}) of f.

The \textbf{graph} of f is the set of all ordered pairs $\left\{  \left(
x,f(x)\right)  ,x\in E\right\}  .$
\end{definition}

Formally one can define the map as the set of pairs (x,f(x))

We will usually reserve the name \textbf{function} when the codomain is a
field ($%
\mathbb{R}
,%
\mathbb{C}
).$

\begin{definition}
The \textbf{preimage} (or \textbf{inverse image}) of a subset $B\subset F$ of
the map $f:E\rightarrow F$ is the subset denoted $f^{-1}\left(  B\right)
\subset E$ such that $\forall x\in f^{-1}\left(  B\right)  :f\left(  x\right)
\in B$
\end{definition}

It is usually denoted :$f^{-1}\left(  B\right)  =\left\{  x\in E:f(x)\in
B\right\}  .$

Notice that it is not necessary for f to have an inverse map.

The following identities are the consequence of the definitions.\ 

For any sets A,B, map : $f:A\rightarrow B$

$f\left(  A\cup B\right)  =f\left(  A\right)  \cup f\left(  B\right)  $

$f\left(  A\cap B\right)  \subset f\left(  A\right)  \cap f\left(  B\right)  $
Beware !

$f^{-1}\left(  A\cup B\right)  =f^{-1}\left(  A\right)  \cup f^{-1}\left(
B\right)  $

$f^{-1}\left(  A\cap B\right)  =f^{-1}\left(  A\right)  \cap f^{-1}\left(
B\right)  $

The previous identities still hold for any family (even infinite) of sets.

$f\left(  A\right)  \subset B\Leftrightarrow A\subset f^{-1}\left(  B\right)
$

$f\left(  f^{-1}\left(  A\right)  \right)  \subset A$

$A\subset B\Rightarrow f\left(  A\right)  \subset f\left(  B\right)  $

$A\subset B\Rightarrow f^{-1}\left(  A\right)  \subset f^{-1}\left(  B\right)
$

\begin{definition}
The \textbf{restriction} $f_{A}$ of a map $f:E\rightarrow F$ to a subset
$A\subset E$ is the map : $f_{A}:A\rightarrow F::\forall x\in A:f_{A}\left(
x\right)  =f(x)$
\end{definition}

\begin{definition}
An \textbf{embedding} of a subset A of a set E in E is a map $\imath
:A\rightarrow E$ such that $\forall x\in A:\imath\left(  x\right)  =x.$
\end{definition}

\begin{definition}
A \textbf{retraction} of a set E on a subset A of E is a map : $\rho
:E\rightarrow A$ such that : $\forall x\in A,\rho\left(  x\right)  =x.$ Then A
is said to be a \textbf{retract} of E
\end{definition}

Retraction is the converse of an embedding. Usually embedding and retraction
maps are morphisms : they preserve the mathematical structures of both A and
E, and x could be seen indifferently as an element of A or an element of E.

Example : the embedding of a vector subspace in a vector space.

\begin{definition}
The \textbf{characteristic function} (or indicator funtion) of the subset A of
the set E is the function denoted : $1_{A}:E\rightarrow\left\{  0,1\right\}  $
with $1_{A}\left(  x\right)  =1$ if $x\in A,1_{A}\left(  x\right)  =0$ if
$x\notin A.$
\end{definition}

\begin{definition}
A set H of maps $f:E\rightarrow F$\ is said to \textbf{separate} E if :

$\forall x,y\in E,x\neq y,\exists f\in H:f\left(  x\right)  \neq f\left(
y\right)  $
\end{definition}

\begin{definition}
If E,K are sets, F a set of maps : $f:E\rightarrow K$ the \textbf{evaluation}
map at $x\in E$ is the map : $\widehat{x}:F\rightarrow K::\widehat{x}\left(
f\right)  =f\left(  x\right)  $
\end{definition}

\begin{definition}
Let I be a set, the \textbf{Kronecker} function is the function :
$\delta:I\times I\rightarrow\left\{  0,1\right\}  ::\delta\left(  i,j\right)
=1$ if i=j, $\delta\left(  i,j\right)  =0$ if i$\neq j$
\end{definition}

When I is a set of indices it is usually denoted $\delta_{j}^{i}=\delta\left(
i,j\right)  $ or $\delta_{ij}$.

\begin{theorem}
There is a set, denoted $F^{E}$ , of all maps with domain E and codomain F
\end{theorem}

\begin{theorem}
There is a unique map $Id_{E}$\ over a set E, called the \textbf{identity},
such that : $Id_{E}:E\rightarrow E::x=Id_{E}(x)$
\end{theorem}

A map f of several variables $\left(  x_{1},x_{2},...x_{p}\right)  $ is just a
map with domain the cartesian products of several sets $E_{1}\times
E_{2}...\times E_{p}$

From a map $f:E_{1}\times E_{2}\rightarrow F$ one can define a map with one
variable by keeping $x_{1}$\ constant, that we will denote $f\left(
x_{1},\cdot\right)  :E_{2}\rightarrow F$

\begin{definition}
The \textbf{canonical projection} of $E_{1}\times E_{2}...\times E_{p}$ onto
$E_{k}$ is the map $\pi_{k}:E_{1}\times E_{2}...\times E_{p}\rightarrow
E_{k}::\pi_{k}\left(  x_{1},x_{2},...x_{p}\right)  =x_{k}$
\end{definition}

\begin{definition}
A map $f:E\times E\rightarrow F$ is \textbf{symmetric} if

$\forall x_{1}\in E,\forall x_{2}\in E::f\left(  x_{1},x_{2}\right)  =f\left(
x_{2},x_{1}\right)  $
\end{definition}

\begin{definition}
A map is \textbf{onto} (or \textbf{surjective}) if its range is equal to its codomain.
\end{definition}

For each element $y\in F$ of the codomain there is at least one element $x\in
E$ of the domain such that : $y=f(x)$

\begin{definition}
A map is \textbf{one-to-one} (or \textbf{injective}) if each element of the
codomain is mapped at most by one element of the domain
\end{definition}

$\left(  \forall y\in F:f\left(  x\right)  =f\left(  x^{\prime}\right)
\Rightarrow x=x^{\prime}\right)  \Leftrightarrow\left(  \forall x\neq
x^{\prime}\in E:f\left(  x\right)  \neq f\left(  x^{\prime}\right)  \right)  $

\begin{definition}
A map is \textbf{bijective} (or\ one-one and onto) if it is both onto and
one-to-one. If so there is an \textbf{inverse map}

$f^{-1}:F\rightarrow E::x=f^{-1}(y):y=f(x)$
\end{definition}

\subsubsection{Composition of maps}

\begin{definition}
The \textbf{composition}, denoted $g\circ f$ , of the maps $f:E\rightarrow
F,g:F\rightarrow G$ is the map :

$g\circ f:E\rightarrow G::x\in E\overset{f}{\rightarrow}y=f(x)\in F\overset
{g}{\rightarrow}z=g(y)=g\circ f(x)\in G$
\end{definition}

\begin{theorem}
The composition of maps is always associative :

$\left(  f\circ g\right)  \circ h=f\circ\left(  g\circ h\right)  $
\end{theorem}

\begin{theorem}
The composition of a map $f:E\rightarrow E$ with the identity gives f :
$f\circ Id_{E}=Id_{E}\circ f=f$
\end{theorem}

\begin{definition}
The \textbf{inverse} of a map $f:E\rightarrow F$ for the composition of maps
is a map denoted $f^{-1}:F\rightarrow E$ such that : $f\circ f^{-1}%
=Id_{E},f^{-1}\circ f=Id_{F}$
\end{definition}

\begin{theorem}
A bijective map has an \textbf{inverse} map for the composition
\end{theorem}

\begin{definition}
If the codomain of the map f is included in its domain, the
\textbf{n-iterated} map of f is the map $f^{n}=f\circ f...\circ f$ (n times)
\end{definition}

\begin{definition}
A map f is said \textbf{idempotent} if $f^{2}=f\circ f=f$.
\end{definition}

\begin{definition}
A map f such that $f^{2}=Id$ is an \textbf{involution.}
\end{definition}

\subsubsection{Sequence}

\begin{definition}
A \textbf{family} \textbf{of elements} of a set E is a map from a set I,
called the \textbf{index set}, to the set E
\end{definition}

\begin{definition}
A \textbf{subfamily} of a family of elements is the restriction of the family
to a subset of the index set
\end{definition}

\begin{definition}
A \textbf{sequence} in the set E is a family of elements of E indexed on the
set of natural numbers $%
\mathbb{N}
.$
\end{definition}

\begin{definition}
A \textbf{subsequence} is the restriction of a sequence to an
\textit{infinite} subset of $%
\mathbb{N}
.$
\end{definition}

\begin{notation}
$\left(  x_{i}\right)  _{i\in I}\in E^{I}$ is a family of elements of E
indexed on I
\end{notation}

\begin{notation}
$\left(  x_{n}\right)  _{n\in%
\mathbb{N}
}\in E^{%
\mathbb{N}
}$ is a sequence of elements in the set E
\end{notation}

Notice that if X is a subset of E then a sequence \textit{in} X is a map $x:%
\mathbb{N}
\rightarrow X$

\begin{definition}
On a set E on which an addition has been defined, the \textbf{series }$\left(
S_{n}\right)  $ is the sequence : $S_{n}=\sum_{p=0}^{n}x_{p}$ where $\left(
x_{n}\right)  _{n\in%
\mathbb{N}
}\in E^{%
\mathbb{N}
}$ is a sequence.
\end{definition}

\subsubsection{Family of sets}

\begin{definition}
A \textbf{family of sets }$\left(  E_{i}\right)  _{i\in I}$\textbf{\ , }over a
set E is a map from a set I to the power set of E
\end{definition}

For each argument i $E_{i}$ is a subset of E : $F:I\rightarrow2^{E}::F\left(
i\right)  =E_{i}$.

The axiom of choice tells that for any family of sets $\left(  E_{i}\right)
_{i\in I}$\textbf{\ }\ there is a map $f:I\rightarrow E$\ which associates an
element f(i) of $E_{i}$\ to each value i of the index : $\exists
f:I\rightarrow E::f\left(  i\right)  \in E_{i}$

If the sets $E_{i}$ are not previously defined as subsets of E (they are not
related), following the previous axioms of the enlarged set theory, they must
belong to a universe U, and then the set $E=\cup_{i\in I}E_{i}$ also belongs
to U and all the $E_{i}$ are subsets of E.

\begin{definition}
The \textbf{cartesian product} $E=%
{\textstyle\prod\limits_{i\in I}}
E_{i}$ of a family of sets is the set of all maps : $f:I\rightarrow\cup_{i\in
I}E_{i}$ such that $\forall i\in I:f\left(  i\right)  \in E_{i}.$ The elements
$f\left(  i\right)  $\ are the \textbf{components} of f.
\end{definition}

This is the extension of the previous definition to a possibly infinite number
of sets.

\begin{definition}
A \textbf{partition} of a set E is a family $\left(  E_{i}\right)  _{i\in I:}$
of sets over E such that :

$\forall i:E_{i}\neq\varnothing$

$\forall i,j:E_{i}\cap E_{i}=\varnothing$

$\cup_{i\in I}E_{i}=E$
\end{definition}

\begin{definition}
A \textbf{refinement} $\left(  A_{j}\right)  _{j\in J}$\ of a partition
$\left(  E_{i}\right)  _{i\in I}$\textbf{\ }over E\ is a partition of E such
that : $\forall j\in J,\exists i\in I:A_{j}\subset E_{i}$
\end{definition}

\begin{definition}
A \textbf{family of filters} over a set E is a family $\left(  F_{i}\right)
_{i\in I}$\textbf{\ }over E such that :

$\forall i:F_{i}\neq\varnothing$

$\forall i,j:\exists k\in I:F_{k}\subset F_{i}\cap F_{j}$
\end{definition}

For instance the \textbf{Fr\'{e}chet filter} is the family over $%
\mathbb{N}
$\ defined by :

$F_{n}=\{p\in%
\mathbb{N}
:p\geq n\}$

\bigskip

\subsection{Binary relations}

\label{Binary relations}

\subsubsection{Definitions}

\begin{definition}
A \textbf{binary relation} R on a set E is a 2 variables propositional
function : $R:E\times E\rightarrow\left\{  T,F\right\}  $ true, false
\end{definition}

\begin{definition}
A binary relation R on the set E is :

\textbf{reflexive} if : $\forall x\in E:R(x,x)=T$

\textbf{symmetric} if : $\forall x,y\in E:R(x,y)\sim R(y,x)$

\textbf{antisymmetric} if : $\forall x,y\in E:\left(  R(x,y)\wedge
R(y,x)\right)  \Rightarrow x=y$

\textbf{transitive} if : $\forall x,y,z\in E:\left(  R(x,y)\wedge
R(y,z)\right)  \Rightarrow R\left(  x,z\right)  $

\textbf{total} if $\vDash\forall x\in E,\forall y\in E,\left(  R(x,y)\vee
R(y,x)\right)  $
\end{definition}

An antisymmetric relation gives 2 dual binary relations ("greater or equal
than" and "smaller or equal than").

Warning ! in a set endowed with a binary relation two elements x,y can be non
comparable : R(x,y), R(y,x) are still defined but both take the value "False".
If the relation is total then either R(x,y) is true or R(y,x) is true, they
cannot be both false.

\subsubsection{Equivalence relation}

\begin{definition}
An \textbf{equivalence relation} is a binary relation which is reflexive,
symmetric and transitive
\end{definition}

It will be usually denoted by $\sim$

\begin{definition}
If R is an equivalence relation on the set E,

- the \textbf{class of equivalence} of an element $x\in E$\ is the subset
denoted $\left[  x\right]  $ \ $\ $of elements $y\in E$ such that $y\sim x$ .

- the \textbf{quotient set} denoted $E/\sim$ is the partition of E whose
elements are the classes of equivalence of E.
\end{definition}

\begin{theorem}
There is a natural bijection from the set of all possible equivalence
relations on E to the set of all partitions of E.
\end{theorem}

So, if E is a finite set with n elements, the number of possible equivalence
relations on E equals the number of distinct partitions of E, which is the nth
\textbf{Bell number} : $B_{n}=\frac{1}{e}\sum_{k=0}^{\infty}\frac{k^{n}}{k!}$

Example : for any map $f:E\rightarrow F$ the relation $x\sim y$ if f(x)=f(y)
is an equivalence relation.

\subsubsection{Order relation}

\begin{definition}
A \textbf{preordered set }(also called a poset) is a set endowed with a binary
relation which is reflexive and transitive
\end{definition}

\begin{definition}
An \textbf{order relation} is a binary relation which is reflexive,
antisymmetric and transitive.
\end{definition}

\begin{definition}
An \textbf{ordered set} (or totally ordered) is a set endowed with an order
relation which is total
\end{definition}

\paragraph{Bounds\newline}

On an ordered set :

\begin{definition}
An \textbf{upper bound} of a subset A of E is an element of E which is greater
than all the elements of A
\end{definition}

\begin{definition}
A \textbf{lower bound} of a subset A of E is an element of E which is smaller
than all the elements of A
\end{definition}

\begin{definition}
A \textbf{bounded subset} A of E is a subset which has both an upper bound and
a lower bound.
\end{definition}

\begin{definition}
A \textbf{maximum} of a subset A of E is an element of A which is also an
upper bound for A
\end{definition}

$m=\max A\Leftrightarrow m\in E,\forall x\in A:m\geq x$

\begin{definition}
A \textbf{minimum} of a subset A of E is an element of A which is also a lower
bound for A
\end{definition}

$m=\min A\Leftrightarrow m\in E,\forall x\in A:m\leq x$

Maximum and minimum, if they exist, are unique.

\begin{definition}
If the set of upper bounds has a minimum, this element is unique and is called
the \textbf{least upper bound} or supremum
\end{definition}

denoted : $s=\sup A=\min\{m\in E:\forall x\in E:m\geq x\}$

\begin{definition}
If the set of lower bounds has a maximum, this element is unique and is called
the \textbf{greatest lower bound} or infimum.
\end{definition}

denoted :$s=\inf A=\max\{m\in E:\forall x\in E:m\leq x\}$

\begin{theorem}
Over $%
\mathbb{R}
$ any non empty subset which has an upper bound has a least upper bound, and
any non empty subset which has a lower bound has a greatest lower bound,
\end{theorem}

If $f:E\rightarrow%
\mathbb{R}
$ is a real valued function, a maximum of f is an element M of E such that
f(M) is a maximum of f(E), and a mimimum of f is an element m of E such that
f(m) is a minimum of f(E)

\begin{axiom}
\textbf{Zorn lemna} : if E is a preordered set such that any subset for which
the order is total has a least upper bound, then E has also a maximum.
\end{axiom}

The Zorn lemna is equivalent to the axiom of choice.

\begin{definition}
A set is \textbf{well-ordered} if it is totally ordered and if any non empty
subset has a minimum. Equivalently if there is no infinite \textit{decreasing} sequence.
\end{definition}

It is then possible to associate each element with an ordinal number (see
below). The axiom of choice is equivalent to the statement that every set can
be well-ordered.

As a consequence let I be any set.\ Thus for any finite subset J of I it is
possible to order the elements of J and one can write \ J=$\left\{
j_{1},j_{2},...j_{n}\right\}  $ with n=card(J).

\begin{definition}
A \textbf{lattice} is a partially ordered set in which any two elements have a
unique supremum (the least upper bound, called their join) and an infimum
(greatest lower bound, called their meet).
\end{definition}

Example : For any set A, the collection of all subsets of A can be ordered via
subset inclusion to obtain a lattice bounded by A itself and the null set. Set
intersection and union interpret meet and join, respectively.

\begin{definition}
A \textbf{monotone} map $f:E\rightarrow F$ between sets E,F endowed with an
ordering is a map which preserves the ordering:
\end{definition}

$\forall x,y\in E,x\leq_{E}y\Rightarrow f(x)\leq_{F}f(y)$

The converse of such a map is an \textbf{order-reflecting} map :

$\forall x,y\in E,f(x)\leq_{F}f(y)\Rightarrow x\leq_{E}y$

\paragraph{Nets\newline}

\begin{definition}
(Wilansky p.39) A \textbf{directed set} is a set E endowed with a binary
relation $\geq$ which is reflexive and transitive, and such that :

$\forall x,y\in E:\exists z:z\geq x,z\geq y$
\end{definition}

\begin{definition}
A\ \textbf{net} in a set F is a map : $f:E\rightarrow F$ where E is a directed set.
\end{definition}

A sequence is a net.

\subsubsection{Cardinality}

\begin{theorem}
Bernstein (Schwartz I p.23) For any two sets E,F either there is an injective
map $f:E\rightarrow F$ or there is an injective map $g:F\rightarrow E$. If
there is an injective map $f:E\rightarrow F$ and an injective map :
$g:F\rightarrow E$ then there is a bijective map : $\varphi:E\rightarrow
F,\varphi^{-1}:F\rightarrow E$
\end{theorem}

\paragraph{Cardinal numbers\newline}

The binary relation between sets E,F : "there is a bijection between E and F"
is an equivalence relation.

\begin{definition}
Two sets have the same \textbf{cardinality} if there is a bijection between them.
\end{definition}

The cardinality of a set is represented by a \textbf{cardinal number}. It will
be denoted card(E) or $\#E.$

The cardinal of $\varnothing$ is 0.

The cardinal of any finite set is the number of its elements.

The cardinal of the set of natural numbers $%
\mathbb{N}
$ , of algebric numbers $%
\mathbb{Z}
$ and of rational numbers $%
\mathbb{Q}
$ is $\aleph_{0}$ (aleph null: hebra\"{\i}c letter).

The cardinal of the set of the subsets of E (its power set $2^{E}$) is
$2^{card(E)}$

The cardinal of $%
\mathbb{R}
$ (and $%
\mathbb{C}
,$ and more generally $%
\mathbb{R}
^{n},n\in%
\mathbb{N}
)$\ is $c=2^{\aleph_{0}},$ called the \textbf{cardinality of the continuum}

It can be proven that : $c^{\aleph_{0}}=c,c^{c}=2^{c}$

\newpage

\section{CATEGORIES}

\bigskip

A set E, and any map $f:E\rightarrow F$ to another set have all the general
properties listed in the previous section.\ Mathematicians consider classes of
sets sharing additional properties : for instance groups are sets endowed with
an internal operation which is associative, has a unity and for which each
element has an inverse. These additional properties define a \textbf{structure
}on the set. There are many different structures, which can be sophisticated,
and the same set can be endowed with different structures. When one considers
maps between sets having the same structure, it is logical to focus on maps
which preserves the structure, they are called \textbf{morphisms}. For
instance a group morphism is a map $f:G\rightarrow H$ between two groups such
that $f\left(  g\cdot g^{\prime}\right)  =f\left(  g\right)  \circ f\left(
g^{\prime}\right)  $\ and $f\left(  1_{G}\right)  =1_{H}$. There is a general
theory which deals with structures and their morphisms : the category theory,
which is now a mandatory part of advanced mathematics.

It deals with general structures and morphisms and provides a nice language to
describe many usual mathematical objects in a unifying way.\ It is also a
powerful tool in some specialized fields. To deal with any kind of structure
it requires the minimum of properties from the objects which are considered.
The drawback is that it leads quickly to very convoluted and abstract
constructions when dealing with precise subjects, that border mathematical
pedantism, without much added value. So, in this book, we use it when, and
only when, it is really helpful and the presentation in this section is
limited to the main definitions and principles, in short to the vocabulary
needed to understand what lies behind the language. They can be found in
Kashirawa-Shapira or Lane.

\bigskip

\subsection{Categories}

\label{Categories}

\subsubsection{Definitions}

\begin{definition}
A \textbf{category} C consists of the following data:

- a set\ $Ob(C)$\ of \textbf{objects}

- for each ordered pair (X,Y) of objects of Ob(C), a set of \textbf{morphisms}
\ hom(X,Y)\ from the domain X to the codomain Y

- a function $\circ$\ called composition between morphisms :

$\circ:\hom(X,Y)\times\hom(Y,Z)\rightarrow\hom(X,Z)$

which must satisfy the following conditions :

Associativity

$f\in\hom(X,Y),g\in\hom(Y,Z),h\in\hom(Z,T)\Rightarrow(f\circ g)\circ
h=f\circ(g\circ h)$

Existence of an identity morphism for each object

$\forall X\in Ob(C)\ ,\exists id_{X}\in\hom(X,X):\forall f\in\hom(X,Y):$

$f\circ id_{X}=f,\forall g\in\hom(Y,X):id_{X}\circ g=g$
\end{definition}

If Ob(C) is a set of a universe U (therefore all the objects belong also to
U), and if for all objects the set $\hom\left(  A,B\right)  $ is isomorphic to
a set of U then the category is said to be a "U-small category". Here
"isomorphic" means that there is a bijective map which is also a morphism.

\bigskip

Remarks :

i) When it is necessary to identify the category one denotes $\hom_{C}\left(
X,Y\right)  $ for $\hom(X,Y)$

ii) The use of "universe" is necessary as in categories it is easy to run into
the problems of "too large sets".

iii) To be consistent with some definitions one shall assume that the set of
morphisms from one object A to another object B can be empty.

iv) A morphism is not necessarily a map $f:X\rightarrow Y.$ Let U be a
universe of sets (the sets are known), C the category defined as : objects =
sets in U, morphisms : $\hom_{C}\left(  X,Y\right)  =\left\{  X\sqsubseteq
Y\right\}  $ meaning the logical proposition $X\sqsubseteq Y$ which is either
true of false. One can check that it meets the conditions to define a category.

As such, the definition of a category brings nothing new to the usual axioms
and definitions of set theory. The concept of category is useful when all the
objects are endowed with some specific structure and the morphisms are the
specific maps related to this structure: we have the category of "sets",
"vector spaces", "manifolds",..It is similar to set theory : \ one can use
many properties of sets without telling what are the elements of the set.

The term "morphism" refer to the specific maps used in the definition of the
category, but, as a rule, we will always reserve the name \textbf{morphism}
for maps between sets endowed with similar structures which "conserve" these
structures.\ And similarly \textbf{isomorphism} for bijective morphism.

\paragraph{Examples\newline}

1. For a given universe U the category U-set is the category with objects the
sets of U and morphisms any map between sets of Ob(U-set). It is necessary to
fix a universe because there is no "Set of sets".

2. The category of groups and their morphisms. The category of vector spaces
over a field K and the K-linear maps.\ The category of topological spaces and
continuous maps. The category of smooth manifolds and smooth maps.

Notice that the morphisms must meet the axioms (so one has to prove that the
composition of linear maps is a linear map). The manifolds and differentiable
maps are not a category as a manifold can be continuous but not
differentiable. The vector spaces over $%
\mathbb{R}
$ (resp.$%
\mathbb{C}
)$ are categories but the vector spaces (over any field) are not a category as
the product of a R-linear map and a C-linear map is not a C-linear map.

3. A \textbf{monoid} is a category with one unique object and a single
morphism (the identity).. It is similar to a set M, a binary relation MxM
associative with unitary element (semi group).

4. A \textbf{simplicial category} has objects indexed on ordinal numbers and
morphisms are order preserving maps.

More generally the category of ordered sets with objects = ordered sets
belonging to a universe, morphisms = order preserving maps.

\subsubsection{Additional definitions about categories}

\begin{definition}
A \textbf{subcategory} C' of the category C has for objects $Ob(C\prime
)\subset Ob(C)$ and for $X,Y\in C^{\prime},\hom_{C^{\prime}}\left(
X,Y\right)  \subset\hom_{C}\left(  X,Y\right)  $

A subcategory is \textbf{full} if $\hom_{C^{\prime}}\left(  X,Y\right)
=\hom_{C}\left(  X,Y\right)  $
\end{definition}

\begin{definition}
If C is a category, the \textbf{opposite category}, denoted C*, has the same
objects as C and for morphisms : $\hom_{C^{\ast}}(X,Y)=\hom_{C}(Y,X)$ with the
composition :

$f\in\hom_{C^{\ast}}(X,Y),g\in\hom_{C^{\ast}}(Y,Z):g\circ^{\ast}f=f\circ g$
\end{definition}

\begin{definition}
A category is

- \textbf{discrete} if all the morphisms are the identity morphisms

- \textbf{finite} it the set of objects and the set of morphisms are finite

- \textbf{connected} if it is non empty and for any pair X,Y of objects there
is a finite sequence of objects $X_{0}=X,X_{1},..X_{n-1},X_{n}=Y$ such that
$\forall i\in\left[  0,n-1\right]  $ at least one of the sets $\hom\left(
X_{i},X_{i+1}\right)  ,\hom\left(  X_{i+1},X_{i}\right)  $ is non empty.
\end{definition}

\begin{definition}
If $\left(  C_{i}\right)  _{i\in I}$ is a family of categories indexed by the
set I

the \textbf{product category} $C=%
{\displaystyle\prod\limits_{i\in I}}
C_{i}$ has

- for objects : $Ob\left(  C\right)  =%
{\displaystyle\prod\limits_{i\in I}}
Ob\left(  C_{i}\right)  $

- for morphisms : $\hom_{C}\left(
{\displaystyle\prod\limits_{j\in I}}
X_{j},%
{\displaystyle\prod\limits_{j\in I}}
Y_{j}\right)  =%
{\displaystyle\prod\limits_{j\in I}}
\hom_{C_{j}}\left(  X_{j},Y_{j}\right)  $

the \textbf{disjoint union} \textbf{category} $\sqcup_{i\in I}C_{i}$ has

- for objects : $Ob\left(  \sqcup C_{i}\right)  =\left\{  \left(
X_{i},i\right)  ,i\in I,X_{i}\in Ob\left(  C_{i}\right)  \right\}  $

- for morphisms : $\hom_{\sqcup C_{i}}\left(  \left(  X_{j},j\right)  ,\left(
Y_{k},k\right)  \right)  =\hom_{C_{j}}\left(  X_{j},Y_{j}\right)  $ if j=k;
=$\varnothing$ if $j\neq k$
\end{definition}

\begin{definition}
A \textbf{pointed category} is a category with the following properties:

- each object X is a set and there is a unique $x\in X$ (called base point)
which is singled : let $x=\imath\left(  X\right)  $

- there are morphisms which preserve x : $\exists f\in\hom\left(  X,Y\right)
:\imath\left(  Y\right)  =f\left(  \imath\left(  X\right)  \right)  $
\end{definition}

Example : the category of vector spaces over a field K with a basis and linear
maps which preserve the basis.

\subsubsection{Initial and terminal objects}

\begin{definition}
An object I is \textbf{initial }in the category C if

$\forall X\in Ob(C),\#\hom(I,X)=1$
\end{definition}

meaning that there is only one morphism going from I to X

\begin{definition}
An object T is \textbf{terminal }in the category C if

$\forall X\in Ob(C),\#\hom(X,T)=1$
\end{definition}

meaning that there is only one morphism going from X to T

\begin{definition}
An object is \textbf{null} (or zero object) \textbf{\ }in the category C if it
is both\ initial and terminal.
\end{definition}

It is usually denoted 0. So if there is a null object, $\forall X,Y$ there is
a morphism $X\rightarrow Y$ given by the composition : $X\rightarrow
0\rightarrow Y$

In the category of groups the null object is the group 1, comprised of the unity.

Example : define the pointed category of n dimensional vector spaces over a
field K, with an identified basis:

- objects : E any n dimensional vector space over a field K, with a singled
basis $\left(  e_{i}\right)  _{i=1}^{n}$

- morphisms: $\hom(E,F)=L(E;F)$ (there is always a linear map F : $f\left(
e_{i}\right)  =f_{i})$

All the objects are null : the morphisms from E to F such that $f\left(
e_{i}\right)  =f_{i}$\ are unique

\subsubsection{Morphisms}

\paragraph{General definitions\newline}

The following definitions generalize, in the language of categories, concepts
which have been around for a long time for structures such as vector spaces,
topological spaces,...

\begin{definition}
An \textbf{endomorphism} is a morphism in a category with domain = codomain :
$f\in\hom(X,X)$
\end{definition}

\begin{definition}
If $f\in\hom(X,Y),g\in\hom(Y,X)$ such that : $f\circ g=Id_{Y}$ then f is the
\textbf{left-inverse} of g, and g is the \textbf{right-inverse} of f
\end{definition}

\begin{definition}
A morphism $f\in\hom(X,Y)$ is an \textbf{isomorphism} if there exists
$g\in\hom(Y,X)$ such that $f\circ g=Id_{Y},g\circ f=Id_{X}$
\end{definition}

\begin{notation}
$X\simeq Y$ when X,Y are two isomorphic objects of some category
\end{notation}

\begin{definition}
An \textbf{automorphism} is an endomorphism which is also an isomorphism
\end{definition}

\begin{definition}
A category is a \textbf{groupoid} if all its morphisms are isomorphisms
\end{definition}

\bigskip

\paragraph{Definitions specific to categories\newline}

\begin{definition}
Two morphisms in a category are \textbf{parallel} if they have same domain and
same codomain.\ They are denoted : $f,g:X\rightrightarrows Y$
\end{definition}

\begin{definition}
A \textbf{monomorphism} $f\in\hom(X,Y)$ is a morphism such that for any pair
of parallel morphisms :

$g_{1},g_{2}\in\hom(Z,X):f\circ g_{1}=f\circ g_{2}\Rightarrow g_{1}=g_{2}$
\end{definition}

Which can be interpreted as f has a left-inverse and so is an injective morphism

\begin{definition}
An \textbf{epimorphism} $f\in\hom(X,Y)$ is a morphism such that for any pair
of parallel morphisms :

$g_{1},g_{2}\in\hom(Y,Z):g_{1}\circ f=g_{2}\circ f\Rightarrow g_{1}=g_{2}$
\end{definition}

Which can be interpreted as f has a right-inverse and so is a surjective morphism

\begin{theorem}
If $f\in\hom(X,Y),g\in\hom(Y,Z)$ and f,g are monomorphisms (resp.epimorphisms,
isomorphisms) then $g\circ f$ is a monomorphism (resp.epimorphism, isomorphism)
\end{theorem}

\begin{theorem}
The morphisms of a category C are a category denoted $\hom\left(  C\right)  $

- Its objects are the morphisms in C : $Ob\left(  \hom\left(  C\right)
\right)  =\left\{  \hom_{C}\left(  X,Y\right)  ,X,Y\in Ob\left(  C\right)
\right\}  $

- Its morphisms are the maps u,v such that :

$\forall X,Y,X^{\prime},Y\in Ob\left(  C\right)  ,\forall f\in\hom\left(
X,Y\right)  ,g\in\hom\left(  X^{\prime},Y\right)  :$

$u\in\hom\left(  X,X^{\prime}\right)  ,v\in\hom\left(  Y,Y^{\prime}\right)
:v\circ f=g\circ u$
\end{theorem}

The maps u,v must share the general characteristics of the maps in C

\paragraph{Diagrams\newline}

Category theory uses diagrams quite often, to describe, by arrows and symbols,
morphisms or maps between sets. A diagram is \textbf{commutative} if any path
following the arrows is well defined (in terms of morphisms).

Example : the following diagram is commutative :

\bigskip

$\ \ X\overset{f}{\rightarrow}Y$

$u\downarrow$ \ \ \ \ \ $\downarrow v$

\ \ $Z\overset{g}{\rightarrow}T$

\bigskip

means :

$g\circ u=v\circ f$

\paragraph{Exact sequence\newline}

Used quite often with a very abstract definition, which gives, in plain language:

\begin{definition}
For a family $\left(  X_{p}\right)  _{p\leq n}$ of objects of a category C and
of morphisms $f_{p}\in\hom_{C}\left(  X_{p},X_{p+1}\right)  $

the sequence : $X_{0}\overset{f_{0}}{\rightarrow}X_{1}..X_{p}\overset{f_{p}%
}{\rightarrow}X_{p+1}....\overset{f_{n-1}}{\rightarrow}X_{n}$ is \textbf{exact
}if%

\begin{equation}
f_{p}\left(  X_{p}\right)  =\ker\left(  f_{p+1}\right)
\end{equation}

\end{definition}

An exact sequence is also called a \textbf{complex}. It can be infinite.

That requires to give some meaning to ker. In the usual cases ker may be
understood as the subset :

if the $X_{p}$ are groups :

$\ker f_{p}=\left\{  x\in X_{p},f_{p}\left(  x\right)  =1_{X_{p+1}}\right\}  $
so $f_{p}\circ f_{p-1}=1$

if the $X_{p}$ are vector spaces :

$\ker f_{p}=\left\{  x\in X_{p},f_{p}\left(  x\right)  =0_{X_{p+1}}\right\}  $
so $f_{p}\circ f_{p-1}=0$

\begin{definition}
A \textbf{short exact sequence} in a category C is : $X\overset{f}%
{\rightarrow}Y\overset{g}{\rightarrow}Z$ where : $f\in\hom_{C}\left(
X,Y\right)  $ is a monomorphism (injective) $,g\in\hom_{C}\left(  Y,Z\right)
$ is an epimorphism (surjective), equivalently iff $f\circ g$ is an isomorphism.
\end{definition}

Then Y is, in some way, the product of Z and f(X)

it is usally written :

$0\rightarrow X\overset{f}{\rightarrow}Y\overset{g}{\rightarrow}Z\rightarrow0$
for abelian groups or vector spaces

$1\rightarrow X\overset{f}{\rightarrow}Y\overset{g}{\rightarrow}Z\rightarrow1$
for the other groups

A short exact sequence $X\overset{f}{\rightarrow}Y\overset{g}{\rightarrow}%
Z$\ \textbf{splits} if either :

\bigskip

$\exists t\in\hom_{C}\left(  Y,X\right)  ::t\circ f=Id_{X}$

\bigskip%

\begin{tabular}
[c]{ccccccc}%
0 & $\rightarrow$ & X & $\overset{f}{\rightarrow}$ & Y & $\overset
{g}{\rightarrow}$ & Z\\
&  &  & $\overset{t}{\longleftarrow}$ &  &  &
\end{tabular}

\bigskip

or

$\exists u\in\hom_{C}\left(  Z,Y\right)  ::g\circ u=Id_{Z}$

\bigskip%

\begin{tabular}
[c]{ccccccc}%
0 & $\rightarrow$ & X & $\overset{f}{\rightarrow}$ & Y & $\overset
{g}{\rightarrow}$ & Z\\
&  &  &  &  & $\overset{u}{\longleftarrow}$ &
\end{tabular}

\bigskip

then :

- for abelian groups or vector spaces : $Y=X\oplus Z$

- for other groups (semi direct product) : $Y=X\ltimes Z$

\bigskip

\subsection{Functors}

\label{Functors}

Functors are roughly maps between categories. They are used to import
structures from a category C to a category C', using a general procedure so
that some properties can be extended immediately.

\subsubsection{Functors}

\begin{definition}
A \textbf{functor} (a covariant functor) F between the categories C and C' is :

a map $F_{o}:Ob(C)\rightarrow Ob(C^{\prime})$

maps $F_{m}:\hom\left(  C\right)  \rightarrow\hom\left(  C^{\prime}\right)
::f\in\hom_{C}(X,Y)\rightarrow F_{m}\left(  f\right)  \in\hom_{C^{\prime}%
}\left(  F_{o}\left(  X\right)  ,F_{o}\left(  Y\right)  \right)  $

such that

$F_{m}\left(  Id_{X}\right)  =Id_{F_{o}\left(  X\right)  }$

$F_{m}\left(  g\circ f\right)  =F_{m}\left(  g\right)  \circ F_{m}\left(
f\right)  $
\end{definition}

\begin{definition}
A \textbf{contravariant} \textbf{functor} F between the categories C and C' is :

a map $F_{o}:Ob(C)\rightarrow Ob(C^{\prime})$

maps $F_{m}:\hom\left(  C\right)  \rightarrow\hom\left(  C^{\prime}\right)
::f\in\hom_{C}(X,Y)\rightarrow F_{m}\left(  f\right)  \in\hom_{C^{\prime}%
}\left(  F_{o}\left(  X\right)  ,F_{o}\left(  Y\right)  \right)  $

such that

$F_{m}\left(  Id_{X}\right)  =Id_{F_{o}\left(  X\right)  }$

$F_{m}\left(  g\circ f\right)  =F_{m}\left(  f\right)  \circ F_{m}\left(
g\right)  $
\end{definition}

\begin{notation}
$F:C\mapsto C^{\prime}$ (with the arrow $\mapsto)$\ \ is a functor F between
the categories C,C'
\end{notation}

Example : the functor which associes to each vector space its dual and to each
linear map its transpose is a functor from the category of vector spaces over
a field K to itself.

A contravariant functor is a covariant functor $C^{\ast}\mapsto C^{\prime\ast
}.$

A functor F induces a functor : $F^{\ast}:C^{\ast}\mapsto C^{\prime\ast}$

A functor $F:C\mapsto Set$ is said to be \textbf{forgetful }(the underlying
structure in C is lost).

\begin{definition}
A \textbf{constant functor} denoted $\Delta_{X}:I\mapsto C$ between the
categories I,C, where $X\in Ob(C)$ is the functor :

$\forall i\in Ob\left(  I\right)  :\left(  \Delta_{X}\right)  _{o}\left(
i\right)  =X$

$\forall i,j\in Ob\left(  I\right)  ,\forall f\in\hom_{I}\left(  i,j\right)
:\left(  \Delta_{X}\right)  _{m}\left(  f\right)  =Id_{X}$
\end{definition}

\paragraph{Composition of functors\newline}

Functors can be composed :

$F:C\mapsto C^{\prime},F^{\prime}:C^{\prime}\mapsto C"$

$F\circ F^{\prime}:C\mapsto C"::\left(  F\circ F^{\prime}\right)  _{o}%
=F_{o}\circ F_{o}^{\prime};\left(  F\circ F^{\prime}\right)  _{m}=F_{m}\circ
F_{m}^{\prime}$

The composition of functors is associative whenever it is defined.

\begin{definition}
A functor F is \textbf{faithful} if

$F_{m}:\hom_{C}(X,Y)\rightarrow\hom_{C^{\prime}}\left(  F_{o}\left(  X\right)
,F_{o}\left(  Y\right)  \right)  $ is injective
\end{definition}

\begin{definition}
A functor F is \textbf{full} if

$F_{m}:\hom_{C}(X,Y)\rightarrow\hom_{C^{\prime}}\left(  F_{o}\left(  X\right)
,F_{o}\left(  Y\right)  \right)  $ is surjective
\end{definition}

\begin{definition}
A functor F is \textbf{fully faithful} if

$F_{m}:\hom_{C}(X,Y)\rightarrow\hom_{C^{\prime}}\left(  F_{o}\left(  X\right)
,F_{o}\left(  Y\right)  \right)  $ is bijective
\end{definition}

These 3 properties are closed by composition of functors.

\begin{theorem}
If $F:C\mapsto C^{\prime}$ is faithful, and if F(f) with $f\in\hom_{C}\left(
X,Y\right)  $ is an epimorphism (resp. a monomorphism) then f is an
epimorphism (resp. a monomorphism)
\end{theorem}

\paragraph{Product of functors\newline}

One defines naturally the product of functors.\ A \textbf{bifunctor}
$F:C\times C^{\prime}\mapsto C"$ is a functor defined over the product CxC',
so that for any fixed $X\in C,X^{\prime}\in C^{\prime}$ $F\left(  X,.\right)
,F\left(  .,X^{\prime}\right)  $ are functors

If C,C' are categories, CxC' their product, the right and left
\textbf{projections} are functors defined obviously :

$L_{o}\left(  X\times X^{\prime}\right)  =X;R_{o}\left(  X,X^{\prime}\right)
=X^{\prime}$

$L_{m}\left(  f\times f^{\prime}\right)  =f;R_{m}\left(  f,f^{\prime}\right)
=f^{\prime}$

They have the universal property : whatever the category D, the functors
$F:D\mapsto C;F^{\prime}:D\mapsto C^{\prime}$ there is a unique functor
$G:D\mapsto C\times C^{\prime}$ such that $L\circ G=F,R\circ G=F^{\prime}$

\subsubsection{Natural transformation}

A natural transformation is a map between functors. The concept is mostly used
to give the conditions that a map must meet to be consistent with structures
over two categories.

\begin{definition}
Let F,G be two functors from the categories C to C'. A \textbf{natural
transformation} $\phi$\ (also called a morphism of functors) denoted
$\phi:F\hookrightarrow G$ is a map : $\phi:Ob(C)\rightarrow\hom_{C^{\prime}%
}\left(  Ob(C^{\prime}),Ob(C^{\prime})\right)  $ such that the following
diagram commutes :
\end{definition}

\bigskip%

\begin{tabular}
[c]{llllll}
& C &  &  & C' & \\
$\ulcorner$ &  & $\urcorner$ & $\ulcorner$ &  & $\ \ \ \ \ \ \ \ \ \ \urcorner
$\\
& X &  & $F_{o}\left(  X\right)  $ & $\overset{\phi\left(  X\right)
}{\rightarrow}$ & $G_{o}(X)$\\
& $\downarrow$ &  & $\downarrow$ &  & $\downarrow$\\
f & $\downarrow$ &  & $\downarrow F_{m}\left(  f\right)  $ &  & $\downarrow
G_{m}\left(  f\right)  $\\
& $\downarrow$ &  & $\downarrow$ &  & $\downarrow$\\
& Y &  & $F_{o}(Y)$ & $\overset{\phi\left(  Y\right)  }{\rightarrow}$ &
$G_{o}(Y)$%
\end{tabular}

\bigskip

$\forall X,Y\in Ob\left(  C\right)  ,\forall f\in\hom_{C}\left(  X,Y\right)
:$

$G_{m}\left(  f\right)  \circ\phi\left(  X\right)  =\phi\left(  Y\right)
\circ F_{m}\left(  f\right)  \in\hom_{C^{\prime}}\left(  F_{o}\left(
X\right)  ,G_{o}\left(  Y\right)  \right)  $

$F_{m}\left(  f\right)  \in\hom_{C^{\prime}}\left(  F_{o}\left(  X\right)
,F_{o}(Y)\right)  $

$G_{m}\left(  f\right)  \in\hom_{C^{\prime}}\left(  G_{o}\left(  X\right)
,G_{o}(Y)\right)  $

$\phi\left(  X\right)  \in\hom_{C^{\prime}}\left(  F_{o}\left(  X\right)
,G_{o}(X)\right)  $

$\phi\left(  Y\right)  \in\hom_{C^{\prime}}\left(  F_{o}\left(  Y\right)
,G_{o}(Y)\right)  $

The components of the transformation are the maps $\phi\left(  X\right)
,\phi\left(  Y\right)  $

If $\forall X\in Ob(C)$\ $\phi\left(  X\right)  $ is inversible then the
functors are said to be \textbf{equivalent}.

Natural transformations can be composed in the obvious way. Thus :

\begin{theorem}
The set of functors from a category C to a category C' is itself a category
denoted $Fc\left(  C,C^{\prime}\right)  .$ Its objects are $Ob\left(
Fc\left(  C,C^{\prime}\right)  \right)  $ any functor $F:C\mapsto C^{\prime}$
and its morphisms are natural transformations : $\hom\left(  F_{1}%
,F_{2}\right)  =\left\{  \phi:F_{1}\hookrightarrow F_{2}\right\}  $
\end{theorem}

\subsubsection{Yoneda lemna}

(Kashirawa p.23)

Let U be a universe, C a category such that all its objects belong to U, and
USet the category of all sets belonging to U and their morphisms.

Let :

Y be the category of contravariant functors $C\mapsto USet$

$Y^{\ast}$ be the category of contravariant functors $C\mapsto USet^{\ast}$

$h_{C}$ be the functor : $h_{C}:C\mapsto Y$ defined by : $h_{C}\left(
X\right)  =\hom_{C}\left(  -,X\right)  .$ To an object X\ of C it associes all
the morphisms of C whith codomain X and domain any set of U.

$k_{C}$ be the functor : $k_{C}:C\mapsto Y^{\ast}$ defined by : $k_{C}\left(
X\right)  =\hom_{C}\left(  F,-\right)  .$ To an object X\ of C it associes all
the morphisms of C whith domain X and codomain any set of U.

So : $Y=Fc\left(  C^{\ast},USet\right)  ,Y^{\ast}=Fc\left(  C^{\ast
},USet^{\ast}\right)  $

\begin{theorem}
Yoneda Lemna

i) For $F\in Y,X\in C:\hom_{Y}\left(  h_{C}\left(  X\right)  ,F\right)  \simeq
F\left(  X\right)  $

ii) For $G\in Y^{\ast},X\in C:\hom_{Y^{\ast}}\left(  k_{C}\left(  X\right)
,G\right)  \simeq G\left(  X\right)  $

Moreover these isomorphisms are functorial with respect to X,F,G : they define
isomorphisms of functors from $C^{\ast}\times Y$ to USet and from $Y^{\ast
\ast}\times C$ to USet.
\end{theorem}

\begin{theorem}
The two functors $h_{C},k_{C}$ are fully faithful
\end{theorem}

\begin{theorem}
A contravariant functor $F:C\mapsto Uset$ is \textbf{representable} if there
are an object X of C, called a representative of F, and an isomorphism
$h_{C}\left(  X\right)  \hookrightarrow F$
\end{theorem}

\begin{theorem}
A covavariant functor $F:C\mapsto Uset$ is representable if there are an
object X of C, called a representative of F, and an isomorphism $k_{C}\left(
X\right)  \hookrightarrow F$
\end{theorem}

\subsubsection{Universal functors}

Many objects in mathematics are defined through an "universal property"
(tensor product, Clifford algebra,...) which can be restated in the language
of categories. It gives the following.

Let : $F:C\mapsto C^{\prime}$ be a functor and X' an object of C'

1. An \textbf{initial morphism} from X' to F is a pair $\left(  A,\phi\right)
\in Ob\left(  C\right)  \times\hom_{C^{\prime}}\left(  X^{\prime},F_{o}\left(
A\right)  \right)  $ such that :

$\forall X\in Ob\left(  C\right)  ,f\in\hom_{C^{\prime}}\left(  X^{\prime
},F_{o}\left(  X\right)  \right)  ,\exists g\in\hom_{C}\left(  A,X\right)
:f=F_{m}\left(  g\right)  \circ\phi$

The key point is that g must be unique, then A is unique up to isomorphism

\bigskip

$\mathbf{X}^{\prime}\rightarrow\rightarrow\mathbf{\phi}\rightarrow F_{o}(A)$
\ \ \ \ \ \ \ \ $\mathbf{A}$

\ $\searrow\;$\ \ \ \ \ \ \ \ \ \ \ $\downarrow$
\ \ \ \ \ \ \ \ \ \ \ \ \ \ \ \ $\downarrow$

\ \ \ \ \ $f$ \ \ \ \ \ \ \ \ \ \ $\downarrow\;F_{m}(g)$ \ \ \ \ $\downarrow
g$

\ \ \ \ \ \ \ $\searrow$ \ \ \ \ \ $\downarrow$%
\ \ \ \ \ \ \ \ \ \ \ \ \ \ \ \ \ $\downarrow$

\ \ \ \ \ \ \ \ \ \ \ \ \ $F_{o}(X)$ \ \ \ \ \ \ \ \ \ \ \ \ $X$

\bigskip

2. A \textbf{terminal morphism} from X' to F is a pair $\left(  A,\phi\right)
\in Ob\left(  C\right)  \times\hom_{C^{\prime}}\left(  F_{o}\left(  A\right)
,X^{\prime}\right)  $ such that :

$\forall X\in Ob(X),f\in\hom_{C^{\prime}}\left(  F(X);X^{\prime}\right)
,\exists g\in\hom_{C}\left(  X,A\right)  :f=\phi\circ F_{m}\left(  g\right)  $

The key point is that g must be unique, then A is unique up to isomorphism

\bigskip

$X$ \ \ \ \ \ \ \ \ \ \ \ \ \ $F_{o}(X)$

$\downarrow$ \ \ \ \ \ \ \ \ \ \ \ \ \ $\downarrow$ \ \ \ \ $\searrow$

$g$ \ $F_{m}(g)\;\;\ \downarrow$\ \ \ \ \ \ \ \ \ $f$

$\downarrow$ \ \ \ \ \ \ \ \ \ \ \ \ \ $\downarrow$
\ \ \ \ \ \ \ \ \ \ \ $\searrow$

$\mathbf{A}$ \ \ \ \ \ \ \ \ \ \ \ $F_{o}(A)\rightarrow\mathbf{\phi
}\rightarrow\mathbf{X}^{\prime}$

\bigskip

3. \textbf{Universal morphism} usually refers to initial morphism.

\newpage

\part{ALGEBRA}

\bigskip

Given a set, the theory of sets provides only a limited number of tools.\ To
go further "mathematical structures" are added on sets, meaning operations,
special collection of sets, maps...which become the playing ground of mathematicians.

Algebra is the branch of mathematics which deals with structures defined by
operations between elements of sets. An algebraic structure consists of one or
more sets closed under one or more operations, satisfying some axioms. The
same set can be given different algebraic structures. Abstract algebra is
primarily the study of algebraic structures and their properties.

To differentiate algebra from other branches of mathematics, one can say that
in algebra there is no concept of limits or "proximity" such that are defined by\ topology.

We will give a long list of definitions of all the basic objects of common
use, and more detailed (but still schematic) study of groups (there is a part
dedicated to Lie groups and Lie algebras) and a detailed study of vector
spaces and Clifford algebras, as they are fundamental for the rest of the book.

\bigskip

\section{USUAL\ ALGEBRAIC\ STRUCTURES}

\bigskip

\subsubsection{Operations}

\begin{definition}
An \textbf{operation} over a set A is a map : $\cdot:A\times A\rightarrow A$
such that :

$\forall x,y\in A,\exists z\in A:z=x\cdot y$.

It is :

- \textbf{associative} if $\forall x,y,z\in A:\left(  x\cdot y\right)  \cdot
z=x\cdot\left(  y\cdot z\right)  $

- \textbf{commutative} if $\forall x,y\in A:x\cdot y=y\cdot x$
\end{definition}

Notice that an operation is always defined over the whole of its domain.

\begin{definition}
An element e of a set A is an \textbf{identity element }for the
operation\textbf{\ }$\cdot$ if : $\forall x\in A:e\cdot x=x\cdot e=x$

An element x of a set A is a :

- \textbf{right-inverse} of y for the operation\textbf{\ }$\cdot$ if : $y\cdot
x=e$

- \textbf{left-inverse} of y for the operation\textbf{\ }$\cdot$ if : $x\cdot
y=e$

- is \textbf{invertible} if it has a right-inverse and a left-inverse (which
then are necessarily equal and called \textbf{inverse})
\end{definition}

\begin{definition}
If there are two operations denoted $+$ and $\ast$ on the same set A, then
$\ast$ is \textbf{distributive} over (say also distributes over) $+$ if:
$\forall x,y,z\in A:x\ast(y+z)=\left(  x\ast y\right)  +\left(  x\ast
z\right)  ,(y+z)\ast x=\left(  y\ast x\right)  +\left(  z\ast x\right)  $
\end{definition}

\begin{definition}
An operation $\cdot$ on a set A is said to be \textbf{closed} in a subset B of
A if $\forall x,y\in B:x\cdot y\in B$
\end{definition}

If E and F are sets endowed with the operations $\cdot,\ast$ the product set
ExF is endowed with an operation in the obvious way :

$\left(  x,x^{\prime}\right)  \symbol{94}\left(  y,y^{\prime}\right)  =\left(
x\cdot y,x^{\prime}\ast y^{\prime}\right)  $

\bigskip

\subsection{From Monoid to fields}

\label{From monoids to fields}

\begin{definition}
A \textbf{monoid} is a set endowed with an associative operation for which it
has an identity element
\end{definition}

but its elements have not necessarily an inverse.

Classical monoids :

$%
\mathbb{N}
:$ natural integers with addition

$%
\mathbb{Z}
:$ the algebraic integers with multiplication

the square nxn matrices with multiplication

\subsubsection{Group}

\begin{definition}
A \textbf{group} $\left(  G,\cdot\right)  $\ is a set endowed G with an
associative operation $\cdot$, for which there is an identity element and
every element has an inverse.
\end{definition}

\begin{theorem}
In a group, the identity element is unique. The inverse of an element is unique.
\end{theorem}

\begin{definition}
A commutative (or \textbf{abelian}) group is a group with a commutative operation.
\end{definition}

\begin{notation}
+ denotes the operation in a commutative group
\end{notation}

\begin{notation}
0 denotes the identity element in a commutative group
\end{notation}

\begin{notation}
$-x$ denotes the inverse of $x$ in a commutative group
\end{notation}

\begin{notation}
1 (or $1_{G}$\ ) denotes the identity element in a non commutative group G
\end{notation}

\begin{notation}
$x^{-1}$ denotes the inverse of $x$ in a non commutative group G
\end{notation}

Classical groups (see the list of classical linear groups in "Lie groups"):

$%
\mathbb{Z}
:$ the algebraic integers with addition

$%
\mathbb{Z}
/k%
\mathbb{Z}
:$the algebraic integers multiples of $k\in%
\mathbb{Z}
$ with addition

the m$\times$p matrices with addition

$%
\mathbb{Q}
:$ rational numbers with addition and multiplication

$%
\mathbb{R}
:$ real numbers with addition and multiplication

$%
\mathbb{C}
:$ complex numbers with addition and multiplication

The trivial group is the group denoted $\left\{  1\right\}  $ with only one element.

A group G is a category, with Ob=the unique element G and morphisms
$\hom\left(  G,G\right)  $

\subsubsection{Ring}

\begin{definition}
A \textbf{ring} is a set endowed with two operations : one called addition,
denoted + for which it is an abelian group, the other denoted $\cdot$ for
which it is a monoid, and $\cdot$ is distributive over $+.$
\end{definition}

Remark : some authors do not require the existence of an identity element
\ for $\cdot$ and then call unital ring a ring with an identity element\ for
$\cdot$ .

If 0=1 (the identity element for + is also the identity element for $\cdot)$
the ring has only one element, said 1 and is called a trivial ring.

Classical rings :

$%
\mathbb{Z}
:$ the algebraic integers with addition and multiplication

the square nxn matrices with addition and multiplication

\paragraph{Ideals\newline}

\begin{definition}
A \textbf{right-ideal} of a ring E is a subset R of E such that :

R is a subgroup of E for addition and $\forall a\in R,\forall x\in E:x\cdot
a\in R$

A \textbf{left-ideal} of a ring E is a subset L of E such that :

L is a subgroup of E for addition and $\forall a\in L,\forall x\in E:a\cdot
x\in L$

A \textbf{two-sided ideal} (or simply an \textbf{ideal}) is a subset which is
both a right-ideal and a left-ideal.
\end{definition}

\begin{definition}
For any element a of the ring E :

the \textbf{principal right-ideal} is the right-ideal : $R=\left\{  x\cdot
a,x\in E\right\}  $

the \textbf{principal left-ideal} is the left-ideal : $L=\left\{  a\cdot
x,x\in E\right\}  $
\end{definition}

\paragraph{Division ring :\newline}

\begin{definition}
A \textbf{division ring} is a ring for which any element \textit{other than 0}
has an inverse for the second operation $\cdot$.\ 
\end{definition}

The difference between a division ring and a field (below) is that $\cdot$ is
not necessarily commutative.

\begin{theorem}
Any finite division ring is also a field.
\end{theorem}

Examples of division rings : the square invertible matrices, quaternions

\paragraph{Quaternions :\newline}

This is a division ring, usually denoted H, built over the real numbers, using
3 special "numbers" i,j,k (similar to the i of complex numbers) with the
multiplication table :

\bigskip

$%
\begin{bmatrix}
1\backslash2 & i & j & k\\
i & -1 & k & -j\\
j & -k & -1 & i\\
k & j & -i & -1
\end{bmatrix}
$

\bigskip

$i^{2}=j^{2}=k^{2}=-1,ij=k=-ji,jk=i=-kj,ki=j=-ik,$

Quaternions numbers are written as : $x=a+bi+cj+dk$ with $a,b,c,d\in%
\mathbb{R}
.$ Addition and multiplication are processed as usual for a,b,c,d and as the
table above for i,j,k. So multiplication is \textit{not} commutative.

$%
\mathbb{R}
,%
\mathbb{C}
$ can be considered as subsets of H (with b=c=d=0 or c=d=0 respectively).

The "real part" of a quaternion number is : $\operatorname{Re}(a+bi+cj+dk)=a$
so $\operatorname{Re}(xy)=\operatorname{Re}(yx)$

The "conjugate" of a quaternion is : $\overline{a+bi+cj+dk}=a-bi-cj-dk$ so
$\operatorname{Re}(x\overline{y})=\operatorname{Re}(y\overline{x})$ ,
$x\overline{x}=a^{2}+b^{2}+c^{2}+d^{2}=\left\Vert x\right\Vert _{R^{4}}^{2}$

\subsubsection{Field}

\paragraph{Definition\newline}

\begin{definition}
A \textbf{field} is a set with two operations (+ addition and $\times$
multiplication) which is an abelian group for +, the non zero elements are an
\textit{abelian} group for $\times$, and multiplication is distributive over addition.
\end{definition}

A field is a commutative division ring.

Remark : an older usage did not require the multiplication to be commutative,
and distinguished commutative fields and non commutative fields.\ It seems now
that fields=commutative fields only. "Old" non commutative fields are now
called division rings.

\paragraph{Classical fields :\newline}

$%
\mathbb{Q}
:$ rational numbers with addition and multiplication

$%
\mathbb{R}
:$ real numbers with addition and multiplication

$%
\mathbb{C}
:$ complex numbers with addition and multiplication

\textbf{Algebraic numbers} : real numbers which are the root of a one variable
polynomial equation with integers coefficients

$x\in A\Leftrightarrow\exists n,\left(  q_{k}\right)  _{k=1}^{n-1},q_{k}\in%
\mathbb{Q}
:x^{n}+\sum_{k=0}^{n-1}q_{k}x^{k}=0$

$%
\mathbb{Q}
\subset A\subset%
\mathbb{R}
$

For $a\in A,a\notin%
\mathbb{Q}
,$ define $A^{\ast}\left(  a\right)  =\left\{  x\in%
\mathbb{R}
:\exists\left(  q_{k}\right)  _{k=1}^{n-1},q_{k}\in Q:x=\sum_{k=0}^{n-1}%
q_{k}a^{k}\right\}  $ then $A^{\ast}\left(  a\right)  $ is a field. It is also
a n dimensional vector space over the field $%
\mathbb{Q}
$

\paragraph{Characteristic :\newline}

\begin{definition}
The \textbf{characteristic} of a field is the smallest integer n such that
1+1+...+1 (n times)=0. If there is no such number the field is said to be of
characteristic 0 .\ 
\end{definition}

All finite fields (with only a finite number of elements), also called
"Gallois fields", have a finite characteristic which is a prime number.

Fields of characteristic 2 are the boolean algebra of computers.

\paragraph{Polynomials\newline}

1. Polynomials are defined on a field (they can also be defined on a ring but
we will not use them) :

\begin{definition}
A \textbf{polynomial} of degree n with p variables on a field K is a function :

$P:K^{p}\rightarrow K::P\left(  X_{1},..,X_{p}\right)  =\sum a_{i_{1}...i_{k}%
}X_{1}^{i_{1}}..X_{p}^{i_{p}},\sum_{j=1}^{p}i_{j}\leq n$

If $\sum_{j=1}^{p}i_{j}=n$ the polynomial is said to be \textbf{homogeneous}.
\end{definition}

\begin{theorem}
The set of polynomials of degree n with p variables over a field K has the
structure of a finite dimensional vector space over K denoted usually
K$_{n}\left[  X_{1},...X_{p}\right]  $
\end{theorem}

The set of polynomials of any degree with k variables has the structure of a
commutative ring, with pointwise multiplication, denoted usually K$\left[
X_{1},...X_{p}\right]  .$ So it is an infinite dimensional commutative algebra.

\begin{definition}
A field is \textbf{algebraically closed} if any polynomial equation (with 1
variable) has at least one solution :
\end{definition}

$\forall n\in%
\mathbb{N}
,\forall a_{0},..a_{n}\in K,\exists x\in K:P\left(  x\right)  =a_{n}%
x^{n}+a_{n-1}x^{n-1}+...+a_{1}x+a_{.0}=0$

$%
\mathbb{R}
$ is not algebraically closed, but $%
\mathbb{C}
$ is closed (this is the main motive to introduce $%
\mathbb{C}
).$

Anticipating on the following, this generalization of a classic theorem.

\begin{theorem}
Homogeneous functions theorem (Kolar p.213): Any smooth function $f:%
{\textstyle\prod\limits_{i=1}^{n}}
E_{i}\rightarrow%
\mathbb{R}
$ where $E_{i},i=1..n$ are finite dimensional real vector spaces, such that :
$\exists a_{i}>0,b\in%
\mathbb{R}
,\forall k\in%
\mathbb{R}
:f\left(  k^{a_{1}}x_{1},..,k^{a_{n}}x_{n}\right)  =k^{b}f\left(
x_{1},..,x_{n}\right)  $ is the sum of polynomials of degree $d_{i}$ in $x_{i}
$ satisfying the relation : $b=\sum_{i=1}^{n}d_{i}a_{i}.$ If there is no such
non negative integer $d_{i}$ then f=0.
\end{theorem}

\paragraph{Complex numbers\newline}

This is the algebraic extension $%
\mathbb{C}
$\ of the real numbers $%
\mathbb{R}
.$ The fundamental theorem of algebra says that any polynomial equation has a
solution over $%
\mathbb{C}
.$

Complex numbers are written : $z=a+ib$ with $a,b\in%
\mathbb{R}
,i^{2}=-1$

The \textbf{real part} of a complex number is :$\operatorname{Re}(a+bi)=a$

The \textbf{imaginary part} of a complex number is $\operatorname{Im}\left(
a+ib\right)  =b.$

The \textbf{conjugate} of a complex number is : $\overline{a+bi}=a-bi$

The \textbf{module} of a complex number is : $\left\vert a+ib\right\vert
=\sqrt{a^{2}+b^{2}}$ and $z\overline{z}=\left\vert z\right\vert ^{2}$

The infinite sum : $\sum_{n=0}^{\infty}\frac{z^{n}}{n!}=\exp z$ always
converges and defines the \textbf{exponential} function. The cos and sin
functions can be defined as :

$\exp z=\left\vert z\right\vert \left(  \cos\theta+i\sin\theta\right)  $

thus any complex number can be written as : $z=\left\vert z\right\vert \left(
\cos\theta+i\sin\theta\right)  =\left\vert z\right\vert e^{i\theta},\theta
\in\left[  0,\pi\right]  $

and $\exp(z_{1}+z_{2})=\exp z_{1}\exp z_{2}.$

Useful identities :

$\operatorname{Re}\left(  zz^{\prime}\right)  =\operatorname{Re}\left(
z\right)  \operatorname{Re}\left(  z^{\prime}\right)  -\operatorname{Im}%
\left(  z\right)  \operatorname{Im}\left(  z^{\prime}\right)  $

$\operatorname{Im}\left(  zz^{\prime}\right)  =\operatorname{Re}\left(
z\right)  \operatorname{Im}\left(  z^{\prime}\right)  +\operatorname{Im}%
\left(  z\right)  \operatorname{Re}\left(  z^{\prime}\right)  $

$\operatorname{Re}\left(  z\right)  =\operatorname{Re}\left(  \overline
{z}\right)  ;\operatorname{Im}\left(  z\right)  =-\operatorname{Im}\left(
\overline{z}\right)  $

$\operatorname{Re}\left(  z\overline{z^{\prime}}\right)  =\operatorname{Re}%
\left(  z\right)  \operatorname{Re}\left(  z^{\prime}\right)
+\operatorname{Im}\left(  z\right)  \operatorname{Im}\left(  z^{\prime
}\right)  $

$\operatorname{Im}\left(  z\overline{z^{\prime}}\right)  =\operatorname{Re}%
\left(  z\right)  \operatorname{Im}\left(  z^{\prime}\right)
-\operatorname{Im}\left(  z\right)  \operatorname{Re}\left(  z^{\prime
}\right)  $

Let be $z=a+ib$ then the complex numbers $\alpha+i\beta$ such that $\left(
\alpha+i\beta\right)  ^{2}=z$ are :

$\alpha+i\beta=\pm\frac{1}{\sqrt{2}}\left(  \sqrt{a+\left\vert z\right\vert
}+i\frac{b}{\sqrt{a+\left\vert z\right\vert }}\right)  =\pm\frac{1}{\sqrt
{2}\sqrt{a+\left\vert z\right\vert }}\left(  a+\left\vert z\right\vert
+ib\right)  =\pm\frac{z+\left\vert z\right\vert }{\sqrt{2}\sqrt{a+\left\vert
z\right\vert }}$

\bigskip

\subsection{From vector spaces to algebras}

\label{From vector spaces to algebras}

\subsubsection{Vector space}

\begin{definition}
A \textbf{vector space E over a field K} is a set with two operations :
addition denoted + for which it is an abelian group, and multiplication by a
scalar : $K\times E\rightarrow E$ which is distributive over addition.
\end{definition}

The elements of vector spaces are \textbf{vectors}.\ And the elements of the
field K are \textbf{scalars}.

Remark : a module over a ring R is a set with the same operations as above.
The properties are not the same.\ Definitions and names differ according to
the authors.

Affine spaces are considered in the section vector spaces.

\subsubsection{Algebra}

\paragraph{Definition\newline}

\begin{definition}
An \textbf{algebra }$\left(  A,\cdot\right)  $\textbf{\ over a field K} is a
set A which is a vector space over K, endowed with an additional internal
operation $\cdot:A\times A\rightarrow A$ with the following properties :

$\cdot$ it is associative : $\forall X,Y,Z\in A:X\cdot\left(  Y\cdot Z\right)
=\left(  X\cdot Y\right)  \cdot Z$

$\cdot$ it is distributive over addition :

$\forall X,Y,Z\in A:X\cdot\left(  Y+Z\right)  =X\cdot Y+X\cdot Z;\left(
Y+Z\right)  \cdot X=Y\cdot X+Z\cdot X$

$\cdot$ it is compatible with scalar multiplication :

$\forall X,Y\in A,\forall\lambda,\mu\in K:\left(  \lambda X\right)
\cdot\left(  \mu Y\right)  =\left(  \lambda\mu\right)  X\cdot Y$

If there is an identity element I for $\cdot$ the algebra is said to be
\textbf{unital}.
\end{definition}

Remark : some authors do not require $\cdot$\ to be associative

An algebra A\ can be made unital by the extension :

$A\rightarrow\widetilde{A}=K\oplus A=\left\{  \left(  k,X\right)  \right\}
,I=\left(  1,0\right)  ,\left(  k,X\right)  =k1+X$

\begin{definition}
A \textbf{subalgebra} of an algebra $\left(  A,\cdot\right)  $ is a subset B
of A which is also an algebra for the same operations
\end{definition}

So it must be closed for the operations of the algebra.

\paragraph{Examples :\newline}

quaternions

square matrices over a field

polynomials over a field

linear endomorphisms over a vector space (with composition)

Clifford algebra (see specific section)

\paragraph{Ideal\newline}

\begin{definition}
A \textbf{right-ideal} of an algebra $\left(  A,\cdot\right)  $ is a
\textit{vector subspace} R of A such that : $\forall a\in R,\forall x\in
E:x\cdot a\in R$

A \textbf{left-ideal} of an algebra $\left(  A,\cdot\right)  $ is a
\textit{vector subspace} L of A: $\forall a\in L,\forall x\in E:a\cdot x\in L$

A \textbf{two-sided ideal} (or simply an \textbf{ideal}) is a subset which is
both a right-ideal and a left-ideal.
\end{definition}

\begin{definition}
An algebra $\left(  A,\cdot\right)  $ is \textbf{simple} if the only two-sided
ideals are 0 and A
\end{definition}

\paragraph{Derivation\newline}

\begin{definition}
A \textbf{derivation} over an algebra $\left(  A,\cdot\right)  $ is a linear
map :

$D:A\rightarrow A$ such that $\forall u,v\in A:D(u\cdot v)=(Du)\cdot
v+u\cdot(Dv)$
\end{definition}

The relation is similar to the Leibniz rule for the derivative of the product
of two scalar functions.

\paragraph{Commutant\newline}

\begin{definition}
The \textbf{commutant, }denoted S', of a subset S of an algebra $\left(
A,\cdot\right)  $, is the set of all elements in A which commute with all the
elements of S for the operation $\cdot$.
\end{definition}

\begin{theorem}
(Thill p.63-64) A commutant is a subalgebra, containing I if A is unital.
\end{theorem}

$S\subset T\Rightarrow T^{\prime}\subset S^{\prime}$

For any subset S, the elements of S commute with each others iff $S\subset
S^{\prime}$

S' is the centralizer (see Groups below) of S for the internal operation
$\cdot.$

\begin{definition}
The \textbf{second commutant} of a subset of an algebra $\left(
A,\cdot\right)  $, is the commutant denoted S"\ of the commutant S' of S
\end{definition}

\begin{theorem}
(Thill p.64)

$S\subset S"$

$S^{\prime}=\left(  S"\right)  ^{\prime}$

$S\subset T\Rightarrow\left(  S^{\prime}\right)  ^{\prime}\subset\left(
T^{\prime}\right)  ^{\prime}$

$X,X^{-1}\in A\Rightarrow X^{-1}\in\left(  X\right)  "$
\end{theorem}

\paragraph{Projection and reflexion\newline}

\begin{definition}
A \textbf{projection} in an algebra $\left(  A,\cdot\right)  $ is a an element
X\ of A such that : $X\cdot X=X$
\end{definition}

\begin{definition}
Two projections X,Y of an algebra $\left(  A,\cdot\right)  $ \ are said to be
\textbf{orthogonal} if $X\cdot Y=0$ (then $Y\cdot X=0)$
\end{definition}

\begin{definition}
Two projections X,Y of a unital algebra $\left(  A,\cdot\right)  $ are said to
be \textbf{complementary} if X+Y=I
\end{definition}

\begin{definition}
A \textbf{reflexion} of a unital algebra $\left(  A,\cdot\right)  $ is an
element X of A such that $X=X^{-1}$
\end{definition}

\begin{theorem}
If X is a reflexion of a unital algebra $\left(  A,\cdot\right)  $ then there
are two complementary projections such that X=P-Q
\end{theorem}

\begin{definition}
An element X of an algebra $\left(  A,\cdot\right)  $ is \textbf{nilpotent} if
$X\cdot X=0$
\end{definition}

\paragraph{*-algebra\newline}

*-algebras (say star algebra) are endowed with an additional operation similar
to conjugation-transpose of matrix algebras.

\begin{definition}
A \textbf{*-algebra }is an algebra $\left(  A,\cdot\right)  $ over a field K,
endowed with an involution : $\ast:A\rightarrow A$ such that :

$\forall X,Y\in A,\lambda\in K:$

$\left(  X+Y\right)  ^{\ast}=X^{\ast}+Y^{\ast}$

$\left(  X\cdot Y\right)  ^{\ast}=Y^{\ast}\cdot X^{\ast}$

$\left(  \lambda X\right)  ^{\ast}=\overline{\lambda}X^{\ast}$ (if the field K
is $%
\mathbb{C}
)$

$\left(  X^{\ast}\right)  ^{\ast}=X$
\end{definition}

\begin{definition}
The \textbf{adjoint} of an element X of a *-algebra is X*
\end{definition}

\begin{definition}
A subset S of a *-algebra is \textbf{stable} if it contains all its adjoints :
$X\in S\Rightarrow X^{\ast}\in S$
\end{definition}

The commutant S' of a stable subset S is stable

\begin{definition}
A \textbf{*-subalgebra} B of A is a stable subalgebra : $B^{\ast}\sqsubseteq
B$
\end{definition}

\begin{definition}
An element X of a *-algebra $\left(  A,\cdot\right)  $ is said to be :

\textbf{normal} if X$\cdot$X*=X*$\cdot$X,

\textbf{self-adjoint} (or hermitian) if X=X*

\textbf{anti self-adjoint} (or antihermitian) if X=-X*

\textbf{unitary} if X$\cdot$X*=X*$\cdot$X=I
\end{definition}

All this terms are consistent with those used for matrices where * is the transpose-conjugation.

\begin{theorem}
A *-algebra is commutative iff each element is normal
\end{theorem}

If the *algebra A is over $%
\mathbb{C}
$\ then :

i) Any element X in A can be written : $X=Y+iZ$ with Y,Z self-adjoint :

$Y=\frac{1}{2}\left(  X+X^{\ast}\right)  ,Z=\frac{1}{2i}\left(  X-X^{\ast
}\right)  $

ii) The subset of self-adjoint elements in A is a real vector space, real form
of the vector space A.

\subsubsection{Lie Algebra}

There is a section dedicated to Lie algebras in the part Lie Groups.

\begin{definition}
A \textbf{Lie algebra} over a field K is a vector space A over K endowed with
a bilinear map called \textbf{bracket} :$\left[  {}\right]  :A\times
A\rightarrow A$

$\forall X,Y,Z\in A,\forall\lambda,\mu\in K:\left[  \lambda X+\mu Y,Z\right]
=\lambda\left[  X,Z\right]  +\mu\left[  Y,Z\right]  $

such that :

$\left[  X,Y\right]  =-\left[  Y,X\right]  $

$\left[  X,\left[  Y,Z\right]  \right]  +\left[  Y,\left[  Z,X\right]
\right]  +\left[  Z,\left[  X,Y\right]  \right]  =0$ (Jacobi identities)
\end{definition}

Notice that a Lie algebra is not an algebra, because the bracket is not
associative.\ But any algebra $\left(  A,\cdot\right)  $ becomes a Lie algebra
with the bracket : $\left[  X,Y\right]  =X\cdot Y-Y\cdot X.$ This is the case
for the linear endomorphisms over a vector space.

\subsubsection{Algebraic structures and categories}

If the sets E and F are endowed with the same algebraic structure a map
$f:E\rightarrow F$ is a \textbf{morphism} (also called homomorphism) if f
preserves the structure = the image of the result of any operation between
elements of E is the result of the same operation in F between the images of
the elements of E.

Groups : $\forall x,y\in E:f\left(  x\ast y\right)  =f\left(  x\right)  \cdot
f\left(  y\right)  $

Ring : $\forall x,y,z\in E:f\left(  \left(  x+y\right)  \ast z\right)
=f\left(  x\right)  \cdot f\left(  z\right)  +f\left(  y\right)  \cdot
f\left(  z\right)  $

Vector space : $\forall x,y\in E,\lambda,\mu\in K:f\left(  \lambda x+\mu
y\right)  =\lambda f\left(  x\right)  +\mu f\left(  y\right)  $

Algebra : $\forall x,y\in A,\lambda,\mu\in K::f\left(  x\ast y\right)
=f\left(  x\right)  \cdot f\left(  y\right)  ;f\left(  \lambda x+\mu y\right)
=\lambda f\left(  x\right)  +\mu f\left(  y\right)  $

Lie algebra : $\forall X,Y\in E:f\left(  \left[  X,Y\right]  _{E}\right)
=\left[  f\left(  X\right)  ,f\left(  Y\right)  \right]  _{F}$

If f is bijective then f is an isomorphism

If E=F then f is an endomorphism

If f is an endomorphism and an isomorphism it is an automorphism

All these concepts are consistent with the morphisms defined in the category theory.

There are many definitions of "homomorphisms", implemented for various
mathematical objects. When only algebraic properties are involved we will
stick to the universal and clear concept of morphism.

There are the categories of Groups, Rings, Fields, Vector Spaces, Algebras
over a field K.

\newpage

\section{GROUPS}

We see here mostly general definitions about groups, and an overview of the
finite groups.\ Topological groups and Lie groups are studied in a dedicated part.

\bigskip

\subsection{Definitions}

\label{Groups Definitions}

\begin{definition}
A \textbf{group} $\left(  G,\cdot\right)  $\ is a set endowed G with an
associative operation $\cdot$, for which there is an identity element and
every element has an inverse.
\end{definition}

In a group, the identity element is unique. The inverse of an element is unique.

\begin{definition}
A commutative (or \textbf{abelian}) group is a group whith a commutative operation
\end{definition}

\begin{definition}
A \textbf{subgroup} of the group $\left(  G,\cdot\right)  $\ is a subset A of
G which is also a group for $\cdot$
\end{definition}

So : $1_{G}\in A,\forall x,y\in A:x\cdot y\in A,x^{-1}\in A$

\subsubsection{Involution}

\begin{definition}
An \textbf{involution} on a group (G,$\cdot)$ is a map : $\ast:G\rightarrow G$
such that :

$\forall g,h\in G:\left(  g^{\ast}\right)  ^{\ast}=g;\left(  g\cdot h\right)
^{\ast}=h^{\ast}\cdot g^{\ast};\left(  1\right)  ^{\ast}=1$
\end{definition}

$\Rightarrow\left(  g^{-1}\right)  ^{\ast}=\left(  g^{\ast}\right)  ^{-1}$

A group endowed with an involution is said to be an involutive group.

Any group has the involution : $\left(  g\right)  ^{\ast}=g^{-1}$ but there
are others

Example : $\left(
\mathbb{C}
,\times\right)  $ with $\left(  z\right)  ^{\ast}=\overline{z}$

\subsubsection{Morphisms}

\begin{definition}
If $\left(  G,\cdot\right)  $ and $\left(  G^{\prime},\ast\right)  $ are
groups a \textbf{morphism} (or homomorphism) is a map : $f:G\rightarrow
G^{\prime}$ such that :

$\forall x,y\in G:f\left(  x\cdot y\right)  =f\left(  x\right)  \ast f\left(
y\right)  $ \ ; $f\left(  1_{G}\right)  =1_{G^{\prime}}$
\end{definition}

$\Rightarrow f\left(  x^{-1}\right)  =f\left(  x\right)  ^{-1}$

The set of such morphisms f is denoted $\hom\left(  G,G^{\prime}\right)  $

The \textbf{category of groups} has objects = groups and morphisms = homomorphisms.\ 

\begin{definition}
The \textbf{kernel} of a morphism $f\in\hom\left(  G,G^{\prime}\right)  $ is
the set :

$\ker f=\left\{  g\in G:f\left(  g\right)  =1_{G^{\prime}}\right\}  $
\end{definition}

\subsubsection{Translations}

\begin{definition}
The \textbf{left-translation} by $a\in\left(  G,\cdot\right)  $ is the map :

$L_{a}:G\rightarrow G::L_{a}x=a\cdot x$
\end{definition}

\begin{definition}
The \textbf{right-translation} by $a\in\left(  G,\cdot\right)  $ is the map :

$R_{a}:G\rightarrow G::R_{a}x=x\cdot a$
\end{definition}

\begin{definition}
The \textbf{conjugation} with respect to $a\in\left(  G,\cdot\right)  $ is the
map :

$Conj_{a}:G\rightarrow G::Conj_{a}x=a\cdot x\cdot a^{-1}$
\end{definition}

So : $L_{x}y=x\cdot y=R_{y}x.$ Translations are bijective maps.

$Conj_{a}x=L_{a}\circ R_{a^{-1}}(x)=R_{a^{-1}}\circ L_{a}(x)$

\begin{definition}
The \textbf{commutator} of two elements $x,y\in\left(  G,\cdot\right)  $ is :

$\left[  x,y\right]  =x^{-1}\cdot y^{-1}\cdot x\cdot y$ \ 
\end{definition}

It is 0 (or 1) for abelian groups.

\subsubsection{Centralizer}

\begin{definition}
The \textbf{normalizer} of a subset A of a group $\left(  G,\cdot\right)  $ is
the set :

$N_{A}=\{x\in G:Conj_{x}(A)=A\}$
\end{definition}

\begin{definition}
The \textbf{centralizer} $Z_{A}$\ of a subset A of a group $\left(
G,\cdot\right)  $ is the set $Z_{A}$\ of elements of G which commute with the
elements of A : $Z_{A}=\{x\in G:\forall a\in A:ax=xa\}$
\end{definition}

\begin{definition}
The \textbf{center} $Z_{G}$ of G is the centralizer of G
\end{definition}

$Z_{A}$ is a subgroup of G.

\subsubsection{Quotient sets}

Cosets are similar to ideals.

\begin{definition}
For a \textit{subgroup H} of a group $\left(  G,\cdot\right)  $ and $a\in G$

The \textbf{right coset} of a (with respect to H) is the set : $H\cdot
a=\left\{  h\cdot a,h\in H\right\}  $

The \textbf{left coset} of a (with respect to H) is the set : $a\cdot
H=\left\{  a\cdot h,h\in H\right\}  $
\end{definition}

The left and right cosets of H may or may not be equal.

\begin{definition}
A subgroup of a group $\left(  G,\cdot\right)  $ is a \textbf{normal subgroup}
if its right-coset is equal to its left coset
\end{definition}

Then for all g in G, gH = Hg, and $\forall x\in G:x\cdot H\cdot x^{-1}\in H.$

If G is abelian any subgroup is normal.

\begin{theorem}
The kernel of a morphism $f\in\hom\left(  G,G^{\prime}\right)  $ is a normal
subgroup.\ Conversely any normal subgroup is the kernel of some morphism.
\end{theorem}

\begin{definition}
A group $\left(  G,\cdot\right)  $\ is \textbf{simple} if the only normal
subgroups are 1 and G itself.
\end{definition}

\begin{theorem}
The left-cosets (resp.right-cosets) of any subgroup H form a partition of G
\end{theorem}

that is, the union of all left cosets is equal to G and two left cosets are
either equal or have an empty intersection.\ 

\bigskip

So a subgroup defines an equivalence relation :

\begin{definition}
The \textbf{quotient set} $G/H$ of a subgroup H of a group $\left(
G,\cdot\right)  $ is the set $G/\sim$ of classes of equivalence $:$ $x\sim
y\Leftrightarrow\exists h\in H:x=y\cdot h$
\end{definition}

\begin{definition}
The \textbf{quotient set} $H\backslash G$ of a subgroup H of a group $\left(
G,\cdot\right)  $ is the set $G/\sim$ of classes of equivalence $:$ $x\sim
y\Leftrightarrow\exists h\in H:x=h\cdot y$
\end{definition}

\bigskip

It is useful to characterize these quotient sets.

The projections give the classes of equivalences denoted $\left[  x\right]  $ :

$\pi_{L}:G\rightarrow G/H:\pi_{L}\left(  x\right)  =\left[  x\right]
_{L}=\left\{  y\in G:\exists h\in H:x=y\cdot h\right\}  =x\cdot H$

$\pi_{R}:G\rightarrow H\backslash G:\pi_{R}\left(  x\right)  =\left[
x\right]  _{R}=\left\{  y\in G:\exists h\in H:x=h\cdot y\right\}  =H\cdot x$

Then :

$x\in H\Rightarrow\pi_{L}\left(  x\right)  =\pi_{R}\left(  x\right)  =\left[
x\right]  =1$

Because the classes of equivalence define a partition of G, by the Zorn lemna
one can pick one element in each class. So we have two families :

For $G/H:\left(  \lambda_{i}\right)  _{i\in I}:\lambda_{i}\in G:\left[
\lambda_{i}\right]  _{L}=\lambda_{i}\cdot H,$

$\forall i,j:\left[  \lambda_{i}\right]  _{L}\cap\left[  \lambda_{j}\right]
_{L}=\varnothing,\cup_{i\in I}\left[  \lambda_{i}\right]  _{L}=G$

For $H\backslash G:\left(  \rho_{j}\right)  _{j\in J}:\rho_{j}\in G:\left[
\rho_{j}\right]  _{R}=H\cdot\rho_{j}$

$\forall i,j:\left[  \rho_{i}\right]  _{R}\cap\left[  \rho_{j}\right]
_{R}=\varnothing,\cup_{j\in J}\left[  \rho_{j}\right]  _{R}=G$

Define the maps :

$\phi_{L}:G\rightarrow\left(  \lambda_{i}\right)  _{i\in I}:\phi_{L}\left(
x\right)  =\lambda_{i}::\pi_{L}\left(  x\right)  =\left[  \lambda_{i}\right]
_{L}$

$\phi_{R}:G\rightarrow\left(  \rho_{j}\right)  _{j\in J}:\phi_{R}\left(
x\right)  =\rho_{j}::\pi_{R}\left(  x\right)  =\left[  \rho_{j}\right]  _{R}$

Then any $x\in G$ can be written as :$x=\phi_{L}\left(  x\right)  \cdot h$ or
$x=h^{\prime}\cdot\phi_{R}\left(  x\right)  $ for unique h,h'$\in H$

\bigskip

\begin{theorem}
G/H=H%
$\backslash$%
G iff H is a normal subgroup.\ If so then G/H=H%
$\backslash$%
G is a group and the sequence 1 $\rightarrow$ H $\rightarrow$ G $\rightarrow$
G/H$\rightarrow$ 1 is exact (in the category of groups, with 1=trivial group
with only one element). The projection $G\rightarrow G/H$ is a morphism with
kernel H.
\end{theorem}

\bigskip

There is a similar relation of equivalence with conjugation:

\begin{theorem}
The relation : $x\sim y\Leftrightarrow x=y\cdot x\cdot y^{-1}\Leftrightarrow
x\cdot y=y\cdot x$ is an equivalence relation over $\left(  G,\cdot\right)  $
which defines a partition of G : $G=\cup_{p\in P}G_{p},p\neq q:G_{p}\cap
G_{q}=\varnothing$ . Each subset $G_{p}$\ of G is a \textbf{conjugation
class}. If G is commutative there is only one subset, G itself
\end{theorem}

as any element commutes with ist powers $x^{n}$\ the conjugation class of $x$
contains at least its powers, including the unity element.

\subsubsection{Product of groups}

\paragraph{Direct product of groups\newline}

\begin{definition}
The \textbf{direct product} of two groups $\left(  G,\times\right)  ,\left(
H,\cdot\right)  $ is the set which is the cartesian product $G\times H$
endowed with the operations :

$\left(  g_{1},h_{1}\right)  \circ\left(  g_{2},h_{2}\right)  =\left(
g_{1}\times g_{2},h_{1}\cdot h_{2}\right)  $

unity : $\left(  1_{G},1_{H}\right)  $

$\left(  g,h\right)  ^{-1}=\left(  g^{-1},h^{-1}\right)  $
\end{definition}

So the direct product deals with \textit{ordered pairs} of elements of G,H

If both groups are abelian then the direct product is abelian and usually the
direct product is called the direct sum $G+H.$

Many groups can be considered as the product of smaller groups.\ The process
is the following. Take two groups G,H and define ; the direct product
$P=G\times H$ and the two subgroups of P : $G^{\prime}=\left(  G,1_{H}\right)
,H^{\prime}=\left(  1_{G},H\right)  .$ G',H' have the properties :

i) G' is isomorphic to G, H' is isomorphic to H

ii) $G^{\prime}\cap H^{\prime}=\left(  1_{G},1_{H}\right)  $

iii) Any element of P can be written as the product of an element of G' and an
element of H'

iv) Any element of G' commutes with any element of H', or equivalently G',H'
are normal subgroups of P

If any group P has two subgroups G,H with these properties, it can be
decomposed in the direct product $G\times H$

If a group is simple its only normal subgroups are trivial, thus it cannot be
decomposed in the direct product of two other groups. Simple groups are the
basic bricks from which other groups can be built.

\paragraph{Semi-direct product of groups\newline}

The semi-direct product of groups is quite different, in that it requires an
additional map to define the operations.

\begin{definition}
If G,T are groups, and $f:G\times T\rightarrow T$ is such that:

i) for every $g\in G$ the map $f\left(  g,.\right)  $ is a group automorphism
on T:

$\forall g\in G,t,t^{\prime}\in T:$

$f\left(  g,t\cdot t^{\prime}\right)  =f\left(  g,t\right)  \cdot f\left(
g,t^{\prime}\right)  ,f\left(  g,1_{T}\right)  =1_{T},f\left(  g,t\right)
^{-1}=f\left(  g,t^{-1}\right)  $

ii) $\forall g,g^{\prime}\in G,t\in T:f\left(  gg^{\prime},t\right)  =f\left(
g,f(g^{\prime},t)\right)  ,f\left(  1_{G},t\right)  =t$

then the set $G\times T$ endowed with the operations :

$\left(  g,t\right)  \times\left(  g^{\prime},t^{\prime}\right)  =\left(
gg^{\prime},f\left(  g,t^{\prime}\right)  \cdot t\right)  $

$\left(  g,t\right)  ^{-1}=\left(  g^{-1},f\left(  g^{-1},t^{-1}\right)
\right)  $

is a group, called the semi-direct product of G and T denoted $G\ltimes_{f}T $
\end{definition}

We can check that the product is associative :

$\left(  \left(  g,t\right)  \times\left(  g^{\prime},t^{\prime}\right)
\right)  \times\left(  g",t"\right)  =\left(  gg^{\prime},f\left(
g,t^{\prime}\right)  \cdot t\right)  \times\left(  g",t"\right)  $

$=\left(  gg^{\prime}g",f\left(  gg^{\prime},t"\right)  \cdot\left(  f\left(
g,t^{\prime}\right)  \cdot t\right)  \right)  $

$\left(  g,t\right)  \times\left(  \left(  g^{\prime},t^{\prime}\right)
\times\left(  g",t"\right)  \right)  =\left(  g,t\right)  \times\left(
g^{\prime}g",f\left(  g^{\prime},t"\right)  \cdot t^{\prime}\right)  $

$=\left(  gg^{\prime}g",f\left(  g,f\left(  g^{\prime},t"\right)  \cdot
t^{\prime}\right)  \cdot t\right)  $

$f\left(  gg^{\prime},t"\right)  \cdot\left(  f\left(  g,t^{\prime}\right)
\cdot t\right)  =f\left(  gg^{\prime},t"\right)  \cdot f\left(  g,t^{\prime
}\right)  \cdot t=f\left(  g,f\left(  g^{\prime},t"\right)  \right)  \cdot
f\left(  g,t^{\prime}\right)  \cdot t=f\left(  g,f\left(  g^{\prime
},t"\right)  \cdot t^{\prime}\right)  \cdot t=f\left(  g,f\left(  g^{\prime
},t"\right)  \cdot t^{\prime}\right)  \cdot t$

and the inverse :

$\left(  g,t\right)  ^{-1}\times\left(  g,t\right)  =\left(  g^{-1},f\left(
g^{-1},t^{-1}\right)  \right)  \times\left(  g,t\right)  =\left(
g^{-1}g,f\left(  g^{-1},t\right)  \cdot f\left(  g^{-1},t^{-1}\right)
\right)  =\left(  g^{-1}g,f\left(  g^{-1},t\cdot t^{-1}\right)  \right)
=\left(  1_{G},1_{T}\right)  $

$\left(  g,t\right)  \times\left(  g,t\right)  ^{-1}=\left(  g,t\right)
\times\left(  g^{-1},f\left(  g^{-1},t^{-1}\right)  \right)  =\left(
gg^{-1},f\left(  g,f\left(  g^{-1},t^{-1}\right)  \right)  \cdot t\right)
=\left(  1_{G},f\left(  gg^{-1},t^{-1}\right)  \cdot t\right)  =\left(
1_{G},f\left(  1_{G},t^{-1}\right)  \cdot t\right)  =\left(  1_{G},t^{-1}\cdot
t\right)  $

\subparagraph{Example : Group of displacements\newline}

If G is a rotation group and T a translation group on a vector space F we have
the group of displacements.\ T is abelian so :

$\left(  R,T\right)  \times\left(  R^{\prime},T^{\prime}\right)  =\left(
R\circ R^{\prime},R\left(  T^{\prime}\right)  +T\right)  $

$\left(  R,T\right)  ^{-1}=\left(  R^{-1},-R^{-1}\left(  T\right)  \right)  $

with the natural action $\tau:G\rightarrow GL\left(  F;F\right)  $ of G on
vectors of F$\equiv T$.

\subsubsection{Generators}

\begin{definition}
A set of \textbf{generators} of a group $\left(  G,\cdot\right)  $ is a family
$\left(  x_{i}\right)  _{i\in I}$\ of elements of G indexed on an ordered set
I such that any element of G can be written uniquely as the product of a
finite ordered subfamily J of $\left(  x_{i}\right)  _{i\in I}$
\end{definition}

$\forall g\in G,\exists J=\left\{  j_{1},...j_{n},..\right\}  \subset
I,:g=x_{j_{1}}\cdot x_{j_{2}}...\cdot x_{j_{n}}..$

The \textbf{rank} of a group is the cardinality of the smallest set of its
generators (if any).

\subsubsection{Action of a group}

Maps involving a group and a set can have special properties, which deserve
definitions because they are frequently used.

\begin{definition}
A \textbf{left-action} of a group $\left(  G,\cdot\right)  $ on a set E is a
map : $\lambda:G\times E\rightarrow E$ such that :

$\forall x\in E,\forall g,g^{\prime}\in G:\lambda\left(  g,\lambda\left(
g^{\prime},x\right)  \right)  =\lambda\left(  g\cdot g^{\prime},x\right)
;\lambda\left(  1,x\right)  =x$
\end{definition}

\begin{definition}
A \textbf{right-action} of a group $\left(  G,\cdot\right)  $ on a set E is a
map : $\rho:E\times G\rightarrow E$ such that :

$\forall x\in E,\forall g,g^{\prime}\in G:\rho\left(  \rho\left(  x,g^{\prime
}),g\right)  \right)  =\rho\left(  x,g^{\prime}\cdot g\right)  ;\rho\left(
x,1\right)  =x$
\end{definition}

Notice that left, right is related to the place of g.

Any subgroup H of G defines left and right actions by restriction of the map
to H.

Any subgroup H of G defines left and right actions on G itself in the obvious way.

All the following definitions are easily adjusted for a right action.

\begin{definition}
The \textbf{orbit }of the action through a$\in G$ of the left-action $\lambda$
of a group $\left(  G,\cdot\right)  $ on a set E is the subset of E denoted
G$\left(  a\right)  =$ $\left\{  \lambda\left(  g,a\right)  ,g\in G\right\}  $
\end{definition}

The relation $y\in G\left(  x\right)  $ is an equivalence relation between
x,y. The classes of equivalence form a partition of G called the
\textbf{orbits} of the action (an orbit = the subset of elements of E which
can be deduced from each other by the action).

The orbits of the left action of a subgroup H on G are the right cosets
defined above.

\begin{definition}
A left-action of a group $\left(  G,\cdot\right)  $ on a set E is

\textbf{transitive} if : $\forall x,y\in E,\exists g\in G:y=\lambda\left(
g,x\right)  $

\textbf{free} if : $\lambda\left(  g,x\right)  =x\Rightarrow g=1$

\textbf{effective} if : $\forall x:\lambda(g,x)=\lambda(h,x)=>g=h$
\end{definition}

\begin{theorem}
For any action : effective $\Leftrightarrow$ free
\end{theorem}

\begin{proof}
free $\Rightarrow$ effective : $\lambda(g,x)=\lambda(h,x)=>\lambda
(g^{-1},\lambda\left(  h,x\right)  )=\lambda(g^{-1}h,x)=\lambda(g^{-1}%
,\lambda\left(  g,x\right)  )=\lambda(1,x)=x\Rightarrow g^{-1}h=1$

effective $\Rightarrow$ free : $\lambda(g,x)=x=\lambda(gg^{-1},x)=g=gg^{-1}$
\end{proof}

\begin{definition}
A subset F of E is \textbf{invariant} by the left-action $\lambda$ of a group
$\left(  G,\cdot\right)  $ on E if : $\forall x\in F,\forall g\in
G:\lambda\left(  g,x\right)  \in F.$
\end{definition}

F is invariant iff it is the union of a collection of orbits. The minimal non
empty invariant sets are the orbits.

\begin{definition}
The \textbf{stabilizer} of an element $a\in E$ with respect to the left-action
$\lambda$ of a group $\left(  G,\cdot\right)  $ on E is the subset of G :
A$\left(  a\right)  =\left\{  g\in G:\lambda\left(  g,a\right)  =a\right\}  $
\end{definition}

It is a subgroup of G also called the \textbf{isotropy subgroup} (with respect
to a). If the action is free the map : $A:E\rightarrow G$ is bijective.

\begin{definition}
Two set E,F are \textbf{equivariant} under the left actions $\lambda
_{1}:G\times E\rightarrow E,\lambda_{2}:G\times F\rightarrow F$ of a group
$\left(  G,\cdot\right)  $ if there is a map : $f:E\rightarrow F$ such that :
$\forall x\in E,\forall g\in G:f\left(  \lambda_{1}\left(  g,x\right)
\right)  =\lambda_{2}\left(  g,f\left(  x\right)  \right)  $
\end{definition}

So if E=F the set is equivariant under the action if :

$\forall x\in E,\forall g\in G:f\left(  \lambda\left(  g,x\right)  \right)
=\lambda\left(  g,f\left(  x\right)  \right)  $

\bigskip

\subsection{Finite groups}

\label{FINITE GROUPS}

A finite group is a group which has a finite number of elements. So, for a
finite group, one can dress the multiplication table, and one can guess that
there a not too many ways to build such a table : mathematicians have strive
for years to establish a classification of finite groups.

\subsubsection{Classification of finite groups}

1. Order:

\begin{definition}
The \textbf{order} of a finite group is the number of its elements. The order
\textit{of an element} \textbf{a} of a finite group is the smallest positive
integer number k with $\mathbf{a}^{k}=1$, where 1 is the identity element of
the group.
\end{definition}

\begin{theorem}
(Lagrange's theorem) The order of a subgroup of a finite group G divides the
order of G. The order of an element \textbf{a} of a finite group divides the
order of that group.
\end{theorem}

\begin{theorem}
If n is the square of a prime, then there are exactly two possible (up to
isomorphism) types of group of order n, both of which are abelian.
\end{theorem}

2. Cyclic groups :

\begin{definition}
A group is \textbf{cyclic} if it is generated by an element : $G=\left\{
a^{p},p\in%
\mathbb{N}
\right\}  .$
\end{definition}

A cyclic group always has at most countably many elements and is commutative.
For every positive integer n there is exactly one cyclic group (up to
isomorphism) whose order is n, and there is exactly one infinite cyclic group
(the integers under addition). Hence, the cyclic groups are the simplest
groups and they are completely classified. They are usually denoted $%
\mathbb{Z}
/p%
\mathbb{Z}
:$ the algebraic numbers multiple of p with addition.

\begin{theorem}
Any finite abelian group can be decomposed as the direct sum (=product) of
cyclic groups
\end{theorem}

3. Simple finite groups :

Simple groups are the sets from which other groups can be built. All
\textit{simple} finite groups have been classified. Up to isomorphisms there
are 4 classes :

- the cyclic groups with prime order : any group of prime order is cyclic and simple.

- the alternating groups of degree at least 5

- the simple Lie groups

- the 26 sporadic simple groups.

\subsubsection{Symmetric groups}

\paragraph{Definitions\newline}

\begin{definition}
A \textbf{permutation} of a finite set E is a bijective map : $p:E\rightarrow
E.$
\end{definition}

With the composition law the set of permutations of E is a group. As all sets
with the same cardinality are in bijection, their group of permutations are
isomorphics. Therefore it is convenient, for the purpose of the study of
permutations, to consider the set (1,2,....n) of integers.

\begin{notation}
$\mathfrak{S}\left(  n\right)  $ is the group of permutation of a set of n
elements, called the \textbf{symmetric group }of order n
\end{notation}

An element s of $\mathfrak{S}\left(  n\right)  $ can be represented as a table
with 2 rows : the first row comprises the integers 1,2..n, the second row
takes the elements s(1),s(2),...s(n).

$\mathfrak{S}\left(  n\right)  $ is a finite group with n! elements. Its
subgroups are permutations groups. It is abelian iff n%
$<$%
2.

Remark : with regard to permutation, two elements of E are always considered
as distinct, even if it happens that, for other reasons, they are identical.
For instance take the set $\left\{  1,1,2,3\right\}  $ with cardinality
4.\ The two first elements are considered as distinct : indeed in abstract set
theory nothing can tell us that two elements are not distinct, so we have 4
objects $\left\{  a,b,c,d\right\}  $ that are numbered as $\left\{
1,2,3,4\right\}  $

\begin{definition}
A \textbf{transposition} is a permutation which exchanges two elements and
keep inchanged all the others.
\end{definition}

A transposition can be written as a couple (a,b) of the two numbers which are transposed.

Any permutation can be written as the composition of transpositions. This
decomposition is not unique, but the parity of the number p of transpositions
necessary to write a given permutation does not depend of the decomposition.
The \textbf{signature} of a permutation is the number $\left(  -1\right)
^{p}=\pm1.$ A permutation is even if its signature is +1, odd if its signature
is -1. The product of two even permutations is even, the product of two odd
permutations is even, and all other products are odd.

The set of all even permutations is called the \textbf{alternating group}
$A_{n}$ (also denoted \ $\mathfrak{A}_{n}$). It is a normal subgroup of
$\mathfrak{S}\left(  n\right)  $, and for $n\geq2$ it has $n!/2$ elements. The
group $\mathfrak{S}\left(  n\right)  $ is the semidirect product of $A_{n}$
and any subgroup generated by a single transposition.

\paragraph{Young diagrams\newline}

For any partition of (1,2,...n) in p subsets, the permutations of
$\mathfrak{S}\left(  n\right)  $\ which preserve globally each of the subset
of the partition constitute a \textbf{class of conjugation}.

Example : the 3 permutations $\left(  1,2,3,4,5\right)  ,\left(
2,1,4,3,5\right)  ,$\ $\left(  1,2,5,3,4\right)  ,$ preserve the subsets
$\left(  1,2\right)  ,\left(  3,4,5\right)  $ and belong to the same class of conjugation.

A class of conjugation, denoted $\lambda,$ is defined by :

- p integers $\lambda_{1}\leq\lambda_{2}...\leq\lambda_{p}$ such that
$\sum_{i=1}^{p}\lambda_{i}=n$

- a partition of (1,2,...n) in p subsets $\left(  i_{1},..i_{\lambda_{k}%
}\right)  $ containing each $\lambda_{k}$ elements taken in (1,2,..n).

The number S(n,p) of different partitions of n in p subsets is a function of n
(the Stirling number of second kind) which is tabulated .

The tool to build classes of conjugation is a Young diagram. A \textbf{Young
diagram} is a table with p rows i=1,2,...p of $\lambda_{k}$ cells each, placed
below each other, left centered. Any permutation of $\mathfrak{S}\left(
n\right)  $\ obtained by filling such a table with distinct numbers 1,2,...n
is called a \textbf{Young tableau}. The standard (or canonical) tableau is
obtained in the natural manner by filling the cells from the left to the right
in each row, and next to the row below with the ordered numbers 1,2,...n.

Given a Young tableau, two permutations belong to the same class of
conjugation if they have the same elements in each row (but not necessarily in
the same cells).

A Young diagram has also q columns of decreasing sizes $\mu_{j},j=1...q$ with :

$\sum_{i=1}^{p}\lambda_{i}=\sum_{j=1}^{q}\mu_{j}=n;n\geq\mu_{j}\geq\mu
_{j+1}\geq1$

If a diagram is read columns by columns one gets another diagram, called its
\textbf{conjugate}$.$

\subsubsection{Symmetric polynomials}

\begin{definition}
A map of n variables over a set E : $f:E^{n}\rightarrow F$ is
\textbf{symmetric} in its variables if it is invariant for any permutation of
the n variables :%

\begin{equation}
\forall\sigma\in\mathfrak{S}\left(  n\right)  ,f\left(  x_{\sigma\left(
1\right)  },...,x_{\sigma\left(  n\right)  }\right)  =f\left(  x_{1}%
,...,x_{n}\right)
\end{equation}

\end{definition}

The set $S_{d}\left[  X_{1},...X_{n}\right]  $ of symmetric polynomials of n
variables and degree d\ has the structure of a finite dimensional vector
space. These polynomials must be homogeneous :

$P\left(  x_{1},..x_{n}\right)  =\sum a_{i_{1}...i_{p}}x_{1}^{i_{1}}%
...x_{n}^{i_{n}},\sum_{j=1}^{n}i_{j}=d,a_{i_{1}...i_{p}}\in F,X_{i}\in F$

The set $S\left[  X_{1},...X_{n}\right]  $ of symmetric polynomials of n
variables and any degree\ has the structure of a graded commutative algebra
with the multiplication of functions.

A basis of the vector space $S\left[  X_{1},...X_{n}\right]  $ is a set of
symmetric polynomials of n variables. Their elements can be labelled by a
partition $\lambda$ of d : $\lambda=\left(  \lambda_{1}\geq\lambda_{2}%
...\geq\lambda_{n}\geq0\right)  ,\sum_{j=1}^{n}\lambda_{j}=d.$ The most usual
bases are the following.

\subparagraph{1. Monomials :\newline}

the basic monomial is $x_{1}^{\lambda_{1}}\cdot x_{2}^{\lambda_{2}}...\cdot
x_{n}^{\lambda_{n}}.$ The symmetric polynomial\ of degree d associated to the
partition $\lambda$\ is :

$H_{\lambda}=\sum_{\sigma\in\mathfrak{S}\left(  n\right)  }x_{\sigma\left(
1\right)  }^{\lambda_{1}}\cdot x_{\sigma\left(  2\right)  }^{\lambda_{2}%
}...\cdot x_{\sigma\left(  n\right)  }^{\lambda_{n}}$

and a basis of S$_{d}\left[  X_{1},...X_{n}\right]  $ is a set of $H_{\lambda}
$\ for each partition $\lambda.$

\subparagraph{2. Elementary symmetric polynomials :\newline}

the p elementary symmetric polynomial is : $E_{p}=\sum_{\left\{
i_{1},...i_{p}\right\}  }x_{i_{1}}\cdot x_{i_{2}}..\cdot x_{i_{p}}$ where the
sum is for all ordered combinations of p indices taken in (1,2,...n): $1\leq
i_{1}<i_{2}..<i_{n}\leq n$. It is a symmetric polynomial of degree p. The
product of two such polynomials $E_{p}\cdot E_{q}$ is still a symmetric
polynomial of degree p+q. So any partition $\lambda$ defines a polynomial :
$H_{\lambda}=%
{\textstyle\prod\limits_{\lambda}}
E_{\lambda_{1}}...E_{\lambda_{q}}\in S_{d}\left[  x_{1},...x_{n}\right]  $ and
a basis is a set of $H_{\lambda}$\ for all partitions $\lambda.$ There is the
identity : $%
{\displaystyle\prod\limits_{n}^{i=1}}
\left(  1+x_{i}t\right)  =\sum_{j=0}^{\infty}E_{j}t^{j}$

\subparagraph{3. Schur polynomials :\newline the Schur polynomial for a
partition $\lambda$ is :}

$S_{\lambda}=\det\left[  x_{j}^{\lambda_{i}+n-i}\right]  _{n\times n}/\Delta$
where : $\Delta=%
{\displaystyle\prod\limits_{i<j}}
\left(  x_{i}-x_{j}\right)  $ (called the discriminant)

$\det\left[  \frac{1}{1-x_{i}y_{j}}\right]  =\left(
{\displaystyle\prod\limits_{i<j}}
\left(  x_{i}-x_{j}\right)  \right)  \left(
{\displaystyle\prod\limits_{i<j}}
\left(  y_{i}-y_{j}\right)  \right)  /%
{\displaystyle\prod\limits_{i,j}}
\left(  1-x_{i}y_{j}\right)  $

\subsubsection{Combinatorics}

Combinatorics is the study of finite structures, and involves counting the
number of such structures. We will just recall basic results in enumerative
combinatorics and signatures.

\paragraph{Enumerative combinatorics\newline}

Enumerative combinatorics deals with problems such as "how many ways to select
n objects among x ? or how many ways to group x objects in n packets ?..."

1. Many enumerative problems can be modelled as following :

Find the number of maps :$f:N\rightarrow X$ where N is a set with n elements,
X a set with x elements and meeting one of the 3 conditions : f injective, f
surjective, or no condition. Moreover any two maps f,f' :

i) are always distinct (no condition)

or are deemed equivalent (counted only once) if

ii) \ Up to a permutation of X : $f\sim f^{\prime}:\exists s_{X}%
\in\mathfrak{S}\left(  x\right)  :f^{\prime}\left(  N\right)  =s_{X}f\left(
N\right)  $

iii) Up to a permutation of N : $f\sim f^{\prime}:\exists s_{N}\in
\mathfrak{S}\left(  n\right)  :f^{\prime}\left(  N\right)  =f\left(
s_{N}N\right)  $

iv) Up to permutations of N and X : $f\sim f^{\prime}:\exists s\in
\mathfrak{S}\left(  x\right)  ,s_{N}\in\mathfrak{S}\left(  n\right)
:f^{\prime}\left(  N\right)  =s_{X}f\left(  s_{N}N\right)  $

These conditions can be paired in 12 ways.

2. Injective maps from N to X:

i) No condition : this is the number of sequences of n distinct elements of X
without repetitions. The formula is : $\frac{x!}{\left(  n-x\right)  !}$

ii) Up to a permutation of X : 1 si $n\leq x$ ,0 if n
$>$
x

iii) Up to a permutation of N : this is the number of subsets of n elements of
X, the \textbf{binomial coefficient} : $C_{x}^{n}=\frac{x!}{n!\left(
x-n\right)  !}=\binom{x}{n}$. If n%
$>$%
x the result is 0.

iv) Up to permutations of N and X : 1 si $n\leq x$ 0 if n%
$>$%
x

3. Surjective maps f from N to X:

i) No condition : the result is $x!S\left(  n,x\right)  $ where S(n,x), called
the Stirling number of the second kind, is the number of ways to partition a
set of n elements in k subsets (no simple formula).

ii) Up to a permutation of X : the result is the Stirling number of the second
kind S(n,x).

iii) Up to a permutation of N: the result is : $C_{x-1}^{n-1}$

iv) Up to permutations of N and X : this is the the number $p_{x}\left(
n\right)  $ of partitions of n in x non zero integers : $\lambda_{1}%
\geq\lambda_{2}...\geq\lambda_{x}>0:\lambda_{1}+\lambda_{2}+..\lambda_{x}=n $

4. No restriction on f :

i) No condition : the result is $x^{n}$

ii) Up to a permutation of X : the result is $\sum_{k=0}^{x}S(n,k)$ where
S(n,k) is the Stirling number of second kind

iii) Up to a permutation of N : the result is : $C_{x+n-1}^{n}=\binom
{x+n-1}{x}$

iv) Up to permutations of N and X : the result is : $p_{x}\left(  n+x\right)
$ where $p_{k}\left(  n\right)  $ is the number of partitions of n in k
integers : $\lambda_{1}\geq\lambda_{2}...\geq\lambda_{k}:\lambda_{1}%
+\lambda_{2}+..\lambda_{k}=n$

5. The number of distributions of n (distinguishable) elements over r
(distinguishable) containers, each containing exactly $k_{i}$ elements, is
given by the \textbf{multinomial coefficients} :

$\binom{n}{k_{1}k_{2}..k_{r}}=\frac{n!}{k_{1}!k_{2}!..k_{r}!}$

They are the coefficients of the polynomial : $\left(  x_{1}+x_{2}%
+...+x_{r}\right)  ^{n}$

6. Stirling's approximation of n! : $n!\approx\sqrt{2\pi n}\left(  \frac{n}%
{e}\right)  ^{n}$

The gamma function : $\Gamma\left(  z\right)  =\int_{0}^{\infty}t^{z-1}%
e^{-t}dt:$ $n!=\Gamma\left(  n+1\right)  $

\paragraph{Signatures\newline}

To compute the signature of any permutation the basic rule is that the parity
of any permutation of integers $\left(  a_{1},a_{2},...,a_{p}\right)
$\ (consecutive or not) is equal to the number of inversions in the
permutation = the number of times that a given number $a_{i}$ comes before
another number $a_{i+r}$ which is smaller than $a_{i}:a_{i+r}<a_{i}$

Example : $\left(  3,5,1,8\right)  $

take 3 :
$>$
1 $\rightarrow+1$

take 5 :
$>$
1 $\rightarrow+1$

take 1 : $\rightarrow0$

take 8 : $\rightarrow0$

take the sum : 1+1=2 $\rightarrow$ signature $\left(  -1\right)  ^{2}=1$

It is most useful to define the function :

\begin{notation}
$\epsilon$ is the function at n variables : $\epsilon:I^{n}\rightarrow\left\{
-1,0,1\right\}  $ where I is a set of n integers, defined by :

$\epsilon\left(  i_{1},...,i_{n}\right)  =0$ if there are two indices which
are identical : $i_{k},i_{l}$ ,$k\neq l$ such that : $i_{k}=i_{l}$

$\epsilon\left(  i_{1},...,i_{n}\right)  =$ the signature of the permutation
of the integers $\left(  i_{1},...,i_{n}\right)  $\ if they are all distinct

$\epsilon\left(  \mathfrak{\sigma}\right)  $ where $\sigma\in\mathfrak{S}%
\left(  n\right)  $ is the signature of the permutation $\sigma$
\end{notation}

So $\epsilon\left(  3,5,1,8\right)  =1;\epsilon\left(  3,5,5,8\right)  =0$

Basic formulas :

reverse order : $\epsilon\left(  a_{p},a_{p-1},...,a_{1}\right)
=\epsilon\left(  a_{1},a_{2},...,a_{p}\right)  \left(  -1\right)
^{\frac{p\left(  p-1\right)  }{2}}$

inversion of two numbers : $\epsilon\left(  a_{1},a_{2},.a_{j}...a_{i}%
..,a_{p}\right)  =\epsilon\left(  a_{1},a_{2},.a_{i}...a_{j}..,a_{p}\right)
\epsilon\left(  a_{i},a_{j}\right)  $

inversion of one number : $\epsilon\left(  i,1,2,3,..i-1,i+1,...p\right)
=\left(  -1\right)  ^{i-1}$

\newpage

\section{VECTOR\ SPACES}

\bigskip

\subsection{Vector spaces}

\label{Vect.SP.Definitions}

\subsubsection{Vector space}

\begin{definition}
A \textbf{vector space E over a field K} is a set with two operations :
addition denoted + for which it is an abelian group, and multiplication by a
scalar (an element of K) : $K\times E\rightarrow E$ which is distributive over addition.
\end{definition}

So : $\forall x,y\in E,\lambda,\mu\in K:$

$\lambda x+\mu y\in E,$

$\lambda\left(  x+y\right)  =\left(  x+y\right)  \lambda=\lambda x+\lambda y$

Elements of a vector space are called \textbf{vectors}. When necessary (and
only when necessary) vectors will be denoted with an upper arrow :
$\overrightarrow{u}$

Warning ! a vector space structure is defined with respect to a given field
(see below for real and complex vector spaces)

\begin{notation}
(E,K) is a set E with the structure of a vector space over a field K
\end{notation}

\subsubsection{Basis}

\begin{definition}
A family of vectors $\left(  v_{i}\right)  _{i\in I}$ of a vector space over a
field K, indexed on a \textit{finite} set I, is \textbf{linearly independant}
if :

$\forall\left(  x_{i}\right)  _{i\in I},x_{i}\in K:\sum_{i\in I}x_{i}%
v_{i}=0\Rightarrow x_{i}=0$
\end{definition}

\begin{definition}
A family of vectors $\left(  v_{i}\right)  _{i\in I}$ of a vector space,
indexed on a set I (finite of infinite) is \textbf{free} if any finite
subfamily is linearly independant.
\end{definition}

\begin{definition}
A \textbf{basis} of a vector space E is a free family of vectors which
generates E.
\end{definition}

Thus for a basis $\left(  e_{i}\right)  _{i\in I}:$

$\forall v\in E,\exists J\subset I,\#J<\infty,\exists\left(  x_{i}\right)
_{i\in J}\in K^{J}:v=\sum_{i\in J}x_{i}e_{i}$

\bigskip

Warning! These complications are needed because without topology there is no
clear definition of the infinite sum of vectors.\ This implies that for any
vector \textit{at most a finite number of components are non zero} (but there
can be an infinite number of vectors in the basis).\ So usually "Hilbertian
bases" are not bases in this general meaning, because vectors can have
infinitely many non zero components.

The method to define a basis is a common trick in algebra. To define some
property on a family indexed on an infinite set $I$, without any tool to
compute operations on an infinite number of arguments, one says that the
property is valid on $I$ if it is valid on all the finite subsets J of $I$. In
analysis there is another way, by using the limit of a sequence and thus the
sum of an infinite number of arguments.

\begin{theorem}
Any vector space has a basis
\end{theorem}

(this theorem requires the axiom of choice).

\begin{theorem}
The set of indices of bases of a vector space have all the same cardinality,
which is the \textbf{dimension} of the vector space.
\end{theorem}

If K is a field, the set $K^{n}$ is a vector space of dimension n, and its
canonical basis are the vectors $\varepsilon_{i}=\left(
0,0,..0,1,0,...0\right)  $.

\subsubsection{Vector subspaces}

\begin{definition}
A \textbf{vector subspace} of a vector space (E,K) is a subset F of E such
that the operations in E are algebraically closed in F :
\end{definition}

$\forall u,v\in F,\forall k,k^{\prime}\in K:ku+k^{\prime}u^{\prime}\in F$

the operations (+,x) being the operations as defined in E.

\paragraph{Linear span\newline}

\begin{definition}
The \textbf{linear span} of the subset S of a vector space E is the
intersection of all the vector subspaces of E which contains S.
\end{definition}

\begin{notation}
Span(S) is the linear span of the subset S of a vector space
\end{notation}

Span(S) is a vector subspace of E, which contains any \textit{finite} linear
combination of vectors of S.

\paragraph{Direct sum\newline}

\begin{definition}
The sum of a family $\left(  E_{i}\right)  _{i\in I}$\ of \textit{vector
subspaces} of E is the linear span of $\left(  E_{i}\right)  _{i\in I}$
\end{definition}

So any vector of the sum is the sum of at most a finite number of vectors of
some of the $E_{i}$

\begin{definition}
The sum of a family $\left(  E_{i}\right)  _{i\in I}$\ of \textit{vector
subspaces} of E is \textbf{direct }and denoted\textbf{\ }$\oplus_{i\in I}%
E_{i}$ if for any finite subfamily J of I :

$\sum_{i\in J}v_{i}=\sum_{i\in J}w_{i},v_{i},w_{i}\in E_{i}\Rightarrow
v_{i}=w_{i}$
\end{definition}

The sum is direct iff the $E_{i}$ have no common vector but 0 :

$\forall j\in I,E_{j}\cap\left(  \sum_{i\in I-j}E_{i}\right)  =\overrightarrow
{0}$

Or equivalently the sum is direct iff the decomposition over each $E_{i}$ is unique:

$\forall v\in E,\exists J\subset I,\#J<\infty,\exists v_{j}$ unique $\in
E_{j}:v=\sum_{j\in J}v_{j}$

If the sum is direct the projections are the maps : $\pi_{i}:\oplus_{j\in
I}E_{j}\rightarrow E_{i}$

This concept is important, and it is essential to understand fully its significance.Warning!

i) If $\oplus_{i\in I}E_{i}=E$\ the sum is direct iff the decomposition of any
vector of E with respect to the $E_{i}$ is unique, but this does not entail
that there is a unique collection of subspaces $E_{i}$ for which we have such
a decomposition. Indeed take any basis : the decomposition with respect to
each vector subspace generated by the vectors of the basis is unique, but with
another basis we have another unique decomposition.

ii) If F is a vector subspace of E there is always a unique subset G\ of E
such that $G=F^{c}$ but G is not a vector subspace (because 0 must be both in
F and G for them to be vector spaces).

iii) There are always vector subspaces G such that : $E=F\oplus G$ but G is
not unique. A way to define uniquely G is by using a bilinear form, then G is
the orthogonal complement (see below) and the projection is the orthogonal projection.

Example : Let $\left(  e_{i}\right)  _{i=1..n}$ be a\ basis of a n dimensional
vector space E. Take F the vector subspace generated by $\left(  e_{i}\right)
_{i=1}^{p}$ and G the vector subspace generated by\ $\left(  e_{i}\right)
_{i=p+1}^{n}$ .\ Obviously $E=F\oplus G$\ . But $G_{a}^{\prime}%
=\{w=a(u+v),u\in G,v\in F\}$ for any fixed $a\in K$ is such that : $E=F\oplus
G_{a}^{\prime}$

\paragraph{Product of vector spaces\newline}

\subparagraph{1. Product of two vector spaces\newline}

\begin{theorem}
If E,F are vectors spaces over the same field K, the product set E$\times$F
can be endowed with the structure of a vector space over K with the operations
: $\left(  u,v\right)  +\left(  u^{\prime},v^{\prime}\right)  =\left(
u+u^{\prime},v+v^{\prime}\right)  ;k\left(  u,v\right)  =\left(  ku,kv\right)
;0=\left(  0,0\right)  $
\end{theorem}

The subsets of E$\times$F : E'=(u,0), F'=(0,v) are vector subspaces of
E$\times$F and we have E$\times$F=E'$\oplus$F$^{\prime}.$

Conversely, if $E_{1},E_{2}$ are \textit{vector subspaces} of E such that
$E=E_{1}\oplus E_{2}$ then to each vector of E can be associated its unique
pair $(u,v)\in E_{1}\times E_{2}.$ Define $E_{1}^{\prime}=(u,0),E_{2}^{\prime
}=(0,v)$ which are vector subspaces of $E_{1}\times E_{2}$ and $E_{1}\times
E_{2}=E_{1}^{\prime}\oplus E_{2}^{\prime}$ but $E_{1}^{\prime}\oplus
E_{2}^{\prime}\simeq E$. So in this case one can see the direct sum as the
product $E_{1}\times E_{2}\simeq E_{1}\oplus E_{2}$

In the converse, it is mandatory that $E=E_{1}\oplus E_{2}$ . Indeed take
E$\times$E, the product is well defined, but not the direct sum (it would be
just E).

In a somewhat pedantic way : a vector subspace $E_{1}$\ of a vector space E
splits in E if : $E=E_{1}\oplus E_{2}$ and $E\simeq E_{1}\times E_{2}$ (Lang p.6)

\subparagraph{2. Infinite product of vector spaces\newline}

This can be generalized to any product of vector spaces $\left(  F_{i}\right)
_{i\in I}$\ over the same field where I is finite. If I is infinite this is a
bit more complicated : first one must assume that all the vector spaces
$F_{i}$ belong to some universe.

One defines : $E_{T}=\cup_{i\in I}F_{i}$ (see set theory). Using the axiom of
choice there are maps : $C:I\rightarrow E_{T}::C(i)=u_{i}\in F_{i}$

One restricts $E_{T}$ to the subset $E$\ of $E_{T}$ comprised of elements such
that only finitely many $u_{i}$ are non zero. E can be endowed with the
structure of a vector space and $E=%
{\textstyle\prod\limits_{i\in I}}
F_{i}$

The identity $E=\oplus_{i\in I}E_{i}$ with $E_{i}=\left\{  u_{j}=0,j\neq i\in
I\right\}  $ does not hold any longer : it would be $E_{T}.$

But if the $F_{i}$ are vector subspaces of some $E=\oplus_{i\in I}F_{i}$ which
have only 0 as common element on can still write $%
{\textstyle\prod\limits_{i\in I}}
F_{i}\simeq\oplus_{i\in I}F_{i}$

\paragraph{Quotient space\newline}

\begin{definition}
The \textbf{quotient space}, denoted E/F, of a vector space E by any of its
vector subspace F is the quotient set $E/\sim$ by the relation of equivalence
: $x,y\in E:x-y\in F\Leftrightarrow x\equiv y\;(\operatorname{mod}F)$
\end{definition}

It is a vector space on the same field. The class [0] contains the vectors of F.

The mapping $E\rightarrow E/F$ that associates to $x\in E$ its class of
equivalence [x] , called the quotient map, is a natural epimorphism, whose
kernel is F. This relationship is summarized by the short exact sequence

$0\rightarrow F\rightarrow E\rightarrow E/F\rightarrow0$

The dimension of E/F is sometimes called the codimension. For finite
dimensional vector spaces : dim(E/F)=dim(E) - dim(F)

If $E=F\oplus F^{\prime}$ then E/F is isomorphic to F'

\paragraph{Graded vector spaces\newline}

\begin{definition}
A I-\textbf{graded vector space} is a vector space E endowed with a family of
filters $\left(  E_{i}\right)  _{i\in I}$\ such that each $E_{i}$\ is a vector
subspace of E and $E=\oplus_{i\in I}E_{i}.$ A vector of E which belongs to a
single $E_{i}$ is said to be an \textbf{homogeneous element}.
\end{definition}

Usually the family is indexed on $%
\mathbb{N}
$ and then the family is decreasing : $E_{n+1}\subset E_{n}.$ The simplest
example is $E_{n}=$ the vector subspace generated by the vectors $\left(
e_{i}\right)  _{i\geq n}$ of a basis. The graded space is $grE=\oplus_{n\in%
\mathbb{N}
}E_{n}/E_{n+1}$

A linear map between two I-graded vector spaces $f:E\rightarrow F$ is called a
graded linear map if it preserves the grading of homogeneous elements:

$\forall i\in I:f\left(  E_{i}\right)  \subset F_{i}$

\paragraph{Cone\newline}

\begin{definition}
A \textbf{cone} with apex a in a \textit{real} vector space E is a non empty
subset C of E such that : $\forall k\geq0,u\in C\Rightarrow k\left(
u-a\right)  \in C$
\end{definition}

A cone C is proper if $C\cap\left(  -C\right)  =0.$ Then there is an order
relation on E by : $X\geq Y\Leftrightarrow X-Y\in C$\ thus :

$X\geq Y\Rightarrow X+Z\geq Y+Z,k\geq0:kX\geq kY$

\begin{definition}
A \textbf{vectorial lattice} is a real vector space E endowed with an order
relation for which it is a lattice :
\end{definition}

$\forall x,y\in E,\exists\sup(x,y),\inf(x,y)$

$x\leq y\Rightarrow\forall z\in E:x+z\leq y+z$

$x\geq0,k\geq0$ $\Rightarrow kx\geq0$

On a vectorial lattice :

- the cone with apex a is the set : $C_{a}=\left\{  v\in E:a\geq v\right\}  $

- the sets :

$x_{+}=\sup(x,0);x_{-}=\sup(-x,0),\left\vert x\right\vert =x_{+}+x_{-}$

$a\leq b:\left[  a,b\right]  =\{x\in E:a\leq x\leq b\}$

\bigskip

\subsection{Linear maps}

\label{Linear map}

\subsubsection{Definitions}

\begin{definition}
A l\textbf{inear map} is a morphism between vector spaces over the same field
K :
\end{definition}

$f\in L\left(  E;F\right)  \Leftrightarrow$

$f:E\rightarrow F::\forall a,b\in K,\forall\overrightarrow{u},\overrightarrow
{v}\in E:g(a\overrightarrow{u}+b\overrightarrow{v})=ag(\overrightarrow
{u})+bg(\overrightarrow{v})\in F$

Warning ! To be fully consistent, the vector spaces E and F \textit{must be
defined over the same field K}. So if E is a real vector space and F a complex
vector space we will not consider as a linear map a map such that :
f(u+v)=f(u)+f(v), f(ku)=kf(u) for any k real. This complication is necessary
to keep simple the more important definition of linear map. It will be of
importance when K=$%
\mathbb{C}
.$

\bigskip

\begin{theorem}
The composition of linear map between vector spaces over the same field is
still a linear map, so vector spaces over a field K with linear maps define a category.
\end{theorem}

\begin{theorem}
The set of linear maps from a vector space to a vector space on the same field
K is a vector space over K
\end{theorem}

\begin{theorem}
If a linear map is bijective then its inverse is a linear map and f is an
\textbf{isomorphism.}
\end{theorem}

\begin{definition}
Two vector spaces over the same field are isomorphic if there is an
isomorphism between them.
\end{definition}

\begin{theorem}
Two vector spaces over the same field\ are isomorphic iff they have the same dimension
\end{theorem}

We will usually denote $E\simeq F$ if the two vector spaces E,F are isomorphic.

\begin{definition}
An \textbf{endomorphism} is a linear map on a vector space
\end{definition}

\begin{theorem}
The set of endomorphisms of a vector space E, endowed with the composition
law, is a unital algebra on the same field.
\end{theorem}

\begin{definition}
An endomorphism which is also an isomorphism is called an
\textbf{automorphism}.
\end{definition}

\begin{theorem}
The set of automorphisms of a vector space E, endowed with the composition
law, is a group denoted GL(E).
\end{theorem}

\begin{notation}
L(E;F) with a semi-colon (;) before the codomain F is the set of linear maps
$\hom\left(  E,F\right)  $.
\end{notation}

\begin{notation}
GL(E;F) is the subset of invertible linear maps
\end{notation}

\begin{notation}
GL(E) is the set of automorphisms over the vector space E
\end{notation}

\begin{definition}
A linear endomorphism such that its k iterated, for some k%
$>$%
0 is null is said to be nilpotent : $f\in L\left(  E;E\right)  :f\circ
f\circ..\circ f=\left(  f\right)  ^{k}=0$
\end{definition}

Let $\left(  e_{i}\right)  _{i\in I}$\ be a basis of E over the field K,
consider the set $K^{I}$ of all maps from I to K : $\tau:I\rightarrow
K::\tau\left(  i\right)  =x_{i}\in K$ .Take the subset $K_{0}^{I}$ of $K^{I}$
such that only a finite number of $x_{i}\neq0.$ This is a vector space over K.

For any basis $\left(  e_{i}\right)  _{i\in I}$ there is a map : $\tau
_{e}:E\rightarrow K_{0}^{I}::\tau_{e}\left(  i\right)  =x_{i}.$ This map is
linear and bijective.\ So E is isomorphic to the vector space $K_{0}^{I}$.
This property is fundamental in that whenever only linear operations over
finite dimensional vector spaces are involved it is equivalent to consider the
vector space $K^{n}$ with a given basis. This is the implementation of a
general method using the category theory : $K_{0}^{I}$ is an object in the
category of vector spaces over K.\ So if there is a functor acting on this
category we can see how it works on $K_{0}^{I}$ and the result can be extended
to other vector spaces.

\begin{definition}
If E,F are two complex vector spaces, an \textbf{antilinear map} is a map
$f:E\rightarrow F$ such that :

$\forall u,v\in E,z\in%
\mathbb{C}
:f\left(  u+v\right)  =f\left(  u\right)  +f\left(  v\right)  ;f\left(
zu\right)  =\overline{z}f\left(  u\right)  $
\end{definition}

Such a map is linear when z is limited to a real scalar.

\subsubsection{Matrix of a linear map}

(see the "Matrices" section below for more)

Let E be a vector space with basis $\left(  e_{i}\right)  _{i=1}^{n}$, F a
vector space with basis $\left(  f_{j}\right)  _{j=1}^{p}$ both on the field
K, and $L\in L\left(  E;F\right)  $.

The matrix of L in these bases is the matrix $\left[  M\right]  =\left[
M_{ij}\right]  _{j=1...n}^{i=1...p}$ with p rows and n columns such that :
$L\left(  e_{i}\right)  =\sum_{j=1}^{p}M_{ij}f_{j}$

So that : $\forall u=\sum_{i=1}^{n}u_{i}e_{i}\in E:L\left(  u\right)
=\sum_{j=1}^{p}\sum_{i=1}^{n}\left(  M_{ij}u_{i}\right)  f_{j}$

\bigskip

$%
\begin{bmatrix}
v_{1}\\
...\\
v_{p}%
\end{bmatrix}
=%
\begin{bmatrix}
M_{11} & .. & M_{1n}\\
.. & .. & ..\\
M_{p1} & .. & M_{pn}%
\end{bmatrix}%
\begin{bmatrix}
u_{1}\\
..\\
u_{n}%
\end{bmatrix}
\Leftrightarrow v=L\left(  u\right)  $

\bigskip

or with the vectors represented as column matrices : $\left[  L\left(
u\right)  \right]  =\left[  M\right]  \left[  u\right]  $

The matrix of the composed map $L\circ L^{\prime}$ is the product of the
matrices MxM' (the dimensions must be consistent).

The matrix is square is dim(E)=dim(F).\ f is an isomorphism iff M is
invertible (det(M) non zero).

\bigskip

\begin{theorem}
A \textbf{change of basis} in a vector space is an endomorphism. Its matrix P
has for columns the components of the new basis expressed in the old basis :
$\overrightarrow{e_{i}}\rightarrow\overrightarrow{E}_{i}=\sum_{j=1}^{n}%
P_{ij}\overrightarrow{e}_{j}$ .The new components $\widetilde{u}_{i}$ of
a\ vector u are given by :%

\begin{equation}
\left[  \widetilde{u}\right]  =\left[  P\right]  ^{-1}\left[  u\right]
\end{equation}

\end{theorem}

\begin{proof}
$\overrightarrow{u}=\sum_{i=1}^{n}u_{i}\overrightarrow{e_{i}}=\sum_{i=1}%
^{n}U_{i}\overrightarrow{E_{i}}\Leftrightarrow\left[  u\right]  =\left[
P\right]  \left[  U\right]  \Leftrightarrow\left[  U\right]  =\left[
P\right]  ^{-1}\left[  u\right]  $
\end{proof}

\begin{theorem}
In a change of basis in E with matrix P, and F with matrix Q, the matrix M of
the map $L\in L\left(  E;F\right)  $ in the new bases becomes :%

\begin{equation}
\left[  \widetilde{M}\right]  =\left[  Q\right]  ^{-1}\left[  M\right]
\left[  P\right]
\end{equation}

\end{theorem}

\begin{proof}
$\overrightarrow{f_{i}}\rightarrow\overrightarrow{F_{i}}=\sum_{j=1}^{p}%
Q_{ij}\overrightarrow{f_{j}}$

$\overrightarrow{v}=\sum_{i=1}^{p}v_{i}\overrightarrow{f_{i}}=\sum_{i=1}%
^{p}V_{i}\overrightarrow{F_{i}}\Leftrightarrow\left[  v\right]  =\left[
Q\right]  \left[  V\right]  \Leftrightarrow\left[  V\right]  =\left[
Q\right]  ^{-1}\left[  v\right]  $

$\left[  v\right]  =\left[  M\right]  \left[  u\right]  =\left[  Q\right]
\left[  V\right]  =\left[  M\right]  \left[  P\right]  \left[  U\right]
\Rightarrow\left[  V\right]  =\left[  Q\right]  ^{-1}\left[  M\right]  \left[
P\right]  \left[  U\right]  $

$\left[  p,1\right]  =\left[  p,p\right]  \times\left[  p,n\right]
\times\left[  n,n\right]  \times\left[  n,1\right]  $

$\Rightarrow\left[  \widetilde{M}\right]  =\left[  Q\right]  ^{-1}\left[
M\right]  \left[  P\right]  $
\end{proof}

If L is an endomorphism then P=Q, and

$\left[  \widetilde{M}\right]  =\left[  P\right]  ^{-1}\left[  M\right]
\left[  P\right]  \Rightarrow\det\widetilde{M}=\det M$

An obvious, but convenient, result : a vector subspace F of E is generated by
a basis of r vectors $f_{j}$, expressed in a basis $e_{i}$ of E by a n$\times
$r matrix $\left[  A\right]  $:

$u\in F\Leftrightarrow u=\sum_{j=1}^{r}x_{j}f_{j}=\sum_{i=1}^{n}\sum_{j=1}%
^{r}x_{i}A_{ji}e_{j}=\sum_{i=1}^{n}u_{i}e_{i}$

so : $u\in F\Leftrightarrow\exists\left[  x\right]  :$ $\left[  u\right]
=\left[  A\right]  \left[  x\right]  $

\subsubsection{Eigen values}

\begin{definition}
An \textbf{eigen vector} of the endomorphism $f\in L\left(  E;E\right)  $ on
(E,K) with \textbf{eigen value} $\lambda\in K$ is a vector $u\neq0$ such that
$f\left(  u\right)  =\lambda u$
\end{definition}

Warning !

i) An eigen vector is non zero, but an eigen value can be zero.\ 

ii) A linear map may have or have not eigen values.

iii) the eigen value must belong to the field K

\begin{theorem}
The eigenvectors of an endomorphism $f\in L\left(  E;E\right)  $ with the same
eigenvalue $\lambda,$ form, with the vector 0, a vector subspace $E_{\lambda}$
of E called an \textbf{eigenspace}.
\end{theorem}

\begin{theorem}
The eigenvectors corresponding to different eigenvalues are linearly independent
\end{theorem}

\begin{theorem}
If $u,\lambda$ are eigen vector and eigen value of f, then, for k%
$>$
0, u and $\lambda^{k}$ are eigen vector and eigen value of $\left(  \circ
f\right)  ^{k}$ (k-iterated map)
\end{theorem}

So f is nilpotent if its only eigen values are 0.

\begin{theorem}
f is injective iff it has no zero eigen value.
\end{theorem}

If E is finite dimensional, the eigen value and vectors are the eigen value
and vectors of its matrix in any basis (see Matrices)

If E is infinite dimensional the definition stands but the main concept is a
bit different : the spectrum of f is the set of scalars $\lambda$\ such that
$\left(  f-\lambda Id\right)  $ has no bounded inverse. So an eigenvalue
belongs to the spectrum but the converse is not true (see Banach spaces).

\subsubsection{Rank of a linear map}

\paragraph{Rank\newline}

\begin{theorem}
The range f(E) of a linear map $f\in L(E;F)$ is a vector subspace of the
codomain F. The \textbf{rank} rank(f) of f is the dimension of $f(E)\subset F$.

$rank(f)=\dim f(E)\leq\dim(F)$

$f\in L(E;F)$ is surjective iff f(E) = F, or equivalently if rank(f) = dimE
\end{theorem}

\begin{proof}
f is surjective iff$\ \forall v\in F,\exists u\in E:f(u)=v$ $\Leftrightarrow
\dim f(E)=\dim F=rank(f)$
\end{proof}

So the map : $\widetilde{f}:E\rightarrow f(E)$ is a linear surjective map L(E;f(E))

\paragraph{Kernel\newline}

\begin{theorem}
The \textbf{kernel,} denoted $\ker\left(  f\right)  ,$ of a linear map $f\in
L(E;F)$ is the set : $\ker\left(  f\right)  =\left\{  u\in E:f\left(
u\right)  =0_{F}\right\}  .$ It is a vector subspace of its domain E and

$\dim\ker(f)\leq\dim E$ and if $\dim\ker(f)=\dim E$ then f=0

f is injective if ker(f)=0
\end{theorem}

\begin{proof}
f is injective iff $\forall u_{1},u_{2}\in E:f(u_{1})=f(u_{2})\Rightarrow
u_{1}=u_{2}\Leftrightarrow\ker\left(  f\right)  =0_{E}$
\end{proof}

So with the quotient space E/ker(f) the map : $\widehat{f}:E/\ker f\rightarrow
F$ is a linear injective map L(E/ker(f);F) (two vectors giving the same result
are deemed equivalent).

\paragraph{Isomorphism\newline}

\begin{theorem}
If $f\in L(E;F)$ then $rank(f)\leq\min\left(  \dim E,\dim F\right)  $ and f is
an isomorphism iff rank(f)=dim(E)=dim(F)
\end{theorem}

\begin{proof}
$g:E/\ker f\rightarrow f(E)$ is a linear bijective map, that is an isomorphism
and we can write : $f\left(  E\right)  \simeq E/\ker\left(  f\right)  $

The two vector spaces have the same dimension thus :

dim(E/ker(f)) = dim E - dimker(f) = dimf(E) =rank(f)

$rank(f)\leq\min\left(  \dim E,\dim F\right)  $ and f is an isomorphism iff rank(f)=dim(E)=dim(F)
\end{proof}

\paragraph{To sum up\newline}

A linear map $f\in L\left(  E;F\right)  $ falls into one of the three
following cases :

i) f is surjective : f(E)=F :

rank$\left(  f\right)  $ = $\dim f\left(  E\right)  =\dim F=\dim E-\dim\ker
f\leq\dim E$

F is "smaller" or equal to E

With dim(E)=n, dim(F)=p the matrix of f is $\left[  f\right]  _{n\times
p},p\leq n$

There is a linear bijection from E/ker(f) to F

ii) f is injective : ker(f)=0

dim E= dimf(E) =rank(f)$\leq\dim F$

(E is "smaller" or equal to F)

With dim(E)=n, dim(F)=p the matrix of f is $\left[  f\right]  _{n\times
p},n\leq p$

There is a linear bijection from E to f(E)

iii) f is bijective :

f(E)=F ,ker(f)=0, dimE=dimF=rank(f)

With dim(E)=dimF=n, the matrix of f is square $\left[  f\right]  _{n\times n}$
and $\det\left[  f\right]  \neq0$

\subsubsection{Multilinear maps}

\begin{definition}
A \textbf{r multilinear map} is a map : $f:E_{1}\times E_{2}\times
..E_{r}\rightarrow F,$ where $\left(  E_{i}\right)  _{i=1}^{r}$ is a family of
r vector spaces, and F a vector space, all over the same field K, which is
linear with respect to each variable
\end{definition}

$\forall u_{i},v_{i}\in E_{i},k_{i}\in K:$

$f\left(  k_{1}u_{1},k_{2}u_{2},...,k_{r}u_{r}\right)  =k_{1}k_{2}%
...k_{r}f(u_{1},u_{2},...u_{r})$

$f(u_{1},u_{2},.,u_{i}+v_{i},..,u_{r})=f(u_{1},u_{2},...u_{i},...,u_{r}%
)+f(u_{1},u_{2},...v_{i},...,u_{r}) $

\begin{notation}
$L^{r}\left(  E_{1},E_{2}...,E_{r};F\right)  $ is the set of r-linear maps
from $E_{1}\times E_{2}...\times E_{r}$ to F
\end{notation}

\begin{notation}
$L^{r}\left(  E;F\right)  $ is the set of r-linear map from E$^{r}$ to F
\end{notation}

Warning !

$E_{1}\times E_{2}$ can be endowed with the structure of a vector space. A
\textit{linear} map $f:E_{1}\times E_{2}\rightarrow F$\ is such that :

$\forall\left(  u_{1},u_{2}\right)  \in E_{1}\times E_{2}:$

$\left(  u_{1},u_{2}\right)  =\left(  u_{1},0\right)  +\left(  0,u_{2}\right)
$ so $f\left(  u_{1},u_{2}\right)  =f\left(  u_{1},0\right)  +f\left(
0,u_{2}\right)  $

that can be written :

$f\left(  u_{1},u_{2}\right)  =f_{1}\left(  u_{1}\right)  +f_{2}\left(
u_{2}\right)  $ with $f_{1}\in L\left(  E_{1};F\right)  ,f_{2}\in L(E_{2};F)$

So : $L\left(  E_{1}\times E_{2};F\right)  \simeq L\left(  E_{1};F\right)
\oplus L\left(  E_{2};F\right)  $

\begin{theorem}
The space $L^{r}\left(  E;F\right)  \equiv L\left(  E;L\left(
E;....L(E;F\right)  \right)  $
\end{theorem}

\begin{proof}
For $f\in L^{2}\left(  E,E;F\right)  $ and u fixed $f_{u}:E\rightarrow
F::f_{u}(v)=f(u,v)$ is a linear map.

Conversely a map : $g\in L\left(  E;L\left(  E;F\right)  \right)  ::g\left(
u\right)  \in L\left(  E;F\right)  $ is equivalent to a bilinear map :
$f\left(  u,v\right)  =g\left(  u\right)  \left(  v\right)  $
\end{proof}

\bigskip

For E n dimensional and F p dimensional the components of the bilinear map
$f\in L^{2}\left(  E;F\right)  $ read :

$f\left(  u,v\right)  =\sum_{i,j=1}^{n}u_{i}v_{j}f(e_{i},e_{j})$ \ \ with :
$f(e_{i},e_{j})=\sum_{k=1}^{p}\left(  F_{kij}\right)  f_{k},F_{kij}\in K$

A bilinear map cannot be represented by a single matrix if F is not
unidimensional (meaning if F is not K). It is a tensor.

\begin{definition}
A r-linear map $f\in L^{r}\left(  E;F\right)  $ is :

\textbf{symmetric} if : $\forall u_{i}\in E,i=1...r,\sigma\in\mathfrak{S}%
\left(  r\right)  :f(u_{1},...,u_{r})=f(u_{\sigma\left(  1\right)
},...,u_{\sigma\left(  r\right)  })$

\textbf{antisymmetric} if : $\forall u_{i}\in E,i=1...r,\sigma\in
\mathfrak{S}\left(  r\right)  :f(u_{1},...,u_{r})=\epsilon\left(
\sigma\right)  f(u_{\sigma\left(  1\right)  },...,u_{\sigma\left(  r\right)
})$
\end{definition}

\subsubsection{Dual of a vector space}

\paragraph{Linear form\newline}

A field K is endowed with the structure of a 1-dimensional vector space over
itself in the obvious way, so one can consider morphisms from a vector space E
to K.

\begin{definition}
A \textbf{linear form} on a vector space (E,K) on the field K is a linear map
valued in K
\end{definition}

A linear form can be seen as a linear function with argument a vector of E and
value in the field K : $\varpi\left(  u\right)  =k$

Warning ! A linear form must be valued in the same field as E.\ A "linear form
on a complex vector space and valued in $%
\mathbb{R}
"$ cannot be defined without a real structure on E.

\paragraph{Dual of a vector space\newline}

\begin{definition}
The \textbf{algebraic dual} of a vector space is the set of its linear form,
which has the structure of a vector space on the same field
\end{definition}

\begin{notation}
E* is the algebraic dual of the vector space E
\end{notation}

The vectors of the dual $\left(  K^{n}\right)  ^{\ast}$ are usually
represented as 1xn matrices (row matrices).

\begin{theorem}
A vector space and its algebraic dual are isomorphic \textit{iff} they are
finite dimensional.\ 
\end{theorem}

\bigskip

This important point deserves some comments.

i) Consider first a finite finite n dimensional vector space E.

For each basis $\left(  e_{i}\right)  _{i=1}^{n}$ the \textbf{dual basis}
$\left(  e^{i}\right)  _{i=1}^{n}$ of the dual $E^{\ast}$ is defined by the
condition : $e^{i}\left(  e_{j}\right)  =\delta_{j}^{i}.$ where $\delta
_{j}^{i}$ is the \textbf{Kronecker'symbol }$\delta_{j}^{i}=1$ if $i=j,=0$ if
not. These conditions define uniquely a basis of the dual, which is indexed on
the same set I.

The map : $L:E\rightarrow E^{\ast}:L\left(  \sum_{i\in I}u_{i}e_{i}\right)
=\sum_{i\in I}u_{i}e^{i}$ is an isomorphism.

In a change of basis in E with matrix P (which has for columns the components
of the new basis expressed in the old basis) :

$e_{i}\rightarrow\widetilde{e}_{i}=\sum_{j=1}^{n}P_{ij}e_{j},$

the dual basis changes as : $e^{i}\rightarrow\widetilde{e}^{i}=\sum_{j=1}%
^{n}Q_{ij}e^{j}$ with $\left[  Q\right]  =\left[  P\right]  ^{-1}$

Warning! This isomorphism is not canonical, even in finite dimensions, in that
it depends of the choice of the basis. In general \textit{there is no natural
transformation which is an isomorphism between a vector space and its dual},
even finite dimensional. So to define an isomorphism one uses a bilinear form
(when there is one).

ii) Consider now an infinite dimensional vector space E over the field K.

Then $\dim\left(  E^{\ast}\right)  >\dim\left(  E\right)  $ . For infinite
dimensional vector spaces the algebraic dual $E^{\ast}$ is a \textit{larger
set} then E.

Indeed if E has the basis $\left(  e_{i}\right)  _{i\in I}$ there is a map :
$\tau_{e}:E\rightarrow K_{0}^{I}::\tau_{e}\left(  i\right)  =x_{i}$ giving the
components of a vector, in the set $K_{0}^{I}$ of maps $I\rightarrow K$ such
that only a finite number of components is non zero and $K_{0}^{I}\simeq E$.
But any map : $\lambda:I\rightarrow K$ gives a linear map $\sum_{i\in
I}\lambda\left(  i\right)  x_{i}$ which is well defined because only a finite
number of terms are non zero, whatever the vector, and can represent a vector
of the dual.\ So the dual $E^{\ast}\simeq K^{I}$ which is larger than
$K_{0}^{I}.$

The condition $\forall i,j\in I:e^{i}\left(  e_{j}\right)  =\delta_{j}^{i}%
$\ still defines a family $\left(  e^{i}\right)  _{i\in I}$ of linearly
independant vectors of the dual E* but this is not a basis of E*. However
there is always a basis of the dual, that we can denote $\left(  e^{i}\right)
_{i\in I^{\prime}}$ with \#I'
$>$
\#I and one can require that $\forall i,j\in I:e^{i}\left(  e_{j}\right)
=\delta_{j}^{i}$

For infinite dimensional vector spaces one considers usually the topological
dual which is the set of continuous forms over E. If E is finite dimensional
the algebraic dual is the same as the topological dual.

\begin{definition}
The \textbf{double dual} $E^{\ast\ast}$ of a vector space is the algebraic
dual of E*. The double dual E** is isomorphic to E iff E is finite dimensional
\end{definition}

There is a natural homomorphism $\phi$ from E into the double dual E**,
defined by the evaluation map : $(\phi(u))(\varpi)=\varpi(u)$ for all $v\in
E,\varpi\in E^{\ast}$. This map $\phi$ is always injective so $E\sqsubseteq
\left(  E^{\ast}\right)  ^{\ast}$; it is an isomorphism if and only if E is
finite-dimensional, and if so then E$\simeq$E**.

\begin{definition}
The \textbf{annihiliator} $S^{\intercal}$ of a vector subspace S of E is the
set : $S^{\intercal}=\left\{  \varphi\in E^{\ast}:\forall u\in S:\varphi
\left(  u\right)  =0\right\}  .$
\end{definition}

It is a vector subspace of E*.\ $E^{\intercal}=0;$ $S^{\intercal}%
+S^{\prime\intercal}\subset\left(  S\cap S^{\prime}\right)  ^{\intercal}$

\paragraph{Transpose of a linear map\newline}

\begin{theorem}
If E,F are vector spaces on the same field, $\forall f\in L(E;F)$ there is a
unique map, called the (algebraic) \textbf{transpose} (called also
dual\textbf{) }and denoted\textbf{\ }$f^{t}\in L(F^{\ast};E^{\ast})$ such that
: $\forall\varpi\in F^{\ast}:f^{t}\left(  \varpi\right)  =\varpi\circ f$
\end{theorem}

The relation $^{t}:L(E;F)\rightarrow L(F^{\ast};E^{\ast})$ is injective
(whence the unicity) but not surjective (because E**$\neq E$ if E is infinite dimensional).

The functor which associes to each vector space its dual and to each linear
map its transpose is a functor from the category of vector spaces over a field
K to itself.

If the linear map f is represented by the matrix A with respect to two bases
of E and F, then $f^{t}$ is represented by the \textit{same} matrix\ with
respect to the dual bases of F* and E*. Alternatively, as f is represented by
A acting on the left on column vectors, $f^{t}$ is represented by the same
matrix acting on the right on row vectors. So if vectors are always
represented as matrix columns the matrix of $f^{t}$ is the transpose of the
matrix of f :

\begin{proof}
$\forall u,\lambda:\left[  \lambda\right]  ^{t}\left[  f^{t}\right]
^{t}\left[  u\right]  =\left[  \lambda\right]  ^{t}\left[  f\right]  \left[
u\right]  \Leftrightarrow\left[  f^{t}\right]  =\left[  f\right]  $
\end{proof}

\subsubsection{Bilinear forms}

\begin{definition}
A \textbf{multilinear form} is a multilinear map defined on vector spaces on a
field K and valued in K.
\end{definition}

So a \textbf{bilinear form} g\ on a vector space E on a field K is a bilinear
map on E valued on K: $g:E\times E\rightarrow K$ is such that :

$\forall u,v,w\in E,k,k^{\prime}\in K:$

$g\left(  ku,k^{\prime}v\right)  =kk^{\prime}g(u,v),$

$g(u+w,v)=g(u,v)+g(u,w),g(u,v+w)=g(u,v)+g(u,w)$

Warning ! A multilinear form must be valued in the same field as E.\ A
"multilinear form on a complex vector space and valued in $%
\mathbb{R}
"$ cannot be defined without a real structure on E.

\paragraph{Symmetric, antisymmetric forms\newline}

\begin{definition}
A multilinear form $g\in L^{r}\left(  E;K\right)  $ is

\textbf{symmetric} if : $\forall\left(  u_{j}\right)  _{j=1}^{r},\sigma
\in\mathfrak{S}\left(  r\right)  :g\left(  u_{\sigma\left(  1\right)
},...u_{\sigma\left(  r\right)  }\right)  =g\left(  u_{1},...,u_{r}\right)  $

\textbf{antisymmetric} if : $\forall\left(  u_{j}\right)  _{j=1}^{r},\sigma
\in\mathfrak{S}\left(  r\right)  :g\left(  u_{\sigma\left(  1\right)
},...u_{\sigma\left(  r\right)  }\right)  =\epsilon\left(  \sigma\right)
g\left(  u_{1},...,u_{r}\right)  $
\end{definition}

Any bilinear symmetric form defines the \textbf{quadratic form} :

$Q:E\rightarrow K::Q\left(  u\right)  =g\left(  u,u\right)  $

Conversely $g\left(  u,v\right)  =\frac{1}{2}\left(  Q\left(  u+v\right)
-Q\left(  u\right)  -Q\left(  v\right)  \right)  $ (called the
\textbf{polarization formula}) defines the bilinear symmetric form g from Q.

\paragraph{Non degenerate bilinear forms\newline}

\begin{definition}
A bilinear symmetric form $g\in L^{2}\left(  E;K\right)  $ is \textbf{non
degenerate} if : $\forall v:g\left(  u,v\right)  =0\Rightarrow u=0$
\end{definition}

Warning ! one can have g(u,v)=0 with u,v non null.

\begin{theorem}
A non degenerate bilinear form on a finite dimensional vector space E on a
field K defines isomorphisms between E and its dual E*:

$\phi_{R}:E\rightarrow E^{\ast}::\phi_{R}\left(  u\right)  \left(  v\right)
=g\left(  u,v\right)  $

$\phi_{L}:E\rightarrow E^{\ast}::\phi_{L}\left(  u\right)  \left(  v\right)
=g\left(  v,u\right)  $
\end{theorem}

This is the usual way to map vectors to forms and vice versa.

$L^{2}\left(  E;K\right)  \equiv L\left(  E;L\left(  E;K\right)  \right)
=L\left(  E;E^{\ast}\right)  $

$\phi_{R},\phi_{L}$ are injective, they are surjective iff E is finite dimensional.

$\phi_{R},\phi_{L}$ are identical if g is symmetric and opposite from each
other if g is skew-symmetric.

Conversely to the linear map $\phi\in L\left(  E;E^{\ast}\right)  $ is
associated the bilinear forms :

$g_{R}\left(  u,v\right)  =\phi\left(  u\right)  \left(  v\right)
;g_{L}\left(  u,v\right)  =\phi\left(  v\right)  \left(  u\right)  $

Remark : it is usual to say that g is non degenerate if $\phi_{R},\phi_{L}\in
L\left(  E;E^{\ast}\right)  $ are isomorphisms.\ The two definitions are
equivalent if E is finite dimensional, but we will need non degeneracy for
infinite dimensional vector spaces.

\paragraph{Matrix representation of a bilinear form\newline}

If E is finite dimensional g is represented in a basis $\left(  e_{i}\right)
_{i=1}^{n}$ by a square matrix nxn $\left[  g_{ij}\right]  =g\left(
e_{i},e_{j}\right)  $ with : $g\left(  u,v\right)  =\left[  u\right]
^{t}\left[  g\right]  \left[  v\right]  $

The matrix $\left[  g\right]  $ is symmetric if g is symmetric, antisymmetric
if g is antisymmetric, and its determinant is non zero iff g is non degenerate.

In a change of basis : the new matrice is%

\begin{equation}
\left[  \widetilde{g}\right]  =\left[  P\right]  ^{t}\left[  g\right]  \left[
P\right]
\end{equation}

where $\left[  P\right]  $ is the matrix with the components of the new basis :

$g(u,v)=\left[  u\right]  ^{t}\left[  g\right]  \left[  v\right]  ,\left[
u\right]  =\left[  P\right]  \left[  U\right]  ,v=\left[  P\right]  \left[
v\right]  $

$\Rightarrow g(u,v)=\left[  U\right]  ^{t}\left[  P\right]  ^{t}\left[
g\right]  \left[  P\right]  \left[  V\right]  \rightarrow\left[  \widetilde
{g}\right]  =\left[  P\right]  ^{t}\left[  g\right]  \left[  P\right]  $

A symmetric matrix has real eigen values, which are non null if the matrix has
a non null determinant. They do not depend on the basis which is used. So :

\begin{definition}
The \textbf{signature}, denoted (p,q) of a non degenerate symmetric bilinear
form g on a n dimensional real vector space is the number p of positive eigen
values and the number q of negative eigen values of its matrix, expressed in
any basis.
\end{definition}

\paragraph{Positive bilinear forms\newline}

\begin{definition}
A bilinear symmetric form g on a real vector space E is

\textbf{positive} if : $\forall u\in E:g(u,u)\geq0$

\textbf{definite positive} if it is positive and $\forall u\in
E:g(u,u)=0\Rightarrow u=0$
\end{definition}

definite positive $\Rightarrow$\ non degenerate .\ The converse is not true

Notice that E must be a real vector space.

\begin{theorem}
(Schwartz I p.175) If the bilinear symmetric form g on a real vector space E
is positive then $\forall u,v\in E$

i) \textbf{Schwarz inequality} : $\left\vert g(u,v)\right\vert \leq
\sqrt{g(u,u)g(v,v)}$

ii) \textbf{Triangular inequality} : $\sqrt{g(u+v,u+v)}\leq\sqrt{g(u,u)}%
+\sqrt{g(v,v)}$

and if g is positive definite, in both cases the equality implies $\exists
k\in%
\mathbb{R}
:v=ku$

iii) Pythagore's theorem :

$\sqrt{g(u+v,u+v)}=\sqrt{g(u,u)}+\sqrt{g(v,v)}\Leftrightarrow g\left(
u,v\right)  =0$
\end{theorem}

\subsubsection{Sesquilinear forms}

\begin{definition}
A \textbf{sesquilinear} form on a complex vector space E is a map $g:E\times
E\rightarrow%
\mathbb{C}
$ linear in the second variable and antilinear in the \textit{first} variable:
\end{definition}

$g\left(  \lambda u,v\right)  =\overline{\lambda}g\left(  u,v\right)  $

$g\left(  u+u^{\prime},v\right)  =g\left(  u,v\right)  +g\left(  u^{\prime
},v\right)  $

So the only difference with a bilinear form is the way it behaves by
multiplication by a complex scalar in the first variable.

Remarks :

i) this is the usual convention in physics.\ One finds also sesquilinear =
linear in the first variable, antilinear in the second variable

ii) if E is a real vector space then a bilinear form is the same as a
sesquilinear form

The definitions for bilinear forms extend to sesquilinear forms. In most of
the results transpose must be replaced by conjugate-transpose.

\paragraph{Hermitian forms\newline}

\begin{definition}
A \textbf{hermitian form} is a sesquilinear form such that :

$\forall u,v\in E:g\left(  v,u\right)  =\overline{g\left(  u,v\right)  }$
\end{definition}

Hermitian forms play the same role in complex vector spaces as the symmetric
bilinear forms in real vector spaces. If E is a real vector space a bilinear
symmetric form is a hermitian form.

The quadratic form associated to an hermitian form is :

$Q:E\rightarrow%
\mathbb{R}
::Q\left(  u,u\right)  =g\left(  u,u\right)  =\overline{g\left(  u,u\right)
}$

\begin{definition}
A \textbf{skew hermitian form} (also called an anti-symmetric sesquilinear
form) is a sesquilinear form such that :

$\forall u,v\in E:g\left(  v,u\right)  =-\overline{g\left(  u,v\right)  }$
\end{definition}

Notice that, on a complex vector space, there are also bilinear form (they
must be C-linear), and symmetric bilinear form

\paragraph{Non degenerate hermitian form\newline}

\begin{definition}
A hermitian form is \textbf{non degenerate} if :

$\forall v\in E:g\left(  u,v\right)  =0\Rightarrow u=0$
\end{definition}

Warning ! one can have g(u,v)=0 with u,v non null.

\begin{theorem}
A non degenerate form on a finite dimensional vector space defines the
\textit{anti-isomorphism} between E and E* :

$\phi_{R}:E\rightarrow E^{\ast}::\phi_{R}\left(  u\right)  \left(  v\right)
=g\left(  u,v\right)  $

$\phi_{L}:E\rightarrow E^{\ast}::\phi_{L}\left(  u\right)  \left(  v\right)
=\overline{g\left(  v,u\right)  }$
\end{theorem}

which are identical if g is hermitian and opposite from each other if g is skew-hermitian.

\paragraph{Matrix representation of a sequilinear form\newline}

If E is finite dimensional a sequilinear form g is represented in a basis
$\left(  e_{i}\right)  _{i=1}^{n}$ by a square matrix nxn $\left[
g_{ij}\right]  =g\left(  e_{i},e_{j}\right)  $ with :%

\begin{equation}
g\left(  u,v\right)  =\overline{\left[  u\right]  }^{t}\left[  g\right]
\left[  v\right]  =\left[  u\right]  ^{\ast}\left[  g\right]  \left[
v\right]
\end{equation}

The matrix $\left[  g\right]  $ is hermitan $\left(  \left[  g\right]
=\overline{\left[  g\right]  }^{t}=\left[  g\right]  ^{\ast}\right)  $ if g is
hermitian, antihermitian $\left(  \left[  g\right]  =-\overline{\left[
g\right]  }^{t}=-\left[  g\right]  ^{\ast}\right)  $ if g is skewhermitian,
and its determinant is non zero iff g is no degenerate.

In a change of basis : the new matrice is%

\begin{equation}
\left[  \widetilde{g}\right]  =\left[  P\right]  ^{\ast}\left[  g\right]
\left[  P\right]
\end{equation}

where $\left[  P\right]  $ is the matrix with the components of the new basis :

$g(u,v)=\left[  u\right]  ^{\ast}\left[  g\right]  \left[  v\right]  ,\left[
u\right]  =\left[  P\right]  \left[  U\right]  ,v=\left[  P\right]  \left[
v\right]  $

$\Rightarrow g(u,v)=\left[  U\right]  ^{\ast}\left[  P\right]  ^{\ast}\left[
g\right]  \left[  P\right]  \left[  V\right]  \rightarrow\left[  \widetilde
{g}\right]  =\left[  P\right]  ^{\ast}\left[  g\right]  \left[  P\right]  $

\paragraph{Positive hermitian forms\newline}

As $g\left(  u,u\right)  \in%
\mathbb{R}
$ for a hermitian form, there are define positive (resp. definite positive)
hermitian forms.

\begin{definition}
A hermitian form g on a complex vector space E is :

\textbf{positive} if: $\forall u\in E:g\left(  u,u\right)  \geq0$

\textbf{definite positive} if $\forall u\in E:g\left(  u,u\right)
\geq0,g\left(  u,u\right)  =0\Rightarrow u=0$
\end{definition}

\begin{theorem}
(Schwartz I p.178) If g is a hermitian, positive form on a complex vector
space E, then $\forall u,v\in E$

Schwarz inequality : $\left\vert g(u,v)\right\vert \leq\sqrt{g(u,u)g(v,v)}$

Triangular inequality : $\sqrt{g(u+v,u+v)}\leq\sqrt{g(u,u)}+\sqrt{g(v,v)}$

and if g is positive definite, in both cases the equality implies $\exists
k\in%
\mathbb{C}
:v=ku$
\end{theorem}

\subsubsection{Adjoint of a map}

\begin{definition}
On a vector space E, endowed with a bilinear symmetric form g if E is real, a
hermitian sesquilinear form g if E is complex, the \textbf{adjoint} of an
endomorphism f with respect to g is the map $f^{\ast}\in L\left(  E;E\right)
$ such that $\forall u,v\in E:g\left(  f\left(  u\right)  ,v\right)  =g\left(
u,f^{\ast}\left(  v\right)  \right)  $
\end{definition}

Warning ! the transpose of a linear map can be defined without a bilinear map,
the adjoint is always defined with respect to a form.

\begin{theorem}
On a vector space E, endowed with a bilinear symmetric form g if E is real, a
hermitian sesquilinear form g if E is complex, which is non degenerate :

i) the adjoint of an endormorphism, if it exists, is unique and $\left(
f^{\ast}\right)  ^{\ast}=f$

ii) If E is finite dimensional any endomorphism has a unique adjoint
\end{theorem}

The matrix of f* is : $[f^{\ast}]=\left[  g\right]  ^{-1}\left[  f\right]
^{t}\left[  g\right]  $ with $\left[  g\right]  $\ \ the matrix of g

\begin{proof}
$\left(  \left[  f\right]  \left[  u\right]  \right)  ^{\ast}\left[  g\right]
\left[  v\right]  =\left[  u\right]  ^{\ast}\left[  g\right]  \left[  f^{\ast
}\right]  \left[  v\right]  \Leftrightarrow\left[  f\right]  ^{\ast}\left[
g\right]  =\left[  g\right]  \left[  f^{\ast}\right]  \Leftrightarrow\lbrack
f^{\ast}]=\left[  g\right]  ^{-1}\left[  f\right]  ^{\ast}\left[  g\right]  $
\end{proof}

And usually $[f^{\ast}]\neq\left[  f\right]  ^{\ast}$

\paragraph{Self-adjoint, orthogonal maps\newline}

\begin{definition}
An endomorphism f on a vector space E, endowed with a bilinear symmetric form
g if E is real, a hermitian sesquilinear form g if E is complex, is:

\textbf{self-adjoint} if it is equal to its adjoint : $f^{\ast}%
=f\Leftrightarrow g\left(  f\left(  u\right)  ,v\right)  =g\left(  u,f\left(
v\right)  \right)  $

\textbf{orthogonal }(real case),\textbf{\ unitary }(complex case) if it
preserves the form : $g\left(  f\left(  u\right)  ,f\left(  v\right)  \right)
=g\left(  u,v\right)  $
\end{definition}

If E is finite dimensional the matrix $\left[  f\right]  $\ of a self adjoint
map f is such that : $\left[  f\right]  ^{\ast}\left[  g\right]  =\left[
g\right]  \left[  f\right]  $

\begin{theorem}
If the form g is non degenerate then for any unitary endomorphism f : $f\circ
f^{\ast}=f^{\ast}\circ f=Id$
\end{theorem}

\begin{proof}
$\forall u,v:g\left(  f\left(  u\right)  ,f\left(  v\right)  \right)
=g\left(  u,v\right)  =g\left(  u,f^{\ast}f\left(  v\right)  \right)
\Rightarrow g\left(  u,(Id-f^{\ast}f\right)  v)=0\Rightarrow f^{\ast}f=Id$
\end{proof}

\begin{definition}
The \textbf{orthogonal group} denoted O(E,g) of a vector space E endowed with
a non degenerate bilinear symmetric form g is the set of its orthogonal
invertible endomorphisms. The \textbf{special orthogonal group} denoted
SO(E,g) is its subgroup comprised of elements with det(f) = 1.

The \textbf{unitary group }denoted U(E,g) on a complex vector space E endowed
with a hermitian sesquilinear form g is the set, denoted U(E,g), of its
unitary invertible\ endomorphisms. The \textbf{special unitary group} denoted
SU(E,g) is its subgroup comprised of elements with det(f) = 1.
\end{definition}

\bigskip

\subsection{Scalar product on a vector space}

Many interesting properties of vector spaces occur when there is some non
degenerate bilinear form defined on them.

\label{Inner product space}

There are 4 mains results : existence of orthonormal basis, partition of the
vector space, orthogonal complement and isomorphism with the dual.

\subsubsection{Definitions}

\begin{definition}
A \textbf{scalar product} on a vector space E on a field K is either a non
degenerate, bilinear symmetric form g, or a non degenerate hermitian
sesquilinear form g. This is an \textbf{inner product} if g is definite positive.
\end{definition}

If g is definite positive then g defines a metric and a norm over E and E is a
normed vector space (see Topology). Moreover if E is complete (which happens
if E is finite dimensional), it is a Hilbert space.\ If K=$%
\mathbb{R}
$\ then E is an \textbf{euclidean space}.

If the vector space is finite dimensional the matrix $\left[  g\right]  $ is
symmetric or hermitian\ and its eigen values are all distinct, real and non
zero. Their signs defines the \textbf{signature} of g, denoted (p,q) for p (+)
and q (-). g is definite positive iff all the eigen values are
$>$%
0.

If $K=%
\mathbb{R}
$ then the p in the signature of g is the maximum dimension of the vector
subspaces where g is definite positive

With E 4 real dimensional and g the Lorentz metric of signature + + + - E is
the Minkowski space of Special Relativity (remark : if the chosen signature is
- - - +, all the following results still stand with the appropriate adjustments).

\begin{definition}
An \textbf{isometry} is a linear map $f\in L(E;F)$ between two vector spaces
$\left(  E,g\right)  $,$\left(  F,h\right)  $ endowed with scalar products,
which preserves the scalar product : $\forall u,v\in E,g\left(  f\left(
u\right)  ,f\left(  v\right)  \right)  =h\left(  u,v\right)  $
\end{definition}

\subsubsection{Induced scalar product}

Let be F a vector subspace, and define the form $h:F\times F\rightarrow
K::\forall u,v\in F:h\left(  u,v\right)  =g\left(  u,v\right)  $ that is the
restriction of g to F. h has the same linearity or anti-linearity as g. If F
is defined by the n$\times$r matrix A ($u\in F\Leftrightarrow\left[  u\right]
=\left[  A\right]  \left[  x\right]  )$, then h has the matrix $\left[
H\right]  =\left[  A\right]  ^{t}\left[  g\right]  \left[  A\right]  .$

If g is definite positive, so is h and (F,h) is endowed with an inner product
induced by g on F

If not, h can be degenerate, because there are vector subspaces of
null-vectors, and its signature is usually different

\subsubsection{Orthonormal basis}

\begin{definition}
In a vector space endowed (E,g) with a scalar product\ :

Two vectors u,v are \textbf{orthogonal} if g(u,v)=0.

A vector u and a subset A are orthogonal if $\forall v\in A,g\left(
u,v\right)  =0.$\ 

Two subsets A and B\ are orthogonal if $\forall u\in A,v\in B,g\left(
u,v\right)  =0$
\end{definition}

\begin{definition}
A basis $\left(  e_{i}\right)  _{i\in I}$ of a vector space (E,g) endowed with
a scalar product is :

\textbf{orthogonal} if : $\forall i\neq j\in I:g\left(  e_{i},e_{j}\right)
=0$

\textbf{orthonormal}\ if $\forall i,j\in I:g\left(  e_{i},e_{j}\right)
=\pm\delta_{ij}$.
\end{definition}

Notice that we do not require $g\left(  e_{i},e_{j}\right)  =1$

\bigskip

\begin{theorem}
A finite dimensional vector space (E,g) endowed with a scalar product has
orthonormal bases. If E is euclidian $g\left(  e_{i},e_{j}\right)
=\delta_{ij}.$ If $K=%
\mathbb{C}
$\ it is always possible to choose the basis such that $g\left(  e_{i}%
,e_{j}\right)  =\delta_{ij}.$
\end{theorem}

\begin{proof}
the matrix $\left[  g\right]  $ is hermitian so it is diagonalizable : there
are matrix $P$ either orthogonal or unitary such that $\left[  g\right]
=\left[  P\right]  ^{-1}\left[  \Lambda\right]  \left[  P\right]  $ with
$\left[  P\right]  ^{-1}=\left[  P\right]  ^{\ast}=\overline{\left[  P\right]
}^{t}$ and $\left[  \Lambda\right]  =Diag(\lambda_{1},...\lambda_{n})$ the
diagonal matrix with the eigen values of P which are all real.

In a change of basis with new components given by $\left[  P\right]  $ , the
form is expressed in the matrix $\left[  \Lambda\right]  $

If $K=%
\mathbb{R}
$ take as new basis $\left[  P\right]  \left[  D\right]  $ with $\left[
D\right]  =Diag\left(  sgn\left(  \lambda_{i}\right)  \sqrt{\left\vert
\lambda_{i}\right\vert }\right)  .$

If $K=%
\mathbb{C}
$ take as new basis $\left[  P\right]  \left[  D\right]  $ with $\left[
D\right]  =Diag\left(  \mu_{i}\right)  ,$ with $\mu_{i}=\sqrt{\left\vert
\lambda_{i}\right\vert }$ if $\lambda_{i}>0,\mu_{i}=i\sqrt{\left\vert
\lambda_{i}\right\vert }$ if $\lambda_{i}<0$
\end{proof}

In an orthonormal basis g takes the following form (expressed in the
components on this basis):

If $K=%
\mathbb{R}
:g\left(  u,v\right)  =\sum_{i=1}^{n}\epsilon_{i}u_{i}v_{i}$ with
$\epsilon_{i}=\pm1$

If $K=%
\mathbb{C}
:g\left(  u,v\right)  =\sum_{i=1}^{n}\overline{u}_{i}v_{i}$

(remember that $u_{i},v_{i}\in K)$

Warning ! the integers p,q of the signature (p,q) are related to the number of
eigen values of each sign, not to the index of the vector of a basis. In an
orthonormal basis we have always $g\left(  e_{i},e_{j}\right)  =\pm1$\ but not
necessarily $g\left(  u,v\right)  =\sum_{i=1}^{p}u_{i}v_{i}-\sum_{i=p+1}%
^{p+q}u_{i}v_{i}$

\begin{notation}
$\eta_{ij}=\pm1$ denotes usually the product $g\left(  e_{i},e_{j}\right)  $
for an orthonormal basis and $\left[  \eta\right]  $ is the diagonal matrix
$\left[  \eta_{ij}\right]  $
\end{notation}

As a consequence (take orthonormal basis in each vector space):

- all complex vector spaces with hermitian non degenerate form and the same
dimension are isometric.

- all real vector spaces with symmetric bilinear form of identical signature
and the same dimension are isometric.

\subsubsection{Time like and space like vectors}

The quantity g(u,u) is always real, it can be
$>$
0,
$<$
0,or 0. The sign does not depend on the basis. So one distinguishes the
vectors according to the sign of g(u,u) :

- \textbf{time-like vectors} : g(u,u)
$<$
0

- \textbf{space-like vectors} : g(u,u)
$>$
0

- \textbf{null vectors} : g(u,u) = 0

Remark : with the Lorentz metric the definition varies with the basic
convention used to define g. The definitions above hold with the signature + +
+ - . With the signature - - - + time-like vectors are such that g(u,u)
$>$
0.

The sign does no change if one takes $u\rightarrow ku,k>0$ so these sets of
vectors are half-cones. The cone of null vectors is commonly called the
light-cone (as light rays are null vectors).

The following theorem is new.

\begin{theorem}
If g has the signature (+p,-q) a vector space (E,g) endowed with a scalar
product is partitioned in 3 subsets :

$E_{+}:$space-like vectors, open, arc-connected if p%
$>$%
1, with 2 connected components if p=1

$E_{-}:$ time-like vectors, open, arc-connected if q%
$>$%
1, with 2 connected components if q=1

$E_{0}:$ null vectors, closed, arc-connected
\end{theorem}

Openness an connectedness are topological concepts, but we place this theorem
here as it fits the story.

\begin{proof}
It is clear that the 3 subsets are disjoint and that their union is E. g being
a continuous map $E_{+}$ is the inverse image of an open set, and $E_{0}$ is
the inverse image of a closed set.

For arc-connectedness we will exhibit a continuous path internal to each
subset. Choose an orthonormal basis $\varepsilon_{i}$ (with p+ and q- even in
the complex case). Define the projections over the first p and the last q
vectors of the basis :

$u=\sum_{i=1}^{n}u^{i}\varepsilon_{i}\rightarrow P_{h}(u)=\sum_{i=1}^{p}%
u^{i}\varepsilon_{i};P_{v}(u)=\sum_{i=p+1}^{n}u^{i}\varepsilon_{i}$

and the real valued functions : $f_{h}(u)=g(P_{h}(u),P_{h}(u));f_{v}%
(u)=g(P_{v}(u),P_{v}(u))$

so : $g(u,u)=f_{h}(u)-f_{v}(u)$

Let be $u_{a},u_{b}\in E_{+}:$

$f_{h}(u_{a})-f_{v}(u_{a})>0,f_{h}(u_{b})-f_{v}(u_{b})>0\Rightarrow
f_{h}(u_{a}),f_{h}(u_{b})>0$

Define the path $x(t)\in E$ with 3 steps:

a) $t=0\rightarrow t=1:x(0)=u_{a}\rightarrow x(1)=\left(  u_{a}^{h},0\right)
$

$x(t):i\leq p:x^{i}(t)=u_{a}^{i};$ if p%
$>$%
1 : $i>p:x^{i}(t)=(1-t)u_{a}^{i}$

$g(x(t),x(t))=f_{h}(u_{a})-(1-t)^{2}f_{v}(u_{a})>f_{h}(u_{a})-f_{v}%
(u_{a})=g(u_{a},u_{a})>0 $

$\Rightarrow x(t)\in E_{+}$

b) $t=1\rightarrow t=2:x(1)=\left(  u_{a}^{h},0\right)  \rightarrow
x(1)=\left(  u_{b}^{h},0\right)  $

$x(t):i\leq p:x^{i}(t)=(t-1)u_{b}^{i}+(2-t)u_{a}^{i}=u_{a}^{i};$ if p%
$>$%
1: $i>p:x^{i}(t)=0$

$g(x(t),x(t))=f_{h}((t-1)u_{b}+(2-t)u_{a})>0\Rightarrow x(t)\in E_{+}$

c) $t=2\rightarrow t=3:x(2)=\left(  u_{b}^{h},0\right)  \rightarrow
x(3)=u_{b}$

$x(t):i\leq p:x^{i}(t)=u_{b}^{i};$ if p%
$>$%
1: $i>p:x^{i}(t)=(t-2)u_{b}^{i}$

$g(x(t),x(t))=f_{h}(u_{b})-(t-2)^{2}f_{v}(u_{b})>f_{h}(u_{b})-f_{v}%
(u_{b})=g(u_{b},u_{b})>0\Rightarrow x(t)\in E_{+} $

So if $u_{a},u_{b}\in E_{+},x(t)\subset E_{+}$ whenever p%
$>$%
1.

For $E_{-}$ we have a similar demonstration.

If q=1 one can see that the two regions t%
$<$%
0 and t%
$>$%
0 cannot be joined : the component along $\varepsilon_{n}$\ must be zero for
some t and then g(x(t),x(t))=0

If $u_{a},u_{b}\in E_{0}\Leftrightarrow f_{h}(u_{a})=f_{v}(u_{a}),f_{h}%
(u_{b})=f_{v}(u_{b})$

The path comprises of 2 steps going through 0 :

a) $t=0\rightarrow t=1:x(t)=(1-t)u_{a}\Rightarrow g(x(t))=(1-t)^{2}%
g(u_{a},u_{a})=0$

b) $t=1\rightarrow t=2:x(t)=(t-1)u_{b}\Rightarrow g(x(t))=(1-t)^{2}%
g(u_{b},u_{b})=0$

This path always exists.
\end{proof}

\bigskip

The partition of E in two disconnected components is crucial, because it gives
the distinction between "past oriented" and "future oriented" time-like
vectors (one cannot go from one region to the other without being in trouble).
This theorem shows that the Lorentz metric is special, in that it is the only
one for which this distinction is possible.

One can go a little further. One can show that there is always a vector
subspace F of dimension $\min(p,q)$ such that all its vectors are null
vectors. In the Minkowski space the only null vector subspaces are 1-dimensional.

\subsubsection{Graham-Schmitt's procedure}

The problem is the following : in a finite dimensional vector space (E,g)
endowed with a scalar product, starting from a given basis $\left(
e_{i}\right)  _{i=1}^{n}$ compute an orthonormal basis $\left(  \varepsilon
_{i}\right)  _{i=1}^{n}$

Find a vector of the basis which is not a null-vector.\ If all the vectors of
the basis are null vectors then g=0 on the vector space.

So let be : $\varepsilon_{1}=\frac{1}{g\left(  e_{1},e_{1}\right)  }e_{1}$

Then by recursion : $\varepsilon_{i}=e_{i}-\sum_{j=1}^{i-1}\frac{g\left(
e_{i},\varepsilon_{j}\right)  }{g\left(  \varepsilon_{j},\varepsilon
_{j}\right)  }\varepsilon_{j}$

All the $\varepsilon_{i}$ are linearly independant.\ They are orthogonal :

$g\left(  \varepsilon_{i},\varepsilon_{k}\right)  =g\left(  e_{i}%
,\varepsilon_{k}\right)  -\sum_{j=1}^{i-1}\frac{g\left(  e_{i},\varepsilon
_{j}\right)  }{g\left(  \varepsilon_{j},\varepsilon_{j}\right)  }g\left(
\varepsilon_{j},\varepsilon_{k}\right)  =g\left(  e_{i},\varepsilon
_{k}\right)  -\frac{g\left(  e_{i},\varepsilon_{k}\right)  }{g\left(
\varepsilon_{k},\varepsilon_{k}\right)  }g\left(  \varepsilon_{k}%
,\varepsilon_{k}\right)  =0$

The only trouble that can occur is if for some i :

$g\left(  e_{i},e_{i}\right)  =g\left(  e_{i},e_{i}\right)  -\sum_{j=1}%
^{i-1}\frac{g\left(  e_{i},\varepsilon_{j}\right)  ^{2}}{g\left(
\varepsilon_{j},\varepsilon_{j}\right)  }=0.$

But from the Schwarz inegality :

$g\left(  e_{i},\varepsilon_{j}\right)  ^{2}\leq g\left(  \varepsilon
_{j},\varepsilon_{j}\right)  g\left(  e_{i},e_{i}\right)  $

and, if g is positive definite, equality can occur only if $\varepsilon_{i}$
is a linear combination of the $\varepsilon_{j}.$

So\ if g is positive definite the procedure always works.

\subsubsection{Orthogonal complement}

\begin{definition}
A vector subspace, denoted $F^{\bot},$ of a vector space E endowed with a
scalar product is an \textbf{orthogonal complement} of a vector subspace F of
E if $F^{\bot}$ is orthogonal to F and $E=F\oplus F^{\perp}.$
\end{definition}

If E is finite dimensional there are always orthogonal vector spaces F' and
$\dim F+\dim F^{\prime}=\dim E$ (Knapp p.50) but we have not necessarily
$E=F\oplus F^{\perp}$ (see below) and they are not necessarily unique if g is
not definite positive.

\begin{theorem}
In a vector space endowed with an inner product the orthogonal complement
always exist and is unique.\ 
\end{theorem}

\begin{proof}
To find the orthogonal complement of a vector subspace F start with a basis of
E such that the first r vectors are a basis of F. Then if there is an
orthonormal basis deduced from $\left(  e_{i}\right)  $ the last n-r vectors
are an orthonormal basis of the unique orthogonal complement of F. If g is not
positive definite there is not such guaranty.
\end{proof}

This theorem is important : if F is a vector subspace there is always a vector
space B such that $E=F\oplus B$ but B is not unique.\ This decomposition is
useful for many purposes, and it is an hindrance when B cannot be defined more
precisely.\ This is just what g does : $A^{\bot}$\ is the orthogonal
projection of A. But the theorem is not true if g is not definite positive.\ 

\bigskip

\subsection{Symplectic vector spaces}

\label{Symplectic vector space}

If the symmetric bilinear form is replaced by an antisymmetric form we get a
symplectic structure. In many ways the results are similar, All symplectic
vector spaces of same dimension are isomorphic. Symplectic spaces are commonly
used in lagrangian mechanics.

\subsubsection{Definitions}

\begin{definition}
A \textbf{symplectic vector space} (E,h) is a real vector space E endowed with
a non degenerate antisymmetric 2-form h called the \textbf{symplectic}
\textbf{form}
\end{definition}

$\forall u,v\in E:h\left(  u,v\right)  =-h\left(  v,u\right)  \in%
\mathbb{R}
$

$\forall u\in E:\forall v\in E:h\left(  u,v\right)  =0\Rightarrow u=0$

\begin{definition}
2 vectors u,v of a symplectic vector space (E,h) are \textbf{orthogonal} if h(u,v)=0.
\end{definition}

\begin{theorem}
The set of vectors orthogonal to all vectors of a vector subspace F of a
symplectic vector space is a vector subspace denoted $F^{\perp}$
\end{theorem}

\begin{definition}
A vector subspace is :

\textbf{isotropic} if $F^{\perp}\subset F$

\textbf{co-isotropic} if $F\subset F^{\perp}$

\textbf{self-orthogonal} if $F^{\perp}=F$
\end{definition}

The 1-dimensional vector subspaces are isotropic

An isotropic vector subspace is included in a self-orthogonal vector subspace

\begin{theorem}
The symplectic form of symplectic vector space (E,h) induces a map $j:E^{\ast
}\rightarrow E::\lambda\left(  u\right)  =h\left(  j\left(  \lambda\right)
,u\right)  $ which is an isomorphism iff E is finite dimensional.
\end{theorem}

\subsubsection{Canonical basis}

The main feature of symplectic vector spaces if that they admit basis in which
any symplectic form is represented by the same matrix. So all symplectic
vector spaces of the same dimension are isomorphic.

\begin{theorem}
(Hofer p.3) A symplectic (E,h) finite dimensional vector space must have an
even dimension n=2m. There are always canonical bases $\left(  \varepsilon
_{i}\right)  _{i=1}^{n}$ such that $h\left(  \varepsilon_{i},\varepsilon
_{j}\right)  =0,\forall\left\vert i-j\right\vert <m,h\left(  \varepsilon
_{i},\varepsilon_{j}\right)  =\delta_{ij},\forall\left\vert i-j\right\vert
>m.$ All finite dimensional symplectic vector space of the same dimension are isomorphic.
\end{theorem}

h reads in any basis : $h\left(  u,v\right)  =\left[  u\right]  ^{t}\left[
h\right]  \left[  v\right]  ,$\ with $\left[  h\right]  =\left[
h_{ij}\right]  $ skew-symmetric and det(h)$\neq0$.

In a canonical basis:%

\begin{equation}
\left[  h\right]  =J_{m}=%
\begin{bmatrix}
0 & I_{m}\\
-I_{m} & 0
\end{bmatrix}
soJ_{m}^{2}=-I_{2m}%
\end{equation}

$h\left(  u,v\right)  =\left[  u\right]  ^{t}\left[  J_{m}\right]  \left[
v\right]  =\sum_{i=1}^{m}\left(  u_{i}v_{i+m}-u_{i+m}v_{i}\right)  $

The vector subspaces \ $E_{1}$\ spanned by $\left(  \varepsilon_{i}\right)
_{i=1}^{m}$, $E_{2}$\ spanned by $\left(  \varepsilon_{i}\right)
_{i=m+1}^{2m}$ are self-orthogonal and $E=E_{1}\oplus E$

\subsubsection{Symplectic maps}

\begin{definition}
A symplectic map (or \textbf{symplectomorphism}) between two symplectic vector
spaces (E$_{1}$,h$_{1}$),(E$_{2}$,h$_{2}$), is a linear map $f\in L\left(
E_{1};E_{2}\right)  $ such that \ $\forall u,v\in E_{1}:h_{2}\left(  f\left(
u\right)  ,f\left(  v\right)  \right)  =h_{1}\left(  u,v\right)  $
\end{definition}

f is injective so $\dim E_{1}\leq\dim E_{2}$

\begin{theorem}
(Hofer p.6) There is always a bijective symplectomorphism between two
symplectic vector spaces (E$_{1}$,h$_{1}$),(E$_{2}$,h$_{2}$) of the same dimension
\end{theorem}

\begin{definition}
A symplectic map (or \textbf{symplectomorphism}) of a symplectic vector space
(E,h) is an endomorphism of E which preserves the symplectic form h : $f\in
L\left(  E;E\right)  :h\left(  f\left(  u\right)  ,f\left(  v\right)  \right)
=h\left(  u,v\right)  $
\end{definition}

\begin{theorem}
The symplectomorphisms over a symplectic vector space (E,h) constitute the
\textbf{symplectic group} Sp(E,h).
\end{theorem}

\textit{In a canonical basis} a symplectomorphism is represented by a
symplectic matrix A which is such that :

$A^{t}J_{m}A=J_{m}$

because : $h\left(  f\left(  u\right)  ,f\left(  v\right)  \right)  =\left(
A\left[  u\right]  \right)  ^{t}J_{m}\left[  A\left[  v\right]  \right]
=\left[  u\right]  ^{t}A^{t}J_{m}A\left[  v\right]  =\left[  u\right]
^{t}J_{m}\left[  v\right]  $

so $\det A=1$

\begin{definition}
The \textbf{symplectic group} Sp(2m) is the linear group of 2m$\times$2m real
matrices A such that : $A^{t}J_{m}A=J_{m}$
\end{definition}

$A\in Sp\left(  2m\right)  \Leftrightarrow A^{-1},A^{t}\in Sp\left(
2m\right)  $

\subsubsection{Liouville form}

\begin{definition}
The \textbf{Liouville form} on a 2m dimensional symplectic vector space (E,h)
is the 2m form : $\varpi=\frac{1}{m!}h\wedge h\wedge..\wedge h$ (m times).
Symplectomorphisms preserve the Liouville form.
\end{definition}

In a canonical basis :

$\varpi=\varepsilon^{1}\wedge\varepsilon^{m+1}\wedge..\wedge\varepsilon
^{m}\wedge\varepsilon^{2m}$

\begin{proof}
Put : $h=\sum_{i=1}^{m}\varepsilon^{i}\wedge\varepsilon^{i+m}=\sum_{i=1}%
^{m}h_{i}$

$h_{i}\wedge h_{j}=0$ if i=j so $\left(  \wedge h\right)  ^{m}=\sum_{\sigma
\in\mathfrak{S}_{m}}h_{\sigma\left(  1\right)  }\wedge h_{\sigma\left(
2\right)  }..\wedge h_{\sigma\left(  m\right)  }$

remind that : $h_{\sigma\left(  1\right)  }\wedge h_{\sigma\left(  2\right)
}=\left(  -1\right)  ^{2\times2}h_{\sigma\left(  2\right)  }\wedge
h_{\sigma\left(  1\right)  }=h_{\sigma\left(  2\right)  }\wedge h_{\sigma
\left(  1\right)  }$

$\left(  \wedge h\right)  ^{m}=m!\sum_{\sigma\in\mathfrak{S}_{m}}h_{1}\wedge
h_{2}..\wedge h_{m}$
\end{proof}

\subsubsection{Complex structure}

\begin{theorem}
A finite dimensional symplectic vector space (E,h) admits a complex structure
\end{theorem}

Take a canonical basis and define :

$J:E\rightarrow E::J\left(  \sum_{i=1}^{m}u_{i}\varepsilon_{i}+v_{i}%
\varphi_{i}\right)  =\sum_{i=1}^{m}\left(  -v_{i}\varepsilon_{i}+u_{i}%
\varphi_{i}\right)  $

So $J^{2}=-Id$ (see below)

It sums up to take as complex basis :$\left(  \varepsilon_{j},i\varepsilon
_{j+m}\right)  _{j=1}^{m}$ with complex components.\ Thus E becomes a
m-dimensional complex vector space.

\bigskip

\subsection{Complex vector spaces}

\label{Complex vector space}

Complex vector spaces are vector spaces over the field $%
\mathbb{C}
$ .\ They share all the properties listed above, but have some specificities
linked to :

- passing from a vector space over $%
\mathbb{R}
$ to a vector space over $%
\mathbb{C}
$ and vice versa

- the definition of the conjugate of a vector

\bigskip

\subsubsection{From complex to real vector space}

In a complex vector space E the restriction of the multiplication by a scalar
to real scalars gives a real vector space, but as a set one must distinguish
the vectors u and iu : we need some rule telling which are "real" vectors and
"imaginary" vectors \textit{in the same set of vectors}.

There is always a solution but it is not unique and depends on a specific map.

\paragraph{Real structure\newline}

\begin{definition}
A \textbf{real structure} on a complex vector space E is a map :
$\sigma:E\rightarrow E$ which is \textit{antilinear} and such that $\sigma
^{2}=Id_{E}:$
\end{definition}

$z\in%
\mathbb{R}
,u\in E,\sigma\left(  zu\right)  =\overline{z}\sigma\left(  u\right)
\Rightarrow\sigma^{-1}=\sigma$

\begin{theorem}
There is always a real structure $\sigma$ on a complex vector space E. Then E
is the direct sum of two real vector spaces : $E=E_{%
\mathbb{R}
}\oplus iE_{%
\mathbb{R}
}$ where $E_{%
\mathbb{R}
}$ , called the \textbf{real kernel} of $\sigma,$\ is the subset of vectors
invariant by $\sigma$
\end{theorem}

\begin{proof}
i) Take any (complex) basis $\left(  e_{j}\right)  _{j\in I}$ of E and define
the map $:\sigma\left(  e_{j}\right)  =e_{j},\sigma\left(  ie_{j}\right)
=-ie_{j}$

$\forall u\in E:u=\sum_{j\in I}z_{j}e_{j}\rightarrow\sigma\left(  u\right)
=\sum_{j\in I}\overline{z}_{j}e_{j}$

$\sigma^{2}\left(  u\right)  =\sum_{j\in I}z_{j}e_{j}=\sigma\left(  u\right)
$

It is antilinear :

$\sigma\left(  (a+ib)u\right)  =\sigma\left(  \sum_{j\in I}(a+ib)z_{j}%
e_{j}\right)  =(a-ib)\sum_{j\in I}\sigma\left(  z_{j}e_{j}\right)
=(a-ib)\sigma\left(  u\right)  $

This structure is not unique and depends on the choice of a basis.

ii) Define $E_{%
\mathbb{R}
}$\ as the subset of vectors of E invariant by $\sigma:E_{%
\mathbb{R}
}=\left\{  u\in E:\sigma\left(  u\right)  =u\right\}  .$

It is not empty : with the real structure above any vector with real
components in the basis $\left(  e_{j}\right)  _{j\in I}$\ belongs to $E_{%
\mathbb{R}
}$

It is a \textit{real vector subspace} of E. Indeed the multiplication by a
real scalar gives : $ku=\sigma\left(  ku\right)  \in E_{%
\mathbb{R}
}.$

iii) Define the maps :

$\operatorname{Re}:E\rightarrow E_{%
\mathbb{R}
}::\operatorname{Re}u=\frac{1}{2}\left(  u+\sigma\left(  u\right)  \right)  $

$\operatorname{Im}:E\rightarrow E_{%
\mathbb{R}
}::\operatorname{Im}u=\frac{1}{2i}\left(  u-\sigma\left(  u\right)  \right)  $

Any vector can be uniquely written with a real and imaginary part : $u\in
E:u=\operatorname{Re}u+i\operatorname{Im}u$ which both belongs to the real
kernel of E. Thus : $E=E_{%
\mathbb{R}
}\oplus iE_{%
\mathbb{R}
}$
\end{proof}

E can be seen as a real vector space with two fold the dimension of E :
$E_{\sigma}=E_{%
\mathbb{R}
}\times iE_{%
\mathbb{R}
}$

\paragraph{Conjugate\newline}

Warning ! The definition of the conjugate of a vector makes sense only iff E
is a complex vector space endowed with a real structure.

\begin{definition}
The \textbf{conjugate of a vector} u on a complex vector space E endowed with
a real structure $\sigma$\ is $\sigma\left(  u\right)  =\overline{u}$
\end{definition}

$\overline{}:E\rightarrow E:\overline{u}=\operatorname{Re}u-i\operatorname{Im}%
u=\frac{1}{2}\left(  u+\sigma\left(  u\right)  \right)  -i\frac{1}{2i}\left(
u-\sigma\left(  u\right)  \right)  =\sigma\left(  u\right)  $

This definition is valid whatever the dimension of E. And as one can see
conjugation is an involution on E, and $\overline{E}=E:$ they are the same
set$.$

\paragraph{Real form\newline}

\begin{definition}
A real vector space F is a \textbf{real form} of a complex vector space E if F
is a real vector subspace of E and there is a real structure $\sigma$ on E for
which F is invariant by $\sigma.$
\end{definition}

Then E can be written as : $E=F\oplus iF$

As any complex vector space has real structures, there are always real forms,
which are not unique.

\subsubsection{From real to complex vector space}

There are two different ways for endowing a real vector space with a complex
vector space structure.

\bigskip

\paragraph{Complexification\newline}

The simplest, and the most usual, way is to enlarge the real vector space
itself (as a set). This is always possible and called
\textbf{complexification.}

\begin{theorem}
For any real vector space E there is a structure $E_{%
\mathbb{C}
}$ of complex vector space on E$\times$E, called the \textbf{complexification}
of E, such that $E_{%
\mathbb{C}
}=E\oplus iE$
\end{theorem}

\begin{proof}
ExE is a real vector space with the usual operations :

$\forall u,v,u^{\prime},v^{\prime}\in E,k\in%
\mathbb{R}
:\left(  u,v\right)  +\left(  u^{\prime},v^{\prime}\right)  =\left(
u+u^{\prime},v+v^{\prime}\right)  ;k(u,v)=(ku,kv)$

We add the operation : $i\left(  u,v\right)  =\left(  -v,u\right)  $. Then :

$z=a+ib\in%
\mathbb{C}
:z\left(  u,v\right)  =\left(  au-vb,av+bu\right)  \in E\times E$

$i\left(  i\left(  u,v\right)  \right)  =i\left(  -v,u\right)  =-\left(
u,v\right)  $

ExE becomes a vector space $E_{%
\mathbb{C}
}$ over $%
\mathbb{C}
$\ . This is obvious if we denote : $\left(  u,v\right)  =u+iv$

The direct sum of two vector spaces can be identified with a product of these
spaces, so $E_{%
\mathbb{C}
}$ is defined as :

$E_{%
\mathbb{C}
}=E\oplus iE\Leftrightarrow\forall u\in E_{%
\mathbb{C}
},\exists v,w$ unique $\in E:u=v+iw$ or $u=\operatorname{Re}%
u+i\operatorname{Im}u$ with $\operatorname{Re}u,\operatorname{Im}u\in E$
\end{proof}

So E and iE are real vector subspaces of $E_{%
\mathbb{C}
}.$

The map : $\sigma:E_{%
\mathbb{C}
}\rightarrow E_{%
\mathbb{C}
}::\sigma\left(  \operatorname{Re}u+i\operatorname{Im}u\right)
=\operatorname{Re}u-i\operatorname{Im}u$ is antilinear and E, iE are real
forms of $E_{%
\mathbb{C}
}.$

The conjugate of a vector of $E_{%
\mathbb{C}
}$ is $\overline{u}=\sigma\left(  u\right)  $

Remark : the complexified is often defined as $E_{%
\mathbb{C}
}=E\otimes_{R}%
\mathbb{C}
$ the tensoriel product being understood as acting over $%
\mathbb{R}
.$ The two definitions are equivalent, but the second is less enlighting...

\begin{theorem}
Any basis $\left(  e_{j}\right)  _{j\in I}$ of a real vector space E is a
basis of the complexified $E_{%
\mathbb{C}
}$ with complex components. $E_{%
\mathbb{C}
}$ has same complex dimension as E.
\end{theorem}

As a set $E_{%
\mathbb{C}
}$ is "larger" than E : indeed it is defined through ExE, the vectors
$e_{i}\in E,$ and $ie_{j}\in E_{%
\mathbb{C}
}$ but $ie_{j}\notin E.$To define a vector in $E_{%
\mathbb{C}
}$ we need two vectors in E.\ However $E_{%
\mathbb{C}
}$ has the same \textit{complex} dimension as the real vector space E :a
complex component needs two real scalars.

\begin{theorem}
Any linear map $f\in L\left(  E;F\right)  $ between real vector spaces has a
unique prolongation $f_{%
\mathbb{C}
}\in L\left(  E_{%
\mathbb{C}
};F_{%
\mathbb{C}
}\right)  $
\end{theorem}

\begin{proof}
i) If $f_{%
\mathbb{C}
}\in L\left(  E_{%
\mathbb{C}
};F_{%
\mathbb{C}
}\right)  $ is a C-linear map : $f_{%
\mathbb{C}
}\left(  u+iv\right)  =f_{%
\mathbb{C}
}\left(  u\right)  +if_{%
\mathbb{C}
}\left(  v\right)  $ and if it is the prolongation of f : $f_{%
\mathbb{C}
}\left(  u\right)  =f\left(  u\right)  ,f_{%
\mathbb{C}
}\left(  v\right)  =f\left(  v\right)  $

ii) $f_{%
\mathbb{C}
}\left(  u+iv\right)  =f\left(  u\right)  +if\left(  v\right)  $ is C-linear
and the obvious prolongation of f.
\end{proof}

If $f\in L\left(  E;E\right)  $\ has $\left[  f\right]  $\ for matrix in the
basis $\left(  e_{i}\right)  _{i\in I}$\ then its extension $f_{c}\in L\left(
E_{%
\mathbb{C}
};E_{%
\mathbb{C}
}\right)  $\ has the \textit{same} matrix in the basis $\left(  e_{i}\right)
_{i\in I}.$. This is exactly what is done to compute the complex eigen values
of a real matrix.

Notice that $L\left(  E_{%
\mathbb{C}
};E_{%
\mathbb{C}
}\right)  \neq\left(  L\left(  E;E\right)  \right)  _{%
\mathbb{C}
}$ which is the set : $\left\{  F=f+ig,f,g\in L(E;E)\right\}  $ of maps from E
to $E_{%
\mathbb{C}
}$

Similarly $\left(  E_{%
\mathbb{C}
}\right)  ^{\ast}=\left\{  F;F(u+iv)=f(u)+if(v),f\in E^{\ast}\right\}  $

and $\left(  E^{\ast}\right)  _{%
\mathbb{C}
}=\left\{  F=f+ig,f,g\in E^{\ast}\right\}  $

\bigskip

\paragraph{Complex structure\newline}

The second way leads to define a complex vector space structure $E_{%
\mathbb{C}
}$ on the \textit{same} set E :

i) the sets are the same : if u is a vector of E it is a vector of $E_{%
\mathbb{C}
}$ and vice versa

ii) the operations (sum and product by a scalar) defined in $E_{%
\mathbb{C}
}$ are closed over $%
\mathbb{R}
$ and $%
\mathbb{C}
$

So the goal is to find a way to give a meaning to the operation : $%
\mathbb{C}
\times E\rightarrow E$ and it would be enough if there is an operation with
$i\times E\rightarrow E$

This is not always possible and needs the definition of a special map.

\begin{definition}
A complex structure on a real vector space is a linear map $J\in L\left(
E;E\right)  $ such that $J^{2}=-Id_{E}$
\end{definition}

\begin{theorem}
A real vector space can be endowed with the structure of a complex vector
space iff there is a complex structure.
\end{theorem}

\begin{proof}
a) the condition is necessary : if E has the structure of a complex vector
space then the map : $J:E\rightarrow E::J\left(  u\right)  =iu$ is well
defined and $J^{2}=-Id$

b) the condition is sufficient : what we need is to define the multiplication
by i such that it is a complex linear operation. Define on E : $iu=J\left(
u\right)  .$ Then $i\times i\times u=-u=J\left(  J\left(  u\right)  \right)
=J^{2}\left(  u\right)  =-u$
\end{proof}

\begin{theorem}
A real vector space has a complex structure iff it has a dimension which is
infinite countable or finite even.
\end{theorem}

\begin{proof}
i) Let us assume that E has a complex structure, then it can be made a complex
vector space and $E=E_{%
\mathbb{R}
}\oplus iE_{%
\mathbb{R}
}.$ The two real vector spaces $E_{%
\mathbb{R}
},iE_{%
\mathbb{R}
}$ are real isomorphic and have same dimension, so $\dim E=2\dim E_{%
\mathbb{R}
}$ is either infinite or even

ii) Pick any basis $\left(  e_{i\in I}\right)  _{i\in I}$ of E. If E is finite
dimensional or countable we can order I according to the ordinal number, and
define the map :

$J\left(  e_{2k}\right)  =e_{2k+1}$

$J\left(  e_{2k+1}\right)  =-e_{2k}$

It meets the condition :

$J^{2}\left(  e_{2k}\right)  =J\left(  e_{2k+1}\right)  =-e_{2k}$

$J^{2}\left(  e_{2k+1}\right)  =-J\left(  e_{2k}\right)  =-e_{2k+1}$

So any vector of $E_{J}$ can be written as :

$u=\sum_{k\in I}u_{k}e_{k}=\sum u_{2k}e_{2k}+\sum u_{2k+1}e_{2k+1}=\sum
u_{2k}e_{2k}-\sum u_{2k+1}J\left(  e_{2k}\right)  =\sum\left(  u_{2k}%
-iu_{2k+1}\right)  e_{2k}=\sum\left(  -iu_{2k}+u_{2k+1}\right)  e_{2k+1}$

A basis of the complex structure is then either $e_{2k}$ or $e_{2k+1}$
\end{proof}

Remark : this theorem can be extended to the case (of scarce usage !) of
uncountable dimensional vector spaces, but this would involve some hypotheses
about the set theory which are not always assumed.

The complex dimension of the complex vector space is half the real dimension
of E if E is finite dimensional, equal to the dimension of E if E has a
countable infinite dimension.

Contrary to the complexification it is not always possible to extend a real
linear map $f\in L\left(  E;E\right)  $ to a complex linear map. It must be
complex linear : $f\left(  iu\right)  =if\left(  u\right)  \Leftrightarrow
f\circ J\left(  u\right)  =J\circ f\left(  u\right)  $ so it must commute with
J : $J\circ f=f\circ J.$ If so then $f\in L\left(  E_{%
\mathbb{C}
};E_{%
\mathbb{C}
}\right)  $ but it is not represented by the same matrix in the complex basis.

\subsubsection{Real linear and complex linear maps}

\paragraph{Real linear maps\newline}

\begin{definition}
Let E,F be two complex vector spaces.\ A map $f:E\rightarrow F$ is
\textbf{real linear} if :

$\forall u,v\in E,\forall k\in%
\mathbb{R}
:f\left(  u+v\right)  =f\left(  u\right)  +f\left(  v\right)  ;f\left(
ku\right)  =kf\left(  u\right)  $
\end{definition}

A real linear map (or R-linear map) is then a complex-linear maps (that is a
linear map according to our definition) iff :

$\forall u\in E:$ $f\left(  iu\right)  =if\left(  u\right)  $

Notice that these properties do not depend on the choice of a real structure
on E or F.

\begin{theorem}
If E is a real vector space, F a complex vector space, a real linear map :
$f:E\rightarrow F$ can be uniquely extended to a linear map : $f_{%
\mathbb{C}
}:E_{%
\mathbb{C}
}\rightarrow F$ where $E_{%
\mathbb{C}
}$ is the complexification of E.
\end{theorem}

\begin{proof}
Define : $f_{%
\mathbb{C}
}\left(  u+iv\right)  =f\left(  u\right)  +if\left(  v\right)  $
\end{proof}

\paragraph{Cauchy identities\newline}

A complex linear map f between complex vector spaces endowed with real
structures, must meet some specific identities, which are called (in the
homolorphic map context) the Cauchy identities.

\begin{theorem}
A linear map $f:E\rightarrow F$ between two complex vector spaces endowed with
real structures can be written :

$f\left(  u\right)  =P_{x}\left(  \operatorname{Re}u\right)  +P_{y}%
(\operatorname{Im}u)+i\left(  Q_{x}\left(  \operatorname{Re}u\right)
+Q_{y}(\operatorname{Im}u)\right)  $ where $P_{x},P_{y},Q_{x},Q_{y}$ are real
linear maps between the real kernels $E_{%
\mathbb{R}
},F_{%
\mathbb{R}
}$ which satisfy the identities%

\begin{equation}
P_{y}=-Q_{x};Q_{y}=P_{x}%
\end{equation}

\end{theorem}

\begin{proof}
Let $\sigma,\sigma^{\prime}$ be the real structures on E,F

Using the sums : $E=E_{%
\mathbb{R}
}\oplus iE_{%
\mathbb{R}
},F=F_{%
\mathbb{R}
}\oplus iF_{%
\mathbb{R}
}$ one can write for any vector u of E :

$\operatorname{Re}u=\frac{1}{2}\left(  u+\sigma\left(  u\right)  \right)  $

$\operatorname{Im}u=\frac{1}{2i}\left(  u-\sigma\left(  u\right)  \right)  $

$f\left(  \operatorname{Re}u+i\operatorname{Im}u\right)  =f\left(
\operatorname{Re}u\right)  +if\left(  \operatorname{Im}u\right)  $

$=\operatorname{Re}f\left(  \operatorname{Re}u\right)  +i\operatorname{Im}%
f\left(  \operatorname{Re}u\right)  +i\operatorname{Re}f\left(
\operatorname{Im}u\right)  -\operatorname{Im}f\left(  \operatorname{Im}%
u\right)  $

$P_{x}\left(  \operatorname{Re}u\right)  =\operatorname{Re}f\left(
\operatorname{Re}u\right)  =\frac{1}{2}\left(  f\left(  \operatorname{Re}%
u\right)  +\sigma^{\prime}\left(  f\left(  \operatorname{Re}u\right)  \right)
\right)  $

$Q_{x}\left(  \operatorname{Re}u\right)  =\operatorname{Im}f\left(
\operatorname{Re}u\right)  =\frac{1}{2i}\left(  f\left(  \operatorname{Re}%
u\right)  -\sigma^{\prime}\left(  f\left(  \operatorname{Re}u\right)  \right)
\right)  $

$P_{y}(\operatorname{Im}u)=-\operatorname{Im}f\left(  \operatorname{Im}%
u\right)  =\frac{i}{2}\left(  f\left(  \operatorname{Im}u\right)
-\sigma^{\prime}f\left(  \operatorname{Im}u\right)  \right)  $

$Q_{y}\left(  \operatorname{Im}u\right)  =\operatorname{Re}f\left(
\operatorname{Im}u\right)  =\frac{1}{2}\left(  f\left(  \operatorname{Im}%
u\right)  +\sigma^{\prime}f\left(  \operatorname{Im}u\right)  \right)  $

So : $f\left(  \operatorname{Re}u+i\operatorname{Im}u\right)  =P_{x}\left(
\operatorname{Re}u\right)  +P_{y}(\operatorname{Im}u)+i\left(  Q_{x}\left(
\operatorname{Re}u\right)  +Q_{y}(\operatorname{Im}u)\right)  $

As f is complex linear :

$f\left(  i\left(  \operatorname{Re}u+i\operatorname{Im}u\right)  \right)
=f\left(  -\operatorname{Im}u+i\operatorname{Re}u\right)  =if\left(
\operatorname{Re}u+i\operatorname{Im}u\right)  $

which gives the identities :

$f\left(  -\operatorname{Im}u+i\operatorname{Re}u\right)  =P_{x}\left(
-\operatorname{Im}u\right)  +P_{y}(\operatorname{Re}u)+i\left(  Q_{x}\left(
-\operatorname{Im}u\right)  +Q_{y}(\operatorname{Re}u\right)  $

$if\left(  \operatorname{Re}u+i\operatorname{Im}u\right)  =iP_{x}\left(
\operatorname{Re}u\right)  +iP_{y}(\operatorname{Im}u)-Q_{x}\left(
\operatorname{Re}u\right)  -Q_{y}(\operatorname{Im}u)$

$P_{x}\left(  -\operatorname{Im}u\right)  +P_{y}(\operatorname{Re}%
u)=-Q_{x}\left(  \operatorname{Re}u\right)  -Q_{y}(\operatorname{Im}u)$

$Q_{x}\left(  -\operatorname{Im}u\right)  +Q_{y}(\operatorname{Re}%
u)=P_{x}\left(  \operatorname{Re}u\right)  +P_{y}(\operatorname{Im}u)$

$P_{y}(\operatorname{Re}u)=-Q_{x}\left(  \operatorname{Re}u\right)  $

$Q_{y}(\operatorname{Re}u)=P_{x}\left(  \operatorname{Re}u\right)  $

$P_{x}\left(  -\operatorname{Im}u\right)  =-Q_{y}(\operatorname{Im}u)$

$Q_{x}\left(  -\operatorname{Im}u\right)  =P_{y}(\operatorname{Im}u)$
\end{proof}

f can then be written : $f\left(  \operatorname{Re}u+i\operatorname{Im}%
u\right)  =\left(  P_{x}-iP_{y}\right)  \left(  \operatorname{Re}u\right)
+\left(  P_{y}+iP_{x}\right)  (\operatorname{Im}u)$

\paragraph{Conjugate of a map\newline}

\begin{definition}
The \textbf{conjugate} of a linear map $f:E\rightarrow F$ between two complex
vector spaces endowed with real structures $\sigma,\sigma^{\prime} $ is the
map : $\overline{f}=\sigma^{\prime}\circ f\circ\sigma$
\end{definition}

so $\overline{f}\left(  u\right)  =\overline{f\left(  \overline{u}\right)  }.$
Indeed the two conjugations are necessary to ensure that $\overline{f}$ is C-linear.

With the previous notations : $\overline{P}_{x}=P_{x},\overline{P}_{y}=-P_{y}
$

\paragraph{Real maps\newline}

\begin{definition}
A linear map $f:E\rightarrow F$ between two complex vector spaces endowed with
real structures is \textbf{real} if it maps a real vector of E to a real
vector of F.
\end{definition}

$\operatorname{Im}u=0\Rightarrow f\left(  \operatorname{Re}u\right)  =\left(
P_{x}-iP_{y}\right)  \left(  \operatorname{Re}u\right)  =P_{x}\left(
\operatorname{Re}u\right)  \Rightarrow P_{y}=Q_{x}=0$

Then $f=\overline{f}$

But conversely a map which is equal to its conjugate is not necessarily real.

\begin{definition}
A multilinear form $f\in L^{r}\left(  E;%
\mathbb{C}
\right)  $ on a complex vector space E, endowed with a real structure $\sigma
$\ is said to be \textbf{real valued} if its value is real whenever it acts on
real vectors.
\end{definition}

A real vector is such that $\sigma\left(  u\right)  =u$

\begin{theorem}
An antilinear map f on a complex vector space E, endowed with a real structure
$\sigma$\ can be uniquely decomposed into two real linear forms.
\end{theorem}

\begin{proof}
Define the real linear forms :

$g\left(  u\right)  =\frac{1}{2}\left(  f\left(  u\right)  +\overline{f\left(
\sigma\left(  u\right)  \right)  }\right)  $

$h\left(  u\right)  =\frac{1}{2i}\left(  f\left(  u\right)  -\overline
{f\left(  \sigma\left(  u\right)  \right)  }\right)  $

$f\left(  u\right)  =g\left(  u\right)  +ih\left(  u\right)  $
\end{proof}

Similarly :

\begin{theorem}
Any sesquilinear form $\gamma$ on a complex vector space E endowed with a real
structure $\sigma$\ can be uniquely defined by a C-bilinear form on E. A
hermitian sesquilinear form $\gamma$ is defined by a C-bilinear form g on E
such that : $g\left(  \sigma u,\sigma v\right)  =\overline{g\left(
v,u\right)  }$
\end{theorem}

\begin{proof}
i) If g is a C-bilinear form on E then : $\gamma\left(  u,v\right)  =g\left(
\sigma u,v\right)  $ defines a sesquilinear form

ii) If g is a C-bilinear form on E such that : $\forall u,v\in E:g\left(
\sigma u,v\right)  =\overline{g\left(  \sigma v,u\right)  }$ then
$\gamma\left(  u,v\right)  =g\left(  \sigma u,v\right)  $ defines a hermitian
sesquilinear form. In a basis with $\sigma\left(  ie_{\alpha}\right)
=-ie_{\alpha}$ g must have components : $g_{\alpha\beta}=\overline
{g_{\beta\alpha}}$

$g\left(  \sigma u,v\right)  =\overline{g\left(  \sigma v,u\right)
}\Leftrightarrow g\left(  \sigma u,\sigma v\right)  =\overline{g\left(
\sigma^{2}v,u\right)  }=\overline{g\left(  v,u\right)  }\Leftrightarrow
\overline{g\left(  \sigma u,\sigma v\right)  }=\overline{g}\left(  u,v\right)
=g\left(  v,u\right)  $

iii) And conversely : $\gamma\left(  \sigma u,v\right)  =g\left(  u,v\right)
$ defines a C-bilinear form on E
\end{proof}

\begin{theorem}
A symmetric bilinear form g on a real vector space E can be extended to a
hermitian, sesquilinear form $\gamma$ on the complexified $E_{%
\mathbb{C}
}$. If g is non degenerate then $\gamma$ is non degenerate. An orthonormal
basis of E for g is an orthonormal basis of $E_{%
\mathbb{C}
}$ for $\gamma.$
\end{theorem}

\begin{proof}
On the complexified $E_{%
\mathbb{C}
}=E\oplus iE$ we define the hermitian, sesquilinear form $\gamma$,
prolongation of g by :

For any $u,v\in E$ :

$\gamma\left(  u,v\right)  =g\left(  u,v\right)  $

$\gamma\left(  iu,v\right)  =-i\gamma\left(  u,v\right)  =-ig\left(
u,v\right)  $

$\gamma\left(  u,iv\right)  =i\gamma\left(  u,v\right)  =ig\left(  u,v\right)
$

$\gamma\left(  iu,iv\right)  =g\left(  u,v\right)  $

$\gamma\left(  u+iv,u^{\prime}+iv^{\prime}\right)  =g\left(  u,u^{\prime
}\right)  +g\left(  v,v^{\prime}\right)  +i\left(  g\left(  u,v^{\prime
}\right)  -g\left(  v,u^{\prime}\right)  \right)  $

$=\overline{\gamma\left(  u^{\prime}+iv^{\prime},u+iv\right)  }$

If $\left(  e_{i}\right)  _{i\in I}$ is an orthonormal basis of E : $g\left(
e_{i},e_{j}\right)  =\eta_{ij}=\pm1$ then $\left(  e_{p}\right)  _{p\in I}$ is
a basis of $E_{%
\mathbb{C}
}$ and it is orthonormal :

$\gamma\left(  \sum_{j\in I}\left(  x_{j}+iy_{j}\right)  e_{j},\sum_{k\in
I}\left(  x_{k}^{\prime}+iy_{k}^{\prime}\right)  e_{k}\right)  $

$=\sum_{j\in I}\eta_{jj}\left(  x_{j}x_{j}^{\prime}+y_{j}y_{j}^{\prime
}+i\left(  x_{j}y_{j}^{\prime}-y_{j}x_{j}^{\prime}\right)  \right)  $

$\gamma\left(  e_{j},e_{k}\right)  =\eta_{jk}$

So the matrix of $\gamma$ in this basis has a non null determinant and
$\gamma$ is not degenerate. It has the same signature as g, but it is always
possible to choose a basis such that $\left[  \gamma\right]  =I_{n}.$
\end{proof}

\begin{theorem}
A symmetric bilinear form g on a real vector space E can be extended to a
symmetric bilinear form $\gamma$ on the complexified $E_{%
\mathbb{C}
}$. If g is non degenerate then $\gamma$ is non degenerate. An orthonormal
basis of E for g is an orthonormal basis of $E_{%
\mathbb{C}
}$ for $\gamma.$
\end{theorem}

\begin{proof}
On the complexified $E_{%
\mathbb{C}
}=E\oplus iE$ we define the form $\gamma$, prolongation of g by :

For any $u,v\in E$ :

$\gamma\left(  u,v\right)  =g\left(  u,v\right)  $

$\gamma\left(  iu,v\right)  =i\gamma\left(  u,v\right)  =ig\left(  u,v\right)
$

$\gamma\left(  u,iv\right)  =i\gamma\left(  u,v\right)  =ig\left(  u,v\right)
$

$\gamma\left(  iu,iv\right)  =-g\left(  u,v\right)  $

$\gamma\left(  u+iv,u^{\prime}+iv^{\prime}\right)  =g\left(  u,u^{\prime
}\right)  -g\left(  v,v^{\prime}\right)  +i\left(  g\left(  u,v^{\prime
}\right)  +g\left(  v,u^{\prime}\right)  \right)  =\gamma\left(  u^{\prime
}+iv^{\prime},u+iv\right)  $

If $\left(  e_{i}\right)  _{i\in I}$ is an orthonormal basis of E : $g\left(
e_{i},e_{j}\right)  =\eta_{ij}=\pm1$ then $\left(  e_{p}\right)  _{p\in I}$ is
a basis of $E_{%
\mathbb{C}
}$ and it is orthonormal :

$\gamma\left(  \sum_{j\in I}\left(  x_{j}+iy_{j}\right)  e_{j},\sum_{k\in
I}\left(  x_{k}^{\prime}+iy_{k}^{\prime}\right)  e_{k}\right)  $

$=\sum_{j\in I}\eta_{jj}\left(  x_{j}x_{j}^{\prime}-y_{j}y_{j}^{\prime
}+i\left(  x_{j}y_{j}^{\prime}+y_{j}x_{j}^{\prime}\right)  \right)  $

$\gamma\left(  e_{j},e_{k}\right)  =\eta_{jk}$

So the matrix of $\gamma$ in this basis has a non null determinant and
$\gamma$ is not degenerate. It has the same signature as g, but it is always
possible to choose a basis such that $\left[  \gamma\right]  =I_{n}.$
\end{proof}

\bigskip

\subsection{Affine Spaces}

\label{Affine space}

Affine spaces are the usual structures of elementary geometry.\ However their
precise definition requires attention.

\subsubsection{Definitions}

\begin{definition}
An \textbf{affine space }$\left(  E,\overrightarrow{E}\right)  $ is a set $E$
with an underlying vector space $\overrightarrow{E}$ over a field K and a map
: $\overset{\rightarrow}{}:E\times E\rightarrow\overrightarrow{E}$ \ such that :

i) $\forall A,B,C\in E:\overrightarrow{AB}+\overrightarrow{BC}+\overrightarrow
{CA}=\overrightarrow{0}$

ii) $\forall A\in E$ fixed the map $\tau_{A}:\overrightarrow{E}\rightarrow
E::\overrightarrow{A\tau_{A}\left(  \overrightarrow{u}\right)  }%
=\overrightarrow{u}$ is bijective
\end{definition}

\begin{definition}
The dimension of an affine space $\left(  E,\overrightarrow{E}\right)  $ is
the dimension of $\overrightarrow{E}.$
\end{definition}

i) $\Rightarrow\forall A,B\in E:\overrightarrow{AB}=-\overrightarrow{BA}$ and
$\overrightarrow{AA}=\overrightarrow{0}$

ii) $\Rightarrow\forall A\in E,\forall\overrightarrow{u}\in\overrightarrow{E}$
\ there is a unique $B\in E:\overrightarrow{AB}=\overrightarrow{u}$

On an affine space the sum of points is the map :

$E\times E\rightarrow E::\overrightarrow{OA}+\overrightarrow{OB}%
=\overrightarrow{OC}$

The result does not depend on the choice of O.

We will usually denote : $\overrightarrow{AB}=\overrightarrow{u}%
\Leftrightarrow B=A+\overrightarrow{u}\Leftrightarrow B-A=\overrightarrow{u}$

An affine space is fully defined with a point O, and a vector space
$\overrightarrow{E}$ :

Define : $E=\left\{  A=\left(  O,\overrightarrow{u}\right)  ,\overrightarrow
{u}\in\overrightarrow{E}\right\}  ,$ $\overrightarrow{\left(
O,\overrightarrow{u}\right)  \left(  O,\overrightarrow{v}\right)
}=\overrightarrow{v}-\overrightarrow{u}$

So a vector space can be endowed with the structure of an affine space by
taking O=$\overrightarrow{0}.$

\paragraph{Frame\newline}

\begin{definition}
A \textbf{frame} in an affine space\textbf{\ }$\left(  E,\overrightarrow
{E}\right)  $ is a pair $\left(  O,\left(  \overrightarrow{e}_{i}\right)
_{i\in I}\right)  $\ of a point $O\in E$ and a basis $\left(  \overrightarrow
{e}_{i}\right)  _{i\in I}$ of $\overrightarrow{E}.$ The \textbf{coordinates}
of a point M of E are the components of the vector $\overrightarrow{OM}$ with
respect to $\left(  \overrightarrow{e}_{i}\right)  _{i\in I}$
\end{definition}

If I is infinite only a finite set of coordinates is non zero.

An affine space \textbf{\ }$\left(  E,\overrightarrow{E}\right)  $ is real if
$\overrightarrow{E}$ is real (the coordinates are real), complex if
$\overrightarrow{E}$ is complex (the coordinates are complex).

\paragraph{Affine subspace\newline}

\begin{definition}
An \textbf{affine subspace} F of E is a pair $\left(  A,\overrightarrow
{F}\right)  $ of a point A of E and a vector subspace $\overrightarrow
{F}\subset\overrightarrow{E}$ with the condition : $\forall M\in
F:\overrightarrow{AM}\in\overrightarrow{F}$
\end{definition}

Thus $A\in F$

The dimension of F is the dimension of $\overrightarrow{F}$

\begin{definition}
A \textbf{line} is a 1-dimensional affine subspace.
\end{definition}

\begin{definition}
A \textbf{hyperplane} passing through A is the affine subspace complementary
of a line passing through A.
\end{definition}

If E is finite dimensional an hyperplane is an affine subspace of dimension n-1.

If $K=%
\mathbb{R}
,%
\mathbb{C}
$\ the \textbf{segment} AB between two points A$\neq$B is the set :

$AB=\left\{  M\in E:\exists t\in\left[  0,1\right]  ,t\overrightarrow
{AM}+\left(  1-t\right)  \overrightarrow{BM}=0\right\}  $

\begin{theorem}
The intersection of a family finite or infinite of affine subspaces is an
affine subspace.\ Conversely given any subset F of an affine space the affine
subspace generated by F is the intersection of all the affine subspaces which
contains F.
\end{theorem}

\begin{definition}
Two affine subspaces are said to be \textbf{parallel} if they share the same
underlying vector subspace $\overrightarrow{F}:\left(  A,\overrightarrow
{F}\right)  //\left(  B,\overrightarrow{G}\right)  \Leftrightarrow
\overrightarrow{F}=\overrightarrow{G}$
\end{definition}

\paragraph{Product of affine spaces\newline}

1. If $\overrightarrow{E},\overrightarrow{F}$ are vector spaces over the same
field, $\overrightarrow{E}\times\overrightarrow{F}$ can be identified with
$\overrightarrow{E}\oplus\overrightarrow{F}.$ Take any point O and define the
affine space : $\left(  O,\overrightarrow{E}\oplus\overrightarrow{F}\right)
.$ It can be identified with the set product of the affine spaces : $\left(
O,\overrightarrow{E}\right)  \times\left(  O,\overrightarrow{F}\right)  .$

2. A real affine space\textbf{\ }$\left(  E,\overrightarrow{E}\right)  $
becomes a complex affine space\textbf{\ }$\left(  E,\overrightarrow{E}_{%
\mathbb{C}
}\right)  $ with the complexification $\overrightarrow{E}_{%
\mathbb{C}
}=\overrightarrow{E}\oplus i\overrightarrow{E}.$

$\left(  E,\overrightarrow{E}_{%
\mathbb{C}
}\right)  $ can be dentified with the product of real affine space $\left(
O,\overrightarrow{E}\right)  \times\left(  O,i\overrightarrow{E}\right)  .$

3. Conversely a complex affine space\textbf{\ }$\left(  E,\overrightarrow
{E}\right)  $ endowed with a real structure can be identified with the product
of two real affine space $\left(  O,\overrightarrow{E}_{%
\mathbb{R}
}\right)  \times\left(  O,i\overrightarrow{E}_{%
\mathbb{R}
}\right)  .$ The "complex plane" is just the affine space $%
\mathbb{C}
\simeq%
\mathbb{R}
\times i%
\mathbb{R}
$

\subsubsection{Affine transformations}

\begin{definition}
The \textbf{translation } by the vector $\overrightarrow{u}\in\overrightarrow
{E}$\ \textbf{\ }on\textbf{\ }of an affine space $\left(  E,\overrightarrow
{E}\right)  $ is the map $\tau:E\rightarrow E::\tau\left(  A\right)
=B::\overrightarrow{AB}=\overrightarrow{u}.$
\end{definition}

\begin{definition}
An \textbf{affine map} $f:E\rightarrow E$ \textbf{\ }on\textbf{\ }an affine
space $\left(  E,\overrightarrow{E}\right)  $ is such that there is a map
$\overrightarrow{f}\in L\left(  \overrightarrow{E};\overrightarrow{E}\right)
$ and :

$\forall M,P\in E:M^{\prime}=f\left(  M\right)  ,P^{\prime}=f\left(  P\right)
:\overrightarrow{M^{\prime}P^{\prime}}=\overrightarrow{f}\left(
\overrightarrow{MP}\right)  $
\end{definition}

If is fully defined by a triple $\left(  O,\overrightarrow{a},\overrightarrow
{f}\right)  ,O\in E,\overrightarrow{a}\in\overrightarrow{F},\overrightarrow
{f}\in L\left(  \overrightarrow{E};\overrightarrow{E}\right)  :$ then
$\overrightarrow{Of\left(  M\right)  }=\overrightarrow{a}+\overrightarrow
{f}\left(  \overrightarrow{OM}\right)  $ so $A=f(O)$ with $\overrightarrow
{OA}=\overrightarrow{a}$

With another point O', the vector $\overrightarrow{a^{\prime}}=\overrightarrow
{O^{\prime}f\left(  O^{\prime}\right)  }$ defines the same map:

\begin{proof}
$\overrightarrow{O^{\prime}f^{\prime}\left(  M\right)  }=\overrightarrow
{a}^{\prime}+\overrightarrow{f}\left(  \overrightarrow{O^{\prime}M}\right)
=\overrightarrow{O^{\prime}O}+\overrightarrow{Of\left(  O^{\prime}\right)
}+\overrightarrow{f}\left(  \overrightarrow{O^{\prime}O}\right)
+\overrightarrow{f}\left(  \overrightarrow{OM}\right)  =\overrightarrow
{O^{\prime}O}+\overrightarrow{Of^{\prime}\left(  M\right)  }$

$\overrightarrow{Of\left(  O^{\prime}\right)  }=\overrightarrow{a}%
+\overrightarrow{f}\left(  \overrightarrow{OO^{\prime}}\right)  $

$\overrightarrow{Of^{\prime}\left(  M\right)  }=\overrightarrow{a}%
+\overrightarrow{f}\left(  \overrightarrow{OO^{\prime}}\right)
+\overrightarrow{f}\left(  \overrightarrow{O^{\prime}O}\right)
+\overrightarrow{f}\left(  \overrightarrow{OM}\right)  =\overrightarrow
{a}+\overrightarrow{f}\left(  \overrightarrow{OM}\right)  =\overrightarrow
{Of\left(  M\right)  }$
\end{proof}

It can be generalized to an affine map between affine spaces : $f:E\rightarrow
F:$

take $(O,O\prime,\overrightarrow{a\prime},\overrightarrow{f})\in E\times
F\times\overrightarrow{F}\times L\left(  \overrightarrow{E};\overrightarrow
{F}\right)  :$ then

$\overrightarrow{O^{\prime}f\left(  M\right)  }=\overrightarrow{a\prime
}+\overrightarrow{f}\left(  \overrightarrow{OM}\right)  \Rightarrow
\overrightarrow{O^{\prime}f\left(  O\right)  }=\overrightarrow{a\prime}$

If E is finite dimensional $\overrightarrow{f}$ is defined by a matrix
$\left[  F\right]  $ in a basis and the coordinates of the image f(M) are
given by the affine relation : $\left[  y\right]  =A+\left[  F\right]  \left[
x\right]  $ with $\overrightarrow{OA}=\sum_{i\in I}a_{i}\overrightarrow{e}%
_{i},$ $\overrightarrow{OM}=\sum_{i\in I}x_{i}\overrightarrow{e}_{i},$
$\overrightarrow{Of\left(  M\right)  }=\sum_{i\in I}y_{i}\overrightarrow
{e}_{i}$

\bigskip

\begin{theorem}
(Berge p.144) A hyperplane in an affine space E over K is defined by f(x)=0
where $f:E\rightarrow K$ is an affine, non constant, map.
\end{theorem}

f is not unique.

\begin{theorem}
Affine maps are the morphisms of the affine spaces over the same field K,
which is a category
\end{theorem}

\begin{theorem}
The set of invertible affine transformations over an affine space $\left(
E,\overrightarrow{E}\right)  $ is a group with the composition law :

$\left(  O,\overrightarrow{a}_{1},\overrightarrow{f}_{1}\right)
\circ(O,\overrightarrow{a}_{2},\overrightarrow{f}_{2})=(O,\overrightarrow
{a}_{1}+\overrightarrow{f}_{1}\left(  \overrightarrow{a}_{2}\right)
,\overrightarrow{f}_{1}\circ\overrightarrow{f}_{2});$

$(O,\overrightarrow{a},\overrightarrow{f})^{-1}=(O,-\overrightarrow{f}%
^{-1}\left(  \overrightarrow{a}\right)  ,\overrightarrow{f}^{-1})$
\end{theorem}

\subsubsection{Convexity}

\paragraph{Barycenter\newline}

\begin{definition}
A set of points $\left(  M_{i}\right)  _{i\in I}$ of an affine space E is said
to be \textbf{independant} if all the vectors $\left(  \overrightarrow
{M_{i}M_{j}}\right)  _{i,j\in I}$ are linearly independant.
\end{definition}

If E has the dimension n at most n+1 points can be independant.

\begin{definition}
A weighted family in an affine space $\left(  E,\overrightarrow{E}\right)  $
over a field K is a family $\left(  M_{i},w_{i}\right)  _{i\in I}$ where
$M_{i}\in E$ and $w_{i}\in K$

The \textbf{barycenter} of a weighted family is the point G such that for each
finite subfamily J of I : $\sum_{i\in J}m_{i}\overrightarrow{GM}_{i}=0.$
\end{definition}

One writes : $G=\sum_{i\in J}m_{i}M_{i}$

In any frame : $\left(  x_{G}\right)  _{i}=\sum_{j\in I}m_{j}\left(
x_{Mj}\right)  _{i}$

\paragraph{Convex subsets\newline}

Convexity is a purely geometric property. However in many cases it provides a
"proxy" for topological concepts.

\begin{definition}
A subset A of an affine space $\left(  E,\overrightarrow{E}\right)  $ is
\textbf{convex} iff the barycenter of any weighted family $\left(
M_{i},1\right)  _{i\in I}$ where $M_{i}\in A$ belongs to A
\end{definition}

\begin{theorem}
A subset A of an affine space $\left(  E,\overrightarrow{E}\right)  $ over $%
\mathbb{R}
$ or $%
\mathbb{C}
$ is convex iff $\forall t\in\left[  0,1\right]  ,\forall M,P\in
A,Q:t\overrightarrow{QM}+\left(  1-t\right)  \overrightarrow{QP}=0,Q\in A$
\end{theorem}

that is if any point in the segment joining M and P is in A.

Thus in $%
\mathbb{R}
$ convex sets are closed intervals $\left[  a,b\right]  $

\begin{theorem}
In an affine space $\left(  E,\overrightarrow{E}\right)  $ :

the empty set $\varnothing$\ and E are convex.

the intersection of any collection of convex sets is convex.

the union of a non-decreasing sequence of convex subsets is a convex set.

if $A_{1},A_{2}$ are convex then $A_{1}+A_{2}$ is convex
\end{theorem}

\begin{definition}
The \textbf{convex hull} of a subset A of an affine space $\left(
E,\overrightarrow{E}\right)  $ is the intersection of all the convex sets
which contains A.\ It is the smallest convex set which contains A.
\end{definition}

\begin{definition}
A convex subset C of a real affine space $\left(  E,\overrightarrow{E}\right)
$ is \textbf{absolutely convex} iff :

$\forall\lambda,\mu\in%
\mathbb{R}
,\left\vert \lambda\right\vert +\left\vert \mu\right\vert \leq1,\forall M\in
C,\forall O:\lambda\overrightarrow{OM}+\mu\overrightarrow{OM}\in C$
\end{definition}

There is a separation theorem which does not require any topological structure
(but uses the Zorn lemna).

\begin{theorem}
Kakutani (Berge p.162): If X,Y are two disjunct convex subset of an affine
space E, there are two convex subsets X',Y' such that :

$X\subset X^{\prime},Y\subset Y^{\prime},X^{\prime}\cap Y^{\prime}%
=\varnothing,X^{\prime}\cup Y^{\prime}=E$
\end{theorem}

\begin{definition}
A point a is an \textbf{extreme point} of a convex subset C of a real affine
space if it does not lie in any open segment of C
\end{definition}

Meaning : $\forall M,P\in C,\forall t\in]0,1[:tM+(1-t)P\neq a$

\paragraph{Convex function\newline}

\begin{definition}
A real valued function $f:A\rightarrow%
\mathbb{R}
$ defined on a convex set A of a real affine space $\left(  E,\overrightarrow
{E}\right)  $ is \textbf{convex} if :

$\forall M,P\in A,\forall t\in\left[  0,1\right]  :t\overrightarrow
{QM}+\left(  1-t\right)  \overrightarrow{QP}=0\Rightarrow f\left(  Q\right)
\leq tf(M)+\left(  1-t\right)  f\left(  P\right)  $

It is stricly convex if $\forall t\in]0,1[:f\left(  Q\right)  <tf(M)+\left(
1-t\right)  f\left(  P\right)  $
\end{definition}

\begin{definition}
A function f is said to be (strictly) \textbf{concave} if -f is (strictly) convex.
\end{definition}

\begin{theorem}
If g is an affine map : $g:A\rightarrow A$ and f is convex, then $f\circ g$ is convex
\end{theorem}

\subsubsection{Homology}

Homology is a branch of abstract algebra.\ We will limit here to the
definitions and results which are related to simplices, which can be seen as
solids bounded by flat faces and straight edges. Simplices appear often in
practical optimization problems : the extremum of a linear function under
linear constraints (what is called a linear program) is on the simplex
delimited by the constraints.

Definitions and results can be found in Nakahara p.110, Gamelin p.171

\paragraph{Simplex\newline}

(plural simplices)

\begin{definition}
A \textbf{k-simplex} denoted $\left\langle A_{0},...A_{k}\right\rangle $ where
$\left(  A_{i}\right)  _{i=0}^{k}$ are k+1 independant points of a n
dimensional real affine space $\left(  E,\overrightarrow{E}\right)  ,$ is the
convex subset:

$\left\langle A_{0},...A_{k}\right\rangle =\{P\in E:P=\sum_{i=0}^{k}t_{i}%
A_{i};0\leq t_{i}\leq1,\sum_{i=0}^{k}t_{i}=1\}$

A \textbf{vertex} (plural vertices) is a 0-simplex (a point)

An \textbf{edge} is a 1-simplex (the segment joining 2 points)

A \textbf{polygon} is a 2-simplex in a 3 dimensional affine space

A \textbf{polyhedron} is a 3-simplex in a 3 dimensional affine space (the
solid delimited by 4 points)

A \textbf{p-face} is a p-simplex issued from a k-simplex.
\end{definition}

So a k-simplex is a convex subset of a k dimensional affine subspace delimited
by straight lines.

A regular simplex is a simplex which is symmetric for some group of affine transformations.

The standard simplex is the n-1-simplex in $%
\mathbb{R}
^{n}$ delimited by the points of coordinates $A_{i}=\left(
0,..0,1,0,...0\right)  $

Remark : the definitions vary greatly, but these above are the most common and
easily understood.\ The term simplex is sometimes replaced by polytope.

\paragraph{Orientation of a k-simplex\newline}

Let be a path connecting any two vertices $A_{i},A_{j}$ of a simplex.\ This
path can be oriented in two ways (one goes from $A_{i}$ to $A_{j}$ or from
$A_{j}$ to $A_{i}).$ So for any path connecting all the vertices, there are
only two possible consistent orientations given by the parity of the
permutation $\left(  A_{i_{0}},A_{i_{1}},...,A_{i_{k}}\right)  $ of $\left(
A_{0},A_{1},...A_{k}\right)  .$ So a k-simplex can be oriented.

\paragraph{Simplicial complex\newline}

Let be $\left(  A_{i}\right)  _{i\in I}$ a family of points in E.\ For any
finite subfamily J one can define the simplex delimited by the points $\left(
A_{i}\right)  _{i\in J}$ denoted $\left\langle A_{i}\right\rangle _{i\in
J}=C_{J}.$ .The set $C=\cup_{J}C_{J}$ is a \textbf{simplicial complex} if :
$\forall J,J^{\prime}:C_{J}\cap C_{J^{\prime}}\subset C$ or is empty

The dimension m of the simplicial complex is the maximum of the dimension of
its simplices.

The \textbf{Euler characteristic} of a n dimensional simplicial complex is :
$\chi\left(  C\right)  =\sum_{r=0}^{n}\left(  -1\right)  ^{r}I_{r}$ where
$I_{r} $ is the number of r-simplices in C (non oriented). It is a
generalization of the Euler Number in 3 dimensions :

Number of vertices - Number of edges + Number of 2-faces = Euler Number

\paragraph{r-chains\newline}

It is intuitive that, given a simplicial complex, one can build many different
simplices by adding or removing vertices.\ This is formalized in the concept
of chain and homology group, which are the basic foundations of algebraic
topology (the study of "shapes" of objects in any dimension).

\subparagraph{1. Definition:\newline}

Let C a simplicial complex, whose elements are simplices, and $C_{r}\left(
C\right)  $ its subset comprised of all r-simplices. $C_{r}\left(  C\right)  $
is a finite set with $I_{r}$ different non oriented elements.

A \textbf{r-chain} is a formal finite linear combination of r-simplices
belonging to the same simplicial complex.\ The set of all r-chains of the
simplicial complex C is denoted $G_{r}\left(  C\right)  :$

$G_{r}\left(  C\right)  =\left\{  \sum_{i=1}^{I_{r}}z_{i}S_{i},S_{i}\in
C_{r}\left(  C\right)  ,z_{i}\in%
\mathbb{Z}
\right\}  ,i=$ index running over all the elements of $C_{r}\left(  C\right)
$

Notice that the coefficients $z_{i}\in%
\mathbb{Z}
.$

\subparagraph{2. Group structure:\newline}

$G_{r}\left(  C\right)  $ is an abelian group with the following operations :

$\sum_{i=1}^{I_{r}}z_{i}S_{i}+\sum_{i=1}^{I_{r}}z_{i}^{\prime}S_{i}=\sum
_{i=1}^{I_{r}}\left(  z_{i}+z_{i}^{\prime}\right)  S_{i}$

$0=\sum_{i=1}^{I_{r}}0S_{i}$

$-S_{i}=$ the same r-simplex with the opposite orientation

The group $G\left(  C\right)  =\oplus_{r}G_{r}\left(  C\right)  $

\subparagraph{3. Border:\newline}

Any r-simplex of the complex can be defined from r+1 independant points.\ If
one point of the simplex is removed we get a r-1-simplex which still belongs
to the complex. The \textbf{border} of the simplex $\left\langle A_{0}%
,A_{1},...A_{r}\right\rangle $ is the r-1-chain :

$\partial\left\langle A_{0},A_{1},...A_{r}\right\rangle =\sum_{k=0}^{r}\left(
-1\right)  ^{k}\left\langle A_{0},A_{1},.,\hat{A}_{k},...A_{r}\right\rangle $
where the point $A_{k}$ has been removed

Conventionnaly : $\partial\left\langle A_{0}\right\rangle =0$

The operator $\partial$ is a morphism $\partial\in\hom\left(  G_{r}\left(
C\right)  ,G_{r-1}\left(  C\right)  \right)  $ and there is the exact sequence :

$0\rightarrow G_{n}(C)\overset{\partial}{\rightarrow}G_{n-1}(C)\overset
{\partial}{\rightarrow}....G_{0}(C)\overset{\partial}{\rightarrow}0$

\subparagraph{3. Cycle:\newline}

A simplex such that $\partial S=0$ is a \textbf{r-cycle}. The set
$Z_{r}\left(  C\right)  =\ker\left(  \partial\right)  $ is the r-cycle
subgroup of $G_{r}\left(  C\right)  $ and $Z_{0}(C)=G_{0}(C)$

Conversely if there is $A\in G_{r+1}\left(  C\right)  $ such that $B=\partial
A\in G_{r}\left(  C\right)  $ then B is called a \textbf{r-border}.\ The set
of r-borders is a subgroup $B_{r}\left(  C\right)  $ of $G_{r}\left(
C\right)  $ and $B_{n}(C)=0$

$B_{r}(C)\subset Z_{r}(C)\subset G_{r}(C)$

\subparagraph{4. Homology group:\newline}

The \textbf{r-homology group} of C is the quotient set : $H_{r}\left(
C\right)  =Z_{r}(C)/B_{r}\left(  C\right)  $

The rth \textbf{Betti number} is $b_{r}\left(  C\right)  =\dim H_{r}\left(
C\right)  $

Euler-Poincar\'{e} theorem : $\chi\left(  C\right)  =\sum_{r=0}^{n}\left(
-1\right)  ^{r}b_{r}\left(  C\right)  $

The situation is very similar to the exact $\left(  \varpi=d\pi\right)  $ and
closed $\left(  d\varpi=0\right)  $ forms on a manifold, and there are strong
relations between the groups of homology and cohomology.

\newpage

\section{TENSORS}

\bigskip

Tensors are mathematical objects defined over a space vector.\ As they are
ubiquituous in mathematics, they deserve a full section. Many of the concepts
presented here hold in vector bundles, due to the functorial nature of tensors constructs.

\bigskip

\subsection{Tensorial product of vector spaces}

\label{Tensorial product of vector spaces}

\subsubsection{Definition}

\paragraph{Universal property\newline}

\begin{definition}
The \textbf{tensorial product} $E\otimes F$ of two vector spaces on the same
field K is defined by the following universal property : there is a map
$\imath:E\times F\rightarrow E\otimes F$ such that for any vector space V and
bilinear map $g:E\times F\rightarrow V$ , there is a unique linear map :
$G:E\otimes F\rightarrow V$ such that $g=G\circ\imath$
\end{definition}

\bigskip%

\begin{tabular}
[c]{lllll}
&  & $g$ &  & \\
$E\times F$ & $\rightarrow$ & $\rightarrow$ & $\rightarrow$ & $V$\\
$\downarrow$ &  &  & $\nearrow$ & \\
$\downarrow\hspace{0in}\imath$ &  & $\nearrow$ & $G$ & \\
$\downarrow$ & $\nearrow$ &  &  & \\
$E\otimes F$ &  &  &  &
\end{tabular}

\bigskip

This definition can be seen as abstract, but it is in fact the most natural
introduction of tensors.\ Let g be a bilinear map so :

$g(u,v)=g\left(  \sum_{i}u_{i}e_{i},\sum_{j}v_{j}f_{j}\right)  =\sum
_{i,j}u_{i}v_{j}g\left(  e_{i},f_{j}\right)  =\sum_{ijk}g_{ijk}u_{i}%
v_{j}\varepsilon_{k}$

it is intuitive to extend the map by linearity to something like : $\sum
_{ijk}G_{ijk}U_{ij}\varepsilon_{k}$ meaning that $U=u\otimes v$

Definition is not proof of existence. So to prove that the tensorial product
does exist the construct is the following :

1. Take the product $E\times F$ with the obvious structure of vector space.

2. Take the equivalence relation : $\left(  x,0\right)  \sim\left(
0,y\right)  \sim0$ and 0 as identity element for addition

3. Define $E\otimes F=E\times F/\sim$

\paragraph{Example}

The set $K_{p}\left[  x_{1},...x_{n}\right]  $\ of polynomials of degree p in
n variables is a vector space over K.

$P_{p}\in K_{p}\left[  x\right]  $ reads : $P_{p}\left(  x\right)  =\sum
_{r=0}^{p}a_{r}x^{r}=\sum_{r=0}^{p}a_{r}e_{r}$ with as basis the monomials :
$e_{r}=x^{r},r=0..p$

Consider the bilinear map :

$f:K_{p}\left[  x\right]  \times K_{q}\left[  y\right]  \rightarrow
K_{p+q}\left[  x,y\right]  ::f\left(  P_{p}\left(  x\right)  ,P_{q}\left(
y\right)  \right)  =P_{p}\left(  x\right)  \times P_{q}\left(  y\right)
=\sum_{r=0}^{p}\sum_{s=0}^{q}a_{r}b_{s}x^{r}y^{s}$

So there is a linear map : $F:K_{p}\left[  x\right]  \otimes K_{q}\left[
y\right]  \rightarrow K_{p+q}\left[  x,y\right]  ::f=F\circ\imath$

$\imath\left(  e_{r},e_{s}\right)  =e_{r}\otimes e_{s}$

$\imath\left(  P_{p}\left(  x\right)  ,P_{q}\left(  y\right)  \right)
=\sum_{r=0}^{p}\sum_{s=0}^{q}a_{r}b_{s}e_{r}\otimes e_{s}$

$\sum_{r=0}^{p}\sum_{s=0}^{q}a_{r}b_{s}x^{r}y^{s}=\sum_{r=0}^{p}\sum_{s=0}%
^{q}a_{r}b_{s}e_{r}\otimes e_{s}$

So $e_{r}\otimes e_{s}=x^{r}y^{s}$

And one can write : $K_{p}\left[  x\right]  \otimes K_{q}\left[  y\right]
=K_{p+q}\left[  x,y\right]  $

\subsubsection{Properties}

\begin{theorem}
The tensorial product $E\otimes F$ of two vector spaces on a field K is a
vector space on K whose vectors are called \textbf{tensors}.
\end{theorem}

\begin{definition}
The bilinear map : $\imath:E\times F\rightarrow E\otimes F::\imath\left(
u,v\right)  =u\otimes v$ is the \textbf{tensor product of vectors}
\end{definition}

with the properties :

$\forall k,k^{\prime}\in K,u,u^{\prime}\in E,v,v^{\prime}\in F$

$(ku+k^{\prime}u^{\prime})\otimes v=ku\otimes v+k^{\prime}u^{\prime}\otimes v$

$u\otimes\left(  kv+k^{\prime}v^{\prime}\right)  =ku\otimes v+k^{\prime
}u\otimes v^{\prime}$

$0\otimes u=u\otimes0=0\in E\otimes F$

But if E=F \textit{it is not commutative} : $u,v\in E,u\otimes v=v\otimes
u\Leftrightarrow\exists k\in K:v=ku$

\begin{theorem}
If $\left(  e_{i}\right)  _{i\in I},\left(  f_{j}\right)  _{j\in J}$ are basis
of E and F, $\left(  e_{i}\otimes f_{j}\right)  _{I\times J}$\ is a basis of
$E\otimes F$ called a tensorial basis$.$
\end{theorem}

So tensors are linear combinations of $e_{i}\otimes f_{j}.$ If E and F are
finite dimensional with dimensions n,p then $E\otimes F$ is finite dimensional
with dimensions nxp.

If $u=\sum_{i\in I}U_{i}e_{i},v=\sum_{j\in J}V_{j}f_{j}:u\otimes
v=\sum_{\left(  i,j\right)  \in I\times J}U_{i}V_{j}e_{i}\otimes f_{j}$

The components of the tensorial product are the sum of all the combinations of
the components of the vectors

If $T\in E\otimes F:T=\sum_{\left(  i,j\right)  \in I\times J}T_{i}{}_{j}%
e_{i}\otimes f_{j}$

A tensor which can be put in the form : $T\in E\otimes F:T=u\otimes v,u\in
E,v\in F$ is said to be \textbf{decomposable}.

Warning ! all tensors are not decomposable : they are sum of decomposable tensors

\begin{theorem}
The vector spaces $E\otimes F,F\otimes E,$ are canonically isomorphic
$E\otimes F\simeq F\otimes E$ and can be identified whenever $E\neq F$

The vector spaces $E\otimes K,E$ are canonically isomorphic $E\otimes K\simeq
E$ and can be identified
\end{theorem}

\subsubsection{Tensor product of more than two vector spaces}

\begin{definition}
The \textbf{tensorial product} $E_{1}\otimes E_{2}...\otimes E_{r}$ of the
vector spaces $\left(  E_{i}\right)  _{i=1}^{r}$\ on the same field K is
defined by the following universal property : there is a multilinear map :
$\imath:E_{1}\times E_{2}...\times E_{r}\rightarrow E_{1}\otimes
E_{2}...\otimes E_{r}$ such that for any vector space S and multilinear map
$f:E_{1}\times E_{2}...\times E_{r}\rightarrow S$ there is a unique linear map
: $F:E_{1}\otimes E_{2}...\otimes E_{r}\rightarrow S$ such that $f=F\circ
\imath$
\end{definition}

The \textbf{order} of a tensor is the number r of vectors spaces.

The multilinear map : $\imath:E_{1}\times E_{2}...\times E_{r}\rightarrow
E_{1}\otimes E_{2}...\otimes E_{r}$ is the tensor product of vectors

If $\left(  e_{ij}\right)  _{j\in J_{i}}$ is a basis of $E_{i}$ then $\left(
e_{1j_{1}}\otimes e_{2j_{2}}...\otimes e_{rj_{r}}..\right)  _{j_{k}\in J_{k}}$
is a basis of $E_{1}\otimes E_{2}...\otimes E_{r}$

In components with : $u_{k}=\sum_{j\in J_{k}}U_{kj}e_{kj}$

$u_{1}\otimes u_{2}...\otimes u_{r}=\sum_{\left(  j_{1},j_{2},...j_{r}\right)
\in J_{1}\times J_{2}...\times J_{r}}U_{1j_{1}}U_{2j_{2}}...U_{rj_{r}%
}e_{1j_{1}}\otimes e_{2j_{2}}...\otimes e_{rj_{r}}$

As each tensor product $E_{i_{1}}\otimes E_{i_{2}}...\otimes E_{i_{k}}$ is
itself a vector space the \textbf{tensorial product of tensors} can be defined.

\begin{theorem}
The tensorial product of tensors is associative, and distributes over direct
sums, even infinite sums :
\end{theorem}

$E\otimes\left(  \oplus_{I}F_{i}\right)  =\oplus_{I}\left(  E\otimes
F_{i}\right)  $

In components :

$T=\sum_{\left(  i_{1},i_{2},...i_{r}\right)  \in I_{1}\times I_{2}...\times
I_{r}}T_{i_{1}i_{2}..i_{r}}e_{1i_{1}}\otimes e_{2i_{2}}...\otimes e_{ri_{r}%
}\in E=E_{1}\otimes E_{2}...\otimes E_{r}$

$S=\sum_{\left(  j_{1},j_{2},...j_{s}\right)  \in J_{1}\times J_{2}...\times
J_{s}}S_{j_{1}j_{2}..j_{s}}f_{1j_{1}}\otimes f_{2j_{2}}...\otimes f_{sj_{s}%
}\in F=F_{1}\otimes F_{2}...\otimes F_{s}$

$T\otimes S$

$=\sum_{\left(  i_{1},...i_{r}\right)  \in I_{1}..\times I_{r}}\sum_{\left(
j_{1},...j_{s}\right)  \in J_{1}..\times J_{s}}T_{i_{1}..i_{r}}S_{j_{1}%
..j_{s}}e_{1i_{1}}...\otimes e_{ri_{r}}\otimes f_{1j_{1}}...\otimes f_{sj_{s}%
}\in E\otimes F$

\subsubsection{Tensorial algebra}

\begin{definition}
The \textbf{tensorial algebra}, denoted $T(E),$ of the vector space E on the
field K is the direct sum $T(E)=\oplus_{n=0}^{\infty}\left(  \otimes
^{n}E\right)  $ of the tensorial products $\otimes^{n}E=E\otimes E...\otimes
E$ where for n=0 : $\otimes^{0}E=K$
\end{definition}

\begin{theorem}
The tensorial algebra of the vector space E on the field K is an algebra on
the field K with the tensor product as internal operation and\ the unity
element is $1\in K.$
\end{theorem}

The elements of $\otimes^{n}E$ are homogeneous tensors of order n.\ Their
components in a basis $\left(  e_{i}\right)  _{i\in I}$ are such that :

$T=\sum_{\left(  i_{1}...i_{n}\right)  }t^{i_{1}...i_{n}}e_{i_{1}}%
\otimes...\otimes e_{i_{n}}$ with the sum over all finite n-sets of indices
$\left(  i_{1}...i_{n}\right)  ,i_{k}\in I$

\begin{theorem}
The tensorial algebra $T(E),$ of the vector space E on the field K has the
universal property : for any algebra A on the field K and linear map
$l:E\rightarrow A$ there is a unique algebra morphism $L:T(E)\rightarrow A$
such that : $l=L\circ\jmath$ where $\jmath:E\rightarrow\otimes^{1}E$
\end{theorem}

\begin{definition}
A derivation D over the algebra $T(E)$ is a map $D:T(E)\rightarrow T(E)$ such
that :

$\forall u,v\in T(E):D\left(  u\otimes v\right)  =D(u)\otimes v+u\otimes D(v)$
\end{definition}

\begin{theorem}
The tensorial algebra $T(E)$ of the vector space E has the universal property
that for any linear map $d:E\rightarrow T(E)$ there is a unique derivation
$D:T(E)\rightarrow T(E)$ such that : $d=D\circ\jmath$ where $\jmath
:E\rightarrow\otimes^{1}E$
\end{theorem}

\subsubsection{Covariant and contravariant tensors}

\begin{definition}
Let be E a vector space and E* its algebraic dual

The tensors of the tensorial product of p copies of E are p
\textbf{contravariant} tensors

The tensors of the tensorial product of q copies of E* are q
\textbf{covariant} tensors

The tensors of the tensorial product of p copies of E and q copies of E* are
mixed, p contravariant,q covariant tensors (or a \textbf{type (p,q)} tensor)
\end{definition}

The tensorial product is not commutative if E=F, so in a mixed product (p,q)
the order between contravariant on one hand, covariant on the other hand,
matters, but not the order between contravariant and covariant. So :

$\underset{q}{\overset{p}{\otimes}}E=\left(  \otimes E\right)  ^{p}%
\otimes\left(  \otimes E^{\ast}\right)  ^{q}=\left(  \otimes E^{\ast}\right)
^{q}\otimes\left(  \otimes E\right)  ^{p}$

\begin{notation}
$\otimes_{q}^{p}E$ is the vector space of type (p,q) tensors over E :

Components of contravariant tensors are denoted with upper index: $a^{ij...m}
$

Components of covariant tensors are denoted with lower index: $a_{ij...m}.$

Components of mixed tensors are denoted with upper and lower indices:
$a_{qr...t}^{ij...m}$

Basis vectors $e_{i}$ of E are denoted with lower index

Basis vectors $e^{i}$ of E* are denoted with upper index.
\end{notation}

The order of the upper indices (resp.lower indices) matters

Notice that a covariant tensor is a multilinear map acting on vectors the
usual way :

If $T=\sum t_{ij}e^{i}\otimes e^{j}$ then $T(u,v)=\sum_{ij}t_{ij}u^{i}v^{j}\in
K$

Similarly a contravariant tensor can be seen as a linear map acting on 1-forms:

If $T=\sum t^{ij}e_{i}\otimes e_{j}$ then $T(\lambda,\mu)=\sum_{ij}%
t^{ij}\lambda_{i}\mu_{j}\in K$

And a mixed tensor is a map acting on vectors and giving vectors (see below)

\paragraph{Isomorphism L(E;E)$\simeq E\otimes E^{\ast}$\newline}

\begin{theorem}
If the vector space E is finite dimensional, there is an isomorphism between
L(E;E) and $E\otimes E^{\ast}$
\end{theorem}

\begin{proof}
Define the bilinear map :

$\lambda:E\times E^{\ast}\rightarrow L(E;E)::\lambda\left(  u,\varpi\right)
\left(  v\right)  =\varpi\left(  u\right)  v$

$\lambda\in L^{2}\left(  E,E^{\ast};L(E;E)\right)  $

From the universal property of tensor product :

\i:$E\times E^{\ast}\rightarrow E\otimes E^{\ast}$

$\exists$ unique $\Lambda\in L\left(  E\otimes E^{\ast};L(E;E)\right)
:\lambda=\Lambda\circ\imath$

$t\in E\otimes E^{\ast}\rightarrow f=\Lambda\left(  t\right)  \in L(E;E)$

Conversely :

$\forall f\in L\left(  E;E\right)  ,\exists f^{\ast}\in L\left(  E^{\ast
};E^{\ast}\right)  :f^{\ast}\left(  \varpi\right)  =\varpi\circ f$

$\exists f\otimes f^{\ast}\in L\left(  E\otimes E^{\ast};E\otimes E^{\ast
}\right)  ::$

$\left(  f\otimes f^{\ast}\right)  \left(  u\otimes\varpi\right)  =f\left(
u\right)  \otimes f^{\ast}\left(  \varpi\right)  =f\left(  u\right)
\otimes\left(  \varpi\circ f\right)  \in E\otimes E^{\ast}$

Pick up any basis of E : $\left(  e_{i}\right)  _{i\in I}$ and its dual basis
$\left(  e^{i}\right)  _{i\in I}$

Define : $T=\sum_{i,j}\left(  f\otimes f^{\ast}\right)  \left(  e_{i}\otimes
e^{j}\right)  \in E\otimes E^{\ast}$

In components : $f\left(  u\right)  =\sum_{ij}g_{i}^{j}u^{i}e_{j}\rightarrow
T(f)=\sum_{ij}g_{i}^{j}e^{i}\otimes e_{j}$
\end{proof}

Warning ! E must be finite dimensional

This isomorphism justifies the notation of matrix elements with upper indexes
(rows, for the contravariant part) and lower indexes (columns, for the
covariant part) : the matrix $A=\left[  a_{j}^{i}\right]  $ is the matrix of
the linear map : $f\in L(E;E)::f(u)=\sum_{i,j}\left(  a_{j}^{i}u^{j}\right)
e_{i}$ which is identified with the mixed tensor in $E\otimes E^{\ast}$ acting
on a vector of E.

\begin{definition}
The \textbf{Kronecker tensor} is $\delta=\sum_{i=1}^{n}e^{i}\otimes e_{i}%
=\sum_{ij}\delta_{j}^{i}e^{i}\otimes e_{j}\in E\otimes E^{\ast}$
\end{definition}

It has the same components in any basis, and corresponds to the identity map
$E\rightarrow E$

\paragraph{The trace operator\newline}

\begin{theorem}
If E is a vector space on the field K there is a unique linear map called the
trace $Tr:E^{\ast}\otimes E\rightarrow K$ such that : $Tr\left(  \varpi\otimes
u\right)  =\varpi\left(  u\right)  $
\end{theorem}

\begin{proof}
This is the consequence of the universal property :

For : $f:E^{\ast}\times E\rightarrow K::f\left(  \varpi,u\right)
=\varpi\left(  u\right)  $

we have : $f=Tr\circ\imath\Leftrightarrow f\left(  \varpi,u\right)  =F\left(
\varpi\otimes u\right)  =\varpi\left(  u\right)  $
\end{proof}

So to any (1,1) tensor S is associated one scalar Tr(S) called the
\textbf{trace} of the tensor, whose value does not depend on a basis.\ In
components it reads :

$S=\sum_{i,j\in I}S_{i}^{j}e^{i}\otimes e_{j}\rightarrow Tr(S)=\sum_{i\in
I}S_{i}^{i}$

\bigskip

\textit{If E is finite dimensional} there is an isomorphism between L(E;E) and
$E\otimes E^{\ast},$ and $E^{\ast}\otimes E\equiv E\otimes E^{\ast}$ .\ So to
any linear map $f\in L\left(  E;E\right)  $ is associated a scalar.\ In a
basis it is $Tr\left(  f\right)  =\sum_{i\in I}f_{ii}.$ This is the geometric
(basis independant) definition of the \textbf{Trace} operator of an endomorphism.

Remark : this is an algebraic definition of the trace operator. This
definition uses the algebraic dual E* which is replaced in analysis by the
topological dual. So there is another definition for the Hilbert spaces, they
are equivalent in finite dimension.

\bigskip

\begin{theorem}
If E is a finite dimensional vector space and$\ f,g\in L(E;E)$ then $Tr(f\circ
g)=Tr(g\circ f)$
\end{theorem}

\begin{proof}
Check with a basis :

$f=\sum_{i\in I}f_{i}^{j}e^{i}\otimes e_{j},g=\sum_{i\in I}g_{i}^{j}%
e^{i}\otimes e_{j}$

$f\circ g=\sum_{i,j,k\in I}f_{k}^{j}g_{i}^{k}e^{i}\otimes e_{j}$

$Tr\left(  f\circ g\right)  =\sum_{i,k\in I}f_{k}^{i}g_{i}^{k}=\sum_{i,k\in
I}g_{k}^{i}f_{i}^{k}$
\end{proof}

\paragraph{Contraction of tensors\newline}

Over mixed tensors there is an additional operation, called
\textbf{contraction}.

Let $T\in\otimes_{q}^{p}E.$ One can take the trace of T over one covariant and
one contravariant component of T (or similarly one contravariant component and
one covariant component of T). The resulting tensor $\in\otimes_{q-1}^{p-1}E$
. The result depends of the choice of the components which are to be
contracted (but not of the basis).

Example :

Let $T=\sum_{ijk}a_{jk}^{i}e_{i}\otimes e^{j}\otimes e^{k}\in\underset
{2}{\overset{1}{\otimes}}E,$ the contracted tensor is

$\sum_{i}\sum_{k}a_{ik}^{i}e_{i}\otimes e^{k}\in\underset{1}{\overset
{1}{\otimes}}E$

$\sum_{i}\sum_{k}a_{ik}^{i}e_{i}\otimes e^{k}\neq\sum_{i}\sum_{k}a_{ki}%
^{i}e_{i}\otimes e^{k}\in\underset{1}{\overset{1}{\otimes}}E$

\paragraph{Einstein summation convention :\newline}

In the product of components of mixed tensors, whenever a index has the same
value in a upper and in a lower position it is assumed that the formula is the
sum of these components. This convention is widely used and most convenient
for the contraction of tensors.

Examples :

$a_{jk}^{i}b_{i}^{l}=\sum_{i}a_{jk}^{i}b_{i}^{l}$

$a^{i}b_{i}=\sum_{i}a^{i}b_{i}$

So with this convention $a_{ik}^{i}=\sum_{i}a_{ik}^{i}$ is the contracted tensor

\paragraph{Change of basis\newline}

Let E a finite dimensional n vector space. So the dual E* is well defined and
is n dimensional.

A basis $\left(  e_{i}\right)  _{i=1}^{n}$ of E and the dual basis $\left(
e^{i}\right)  _{i=1}^{n}$ of E*

In a change of basis : $f_{i}=\sum_{j=1}^{n}P_{i}^{j}e_{j}$ the components of
tensors change according to the following rules :

$\left[  P\right]  =\left[  Q\right]  ^{-1}$

- the contravariant components are multiplied by Q (as for vectors)

- the covariant components are multiplied by P (as for forms)

$T=\sum_{i_{1}...i_{p}}\sum_{j_{1}....j_{q}}t_{j_{1}...j_{q}}^{i_{1}...i_{p}%
}e_{i_{1}}\otimes e_{i_{2}}...\otimes e_{i_{p}}\otimes e^{j_{1}}%
\otimes...\otimes e^{j_{q}}$

$\rightarrow$

$T=\sum_{i_{1}...i_{p}}\sum_{j_{1}....j_{q}}\widetilde{t}_{j_{1}...j_{q}%
}^{i_{1}...i_{p}}f_{i_{1}}\otimes f_{i_{2}}...\otimes f_{i_{p}}\otimes
f^{j_{1}}\otimes...\otimes f^{j_{q}}$

with :

$\widetilde{t}_{j_{1}...j_{q}}^{i_{1}...i_{p}}=\sum_{k_{1}...k_{p}}\sum
_{l_{1}....l_{q}}t_{l_{1}...l_{q}}^{k_{1}...k_{p}}Q_{k_{1}}^{i_{1}}%
...Q_{k_{p}}^{i_{p}}P_{j_{1}}^{l_{1}}...P_{j_{q}}^{l_{q}} $

\paragraph{Bilinear forms\newline}

Let E a finite dimensional n vector space. So the dual E* is well defined and
is n dimensional. Let $\left(  e_{i}\right)  _{i=1}^{n}$ be a basis of E with
its the dual basis $\left(  e^{i}\right)  _{i=1}^{n}$ of E*.

1. Bilinear forms: $g:E\times E\rightarrow K$ can be seen as tensors :
$G:E^{\ast}\otimes E^{\ast}:$

$g\left(  u,v\right)  =\sum_{ij}g_{ij}u^{i}u^{j}\rightarrow G=\sum_{ij}%
g_{ij}e^{i}\otimes e^{j}$

Indeed in a change of basis \ the components of the 2 covariant tensor
$G=\sum_{ij}g_{ij}e^{i}\otimes e^{j}$ change as :

$G=\sum_{ij}\widetilde{g}_{ij}f^{i}\otimes f^{j}$ with $\widetilde{g}%
_{ij}=\sum_{kl}g_{kl}P_{i}^{k}P_{j}^{l}$ so $\left[  \widetilde{g}\right]
=\left[  P\right]  ^{t}\left[  g\right]  \left[  P\right]  $ is transformed
according to the rules for bilinear forms.

Similarly let be $\left[  g\right]  ^{-1}=\left[  g^{ij}\right]  $ and
$H=\sum_{ij}g^{ij}e_{i}\otimes e_{j}$ .\ H is a 2 contravariant tensor
$H\in\otimes^{2}E$

2. Let E be a a n-dimensional vector space over $%
\mathbb{R}
$\ endowed with a bilinear symmetric form g, non degenerate (but not
necessarily definite positive). Its matrix is $\left[  g\right]  =\left[
g_{ij}\right]  $ and $\left[  g\right]  ^{-1}=\left[  g^{ij}\right]  $

A contravariant tensor is "lowered" by contraction with the 2 covariant tensor
$G=\sum_{ij}g_{ij}e^{i}\otimes e^{j}$:

$T=\sum_{i_{1}...i_{p}}\sum_{j_{1}....j_{q}}t_{j_{1}...j_{q}}^{i_{1}...i_{p}%
}e_{i_{1}}\otimes e_{i_{2}}...\otimes e_{i_{p}}\otimes e^{j_{1}}%
\otimes...\otimes e^{j_{q}}$

$\rightarrow\widetilde{T}=\sum_{i_{2}...i_{p}}\sum_{j_{1}....j_{q+1}}%
\sum_{i_{1}}g_{j_{q+1}i_{1}}t_{j_{1}...j_{q}}^{i_{1}...i_{p}}e_{i_{2}%
}...\otimes e_{i_{p}}\otimes e^{j_{1}}\otimes...\otimes e^{j_{q+1}}$

so $T\in\otimes_{q}^{p}\rightarrow\widetilde{T}\in\otimes_{q+1}^{p-1}$

This operation can be done on any (or all) contravariant components (it
depends of the choice of the component) and the result does not depend of the basis.

Similarly a contravariant tensor is "lifted" by contraction with the 2
covariant tensor $H=\sum_{ij}g^{ij}e_{i}\otimes e_{j}$ :

$T=\sum_{i_{1}...i_{p}}\sum_{j_{1}....j_{q}}t_{j_{1}...j_{q}}^{i_{1}...i_{p}%
}e_{i_{1}}\otimes e_{i_{2}}...\otimes e_{i_{p}}\otimes e^{j_{1}}%
\otimes...\otimes e^{j_{q}}$

$\rightarrow\widetilde{T}=\sum_{i_{1}...i_{p+1}}\sum_{j_{2}....j_{q}}%
\sum_{j_{1}}g^{i_{p+1}j_{1}}t_{j_{1}...j_{q}}^{i_{1}...i_{p}}e_{i_{1}%
}...\otimes e_{i_{p+1}}\otimes e^{j_{2}}\otimes...\otimes e^{j_{q}}$

so $T\in\otimes_{q}^{p}\rightarrow\widetilde{T}\in\otimes_{q-1}^{p+1}$

These operations are just the generalization of the isomorphism $E\simeq
E^{\ast}$ using a bilinear form.

\paragraph{Derivation\newline}

The tensor product of any mixed tensor defines the algebra of tensors over a
vector space E :

\begin{notation}
$\otimes E=\oplus_{r,s=0}^{\infty}\left(  \otimes_{s}^{r}E\right)  $ is the
algebra of all tensors over E
\end{notation}

\begin{theorem}
The tensorial algebra $\otimes E$ of the vector space E on the field K is an
algebra on the field K with the tensor product as internal operation and\ the
unity element is $1\in K.$
\end{theorem}

\begin{definition}
A \textbf{derivation on the tensorial algebra} $\otimes E\ $\ is a linear map
$D:\otimes E\rightarrow\otimes E$ such that :

i) it preserves the tensor type : $\forall r,s,T\in\otimes_{s}^{r}%
E:DT\in\otimes_{s}^{r}E$

ii) it follows the Leibnitz rule for tensor product :

$\forall S,T\in\otimes E:D\left(  S\otimes T\right)  =D(S)\otimes T+S\otimes
D(T)$

iii) it commutes with the trace operator.
\end{definition}

So it will commute with the contraction of tensors.

A derivation on the tensorial algebra is a derivation as defined previously
(see Algebras) with the i),iii) additional conditions.

\begin{theorem}
The set of all derivations on $\otimes E$\ is a vector space and a Lie algebra
with the bracket : $\left[  D,D^{\prime}\right]  =D\circ D^{\prime}-D^{\prime
}\circ D.$
\end{theorem}

\begin{theorem}
(Kobayashi p.25) If E is a finite dimensional vector space, the Lie algebra of
derivations on $\otimes E$\ is isomorphic to the Lie algebra of endomorphisms
on E.\ This isomorphism is given by assigning to each derivation its value on E.
\end{theorem}

So given an endomorphism $f\in L(E;E)$ there is a unique derivation D on
$\otimes E$ such that :

$\forall u\in E,\varpi\in E^{\ast}:Du=f\left(  u\right)  ,D\left(
\varpi\right)  =-f^{\ast}\left(  \varpi\right)  $ where f* is the dual of f
and we have $\forall k\in K:D\left(  k\right)  =0$

\bigskip

\subsection{Algebras of symmetric and antisymmetric tensors}

\label{algebras of symmetric and antisymmetric tensors}

\begin{notation}
For any finite set I of indices:

$\left(  i_{1},i_{2},...i_{n}\right)  $ is any subset of n indexes chosen in
I, two subsets deduced by permutation are considered distinct

$\sum_{\left(  i_{1},i_{2},...i_{n}\right)  }$ is the sum over all
permutations of n indices in I

$\left\{  i_{1},...i_{n}\right\}  $ is any strictly ordered permutation of n
indices in I: $i_{1}<..<i_{n}$

$\sum_{\left\{  i_{1},...i_{n}\right\}  }$ is the sum over all ordered
permutations of n indices chosen in I

$\left[  i_{1},i_{2},...i_{n}\right]  $ is any set of n indexes in I such
that: $i_{1}\leq i_{2}\leq..\leq i_{n}$

$\sum_{\left[  i_{1},i_{2},...i_{n}\right]  }$ is the sum over all distinct
such sets of indices chosen in I
\end{notation}

We remind the notations:

$\mathfrak{S(n)}$ is the symmetric group of permutation of n indexes

$\sigma\left(  i_{1},i_{2},...i_{n}\right)  =\left(  \sigma\left(
i_{1}\right)  ,\sigma\left(  i_{2}\right)  ,...\sigma\left(  i_{n}\right)
\right)  $ is the image of the set $\left(  i_{1},i_{2},...i_{n}\right)  $ by
$\sigma\in\mathfrak{S(n)}$

$\epsilon\left(  \sigma\right)  $ where $\sigma\in\mathfrak{S(n)}$ is the
signature of $\sigma$

Permutation is a set operation, without respect for the possible equality of
some of the elements of the set. So $\left\{  a,b,c\right\}  $ and $\left\{
b,a,c\right\}  $ are two distinct permutations of the set even if it happens
that a=b.

\subsubsection{Algebra of symmetric tensors}

\paragraph{Symmetric tensors\newline}

\begin{definition}
On a vector space E the \textbf{symmetrisation operator }or symmetrizer is the
map :

$s_{r}:E^{r}\rightarrow\overset{r}{\otimes}E::s_{r}\left(  u_{1}%
,..,u_{r}\right)  =\sum_{\sigma\in\mathfrak{S}\left(  r\right)  }%
u_{\sigma\left(  1\right)  }\otimes..\otimes u_{\sigma\left(  r\right)  }$
\end{definition}

It is a multilinear symmetric map : $s_{r}\in L^{r}\left(  E;\otimes
^{r}E\right)  $

$s_{r}(u_{\sigma\left(  1\right)  },u_{\sigma\left(  2\right)  }%
,...,u_{\sigma\left(  r\right)  })=\sum_{\sigma^{\prime}\in\mathfrak{S}\left(
r\right)  }u_{\sigma^{\prime}\sigma\left(  1\right)  }\otimes..\otimes
u_{\sigma^{\prime}\sigma\left(  r\right)  }$

$=\sum_{\theta\in\mathfrak{S}\left(  r\right)  }u_{\theta\left(  1\right)
}\otimes..\otimes u_{\theta\left(  r\right)  }=s_{r}(u_{1},u_{2},...,u_{r})$

So there is a unique linear map : $S_{r}:\overset{r}{\otimes}E\rightarrow
\overset{r}{\otimes}E:$ such that : $s_{r}=S_{r}\circ\imath$ with
$\imath:E^{r}\rightarrow\overset{r}{\otimes}E$

For any tensor $T=\sum_{\left(  i_{1}...i_{r}\right)  }t^{i_{1}...i_{r}%
}e_{i_{1}}\otimes...\otimes e_{i_{r}}\in\overset{r}{\otimes}E:$

$S_{r}\left(  T\right)  =\sum_{\left(  i_{1}...i_{r}\right)  }t^{i_{1}%
...i_{r}}S_{r}\left(  e_{i_{1}}\otimes...\otimes e_{i_{r}}\right)
=\sum_{\left(  i_{1}...i_{r}\right)  }t^{i_{1}...i_{r}}S_{r}\circ\imath\left(
e_{i_{1}},..,e_{i_{r}}\right)  $

$=\sum_{\left(  i_{1}...i_{r}\right)  }t^{i_{1}...i_{r}}s_{r}\left(  e_{i_{1}%
},..,e_{i_{r}}\right)  =\sum_{\left(  i_{1}...i_{r}\right)  }t^{i_{1}...i_{r}%
}\sum_{\sigma\in\mathfrak{S}\left(  r\right)  }e_{i_{1}}\otimes...\otimes
e_{i_{r}}$

$S_{r}\left(  T\right)  =\sum_{\left(  i_{1}...i_{r}\right)  }t^{i_{1}%
...i_{r}}\sum_{\sigma\in\mathfrak{S}\left(  r\right)  }e_{\sigma\left(
i_{1}\right)  }\otimes..\otimes e_{\sigma\left(  i_{r}\right)  }$

\begin{definition}
A \textbf{symmetric r tensor} is a tensor T such that $S_{r}\left(  T\right)
=r!T$
\end{definition}

In a basis a symmetric r tensor reads :

$T=\sum_{\left(  i_{1}...i_{r}\right)  }t^{i_{1}...i_{r}}e_{i_{1}}%
\otimes...\otimes e_{i_{r}},$

where $t^{i_{1}...i_{r}}=t^{\sigma\left(  i_{1}...i_{r}\right)  }$ with
$\sigma$ is any permutation of the set of r-indices. Thus a symmetric tensor
is uniquely defined by a set of components $t^{i_{1}...i_{r}}$ for all ordered
indices $\left[  i_{1}...i_{r}\right]  $.

Example :

$T=t^{111}e_{1}\otimes e_{1}\otimes e_{1}+t^{112}e_{1}\otimes e_{1}\otimes
e_{2}+t^{121}e_{1}\otimes e_{2}\otimes e_{1}+t^{122}e_{1}\otimes e_{2}\otimes
e_{2}$

$+t^{211}e_{2}\otimes e_{1}\otimes e_{1}+t^{212}e_{2}\otimes e_{1}\otimes
e_{2}+t^{221}e_{2}\otimes e_{2}\otimes e_{1}+t^{222}e_{2}\otimes e_{2}\otimes
e_{2}$

$S_{3}\left(  T\right)  =6t^{111}e_{1}\otimes e_{1}\otimes e_{1}+6t^{222}%
e_{2}\otimes e_{2}\otimes e_{2}$

$+2\left(  t^{112}+t^{121}+t^{211}\right)  \left(  e_{1}\otimes e_{1}\otimes
e_{2}+e_{1}\otimes e_{2}\otimes e_{1}+e_{2}\otimes e_{1}\otimes e_{1}\right)
$

$+2\left(  t^{122}+t^{212}+t^{221}\right)  \left(  e_{1}\otimes e_{2}\otimes
e_{2}+e_{2}\otimes e_{1}\otimes e_{2}+e_{2}\otimes e_{2}\otimes e_{1}\right)
$

If the tensor is symmetric : $t^{112}=t^{121}=t^{211},t^{122}=t^{212}=t^{221}
$ and

$S_{3}\left(  T\right)  =6\{t^{111}e_{1}\otimes e_{1}\otimes e_{1}%
+t^{112}\left(  e_{1}\otimes e_{1}\otimes e_{2}+e_{1}\otimes e_{2}\otimes
e_{1}+e_{2}\otimes e_{1}\otimes e_{1}\right)  $

$+t^{122}\left(  e_{1}\otimes e_{2}\otimes e_{2}+e_{2}\otimes e_{1}\otimes
e_{2}+e_{2}\otimes e_{2}\otimes e_{1}\right)  +t^{222}e_{2}\otimes
e_{2}\otimes e_{2}\}$

$S_{3}\left(  T\right)  =6T$

\begin{notation}
$\odot^{r}E$ is the set of symmetric r-contravariant tensors on E
\end{notation}

\begin{notation}
$\odot_{r}E^{\ast}$ is the set of symmetric r-covariant tensors on E
\end{notation}

\begin{theorem}
The set $\odot^{r}E$ of symmetric r-contravariant tensors is a vector subspace
of $\otimes^{r}E.$
\end{theorem}

If $\left(  e_{i}\right)  _{i\in I}$\ \ is a basis of E, with I an ordered set,

$\otimes e_{j_{1}}\otimes e_{j_{2}}...\otimes e_{j_{r}},j_{1}\leq j_{2}..\leq
j_{r}$

$\equiv\left(  \otimes e_{i_{1}}\right)  ^{r_{1}}\otimes\left(  \otimes
e_{i_{2}}\right)  ^{r_{2}}\otimes...\left(  \otimes e_{i_{k}}\right)  ^{r_{k}%
},i_{1}<i_{2}..<i_{k},\sum_{l=1}^{k}r_{l}=r$

is a basis of $\odot^{r}E$

If E is n-dimensional $\dim\odot^{r}E=C_{n-1+r}^{n-1}$

The symmetrizer is a multilinear symmetric map : $s_{r}:E^{r}\rightarrow
\odot^{r}E::s_{r}\in L^{r}\left(  E;\odot^{r}E\right)  $

\begin{theorem}
For any multilinear symmetric map $f\in L^{r}\left(  E;E^{\prime}\right)  $
there is a unique linear map $F\in L\left(  \overset{r}{\otimes}E;E^{\prime
}\right)  $ such that : $F\circ s_{r}=r!f$
\end{theorem}

\begin{proof}
$\forall u_{i}\in E,i=1...r,\sigma\in\mathfrak{S}_{r}:f(u_{1},u_{2}%
,...,u_{r})=f(u_{\sigma\left(  1\right)  },u_{\sigma\left(  2\right)
},...,u_{\sigma\left(  r\right)  })$

There is a unique linear map : $F\in L\left(  \overset{r}{\otimes}E;E^{\prime
}\right)  $ such that : $f=F\circ\imath$

$F\circ s_{r}(u_{1},u_{2},...,u_{r})=\sum_{\sigma\in\mathfrak{s}_{r}}F\left(
u_{\sigma\left(  1\right)  }\otimes..\otimes u_{\sigma\left(  r\right)
}\right)  $

$=\sum_{\sigma\in\mathfrak{S}\left(  r\right)  }F\circ i\left(  u_{\sigma
\left(  1\right)  },..,u_{\sigma\left(  r\right)  }\right)  =\sum_{\sigma
\in\mathfrak{S}\left(  r\right)  }f\left(  u_{\sigma\left(  1\right)
},..,u_{\sigma\left(  r\right)  }\right)  $

$=\sum_{\sigma\in\mathfrak{S}\left(  r\right)  }f\left(  u_{1},..,u_{r}%
\right)  =r!f\left(  u_{1},..,u_{r}\right)  $
\end{proof}

\paragraph{Symmetric tensorial product\newline}

\bigskip

The tensorial product of two symmetric tensors is not necessarily symmetric
so, in order to have an internal operation for $\odot^{r}E$ one defines :

\begin{definition}
The symmetric tensorial product of r vectors of E, denoted by $\odot$ , is the
map :

$\odot:E^{r}\rightarrow\odot^{r}E::u_{1}\odot u_{2}..\odot u_{r}=\sum
_{\sigma\in\mathfrak{S}\left(  r\right)  }u_{\sigma\left(  1\right)  }%
\otimes..\otimes u_{\sigma\left(  r\right)  }$

$=s_{r}\left(  u_{1},..,u_{r}\right)  =S_{r}\circ\imath\left(  u_{1}%
,..,u_{r}\right)  $
\end{definition}

notice that there is no r!

\begin{theorem}
The symmetric tensorial product of r vectors is a multilinear, distributive
over addition, \textit{symmetric} map
\end{theorem}

$u_{\sigma\left(  1\right)  }\odot u_{\sigma\left(  2\right)  }..\odot
u_{\sigma\left(  r\right)  }=u_{1}\odot u_{2}..\odot u_{r}$

$\left(  \lambda u+\mu v\right)  \odot w=\lambda u\odot w+\mu v\odot w$

Examples:

$u\odot v=u\otimes v+v\otimes u$

$u_{1}\odot u_{2}\odot u_{3}=$

$u_{1}\otimes u_{2}\otimes u_{3}+u_{1}\otimes u_{3}\otimes u_{2}+u_{2}\otimes
u_{1}\otimes u_{3}+u_{2}\otimes u_{3}\otimes u_{1}+u_{3}\otimes u_{1}\otimes
u_{2}+u_{3}\otimes u_{2}\otimes u_{1}$

$u\odot u\odot v$

$=u\otimes u\otimes v+u\otimes v\otimes u+u\otimes u\otimes v+u\otimes
v\otimes u+v\otimes u\otimes u+v\otimes u\otimes u$

$=2\left(  u\otimes u\otimes v+u\otimes v\otimes u+v\otimes u\otimes u\right)
$

\begin{theorem}
If $\left(  e_{i}\right)  _{i\in I}$\ \ is a basis of E, with I an ordered
set, the set of ordered products $e_{i_{1}}\odot e_{i_{2}}\odot...\odot
e_{i_{r}},i_{1}\leq i_{2}..\leq i_{r}$ is a basis of $\odot^{r}E$
\end{theorem}

\begin{proof}
Let $T=\sum_{\left(  i_{1}...i_{r}\right)  }t^{i_{1}...i_{r}}e_{i_{1}}%
\otimes...\otimes e_{i_{r}}\in\odot^{r}E$

$S_{r}\left(  T\right)  =\sum_{\left(  i_{1}...i_{r}\right)  }t^{i_{1}%
...i_{r}}\sum_{\sigma\in\mathfrak{S}\left(  r\right)  }e_{\sigma\left(
i_{1}\right)  }\otimes..\otimes e_{\sigma\left(  i_{r}\right)  }=\sum_{\left(
i_{1}...i_{r}\right)  }t^{i_{1}...i_{r}}e_{i_{1}}\odot..\odot e_{i_{r}}=r!T$

For any set $\left(  i_{1}...i_{r}\right)  $ let $\left[  j_{1},j_{2}%
,...j_{r}\right]  =\sigma\left(  i_{1}...i_{r}\right)  $ with $j_{1}\leq
j_{2}..\leq j_{r}$

$t^{i_{1}...i_{r}}e_{i_{1}}\odot..\odot e_{i_{r}}=t^{j_{1}...j_{r}}e_{j_{1}%
}\odot..\odot e_{j_{r}}$

$S_{r}\left(  T\right)  =\sum_{\left[  j_{1}...j_{r}\right]  }s_{\left[
j_{1}...j_{r}\right]  }t^{j_{1}...j_{r}}e_{j_{1}}\odot..\odot e_{j_{r}}=r!T$

$T=\sum_{\left[  j_{1}...j_{r}\right]  }\frac{1}{r!}s_{\left[  j_{1}%
...j_{r}\right]  }t^{j_{1}...j_{r}}e_{j_{1}}\odot..\odot e_{j_{r}}$
\end{proof}

$s_{\left[  j_{1}...j_{r}\right]  }$ depends on the number of identical
indices in $\left[  j_{1}...j_{r}\right]  $

Example :

$T=\{t^{111}e_{1}\otimes e_{1}\otimes e_{1}+t^{112}\left(  e_{1}\otimes
e_{1}\otimes e_{2}+e_{1}\otimes e_{2}\otimes e_{1}+e_{2}\otimes e_{1}\otimes
e_{1}\right)  $

$+t^{122}\left(  e_{1}\otimes e_{2}\otimes e_{2}+e_{2}\otimes e_{1}\otimes
e_{2}+e_{2}\otimes e_{2}\otimes e_{1}\right)  +t^{222}e_{2}\otimes
e_{2}\otimes e_{2}\}$

$=\frac{1}{6}t^{111}e_{1}\odot e_{1}\odot e_{1}+\frac{1}{2}t^{112}e_{1}\odot
e_{1}\odot e_{2}+\frac{1}{2}t^{122}e_{1}\odot e_{2}\odot e_{2}+\frac{1}%
{6}t^{222}e_{2}\odot e_{2}\odot e_{2}$

\bigskip

The symmetric tensorial product is generalized for symmetric tensors :

a) define for vectors :

$\left(  u_{1}\odot...\odot u_{p}\right)  \odot\left(  u_{p+1}\odot...\odot
u_{p+q}\right)  =\sum_{\sigma\in\mathfrak{S}\left(  p+q\right)  }%
u_{\sigma\left(  1\right)  }\otimes u_{\sigma\left(  2\right)  }...\otimes
u_{\sigma\left(  +q\right)  }=u_{1}\odot...\odot u_{p}\odot u_{p+1}%
\odot...\odot u_{p+q}$

This product is commutative

b) so for any symmetric tensors :

$T=\sum_{\left[  i_{1}...i_{p}\right]  }t^{i_{1}...i_{p}}e_{i_{1}}\odot
e_{i_{2}}\odot...e_{i_{p}},U=\sum_{\left[  i_{1}...j_{q}\right]  }%
u^{j_{1}...j_{q}}e_{j_{1}}\odot e_{j_{2}}\odot...e_{j_{q}}$

$T\odot U=\sum_{\left[  i_{1}...i_{p}\right]  }\sum_{\left[  i_{1}%
...j_{q}\right]  }t^{i_{1}...i_{p}}u^{j_{1}...j_{q}}e_{i_{1}}\odot e_{i_{2}%
}\odot...e_{i_{p}}\odot e_{j_{1}}\odot e_{j_{2}}\odot...e_{j_{q}}$

$T\odot U=\sum_{\left[  k_{1},..k_{p+q}\right]  _{p+q}}\sum_{\left[
i_{1}...i_{p}\right]  ,\left[  i_{1}...j_{q}\right]  \subset\left[
k_{1},..k_{p+q}\right]  }t^{i_{1}...i_{p}}u^{j_{1}...j_{q}}e_{k_{1}}\odot
e_{k_{2}}...\odot e_{k_{p+q}}$

\begin{theorem}
The symmetric tensorial product of symmetric tensors is a bilinear,
associative, commutative, distributive over addition, map :

$\odot:\odot^{p}E\times\odot^{q}E\rightarrow\odot^{p+q}E$
\end{theorem}

\begin{theorem}
If E is a vector space over the field K, the set $\odot E=\oplus_{r=0}%
^{\infty}\odot^{r}E\subset T\left(  E\right)  ,$ with $\odot^{0}E=K,\odot
^{1}E=E$ is, with symmetric tensorial product, a graded unital algebra over K,
called the \textbf{symmetric algebra S(E)}
\end{theorem}

Notice that $\odot E\subset T(E)$

There are the algebra isomorphisms :

$\hom\left(  \odot^{r}E,F\right)  \simeq L_{s}^{r}\left(  E^{r};F\right)  $
symmetric multilinear maps

$\odot^{r}E^{\ast}\simeq\left(  \odot^{r}E\right)  ^{\ast}$

\paragraph{Algebraic definition\newline}

There is another definition, more common in pure algebra (Knapp p.645). The
\textbf{symmetric algebra S(E)} is the quotient set :

$S(E)=$

$T(E)/\left(  \text{two-sided ideal generated by the tensors }u\otimes
v-v\otimes u\text{ with }u,v\in E\right)  $

The tensor product translates in a symmetric tensor product $\odot$ which
makes S(E) an algebra.

With this definition the elements of S(E) are not tensors (but classes of
equivalence) so in practical calulations it is rather confusing. In this book
we will only consider symmetric tensors (and not bother with the quotient space).

\subsubsection{The set of antisymmetric tensors}

\paragraph{Antisymmetric tensors\newline}

\begin{definition}
On a vector space E the \textbf{antisymmetrisation operator }or
antisymmetrizer is the map :

$a_{r}:E^{r}\rightarrow\overset{r}{\otimes}E::a_{r}\left(  u_{1}%
,..,u_{r}\right)  =\sum_{\sigma\in\mathfrak{S}\left(  r\right)  }%
\epsilon\left(  \sigma\right)  u_{\sigma\left(  1\right)  }\otimes..\otimes
u_{\sigma\left(  r\right)  }$
\end{definition}

The antisymmetrizer is an antisymmetric multilinear map : $a_{r}\in
L^{r}\left(  E;\otimes^{r}E\right)  $

$a_{r}\left(  u_{\sigma\left(  1\right)  },u_{\sigma\left(  2\right)
},...,u_{\sigma\left(  r\right)  }\right)  =\sum_{\sigma^{\prime}%
\in\mathfrak{S}\left(  r\right)  }\epsilon\left(  \sigma^{\prime}\right)
u_{\sigma^{\prime}\sigma\left(  1\right)  }\otimes..\otimes u_{\sigma^{\prime
}\sigma\left(  r\right)  }$

$=\sum_{\sigma\sigma^{\prime}\in\mathfrak{S}\left(  r\right)  }\epsilon\left(
\sigma\right)  \epsilon\left(  \sigma\sigma^{\prime}\right)  u_{\sigma
^{\prime}\sigma\left(  1\right)  }\otimes..\otimes u_{\sigma^{\prime}%
\sigma\left(  r\right)  }$

$=\epsilon\left(  \sigma\right)  \sum_{\theta\in\mathfrak{S}\left(  r\right)
}\epsilon\left(  \theta\right)  u_{\theta\left(  1\right)  }\otimes..\otimes
u_{\theta\left(  r\right)  }=\epsilon\left(  \sigma\right)  a_{r}(u_{1}%
,u_{2},...,u_{r})$

It is a multilinear map so there is a unique linear map : $A_{r}:\overset
{r}{\otimes}E\rightarrow\overset{r}{\otimes}E:$ such that : $a_{r}=A_{r}%
\circ\imath$ with $\imath:E^{r}\rightarrow\overset{r}{\otimes}E$

For any tensor $T\in\overset{r}{\otimes}E:$

$A_{r}\left(  T\right)  =\sum_{\left(  i_{1}...i_{r}\right)  }t^{i_{1}%
...i_{r}}A_{r}\left(  e_{i_{1}}\otimes...\otimes e_{i_{r}}\right)
=\sum_{\left(  i_{1}...i_{r}\right)  }t^{i_{1}...i_{r}}A_{r}\circ\imath\left(
e_{i_{1}},...,e_{i_{r}}\right)  $

$=\sum_{\left(  i_{1}...i_{r}\right)  }t^{i_{1}...i_{r}}a_{r}\left(  e_{i_{1}%
},...,e_{i_{r}}\right)  =\sum_{\left(  i_{1}...i_{r}\right)  }t^{i_{1}%
...i_{r}}\sum_{\sigma\in\mathfrak{S}\left(  r\right)  }\epsilon\left(
\sigma\right)  e_{\sigma\left(  i_{1}\right)  }\otimes..\otimes e_{\sigma
\left(  i_{r}\right)  }$

$a_{r}\left(  e_{1},..,e_{r}\right)  =A_{r}\circ\imath\left(  e_{1}%
,..,e_{r}\right)  =A_{r}\left(  e_{1}\otimes..\otimes e_{r}\right)
=\sum_{\sigma\in\mathfrak{S}\left(  r\right)  }\epsilon\left(  \sigma\right)
e_{\sigma\left(  1\right)  }\otimes..\otimes e_{\sigma\left(  r\right)  }$

\begin{definition}
An \textbf{antisymmetric r tensor} is a tensor T such that $A_{r}\left(
T\right)  =r!T$
\end{definition}

In a basis a r antisymmetric tensor $T=\sum_{\left(  i_{1}...i_{r}\right)
}t^{i_{1}...i_{r}}e_{i_{1}}\otimes...\otimes e_{i_{r}}$ is such that :

$t^{i_{1}...i_{r}}=\epsilon\left(  \sigma\right)  t^{\sigma\left(
i_{1}...i_{r}\right)  }$ $\Leftrightarrow i_{1}<i_{2}..<i_{k}:t^{\sigma\left(
i_{1}...i_{r}\right)  }=\epsilon\left(  \sigma\left(  i_{1},...i_{r}\right)
\right)  t^{i_{1}...i_{r}}$

where $\sigma$ is any permutation of the set of r-indices. It implies that
$t^{i_{1}...i_{r}}=0$ whenever two of the indices have the same value.

Thus an antisymmetric tensor is uniquely defined by a set of components
$t^{i_{1}...i_{r}}$ for all ordered indices $\left\{  i_{1}...i_{r}\right\}  $

$T=\sum_{\left\{  i_{1}...i_{r}\right\}  }t^{i_{1}...i_{r}}\left(
\sum_{\sigma\in\mathfrak{S}\left(  r\right)  }\epsilon\left(  \sigma\right)
e_{\sigma\left(  i_{1}\right)  }\otimes...\otimes e_{\sigma\left(
i_{r}\right)  }\right)  $

\begin{notation}
$\Lambda^{r}E$ is the set of antisymmetric r-contravariant tensors on E
\end{notation}

\begin{notation}
$\Lambda_{r}E^{\ast}$ is the set of antisymmetric r-covariant tensors on E
\end{notation}

\begin{theorem}
The set of antisymmetric r-contravariant tensors $\Lambda^{r}E$ is a vector
subspace of $\otimes^{r}E.$
\end{theorem}

A basis of the vector subspace $\Lambda^{r}E$ is : $e_{i_{1}}\otimes e_{i_{2}%
}\otimes...\otimes e_{i_{r}},i_{1}<i_{2}..<i_{n}$

If E is n-dimensional $\dim\Lambda^{r}E=C_{n}^{r}$ and :

i) there is no antisymmetric tensor of order r%
$>$%
N

ii) $\dim\Lambda^{n}E=1$ so all antisymmetric n-tensors are proportionnal

iii) $\Lambda^{n-r}E\simeq\Lambda^{r}E$ : they are isomorphic vector spaces

\begin{theorem}
For any multilinear antisymmetric map $f\in L^{r}\left(  E;E^{\prime}\right)
$ there is a unique linear map $F\in L\left(  \overset{r}{\otimes}E;E^{\prime
}\right)  $ such that : $F\circ a_{r}=r!f$
\end{theorem}

\begin{proof}
$\forall u_{i}\in E,i=1...r,\sigma\in\mathfrak{S}_{r}:f(u_{1},u_{2}%
,...,u_{r})=\epsilon\left(  \sigma\right)  f(u_{\sigma\left(  1\right)
},u_{\sigma\left(  2\right)  },...,u_{\sigma\left(  r\right)  })$

There is a unique linear map : $F\in L\left(  \overset{r}{\otimes}E;E^{\prime
}\right)  $ such that : $f=F\circ\imath$

$F\circ a_{r}(u_{1},u_{2},...,u_{r})=\sum_{\sigma\in\mathfrak{S}\left(
r\right)  }\epsilon\left(  \sigma\right)  F\left(  u_{\sigma\left(  1\right)
}\otimes..\otimes u_{\sigma\left(  r\right)  }\right)  $

$=\sum_{\sigma\in\mathfrak{S}\left(  r\right)  }\epsilon\left(  \sigma\right)
F\circ i\left(  u_{\sigma\left(  1\right)  },..,u_{\sigma\left(  r\right)
}\right)  =\sum_{\sigma\in\mathfrak{S}\left(  r\right)  }\epsilon\left(
\sigma\right)  f\left(  u_{\sigma\left(  1\right)  },..,u_{\sigma\left(
r\right)  }\right)  $

$=\sum_{\sigma\in\mathfrak{S}\left(  r\right)  }f\left(  u_{1},..,u_{r}%
\right)  =r!f\left(  u_{1},..,u_{r}\right)  $
\end{proof}

\paragraph{Exterior product\newline}

\bigskip

The tensor product of 2 antisymmetric tensor is not necessarily antisymmetric
so, in order to have an internal operation for $\Lambda^{r}E$ one defines :

\begin{definition}
The \textbf{exterior product} (or \textbf{wedge product}) of r vectors is the
map :

$\wedge:E^{r}\rightarrow\Lambda^{r}E::u_{1}\wedge u_{2}..\wedge u_{r}%
=\sum_{\sigma\in\mathfrak{S}\left(  r\right)  }\epsilon\left(  \sigma\right)
u_{\sigma\left(  1\right)  }\otimes..\otimes u_{\sigma\left(  r\right)
}=a_{r}\left(  u_{1},..,u_{r}\right)  $
\end{definition}

notice that there is no r!

\begin{theorem}
The exterior product of vectors is a multilinear, antisymmetric map , which is
distributive over addition
\end{theorem}

$u_{\sigma\left(  1\right)  }\wedge u_{\sigma\left(  2\right)  }..\wedge
u_{\sigma\left(  r\right)  }=\epsilon\left(  \sigma\right)  u_{1}\wedge
u_{2}..\wedge u_{r}$

$\left(  \lambda u+\mu v\right)  \wedge w=\lambda u\wedge w+\mu v\wedge w$

Moreover :

$u_{1}\wedge u_{2}..\wedge u_{r}=0\Leftrightarrow$ the vectors are linearly dependant

$u\wedge v=0\Leftrightarrow\exists k\in K:u=kv$

Examples :

$u\wedge v=u\otimes v-v\otimes u$

$u_{1}\wedge u_{2}\wedge u_{3}=$

$u_{1}\otimes u_{2}\otimes u_{3}-u_{1}\otimes u_{3}\otimes u_{2}-u_{2}\otimes
u_{1}\otimes u_{3}+u_{2}\otimes u_{3}\otimes u_{1}+u_{3}\otimes u_{1}\otimes
u_{2}-u_{3}\otimes u_{2}\otimes u_{1}$

$u_{1}\wedge u_{1}\wedge u_{3}=0$

\begin{theorem}
The set of antisymmetric tensors : $e_{i_{1}}\wedge e_{i_{2}}\wedge
...e_{i_{r}},i_{1}<i_{2}..<i_{r},$ is a basis of $\Lambda^{r}E$
\end{theorem}

\begin{proof}
Let $T=\sum_{\left(  i_{1}...i_{r}\right)  }t^{i_{1}...i_{r}}e_{i_{1}}%
\otimes...\otimes e_{i_{r}}\in\Lambda^{r}E$

$A_{r}\left(  T\right)  =\sum_{\left(  i_{1}...i_{r}\right)  }t^{i_{1}%
...i_{r}}\sum_{\sigma\in\mathfrak{s}_{r}}\epsilon\left(  \sigma\right)
e_{\sigma\left(  i_{1}\right)  }\otimes..\otimes e_{\sigma\left(
i_{r}\right)  }=\sum_{\left(  i_{1}...i_{r}\right)  }t^{i_{1}...i_{r}}%
e_{i_{1}}\wedge..\wedge e_{i_{r}}=r!T$

For any set $\left(  i_{1}...i_{r}\right)  $ let $\left\{  j_{1}%
,j_{2},...j_{r}\right\}  =\sigma\left(  i_{1}...i_{r}\right)  $ with
$j_{1}<j_{2}..<j_{r}$

$t^{i_{1}...i_{r}}e_{i_{1}}\wedge..\wedge e_{i_{r}}=t^{j_{1}...j_{r}}e_{j_{1}%
}\wedge..\wedge e_{j_{r}}$

$A_{r}\left(  T\right)  =\sum_{\left\{  j_{1}...j_{r}\right\}  }%
r!t^{j_{1}...j_{r}}e_{j_{1}}\wedge..\wedge e_{j_{r}}=r!T$

$T=\sum_{\left\{  j_{1}...j_{r}\right\}  }t^{j_{1}...j_{r}}e_{j_{1}}%
\wedge..\wedge e_{j_{r}}$
\end{proof}

\bigskip

The exterior product is generalized between antisymmetric tensors:

a) define for vectors :

$\left(  u_{1}\wedge...\wedge u_{p}\right)  \wedge\left(  u_{p+1}%
\wedge...\wedge u_{p+q}\right)  =\sum_{\sigma\in\mathfrak{s}_{p+q}}%
\epsilon\left(  \sigma\right)  u_{\sigma\left(  1\right)  }\otimes
u_{\sigma\left(  2\right)  }...\otimes u_{\sigma\left(  p+q\right)  }%
=u_{1}\wedge...\wedge u_{p}\wedge u_{p+1}\wedge...\wedge u_{p+q}$

Notice that the exterior product of vectors is anticommutative, but the
exterior product of tensors\textit{\ is not anticommutative} :

$\left(  u_{1}\wedge...\wedge u_{p}\right)  \wedge\left(  u_{p+1}%
\wedge...\wedge u_{p+q}\right)  =\left(  -1\right)  ^{pq}\left(  u_{p+1}%
\wedge...\wedge u_{p+q}\right)  \wedge\left(  u_{1}\wedge...\wedge
u_{p}\right)  $

b) so for any antisymmetric tensor :

$T=\sum_{\left\{  i_{1}...i_{p}\right\}  }t^{i_{1}...i_{p}}e_{i_{1}}\wedge
e_{i_{2}}\wedge...e_{i_{p}},U=\sum_{\left\{  i_{1}...j_{q}\right\}  }%
u^{j_{1}...j_{q}}e_{j_{1}}\wedge e_{j_{2}}\wedge...e_{j_{q}}$

$T\wedge U=\sum_{\left\{  i_{1}...i_{p}\right\}  }\sum_{\left\{  i_{1}%
...j_{q}\right\}  }t^{i_{1}...i_{p}}u^{j_{1}...j_{q}}e_{i_{1}}\wedge e_{i_{2}%
}\wedge...e_{i_{p}}\wedge e_{j_{1}}\wedge e_{j_{2}}\wedge...e_{j_{q}}$

$T\wedge U=\frac{1}{p!q!}\sum_{\left(  i_{1}...i_{p}\right)  }\sum_{\left(
i_{1}...j_{q}\right)  }t^{i_{1}...i_{p}}u^{j_{1}...j_{q}}e_{i_{1}}\wedge
e_{i_{2}}\wedge...e_{i_{p}}\wedge e_{j_{1}}\wedge e_{j_{2}}\wedge...e_{j_{q}}$

$e_{i_{1}}\wedge e_{i_{2}}\wedge...e_{i_{p}}\wedge e_{j_{1}}\wedge e_{j_{2}%
}\wedge...e_{j_{q}}=\epsilon\left(  i_{1},..i_{p},j_{1},...j_{q}\right)
e_{k_{1}}\wedge e_{k_{2}}\wedge..\wedge e_{k_{p+q}}$

where $\left(  k_{1},...k_{p+q}\right)  $ is the ordered set of indices :
$\left(  i_{1},..i_{p},j_{1},...j_{q}\right)  $

Expressed in the basis $e_{i_{1}}\wedge e_{i_{2}}...\wedge e_{i_{p+q}}$ of
$\Lambda^{p+q}E:$

$T\Lambda S=$

$\sum_{\{j_{1},..j_{p+q}\}_{p+q}}\left(  \sum_{\{j_{1},..j_{p}\},\left\{
j_{p+1},...j_{p+q}\right\}  \subset\{i_{1},..i_{p+q}\}}\epsilon\left(
j_{1},..j_{p},j_{p+1},...j_{p+q}\right)  T^{\{j_{1},..j_{p}\}}S^{\left\{
j_{p+1},...j_{p+q}\right\}  }\right)  $

$e_{j_{1}}\wedge e_{j_{2}}...\wedge e_{j_{p+q}}$

or with

$\left\{  A\right\}  =\{j_{1},..j_{p}\},\left\{  B\right\}  =\left\{
j_{p+1},...j_{p+q}\right\}  ,\left\{  C\right\}  =\left\{  j_{1}%
,..j_{p},j_{p+1},...j_{p+q}\right\}  =\left\{  \left\{  A\right\}
\cup\left\{  B\right\}  \right\}  $

$\left\{  B\right\}  =\left\{  j_{p+1},...j_{p+q}\right\}  =\left\{
C/\left\{  A\right\}  \right\}  $

$T\Lambda S=\sum_{\left\{  C\right\}  _{p+q}}\left(  \sum_{\{A\}_{p}}%
\epsilon\left(  \left\{  A\right\}  ,\left\{  C/\left\{  A\right\}  \right\}
\right)  T^{\{A\}}S^{\left\{  C/\left\{  A\right\}  \right\}  }\right)  \wedge
e_{\left\{  C\right\}  }$

\begin{theorem}
The wedge product of antisymmetric tensors is a multilinear, distributive over
addition, associative map :

$\wedge:\Lambda^{p}E\times\Lambda^{q}E\rightarrow\Lambda^{p+q}E$
\end{theorem}

Moreover:

$T\wedge U=\left(  -1\right)  ^{pq}U\wedge T$

$k\in K:T\wedge k=kT$

\begin{theorem}
For the vector space E over the field K, the set denoted : $\Lambda
E=\oplus_{n=0}^{\dim E}\Lambda^{n}E$ with $\Lambda^{0}E=K$ is, with the
exterior product, a graded unital algebra (the identity element is 1$\in K)$
over K
\end{theorem}

$\dim\Lambda E=2^{\dim E}$

There are the algebra isomorphisms :

$\hom\left(  \Lambda^{r}E,F\right)  \simeq L_{A}^{r}\left(  E^{r};F\right)  $
antisymmetric multilinear maps

$\Lambda^{r}E^{\ast}\simeq\left(  \Lambda^{r}E\right)  ^{\ast}$

The elements T of $\Lambda E$ which can be written as : $T=u_{1}\wedge
u_{2}...\wedge u_{r}$ are homogeneous.

\begin{theorem}
An antisymmetric tensor is homogeneous iff $T\wedge T=0$
\end{theorem}

Warning ! usually $T\wedge T\neq0$

\begin{theorem}
On a finite dimensional vector space E on a field K there is a unique map,
called \textbf{determinant} :

$\det:L\left(  E;E\right)  \rightarrow K$ such that

$\forall u_{1},u_{2},...u_{n}\in E:f\left(  u_{1}\right)  \wedge f\left(
u_{2}\right)  ...\wedge f\left(  u_{n}\right)  =\left(  \det f\right)
u_{1}\wedge u_{2}...\wedge u_{n}$
\end{theorem}

\begin{proof}
$F=a_{r}\circ f:E^{n}\rightarrow\Lambda^{n}E::F\left(  u_{1},...,u_{n}\right)
=f\left(  u_{1}\right)  \wedge f\left(  u_{2}\right)  ...\wedge f\left(
u_{n}\right)  $

is a multilinear, antisymmetric map. So there is a unique linear map
$D:\Lambda^{n}E\rightarrow\Lambda^{n}E$ such that

$\det\circ a_{r}=n!F$

$F\left(  u_{1},...,u_{n}\right)  =f\left(  u_{1}\right)  \wedge f\left(
u_{2}\right)  ...\wedge f\left(  u_{n}\right)  =\frac{1}{n!}D\left(
u_{1}\wedge...\wedge u_{n}\right)  $

As all the n-antisymmetric tensors are proportional, $D\left(  u_{1}%
\wedge...\wedge u_{n}\right)  =k\left(  f\right)  \left(  u_{1}\wedge...\wedge
u_{n}\right)  $ with $k:L(E;E)\rightarrow K.$
\end{proof}

\paragraph{Algebraic definition\newline}

There is another definition which is common in algebra (see Knapp p.651). The
algebra $A\left(  E\right)  $ is defined as the quotient set
:$A(E)=T(E)/\left(  I\right)  $ where $I=$two-sided ideal generated by the
tensors of the kind $u\otimes v+v\otimes u$ with $u,v\in E.$ The interior
product of T(E), that is the tensor product, goes in $A\left(  E\right)  $ as
an interior product denoted $\wedge$ and called wedge product, with which
$A\left(  E\right)  $ is an algebra. It is computed differently :

$u_{1}\wedge..\wedge u_{r}=\frac{1}{r!}\sum_{\sigma\in\mathfrak{s}_{n}%
}\epsilon\left(  \sigma\left(  1,..r\right)  \right)  u_{\sigma\left(
1\right)  }\otimes..\otimes u_{\sigma\left(  r\right)  }$

The properties are the same than above, but $A^{r}\left(  E\right)  $
\textit{is not a subset} of $\overset{r}{\otimes}E.$ So the exterior product
of two elements of A(E) is more complicated and not easily linked to the
tensorial product.

In this book we will only consider antisymmetric tensors (and not bother with
the quotient space).

\subsubsection{Exterior algebra}

All the previous material can be easily extended to the dual E* of a vector
space, but the exterior algebra $\Lambda E^{\ast}$ is by far more widely used
than $\Lambda E$ and has some specific properties.

\paragraph{r-forms\newline}

\begin{definition}
The \textbf{exterior algebra} (also called Grassman algebra) of a vector space
E is the algebra $\Lambda E^{\ast}=\Lambda\left(  E^{\ast}\right)  =\left(
\Lambda E\right)  ^{\ast}.$
\end{definition}

So $\Lambda E^{\ast}=\oplus_{r=0}^{\dim E}\Lambda_{r}E^{\ast}$ and
$\Lambda_{0}E^{\ast}=K,\Lambda_{1}E^{\ast}=E^{\ast}$ (all indices down)

The tensors of $\Lambda_{r}E^{\ast}$ are called \textbf{r-forms} : they are
antisymmetric multilinear functions $E^{r}\rightarrow K$

\bigskip

In the following E is a n-dimensional vector space with basis $\left(
e_{i}\right)  _{i=1}^{n},$ and the dual basis $\left(  e^{i}\right)
_{i=1}^{n}$ of E*:$e^{i}\left(  e_{j}\right)  =\delta_{j}^{i}$

So $\varpi\in\Lambda_{r}E^{\ast}$ can be written equivalently :

i) $\varpi=\sum_{\left\{  i_{1}...i_{r}\right\}  }\varpi_{i_{1}...i_{r}%
}e^{i_{1}}\wedge e^{i_{2}}\wedge...\wedge e^{i_{r}}$ with ordered indices

ii) $\varpi=\frac{1}{r!}\sum_{\left(  i_{1}...i_{r}\right)  }\varpi
_{i_{1}...i_{r}}e^{i_{1}}\wedge e^{i_{2}}\wedge...\wedge e^{i_{r}}$ with non
ordered indices

ii) $\varpi=\sum_{\left(  i_{1}...i_{r}\right)  }\varpi_{i_{1}...i_{r}%
}e^{i_{1}}\otimes e^{i_{2}}\otimes...\otimes e^{i_{r}}$ with non ordered indices

\begin{proof}
$\varpi=\sum_{\left\{  i_{1}...i_{r}\right\}  }\varpi_{i_{1}...i_{r}}e^{i_{1}%
}\wedge e^{i_{2}}\wedge...\wedge e^{i_{r}}$

$=\sum_{\left\{  i_{1}...i_{r}\right\}  }\varpi_{i_{1}...i_{r}}\sum_{\sigma
\in\mathfrak{S}\left(  r\right)  }\epsilon\left(  \sigma\right)
e^{i_{\sigma\left(  1\right)  }}\otimes e^{i_{\sigma\left(  2\right)  }%
}\otimes...\otimes e^{i_{\sigma\left(  r\right)  }}$

$=\sum_{\left\{  i_{1}...i_{r}\right\}  }\sum_{\sigma\in\mathfrak{S}\left(
r\right)  }\varpi_{i_{\sigma\left(  1\right)  }...i_{\sigma\left(  r\right)
}}e^{i_{\sigma\left(  1\right)  }}\otimes e^{i_{\sigma\left(  2\right)  }%
}\otimes...\otimes e^{i_{\sigma\left(  r\right)  }}$

$=\sum_{\sigma\in\mathfrak{S}\left(  r\right)  }\sum_{\left\{  i_{1}%
...i_{r}\right\}  }\varpi_{i_{\sigma\left(  1\right)  }...i_{\sigma\left(
r\right)  }}e^{i_{\sigma\left(  1\right)  }}\otimes e^{i_{\sigma\left(
2\right)  }}\otimes...\otimes e^{i_{\sigma\left(  r\right)  }}$

$=\sum_{\left(  i_{1}...i_{r}\right)  }\varpi_{i_{1}...i_{r}}e^{i_{1}}\otimes
e^{i_{2}}\otimes...\otimes e^{i_{r}}$

$\varpi=\sum_{\left(  i_{1}...i_{r}\right)  }\varpi_{i_{1}...i_{r}}e^{i_{1}%
}\wedge e^{i_{2}}\wedge...\wedge e^{i_{r}}$

$=\sum_{\left\{  i_{1}...i_{r}\right\}  }\sum_{\sigma\in\mathfrak{S}\left(
r\right)  }\varpi_{i_{\sigma\left(  1\right)  }...i_{\sigma\left(  r\right)
}}e^{i_{\sigma\left(  1\right)  }}\wedge e^{i_{\sigma\left(  2\right)  }%
}\wedge...\wedge e^{i_{\sigma\left(  r\right)  }}$

$=\sum_{\left\{  i_{1}...i_{r}\right\}  }\sum_{\sigma\in\mathfrak{S}\left(
r\right)  }\varpi_{i_{\sigma\left(  1\right)  }...i_{\sigma\left(  r\right)
}}\epsilon\left(  \sigma\right)  e^{i_{1}}\wedge e^{i_{2}}\wedge...\wedge
e^{i_{r}}$

$=\sum_{\left\{  i_{1}...i_{r}\right\}  }\sum_{\sigma\in\mathfrak{S}\left(
r\right)  }\varpi_{i_{1}...i_{r}}e^{i_{1}}\wedge e^{i_{2}}\wedge...\wedge
e^{i_{r}}$

$=r!\sum_{\left\{  i_{1}...i_{r}\right\}  }\varpi_{i_{1}...i_{r}}e^{i_{1}%
}\wedge e^{i_{2}}\wedge...\wedge e^{i_{r}}$
\end{proof}

\bigskip

In a change of basis : $f_{i}=\sum_{j=1}^{n}P_{i}^{j}e_{j}$ the components of
antisymmetric tensors change according to the following rules :

$\varpi=\sum_{\left(  i_{1}...i_{r}\right)  }\varpi_{i_{1}...i_{r}}e^{i_{1}%
}\otimes e^{i_{2}}\otimes...\otimes e^{i_{r}}\rightarrow$

$\varpi=\sum_{\left(  i_{1}...i_{r}\right)  }\widetilde{\varpi}_{i_{1}%
...i_{r}}f^{i_{1}}\otimes f^{i_{2}}\otimes...\otimes f^{i_{r}}=\sum_{\left\{
i_{1}...i_{r}\right\}  }\widetilde{\varpi}_{i_{1}...i_{r}}f^{i_{1}}\wedge
f^{i_{2}}\wedge...\wedge f^{i_{r}}$

with $\widetilde{\varpi}_{i_{1}...i_{r}}=\sum_{\left(  j_{1}....j_{r}\right)
}\varpi_{j_{1}...j_{r}}P_{i_{1}}^{j_{1}}...P_{i_{r}}^{j_{r}}=\sum_{\left\{
j_{1}....j_{r}\right\}  }\epsilon\left(  \sigma\right)  \varpi_{j_{1}...j_{r}%
}P_{i_{1}}^{\sigma\left(  j_{1}\right)  }...P_{i_{r}}^{\sigma\left(
j_{r}\right)  }$%

\begin{equation}
\widetilde{\varpi}_{i_{1}...i_{r}}=\sum_{\left\{  j_{1}....j_{r}\right\}
}\varpi_{j_{1}...j_{r}}\det\left[  P\right]  _{i_{1}...i_{r}}^{j_{1}...j_{r}}%
\end{equation}

where $\det\left[  P\right]  _{i_{1}...i_{r}}^{j_{1}...j_{r}}$ is the
determinant of the matrix with r column $\left(  i_{1},..i_{r}\right)  $
comprised each of the components $\left(  j_{1}...j_{r}\right)  $ of the new
basis vectors

\paragraph{Interior product\newline}

\bigskip

The value of a r-form over r vectors of E is :

$\varpi=\sum_{\left(  i_{1}...i_{r}\right)  }\varpi_{i_{1}...i_{r}}e^{i_{1}%
}\otimes e^{i_{2}}\otimes...\otimes e^{i_{r}}$

$\varpi\left(  u_{1},...,u_{r}\right)  =\sum_{\left(  i_{1}...i_{r}\right)
}\varpi_{i_{1}...i_{r}}e^{i_{1}}\left(  u_{1}\right)  e^{i_{2}}\left(
u_{2}\right)  ...e^{i_{r}}\left(  u_{r}\right)  $

$\varpi\left(  u_{1},...,u_{r}\right)  =\sum_{\left(  i_{1}...i_{r}\right)
}\varpi_{i_{1}...i_{r}}u_{1}^{i_{1}}u_{2}^{i_{2}}...u_{r}^{i_{r}}$

The value of the exterior product of a p-form and a q-form $\varpi\wedge\pi$
for p+q vectors is given by the formula (Kolar p.62):

$\varpi\Lambda\pi\left(  u_{1},...,u_{p+q}\right)  =\frac{1}{p!q!}\sum
_{\sigma\in\mathfrak{S}\left(  p+q\right)  }\epsilon\left(  \sigma\right)
\varpi\left(  u_{\sigma\left(  1\right)  },...u_{\sigma\left(  p\right)
}\right)  \pi\left(  u_{\sigma\left(  p+1\right)  },...u_{\sigma\left(
p+q\right)  }\right)  $

If $r=\dim E:\varpi_{i_{1}...i_{n}}=\epsilon\left(  i_{1},...i_{n}\right)
\varpi_{12...n}$

$\varpi\left(  u_{1},...,u_{n}\right)  =\varpi_{12...n}\sum_{\sigma
\in\mathfrak{S}\left(  n\right)  }\epsilon\left(  \sigma\right)  u_{1}%
^{\sigma\left(  1\right)  }u_{2}^{\sigma\left(  2\right)  }...u_{n}%
^{\sigma\left(  n\right)  }=\varpi_{12...n}\det\left[  u_{1},u_{2}%
,...u_{n}\right]  $

This is the determinant of the matrix with columns the components of the
vectors u

\begin{definition}
The \textbf{interior product} of a r form $\varpi\in\Lambda_{r}E^{\ast}$ and a
vector $u\in E,$ denoted $i_{u}\varpi,$ is the r-1-form :%

\begin{equation}
i_{u}\varpi=\sum_{\{i_{1}...i_{r}\}}\sum_{k=1}^{r}(-1)^{k-1}u^{i_{k}}%
\varpi_{\{i_{1}...i_{r}\}}e^{\{i_{1}}\Lambda...\Lambda\widehat{e^{i_{k}}%
}...\Lambda e^{i_{p}\}}%
\end{equation}

where \symbol{94} means that the vector shall be omitted, with $\left(
e^{i}\right)  _{i\in I}$ a basis of E*.
\end{definition}

For u fixed the map : $i_{u}:\Lambda_{r}E^{\ast}\rightarrow\Lambda
_{r-1}E^{\ast}$ is linear : $i_{u}\in L\left(  \Lambda E;\Lambda E\right)  $

$i_{u}\circ i_{v}=-i_{v}\circ i_{u}$

$i_{u}\circ i_{u}=0$

$i_{u}\left(  \lambda\wedge\mu\right)  =\left(  i_{u}\lambda\right)  \wedge
\mu+\left(  -1\right)  ^{\deg\lambda}\lambda\wedge\mu$

\paragraph{Orientation of a vector space\newline}

For any n dimensional vector space E a basis can be chosen and its vectors
labelled $e_{1},...e_{n}$. One says that there are two possible orientations :
direct and indirect according to the value of the signature of any permutation
of these vectors. A vector space is orientable if it is possible to compare
the orientation of two different bases.

A change of basis is defined by an endomorphism $f\in GL\left(  E;E\right)  .$
Its determinant is such that :%

\begin{equation}
\forall u_{1},u_{2},...u_{n}\in E:f\left(  u_{1}\right)  \wedge f\left(
u_{2}\right)  ...\wedge f\left(  u_{n}\right)  =\left(  \det f\right)
u_{1}\wedge u_{2}...\wedge u_{n}%
\end{equation}

So if E is a real vector space det(f) is a non null real scalar, and two bases
have the same orientation if det(f)
$>$
0.

If E is a complex vector space, it has a real structure such that : $E=E_{%
\mathbb{R}
}\oplus iE_{%
\mathbb{R}
}.$ So take any basis $\left(  e_{i}\right)  _{i=1}^{n}$ of $E_{%
\mathbb{R}
}$\ and say that the basis : $\left(  e_{1},ie_{1},e_{2},ie_{2},...e_{n}%
,ie_{n}\right)  $ is direct. It does not depend on the choice of $\left(
e_{i}\right)  _{i=1}^{n}$ and is called the canonical orientation of E.

To sum up :

\begin{theorem}
All finite dimensional vector spaces over $%
\mathbb{R}
$ or $%
\mathbb{C}
$\ are orientable.
\end{theorem}

\paragraph{Volume}

\begin{definition}
A volume form on a n dimensional vector space $\left(  E,g\right)  $ with
scalar product is a n-form $\varpi$ such that its value on any direct
orthonormal basis is 1.
\end{definition}

\begin{theorem}
In any direct basis $\left(  e^{i}\right)  _{i=1}^{n}$ a volume form is%

\begin{equation}
\varpi=\sqrt{\left\vert \det g\right\vert }e_{1}\wedge e_{2}..\wedge e_{n}%
\end{equation}

In any orthonormal basis $\left(  \varepsilon_{i}\right)  _{i=1}^{n}$
$\varpi=\varepsilon_{1}\wedge\varepsilon_{2}..\wedge\varepsilon_{n}$
\end{theorem}

\begin{proof}
$\left(  E,g\right)  $ is\ endowed with a bilinear symmetric form g, non
degenerate (but not necessarily definite positive)$.$

In $\left(  e^{i}\right)  _{i=1}^{n}$ g has for matrix is $\left[  g\right]
=\left[  g_{ij}\right]  .$ $g_{ij}=g\left(  e_{i},e_{j}\right)  $

Let $\varepsilon_{i}=\sum_{j}P_{i}^{j}e_{j}$ then g has for matrix in
$\varepsilon_{i}:\left[  \eta\right]  =\left[  P\right]  ^{\ast}\left[
g\right]  \left[  P\right]  $ with $\eta_{ij}=\pm\delta_{ij}$

The value of $\varpi\left(  \varepsilon_{1},\varepsilon_{2},...\varepsilon
_{n}\right)  =\varpi_{12...n}\det\left[  \varepsilon_{1},\varepsilon
_{2},...\varepsilon_{n}\right]  =\varpi_{12...n}\det\left[  P\right]  $

But : $\det\left[  \eta\right]  =\det\left(  \left[  P\right]  ^{\ast}\left[
g\right]  \left[  P\right]  \right)  =\left\vert \det\left[  P\right]
\right\vert ^{2}\det\left[  g\right]  =\pm1$ depending on the signature of g

If E is a real vector space, then $\det\left[  P\right]  >0$ as the two bases
are direct. So : $\det\left[  P\right]  =1/\sqrt{\left\vert \det g\right\vert
}$ and :

$\varpi=\sqrt{\left\vert \det g\right\vert }e_{1}\wedge e_{2}..\wedge
e_{n}=\varepsilon_{1}\wedge\varepsilon_{2}..\wedge\varepsilon_{n}$

If E is a complex vector space $\left[  g\right]  =\left[  g\right]  ^{\ast}$
and $\det\left[  g\right]  ^{\ast}=\overline{\det\left[  g\right]  }%
=\det\left[  g\right]  $ so $\det\left[  g\right]  $ is real. It is always
possible to choose an orthonormal basis such that : $\eta_{ij}=\delta_{ij}$ so
we can still take $\varpi=\sqrt{\left\vert \det g\right\vert }e_{1}\wedge
e_{2}..\wedge e_{n}=\varepsilon_{1}\wedge\varepsilon_{2}..\wedge
\varepsilon_{n}$
\end{proof}

\begin{definition}
The \textbf{volume} spanned by n vectors $\left(  u_{1},...,u_{n}\right)  $ of
a real n dimensional vector space $\left(  E,g\right)  $ with scalar product
endowed with the volume form $\varpi$ is $\varpi\left(  u_{1},...,u_{n}%
\right)  $
\end{definition}

It is null if the vectors are linearly dependant.

Maps of the special orthogonal group SO(E,g) preserve both g and the
orientation, so they preserve the volume.

\bigskip

\subsection{Tensorial product of maps}

\label{Functors on vector spaces}

\subsubsection{Tensorial product of maps}

\bigskip

\paragraph{Maps on contravariant or covariant tensors:\newline}

The following theorems are the consequences of the universal property of the
tensorial product, implemented to the vector spaces of linear maps.

\begin{theorem}
For any vector spaces $E_{1},E_{2},F_{1},F_{2}$ on the same field, $\forall
f_{1}\in L\left(  E_{1};F_{1}\right)  ,f_{2}\in L(E_{2};F_{2})$ there is a
unique map denoted $f_{1}\otimes f_{2}\in L\left(  E_{1}\otimes F_{1}%
;E_{2}\otimes F_{2}\right)  $ such that : $\forall u\in E_{1},v\in
F_{1}:\left(  f_{1}\otimes f_{2}\right)  \left(  u\otimes v\right)
=f_{1}\left(  u\right)  \otimes f_{2}\left(  v\right)  $
\end{theorem}

\begin{theorem}
For any vector spaces E,F on the same field,

$\forall r\in%
\mathbb{N}
$. $\forall f\in L\left(  E;F\right)  $

i) there is a unique map denoted $\otimes^{r}f\in L\left(  \otimes
^{r}E;\otimes^{r}F\right)  $ such that :

$\forall u_{k}\in E,k=1...r:\left(  \otimes^{r}f\right)  \left(  u_{1}\otimes
u_{2}...\otimes u_{r}\right)  =f\left(  u_{1}\right)  \otimes f\left(
u_{2}\right)  ...\otimes f\left(  u_{r}\right)  $

ii) there is a unique map denoted $\otimes_{s}f^{t}\in L\left(  \otimes
^{s}F^{\ast};\otimes^{s}E^{\ast}\right)  $ such that :

$\forall\lambda_{k}\in F^{\ast},k=1...s:\left(  \otimes_{s}f^{t}\right)
\left(  \lambda_{1}\otimes\lambda_{2}...\otimes\lambda_{r}\right)
=f^{t}\left(  \lambda_{1}\right)  \otimes f^{t}\left(  \lambda_{2}\right)
...\otimes f^{t}\left(  \lambda_{r}\right)  =\left(  \lambda_{1}\circ
f\right)  \otimes\left(  \lambda_{2}\circ f\right)  ...\otimes\left(
\lambda_{s}\circ f\right)  $
\end{theorem}

\paragraph{Maps on mixed tensors\newline}

If f is inversible : $f^{-1}\in L\left(  F;E\right)  $ and $\left(
f^{-1}\right)  ^{t}\in L\left(  E^{\ast};F^{\ast}\right)  .$ So to extend a
map from $L\left(  E;F\right)  $ to $L\left(  \otimes_{s}^{r}E;\otimes_{s}%
^{r}F\right)  $ an inversible map $f\in GL\left(  E;F\right)  $\ is required.

Take as above :

$E_{1}=\otimes^{r}E,E_{2}=\otimes^{s}E^{\ast},F_{1}=\otimes^{r}F,F_{2}%
=\otimes_{s}F^{\ast}=\otimes^{s}F^{\ast},$

$f\in L\left(  E_{1};F_{1}\right)  ,\otimes^{r}f\in L\left(  \otimes
^{r}E;\otimes^{r}F\right)  $

$f^{-1}\in L(F_{2};E_{2}),\otimes^{s}\left(  f^{-1}\right)  ^{t}=L\left(
\otimes^{s}E^{\ast};\otimes^{s}F^{\ast}\right)  $

There is a unique map : $\left(  \otimes^{r}f\right)  \otimes\left(
\otimes^{s}\left(  f^{-1}\right)  ^{t}\right)  \in L\left(  \otimes_{s}%
^{r}E;\otimes_{s}^{r}F\right)  $ such that :

$\forall u_{k}\in E,\lambda_{l}\in E^{\ast},k=..r,l=..s:$

$\left(  \otimes^{r}f\right)  \otimes\left(  \otimes^{s}\left(  f^{-1}\right)
^{t}\right)  \left(  \left(  u_{1}\otimes u_{2}...\otimes u_{r}\right)
\otimes\left(  \lambda_{1}\otimes\lambda_{2}...\otimes\lambda_{s}\right)
\right)  =f\left(  u_{1}\right)  \otimes f\left(  u_{2}\right)  ...\otimes
f\left(  u_{r}\right)  \otimes f\left(  \lambda_{1}\right)  \otimes f\left(
\lambda_{2}\right)  ...\otimes f\left(  \lambda_{r}\right)  $

This can be done for any r,s and from a map $f\in L\left(  E;F\right)
$\ build a family of linear maps $\otimes_{s}^{r}f=\left(  \otimes
^{r}f\right)  \otimes\left(  \otimes^{s}\left(  f^{-1}\right)  ^{t}\right)
\in L\left(  \otimes_{s}^{r}E;\otimes_{s}^{r}F\right)  $ such that the maps
commute with the trace operator and preserve the tensorial product :

$S\in\otimes_{s}^{r}E,T\in\otimes_{s^{\prime}}^{r^{\prime}}E:F_{s+s^{\prime}%
}^{r+r^{\prime}}\left(  S\otimes T\right)  =F_{s}^{r}\left(  S\right)  \otimes
F_{s^{\prime}}^{r^{\prime}}\left(  T\right)  $

\paragraph{These results are summarized in the following theorem\newline}

\begin{theorem}
(Kobayashi p.24) For any vector spaces E,F on the same field, there is an
isomorphism between the isomorphisms in L(E;F) and the isomorphisms of
algebras $L(\otimes E;\otimes F)$ which preserves the tensor type and commute
with contraction. So there is a unique extension of an isomorphism $f\in
L(E,F)$ to a linear bijective map $F\in L\left(  \otimes E;\otimes F\right)  $
such that $F\left(  S\otimes T\right)  =F\left(  S\right)  \otimes F\left(
T\right)  ,$ F preserves the type and commutes with the contraction. And this
extension can be defined independantly of the choice of bases.
\end{theorem}

Let E be a vector space and G a subgroup of GL(E;E). Then any fixed f in G is
an isomorphism and can be extended to a unique linear bijective map $F\in
L\left(  \otimes E;\otimes E\right)  $ such that $F\left(  S\otimes T\right)
=F\left(  S\right)  \otimes F\left(  T\right)  ,$ F preserves the type and
commutes with the contraction. For $F\left(  T,1\right)  :\otimes
E\rightarrow\otimes E$ we have a linear map.

\subsubsection{Tensorial product of bilinear forms}

\paragraph{Bilinear form on $\otimes^{r}E$\newline}

\begin{theorem}
A bilinear symmetric form on a finite n dimensional vector space E over the
field K can be extended to a bilinear symmetric form : $G_{r}:\otimes
^{r}E\times\otimes^{r}E\rightarrow K::G_{r}=\otimes^{r}g$
\end{theorem}

\begin{proof}
$g\in E^{\ast}\otimes E^{\ast}$ . g reads in in a basis $\left(  e^{i}\right)
_{i=1}^{n}$ of E* : $g=\sum_{i,j=1}^{n}g_{ij}e^{i}\otimes e^{j}$

The r tensorial product of g : $\otimes^{r}g\in\otimes^{2r}E^{\ast}$ reads :

$\otimes^{r}g=\sum_{i_{1}...i_{2}=1}^{n}g_{i_{1}i_{2}}...g_{i_{2r-1}i_{2r}%
}e^{i_{1}}\otimes e^{i_{2}}...\otimes e^{i_{2r}}$

It acts on tensors $U\in\otimes^{2r}E:\otimes^{r}g\left(  U\right)
=\sum_{i_{1}...i_{2}=1}^{n}g_{i_{1}i_{2}}...g_{i_{2r-1}i_{2r}}U^{i_{1}%
...i_{2r}} $

Take two r contravariant tensors S,T$\in\otimes^{r}E$ then

$\otimes^{r}g\left(  S\otimes T\right)  =\sum_{i_{1}...i_{2}=1}^{n}%
g_{i_{1}i_{2}}...g_{i_{2r-1}i_{2r}}S^{i_{1}...i_{r}}T^{i_{r+1}...i_{2r}} $

From the properties of the tensorial product :

$\otimes^{r}g\left(  \left(  kS+k^{\prime}S^{\prime}\right)  \otimes T\right)
=k\otimes^{r}g\left(  S\otimes T\right)  +k^{\prime}\otimes^{r}g\left(
S^{\prime}\otimes T\right)  $

So it can be seen as a bilinear form acting on $\otimes^{r}E.$ Moreover it is symmetric:

$G_{r}\left(  S,T\right)  =\otimes^{r}g\left(  S\otimes T\right)
=G_{r}\left(  T,S\right)  $
\end{proof}

\paragraph{Bilinear form on $L^{r}(E;E)$\newline}

\begin{theorem}
A bilinear symmetric form on a finite n dimensional vector space E over the
field K can be extended to a bilinear symmetric form : $B_{r}:\otimes
^{r}E^{\ast}\times\otimes^{r}E\rightarrow K::B_{r}=\otimes^{r}g^{\ast}\otimes
g$
\end{theorem}

\begin{proof}
The vector space of r linear maps L$^{r}$(E;E) is isomorphic to the tensorial
subspace : $\otimes^{r}E^{\ast}\otimes E$

We define a bilinear symmetric form on L$^{r}$(E;E) as follows :

$\varphi,\psi\in L^{r}(E;E)$:$B_{r}\left(  \varphi,\psi\right)  =B_{r}\left(
\varphi\otimes\psi\right)  $

with : $B_{r}=\otimes^{r}g^{\ast}\otimes g=\sum_{i_{1}...i_{2}=1}^{n}%
g^{i_{1}i_{2}}...g^{i_{2r-1}i_{2r}}g_{j_{1}j_{2}}e_{i_{1}}\otimes...\otimes
e_{i_{2r}}\otimes e^{j_{1}}\otimes e^{j_{2}} $

This is a bilinear form, and it is symmetric because g is symmetric.
\end{proof}

Notice that if E is a complex vector space and g is hermitian we do not have a
hermitian scalar product.

\subsubsection{Hodge duality}

Hodge duality is a special case of the previous construct : if the tensors are
anti-symmetric then we get the determinant. However we will extend the study
to the case of hermitian maps.

Remind that a vector space (E,g) on a field K is endowed with a scalar product
if g is either a non degenerate, bilinear symmetric form, or a non degenerate
hermitian form.

\paragraph{Scalar product of r-forms\newline}

\begin{theorem}
If (E,g) is a finite dimensional vector space endowed with a scalar product,
then the map :

$G_{r}:\Lambda_{r}E^{\ast}\times\Lambda_{r}E^{\ast}\rightarrow%
\mathbb{R}
::$

$G_{r}\left(  \lambda,\mu\right)  =\sum_{\left\{  i_{1}..i_{r}\right\}
\left\{  j_{1}..j_{r}\right\}  }\overline{\lambda}_{i_{1}..i_{r}}\mu
_{j_{1}...j_{r}}\det\left[  g^{-1}\right]  ^{\left\{  i_{1}..i_{r}\right\}
,\left\{  j_{1}..j_{r}\right\}  }$

is a non degenerate hermitian form and defines a scalar product which does not
depend on the basis. It is definite positive if g is definite positive
\end{theorem}

In the matrix $\left[  g^{-1}\right]  $\ one takes the elements $g^{i_{k}%
j_{l}} $ with $i_{k}\in\left\{  i_{1}..i_{r}\right\}  ,j_{l}\in\left\{
j_{1}..j_{r}\right\}  $

$G_{r}\left(  \lambda,\mu\right)  =\sum_{\left\{  i_{1}...i_{r}\right\}
}\overline{\lambda}_{\{i_{1}...i_{r}\}}\sum_{j_{1}...j_{r}}g^{i_{1}j_{1}%
}...g^{i_{r}j_{r}}\mu_{j_{1}...j_{r}}$

$=\sum_{\left\{  i_{1}...i_{r}\right\}  }\overline{\lambda}_{\{i_{1}%
...i_{r}\}}\mu^{\left\{  i_{1}i_{2}...i_{r}\right\}  }$

where the indexes are lifted and lowered with g.

In an orthonormal basis : $G_{r}\left(  \lambda,\mu\right)  =\sum_{\left\{
i_{1}..i_{r}\right\}  \left\{  j_{1}..j_{r}\right\}  }\overline{\lambda
}_{i_{1}..i_{r}}\mu_{j_{1}...j_{r}}\eta^{i_{1}j_{1}}...\eta^{i_{r}j_{r}}$

This is the application of the first theorem of the previous subsection, where
the formula for the determinant is used.

For $r=1$ one gets the usual bilinear symmetric form over $E^{\ast}:$

$G_{1}\left(  \lambda,\mu\right)  =\sum_{ij}\overline{\lambda}_{i}\mu
_{j}g^{ij}$

\begin{theorem}
For a vector u fixed in $\left(  E,g\right)  $, the map : $\lambda\left(
u\right)  :\Lambda_{r}E\rightarrow\Lambda_{r+1}E::\lambda\left(  u\right)
\mu=u\wedge\mu$ has an adjoint with respect to the scalar product of forms :
$G_{r+1}\left(  \lambda\left(  u\right)  \mu,\mu^{\prime}\right)
=G_{r}\left(  \mu,\lambda^{\ast}\left(  u\right)  \mu^{\prime}\right)  $ which is

$\lambda^{\ast}\left(  u\right)  :\Lambda_{r}E\rightarrow\Lambda
_{r-1}E::\lambda^{\ast}\left(  u\right)  \mu=i_{u}\mu$
\end{theorem}

It suffices to compute the two quantities.

\paragraph{Hodge duality\newline}

g can be used to define the isomorphism $E\simeq E^{\ast}.$ Similarly this
scalar product can be used to define the isomorphism $\Lambda_{r}%
E\simeq\Lambda_{n-r}E$

\begin{theorem}
If (E,g) is a n dimensional vector space endowed with a scalar product with
the volume form $\varpi_{0}$, then the map :

$\ast:\Lambda_{r}E^{\ast}\rightarrow\Lambda_{n-r}E$ defined by the condition

$\forall\mu\in\Lambda_{r}E^{\ast}:\ast\lambda_{r}\wedge\mu=G_{r}\left(
\lambda,\mu\right)  \varpi_{0}$

is an anti-isomorphism
\end{theorem}

A direct computation gives the value of the Hodge dual $\ast\lambda$ in the
basis $\left(  e^{i}\right)  _{i=1}^{n}$ of E*:

$\ast\left(  \sum_{\left\{  i_{1}...i_{r}\right\}  }\lambda_{\{i_{1}%
...i_{r}\}}e^{i_{1}}\wedge..\wedge e^{i_{r}}\right)  $

$=\sum_{\left\{  i_{1}..i_{n-r}\right\}  \left\{  j_{1}..j_{r}\right\}
}\epsilon\left(  j_{1}..j_{r},i_{1},...i_{n-r}\right)  \overline{\lambda
}^{j_{1}...j_{r}}\sqrt{\left\vert \det g\right\vert }e^{i_{1}}\wedge e^{i_{2}%
}...\wedge e^{i_{n-r}}$

With $\epsilon=sign\det\left[  g\right]  $ (which is always real)

For r=0:

$\ast\lambda=\overline{\lambda}\varpi_{0}$

For r=1 :

$\ast\left(  \sum_{i}\lambda_{i}e^{i}\right)  =\sum_{j=1}^{n}\left(
-1\right)  ^{j+1}g^{ij}\overline{\lambda}_{j}\sqrt{\left\vert \det
g\right\vert }e^{1}\wedge..\widehat{e^{j}}\wedge...\wedge e^{n}$

For r=n-1:

$\ast\left(  \sum_{i=1}^{n}\lambda_{1..\widehat{i}...n}e^{1}\wedge
..\widehat{e^{i}}\wedge...\wedge e^{n}\right)  =\sum_{i=1}^{n}\left(
-1\right)  ^{i-1}\overline{\lambda}^{1..\widehat{i}...n}\sqrt{\left\vert \det
g\right\vert }e^{i}$

For r=n:

$\ast\left(  \lambda e^{1}\wedge.....\wedge e^{n}\right)  =\epsilon\frac
{1}{\sqrt{\left\vert \det g\right\vert }}\overline{\lambda}$

The usual cross product of 2 vectors in an 3 dimensional euclidean vector
space can be defined as $u\times v=\ast\left(  a\wedge b\right)  $ where the
algebra $\Lambda^{r}E$ is used

The inverse of the map * is :

$\ast^{-1}\lambda_{r}=\epsilon(-1)^{r\left(  n-r\right)  }\ast\lambda
_{r}\Leftrightarrow\ast\ast\lambda_{r}=\epsilon(-1)^{r\left(  n-r\right)
}\lambda_{r}$

$G_{q}(\lambda,\ast\mu)=G_{n-q}(\ast\lambda,\mu)$

$G_{n-q}(\ast\lambda,\ast\mu)=G_{q}(\lambda,\mu)$

Contraction is an operation over $\Lambda E^{\ast}.$ It is defined, on a real
n dimensional vector space by:

$\lambda\in\Lambda_{r}E,\mu\in\Lambda_{q}E:\lambda\vee\mu=\epsilon\left(
-1\right)  ^{p+(r-q)n}\ast\left(  \lambda\wedge\ast\mu\right)  \in\wedge
_{r-q}E^{\ast}$

It is distributive over addition and not associative

$\ast\left(  \lambda\vee\mu\right)  =\epsilon\left(  -1\right)  ^{(r-q)n}%
\epsilon(-1)^{\left(  r-q\right)  \left(  n-\left(  r-q\right)  \right)
}\left(  \lambda\wedge\ast\mu\right)  =\left(  -1\right)  ^{q^{2}+r^{2}%
}\left(  \lambda\wedge\ast\mu\right)  $

$\lambda\vee\left(  \lambda\vee\mu\right)  =0$

$\lambda\in E^{\ast},\mu\in\Lambda_{q}E:$

$\ast\left(  \lambda\wedge\mu\right)  =\left(  -1\right)  ^{q}\lambda\vee
\ast\mu$

$\ast\left(  \lambda\vee\mu\right)  =\left(  -1\right)  ^{q-1}\lambda
\wedge\ast\mu$

\subsubsection{Tensorial Functors}

\begin{theorem}
The vector spaces over a field K with their morphisms form a category
$\mathfrak{V}$.
\end{theorem}

The vector spaces isomorphic to some vector space E form a subcategory
$\mathfrak{V}_{E}$

\begin{theorem}
The functor $\mathfrak{D:V\mapsto V}$ which associates :

to each vector space E its dual : $\mathfrak{D}(E)=E^{\ast}$

to each linear map $f:E\rightarrow F$ its dual : $f^{\ast}:F^{\ast}\rightarrow
E^{\ast}$

is contravariant : $\mathfrak{D}\left(  f\circ g\right)  =\mathfrak{D}\left(
g\right)  \circ\mathfrak{D}\left(  f\right)  $
\end{theorem}

\begin{theorem}
The r-tensorial power of vector spaces is a faithful covariant functor
$\mathfrak{T}^{r}\mathfrak{:V\mapsto V}$
\end{theorem}

$\mathfrak{T}^{r}\left(  E\right)  =\otimes^{k}E$

$f\in L\left(  E;F\right)  :\mathfrak{T}^{r}\left(  f\right)  =\otimes^{r}f\in
L\left(  \otimes^{r}E;\otimes^{r}F\right)  $

$\mathfrak{T}^{r}\left(  f\circ g\right)  =\mathfrak{T}^{r}\left(  f\right)
\circ\mathfrak{T}^{r}\left(  g\right)  =\left(  \otimes^{r}f\right)
\circ\left(  \otimes^{r}g\right)  $

\begin{theorem}
The s-tensorial power of dual vector spaces is a faithful contravariant
functor $\mathfrak{T}_{s}\mathfrak{:V\mapsto V}$
\end{theorem}

$\mathfrak{T}_{s}\left(  E\right)  =\otimes_{s}E=\otimes^{s}E^{\ast}$

$f\in L\left(  E;F\right)  :\mathfrak{T}_{s}\left(  f\right)  =\otimes_{s}f\in
L\left(  \otimes_{s}F^{\ast};\otimes_{s}E^{\ast}\right)  $

$\mathfrak{T}_{s}\left(  f\circ g\right)  =\mathfrak{T}_{s}\left(  g\right)
\circ\mathfrak{T}_{s}\left(  f\right)  =\left(  \otimes_{s}f^{\ast}\right)
\circ\left(  \otimes_{s}g^{\ast}\right)  $

\begin{theorem}
The (r,c)-tensorial product of vector spaces is a faithful bi-functor :
$\mathfrak{T}_{s}^{r}\mathfrak{:V}_{E}\mathfrak{\mapsto V}_{E}$
\end{theorem}

The following functors are similarly defined:

the covariant functors $\mathfrak{T}_{S}^{r}\mathfrak{:V\mapsto V::T}_{S}%
^{r}\left(  E\right)  =\odot^{r}E$ for symmetric r tensors

the covariant functors $\mathfrak{T}_{A}^{r}\mathfrak{:V\mapsto V::T}_{A}%
^{r}\left(  E\right)  =\wedge^{r}E$ for antisymmetric r contravariant tensors

the contravariant functors $\mathfrak{T}_{As}\mathfrak{:V\mapsto V::T}%
_{As}\left(  E\right)  =\wedge_{s}E$ for antisymmetric r covariant tensors

\begin{theorem}
Let $\mathfrak{A}$\ be the category of algebras over the field K. The functor
$\mathfrak{T:V\mapsto A}$\ defined as :

$\mathfrak{V}\left(  E\right)  =\otimes E=\sum_{r,s=0}^{\infty}\otimes_{s}%
^{r}E$

$\forall f\in L\left(  E;F\right)  :\mathfrak{T}\left(  f\right)  \in
\hom\left(  \otimes E;\otimes F\right)  =L\left(  \otimes E;\otimes F\right)
$

is faithful : there is a unique map $\mathfrak{T}\left(  f\right)  \in
L\left(  \otimes E;\otimes F\right)  $ such that :

$\forall u\in E,v\in F:\mathfrak{T}\left(  f\right)  \left(  u\otimes
v\right)  =f\left(  u\right)  \otimes f\left(  v\right)  $
\end{theorem}

\subsubsection{Invariant and equivariant tensors}

Let E be a vector space, GL(E) the group of linear inversible endomorphisms, G
a subgroup of GL(E).

The action of $g\in G$ on E is : $f\left(  g\right)  :E\rightarrow E$ and we
have the dual action: $f^{\ast}\left(  g\right)  :E^{\ast}\rightarrow E^{\ast
}::f^{\ast}\left(  g\right)  \lambda=\lambda\circ f\left(  g^{-1}\right)  $

This action induces an action $F_{s}^{r}\left(  g\right)  :\otimes_{s}%
^{r}E\rightarrow\otimes_{s}^{r}E$ with

$F_{s}^{r}\left(  g\right)  =\left(  \otimes^{r}f\left(  g\right)  \right)
\otimes\left(  \otimes^{s}\left(  f\left(  g\right)  \right)  ^{\ast}\right)
$

\paragraph{Invariant tensor\newline}

A tensor T$\in\otimes_{s}^{r}E$ is said to be invariant by G if : $\forall
g\in G:F_{s}^{r}\left(  g\right)  T=T$

\begin{definition}
The elementary invariant tensors of rank r of a finite dimensional vector
space E are the tensors $T\in\otimes_{r}^{r}E$ with components :

$T_{j_{1}..j_{r}}^{i_{1}..i_{r}}=\sum_{\sigma\in\mathfrak{S}\left(  r\right)
}C_{\sigma}\delta_{j_{1}}^{\sigma\left(  i_{1}\right)  }\delta_{j_{2}}%
^{\sigma\left(  i_{2}\right)  }...\delta_{j_{r}}^{\sigma\left(  i_{r}\right)
}$
\end{definition}

\begin{theorem}
Invariant tensor theorem (Kolar p.214): On a finite dimensional vector space
E, any tensor $T\in\otimes_{s}^{r}E$ invariant by the action of GL(E) is zero
if r $\neq s.$\ If r=s it is a linear combination of the elementary invariant
tensors of rank r
\end{theorem}

\begin{theorem}
Weyl (Kolar p.265) : The linear space of all linear maps $\otimes^{k}%
\mathbb{R}
^{m}\rightarrow%
\mathbb{R}
$ invariant by the orthogonal group O($%
\mathbb{R}
,$m) is spanned by the elementary invariants tensors if k is even, and 0 if k
is odd.
\end{theorem}

\paragraph{Equivariant map\newline}

A map : $f:\otimes E\rightarrow\otimes E$ is said to be equivariant by the
action of GL(E) if :

$\forall g\in G,T\in\otimes_{s}^{r}E:f\left(  F_{s}^{r}\left(  g\right)
T\right)  =F_{s}^{r}\left(  g\right)  f\left(  T\right)  $

\begin{theorem}
(Kolar p.217) : all smooth GL(E) equivariant maps (not necessarily linear) :

i) $\wedge^{r}E\rightarrow\wedge^{r}E$ are multiples of the identity

ii) $\otimes^{r}E\rightarrow\odot^{r}E$ are multiples of the symmetrizer

iii) $\otimes^{r}E\rightarrow\wedge^{r}E$ are multiples of the antisymmetrizer

iv) $\wedge^{r}E\rightarrow\otimes^{r}E$ or \ $\odot^{r}E\rightarrow
\otimes^{r}E$ are multiples of the inclusion
\end{theorem}

\subsubsection{Invariant polynomials}

\paragraph{Invariant maps\newline}

\begin{definition}
Let E be a vector space on a field K, G a subgroup of GL(E), f a map
:$f:E^{r}\times E^{\ast s}\rightarrow K$ with $r,s\in%
\mathbb{N}
$

f is said to be invariant by G if :

$\forall g\in G,\forall\left(  u_{i}\right)  _{i=1..r}\in E,\forall\left(
\lambda_{j}\right)  _{j=1}^{s}\in E^{\ast}:$

$f\left(  \left(  gu_{1},..gu_{r}\right)  ,\left(  g^{-1}\lambda_{1}\right)
,...\left(  g^{-1}\lambda_{s}\right)  \right)  =f\left(  u_{1},..u_{r}%
,\lambda_{1},..\lambda_{s}\right)  $
\end{definition}

\begin{theorem}
Tensor evaluation theorem (Kolar p.223) Let E a finite dimensional real vector
space. A smooth map $f:E^{r}\times E^{\ast s}\rightarrow%
\mathbb{R}
$ (not necessarily linear) is invariant by GL(E) iff $\exists F\in C_{\infty
}\left(
\mathbb{R}
^{rs};%
\mathbb{R}
\right)  $ such that for all i,j:

$\forall\left(  u_{i}\right)  _{i=1..r}\in E,\forall\left(  \lambda
_{j}\right)  _{j=1}^{s}\in E^{\ast}:f\left(  u_{1},..u_{r},\lambda
_{1},..\lambda_{s}\right)  =F\left(  ..\lambda_{i}\left(  u_{j}\right)
...\right)  $
\end{theorem}

As an application, all smooth GL(E) equivariant maps

$f:E^{r}\times E^{\ast s}\rightarrow E$ are of the form :

$f\left(  u_{1},..u_{r},\lambda_{1},..\lambda_{s}\right)  =\sum_{\beta=1}%
^{k}F_{\beta}\left(  ..\lambda_{i}\left(  u_{j}\right)  ...\right)  u_{\beta}$
where $F_{\beta}\left(  ..\lambda_{i}\left(  u_{j}\right)  ...\right)  \in
C_{\infty}\left(
\mathbb{R}
^{rs};%
\mathbb{R}
\right)  $

$f:E^{r}\times E^{\ast s}\rightarrow E^{\ast}$ are of the form :

$f\left(  u_{1},..u_{r},\lambda_{1},..\lambda_{s}\right)  =\sum_{\beta=1}%
^{l}F_{\beta}\left(  ..\lambda_{i}\left(  u_{j}\right)  ...\right)
\lambda_{\beta}$ where $F_{\beta}\left(  ..\lambda_{i}\left(  u_{j}\right)
...\right)  \in C_{\infty}\left(
\mathbb{R}
^{rs};%
\mathbb{R}
\right)  $

\paragraph{Polynomials on a vector space\newline}

\begin{definition}
A map $f:V\rightarrow W$ between two finite dimensional vector spaces on a
field K\ is said to be polynomial if in its coordinate expression in any bases
: $f_{i}\left(  x_{1,...}x_{j},..x_{m}\right)  =y_{i}$ are polynomials in the
$x_{j}.$
\end{definition}

i) Then f reads : $f_{i}=f_{i0}+f_{i1}+..+f_{ir}$ where $f_{ik}$ , called a
homogeneous component, is, for each component, a monomial of degree k in the
components : $f_{ik}=x_{1}^{\alpha_{1}}x_{2}^{\alpha_{2}}..x_{m}^{\alpha_{m}%
},\alpha_{1}+...\alpha_{m}=k$

ii) let $f:V\rightarrow K$ be a homogeneous polynomial map of degree r. The
\textbf{polarization} of f is defined as $P_{r}$ such that $r!P_{r}\left(
u_{1},...,u_{r}\right)  $ is the coefficient of $t_{1}t_{2}..t_{r}$ in
$f\left(  t_{1}u_{1}+t_{2}u_{2}+...+t_{r}u_{r}\right)  $

$P_{r}$ is a r linear symmetric map : $P_{r}\in L^{r}\left(  V;K\right)  $

Conversely if $P_{r}$ is a r linear symmetric map a homogeneous polynomial map
of degree r is defined by : $f(u)=P_{r}(u,u,..,u)$

iii) by the universal property of the tensor product, the r linear symmetric
map $P_{r}$ induces a unique map : $\widehat{P}_{r}:\odot^{r}V\rightarrow K$
such that :

$P_{r}\left(  u_{1},...,u_{r}\right)  =\widehat{P}_{r}\left(  u_{1}%
\otimes...\otimes u_{r}\right)  $

iv) So if f is a polynomial map of degree r : $f:V\rightarrow K$ there is a
linear map : $P:\odot_{k=0}^{k=r}V\rightarrow K$ given by the sum of the
linear maps $\widehat{P}_{r}.$

\paragraph{Invariant polynomial\newline}

Let V a finite dimensional vector space on a field K, G a group with action on
V : $\rho:G\rightarrow L(E;E)$

A map : $f:V\rightarrow K$ $\ $is said to be invariant by this action if :

$\forall g\in G,\forall u\in V:f\left(  \rho\left(  g\right)  u\right)
=f\left(  u\right)  $

Similarly a map $f:V^{r}\rightarrow K$ is invariant if :

$\forall g\in G,\forall u\in V:f\left(  \rho\left(  g\right)  u_{1}%
,..\rho\left(  g\right)  u_{r}\right)  =f\left(  u_{1},..,u_{r}\right)  $

A polynomial $f:V\rightarrow K$ is invariant iff each of its homogeneous
components $f_{k}$ is invariant

An invariant polynomial induces by polarizarion a r linear symmetric invariant
map, and conversely a r linear, symmetric, invariant map induces an invariant polynomial.

\begin{theorem}
(Kolar p.266) Let $f:%
\mathbb{R}
\left(  m\right)  \rightarrow%
\mathbb{R}
$ a polynomial map from the vector space $%
\mathbb{R}
\left(  m\right)  $ of square m$\times$m real matrices to $%
\mathbb{R}
$ such that : f(OM)=M for any orthogonal matrix $O\in O\left(
\mathbb{R}
,m\right)  $ . Then there is a polynomial map $F:%
\mathbb{R}
\left(  m\right)  \rightarrow%
\mathbb{R}
$ such that : $f(M)=F(M^{t}M)$
\end{theorem}

\newpage

\section{MATRICES}

\bigskip

\subsection{Operations with matrices}

\label{Operations with matrices}

\subsubsection{Definitions}

\begin{definition}
A r$\times$c \textbf{matrix} over a field K is a table A of scalars from K
arranged in r rows and c columns, indexed as : $a_{ij}$ i=1...r, j=1...c (the
fist index is for rows, the second is for columns).
\end{definition}

We will use also the tensor like indexes : $a_{j}^{i},$ up=row, low=column.
When necessary a matrix is denoted within brackets : $A=\left[  a_{ij}\right]
$

When r=c we have the set of \textbf{square r-matrices} over K

\begin{notation}
$K\left(  r,c\right)  $ is the set of r$\times$c matrices over the field K.

K(r) is the set of square r-matrices over the field K
\end{notation}

\subsubsection{Basic operations}

\begin{theorem}
With addition and multiplication by a scalar the set K(r,c) is a vector space
over K, with dimension rc.
\end{theorem}

$A,B\in K\left(  r,c\right)  :A+B=\left[  a_{ij}+b_{ij}\right]  $

$A\in K\left(  r,c\right)  ,k\in K:kA=\left[  ka_{ij}\right]  $

\begin{definition}
The \textbf{product of matrices }is the operation\textbf{\ :}

$K\left(  c,s\right)  \times K\left(  c,s\right)  \rightarrow K\left(
r,s\right)  ::AB=\left[  \sum_{k=1}^{c}a_{ik}b_{kj}\right]  $
\end{definition}

When defined the product distributes over addition and multiplication by a
scalar and is associative :

$A\left(  B+C\right)  =AB+AC$

$A\left(  kB\right)  =kAB$

$\left(  AB\right)  C=A\left(  BC\right)  $

The product \textit{is not commutative}.

The identity element for multiplication is the \textbf{identity matrix} :
$I_{r}=\left[  \delta_{ij}\right]  $

\begin{theorem}
With these operations the set K(r) of square r-matrices over K is a ring and a
unital algebra over K.\ 
\end{theorem}

\begin{definition}
The \textbf{commutator} of 2 matrices is : $\left[  A,B\right]  =AB-BA.$
\end{definition}

\begin{theorem}
With the commutator as bracket K(r) is a Lie algebra.
\end{theorem}

\begin{notation}
GK(r) is the group of square invertible (for the product) r-matrices.
\end{notation}

When a matrix has an inverse, denoted A$^{-1}$ , it is unique and a right and
left inverse : $AA^{-1}=A^{-1}A=I_{r}$ and $\left(  AB\right)  ^{-1}%
=B^{-1}A^{-1}$

\begin{definition}
The \textbf{diagonal} of a squared matrix A is the set of elements : $\left\{
a_{11},a_{22},...,a_{rr}\right\}  $
\end{definition}

A square matrix is diagonal if all its elements =0 but for the diagonal.

A diagonal matrix is commonly denoted as $Diag\left(  m_{1},...m_{r}\right)  $
with $m_{i}=a_{ii}$

Remark : the diagonal is also called the "main diagonal", with reverse
diagonal = the set of elements : $\left\{  a_{r1},a_{r-12},...,a_{1r}\right\}
$

\begin{theorem}
The set of diagonal matrices is a commutative subalgebra of K(r).
\end{theorem}

A diagonal matrix is invertible if there is no zero on its diagonal.

\begin{definition}
A \textbf{triangular} matrix is a square matrix A such that : $a_{ij}=0$
whenever i%
$>$%
j .\ Also called upper triangular (the non zero elements are above the
diagonal).\ A lower triangular matrix is such that $A^{t}$ is upper triangular
(the non zero elements are below the diagonal)
\end{definition}

\subsubsection{Transpose}

\begin{definition}
The \textbf{transpose} of a matrix $A=\left[  a_{ij}\right]  \in K\left(
r,c\right)  $ is the matrix $A^{t}=\left[  a_{ji}\right]  \in K(c,r)$
\end{definition}

Rows and columns are permuted:

\bigskip

$A=%
\begin{bmatrix}
a_{11} & ... & a_{1c}\\
... &  & ...\\
a_{r1} & ... & a_{rc}%
\end{bmatrix}
\rightarrow A^{t}=%
\begin{bmatrix}
a_{11} & ... & a_{r1}\\
... &  & ...\\
a_{1c} & ... & a_{rc}%
\end{bmatrix}
$

\bigskip

Remark : there is also the old (and rarely used nowodays) notation $^{t}A$

For $A,B\in K\left(  r,c\right)  ,k,k^{\prime}\in K:$

$\left(  kA+k^{\prime}B\right)  ^{t}=kA^{t}+k^{\prime}B^{t}$

$\left(  AB\right)  ^{t}=B^{t}A^{t}$

$\left(  A_{1}A_{2}...A_{n}\right)  ^{t}=A_{n}^{t}A_{n-1}^{t}...A_{1}^{t}$

$\left(  A^{t}\right)  ^{-1}=\left(  A^{-1}\right)  ^{t}$

\begin{definition}
A square matrix A is :

\textbf{symmetric} if $A=A^{t}$

\textbf{skew-symmetric} (or antisymmetric) if $A=-A^{t}$

\textbf{orthogonal} if : $A^{t}=A^{-1}$
\end{definition}

\begin{notation}
O(r,K) is the set of orthogonal matrix in K(r)
\end{notation}

So $A\in O(r,K)\Rightarrow A^{t}=A^{-1},AA^{t}=A^{t}A=I_{r}$

Notice that O(r,K) is not an algebra: the sum of two orthogonal matrices is
generally not orthogonal.

\subsubsection{Adjoint}

\begin{definition}
The \textbf{adjoint} of a matrix $A=\left[  a_{ij}\right]  \in%
\mathbb{C}
\left(  r,c\right)  $ is the matrix $A^{\ast}=\left[  \overline{a}%
_{ji}\right]  \in K(c,r)$
\end{definition}

Rows and columns are permuted and the elements are conjugated :

\bigskip

$A=%
\begin{bmatrix}
a_{11} & ... & a_{1c}\\
... &  & ...\\
a_{r1} & ... & a_{rc}%
\end{bmatrix}
\rightarrow A^{\ast}=%
\begin{bmatrix}
\overline{a}_{11} & ... & \overline{a}_{r1}\\
... &  & ...\\
\overline{a}_{1c} & ... & \overline{a}_{rc}%
\end{bmatrix}
$

\bigskip

Remark : the notation varies according to the authors

For $A,B\in%
\mathbb{C}
\left(  r,c\right)  ,k,k^{\prime}\in K:$

$\left(  kA+k^{\prime}B\right)  ^{\ast}=\overline{k}A^{\ast}+\overline
{k}^{\prime}B^{\ast}$

$\left(  AB\right)  ^{\ast}=B^{\ast}A^{\ast}$

$\left(  A_{1}A_{2}...A_{n}\right)  ^{\ast}=A_{n}^{\ast}A_{n-1}^{\ast}%
...A_{1}^{t}$

$\left(  A^{\ast}\right)  ^{-1}=\left(  A^{-1}\right)  ^{\ast}$

\begin{definition}
A square matrix A is

\textbf{hermitian} if $A=A^{\ast}$

\textbf{skew-hermitian} if $A=-A^{\ast}$

\textbf{unitary} if : $A^{\ast}=A^{-1}$

\textbf{normal} if $AA^{\ast}=A^{\ast}A$
\end{definition}

\begin{notation}
U(r) is the group of unitary matrices
\end{notation}

So $A\in U(r)\Rightarrow A^{\ast}=A^{-1},AA^{\ast}=A^{\ast}A=I_{r}$

U(r) is a group but not an algebra: the sum of two unitary is generally not unitary

\begin{theorem}
The real symmetric, real antisymmetric, real orthogonal, complex hermitian,
complex antihermitian, unitary matrices are normal.
\end{theorem}

Remark : $%
\mathbb{R}
\left(  r\right)  $ is a subset of $%
\mathbb{C}
\left(  r\right)  .$ Matrices in $%
\mathbb{C}
\left(  r\right)  $ with real elements are matrices in $%
\mathbb{R}
\left(  r\right)  .$ So hermitian becomes symmetric, skew-hermitian becomes
skew-symmetric, unitary becomes orthogonal, normal becomes $AA^{t}=A^{t}A.$
Any theorem for $%
\mathbb{C}
\left(  r\right)  $ can be implemented for $%
\mathbb{R}
\left(  r\right)  $ with the proper adjustments.

\subsubsection{Trace}

\begin{definition}
The \textbf{trace} of a square matrix $A\in K\left(  r\right)  $ is the sum of
its diagonal elements
\end{definition}

$Tr\left(  A\right)  =\sum_{i=1}^{r}a_{ii}$

It is the trace of the linear map whose matrix is A

$Tr:K(r)\rightarrow K$ is a linear map $Tr\in K\left(  r\right)  ^{\ast}$

$Tr\left(  A\right)  =Tr\left(  A^{t}\right)  $

$Tr\left(  A\right)  =\overline{Tr\left(  A^{\ast}\right)  }$

$Tr\left(  AB\right)  =Tr\left(  BA\right)  \Rightarrow Tr(ABC)=Tr\left(
BCA\right)  =Tr\left(  CAB\right)  $

$Tr\left(  A^{-1}\right)  =\left(  Tr\left(  A\right)  \right)  ^{-1}$

$Tr\left(  PAP^{-1}\right)  =Tr\left(  A\right)  $

Tr(A)= sum of the eigenvalues of A

$Tr(A^{k})=$ sum of its $\left(  \text{eigenvalues}\right)  ^{k}$

If A is symmetric and B skew-symmetric then Tr(AB)=0

$Tr\left(  \left[  A,B\right]  \right)  =0$ where $\left[  A,B\right]  =AB-BA$

\begin{definition}
The Frobenius norm (also called the Hilbert-Schmidt norm) is the map :
$K\left(  r,c\right)  \rightarrow%
\mathbb{R}
::Tr\left(  AA^{\ast}\right)  =Tr\left(  A^{\ast}A\right)  $
\end{definition}

Whenever $A\in%
\mathbb{C}
\left(  r,c\right)  :AA^{\ast}\in%
\mathbb{C}
\left(  r,r\right)  ,A^{\ast}A\in%
\mathbb{C}
\left(  c,c\right)  $ are square matrix, so $Tr\left(  AA^{\ast}\right)  $ and
$Tr\left(  A^{\ast}A\right)  $ are well defined

$Tr\left(  AA^{\ast}\right)  =\sum_{i=1}^{r}\left(  \sum_{j=1}^{c}%
a_{ij}\overline{a}_{ij}\right)  =\sum_{j=1}^{c}\left(  \sum_{i=1}^{r}%
\overline{a}_{ij}a_{ij}\right)  =\sum_{i=1}^{r}\sum_{j=1}^{c}\left\vert
a_{ij}\right\vert ^{2}$

\subsubsection{Permutation matrices}

\begin{definition}
A \textbf{permutation matrix} is a square matrix $P\in K\left(  r\right)  $
which has on each row and column all elements =0 but one =1
\end{definition}

$\forall i,j:P_{ij}=0$ but for one unique couple $\left(  I,J\right)
:P_{IJ}=1 $

It implies that $\forall i,j:\sum_{c}P_{ic}=\sum_{r}P_{rj}=1$

\begin{theorem}
The set P(K,r) of permutation matrices is a subgroup of the orthogonal
matrices O(K,r).
\end{theorem}

The right multiplication of a matrix A by a permutation matrix is a
permutation of the rows of A

The left multiplication of a matrix A by a permutation matrix is a permutation
of the columns of A

So given a permutation $\sigma\in\mathfrak{S}\left(  r\right)  $ of $\left(
1,2,...r\right)  $ the matrix :

$S\left(  \sigma\right)  =P:\left[  P_{ij}\right]  =\delta_{\sigma\left(
j\right)  j}$

is a permutation matrix (remark : one can also take $P_{ij}=\delta
_{i\sigma\left(  i\right)  }$ but it is less convenient) and this map :
$S:\mathfrak{S}\left(  r\right)  \rightarrow P\left(  K,r\right)  $ is a group
isomorphism : $P_{S\left(  \sigma\circ\sigma^{\prime}\right)  }=P_{S\left(
\sigma\right)  }P_{S\left(  \sigma^{\prime}\right)  }$

The identity matrix is the only diagonal permutation matrix.

As any permutation of a set can be decomposed in the product of
transpositions, any permutation matrix can be decomposed in the product of
elementary permutation matrices which transposes two columns (or two rows).

\subsubsection{Determinant}

\begin{definition}
The \textbf{determinant} of a square matrix $A\in K\left(  r\right)  $ is the quantity:%

\begin{equation}
\det A=\sum_{\sigma\in\mathfrak{S}\left(  r\right)  }\varepsilon\left(
\sigma\right)  a_{1\sigma\left(  1\right)  }a_{2\sigma\left(  2\right)
}...a_{n\sigma\left(  n\right)  }=\sum_{\sigma\in\mathfrak{S}\left(  r\right)
}\epsilon\left(  \sigma\right)  a_{\sigma\left(  1\right)  1}a_{\sigma\left(
2\right)  2}...a_{\sigma\left(  n\right)  n}%
\end{equation}

\end{definition}

$\det A^{t}=\det A$

$\det A^{\ast}=\overline{\det A}$ so the determinant of a Hermitian matrix is real

$\det\left(  kA\right)  =k^{r}\det A$ (Beware !)

$\det\left(  AB\right)  =\det\left(  A\right)  \det\left(  B\right)
=\det\left(  BA\right)  $

$\exists A^{-1}\Leftrightarrow\det A\neq0$ and then $\det A^{-1}=\left(  \det
A\right)  ^{-1}$

The determinant of a permutation matrix is equal to the signature of the
corresponding permutation

For $K=%
\mathbb{C}
$\ the determinant of a matrix is equal to the product of its eigen values

As the product of a matrix by a permutation matrix is the matrix with permuted
rows or columns, the determinant of the matrix with permuted rows or columns
is equal to the determinant of the matrix $\times$ the signature of the permutation.

The determinant of a triangular matrix is the product of the elements of its diagonal

\begin{theorem}
Sylvester's determinant theorem :

Let $A\in K\left(  r,c\right)  ,B\in K(c,r),X\in GL(K,r)$ then :

$\det\left(  X+AB\right)  =\det X\det\left(  I_{c}+BX^{-1}A\right)  $

so with X=I: $\det\left(  I+AB\right)  =\det\left(  I_{c}+BA\right)  $
\end{theorem}

\bigskip

\subparagraph{Computation of a determinant : \newline}

Determinant is the unique map : $D:K\left(  r\right)  \rightarrow K$ with the
following properties :

a) For any permutation matrix P, $D\left(  P\right)  =$ signature of the
corresponding permutation

b) $D\left(  AP\right)  =D\left(  P\right)  D\left(  A\right)  =D\left(
A\right)  D\left(  P\right)  $ where P is a permutation matrix

Moreover D has the following linear property :

D(A')=kD(A)+D(A') where A' is (for any i) the matrix

$A=\left[  A_{1},A_{2},...A_{r}\right]  \rightarrow A^{\prime}=\left[
A_{1},A_{2},.,A_{i-1},B,A_{i+1}..A_{r}\right]  $

where $A_{i}$ is the i column of A, B is rx1 matrix and k a scalar

So for $A\in K\left(  r\right)  $ and A' the matrix obtained from A by adding
to the row i a scalar multiple of another row i' :$\det A=\det A^{\prime}.$

There is the same result with columns (but one cannot mix rows and columns in
the same operation).\ This is the usual way to compute determinants, by
gaussian elimination : by successive applications of the previous rules one
strives to get a triangular matrix.

There are many results for the determinants of specific matrices. Many
Internet sites offer results and software for the computation.

\bigskip

\begin{definition}
The (i,j) \textbf{minor} of a square matrix A=$\left[  a_{ij}\right]  \in
K\left(  r\right)  $ is the determinant of the (r-1,r-1) matrix denoted
$A_{ij} $ deduced from A by removing the row i and the column j.
\end{definition}

\begin{theorem}
$\det A=\sum_{i=1}^{r}\left(  -1\right)  ^{i+j}a_{ij}\det A_{ij}=\sum
_{j=1}^{r}\left(  -1\right)  ^{i+j}a_{ij}\det A_{ij}$
\end{theorem}

The row i or the column j are arbitrary. It gives a systematic way to compute
a determinant by a recursive calculus.

This formula is generalized in the \textbf{Laplace's development} :

For any sets of p ordered indices

$I=\left\{  i_{1},i_{2},...i_{p}\right\}  \subset\left(  1,2,...r\right)
,J=\left\{  j_{1},j_{2},...j_{p}\right\}  \subset\left(  1,2,...r\right)  $

Let us denote $\left[  A^{c}\right]  _{J}^{I}$ the matrices deduced from A by
removing all rows with indexes in I, and all columns with indexes in J

Let us denote $\left[  A\right]  _{J}^{I}$ the matrices deduced from A by
keeping only the rows with indexes in I, and the columns with indexes in J

Then :

$\det A=\sum_{\left(  j_{1},...,j_{p}\right)  }\left(  -1\right)
^{i_{1}+i_{2}+...+i_{p}+j_{1}+...+j_{p}}\left(  \det\left[  A\right]
_{\left\{  j_{1},..,j_{p}\right\}  }^{\left\{  i_{1},...i_{p}\right\}
}\right)  \left(  \det\left[  A^{c}\right]  _{\left(  j_{1},..,j_{p}\right)
}^{\left\{  i_{1},...i_{p}\right\}  }\right)  $

The cofactor of a square matrix A=$\left[  a_{ij}\right]  \in K\left(
r\right)  $ is the quantity $\left(  -1\right)  ^{i+j}\det A_{ij}$ where $\det
A_{ij}$ is the minor.

The matrix of cofactors is the matrix : $C\left(  A\right)  =\left[  \left(
-1\right)  ^{i+j}\det A_{ij}\right]  $ and : $A^{-1}=\frac{1}{\det A}C\left(
A\right)  ^{t}$ So

\begin{theorem}
The elements $\left[  A^{-1}\right]  _{ij}$ of $A^{-1}$ are given by the
formula :%

\begin{equation}
\left[  A^{-1}\right]  _{ij}=\frac{1}{\det A}(-1)^{i+j}\det\left[
A_{ji}\right]
\end{equation}

where $A_{ij}$ is the (r-1,r-1) matrix denoted $A_{ij}$ deduced from A by
removing the row i and the column j.
\end{theorem}

Beware of the inverse order of indexes on the right hand side!

\subsubsection{Kronecker's product}

Also called tensorial product of matrices

For $A\in K\left(  m,n\right)  ,B\left(  p,q\right)  $ , $C=A\otimes B\in
K\left(  mp,nq\right)  $ is the matrix built as follows by associating to each
element $\left[  a_{ij}\right]  $\ one block equal to $\left[  a_{ij}\right]
B$

The useful relation is : $\left(  A\otimes B\right)  \times\left(  C\otimes
D\right)  =AC\otimes BD$\ 

Thus : $\left(  A_{1}\otimes...\otimes A_{p}\right)  \times\left(
B_{1}\otimes...B_{p}\right)  =A_{1}B_{1}\otimes...\otimes A_{p}B_{p}$

If the matrices are square the Kronecker product of two symmetric matrices is
still symmetric, the Kronecker product of two hermitian matrices is still hermitian.

\bigskip

\subsection{Eigen values}

\label{Eigen values}

There are two ways to see a matrix : as a vector in the vector space of
matrices, and as the representation of a map in $K^{n}.$

A matrix in K(r,c) can be seen as tensor in $K^{r}\otimes\left(  K^{c}\right)
^{\ast}$ so a morphism in K(r,c) is a 4th order tensor. As this is not the
most convenient way to work, usually matrices are seen as representations of
maps, either linear maps or bilinear forms.

\subsubsection{Canonical isomorphims}

1. The set $K^{n}$ has an obvious n-dimensional vector space structure, with
canonical basis $\varepsilon_{i}=\left(  0,0,..,0,1,0,..0\right)  $

Vectors are represented as nx1 column matrices

$\left(  K^{n}\right)  ^{\ast}$ has the basis $\varepsilon_{i}=\left(
0,0,..,0,1,0,..0\right)  $ with vectors represented as 1xn row matrices

So the action of a form on a vector is given by : $\left[  x\right]  \in
K^{n},\left[  \varpi\right]  \in K^{n\ast}:\varpi\left(  x\right)  =\left[
\varpi\right]  \left[  x\right]  $

2. To any matrix $A\in$ $K(r,c)$ is associated a \textbf{linear map} with the
obvious definition :

$K(r,c)\rightarrow L\left(  K^{c};K^{r}\right)  ::L_{A}\left(  x\right)
=\left[  A\right]  \left[  x\right]  $

$\left(  r,1\right)  =\left(  r,c\right)  \left(  c,1\right)  $ Beware of the dimensions!

The rank of A is the rank of $L_{A}$.

Similarly for the dual map

$K(r,c)\rightarrow L\left(  \left(  K^{r}\right)  ^{\ast};\left(
K^{c}\right)  ^{\ast}\right)  ::L_{A}^{\ast}\left(  x\right)  =\left[
x\right]  ^{t}\left[  A\right]  $

Warning !

The map : $K(r,c)\rightarrow L\left(  K^{c};K^{r}\right)  $ is basis
dependant. With another basis we would have another map. And the linear map
$L_{A}$ is represented by another matrix in another basis.

If r=c, in a change of basis $e_{i}=\sum_{j}P_{i}^{j}\varepsilon_{j}$ the new
matrix of a is : $B=P^{-1}AP$. Conversely, for $A,B,P\in$ $K(r)$ such that :
$B=P^{-1}AP$ the matrices A and B are said to be similar : they represent the
same linear map $L_{A}$. Thus they have same determinant, rank, eigen values.

3. Similarly to each square matrix $A\in K\left(  r\right)  $ is associated a
\textbf{bilinear form} $B_{A}$ whose matrix is A in the canonical basis.

$K\left(  r\right)  \rightarrow L^{2}\left(  K^{r};K\right)  ::B_{A}\left(
x,y\right)  =\left[  y\right]  ^{t}A\left[  x\right]  $

and if $K=%
\mathbb{C}
$ a \textbf{sequilinear form} B$_{A}$ defined by : $B_{A}\left(  x,y\right)
=\left[  y\right]  ^{\ast}A\left[  x\right]  $

$B_{A}$ is symmetric (resp.skew symmetric, hermitian, skewhermitian) is A is
symmetric (resp.skew symmetric, hermitian, skewhermitian)

To the unitary matrix $I_{r}$ is associated the canonical bilinear form :
$B_{I}\left(  x,y\right)  =\sum_{i=1}^{r}x_{i}y_{i}=\left[  x\right]
^{t}\left[  y\right]  .$ The canonical basis is orthonormal. And the
associated isomorphism $K^{r}\rightarrow K^{r\ast}$ is just passing from
column vectors to rows vectors. With respect to this bilinear form the map
associated to a matrix A is orthogonal if A is orthogonal.

If $K=%
\mathbb{C}
,$ to the unitary matrix $I_{r}$ is associated the canonical hermitian form :
$B_{I}\left(  x,x\right)  =\sum_{i=1}^{r}\overline{x}_{i}y_{i}=\left[
x\right]  ^{\ast}\left[  y\right]  .$With respect to this hermitian form the
map associated to a matrix A is unitary if A is unitary.

Remark : the property for a matrix to be symmetric (or hermitian) is not
linked to the associated linear map, but to the associated bilinear or
sesquilinear map. It is easy to check that if a linear map is represented by a
symmetric matrix in a basis, this property is not conserved in a change of basis.

4. A matrix in $%
\mathbb{R}
\left(  r\right)  $ can be considered as a matrix in $%
\mathbb{C}
\left(  r\right)  $ with real elements. As a matrix A in $%
\mathbb{R}
\left(  r\right)  $ is associated a linear map $L_{A}\in L\left(
\mathbb{R}
^{r};%
\mathbb{R}
^{r}\right)  .$ As a matrix in $%
\mathbb{C}
\left(  r\right)  $ is associated $M_{A}\in L\left(
\mathbb{C}
^{r};%
\mathbb{C}
^{r}\right)  $ which is the complexified of the map L$_{A}$ in the
complexified of $%
\mathbb{R}
^{r}.$ $L_{A}$ and $M_{A}$ have same value for real vectors, and same matrix.
It works only with the classic complexification (see complex vector spaces),
and\ not with complex structure.

\begin{definition}
A matrix $A\in%
\mathbb{R}
\left(  r\right)  $ is \textbf{definite positive} if

$\forall\left[  x\right]  \neq0:\left[  x\right]  ^{t}A\left[  x\right]  >0$

An hermitian matrix A is definite positive if $\forall\left[  x\right]
\neq0:\left[  x\right]  ^{\ast}A\left[  x\right]  >0$
\end{definition}

\subsubsection{Eigen values}

\begin{definition}
The \textbf{eigen values} $\lambda$\ of a square matrix $A\in K\left(
r\right)  $ are the eigen values of its associated linear map $L_{A}\in
L\left(  K^{r};K^{r}\right)  $
\end{definition}

So there is the equation : $A\left[  x\right]  =\lambda\left[  x\right]  $ and
the vectors $\left[  x\right]  \in K^{r}$ meeting this relation are the
\textbf{eigen vectors} of A with respect to $\lambda$

\begin{definition}
The \textbf{characteristic equation} of a matrix A$\in K\left(  r\right)  $ is
the polynomial equation of degree r over K in $\lambda:$

$\det\left(  A-\lambda I_{r}\right)  =0$ reads : $\sum_{i=0}^{r}\lambda
^{i}P_{i}=0$
\end{definition}

$A\left[  x\right]  =\lambda\left[  x\right]  $ is a set of r linear equations
with respect to x, so the eigen values of A are the solutions of $\det\left(
A-\lambda I_{r}\right)  =0$

The coefficient of degree 0 is just detA : $P_{0}=\det A$

If the field K is algebraically closed then this equation has always a
solution. So matrices in $%
\mathbb{R}
\left(  r\right)  $ can have no (real) eigen value and matrices in $%
\mathbb{C}
\left(  r\right)  $ have r eigen values (possibly identical). And similarly
the associated real linear maps can have no (real) eigen value and complex
linear maps have r eigen values (possibly identical)

As any $A\in%
\mathbb{R}
\left(  r\right)  $ can be considered as the same matrix (with real elements)
in $%
\mathbb{C}
\left(  r\right)  $ it has always r eigen values (possibly complex and
identical) and the corresponding eigen vectors can have complex components in
$%
\mathbb{C}
^{r}.$ These eigen values and eigen vectors are associated to the complexified
$M_{A}$\ of the real linear map L$_{A}$ and not to L$_{A}$.

The matrix has no zero eigen value iff the associated linear form is
injective. The associated bilinear form is non degenerate iff there is no zero
eigen value, and definite positive iff all the eigen values are
$>$%
0 .

If all eigen values are real the (non ordered) sequence of signs of the eigen
values is the \textbf{signature} of the matrix.

\begin{theorem}
Hamilton-Cayley's Theorem: Any square matrix is a solution of its
characteristic equation : $\sum_{i=0}^{r}A^{i}P_{i}=0$
\end{theorem}

\begin{theorem}
Any symmetric matrix $A\in%
\mathbb{R}
\left(  r\right)  $ has real eigen values

Any hermitian matrix $A\in%
\mathbb{C}
\left(  r\right)  $\ has real eigen values
\end{theorem}

\subsubsection{Diagonalization}

The eigen spaces $E_{\lambda}$ (set of eigen vectors corresponding to the same
eigen value $\lambda$) are independant. Let be $\dim E_{\lambda}=d_{\lambda}$
so $\sum_{\lambda}d_{\lambda}\leq r$

The matrix A is said to be \textbf{diagonalizable} iff $\sum_{\lambda
}d_{\lambda}=r.$ If it is so $K^{r}=\oplus_{\lambda}E_{\lambda}$ and there is
a basis $\left(  e_{i}\right)  _{i=1}^{r}$ of $K^{r}$ such that the linear map
$L_{A}$ associated with A is expressed in a diagonal matrix D=Diag$\left(
\lambda_{1},...\lambda_{r}\right)  $ (several $\lambda$ can be identical).

Matrices are not necessarily diagonalizable. The matrix A is diagonalizable
iff $m_{\lambda}=d_{\lambda}$ where $m_{\lambda}$ is the order of multiplicity
of $\lambda$ in the characteristic equation. Thus if there are r distincts
eigen values the matrix is diagonalizable.

Let be P the matrix whose columns are the components of the eigen vectors (in
the canonical basis), P is also the matrix of the new basis : $e_{i}=\sum
_{j}P_{i}^{j}\varepsilon_{j}$\ and the new matrix of $L_{A}$ is :
$D=P^{-1}AP\Leftrightarrow A=PDP^{-1}.$ The basis $\left(  e_{i}\right)  $ is
not unique : the vectors $e_{i}$ are defined up to a scalar, and the vectors
can be permuted.

Let be $A,P,Q,D,D^{\prime}\in K\left(  r\right)  ,D,D^{\prime}$ diagonal such
that : $A=PDP^{-1}=QD^{\prime}Q^{-1}$ then there is a permutation matrix $\pi$
such that : $D^{\prime}=\pi D\pi^{t};P=Q\pi$

\begin{theorem}
Normal matrices admit a complex diagonalization
\end{theorem}

\begin{proof}
Let K=$%
\mathbb{C}
.$ the Schur decomposition theorem states that any matrix A can be written as
: $A=U^{\ast}TU$ where U is unitary ($UU^{\ast}=I)$ and T is a triangular
matrix whose diagonal elements are the eigen values of A.

T is a diagonal matrix iff A is normal : AA*=A*A. So A can be written as :
$A=U^{\ast}DU$ iff it is normal. The diagonal elements are the eigen values of A.
\end{proof}

Real symmetric matrices can be written : $A=P^{t}DP$ with P orthogonal :
$P^{t}P=PP^{t}=I.$ The eigen vectors are real and orthogonal for the canonical
bilinear form

Hermitian matrices can be written : $A=U^{\ast}DU$ (also called Takagi's
decomposition) where U is unitary.

\bigskip

\subsection{Matrix calculus}

\label{Matrix calculus}

\subsubsection{Decomposition}

The decomposition of a matrix A\ is a way to write A as the product of
matrices with interesting properties.

\paragraph{Singular values\newline}

\begin{theorem}
Any matrix $A\in K(r,c)$ can be written as : $A=VDU$ where V,U are unitary and
D is the matrix :
\end{theorem}

\bigskip

$D=%
\begin{bmatrix}
diag(\sqrt{\lambda_{1}},...\sqrt{\lambda_{c}})\\
0_{(r-c)\times c}%
\end{bmatrix}
_{r\times c}$

\bigskip

with $\lambda_{i}$ the eigen values of $A^{\ast}A$

(as $A^{\ast}A$ is hermitian its eigen values are real, and it is easy to
check that $\lambda_{i}\geq0)$

If K=$%
\mathbb{R}
$ the theorem stands and V,U are orthogonal.

Remark : the theorem is based on the study of the eigen values and vectors of
A*A and AA*.

\begin{definition}
A scalar $\lambda\in K$ is a \textbf{singular value} for $A\in K(r,c)$ if
there are vectors $\left[  x\right]  \in K^{c},\left[  y\right]  \in K^{r}$
such that :

$A\left[  x\right]  =\lambda\left[  y\right]  $ and $A^{\ast}\left[  y\right]
=\lambda\left[  x\right]  $
\end{definition}

\paragraph{Jordan's decomposition\newline}

\begin{theorem}
Any matrix $A\in K(r)$ can be uniquely written as : $A=S+N$ where S is
diagonalizable, N\ is nilpotent (there is k$\in%
\mathbb{N}
:N^{k}=0),$and $SN=NS$ .Furthermore there is a polynomial such that :
$S=\sum_{j=1}^{p}a_{j}A^{j}$
\end{theorem}

\paragraph{Schur's decomposition\newline}

\begin{theorem}
Any matrix $A\in K(r)$ can be written as : $A=U^{\ast}TU$ where U is unitary
($UU^{\ast}=I)$ and T is a triangular matrix whose diagonal elements are the
eigen values of A.
\end{theorem}

T is a diagonal matrix iff A is normal : AA*=A*A. So A can be written as :
$A=U^{\ast}DU$ iff it is normal (see Diagonalization).

\paragraph{With triangular matrices\newline}

\begin{theorem}
Lu decomposition : Any square matrix $A\in K(r)$ can be written : $A=LU$ with
L lower triangular and U upper triangular
\end{theorem}

\begin{theorem}
QR decomposition : any matrix $A\in%
\mathbb{R}
(r,c)$ can be written : $A=QR$ with Q orthogonal and R upper triangular
\end{theorem}

\begin{theorem}
Cholesky decomposition : any symmetric positive definite matrix can be
uniquely written $A=T^{t}T$ where T is triangular with positive diagonal entries
\end{theorem}

\paragraph{Spectral decomposition\newline}

Let be $\lambda_{k},k=1...p$ the eigen values of A$\in%
\mathbb{C}
\left(  n\right)  $ with multiplicity $m_{k},$ A diagonalizable with
$A=PDP^{-1}$

$B_{k}$ the matrix deduced from D by putting 1 for all diagonal terms related
to $\lambda_{k}$\ and 0 for all the others and $E_{k}=PB_{k}P^{-1}$

Then $A=\sum_{k=1}^{p}\lambda_{k}E_{k}$ and :

$E_{j}^{2}=E_{j};E_{i}E_{j}=0,i\neq j$

$\sum_{k=1}^{p}E_{k}=I$

rank E$_{k}=m_{k}$

$\left(  \lambda_{k}I-A\right)  E_{k}=0$

$A^{-1}=\sum\lambda_{k}^{-1}E_{k}$

A matrix commutes with A iff it commutes with each $E_{k}$

If A is normal then the matrices $E_{k}$ are hermitian

\paragraph{Other\newline}

\begin{theorem}
Any non singular real matrix $A\in%
\mathbb{R}
\left(  r\right)  $ can be written A=CP (or A=PC) where C is symmetric
definite positive and P orthogonal
\end{theorem}

\subsubsection{Block calculus}

Quite often matrix calculi can be done more easily by considering
sub-matrices, called blocks.

The basic identities are :

\bigskip

$%
\begin{bmatrix}
A_{np} & B_{nq}\\
C_{rp} & D_{rq}%
\end{bmatrix}%
\begin{bmatrix}
A_{pn^{\prime}}^{\prime} & B_{pp^{\prime}}^{\prime}\\
C_{qn^{\prime}}^{\prime} & D_{qp^{\prime}}^{\prime}%
\end{bmatrix}
=%
\begin{bmatrix}
A_{np}A_{pn^{\prime}}^{\prime}+B_{nq}C_{qn^{\prime}}^{\prime} & A_{np}%
B_{pp^{\prime}}^{\prime}+B_{nq}D_{qp^{\prime}}^{\prime}\\
C_{rp}A_{pn^{\prime}}^{\prime}+D_{rq}C_{qn^{\prime}}^{\prime} & C_{rp}%
B_{pp^{\prime}}^{\prime}+D_{rq}D_{qp^{\prime}}^{\prime}%
\end{bmatrix}
$

\bigskip

so we get nicer results if some of the blocks are 0.

\bigskip

Let be M=$%
\begin{bmatrix}
A & B\\
C & D
\end{bmatrix}
;A(m,m);B(m,n);C(n,m);D(n,n)$

\bigskip

Then :

$\det M=\det(A)\det(D-CA^{-1}B)=\det(D)\det(A-BD^{-1}C)$

If A=I,D=I:det(M)=det(I$_{mm}-BC)=det(I_{nn}-CB)$

$P[n,n]=D-CA^{-1}B$ and $Q[m,m]=A-BD^{-1}C$ are respectively the Schur
Complements of A and D in M.

\bigskip

$M^{-1}=%
\begin{bmatrix}
Q^{-1} & -Q^{-1}BD^{-1}\\
-D^{-1}CQ^{-1} & D^{-1}\left(  I+CQ^{-1}BD^{-1}\right)
\end{bmatrix}
$

\bigskip

\subsubsection{Complex and real matrices}

Any matrix $A\in%
\mathbb{C}
\left(  r,c\right)  $ can be written as : $A=\operatorname{Re}%
A+i\operatorname{Im}A$ where $\operatorname{Re}A,\operatorname{Im}A\in%
\mathbb{R}
\left(  r,c\right)  $

For square matrices $M\in%
\mathbb{C}
\left(  n\right)  $ it can be useful to introduce :

\bigskip

$Z\left(  M\right)  =%
\begin{bmatrix}
\operatorname{Re}M & -\operatorname{Im}M\\
\operatorname{Im}M & \operatorname{Re}M
\end{bmatrix}
\in%
\mathbb{R}
\left(  2n\right)  $

\bigskip

It is the real representation of GL(n,$%
\mathbb{C}
)$ in GL(2n;$%
\mathbb{R}
)$

and :

Z(MN)=Z(M)Z(N)

Z(M*)=Z(M)*

TrZ(M)=2ReTrM

detZ(M)=%
$\vert$%
detM%
$\vert$%
${{}^2}$%

\subsubsection{Pauli's matrices}

They are (with some differences according to authors and usages) the matrices
in $%
\mathbb{C}
\left(  2\right)  :$

\bigskip

$\sigma_{0}=%
\begin{bmatrix}
1 & 0\\
0 & 1
\end{bmatrix}
;\sigma_{1}=%
\begin{bmatrix}
0 & 1\\
1 & 0
\end{bmatrix}
;\sigma_{2}=%
\begin{bmatrix}
0 & -i\\
i & 0
\end{bmatrix}
;\sigma_{3}=%
\begin{bmatrix}
1 & 0\\
0 & -1
\end{bmatrix}
;$

\bigskip

the multiplication tables are:

$i,j=1,2,3:\sigma_{i}\sigma_{j}+\sigma_{j}\sigma_{i}=2\delta_{ij}\sigma_{0}$

$\sigma_{1}\sigma_{2}=i\sigma_{3}$

$\sigma_{2}\sigma_{3}=i\sigma_{1}$

$\sigma_{3}\sigma_{1}=i\sigma_{2}$

that is :

i,j=1,2,3 : $\sigma_{i}\sigma_{j}=\sigma_{i}\sigma_{j}\sigma_{0}=\sigma
_{i}\sigma_{0}\sigma_{j}=\sigma_{0}\sigma_{i}\sigma_{j}=i\epsilon\left(
i,j,k\right)  \sigma_{k}$

i,j,k=1,2,3 : $\sigma_{i}\sigma_{j}\sigma_{k}=i\epsilon\left(  i,j,k\right)
\sigma_{0}$

\subsubsection{Matrix functions}

$%
\mathbb{C}
\left(  r\right)  $ is a finite dimensional vector space, thus a normed vector
space and a Banach vector space (and a C* algebra). All the noms are
equivalent.\ The two most common are :

i) the Frobenius norm (also called the Hilbert-Schmidt norm):

$\left\Vert A\right\Vert _{HS}=Tr\left(  A^{\ast}A\right)  =\sum
_{ij}\left\vert a_{ij}\right\vert ^{2}$

ii) the usual norm on $L\left(
\mathbb{C}
^{n};%
\mathbb{C}
^{n}\right)  :\left\Vert A\right\Vert _{2}=\inf_{\left\Vert u\right\Vert
=1}\left\Vert Au\right\Vert $

$\left\Vert A\right\Vert _{2}\leq\left\Vert A\right\Vert _{HS}\leq n\left\Vert
A\right\Vert _{2}$

\paragraph{Exponential\newline}

\begin{theorem}
The series : $\exp A=\sum_{0}^{\infty}\frac{A^{n}}{n!}$ converges
always\newline
\end{theorem}

$\exp0=I$

$\left(  \exp A\right)  ^{-1}=\exp\left(  -A\right)  $

$\exp\left(  A\right)  \exp\left(  B\right)  =\exp\left(  A+B\right)  $ iff
AB=BA \ \ Beware !

$\left(  \exp A\right)  ^{t}=\exp\left(  A^{t}\right)  $

$\left(  \exp A\right)  ^{\ast}=\exp\left(  A^{\ast}\right)  $

$\det(\exp A)=\exp(Tr\left(  A\right)  )$

The map $t\in%
\mathbb{R}
\rightarrow\exp\left(  tA\right)  $ defines a 1-parameter group.\ It is
differentiable :

$\frac{d}{dt}\left(  \exp tA\right)  |_{t=\tau}=\left(  \exp\tau A\right)
A=A\exp\tau A$

$\frac{d}{dt}\left(  \exp tA\right)  |_{t=0}=A$

Conversely if $f:%
\mathbb{R}
_{+}\rightarrow%
\mathbb{C}
\left(  r\right)  $ is a continuous homomorphism then $\exists A\in%
\mathbb{C}
\left(  r\right)  :f\left(  t\right)  =\exp tA$

Warning !

The map $t\in%
\mathbb{R}
\rightarrow\exp A\left(  t\right)  $ where the matrix A(t) depends on t has no
simple derivative.\ We \textit{do not} have : $\frac{d}{dt}\left(  \exp
A\left(  t\right)  \right)  =A^{\prime}(t)\exp A(t)$

\begin{theorem}
(Taylor 1 p.19) If A is a n$\times$n complex matrix and $\left[  v\right]  $ a
n$\times$1 matrix, then :

$\forall t\in%
\mathbb{R}
:\left(  \exp t\left[  A\right]  \right)  \left[  v\right]  =\sum_{j=1}%
^{n}\left(  \exp\lambda_{j}t\right)  \left[  w_{j}\left(  t\right)  \right]  $

where : $\lambda_{j}$ are the eigen values of A, $\left[  w_{j}\left(
t\right)  \right]  $ is a polynomial in t, valued in $%
\mathbb{C}
\left(  n,1\right)  $
\end{theorem}

If A is diagonalizable then the $\left[  w_{j}\left(  t\right)  \right]  =Cte$

\begin{theorem}
Integral formulation: If all the eigen value of A are in the open disc
$\left\vert z\right\vert <r$ then $\exp A=\dfrac{1}{2i\pi}\int_{C}%
(zI-A)^{-1}e^{z}dz$ with C any closed curve around the origin and included in
the disc
\end{theorem}

The inverse function of exp is the logarithm : $\exp\left(  \log\left(
A\right)  \right)  =A.$ It is usally a multivalued function (as for the
complex numbers).

$\log(BAB^{-1})=B(\log A)B^{-1}$

$\log(A^{-1})=-\log A$

If A has no zero or negative eigen values : $\log A=\int_{-\infty}%
^{0}[(s-A)^{-1}-(s-1)^{-1}]ds$

Cartan's decomposition : Any invertible matrix $A\in%
\mathbb{C}
\left(  r\right)  $ can be uniquely written : $A=P\exp Q$ with :

$P=A\exp\left(  -Q\right)  ;P^{t}P=I$

$Q=\dfrac{1}{2}\log(A^{t}A);Q^{t}=Q$

P,Q are real if A is real

\paragraph{Analytic functions\newline}

\begin{theorem}
Let $f:%
\mathbb{C}
\rightarrow%
\mathbb{C}
$ any holomorphic function on an open disc $\left\vert z\right\vert <r$ then :
$f\left(  z\right)  =\sum_{n=0}^{\infty}a_{n}z^{n}$ and the series : $f\left(
A\right)  =\sum_{n=0}^{\infty}a_{n}A^{n}$ converges for any square matrix A
such that $\left\Vert A\right\Vert <r$\newline
\end{theorem}

With the Cauchy's integral formula, for any closed curve C circling x and
contained within the disc, it holds:

$f\left(  x\right)  =\frac{1}{2i\pi}\int_{C}\frac{f\left(  z\right)  }{z-x}dz$
then : $f\left(  A\right)  =\frac{1}{2i\pi}\int_{C}f\left(  z\right)  \left(
zI-A\right)  ^{-1}dz$ where C is any closed curve enclosing all the eigen
values of A and contained within the disc

If $\left\Vert A\right\Vert <1:\sum_{p=0}^{\infty}(-1)^{p}A^{p}=(I+A)^{-1}$

\paragraph{Derivative\newline}

There are useful formulas for the derivative of functions of a matrix
depending on a variable.

\subparagraph{1. Determinant:\newline}

\begin{theorem}
let $A=\left[  a_{ij}\right]  \in%
\mathbb{R}
\left(  n\right)  ,$ then%

\begin{equation}
\frac{d\det A}{da_{ij}}=(-1)^{i+j}\det\left[  A_{\left\{  1...n\backslash
j\right\}  }^{\left\{  1...n\backslash i\right\}  }\right]  =\left[
A^{-1}\right]  _{i}^{j}\det A
\end{equation}

\end{theorem}

\begin{proof}
we have $\left[  A^{-1}\right]  _{ij}=\frac{1}{\det A}(-1)^{i+j}\det\left[
A_{ji}\right]  $ where $\left[  A^{-1}\right]  _{ij}$ is the element of
$A^{-1}$ and $\det\left[  A_{ji}\right]  $ the minor.
\end{proof}

Beware reversed indices!

\begin{theorem}
If $%
\mathbb{R}
\rightarrow%
\mathbb{R}
\left(  n\right)  ::A\left(  x\right)  =\left[  a_{ij}\left(  x\right)
\right]  ,$ A invertible then%

\begin{equation}
\frac{d\det A}{dx}=\left(  \det A\right)  Tr\left(  \frac{dA}{dx}\left(
A^{-1}\right)  \right)
\end{equation}

\end{theorem}

\begin{proof}
Schur's decomposition : $A=UTU^{\ast},UU^{\ast}=I,T$ triangular

let be : $A^{\prime}=U^{\prime}TU^{\ast}+UT^{\prime}U^{\ast}+UT\left(
U^{\ast}\right)  ^{\prime}$

the derivative of U :$\left[  u_{j}^{i}(x)\right]  =U^{\ast}\rightarrow\left(
U^{\ast}\right)  ^{\prime}=\overline{u}_{i}^{j}(x)^{\prime}\rightarrow\left(
U^{\ast}\right)  ^{\prime}=\left(  U^{\prime}\right)  ^{\ast}$

$A^{\prime}A^{-1}=U^{\prime}TU^{\ast}UT^{-1}U^{\ast}+UT^{\prime}U^{\ast
}UT^{-1}U^{\ast}+UT\left(  U^{\ast}\right)  ^{\prime}UT^{-1}U^{\ast}$

$=U^{\prime}U^{\ast}+UT^{\prime}T^{-1}U^{\ast}+UT\left(  U^{\ast}\right)
^{\prime}UT^{-1}U^{\ast}$

$Tr(A^{\prime}A^{-1})=Tr\left(  U^{\prime}U^{\ast}\right)  +Tr\left(
T^{\prime}T^{-1}\right)  +Tr\left(  \left(  UT\left(  U^{\ast}\right)
^{\prime}\right)  \left(  UT^{-1}U^{\ast}\right)  \right)  $

$Tr\left(  \left(  UT\left(  U^{\ast}\right)  ^{\prime}\right)  \left(
UT^{-1}U^{\ast}\right)  \right)  =Tr\left(  \left(  UT^{-1}U^{\ast}\right)
\left(  UT\left(  U^{\ast}\right)  ^{\prime}\right)  \right)  =Tr\left(
U\left(  U^{\ast}\right)  ^{\prime}\right)  $

$UU^{\ast}=I\Rightarrow U^{\prime}U^{\ast}+U\left(  U^{\ast}\right)  ^{\prime
}=0$

$Tr(A^{\prime}A^{-1})=Tr\left(  T^{\prime}T^{-1}\right)  $

$\Theta=T^{-1}$ is triangular with diagonal such that :$\theta_{k}^{i}%
t_{j}^{k}=\delta_{j}^{i}\Rightarrow\theta_{k}^{i}t_{i}^{k}=1=\sum_{k=i}%
^{n}\theta_{k}^{i}t_{i}^{k}=\theta_{i}^{i}t_{i}^{k}$

so $\theta_{i}^{i}=1/$eigen values of A

$Tr(A^{\prime}A^{-1})=Tr\left(  T^{\prime}T^{-1}\right)  =\sum_{i=1}^{n}%
\frac{\lambda_{i}^{^{\prime}}}{\lambda_{i}}=\sum_{i}\left(  \ln\lambda
_{i}\right)  ^{\prime}=\left(  \sum_{i}\ln\lambda_{i}\right)  ^{\prime}$

$=\left(  \ln%
{\displaystyle\prod\limits_{i}}
\lambda_{i}\right)  ^{\prime}=\left(  \ln\det A\right)  ^{\prime}$
\end{proof}

\subparagraph{2. Inverse:\newline}

\begin{theorem}
If $K=\left[  k_{pq}\right]  \in%
\mathbb{R}
\left(  n\right)  ,$ is an invertible matrix, then%

\begin{equation}
\frac{dk_{pq}}{dj_{rs}}=-k_{pr}k_{sq}\text{ }with\text{ }J=K^{-1}=\left[
j_{pq}\right]
\end{equation}

\end{theorem}

\begin{proof}
Use : $K_{\lambda}^{\gamma}J_{\mu}^{\lambda}=\delta_{\mu}^{\gamma}$

$\Rightarrow\left(  \frac{\partial}{\partial J_{\beta}^{\alpha}}K_{\lambda
}^{\gamma}\right)  J_{\mu}^{\lambda}+K_{\lambda}^{\gamma}\left(
\frac{\partial}{\partial J_{\beta}^{\alpha}}J_{\mu}^{\lambda}\right)  =0$

$=\left(  \frac{\partial}{\partial J_{\beta}^{\alpha}}K_{\lambda}^{\gamma
}\right)  J_{\mu}^{\lambda}+K_{\lambda}^{\gamma}\delta_{\alpha}^{\lambda
}\delta_{\beta}^{\mu}=\left(  \frac{\partial}{\partial J_{\beta}^{\alpha}%
}K_{\lambda}^{\gamma}\right)  J_{\mu}^{\lambda}+K_{\alpha}^{\gamma}%
\delta_{\beta}^{\mu}$

$=\left(  \frac{\partial}{\partial J_{\beta}^{\alpha}}K_{\lambda}^{\gamma
}\right)  J_{\mu}^{\lambda}K_{\nu}^{\mu}+K_{\alpha}^{\gamma}\delta_{\beta
}^{\mu}K_{\nu}^{\mu}=\left(  \frac{\partial}{\partial J_{\beta}^{\alpha}%
}K_{\nu}^{\gamma}\right)  +K_{\alpha}^{\gamma}K_{\nu}^{\beta} $

$\Rightarrow\frac{\partial}{\partial J_{l}^{k}}K_{j}^{i}=-K_{k}^{i}K_{j}^{l}$
\end{proof}

As $%
\mathbb{C}
\left(  r\right)  $ is a normed algebra the derivative with respect to a
matrix (and not only with respect to its elements) is defined :

$\varphi:%
\mathbb{C}
\left(  r\right)  \rightarrow%
\mathbb{C}
\left(  r\right)  ::\varphi\left(  A\right)  =\left(  I_{r}+A\right)  ^{-1}$
then $\frac{d\varphi}{dA}=-A$

\paragraph{Matrices of $SO\left(
\mathbb{R}
,p,q\right)  $\newline}

These matrices are of some importance in physics, because the Lorentz group of
Relativity is just $SO(%
\mathbb{R}
,3,1).$

$SO(%
\mathbb{R}
,p,q)$ is the group of nxn real matrices with n=p+q such that :

detM = 1

$A^{t}\left[  I_{p,q}\right]  A=I_{n\times n}$ where $\left[  I_{p,q}\right]
=%
\begin{bmatrix}
I_{p\times p} & 0\\
0 & I_{q\times q}%
\end{bmatrix}
$

Any matrix of $SO\left(
\mathbb{R}
,p,q\right)  $ has a Cartan decomposition, so can be uniquely written as :

$A=\left[  \exp p\right]  \left[  \exp l\right]  $ with

$\left[  p\right]  =%
\begin{bmatrix}
0 & P_{p\times q}\\
P_{q\times p}^{t} & 0
\end{bmatrix}
,\left[  l\right]  =%
\begin{bmatrix}
M_{p\times p} & 0\\
0 & N_{q\times q}%
\end{bmatrix}
,M=-M^{t},N=-N^{t}$

(or as $A=\left[  \exp l^{\prime}\right]  \left[  \exp p^{\prime}\right]  $
with similar p',l' matrices).

The matrix $\left[  l\right]  $ is block diagonal antisymmetric.

This theorem is new.

\begin{theorem}
$\exp p=%
\begin{bmatrix}
I_{p} & 0\\
0 & I_{q}%
\end{bmatrix}
+%
\begin{bmatrix}
H\left(  \cosh D-I_{q}\right)  H^{t} & H(\sinh D)U^{t}\\
U(\sinh D)H^{t} & U(\cosh D-I_{q})U^{t}%
\end{bmatrix}
$ with $H_{p\times q}$ such that : $H^{t}H=I_{q},P=HDU^{t}$ where D is a real
diagonal q$\times$q matrix and U is a q$\times$q real orthogonal matrix.
\end{theorem}

\begin{proof}
We assume that p
$>$
q

The demonstration is based upon the decomposition of $\left[  P\right]
_{p\times q}$ using the singular values decomposition.\ 

P reads :$P=VQU^{t}$ where :

$Q=%
\begin{bmatrix}
D_{q\times q}\\
0_{(p-q)\times q}%
\end{bmatrix}
_{p\times q};D=diag(d_{k})_{k=1...q};d_{k}\geq0$

$\left[  V\right]  _{p\times p}$,$\left[  U\right]  _{q\times q}$ are orthogonal

Thus : $PP^{t}=V%
\begin{bmatrix}
D^{2}\\
0_{(p-q)\times q}%
\end{bmatrix}
V^{t};P^{t}P=UD^{2}U^{t}$

The eigen values of $PP^{t}$ are $\left\{  d_{1}^{2},...d_{q}^{2}%
,0,..,0\right\}  $ and of $P^{t}P:\left\{  d_{1}^{2},...d_{q}^{2}\right\}  .$
The decomposition is not unique.

Notice that we are free to choose the sign of $d_{k},$ the choice $d_{k}\geq0$
is just a convenience.

So :

$\left[  p\right]  =%
\begin{bmatrix}
0 & P\\
P^{t} & 0
\end{bmatrix}
=%
\begin{bmatrix}
V & 0\\
0 & U
\end{bmatrix}%
\begin{bmatrix}
0 & Q\\
Q^{t} & 0
\end{bmatrix}%
\begin{bmatrix}
V^{t} & 0\\
0 & U^{t}%
\end{bmatrix}
=\left[  k\right]
\begin{bmatrix}
0 & Q\\
Q^{t} & 0
\end{bmatrix}
\left[  k\right]  ^{t}$

with : $k=%
\begin{bmatrix}
V & 0\\
0 & U
\end{bmatrix}
:\left[  k\right]  \left[  k\right]  ^{t}=I_{p\times p}$

and :

$\exp\left[  p\right]  =\left[  k\right]  \left(  \exp%
\begin{bmatrix}
0 & Q\\
Q^{t} & 0
\end{bmatrix}
\right)  \left[  k\right]  ^{t}$

$%
\begin{bmatrix}
0 & Q\\
Q^{t} & 0
\end{bmatrix}
^{2m}=%
\begin{bmatrix}
D^{2m} & 0 & 0\\
0 & 0 & 0\\
0 & 0 & D^{2m}%
\end{bmatrix}
;m>0$

$%
\begin{bmatrix}
0 & Q\\
Q^{t} & 0
\end{bmatrix}
^{2m+1}=%
\begin{bmatrix}
0 & 0 & D^{2m+1}\\
0 & 0 & 0\\
D^{2m+1} & 0 & 0
\end{bmatrix}
$

thus :

$\exp%
\begin{bmatrix}
0 & Q\\
Q^{t} & 0
\end{bmatrix}
$

$=I_{p+q}+\sum_{m=0}^{\infty}\frac{1}{\left(  2m+1\right)  !}%
\begin{bmatrix}
0 & 0 & D^{2m+1}\\
0 & 0 & 0\\
D^{2m+1} & 0 & 0
\end{bmatrix}
+\sum_{m=1}^{\infty}\frac{1}{\left(  2m\right)  !}%
\begin{bmatrix}
D^{2m} & 0 & 0\\
0 & 0 & 0\\
0 & 0 & D^{2m}%
\end{bmatrix}
$

$=I_{p+q}+%
\begin{bmatrix}
0 & 0 & \sinh D\\
0 & 0 & 0\\
\sinh D & 0 & 0
\end{bmatrix}
+%
\begin{bmatrix}
\cosh D & 0 & 0\\
0 & 0 & 0\\
0 & 0 & \cosh D
\end{bmatrix}
-%
\begin{bmatrix}
I_{q} & 0 & 0\\
0 & 0 & 0\\
0 & 0 & I_{q}%
\end{bmatrix}
$

with : $\cosh D=diag(\cosh d_{k});\sinh D=diag(\sinh d_{k})$

And :

$\exp p=%
\begin{bmatrix}
V & 0\\
0 & U
\end{bmatrix}%
\begin{bmatrix}
\cosh D & 0 & \sinh D\\
0 & I_{p-q} & 0\\
\sinh D & 0 & \cosh D
\end{bmatrix}%
\begin{bmatrix}
V^{t} & 0\\
0 & U^{t}%
\end{bmatrix}
$

In order to have some unique decomposition write :

$\exp p=%
\begin{bmatrix}
V%
\begin{bmatrix}
\cosh D & 0\\
0 & I_{p-q}%
\end{bmatrix}
V^{t} & V%
\begin{bmatrix}
\sinh D\\
0
\end{bmatrix}
U^{t}\\
U%
\begin{bmatrix}
\sinh D & 0
\end{bmatrix}
V^{t} & U(\cosh D)U^{t}%
\end{bmatrix}
$

Thus with the block matrices $V_{1}$\ (q,q) and $V_{3}$ (p-q,q)

$V=%
\begin{bmatrix}
V_{1} & V_{2}\\
V_{3} & V_{4}%
\end{bmatrix}
\in O(%
\mathbb{R}
,p)$

$V^{t}V=VV^{t}=I_{p}$

$%
\begin{bmatrix}
V_{1}V_{1}^{t}+V_{2}V_{2}^{t} & V_{1}V_{3}^{t}+V_{2}V_{4}^{t}\\
V_{3}V_{1}^{t}+V_{4}V_{2}^{t} & V_{3}V_{3}^{t}+V_{4}V_{4}^{t}%
\end{bmatrix}
=%
\begin{bmatrix}
V_{1}^{t}V_{1}+V_{3}^{t}V_{3} & V_{2}^{t}V_{1}+V_{4}^{t}V_{3}\\
V_{1}^{t}V_{2}+V_{3}^{t}V_{4} & V_{2}^{t}V_{2}+V_{4}^{t}V_{4}%
\end{bmatrix}
=%
\begin{bmatrix}
I_{q} & 0\\
0 & I_{p-q}%
\end{bmatrix}
$

So :

$V%
\begin{bmatrix}
\cosh D+I_{q} & 0\\
0 & I_{p-q}%
\end{bmatrix}
V^{t}=%
\begin{bmatrix}
V_{2}V_{2}^{t}+V_{1}(\cosh D)V_{1}^{t} & V_{2}V_{4}^{t}+V_{1}(\cosh
D)V_{3}^{t}\\
V_{4}V_{2}^{t}+V_{3}\left(  \cosh D\right)  V_{1}^{t} & V_{4}V_{4}^{t}%
+V_{3}(\cosh D)V_{3}^{t}%
\end{bmatrix}
\allowbreak$

$=I_{p}+%
\begin{bmatrix}
V_{1}\\
V_{3}%
\end{bmatrix}
\left(  \cosh D-I_{q}\right)
\begin{bmatrix}
V_{1}^{t} & V_{3}^{t}%
\end{bmatrix}
$

$V%
\begin{bmatrix}
\sinh D\\
0
\end{bmatrix}
U^{t}=%
\begin{bmatrix}
V_{1}(\sinh D)U^{t}\\
V_{3}(\sinh D)U^{t}%
\end{bmatrix}
=%
\begin{bmatrix}
V_{1}\\
V_{3}%
\end{bmatrix}
(\sinh D)U^{t}$

$U%
\begin{bmatrix}
\sinh D & 0
\end{bmatrix}
V^{t}=U(\sinh D)%
\begin{bmatrix}
V_{1}^{t} & V_{3}^{t}%
\end{bmatrix}
$

$\exp p=%
\begin{bmatrix}
I_{p}+%
\begin{bmatrix}
V_{1}\\
V_{3}%
\end{bmatrix}
\left(  \cosh D-I_{q}\right)
\begin{bmatrix}
V_{1}^{t} & V_{3}^{t}%
\end{bmatrix}
&
\begin{bmatrix}
V_{1}\\
V_{3}%
\end{bmatrix}
(\sinh D)U^{t}\\
U(\sinh D)%
\begin{bmatrix}
V_{1}^{t} & V_{3}^{t}%
\end{bmatrix}
& U(\cosh D)U^{t}%
\end{bmatrix}
$

Let us denote $H=%
\begin{bmatrix}
V_{1}\\
V_{3}%
\end{bmatrix}
$

H is a pxq matrix with rank q : indeed if not the matrix V would not be
regular. Moreover :

$V^{t}V=VV^{t}=I_{p}\Rightarrow V_{1}^{t}V_{1}+V_{3}^{t}V_{3}=I_{q}%
\Leftrightarrow H^{t}H=I_{q}$

And :

$\exp p=%
\begin{bmatrix}
I_{p}+H\left(  \cosh D-I_{q}\right)  H^{t} & H(\sinh D)U^{t}\\
U(\sinh D)H^{t} & U(\cosh D)U^{t}%
\end{bmatrix}
$

The number of parameters are here just pq and as the Cartan decomposition is a
diffeomorphism the decomposition is unique.

H, D and U are related to P and p by :

$P=V%
\begin{bmatrix}
D\\
0
\end{bmatrix}
U^{t}=HDU^{t}$

$p=%
\begin{bmatrix}
0 & P\\
P^{t} & 0
\end{bmatrix}
=%
\begin{bmatrix}
H & 0\\
0 & U
\end{bmatrix}
_{\left(  p+q,2q\right)  }%
\begin{bmatrix}
0 & D\\
D & 0
\end{bmatrix}
_{\left(  2q,2q\right)  }%
\begin{bmatrix}
H & 0\\
0 & U
\end{bmatrix}
_{(2q,p+q)}^{t}$
\end{proof}

With this decomposition it is easy to compute the powers of exp(p)

$k\in Z:\left(  \exp p\right)  ^{k}=\exp(kp)=%
\begin{bmatrix}
I_{p}+H\left(  \cosh kD-I_{q}\right)  H^{t} & H(\sinh kD)U^{t}\\
U(\sinh kD)H^{t} & U(\cosh kD)U^{t}%
\end{bmatrix}
$

Notice that : $\exp(kp)=\exp\left(
\begin{bmatrix}
0 & kP\\
kP^{t} & 0
\end{bmatrix}
\right)  $

so with the same singular values decomposition the matrix D' :

$\left(  kP\right)  ^{t}\left(  kP\right)  =D^{\prime2}=k^{2}D,$

$kP=(\frac{k}{\left\vert k\right\vert }V)D^{\prime}U^{t}=\left(  \epsilon
V\right)  \left(  \left\vert k\right\vert D\right)  U^{t}$

$\left(  \exp p\right)  ^{k}=\exp(kp)=%
\begin{bmatrix}
I_{p}+H\left(  \cosh kD-I_{q}\right)  H^{t} & H(\sinh kD)U^{t}\\
U(\sinh kD)H^{t} & U(\cosh kD)U^{t}%
\end{bmatrix}
$

In particular with k= -1 :

$\left(  \exp p\right)  ^{-1}=%
\begin{bmatrix}
I_{p}+H\left(  \cosh D\right)  H^{t}-HH^{t} & -H\left(  \sinh D\right)
U^{t}\\
-U\left(  \sinh D\right)  H^{t} & U\left(  \cosh D\right)  U^{t}%
\end{bmatrix}
)$

For the Lorentz group the decomposition reads :

H is a vector 3x1 matrix : $H^{t}H=1,$ D is a scalar, U=$\left[  1\right]  ,$

$l=%
\begin{bmatrix}
M_{3\times3} & 0\\
0 & 0
\end{bmatrix}
,M=-M^{t}$ thus $\exp l=%
\begin{bmatrix}
R & 0\\
0 & 1
\end{bmatrix}
$ where $R\in SO\left(
\mathbb{R}
,3\right)  $

$A\in SO\left(  3,1,%
\mathbb{R}
\right)  :$

$A=\exp p\exp l=%
\begin{bmatrix}
I_{3}+\left(  \cosh D-1\right)  HH^{t} & (\sinh D)H\\
(\sinh D)H^{t} & \cosh D
\end{bmatrix}%
\begin{bmatrix}
R & 0\\
0 & 1
\end{bmatrix}
$

\newpage

\section{CLIFFORD\ ALGEBRA}

Mathematical objects such as "spinors" and spin representations are frequently
met in physics. The great variety of definitions, sometimes clever but varying
greatly and too focused on a pretense of simplicity, gives a confusing idea of
this field. In fact the unifiying concept which is the base of all these
mathematical objects is the Clifford algebra. This is a special structure,
involving a vector space, a symmetric bilinear form and a field, which is more
than an algebra and distinct from a Lie algebra. It introduces a new operation
- the product of vectors - which can be seen as disconcerting at first, but
when the structure is built in a coherent way, step by step, we feel much more
comfortable with all its uses in the other fields, such as representation
theory of groups, fiber bundles and functional analysis.

\bigskip

\subsection{Main operations in a Clifford algebra}

\label{Clifford Definition}

\subsubsection{Definition of the Clifford algebra}

\begin{definition}
Let F be a vector space over the field K (of characteristic $\neq2)$\ endowed
with a \textit{symmetric bilinear non degenerate} \textit{form} g (valued in
the field K). The \textbf{Clifford algebra} Cl(F,g) and the canonical map
$\imath:F\rightarrow Cl(F,g)$ are defined by the following universal property
: for any associative algebra A over K (with internal product $\times$\ and
unit e) and K-linear map $f:F\rightarrow A$ such that :

$\forall v,w\in F:f\left(  v\right)  \times f\left(  w\right)  +f\left(
w\right)  \times f\left(  v\right)  =2g\left(  v,w\right)  \times e$

there exists a unique algebra morphism : $\varphi:Cl(F,g)\rightarrow A$ such
that $f=\varphi\circ\imath$
\end{definition}

\bigskip%

\begin{tabular}
[c]{lllll}
&  & $f$ &  & \\
$F$ & $\rightarrow$ & $\rightarrow$ & $\rightarrow$ & $A$\\
$\downarrow$ &  &  & $\nearrow$ & \\
$\downarrow\hspace{0in}\imath$ &  & $\nearrow$ & $\varphi$ & \\
$\downarrow$ & $\nearrow$ &  &  & \\
$Cl(F,g)$ &  &  &  &
\end{tabular}

\bigskip

The Clifford algebra includes the scalar K and the vectors of F (so we
identify $\imath\left(  u\right)  $ with $u\in F$ and $\imath\left(  k\right)
$ with $k\in K)$

Remarks :

i) There is also the definition $f\left(  v\right)  \times f\left(  w\right)
+f\left(  w\right)  \times f\left(  v\right)  +2g\left(  v,w\right)  \times
e=0$ which sums up to take the opposite for g (careful about the signature
which is important).

ii) F can be a real or a complex vector space, but \textit{g must be symmetric
: }it does not work with a hermitian sesquilinear form.

iii) It is common to define a Clifford algebra through a quadratic form : any
quadratic form gives a bilinear symmetric form by polarization, and as a
bilinear symmetric form is necessary for most of the applications, we can
easily jump over this step.

iv) This is an algebraic definition, which encompasses the case of infinite
dimensional vector spaces. However in infinite dimension, the Clifford
algebras are usually defined over Hilbert spaces $\left(  H,g\right)  .$ The
Clifford algebra Cl(H,g) is then the quotient of the tensorial algebra
$\sum_{n=0}^{\infty}\otimes^{n}H$\ by the two-sided ideal generated by the
elements of the form $u\otimes v+v\otimes v-2g\left(  u,v\right)  .$\ Most of
the results presented here can be generalized, moreover Cl(H,g) is a Hilbert
algebra (on this topic see de la Harpe). In the following the vector space V
will be finite dimensional.

A definition is not a proof of existence. Happily :

\begin{theorem}
There is always a Clifford algebra, isomorphic, as vector space, to the
algebra $\Lambda F$ of antisymmetric tensors with the exterior product$.$
\end{theorem}

The isomorphism follows the determination of the bases (see below)

\subsubsection{Algebra structure}

\begin{definition}
The \textbf{internal product }of Cl(F,g) is denoted by a dot $\cdot$\ .\ It is
such that :%

\begin{equation}
\forall v,w\in F:v\cdot w+w\cdot v=2g\left(  v,w\right)
\end{equation}

\end{definition}

\begin{theorem}
With this internal product $\left(  Cl(F,g),\cdot\right)  $ is a unital
algebra on the field K, with unity element the scalar $1\in K$
\end{theorem}

Notice that a Clifford algebra is an algebra but is more than that because of
this fundamental relation (valid only for vectors of F, not for any element of
the Clifford algebra).

$\Rightarrow\forall u,v\in F:u\cdot v\cdot u=-g\left(  u,u\right)  v+2g\left(
u,v\right)  u\in F$

\begin{definition}
The \textbf{homogeneous} elements of degree r of Cl(F,g) are elements which
can be written as product of r vectors of F
\end{definition}

$w=u_{1}\cdot u_{2}...\cdot u_{r}.$ The homogeneous elements of degree n=dimF
are called pseudoscalars (there are also many denominations for various
degrees and dimensions, but they are only complications).

\begin{theorem}
(Fulton p.302) The set of elements :

$\left\{  1,e_{i_{1}}\cdot...\cdot e_{i_{k}},1\leq i_{1}<i_{2}...<i_{k}%
\leq\dim F,k=1...2^{\dim F}\right\}  $\ where $\left(  e_{i}\right)
_{i=1}^{\dim F} $\ is an orthonormal basis of F, is a basis of the Clifford
algebra Cl(F,g) which is a vector space over K of dimCl(F,g)=$2^{\dim F}$
\end{theorem}

Notice that the basis of Cl(F,g) must have the basis vector 1 to account for
the scalars.

So any element of Cl(F,g) can be expressed as :

$w=\sum_{k=0}^{2^{\dim F}}\sum_{\left\{  i_{1},...,i_{k}\right\}  }w_{\left\{
i_{1},...,i_{k}\right\}  }e_{i_{1}}\cdot...\cdot e_{i_{k}}=\sum_{k=0}^{2^{\dim
F}}\sum_{I_{k}}w_{I_{k}}e_{i_{1}}\cdot...\cdot e_{i_{k}}$

Notice that $w_{0}\in K$

A bilinear symmetric form is fully defined by an orthonormal basis.\ They will
always be used in a Clifford algebra.

The isomorphism of vector spaces with the algebra $\Lambda F$\ reads :

$e_{i_{1}}\cdot e_{i_{2}}\cdot..e_{i_{k}}\in Cl(F,g)\leftrightarrow e_{i_{1}%
}\wedge e_{i_{2}}\wedge..e_{i_{k}}\in\Lambda F$

This isomorphism does not depend of the choice of the orthonormal basis

Warning ! this an isomorphism of vector spaces, not of algebras : the product
$\cdot$\ does not correspond to the product $\wedge$

\begin{theorem}
\textit{For an orthonormal basis} $\left(  e_{i}\right)  :$%

\begin{equation}
e_{i}\cdot e_{j}+e_{j}\cdot e_{i}=2\eta_{ij}\text{ where }\eta_{ij}=g\left(
e_{i},e_{j}\right)  =0,\pm1
\end{equation}

\end{theorem}

so

$i\neq j:e_{i}\cdot e_{j}=-e_{j}\cdot e_{i}$

$e_{i}\cdot e_{i}=\pm1$

$e_{p}\cdot e_{q}\cdot e_{i}-e_{i}\cdot e_{p}\cdot e_{q}=2\left(  \eta
_{iq}e_{p}-\eta_{ip}e_{q}\right)  $

\begin{theorem}
For any permutation of the \textit{ordered} set of indices

$\left\{  i_{1},...,i_{n}\right\}  :e_{\sigma\left(  i_{1}\right)  .}\cdot
e_{\sigma\left(  i_{2}\right)  }..\cdot e_{\sigma\left(  i_{r}\right)
}=\epsilon\left(  \sigma\right)  e_{i_{1}}\cdot e_{i_{2}}..\cdot e_{i_{r}}$
\end{theorem}

Warning ! it works for orthogonal vectors, not for any vector and the indices
must be different

\subsubsection{Involutions}

\begin{definition}
The \textbf{principal involution} in Cl(F,g) denoted $\alpha
:Cl(F,g)\rightarrow Cl(F,g)$ acts on homogeneous elements as : $\alpha\left(
v_{1.}\cdot v_{2}..\cdot v_{r}\right)  =\left(  -1\right)  ^{r}\left(
v_{1}\cdot v_{2}..\cdot v_{r}\right)  $
\end{definition}

$\alpha\left(  \sum_{k=0}^{2^{\dim F}}\sum_{\left\{  i_{1},...,i_{k}\right\}
}w_{\left\{  i_{1},...,i_{k}\right\}  }e_{i_{1}}\cdot...\cdot e_{i_{k}%
}\right)  $

$=\sum_{k=0}^{2^{\dim F}}\sum_{\left\{  i_{1},...,i_{k}\right\}  }\left(
-1\right)  ^{k}w_{\left\{  i_{1},...,i_{k}\right\}  }e_{i_{1}}\cdot...\cdot
e_{i_{k}}$

It has the properties:

$\alpha\circ\alpha=Id,$

$\forall w,w^{\prime}\in Cl\left(  F,g\right)  :\alpha\left(  w\cdot
w^{\prime}\right)  =\alpha\left(  w\right)  \cdot\alpha\left(  w^{\prime
}\right)  $

It follows that Cl(F,g) is the direct sum of the two eigen spaces with eigen
value $\pm1$ for $\alpha.$

\begin{theorem}
The set $Cl_{0}\left(  F,g\right)  $\ of elements of a Clifford algebra
Cl(F,g) which are invariant by the principal involution is a subalgebra and a
Clifford algebra.
\end{theorem}

$Cl_{0}\left(  F,g\right)  =\left\{  w\in Cl\left(  F,g\right)  :\alpha\left(
w\right)  =w\right\}  $

Its elements are the sum of homogeneous elements which are themselves product
of an even number of vectors.

As a vector space its basis is $1,e_{i_{1}}\cdot e_{i_{2}}\cdot..e_{i_{2k}%
}:i_{1}<i_{2}...<i_{2k}$

\begin{theorem}
The set $Cl_{1}\left(  F,g\right)  $ of elements w of a Clifford algebra
Cl(F,g) such that $\alpha\left(  w\right)  =-w$ is a vector subspace of Cl(F,g)
\end{theorem}

$Cl_{1}\left(  F,g\right)  =\left\{  w\in Cl\left(  F,g\right)  :\alpha\left(
w\right)  =-w\right\}  $

It\ is not a subalgebra. As a vector space its basis is

$e_{i_{1}}\cdot e_{i_{2}}\cdot..e_{i_{2k+1}}:i_{1}<i_{2}...<i_{2k+1}$

$Cl_{0}\cdot Cl_{0}\subset Cl_{0},Cl_{0}\cdot Cl_{1}\subset Cl_{1},Cl_{1}\cdot
Cl_{0}\subset Cl_{1},Cl_{1}\cdot Cl_{1}\subset Cl_{0}$

so Cl(F,g) is a $%
\mathbb{Z}
$/2 graded algebra.\newline

\begin{definition}
The \textbf{transposition }on Cl(F,g) is the involution which acts on
homogeneous elements by : $\left(  v_{1}\cdot v_{2}...\cdot v_{r}\right)
^{t}=\left(  v_{r}\cdot v_{r-1}...\cdot v_{1}\right)  .$
\end{definition}

$\left(  \sum_{k=0}^{2^{\dim F}}\sum_{\left\{  i_{1},...,i_{k}\right\}
}w_{\left\{  i_{1},...,i_{k}\right\}  }e_{i_{1}}\cdot...\cdot e_{i_{k}%
}\right)  ^{t}$

$=\sum_{k=0}^{2^{\dim F}}\sum_{\left\{  i_{1},...,i_{k}\right\}  }\left(
-1\right)  ^{\frac{k\left(  k-1\right)  }{2}}w_{\left\{  i_{1},...,i_{k}%
\right\}  }e_{i_{1}}\cdot...\cdot e_{i_{k}}$

\subsubsection{Scalar product on the Clifford algebra}

\begin{theorem}
A non degenerate bilinear symmetric form g on a vector space F can be extended
in a non degenerate bilinear symmetric form G on Cl(F,g).
\end{theorem}

Define G by :

$i_{1}<i_{2}..<i_{k},j_{1}<j_{2}..<j_{k}:G\left(  e_{i_{1}}\cdot e_{i_{2}%
}\cdot..e_{i_{k}},e_{j_{1}}\cdot e_{j_{2}}\cdot..e_{j_{l}}\right)  $

$=\delta_{kl}g\left(  e_{i_{1}},e_{j_{1}}\right)  \times...g\left(  e_{i_{k}%
},e_{j_{k}}\right)  =\delta_{kl}\eta_{i_{1}j_{1}}...\eta_{i_{k}j_{k}}$

for any orthonormal basis of F. This is an orthonormal basis of Cl(F,g) for G.

G does not depend on the choice of the basis. It is not degenerate.

$k,l\in K:G(k,l)=-kl$

$u,v\in F:G\left(  u,v\right)  =g(u,v)$

$w=\sum_{i<j}w_{ij}e_{i}\cdot e_{j},w^{\prime}=\sum_{i<j}w_{ij}^{\prime}%
e_{i}\cdot e_{j}:G\left(  w,w^{\prime}\right)  =\sum_{i<j}w_{ij}w_{ij}%
^{\prime}\eta_{ii}\eta_{jj}$

$u,v\in Cl\left(  F,g\right)  :G\left(  u,v\right)  =\left\langle u\cdot
v^{t}\right\rangle $ where $\left\langle u\cdot v^{t}\right\rangle $ is the
scalar component of $u\cdot v^{t}$

The transpose is the adjoint of the left and right Clifford product in the
meaning :

$a,u,v\in Cl\left(  F,g\right)  :G\left(  a\cdot u,v\right)  =G\left(
u,a^{t}\cdot v\right)  ;G\left(  u\cdot a,v\right)  =G\left(  u,v\cdot
a^{t}\right)  $

\subsubsection{Volume element}

\paragraph{Volume element\newline}

\begin{definition}
A volume element of the Clifford algebra Cl(F,g) is an element $\varpi$\ such
that $\varpi\cdot\varpi=1$
\end{definition}

Let F be n dimensional and $\left(  e_{i}\right)  _{i=1}^{n}$\ an orthonormal
basis of (F,g) with $K=%
\mathbb{R}
,%
\mathbb{C}
$.

The element : $e_{0}=e_{1}\cdot e_{2}...\cdot e_{n}\in Cl(F,g)$ does not
depend on the choice of the orthonormal basis.

It has the properties :

$e_{0}\cdot e_{0}=\left(  -1\right)  ^{\frac{n\left(  n-1\right)  }{2}+q}%
=\pm1$

$e_{0}\cdot e_{0}=+1$ if p-q=0,1 mod 4

$e_{0}\cdot e_{0}=-1$ if p-q=2,3 mod 4

where p,q is the signature of g if $K=%
\mathbb{R}
$.\ If $K=%
\mathbb{C}
$ then q=0 and p=n

Thus if $K=%
\mathbb{C}
$ there is always a volume element $\varpi$\ of Cl(F,g), which does not depend
of a basis$.$ It is defined up to sign by: $\varpi=e_{0}$ if $e_{0}\cdot
e_{0}=1,$ and $\varpi=ie_{0}$ if $e_{0}\cdot e_{0}=-1$

If $K=%
\mathbb{R}
$ and $e_{0}\cdot e_{0}=1$ there is always a volume element $\varpi$\ of
Cl(F,g), which does not depend of a basis, such that $\varpi\cdot\varpi=1. $
It is defined up to sign by: $\varpi=e_{0}$

\begin{theorem}
If the Clifford algebra Cl(F,g) has a volume element $\varpi$\ then $\varpi$
commute with any element of $Cl_{0}\left(  F,g\right)  $
\end{theorem}

\begin{proof}
Let $\varpi=ae_{1}\cdot e_{2}...\cdot e_{n}$ with $a=\pm1$ or $\pm i$

$v\in F:v=\sum_{j=1}^{n}v_{j}e_{j}$

$\varpi\cdot v=\sum_{j=1}^{n}v_{j}ae_{1}\cdot e_{2}...\cdot e_{n}\cdot
e_{j}=a\sum_{j=1}^{n}v_{j}\left(  -1\right)  ^{n-j}\eta_{jj}e_{1}\cdot
e_{2}..\widehat{e_{j}}.\cdot e_{n}\cdot e_{j}$

$v\cdot\varpi=\sum_{j=1}^{n}v_{j}ae_{j}\cdot e_{1}\cdot e_{2}...\cdot
e_{n}\cdot e_{j}=a\sum_{j=1}^{n}v_{j}\left(  -1\right)  ^{j-1}\eta_{jj}%
e_{1}\cdot e_{2}..\widehat{e_{j}}.\cdot e_{n}\cdot e_{j}$

$\varpi\cdot v=\left(  -1\right)  ^{n-j}\left(  -1\right)  ^{1-j}v\cdot
\varpi=\left(  -1\right)  ^{n+1}v\cdot\varpi$

$w=v_{1}\cdot...\cdot v_{k}$

$\varpi\cdot w=\varpi\cdot v_{1}\cdot v_{2}...\cdot v_{k}=\left(  -1\right)
^{n+1}v_{1}\cdot\varpi\cdot v_{2}...\cdot v_{k}=\left(  -1\right)  ^{\left(
n+1\right)  k}v_{1}\cdot v_{2}...\cdot v_{k}\cdot\varpi=\left(  -1\right)
^{\left(  n+1\right)  k}w\cdot\varpi$
\end{proof}

\paragraph{Decomposition of Cl(F,g)\newline}

\begin{theorem}
If the Clifford algebra Cl(F,g) has a volume element $\varpi$\ then Cl(F,g) is
the direct sum of two subalgebras :

$Cl\left(  F,g\right)  =Cl^{+}\left(  F,g\right)  \oplus Cl^{-}\left(
F,g\right)  $

$Cl^{+}\left(  F,g\right)  =\left\{  w\in Cl(F,g):\varpi\cdot w=w\right\}  ,$

$Cl^{-}\left(  F,g\right)  =\left\{  w\in Cl(F,g):\varpi\cdot w=-w\right\}  $
\end{theorem}

\begin{proof}
The map : $w\rightarrow\varpi\cdot w$ is a linear map on Cl(F,g) and
$\varpi\cdot\left(  \varpi\cdot w\right)  =w$ so it has $\pm1$ as eigen values.

$Cl\left(  F,g\right)  =Cl^{+}\left(  F,g\right)  \oplus Cl^{-}\left(
F,g\right)  $ because they are eigen spaces for different eigen values

if $w,w^{\prime}\in Cl^{-}\left(  4,C\right)  ,\varpi\cdot w\cdot w^{\prime
}=-w\cdot w^{\prime}\Leftrightarrow w\cdot w^{\prime}\in Cl^{-}\left(
F,g\right)  :$ $Cl_{0}^{-}\left(  F,g\right)  $ is a subalgebra

if $w,w^{\prime}\in Cl^{+}\left(  4,C\right)  ,\varpi\cdot w\cdot w^{\prime
}=w\cdot w^{\prime}\Leftrightarrow w\cdot w^{\prime}\in Cl^{+}\left(
F,g\right)  :$ $Cl_{0}^{+}\left(  F,g\right)  $ is a subalgebra
\end{proof}

So any element w\ of Cl(F,g) can be written : $w=w^{+}+w^{-}$ with $w^{+}\in
Cl^{+}\left(  F,g\right)  ,w^{-}\in Cl^{-}\left(  F,g\right)  $

Moreover :

$Cl_{0}^{+}\left(  F,g\right)  =Cl_{0}\left(  F,g\right)  \cap Cl^{+}\left(
F,g\right)  $ and $Cl_{0}^{-}\left(  F,g\right)  =Cl_{0}\left(  F,g\right)
\cap Cl^{-}\left(  F,g\right)  $ are subalgebras

and $Cl_{0}\left(  F,g\right)  =Cl_{0}^{+}\left(  F,g\right)  \oplus
Cl_{0}^{-}\left(  F,g\right)  $

$Cl_{1}^{+}\left(  F,g\right)  =Cl_{1}\left(  F,g\right)  \cap Cl^{+}\left(
F,g\right)  $ and $Cl_{1}^{-}\left(  F,g\right)  =Cl_{1}\left(  F,g\right)
\cap Cl^{-}\left(  F,g\right)  $ are vector subspaces

and $Cl_{1}\left(  F,g\right)  =Cl_{1}^{+}\left(  F,g\right)  \oplus
Cl_{1}^{-}\left(  F,g\right)  $ (but they are not subalgebras)

\paragraph{Projection operators\newline}

\begin{definition}
If the Clifford algebra Cl(F,g) has a volume element $\varpi$\ the projectors
on the subalgebras $Cl^{+}\left(  F,g\right)  ,Cl^{-}\left(  F,g\right)  $ are :

$p_{+}:Cl(F,g)\rightarrow Cl^{+}(F,g)::p_{+}=\frac{1}{2}\left(  1+\varpi
\right)  $ (also called the creation operator)

$p_{-}:Cl(F,g)\rightarrow Cl^{-}(F,g)::p_{-}=\frac{1}{2}\left(  1-\varpi
\right)  $ (also called the annihiliation operator)
\end{definition}

$\forall w\in Cl(F,g):p_{+}\cdot w=w^{+},p_{-}\cdot w=w^{-};$

$p_{+}^{2}=p_{+};p_{-}^{2}=p_{-}$

They are orthogonal in the meaning :

$p_{+}\cdot p_{-}=p_{-}\cdot p_{+}=0,p_{+}+p_{-}=1$

$p_{+}\cdot w^{-}=0;p_{-}\cdot w^{+}=0$

\subsubsection{Complex and real Clifford algebras}

\begin{theorem}
The complexified $Cl_{c}\left(  F,g\right)  $ of a Clifford algebra on a real
vector space F is the Clifford algebra $Cl\left(  F_{%
\mathbb{C}
},g_{%
\mathbb{C}
}\right)  $ on the complexified of F, endowed with the complexified of g
\end{theorem}

\begin{proof}
A real vector space can be complexified $F_{%
\mathbb{C}
}=F\oplus iF.$ The symmetric bilinear form g on F can be extended to a
symmetric bilinear form $g_{%
\mathbb{C}
}$ on the complexified $F_{%
\mathbb{C}
}$, \ it is non degenerate if g is non degenerate. An orthonormal basis
$\left(  e_{i}\right)  _{i\in I}$ of F for g is an orthonormal basis of $F_{%
\mathbb{C}
}$ for $g_{%
\mathbb{C}
}$ (see Complex vector spaces).

The basis of $Cl\left(  F_{%
\mathbb{C}
},g_{%
\mathbb{C}
}\right)  $ is comprised of

$\left\{  1,e_{i_{1}}\cdot...\cdot e_{i_{k}},1\leq i_{1}<i_{2}...<i_{k}%
\leq\dim F,k=1...2^{\dim F}\right\}  $ with complex components.

Conversely the complexified of $Cl\left(  F,g\right)  $ is the complexified of
the vector space structure.\ It has the same basis as above with complex components.
\end{proof}

A symmetric bilinear form on a complex vector space has no signature, the
orthonormal basis can always be chosen such that $g\left(  e_{j},e_{k}\right)
=\delta_{jk}.$

\begin{theorem}
A real structure $\sigma$ on a complex vector space (F,g) can be extended to a
real structure on Cl(F,g)
\end{theorem}

\begin{proof}
$\sigma:F\rightarrow F$ is antilinear and $\sigma^{2}=I.$

Define on $Cl\left(  F,g\right)  :$

$\sigma\left(  u_{1}\cdot...\cdot u_{r}\right)  =\sigma\left(  u_{1}\right)
\cdot..\cdot\sigma\left(  u_{r}\right)  $

$\sigma\left(  aw+bw^{\prime}\right)  =\overline{a}\sigma\left(  w\right)
+\overline{b}\sigma\left(  w^{\prime}\right)  $

$\Rightarrow$

$\sigma\left(  iu_{1}\cdot...\cdot u_{r}\right)  =-i\sigma\left(  u_{1}%
\cdot...\cdot u_{r}\right)  ;$

$\sigma^{2}\left(  u_{1}\cdot...\cdot u_{r}\right)  =\sigma\left(
\sigma\left(  u_{1}\right)  \cdot..\cdot\sigma\left(  u_{r}\right)  \right)
=u_{1}\cdot...\cdot u_{r}$

$\sigma^{2}\left(  aw+bw^{\prime}\right)  =\sigma\left(  \overline{a}%
\sigma\left(  w\right)  +\overline{b}\sigma\left(  w^{\prime}\right)  \right)
=aw+bw^{\prime}$
\end{proof}

\begin{definition}
On a complex Clifford algebra Cl(F,g) endowed with a real structure $\sigma$
the \textbf{conjugate} of an element w is $\overline{w}=\sigma\left(
w\right)  $
\end{definition}

\bigskip

\subsection{Pin and Spin groups}

\label{Pin and spin groups}

\subsubsection{Adjoint map}

\begin{theorem}
In a Clifford algebra any element which is the product of non null norm
vectors has an inverse for $\cdot$:
\end{theorem}

$\left(  u_{1}\cdot...\cdot u_{k}\right)  ^{-1}=\left(  u_{1}\cdot...\cdot
u_{k}\right)  ^{t}/%
{\displaystyle\prod\limits_{r=1}^{k}}
g\left(  u_{r},u_{r}\right)  $

\begin{proof}
$\left(  u_{1}\cdot...\cdot u_{k}\right)  ^{t}\cdot\left(  u_{1}\cdot...\cdot
u_{k}\right)  =u_{k}\cdot...u_{1}\cdot u_{1}...\cdot u_{k}=%
{\displaystyle\prod\limits_{r=1}^{k}}
g\left(  u_{r},u_{r}\right)  $
\end{proof}

\begin{definition}
The \textbf{adjoint map}, denoted \textbf{Ad} , is the map :

$\mathbf{Ad}:GCl(F,g)\times Cl(F,g)\rightarrow Cl(F,g)::\mathbf{Ad}%
_{w}u=\alpha\left(  w\right)  \cdot u\cdot w^{-1}$

Where GCl(F,g) is the group of invertible elements of the Clifford algebra Cl(F,g)
\end{definition}

\begin{theorem}
The adjoint map \textbf{Ad} is a $\left(  GCl(F,g),\cdot\right)  $ group automorphism
\end{theorem}

If $w,w^{\prime}\in GCl(F,g):\mathbf{Ad}_{w}\circ\mathbf{Ad}_{w^{\prime}%
}=\mathbf{Ad}_{w\cdot w^{\prime}}$

\begin{definition}
The \textbf{Clifford group} is the set :

$P=\left\{  w\in GCl\left(  F,g\right)  :\mathbf{Ad}_{w}\left(  F\right)
\subset F\right\}  $
\end{definition}

\begin{theorem}
The map : $\mathbf{Ad:}P\mathbf{\rightarrow}O\left(  F,g\right)  $ where
O(F,g) is the orthogonal group, is a surjective morphism of groups
\end{theorem}

\begin{proof}
If $w\in P$ then $\forall u,v\in F:g\left(  \mathbf{Ad}_{w}u,\mathbf{Ad}%
_{w}v\right)  =g\left(  u,v\right)  $ so $\mathbf{Ad}_{w}\in O\left(
F,g\right)  $

Over (F,g) a reflexion of vector u with $g(u,u)\neq0$ is the orthogonal map :

$R\left(  u\right)  :F\rightarrow F::$ $R\left(  u\right)  x=x-2\frac
{g(x,u)}{g(u,u)}u$ and $x,u\in F:\mathbf{Ad}_{u}x=R\left(  u\right)  x$

According to a classic theorem by Cartan-Dieudonn\'{e}, any orthogonal linear
map over a n-dimensional vector space can be written as the product of at most
n reflexions.\ Which reads :

$\forall h\in O\left(  F,g\right)  ,\exists u_{1},...u_{k}\in F,k\leq\dim F:$

$h\mathbf{=}R(u_{1})\circ...R(u_{k})=\mathbf{Ad}_{u_{1}}\circ...\circ
\mathbf{Ad}_{u_{r}}=\mathbf{Ad}_{u_{1}\cdot...\cdot u_{k}}=\mathbf{Ad}%
_{w},w\in P$

$\mathbf{Ad}_{w\cdot w^{\prime}}=\mathbf{Ad}_{u_{1}\cdot...\cdot u_{k}}%
\circ\mathbf{Ad}_{u_{1}^{\prime}\cdot...\cdot u_{l}^{\prime}}=\mathbf{Ad}%
_{u_{1}}\circ...\circ\mathbf{Ad}_{u_{r}}\circ\mathbf{Ad}_{u_{1}^{\prime}}%
\circ...\circ\mathbf{Ad}_{u_{l}^{\prime}}=h\circ h^{\prime}$

Thus the map : $\mathbf{Ad:}P\mathbf{\rightarrow}O\left(  F,g\right)  $ is a
surjective homomorphism and $\left(  \mathbf{Ad}\right)  ^{-1}\left(  O\left(
F,g\right)  \right)  $ is the subset of P comprised of homogeneous elements of
Cl(F,g), products of vectors $u_{k}$ with $g(u_{k},u_{k})\neq0$
\end{proof}

\subsubsection{Pin group}

\begin{definition}
The \textbf{Pin group} of Cl(F,g) is the set :

$Pin\left(  F,g\right)  =\left\{  w\in Cl(F,g),w=u_{1}\cdot...\cdot
u_{r},u_{k}\in F,g(u_{k},u_{k})=\pm1,r\geq0\right\}  $

with the product $\cdot$ as operation
\end{definition}

If $w\in Pin(F,g)$ then : $w^{-1}=\left(  u_{r}/g\left(  u_{r},u_{r}\right)
\right)  \cdot...\cdot\left(  u_{1}/g\left(  u_{1},u_{1}\right)  \right)  \in
Pin(F,g)$

$\forall v\in F:\mathbf{Ad}_{w}v=\alpha\left(  w\right)  \cdot v\cdot
w^{-1}=\left(  -1\right)  ^{r}u_{1}\cdot...u_{r}\cdot v\cdot\left(
u_{r}/g\left(  u_{r},u_{r}\right)  \right)  \cdot...\cdot\left(
u_{1}/g\left(  u_{1},u_{1}\right)  \right)  $

$u_{r}\cdot v\cdot\left(  u_{r}/g\left(  u_{r},u_{r}\right)  \right)
=\frac{1}{g\left(  u_{r},u_{r}\right)  }\left(  u_{r}\cdot v\cdot
u_{r}\right)  =\frac{1}{g\left(  u_{r},u_{r}\right)  }\left(  -g\left(
u_{r},u_{r}\right)  v+2g\left(  u_{r},v\right)  u_{r}\right)  \in F$

The scalars $\pm1$ belong to the Pin group. The identity is + 1

$\forall u,v\in F,w\in Pin\left(  F,g\right)  :g\left(  \mathbf{Ad}%
_{w}u,\mathbf{Ad}_{w}v\right)  =g\left(  u,v\right)  $

\begin{theorem}
\textbf{Ad} is a surjective group morphism :

$\mathbf{Ad:}\left(  Pin(F,g),\cdot\right)  \rightarrow\left(  O\left(
F,g\right)  ,\circ\right)  $

and O(F,g) is isomorphic to Pin(F,g)/$\left\{  +1,-1\right\}  $
\end{theorem}

\begin{proof}
It is the restriction of the map $\mathbf{Ad:}P\mathbf{\rightarrow}O\left(
F,g\right)  $ to Pin(F,g)

For any $h\in O\left(  F,g\right)  $ there are two elements $\left(
w,-w\right)  $ of Pin(F,g) such that : $\mathbf{Ad}_{w}=h$
\end{proof}

\begin{theorem}
There is an action of O(F,g) on Cl(F,g) :

$\lambda:O(F,g)\times Cl(F,g)\rightarrow Cl(F,g)::\lambda\left(  h,w\right)
=Ad_{s}w$

where $s\in Pin(F,g):Ad_{s}=h$
\end{theorem}

\begin{theorem}
(Cl(F,g),\textbf{Ad}) is a representation of Pin(F,g)$:$
\end{theorem}

\begin{proof}
For any s in Pin(F,g) the map \textbf{Ad}$_{s}$\ is linear on Cl(F,g) :

$\mathbf{Ad}_{s}\left(  kw+k^{\prime}w^{\prime}\right)  =\alpha\left(
s\right)  \cdot\left(  kw+k^{\prime}w^{\prime}\right)  \cdot s^{-1}%
=k\alpha\left(  s\right)  \cdot w\cdot s^{-1}+k^{\prime}\alpha\left(
s\right)  \cdot w^{\prime}\cdot s^{-1}$

and \textbf{Ad}$_{s}$\textbf{Ad}$_{s^{\prime}}=$\textbf{Ad}$_{ss^{\prime}},
$\textbf{Ad}$_{1}=Id_{F}$
\end{proof}

\begin{theorem}
(F,\textbf{Ad}) is a representation of Pin(F,g)
\end{theorem}

This is the restriction of the representation on Cl(F,g)

\subsubsection{Spin group}

\begin{definition}
The \textbf{Spin group} of Cl(F,g) is the set :

$Spin\left(  F,g\right)  =\left\{  w\in Cl\left(  F,g\right)  :w=w_{1}%
\cdot...\cdot w_{2r},w_{k}\in F,g(w_{k},w_{k})=\pm1,r\geq0\right\}  $

with the product $\cdot$ as operation
\end{definition}

So $Spin\left(  F,g\right)  =Pin\left(  F,g\right)  \cap Cl_{0}\left(
F,g\right)  $

$\forall u,v\in F,w\in Spin\left(  F,g\right)  :g\left(  \mathbf{Ad}%
_{w}u,\mathbf{Ad}_{w}v\right)  =g\left(  u,v\right)  $

The scalars $\pm1$ belong to the Spin group. The identity is + 1

\begin{theorem}
\textbf{Ad} is a surjective group morphism :

$\mathbf{Ad:}\left(  Spin(F,g),\cdot\right)  \rightarrow\left(  SO\left(
F,g\right)  ,\circ\right)  $

and SO(F,g) is isomorphic to Spin(F,g)/$\left\{  +1,-1\right\}  $
\end{theorem}

\begin{proof}
It is the restriction of the map $\mathbf{Ad:}P\mathbf{\rightarrow S}O\left(
F,g\right)  $ to Spin(F,g)

For any $h\in SO\left(  F,g\right)  $ there are two elements $\left(
w,-w\right)  $ of Spin(F,g) such that : $\mathbf{Ad}_{w}=h$
\end{proof}

\begin{theorem}
There is an action of SO(F,g) on Cl(F,g) :

$\lambda:SO(F,g)\times Cl(F,g)\rightarrow Cl(F,g)::\lambda\left(  h,u\right)
=w\cdot u\cdot w^{-1}$ where $w\in Spin(F,g):\mathbf{Ad}_{w}=h$
\end{theorem}

\begin{theorem}
(Cl(F,g),\textbf{Ad}) is a representation of Spin(F,g)$:\mathbf{Ad}%
_{s}w=s\cdot w\cdot s^{-1}=s\cdot w\cdot s^{t}$
\end{theorem}

\begin{proof}
This is the restriction of the representation of Pin(F,g)
\end{proof}

\begin{theorem}
(F,\textbf{Ad}) is a representation of Spin(F,g)
\end{theorem}

\begin{proof}
This is the restriction of the representation on Cl(F,g)
\end{proof}

\begin{definition}
The group $Spin_{c}\left(  F,g\right)  $ is the subgroup of $Cl_{c}\left(
F,g\right)  $ comprised of elements : $S=zs$ where z is a module 1 complex
scalar, and $s\in Spin(F,g).$ It is a subgroup of the group $Spin\left(
F_{c},g_{c}\right)  .$
\end{definition}

\subsubsection{Characterization of Spin(F,g) and Pin(F,g)}

We need here results which can be found in the Part Lie groups. We assume that
the vector space F is finite n dimensional.

\paragraph{Lie Groups\newline}

\begin{theorem}
The groups Pin(F,g) and Spin(F,g) are Lie groups
\end{theorem}

\begin{proof}
O(F,g) is a Lie group and $Pin(F,g)=O(F,g)\times\left\{  +1,-1\right\}  $
\end{proof}

Any element of Pin(F,g) reads in an orthonormal basis of F:

$s=\sum_{k=0}^{n}\sum_{\left\{  i_{1},..i_{k}\right\}  }S_{i_{1}...i_{2k}%
}e_{i_{1}}\cdot e_{i_{2}}\cdot..e_{i_{k}}=\sum_{k=0}^{n}\sum_{I_{k}}S_{I_{k}%
}E_{I_{k}}$ with $S_{I_{k}}\in K,$

where the components $S_{I_{k}}$ are not independant because the generator
vectors must have norm $\pm1$.

Any element of Spin(F,g) reads :

$s=\sum_{k=0}^{N}\sum_{\left\{  i_{1},..i_{2k}\right\}  }S_{i_{1}...i_{2k}%
}e_{i_{1}}\cdot e_{i_{2}}\cdot..e_{i_{2k}}=\sum_{k=0}^{N}\sum_{I_{k}}S_{I_{k}%
}E_{I_{k}}$ with $S_{I_{k}}\in K,N\leq n/2$

with the same remark.

So Pin(F,g) and Spin(F,g) are not vector spaces, but manifolds embedded in the
vector space Cl(F,g) : they are Lie groups.

Pin(F,g) and Spin(F,g) are respectively a double cover, \textit{as manifold},
of O(F,g) and SO(F,g). However the latter two groups may be not connected and
in these cases Pin(F,g) and Spin(F,g) are not a double cover as Lie group.

\begin{theorem}
The Lie algebra $T_{1}Pin(F,g)$ is isomorphic to the Lie algebra o(F,g) of O(F,g)

The Lie algebra $T_{1}Spin(F,g)$ is isomorphic to the Lie algebra so(F,g) of SO(F,g)
\end{theorem}

\begin{proof}
O(F,g) is isomorphic to $Pin(F,g)/\left\{  +1,-1\right\}  .$ The subgroup
$\left\{  +1,-1\right\}  $ is a normal, abelian subgroup of Pin(F,g). So the
derivative of the map $h:Pin(F,g)\rightarrow O(F,g)$ is a morphism of Lie
algebra with kernel the Lie algebra of $\left\{  +1,-1\right\}  $ which is 0
because the group is abelian.\ So h'(1) is an isomorphism (see Lie groups).
Similarly for\ $T_{1}Spin(F,g)$
\end{proof}

\paragraph{Component expressions of the Lie algebras\newline}

\begin{theorem}
The Lie algebra of Pin(F,g) is a subset of Cl(F,g).
\end{theorem}

\begin{proof}
With the formula above, for any map $s:%
\mathbb{R}
\rightarrow Pin(F,g)$ : $s\left(  t\right)  =\sum_{k=0}^{n}\sum_{\left\{
i_{1},..i_{k}\right\}  }S_{i_{1}...i_{2k}}\left(  t\right)  e_{i_{1}}\cdot
e_{i_{2}}\cdot..e_{i_{k}}$ and its derivative reads \ $\frac{d}{dt}s\left(
t\right)  |_{t=0}=\sum_{k=0}^{n}\sum_{\left\{  i_{1},..i_{k}\right\}  }%
\frac{d}{dt}S_{i_{1}...i_{2k}}\left(  t\right)  |_{t=0}e_{i_{1}}\cdot
e_{i_{2}}\cdot..e_{i_{k}}$ that is an element of Cl(F,g)
\end{proof}

Because $h^{\prime}(1):T_{1}Pin(F,g)\rightarrow o(F,g)$ is an isomorphism, for
any vector $\overrightarrow{\kappa}\in o\left(  F,g\right)  $ there is an
element $\upsilon\left(  \overrightarrow{\kappa}\right)  =h^{\prime}%
(1)^{-1}\overrightarrow{\kappa}$ of Cl(F,g)$.$ Our objective here is to find
the expression of $\upsilon\left(  \overrightarrow{\kappa}\right)  $\ in the
basis of Cl(F,g).

\begin{lemma}
$\forall u\in F:\upsilon\left(  \overrightarrow{\kappa}\right)  \cdot
u-u\cdot\upsilon\left(  \overrightarrow{\kappa}\right)  =\left[  J\left(
\overrightarrow{\kappa}\right)  \right]  u$ where $J\left(  \overrightarrow
{\kappa}\right)  $ is the matrix of $\overrightarrow{\kappa}$ in the standard
representation $\left(  F,\jmath\right)  $ of o(F,g)
\end{lemma}

\begin{proof}
i) An endomorphism f of O(F,g) is represented in an orthonormal basis by a
n$\times$n matrix $\left[  f\right]  $\ such that : $\left[  f\right]
^{t}\left[  \eta\right]  \left[  f\right]  =\left[  \eta\right]  .$ Thus if f
= h(s) is such that : $\mathbf{Ad}_{s}u=s\cdot u\cdot s^{-1}=h(s)u$ in
components it reads : $s\cdot u\cdot s^{-1}=\left[  h(s)\right]  u$

ii) By derivation with respect to s at s=1 $\left(  \mathbf{Ad}\right)
^{\prime}|_{s=1}:T_{1}Pin(F,g)\rightarrow o(F,g)$ reads :

$\left(  \mathbf{Ad}\right)  ^{\prime}|_{s=1}\sigma\left(  \overrightarrow
{\kappa}\right)  =\left[  h^{\prime}(1)\right]  \overrightarrow{\kappa
}=\left[  J\left(  \overrightarrow{\kappa}\right)  \right]  u$

where $\left[  J\left(  \overrightarrow{\kappa}\right)  \right]  $ is a nxn
matrix such that : $\left[  \eta\right]  \left[  J\left(  \overrightarrow
{\kappa}\right)  \right]  ^{t}+\left[  J\left(  \overrightarrow{\kappa
}\right)  \right]  \left[  \eta\right]  =0$

ii) On the other hand the derivation of the product $\mathbf{Ad}_{s}u=s\cdot
u\cdot s^{-1}$\ with respect to s at s=t gives :

$\left(  \mathbf{Ad}_{s}u\right)  ^{\prime}|_{s=t}\xi_{t}=\xi_{t}\cdot u\cdot
t^{-1}-t\cdot u\cdot t^{-1}\cdot\xi_{t}\cdot t^{-1}$

For t=1 and $\xi_{t}=\upsilon\left(  \overrightarrow{\kappa}\right)  :\left(
\mathbf{Ad}_{s}u\right)  ^{\prime}|_{s=1}\upsilon\left(  \overrightarrow
{\kappa}\right)  =\upsilon\left(  \overrightarrow{\kappa}\right)  \cdot
u-u\cdot\upsilon\left(  \overrightarrow{\kappa}\right)  $
\end{proof}

\begin{theorem}
The isomophism of Lie algebras : $\upsilon:so(F,g)\rightarrow T_{1}Spin(F,g)$
reads :%

\begin{equation}
\upsilon\left(  \overrightarrow{\kappa}\right)  =\sum_{ij}\left[
\upsilon\right]  _{j}^{i}e_{i}\cdot e_{j}\text{ with }\left[  \upsilon\right]
=\frac{1}{4}\left[  J\left(  \overrightarrow{\kappa}\right)  \right]  \left[
\eta\right]
\end{equation}
\ 

where $\left[  J\left(  \overrightarrow{\kappa}\right)  \right]  $ is the
matrix of $\overrightarrow{\kappa}$ in the standard representation of so(F,g)
\end{theorem}

\begin{proof}
We have also $\upsilon\left(  \overrightarrow{\kappa}\right)  =\sum_{k=0}%
^{N}\sum_{I_{k}}s_{I_{k}}E_{I_{k}}$ where $s_{I_{k}}$ are fixed scalars
(depending on the bases and $\overrightarrow{\kappa}$).

Thus : $\sum_{k=0}^{N}\sum_{I_{k}}s_{I_{k}}\left(  E_{I_{k}}\cdot u-u\cdot
E_{I_{k}}\right)  =\left[  J\right]  u$ and taking $u=e_{i}:$

$\forall i=1..n:\sum_{k=0}^{N}\sum_{I_{k}}s_{I_{k}}\left(  E_{I_{k}}\cdot
e_{i}-e_{i}\cdot E_{I_{k}}\right)  =\left[  J\right]  \left(  e_{i}\right)
=\sum_{j=1}^{n}\left[  J\right]  _{i}^{j}e_{j}$

$I_{k}=\left\{  i_{1},...,i_{2k}\right\}  :$

$\left(  E_{I_{k}}\cdot e_{i}-e_{i}\cdot E_{I_{k}}\right)  =e_{i_{1}}\cdot
e_{i_{2}}\cdot..e_{i_{2k}}\cdot e_{i}-e_{i}\cdot e_{i_{1}}\cdot e_{i_{2}}%
\cdot..e_{i_{2k}}$

If $i\notin I_{k}:E_{I_{k}}\cdot e_{i}-e_{i}\cdot E_{I_{k}}=2\left(
-1\right)  ^{l+1}e_{i_{1}}\cdot e_{i_{2}}\cdot..e_{i_{2k}}\cdot e_{i}$

If $i\in I_{k},i=i_{i}:E_{I_{k}}\cdot e_{i}=\left(  -1\right)  ^{2k-l}%
\eta_{ii}e_{i_{1}}\cdot e_{i_{2}}..\cdot\widehat{e_{i_{l}}}..e_{i_{2k}}%
,e_{i}\cdot E_{I_{k}}=\left(  -1\right)  ^{l-1}\eta_{ii}e_{i_{1}}\cdot
e_{i_{2}}..\cdot\widehat{e_{i_{l}}}..e_{i_{2k}}$

so : $E_{I_{k}}\cdot e_{i}-e_{i}\cdot E_{I_{k}}=2\left(  -1\right)  ^{l}%
\eta_{ii}e_{i_{1}}\cdot e_{i_{2}}..\cdot\widehat{e_{i_{l}}}..e_{i_{2k}}$

So : $s_{I_{k}}=0$ for $k\neq1$ and for k=1 : $I_{1pq}=\left\{  e_{p}%
,e_{q}\right\}  ,p<q:$

$\sum_{p<q}s_{pq}\left(  e_{p}\cdot e_{q}\cdot e_{i}-e_{i}\cdot e_{p}\cdot
e_{q}\right)  =\sum_{i<j}\left(  -2s_{ij}\eta_{ii}e_{j}\right)  =\sum
_{j=1}^{n}\left[  J\right]  _{i}^{j}e_{j}$

i%
$<$%
j : $s_{ij}=-\frac{1}{2}\eta_{ii}\left[  J\right]  _{i}^{j}$

$\upsilon\left(  \overrightarrow{\kappa}\right)  =-\frac{1}{2}\sum_{i<j}%
\eta_{ii}\left[  J\left(  \overrightarrow{\kappa}\right)  \right]  _{i}%
^{j}e_{i}\cdot e_{j}$

$\left[  \eta\right]  \left[  J\right]  ^{t}+\left[  J\right]  \left[
\eta\right]  =0\Rightarrow\left[  J\right]  _{i}^{j}=-\eta_{ii}\eta
_{jj}\left[  J\right]  _{j}^{i}$

so the formula is consistent if we replace i by j :

$\upsilon\left(  \overrightarrow{\kappa}\right)  =-\frac{1}{2}\sum_{j<i}%
\eta_{jj}\left[  J\right]  _{j}^{i}e_{j}\cdot e_{i}$

$\upsilon\left(  \overrightarrow{\kappa}\right)  =-\frac{1}{4}\left(
\sum_{i<j}\eta_{ii}\left[  J\right]  _{i}^{j}e_{i}\cdot e_{j}+\sum_{j<i}%
\eta_{jj}\left[  J\right]  _{j}^{i}e_{j}\cdot e_{i}\right)  $

$=-\frac{1}{4}\left(  \sum_{i<j}\eta_{ii}\left[  J\right]  _{i}^{j}e_{i}\cdot
e_{j}-\sum_{j<i}\eta_{ii}\left[  J\right]  _{i}^{j}e_{j}\cdot e_{i}\right)  $

$=-\frac{1}{4}\left(  \sum_{i<j}\eta_{ii}\left[  J\right]  _{i}^{j}e_{i}\cdot
e_{j}+\sum_{j<i}\eta_{ii}\left[  J\right]  _{i}^{j}e_{i}\cdot e_{j}\right)  $

$=-\frac{1}{4}\left(  \sum_{i,j}\eta_{ii}\left[  J\right]  _{i}^{j}e_{i}\cdot
e_{j}-\sum_{i}\eta_{ii}\left[  J\right]  _{i}^{i}e_{i}\cdot e_{i}\right)  $

$=-\frac{1}{4}\left(  \sum_{i,j}\eta_{ii}\left[  J\right]  _{i}^{j}e_{i}\cdot
e_{j}-Tr\left(  \left[  J\right]  \right)  \right)  $

$=-\frac{1}{4}\left(  \sum_{i,j}\eta_{ii}\left[  J\right]  _{i}^{j}e_{i}\cdot
e_{j}\right)  $ because J is traceless.

$\upsilon\left(  \overrightarrow{\kappa}\right)  =-\frac{1}{4}\left(
\sum_{i,j}\eta_{ii}\left[  J\left(  \overrightarrow{\kappa}\right)  \right]
_{i}^{j}e_{i}\cdot e_{j}\right)  $

If we represent the components of $\upsilon\left(  \overrightarrow{\kappa
}\right)  $ in a matrix $\left[  \upsilon\right]  $\ nxn :

$\upsilon\left(  \overrightarrow{\kappa}\right)  =\sum_{ij}$ $\left[
\upsilon\right]  _{j}^{i}e_{i}\cdot e_{j}=-\frac{1}{4}\left(  \sum_{i,j}%
\eta_{ii}\left[  J\right]  _{i}^{j}e_{i}\cdot e_{j}\right)  $

$\left[  \upsilon\right]  _{j}^{i}=-\frac{1}{4}\left(  \left[  J\right]
\left[  \eta\right]  \right)  _{i}^{j}\Leftrightarrow\left[  \upsilon\right]
=-\frac{1}{4}\left(  \left[  J\right]  \left[  \eta\right]  \right)
^{t}=-\frac{1}{4}\left[  \eta\right]  \left[  J\right]  ^{t}=\frac{1}%
{4}\left[  J\right]  \left[  \eta\right]  $
\end{proof}

$\upsilon$ is an isomorphism of Lie algebras, so :

$\upsilon\left(  \left[  \overrightarrow{\kappa},\overrightarrow
{\kappa^{\prime}}\right]  \right)  =\left[  \upsilon\left(  \overrightarrow
{\kappa}\right)  ,\upsilon\left(  \overrightarrow{\kappa^{\prime}}\right)
\right]  =\upsilon\left(  \overrightarrow{\kappa}\right)  \cdot\upsilon\left(
\overrightarrow{\kappa^{\prime}}\right)  -\upsilon\left(  \overrightarrow
{\kappa^{\prime}}\right)  \cdot\upsilon\left(  \overrightarrow{\kappa}\right)
$

\bigskip

For Cl$\left(
\mathbb{R}
,3,1\right)  $ we have more precise results, involving the Spin group, which
lead to the definition of relativist motion (see Dutailly 2014).

\bigskip

\paragraph{Derivatives of the translation and adjoint map\newline}

The translations on Pin(F,g) are : $s,t\in Pin(F,g):L_{s}t=s\cdot
t,R_{s}t=t\cdot s$

The derivatives with respect to t are :

$L_{s}^{\prime}t\left(  \xi_{t}\right)  =s\cdot\xi_{t},R_{s}^{\prime}t\left(
\xi_{t}\right)  =\xi_{t}\cdot s$ with $\xi_{t}\in T_{t}Pin(F,g)$

With : $\xi_{t}=L_{t}^{\prime}\left(  1\right)  \upsilon\left(
\overrightarrow{\kappa}\right)  =R_{t}^{\prime}\left(  1\right)
\upsilon\left(  \overrightarrow{\kappa}\right)  =t\cdot\upsilon\left(
\overrightarrow{\kappa}\right)  =\upsilon\left(  \overrightarrow{\kappa
}\right)  \cdot t$

\begin{theorem}
The adjoint map Ad : Pin(F,g) $\rightarrow GL\left(  T_{1}Pin(F,g),T_{1}%
Pin(F,g)\right)  $ is the map \textbf{Ad}
\end{theorem}

\begin{proof}
As a Lie group the adjoint map of Pin(F,g) is the derivative of $s\cdot x\cdot
s^{-1}$ with respect to x at x=1:

$Ad:T_{1}Pin(F,g)\rightarrow%
\mathcal{L}%
\left(  T_{1}Pin(F,g);T_{1}Pin(F,g)\right)  ::Ad_{s}=(s\cdot x\cdot
s^{-1})^{\prime}|_{x=1}=L_{s}^{\prime}(s^{-1})\circ R_{s^{-1}}^{\prime
}(1)=R_{s^{-1}}^{\prime}(s)\circ L_{s}^{\prime}(1)$

$Ad_{s}\upsilon\left(  \overrightarrow{\kappa}\right)  =s\cdot\upsilon\left(
\overrightarrow{\kappa}\right)  \cdot s^{-1}=\mathbf{Ad}_{s}\upsilon\left(
\overrightarrow{\kappa}\right)  $
\end{proof}

Using : $\left(  \mathbf{Ad}_{s}u\right)  ^{\prime}|_{s=t}\xi_{t}=\xi_{t}\cdot
u\cdot t^{-1}-t\cdot u\cdot t^{-1}\cdot\xi_{t}\cdot t^{-1}$

and $\xi_{t}=L_{t}^{\prime}\left(  1\right)  \upsilon\left(  \overrightarrow
{\kappa}\right)  =t\cdot\upsilon\left(  \overrightarrow{\kappa}\right)  :$

$\left(  \mathbf{Ad}_{s}u\right)  ^{\prime}|_{s=t}t\cdot\upsilon\left(
\overrightarrow{\kappa}\right)  =\mathbf{Ad}_{t}\left(  \upsilon\left(
\overrightarrow{\kappa}\right)  \cdot u-u\cdot\upsilon\left(  \overrightarrow
{\kappa}\right)  \right)  $

\bigskip

\subsection{Classification of Clifford algebras}

\label{Classification of Clifford algebras}

\subsubsection{Morphisms of Clifford algebras}

\begin{definition}
A \textbf{Clifford algebra morphism} between the Clifford algebras
$Cl(F_{1},g_{1}),Cl\left(  F_{2},g_{2}\right)  $ on the same field K is an
algebra morphism

$F:Cl\left(  F_{1},g_{1}\right)  \rightarrow Cl\left(  F_{2},g_{2}\right)  $
\end{definition}

Which means that :

$\forall w,w^{\prime}\in F_{1},\forall k,k^{\prime}\in K:$

$F\left(  kw+k^{\prime}w^{\prime}\right)  =kF\left(  w\right)  +k^{\prime
}F(w^{\prime}),F\left(  1\right)  =1,F\left(  w\cdot w^{\prime}\right)
=F\left(  w\right)  \cdot F\left(  w^{\prime}\right)  $

It entails that :

$F\left(  u\cdot v+v\cdot u\right)  =F\left(  u\right)  \cdot F\left(
v\right)  +F\left(  v\right)  \cdot F\left(  u\right)  =2g_{2}\left(  F\left(
u\right)  ,F\left(  v\right)  \right)  $

$=F\left(  2g_{1}\left(  u,v\right)  \right)  =2g_{1}\left(  u,v\right)  $

so F must preserve the scalar product : $g_{2}\left(  F\left(  u\right)
,F\left(  v\right)  \right)  =g_{1}\left(  u,v\right)  $

\begin{theorem}
For any vector spaces over the same field, endowed with bilinear symmetric
forms $\left(  F_{1},g_{1}\right)  ,\left(  F_{2},g_{2}\right)  $\ every
linear maps \ $f\in L(F_{1};F_{2})$ which preserves the scalar product\ can be
extended to a morphism $F$ over the Clifford algebras such that the diagram
commutes :
\end{theorem}

\bigskip%

\begin{tabular}
[c]{lll}%
$\left(  F_{1},g_{1}\right)  $ & $\overset{Cl_{1}}{\rightarrow}$ & $Cl\left(
F_{1},g_{1}\right)  $\\
$\downarrow$ &  & $\downarrow$\\
$\downarrow f$ &  & $\downarrow F$\\
$\downarrow$ &  & $\downarrow$\\
$\left(  F_{2},g_{2}\right)  $ & $\overset{Cl_{2}}{\rightarrow}$ & $Cl\left(
F_{2},g_{2}\right)  $%
\end{tabular}

\bigskip

\begin{proof}
$F:Cl(F_{1},g_{1})\rightarrow Cl\left(  F_{2},g_{2}\right)  $ is defined as
follows :

$\forall k,k^{\prime}\in K,\forall u,v\in F_{1}:$

$F\left(  k\right)  =k,F\left(  u\right)  =f\left(  u\right)  ,F\left(
ku+k^{\prime}v\right)  =kf\left(  u\right)  +k^{\prime}f\left(  v\right)
,F\left(  u\cdot v\right)  =f\left(  u\right)  \cdot f\left(  v\right)  $

and as a consequence :

$F\left(  u\cdot v+v\cdot u\right)  =f\left(  u\right)  \cdot f\left(
v\right)  +f\left(  v\right)  \cdot f\left(  u\right)  =2g_{2}\left(  f\left(
u\right)  ,f\left(  v\right)  \right)  =2g_{1}\left(  u,v\right)  =F\left(
2g_{1}\left(  u,v\right)  \right)  $
\end{proof}

\begin{theorem}
Clifford algebras on a field K and their morphisms constitute a category
$\mathfrak{Cl}_{K}.$
\end{theorem}

The product of Clifford algebras morphisms is a Clifford algebra morphism.

Vector spaces (V,g) on the same field K endowed with a symmetric bilinear form
g, and linear maps f \textit{which preserve this form}, constitute a category,
denoted $\mathfrak{V}_{B}$

$f\in\hom_{\mathfrak{V}_{B}}\left(  \left(  F_{1},g_{1}\right)  ,\left(
F_{2},g_{2}\right)  \right)  $

$\Leftrightarrow f\in L\left(  V_{1};V_{2}\right)  ,\forall u,v\in F_{1}%
:g_{2}\left(  f\left(  u\right)  ,f\left(  v\right)  \right)  =g_{1}\left(
u,v\right)  $

\begin{theorem}
$\mathfrak{TCl:V}_{B}\mapsto\mathfrak{Cl}_{K}$ is a functor from the category
of vector spaces over K endowed with a symmetric bilinear form, to the
category of Clifford algebras over K.
\end{theorem}

$\mathfrak{TCl:V}_{B}\mapsto\mathfrak{Cl}_{K}$ associates :

to each object (F,g)\ of $\mathfrak{V}_{B}$ its Clifford algebra Cl(F,g) :

$\mathfrak{TCl:}\left(  F,g\right)  \mapsto Cl\left(  F,g\right)  $

to each morphism of vector spaces a morphism of Clifford algebras :

$\mathfrak{TCl:}f\in\hom_{\mathfrak{V}_{B}}\left(  \left(  F_{1},g_{1}\right)
,\left(  F_{2},g_{2}\right)  \right)  \mapsto F\in\hom_{\mathfrak{Cl}_{K}%
}\left(  \left(  F_{1},g_{1}\right)  ,\left(  F_{2},g_{2}\right)  \right)  $

\begin{definition}
An isomorphism of Clifford algebras is a morphism which is also a bijective
map. Two Clifford algebras which are linked by an isomorphism are said to be
isomorphic. An automorphism of Clifford algebra is a Clifford isomorphism\ on
the same Clifford algebra.
\end{definition}

\begin{theorem}
The only Clifford automorphisms of Clifford algebras are the changes of
orthonormal basis, which can be characterized as :

$f:Cl(V,g)\rightarrow Cl(V,g)::f\left(  w\right)  =\mathbf{Ad}_{s}w$ for $s\in
Pin(V,g)$
\end{theorem}

\begin{proof}
f must preserve the scalar product on F, so it belongs to O(F,g) : $\forall
u\in F:f\left(  u\right)  =Ad_{s}u.$ Take $u,v\in F:f\left(  u\cdot v\right)
=f\left(  u\right)  \cdot f(v)=Ad_{s}u\cdot Ad_{s}v=s\cdot u\cdot v\cdot
s^{-1}=Ad_{s}\left(  u\cdot v\right)  $
\end{proof}

By picking an orthonormal basis in each Clifford algebra one deduces :

\begin{theorem}
All Clifford algebras Cl(F,g) where F is a complex n dimensional vector space
are isomorphic.

All Clifford algebras Cl(F,g) where F is a real n dimensional vector space and
g have the same signature, are isomorphic.
\end{theorem}

\begin{notation}
$Cl\left(
\mathbb{C}
,n\right)  $ is the common structure of Clifford algebras over a n dimensional
complex vector space
\end{notation}

\begin{notation}
$Cl\left(
\mathbb{R}
,p,q\right)  $ is the common structure of Clifford algebras over a real vector
space endowed with a bilinear symmetric form of signature (+ p, - q).
\end{notation}

The common structure $Cl\left(
\mathbb{C}
,n\right)  $\ is the Clifford algebra $Cl\left(
\mathbb{C}
^{n},g\right)  $ over $%
\mathbb{C}
$ \ endowed with the canonical \textit{bilinear} (Beware !) form : $g\left(
u,v\right)  =\sum_{i=1}^{n}\left(  u_{i}\right)  ^{2},u_{i}\in%
\mathbb{C}
$

The common structure $Cl\left(
\mathbb{R}
,p,q\right)  $ is the Clifford algebra $Cl\left(
\mathbb{R}
^{n},g\right)  $ over $%
\mathbb{R}
$ with p+q=n\ endowed with a bilinear form with signature p + and q -.

Warning !

i) The traditional notation (p,q) is confusing when we compare the signatures.
(p,q) means always p + and q -\ so the bilinear forms for $Cl\left(
\mathbb{R}
,p,q\right)  $\ and $Cl\left(
\mathbb{R}
,q,p\right)  $\ have opposite signs. But the expression of the bilinear form
depends on the basis which is used.\ As we are in $%
\mathbb{R}
^{n}$ it is customary to use the canonical basis $\left(  e_{j}\right)
_{j=1}^{n}.$ So, with the same basis, the meaning is the following :

$Cl\left(
\mathbb{R}
,p,q\right)  $ corresponds to : $g\left(  u,v\right)  =\sum_{i=1}^{p}\left(
u_{i}\right)  ^{2}-\sum_{i=p+1}^{n}\left(  u_{i}\right)  ^{2},$ that is p + ,
q -

$Cl\left(
\mathbb{R}
,q,p\right)  $ corresponds to : $\widetilde{g}\left(  u,v\right)  =-\sum
_{i=1}^{p}\left(  u_{i}\right)  ^{2}+\sum_{i=p+1}^{n}\left(  u_{i}\right)
^{2}=-g\left(  u,v\right)  $ that is q + , p -

Of course the custom to index the basis from 0 to 3 in physics does not change
anything to this remark.

ii) The algebras $Cl(%
\mathbb{R}
,p,q)$ and $Cl(%
\mathbb{R}
,q,p)$ are \textit{not} isomorphic if $p\neq q$ . However :

\begin{theorem}
For any $n,p,q\geq0$ we have the algebras isomorphisms :

$Cl\left(
\mathbb{R}
,p,q\right)  \simeq Cl_{0}\left(
\mathbb{R}
,p+1,q\right)  \simeq Cl_{0}\left(
\mathbb{R}
,q,p+1\right)  $

$Cl_{0}\left(
\mathbb{R}
,p,q\right)  \simeq Cl_{0}\left(
\mathbb{R}
,q,p\right)  $

$Cl(%
\mathbb{R}
,0,p)\simeq Cl(%
\mathbb{R}
,p,0$

$Cl_{0}\left(
\mathbb{C}
,n\right)  \simeq Cl(%
\mathbb{C}
,n-1)$
\end{theorem}

\begin{theorem}
The complexified $Cl_{c}\left(
\mathbb{R}
,p,q\right)  $ is isomorphic to $Cl\left(
\mathbb{C}
,p+q\right)  $
\end{theorem}

\begin{proof}
Let g be the symmetric bilinear form of signature (p,q) on $%
\mathbb{R}
^{p+q}$

$Cl_{c}\left(
\mathbb{R}
^{p+q},g\right)  \equiv Cl\left(
\mathbb{C}
^{p+q},g_{%
\mathbb{C}
}\right)  \simeq Cl\left(
\mathbb{C}
,p+q\right)  $ because all the Clifford algebras on \ $%
\mathbb{C}
^{p+q}$\ are isomorphic.
\end{proof}

\begin{theorem}
There are real algebra morphisms : $T:Cl\left(
\mathbb{R}
,p,q\right)  \rightarrow Cl\left(
\mathbb{C}
,p+q\right)  $ and $T^{\prime}:Cl\left(
\mathbb{R}
,q,p\right)  \rightarrow Cl\left(
\mathbb{C}
,p+q\right)  $ with $T^{\prime}\left(  e_{j}\right)  =i\eta_{jj}T\left(
e_{j}\right)  $ with $\eta_{jj}=+1$ for j=1...p, -1 for j=p+1...n
\end{theorem}

\begin{proof}
Let $\left(  e_{i}\right)  _{i=1}^{n}$ be the canonical orthonormal basis of $%
\mathbb{R}
^{n}$ with n = p + q, $V\left(  p,q\right)  $\ $%
\mathbb{R}
^{n}$ endowed with the scalar product of signature (p,q) with bilinear form :
$g\left(  u,v\right)  =\sum_{j=1}^{p}u_{j}v_{j}-$\ $\sum_{j=p+1}^{n}u_{j}%
v_{j}$

$\left(  e_{i}\right)  _{i=1}^{n}$ is the canonical basis of $%
\mathbb{C}
^{n}$ with complex components. $Cl\left(
\mathbb{C}
,p+q\right)  $ is a complex Clifford algebra, but it is also a real algebra
(not Clifford) with basis :

$\widetilde{e}_{j}=e_{j},j=1..n$

$\widetilde{e}_{j+1}=ie_{j},j=1..n$

and the product :

$\widetilde{e}_{j}\cdot\widetilde{e}_{n+k}=\widetilde{e}_{n+j}\cdot
\widetilde{e}_{k}=i\left(  e_{j}\cdot e_{k}\right)  \in Cl\left(
\mathbb{C}
,p+q\right)  $

Define the real linear map :

$\tau:V\left(  p,q\right)  \rightarrow Cl\left(
\mathbb{C}
,p+q\right)  $

$\tau\left(  e_{j}\right)  =\widetilde{e}_{j}=e_{j}$ for j = 1 ...p

$\tau\left(  e_{j}\right)  =\widetilde{e}_{j+1}=ie_{j}$ for j = p + 1 ...n

This is a real linear map from the real vector space $V\left(  p,q\right)  $
into the real algebra $Cl\left(
\mathbb{C}
,p+q\right)  .$ It is not surjective.

$\tau\left(  u\right)  \cdot\tau\left(  v\right)  =\left(  \sum_{j=1}^{p}%
u_{j}e_{j}+i\sum_{j=p+1}^{n}u_{j}e_{j}\right)  \cdot\left(  \sum_{j=1}%
^{p}v_{j}e_{j}+i\sum_{j=p+1}^{n}v_{j}e_{j}\right)  $

$=\sum_{j,k=1}^{p}u_{j}v_{k}e_{j}\cdot e_{k}+i\sum_{j=1}^{p}\sum_{k=p+1}%
^{n}u_{j}v_{k}e_{j}\cdot e_{k}$

$+i\sum_{j=p+1}^{n}\sum_{k=1}^{p}u_{j}v_{k}e_{j}\cdot e_{k}-\sum_{j,k=p+1}%
^{n}u_{j}v_{k}e_{j}\cdot e_{k}$

$\tau\left(  v\right)  \cdot\tau\left(  u\right)  =\sum_{j,k=1}^{p}v_{j}%
u_{k}e_{j}\cdot e_{k}+i\sum_{j=1}^{p}\sum_{k=p+1}^{n}v_{j}u_{k}e_{j}\cdot
e_{k}$

$+i\sum_{j=p+1}^{n}\sum_{k=1}^{p}v_{j}u_{k}e_{j}\cdot e_{k}-\sum_{j,k=p+1}%
^{n}v_{j}u_{k}e_{j}\cdot e_{k}$

In $Cl\left(
\mathbb{C}
,p+q\right)  :e_{j}\cdot e_{k}+e_{k}\cdot e_{j}=\delta_{jk}$

$\tau\left(  u\right)  \cdot\tau\left(  v\right)  +\tau\left(  v\right)
\cdot\tau\left(  u\right)  =2\left(  \sum_{j=1}^{p}u_{j}v_{j}-\sum_{j=p+1}%
^{n}u_{j}v_{j}\right)  =2g\left(  u,v\right)  $

Thus by the universal property of Clifford algebras, there is a real algebra
morphism : $T:Cl\left(
\mathbb{R}
,p,q\right)  \rightarrow Cl\left(
\mathbb{C}
,p+q\right)  $ such that : \ $\tau=T\circ\imath$ with $\imath:V\left(
p,q\right)  \rightarrow Cl\left(
\mathbb{R}
,p,q\right)  $

The image $T\left(  Cl\left(
\mathbb{R}
,p,q\right)  \right)  $ is a real subalgebra of $Cl\left(
\mathbb{C}
,p+q\right)  .$

Now assume that we take for scalar product on V(q,p) :

$g^{\prime}\left(  u,v\right)  =-\sum_{j=1}^{p}u_{j}v_{j}+$\ $\sum_{j=p+1}%
^{n}u_{j}v_{j}=-g\left(  u,v\right)  $

Define the real linear map :

$\tau^{\prime}:V(q,p)\rightarrow Cl\left(
\mathbb{C}
,p+q\right)  $

$\tau^{\prime}\left(  e_{j}\right)  =i\widetilde{e}_{j}=e_{j}$ for j = 1 ...p

$\tau^{\prime}\left(  e_{j}\right)  =\widetilde{e}_{j+1}=ie_{j}$ for j = p + 1 ...n

We have :

$\tau^{\prime}\left(  u\right)  \cdot\tau^{\prime}\left(  v\right)
+\tau^{\prime}\left(  v\right)  \cdot\tau^{\prime}\left(  u\right)  =2\left(
-\sum_{j=1}^{p}u_{j}v_{j}+\sum_{j=p+1}^{n}u_{j}v_{j}\right)  =-2g\left(
u,v\right)  $

Thus there is a real algebra morphism : $T^{\prime}:Cl\left(
\mathbb{R}
,q,p\right)  \rightarrow Cl\left(
\mathbb{C}
,p+q\right)  $ such that : \ $\tau^{\prime}=T^{\prime}\circ\imath^{\prime}$
with $\imath^{\prime}:V(q,p)\rightarrow Cl\left(
\mathbb{R}
,p,q\right)  $ .The image $T\left(  Cl\left(
\mathbb{R}
,q,p\right)  \right)  $ is a real subalgebra of $Cl\left(
\mathbb{C}
,p+q\right)  $

In $Cl(%
\mathbb{C}
,p+q):T^{\prime}\left(  e_{j}\right)  =i\eta_{ii}T\left(  e_{j}\right)  $ with
$\eta_{ii}=+1$ for j=1...p, -1 for j=p+1...n
\end{proof}

\subsubsection{Representation of a Clifford algebra}

\paragraph{Algebraic representations\newline}

\begin{definition}
An \textbf{algebraic representation} of a Clifford algebra Cl(F,g) over a
field K is the couple $\left(  A,\rho\right)  $ of an algebra $\left(
A,\circ\right)  $ on the field K and a map : $\rho:Cl\left(  F,g\right)
\rightarrow A $ which is an algebra morphism :
\end{definition}

$\forall X,Y\in Cl(F,g),k,k^{\prime}\in K:$

$\rho\left(  kX+k^{\prime}Y\right)  =k\rho(X)+k^{\prime}\rho(Y),$

$\rho\left(  X\cdot Y\right)  =\rho(X)\circ\rho(Y),\rho\left(  1\right)
=I_{A}$

(with $\circ$\ as internal operation in A, and A is required to be unital with
unity element I) .

No scalar product is required on A, which is an "ordinary" unital algebra (so
this is different from the previous case). The algebra is usually a set of
matrices, or of couple of matrices (see below).

A, F must be on the same field K.\ A and Cl(F,g) must have the same dimension
$2^{\dim F}.$

A morphism of Clifford algebras $\rho:Cl\left(  F_{1},g_{1}\right)
\rightarrow Cl\left(  F_{2},g_{2}\right)  $ gives an algebraic representation
$\left(  Cl\left(  F_{2},g_{2}\right)  ,\rho\right)  $ of $Cl\left(
F_{1},g_{1}\right)  .$ In particular $\left(  Cl(F,g),\tau\right)  $ where
$\tau$\ is any automorphism is an algebraic representation of Cl(F,g) on itself.

If there is an isomorphism : $\tau:Cl\left(  F_{1},g_{1}\right)  \rightarrow
Cl\left(  F_{2},g_{2}\right)  $ between the Clifford algebras $Cl\left(
F_{1},g_{1}\right)  ,Cl\left(  F_{2},g_{2}\right)  $ then, for any algebraic
representation $\left(  A,\rho\right)  $ of $Cl\left(  F_{2},g_{2}\right)  $ ,
$\left(  A,\rho\circ\tau\right)  $ is a representation of $Cl\left(
F_{1},g_{1}\right)  .$

If $u\in Cl\left(  F,g\right)  $ is invertible then $\rho\left(  u\right)  $
is invertible and

$\rho\left(  u^{-1}\right)  =\rho\left(  u\right)  ^{-1}=\rho\left(
-u/g\left(  u,u\right)  \right)  =-\rho\left(  u\right)  /g\left(  u,u\right)
$

The image $\rho\left(  Cl(F,g)\right)  $ is a subalgebra of A. If $\rho$ is
injective $\left(  \rho\left(  Cl(F,g)\right)  ,\rho\right)  $ is a faithful representation.

The images $\rho\left(  Pin(F,g)\right)  ,\rho\left(  Spin(F,g)\right)  $ are
subgroups of the group GA of invertible elements of A, and $\rho$ is a group morphism.

\begin{definition}
An algebraic representation $\left(  A,\rho\right)  $ of a Clifford algebra
Cl(F,g) over a field K is \textbf{faithful} if $\rho$ is bijective.
\end{definition}

\begin{definition}
If $\left(  A,\rho\right)  $\ is an algebraic representation $\left(
A,\rho\right)  $ of a Clifford algebra Cl(F,g) over a field K, a subalgebra A'
of A is \textbf{invariant} if

$\forall w\in Cl(F,g),\forall a\in A^{\prime}:\rho\left(  w\right)  a\in
A^{\prime}$
\end{definition}

\begin{definition}
An algebraic representation $\left(  A,\rho\right)  $ of a Clifford algebra
Cl(F,g) over a field K is \textbf{irreducible} if there is no subalgebra A' of
A which is invariant by $\rho$
\end{definition}

\bigskip

\paragraph{Geometric representation\newline}

\begin{definition}
A \textbf{geometric representation} of a Clifford algebra Cl(F,g) over a field
K is a couple $\left(  V,\rho\right)  $ of a vector space V on the field K and
a map : $\rho:Cl\left(  F,g\right)  \rightarrow L\left(  V;V\right)  $ which
is an algebra morphism :
\end{definition}

$\forall w,w^{\prime}\in Cl(F,g),k,k^{\prime}\in K:\rho\left(  kw+k^{\prime
}w^{\prime}\right)  =k\rho(w)+k^{\prime}\rho(w^{\prime}),$

$\rho\left(  w\cdot w^{\prime}\right)  =\rho(w)\circ\rho(w^{\prime}%
),\rho\left(  1\right)  =Id_{V}$

Notice that the internal operation in L(V;V) is the composition of maps, and
L(V;V) is always unital.

The vectors of V are called \textbf{spinors} (and not the elements
$\rho\left(  w\right)  $\ of the representation).\ When the dimension of F is
4 they are called \textbf{Dirac spinors}.\ On $Cl\left(
\mathbb{C}
,4\right)  $ there is a volume element denoted $e_{5},$ the Dirac spinors can
be decomposed in the sum of two spinors in $%
\mathbb{C}
^{2}$\ , called half spinors or \textbf{Weyl's spinors}, the left handed
spinors corresponding to $Cl_{-}\left(
\mathbb{C}
,4\right)  .$

By restriction of $\rho$ we have representations (in the usual meaning - see
Lie groups) $\left(  V,\rho\right)  $ :

i) of the groups Pin(F,g) and Spin(F,g).

ii) of the Lie algebras $T_{1}Pin\left(  F,g\right)  ,T_{1}Spin\left(
F,g\right)  $ (which are vector subspaces of Cl(F,g)). The range is the Lie
algebra of L(V;V) with bracket $f\circ g-g\circ f.$ This is the application of
a general theorem and $\rho^{\prime}\left(  1\right)  \equiv\rho$ because
$\rho$ is linear.

Remarks :

i) A geometric representation is a special algebraic representation, where a
vector space V has been specified and the algebra is L(V;V). Thus any
geometric representation gives an algebraic representation on L(V;V).

ii) However the converse is more complicated. If $(A,\rho)$ is an algebraic
representation of a Clifford algebra Cl over a field K, and A is an algebra of
m$\times$m matrices over the same field K, then $\left(  K^{m},\left[
\rho\right]  \right)  $ is a geometric representation of Cl, by taking the
endomorphisms associated to the matrices. It is always possible to replace
$K^{m}$ by any vector space V on K, and in particular one can choose any basis
of V in which the matrices are expressed. They will give equivalent representations.

But if the fields are different the operation is not always possible (see below).

\bigskip

\paragraph{Real and complex representations\newline}

If Cl(F,g) is a real Clifford algebra and A a complex algebra with a real
structure : $A=A_{%
\mathbb{R}
}\oplus iA_{%
\mathbb{R}
}$\ , an algebraic representation of Cl(F,g) is a real representation where
elements $X\in A_{%
\mathbb{R}
}$ and $iX\in iA_{%
\mathbb{R}
}$ are deemed different.

If Cl(F,g) is a complex algebra then A must be complex, possibly through a
complex structure on A (usually by complexification : $A\rightarrow A_{%
\mathbb{C}
}=A\oplus iA).$

\begin{theorem}
If Cl(F,g) is a real Clifford algebra and $\left(  A,\rho\right)  $ a complex
representation of the complexified Clifford algebra $Cl_{c}\left(  F,g\right)
$ then $\left(  A,\rho\circ T\right)  $ is a real representation of Cl(F,g),
where $T:Cl(F,g)\rightarrow Cl_{c}\left(  F,g\right)  $ is a real linear morphism
\end{theorem}

\begin{proof}
There is a real morphism of algebras $T:Cl\left(
\mathbb{R}
,p,q\right)  \rightarrow Cl\left(
\mathbb{C}
,p+q\right)  .$ All Clifford algebras of same dimension and signature being
isomorphic, this morphism stands for $T:Cl(F,g)\rightarrow Cl_{c}\left(
F,g\right)  $ and $\rho\circ T$ is by restriction a real morphism of algebras.
\end{proof}

So, if $\left(  A,\rho\right)  $ is a complex representation of $Cl_{c}\left(
F,g\right)  $ on a complex algebra of $n\times n$\ matrices, from which we
deduce a complex geometric representation $\left(
\mathbb{C}
^{n},\rho\right)  ,$ then for each $w\in Cl\left(  F,g\right)  $ the element
$\rho\circ T\left(  w\right)  $ is the matrix of a complex linear map, but
$\rho\circ T:Cl(F,g)\rightarrow A$ is real linear.

\bigskip

\paragraph{The generators of a representation\newline}

\begin{definition}
The \textbf{generators }of an algebraic representation $\left(  A,\rho\right)
$ of the Clifford algebra Cl(F,g), for the vector space F with orthonormal
basis $\left(  e_{i}\right)  _{i=1}^{n}$, are the images :

$\left(  \gamma_{i}\right)  _{i=0}^{n}:\gamma_{i}=\rho\left(  e_{i}\right)
,i=1..n,\gamma_{0}=\rho\left(  1\right)  $
\end{definition}

They meet the conditions :%

\begin{equation}
\forall i,j=0...n:\gamma_{i}\gamma_{j}+\gamma_{j}\gamma_{i}=2\eta_{ij}%
\gamma_{0}%
\end{equation}

Moreover :

each generator is invertible in A : $\gamma_{i}^{-1}=-\eta_{ii}\gamma_{i}$

$\gamma_{0}$ is the unity in A

$\rho$ is injective iff all the generators are distinct

Conversely :

\begin{theorem}
A family $\left(  \gamma_{i}\right)  _{i=0}^{n}$ of elements of the unital
algebra $\left(  A,\circ\right)  $ on the field K \ which meet the conditions :

i) $\gamma_{0}$ is the unity in A

ii) all the $\gamma_{i}$ are invertible in A

iii) $\forall i,j=0...n:\gamma_{i}\circ\gamma_{j}+\gamma_{j}\circ\gamma
_{i}=2\eta_{ij}\gamma_{0}$ with $\eta_{ij}=\pm\delta_{ij}$

define a unique algebraic representation $\left(  A,\rho\right)  $ of the
Clifford algebra $Cl\left(  K,p,q\right)  $ where p,q is the signature of
$\eta_{ij}$
\end{theorem}

\begin{proof}
Define F as the vector space $K^{n}$ endowed with the symmetric bilinear form
of signature (p,q) and $\left(  e_{i}\right)  _{i=1}^{n}$ as an orthonormal basis.

Any element of $Cl\left(  K,p,q\right)  $ reads : $w=\sum_{k=0}^{2^{\dim F}%
}\sum_{\left\{  i_{1},...,i_{k}\right\}  }w_{\left\{  i_{1},...,i_{k}\right\}
}e_{i_{1}}\cdot...\cdot e_{i_{k}}$

Define : i=1...n : $\rho\left(  e_{i}\right)  =\gamma_{i},\rho\left(
1\right)  =\gamma_{0}=1_{A}$

Define $\rho\left(  w\right)  =\sum_{k=0}^{2^{\dim F}}\sum_{\left\{
i_{1},...,i_{k}\right\}  }w_{\left\{  i_{1},...,i_{k}\right\}  }\gamma_{i_{1}%
}\circ...\circ\gamma_{i_{k}}$

$e_{i}$ is invertible in $Cl\left(  K,p,q\right)  $, thus $\gamma_{i}$ must be invertible

$e_{i}\cdot e_{j}+e_{j}\cdot e_{i}=2\eta_{ij}$thus we must have : $\gamma
_{i}\circ\gamma_{j}+\gamma_{j}\circ\gamma_{i}=2\eta_{ij}\gamma_{0}$
\end{proof}

All irreducible representations of Clifford algebras are on sets of rxr
matrices with $r=2^{k}.$\ So a practical way to find a set of generators is to
start with 2$\times$2 matrices and extend the scope by Kronecker product :

Pick four 2$\times$2 matrices $E_{j}$ such that : $E_{i}E_{j}+E_{j}E_{i}%
=2\eta_{ij}I_{2},E_{0}=I_{2}$ (the Dirac matrices are usually adequate)

Compute : $F_{ij}=E_{i}\otimes E_{j}$

Then : $F_{ij}F_{kl}=E_{i}E_{k}\otimes E_{j}E_{l}$

With some good choices of combination by recursion one gets the right
$\gamma_{i}$

The Kronecker product preserves the symmetry and the hermicity, so if one
starts with $E_{j}$ having these properties the $\gamma_{i}$ will have it.

\bigskip

\paragraph{Equivalence of representations\newline}

For a given representation of a Clifford algebra there is a unique set of
generators. A given set of generators on an algebra defines a unique
representation of the generic Clifford algebra Cl(K,p,q).\ But a
representation is not unique. It is of particular interest to tell when two
representations are "equivalent", and to know the relations between such
representations. The generators are a useful tool for this purpose.

\bigskip

\subparagraph{Algebraic representations\newline}

\begin{definition}
Two algebraic representations $\left(  A_{1},\rho_{1}\right)  ,\left(
A_{2},\rho_{2}\right)  $ of a Clifford algebra Cl(F,g) are said to be
\textbf{equivalent} if there are :

i) a bijective algebra morphism $\phi:A_{1}\rightarrow A_{2}$\ 

ii) an automorphism $\tau:Cl\left(  F,g\right)  \rightarrow Cl\left(
F,g\right)  $\ 

such that $:\phi\circ\rho_{1}=\rho_{2}\circ\tau$
\end{definition}

$\bigskip$%

\begin{tabular}
[c]{lllll}
&  & $\tau$ &  & \\
& Cl(F,g) & $\rightarrow$ & Cl(F,g) & \\
$\rho_{1}$ & \multicolumn{1}{c}{$\downarrow$} &  &
\multicolumn{1}{c}{$\downarrow$} & $\rho_{2}$\\
& \multicolumn{1}{c}{$A_{1}$} & $\rightarrow$ & \multicolumn{1}{c}{$A_{2}$} &
\\
&  & $\phi$ &  &
\end{tabular}

\bigskip

$A_{1},A_{2},F$ must be on the same field K, $A_{1},A_{2}$ must have the same
dimension. $\phi$ must be such that : $\phi\left(  X\circ Y\right)
=\phi\left(  X\right)  \circ\phi\left(  Y\right)  $ thus the right or left
translation $L_{U}\left(  X\right)  =U\circ X,R_{U}\left(  X\right)  =X\circ
U$ are not automorphisms.

Two representations are not necessarily equivalent.

All the representations equivalent to $\left(  A,\rho\right)  $\ are generated
by :

$\phi\in GL\left(  A;\widetilde{A}\right)  ,\tau=\mathbf{Ad}_{s},s\in
Pin\left(  F,g\right)  :$

$\widetilde{e}_{j}=\mathbf{Ad}_{s}\left(  e_{j}\right)  $

$\widetilde{\gamma}_{j}=\phi\circ\rho\circ\mathbf{Ad}_{s^{-1}}\left(
e_{j}\right)  =\phi\circ\rho\circ\left(  s^{-1}\cdot e_{j}\cdot s\right)
\mathbf{=}\phi\left(  \rho\left(  s^{-1}\right)  \gamma_{j}\rho\left(
s\right)  \right)  $

$\mathbf{=}\phi\left(  \rho\left(  s\right)  \right)  ^{-1}\circ\phi\left(
\gamma_{j}\right)  \circ\phi\left(  \rho\left(  s\right)  \right)
=U^{-1}\circ\phi\left(  \gamma_{j}\right)  \circ U$

In particular, on the same algebra A, all the equivalent representations are
defined by conjugation with a fixed invertible element U ($\widetilde
{A}=A,\phi=Id).$

An involution : $\ast:A\rightarrow A$ such that $\left(  X\circ Y\right)
^{\ast}=Y^{\ast}\circ X^{\ast}$ is not an automorphism, so one cannot define
an equivalent representation by taking the conjugate, the transpose or the
adjoint of matrices.\ But we have the following :

\begin{theorem}
If Cl(F,g) is a complex Clifford algebra, with real structure $\sigma,$ A a
complex algebra endowed with a real structure $\theta$, then to any algebraic
representation $\left(  A,\rho\right)  $ is associated the
\textbf{contragredient representation} : $\left(  A,\overline{\rho}\right)  $
with $\overline{\rho}=\theta\circ\rho\circ\sigma$
\end{theorem}

\begin{proof}
The conjugate of $w\in Cl\left(  F,g\right)  $ is $\overline{w}=\sigma\left(
w\right)  $ and $\sigma$ is antilinear and compatible with the product.

$\theta$ is antilinear on A, compatible with the product, the conjugate of
$X\in A$ is $\overline{X}=\theta\left(  X\right)  $

We have a complex linear morphism $\overline{\rho}:Cl\left(  F,g\right)
\rightarrow A$

$\overline{\rho}\left(  kw\right)  =\theta\circ\rho\circ\sigma\left(
kw\right)  =k\theta\circ\rho\circ\sigma\left(  w\right)  =k\overline{\rho
}\left(  w\right)  $
\end{proof}

The two representations are not equivalent.

If $\left(  A,\rho\right)  $ is a representation on an algebra of matrices,
and if the algebra is closed under transposition, the transpose $\gamma
_{j}^{t}$ of the generators still meet the requirements, so we have a non
equivalent representation. And similarly for complex conjugation.

\bigskip

\subparagraph{Geometric representations\newline}

For a geometric representation a morphism such that : $\phi:L(V_{1}%
;V_{1})\rightarrow L\left(  V_{2};V_{2}\right)  $ is not very informative.
This leads to:

\begin{definition}
An \textbf{interwiner} between two geometric representations $(V_{1},\rho
_{1}),\left(  V_{2},\rho_{2}\right)  $ of a Clifford algebra Cl(F,g) is a
linear map $\phi:V_{1}\rightarrow V_{2}$\ such that $\forall w\in Cl\left(
F,g\right)  :\phi\circ\rho_{1}\left(  w\right)  =\rho_{2}\left(  w\right)
\circ\phi\in L\left(  V_{1};V_{2}\right)  $
\end{definition}

\begin{definition}
Two geometric representations of a Clifford algebra Cl(F,g) are said to be
\textbf{equivalent} if there is a bijective interwiner.
\end{definition}

In two equivalent geometric \ representations $(V_{1},\rho_{1}),\left(
V_{2},\rho_{2}\right)  $ the vector spaces must have the same dimension and be
on the same field. \ Conversely two vector spaces with the same dimension on
the same field are isomorphic, so $\left(  V_{1},\rho_{1}\right)  $\ give the
equivalent representation $\left(  V_{2},\rho_{2}\right)  $ by : $\rho
_{2}\left(  w\right)  =\phi\circ\rho_{1}\left(  w\right)  \circ\phi^{-1}.$ In
particular we are free to define a basis in V.

All the equivalent representation to $\left(  V,\rho\right)  $ on the same
vector space are deduced by conjugation by a fixed automorphism $\phi$ of V :
$\left(  V,\widetilde{\rho}\right)  \simeq\left(  V,\rho\right)
::\widetilde{\rho}\left(  w\right)  =\phi\circ\rho\left(  w\right)  \circ
\phi^{-1},\phi\in GL\left(  V;V\right)  $

\begin{theorem}
If Cl(F,g) is a complex Clifford algebra, with real structure $\sigma,$ V a
complex vector space endowed with a real structure $\theta$, then to any
geometric representation $\left(  V,\rho\right)  $ is associated the
\textbf{contragredient representation} : $\left(  V,\overline{\rho}\right)  $
with $\overline{\rho}=\theta\circ\rho\circ\sigma$
\end{theorem}

The conjugate of a vector $X\in V$ is $\overline{X}=\theta\left(  X\right)  .$
The demonstration is the same as above.\ The two representations are not equivalent.

Similarly :

If $\left(  V,\rho\right)  $ is a geometric representation of Cl(F,g), then
$\left(  V^{\ast},\rho^{t}\right)  $ is a non equivalent representation of
Cl(F,g), with same generators. $\rho\left(  w\right)  $ and $\rho^{t}\left(
w\right)  =\left(  \rho\left(  V\right)  \right)  ^{t}\in L\left(  V^{\ast
};V^{\ast}\right)  $ are represented by the same matrix in a basis of V and
its dual in $V^{\ast}.$

If $\left(  V,\rho\right)  $ is a geometric representation of Cl(F,g), then
$\left(  V,\rho^{t}\right)  $ is a non equivalent representation of Cl(F,g),
whose generators are the transpose of the generators of $\left(
V,\rho\right)  $\ . Clearly $\widetilde{\gamma}_{i}=\left[  \gamma_{i}\right]
^{t}$ meets all the required conditions, so $\left(  L\left(  V;V\right)
,\rho^{t}\right)  $ is an algebraic representation of Cl(F,g), and $\left(
V,\rho^{t}\right)  $ is a geometric representation.

If $\left(  V,\rho\right)  $ is a geometric representation of Cl(F,g),
$\tau=\mathbf{Ad}_{s}$ an automorphism of Cl(F,g), an equivalent algebraic
representation to \ $\left(  L(V;V),\rho\circ\tau\right)  $ is $\left(
L(V;V),\phi\circ\rho\right)  $\ with $\phi\in L\left(  V;V\right)  :\phi
\circ\rho=\rho\circ\tau\Leftrightarrow\phi=\rho\circ\tau\circ\rho^{-1}.$ The
new generators are :

$\widetilde{\gamma}_{j}=\phi\left(  \gamma_{j}\right)  =\rho\circ\tau\left(
e_{j}\right)  =\rho\left(  s\cdot e_{j}\cdot s^{-1}\right)  =\rho\left(
s\right)  \gamma_{j}\rho\left(  s\right)  ^{-1}\Rightarrow\forall w\in
Cl(F,g):\widetilde{\gamma}\left(  w\right)  =\rho\left(  s\right)
\gamma\left(  w\right)  \rho\left(  s\right)  ^{-1}$

This is equivalent to a change of basis in V by $\rho\left(  s\right)  ^{-1}.$

\bigskip

\paragraph{A classic representation\newline}

A Clifford algebra Cl(F,g) has a geometric representation on the algebra
$\Lambda F^{\ast}$ of linear forms on F

Consider the maps with $u\in V:$

$\lambda\left(  u\right)  :\Lambda_{r}F^{\ast}\rightarrow\Lambda_{r+1}F^{\ast
}::\lambda\left(  u\right)  \mu=u\wedge\mu$

$i_{u}:\Lambda_{r}F^{\ast}\rightarrow\Lambda_{r-1}F^{\ast}::i_{u}\left(
\mu\right)  =\mu\left(  u\right)  $

The map : $\Lambda F^{\ast}\rightarrow\Lambda F^{\ast}::\widetilde{\rho
}\left(  u\right)  =\lambda\left(  u\right)  -i_{u}$ is such that :

$\widetilde{\rho}\left(  u\right)  \circ\widetilde{\rho}\left(  v\right)
+\widetilde{\rho}\left(  v\right)  \circ\widetilde{\rho}\left(  u\right)
=2g\left(  u,v\right)  Id$

thus there is a map : $\rho:Cl(F,g)\rightarrow\Lambda F^{\ast}$ such that :
$\rho\cdot\imath=\widetilde{\rho}$ and $\left(  \Lambda F^{\ast},\rho\right)
$ is a geometric representation of Cl(F,g). It is reducible.

\subsubsection{Classification of Clifford algebras}

\paragraph{Complex algebras\newline}

\begin{theorem}
The unique faithful irreducible algebraic representation of the complex
Clifford algebra $Cl(%
\mathbb{C}
,n)$ is over an algebra \ of matrices of complex numbers
\end{theorem}

The algebra A depends on n :

If n=2m : $A=%
\mathbb{C}
(2^{m}):$ the square matrices $2^{m}\times2^{m}$ (we get the dimension
$2^{2m}$ as vector space)

If n=2m+1 : $A=%
\mathbb{C}
(2^{m})\oplus%
\mathbb{C}
(2^{m})\simeq%
\mathbb{C}
(2^{m})\times%
\mathbb{C}
(2^{m}):$ couples (A,B) of square matrices $2^{m}\times2^{m}$ (the vector
space has the dimension $2^{2m+1}$). A and B are two independant matrices.

The representation is faithful so there is a bijective correspondance between
elements of the Clifford algebra and matrices.

The internal operations on A are the addition, multiplication by a complex
scalar and product of matrices. When there is a couple of matrices each
operation is performed independantly on each component (as in the product of a
vector space):

$\forall\left(  \left[  A\right]  ,\left[  B\right]  \right)  ,\left(  \left[
A^{\prime}\right]  ,\left[  B^{\prime}\right]  \right)  \in A,k\in%
\mathbb{C}
$

$\left(  \left[  A\right]  ,\left[  B\right]  \right)  +\left(  \left[
A^{\prime}\right]  ,\left[  B^{\prime}\right]  \right)  =\left(  \left[
A\right]  +\left[  A^{\prime}\right]  ,\left[  B\right]  +\left[  B^{\prime
}\right]  \right)  $

$k\left(  \left[  A\right]  ,\left[  B\right]  \right)  =\left(  k\left[
A\right]  ,k\left[  B\right]  \right)  $

The map $\rho$\ is an isomorphism of algebras : $\forall w,w^{\prime}\in Cl(%
\mathbb{C}
,n),z,z^{\prime}\in%
\mathbb{C}
:$

$\rho\left(  w\right)  =\left[  A\right]  $ or $\rho\left(  w\right)  =\left(
\left[  A\right]  ,\left[  B\right]  \right)  $

$\rho\left(  zw+z^{\prime}w^{\prime}\right)  =z\rho\left(  w\right)
+z^{\prime}\rho\left(  w^{\prime}\right)  =z\left[  A\right]  +z^{\prime
}\left[  A^{\prime}\right]  $ or $=\left(  z\left[  A\right]  +z^{\prime
}\left[  A^{\prime}\right]  ,z\left[  B\right]  +z^{\prime}\left[  B^{\prime
}\right]  \right)  $

$\rho\left(  w\cdot w^{\prime}\right)  =\rho\left(  w\right)  \cdot\rho\left(
w^{\prime}\right)  =\left[  A\right]  \left[  B\right]  $ or $=\left(  \left[
A\right]  \left[  A^{\prime}\right]  ,\left[  B\right]  \left[  B^{\prime
}\right]  \right)  $

In particular :

$Cl(%
\mathbb{C}
,0)\simeq%
\mathbb{C}
;Cl(%
\mathbb{C}
,1)\simeq%
\mathbb{C}
\oplus%
\mathbb{C}
;Cl(%
\mathbb{C}
,2)\simeq%
\mathbb{C}
\left(  4\right)  $

\paragraph{Real Clifford algebras\newline}

\begin{theorem}
The unique faithful irreducible algebraic representation of the Clifford
algebra Cl($%
\mathbb{R}
,p,q)$ is over an algebra of matrices
\end{theorem}

(Husemoller p.161) The matrices algebras are over a field K'\ ($%
\mathbb{C}
,%
\mathbb{R}
)$ or the division ring H of quaternions with the following rules :

\bigskip

$%
\begin{bmatrix}
\left(  p-q\right)  \operatorname{mod}8 & Matrices & \left(  p-q\right)
\operatorname{mod}8 & Matrices\\
0 &
\mathbb{R}
\left(  2^{m}\right)  & 0 &
\mathbb{R}
\left(  2^{m}\right) \\
1 &
\mathbb{R}
\left(  2^{m}\right)  \oplus%
\mathbb{R}
\left(  2^{m}\right)  & -1 &
\mathbb{C}
\left(  2^{m}\right) \\
2 &
\mathbb{R}
\left(  2^{m}\right)  & -2 & H\left(  2^{m-1}\right) \\
3 &
\mathbb{C}
\left(  2^{m}\right)  & -3 & H\left(  2^{m-1}\right)  \oplus H\left(
2^{m-1}\right) \\
4 & H\left(  2^{m-1}\right)  & -4 & H\left(  2^{m-1}\right) \\
5 & H\left(  2^{m-1}\right)  \oplus H\left(  2^{m-1}\right)  & -5 &
\mathbb{C}
\left(  2^{m}\right) \\
6 & H\left(  2^{m-1}\right)  & -6 &
\mathbb{R}
\left(  2^{m}\right) \\
7 &
\mathbb{C}
\left(  2^{m}\right)  & -7 &
\mathbb{R}
\left(  2^{m}\right)  \oplus%
\mathbb{R}
\left(  2^{m}\right)
\end{bmatrix}
$

\bigskip

On H matrices are defined similarly as over a field, with the non
commutativity of product.

Remark : the division ring of quaternions can be built as $Cl_{0}\left(
\mathbb{R}
,0,3\right)  $

$H\oplus H,%
\mathbb{R}
\oplus%
\mathbb{R}
:$ take couples of matrices as above.

The representation is faithful so there is a bijective correspondance between
elements of the Clifford algebra and of matrices. The dimension of the
matrices in the table must be adjusted to n=2m or 2m+1 so that $\dim_{%
\mathbb{R}
}A=2^{n}$

The internal operations on A are performed as above when A is a direct product
of group of matrices.

$\rho$ is a real isomorphism, meaning that $\rho\left(  kw\right)
=k\rho\left(  w\right)  $ only if $k\in%
\mathbb{R}
$ even if the matrices are complex.

There are the following isomorphisms of algebras :

$Cl(%
\mathbb{R}
,0)\simeq%
\mathbb{R}
;Cl(%
\mathbb{R}
,1,0)\simeq%
\mathbb{R}
\oplus%
\mathbb{R}
;Cl(%
\mathbb{R}
,0,1)\simeq%
\mathbb{C}
$

$Cl(%
\mathbb{R}
,3,1)\simeq%
\mathbb{R}
\left(  4\right)  ,Cl(%
\mathbb{R}
,1,3)\simeq H\left(  2\right)  $

When the Clifford algebra is real, and is represented by a set of real
$2^{m}\times2^{m}$ matrices there is a geometric representation on $%
\mathbb{R}
^{2m}.$ The vectors of $%
\mathbb{R}
^{2m}$\ in such a representation are the \textbf{Majorana spinors}.

\subsubsection{Classification of Pin and Spin groups}

Pin and Spin are subset of the respective Clifford algebras, so the previous
algebras morphisms entail group morphisms with the invertible elements of the
algebras. Moreover, groups of matrices are well known and themselves
classified. So what matters here is the group morphism with these classical
groups. The respective classical groups involved are the orthogonal groups
O(K,p,q) for Pin and the special orthogonal groups SO(K,p,q) for Spin. A key
point is that to one element of O(K,p,q) or SO(K,p,q) correspond two elements
of Pin or Spin.\ This topic is addressed through the formalism of "cover" of a
manifold (see Differential geometry) and the results about the representations
of the Pin and Spin groups are presented in the Lie group part (Linear groups) .

\begin{theorem}
All Pin(F,g),Spin(F,g) groups, where F is a complex n dimensional vector space
are group isomorphic.

All Pin(F,g),Spin(F,g) where F is a real n dimensional vector space and g has
the same signature, are group isomorphic.
\end{theorem}

\begin{notation}
$Pin\left(
\mathbb{C}
,n\right)  ,Spin\left(
\mathbb{C}
,n\right)  $ are the common structure of the Pin and Spin groups over a n
dimensional complex vector space
\end{notation}

\begin{notation}
$Pin\left(
\mathbb{R}
,p,q\right)  ,Spin\left(
\mathbb{R}
,p,q\right)  $ are the common structure of the Pin and Spin group over a real
vector space endowed with a bilinear symmetric form of signature (+ p, - q).
\end{notation}

\newpage

\part{\textbf{ANALYSIS}}

\bigskip

\bigskip

Analysis is a very large area of mathematics.\ It adds to the structures and
operations of algebra the concepts of "proximity" and "limit".\ Its key
ingredient is topology, a way to introduce these concepts in a very general
but rigorous manner, to which is dedicated the first section. It is mainly a
long, but by far not exhaustive, list of definitions and results which are
necessary for a basic understanding of the rest of the book. The second
section is dedicated to the theory of measure, which is the basic tool for
integrals, with a minimum survey of probability theory. The third and fourth
sections are dedicated to analysis on sets endowed with a vector space
structure, mainly Banach spaces and algebras, which lead to Hilbert spaces and
the spectral theory.

\bigskip

\section{GENERAL\ TOPOLOGY}

\bigskip

Topology can be understood with two different, related, meanings. Initially it
has been an extension of geometry, starting with Euler, Listing and pursued by
Poincar\'{e}, to study qualitative properties of objects without referring to
a vector space structure. Today this part is understood as algebraic topology,
of which some basic elements are presented below.

The second meaning, called "general topology", is the mathematical way to
define "proximity" and "limit", and is the main object of this section. It has
been developped in the beginning of the XX$%
{{}^\circ}%
$ century by Cantor, as an extension of the set theory, and developped with
metrics over a set by Fr\'{e}chet, Hausdorff and many others. General topology
is still often introduced through metric spaces. But, when the basic tools
such as open, compact,... have been understood, they are often easier to use,
with a much larger scope. So we start with these general concepts. Metric
spaces bring additional properties. Here also it has been usual to focus on
definite positive metrics, but many results still hold with semi-metrics which
are common.

This is a vast area, so there are many definitions, depending on the authors
and the topic studied. We give only the most usual, which can be useful, and
often a prerequisite, in advanced mathematics. We follow mainly Wilansky,
Gamelin and Schwartz (tome 1). The reader can also usefully consult the tables
of theorems in Wilansky.

\bigskip

\subsection{Topological space}

\label{Topological spaces}

In this subsection topological concepts are introduced without any metric.
They all come from the definition of a special collection of subsets, the open subsets.

\subsubsection{Topology}

\paragraph{Open subsets\newline}

\begin{definition}
A \textbf{topological space} is a set E, endowed with a collection
$\Omega\subset2^{E}$ of subsets called \textbf{open} subsets such that :

$E\in\Omega,\varnothing\in\Omega$

$\forall I:O_{i}\in\Omega,\cup_{i\in I}O_{i}\in\Omega$

$\forall I,cardI<\infty:O_{i}\in\Omega,\cap_{i\in I}O_{i}\in\Omega$
\end{definition}

The key points are that every (even infinite) union of open sets is open, and
every \textit{finite intersection} of open sets is open.

The power set $2^{E}$ is the set of subsets of E, so $\Omega\subset2^{E}.$
Quite often the open sets are not defined by a family of sets, meaning a map :
$I\rightarrow2^{E}$

Example : in $%
\mathbb{R}
$ the open sets are generated by the open intervals ]a,b[ (a and b
excluded)$.$

\paragraph{Topology\newline}

The \textbf{topology} on E is just the collection $\Omega$ of its open
subsets, and a topological space will be denoted $\left(  E,\Omega\right)  .$
Different collections define different topologies (but they can be equivalent
: see below). There are many different topologies on the same set : there is
always $\Omega_{0}=\left\{  \varnothing,E\right\}  $ and $\Omega_{\infty
}=2^{E}$ (called the \textbf{discrete topology}).

When $\Omega_{1}\subset\Omega_{2}$ the topology defined by $\Omega_{1}$ is
said to be \textbf{thinner} (or \textbf{stronger}) then $\Omega_{2},$ and
$\Omega_{2}$ \textbf{coarser} (or \textbf{weaker}) than $\Omega_{1}$. The
issue is usually to find the "right" topology, meaning a collection of open
subsets which is not too large, but large enough to bring interesting properties.

\paragraph{Closed subsets\newline}

\begin{definition}
A subset A of a topological space $\left(  E,\Omega\right)  $ is
\textbf{closed} if $A^{c}$ is open.
\end{definition}

\begin{theorem}
In a topological space :

$\varnothing,E$ are closed,

any intersection of closed subsets is closed,

any \textit{finite union} of closed subsets is closed.
\end{theorem}

A topology can be similarly defined by a collection of closed subsets.

\paragraph{Relative topology\newline}

\begin{definition}
If X is a subset of the topological space $\left(  E,\Omega\right)  $ the
\textbf{relative topology} (or induced topology) in X inherited from E is
defined by taking as open subsets of X : $\Omega_{X}=\left\{  O\cap
X,O\in\Omega\right\}  $. Then $\left(  X,\Omega_{X}\right)  $ is a topological
space, and the subsets of $\Omega_{X}$ are said to be \textbf{relatively open}
in X.
\end{definition}

But they are not necessarily open in E : indeed X can be any subset and one
cannot know if $O\cap X$ is open or not in E.

\subsubsection{Neighborhood}

Topology is the natural way to define what is "close" to a point.

\bigskip

\paragraph{Neighborhood\newline}

\begin{definition}
A \textbf{neighborhood} of a point x in a topological space (E,$\Omega)$ is a
subset n(x) of E which contains an open subset containing x:

$\exists O\in\Omega:O\subset n(x),x\in O$
\end{definition}

Indeed a neighborhood is just a convenient, and abbreviated, way to say : "a
subset which contains on open subset which contains x".

\begin{notation}
n(x) is a neighborhood of a point x of the topological space (E,$\Omega)$
\end{notation}

\begin{definition}
A point x of a subset X in a topological space (E,$\Omega)$ is
\textbf{isolated} in X if there is a neighborhood n(x) of x such that
$n\left(  x\right)  \cap X=\left\{  x\right\}  $
\end{definition}

\bigskip

\paragraph{Interior, exterior\newline}

\begin{definition}
A point x is an \textbf{interior point} of a subset $X$ of the topological
space (E,$\Omega)$\ if X is a neighborhood of x. The \textbf{interior
}$\overset{\circ}{X}$ of X is the set of its interior points, or
equivalently,the largest open subset contained in X (the union of all open
sets contained in X) . The \textbf{exterior }$\overset{\circ}{\left(
X^{c}\right)  }$ of X is the interior of its complement, or equivalently,the
largest open subset which does not intersect X (the union of all open sets
which do not intersect X)
\end{definition}

\begin{notation}
$\overset{\circ}{X}$ is the interior of the set X
\end{notation}

\begin{theorem}
$\overset{\circ}{X}$ is an open subset : $\overset{\circ}{X}\sqsubseteq X$ and
$\overset{\circ}{X}=X$ iff X is open.
\end{theorem}

\bigskip

\paragraph{Closure\newline}

\begin{definition}
A point x is \textbf{adherent} to a subset $X$ of the topological space
(E,$\Omega)$ if each of its neighborhoods meets X. The \textbf{closure
}$\overline{X}$ of X is the set of the points which are adherent to X or,
equivalently, the smallest closed subset which contains X (the intersection of
all closed subsets which contains X)
\end{definition}

\begin{notation}
$\overline{X}$ is the closure of the subset X
\end{notation}

\begin{theorem}
$\overline{X}$ is a closed subset : $X\sqsubseteq\overline{X}$ and
$\overline{X}=X$ iff X is closed.
\end{theorem}

\begin{definition}
A subset X of the topological space $\left(  E,\Omega\right)  $ is
\textbf{dense} in E if its closure is E : $\overline{X}=E$
\end{definition}

$\Leftrightarrow\forall\varpi\in\Omega,\varpi\cap X\neq\varnothing$

\bigskip

\paragraph{Border\newline}

\begin{definition}
A point x is a \textbf{boundary point} of a subset X of the topological space
(E,$\Omega)$ if each of its neighborhoods meets both X and X$^{c}.$The
\textbf{border} $\partial X\ $\ of X is the set of its boundary points.
\end{definition}

\begin{notation}
$\partial X$ is the border (or boundary) of the set X
\end{notation}

Another common notation is $\overset{\cdot}{X}=\partial X$

\begin{theorem}
$\partial X$ is a closed subset
\end{theorem}

\bigskip

The relation between interior, border, exterior and closure is summed up in
the following theorem:

\begin{theorem}
If X is a subset of a topological space (E,$\Omega)$\ then :

$\overline{X}=\overset{\circ}{X}\cup\partial X=\left(  \overset{\circ}{\left(
X^{c}\right)  }\right)  ^{c}$

$\overset{\circ}{X}\cap\partial X=\varnothing$

$\overset{\circ}{\left(  X^{c}\right)  }\cap\partial X=\varnothing$

$\partial X=\overline{X}\cap\overline{\left(  X^{c}\right)  }=\partial\left(
X^{c}\right)  $
\end{theorem}

\subsubsection{Base of a topology}

A topology is not necessarily defined by a family of subsets. The base of a
topology is just a way to define a topology through a family of subsets, and
it gives the possibility to precise the thinness of the topology by the
cardinality of the family.

\paragraph{Base of a topology\newline}

\begin{definition}
A \textbf{base} of a topological space $\left(  E,\Omega\right)  $ is a family
$\left(  B_{i}\right)  _{i\in I}$ of subsets of E such that : $\forall
O\in\Omega,\exists J\subset I:O=\cup_{j\in J}B_{j}$
\end{definition}

\begin{theorem}
(Gamelin p.70) A family $\left(  B_{i}\right)  _{i\in I}$\ of subsets of E is
a base of the topological space $\left(  E,\Omega\right)  $ iff

$\forall x\in E,\exists i\in I:x\in B_{i}$

$\forall i,j\in I:x\in B_{i}\cap B_{j}\Rightarrow\exists k\in I:x\in
B_{k},B_{k}\subset B_{i}\cap B_{j}$
\end{theorem}

\begin{theorem}
(Gamelin p.70) A family $\left(  B_{i}\right)  _{i\in I}$\ of open subsets of
$\Omega$ is a base of the topological space $\left(  E,\Omega\right)  $ iff

$\forall x\in E,\forall n\left(  x\right)  $ neighborhood of x, $\exists i\in
I:x\in B_{i},B_{i}\subset n(x)$
\end{theorem}

\paragraph{Countable spaces\newline}

The word "countable" in the following can lead to some misunderstanding. It
does not refer to the number of elements of the topological space but to the
cardinality of a base used to define the open subsets. It is clear that a
topology is stronger if it has more open subsets, but too many opens make
difficult to deal with them.\ Usually the "right size" is a countable base.

\bigskip

\subparagraph{Basic definitions\newline}

\begin{definition}
A topological space is

\textbf{first countable} if each of its points has a neighborhood with a
countable base.

\textbf{second countable} if it has a countable base.
\end{definition}

Second countable$\Rightarrow$First countable

In a second countable topological space there is a family $\left(
B_{n}\right)  _{n\in%
\mathbb{N}
}$ of subsets which gives, by union and finite intersection, all the open
subsets of $\Omega.$

\bigskip

\subparagraph{Open cover\newline}

The "countable" property appears quite often through the use of open covers,
where it is useful to restrict their size.

\begin{definition}
An \textbf{open cover} of a topological space $\left(  E,\Omega\right)  $ is a family

$\left(  O_{i}\right)  _{i\in I},O_{i}\subset\Omega$\ of open subsets whose
union is E. A \textbf{subcover} is a subfamily of an open cover which is still
an open cover. A \textbf{refinement} of an open cover is a family $\left(
F_{j}\right)  _{j\in J}$\ of subsets of E whose union is E and such that each
member is contained in one of the subset of the cover : $\forall j\in
J,\exists i\in I:F_{j}\sqsubseteq O_{i}$
\end{definition}

\begin{theorem}
Lindel\"{o}f (Gamelin p.71) If a topological space is second countable then
every open cover has a countable open subcover.
\end{theorem}

\bigskip

\subparagraph{Separable space\newline}

\begin{definition}
A topological space $\left(  E,\Omega\right)  $ is \textbf{separable} if there
is a countable subset of E which is dense in E.
\end{definition}

\begin{theorem}
(Gamelin p.71) A second countable topological space is separable
\end{theorem}

When a subset is dense it is often possible to extend a property to the set.

\subsubsection{Separation}

It is useful to have not too many open subsets, but it is also necessary to
have not too few in order to be able to "distinguish" points. They are
different definitions of this concept.\ They are often labeled by a T from the
german "Trennung"=separation. By far the most common is the "Hausdorff" property.

\begin{definition}
(Gamelin p.73) A topological space $\left(  E,\Omega\right)  $ is

\textbf{Hausdorff} (or T2) if for any pair x,y of distinct points of E there
are open subsets O,O' such that $x\in O,y\in O^{\prime},O\cap O^{\prime
}=\varnothing$

\textbf{regular} if for any pair of a closed subset X and a point $y\notin X$
there are open subsets O,O' such that $X\subset O,y\in O^{\prime},O\cap
O^{\prime}=\varnothing$

\textbf{normal} if for any pair of closed disjoint subsets X,Y $X\cap
Y=\varnothing$ there are open subsets O,O' such that $X\subset O,Y\subset
O^{\prime},O\cap O^{\prime}=\varnothing$

\textbf{T1} if a point is a closed set.

\textbf{T3} if it is T1 and regular

\textbf{T4} if it is T1 and normal
\end{definition}

The definitions for regular and normal can vary in the litterature (but
Hausdorff is standard). See Wilansky p.46 for more.

\begin{theorem}
(Gamelin p.73) T4 $\Rightarrow$ T3 $\Rightarrow$ T2 $\Rightarrow$ T1
\end{theorem}

\begin{theorem}
(Gamelin p.74) A topological space $\left(  E,\Omega\right)  $ is normal iff
for any closed subset X and open set O containing X there is an open subset O'
such that $\overline{O^{\prime}}\subset O$ and $X\subset O^{\prime}$
\end{theorem}

\begin{theorem}
(Thill p.84) A topological space is regular iff it is homeomorphic to a
subspace of a compact Hausdorff space
\end{theorem}

\begin{theorem}
Urysohn (Gamelin p.75): If X,Y are disjoint subsets of a normal topological
space $\left(  E,\Omega\right)  $\ there is a continuous function
$f:E\rightarrow\left[  0,1\right]  \subset%
\mathbb{R}
$ such that f(x)=0 on X and f(x)=1 on Y.
\end{theorem}

\begin{theorem}
Tietze (Gamelin p.76): If F is a closed subset of a normal topological space
$\left(  E,\Omega\right)  $ $\varphi:F\rightarrow%
\mathbb{R}
$ bounded continuous, then there is $\Phi:E\rightarrow%
\mathbb{R}
$ bounded continuous such that $\Phi=\varphi$ over F.
\end{theorem}

\bigskip

Remarks :

1) "separable" is a concept which is not related to separation (see base of a topology).

2) it could seem strange to consider non Hausdorff space.\ In fact usually
this is the converse which happens : one wishes to consider as "equal" two
different objects which share basic properties (for instance functions which
are almost everywhere equal) : thus one looks for a topology that does not
distinguish these objects.\ Another classic solution is to build a quotient
space through an equivalence relation.

\subsubsection{Compact}

Compact is a topological mean to say that a set is not "too large". The other
useful concept is locally compact, which means that "bounded" subsets are compact.

\begin{definition}
A topological space $\left(  E,\Omega\right)  $ is :

\textbf{compact} if for any open cover there is a finite open subcover.\ 

\textbf{countably} \textbf{compact} if for any countable open cover there is a
finite open subcover.

\textbf{locally} \textbf{compact} if each point has a compact neighborhood

\textbf{compactly generated} if a subset X of E is closed in E iff $X\cap K$
is closed for any compact K in E. We have the equivalent for open subsets.
\end{definition}

In a second countable space an open cover has a countable subcover
(Lindel\"{o}f theorem). Here it is finite.

\begin{definition}
A subset X of topological space $\left(  E,\Omega\right)  $ is :

\textbf{compact in E} if for any open cover of X there is a finite subcover of X

\textbf{relatively compact} if its closure is compact
\end{definition}

\begin{definition}
A \textbf{Baire space} is a topological space where the intersection of any
sequence of dense subsets is dense
\end{definition}

\bigskip

Compact $\Rightarrow$ countably compact

Compact $\Rightarrow$ locally compact

Compact, locally compact, first countable spaces are compactly generated.

\begin{theorem}
(Gamelin p.83) Any compact topological space is locally compact. Any discrete
set is locally compact.\ Any non empty open subset of $%
\mathbb{R}
^{n}$ is locally compact.
\end{theorem}

\begin{theorem}
(Gamelin p.79) Any finite union of compact subsets is compact.
\end{theorem}

\begin{theorem}
(Gamelin p.79) A closed subset of a compact topological space is compact.
\end{theorem}

\begin{theorem}
(Wilansky p.81) A topological space $\left(  E,\Omega\right)  $ is compact iff
for any family $\left(  X_{i}\right)  _{i\in I}$ of subsets for wich
$\cap_{i\in I}X_{i}=\varnothing$ there is a finite subfamily J for which
$\cap_{i\in J}X_{i}=\varnothing$
\end{theorem}

\begin{theorem}
(Wilansky p.82) The image of a compact subset by a continuous map is compact
\end{theorem}

\begin{theorem}
(Gamelin p.80) If X is a compact subset of a Hausdorff topological space
$\left(  E,\Omega\right)  $ :

- X is closed

- $\forall y\notin X$ there are open subsets O,O' such that : $y\in O,X\subset
O^{\prime},O\cap O^{\prime}=\varnothing$
\end{theorem}

\begin{theorem}
(Wilansky p.83) A compact Hausdorff space is normal and regular.
\end{theorem}

\begin{theorem}
(Gamelin p.85) A locally compact Hausdorff space is regular
\end{theorem}

\begin{theorem}
(Wilansky p.180) A locally compact, regular topological space is a Baire space
\end{theorem}

\paragraph{Compactification\newline}

(Gamelin p.84)

This is a general method to build a compact space from a locally compact
Hausdorff topological space $\left(  E,\Omega\right)  $.\ Define
$F=E\cup\left\{  \infty\right\}  $ where $\infty$ is any point (not in
E).\ There is a unique topology for F such that F is compact and the topology
inherited in E from F coincides with the topology of E.\ The open subsets O in
F are either open subsets of E or O such that $\infty\in O$ and E%
$\backslash$%
O is compact in E.

\subsubsection{Paracompact spaces}

The most important property of paracompact spaces is that they admit a
\textbf{partition of unity}, with which it is possible to extend local
constructions to global constructions over E.\ This is a mandatory tool in
differential geometry.

\begin{definition}
A family $\left(  X_{i}\right)  _{i\in I}$ of subsets of a topological space
$\left(  E,\Omega\right)  $ is :

\textbf{locally finite} if every point has a neighborhood which intersects
only finitely many elements of the family.

$\sigma$\textbf{-locally finite }if it is the union of countably many locally
finite families.
\end{definition}

\begin{definition}
A topological space $\left(  E,\Omega\right)  $ is \textbf{paracompact} if any
open cover of E \ has a refinement which is locally finite.
\end{definition}

\begin{theorem}
(Wilansky p.191) The union of a locally finite family of closed sets is closed
\end{theorem}

\begin{theorem}
(Bourbaki) Every compact space is paracompact.\ Every closed subspace of a
paracompact space is paracompact.
\end{theorem}

\begin{theorem}
(Bourbaki) Every paracompact space is normal
\end{theorem}

Warning ! an infinite dimensional Banach space may not be paracompact

\begin{theorem}
(Nakahara p.206) For any paracompact Hausdorff topological space $\left(
E,\Omega\right)  $ and open cover $\left(  O_{i}\right)  _{i\in I},$ there is
a family $\left(  f_{j}\right)  _{j\in J}$ of continuous functions
$f_{j}:E\rightarrow\left[  0,1\right]  \subset%
\mathbb{R}
$ such that :

- $\forall j\in J,\exists i\in I:$ support$\left(  f_{i}\right)  \subset
O_{j}$

- $\forall x\in E,\exists n\left(  x\right)  ,\exists K\subset
J,card(K)<\infty:\forall y\in n\left(  x\right)  :$

$\forall j\in J\backslash K:f_{j}\left(  y\right)  =0,\sum_{j\in K}%
f_{j}\left(  y\right)  =1$
\end{theorem}

\subsubsection{Connected space}

Connectedness is related to the concepts of "broken into several parts". This
is a global property, which is involved in many theorems about unicity of a result.

\begin{definition}
(Schwartz I p.87) A topological space $\left(  E,\Omega\right)  $ is
\textbf{connected} if it does not admit a partition into two subsets (other
than E and $\varnothing)$ which are both closed or both open, or equivalently
if there are no subspace (other than E and $\varnothing)$\ which are both
closed and open. A subset X of a topological space $\left(  E,\Omega\right)  $
is connected if it is connected in the induced topology.
\end{definition}

So if X is not connected (say disconnected) in E if there are two subspaces of
E, both open or closed in E, such that $X=\left(  X\cap A\right)  \cup\left(
X\cap B\right)  ,A\cap B\cap X=\varnothing.$

\begin{definition}
A topological space $\left(  E,\Omega\right)  $ is \textbf{locally connected}
if for each point x and each open subset O\ which contains x, there is a
connected open subset O' such that $x\in O^{\prime},O^{\prime}\sqsubseteq O $\ .
\end{definition}

\begin{definition}
The \textbf{connected component} C(x) of a point x of E is the \textit{union}
of all the connected subsets which contains x.
\end{definition}

It is the largest connected subset of E which contains x. So $x\sim y$ if
$C\left(  x\right)  =C\left(  y\right)  $ is a relation of equivalence which
defines a partition of E.\ The classes of equivalence of this relation are the
connected components of E. They are disjoint, connected subsets of E and their
union is E. Notice that the components are not necessarily open or closed. If
E is connected it has only one component.

\begin{theorem}
The only connected subsets of $%
\mathbb{R}
$\ are the intervals

$\left[  a,b\right]  ,]a,b[,[a,b[,]a,b]$ (usually denoted $\left\vert
a,b\right\vert )$ a and b can be $\pm\infty$
\end{theorem}

\begin{theorem}
(Gamelin p.86) It $\left(  X_{i}\right)  _{i\in I}$ is a family of connected
subsets of a topological space $\left(  E,\Omega\right)  $\ such that $\forall
i,j\in I:X_{i}\cap X_{j}\neq\varnothing$ then $\cup_{i\in I}X_{i}$\ is connected
\end{theorem}

\begin{theorem}
(Gamelin p.86) The image of a connected subset by a continuous map is connected
\end{theorem}

\begin{theorem}
(Wilansky p.70) If X is connected in E, then its closure $\overline{X}$ is
connected in E
\end{theorem}

\begin{theorem}
(Schwartz I p.91) If X is a connected subset of a topological space $\left(
E,\Omega\right)  $ , Y a subset of E such that $X\cap\overset{\circ}{Y}%
\neq\varnothing$ and $X\cap\left(  \overline{Y}\right)  ^{c}\neq\varnothing$
then $X\cap\partial Y\neq\varnothing$
\end{theorem}

\begin{theorem}
(Gamelin p.88) Each connected component of a topological space is closed. Each
connected component of a locally connected space is both open and closed.
\end{theorem}

\subsubsection{Path connectedness}

Path connectedness is a stronger form of connectedness.

\begin{definition}
A \textbf{path }on a topological space E is a continuous map : $c:J\rightarrow
E$ from a connected subset J of $%
\mathbb{R}
$\ to E. The codomain C=$\left\{  c(t),t\in J\right\}  $ of c is a subset of
E, which is a \textbf{curve}.
\end{definition}

The same curve can be described using different paths, called
\textbf{parametrisation}. Take $f:J^{\prime}\rightarrow J$ where J' is another
interval of $%
\mathbb{R}
$\ and f is any bijective continuous map, then : $c^{\prime}=c\circ
f:J^{\prime}\rightarrow E$ is another path with image C.

A\ path from a point x of E to a point y of E is a path such that $x\in C,y\in
C$

\begin{definition}
Two points x,y of a topological space $\left(  E,\Omega\right)  $ are
\textbf{path-connected} (or arc-connected) if there is a path from x to y.

A subset X of E is path-connected if any pair of its points are path-connected.

The \textbf{path-connected component} of a point x of E is the set of the
points of E which are path-connected to x.
\end{definition}

$x\sim y$ if x and y are path-connected is a relation of equivalence which
defines a partition of E.\ The classes of equivalence of this relation are the
path-connected components of E.

\begin{definition}
A topological space $\left(  E,\Omega\right)  $ is \textbf{locally
path-connected} if for any point x and neighborhood n(x) of x there is a
neighborhood n'(x) included in n(x) which is path-connected.
\end{definition}

\begin{theorem}
(Schwartz I p.91) if X is a subset of a topological space $\left(
E,\Omega\right)  $, any path from $a\in\overset{\circ}{X}$ to $b\in
\overset{\circ}{\left(  X^{c}\right)  }$ meets $\partial X$
\end{theorem}

\begin{theorem}
(Gamelin p.90) If a subset X of a topological space $\left(  E,\Omega\right)
$ is path connected then it is connected.
\end{theorem}

\begin{theorem}
(Gamelin p.90) Each connected component of a topological space $\left(
E,\Omega\right)  $ is the union of path-connected components of E.
\end{theorem}

\begin{theorem}
(Schwartz I p.97) A path-connected topological space is locally
path-connected. A connected, locally path-connected topological space, is
path-connected. The connected components of a locally path-connected
topological space are both open and closed, and are path connected.
\end{theorem}

\subsubsection{Limit of a sequence}

\begin{definition}
A point $x\in E$ is an \textbf{accumulation point} (or cluster) of the
sequence $\left(  x_{n}\right)  _{n\in%
\mathbb{N}
}$ in the topological space $\left(  E,\Omega\right)  $ if for any
neighborhood n(x) and any N there is $p>N$\ such that $x_{p}\in n\left(
x\right)  $
\end{definition}

A neighborhood of x contains infinitely many $x_{n}$

\begin{definition}
A point $x\in E$ is a \textbf{limit} of the sequence $\left(  x_{n}\right)
_{n\in%
\mathbb{N}
}$ in the topological space $\left(  E,\Omega\right)  $ if for any
neighborhood n(x) of x there is N such that $\forall n\geq N:x_{n}\in n\left(
x\right)  .$ Then $\left(  x_{n}\right)  _{n\in%
\mathbb{N}
}$ \textbf{converges} to x and this property is denoted $x=\lim_{n\rightarrow
\infty}x_{n}$.
\end{definition}

There is a neighborhood of x which contains \textit{all} the $x_{n}$ for n%
$>$%
N

\begin{definition}
A sequence $\left(  x_{n}\right)  _{n\in%
\mathbb{N}
}$ in the topological space $\left(  E,\Omega\right)  $ is \textbf{convergent}
if it admits at least one limit.
\end{definition}

So a limit is an accumulation point, but the converse is not aways true. And a
limit is not necessarily unique.

\begin{theorem}
(Wilansky p.47) The limit of a convergent sequence in a Hausdorff topological
space $\left(  E,\Omega\right)  $ is unique. Conversely if the limit of any
convergent sequence in a topological space $\left(  E,\Omega\right)  $ is
unique then $\left(  E,\Omega\right)  $\ is Hausdorff.
\end{theorem}

\begin{theorem}
(Wilansky p.27) The limit (s) of a convergent sequence $\left(  x_{n}\right)
_{n\in%
\mathbb{N}
}$ in the subset X of a topological space $\left(  E,\Omega\right)  $ belong
to the closure of X :

$\forall\left(  x_{n}\right)  _{n\in%
\mathbb{N}
}\in X^{%
\mathbb{N}
}:$ $\lim_{n\rightarrow\infty}x_{n}\in\overline{X}$

Conversely if the topological space $\left(  E,\Omega\right)  $ is
first-countable then any point adherent to a subset X of E is the limit of a
sequence in X.
\end{theorem}

As a consequence :

\begin{theorem}
A subset X of the topological space $\left(  E,\Omega\right)  $ is closed if
the limit of any convergent sequence in X belongs to X
\end{theorem}

This is the usual way to prove that a subset is closed. Notice that the
condition is sufficient and not necessary if E is not first countable.

\begin{theorem}
Weierstrass-Bolzano (Schwartz I p.75): In a compact topological space every
sequence has a accumulation point.
\end{theorem}

\begin{theorem}
(Schwartz I p.77) A sequence in a compact topological space converges to a iff
a is its unique accumulation point.
\end{theorem}

\subsubsection{Product topology}

\begin{definition}
(Gamelin p.100) If $\left(  E_{i},\Omega_{i}\right)  _{i\in I}$ is a family of
topological spaces, the \textbf{product topology} on $E=%
{\textstyle\prod\limits_{i\in I}}
E_{i}$ is defined by the collection of open sets :

If I is finite : $\Omega=%
{\textstyle\prod\limits_{i\in I}}
\Omega_{i}$

If I is infinite : $\Omega=%
{\textstyle\prod\limits_{i\in I}}
\varpi_{i}$ , $\varpi_{i}\subset E_{i}$ such that $\exists J$ finite $\subset
I:i\in J:\varpi_{i}\subset\Omega_{i}$
\end{definition}

So the open sets of E are the product of a finite number of open sets, and the
other components are any subsets.

The projections are the maps : $\pi_{i}:E\rightarrow E_{i}$

The product topology is the smallest $\Omega$\ for which the projections are
continuous maps

\begin{theorem}
(Gamelin p.100-103) If $\left(  E_{i}\right)  _{i\in I}$ is a family of
topological spaces $\left(  E_{i},\Omega_{i}\right)  $ and $E=%
{\textstyle\prod\limits_{i\in I}}
E_{i}$ their product endowed with the product topology, then:

i) E is Hausdorff iff the $\left(  E_{i},\Omega_{i}\right)  $ are Hausdorff

ii) E is connected iff the $\left(  E_{i},\Omega_{i}\right)  $ are connected

iii) E is compact iff the $\left(  E_{i},\Omega_{i}\right)  $ are compact
(Tychonoff's theorem)

iv) If I is finite, then E is regular iff the $\left(  E_{i},\Omega
_{i}\right)  $ are regular

v) If I is finite, then E is normal iff the $\left(  E_{i},\Omega_{i}\right)
$ are normal

vi) If I is finite, then a sequence in E is convergent iff each of its
components is convergent

vii) If I is finite and the $\left(  E_{i},\Omega_{i}\right)  $ are secound
countable then E is secound countable
\end{theorem}

\begin{theorem}
(Wilansky p.101) An uncountable product of non discrete space cannot be first countable.
\end{theorem}

Remark : the topology defined by taking only products of open subsets in all
$E_{i}$ (called the box topology) gives too many open sets if I is infinite
and the previous results are no longer true.

\subsubsection{Quotient topology}

An equivalence relation on a space E is just a partition of E, and the
quotient set E'=E/$\sim$ the set of of its classes of equivalence (so each
element is itself a subset). The key point is that E' is not necessarily
Hausdorff, and it happens only if the classes of equivalence are closed
subsets of E.

\begin{definition}
(Gamelin p.105) Let $\left(  E,\Omega\right)  $ be a topological space, $\sim$
and an equivalence relation on E, $\pi:E\rightarrow E^{\prime}$ the projection
on the quotient set E'=E/$\sim$. The \textbf{quotient topology} on E' is
defined by taking as open sets $\Omega^{\prime}$ in E' : $\Omega^{\prime
}=\left\{  O^{\prime}\subset E^{\prime}:\pi^{-1}\left(  O^{\prime}\right)
\in\Omega\right\}  $
\end{definition}

So $\pi$\ is continuous and this is the coarsest (meaning the smallest
$\Omega^{\prime})$\ topology for which $\pi$\ is continuous.\ 

The quotient topology is the final topology with respect to the projection
(see below).

\begin{theorem}
(Gamelin p.107) The quotient set E' of a topological space (E,$\Omega)$
endowed with the quotient topology is :

i) connected if E is connected

ii) path-connected if E is path-connected

iii) compact if E is compact

iv) Hausdorff iff E is Hausdorff and each equivalence class is closed in E
\end{theorem}

The property iv) is used quite often.

\begin{theorem}
(Gamelin p.105) Let $\left(  E,\Omega\right)  $ be a topological space,
E'=E/$\sim$ the quotient set endowed with the quotient topology,
$\pi:E\rightarrow E^{\prime}$\ the projection, F a topological space

i) a map $\varphi:E^{\prime}\rightarrow F$ \ is continuous iff $\varphi
\circ\pi$ is continuous

ii) If a continuous map $f:E\rightarrow F$ \ is such that f is constant on
each equivalence class, then there is a continuous map : $\varphi:E^{\prime
}\rightarrow F$ such that $f=\varphi\circ\pi$
\end{theorem}

A map $f:E\rightarrow F$ \ is called a quotient map if F is endowed with the
quotient topology (Wilansky p.103).

Let $f:E\rightarrow F$ be a continuous map between compact, Hausdorff,
topological spaces E,F. Then $a\sim b\Leftrightarrow f\left(  a\right)
=f\left(  b\right)  $ is an equivalence relation over E and E/$\sim$ is
homeomorphic to F.

\bigskip

\subsection{Maps on topological spaces}

\label{Continuous maps}

\subsubsection{Support of a function}

\begin{definition}
The \textbf{support} of the function $f:E\rightarrow K$ from a topological
space (E,$\Omega)$ to a field K is the subset of E : Supp(f)=$\overline
{\left\{  x\in E:f(x)\neq0\right\}  }$ or equivalently the complement of the
largest open set where f(x) is zero.
\end{definition}

This is a closed subset of the domain of f

\begin{notation}
Supp(f) is the support of the function f.
\end{notation}

Warning ! f(x) can be zero in the support, it is necessarily zero outside the support.

\subsubsection{Continuous map}

\paragraph{Definitions\newline}

\begin{definition}
A map $f:E\rightarrow F$ between two topological spaces $\left(
E,\Omega\right)  ,\left(  F,\Omega^{\prime}\right)  :$

i) \textbf{converges} to $b\in F$ when x converges to $a\in E$ if for any open
O' in F such that $b\in O^{\prime}$ there is an open O in E such that $a\in O$
and $\forall x\in O:f\left(  x\right)  \in O^{\prime}$

ii).is \textbf{continuous in }$a\in E$ if for any open O' in F such that
$f\left(  a\right)  \in O^{\prime}$ there is an open O in E such that $a\in O$
and $\forall x\in O:f\left(  x\right)  \in O^{\prime}$

iii) is \textbf{continuous over a subset X of E} if it is continuous in any
point of X
\end{definition}

f converges to a is denoted : $f(x)\rightarrow b$ when $x\rightarrow a$ or
equivalently : $\lim_{x\rightarrow a}f\left(  x\right)  =b$

if f is continuous in a, it converges towards b = f(a), and conversely if f
converges towards b then one can define by continuity f in a by f(a) = b.

\begin{notation}
$C_{0}\left(  E;F\right)  $\ is the set of continuous maps from E to F
\end{notation}

\begin{definition}
A map $f:X\rightarrow F$ from a closed subset X of a topological space E to a
topological space F is \textbf{semi-continuous} in $a\in\partial X$ if, for
any open O' in F such that $f\left(  a\right)  \in O^{\prime}$ , there is an
open O in E such that $a\in O$ and $\forall x\in O\cap X:f\left(  x\right)
\in O^{\prime}$
\end{definition}

Which is, in the language of topology, the usual $f\rightarrow b$ when
$x\rightarrow a_{+}$

\begin{definition}
A function $f:E\rightarrow%
\mathbb{C}
$ \ from a topological space $\left(  E,\Omega\right)  $ to $%
\mathbb{C}
$\ \textbf{vanishes at infinity} if : $\forall\varepsilon>0,\exists K$ compact
$:\forall x\in K:\left\vert f(x)\right\vert <\varepsilon$
\end{definition}

Which is, in the language of topology, the usual $f\rightarrow0$ when
$x\rightarrow\infty$

\paragraph{Properties of continuous maps\newline}

\begin{theorem}
The composition of continuous maps is a continuous map
\end{theorem}

if $f:E\rightarrow F$ , $g:F\rightarrow G$ then $g\circ f$ is continuous.

\begin{theorem}
The topological spaces and continuous maps constitute a category
\end{theorem}

\begin{theorem}
If the map $f:E\rightarrow F$ between two topological spaces is continuous in
a, then for any sequence $\left(  x_{n}\right)  _{n\in%
\mathbb{N}
}$\ in E which converges to a : $f(x_{n})\rightarrow f(a)$

The converse is true only if E is first countable.\ Then f is continuous in a
iff for any sequence $\left(  x_{n}\right)  _{n\in%
\mathbb{N}
}$\ in E which converges to a : $f(x_{n})\rightarrow f(a).$
\end{theorem}

\begin{theorem}
The map $f:E\rightarrow F$ between two topological spaces is
\textbf{continuous over E} if the preimage of any open subset of F is an open
subset of E : $\forall O^{\prime}\in\Omega^{\prime},f^{-1}\left(  O^{\prime
}\right)  \in\Omega$
\end{theorem}

This a fundamental property of continuous maps.

\begin{theorem}
If the map $f:E\rightarrow F$ between two topological spaces $\left(
E,\Omega\right)  ,\left(  F,\Omega^{\prime}\right)  $ is continuous over E
then :

i) if $X\subset E$ is compact in E, then f(X) is a compact in F

ii) if $X\subset E$ is connected in E, then f(X) is connected in F

iii) if $X\subset E$ is path-connected in E, then f(X) is path-connected in F

iv) if E is separable, then f(E) is separable

v) if Y is open in F, then $f^{-1}\left(  Y\right)  $\ is open in E

vi) if Y is closed in F, then $f^{-1}\left(  Y\right)  $\ is closed in E

vii) if X is dense in E and f surjective, then f(X) is dense in F

viii) the graph of $f=\left\{  \left(  x,f(x)\right)  ,x\in E\right\}  $ is
closed in E$\times$F
\end{theorem}

\begin{theorem}
If $f\in C_{0}\left(  E;%
\mathbb{R}
\right)  $ and E is a non empty, compact topological space, then f has a
maximum and a minimum.
\end{theorem}

\begin{theorem}
(Wilansky p.57) If $f,g\in C_{0}\left(  E;F\right)  $ E,F Hausdorff
topological spaces and f(x) = g(x) for any x in a dense subset X of E, then f
= g in E.
\end{theorem}

\begin{theorem}
(Gamelin p.100) If $\left(  E_{i}\right)  _{i\in I}$ is a family of
topological spaces $\left(  E_{i},\Omega_{i}\right)  $ and $E=%
{\textstyle\prod\limits_{i\in I}}
E_{i}$ their product endowed with the product topology, then:

i) The projections : $\pi_{i}:E\rightarrow E_{i}$ are continuous

ii) If F is a topological space, a map $\varphi:E\rightarrow F$ is continuous
iff $\forall i\in I,\pi_{i}\circ\varphi$ is continuous
\end{theorem}

\begin{theorem}
(Wilansky p.53) A map $f:E\rightarrow F$ between two topological spaces E,F is
continuous iff $\ \forall X\subset E:f\left(  \overline{X}\right)
\subset\overline{f\left(  X\right)  }$
\end{theorem}

\begin{theorem}
(Wilansky p.57) The characteristic function of a subset which is both open and
closed is continuous
\end{theorem}

\paragraph{Algebraic topological spaces\newline}

Whenever there is some algebraic structure on a set E, and a topology on E,
the two structures are said to be consistent if the operations defined over E
in the algebraic structure are continuous. So we have topological groups,
topological vector spaces,...which themselves define Categories.

Example : a group $\left(  G,\cdot\right)  $ is a topological group if :
$\cdot:G\times G\rightarrow G,$ $G\rightarrow G::g^{-1}$\ are continuous

\subsubsection{Topologies defined by maps}

\paragraph{Compact-open topology\newline}

\begin{definition}
(Husemoller p.4) The \textbf{compact-open topology} on the set $C_{0}(E;F)$ of
all continuous maps between two topological spaces $\left(  E,\Omega\right)  $
and $\left(  F,\Omega^{\prime}\right)  $ is defined by the base of open
subsets : $\left\{  \varphi:\varphi\in C_{0}(E;F),\varphi\left(  K\right)
\subset O^{\prime}\right\}  $ where K is a compact subset of E and O' an open
subset of F.
\end{definition}

\paragraph{Weak, final topology\newline}

This is the implementation in topology of a usual mathematical trick : to pull
back or to push forward a structure from a space to another. These two
procedures are inverse from each other.\ They are common in functional analysis.

\begin{definition}
Let E be a set, $\Phi$ a family $\left(  \varphi_{i}\right)  _{i\in I}$\ of
maps : $\varphi_{i}:E\rightarrow F_{i}$ where $\left(  F_{i},\Omega
_{i}\right)  $\ are topological spaces. The \textbf{weak topology} on E with
respect to $\Phi$\ is defined by the collection of open subsets in E :
$\Omega=\cup_{i\in I}\left\{  \varphi_{i}^{-1}\left(  \varpi_{i}\right)
,\varpi_{i}\in\Omega_{i}\right\}  $
\end{definition}

So the topology on $\left(  F_{i}\right)  _{i\in I}$ is "pulled-back" on E.

\begin{definition}
Let F be a set, $\Phi$ a family $\left(  \varphi_{i}\right)  _{i\in I}$\ of
maps : $\varphi_{i}:E_{i}\rightarrow F$ where $\left(  E_{i},\Omega
_{i}\right)  $\ are topological spaces. The \textbf{final topology} on F with
respect to $\Phi$\ is defined by the collection of open subsets in F :
$\Omega^{\prime}=\cup_{i\in I}\left\{  \varphi_{i}\left(  \varpi_{i}\right)
,\varpi_{i}\in\Omega_{i}\right\}  $
\end{definition}

So the topology on $\left(  E_{i}\right)  _{i\in I}$ is "pushed-forward" on F.

\bigskip

In both cases, this is the coarsest topology for which all the maps
$\varphi_{i}$ are continuous.

They have the universal property :

Weak topology : given a topological space G, a map $g:G\rightarrow E$ is
continuous iff all the maps $\varphi_{i}\circ g$ are continuous (Thill p.251)

Final topology : given a topological space G, a map $g:F\rightarrow G$ is
continuous iff all the maps $g\circ\varphi_{i}$ are continuous.

\begin{theorem}
(Thill p.251) If E is endowed by the weak topology induced by the family
$\left(  \varphi_{i}\right)  _{i\in I}$\ of maps : $\varphi_{i}:E\rightarrow
F_{i}$ , a sequence $\left(  x_{n}\right)  _{n\in%
\mathbb{N}
}$ in E converges to x iff $\forall i\in I:f_{i}\left(  x_{n}\right)
\rightarrow f\left(  x\right)  $
\end{theorem}

\begin{theorem}
(Wilansky p.94) The weak topology is Hausdorff iff $\Phi$ is separating over E.
\end{theorem}

Which means $\forall$x$\neq y,\exists i\in I:\varphi_{i}\left(  x\right)
\neq\varphi_{i}\left(  y\right)  $

\begin{theorem}
(Wilansky p.94) The weak topology is semi-metrizable if $\Phi$ is a sequence
of maps to semi-metrizable spaces. The weak topology is metrizable iff $\Phi$
is a sequence of maps to metrizable spaces which is separating over E
\end{theorem}

\subsubsection{Homeomorphism}

\begin{definition}
A \textbf{homeomorphism} is a bijective and continuous\ map $f:E\rightarrow F
$\ between two topological spaces E,F such that its inverse $f^{-1}$ is continuous.
\end{definition}

\begin{definition}
A \textbf{local} \textbf{homeomorphism} is a map $f:E\rightarrow F$\ between
two topological spaces E,F such that for each $a\in E$\ there is a
neighborhood n(a) and a neighborhood n(b) of b=f(a) and the restriction of
$f:n(a)\rightarrow n(b)$ is a homemomorphism.
\end{definition}

The homeomorphisms are the isomorphisms of the category of topological spaces.

\begin{definition}
Two topological spaces are \textbf{homeomorphic} if there is a homeomorphism
between them.
\end{definition}

Homeomorphic spaces share the same topological properties. Equivalently a
topological property is a property which is preserved by homeomorphism. Any
property than can be expressed in terms of open and closed sets is
topological. Examples : if E and F are homeomorphic, E is connected iff F is
connected, E is compact iff F is compact, E is Hausdorff iff F is Hausdorff,...

Warning ! this is true for a global homeomorphism, not a local homeomorphism

\begin{definition}
The topologies defined by the collections of open subsets $\Omega
,\Omega^{\prime}$ on the same set E are \textbf{equivalent} if there is an
homeomorphism between $\left(  E,\Omega\right)  $ and $\left(  E,\Omega
^{\prime}\right)  .$
\end{definition}

So, for all topological purposes, it is equivalent to take $\left(
E,\Omega\right)  $ or $\left(  E,\Omega^{\prime}\right)  $

\begin{theorem}
(Wilansky p.83) If $f\in C_{0}\left(  E;F\right)  $ is one to one, E compact,
F Hausdorff then f is a homeomorphism of E and f(E)
\end{theorem}

\begin{theorem}
(Wilansky p.68) Any two non empty convex open sets of $%
\mathbb{R}
^{m}$ are homeomorphic
\end{theorem}

\subsubsection{Open and closed maps}

It would be handy if the image of an open set by a map would be an open set,
but this is the contrary which happens with a continuous map. This leads to
the following definitions :

\begin{definition}
A map $f:E\rightarrow F$ between two topological spaces is :

an \textbf{open map}, if the image of an open subset is open

a \textbf{closed map}, if the image of a closed subset is closed
\end{definition}

The two properties are distinct : a map can be open and not closed (and vice versa).

Every homeomorphism is open and closed.

\begin{theorem}
(Wilansky p.58) A bijective map is open iff its inverse is continuous.
\end{theorem}

\begin{theorem}
The composition of two open maps is open; the composition of two closed maps
is closed.
\end{theorem}

\begin{theorem}
(Schwartz II p.190) A local homeomorphism is an open map.
\end{theorem}

\begin{theorem}
A map $f:E\rightarrow F$ between two topological spaces is :

open iff $\forall X\subset E:$ $f(\overset{\circ}{X})\subset\overset{\circ
}{\left(  f(X)\right)  }$

closed iff $\forall X\subset E:$ $\overline{f(X)}\subset f(\overline{X})$
\end{theorem}

\begin{theorem}
(Wilansky p.103) Any continuous open surjective map $f:E\rightarrow F$\ is a
quotient map. Any continuous closed surjective map $f:E\rightarrow F$\ is a
quotient map.
\end{theorem}

meaning that F has the quotient topology. They are the closest thing to a homeomorphism.

\begin{theorem}
(Thill p.253) If $f:E\rightarrow F$ is a continuous closed map from a compact
space E to a Hausdorff space, if f is injective it is an embedding, if is
bijective it is a homeomorphism.
\end{theorem}

\subsubsection{Proper maps}

This is the same purpose as above : remedy to the defect of continuous maps
that the image of a compact space is compact.

\begin{definition}
A map $f:E\rightarrow F$ between two topological spaces is a \textbf{proper
map} (also called a compact map) is the preimage of a compact subset of F is a
compact subset of E.
\end{definition}

\begin{theorem}
A continuous map $f\in C_{0}\left(  E;F\right)  $ is proper if it is a closed
map and the pre-image of every point in F is compact.
\end{theorem}

\begin{theorem}
Closed map lemma: Every continuous map $f\in C_{0}\left(  E;F\right)  $ from a
compact space E to a Hausdorff space F is closed and proper.
\end{theorem}

\begin{theorem}
A continuous function between locally compact Hausdorff spaces which is proper
is also closed.
\end{theorem}

\begin{theorem}
A topological space is compact iff the maps from that space to a single point
are proper.
\end{theorem}

\begin{theorem}
If $f\in C_{0}\left(  E;F\right)  $ is a proper continuous map and F is a
compactly generated Hausdorff space, then f is closed.
\end{theorem}

this includes Hausdorff spaces which are either first-countable or locally compact

\bigskip

\subsection{Metric and Semi-metric spaces}

\label{Metric spaces}

The existence of a metric on a set is an easy way to define a topology and,
indeed, this is still the way it is taught usually. Anyway a metric brings
more properties. Most of the properties extend to semi-metric.

\subsubsection{Metric and Semi-metric spaces}

\paragraph{Semi-metric, Metric\newline}

\begin{definition}
A \textbf{semi-metric} (or pseudometric\textbf{)} on a set E is a map :
$d:E\times E\rightarrow%
\mathbb{R}
$ which is symmetric, positive and such that :

d(x,x)=0, $\forall x,y,z\in E:d\left(  x,z\right)  \leq d\left(  x,y\right)
+d\left(  y,z\right)  $
\end{definition}

\begin{definition}
A \textbf{metric} on a set E is a definite positive semi-metric :

$d\left(  x,y\right)  =0\Leftrightarrow x=y$
\end{definition}

Examples :

i) on a real vector space a bilinear definite positive form defines a metric :
$d\left(  x,y\right)  =g\left(  x-y,x-y\right)  $

ii) a real affine space whose underlying vector space is endowed with a
bilinear definite positive form :

$d\left(  A,B\right)  =g\left(  \overrightarrow{AB},\overrightarrow
{AB},\right)  $

iii) on any set there is the discrete metric : d(x,y)=0 if x=y, d(x,y)=1 otherwise

\bigskip

\begin{definition}
If the set E is endowed with a semi-metric d:

a \textbf{Ball} is the set $B\left(  a,r\right)  =\left\{  x\in E:d\left(
a,x\right)  <r\right\}  $ with r
$>$
0

the \textbf{diameter} of a subset X of is diam = $\sup_{x,y\in X}d(x,y)$

the \textbf{distance} between a subset X and a point a is : $\delta\left(
a,X\right)  =\inf d(x,a)_{x\in X}$

the distance between 2 subsets X,Y is : $\delta(A,B)=\inf d(x,y)_{x\in X,y\in
Y}$
\end{definition}

\begin{definition}
If the set E is endowed with a semi-metric d, a subset X of E is:

\textbf{bounded} if $\exists R\subset%
\mathbb{R}
:\forall x,y\in X:d\left(  x,y\right)  \leq R\Leftrightarrow diam(X)<\infty$

\textbf{totally bounded} if $\forall r>0$ there is a finite number of balls of
radius r which cover X.\ 
\end{definition}

totally bounded $\Rightarrow$ bounded

\paragraph{Topology on a semi-metric space\newline}

One of the key differences between semi metric and metric spaces is that a
semi metric space is usually not Hausdorff.

\subparagraph{Topology\newline}

\begin{theorem}
A semi-metric on a set E induces a topology whose base are the open balls :
$B\left(  a,r\right)  =\left\{  x\in E:d\left(  a,x\right)  <r\right\}  $ with
r
$>$
0
\end{theorem}

The open subsets of E are generated by the balls, through union and finite intersection.

\begin{definition}
A \textbf{semi-metric space} (E,d) is a set E endowed with the topology
denoted (E,d) defined by its semi-metric. It is a \textbf{metric space} if d
is a metric.
\end{definition}

\subparagraph{Neighborhood\newline}

\begin{theorem}
A neighborhood of the point x of a semi-metric space (E,d) is any subset of E
that contains an open ball B(x,r).
\end{theorem}

\begin{theorem}
(Wilansky p.19) If X is a subset of the semi-metric space (E,d), then
$x\in\overline{X}$ iff $\delta(x,X)=0$
\end{theorem}

\subparagraph{Equivalent topology\newline}

\begin{theorem}
(Gamelin p.27) The topology defined on a set E by two semi-metrics d,d' are
equivalent iff the identity map $(E,d)\rightarrow(E,d^{\prime})$ is an homeomorphism
\end{theorem}

\begin{theorem}
A semi-metric d induces in any subset X of E an equivalent topology defined by
the restriction of d to X.
\end{theorem}

Example : If d is a semi metric, $\min\left(  d,1\right)  $ is a semi metric
equivalent to d.

\subparagraph{Base of the topology\newline}

\begin{theorem}
(Gamelin p.72) A metric space is first countable
\end{theorem}

\begin{theorem}
(Gamelin p.24, Wilansky p.76\ ) A metric or semi-metric space is separable iff
it is second countable.\ 
\end{theorem}

\begin{theorem}
(Gamelin p.23) A subset of a separable metric space is separable
\end{theorem}

\begin{theorem}
(Gamelin p.23) A totally bounded metric space is separable
\end{theorem}

\begin{theorem}
(Gamelin p.25) A compact metric space is separable and second countable
\end{theorem}

\begin{theorem}
(Wilansky p.128) A totally bounded semi-metric space is second countable and
so is separable
\end{theorem}

\begin{theorem}
(Kobayashi I p.268) A connected, locally compact, metric space is second
countable and separable
\end{theorem}

\subparagraph{Separability\newline}

\begin{theorem}
(Gamelin p.74) A metric space is a T4 topological space, so it is a normal,
regular, T1 and Hausdorff topological space
\end{theorem}

\begin{theorem}
(Wilansky p.62) A semi-metric space is normal and regular
\end{theorem}

\subparagraph{Compactness\newline}

\begin{theorem}
(Wilansky p.83) A compact subset of a semi-metric space is bounded
\end{theorem}

\begin{theorem}
(Wilansky p.127) A countably compact semi-metric space is totally bounded
\end{theorem}

\begin{theorem}
(Gamelin p.20) In a metric space E, the following properties are equivalent
for any subset X of E :

i) X is compact

ii) X is closed and totally bounded

iii) every sequence in X has an accumulation point (Weierstrass-Bolzano)

iv) every sequence in X has a convergent subsequence
\end{theorem}

Warning ! in a metric space a subset closed and bounded is not necessarily compact

\begin{theorem}
Heine-Borel: A subset X of $%
\mathbb{R}
^{m}$ is closed and bounded iff it is compact
\end{theorem}

\begin{theorem}
(Gamelin p.28) A metric space (E,d) is compact iff every continuous function
$f:E\rightarrow%
\mathbb{R}
$ is bounded
\end{theorem}

\subparagraph{Paracompactness\newline}

\begin{theorem}
(Wilansky p.193) A semi-metric space has a $\sigma-$locally finite base for
its topology.
\end{theorem}

\begin{theorem}
(Bourbaki, Lang p.34) A metric space is paracompact
\end{theorem}

\subparagraph{Convergence of a sequence\newline}

\begin{theorem}
A sequence $\left(  x_{n}\right)  _{n\in%
\mathbb{N}
}$ in a semi-metric space (E,d) converges to the limit x iff $\forall
\varepsilon>0,\exists N\in%
\mathbb{N}
:\forall n>N:d\left(  x_{n},x\right)  <\varepsilon$
\end{theorem}

\begin{theorem}
(Schwartz I p.77) In a metric space a sequence is convergent iff it has a
unique point of accumulation
\end{theorem}

The limit is unique \textit{if\ d is a metric}.

\subparagraph{Product of semi-metric spaces\newline}

There are different ways to define a metric on the product of a finite number
of metric spaces $E=E_{1}\times E_{2}\times...\times E_{n}$

The most usual ones for $x=\left(  x_{1},...,x_{n}\right)  $ are : the
euclidean metric : $d\left(  x,y\right)  =\left(  \sum_{i=1}^{n}d_{i}\left(
x_{i},y_{i}\right)  ^{2}\right)  $ and the max metric : $d\left(  x,y\right)
=\max d_{i}\left(  x_{i},y_{i}\right)  $

With these metrics (E,d) is endowed with the product topology (cf.above).

\paragraph{Semi-metrizable and metrizable spaces\newline}

\subparagraph{Definitions\newline}

\begin{definition}
A topological space $\left(  E,\Omega\right)  $ is said to be
\textbf{semi-metrizable} if there is a semi-metric d on E such that the
topologies $\left(  E,\Omega\right)  ,\left(  E,d\right)  $ are equivalent. A
topological space $\left(  E,\Omega\right)  $ is said to be
\textbf{metrizable} if there is a metric d on E such that the topologies
$\left(  E,\Omega\right)  ,\left(  E,d\right)  $ are equivalent.
\end{definition}

\subparagraph{Conditions for semi-metrizability\newline}

\begin{theorem}
Nagata-Smirnov( Wilansky p.198) A topological space is semi-metrizable iff it
is regular and has a $\sigma-$locally finite base.
\end{theorem}

\begin{theorem}
Urysohn (Wilansky p.185): A second countable regular topological space is semi-metrizable
\end{theorem}

\begin{theorem}
(Wilansky p.186) A separable topological space is semi-metrizable iff it is
second countable and regular.
\end{theorem}

\begin{theorem}
(Schwartz III p.428) A compact or locally compact topological space is semi-metrizable.
\end{theorem}

\begin{theorem}
(Schwartz III p.427) A topological space $\left(  E,\Omega\right)  $ is
semi-metrizable iff : $\forall a\in E,\forall n\left(  a\right)  ,\exists f\in
C_{0}(E;%
\mathbb{R}
^{+}):f(a)>0,x\in n(a)^{c}:f(x)=0$
\end{theorem}

\subparagraph{Conditions for metrizability\newline}

\begin{theorem}
(Wilansky p.186) A second countable T3 topological space is metrizable
\end{theorem}

\begin{theorem}
(Wilansky p.187) A compact Hausdorff space is metrizable iff it is second-countable
\end{theorem}

\begin{theorem}
Urysohn (Wilansky p.187) A topological space is separable and metrizable iff
it is T3 and second-countable.
\end{theorem}

\begin{theorem}
Nagata-Smirnov: A topological space is metrizable iff it is regular, Hausdorff
and has a $\sigma$-locally finite base.
\end{theorem}

A $\sigma$-locally finite base is a base which is a union of countably many
locally finite collections of open sets.

\paragraph{Pseudo-metric spaces\newline}

Some sets (such that the Fr\'{e}chet spaces) are endowed with a family of
semi-metrics, which have some specific properties. In particular they can be Hausdorff.

\begin{definition}
A \textbf{pseudo-metric space} is a set endowed with a family $\left(
d_{i}\right)  _{i\in I}$ such that each $d_{i}$ is a semi-metric on E and

$\forall J\subset I,\exists k\in I:\forall j\in J:d_{k}\geq d_{j}$
\end{definition}

\begin{theorem}
(Schwartz III p.426) On a pseudo-metric space $(E,\left(  d_{i}\right)  _{i\in
I}$), the collection $\Omega$ of open sets defined by :

$O\in\Omega\Leftrightarrow\forall x\in O,\exists r>0,\exists i\in
I:B_{i}\left(  x,r\right)  \subset O$

where $B_{i}(a,r)=\{x\in E:d_{i}(a,x)<r\}$

is the base for a topology.
\end{theorem}

\begin{theorem}
(Schwartz III p.427) A pseudometric space $\left(  E,\left(  d_{i}\right)
_{i\in I}\right)  $ is Hausdorff iff $\forall x\neq y\in E,\exists i\in
I:d_{i}(x,y)>0$
\end{theorem}

\begin{theorem}
(Schwartz III p.440) A map $f:E\rightarrow F$ from a topological space
$\left(  E,\Omega\right)  $ to a pseudo-metric space $\left(  F,\left(
d_{i}\right)  _{i\in I}\right)  $ is continuous at $a\in E$ if :
$\forall\varepsilon>0,\exists\varpi\in E:\forall x\in\varpi,\forall i\in
I:d_{i}\left(  f\left(  x\right)  ,f\left(  a\right)  \right)  <\varepsilon$
\end{theorem}

\begin{theorem}
Ascoli (Schwartz III p.450)\ A family $\left(  f_{k}\right)  _{k\in K}$ of
maps : $f_{k}:E\rightarrow F$ from a topological space $\left(  E,\Omega
\right)  $ to a pseudo-metric space $\left(  F,\left(  d_{i}\right)  _{i\in
I}\right)  $ is \textbf{equicontinuous} at $a\in E$ if :

$\forall\varepsilon>0,\exists\varpi\in\Omega:\forall x\in\varpi,\forall i\in
I,\forall k\in K:d_{i}\left(  f_{k}\left(  x\right)  ,f\left(  a\right)
\right)  <\varepsilon$

Then the closure $\digamma$ of $\left(  f_{k}\right)  _{k\in K}$ in F$^{E}$
(with the topology of simple convergence) is still equicontinuous at a. All
maps in $\digamma$\ are continuous at a, the limit of every convergent
sequence of maps in $\digamma$\ is continuous at a.

If a sequence $\left(  f_{n}\right)  _{n\in%
\mathbb{N}
}$ of continuous maps on E, is equicontinuous and converges to a continuous
map f on a dense subset of E, then it converges to f in E and uniformly on any
compact of E.
\end{theorem}

\begin{definition}
A topological space $\left(  E,\Omega\right)  $ is pseudo-metrizable if it is
homeomorphic to a space endowed with a family of pseudometrics
\end{definition}

\begin{theorem}
(Schwartz III p.433) A pseudo-metric space $\left(  E,\left(  d_{i}\right)
_{i\in I}\right)  $ such that the set I is countable is metrizable.
\end{theorem}

\subsubsection{Maps on a semi-metric space}

\paragraph{Continuity\newline}

\begin{theorem}
A map $f:E\rightarrow F$ between semi-metric space (E,d),(F,d') is continuous
in $a\in E$ iff $\forall\varepsilon>0,\exists\eta>0:\forall d\left(
x,a\right)  <\eta,d^{\prime}\left(  f(x),f(a)\right)  <\varepsilon$
\end{theorem}

\begin{theorem}
On a semi-metric space (E,d) the map $d:E\times E\rightarrow%
\mathbb{R}
$ is continuous
\end{theorem}

\paragraph{Uniform continuity\newline}

\begin{definition}
A map $f:E\rightarrow F$ between the semi-metric spaces (E,d),(F,d') is
\textbf{uniformly continuous} if

$\forall\varepsilon>0,\exists\eta>0:\forall x,y\in E:d\left(  x,y\right)
<\eta,d^{\prime}\left(  f(x),f(y)\right)  <\varepsilon$
\end{definition}

\begin{theorem}
(Wilansky p.59) A map f uniformly continuous is continuous (but the converse
is not true)
\end{theorem}

\begin{theorem}
(Wilansky p.219) A subset X of a semi-metric space is bounded iff any
uniformly continuous real function on X is bounded
\end{theorem}

\begin{theorem}
(Gamelin p.26, Schwartz III p.429) A continuous map $f:E\rightarrow F$ between
the semi-metric spaces E,F where E is compact is uniformly continuous
\end{theorem}

\begin{theorem}
(Gamelin p.27) On a semi-metric space (E,d), $\forall a\in E$ the map
$d\left(  a,.\right)  :E\rightarrow%
\mathbb{R}
$ is uniformly continuous
\end{theorem}

\paragraph{Uniform convergence of sequence of maps\newline}

\begin{definition}
The sequence of maps : $f_{n}:E\rightarrow F$ where (F,d) is a semi-metric
space \textbf{converges uniformly} to $f:E\rightarrow F$ if :

$\forall\varepsilon>0,\exists N\in%
\mathbb{N}
:\forall x\in E,\forall n>N:d\left(  f_{n}\left(  x\right)  ,f\left(
x\right)  \right)  <\varepsilon$
\end{definition}

Convergence uniform $\Rightarrow$\ convergence but the converse is not true

\begin{theorem}
(Wilansky p.55) If the sequence of maps : $f_{n}:E\rightarrow F$ ,where E is a
topological space and F is a semi-metric space, converges uniformly to $f$
then :

i) if the maps $f_{n}$ are continuous at a, then f is continuous at a.

ii) If the maps $f_{n}$ are continuous in E, then f is continuous in E
\end{theorem}

\paragraph{Lipschitz map\newline}

\begin{definition}
A map $f:E\rightarrow F$ between the semi-metric spaces (E,d),(F,d') is

i) a \textbf{globally} \textbf{Lipschitz} (or H\"{o}lder continuous) map of
order a%
$>$%
0 if

$\exists k\geq0:\forall x,y\in E:d^{\prime}(f(x),f(y))\leq k\left(
d(x,y)\right)  ^{a}$

ii) a \textbf{locally} \textbf{Lipschitz} map of order a%
$>$%
0 if

$\forall x\in E,\exists n(x),\exists k\geq0:\forall y\in n(x):d^{\prime
}(f(x),f(y))\leq k\left(  d(x,y)\right)  ^{a}$

iii) a \textbf{contraction} if

$\exists k,0<k<1:\forall x,y\in E:d(f(x),f(y))\leq kd(x,y)$

iv) an \textbf{isometry }if

$\forall x,y\in E:d^{\prime}(f(x),f(y))=d(x,y)$
\end{definition}

\paragraph{Embedding of a subset\newline}

It is a way to say that a subset contains enough information so that a
function can be continuously extended from it.

\begin{definition}
(Wilansky p.155) A subset X of a topological set E is said to be
\textbf{C-embedded} in E if every continuous real function on X can be
extended to a real continuous function on E.
\end{definition}

\begin{theorem}
(Wilansky p.156) Every closed subset of a normal topological space E is C-embedded.
\end{theorem}

\begin{theorem}
(Schwartz 2 p.443) Let E be a metric space, X a closed subset of E,
$f:X\rightarrow%
\mathbb{R}
$ a continuous map on X, then there is a map $F:E\rightarrow%
\mathbb{R}
$ continuous on E, such that : $\forall x\in X:$

$F(x)=f(x),\sup_{x\in E}F\left(  x\right)  =\sup_{y\in X}f\left(  y\right)
,\inf_{x\in E}F\left(  x\right)  =\inf_{y\in X}f\left(  y\right)  $
\end{theorem}

\subsubsection{Completeness}

Completeness is an important property for infinite dimensional vector spaces
as it is the only way to assure some fundamental results (such that the
inversion of maps) through the fixed point theorem.

\paragraph{Cauchy sequence\newline}

\begin{definition}
A sequence $\left(  x_{n}\right)  _{n\in%
\mathbb{N}
}$ in a semi-metric space (E,d) is a \textbf{Cauchy sequence} if :

$\forall\varepsilon>0,\exists N\in%
\mathbb{N}
:\forall n,m>N:d\left(  x_{n},x_{m}\right)  <\varepsilon$
\end{definition}

Any convergent sequence is a Cauchy sequence. But the converse is not always true.

Similarly a sequence of maps $f_{n}:E\rightarrow F$ where (F,d) is a
semi-metric space, is a Cauchy sequence of maps if :

$\forall\varepsilon>0,\exists N\in%
\mathbb{N}
:\forall x\in E,\forall n,m>N:d\left(  f_{n}\left(  x\right)  ,f_{m}\left(
x\right)  \right)  <\varepsilon$

\begin{theorem}
(Wilansky p.171) A Cauchy sequence which has a convergent subsequence is convergent
\end{theorem}

\begin{theorem}
(Gamelin p.22) Every sequence in a totally bounded metric space has a Cauchy subsequence
\end{theorem}

\paragraph{Definition of complete semi-metric space\newline}

\begin{definition}
A semi-metric space (E,d) is \textbf{complete} if any Cauchy sequence converges.
\end{definition}

Examples of complete metric spaces:

- Any finite dimensional vector space endowed with a metric

- The set of continuous, bounded real or complex valued functions over a
metric space

- The set of linear continuous maps from a normed vector space E to a normed
complete vector space F

\paragraph{Properties of complete semi-metric spaces\newline}

\begin{theorem}
(Wilansky p.169) A semi-metric space is compact iff it is complete and totally bounded
\end{theorem}

\begin{theorem}
(Wilansky p.171) A closed subset of a complete metric space is complete.
Conversely a complete subset of a metric space is closed.(untrue for
semi-metric spaces)
\end{theorem}

\begin{theorem}
(Wilansky p.172) The countable product of complete spaces is complete
\end{theorem}

\begin{theorem}
(Schwartz I p.96) Every compact metric space is complete (the converse is not true)
\end{theorem}

\begin{theorem}
(Gamelin p.10) If $\left(  f_{n}\right)  _{n\in%
\mathbb{N}
}$ is a Cauchy sequence of maps$\ f_{n}:E\rightarrow F$ in a complete matric
space F, then there is a map : $f:E\rightarrow F$ such that $f_{n}$ converges
uniformly to f on E.
\end{theorem}

\begin{theorem}
Every increasing sequence on $%
\mathbb{R}
$ with an upper bound converges. Every decreasing sequence on $%
\mathbb{R}
$ with a lower bound converges
\end{theorem}

\paragraph{Baire spaces\newline}

\begin{theorem}
(Wilansky p.178) A complete semi metric space is a Baire space
\end{theorem}

\begin{theorem}
(Doob, p.4) If $f:X\rightarrow F$ is a uniformly continuous map on a dense
subset X of a metric space E to a complete metric space F, then f has a unique
uniformly continuous extension to E.
\end{theorem}

\begin{theorem}
Baire Category (Gamelin p.11): If $\left(  X_{n}\right)  _{n\in%
\mathbb{N}
}$ is a family of dense open subsets of the complete metric space (E,d), then
$\cap_{n=1}^{\infty}X_{n}$ is dense in E.
\end{theorem}

\begin{theorem}
A metric space (E,d) is complete iff every decreasing sequence of non-empty
closed subsets of E, with diameters tending to 0, has a non-empty intersection.
\end{theorem}

\paragraph{Fixed point\newline}

\begin{theorem}
Contraction mapping theorem (Schwartz I p.101): If $f:E\rightarrow E$ is a
contraction over a complete metric space then it has a unique \textbf{fixed
point} a : $\exists a\in E:f\left(  a\right)  =a$

Furthermore if $f:E\times T\rightarrow E$ is continuous with respect to $t\in
T,$ a topological space, and

$\exists1>k>0:$ $\forall x,y\in E,t\in T:d\left(  f\left(  x,t\right)
,f\left(  y,t\right)  \right)  \leq kd\left(  x,y\right)  $

then there is a unique fixed point a(t) and $a:T\rightarrow E$ is continuous
\end{theorem}

The point a can be found by iteration starting from any point b :
$b_{n+1}=f\left(  b_{n}\right)  \Rightarrow a=\lim_{n\rightarrow\infty}b_{n}$
and we have the estimate : $d\left(  b_{n},a\right)  \leq\frac{k^{n}}%
{k-1}d\left(  b,f\left(  b\right)  \right)  $. So, if f is not a contraction,
but if one of its iterated is a contraction, the theorem still holds.

This theorem is fundamental, for instance it is the key to prove the existence
of solutions for differential equations, and it is a common way to compute solutions.

\begin{theorem}
Brower: In $%
\mathbb{R}
^{n}$\ , n$\geq$1 any continuous map $f:B\left(  0,1\right)  \rightarrow
B\left(  0,1\right)  $ (closed balls) has a fixed point.
\end{theorem}

With the generalization : every continuous function from a convex compact
subset K of a Banach space to K itself has a fixed point

\paragraph{Completion\newline}

Completeness \textit{is not a topological property} : it is not preserved by
homeomorphism. A topological space homeomorphic to a separable complete metric
space is called a Polish space. But a metric space which is not complete can
be completed : it is enlarged so that, with the same metric, any Cauchy
sequence converges.

\begin{definition}
(Wilansky p.174) A completion of a semi-metric space (E,d) is a pair $\left(
\overline{E},\imath\right)  $ of a complete semi-metric space $\overline{E}$
and an isometry \i\ from E to a dense subset of $\overline{E}$

A completion of a metric space (E,d) is a pair $\left(  \overline{E}%
,\imath\right)  $ of a complete metric space $\overline{E}$ and an isometry
\i\ from E to a dense subset of $\overline{E}$
\end{definition}

\begin{theorem}
(Wilansky p.175) A semi-metric space has a completion. A metric space has a
completion, unique up to an isometry.
\end{theorem}

The completion of a metric space $\left(  \overline{E},\imath\right)  $ has
the universal property that for any complete space (F,d') and uniformly
continuous map $f:E\rightarrow F$ then there is a unique uniformly continuous
function f' from $\overline{E}$ to F, which extends f.

The set of real number $%
\mathbb{R}
$ is the completion of the set of rational numbers $%
\mathbb{Q}
.$ So $%
\mathbb{R}
^{n},%
\mathbb{C}
^{n}$ are complete metric spaces for any fixed n, but not $%
\mathbb{Q}
.$

If the completion procedure is applied to a normed vector space, the result is
a Banach space containing the original space as a dense subspace, and if it is
applied to an inner product space, the result is a Hilbert space containing
the original space as a dense subspace.

\bigskip

\subsection{Algebraic topology}

\label{Algebraic topology}

Algebraic topology deals with the shape of objects, where two objects are
deemed to have the same shape if one can pass from one to the other by a
continuous deformation (so it is purely topological, without metric). The
tools which have been developped for this purpose have found many other useful
application in other fields. They highlight some fundamental properties of
topological spaces (topological invariants) so, whenever we look for some
mathematical objects which "look alike" in some way, they give a quick way to
restrict the scope of the search. For instance two manifolds which are not
homotopic cannot be homeomorphic.

We will limit the scope at a short view of homotopy and covering spaces, with
an addition for the Hopf-Rinow theorem. The main concept is that of simply
connected spaces.

\subsubsection{Homotopy}

The basic idea of homotopy theory is that the kind of curves which can be
drawn on a set, notably of loops, is in someway a characteristic of the set
itself. It is studied by defining a group structure on loops which can be
deformed continuously.

\paragraph{Homotopic paths\newline}

This construct is generalized below, but it is very common and useful to
understand the concept. A curve can be continuously deformed. Two curves are
homotopic if they coincide in a continuous transformation. The precise
definition is the following:

\begin{definition}
Let $\left(  E,\Omega\right)  $ be a topological space, P the set P(a,b) of
continuous maps $f\in C_{0}\left(  \left[  0,1\right]  ;E\right)  :f\left(
0\right)  =a,f\left(  1\right)  =b$

The paths $f_{1},f_{2}\in P\left(  a,b\right)  $ are \textbf{homotopic} if

$\exists F\in C_{0}([0,1]\times\lbrack0,1];E)$ such that :

$\forall s\in\lbrack0,1]:F(s,0)=f_{1}(s),F(s,1)=f_{2}(s),$

$\forall t\in\lbrack0,1]:F(0,t)=a,F(1,t)=b$
\end{definition}

$f_{1}\sim f_{2}$ is an equivalence relation. It does not depend on the
parameter :

$\forall\varphi\in C_{0}\left(  \left[  0,1\right]  ;\left[  0,1\right]
\right)  ,\varphi\left(  0\right)  =0,\varphi\left(  1\right)  =1:f_{1}\sim
f_{2}\Rightarrow f_{1}\circ\varphi\sim f_{2}\circ\varphi$

The quotient space P(a,b)/$\sim$ is denoted $\left[  P\left(  a,b\right)
\right]  $ and the classes of equivalences $\left[  f\right]  .$

Example : all the paths with same end points (a,b) in a convex subset of $%
\mathbb{R}
^{n}$ are homotopic.

The key point is that not any curve can be similarly transformed in each
other. In $%
\mathbb{R}
^{3}$\ curves which goes through a tore or envelop it are not homotopic.

\paragraph{Fundamental group\newline}

The set $\left[  P\left(  a,b\right)  \right]  $ is endowed with the operation
$\cdot$\ :

If $a,b,c\in E,f\in P\left(  a,b\right)  ,g\in P\left(  b,c\right)  $ define
the product

$f\cdot g$ : $P\left(  a,b\right)  \times P\left(  b,c\right)  \rightarrow
P\left(  a,c\right)  $ by :

$0\leq s\leq1/2\leq:f\cdot g(s)=f(2s),$

$1/2\leq s\leq1:f\cdot g(s)=g(2s-1)$

This product is associative.

Define the inverse : $\left(  f\right)  ^{-1}\left(  s\right)  =f\left(
1-s\right)  \Rightarrow\left(  f\right)  ^{-1}\circ f\in P\left(  a,a\right)
$

This product is defined over $\left[  P\left(  a,b\right)  \right]  $ : If
$f_{1}\sim f_{2}$ $,g_{1}\sim g_{2}$ then $f_{1}\cdot g_{1}\sim f_{2}\cdot
g_{2},\left(  f_{1}\right)  ^{-1}\sim\left(  f_{2}\right)  ^{-1}$

\begin{definition}
A \textbf{loop} is a path which begins and ends at the same point called
\textbf{the base point}.
\end{definition}

The product of two loops with same base point is well defined, as is the
inverse, and the identity element (denoted $\left[  0\right]  )$ is the
constant loop $f\left(  t\right)  =a.\ $So the set of loops with same base
point is a group with $\cdot.$ (it is not commutative).

\begin{definition}
The \textbf{fundamental group} at a point a, denoted $\pi_{1}\left(
E,a\right)  ,$ of a topological space E, is the set of homotopic loops with
base point a, endowed with the product of loops.
\end{definition}

$\pi_{1}\left(  E,a\right)  =\left(  \left[  P\left(  a,a\right)  \right]
,\cdot\right)  $

Let a,b$\in E$ such that there is a path f from a to b.\ Then :

$f_{\ast}:\pi_{1}\left(  E,a\right)  \rightarrow\pi_{1}\left(  E,b\right)
::f_{\ast}\left(  \left[  \gamma\right]  \right)  =\left[  f\right]
\cdot\left[  \gamma\right]  \cdot\left[  f\right]  ^{-1}$

is a group isomorphism. So :

\begin{theorem}
The fundamental groups $\pi_{1}\left(  E,a\right)  $ whose base point a belong
to the same path-connected component of E are isomorphic.
\end{theorem}

\begin{definition}
The fundamental group of a path-connected topological space E, denoted
$\pi_{1}\left(  E\right)  $ , is the common group structure of its fundamental
groups $\pi_{1}\left(  E,a\right)  $
\end{definition}

\begin{theorem}
The fundamental groups of homeomorphic topological spaces are isomorphic.
\end{theorem}

So the fundamental group is a pure topological concept, and this is a way to
check the homeomorphism of topological spaces. One of the consequences is the
following :

\begin{theorem}
(Gamelin p.123) If E,F are two topological spaces, $f:E\rightarrow F$ a
homeomorphism such that $f\left(  a\right)  =b,$ then there is an isomorphism
$\varphi:\pi_{1}\left(  E,a\right)  \rightarrow\pi_{1}\left(  F,b\right)  $
\end{theorem}

\paragraph{Simply-connectedness\newline}

If $\pi_{1}\left(  E\right)  \sim\left[  0\right]  $ the group is said
\textbf{trivial} : every loop can be continuously deformed to coincide with
the point a.

\begin{definition}
A path-connected topological group E is \textbf{simply connected} if its
fundamental group is trivial : $\pi_{1}\left(  E\right)  \sim\left[  0\right]
$
\end{definition}

Roughly speaking a space is simply connected if there is no "hole" in it.

\begin{definition}
A topological space E is \textbf{locally simply connected} if any point has a
neighborhood which is simply connected
\end{definition}

\begin{theorem}
(Gamelin p.121) The product of two simply connected spaces is simply connected
\end{theorem}

\begin{theorem}
A convex subset of $%
\mathbb{R}
^{n}$ is simply connected.\ The sphere $S^{n}$ (in $%
\mathbb{R}
^{n+1}$)\ is simply connected for n%
$>$%
1 (the circle is not).
\end{theorem}

\paragraph{Homotopy of maps\newline}

Homotopy can be generalized from paths to maps as follows:

\begin{definition}
Two continuous maps $f,g\in C_{0}\left(  E;F\right)  $ between the topological
spaces E,F are homotopic if there is a continuous map : $F:E\times\left[
0,1\right]  \rightarrow F$ such that : $\forall x\in E:F\left(  x,0\right)
=f\left(  x\right)  ,F\left(  x,1\right)  =g\left(  x\right)  $
\end{definition}

Homotopy of maps is an equivalence relation, which is compatible with the
composition of maps.

\paragraph{Homotopy of spaces\newline}

\begin{definition}
Two topological spaces E,F are \textbf{homotopic} if there are maps
$f:E\rightarrow F,g:F\rightarrow E,$ such that $f\circ g$ is homotopic to the
identity on E and $g\circ f$ is homotopic to the identity on F.
\end{definition}

Homeomorphic spaces are homotopic, but the converse is not always true.

Two spaces are homotopic if they can be transformed in each other by a
continuous transformation : by bending, shrinking and expending.

\begin{theorem}
If two topological spaces E,F are homotopic then if E is path-connected, F is
path connected and their fundamental group are isomorphic. Thus if E is simply
connected, F is simply connected
\end{theorem}

The topologic spaces which are homotopic, with homotopic maps as morphisms,
constitute a category.

\begin{definition}
A topological space is \textbf{contractible} if it is homotopic to a point
\end{definition}

The sphere is not contractible.

\begin{theorem}
(Gamelin p.140) A contractible space is simply connected.
\end{theorem}

More generally, a map $f\in C_{0}\left(  E;X\right)  $ between a topological
space E and its subset X, is a continuous retract if $\forall x\in X:f\left(
x\right)  =x$ and then X is a retraction of E. \ E is retractible into X if
there is a continuous retract (called a deformation retract) which is
homotopic to the identity map on E.

If the subset X of the topological space E, is a continuous retraction of E
and is simply connected, then E is simply connected.

\paragraph{Extension\newline}

\begin{definition}
Two continuous maps $f,g\in C_{0}\left(  E;F\right)  $ between the topological
spaces E,F are homotopic relative to the subset $X\subset E$ if there is a
continuous map : $F:E\times\left[  0,1\right]  \rightarrow F$ such that :
$\forall x\in E:F\left(  x,0\right)  =f\left(  x\right)  ,F\left(  x,1\right)
=g\left(  x\right)  $ and $\forall t\in\left[  0,1\right]  ,x\in X:F\left(
x,t\right)  =f\left(  x\right)  =g\left(  x\right)  $
\end{definition}

One gets back the homotopy of paths with E=$\left[  0,1\right]  ,$ $X=\left\{
a,b\right\}  .$

This leads to the extension to homotopy of higher orders, by considering the
homotopy of maps between n-cube $\left[  0,1\right]  ^{r}$\ in $%
\mathbb{R}
^{r}$ and a topological space E, with the fixed subset the boundary
$\partial\left[  0,1\right]  ^{r}$ (all of its points such at least one
$t_{i}=0$ or 1). The homotopy groups of order $\pi_{r}\left(  E,a\right)  $
are defined by proceeding as above.\ They are abelian for r%
$>$%
1.

\subsubsection{Covering spaces}

A fibered manifold (see the Fiber bundle part) is basically a pair of
manifolds (M,E) where E is projected on M.\ Covering spaces can be seen as a
generalization of this concept to topological spaces.

\paragraph{Definitions\newline}

\bigskip

1. The definition varies according to the authors.\ This is the most general.

\begin{definition}
Let $\left(  E,\Omega\right)  ,\left(  M,\Theta\right)  $ two topological
spaces and a continuous surjective map : $\pi:E\rightarrow M$

An open subset U of M is \textbf{evenly covered} by E if :

$\pi^{-1}\left(  U\right)  $\ is the disjoint union of open subsets of E :

$\pi^{-1}\left(  U\right)  =\cup_{i\in I}O_{i};O_{i}\in\Omega;\forall i,j\in
I:O_{i}\cap O_{j}=\varnothing$

and $\pi$ is an homeomorphism on each $O_{i}\rightarrow\pi\left(
O_{i}\right)  $
\end{definition}

The O$_{i}$ are called the \textbf{sheets}. If U is connected they are the
connected components of $\pi^{-1}\left(  U\right)  $

\begin{definition}
$E\left(  M,\pi\right)  $ is a \textbf{covering space} if any point of M has a
neighborhood which is evenly covered by E
\end{definition}

E is the \textbf{total space} , $\pi$ the \textbf{covering map}, M the
\textbf{base space}, $\pi^{-1}\left(  x\right)  $ the \textbf{fiber} over
$x\in M$

Thus E and M share all local topological properties : if M is locally
connected so is E.

Example : M=$%
\mathbb{R}
,E=S_{1}$ the unit circle, $\pi:S_{1}\rightarrow M::\pi\left(  \left(  \cos
t,\sin t\right)  \right)  =t$

$\pi$ is a local homeomorphism : each x in M has a neigborhood which is
homeomorphic to a neighborhood $n\left(  \pi^{-1}\left(  x\right)  \right)  .$

\bigskip

2. Order of a covering:

If M is connected every x in M has a neighborhood n(x) such that $\pi
^{-1}\left(  n\left(  x\right)  \right)  $\ is homeomorphic to n(x)xV where V
is a discrete space (Munkres). The cardinality of V is called the
\textbf{degree r of the cover} : E is a double-cover of M if r=2. From the
topological point of view E is r "copies" of M piled over M. This is in stark
contrast with a fiber bundle E which is locally the "product" of M and a
manifold V : so we can see a covering space as a fiber bundle with typical
fiber a discrete space V (but of course the maps cannot be differentiable).

\bigskip

3. Isomorphims of fundamental groups:

\begin{theorem}
Munkres: In a covering space $E\left(  M,\pi\right)  ,$ if M is connected and
the order is r%
$>$%
1 then there is an isomorphism between the fundamental groups : $\widetilde
{\pi}:\pi_{1}\left(  E,a\right)  \rightarrow\pi_{1}\left(  M,\pi\left(
a\right)  \right)  $
\end{theorem}

\paragraph{Fiber preserving maps\newline}

\begin{definition}
A map : $f:E_{1}\rightarrow E_{2}$ between two covering spaces $E_{1}\left(
M,\pi_{1}\right)  ,E_{2}\left(  M,\pi_{2}\right)  $ is \textbf{fiber
preserving} if : $\pi_{2}\circ f=\pi_{1}$
\end{definition}

\bigskip%

\begin{tabular}
[c]{lllll}%
$E_{1}$ & $\rightarrow$ & $f$ & $\rightarrow$ & $E_{2}$\\
\multicolumn{1}{r}{$\pi_{1}$} & $\searrow$ &  & $\swarrow$ & $\pi_{2}$\\
&  & $E$ &  &
\end{tabular}

\bigskip

If f is an homeomorphisme then the covers are said \textbf{equivalent}.

\paragraph{Lifting property\newline}

\begin{theorem}
(Munkres) In a covering space $E\left(  M,\pi\right)  ,$ if $\gamma:\left[
0,1\right]  \rightarrow M$ is a path, then there exists a unique path
$\Gamma:\left[  0,1\right]  \rightarrow E$ such that $\pi\circ\Gamma=\gamma$
\end{theorem}

The path $\Gamma$ is called the \textbf{lift} of $\gamma$.

If x and y are two points in E connected by a path, then that path furnishes a
bijection between the fiber over x and the fiber over y via the lifting property.

If $\varphi:N\rightarrow M$ is a continuous map in a simply connected
topological space N, fix y$\in N,a\in\pi^{-1}\left(  \varphi\left(  a\right)
\right)  $ in E, then there is a unique continuous map $\Phi:N\rightarrow E$
such that $\varphi=\pi\circ\Phi$ .

\paragraph{Universal cover\newline}

\begin{definition}
A covering space $E\left(  M,\pi\right)  $ is a \textbf{universal cover} if E
is connected and simply connected
\end{definition}

If M is simply connected and E connected then $\pi$\ is bijective

The meaning is the following : let $E^{\prime}\left(  M,\pi^{\prime}\right)  $
another covering space of M such that E' is connected.\ Then there is a map :
$f:E\rightarrow E^{\prime}$ such that $\pi=\pi^{\prime}\circ f$

A universal cover is unique : if we fix a point x in M,\ there is a unique f
such that $\pi\left(  a\right)  =x,\pi^{\prime}\left(  a^{\prime}\right)
=x,\pi=\pi^{\prime}\circ f$

\subsubsection{Geodesics}

This is a generalization of the topic studied on manifolds.

1. Let $\left(  E,d\right)  $ be a metric space. A path on E is a continuous
injective map from an interval $\left[  a,b\right]  \subset%
\mathbb{R}
$ to E. If $\left[  a,b\right]  $ is bounded then the set $C_{0}\left(
\left[  a,b\right]  ;E\right)  $ is a compact connected subset. The curve
generated by $p\in C_{0}\left(  \left[  a,b\right]  ;E\right)  ,$ denoted
$p\left[  a,b\right]  ,$ is a connected, compact subset of E.

2. The \textbf{length of a curve} $p\left[  a,b\right]  $ is defined as :
$\ell\left(  p\right)  =\sup\sum_{k=1}^{n}d(p\left(  t_{k+1}\right)
),p(t_{k})) $ for any increasing sequence $\left(  t_{n}\right)  _{n\in%
\mathbb{N}
}$ in [a,b]

The curve is said to be \textbf{rectifiable} if $\ell\left(  p\right)
<\infty.$

3. The length is inchanged by any change of parameter $p\rightarrow
\widetilde{p}=p\circ\varphi$ where $\varphi$ is order preserving.

The path is said to be at constant speed v if there is a real scalar v such
that : $\forall t,t^{\prime}\in\left[  a,b\right]  :\ell\left(  p\left[
t,t^{\prime}\right]  \right)  =v\left\vert t-t^{\prime}\right\vert $

If the curve is rectifiable it is always possible to choose a path at constant
speed 1 by :$\varphi\left(  t\right)  =\ell\left(  p(t)\right)  $

4. A \textbf{geodesic} on E is a curve such that there is a path $p\in
C_{0}\left(  I;E\right)  ,$ with $I$ some interval of $%
\mathbb{R}
$ , such that :

$\forall t,t^{\prime}\in I:d\left(  p\left(  t\right)  ,p\left(  t^{\prime
}\right)  \right)  =\left\vert t^{\prime}-t\right\vert $

5. A subset X is said \textbf{geodesically convex} if there is a geodesic
which joins any pair of its points.

6. Define over E the new metric $\delta$, called internally metric, by :

$\delta:E\times E\rightarrow%
\mathbb{R}
::$

$\delta\left(  x,y\right)  =\inf_{c}\ell\left(  c\right)  ,p\in C_{0}(\left[
0,1\right]  ;E):p(0)=x,p(1)=y,\ell\left(  c\right)  <\infty$

$\delta\geq d$ and (E,d) is said to be an \textbf{internally metric} space if
$d=\delta$

A geodesically convex set is internally metric

A riemanian manifold is an internal metric space (with $p\in C_{1}(\left[
0,1\right]  ;E))$

If $\left(  E,d\right)  ,\left(  F,d^{\prime}\right)  $ are metric spaces and
D is defined on ExF as

$D\left(  \left(  x_{1},y_{1}\right)  ,\left(  x_{2},y_{2}\right)  \right)
=\sqrt{d\left(  x_{1},x_{2}\right)  ^{2}+d^{\prime}\left(  y_{1},y_{2}\right)
^{2}}$

then $\left(  E\times F,D\right)  $ is internally metric space iff E and F are
internally metric spaces

A curve is a geodesic iff its projections are geodesic

7. The main result is the following:

\begin{theorem}
Hopf-Rinow : If (E,d) is an internally metric, complete, locally compact space then:

- every closed bounded subset is compact

- E is geodesically convex

Furthermore if, in the neighborhood of any point, any close curve is homotopic
to a point (it is semi-locally simply connected) then every close curve is
homotopic either\ to a point or a geodesic
\end{theorem}

It has been proven (Atkin) that the theorem is false for an infinite
dimensional vector space (which is not, by the way, locally compact).

\newpage

\section{MEASURE}

A measure is roughly the generalization of the concepts of "volume" or
"surface" for a topological space. There are several ways to introduce
measures :

- the first, which is the most general and easiest to understand, relies on
set functions \ So roughly a measure on a set E is a map $\mu:S\rightarrow%
\mathbb{R}
$ where S is a set of subsets of E (a $\sigma-$algebra). We do not need a
topology and the theory, based upon the ZFC model of sets, is quite
straightforward. From a measure we can define integral $\int f\mu$, which are
linear functional on the set C(E;$%
\mathbb{R}
).$

- the "Bourbaki way" goes the other way around, and is based upon Radon
measures. It requires a topology, and, from my point of view, is more convoluted.

So we will follow the first way. Definitions and results can be found in Doob
and Schwartz (tome 2).

\bigskip

\subsection{Measurable spaces}

\label{Measurable spaces}

\subsubsection{Limit of sequence of subsets}

(Doob p.8)

\begin{definition}
A sequence $\left(  A_{n}\right)  _{n\in%
\mathbb{N}
}$ of subsets in E is :

\textbf{monotone increasing} if : $A_{n}\sqsubseteq A_{n+1}$

\textbf{monotone decreasing} if : $A_{n+1}\sqsubseteq A_{n}$
\end{definition}

\begin{definition}
The \textbf{superior limit} of a sequence $\left(  A_{n}\right)  _{n\in%
\mathbb{N}
}$ of subsets in E is the subset :

$\lim\sup A_{n}=\cap_{k=1}^{\infty}\cup_{j=k}^{\infty}A_{n}$
\end{definition}

It is the set of point in $A_{n}$ for an infinite number of n

\begin{definition}
The \textbf{inferior limit} of a sequence $\left(  A_{n}\right)  _{n\in%
\mathbb{N}
}$ of subsets in E is the subset :

$\lim\inf A_{n}=\cup_{k=1}^{\infty}\cap_{j=k}^{\infty}A_{n}$
\end{definition}

It is the set of point in $A_{n}$ for all but a finite number of n

\begin{theorem}
Every sequence $\left(  A_{n}\right)  _{n\in%
\mathbb{N}
}$ of subsets in E has a superior and an inferior limit and :

$\lim\inf A_{n}\sqsubseteq\lim\sup A_{n}$
\end{theorem}

\begin{definition}
A sequence $\left(  A_{n}\right)  _{n\in%
\mathbb{N}
}$ of subsets in E \textbf{converges} if its superior and inferior limits are
equal, and\ then its limit is:

$\lim_{n\rightarrow\infty}A_{n}=\lim\sup A_{n}=\lim\inf A_{n}$
\end{definition}

\begin{theorem}
A monotone increasing sequence of subsets converges to their union
\end{theorem}

\begin{theorem}
A monotone decreasing sequence of subsets converges to their intersection
\end{theorem}

\begin{theorem}
If $B_{p}$ is a subsequence of a sequence $\left(  A_{n}\right)  _{n\in%
\mathbb{N}
}$ then $B_{p}$ converges iff $\left(  A_{n}\right)  _{n\in%
\mathbb{N}
}$ converges
\end{theorem}

\begin{theorem}
If $\left(  A_{n}\right)  _{n\in%
\mathbb{N}
}$ converges to A, then $\left(  A_{n}^{c}\right)  _{n\in%
\mathbb{N}
}$ converges to $A^{c}$
\end{theorem}

\begin{theorem}
If $\left(  A_{n}\right)  _{n\in%
\mathbb{N}
},\left(  B_{n}\right)  _{n\in%
\mathbb{N}
}$ converge respectively to A,B, then $\left(  A_{n}\cup B_{n}\right)  _{n\in%
\mathbb{N}
}$,$\left(  A_{n}\cap B_{n}\right)  _{n\in%
\mathbb{N}
}$ converge respectively to $A\cup B,A\cap B$
\end{theorem}

\begin{theorem}
$\left(  A_{n}\right)  _{n\in%
\mathbb{N}
}$ converges to A iff the sequence of indicator functions $\left(  1_{A_{n}%
}\right)  _{n\in%
\mathbb{N}
}$ converges to 1$_{A}$
\end{theorem}

\paragraph{Extension of $%
\mathbb{R}
$\newline}

The compactification of $%
\mathbb{R}
$ leads to define :

$%
\mathbb{R}
_{+}=\left\{  r\in%
\mathbb{R}
,r\geq0\right\}  ,\overline{%
\mathbb{R}
}_{+}=%
\mathbb{R}
_{+}\cup\left\{  \infty\right\}  ,\overline{%
\mathbb{R}
}=%
\mathbb{R}
\cup\left\{  +\infty,-\infty\right\}  $

$\overline{%
\mathbb{R}
}$ is compact .

\begin{definition}
If $\left(  x_{n}\right)  _{n\in%
\mathbb{N}
}$ is a sequence of real scalar on $\overline{%
\mathbb{R}
}$

the \textbf{limit superior} is : $\lim\sup\left(  x_{n}\right)  =\lim
_{n\rightarrow\infty}\sup_{p\geq n}\left(  x_{p}\right)  $

the \textbf{limit inferior} is : $\ \lim\inf\left(  x_{n}\right)
=\lim_{n\rightarrow\infty}\inf_{p\geq n}\left(  x_{p}\right)  $
\end{definition}

\begin{theorem}
$\lim\inf\left(  x_{n}\right)  \leq\lim\sup\left(  x_{n}\right)  $ and are
equal if the sequence converges (possibly at infinity).
\end{theorem}

Warning ! this is different from the least upper bound \textbf{:}

$\sup A=\min\{m\in E:\forall x\in E:m\geq x\}$

and the greatest lower bound $\inf A=\max\{m\in E:\forall x\in E:m\leq x\}.$

\subsubsection{Measurable spaces}

\begin{definition}
A collection S of subsets of E is an \textbf{algebra} if :

\begin{definition}
$\varnothing\in S$

If $A\in S$ then $A^{c}\in S$ so $E\in S$

S is closed under finite union and finite intersection
\end{definition}
\end{definition}

\begin{definition}
A $\sigma-$\textbf{algebra} is an algebra which contains the limit of any
monotone sequence of its elements.
\end{definition}

The smallest $\sigma-$algebra is S=$\left\{  \varnothing,E\right\}  ,$ the
largest is $S=2^{E}$

\begin{definition}
A \textbf{measurable space} (E,S) is a set E endowed with a $\sigma-$algebra
S. Every subset which belongs to S is said to be \textbf{measurable}.
\end{definition}

\bigskip

Take any collection S of subsets of E, it is always possible to enlarge S in
order to get a $\sigma-$algebra.

The smallest of the $\sigma-$algebras which include S will be denoted
$\sigma\left(  S\right)  .$

If $\left(  S_{i}\right)  _{i=1}^{n}$ is a finite collection of subsets of
$2^{E}$ then

$\sigma\left(  S_{1}\times S_{2}..\times S_{n}\right)  =\sigma\left(
\sigma\left(  S_{1}\right)  \times\sigma\left(  S_{2}\right)  \times
...\sigma\left(  S_{n}\right)  \right)  $

If $\left(  E_{i},S_{i}\right)  $ i=1..n are measurable spaces, then $\left(
E_{1}\times E_{2}\times...E_{n},S\right)  $ with $S=\sigma\left(  S_{1}\times
S_{2}\times...\times S_{n}\right)  $ is a measurable space

Warning ! $\sigma\left(  S_{1}\times S_{2}\times...\times S_{n}\right)  $ \ is
by far larger than $S_{1}\times S_{2}\times...\times S_{n}.$ If $E_{1}=E_{2}=%
\mathbb{R}
$ then S encompasses not only "rectangles" but almost any area in $%
\mathbb{R}
^{2}$

\bigskip

Notice that in all these definitions there is no reference to a
topology.\ However usually a $\sigma-$algebra is defined with respect to a
given topology, meaning a collection of open subsets.

\begin{definition}
A topological space $\left(  E,\Omega\right)  $ has a unique $\sigma-$algebra
$\sigma\left(  \Omega\right)  ,$ called its \textbf{Borel algebra,} which is
generated either by the open or the closed subsets.
\end{definition}

So a topological space can always be made a measurable space.

\subsubsection{Measurable functions}

A measurable function is different from an integrable function. They are
really different concepts. Almost every map is measurable.

\begin{theorem}
If $\left(  F,S^{\prime}\right)  $ is a measurable space, f a map:
$f:E\rightarrow F$ then the collection of subsets $\left(  f^{-1}\left(
A^{\prime}\right)  ,A^{\prime}\in S^{\prime}\right)  $ is a $\sigma$ -algebra
in E denoted $\sigma\left(  f\right)  $
\end{theorem}

\begin{definition}
A map $f:E\rightarrow F$ between the measurable spaces $\left(  E,S\right)
,\left(  F,S^{\prime}\right)  $ is \textbf{measurable} if $\sigma\left(
f\right)  \sqsubseteq S$
\end{definition}

\begin{definition}
A \textbf{Baire map} is a measurable map $f:E\rightarrow F$ between
topological spaces endowed with their Borel algebras.
\end{definition}

\begin{theorem}
Every continuous map is a Baire map.
\end{theorem}

(Doob p.56)

\begin{theorem}
The composed $f\circ g$ of measurable maps is a measurable map.\ 
\end{theorem}

The category of measurable spaces as for objects measurable spaces and for
morphisms measurable maps

\begin{theorem}
If $\left(  f_{n}\right)  _{n\in%
\mathbb{N}
}$\ is a sequence of measurable maps $f_{n}:E\rightarrow F,$with $\left(
E,S\right)  ,\left(  F,S^{\prime}\right)  $ measurable spaces, such that
$\forall x\in E,\exists\lim_{n\rightarrow\infty}f_{n}\left(  x\right)
=f\left(  x\right)  ,$ then f is measurable
\end{theorem}

\begin{theorem}
If $\left(  f_{n}\right)  _{n\in%
\mathbb{N}
}$ is a sequence of measurable functions : $f_{n}:E\rightarrow\overline{%
\mathbb{R}
}$

then the functions : $\lim\sup f_{n}=\inf_{j>i}\sup_{n>j}f_{n};\lim\inf
_{n}f=\sup_{j>i}\inf f_{n}$ are measurable
\end{theorem}

\begin{theorem}
If for i=1...n: $f_{i}:E\rightarrow F_{i}$ with $\left(  E,S\right)
,(F_{i},S_{i}^{\prime})\ $measurables spaces then the map: $f=\left(
f_{1},f_{2},..f_{n}\right)  :E\rightarrow F_{1}\times F_{2}...\times F_{n}$ is
measurable iff each $f_{i}$ is measurable.
\end{theorem}

\begin{theorem}
If the map $f:E_{1}\times E_{2}\rightarrow F,$ between measurable spaces is
measurable, then for each $x_{1}$ fixed the map : $f_{x_{1}}:x_{1}\times
E_{2}\rightarrow F::f_{x_{1}}\left(  x_{2}\right)  =f(x_{1},x_{2})$ is measurable
\end{theorem}

\bigskip

\subsection{Measured spaces}

\label{Measured spaces}

A measure is a function acting on \textit{subsets} : $\mu:S\rightarrow%
\mathbb{R}
$ with some minimum properties.

\subsubsection{Definition of a measure}

\begin{definition}
Let (E,S) a measurable space, $A=\left(  A_{i}\right)  _{i\in I},A_{i}\in S$ a
family, a function\ $\mu:S\rightarrow\overline{%
\mathbb{R}
}$ is said :

\textbf{I-subadditive} if : $\mu\left(  \cup_{i\in I}A_{i}\right)  \leq
\sum_{i\in I}\mu\left(  A_{i}\right)  $ for any family $A$

\textbf{I-additive} if : $\mu\left(  \cup_{i\in I}A_{i}\right)  =\sum_{i\in
I}\mu\left(  A_{i}\right)  $ for any family $A$ of disjointed subsets in S:
$\forall i,j\in I:A_{i}\cap A_{j}=\varnothing$

\textbf{finitely subadditive} if it is I-subadditive for any finite family A

\textbf{finitely additive} if it is I-additive for any finite family A

\textbf{countably subadditive} if it is I-subadditive for any countable family A

\textbf{countably additive} if it is I-additive for any countable family A
\end{definition}

\begin{definition}
A \textbf{measure} on the measurable space (E,S) is a map $\mu:S\rightarrow
\overline{%
\mathbb{R}
}_{+}$ which is \textit{countably} additive. Then $\left(  E,S,\mu\right)  $
is a \textbf{measured space}.
\end{definition}

So a measure has the properties :

$\forall A\in S:0\leq\mu\left(  A\right)  \leq\infty$

$\mu\left(  \varnothing\right)  =0$

For any countable \textit{disjointed} family $\left(  A_{i}\right)  _{i\in I}$
of subsets in S :

$\mu\left(  \cup_{i\in I}A_{i}\right)  =\sum_{i\in I}\mu\left(  A_{i}\right)
$ (possibly both infinite)

Notice that here a measure - without additional name - is always positive, but
can take infinite value. It is necessary to introduce infinite value because
the value of a measure on the whole of E is often infinite (think to the
Lebesgue measure).

\begin{definition}
A \textbf{Borel measure} is a measure on a topological space with its Borel algebra.
\end{definition}

\begin{definition}
A \textbf{signed-measure} on the measurable space (E,S) is a map
$\mu:S\rightarrow\overline{%
\mathbb{R}
}$ which is \textit{countably} additive. Then $\left(  E,S,\mu\right)  $ is a
\textbf{signed} \textbf{measured space}.
\end{definition}

So a signed measure can take negative value. Notice that a signed measure can
take the values $\pm\infty.$

An outer \textbf{measure} on a set E is a map: $\lambda:2^{E}\rightarrow
\overline{%
\mathbb{R}
}_{+}$ which is countably \textit{subadditive, }monotone increasing and such
that\textit{\ }$\lambda\left(  \varnothing\right)  =0$

So the key differences with a measure is that : there is no $\sigma-$algebra
and $\lambda$ is only countably subadditive (and not additive)

\paragraph{Finite measures\newline}

\begin{definition}
A measure on E is \textbf{finite} if $\mu\left(  E\right)  <\infty$ so it
takes only finite positive values : $\mu:S\rightarrow%
\mathbb{R}
_{+}$
\end{definition}

A finite signed measure is a signed measure that takes only finite values :
$\mu:S\rightarrow%
\mathbb{R}
$

\begin{definition}
A \textbf{locally finite measure} is a Borel measure which is finite on any compact.
\end{definition}

A finite measure is locally finite but the converse is not true.

\begin{definition}
A measure on E is $\sigma-$\textbf{finite} if E is the countable union of
subsets of finite measure. Accordingly a set is said to be $\sigma-$finite if
it is the countable union of subsets of finite measure.
\end{definition}

\paragraph{Regular measure\newline}

\begin{definition}
A Borel measure $\mu$ on a topological space E is

\textbf{inner regular} if it is locally finite and $\mu\left(  A\right)
=\sup_{K}\mu\left(  K\right)  $ where K is a compact $K\sqsubseteq A.$

\textbf{outer regular} if $\mu\left(  A\right)  =\inf_{O}\mu\left(  O\right)
$ where O is an open subset $A\sqsubseteq O.$

\textbf{regular} if it is both inner and outer regular.
\end{definition}

\begin{theorem}
(Thill p.254) An inner regular measure $\mu$ on a Hausdorff space such that
$\mu\left(  E\right)  =1$ is regular.
\end{theorem}

\begin{theorem}
(Neeb p.43) On a locally compact topological space, where every open subset is
the countable union of compact subsets, every locally finite Borel measure is
inner regular.
\end{theorem}

\paragraph{Separable measure\newline}

(see Stochel)

A separable measure is a measure $\mu$\ on a measured space $\left(
E,S,\mu\right)  $ such that there is a countable family $\left(  S_{i}\right)
_{i\in I}\in S^{I}$ with the property :

$\forall s\in S,\mu\left(  s\right)  <\infty,\forall\varepsilon>0,\exists i\in
I:\mu\left(  S_{i}\Delta s\right)  <\varepsilon$

$\Delta$ is the symmetric difference : $S_{i}\Delta s=\left(  S_{i}\cup
s\right)  \backslash\left(  S_{i}\cap s\right)  $

If $\mu$\ \ is separable, then $L^{2}\left(  E,S,\mu,%
\mathbb{C}
\right)  $ is a separable Hilbert space. Conversely, if $L^{2}\left(  E,S,\mu,%
\mathbb{C}
\right)  $ is a separable Hilbert space and \ $\mu$ is $\sigma-$finite, then
$\mu$ is separable

\subsubsection{Radon measures}

Radon measures are a class of measures which have some basic useful properties
and are often met in Functional Analysis.

\begin{definition}
A \textbf{Radon measure} is a Borel, locally finite, regular, signed measure
on a topological Hausdorff locally compact space
\end{definition}

So : if $\left(  E,\Omega\right)  $ is a topological Hausdorff locally compact
space with its Borel algebra S, a Radon measure $\mu$ has the following
properties :

it is locally finite : $\left\vert \mu\left(  K\right)  \right\vert <\infty$
for any compact K of E

it is regular :

$\forall X\in S:\mu\left(  X\right)  =\inf\left(  \mu\left(  Y\right)
,X\sqsubseteq Y,Y\in\Omega\right)  $

$\left(  \left(  \forall X\in\Omega\right)  \vee\left(  X\in S\right)
\right)  \&\left(  \mu\left(  X\right)  <\infty\right)  :\mu\left(  X\right)
=\sup\left(  \mu\left(  K\right)  ,K\sqsubseteq X,K\text{ compact}\right)  $

The Lebesgue measure on $%
\mathbb{R}
^{m}$ is a Radon measure.

Remark : There are other definitions : this one is the easiest to understand
and use.

One useful theorem:

\begin{theorem}
(Schwartz III p.452) Let $\left(  E,\Omega\right)  $ a topological Hausdorff
locally compact space, $\left(  O_{i}\right)  _{i\in I}$ an open cover of E,
$\left(  \mu_{i}\right)  _{i\in I}$ a family of Radon measures defined on each
$O_{i}.$ If on each non empty intersection $O_{i}\cap O_{j}$ we have $\mu
_{i}=\mu_{j}$ then there is a unique measure $\mu$ defined on the whole of E
such that $\mu=\mu_{i}$ on each $O_{i}.$
\end{theorem}

\subsubsection{Lebesgue measure}

(Doob p.47)

So far measures have been reviewed through their properties. The Lebesgue
measure is the basic example of a measure on the set of real numbers, and from
there is used to compute integrals of functions. Notice that the Lebesgue
measure is not, by far, the unique measure that can be defined on $%
\mathbb{R}
$, but it has remarquable properties listed below.

\paragraph{Lebesgue measure on $%
\mathbb{R}
$\newline}

\begin{definition}
The \textbf{Lebesgue measure} on $%
\mathbb{R}
$\ denoted dx is the only complete, locally compact, translation invariant,
positive Borel measure, such that $dx\left(  ]a,b]\right)  =b-a$ for any
interval in $%
\mathbb{R}
.$ It is regular and $\sigma-$finite.
\end{definition}

It is built as follows.

1. $%
\mathbb{R}
$ is a metric space, thus a measurable space with its Borel algebra S

Let $F:%
\mathbb{R}
\rightarrow%
\mathbb{R}
$ be an increasing right continuous function, define

$F\left(  \infty\right)  =\lim_{x\rightarrow\infty}F\left(  x\right)
,F\left(  -\infty\right)  =\lim_{x\rightarrow-\infty}F\left(  x\right)  $

2. For any semi closed interval ]a,b] define the set function :

$\lambda\left(  ]a,b]\right)  =F(b)-F(a)$

then $\lambda$ has a unique extension as a complete measure on $\left(
\mathbb{R}
,S\right)  $ finite on compact subsets

3. Conversely if $\mu$ is a measure on $\left(
\mathbb{R}
,S\right)  $ finite on compact subsets there is an increasing right continuous
function F, defined up to a constant, such that : $\mu\left(  ]a,b]\right)
=F(b)-F(a)$

4. If $F\left(  x\right)  =x$ the measure is the Lebesgue measure, also called
the Lebesgue-Stieljes measure, and denoted dx. It is the usual measure on $%
\mathbb{R}
.$

5. If $\mu$ is a probability then F is the distribution function.

\paragraph{Lebesgue measure on $%
\mathbb{R}
^{n}$\newline}

The construct can be extended to $%
\mathbb{R}
^{n}$ :

For functions $F:%
\mathbb{R}
^{n}\rightarrow%
\mathbb{R}
$ define the operators

$D_{j}\left(  ]a,b]F\right)  (x_{1},x_{2},..x_{j-1},x_{j+1},...x_{n})$

$=F(x_{1},x_{2},..x_{j-1},b,x_{j+1},...x_{n})-F(x_{1},x_{2},..x_{j-1}%
,a,x_{j+1},...x_{n}) $

Choose F such that it is right continuous in each variable and :

$\forall a_{j}<b_{j}:%
{\textstyle\prod\limits_{j=1}^{n}}
D_{j}\left(  ]a_{j},b_{j}]F\right)  \geq0$

The measure of an hypercube is then defined as the difference of F between its faces.

\begin{theorem}
The Lebesgue measure on $%
\mathbb{R}
^{n}$ is the tensorial product $dx=dx_{1}\otimes...\otimes dx_{n}$ of the
Lebesgue measure on each component $x_{k}.$
\end{theorem}

So with the Lebesgue measure the measure of any subset of $%
\mathbb{R}
^{n}$ which is defined as disjointed union of hypercubes can be computed.\ Up
to a multiplicative constant the Lebesgue measure is "the volume" enclosed in
an open subset of $%
\mathbb{R}
^{n}$ . To go further and compute the Lebesgue measure of any set on $%
\mathbb{R}
^{n}$\ the integral on manifolds is used.

\subsubsection{Properties of measures}

\paragraph{A measure is order preserving on subsets\newline}

\begin{theorem}
(Doob p.18) A measure $\mu$ on a measurable space $\left(  E,S\right)  $ \ is :

i) countably subadditive:

$\mu\left(  \cup_{i\in I}A_{i}\right)  \leq\sum_{i\in I}\mu\left(
A_{i}\right)  $ for any countable family $\left(  A_{i}\right)  _{i\in
I},A_{i}\in S$ of subsets in S

ii) order preserving :

$A,B\in S,A\sqsubseteq B\Rightarrow\mu\left(  A\right)  \leq\mu\left(
B\right)  $

$\mu\left(  \varnothing\right)  =0$
\end{theorem}

\paragraph{Extension of a finite additive function on an algebra:}

\begin{theorem}
Hahn-Kolmogorov (Doob p.40)\ There is a unique extension of a
finitely-additive function $\mu_{0}:S_{0}\rightarrow%
\mathbb{R}
_{+}$ on an algebra $S_{0}$ on a set E into a measure on $\left(
E,\sigma\left(  S_{0}\right)  \right)  .$
\end{theorem}

\paragraph{Value of a measure on a sequence of subsets\newline}

\begin{theorem}
Cantelli (Doob p.26) For a sequence $\left(  A_{n}\right)  _{n\in%
\mathbb{N}
}$ of subsets $A_{n}\in S$ of the measured space $\left(  E,S,\mu\right)  $ :

i) $\mu\left(  \lim\inf A_{n}\right)  \leq\lim\inf\mu\left(  A_{n}\right)
\leq\lim\sup\mu\left(  A_{n}\right)  $

ii) if $\sum_{n}\mu\left(  A_{n}\right)  <\infty$ then $\mu\left(  \lim\sup
A_{n}\right)  =0$

iii) if $\mu$ is finite then $\lim\sup\mu\left(  A_{n}\right)  \leq\mu\left(
\lim\sup A_{n}\right)  $
\end{theorem}

\begin{theorem}
(Doob p.18) For a map $\mu:S\rightarrow%
\mathbb{R}
_{+}$ on a measurable space $\left(  E,S\right)  ,$ the following conditions
are equivalent :

i) $\mu$ is a finite measure

ii) For any disjoint sequence $\left(  A_{n}\right)  _{n\in%
\mathbb{N}
}$ in S : $\mu\left(  \cup_{n\in%
\mathbb{N}
}A_{n}\right)  =\sum_{n\in%
\mathbb{N}
}\mu\left(  A_{n}\right)  $

iii) For any increasing sequence $\left(  A_{n}\right)  _{n\in%
\mathbb{N}
}$ in S with $\lim A_{n}=A$ : $\lim\mu\left(  A_{n}\right)  =\mu\left(
A\right)  $

iv) For any decreasing sequence $\left(  A_{n}\right)  _{n\in%
\mathbb{N}
}$ in S with $\lim A_{n}=\varnothing$ : $\lim\mu\left(  A_{n}\right)  =0$
\end{theorem}

\paragraph{Tensorial product of measures\newline}

\begin{theorem}
(Doob p.48) If $\left(  E_{i},S_{i},\mu_{i}\right)  _{i=1}^{n}$ are measured
spaces and $\mu_{i}$ are $\sigma-$finite measures then there is a unique
measure $\mu$ on (E,S) :

$E=%
{\textstyle\prod\limits_{i=1}^{n}}
E_{i},S=\sigma\left(  S_{1}\times S_{2}..\times S_{n}\right)  =\sigma\left(
\sigma\left(  S_{1}\right)  \times\sigma\left(  S_{2}\right)  \times
...\sigma\left(  S_{n}\right)  \right)  $

such that : $\forall\left(  A_{i}\right)  _{i=1}^{n},A_{i}\in S_{i}:\mu\left(
%
{\textstyle\prod\limits_{i=1}^{n}}
A_{i}\right)  =%
{\textstyle\prod\limits_{i=1}^{n}}
\mu_{i}\left(  A_{i}\right)  $
\end{theorem}

$\mu$ is the \textbf{tensorial product of the measures} $\mu=\mu_{1}\otimes
\mu_{2}...\otimes\mu_{n}$ (also denoted as a product $\mu=\mu_{1}\times\mu
_{2}...\times\mu_{n})$

\paragraph{Sequence of measures\newline}

\begin{definition}
A sequence of measures or signed measures $\left(  \mu_{n}\right)  _{n\in%
\mathbb{N}
}$ on the measurable space (E,S) converges to a limit $\mu$ if $\forall A\in
S,\exists\mu\left(  A\right)  =\lim\mu_{n}\left(  A\right)  .$
\end{definition}

\begin{theorem}
Vitali-Hahn-Saks (Doob p.30) The limit $\mu$\ of a convergent sequence of
measures $\left(  \mu_{n}\right)  _{n\in%
\mathbb{N}
}$ on the measurable space (E,S) is a measure if each of the $\mu_{n}$ is
finite or if the sequence is increasing.

The limit $\mu$\ of a convergent sequence of signed measures $\left(  \mu
_{n}\right)  _{n\in%
\mathbb{N}
}$ on the measurable space (E,S) is a signed measure if each of the $\mu_{n} $
is finite..
\end{theorem}

\paragraph{Pull-back, push forward of a measure\newline}

\begin{definition}
Let $\left(  E_{1},S_{1}\right)  $, $(E_{2},S_{2})$ be measurable spaces,
$F:E_{1}\rightarrow E_{2}$ a measurable map such that $F^{-1}$ is\ measurable.

the \textbf{push forward} (or image) by F of the measure $\mu_{1}$ on $E_{1} $
is the measure on $(E_{2},S_{2})$ denoted $F_{\ast}\mu_{1}$ defined by :
$\forall A_{2}\in S_{2}:F_{\ast}\mu_{1}\left(  A_{2}\right)  =\mu_{1}\left(
F^{-1}\left(  A_{2}\right)  \right)  $

the \textbf{pull back} by F of the measure $\mu_{2}$ on $E_{2}$ is the measure
on $(E_{1},S_{1})$ denoted $F^{\ast}\mu_{2}$ \ defined by : $\forall A_{1}\in
S_{1}:F^{\ast}\mu_{2}\left(  A_{1}\right)  =\mu_{2}\left(  F\left(
A_{1}\right)  \right)  $
\end{definition}

\begin{definition}
(Doob p.60) If $f_{1},...f_{n}:E\rightarrow F$ are measurable maps from the
measured space $\left(  E,S,\mu\right)  $ into the measurable space (F,S') and
f is the map $f:E\rightarrow F^{n}:f=\left(  f_{1},f_{2},..f_{n}\right)  $
then $f_{\odot}\mu$ \ is called the \textbf{joint measure}. The
i\ \textbf{marginal distribution} is defined as $\forall A^{\prime}\in
S^{\prime}:\mu_{i}\left(  A^{\prime}\right)  =\mu\left(  f_{i}^{-1}\left(
\pi_{i}^{-1}\left(  A^{\prime}\right)  \right)  \right)  $ where $\pi
_{i}:F^{n}\rightarrow F$ is the i projection.
\end{definition}

\subsubsection{Almost eveywhere property}

\begin{definition}
A \textbf{null set} of a measured space $\left(  E,S,\mu\right)  $ is a set
A$\in S:\mu\left(  A\right)  =0.$ A property which is satisfied everywhere in
E but in a null set is said to be $\mu-$ everywhere satisfied (or
\textbf{almost everywhere satisfied}).
\end{definition}

\begin{definition}
The \textbf{support} of a Borel measure $\mu$, denoted Supp$\left(
\mu\right)  ,$ is the complement of the union of all the null open subsets The
support of a measure is a closed subset.
\end{definition}

\paragraph{Completion of a measure\newline}

It can happen that A is a null set and that $\exists B\subset A,B\notin S$ so
B is not measurable.

\begin{definition}
A measure is said to be \textbf{complete} if any subset of a null set is null.\ 
\end{definition}

\begin{theorem}
(Doob p.37) There is always a unique extension of the $\sigma-$algebra S of a
measured space such that the measure is complete (and identical for any subset
of S).
\end{theorem}

Notice that the tensorial product of complete measures is not necessarily complete

\paragraph{Applications to maps\newline}

\begin{theorem}
(Doob p.57) If the maps $f,g:E\rightarrow F$ from the complete measured space
$\left(  E,S,\mu\right)  $ to the measurable space (F,S') are almost eveywhere
equal, then if f is measurable then g is measurable.
\end{theorem}

\begin{theorem}
Egoroff (Doob p.69) If the sequence $\left(  f_{n}\right)  _{n\in%
\mathbb{N}
}$ of measurable maps $f_{n}:E\rightarrow F$ from the finite measured space
$\left(  E,S,\mu\right)  $ to the metric space $\left(  F,d\right)  $ is
almost everywhere convergent in E to f, then $\forall\varepsilon>0,\exists
A_{\varepsilon}\in S,\mu\left(  E\backslash A_{\varepsilon}\right)
<\varepsilon$ such that $\left(  f_{n}\right)  _{n\in%
\mathbb{N}
}$ is uniformly convergent in A$_{\varepsilon}$ .
\end{theorem}

\begin{theorem}
Lusin (Doob p.69) For every measurable map $f:E\rightarrow F$ from a complete
metric space $\left(  E,S,\mu\right)  $ endowed with a finite mesure $\mu$ to
a metric space F, then $\forall\varepsilon>0$ there is a compact
A$_{\varepsilon}$ , $\mu\left(  E\backslash A_{\varepsilon}\right)
<\varepsilon,A_{\varepsilon}$ such that f is continuous in A$_{\varepsilon}$ .
\end{theorem}

\subsubsection{Decomposition of signed measures}

Signed measures can be decomposed in a positive and a negative measure.
Moreover they can be related to a measure (specially the Lebesgue measure)
through a procedure similar to the differentiation.

\paragraph{Decomposition of a signed measure\newline}

\begin{theorem}
Jordan decomposition (Doob p.145): If $\left(  E,S,\mu\right)  $ is a signed
measure space,

define : $\forall A\in S:\mu_{+}\left(  A\right)  =\sup_{B\subset A}\mu\left(
B\right)  ;\mu_{-}\left(  A\right)  =-\inf_{B\subset A}\mu\left(  B\right)  $

then :

i) $\mu_{+},\mu_{-}$\ are positive measures on (E,S) such that $\mu=\mu
_{+}-\mu_{-}$

ii) $\mu_{+}$ is finite if $\mu<\infty,$

iii) $\mu_{-}$ is finite if $\mu>-\infty$

iv) $\left\vert \mu\right\vert =\mu_{+}+\mu_{-}$ is a positive measure on
(E,S) called the \textbf{total variation} of the measure

v) If there are measures $\lambda_{1},\lambda_{2}$ such that $\mu=\lambda
_{1}-\lambda_{2}$ then $\mu_{+}\leq\lambda_{1},\mu_{-}\leq\lambda_{2}$

vi) (Hahn decomposition) There are subsets $E_{+},E_{-}$ unique up to a null
subset, such that :

$E=E_{+}\cup E_{-};E_{+}\cap E_{-}=\varnothing$

$\forall A\in S:\mu_{+}\left(  A\right)  =\mu\left(  A\cap E_{+}\right)
,\mu_{-}\left(  A\right)  =\mu\left(  A\cap E_{-}\right)  $
\end{theorem}

The decomposition is not unique.

\paragraph{Complex measure\newline}

\begin{theorem}
If $\mu,\nu$ are signed measure on (E,S), then $\mu+i\nu$ is a measure valued
in $%
\mathbb{C}
,$ called a \textbf{complex measure}.

Conversely any complex measure can be uniquely decomposed as $\mu+i\nu$ where
$\mu,\nu$ are real signed measures.
\end{theorem}

\begin{definition}
A signed or complex measure $\mu$ is said to be finite if $\left\vert
\mu\right\vert $ is finite.
\end{definition}

\paragraph{Absolute continuity of a measure\newline}

\begin{definition}
If $\lambda$ is a positive measure on the measurable space (E,S), $\mu$ a
signed measure on (E,S):

i) $\mu$ is \textbf{absolutely continuous} relative to $\lambda$ if $\mu$ (or
equivalently $\left\vert \mu\right\vert )$\ vanishes on null sets of
$\lambda.$

ii) $\mu$ is \textbf{singular} relative to $\lambda$ if there is a null set A
for $\lambda$ such that $\left\vert \mu\right\vert \left(  A^{c}\right)  =0$

iii) if $\mu$ is absolutely continuous (resp.singular) then $\mu_{+},\mu_{-}$
are absolutely continuous (resp.singular)
\end{definition}

Thus with $\lambda=dx$ the Lebesgue measure, a singular measure can take non
zero value for finite sets of points in $%
\mathbb{R}
.$ And an absolutely continuous measure is the product of a function and the
Lebesgue measure.

\begin{theorem}
(Doob p.147) A signed measure $\mu$ on the measurable space (E,S) is
absolutely continuous relative to the finite measure $\lambda$ on (E,S) iff :

$\lim_{\lambda\left(  A\right)  \rightarrow0}\mu\left(  A\right)  =0$
\end{theorem}

\begin{theorem}
Vitali-Hahn-Saks (Doob p.147) If the sequence $\left(  \mu_{n}\right)  _{n\in%
\mathbb{N}
}$ of measures on (E,S), absolutely continuous relative to a finite measure
$\lambda,$ converges to $\mu$ then $\mu$ is a measure and it is also
absolutely continuous relative to $\lambda$
\end{theorem}

\begin{theorem}
Lebesgue decomposition (Doob p.148) A signed measure $\mu$\ on a measured
space $\left(  E,S,\lambda\right)  $ can be uniquely decomposed as : $\mu
=\mu_{c}+\mu_{s}$ where $\mu_{c}$ is a signed measure absolutely continuous
relative to $\lambda$ and $\mu_{s}$ is a signed measure singular relative to
$\lambda$
\end{theorem}

\paragraph{Radon\textbf{-}Nikodym derivative\newline}

\begin{theorem}
Radon-Nicodym (Doob p.150)\ For every finite signed measure $\mu$ on the
finite measured space $\left(  E,S,\lambda\right)  $, there is an integrable
function $f:E\rightarrow%
\mathbb{R}
$ uniquely defined up to null $\lambda$\ subsets, such that for the absolute
continuous component $\mu_{c}$ of $\mu:$ $\mu_{c}\left(  A\right)  =\int
_{A}f\lambda$ . For a scalar c such that $\mu_{c}\geq c\lambda$ (resp.$\mu
_{c}\leq c\lambda)$ then $f\geq c$ (resp.$f\leq c)$ almost everywhere
\end{theorem}

f is the \textbf{Radon-Nikodym derivative} (or density) of $\mu_{c}$ with
respect to $\lambda$

There is a useful extension if $E=%
\mathbb{R}
:$

\begin{theorem}
(Doob p.159) Let $\lambda,\mu$ be locally finite measures on $%
\mathbb{R}
,\lambda$ complete, a closed interval I containing x, then

$\forall x\in%
\mathbb{R}
:\varphi\left(  x\right)  =\lim_{I\rightarrow x}\frac{\mu\left(  I\right)
}{\lambda\left(  I\right)  }$ exists and is an integrable function on $%
\mathbb{R}
$ almost $\lambda$ everywhere finite

$\forall X\in S:\mu_{c}\left(  X\right)  =\int_{X}\varphi\lambda$ where
$\mu_{c}$ is the absolutely continuous component of $\mu$ relative to
$\lambda$
\end{theorem}

$\varphi$\ is denoted : $\varphi\left(  x\right)  =\frac{d\mu}{d\lambda
}\left(  x\right)  $

\subsubsection{Kolmogorov extension of a measure}

These concepts are used in stochastic processes. The Kolmogorov extension can
be seen as the tensorial product of an infinite number of measures.

Let (E,S,$\mu$) a measured space and I any set. $E^{I}$ is the set of maps :
$\varpi:I\rightarrow E$ . The purpose is to define a measure on the set
$E^{I}.$

Any finite subset J of I of cardinality n can be written $J=\left\{
j_{1},j_{2},..j_{n}\right\}  $

Define for $\varpi:I\rightarrow E$ \ the map:

$\varpi_{J}:J\rightarrow E^{n}::\varpi_{J}=\left(  \varpi\left(  j_{1}\right)
,\varpi\left(  j_{2}\right)  ,..\varpi\left(  j_{n}\right)  \right)  \in
E^{n}$

For each n there is a $\sigma-$algebra : $S_{n}=\sigma\left(  S^{n}\right)  $
and for each $A_{n}\in S_{n}$ the condition $\varpi_{J}\in A_{n}$ defines a
subset\ of $E^{I}$ : all maps $\varpi\in E$ such that $\varpi_{J}\in A_{n}. $
If, for a given J, $A_{n}$ varies in $S_{n}$ one gets an algebra $\Sigma_{J}.$
The union of all these algebras is an algebra $\Sigma_{0}$ in $E^{I}$ but
usually not a $\sigma-$algebra. Each of its subsets can be expressed as the
combination of $\Sigma_{J}$\ , with J finite.

However it is possible to get a measure on $E^{I}:$ this is the Kolmogorov extension.

\bigskip

\begin{theorem}
Kolmogorov (Doob p.61) If E is a complete metric space with its Borel algebra,
$\lambda:\Sigma_{0}\rightarrow%
\mathbb{R}
_{+}$ a function countably additive on each $\Sigma_{J},$ then $\lambda$ has
an extension in a measure $\mu$ on $\sigma\left(  \Sigma_{0}\right)  .$
\end{theorem}

Equivalently :

If for any finite subset J of n elements of I there is a finite measure
$\mu_{J}$ on (E$^{n}$,S$^{n})$ such that :

$\forall s\in\mathfrak{S}\left(  n\right)  ,\mu_{J}=\mu_{s(J)}$ : it is symmetric

$\forall K\subset I,card(K)=p<\infty,\forall A_{j}\in S:$

$\mu_{J}\left(  A_{1}\times A_{2}..\times A_{n}\right)  \mu_{K}\left(
E^{p}\right)  =\mu_{J\cup K}\left(  A_{1}\times A_{2}..\times A_{n}\times
E^{p}\right)  $

then there is a $\sigma-$algebra $\Sigma,$\ and a measure $\mu$ such that :

$\mu_{J}\left(  A_{1}\times A_{2}..\times A_{n}\right)  =\mu\left(
A_{1}\times A_{2}..\times A_{n}\right)  $

Thus if there are marginal measures $\mu_{J},$ meeting reasonnable
requirements, $\left(  E^{I},\Sigma,\mu\right)  $ is a measured space.

\bigskip

\subsection{Integral}

\label{Integral}

Measures act on subsets.\ Integrals act on functions. Here integral are
integral of \textit{real functions} defined on a measured space.\ We will see
later integral of r-forms on r-dimensional manifolds, which are a different breed.

\subsubsection{Definition}

\begin{definition}
A \textbf{step function} on a measurable space (E,S) is a map :
$f:E\rightarrow%
\mathbb{R}
_{+}$ defined by a disjunct family $\left(  A_{i},y_{i}\right)  _{i\in I}$ where

$A_{i}\in S,y_{i}\in%
\mathbb{R}
_{+}$: $\forall x\in E:f\left(  x\right)  =\sum_{i\in I}y_{i}1_{A_{i}}\left(
x\right)  $
\end{definition}

\begin{definition}
The integral of a step function on a measured space $(E,S,\mu)$ is : $\int
_{E}f\mu=\sum_{i\in I}y_{i}\mu\left(  A_{i}\right)  $
\end{definition}

\begin{definition}
The integral of a measurable positive function $f:E\rightarrow\overline{%
\mathbb{R}
}_{+}$ on a measured space $(E,S,\mu)$ is :

$\int_{E}f\mu=\sup\int_{E}g\mu$ for all step functions g such that $g\leq f$
\end{definition}

Any measurable function $f:E\rightarrow\overline{%
\mathbb{R}
}$ can always be written as : $f=f_{+}-f_{-}$ with $f_{+},f_{-}:E\rightarrow
\overline{%
\mathbb{R}
}_{+}$ measurable such that they do not take $\infty$ values on the same set.

The integral of a measurable function $f:E\rightarrow\overline{%
\mathbb{R}
}$ on a measured space (E,S,$\mu$) is :

$\int_{E}f\mu=\int_{E}f_{+}\mu-\int_{E}f_{-}\mu$

\begin{definition}
A function $f:E\rightarrow\overline{%
\mathbb{R}
}$ is \textbf{integrable} if $\left\vert \int_{E}f\mu\right\vert <\infty$ and
$\int_{E}f\mu$ is the \textbf{integral} of f over E with respect to $\mu$
\end{definition}

Notice that the integral can be defined for functions which take infinite values.

A function $f:E\rightarrow\overline{%
\mathbb{C}
}$ is integrable iff its real part and its imaginary part are integrable and
$\int_{E}f\mu=\int_{E}\left(  \operatorname{Re}f\right)  \mu+i\int_{E}\left(
\operatorname{Im}f\right)  \mu$

Warning ! $\mu$ is a real measure, and this is totally different from the
integral of a function over a complex variable

The integral of a function on a measurable subset A of E is :

$\int_{A}f\mu=\int_{E}\left(  f\times1_{A}\right)  \mu$

The \textbf{Lebesgue integral} denoted $\int fdx$ is the integral with $\mu=
$\ the Lebesgue measure dx on $%
\mathbb{R}
$.

Any \textbf{Riemann} \textbf{integrable} function is Lebesgue integrable, and
the integrals are equal.\ But the converse is not true. A function is Riemann
integrable iff it is continuous but for a set of Lebesgue null measure.

\subsubsection{Properties of the integral}

The spaces of integrable functions are studied in the Functional analysis part.

\begin{theorem}
The set of real (resp.complex) integrable functions on a measured space
(E,S,$\mu)$ is a real (resp.complex) vector space and the integral is a linear map.
\end{theorem}

if f,g are integrable functions $f:E\rightarrow\overline{%
\mathbb{C}
},$ a,b constant scalars then af+bg is integrable and $\int_{E}\left(
af+bg\right)  \mu=a\int_{E}f\mu+b\int_{E}g\mu$

\begin{theorem}
If is an integrable function $f:E\rightarrow\overline{%
\mathbb{C}
}$ on a measured space (E,S,$\mu)$ then : $\lambda\left(  A\right)  =\int
_{A}f\mu$ is a measure on (E,S).

If $f\geq0$ and g is measurable\ then $\int_{E}g\lambda=\int_{E}gf\mu$
\end{theorem}

\begin{theorem}
Fubini (Doob p.85) If $\left(  E_{1},S_{1},\mu_{1}\right)  ,\left(
E_{2},S_{2},\mu_{2}\right)  $ are $\sigma-$finite\ measured spaces,
$f:E_{1}\times E_{2}\rightarrow\overline{%
\mathbb{R}
}_{+}$ \ an integrable function on $\left(  E_{1}\times E_{2},\sigma\left(
S_{1}\times S_{2}\right)  ,\mu_{1}\otimes\mu_{2}\right)  $ then :

i) for almost all $x_{1}\in E_{1},$ the function f$\left(  x_{1},.\right)
:E_{2}\rightarrow\overline{%
\mathbb{R}
}_{+}$ is $\mu_{2}$\ integrable

ii) $\forall x_{1}\in E_{1},$ the function $\int_{\left\{  x_{1}\right\}
\times E_{2}}f\mu_{2}:E_{1}\rightarrow\overline{%
\mathbb{R}
}_{+}$ \ is $\mu_{1}$\ integrable

iii) $\int_{E_{1}\times E_{2}}f\mu_{1}\otimes\mu_{2}=\int_{E_{1}}\mu
_{1}\left(  \int_{\left\{  x_{1}\right\}  \times E_{2}}f\mu_{2}\right)
=\int_{E_{2}}\mu_{2}\left(  \int_{E_{1}\times\left\{  x_{2}\right\}  }f\mu
_{1}\right)  $
\end{theorem}

\begin{theorem}
(Jensen's inequality) (Doob p.87)

Let : $\left[  a,b\right]  \subset%
\mathbb{R}
,\varphi:\left[  a,b\right]  \rightarrow%
\mathbb{R}
$ an integrable convex function, semi continuous in a,b, $\left(
E,S,\mu\right)  $ a finite measured space, f an integrable function
$f:E\rightarrow\left[  a,b\right]  ,$

then $\varphi\left(  \int_{E}f\mu\right)  \leq\int_{E}\left(  \varphi\circ
f\right)  \mu$
\end{theorem}

The result holds if f,$\varphi$ are not integrable but are lower bounded

\begin{theorem}
If f is a function $f:E\rightarrow\overline{%
\mathbb{C}
}$ on a measured space (E,S,$\mu)$ :

i) If $f\geq0$ almost everywhere and $\int_{E}f\mu=0$ then f=0 almost everywhere

ii) If f is integrable then $\left\vert f\right\vert <\infty$ almost everywhere

iii) if $f\geq0,c\geq0:$ $\int_{E}f\mu\geq c\mu\left(  \left\{  \left\vert
f\right\vert \geq c\right\}  \right)  $

iv) If f is measurable, $\varphi:%
\mathbb{R}
_{+}\rightarrow%
\mathbb{R}
_{+}$ monotone increasing, $c\in%
\mathbb{R}
_{+}$ then

$\int_{E}\left\vert f\right\vert \varphi\mu\geq\int_{\left\vert f\right\vert
\geq c}\varphi\left(  c\right)  f\mu=\varphi\left(  c\right)  \mu\left(
\left\{  \left\vert f\right\vert \geq c\right\}  \right)  $
\end{theorem}

\begin{theorem}
(Lieb p.26) Let $\nu$\ be a Borel measure on $\overline{%
\mathbb{R}
}_{+}$ such that $\forall t\geq0:\phi\left(  t\right)  =\nu\left(
\lbrack0,t)\right)  <\infty,\left(  E,S,\mu\right)  $ a $\sigma-$finite
measured space, $f:E\rightarrow%
\mathbb{R}
_{+}$ integrable, then :

$\int_{E}\phi\left(  f\left(  x\right)  \right)  \mu\left(  x\right)
=\int_{0}^{\infty}\mu\left(  \left\{  f\left(  x\right)  >t\right\}  \right)
\nu\left(  t\right)  $

$\forall p>0\in%
\mathbb{N}
:\int_{E}\left(  f\left(  x\right)  \right)  ^{p}\mu\left(  x\right)
=p\int_{0}^{\infty}t^{p-1}\mu\left(  \left\{  f\left(  x\right)  >t\right\}
\right)  \nu\left(  t\right)  $

$f\left(  x\right)  =\int_{0}^{\infty}1_{\left\{  f>x\right\}  }dx$
\end{theorem}

\begin{theorem}
Beppo-Levi (Doob p.75) If $\left(  f_{n}\right)  _{n\in%
\mathbb{N}
}$ is an increasing sequence of measurable functions $f_{n}:E\rightarrow
\overline{%
\mathbb{R}
}_{+}$ on a measured space $(E,S,\mu),$which converges to f then :
$\lim_{n\rightarrow\infty}\int_{E}f_{n}\mu=\int_{E}f\mu$
\end{theorem}

\begin{theorem}
Fatou (Doob p.82) If $\left(  f_{n}\right)  _{n\in%
\mathbb{N}
}$ is a sequence of measurable functions $f_{n}:E\rightarrow\overline{%
\mathbb{R}
}_{+}$ on a measured space $(E,S,\mu)$ and $f=\lim\inf f_{n}$

then $\int_{E}f\mu\leq\lim\inf\left(  \int_{E}f_{n}\mu\right)  $

If the functions $f$, $f_{n}$ are integrable and $\int_{E}f\mu=\lim
_{n\rightarrow\infty}\int_{E}f_{n}\mu$

then $\lim_{n\rightarrow\infty}\int_{E}\left\vert f-f_{n}\right\vert \mu=0 $
\end{theorem}

\begin{theorem}
Dominated convergence Lebesgue's theorem (Doob p.83) If $\left(  f_{n}\right)
_{n\in%
\mathbb{N}
}$ is a sequence of measurable functions $f_{n}:E\rightarrow\overline{%
\mathbb{R}
}_{+}$ on a measured space $(E,S,\mu)$ if there is an integrable function g on
$\left(  E,S,\mu\right)  $ such that $\forall x\in E,\forall n:$ $\left\vert
f_{n}\left(  x\right)  \right\vert \leq g\left(  x\right)  $, and
$f_{n}\rightarrow f$ almost everywhere then : $\lim_{n\rightarrow\infty}%
\int_{E}f_{n}\mu=\int_{E}f\mu$
\end{theorem}

\subsubsection{Pull back and push forward of a Radon measure}

This is the principle behind the change of variable in an integral.

\begin{definition}
If $\mu$ is a Radon measure on a topological space E endowed with its Borel
$\sigma-$algebra, a \textbf{Radon integral} is the integral $\ell\left(
\varphi\right)  =\int\varphi\mu$ for an integrable function : $\varphi
:E\rightarrow\overline{%
\mathbb{R}
}.$ $\ell$ is a linear functional on the functions $C\left(  E;\overline{%
\mathbb{R}
}\right)  $
\end{definition}

The set of linear functional on a vector space (of functions) is a vector
space which can be endowed with a norm (see Functional Analysis).

\begin{definition}
Let $E_{1},E_{2}$ be two sets, K a field and a map : $F:E_{1}\rightarrow
E_{2}$

The \textbf{pull back of a function }$\varphi_{2}:E_{2}\rightarrow K$ is the
map : $F^{\ast}:C\left(  E_{2};K\right)  \rightarrow C\left(  E_{1};K\right)
::F^{\ast}\varphi_{2}=\varphi_{2}\circ F$

The \textbf{push forward of a function }$\varphi_{1}:E_{1}\rightarrow K$ is
the map $F_{\ast}:C\left(  E_{1};K\right)  \rightarrow C\left(  E_{2}%
;K\right)  ::F_{\ast}\varphi_{1}=F\circ\varphi_{1}$
\end{definition}

\begin{theorem}
(Schwartz III p.535) Let $\left(  E_{1},S_{1}\right)  ,\left(  E_{2}%
,S_{2}\right)  $ be two topological Hausdorff locally compact spaces with
their Borel algebra,a \textit{continuous} map $F:E_{1}\rightarrow E_{2}$ .

i) let $\mu$ be a Radon measure in $E_{1},$ $\ell\left(  \varphi_{1}\right)
=\int_{E_{1}}\varphi_{1}\mu$ be the Radon integral

If F is a compact (proper) map, then there is a Radon measure on E$_{2},$
called the \textbf{push forward} of $\mu$\ and denoted $F_{\ast}\mu,$ such
that :

$\varphi_{2}\in C\left(  E_{2};%
\mathbb{R}
\right)  $ is $F_{\ast}\mu$ integrable iff $F^{\ast}\varphi_{2}$ is $\mu$
integrable and

$F_{\ast}\ell\left(  \varphi_{2}\right)  =\int_{E_{2}}\varphi_{2}\left(
F_{\ast}\mu\right)  =\ell\left(  F^{\ast}\varphi_{2}\right)  =\int_{E_{1}%
}\left(  F^{\ast}\varphi_{2}\right)  \mu$

ii) let $\mu$ be a Radon measure in $E_{2},$ $\ell\left(  \varphi_{2}\right)
=\int_{E_{2}}\varphi_{2}\mu$ be the Radon integral,

If F is an open map, then there is a Radon measure on E$_{1},$ called the
\textbf{pull back} of $\mu$ and denoted $F^{\ast}\mu$ such that :

$\varphi_{1}\in C\left(  E_{1};%
\mathbb{R}
\right)  $ is $F^{\ast}\mu$ integrable iff $F_{\ast}\varphi_{1}$ is $\mu$
integrable and

$F^{\ast}\ell\left(  \varphi_{1}\right)  =\int_{E_{1}}\varphi_{1}\left(
F^{\ast}\mu\right)  =\ell\left(  F_{\ast}\varphi_{1}\right)  =\int_{E_{2}%
}\left(  F_{\ast}\varphi_{2}\right)  \mu$

Moreover :

i) the maps $F_{\ast},F^{\ast}$ when defined, are linear on measures and functionals

ii) The support of the measures are such that :

$Supp\left(  F_{\ast}\ell\right)  \subset F\left(  Supp(\ell)\right)
,Supp\left(  F^{\ast}\ell\right)  \subset F^{-1}\left(  Supp(\ell)\right)  $

iii) The norms of the functionals :$\left\Vert F_{\ast}\ell\right\Vert
=\left\Vert F^{\ast}\ell\right\Vert =\left\Vert \ell\right\Vert \leq\infty$

iv) $F_{\ast}\mu,F^{\ast}\mu$ are positive iff $\mu$ is positive

v) If $\left(  E_{3},S_{3}\right)  $ is also a topological Hausdorff locally
compact space and $G:E_{2}\rightarrow E_{3},$ then, when defined :

$\left(  F\circ G\right)  _{\ast}\mu=F_{\ast}\left(  G_{\ast}\mu\right)  $

$\left(  F\circ G\right)  ^{\ast}\mu=G^{\ast}\left(  F^{\ast}\mu\right)  $

If F is an homeomorphism then the push forward and the pull back are inverse
operators :

$\left(  F^{-1}\right)  ^{\ast}\mu=F_{\ast}\mu,\left(  F^{-1}\right)  _{\ast
}\mu=F^{\ast}\mu$
\end{theorem}

Remark : the theorem still holds if $E_{1},E_{2}$\ are the countable union of
compact subsets, F is measurable and $\mu$\ is a positive finite
measure.\ Notice that there are conditions attached to the map F.

\paragraph{Change of variable in a multiple integral\newline}

An application of this theorem is the change of variable in multiple integrals
(in anticipation of the next part). The Lebesgue measure dx on $%
\mathbb{R}
^{n}$ can be seen as the tensorial product of the measures $dx^{k},k=1...n$
which reads : $dx=dx^{1}\otimes...\otimes dx^{n}$ or more simply :
$dx=dx^{1}..dx^{n}$ so that the integral $\int_{U}fdx$\ of $f\left(
x^{1},...x^{n}\right)  $ over a subset U is by Fubini's theorem computed by
taking the successive integral over the variables $x^{1},...x^{n}.$ Using the
definition of the Lebesgue measure we have the following theorem.

\begin{theorem}
(Schwartz IV p.71) Let $U,V$ be two open subsets of $%
\mathbb{R}
^{n},F:U\rightarrow V$ a diffeomorphism, x coordinates in U, y coordinates in
V, $y^{i}=F^{i}\left(  x^{1},...x^{n}\right)  $ then :

$\varphi_{2}\in C\left(  V;%
\mathbb{R}
\right)  $ is Lebesgue integrable iff $F^{\ast}\varphi_{2}$ is Lebesgue
integrable and

$\int_{V}\varphi_{2}\left(  y\right)  dy=\int_{U}\varphi_{2}\left(  F\left(
x\right)  \right)  \left\vert \det\left[  F^{\prime}\left(  x\right)  \right]
\right\vert dx$

$\varphi_{1}\in C\left(  U;%
\mathbb{R}
\right)  $ is Lebesgue integrable iff $F_{\ast}\varphi_{1}$ is Lebesgue
integrable and

$\int_{U}\varphi_{1}\left(  x\right)  dx=\int_{V}\varphi_{1}\left(
F^{-1}\left(  y\right)  \right)  \left\vert \det\left[  F^{\prime}\left(
y\right)  \right]  ^{-1}\right\vert dy$

So :

$F_{\ast}dx=dy=\left\vert \det\left[  F^{\prime}\left(  x\right)  \right]
\right\vert dx$

$F^{\ast}dy=dx=\left\vert \det\left[  F^{\prime}\left(  y\right)  \right]
^{-1}\right\vert dy$
\end{theorem}

This formula is the basis for any change of variable in a multiple
integral.\ We use $dx,dy$ to denote the Lebesgue measure for clarity but
\textit{there is only one measure on }$%
\mathbb{R}
^{n}$ which applies to different real scalar variables.\ For instance in $%
\mathbb{R}
^{3}$ when we go from cartesian coordinates (the usual x,y,z) to spherical
coordinates : $x=r\cos\theta\cos\varphi;y=r\sin\theta\cos\varphi
;z=r\sin\varphi$ the new variables are real scalars $\left(  r,\theta
,\varphi\right)  $ subject to the Lebesgue measure which reads $drd\theta
d\varphi$ and

$\int_{U}\varpi(x,y,z)dxdydz$

$=\int_{V(r,\theta,\varphi)}\varpi(r\cos\theta\cos\varphi,r\sin\theta
\cos\varphi,r\sin\varphi)\left\vert r^{2}\cos\varphi\right\vert drd\theta
d\varphi$

The presence of the absolute value in the formula is due to the fact that the
Lebesgue measure is positive : the measure of a set must stay positive when we
use one variable or another.

\bigskip

\subsection{Probability}

\label{Probability}

Probability is a full branch of mathematics, which relies on measure theory,
thus its place here.

\begin{definition}
A \textbf{probability space} is a measured space $\left(  \Omega,S,P\right)  $
endowed with a measure P called a \textbf{probability} such that $P(\Omega)=1$.\ 
\end{definition}

So all the results above can be fully extended to a probability space, and we
have many additional definitions and results. The presentation is limited to
the basic concepts.

\subsubsection{Definitions}

Some adjustments in vocabulary are common in probability :

1. An element of a $\sigma-$algebra S is called an \textbf{event} : basically
it represents the potential occurence of some phenomena.\ Notice that an event
is usually not a single point in S, but a subset. A subset of S or a
subalgebra of S can be seen as the "simultaneous" realization of events.

2. A measurable map $X:\Omega\rightarrow F$ with F usually a discrete space or
a metric space endowed with its Borel algebra, is called a \textbf{random
variable} (or \textbf{stochastic variable}). So the events $\varpi$\ occur in
$\Omega$ and the value $X\left(  \varpi\right)  $ is in F.

3. two random variables X,Y are \textbf{equal almost surely} if $P\left(
\left\{  \varpi:X\left(  \varpi\right)  \neq Y\left(  \varpi\right)  \right\}
\right)  =0$ so they are equal almost everywhere

4. If X is a real valued random variable :

its \textbf{distribution function} is the map : $F:%
\mathbb{R}
\rightarrow\left[  0,1\right]  $ defined as :

$F\left(  x\right)  =P\left(  \varpi\in\Omega:X\left(  \varpi\right)  \leq
x\right)  $

Its \textbf{expected value} is $E\left(  X\right)  =\int_{\Omega}XP$ this is
its "average" value

its \textbf{moment of order r} is : $\int_{\Omega}\left(  X-E(X)\right)  ^{r}P
$ the moment of order 2 is the \textbf{variance}

the Jensen's inequality reads : for $1\leq p:\left(  E\left(  \left\vert
X\right\vert \right)  \right)  ^{p}\leq E\left(  \left\vert X\right\vert
^{p}\right)  $

and for X valued in [a,b] and\ any function $\varphi:\left[  a,b\right]
\rightarrow%
\mathbb{R}
$ integrable convex, semi continuous in a,b : $\varphi\left(  E\left(
X\right)  \right)  \leq E\left(  \varphi\circ X\right)  $

5. If $\Omega=%
\mathbb{R}
$ then, according to Radon-Nikodym, there is a \textbf{density function}
defined as the derivative relative to the Lebesgue measure :

$\rho\left(  x\right)  =\lim_{I\rightarrow x}\frac{P(I)}{dx(I)}=\lim
_{h_{1},h_{2}\rightarrow0_{+}}\frac{1}{h_{1}+h_{2}}\left(  F(x+h_{1}%
)-F(x-h_{2})\right)  $

where I is an interval containing I, $h_{1},h_{2}>0$

and the absolutely continuous component of P is such that :

$P_{c}\left(  \varpi\right)  =\int_{\varpi}\rho\left(  x\right)  dx$

\subsubsection{Independant sets}

\paragraph{Independant events\newline}

\begin{definition}
The \textit{events} $A_{1},A_{2},...A_{n}\in S$ of a probability space
$(\Omega,S,P)$\ are \textbf{independant} if :

$P\left(  B_{1}\cap B_{2},..\cap B_{n}\right)  =P\left(  B_{1}\right)
P\left(  B_{2}\right)  ..P\left(  B_{n}\right)  $ \ 

where for any i : $B_{i}=A_{i}$ or $B_{i}=A_{i}^{c}$

A family $\left(  A_{i}\right)  _{i\in I}$ of events are independant if any
finite subfamily is independant
\end{definition}

\begin{definition}
Two $\sigma-$algebra $S_{1},S_{2}$ are independant if any pair of subsets
$\left(  A_{1},A_{2}\right)  \in S_{1}\times S_{2}$ are independant.
\end{definition}

If a collection of $\sigma-$algebras $\left(  S_{i}\right)  _{i=1}^{n}$ are
independant then $\sigma\left(  S_{i}\times S_{j}\right)  ,\sigma\left(
S_{k}\times S_{l}\right)  $ are independant for i,j,k,l distincts

\paragraph{Conditional probability\newline}

\begin{definition}
On a probability space $(\Omega,S,P)$, if $A\in S,P\left(  A\right)  \neq0$
then $P\left(  B|A\right)  =\frac{P\left(  B\cap A\right)  }{P\left(
A\right)  }$ defines a new probability on $(\Omega,S)$ called
\textbf{conditional probability} (given A).Two events are independant iff
$P\left(  B|A\right)  =P\left(  B\right)  $
\end{definition}

\paragraph{Independant random variables\newline}

\begin{definition}
Let $(\Omega,S,P)$ a probability space, (F,S') a measurable space, a family of
random variables $\left(  X_{i}\right)  _{i\in I},X_{i}:\Omega\rightarrow F$
are \textbf{independant} if for any finite $J\subset I$ the $\sigma-$algebras
$\left(  \sigma\left(  X_{j}\right)  \right)  _{j\in J}$ are independant
\end{definition}

remind that $\sigma\left(  X_{j}\right)  =X_{j}^{-1}\left(  S^{\prime}\right)
)$

Equivalently :

$\forall\left(  A_{j}\right)  _{j\in J}\in S^{\prime J},$ $P\left(  \cap_{j\in
J}X_{j}^{-1}\left(  A_{j}\right)  \right)  =%
{\textstyle\prod\limits_{j\in J}}
P\left(  X_{j}^{-1}\left(  A_{j}\right)  \right)  $

usually denoted : $P\left(  \left(  X_{j}\in A_{j}\right)  _{j\in J}\right)  =%
{\textstyle\prod\limits_{j\in J}}
P\left(  X_{j}\in A_{j}\right)  $

\paragraph{The 0-1 law\newline}

The basic application of the theorems on sequence of sets give the following theorem:

\begin{theorem}
\textbf{the 0-1 law}: Let in the probability space $(\Omega,S,P)$:

$\left(  U_{n}\right)  _{n\in%
\mathbb{N}
}$ an increasing sequence of $\sigma-$algebras of measurable subsets,

$\left(  V_{n}\right)  _{n\in%
\mathbb{N}
}$ a decreasing sequence of $\sigma-$algebras of measurable subsets with
$V_{1}\subset\sigma\left(  \cup_{n\in%
\mathbb{N}
}U_{n}\right)  $

If, for each n, $U_{n},V_{n}$ are independant, then $\cap_{n\in%
\mathbb{N}
}V_{n}$ contains only null subsets and their complements
\end{theorem}

Applications : let the sequence of independant random variables $\left(
X_{n}\right)  _{n\in%
\mathbb{N}
},$ $X_{n}\in%
\mathbb{R}
,$ take $U_{n}=\sigma\left(  X_{1},...X_{n}\right)  ,V_{n}=\sigma\left(
X_{n+1},...\right)  .$ The series $\sum_{n}X_{n}$ converges either almost
everywhere or almost nowhere. The random variables $\lim\sup\frac{1}{n}\left(
\sum_{m=1}^{n}X_{m}\right)  ,\lim\inf\frac{1}{n}\left(  \sum_{m=1}^{n}%
X_{m}\right)  $ are almost everywhere constant (possibly infinite). Thus :

\begin{theorem}
On a probability space $(\Omega,S,P)$ for every sequence of independant random
real variables $\left(  X_{n}\right)  _{n\in%
\mathbb{N}
},$ the series $\frac{1}{n}\left(  \sum_{m=1}^{n}X_{m}\right)  $ converges
almost everywhere to a constant or almost nowhere
\end{theorem}

\begin{theorem}
Let $\left(  A_{n}\right)  $ a sequence of Borel subsets in $%
\mathbb{R}
$ with a probability P :

$P\left(  \lim\sup\left(  X_{n}\in A_{n}\right)  \right)  =0$ or 1 . This is
the probability that $X_{n}\in A_{n}$ infinitely often

$P\left(  \lim\inf\left(  X_{n}\in A_{n}\right)  \right)  =0$ or 1. This is
the probability that $X_{n}\in A_{n}^{c}$ only finitely often
\end{theorem}

\subsubsection{Conditional expectation of random variables}

The conditional probability is a measure acting on subsets.\ Similarly the
conditional expectation of a random variable is the integral of a random
variable using a conditional probablity.

Let $(\Omega,S,P)$ be a probability space and s a sub$\ \sigma-$algebra of
S.\ Thus the subsets in s are S measurable sets.\ The restriction $P_{s}$ of P
to s is a finite measure on $\Omega$.

\begin{definition}
On a probability space $(\Omega,S,P)$, the \textbf{conditional expectation} of
a random variable $X:\Omega\rightarrow F$ given a sub$\ \sigma-$algebra
$s\subset S$ is a random variable $Y:\Omega\rightarrow F$ denoted $E(X|s)$
meeting the two requirements:

i) Y is s measurable and $P_{s}$ integrable

ii) $\forall\varpi\in s:\int_{\varpi}YP_{s}=\int_{\varpi}XP$
\end{definition}

Thus X defined on $(\Omega,S,P)$ is replaced by Y defined on $(\Omega
,s,P_{s})$ with the condition that X,Y have the same expectation value on
their common domain, which is s.

Y is not unique : any other function which is equal to Y almost everywhere but
on P null subsets of s meets the same requirements.

With s=A it gives back the previous definition of P(B%
$\vert$%
A)=E(1$_{B}|A)$

\begin{theorem}
(Doob p.183) If s is a sub$\ \sigma-$algebra of S on a probability space
$(\Omega,S,P)$, we have the following relations for the conditional
expectations of random variables $X,Z:\Omega\rightarrow F$

i) If X=Z almost everywhere then $E(X|s)=E(Z|s)$ almost everywhere

ii) If a,b are real constants and X,Z are real random variables :

$E(aX+bZ|s)=aE(X|s)+bE(Z|s)$

iii) If $F=%
\mathbb{R}
:X\leq Z\Rightarrow E(X|s)\leq E(Z|s)$ and $E(X|s)\leq E(\left\vert
X\right\vert |s)$

iv) if X is a constant function : $E\left(  X|s\right)  =X$

v) If $S^{\prime}\subset S$ then $E\left(  E\left(  X|S^{\prime}\right)
|S\right)  =E\left(  E\left(  X|S\right)  |S^{\prime}\right)  =E\left(
X|S^{\prime}\right)  $
\end{theorem}

\begin{theorem}
Bepo-Levi (Doob p.183) If s is a sub$\ \sigma-$algebra of S on a probability
space $(\Omega,S,P)$ and $\left(  X_{n}\right)  _{n\in%
\mathbb{N}
}$ an increasing sequence of positive random variables with integrable limit,
then : $\lim E\left(  X_{n}|s\right)  =E\left(  \lim X_{n}|s\right)  $
\end{theorem}

\begin{theorem}
Fatou (Doob p.184) If s is a sub$\ \sigma-$algebra of S on a probability space
$(\Omega,S,P)$ and $\left(  X_{n}\right)  _{n\in%
\mathbb{N}
}$ is a sequence of positive integrable random variables with $X=\lim\inf
X_{n}$ integrable then :

$E\left(  X|s\right)  \leq\lim\inf E\left(  X_{n}|s\right)  $ almost everywhere

$\lim E\left(  \left\vert X-X_{n}\right\vert |s\right)  =0$ almost everywhere
\end{theorem}

\begin{theorem}
Lebesgue (Doob p.184) If s is a sub$\ \sigma-$algebra of S on a probability
space $(\Omega,S,P)$ and $\left(  X_{n}\right)  _{n\in%
\mathbb{N}
}$ a sequence of real random variables such that there is an integrable
function g : with $\forall n,\forall x\in E:\left\vert X_{n}\left(  x\right)
\right\vert \leq g\left(  x\right)  $ and $X_{n}\rightarrow X$ almost
everywhere then : $\lim E\left(  X_{n}|s\right)  =E\left(  \lim X_{n}%
|s\right)  $
\end{theorem}

\begin{theorem}
Jensen (Doob p.184) If $\left[  a,b\right]  \subset%
\mathbb{R}
,\varphi:\left[  a,b\right]  \rightarrow%
\mathbb{R}
$ is an integrable convex function, semi continuous in a,b,

X is a real random variable with range in [a,b] on a probability space
$(\Omega,S,P)$,s is a sub$\ \sigma-$algebra of S

then $\varphi\left(  E\left(  X|s\right)  \right)  \leq E\left(
\varphi\left(  X\right)  |s\right)  $
\end{theorem}

\subsubsection{Stochastic process}

\paragraph{The problem\newline}

Consider the simple case of coin tossing.\ For one shot the set of events is
$\Omega=\left\{  H,T\right\}  $ (for "head" and "tail") with $H\cap
T=\varnothing,H\cup T=E,P(H)=P(T)=1/2$ and the variable is $X=0,1.$ For n
shots the set of events must be extended : $\Omega_{n}=\left\{
HHT...T,...\right\}  $ and the value of the realization of the shot n is :
$X_{n}\in\left\{  0,1\right\}  .$ Thus one can consider the family $\left(
X_{p}\right)  _{p=1}^{n}$ which is some kind of random variable, but the
\textit{set of events depend on n}, and \textit{the probability law depends on
n}, and could also depends on the occurences of the $X_{p}$ if the shots are
not independant.

\begin{definition}
A \textbf{stochastic process} is a random variable $X=\left(  X_{t}\right)
_{t\in T}$ on a probability space $\left(  \Omega,S,P\right)  $ such that
$\forall t\in T,X_{t}$ is a random variable on a probability space $\left(
\Omega_{t},S_{t},P_{t}\right)  $ valued in a measurable space (F,S')\ with
$X_{t}^{-1}\left(  F\right)  =\Omega_{t}$
\end{definition}

i) T is any set ,which can be uncountable, but is well ordered so for any
finite subspace J of T we can write : $J=\left\{  j_{1},j_{2},..,j_{n}%
\right\}  $

ii) so far no relation is assumed between the spaces $\left(  \Omega_{t}%
,S_{t},P_{t}\right)  _{t\in T}$

iii) $X_{t}^{-1}\left(  F\right)  =E_{t}\Rightarrow P_{t}\left(  X_{t}\in
F\right)  =1$

If T is infinite, given each element, there is no obvious reason why there
should be a stochastic process, and how to build $\Omega$,S,P.

\paragraph{The measurable space $(\Omega,S)$\newline}

The first step is to build a measurable space $(\Omega,S)$:

1. $\Omega=%
{\textstyle\prod\limits_{t\in T}}
\Omega_{t}$ which always exists and is the set of all \textit{maps}
$\phi:T\rightarrow\cup_{t\in T}\Omega_{t}$ such that $\forall t:\phi\left(
t\right)  \in\Omega_{t}$

2. Let us consider the subsets of $\Omega$ of the type : $A_{J}=%
{\textstyle\prod\limits_{t\in T}}
A_{t}$ where $A_{t}\in S_{t}$ and all but a finite number $t\in J$ of which
are equal to $\Omega_{t}$ (they are called \textbf{cylinders}). For a given J
and varying $\left(  A_{t}\right)  _{t\in J} $ in $S_{t}$ we get a $\sigma
-$algebra denoted $S_{J}.$ It can be shown that the union of all these
algebras for all finite J generates a $\sigma-$algebra $S$

3. We can do the same with F : define $A_{J}^{\prime}=%
{\textstyle\prod\limits_{t\in T}}
A_{t}^{\prime}$ where $A_{t}^{\prime}\in S_{t}^{\prime}$ and all but a finite
number $t\in J$ of which are equal to $F.$ The preimage of such $A_{J}%
^{\prime}$ by X is such that $X_{t}^{-1}\left(  A_{t}^{\prime}\right)  \in
S_{t}$ and for $t\in J:X_{t}^{-1}\left(  F\right)  =\Omega_{t}$ so
$X^{-1}\left(  A_{J}^{\prime}\right)  =A_{J}\in S_{J}.$ And the $\sigma
-$algebra generated by the $S_{J}^{\prime}$ has a preimage in S. S is the
smallest $\sigma-$algebra for which all the $\left(  X_{t}\right)  _{t\in T}$
are simultaneouly measurable (meaning that the map X is measurable).

4. The next step, finding P, is less obvious.\ There are many constructs based
upon relations between the $\left(  \Omega_{t},S_{t},P_{t}\right)  ,$ we will
see some of them later.\ There are 2 general results.

\paragraph{The Kolmogov extension\newline}

This is one implementation of the extension with $(\Omega,S)$ as presented above

\begin{theorem}
Given a family $\left(  X_{t}\right)  _{t\in T}$ of random variables on
probability spaces $\left(  \Omega_{t},S_{t},P_{t}\right)  $ valued in a
measurable space (F,S')\ with $X_{t}^{-1}\left(  F\right)  =\Omega_{t},$ if
all the $\Omega_{t}$ and F are complete metric spaces with their Borel
algebras, and if for any finite subset J there are marginal probabilities
$P_{J}$ defined on $\left(  \Omega,S_{J}\right)  ,S_{J}\subset S$ such that :

$\forall s\in\mathfrak{S}\left(  n\right)  ,P_{J}=P_{s(J)}$ the marginal
probabilities $P_{J}$ do not depend on the order of J

$\forall J,K\subset I,card(J)=n,card(K)=p<\infty,\forall A_{j}\in S^{\prime}:$

$P_{J}\left(  X_{j_{1}}^{-1}\left(  A_{1}\right)  \times X_{j_{2}}^{-1}\left(
A_{2}\right)  ..\times X_{j_{n}}^{-1}\left(  A_{n}\right)  \right)  $

$=P_{J\cup K}\left(  X_{j_{1}}^{-1}\left(  A_{1}\right)  \times X_{j_{2}}%
^{-1}\left(  A_{2}\right)  ..\times X_{j_{n}}^{-1}\left(  A_{n}\right)  \times
E^{p}\right)  $

then there is a $\sigma-$algebra $S$ on $\Omega=%
{\textstyle\prod\limits_{t\in T}}
\Omega_{t},$\ a probability P such that :

$P_{J}\left(  X_{j_{1}}^{-1}\left(  A_{1}\right)  \times X_{j_{2}}^{-1}\left(
A_{2}\right)  ..\times X_{j_{n}}^{-1}\left(  A_{n}\right)  \right)  $

$=P\left(  X_{j_{1}}^{-1}\left(  A_{1}\right)  \times X_{j_{2}}^{-1}\left(
A_{2}\right)  ..\times X_{j_{n}}^{-1}\left(  A_{n}\right)  \right)  $
\end{theorem}

These conditions are reasonable, notably in physics : if for any finite J,
there is a stochastic process $\left(  X_{t}\right)  _{t\in J}$\ then one can
assume that the previous conditions are met and say that there is a stochastic
process $\left(  X_{t}\right)  _{t\in T}$ with some probability P, usually not
fully explicited, from which all the marginal probability $P_{J}$ are deduced.

\paragraph{Conditional expectations\newline}

The second method involves conditional expectation of random variables, with
$(\Omega,S)$ as presented above .

\begin{theorem}
Given a family $\left(  X_{t}\right)  _{t\in T}$ of random variables on
probability spaces $\left(  \Omega_{t},S_{t},P_{t}\right)  $ valued in a
measurable space (F,S')\ with $X_{t}^{-1}\left(  F\right)  =\Omega_{t}.$ If,
for any finite subset J of T, $S_{J}\subset S,\Omega_{J}=%
{\textstyle\prod\limits_{j\in I}}
\Omega_{j}$ and the map $X_{J}=\left(  X_{j_{1}},X_{j_{2}},..X_{j_{n}}\right)
:\Omega_{J}\rightarrow F^{J},$ there is a probability $P_{J}$ on $\left(
\Omega_{J},S_{J}\right)  $ and a conditional expectation $Y_{J}=E\left(
X_{J}|S_{J}\right)  $ then there is a probability on $(\Omega,S)$ such that :
$\forall\varpi\in S_{J}:\int_{\varpi}Y_{J}P_{J}=\int_{\varpi}XP$
\end{theorem}

This result if often presented (Tulc\'{e}a) with $T=%
\mathbb{N}
$ and

$P_{J}=P\left(  X_{n}=x_{n}|X_{1}=x_{1},...X_{n-1}=x_{n-1}\right)  $ which are
the transition probabilities.

\subsubsection{Martingales}

Martingales are classes of stochastic processes. They precise the relation
between the probability spaces $\left(  \Omega_{t},S_{t},P_{t}\right)  _{t\in
T}$

\begin{definition}
A \textbf{filtered probability space }$\left(  \Omega,S,\left(  S_{i}\right)
_{i\in I},\left(  X_{i}\right)  _{i\in I},P\right)  $ is a probability space
$(\Omega,S)$ , an ordered set I, and a family $\left(  S_{i}\right)  _{i\in
I}$ where $S_{i}$ is a $\sigma-$subalgebra of S such that : $S_{i}\sqsubseteq
S_{j}$ whenever $i<j$.
\end{definition}

\begin{definition}
A \textbf{filtered stochastic process} on a filtered probability space is a
family $\left(  X_{i}\right)  _{i\in I}$ of random variables $X_{i}%
:\Omega\rightarrow F$ such that each $X_{i}$ is measurable in $\left(
\Omega,S_{i}\right)  $.
\end{definition}

\begin{definition}
A filtered stochastic process is a \textbf{Markov process} if :

$\forall i<j,A\subset F:P\left(  X_{j}\in A|S_{i}\right)  =P\left(  X_{j}\in
A|X_{i}\right)  $ almost everywhere
\end{definition}

So the probability at the step j depends only of the state $X_{i}$ meaning the
last one

\begin{definition}
A filtered stochastic process is a \textbf{martingale} if $\forall
i<j:X_{i}=E\left(  X_{j}|S_{i}\right)  $ almost everywhere
\end{definition}

That means that the future is totally conditionned by the past.

Then the function : $I\rightarrow F::E(X_{i})$ is constant almost everywhere

If $I=%
\mathbb{N}
$ the condition $X_{n}=E\left(  X_{n+1}|S_{n}\right)  $ is sufficient

A useful application of the theory is the following :

\begin{theorem}
Kolomogorov: Let $\left(  X_{n}\right)  $ a sequence of independant real
random variables on $(\Omega,S,P)$ with the same distribution law, then if
$X_{1}$ is integrable : $\lim_{n\rightarrow\infty}\left(  \sum_{p=1}^{n}%
X_{p}\right)  /n=E\left(  X_{1}\right)  $
\end{theorem}

\newpage

\section{BANACH\ SPACES}

The combination of an algebraic structure and a topologic structure on the
same set gives rise to new properties. Topological groups are \ studied in the
part "Lie groups". Here we study the other major algebraic structure : vector
spaces, which include algebras. A key feature of vector spaces is that all
n-dimensional vector spaces are agebraically isomorphs and homeomorphs : all
the topologies are equivalent and metrizable. Thus most of their properties
stem from their algebraic structure.\ The situation is totally different for
the infinite dimensional vector spaces.\ And the most useful of them are the
complete normed vector spaces, called Banach spaces which are the spaces
inhabited by many functions. Among them we have the Banach algebras and the
Hilbert spaces.

\bigskip

\subsection{Topological vector spaces}

\label{Topological vector space}

\subsubsection{Definitions}

\begin{definition}
A \textbf{topological vector space} is a vector space endowed with a topology
such that the operations (linear combination of vectors and scalars) are continuous.
\end{definition}

\begin{theorem}
(Wilansky p.273, 278) A topological vector space is regular and connected
\end{theorem}

\paragraph{Finite dimensional vector spaces\newline}

\begin{theorem}
Every Hausdorff n-dimensional topological vector space over a field K is
isomorphic (algebraically) and homeomorphic (topologically) to $K^{n}$.
\end{theorem}

So on a finite dimensional Haussdorff topological space all the topologies are
equivalent to the topology defined by a norm (see below) and are metrizable.
In the following all n-dimensional vector spaces will be endowed with their
unique normed topology if not stated otherwise. Conversely we have the
fundamental theorem:

\begin{theorem}
A Hausdorff topological vector space is finite-dimensional if and only if it
is locally compact.
\end{theorem}

And we have a less obvious result :

\begin{theorem}
(Schwartz II p.97) If there is an homeomorphism between open sets of two
finite dimensional vector spaces E,F on the same field, then dimE=dimF
\end{theorem}

\paragraph{Vector subspace\newline}

\begin{theorem}
A vector subspace F of a topological vector space is itself a topological
vector space.
\end{theorem}

\begin{theorem}
A finite dimensional vector subspace F is always closed in a topological
vector space E.\ 
\end{theorem}

\begin{proof}
A finite dimensional vector space is defined by a finite number of linear
equations, which constitute a continuous map and F is the inverse image of 0.
\end{proof}

Warning ! If F is infinite dimensional it can be open or closed, or neither of
the both.

\begin{theorem}
If F is a vector subspace of E, then the quotient space E/F is Hausdorff iff F
is closed in E.\ In particular E is Hausdorff iff the subset \{0\} is closed.
\end{theorem}

This is the application of the general theorem on quotient topology.

Thus if E is not Hausdorff E can be replaced by the set E/F where F is the
closure of $\left\{  0\right\}  .$ For instance functions which are almost
everywhere equal are taken as equal in the quotient space and the latter
becomes Hausdorff.

\begin{theorem}
(Wilansky p.274) The closure of a vector subspace is still a vector subspace.
\end{theorem}

\paragraph{Bounded vector space\newline}

Without any metric it is still possible to define some kind of "bounded
subsets". The definition is consistent with the usual one when there is a semi-norm.

\begin{definition}
A subset X of a topological vector space over a field K is bounded if for any
n(0) neighborhood of 0 there is $k\in K$ such that $X\subset kn(0)$
\end{definition}

\paragraph{Product of topological vector spaces\newline}

\begin{theorem}
The product (possibly infinite) of topological vector spaces, endowed with its
vector space structure and the product topology, is a topological vector space.
\end{theorem}

This is the direct result of the general theorem on the product topology.

Example : the space of real functions : $f:%
\mathbb{R}
\rightarrow%
\mathbb{R}
$ can be seen as $%
\mathbb{R}
^{%
\mathbb{R}
}$ and is a topological vector space

\paragraph{Direct sum\newline}

\begin{theorem}
The direct sum $\oplus_{i\in I}E_{i}$ (finite or infinite) of vector subspaces
of a vector space E is a topological vector space.
\end{theorem}

\begin{proof}
It is algebraically isomorphic to their product $\widetilde{E}=%
{\textstyle\prod\limits_{i\in I}}
E_{i}.$ Endowed with the product topology $\widetilde{E}$ is a topological
vector space, and the projections $\pi_{i}:\widetilde{E}\rightarrow E_{i}$ are
continuous. So the direct sum E is a topological vector space homeomorphic to
$\widetilde{E}$.
\end{proof}

This, obvious, result is useful because it is possible to part a vector space
without any reference to a basis. A usual case if of a topological space which
splits. Algebraically $E=E_{1}\oplus E_{2}$ and it is isomorphic to $\left(
E_{1},0\right)  \times\left(  0,E_{2}\right)  \subset E\times E.$ 0 is closed
in E

\subsubsection{Linear maps on topological vector spaces}

The key point is that, in an infinite dimensional vector space, there are
linear maps which are not continuous. So it is necessary to distinguish
continuous linear maps, and this holds also for the dual space.

\paragraph{Continuous linear maps\newline}

\begin{theorem}
A linear map $f\in L(E;F)$ is continuous if the vector spaces E,F are on the
same field and finite dimensional.

A multilinear map $f\in L^{r}(E_{1},E_{2},..E_{r};F)$ is continuous if the
vector spaces $\left(  E_{i}\right)  _{i=1}^{r}$,F are on the same field and
finite dimensional.
\end{theorem}

\begin{theorem}
A linear map $f\in L(E;F)$ is continuous on the topological vector spaces E,F
iff it is continuous at 0 in E.

A multilinear map $f\in L^{r}(E_{1},E_{2},..E_{r};F)$ is continuous if it is
continuous at $\left(  0,..,0\right)  $ in $E_{1}\times E_{2}\times..E_{r}$.
\end{theorem}

\begin{theorem}
The kernel of a linear map $f\in L\left(  E;F\right)  $ between topological
vector space is either closed or dense in E. It is closed if f is continuous.
\end{theorem}

\begin{notation}
$%
\mathcal{L}%
\left(  E;F\right)  $ is the set of continuous linear map between topological
vector spaces E,F on the same field
\end{notation}

\begin{notation}
$G%
\mathcal{L}%
\left(  E;F\right)  $ is the set of continuous invertible linear map, with
continuous inverse, between topological vector spaces E,F on the same field
\end{notation}

\begin{notation}
$%
\mathcal{L}%
^{r}(E_{1},E_{2},...E_{r};F)$ is the set of continuous r-linear maps in
L$^{r}\left(  E_{1},E_{2},...E_{r};F\right)  $
\end{notation}

Warning ! The inverse of an invertible continuous map is not necessarily continuous.

\paragraph{Compact maps\newline}

Compact maps (also called proper maps) are defined for any topological space,
with the meaning that it maps compact sets to compact sets. However, because
compact sets are quite rare in infinite dimensional vector spaces, the
definition is extended as follows.

\begin{definition}
(Schwartz 2 p.58) A linear map $f\in L\left(  E;F\right)  $\ between
topological vector spaces E,F is said to be \textbf{compact} if the closure
$\overline{f\left(  X\right)  }$\ in F of the image of a bounded subset X of E
is compact in F.
\end{definition}

So compact maps "shrink" a set.

\begin{theorem}
(Schwartz 2.p.59) A compact map is continuous.
\end{theorem}

\begin{theorem}
(Schwartz 2.p.59) A continuous linear map $f\in%
\mathcal{L}%
\left(  E;F\right)  $\ between topological vector spaces E,F such that f(E) is
\ finite dimensional is compact.
\end{theorem}

\begin{theorem}
(Schwartz 2.p.59) The set of compact maps is a subspace of $%
\mathcal{L}%
\left(  E;F\right)  $\ . It is a two-sided ideal of the algebra
$\mathcal{L}$%
(E;E)
\end{theorem}

Thus the identity map in $%
\mathcal{L}%
\left(  E;E\right)  $ is compact iff E is finite dimensional.

\begin{theorem}
Riesz (Schwartz 2.p.66) : If $\lambda\neq0$\ is an eigen value of the compact
linear endomorphism f on a topological vector space E, then the vector
subspace $E_{\lambda}$ of corresponding eigen vectors is finite dimensional.
\end{theorem}

\paragraph{Dual vector space\newline}

As a consequence a linear form : $\varpi:E\rightarrow K$ is not necessarily continuous.

\begin{definition}
The vector space of continuous linear forms on a topological vector space E is
called its \textbf{topological dual}
\end{definition}

\begin{notation}
E' is the topological dual of a topological vector space E
\end{notation}

So $E^{\ast}=L\left(  E;K\right)  $ and $E^{\prime}=%
\mathcal{L}%
\left(  E;K\right)  $

The topological dual E' is \textit{included} in the algebraic dual E*, and
they are identical iff E is finite dimensional.

The topological bidual $\left(  E^{\prime}\right)  ^{\prime}$ may be or not
isomorphic to E if E is infinite dimensional.

\begin{definition}
The map: $\imath:E\rightarrow\left(  E^{\prime}\right)  ^{\prime}%
::\imath\left(  u\right)  \left(  \varpi\right)  =\varpi\left(  u\right)  $
between E and its topological bidual $\left(  E^{\prime}\right)  ^{\prime}$ is
linear and injective.

If it is also surjective then E is said to be \textbf{reflexive} and $\left(
E^{\prime}\right)  ^{\prime}$ is isomorphic to E.
\end{definition}

The map \i, called the \textbf{evaluation map,} is met quite often in this
kind of problems.

\begin{theorem}
The transpose of a linear continuous map : $f\in%
\mathcal{L}%
\left(  E;F\right)  $ is the continuous linear map : $f^{t}\in%
\mathcal{L}%
\left(  F^{\prime};E^{\prime}\right)  ::\forall\varpi\in F^{\prime}:f^{\prime
}\left(  \varpi\right)  =\varpi\circ f$
\end{theorem}

\begin{proof}
The transpose of a linear map $f\in L\left(  E;F\right)  $ is : $f^{t}\in
L(F^{\ast};E^{\ast})::\forall\varpi\in F^{\ast}:f^{t}\left(  \varpi\right)
=\varpi\circ f$

If f is continuous by restriction of F* to F' : $\forall\varpi\in F^{\prime
}:f^{\prime}\left(  \varpi\right)  =\varpi\circ f$ is a continuous map
\end{proof}

\begin{theorem}
Hahn-Banach (Bratelli 1 p.66) If C is a closed convex subset of a real locally
convex topological Hausdorff vector space E, and $p\notin C$ then there is a
continuous affine map : $f:E\rightarrow%
\mathbb{R}
$ such that $f(p)>1$ and $\forall x\in C:f\left(  x\right)  \leq1$
\end{theorem}

This is one of the numerous versions of this theorem.

\subsubsection{Tensor algebra}

Tensor, tensor products and tensor algebras have been defined without any
topology involved.\ All the definitions and results in the Algebra part can be
fully translated by taking continuous linear maps (instead of simply linear maps).

Let be E,F vector spaces over a field K.\ Obviously the map $\imath:E\times
F\rightarrow E\otimes F$ is continuous. So the universal property of the
tensorial product can be restated as : for every topological space S and
continuous bilinear map $f:E\times F\rightarrow S$ there is a unique
continuous linear map : $\widehat{f}:E\otimes F\rightarrow S$ such that
$f=\widehat{f}\circ\imath$

Covariant tensors must be defined in the topological dual $E^{\prime}.$
However the isomorphism between L(E;E) and $E\otimes E^{\ast}$ holds only if E
is finite dimensional so, in general,
$\mathcal{L}$%
$\left(  E;E\right)  $ is not isomorphic to $E\otimes E^{\prime}.$

\subsubsection{Affine topological space}

\begin{definition}
A \textbf{topological affine space} E is an affine space E with an underlying
topological vector space $\overrightarrow{E}$ such that the map :
$\overrightarrow{}:E\times E\rightarrow\overrightarrow{E}$ is continuous.
\end{definition}

So the open subsets in an affine topological space E can be deduced by
translation from the collection of open subsets at any given point of E.

An affine subspace is closed in E iff its underlying vector subspace is closed
in $\overrightarrow{E}.$ So :

\begin{theorem}
A finite dimensional affine subspace is closed.
\end{theorem}

Convexity plays an important role for topological affine spaces.\ In many ways
convex subsets behave like compact subsets.

\begin{definition}
A topological affine space $\left(  E,\Omega\right)  $ is \textbf{locally
convex} if there is a base of the topology comprised of convex subsets.\ 
\end{definition}

Such a base is a family C of open absolutely convex subsets $\varpi$
containing a point O :

$\forall\varpi\in C,M,N\in\varpi,\lambda,\mu\in K:\left\vert \lambda
\right\vert +\left\vert \mu\right\vert \leq1:\lambda M+\mu N\in\varpi$

and such that every neighborhood of O contains a element $k\varpi$ for some
$k\in K,\varpi\in C$

A locally convex space has a family of pseudo-norms and conversely (see below).

\begin{theorem}
(Berge p.262) The closure of a convex subset of a topological affine space is
convex. The interior of a convex subset of a topological affine space is convex.
\end{theorem}

\begin{theorem}
Schauder (Berge p.271) If f is a continuous map $f:C\rightarrow C$ where C is
a non empty compact convex subset of a locally convex affine topological
space, then there is $a\in C:f\left(  a\right)  =a$
\end{theorem}

\begin{theorem}
An affine map f is continuous iff its underlying linear map $\overrightarrow
{f}$ is continuous.
\end{theorem}

\begin{theorem}
Hahn-Banach theorem (Schwartz) : For every non empty convex subsets X,Y of a
topological affine space E over $%
\mathbb{R}
$, $X$ open subset, such that $X\cap Y=\varnothing$ , there is a closed
hyperplane H which does not meet X or Y.
\end{theorem}

A hyperplane H in an affine space is defined by an affine scalar equation
f(x)=0. If $f:E\rightarrow K$\ is continuous then H is closed and $f\in
E^{\prime}.$

So the theorem can be restated :

\begin{theorem}
For every non empty convex subsets X,Y of a topological affine space $\left(
E,\overrightarrow{E}\right)  $ over $%
\mathbb{C}
$, X open subset, such that $X\cap Y=\varnothing$ , there is a linear map
$\overrightarrow{f}\in\overrightarrow{E}^{\prime}$\ , $c\in%
\mathbb{R}
$\ such that \ for any $O\in E$:

$\forall x\in X,y\in Y:\operatorname{Re}\overrightarrow{f}\left(
\overrightarrow{Ox}\right)  <c<\operatorname{Re}\overrightarrow{f}\left(
\overrightarrow{Oy}\right)  $
\end{theorem}

\bigskip

\subsection{Normed vector spaces}

\label{Normed vector space}

\subsubsection{Norm on a vector space}

A topological vector space can be endowed with a metric, and thus becomes a
metric space.\ But an ordinary metric does not reflect the algebraic
properties, so what is useful is a norm.

\begin{definition}
A \textbf{semi-norm} on a vector space E over the field K (which is either $%
\mathbb{R}
$ or $%
\mathbb{C}
)$ is a function :$\left\Vert {}\right\Vert :E\rightarrow%
\mathbb{R}
_{+}$ such that :

$\forall u,v\in E,k\in K:$

$\left\Vert u\right\Vert \geq0;$

$\left\Vert ku\right\Vert =\left\vert k\right\vert \left\Vert u\right\Vert $
where $\left\vert k\right\vert $ is either the absolute value or the module of k

$\left\Vert u+v\right\Vert \leq\left\Vert u\right\Vert +\left\Vert
v\right\Vert $
\end{definition}

\begin{definition}
A vector space endowed with a semi norm is a \textbf{semi-normed vector space}
\end{definition}

\begin{theorem}
A semi-norm is a continuous convex map.
\end{theorem}

\begin{definition}
A \textbf{norm} on a vector space E is a semi-norm such that :

$\left\Vert u\right\Vert =0\Rightarrow u=0$
\end{definition}

\begin{definition}
If E is endowed with a norm $\left\Vert {}\right\Vert $ it is a \textbf{normed
vector space} $\left(  E,\left\Vert {}\right\Vert \right)  $
\end{definition}

The usual norms are :

i) $\left\Vert u\right\Vert =\sqrt{g(u,u)}$ where g is a definite positive
symmetric (or hermitian) form

ii) $\left\Vert u\right\Vert =\max_{i}\left\vert u_{i}\right\vert $ where
$u_{i}$\ are the components relative to a basis

iii) $\left\Vert k\right\Vert =\left\vert k\right\vert $ is a norm on K with
its vector space structure.

iv) On $%
\mathbb{C}
^{n}$ we have the norms :

$\left\Vert X\right\Vert _{p}=\sum_{k=1}^{n}c_{k}\left\vert x_{k}\right\vert
^{p}$ for p%
$>$%
0$\in%
\mathbb{N}
$

$\left\Vert X\right\Vert _{\infty}=\sup_{k=1..n}\left\vert x_{k}\right\vert $

with the fixed scalars : $\left(  c_{k}\right)  _{k=1}^{n},c_{k}>0\in%
\mathbb{R}
$

The \textbf{inequalities of H\"{o}lder-Minkovski} give :

$\forall p\geq1:\left\Vert X+Y\right\Vert _{p}\leq\left\Vert X\right\Vert
_{p}+\left\Vert Y\right\Vert _{p}$

and if $p<\infty$ then $\left\Vert X+Y\right\Vert _{p}=\left\Vert X\right\Vert
_{p}+\left\Vert Y\right\Vert _{p}\Rightarrow\exists a\in%
\mathbb{C}
:Y=aX$

\subsubsection{Topology on a semi-normed vector space}

A semi-norm defines a semi-metric by : $d\left(  u,v\right)  =\left\Vert
u-v\right\Vert $ but the converse is not true. There are vector spaces which
are metrizable but not normable (see Fr\'{e}chet spaces). So every result and
definition for semi-metric spaces hold for semi-normed vector space.

\begin{theorem}
A semi-norm (resp.norm) defines by restriction a semi-norm (resp.norm) on
every vector subspace.
\end{theorem}

\begin{theorem}
On a vector space E two semi-norms $\left\Vert {}\right\Vert _{1},\left\Vert
{}\right\Vert _{2}$ are \textbf{equivalent} if they define the same topology.
It is necessary and sufficient that :

$\exists k,k^{\prime}>0:\forall u\in E:\left\Vert u\right\Vert _{1}\leq
k\left\Vert u\right\Vert _{2}\Leftrightarrow\left\Vert u\right\Vert _{2}\leq
k^{\prime}\left\Vert u\right\Vert _{1}$
\end{theorem}

\begin{proof}
The condition is necessary. If $B_{1}\left(  0,r\right)  $ is a ball centered
at 0, open for the topology 1, and if the topology are equivalent then there
is ball $B_{2}\left(  0,r_{2}\right)  \subset B_{1}\left(  0,r\right)  $ so
$\left\Vert u\right\Vert _{2}\leq r_{2}\Rightarrow\left\Vert u\right\Vert
_{1}\leq r=kr_{2}.$ And similarly for a ball $B_{2}\left(  0,r\right)  .$

The condition is sufficient. Every ball $B_{1}\left(  0,r\right)  $ contains a
ball $B_{2}\left(  0,\frac{r}{k^{\prime}}\right)  $ and vice versa.
\end{proof}

The theorem is still true for norms.

\begin{theorem}
On a finite dimensional vector space all the norms are equivalent.
\end{theorem}

\begin{theorem}
The product E=$%
{\textstyle\prod\limits_{i\in I}}
E_{i}$ of a finite number of semi-normed vector spaces on a field K is still a
semi-normed vector space with one of the equivalent semi-norm :

$\left\Vert {}\right\Vert _{E}=\max\left\Vert {}\right\Vert _{E_{i}}$

$\left\Vert {}\right\Vert _{E}=\left(  \sum_{i\in I}\left\Vert {}\right\Vert
_{E_{i}}^{p}\right)  ^{1/p},1\leq p<\infty$
\end{theorem}

The product of an infinite number of normed vector spaces is not a normable
vector space.

\begin{theorem}
(Wilansky p.268) Every first countable topological vector space is semi-metrizable
\end{theorem}

\begin{theorem}
A topological vector space is normable iff it is Hausdorff and has a convex
bounded neighborhood of 0.
\end{theorem}

\begin{theorem}
(Schwartz I p.72) A subset of a finite dimensional vector space is compact iff
it is bounded and closed.
\end{theorem}

Warning ! This is false in an infinite dimensional normed vector space.

\begin{theorem}
(Wilansky p.276) If a semi-normed vector space has a totally bounded
neighborhood of 0 it has a dense finite dimensional vector subspace.
\end{theorem}

\begin{theorem}
(Wilansky p.271) A normed vector space is locally compact iff it is finite dimensional
\end{theorem}

\subsubsection{Linear maps}

The key point is that a norm can be assigned to every continuous linear map.

\paragraph{Continuous linear maps\newline}

\begin{theorem}
If E,F are semi-normed vector spaces on the same field, an $f\in L\left(
E;F\right)  $ \ then the following are equivalent:

i) f is continuous

ii) $\exists k\geq0:\forall u\in E:\left\Vert f\left(  u\right)  \right\Vert
_{F}\leq k\left\Vert u\right\Vert _{E}$

iii) f is uniformly continuous and globally Lipschitz of order 1
\end{theorem}

So it is equivalently said that f is bounded.

\begin{theorem}
Every linear map $f\in L\left(  E;F\right)  $ from a finite dimensional vector
space E to a normed vector space F, both on the same field, is uniformly
continuous and Lipschitz of order 1
\end{theorem}

If E,F are semi-normed vector spaces on the same field f is said to be
"bounded below" if : $\exists k\geq0:\forall u\in E:\left\Vert f\left(
u\right)  \right\Vert _{F}\geq k\left\Vert u\right\Vert _{E}$

\paragraph{Space of linear maps\newline}

\begin{theorem}
The space
$\mathcal{L}$%
(E;F) of continuous linear maps on the semi-normed vector spaces E,F on the
same field is a semi-normed vector space with the semi-norm : $\left\Vert
f\right\Vert _{%
\mathcal{L}%
\left(  E;F\right)  }=\sup_{\left\Vert u\right\Vert \neq0}\frac{\left\Vert
f\left(  u\right)  \right\Vert _{F}}{\left\Vert u\right\Vert _{E}}%
=\sup_{\left\Vert u\right\Vert _{E}=1}\left\Vert f\left(  u\right)
\right\Vert _{F}$

The semi-norm $\left\Vert {}\right\Vert _{%
\mathcal{L}%
\left(  E;F\right)  }$ has the following properties :

i) $\forall u\in E:\left\Vert f\left(  u\right)  \right\Vert \leq\left\Vert
f\right\Vert _{%
\mathcal{L}%
\left(  E;F\right)  }\left\Vert u\right\Vert _{E}$

ii) If E=F $\left\Vert Id\right\Vert _{E}=1$

iii) (Schwartz I p.107) In the composition of linear continuous maps :
$\left\Vert f\circ g\right\Vert \leq\left\Vert f\right\Vert \left\Vert
g\right\Vert $

iv) If $f\in%
\mathcal{L}%
(E;E)$ then its iterated $f^{n}\in%
\mathcal{L}%
(E;E) $ and $\left\Vert f^{n}\right\Vert =\left\Vert f\right\Vert ^{n}$
\end{theorem}

\paragraph{Dual\newline}

\begin{theorem}
The topological dual E' of the semi-normed vector spaces E is a semi-normed
vector space with the semi-norm : $\left\Vert f\right\Vert _{E^{\prime}}%
=\sup_{\left\Vert u\right\Vert \neq0}\frac{\left\vert f\left(  u\right)
\right\vert }{\left\Vert u\right\Vert _{E}}=\sup_{\left\Vert u\right\Vert
_{E}=1}\left\vert f\left(  u\right)  \right\vert $
\end{theorem}

This semi-norm defines a topology on E' called the \textbf{strong topology}.

\begin{theorem}
Banach lemna (Taylor 1 p.484): A linear form $\varpi\in F^{\ast}\ $on a a
vector subspace F of a semi-normed vector space E on a field K, such that :
$\forall u\in F$ $\left\vert \varpi\left(  u\right)  \right\vert
\leq\left\Vert u\right\Vert $ can be extended in a map $\widetilde{\varpi}\in
E^{\prime}$ such that $\forall u\in E:\left\vert \widetilde{\varpi}\left(
u\right)  \right\vert \leq\left\Vert u\right\Vert $
\end{theorem}

The extension is not necessarily unique. It is continuous. Similarly :

\begin{theorem}
Hahn-Banach (Wilansky p.269): A linear form $\varpi\in F^{\prime}\ $continuous
on a vector subspace F of a semi-normed vector space E on a field K can be
extended in a continuous map $\widetilde{\varpi}\in E^{\prime}$ such that
$\left\Vert \widetilde{\varpi}\right\Vert _{E^{\prime}}=\left\Vert
\varpi\right\Vert _{F^{\prime}}$
\end{theorem}

\begin{definition}
In a semi normed vector space E a \textbf{tangent functional} at $u\in E$ is a
1 form $\varpi\in E^{\prime}:\varpi\left(  u\right)  =\left\Vert
\varpi\right\Vert \left\Vert u\right\Vert $
\end{definition}

Using the Hahan-Banach theorem one can show that there are always non unique
tangent functionals.

\paragraph{Multilinear maps\newline}

\begin{theorem}
If $\left(  E_{i}\right)  _{i=1}^{r}$,F are semi-normed vector spaces on the
same field, and f$\in L^{r}\left(  E_{1},E_{2},..E_{r};F\right)  $ \ then the
following are equivalent:

i) f is continuous

ii) $\exists k\geq0:\forall\left(  u_{i}\right)  _{i=1}^{r}\in E:\left\Vert
f\left(  u_{1},..,u_{r}\right)  \right\Vert _{F}\leq k%
{\textstyle\prod\limits_{i=1}^{r}}
\left\Vert u_{i}\right\Vert _{E_{i}}$
\end{theorem}

Warning ! a multilinear map is never uniformly continuous.

\begin{theorem}
If $\left(  E_{i}\right)  _{i=1}^{r}$,F are semi-normed vector spaces on the
same field, the vector space of continuous r linear maps $f\in L^{r}\left(
E_{1},E_{2},..E_{r};F\right)  $ is a semi-normed vector space on the same
field with the norm :

$\left\Vert f\right\Vert _{%
\mathcal{L}%
^{r}}=\sup_{\left\Vert u\right\Vert _{i}\neq0}\frac{\left\Vert f\left(
u_{1},..,u_{r}\right)  \right\Vert _{F}}{\left\Vert u_{1}\right\Vert
_{1}...\left\Vert u_{r}\right\Vert _{r}}=\sup_{\left\Vert u_{i}\right\Vert
_{E_{i}}=1}\left\Vert f\left(  u_{1},..,u_{r}\right)  \right\Vert _{F}$
\end{theorem}

So : $\forall\left(  u_{i}\right)  _{i=1}^{r}\in E:\left\Vert f\left(
u_{1},..,u_{r}\right)  \right\Vert _{F}\leq\left\Vert f\right\Vert _{%
\mathcal{L}%
^{r}}%
{\textstyle\prod\limits_{i=1}^{r}}
\left\Vert u_{i}\right\Vert _{E_{i}}$

\begin{theorem}
(Schwartz I p.119) If E,F are semi-normed spaces,the map : $%
\mathcal{L}%
\left(  E.F\right)  \times E\rightarrow F::\varphi\left(  f,u\right)
=f\left(  u\right)  $ is bilinear continuous with norm 1
\end{theorem}

\begin{theorem}
(Schwartz I p.119) If E,F,G are semi-normed vector spaces then the composition
of maps : $%
\mathcal{L}%
\left(  E;F\right)  \times%
\mathcal{L}%
\left(  F;G\right)  \rightarrow%
\mathcal{L}%
\left(  E;G\right)  ::\circ\left(  f,g\right)  =g\circ f$ is bilinear,
continuous and its norm is 1
\end{theorem}

\subsubsection{Family of semi-norms\newline}

A family of semi-metrics on a topological space can be useful because its
topology can be Haussdorff (but usually is not semi-metric). Similarly on
vector spaces :

\begin{definition}
A \textbf{pseudo-normed space} is a vector space endowed with a family
$\left(  p_{i}\right)  _{i\in I}$ of semi-norms such that for any finite
subfamily J :

$\exists k\in I:\forall j\in J:p_{j}\leq p_{k}$
\end{definition}

\begin{theorem}
(Schwartz III p.435) A pseudo-normed vector space $(E,\left(  p_{i}\right)
_{i\in I})$ is a\ topological vector space with the base of open balls :

$B\left(  u\right)  =\cap_{j\in J}B_{j}\left(  u,r_{j}\right)  $ with
$B_{j}\left(  u,r_{j}\right)  =\left\{  v\in E:p_{j}\left(  u-v\right)
<r_{j}\right\}  $ ,

for every finite subset J of I and family $\left(  r_{j}\right)  _{j\in J}%
$,$r_{j}>0$
\end{theorem}

It works because all the balls $B_{j}\left(  u,r_{j}\right)  $ are convex
subsets, and the open balls $B\left(  u\right)  $ are convex subsets.

The functions $p_{i}$ must satisfy the usual conditions of semi-norms.

\begin{theorem}
A pseudo-normed vector space $(E,\left(  p_{i}\right)  _{i\in I})$ is
Hausdorff iff $\ $

$\forall u\neq0\in E,\exists i\in I:p_{i}\left(  u\right)  >0$
\end{theorem}

\begin{theorem}
A countable family of seminorms on a vector space defines a semi-metric on E
\end{theorem}

It is defined by : $d\left(  x,y\right)  =\sum_{n=0}^{\infty}\frac{1}{2^{n}%
}\frac{p_{n}\left(  x-y\right)  }{1+p_{n}\left(  x-y\right)  }.$

If E is Hausdorff then this pseudo-metric is a metric.

However usualy a pseudo-normed space is not normable.

\begin{theorem}
(Schwartz III p.436) A linear map between pseudo-normed spaces is continuous
if it is continuous at 0. It is then uniformly continuous and Lipschitz.
\end{theorem}

\begin{theorem}
A topological vector space is locally convex iff its topology can be defined
by a family of semi-norms.
\end{theorem}

\subsubsection{Weak topology}

Weak topology is defined for general topological spaces. The idea is to use a
collection of maps $\varphi_{i}:E\rightarrow F$ where $F$ is a topological
space to pull back a topology on E such that every $\varphi_{i}$ is continuous.

This idea can be implemented for a topological vector space and its dual. It
is commonly used when the vector space has already an initial topology,
usually defined from a semi-norm. Then another topology can be defined, which
is weaker than the initial topology and this is useful when the normed
topology imposes too strict conditions. This is easily done by using families
of semi-norms as above. For finite dimensional vector spaces the weak and the
"strong" (usual) topologies are equivalent.

\paragraph{Weak-topology\newline}

\begin{definition}
The \textbf{weak-topology} on a topological vector space E is the topology
defined by the family of semi-norms on E:

$\left(  p_{\varpi}\right)  _{\varpi\in E^{\prime}}:\forall u\in E:p_{\varpi
}\left(  u\right)  =\left\vert \varpi\left(  u\right)  \right\vert $
\end{definition}

It sums up to take as collection of maps the continuous (as defined by the
initial topology on E) linear forms on E.

\begin{theorem}
The weak topology is Hausdorff
\end{theorem}

\begin{proof}
It is Hausdorff if E' is separating : if $\forall u\neq v\in E,\exists
\varpi\in E^{\prime}:\varpi\left(  u\right)  \neq\varpi\left(  v\right)  $ and
this is a consequence of the Hahn-Banach theorem
\end{proof}

\begin{theorem}
A sequence $\left(  u_{n}\right)  _{n\in%
\mathbb{N}
}$ in a topological space E \textbf{converges weakly} to u if $:\forall
\varpi\in E^{\prime}:\varpi\left(  u_{n}\right)  \rightarrow\varpi\left(
u\right)  .$
\end{theorem}

convergence (with the initial topology in E) $\Rightarrow$\ weak convergence
(with the weak topology in E)

So the criterium for convergence is weaker, and this is one of the main
reasons for using this topology.

\begin{theorem}
If E is a semi-normed vector space, then the weak-topology on E is equivalent
to the topology of the semi-norm :

$\left\Vert u\right\Vert _{W}=\sup_{\left\Vert \varpi\right\Vert _{E^{\prime}%
}=1}\left\vert \varpi\left(  u\right)  \right\vert $
\end{theorem}

The weak norm $\left\Vert u\right\Vert _{W}$\ and the initial norm $\left\Vert
u\right\Vert $ are not equivalent if E is infinite dimensional (Wilansky p.270).

\begin{theorem}
(Banach-Alaoglu): if E is a normed vector space, then the closed unit ball E
is compact with respect to the weak topology iff E is reflexive.
\end{theorem}

This is the application of the same theorem for the *weak topology to the bidual.

\paragraph{*weak-topology\newline}

\begin{definition}
The \textbf{*weak-topology} on the topological dual E' of a topological vector
space E is the topology defined by the family of semi-norms on E':

$\left(  p_{u}\right)  _{u\in E}:\forall\varpi\in E^{\prime}:p_{u}\left(
\varpi\right)  =\left\vert \varpi\left(  u\right)  \right\vert $
\end{definition}

It sums up to take as collection of maps the evaluation maps given by vectors
of E.

\begin{theorem}
The *weak topology is Hausdorff
\end{theorem}

\begin{theorem}
(Wilansky p.274) With the *weak-topology E' is $\sigma-$compact, normal
\end{theorem}

\begin{theorem}
(Thill p.252) A sequence $\left(  \varpi_{n}\right)  _{n\in%
\mathbb{N}
}$ in the topological dual E' of a topological space E \textbf{converges
weakly} to u if $:\forall u\in E:\varpi_{n}\left(  u\right)  \rightarrow
\varpi\left(  u\right)  .$
\end{theorem}

convergence (with the initial topology in E') $\Rightarrow$\ weak convergence
(with the weak topology in E')

This is the topology of pointwise convergence (Thill p.252)

\begin{theorem}
If E is a semi-normed vector space, then the weak-topology on E' is equivalent
to the topology of the semi-norm :

$\left\Vert \varpi\right\Vert _{W}=\sup_{\left\Vert u\right\Vert _{E}%
=1}\left\vert \varpi\left(  u\right)  \right\vert $
\end{theorem}

The weak norm $\left\Vert \varpi\right\Vert _{W}$\ and the initial norm
$\left\Vert \varpi\right\Vert _{E^{\prime}}$ are not equivalent if E is
infinite dimensional.

\begin{theorem}
Banach-Alaoglu (Wilansky p.271): If E is a semi-normed vector space, then the
closed unit ball in its topological dual E' is a compact Hausdorff subset with
respect to the *-weak topology.
\end{theorem}

Remark : in both cases one can complicate the definitions by taking only a
subset of E' (or E), or extend E' to the algebraic dual E*.\ See Bratelli (1
p.162) and Thill.

\subsubsection{Fr\'{e}chet space}

Fr\'{e}chet spaces have a somewhat complicated definition.\ However they are
very useful, as they share many (but not all) properties of the Banach spaces.

\begin{definition}
A \textbf{Fr\'{e}chet space} is a Hausdorff, complete, topological vector
space, endowed with a countable family $\left(  p_{n}\right)  _{n\in%
\mathbb{N}
}$\ of semi-norms. So it is locally convex and metric.
\end{definition}

The metric is : $d\left(  x,y\right)  =\sum_{n=0}^{\infty}\frac{1}{2^{n}}%
\frac{p_{n}\left(  x-y\right)  }{1+p_{n}\left(  x-y\right)  }$

And because it is Hausdorff : $\forall u\neq0\in E,\exists n\in%
\mathbb{N}
:p_{n}\left(  u\right)  >0$

\begin{theorem}
A closed vector subspace of a Fr\'{e}chet space is a Fr\'{e}chet space.
\end{theorem}

\begin{theorem}
(Taylor 1 p.482) The quotient of a Fr\'{e}chet space by a closed subspace is a
Fr\'{e}chet space.
\end{theorem}

\begin{theorem}
The direct sum of a finite number of Fr\'{e}chet spaces is a Fr\'{e}chet space.
\end{theorem}

\begin{theorem}
(Taylor 1 p.481) A sequence $\left(  u_{n}\right)  _{n\in%
\mathbb{N}
}$ converges in a Fr\'{e}chet space $\left(  E,\left(  p_{n}\right)  _{n\in%
\mathbb{N}
}\right)  $ iff $\forall m\in%
\mathbb{N}
:p_{m}\left(  u_{n}-u\right)  \rightarrow_{n\rightarrow\infty}0$
\end{theorem}

\paragraph{Linear functions on Fr\'{e}chet spaces\newline}

\begin{theorem}
(Taylor 1 p.491) For every linear map $f\in L\left(  E;F\right)  $ between
Fr\'{e}chet vector spaces :

i) (open mapping theorem) If f\ is continuous and surjective then any
neighborhood of 0 in E is mapped onto a neighborhood of 0 in F (f is open)

ii) If f is continuous and bijective then $f^{-1}$is continuous

iii) (closed graph theorem) if the graph of $f=$ $\left\{  \left(
u,f(u)\right)  ,u\in E\right\}  $ is closed in E$\times$F then f is continuous.
\end{theorem}

\begin{theorem}
(Taylor I p.297) For any bilinear map : $B:E\times F\rightarrow%
\mathbb{C}
$ on two complex Fr\'{e}chet spaces $\left(  E,\left(  p_{n}\right)  _{n\in%
\mathbb{N}
}\right)  ,\left(  F,\left(  q_{n}\right)  _{n\in%
\mathbb{N}
}\right)  $ which is separately continuous on each variable, there are $C\in%
\mathbb{R}
,\left(  k,l\right)  \in%
\mathbb{N}
^{2}:$

$\forall\left(  u,v\right)  \in E\times F:\left\vert B\left(  u,v\right)
\right\vert \leq Cp_{k}\left(  u\right)  q_{l}\left(  v\right)  $
\end{theorem}

\begin{theorem}
(Zuily p.59) If a sequence $\left(  f_{m}\right)  _{m\in%
\mathbb{N}
}$ of continuous maps between two Fr\'{e}chet spaces $\left(  E,p_{n}\right)
,\left(  F,q_{n}\right)  $ is such that :

$\forall u\in E,\exists v\in F:f_{m}\left(  u\right)  _{m\rightarrow\infty
}\rightarrow v$

then there is a map : $f\in%
\mathcal{L}%
\left(  E;F\right)  $ such that :

i) $f_{m}\left(  u\right)  _{m\rightarrow\infty}\rightarrow f(u)$

ii) for any compact K in E, any $n\in%
\mathbb{N}
:$

$\lim_{m\rightarrow\infty}\sup_{u\in K}q_{n}\left(  f_{m}\left(  u\right)
-f\left(  u\right)  \right)  =0.$

If $\left(  u_{m}\right)  _{m\in%
\mathbb{N}
}$ is a sequence in E which converges to u then $\left(  f_{m}\left(
u_{m}\right)  \right)  _{m\in%
\mathbb{N}
}$ converges to f(u).
\end{theorem}

This theorem is important, because it gives a simple rule for the convergence
of sequence of linear maps. It holds in Banach spaces (which are Fr\'{e}chet spaces).

The space
$\mathcal{L}$%
(E;F) of continuous linear maps between Fr\'{e}chet spaces E, F is usually not
a Fr\'{e}chet space. The topological dual of a Fr\'{e}chet space is not
necessarily a Fr\'{e}chet space. However we have the following theorem.

\begin{theorem}
Let $\left(  E_{1},\Omega_{1}\right)  ,\left(  E_{2},\Omega_{2}\right)  $ two
Fr\'{e}chet spaces with their open subsets, if $E_{2}$ is dense in $E_{1},$
then $E_{1}^{\prime}\subset E_{2}^{\prime}$
\end{theorem}

\begin{proof}
$E_{1}^{\ast}\subset E_{2}^{\ast}$ because by restriction any linear map on
$E_{1}$ is linear on $E_{2}$

take $\lambda\in E_{1}^{\prime},a\in E_{2}$ so $a\in E_{1}$

$\lambda$ continuous on $E_{1}$ at a $\Rightarrow\forall\varepsilon
>0:\exists\varpi_{1}\in\Omega_{1}:\forall u\in\varpi_{1}:\left\vert
\lambda\left(  u\right)  -\lambda\left(  a\right)  \right\vert \leq
\varepsilon$

take any u in $\varpi_{1},u\in\overline{E}_{2},E_{2}$ second countable, thus
first countable $\Rightarrow\exists\left(  v_{n}\right)  ,v_{n}\in E_{2}%
:v_{n}\rightarrow u$

So any neighborhood of u contains at least two points $w,w^{\prime}$ in
$E_{2}$

So there are w$\neq$w'$\in\varpi_{1}\cap E_{2}$

$E_{2}$ is Hausdorff $\Rightarrow\exists\varpi_{2},\varpi_{2}^{\prime}%
\in\Omega_{2}:w\in\varpi_{2},w^{\prime}\in\varpi_{2}^{\prime},\varpi_{2}%
\cap\varpi_{2}^{\prime}=\varnothing$

So there is $\varpi_{2}\in\Omega_{2}:\varpi_{2}\subset\varpi_{1}$

and $\lambda$ is continuous at a for $E_{2}$
\end{proof}

\subsubsection{Affine spaces}

All the previous material extends to affine spaces.

\begin{definition}
An affine space $\left(  E,\overrightarrow{E}\right)  $ is semi-normed if its
underlying vector space $\overrightarrow{E}$ is normed.\ The semi-norm defines
uniquely a semi-metric :

$d\left(  A,B\right)  =\left\Vert \overrightarrow{AB}\right\Vert $
\end{definition}

\begin{theorem}
The closure and the interior of a convex subset of a semi-normed affine space
are convex.
\end{theorem}

\begin{theorem}
Every ball B(A,r) of a semi-normed affine space is convex.
\end{theorem}

\begin{theorem}
A map $f:E\rightarrow F$ valued in an affine normed space F is
\textbf{bounded} if for a point $O\in F$ : $\sup_{x\in E}\left\Vert
f(x)-O\right\Vert _{\overrightarrow{F}}<\infty.$ This property does not depend
on the choice of O.
\end{theorem}

\begin{theorem}
(Schwartz I p.173) A hyperplane of a normed affine space E is either closed or
dense in E. It is closed if it is defined by a continuous affine map.
\end{theorem}

\bigskip

\subsection{Banach spaces}

\label{Banach spaces}

For many applications a complete topological space is required, thanks to the
fixed point theorem. So for vector spaces there are Fr\'{e}chet spaces and
Banach spaces.\ The latter is the structure of choice, whenever it is
available, because it is easy to use and brings several useful tools such as
series, analytic functions and one parameter group of linear maps. Moreover
all classic calculations on series, usually done with scalars, can readily be
adaptated to Banach vector spaces.

Banach spaces are named after the Polish mathematician Stefan Banach who
introduced them in 1920--1922 along with Hans Hahn and Eduard Helly

\subsubsection{Banach Spaces}

\paragraph{Definitions\newline}

\begin{definition}
A \textbf{Banach vector space} is a complete normed vector space over a
topologically complete field K
\end{definition}

usually K= $%
\mathbb{R}
$ or $%
\mathbb{C}
$

\begin{definition}
A \textbf{Banach affine space} is a complete normed affine space over a
topologically complete field K
\end{definition}

Usually a "Banach space" is a Banach vector space.

Any finite dimensional vector space is complete. So it is a Banach space when
it is endowed with any norm.

A normed vector space can be completed. If the completion procedure is applied
to a normed vector space, the result is a Banach space containing the original
space as a dense subspace, and if it is applied to an inner product space, the
result is a Hilbert space containing the original space as a dense subspace.
So for all practical purposes the completed space can replace the initial one.

\paragraph{Subspaces\newline}

The basic applications of general theorems gives:

\begin{theorem}
A \textit{closed} vector subspace of a Banach vector space is a Banach vector space
\end{theorem}

\begin{theorem}
Any finite dimensional vector subspace of a Banach vector space is a Banach
vector space
\end{theorem}

\begin{theorem}
If F is a closed vector subspace of the Banach space E then E/F is still a
Banach vector space
\end{theorem}

It can be given (Taylor I p.473) the norm :$\left\Vert u\right\Vert
_{E/F}=\lim_{v\in F,v\rightarrow0}\left\Vert u-v\right\Vert _{E}$

\paragraph{Series on a Banach vector space\newline}

Series must be defined on sets endowed with an addition, so many important
results are on Banach spaces. Of course they hold for series defined on $%
\mathbb{R}
$ or $%
\mathbb{C}
.$ First we define three criteria for convergence.

\bigskip

\subparagraph{Absolutely convergence\newline}

\begin{definition}
A series $\sum_{n\in%
\mathbb{N}
}u_{n}$ on a semi-normed vector space E is \textbf{absolutely convergent} if
the series $\sum_{n\in%
\mathbb{N}
}\left\Vert u_{n}\right\Vert $ converges.
\end{definition}

\begin{theorem}
(Schwartz I p.123) If the series $\sum_{n\in%
\mathbb{N}
}u_{n}$ on a Banach E is \textbf{absolutely convergent} then :

i) $\sum_{n\in%
\mathbb{N}
}u_{n}$ converges in E

ii) $\left\Vert \sum_{n\in%
\mathbb{N}
}u_{n}\right\Vert \leq\sum_{n\in%
\mathbb{N}
}\left\Vert u_{n}\right\Vert $

iii) If $\varphi:%
\mathbb{N}
\rightarrow%
\mathbb{N}
$ is any bijection, the series $\sum_{n\in%
\mathbb{N}
}u_{\varphi\left(  n\right)  }$ is also absolutely convergent and $\lim
\sum_{n\in%
\mathbb{N}
}u_{\varphi\left(  n\right)  }=\lim\sum_{n\in%
\mathbb{N}
}u_{n}$
\end{theorem}

\bigskip

\subparagraph{Commutative convergence\newline}

\begin{definition}
A series $\sum_{i\in I}u_{i}$\ on a topological vector space E, where I is a
countable set, is \textbf{commutatively} \textbf{convergent} if there is $u\in
E$ such that for every bijective map $\varphi$ on I : $\lim\sum_{n}%
u_{\varphi\left(  j_{n}\right)  }=u$
\end{definition}

Then on a Banach : absolute convergence $\Rightarrow$\ commutative convergence

Conversely :

\begin{theorem}
(Neeb p.21) A series on a Banach\ which is commutatively convergent is
absolutely convergent.
\end{theorem}

Commutative convergence enables to define quantities such as $\sum_{i\in
I}u_{i}$ for any set.

\bigskip

\subparagraph{Summable family\newline}

\begin{definition}
(Neeb p.21) \ A family $\left(  u_{i}\right)  _{i\in I}$ of vectors on a
semi-normed vector space E is said to be \textbf{summable} with sum u if :

$\forall\varepsilon>0,\exists J\subset I,card(J)<\infty:\forall K\subset
J:\left\Vert \left(  \sum_{i\in K}u_{i}\right)  -x\right\Vert <\varepsilon$

then one writes : $u=\sum_{i\in I}u_{i}$ .
\end{definition}

\begin{theorem}
(Neeb p.25) If a family $\left(  u_{i}\right)  _{i\in I}$ of vectors in the
Banach E is summable, then only countably many $u_{i}$ are non zero
\end{theorem}

So for a countable set I, E Banach

summability $\Leftrightarrow$\ commutative convergence $\Leftrightarrow$
absolute convergence $\Rightarrow$ convergence in the usual meaning, but the
converse is not true.

\bigskip

\subparagraph{Image of a series by a continuous linear map\newline}

\begin{theorem}
(Schwartz I p.128) For every continuous map $L\in%
\mathcal{L}%
\left(  E;F\right)  $ between normed vector spaces : if the series $\sum_{n\in%
\mathbb{N}
}u_{n}$ on E is convergent then the series $\sum_{n\in%
\mathbb{N}
}L\left(  u_{n}\right)  $ on F is convergent and $\sum_{n\in%
\mathbb{N}
}L\left(  u_{n}\right)  =L\left(  \sum_{n\in%
\mathbb{N}
}u_{n}\right)  .$

If E,F are Banach, then the theorem holds for absolutely convergent (resp.
commutatively convergent) and :

$\sum_{n\in%
\mathbb{N}
}\left\Vert L\left(  u_{n}\right)  \right\Vert \leq\left\Vert L\right\Vert
\sum_{n\in%
\mathbb{N}
}\left\Vert u_{n}\right\Vert $
\end{theorem}

\bigskip

\subparagraph{Image of 2 series by a continuous bilinear map\newline}

\begin{theorem}
(Schwartz I p.129) For every continuous bilinear map $B\in%
\mathcal{L}%
^{2}\left(  E,F;G\right)  $ between the Banach spaces E,F,G, if the series
$\sum_{i\in I}u_{i}$ on E, $\sum_{j\in J}v_{j}$ on F, for I,J countable sets,
are both absolutely convergent, then the series on G : $\sum_{\left(
i,j\right)  \in I}B\left(  u_{i},v_{j}\right)  $ is absolutly convergent and
$\sum_{\left(  i,j\right)  \in I\times J}B\left(  u_{i},v_{j}\right)
=B\left(  \sum_{i\in I}u_{i},\sum_{j\in J}v_{j}\right)  $
\end{theorem}

\begin{theorem}
Abel criterium (Schwartz I p.134) For every continuous bilinear map $B\in%
\mathcal{L}%
^{2}\left(  E,F;G\right)  $ between the Banach spaces E,F,G on the field K, if :

i) the sequence $\left(  u_{n}\right)  _{n\in%
\mathbb{N}
}$ on E converges to 0 and is such that the series

$\sum_{p=0}^{\infty}\left\Vert u_{p+1}-u_{p}\right\Vert $ converges,

ii) the sequence $\left(  v_{n}\right)  _{n\in%
\mathbb{N}
}$ on F is such that $\exists k\in K:\forall m,n:\left\Vert \sum_{p=m}%
^{n}v_{p}\right\Vert \leq k,$

\begin{theorem}
then the series: $\sum_{n\in%
\mathbb{N}
}B\left(  u_{n},v_{n}\right)  $ converges to S, and

$\left\Vert S\right\Vert \leq\left\Vert B\right\Vert \left(  \sum
_{p=0}^{\infty}\left\Vert u_{p+1}-u_{p}\right\Vert \right)  \left(  \left\Vert
\sum_{p=0}^{\infty}v_{p}\right\Vert \right)  $

$\left\Vert \sum_{p>n}B\left(  u_{p},v_{p}\right)  \right\Vert \leq\left\Vert
B\right\Vert \left(  \sum_{p=n+1}^{\infty}\left\Vert u_{p+1}-u_{p}\right\Vert
\right)  \left(  \sup_{p>n}\left\Vert \sum_{m=p}^{\infty}v_{m}\right\Vert
\right)  $
\end{theorem}
\end{theorem}

The last theorem covers all the most common criteria for convergence of series.

\subsubsection{Continuous linear maps}

It is common to say "operator" for a "continuous linear map" on a Banach
vector space.

\paragraph{Properties of continuous linear maps on Banach spaces\newline}

\begin{theorem}
For every linear map $f\in L\left(  E;F\right)  $ between Banach vector spaces

i) open mapping theorem (Taylor 1 p.490): If f\ is continuous and surjective
then any neighborhood of 0 in E is mapped onto a neighborhood of 0 in F (f is open)

ii) closed graph theorem (Taylor 1 p.491): if the graph of f = $\left\{
\left(  u,f(u)\right)  ,u\in E\right\}  $ is closed in E$\times$F then f is continuous.

iii) (Wilansky p.276) if f is continuous and injective then it is a homeomorphism

iv) (Taylor 1 p.490) If f is continuous and bijective then $f^{-1}$is continuous

v) (Schwartz I p.131) If f,g are continuous, f invertible and $\left\Vert
g\right\Vert <\left\Vert f^{-1}\right\Vert ^{-1}$ then f+g is invertible and
$\left\Vert \left(  f+g\right)  ^{-1}\right\Vert \leq\frac{1}{\left\Vert
f^{-1}\right\Vert ^{-1}-\left\Vert g\right\Vert }$
\end{theorem}

\begin{theorem}
(Rudin) For every linear map $f\in L\left(  E;F\right)  $ between Banach
vector spaces and sequence $\left(  u_{n}\right)  _{n\in%
\mathbb{N}
}$ in E:

i) If f is continuous then for every sequence $\left(  u_{n}\right)  _{n\in%
\mathbb{N}
}$ in E :

$u_{n}\rightarrow u$ $\Rightarrow$ $f\left(  u_{n}\right)  \rightarrow f(u)$

ii) Conversely if for every sequence $\left(  u_{n}\right)  _{n\in%
\mathbb{N}
}$ in E which converges to 0 : $f\left(  u_{n}\right)  \rightarrow v$ then
$v=0 $ and f is continuous.
\end{theorem}

\begin{theorem}
(Wilansky p.273) If $\left(  \varpi_{n}\right)  _{n\in%
\mathbb{N}
}$ is a sequence in the topological dual E' of a Banach space such that
$\forall u\in E$ the set $\left\{  \varpi_{n}\left(  u\right)  ,n\in%
\mathbb{N}
\right\}  $ is bounded, then the set $\left\{  \left\Vert \varpi
_{n}\right\Vert ,n\in%
\mathbb{N}
\right\}  $ is bounded
\end{theorem}

\begin{theorem}
(Schwartz I p.109) If $f\in%
\mathcal{L}%
(E_{0};F)$ is a continuous linear map from a\ dense subspace $E_{0}$ of a
normed vector space to a Banach vector space F, then there is a unique
continuous map $\widetilde{f}:E\rightarrow F$ which extends f, $\widetilde
{f}\in%
\mathcal{L}%
(E;F)$ and $\left\Vert \widetilde{f}\right\Vert =\left\Vert f\right\Vert $
\end{theorem}

If F is a vector subspace, the annihiliator $F^{\intercal}$ of F is the set :

$\left\{  \varpi\in E^{\prime}:\forall u\in F:\varpi\left(  u\right)
=0\right\}  $

\begin{theorem}
Closed range theorem (Taylor 1 p.491): For every linear map $f\in L\left(
E;F\right)  $ between Banach vector spaces : $\ker f^{t}=f\left(  E\right)
^{\intercal}.$ Moreover if f(E) is closed in F then $f^{t}\left(  F^{\prime
}\right)  $ is closed in E' and $f^{t}\left(  F^{\prime}\right)  =\left(  \ker
f\right)  ^{\intercal}$
\end{theorem}

\paragraph{Properties of the set of linear continuous maps\newline}

\begin{theorem}
(Schwartz I p.115) The set of continuous linear maps
$\mathcal{L}$%
(E;F) between a normed vector space and a Banach vector space F on the same
field is a Banach vector space
\end{theorem}

\begin{theorem}
(Schwartz I p.117) The set of continuous multilinear maps

$%
\mathcal{L}%
^{r}(E_{1},E_{2},..E_{r};F)$ between normed vector spaces $\left(
E_{i}\right)  _{i=1}^{r}$ and a Banach vector space F on the same field is a
Banach vector space
\end{theorem}

\begin{theorem}
if E,F are Banach :
$\mathcal{L}$%
(E;F) is Banach
\end{theorem}

\begin{theorem}
The topological dual E' of a Banach vector space is a Banach vector space
\end{theorem}

A Banach vector space may be not reflexive : the bidual (E')' is not
necessarily isomorphic to E.

\begin{theorem}
(Schwartz II p.81) The sets of invertible continuous linear maps G%
$\mathcal{L}$%
(E;F),G%
$\mathcal{L}$%
(F;E) between the Banach vector spaces E,F are \textit{open} subsets in
$\mathcal{L}$%
(E;F),%
$\mathcal{L}$%
(F;E), thus they are normed vector spaces but not complete. The map $\Im:G%
\mathcal{L}%
(E;F)\rightarrow G%
\mathcal{L}%
(F;E)::\Im(f)=f^{-1}$ is an homeomorphism (bijective, continuous as its inverse).
\end{theorem}

$\left\Vert f\circ f^{-1}\right\Vert =\left\Vert Id\right\Vert =1\leq
\left\Vert f\right\Vert \left\Vert f^{-1}\right\Vert \leq\left\Vert
\Im\right\Vert ^{2}\left\Vert f\right\Vert \left\Vert f^{-1}\right\Vert $

$\Rightarrow\left\Vert \Im\right\Vert \geq1,\left\Vert f^{-1}\right\Vert
\geq1/\left\Vert f\right\Vert $

\begin{theorem}
The set G%
$\mathcal{L}$%
(E;E) of invertible endomorphisms on a Banach vector space is a topological
group with compose operation and the metric associated to the norm, open
subset in
$\mathcal{L}$%
$\left(  E;E\right)  .$
\end{theorem}

Notice that an "invertible map f in G%
$\mathcal{L}$%
(E;F)"\ means that f$^{-1}$\ must also be a continuous map, and for this it is
sufficient that f is continuous and bijective .

\begin{theorem}
(Neeb p.141) If X is a compact topological space, endowed with a Radon measure
$\mu,$ E,F are Banach vector spaces, then:

i) for every continuous map : $f\in C_{0}\left(  X;E\right)  $ there is a
unique vector U in E such that :

$\forall\lambda\in E^{\prime}:\lambda\left(  U\right)  =\int_{X}\lambda\left(
f\left(  x\right)  \right)  \mu\left(  x\right)  $ and we write : $U=\int
_{X}f\left(  x\right)  \mu\left(  x\right)  $

ii) for every continuous map :

$L\in%
\mathcal{L}%
\left(  E;F\right)  $ : $L\left(  \int_{X}f\left(  x\right)  \mu\left(
x\right)  \right)  =\int_{X}\left(  L\circ f\left(  x\right)  \right)
\mu\left(  x\right)  $
\end{theorem}

\paragraph{Spectrum of a map\newline}

A scalar $\lambda$ if an eigen value for the endomorphism $f\in%
\mathcal{L}%
\left(  E;E\right)  $ if there is a vector u such that $f(u)=\lambda u,$ so
$f-\lambda I$ cannot be inversible. On infinite dimensional topological vector
space the definition is enlarged as follows.

\begin{definition}
For every linear continuous endomorphism f on a topological vector space E on
a field K,

i) the \textbf{spectrum }$Sp\left(  f\right)  $ of f is the subset of the
scalars $\lambda\in K$ such that $\left(  f-\lambda Id_{E}\right)  $ has no
inverse in
$\mathcal{L}$%
$\left(  E;E\right)  .$

ii) the \textbf{resolvent} \textbf{set} $\rho\left(  f\right)  $\ of f is the
complement of the spectrum

iii) the map: $R:K\rightarrow L\left(  E;E\right)  ::R\left(  \lambda\right)
=$ $\left(  \lambda Id-f\right)  ^{-1}$ is called the \textbf{resolvent} of f.
\end{definition}

If $\lambda$ is an eigen value of f, it belongs to the spectrum, but the
converse is no true. If $f\in G%
\mathcal{L}%
\left(  E;E\right)  $ then $0\notin Sp(f).$

\begin{theorem}
The spectrum of a continuous endomorphism f on a complex Banach vector space E
is a non empty compact subset of $%
\mathbb{C}
$\ bounded by $\left\Vert f\right\Vert $
\end{theorem}

\begin{proof}
It is a consequence of general theorems on Banach algebras :
$\mathcal{L}$%
(E;E) is a Banach algebra, so the spectrum is a non empty compact, and is
bounded by the spectral radius, which is $\leq\left\Vert f\right\Vert $\ 
\end{proof}

\begin{theorem}
(Schwartz 2 p.69) The set of eigen values of a compact endomorphism on a
Banach space is either finite, or countable in a sequence convergent to 0
(which is or not an eigen value).
\end{theorem}

\begin{theorem}
(Taylor 1 p.493) If f is a continuous endomorphism on a complex Banach space:

$\left\vert \lambda\right\vert >\left\Vert f\right\Vert \Rightarrow\lambda
\in\rho\left(  f\right)  .$ In particular if $\left\Vert f\right\Vert <1$ then
$Id-f$ is invertible and $\sum_{n=0}^{\infty}f^{n}=\left(  Id-f\right)  ^{-1}$

If $\lambda_{0}\in\rho\left(  f\right)  $ then : $R\left(  \lambda\right)
=R\left(  \lambda_{0}\right)  \sum_{n=0}^{\infty}R\left(  \lambda_{0}\right)
^{n}\left(  \lambda-\lambda_{0}\right)  ^{n}$

If $\lambda_{1},\lambda_{2}\in\rho\left(  f\right)  $ then : $R\left(
\lambda_{1}\right)  -R\left(  \lambda_{2}\right)  =\left(  \lambda_{1}%
-\lambda_{2}\right)  R\left(  \lambda_{1}\right)  \circ R\left(  \lambda
_{2}\right)  $
\end{theorem}

\paragraph{Compact maps\newline}

\begin{theorem}
(Schwartz 2 p.60) If f is a continuous compact map $f\in%
\mathcal{L}%
\left(  E;F\right)  $ between a reflexive Banach vector space E and a
topological vector space f, then the closure in F $\overline{f\left(
B(0,1)\right)  }$\ of the image by f of the unit ball B(0,1) in E is compact
in F.
\end{theorem}

\begin{theorem}
(Taylor 1 p.496) The transpose of a compact map is compact.
\end{theorem}

\begin{theorem}
(Schwartz 2 p.63) If $\left(  f_{n}\right)  _{n\in%
\mathbb{N}
}$ is a sequence of linear continuous maps of finite rank between Banach
vector spaces, which converges to f, then f is a compact map.
\end{theorem}

\begin{theorem}
(Taylor 1 p.495)The set of compact linear maps between Banach vector
spaces\ is a closed vector subspace of the space of continuous linear maps $%
\mathcal{L}%
\left(  E;F\right)  .$
\end{theorem}

\begin{theorem}
(Taylor 1 p.499) The spectrum Sp(f) of a compact endomorphism $f\in%
\mathcal{L}%
\left(  E;E\right)  $ on a complex Banach space has only 0 as point of
accumulation, and all $\lambda\neq0\in Sp(f)$ are eigen values of f.
\end{theorem}

\paragraph{Fredholm operators\newline}

Fredholm operators are "proxy" for isomorphisms. Their main feature is the index.

\begin{definition}
(Taylor p.508) A continuous linear map $f\in%
\mathcal{L}%
\left(  E;F\right)  $\ between Banach vector spaces E,F is said to be a
\textbf{Fredholm operator} if $\ker f$ and $F/f(E)$ are finite dimensional.
Equivalentely if there exists $g\in%
\mathcal{L}%
\left(  F;E\right)  $ such that : $Id_{E}-g\circ f$ and $Id_{F}-f\circ g$ are
continuous and compact. The \textbf{index} of f is : $Index(f)=\dim\ker f-\dim
F/f\left(  E\right)  =\dim\ker f-\dim\ker f^{t}$
\end{definition}

\begin{theorem}
(Taylor p.508) The set Fred(E;F) of Fredholm operators is an open vector
subspace of $%
\mathcal{L}%
\left(  E;F\right)  .$ The map : $Index:Fred(E;F)\rightarrow%
\mathbb{Z}
$ is constant on each connected component of Fred(E;F).
\end{theorem}

\begin{theorem}
(Taylor p.508) The compose of two Fredholm operators is Fredholm : If $f\in
Fred(E;F),g\in Fred(F;G)$ then $g\circ f\in Fred(E;G)$ and Index(gf)=Index(f)
+ Index(g). If f is Fredholm and g compact then f+g is Fredholm and Index(f+g)=Index(f)
\end{theorem}

\begin{theorem}
(Taylor p.508) The transpose $f^{t}$ of a Fredholm operator f is Fredholm and
$Index(f^{t})=-Index\left(  f\right)  $
\end{theorem}

\subsubsection{Analytic maps on Banach spaces}

With the vector space structure of
$\mathcal{L}$%
(E;E) one can define any linear combination of maps. But in a Banach space one
can go further and define "functions" of an endomorphism.

\paragraph{Exponential of a linear map\newline}

\begin{theorem}
The \textbf{exponential} of a continuous linear endomorphism $f\in%
\mathcal{L}%
\left(  E;E\right)  $\ on a Banach space E is the continuous linear map :
$\exp f=\sum_{n=0}^{\infty}\frac{1}{n!}f^{n}$ where $f^{n}$ is the n iterated
of f and $\left\Vert \exp f\right\Vert \leq\exp\left\Vert f\right\Vert $
\end{theorem}

\begin{proof}
$\forall u\in F,$ the series $\sum_{n=0}^{\infty}\frac{1}{n!}f^{n}\left(
u\right)  $ converges absolutely :

$\sum_{n=0}^{N}\frac{1}{n!}\left\Vert f^{n}\left(  u\right)  \right\Vert
\leq\sum_{n=0}^{N}\frac{1}{n!}\left\Vert f^{n}\right\Vert \left\Vert
u\right\Vert =\sum_{n=0}^{N}\frac{1}{n!}\left\Vert f\right\Vert ^{n}\left\Vert
u\right\Vert \leq\left(  \exp\left\Vert f\right\Vert \right)  \left\Vert
u\right\Vert $

we have an increasing bounded sequence on $%
\mathbb{R}
$ which converges.

and $\left\Vert \sum_{n=0}^{\infty}\frac{1}{n!}f^{n}\left(  u\right)
\right\Vert \leq\left(  \exp\left\Vert f\right\Vert \right)  \left\Vert
u\right\Vert $ so exp is continuous with $\left\Vert \exp f\right\Vert
\leq\exp\left\Vert f\right\Vert $
\end{proof}

A simple computation as above brings (Neeb p.170):

$f\circ g=g\circ f\Rightarrow\exp(f+g)=(\exp f)\circ(\exp g)$

$\exp(-f)=(\exp f)^{-1}$

$\exp(f^{t})=(\exp f)^{t}$

$g\in G%
\mathcal{L}%
\left(  E;E\right)  :\exp\left(  g^{-1}\circ f\circ g\right)  =g^{-1}%
\circ\left(  \exp f\right)  \circ g$

If E,F are finite dimensional : $\det(\exp f)=\exp(Trace\left(  f\right)  )$

If E is finite dimensional the inverse log of exp is defined as :

$\left(  \log f\right)  \left(  u\right)  =\int_{-\infty}^{0}[(s-f)^{-1}%
-(s-1)^{-1}]\left(  u\right)  ds$ if f has no eigen value $\leq0$

Then : $\log(g\circ f\circ g^{-1})=g\circ(\log f)\circ g^{-1}$

$\log(f^{-1})=-\log f$

\paragraph{Holomorphic groups\newline}

\begin{theorem}
If f is a continuous linear endomorphism $f\in%
\mathcal{L}%
\left(  E;E\right)  $\ on a complex Banach space E then the map : $\exp
zf=\sum_{n=0}^{\infty}\frac{z^{n}}{n!}f^{n}\in%
\mathcal{L}%
\left(  E;E\right)  $ and defines the holomorphic group : $U:%
\mathbb{C}
\rightarrow%
\mathcal{L}%
\left(  E;E\right)  ::U(z)=\exp zf$ with $U\left(  z_{2}\right)  \circ
U\left(  z_{1}\right)  =U\left(  z_{1}+z_{2}\right)  ,U\left(  0\right)  =Id$

U is holomorphic on $%
\mathbb{C}
$ and $\frac{d}{dz}\left(  \exp zf\right)  |_{z=z_{0}}=f\circ\exp z_{0}f$
\end{theorem}

\begin{proof}
i) The previous demonstration can be generalized in a complex Banach space for
$\sum_{n=0}^{\infty}\frac{z^{n}}{n!}f^{n}$

Then, for any continuous endomorphism f we have a map :

$\exp zf=\sum_{n=0}^{\infty}\frac{z^{n}}{n!}f^{n}\in%
\mathcal{L}%
\left(  E;E\right)  $

ii) $\exp zf\left(  u\right)  =\exp f\left(  zu\right)  ,$ $z_{1}f,z_{2}f$
commutes so :

$\exp(z_{1}f)\circ\exp(z_{2}f)=\exp(z_{1}+z_{2})f$

iii) $\frac{1}{z}\left(  U(z)-I\right)  -f=\sum_{n=1}^{\infty}\frac{z^{n-1}%
}{n!}f^{n}-f=f\circ\left(  \sum_{n=1}^{\infty}\frac{z^{n-1}}{n!}%
f^{n-1}-Id\right)  =f\circ\left(  \exp zf-Id\right)  $

$\left\Vert \frac{1}{z}\left(  U(z)-I\right)  -f\right\Vert \leq\left\Vert
f\right\Vert \left\Vert \left(  \exp zf-Id\right)  \right\Vert $

$\lim_{z\rightarrow0}\left\Vert \frac{1}{z}\left(  U(z)-I\right)
-f\right\Vert \leq\lim_{z\rightarrow0}\left\Vert f\right\Vert \left\Vert
\left(  \exp zf-Id\right)  \right\Vert =0$

Thus U is holomorphic at z=0 with $\frac{dU}{dz}|_{z=0}=f$

iv) $\frac{1}{h}\left(  U\left(  z+h\right)  -U\left(  z\right)  \right)
-f\circ U\left(  z\right)  =\frac{1}{h}\left(  U\left(  h\right)  -I\right)
\circ U\left(  z\right)  -f\circ U\left(  z\right)  =\left(  \frac{1}%
{h}\left(  U\left(  h\right)  -I\right)  -f\right)  \circ U\left(  z\right)  $

$\left\Vert \frac{1}{h}\left(  U\left(  z+h\right)  -U\left(  z\right)
\right)  -f\circ U\left(  z\right)  \right\Vert \leq\left\Vert \left(
\frac{1}{h}\left(  U\left(  h\right)  -I\right)  -f\right)  \right\Vert
\left\Vert U\left(  z\right)  \right\Vert $

$\lim_{h\rightarrow0}\left\Vert \frac{1}{h}\left(  U\left(  z+h\right)
-U\left(  z\right)  \right)  -f\circ U\left(  z\right)  \right\Vert $

$\leq\lim_{h\rightarrow0}\left\Vert \left(  \frac{1}{h}\left(  U\left(
h\right)  -I\right)  -f\right)  \right\Vert \left\Vert U\left(  z\right)
\right\Vert =0$

So U is holomorphic on $%
\mathbb{C}
$ and $\frac{d}{dz}\left(  \exp tf\right)  |_{z=z_{0}}=f\circ\exp z_{0}f$
\end{proof}

For every endomorphism $f\in%
\mathcal{L}%
\left(  E;E\right)  $\ on a complex or real Banach space E then the map :
$\exp tf:%
\mathbb{R}
\rightarrow%
\mathcal{L}%
\left(  E;E\right)  $ defines a one parameter group and U(t)=exptf is smooth
and $\frac{d}{dt}\left(  \exp tf\right)  |_{t=t_{0}}=f\circ\exp t_{0}f$

\paragraph{Map defined through a holomorphic map\newline}

The previous procedure can be generalized. This is an application of the
spectral theory (see the dedicated section).

\begin{theorem}
(Taylor 1 p.492) Let $\varphi:\Omega\rightarrow%
\mathbb{C}
$ be a holomorphic map on a bounded region, with smooth border, of $%
\mathbb{C}
$ and $f\in%
\mathcal{L}%
\left(  E;E\right)  $ a continuous endomorphism on the complex Banach E.

i) If $\Omega$ contains the spectrum of f, the following map is a continuous
endomorphism on E:

$\Phi\left(  \varphi\right)  \left(  f\right)  =\frac{1}{2i\pi}\int
_{\partial\Omega}\varphi\left(  \lambda\right)  \left(  \lambda I-f\right)
^{-1}d\lambda\in%
\mathcal{L}%
\left(  E;E\right)  $

ii) If $\varphi\left(  \lambda\right)  =1$ then $\Phi\left(  \varphi\right)
\left(  f\right)  =Id$

iii) If $\varphi\left(  \lambda\right)  =\lambda$ then $\Phi\left(
\varphi\right)  \left(  f\right)  =f$

iv) If $\varphi_{1},\varphi_{2}$ are both holomorphic on $\Omega,$ then :

$\Phi\left(  \varphi_{1}\right)  \left(  f\right)  \circ\Phi\left(
\varphi_{2}\right)  \left(  f\right)  =\Phi\left(  \varphi_{1}\times
\varphi_{2}\right)  \left(  f\right)  $
\end{theorem}

\subsubsection{One parameter group}

The main purpose is the study of the differential equation $\frac{dU}%
{dt}=SU\left(  t\right)  $ where $U\left(  t\right)  ,S\in%
\mathcal{L}%
\left(  E;E\right)  .$ S is the infinitesimal generator of U. If S is
continuous then the general solution is $U\left(  t\right)  =\exp tS$ but as
it is not often the case we have to distinguish norm and weak topologies. On
this topic we follow Bratelli (I p.161). See also Spectral theory on the same
topic for unitary groups on Hilbert spaces.

\paragraph{Definition\newline}

\begin{definition}
A \textbf{one parameter group} of operators on a Banach vector space E is a
map : $U:%
\mathbb{R}
\rightarrow%
\mathcal{L}%
(E;E)$ such that :

$U\left(  0\right)  =Id,U\left(  s+t\right)  =U\left(  s\right)  \circ
U\left(  t\right)  $
\end{definition}

the family U(t)\ has the structure of an abelian group, isomorphic to $%
\mathbb{R}
.$

\begin{definition}
A \textbf{one parameter semi-group} of operators on a Banach vector space E is
a map : $U:%
\mathbb{R}
_{+}\rightarrow%
\mathcal{L}%
(E;E)$ such that :

$U\left(  0\right)  =Id,U\left(  s+t\right)  =U\left(  s\right)  \circ
U\left(  t\right)  $
\end{definition}

the family U(t) has the structure of a monoid (or semi-group)

So we denote T=$%
\mathbb{R}
$ or $%
\mathbb{R}
_{+}$

Notice that U(t) (the value at t) \textit{must be continuous}. The continuity
conditions below do not involve U(t) but the map $U:T\rightarrow%
\mathcal{L}%
\left(  E;E\right)  .$

\paragraph{Norm topology\newline}

\begin{definition}
(Bratelli p.161) A one parameter (semi) group U of continuous operators on E
is said to be \textbf{uniformly continuous} if one of the equivalent
conditions is met:

i) $\lim_{t\rightarrow0}\left\Vert U(t)-Id\right\Vert =0$

ii) $\exists S\in%
\mathcal{L}%
\left(  E;E\right)  :\lim_{t\rightarrow0}\left\Vert \frac{1}{t}\left(
U(t)-I\right)  -S\right\Vert =0$

iii) $\exists S\in%
\mathcal{L}%
\left(  E;E\right)  :U(t)=\sum_{n=0}^{\infty}\frac{t^{n}}{n!}S^{n}=\exp tS$

S is the \textbf{infinitesimal\ generator} of U and one writes $\frac{dU}%
{dt}=SU(t)$
\end{definition}

A uniformly continuous one parameter semi group U can be extended to T=$%
\mathbb{R}
$ such that $\left\Vert U(t)\right\Vert \leq\exp\left(  \left\vert
t\right\vert \left\Vert S\right\Vert \right)  $

If these conditions are met the problem is solved. And conversely a one
parameter (semi) group of continuous operators is uniformly continuous iff its
generator is continuous.

\paragraph{Weak topology\newline}

\begin{definition}
(Bratelli p.164) \ A one parameter (semi) group U of continuous operators on
the banach vector space E on the field K is said to be \textbf{weakly
continuous} if $\forall\varpi\in E^{\prime}$ the map $\phi_{\varpi}:T\times
E\rightarrow K::\phi_{\varpi}\left(  t,u\right)  =\varpi\left(  U(t)u\right)
$ is such that :

$\forall t\in T:\phi_{\varpi}\left(  t,.\right)  :E\rightarrow K$ is continuous

$\forall u\in E:\phi_{\varpi}\left(  .,u\right)  :T\rightarrow K$ is continuous
\end{definition}

So one can say that U is continuous in the weak topology on E.

Similarly a one parameter group U on E' : $U:%
\mathbb{R}
\rightarrow%
\mathcal{L}%
(E^{\prime};E^{\prime})$ is continuous in the *weak topology if $\forall u\in
E$ the map $\phi_{u}:T\times E^{\prime}\rightarrow K::\phi_{u}\left(
t,\varpi\right)  =U(t)\left(  \varpi\right)  \left(  u\right)  $ is such that :

$\forall t\in T:\phi_{u}\left(  t,.\right)  :E^{\prime}\rightarrow K$ is continuous

$\forall\varpi\in E^{\prime}:\phi_{u}\left(  .,\varpi\right)  :T\rightarrow K$
is continuous

\begin{theorem}
(Bratelli p.164-165) If a one parameter (semi) group U of operators on E is
weakly continuous then :

i) $\forall u\in E:\psi_{u}:T\rightarrow E:\psi_{u}\left(  t\right)  =U(t)u$
is continuous in the norm of E

ii) $\exists M\geq1,\exists\beta\geq\inf_{t>0}\frac{1}{t}\ln\left\Vert
U\left(  t\right)  \right\Vert :\left\Vert U\left(  t\right)  \right\Vert \leq
M\exp\beta t$

iii) for any complex borelian measure $\mu$ on T such that $\int_{T}e^{\beta
t}\left\vert \mu\left(  t\right)  \right\vert <\infty$ the map :

$U_{\mu}:E\rightarrow E::$ $U_{\mu}\left(  u\right)  =\int_{T}U\left(
t\right)  \left(  u\right)  \mu\left(  t\right)  $ belongs to $%
\mathcal{L}%
\left(  E;E\right)  $
\end{theorem}

The main difference with the uniformly continuous case is that the
infinitesimal generator does not need to be defined over the whole of E.

\begin{theorem}
(Bratelli p.165-166) A map $S\in L(D(S);E)$ with domain $D(S)\subset E$ is the
infinitesimal generator of the weakly continuous one parameter (semi) group U
on a Banach\ E if :

$\forall u\in D(S),\exists v\in E:\forall\varpi\in E^{\prime}:\varpi\left(
v\right)  =\lim_{t\rightarrow0}\frac{1}{t}\varpi\left(  \left(
U(t)-Id\right)  u\right)  $

then :

i) $\forall u\in E:T\rightarrow E::U(t)u$ is continuous in the norm of E

ii) D(S) is dense in E in the weak topology

iii) $\forall u\in D\left(  S\right)  :S\circ U\left(  t\right)  u=U\left(
t\right)  \circ Su$

iii) if $\operatorname{Re}\lambda>\beta$ then the range of $\left(  \lambda
Id-S\right)  ^{-1}=E$ and

$\forall u\in D(S):\left\Vert \left(  \lambda Id-S\right)  u\right\Vert \geq
M^{-1}\left(  \operatorname{Re}\lambda-\beta\right)  \left\Vert u\right\Vert $

iv) the resolvent $\left(  \lambda Id-S\right)  ^{-1}$ is given by the Laplace
transform :

$\forall\lambda:\operatorname{Re}\lambda>\beta,\forall u\in E:\left(  \lambda
Id-S\right)  ^{-1}u=\int_{0}^{\infty}e^{-\lambda t}U(t)udt$
\end{theorem}

Notice that $\frac{d}{dt}U(t)u=Su$ only if $u\in D\left(  S\right)  .$ The
parameters $\beta,M$ refer to the previous theorem.

The following theorem gives a characterization of the linear endomorphisms S
defined on a subset D(S) of a Banach space which can be an infintesimal generator.

\begin{theorem}
Hille-Yoshida (Bratelli p.171): Let $S\in L(D(S);E),D(S)\sqsubseteq E,$ E
Banach vector space, then the following conditions are equivalent :

i) S is the infinitesimal generator of a weakly continuous semi group U in E
and U(t) is a contraction

ii) D(S) is dense in E and S closed in D(S) (in the weak topology) and

$\forall u\in D(S),\forall\alpha\geq0:\left\Vert \left(  Id-\alpha S\right)
u\right\Vert \geq\left\Vert u\right\Vert $ and for some $\alpha>0: $ the range
of $\left(  Id-\alpha S\right)  ^{-1}=E$

If so then U is defined by :

$\forall u\in D(S):U(t)u=\lim_{\varepsilon\rightarrow0}\exp\left(  tS\left(
Id-\varepsilon S\right)  ^{-1}\right)  u=\lim_{n\rightarrow\infty}\left(
I-\frac{1}{n}tS\right)  ^{-n}u$

where the exponential is defined by power series expansion.\ The limits exist
in compacts in the weak topology uniformly for t, and if u is in the norm
closure of D(S) the limits exist in norm.
\end{theorem}

\bigskip

\subsection{Normed algebras}

\label{Normed algebras}

Algebras are vector spaces with an internal operation. Their main algebraic
properties are seen in the Algebra part. To add a topology the most natural
way is to add a norm and one has a normed algebra and, if it is complete, a
Banach algebra. Several options are common : assumptions about the norm and
the internal operation on one hand, the addition of an involution (copied from
the adjoint of matrices) on the other hand, and both lead to distinguish
several classes of normed algebras, notably C*-algebras.

In this section we review the fundamental properties of normed algebras, their
representation is seen in the Spectral theory section. We use essentially the
comprehensive study of M.Thill. We strive to address as many subjects as
possible, while staying simple and practical. Much more can be found in
M.Thill's study. Bratelli gives an in depth review of the dynamical aspects,
more centered on the C*algebras and one parameter groups.

\subsubsection{Definitions}

\paragraph{Algebraic structure\newline}

This is a reminder of definitions from the Algebra part.

1. Algebra:

An algebra A is a vector space on a field K (it will be $%
\mathbb{C}
,$ if K=$%
\mathbb{R}
$ the adjustments are obvious$)$ endowed with an internal operation (denoted
as multiplication $XY$ with inverse $X^{-1})$ which is associative,
distributive over addition and compatible with scalar multiplication.
\textit{We assume that it is unital}, with unity element denoted I.

An algebra is commutative if XY=YX for every element.

2. Commutant:

The commutant\textbf{, }denoted S', of a subset S of an algebra A is the set
of all elements in A which commute with all the elements of S for the internal
operation. This is a subalgebra, containing I. The second commutant, denoted
S", is the commutant of S'.

3. Projection and reflexion:

An element X of A is a projection if $XX=X,$ a reflexion if $X=X^{-1},$
nilpotent if $X\cdot X=0$

4. Star algebra:

A *algebra is endowed with an involution such that :

$\left(  X+Y\right)  ^{\ast}=X^{\ast}+Y^{\ast};\left(  X\cdot Y\right)
^{\ast}=Y^{\ast}\cdot X^{\ast};\left(  \lambda X\right)  ^{\ast}%
=\overline{\lambda}X^{\ast};\left(  X^{\ast}\right)  ^{\ast}=X$

Then the adjoint of an element X is X*

An element X of a *-algebra is : normal if XX* = X*X, self-adjoint (or
hermitian) if X = X*, anti self-adjoint (or antihermitian) if X= - X*, unitary
if XX* = X*X =I

The subset of self-adjoint elements in A is a real vector space, real form of
the vector space A.

\paragraph{Topological structures\newline}

For the sake of simplicity we will make use only of :

- normed algebra, normed *-algebra

- Banach algebra, Banach *-algebra, C*-algebra

\begin{definition}
A \textbf{topological algebra} is a topological vector space such that the
internal operation is continuous.
\end{definition}

\begin{definition}
A \textbf{normed algebra} is a normed vector space endowed with the structure
of a topological algebra with the topology induced by the norm $\left\Vert
{}\right\Vert $, \textit{with the additional properties} that : $\left\Vert
XY\right\Vert \leq\left\Vert X\right\Vert \left\Vert Y\right\Vert ,\left\Vert
I\right\Vert =1.$
\end{definition}

Notice that each element in A must have a finite norm.

There is always an equivalent norm such that $\left\Vert I\right\Vert =1$

A normed algebra is a rich structure, so much so that if we go further we fall
in known territories :

\begin{theorem}
Gel'fand-Mazur (Thill p.40) A normed algebra which is also a division ring
(each element has an inverse) is isomorphic to $%
\mathbb{C}
$
\end{theorem}

\begin{definition}
A normed *-algebra is a normed algebra and a *algebra such that the involution
is continuous. We will require also that :

$\forall X\in A:\left\Vert X^{\ast}\right\Vert =\left\Vert X\right\Vert $ and
$\left\Vert X\right\Vert ^{2}=\left\Vert X^{\ast}X\right\Vert $
\end{definition}

(so a normed *algebra is a pre C*-algebra in Thill's nomenclature)

\begin{theorem}
(Thill p.120) In a normed *-algebra :

i) $\left\Vert I\right\Vert =1$

ii) the map : $X\rightarrow X^{\ast}X$ is continuous in X = 0

iii) if the sequence $\left(  X_{n}\right)  _{n\in%
\mathbb{N}
}$ converges to 0, then the sequence $\left(  X_{n}^{\ast}\right)  _{n\in%
\mathbb{N}
}$\ is bounded
\end{theorem}

\begin{definition}
A \textbf{Banach algebra} is a normed algebra which is complete with the norm topology.
\end{definition}

It is always possible to complete a normed algebra to make it a Banach algebra.

\begin{theorem}
(Thill p.12) A Banach algebra is isomorphic and homeomorphic to the space of
continuous endomorphisms on a Banach space.
\end{theorem}

Take A as vector space and the maps : $\rho:A\rightarrow%
\mathcal{L}%
\left(  A;A\right)  ::\rho\left(  X\right)  Y=XY$ this is the left regular
representation of A on itself.

\begin{definition}
A \textbf{Banach *-algebra} is a Banach algebra which is endowed with a
continuous involution such that $\left\Vert XY\right\Vert \leq\left\Vert
X\right\Vert \left\Vert Y\right\Vert .$
\end{definition}

\begin{definition}
A \textbf{C*-algebra} is a Banach *-algebra with a continuous involution *
such that $\left\Vert X^{\ast}\right\Vert =\left\Vert X\right\Vert $ and
$\left\Vert X\right\Vert ^{2}=\left\Vert X^{\ast}X\right\Vert $
\end{definition}

The results for series seen in Banach vector space still hold, but the
internal product opens additional possibilities. The main theorem is the following:

\begin{theorem}
Mertens (Thill p.53): If the series in a Banach algebra, $\sum_{n\in%
\mathbb{N}
}X_{n}$ is absolutely convergent$,\sum_{n\in%
\mathbb{N}
}Y_{n}$ is convergent, then the series (called the Cauchy product) $\sum_{n\in%
\mathbb{N}
}Z_{n}=\sum_{n\in%
\mathbb{N}
}\left(  \sum_{k=0}^{n}X_{k}Y_{n-k}\right)  $ converges and

$\sum_{n\in%
\mathbb{N}
}Z_{n}=\left(  \sum_{n\in%
\mathbb{N}
}X_{n}\right)  \left(  \sum_{n\in%
\mathbb{N}
}Y_{n}\right)  $
\end{theorem}

\paragraph{Examples\newline}

\begin{theorem}
On a Banach vector space the set $%
\mathcal{L}%
\left(  E;E\right)  $ is a Banach algebra with composition of maps as internal
product.\ If E is a Hilbert space $%
\mathcal{L}%
\left(  E;E\right)  $ is a C*-algebra
\end{theorem}

\begin{theorem}
The set $%
\mathbb{C}
\left(  r\right)  $ of square complex matrices is a finite dimensional
C*-algebra with the norm $\left\Vert M\right\Vert =\frac{1}{r}Tr\left(
MM^{\ast}\right)  $
\end{theorem}

Spaces of functions (see the Functional analysis part for more) :

Are commutative C*-algebra with pointwise multiplication and the norm :
$\left\Vert f\right\Vert =\max\left\vert f\right\vert $

i) The set $C_{b}\left(  E;%
\mathbb{C}
\right)  $ of bounded functions

ii) if E Hausdorff, the set $C_{0b}\left(  E;%
\mathbb{C}
\right)  $ of bounded continuous functions

iii) if E Hausdorff, locally compact, the set $C_{0v}\left(  E;%
\mathbb{C}
\right)  $ of continuous functions vanishing at infinity.

If is E Hausdorff, locally compact, the set $C_{0c}\left(  E;%
\mathbb{C}
\right)  $ of continuous functions with compact support with the norm :
$\left\Vert f\right\Vert =\max\left\vert f\right\vert $ is a normed *-algebra
which is dense in $C_{0\nu}\left(  E;%
\mathbb{C}
\right)  $

\subsubsection{Morphisms}

\begin{definition}
An \textbf{algebra morphism} between the topological algebras A,B is a
continuous linear map $f\in%
\mathcal{L}%
\left(  A;B\right)  $\ such that:

$f\left(  XY\right)  =f\left(  X\right)  \cdot f\left(  Y\right)  ,f\left(
I_{A}\right)  =I_{B}$
\end{definition}

\begin{definition}
A \textbf{*-algebra morphism} between the topological *-algebras A,B is an
algebra morphism f such that $f\left(  X\right)  ^{\ast}=f\left(  X^{\ast
}\right)  $
\end{definition}

As usual a morphism which is bijective and whose inverse map is also a
morphism is called an isomorphism.

When the algebras are normed, a map which preserves the norm is an isometry.
It is necessarily continuous.

A *-algebra isomorphism between C*-algebras is necessarily an isometry, and
will be called a C*-algebra isomorphism.

\begin{theorem}
(Thill p.48) A map $f\in L(A;B)$ between a Banach *-algebra A, and a normed
*-algebra B, such that : $f\left(  XY\right)  =f\left(  X\right)  \cdot
f\left(  Y\right)  ,f\left(  I\right)  =I$ and $f\left(  X\right)  ^{\ast
}=f\left(  X^{\ast}\right)  $ is continuous, and a *-algebra morphism
\end{theorem}

\begin{theorem}
(Thill p.46) A *morphism f from a C*-algebra A to a normed *-algebra B :

i) is contractive ($\left\Vert f\right\Vert \leq1)$

ii) f(A) is a C*-algebra

iii) $A/\ker f$ is a C*-algebra

iv) if f is injective, it is an isometry

v) f factors in a C*-algebra isomorphism $A/\ker f$ $\rightarrow f\left(
A\right)  $
\end{theorem}

\subsubsection{Spectrum}

The spectrum of an element of an algebra is an extension of the eigen values
of an endomorphism. This is the key tool in the study of normed algebras.

\paragraph{Invertible elements\newline}

"Invertible" will always mean "invertible for the internal operation".

\begin{theorem}
The set G(A) of invertible elements of a topological algebra is a topological group
\end{theorem}

\begin{theorem}
(Thill p.38, 49) In a Banach algebra A, the set G(A) of invertible elements is
an open subset and the map $X\rightarrow X^{-1}$ is continuous.

If the sequence $\left(  X_{n}\right)  _{n\in%
\mathbb{N}
}$ in G(A) converges to X, then the sequence $\left(  X_{n}^{-1}\right)
_{n\in%
\mathbb{N}
}$ converges to $X^{-1}$ iff it is bounded.

The border $\partial G\left(  A\right)  $ is the set of elements X such that
there are sequences $\left(  Y_{n}\right)  _{n\in%
\mathbb{N}
},\left(  Z_{n}\right)  _{n\in%
\mathbb{N}
}$ in A such that :

$\left\Vert Y_{n}\right\Vert =1,\left\Vert Z_{n}\right\Vert =1,XY_{n}%
\rightarrow0,Z_{n}X\rightarrow0$
\end{theorem}

\paragraph{Spectrum\newline}

\begin{definition}
For every element X of an algebra A on a field K:

i) the \textbf{spectrum }$Sp\left(  X\right)  $ of X is the subset of the
scalars $\lambda\in K$ such that $\left(  f-\lambda Id_{E}\right)  $ has no
inverse in A$.$

ii) the \textbf{resolvent} \textbf{set} $\rho\left(  f\right)  $\ of X is the
complement of the spectrum

iii) the map: $R:K\rightarrow A::R\left(  \lambda\right)  =$ $\left(  \lambda
Id-X\right)  ^{-1}$ is called the \textbf{resolvent} of X.
\end{definition}

As we have assumed that K=$%
\mathbb{C}
$ the spectrum is in $%
\mathbb{C}
.$

Warning ! the spectrum is \textit{relative to an algebra} A, and the inverse
must be \textit{in the algebra} :

i) is A is a normed algebra then we must have $\left\Vert \left(  X-\lambda
I\right)  ^{-1}\right\Vert <\infty$

ii) When one considers the spectrum in a subalgebra and when necessary we will
denote Sp$_{A}\left(  X\right)  .$

\paragraph{Spectral radius\newline}

The interest of the spectral radius is that, in a Banach algebra :
$\max\left(  \left\vert \lambda\right\vert ,\lambda\in Sp(X)\right)
=r_{\lambda}\left(  X\right)  $ (Spectral radius formula)

\begin{definition}
The \textbf{spectral radius} of an element X of a normed algebra is the real scalar:

$r_{\lambda}\left(  X\right)  =\inf\left\Vert X^{n}\right\Vert ^{1/n}%
=\lim_{n\rightarrow\infty}\left\Vert X^{n}\right\Vert ^{1/n}$
\end{definition}

\begin{theorem}
(Thill p.35, 40, 41)

$r_{\lambda}\left(  X\right)  \leq\left\Vert X\right\Vert $

$k\geq1:r_{\lambda}\left(  X^{k}\right)  =\left(  r_{\lambda}\left(  X\right)
\right)  ^{k}$

$r_{\lambda}\left(  XY\right)  =r_{\lambda}\left(  YX\right)  $

If $XY=YX$ :

$r_{\lambda}\left(  XY\right)  \leq r_{\lambda}\left(  X\right)  r_{\lambda
}\left(  Y\right)  $

$r_{\lambda}\left(  X+Y\right)  \leq r_{\lambda}\left(  X\right)  +r_{\lambda
}\left(  Y\right)  ;r_{\lambda}\left(  X-Y\right)  \leq\left\vert r_{\lambda
}(X)-r_{\lambda}\left(  Y\right)  \right\vert $
\end{theorem}

\begin{theorem}
(Thill p.36) For every element X of a Banach algebra the series $f\left(
z\right)  =\sum_{n=0}^{\infty}z^{n}X^{n}$ \ converges absolutely for
$\left\vert z\right\vert <1/r_{\lambda}\left(  X\right)  $ and it converges
nowhere for $\left\vert z\right\vert >1/r_{\lambda}\left(  X\right)  .$ The
radius of convergence is $1/r_{\lambda}\left(  X\right)  $
\end{theorem}

\begin{theorem}
(Thill p.60) For $\mu>r_{\lambda}\left(  X\right)  ,$ the \textbf{Cayley
transform} of X : $C_{\mu}(X)=\left(  X-\mu iI\right)  \left(  X+\mu
iI\right)  ^{-1}$ of every self adjoint element of a Banach *-algebra\ is unitary
\end{theorem}

\paragraph{Structure of the spectrum\newline}

\begin{theorem}
(Thill p.40) In a normed algebra the spectrum is never empty.
\end{theorem}

\begin{theorem}
(Thill p.39, 98) In a Banach algebra :

- the spectrum is a non empty compact in $%
\mathbb{C}
,$ bounded by $r_{\lambda}\left(  X\right)  \leq\left\Vert X\right\Vert
:\max\left(  \left\vert \lambda\right\vert ,\lambda\in Sp(X)\right)
=r_{\lambda}\left(  X\right)  $

- the spectrum of a reflexion X is Sp(X)=(-1,+1)
\end{theorem}

\begin{theorem}
(Thill p.34) In a *-algebra :

Sp(X*)=$\overline{Sp(X)}$

for every normal element X : $r_{\lambda}\left(  X\right)  =\left\Vert
X\right\Vert $
\end{theorem}

\begin{theorem}
(Thill p.41) In a Banach *-algebra:

$r_{\lambda}\left(  X\right)  =r_{\lambda}\left(  X^{\ast}\right)
,Sp(X^{\ast})=\overline{Sp(X)}$
\end{theorem}

\begin{theorem}
(Thill p.60) In a C*-algebra the spectrum of an unitary element is contained
in the unit circle
\end{theorem}

\begin{theorem}
(Thill p.33) For every element : $\left(  Sp(XY)\right)  \backslash0=\left(
Sp(YX\right)  )\backslash0$
\end{theorem}

\begin{theorem}
(Thill p.73) In a Banach algebra if $XY=YX$ then :

$Sp\left(  XY\right)  \subset Sp\left(  X\right)  Sp\left(  Y\right)
;Sp\left(  X+Y\right)  \subset\left\{  Sp\left(  X\right)  +Sp\left(
Y\right)  \right\}  $
\end{theorem}

\begin{theorem}
(Thill p.32, 50, 51) For every normed algebra, B subalgebra of A, $X\in B$

$Sp_{A}(X)\sqsubseteq Sp_{B}\left(  X\right)  $

(Silov) if B is complete or has no interior : $\partial Sp_{B}\left(
X\right)  \subset\partial Sp_{A}\left(  X\right)  $
\end{theorem}

\begin{theorem}
(Thill p.32, 48) If $f:A\rightarrow B$ is an algebra morphism then :

$Sp_{B}(f(X))\subset Sp_{A}(X)$

$r_{\lambda}\left(  f\left(  X\right)  \right)  \leq r_{\lambda}\left(
X\right)  $
\end{theorem}

\begin{theorem}
(Rational Spectral Mapping theorem) (Thill p.31) For every element X in an
algebra A, the rational map :

$Q:A\rightarrow A::Q\left(  X\right)  =%
{\displaystyle\prod\limits_{k}}
\left(  X-\alpha_{k}I\right)
{\displaystyle\prod\limits_{l}}
\left(  X-\beta_{l}I\right)  ^{-1}$

where all $\alpha_{k}\neq\beta_{k}$ , $\beta_{k}\notin Sp(X)$ is such that : Sp(Q(X))=Q(Sp(X))
\end{theorem}

\paragraph{Pt\`{a}k function\newline}

\begin{definition}
On a normed *-algebra A the Pt\`{a}k function is :

$r_{\sigma}:A\rightarrow%
\mathbb{R}
_{+}::r_{\sigma}\left(  X\right)  =\sqrt{r_{\lambda}\left(  X^{\ast}X\right)
}$
\end{definition}

\begin{theorem}
(Thill p.43, 44,120) The Pt\`{a}k function has the following properties :

$r_{\sigma}\left(  X\right)  \leq\sqrt{\left\Vert X^{\ast}X\right\Vert }$

$r_{\sigma}\left(  X^{\ast}\right)  =r_{\sigma}\left(  X\right)  $

$r_{\sigma}\left(  X^{\ast}X\right)  =r_{\sigma}\left(  X\right)  ^{2}$

If X is hermitian : $r_{\lambda}\left(  X\right)  =r_{\sigma}\left(  X\right)
$

If X is normal : $r_{\sigma}\left(  X\right)  =\left\Vert X\right\Vert $ and
in a Banach *-algebra: $r_{\lambda}\left(  X\right)  \geq r_{\sigma}\left(
X\right)  $

the map $r_{\sigma}$\ is continuous at 0 and bounded in a neighborhood of 0
\end{theorem}

\begin{definition}
(Thill p.119,120) A map $f:A\rightarrow E$\ from a normed *-algebra A to a
normed vector space F is $\sigma-$\textbf{contractive} if $\left\Vert f\left(
X\right)  \right\Vert \leq r_{\sigma}\left(  X\right)  .$
\end{definition}

Then it is continuous.

\paragraph{Hermitian algebra\newline}

For any element : $Sp(X^{\ast})=\overline{Sp(X)}$ so for a self-adjoint X :
$Sp(X)=\overline{Sp(X)}$ but it does not imply that each element of the
spectrum is real.

\begin{definition}
A \textbf{*}-algebra is said to be \textbf{hermitian} if all its self-adjoint
elements have a real spectrum
\end{definition}

\begin{theorem}
(Thill p.57) A closed *-algebra of a hermitian algebra is hermitian. A
C*-algebra is hermitian.
\end{theorem}

\begin{theorem}
(Thill p.56, 88) For a Banach *-algebra A the following conditions are
equivalent :

i) A is hermitian

ii) $\forall X\in A:X=X^{\ast}:i\notin Sp(X)$

iii) $\forall X\in A:r_{\lambda}\left(  X\right)  \leq r_{\sigma}\left(
X\right)  $

iv) $\forall X\in A:XX^{\ast}=X^{\ast}X\Rightarrow r_{\lambda}\left(
X\right)  =r_{\sigma}\left(  X\right)  $

v) $\forall X\in A:XX^{\ast}=X^{\ast}X\Rightarrow r_{\lambda}\left(  X\right)
\leq\left\Vert X^{\ast}X\right\Vert ^{1/2}$

vi) $\forall X\in A:$ unitary $\Rightarrow Sp(X)$ is contained in the unit circle

vii) Shirali-Ford: $\forall X\in A:X^{\ast}X\geq0$
\end{theorem}

\paragraph{Positive elements\newline}

\begin{definition}
On a *-algebra the set of positive elements denoted $A^{+}$ is the set of
self-adjoint elements with positive spectrum
\end{definition}

$A^{+}=\left\{  X\geq0\right\}  =\left\{  X\in A:X=X^{\ast},Sp\left(
X\right)  \subset\lbrack0,\infty\lbrack\right\}  $

$A^{+}$ is a cone in A

Then the set $A^{+}$ is well ordered by : $X\geq Y\Leftrightarrow
X-Y\geq0\Leftrightarrow X-Y\in A^{+}$

\begin{theorem}
(Thill p.100) In a C*algebra A :

i) $A^{+}$ is a convex and closed cone

ii) $\forall X\in A:X^{\ast}X\geq0$
\end{theorem}

\begin{theorem}
(Thill p.85) If $f:A\rightarrow B$ is a *-morphism : $X\in A^{+}\Rightarrow
f\left(  X\right)  \in B^{+}$
\end{theorem}

\paragraph{Square root\newline}

\begin{definition}
In an algebra A an element Y is the \textbf{square root} of X, denoted
$X^{1/2},$ if $Y^{2}=X$ and $Sp\left(  Y\right)  \subset%
\mathbb{R}
_{+}$
\end{definition}

\begin{theorem}
(Thill p.51) The square root $X^{1/2}$ of X, when it exists, belongs to the
closed subalgebra generated by X.\ If X=X* then $\left(  X^{1/2}\right)
=\left(  X^{1/2}\right)  ^{\ast}$
\end{theorem}

\begin{theorem}
(Thill p.55) In a Banach algebra every element X such that $Sp(X)\subset
]0,\infty\lbrack$ has a unique square root such that $Sp(X^{1/2}%
)\subset]0,\infty\lbrack$.
\end{theorem}

\begin{theorem}
(Thill p.62) In a Banach *-algebra every invertible positive element X has a
unique positive square root which is also invertible.
\end{theorem}

\begin{theorem}
(Thill p.100,101) In a C*-algebra every positive element X has a unique
positive square root. Conversely if there is Y such that X=Y%
${{}^2}$
or X=Y*Y then X is positive.
\end{theorem}

\begin{definition}
(Thill p.100) In a C*algebra A the \textbf{absolute value} of any element X is
$\left\vert X\right\vert =\left(  X^{\ast}X\right)  ^{1/2}$
\end{definition}

\begin{theorem}
(Thill p.100, 102, 103) In a C*algebra A :

i) the absolute value $\left\vert X\right\vert =\left(  X^{\ast}X\right)
^{1/2}$ lies in the closed *-subalgebra generated by X and :

$\left\Vert \left\vert X\right\vert \right\Vert =\left\Vert X\right\Vert
,\left\vert X\right\vert \leq\left\vert Y\right\vert \Rightarrow\left\Vert
X\right\Vert \leq\left\Vert Y\right\Vert $

ii) for every self-adjoint element X :

$-\left\Vert X\right\Vert I\leq X\leq\left\Vert X\right\Vert I,-\left\vert
X\right\vert \leq X\leq+\left\vert X\right\vert ,0\leq X\leq Y\Rightarrow
\left\Vert X\right\Vert \leq\left\Vert Y\right\Vert $

iii) any self-adjoint element X has a unique decomposition : $X_{+}=\frac
{1}{2}\left(  \left\vert X\right\vert +X\right)  ,X_{-}=\frac{1}{2}\left(
\left\vert X\right\vert -X\right)  $

such that $X=X_{+}-X_{-};X_{+},X_{-}\geq0;X_{+}X_{-}=X_{-}X_{+}=0$

iv) every invertible element X has a unique polar decomposition : X=UP with
$P=\left\vert X\right\vert ,UU^{\ast}=I$

v) If f is an *-homomorphism between C*-algebras : $f\left(  \left\vert
X\right\vert \right)  =\left\vert f\left(  X\right)  \right\vert $
\end{theorem}

\subsubsection{Linear functionals}

\begin{definition}
A \textbf{linear functional} on a topological algebra A is an element of its
algebric dual A*
\end{definition}

\begin{definition}
In a *-algebra A\ a linear functional $\varphi$ is :

i) \textbf{hermitian} if $\forall X\in A:\varphi\left(  X^{\ast}\right)
=\overline{\varphi\left(  X\right)  }$

ii) \textbf{positive} if $\forall X\in A:\varphi\left(  X^{\ast}X\right)
\geq0$

The \textbf{variation} of a positive linear functional is :

$v\left(  \varphi\right)  =\inf_{X\in A}\left\{  \gamma:\left\vert
\varphi\left(  X\right)  \right\vert ^{2}\leq\gamma\varphi\left(  X^{\ast
}X\right)  \right\}  $.

If it is finite then $\left\vert \varphi\left(  X\right)  \right\vert ^{2}\leq
v\left(  \varphi\right)  \varphi\left(  X^{\ast}X\right)  $

iii) \textbf{weakly continuous} if for every self-adjoint element X the map
$Y\in A\rightarrow\varphi\left(  Y^{\ast}XY\right)  $ is continuous

iv) a \textbf{quasi-state }if it is positive, weakly continuous, and $v\left(
\varphi\right)  \leq1.$The set of quasi-states is denoted QS(A)

iv) a \textbf{state }if it is a quasi-state and $v\left(  \varphi\right)
=1.$The set of states is denoted S(A).

v) a \textbf{pure state} if it is an extreme point of S(A). The set of pure
states is denoted PS(A).
\end{definition}

\begin{theorem}
(Thill p.139,140) QS(A) is the closed convex hull of $PS(A)\cup0$, and a
compact Hausdorff space in the *weak topology.
\end{theorem}

\begin{definition}
In a *-algebra\ a positive linear functional $\varphi_{2}$ is
\textbf{subordinate} to a positive linear fonctional $\varphi_{1}$ if
$\forall\lambda\geq0:\lambda\varphi_{2}-\varphi_{1}$ is a positive linear
fonctional. A positive linear functional $\varphi$\ is \textbf{indecomposable}
if any other positive linear functional subordinate to $\varphi$\ is a
multiple of $\varphi$
\end{definition}

\begin{theorem}
(Thill p.139,141,142, 144, 145, 151) On a normed *-algebra A:

The variation of a positive linear functional $\varphi$ is finite and given by
$v\left(  \varphi\right)  =\varphi\left(  I\right)  $.

A quasi-state $\varphi$ is continuous, $\sigma-$contractive ($\left\Vert
\varphi\left(  X\right)  \right\Vert \leq r_{\sigma}\left(  X\right)  $) and hermitian

A state $\varphi$ is continuous and $\forall X\in A_{+}:\varphi\left(
X\right)  \geq0$,$\forall X\in A:$ $\sqrt{\psi\left(  X^{\ast}X\right)  }\leq
r_{\sigma}\left(  X\right)  $.

A state is pure iff it is indecomposable
\end{theorem}

\begin{theorem}
(Thill p.142,144,145) On a Banach *-algebra :

A positive linear functional is continuous

A state $\varphi$ (resp a pure state) on a closed *-subalgebra can be extended
to a state (resp. a pure state) if $\varphi$ is hermitian
\end{theorem}

\begin{theorem}
(Thill p.145) On a C*-algebra:

A positive linear functional is continuous and $v\left(  \varphi\right)
=\left\Vert \varphi\right\Vert $

A state is a continuous linear functional such that $\left\Vert \varphi
\right\Vert =\varphi\left(  I\right)  =1$ . Then it is hermitian and $v\left(
\varphi\right)  =\left\Vert \varphi\right\Vert $
\end{theorem}

\begin{theorem}
(Thill p.146) If E is a locally compact Hausdorff topological space, for every
state $\varphi$ in $C_{\nu}\left(  E;%
\mathbb{C}
\right)  $ there is a unique inner regular Borel probability measure $P$ on E
such that : $\forall f\in C_{\nu}\left(  E;%
\mathbb{C}
\right)  :\varphi\left(  f\right)  =\int_{E}fP$
\end{theorem}

\begin{theorem}
If $\varphi$\ is a positive linear functional on a *-algebra A, then
$\left\langle X,Y\right\rangle =\varphi\left(  Y^{\ast}X\right)  $ defines a
sesquilinear form on A, called a Hilbert form.
\end{theorem}

\paragraph{Multiplicative linear functionals\newline}

\begin{definition}
A \textbf{multiplicative linear functional} on a topological algebra is an
element of the algebraic dual A' : $\varphi\in L(A;%
\mathbb{C}
)$ such that $\varphi\left(  XY\right)  =\varphi\left(  X\right)
\varphi\left(  Y\right)  $ and $\varphi\neq0$
\end{definition}

$\Rightarrow\varphi\left(  I\right)  =1$

\begin{notation}
$\Delta\left(  A\right)  $ is the set of multiplicative linear functionals on
an algebra A.
\end{notation}

It is also sometimes denoted $\widehat{A}.$

\begin{theorem}
(Thill p.72) The set of multiplicative linear functional is not empty :
$\Delta\left(  A\right)  \neq\varnothing$
\end{theorem}

\begin{definition}
For X fixed in an algebra A, the \textbf{Gel'fand transform} of X is the map :
$\widehat{X}:\Delta\left(  A\right)  \rightarrow%
\mathbb{C}
::\widehat{X}\left(  \varphi\right)  =\varphi\left(  X\right)  $ and the map
$\symbol{94}:A\rightarrow C\left(  \Delta\left(  A\right)  ;%
\mathbb{C}
\right)  $ is the \textbf{Gel'fand transformation.}
\end{definition}

The Gel'fand transformation is a morphism of algebras.

Using the Gel'fand transformation $\Delta\left(  A\right)  \subset A^{\prime}
$ can be endowed with the *weak topology, called Gel'fand topology. With this
topology $\overline{\Delta\left(  A\right)  }$ is compact Hausdorff and
$\overline{\Delta\left(  A\right)  }\sqsubseteq\Delta\left(  A\right)  \cup0$

\begin{theorem}
(Thill p.68) For every topological algebra A, and $X\in A$\ : $\widehat
{X}\left(  \Delta\left(  A\right)  \right)  \subset Sp\left(  X\right)  $
\end{theorem}

\begin{theorem}
(Thill p.67, 68, 75) In a Banach algebra A:

i) a multiplicative linear functional is continuous with norm $\left\Vert
\varphi\right\Vert \leq1$

ii) the Gel'fand transformation is a contractive morphism in $C_{0v}\left(
\Delta\left(  A\right)  ;%
\mathbb{C}
\right)  $

iii) $\Delta\left(  A\right)  $ is compact Hausdorff in the Gel'fand topology
\end{theorem}

\begin{theorem}
(Thill p.70, 71) In a commutative Banach algebra A:

i) for every element $X\in A:\ \widehat{X}\left(  \Delta\left(  A\right)
\right)  =Sp\left(  X\right)  $

ii) (Wiener) An element X of A is not invertible iff $\exists\varphi\in
\Delta\left(  A\right)  :$ $\widehat{X}\left(  \varphi\right)  =0$
\end{theorem}

\begin{theorem}
Gel'fand - Naimark (Thill p.77) The Gel'fand transformation is a C*-algebra
isomorphism between A and $C_{0v}\left(  \Delta\left(  A\right)  ;%
\mathbb{C}
\right)  ,$ the set of continuous, vanishing at infinity, functions on
$\Delta\left(  A\right)  .$
\end{theorem}

\begin{theorem}
(Thill p.79) For any Hausdorff, locally compact topological space,
$\Delta\left(  C_{0v}\left(  E;%
\mathbb{C}
\right)  \right)  $ is homeomorphic to E.
\end{theorem}

The homeomorphism is : $\delta:E\rightarrow\Delta\left(  C_{0v}\left(  E;%
\mathbb{C}
\right)  \right)  ::\delta_{x}\left(  f\right)  =f\left(  x\right)  $

\newpage

\section{HILBERT\ SPACES}

\label{Hilbert Spaces}

\subsection{Hilbert spaces}

\subsubsection{Definition}

\begin{definition}
A complex \textbf{Hilbert space} (H,g) is a complex Banach vector space H
whose norm is induced by a positive definite hermitian form g. A real Hilbert
space (H,g) is a real Banach vector space H whose norm is induced by a
positive definite symmetric form g.
\end{definition}

As a real hermitian form is a symmetric form we will consider only complex
Hilbert space, all results can be easily adjusted to the real case.

The hermitian form g will be considered as \textit{antilinear in the first
variable}, so :

$g(x,y)=\overline{g(y,x)}$

$g(x,ay+bz)=ag(x,y)+bg(x,z)$

$g(ax+by,z)=\overset{-}{a}g(x,z)+\overset{-}{b}g(y,z)$

$g(x,\overset{-}{x})\geq0$

$g(x,x)=0\Rightarrow x=0$

g is continuous. It induces a norm on H :$\left\Vert x\right\Vert
=\sqrt{g(x,x)}$

\begin{definition}
A \textbf{pre-Hilbert} space is a complex normed vector space whose norm is
induced by a positive definite hermitian form
\end{definition}

A normed space can always be "completed" to become a complete space.

\begin{theorem}
(Schwartz 2 p.9) If E is a separable complex vector space endowed with a
definite positive sesquilinear form g, then its completion is a Hilbert space
with a sesquilinear form which is the extension of g.
\end{theorem}

\bigskip

But it is not always possible to deduce a sesquilinear form from a norm.

Let E be a vector space on the field K with a semi-norm $\left\Vert
{}\right\Vert .$ This semi norm is induced by :

- a sequilinear form iff $K=%
\mathbb{C}
$ and

$g\left(  x,y\right)  =\frac{1}{4}\left(  \left\Vert x+y\right\Vert
^{2}-\left\Vert x-y\right\Vert ^{2}+i\left(  \left\Vert x+iy\right\Vert
^{2}-\left\Vert x-iy\right\Vert ^{2}\right)  \right)  $

is a sequilinear form (not necessarily definite positive).

- a symmetric bilinear forms iff $K=%
\mathbb{R}
$ and

$g\left(  x,y\right)  =\frac{1}{2}\left(  \left\Vert x+y\right\Vert
^{2}-\left\Vert x\right\Vert ^{2}-\left\Vert y\right\Vert ^{2}\right)  $\ 

is a symmetric bilinear form (not necessarily definite positive).

And then the form g is necessarily unique.

Similarly not any norm can lead to a Hilbert space : in $%
\mathbb{R}
^{n}$ it is possible only with the euclidian norm.

\begin{theorem}
(Schwartz 2 p.21) Every closed vector subspace of a Hilbert space is a Hilbert space
\end{theorem}

Warning! a vector subspace is not necessarily closed if H is infinite dimensional

\subsubsection{Projection}

\begin{theorem}
(Neeb p.227) For any vectors x,y in a Hilbert space (H,g) the map :
$P_{xy}:H\rightarrow H::P_{xy}\left(  u\right)  =g\left(  y,u\right)  x$ is a
continuous operator with the properties :

$P_{xy}=P_{yx}^{\ast}$

$\forall X,Y\in%
\mathcal{L}%
\left(  H;H\right)  :P_{Xx,Yy}=XP_{xy}Y^{\ast}$
\end{theorem}

\begin{theorem}
(Schwartz 2 p.11) For every closed convex non empty subset F of a Hilbert
space (H,g):

i) $\forall u\in H,\forall v\in F:\operatorname{Re}g\left(  u-v,u-v\right)
\leq0$

ii) for any $u\in H$ there is a unique $v\in F$ such that : $\left\Vert
u-v\right\Vert =\min_{w\in F}\left\Vert u-w\right\Vert $

iii) the map $\pi_{F}:H\rightarrow F::\pi_{F}\left(  u\right)  =v$ , called
the \textbf{projection} on F, is continuous.
\end{theorem}

\begin{theorem}
(Schwartz 2 p.13) For every closed convex family $\left(  F_{n}\right)  _{n\in%
\mathbb{N}
}$ of subsets of a Hilbert space (H,g), such that their intersection F is non
empty, and every vector $u\in H,$\ the sequence $\left(  v_{n}\right)  _{n\in%
\mathbb{N}
}$\ of the projections of u on each $F_{n}$ converges to the projection v of u
on F and $\left\Vert u-v_{n}\right\Vert \rightarrow\left\Vert u-v\right\Vert $
\end{theorem}

\begin{theorem}
(Schwartz 2 p.15) For every closed convex family $\left(  F_{n}\right)  _{n\in%
\mathbb{N}
}$ of subsets of a Hilbert space (H,g), with union F, and every vector $u\in
H,$\ the sequence $\left(  v_{n}\right)  _{n\in%
\mathbb{N}
}$\ of the projections of u on each $F_{n}$ converges to the projection v of u
on the closure of F and $\left\Vert u-v_{n}\right\Vert \rightarrow\left\Vert
u-v\right\Vert .$
\end{theorem}

\begin{theorem}
(Schwartz 2 p.18) For every closed vector subspace F of a Hilbert space (H,g)
there is a unique projection $\pi_{F}:H\rightarrow F\in%
\mathcal{L}%
\left(  F;H\right)  $ . If $F\neq\left\{  0\right\}  $\ then $\left\Vert
\pi_{F}\right\Vert =1$
\end{theorem}

If $u\in F:\pi_{F}\left(  u\right)  =u$

\begin{theorem}
There is a bijective correspondance between the projections on a Hilbert space
(H,g), meaning the operators

$P\in%
\mathcal{L}%
\left(  H;H\right)  :P^{2}=P,P=P^{\ast}$

and the closed vector subspaces $H_{P}$ of H.\ And P is the orthogonal
projection on $H_{P}$
\end{theorem}

\begin{proof}
i) If P is a projection, it has the eigen values 0,1 with eigen vector spaces
$H_{0},H_{1}.$ They are closed as preimage of 0 by the continuous maps :
$P\psi=0,\left(  P-Id\right)  \psi=0$

Thus : $H=H_{0}\oplus H_{1}$

Take $\psi\in H:\psi=\psi_{0}+\psi_{1}$

$g\left(  P\left(  \psi_{0}+i\psi_{1}\right)  ,\psi_{0}+\psi_{1}\right)
=g\left(  i\psi_{1},\psi_{0}+\psi_{1}\right)  =-ig\left(  \psi_{1},\psi
_{0}\right)  -ig\left(  \psi_{1},\psi_{1}\right)  $

$g\left(  \psi_{0}+i\psi_{1},P\left(  \psi_{0}+\psi_{1}\right)  \right)
=g\left(  \psi_{0}+i\psi_{1},\psi_{1}\right)  =g\left(  \psi_{0},\psi
_{1}\right)  -ig\left(  \psi_{1},\psi_{1}\right)  $

$P=P^{\ast}\Rightarrow-ig\left(  \psi_{1},\psi_{0}\right)  =g\left(  \psi
_{0},\psi_{1}\right)  $

$g\left(  P\left(  \psi_{0}-i\psi_{1}\right)  ,\psi_{0}+\psi_{1}\right)
=g\left(  -i\psi_{1},\psi_{0}+\psi_{1}\right)  =ig\left(  \psi_{1},\psi
_{0}\right)  +ig\left(  \psi_{1},\psi_{1}\right)  $

$g\left(  \psi_{0}-i\psi_{1},P\left(  \psi_{0}+\psi_{1}\right)  \right)
=g\left(  \psi_{0}-i\psi_{1},\psi_{1}\right)  =g\left(  \psi_{0},\psi
_{1}\right)  +ig\left(  \psi_{1},\psi_{1}\right)  $

$P=P^{\ast}\Rightarrow ig\left(  \psi_{1},\psi_{0}\right)  =g\left(  \psi
_{0},\psi_{1}\right)  $

$\left\langle \psi_{0},\psi_{1}\right\rangle =0$ so $H_{0},H_{1}$ are orthogonal

P has norm 1 thus $\forall u\in H_{1}:\left\Vert P\left(  \psi-u\right)
\right\Vert \leq\left\Vert \psi-u\right\Vert \Leftrightarrow\left\Vert
\psi_{1}-u\right\Vert \leq\left\Vert \psi-u\right\Vert $ and $\min_{u\in
H_{1}}\left\Vert \psi-u\right\Vert =\left\Vert \psi_{1}-u\right\Vert $

So P is the orthogonal projection on $H_{1}$ and is necessarily unique.

ii) Conversely any orthogonal projection P on a closed vector space meets the
properties : continuity, and $P^{2}=P,P=P^{\ast}$
\end{proof}

\subsubsection{Orthogonal complement}

2 vectors u,v are orthogonal if g(u,v)=0

\begin{definition}
The \textbf{orthogonal complement} $F^{\bot}$ of a \textit{vector subspace} F
of a Hilbert space H is the set of all vectors which are orthogonal to vectors
of F.
\end{definition}

\begin{theorem}
(Schwartz 2 p.17) The orthogonal complement $F^{\bot}$ of a \textit{vector
subspace} F of a Hilbert space H is a closed vector subspace of H, which is
also a Hilbert space and we have : $H=F\oplus F^{\bot},F^{\bot\bot}%
=\overline{F}$ (closure of F)
\end{theorem}

\begin{theorem}
(Schwartz 2 p.19) For every finite family $\left(  F_{i}\right)  _{i\in I}$ of
closed vector subspaces of a Hilbert space H :$(\cup_{i}F_{i})^{\bot}=\cap
_{i}F_{i}^{\bot};(\cap_{i}F_{i})^{\bot}=\overline{(\cup_{i}F_{i}^{\bot})}$ (closure)
\end{theorem}

\begin{theorem}
(Schwartz 2 p.16) A vector subspace F of a Hilbert space H is dense in H iff
its orthogonal complement is 0
\end{theorem}

If S is a \textit{subset} of H then the orthogonal complement of S is the
orthogonal complement of the linear span of S (intersection of all the vector
subspaces containing S). It is a closed vector subspace, which is also a
Hilbert space.

\subsubsection{Quotient space}

\begin{theorem}
(Schwartz 2 p.21) For every closed vector subspace F of a Hilbert space the
quotient space H/F is a Hilbert space and the projection $\pi_{F}:F^{\perp
}\rightarrow H/F$ is a Hilbert space isomorphism.
\end{theorem}

\subsubsection{Hilbert sum of Hilbert spaces}

\begin{theorem}
(Neeb p.23, Schwartz 2 p.34) The \textbf{Hilbert sum,} denoted\textbf{\ }%
$H=\oplus_{i\in I}H_{i}$ of a family $\left(  H_{i},g_{i}\right)  _{i\in I}$
of Hilbert spaces is the subset of families $\left(  u_{i}\right)  _{i\in I},$
$u_{i}\in H_{i}$\ such that : $\sum_{i\in I}g_{i}\left(  u_{i},u_{i}\right)
<\infty.$ For every family $\left(  u_{i}\right)  _{i\in I}\in H$ ,
$\sum_{i\in I}u_{i}$ is summable and H has the structure of a Hilbert space
with the scalar product : $g\left(  u,v\right)  =\sum_{i\in I}g_{i}\left(
u_{i},v_{i}\right)  .$ The vector subspace generated by the $H_{i}$ is dense
in $H.$
\end{theorem}

The sums are understood as :

$\left(  u_{i}\right)  _{i\in I}\in H\Leftrightarrow\forall J\subset
I,card(J)<\infty:\sum_{i\in J}g_{i}\left(  u_{i},u_{i}\right)  <\infty$

and :

$\exists u:\forall\varepsilon>0,\forall J\subset I:card(J)<\infty,\sqrt
{\sum_{i\notin J}\left\Vert u_{i}\right\Vert _{H_{i}}^{2}}<\varepsilon,$

$\forall K:J\subset K\subset I:\left\Vert u-\sum_{i\in K}u_{i}\right\Vert
<\varepsilon$

which implies that for any family of H only \textit{countably} many $u_{i}%
$\ are non zero.

So this is significantly different from the usual case.

The vector subspace generated by the $H_{i}$ comprises any family $\left(
u_{i}\right)  _{i\in I}$ such that only \textit{finitely} many $u_{i}$\ are
non zero.

\begin{definition}
For a complete field K (=$%
\mathbb{R}
,%
\mathbb{C}
)$ and any set I, the set $\ell%
{{}^2}%
\left(  I\right)  $ is the set of families $\left(  x_{i}\right)  _{i\in I}$
over K such that :

$\left(  \sup_{J\subset I}\sum_{i\in J}\left\vert x_{i}\right\vert
^{2}\right)  <\infty$ for any countable subset J of I.

$\ell^{2}\left(  I\right)  $ is a Hilbert space with the sesquilinear form :
$\left\langle x,y\right\rangle =\sum_{i\in I}\overline{x}_{i}y_{i}$
\end{definition}

\begin{theorem}
(Schwartz 2 p.37) $\ell%
{{}^2}%
\left(  I\right)  ,\ell^{2}\left(  I^{\prime}\right)  $ are isomorphic iff I
and I' have the same cardinality.
\end{theorem}

\bigskip

\subsection{Hilbertian basis}

\subsubsection{Definition}

\begin{definition}
A family $\left(  e_{i}\right)  _{i\in I}$\ of vectors of a Hilbert space
(H,g) is \textbf{orthormal} is $\forall i,j\in I:g(e_{i},e_{j})=\delta_{ij}$
\end{definition}

\begin{theorem}
(Schwartz 2 p.42) For any orthonormal family the map : $\ell^{2}\left(
I\right)  \rightarrow H::y=\sum_{i}x_{i}e_{i}$ is an isomorphism of vector
space from $\ell^{2}\left(  I\right)  $ to the closure $\overline{L}$ of the
linear span L of $\left(  e_{i}\right)  _{i\in I}$ and

Perceval inequality$:\forall x\in H:\sum_{i\in I}\left\vert g\left(
e_{i},x\right)  \right\vert ^{2}\leq\left\Vert x\right\Vert ^{2}$

Perceval equality$\ \ \ :\forall x\in\overline{L}:\sum_{i\in I}\left\vert
g\left(  e_{i},x\right)  \right\vert ^{2}=\left\Vert x\right\Vert ^{2}%
,\sum_{i\in I}g\left(  e_{i},x\right)  e_{i}=x$
\end{theorem}

\begin{definition}
A \textbf{Hilbertian basis} of a Hilbert space (H,g) is an orthonormal family
$\left(  e_{i}\right)  _{i\in I}$ such that the linear span of the family is
dense in H. Equivalently if the only vector orthogonal to the family is 0.
\end{definition}

\begin{equation}
\forall x\in H:\sum_{i\in I}g\left(  e_{i},x\right)  e_{i}=x,\sum_{i\in
I}\left\vert g\left(  e_{i},x\right)  \right\vert ^{2}=\left\Vert x\right\Vert
^{2}%
\end{equation}

\begin{equation}
\forall x,y\in H:\sum_{i\in I}\overline{g\left(  e_{i},x\right)  }g\left(
e_{i},y\right)  =g\left(  x,y\right)
\end{equation}

$\forall\left(  x_{i}\right)  _{i\in I}\in\ell^{2}\left(  I\right)  $ (which
means $\left(  \sup_{J\subset I}\sum_{i\in J}\left\vert x_{i}\right\vert
^{2}\right)  <\infty$ for any countable subset J of I) then : $\sum_{i\in
I}x_{i}e_{i}=x\in H$ and $\left(  x_{i}\right)  _{i\in I}$\ is the unique
family such that $\sum_{i\in I}x_{i}e_{i}=x$.

The quantities $g\left(  e_{i},x\right)  $ are the Fourier coefficients.

Conversely :

\begin{theorem}
A family $\left(  e_{i}\right)  _{i\in I}$\ of vectors of of a Hilbert space
(H,g) is a Hilbert basis of H iff :

$\forall x\in H:\sum_{i\in I}\left\vert g\left(  e_{i},x\right)  \right\vert
^{2}=\left\Vert x\right\Vert ^{2}$
\end{theorem}

Warning ! As vector space, a Hilbert space has bases, for which only a finite
number of components are non zero. In a Hilbert basis there can be countably
non zero components. So the two kinds of bases are not equivalent if H is
infinite dimensional.

\begin{theorem}
(Schwartz 2 p.44) A Hilbert space has always a Hilbertian basis.\ All the
Hilbertian bases of a Hilbert space have the same cardinality.
\end{theorem}

\begin{theorem}
A Hilbert space is \textbf{separable} iff it has a Hilbert basis which is at
most countable.
\end{theorem}

Separable here is understood in the meaning of general topology.

\begin{theorem}
(Lang p.37) For every non empty closed disjoint subsets X,Y of a separable
Hilbert space H there is a smooth function $f:H\rightarrow\left[  0,1\right]
$ such that f(x) = 0 on X and f(x) = 1 on Y.
\end{theorem}

\subsubsection{Ehrardt-Schmidt procedure}

It is the extension of the Graham Schmidt procecure to Hilbert spaces. Let
$\left(  u_{n}\right)  _{n=1}^{N}$ be independant vectors in a Hilbert space
(H,g). Define :

$v_{1}=u_{1}/\left\Vert u_{1}\right\Vert $

$w_{2}=u_{2}-g\left(  u_{2},v_{1}\right)  v_{1}$ and $v_{2}=w_{2}/\left\Vert
w_{2}\right\Vert $

$w_{p}=u_{p}-\sum_{q=1}^{p-1}g\left(  u_{p},v_{q}\right)  v_{q}$ and
$v_{p}=w_{p}/\left\Vert w_{p}\right\Vert $

then the vectors $\left(  u_{n}\right)  _{n=1}^{N}$ are orthonormal.

\subsubsection{Conjugate}

The conjugate of a vector can be defined if we have a real structure on the
complex Hilbert space H, meaning an anti-linear map :$\sigma:H\rightarrow H$
such that $\sigma^{2}=Id_{H}.$ Then the conjugate of u is $\sigma\left(
u\right)  .$

The simplest way to define a real structure is by choosing a Hilbertian basis
which is stated as real, then the conjugate $\overline{u}$ of $u=\sum_{i\in
I}x_{i}e_{i}\in H$ is $\overline{u}=\sum_{i\in I}\overline{x}_{i}e_{i}.$

So we must keep in mind that conjugation is always with respect to some map,
and practically to some Hermitian basis.

\bigskip

\subsection{Operators}

\label{Operators on Hilbert spaces}

\subsubsection{General properties}

Linear endomorphisms on a Hilbert space are commonly called \textbf{operators}
(in physics notably).

\begin{theorem}
The set of continuous linear maps
$\mathcal{L}$%
(H;H') between Hilbert spaces on the field K is a Banach vector space on the
field K.
\end{theorem}

\begin{theorem}
(Schwartz 2 p.20) Any continuous linear map $f\in%
\mathcal{L}%
\left(  F;G\right)  $ from the subspace F of a separable pre-Hilbert space E,
to a complete topological vector space G can be extended to a continuous
linear map $\widetilde{f}\in%
\mathcal{L}%
\left(  E;G\right)  .$ If G is a normed space then $\left\Vert \widetilde
{f}\right\Vert _{%
\mathcal{L}%
(E;G)}=\left\Vert f\right\Vert _{%
\mathcal{L}%
(F;G)}$
\end{theorem}

The conjugate $\overline{f}$ (with respect to a real structure on H) of a
linear endomorphism over a Hilbert space H is defined as : $\overline
{f}:H\rightarrow H::\overline{f}\left(  u\right)  =\overline{f\left(
\overline{u}\right)  }$

\paragraph{Dual\newline}

One of the most important feature of Hilbert spaces is that there is an
anti-isomorphism with the dual.

\begin{theorem}
(Riesz) Let (H,g) be a complex Hilbert space, H' its topological dual. There
is a continuous anti-isomorphism $\tau:H^{\prime}\rightarrow H$\ such that :%

\begin{equation}
\forall\lambda\in H^{\prime},\forall u\in H:g\left(  \tau\left(
\lambda\right)  ,u\right)  =\lambda\left(  u\right)
\end{equation}

(H',g*) is a Hilbert space with the hermitian form : $g^{\ast}\left(
\lambda,\mu\right)  =g\left(  \tau\left(  \mu\right)  ,\tau\left(
\lambda\right)  \right)  $ and $\left\Vert \tau\left(  \mu\right)  \right\Vert
_{H}=\left\Vert \mu\right\Vert _{H^{\prime}},\left\Vert \tau^{-1}\left(
u\right)  \right\Vert _{H^{\prime}}=\left\Vert u\right\Vert _{H}$
\end{theorem}

\begin{theorem}
(Schwartz 2 p.27) A Hilbert space is reflexive : $\left(  H^{\prime}\right)
^{\prime}=H$
\end{theorem}

So :

for any $\varpi\in H^{\prime}$ there is a unique $\tau\left(  \varpi\right)
\in H$ such that : $\forall u\in H:g\left(  \tau\left(  \varpi\right)
,u\right)  =\varpi\left(  u\right)  $ and conversely for any $u\in H$ there is
a unique $\tau^{-1}\left(  u\right)  \in H^{\prime}$ such that : $\forall v\in
H:g\left(  u,v\right)  =\tau^{-1}\left(  u\right)  \left(  v\right)  $

$\tau\left(  z\varpi\right)  =\overline{z}\varpi,\tau^{-1}\left(  zu\right)
=\overline{z}u$

These relations are usually written in physics with the \textbf{bra-ket
notation} :

a vector $u\in H$ is written $|u>$ (ket)

a form $\varpi\in H^{\prime}$ is written $<\varpi|$ (bra)

the inner product of two vectors u,v is written $\left\langle u|v\right\rangle
$

the action of the form $\varpi$ on a vector u \ is written : $<\varpi||u>$ so
$<\varpi|$ can be identified with $\tau\left(  \varpi\right)  \in H$ such that
: $\left\langle \tau\left(  \varpi\right)  |u\right\rangle =<\varpi||u>$

As a consequence :

\begin{theorem}
For every continuous sesquilinear map $B:H\times H\rightarrow%
\mathbb{C}
$ in the Hilbert space (H,g), there is a unique continuous endomorphism $A\in%
\mathcal{L}%
\left(  H;H\right)  $ such that $B\left(  u,v\right)  =g\left(  Au,v\right)  $
\end{theorem}

\begin{proof}
Keep u fixed in H.\ The map : $B_{u}:H\rightarrow K::B_{u}\left(  v\right)
=B\left(  u,v\right)  $ is continuous linear, so $\exists\lambda_{u}\in
H^{\prime}:B\left(  u,v\right)  =\lambda_{u}\left(  v\right)  $

Define : $A:H\rightarrow H::A\left(  u\right)  =\tau\left(  \lambda
_{u}\right)  \in H:B\left(  u,v\right)  =g\left(  Au,v\right)  $
\end{proof}

\paragraph{Adjoint of a linear map\newline}

\begin{theorem}
(Schwartz 2 p.44) For every map $f\in%
\mathcal{L}%
(H_{1};H_{2})$ between the Hilbert spaces $(H_{1},g_{1}),(H_{2},g_{2})$ on the
field K there is a map $f^{\ast}\in%
\mathcal{L}%
(H_{2};H_{1})$ called the \textbf{adjoint} of f such that :%

\begin{equation}
\forall u\in H_{1},v\in H_{2}:g_{1}\left(  u,f^{\ast}v\right)  =g_{2}(fu,v)
\end{equation}

The map $^{\ast}:%
\mathcal{L}%
(H_{1};H_{2})\rightarrow%
\mathcal{L}%
(H_{2};H_{1})$ is antilinear, bijective, continuous, isometric and
$f^{\ast\ast}=f,\left(  f\circ h\right)  ^{\ast}=h^{\ast}\circ f^{\ast}$

If f is invertible, then f* is invertible and $\left(  f^{-1}\right)  ^{\ast
}=\left(  f^{\ast}\right)  ^{-1}$
\end{theorem}

There is a relation between transpose and adjoint : $f^{\ast}\left(  v\right)
=\overline{f^{t}\left(  \overline{v}\right)  }$

$f\left(  H_{1}\right)  ^{\perp}=f^{\ast-1}\left(  H_{1}\right)
,f^{-1}\left(  0\right)  ^{\perp}=\overline{f^{\ast}\left(  H_{2}\right)  }$

\begin{theorem}
(Schwartz 2 p.47) A map $f\in%
\mathcal{L}%
(H_{1};H_{2})$ between the Hilbert spaces $(H_{1},g_{1}),(H_{2},g_{2})$ on the
field K\ is injective iff $f^{\ast}(H_{2})$ is dense in $H_{1}$. Conversely
$f(H_{1})$ is dense in $H_{2}$ iff f is injective
\end{theorem}

\paragraph{Compact operators\newline}

A continuous linear map $f\in%
\mathcal{L}%
\left(  E;F\right)  $\ between Banach spaces E,F is compact if the the closure
$\overline{f\left(  X\right)  } $\ of the image of a bounded subset X of E is
compact in F.

\begin{theorem}
(Schwartz 2 p.63) A map $f\in%
\mathcal{L}%
\left(  E;H\right)  $ between a Banach space E and a Hilbert space H is
compact iff it is the limit of a sequence $\left(  f_{n}\right)  _{n\in%
\mathbb{N}
}$ of finite rank continuous maps in $%
\mathcal{L}%
\left(  E;H\right)  $
\end{theorem}

\begin{theorem}
(Schwartz 2 p.64) The adjoint of a compact map between Hilbert spaces is compact.
\end{theorem}

\paragraph{Hilbert sum of endomorphisms\newline}

\begin{theorem}
(Thill p.124) For a family of Hilbert space $\left(  H_{i}\right)  _{i\in I}$,
a family of operators $\left(  X_{i}\right)  _{i\in I}:$ $X_{i}\in%
\mathcal{L}%
\left(  H_{i};H_{i}\right)  $ ,if $\sup_{i\in I}\left\Vert X_{i}\right\Vert
_{H_{i}}<\infty$ there is a continuous operator $\oplus_{i\in I}X_{i}$ on
$\oplus_{i\in I}H_{i}$ with norm : $\left\Vert \oplus_{i\in I}X_{i}\right\Vert
=\sup_{i\in I}\left\Vert X_{i}\right\Vert _{H_{i}},$ called the
\textbf{Hilbert sum of the operators}, defined by : $\left(  \oplus_{i\in
I}X_{i}\right)  \left(  \oplus_{i\in I}u_{i}\right)  =\oplus_{i\in I}%
X_{i}\left(  u_{i}\right)  $
\end{theorem}

\paragraph{Topologies on
$\mathcal{L}$%
(H;H)\newline}

On the space
$\mathcal{L}$%
(H;H) of continuous endomorphisms of a Hilbert space H, we have the topologies :

i) Norm topology, induced by $\left\Vert f\right\Vert =\max_{\left\Vert
u\right\Vert =1}g\left(  u,u\right)  $

ii) Strong operator topology, induced by the semi-norms : $u\in H:p_{u}\left(
X\right)  =\left\Vert Xu\right\Vert $

iii) Weak operator topology, induced by the functionals : $%
\mathcal{L}%
(H;H)\rightarrow%
\mathbb{C}
::u,v\in H:p_{u,v}\left(  X\right)  =g\left(  u,Xv\right)  $

iv) $\sigma-$strong topology, induced by the semi-norms :

$p_{U}\left(  X\right)  =\sqrt{\sum_{i\in I}\left\Vert Xu_{i}\right\Vert ^{2}%
},U=\left(  u_{i}\right)  _{i\in I}:\sum_{i\in I}\left\Vert u_{i}\right\Vert
^{2}<\infty$

iv) $\sigma-$weak topology, induced by the functionals :$%
\mathcal{L}%
(H;H)\rightarrow%
\mathbb{C}
::$

$p_{UV}\left(  X\right)  =\sum_{i\in I}g\left(  u_{i},Xv_{i}\right)
,U,V:\sum_{i\in I}\left\Vert u_{i}\right\Vert ^{2}<\infty,\sum_{i\in
I}\left\Vert v_{i}\right\Vert ^{2}<\infty$

Weak operator topology
$<$
Strong operator topology
$<$
Norm topology

$\sigma-$weak toplogy
$<$
$\sigma-$strong toplogy
$<$
Norm topology

Weak operator topology
$<$
$\sigma-$weak toplogy

Strong operator topology
$<$
$\sigma-$strong toplogy

The $\sigma-$weak topology is the *weak topology induced by the trace class operators.

\subsubsection{C*-algebra of continuous endomorphisms}

With the map * which associates to each endomorphism its adjoint, the space
$\mathcal{L}$%
(H;H) of endomorphisms on a Hilbert space over a field K is a C*-algebra over K.

So all the results seen for general C*-algebra can be fully implemented, with
some simplifications and extensions.

\paragraph{General properties\newline}

For every endomorphism $f\in%
\mathcal{L}%
(H;H)$ on a Hilbert space on the field K :

$f^{\ast}\circ f$ is hermitian, and positive

$\overline{\exp f}=\exp\overline{f}$

$\exp f^{\ast}=(\exp f)^{\ast}$

The absolute value of f is : $\left\vert f\right\vert =\left(  f^{\ast
}f\right)  ^{1/2}$

$\left\Vert \left\vert f\right\vert \right\Vert =\left\Vert f\right\Vert
,\left\vert f\right\vert \leq\left\vert g\right\vert \Rightarrow\left\Vert
f\right\Vert \leq\left\Vert g\right\Vert $

The set G%
$\mathcal{L}$%
(H;H) of invertible operators is an open subset of $%
\mathcal{L}%
(H;H)$ and the map $f\rightarrow f^{-1}$ is continuous.

The set of \textbf{unitary} endomorphism f in a Hilbert space : $f\in%
\mathcal{L}%
\left(  H;H\right)  :f^{\ast}=f^{-1}$ is a closed subgroup of G%
$\mathcal{L}$%
(H;H).

Warning ! we must have both : f inversible and $f^{\ast}=f^{-1}.$ $\ $The
condition $f^{\ast}\circ f=Id$ is not sufficient.

Every invertible element f has a unique polar decomposition : f=UP with
$P=\left\vert f\right\vert ,UU^{\ast}=I$

\begin{theorem}
Trotter Formula (Neeb p.172) If f,h are continuous operators in a Hilbert
space over the field K, then : $\forall k\in K:e^{k(f+h)}=\lim_{n\rightarrow
\infty}(e^{k\dfrac{f}{n}}e^{k\dfrac{h}{n}})^{n}$
\end{theorem}

\begin{theorem}
(Schwartz 2 p.50) If $f\in%
\mathcal{L}%
(H;H)$ on a complex Hilbert space (H,g) :

$\frac{1}{2}\left\Vert f\right\Vert \leq\sup_{\left\Vert u\right\Vert \leq
1}\left\vert g\left(  u,fu\right)  \right\vert \leq\left\Vert f\right\Vert $
\end{theorem}

\begin{definition}
An operator $f\in%
\mathcal{L}%
(H;H)$ on a Hilbert space (H,g) is \textbf{self adjoint} (or hermitian) if f = f*.\ Then

$\forall u,v\in H,:g\left(  u,fv\right)  =g(fu,v)$

$\left\Vert f\right\Vert =\sup_{\left\Vert u\right\Vert \leq1}\left\vert
g\left(  u,fu\right)  \right\vert $
\end{definition}

\begin{theorem}
(Thill p.104) For a map $f\in%
\mathcal{L}%
(H;H)$ on a Hilbert space (H,g) the following conditions are equivalent :

i) f is hermitian positive : $f\geq0$

ii) $\forall u\in H:g\left(  u,fu\right)  \geq0$
\end{theorem}

\paragraph{Spectrum\newline}

\begin{theorem}
The spectrum of an endomorphism f on a Hilbert space H is a non empty compact
in $%
\mathbb{C}
,$ bounded by $r_{\lambda}\left(  f\right)  \leq\left\Vert f\right\Vert $

$Sp(f^{\ast})=\overline{Sp(f)}$

If f is self-adjoint then its eigen values $\lambda$\ are real and
$-\left\Vert f\right\Vert \leq\lambda\leq\left\Vert f\right\Vert $

The spectrum of an unitary element is contained in the unit circle
\end{theorem}

\begin{theorem}
Riesz (Schwartz 2 p.68) The set of eigen values of a compact normal
endomorphism f on a Hilbert space H on the field K is either finite, or
countable in a sequence convergent to 0 (which is or not an eigen value). It
is contained in a disc of radius $\left\Vert f\right\Vert $\ . If $\lambda$ is
eigen value for f, then $\overline{\lambda}$ is eigen value for f*. If $K=%
\mathbb{C}
,$ or if $K=%
\mathbb{R}
$ and f is symmetric, then at least one eigen value is equal to $\left\Vert
f\right\Vert .$ For each eigen value $\lambda$\ , except possibly for 0, the
eigen space $H_{\lambda}$\ is finite dimensional. The eigen spaces are
orthonormal for distinct eigen values. H is the direct Hilbert sum of the
$H_{\lambda}$ : f can be written $u=\sum_{\lambda}u_{\lambda}\rightarrow
fu=\sum_{\lambda}\lambda u_{\lambda}$ and f* : $f^{\ast}u=\sum_{\lambda
}\overline{\lambda}u_{\lambda}$

Conversely if $\left(  H_{\lambda}\right)  _{\lambda\in\Lambda}$ is a family
of closed, finite dimensional, orthogonal vector subspaces, with direct
Hilbert sum H, then the operator $u=\sum_{\lambda}u_{\lambda}\rightarrow
fu=\sum_{\lambda}\lambda u_{\lambda}$ is normal and compact
\end{theorem}

\paragraph{Hilbert-Schmidt operator\newline}

This is the way to extend the definition of the trace operator to Hilbert spaces.

\begin{theorem}
(Neeb p.228) For every map $f\in%
\mathcal{L}%
(H;H)$ of a Hilbert space (H,g), and Hilbert basis $\left(  e_{i}\right)
_{i\in I}$\ of H, the quantity $\left\Vert f\right\Vert _{HS}=\sqrt{\sum_{i\in
I}g\left(  fe_{i},fe_{i}\right)  }$ does not depend of the choice of the
basis.\ If $\left\Vert f\right\Vert _{HS}<\infty$ then f is said to be a
\textbf{Hilbert-Schmidt operator}.
\end{theorem}

\begin{notation}
HS(H) is the set of Hilbert-Schmidt operators on the Hilbert space H.
\end{notation}

\begin{theorem}
(Neeb p.229) Hilbert Schmidt operators are compact
\end{theorem}

\begin{theorem}
(Neeb p.228) For every Hilbert-Schmidt operators $f,h\in HS(H)$ on a Hilbert
space (H,g):

$\left\Vert f\right\Vert \leq\left\Vert f\right\Vert _{HS}=\left\Vert f^{\ast
}\right\Vert _{HS}$

$\left\langle f,h\right\rangle =\sum_{i\in I}g\left(  e_{i},f^{\ast}\circ
h\left(  e_{i}\right)  \right)  $ does not depend on the basis, converges and
gives to HS(H) the structure of a Hilbert space such that $\left\Vert
f\right\Vert _{HS}=\sqrt{\left\langle f,f\right\rangle }$

$\left\langle f,h\right\rangle =\left\langle h^{\ast},f^{\ast}\right\rangle $

If $f_{1}\in%
\mathcal{L}%
\left(  H;H\right)  ,f_{2},f_{3}\in HS\left(  H\right)  $ then :

$f_{1}\circ f_{2},f_{1}\circ f_{3}\in HS\left(  H\right)  ,\left\Vert
f_{1}\circ f_{2}\right\Vert _{HS}\leq\left\Vert f_{1}\right\Vert \left\Vert
f_{2}\right\Vert _{HS},\left\langle f_{1}\circ f_{2},f_{3}\right\rangle
=\left\langle f_{2},f_{1}^{\ast}f_{3}\right\rangle $
\end{theorem}

\paragraph{Trace\newline}

\begin{definition}
(Neeb p.230) A Hilbert-Schmidt endomorphism X on a Hilbert space H is
\textbf{trace class} if

$\left\Vert X\right\Vert _{T}=\sup\left\{  \left\vert \left\langle
X,Y\right\rangle \right\vert ,Y\in HS(H),\left\Vert Y\right\Vert
\leq1\right\}  <\infty$
\end{definition}

\begin{notation}
T(H) is the set of trace class operators on the Hilbert space H
\end{notation}

\begin{theorem}
(Neeb p.231) $\left\Vert X\right\Vert _{T}$ is a norm on T(H) and
T(H)$\sqsubseteq HS(H)$ is a Banach vector space with $\left\Vert X\right\Vert
_{T}$
\end{theorem}

\begin{theorem}
(Neeb p.230) A trace class operator X on a Hilbert space H has the following properties:

$\left\Vert X\right\Vert _{HS}\leq\left\Vert X\right\Vert _{T}=\left\Vert
X^{\ast}\right\Vert _{T}$

If $X\in%
\mathcal{L}%
\left(  H;H\right)  ,Y\in T\left(  H\right)  $ then : $XY\in T\left(
H\right)  ,\left\Vert XY\right\Vert _{T}\leq\left\Vert X\right\Vert \left\Vert
Y\right\Vert _{T}$

If $X,Y\in HS(H)$ then $XY\in T\left(  H\right)  $
\end{theorem}

\begin{theorem}
(Taylor 1 p.502) A continuous endomorphism X on a Hilbert space is trace class
iff it is compact and the set of eigen values of $\left(  X^{\ast}X\right)
^{1/2}$ is summable.
\end{theorem}

\begin{theorem}
(Neeb p.231) For any trace class operator X on a Hilbert space (H,g) and any
Hilbertian basis $\left(  e_{i}\right)  _{i\in I}$ of H, the sum $\sum_{i\in
I}g\left(  e_{i},Xe_{i}\right)  $ converges absolutely and :%

\begin{equation}
\sum_{i\in I}g\left(  e_{i},Xe_{i}\right)  =Tr(X)
\end{equation}

is the trace of X. It has the following properties:

i) $\left\vert Tr(X)\right\vert \leq\left\Vert X\right\Vert _{T}$

ii) Tr(X) does not depend on the choice of a basis, and is a linear continuous
functional on T(H)

iii) For $X,Y\in HS(H):Tr(XY)=Tr(YX),\left\langle X,Y\right\rangle
=Tr(XY^{\ast})$

iv) For $X\in T(H)$ the map : $%
\mathcal{L}%
\left(  H;H\right)  \rightarrow%
\mathbb{C}
::Tr(YX)$ is continuous, and Tr(XY)=Tr(YX).

v) $\left\Vert X\right\Vert _{T}\leq$ $\sum_{i,j\in I}\left\vert g\left(
e_{i},Xe_{j}\right)  \right\vert $

vi) The space of continuous, finite rank, endomorphims on H is dense in T(H)
\end{theorem}

For H finite dimensional the trace coincides with the usual operator.

\paragraph{Irreducible operators\newline}

\begin{definition}
A continuous linear endomorphism on a Hilbert space H is \textbf{irreducible}
if the only invariant closed subspaces are 0 and H.\ A set of operators is
invariant if each of its operators is invariant.
\end{definition}

\begin{theorem}
(Lang p.521) If S is an irreducible set of continuous linear endomorphism on a
Hilbert space H and f is a self-adjoint endomorphism of H commuting with all
elements of S, then f=kId for some scalar k.
\end{theorem}

\begin{theorem}
(Lang p.521) If S is an irreducible set of continuous linear endomorphism on a
Hilbert space H and f is a normal endomorphism of H, commuting as its adjoint
f*, with all elements of S, then f=kId for some scalar k.
\end{theorem}

\paragraph{Ergodic theorem\newline}

In mechanics a system is ergodic if the set of all its invariant states (in
the configuration space) has either a null measure or is equal to the whole of
the configuration space. Then it can be proven that the system converges to a
state which does not depend on the initial state and is equal to the averadge
of possible states. As the dynamic of such systems is usually represented by a
one parameter group of operators on Hilbert spaces, the topic has received a
great attention.

\begin{theorem}
Alaoglu-Birkhoff (Bratelli 1 p.378) Let $\mathfrak{U}$ be a set of linear
continuous endomorphisms on a Hilbert space H, such that : $\forall
U\in\mathfrak{U:}\left\Vert U\right\Vert \leq1,\forall U_{1},U_{2}%
\in\mathfrak{U:}U_{1}\circ U_{2}\in\mathfrak{U}$ and V the subspace of vectors
invariant by all U: $V=\left\{  u\in H,\forall U\in\mathfrak{U:}Uu=u\right\}
.$

Then the orthogonal projection $\pi_{V}:H\rightarrow V$ belongs to the closure
of the convex hull of $\mathfrak{U.}$
\end{theorem}

\begin{theorem}
For every unitary operator U on a Hilbert space H : $\forall u\in
H:\lim_{n\rightarrow\infty}\frac{1}{n+1}\sum_{p=0}^{n}U^{p}u=Pu$ where P is
the orthogonal projection on the subspaceV of invariant vectors $u\in V:Uu=u$
\end{theorem}

\begin{proof}
Take $\mathfrak{U=}$ the algebra generated by U in
$\mathcal{L}$%
$\left(  H;H\right)  $
\end{proof}

\subsubsection{Unbounded operators}

In physics it is necessary to work with linear maps which are not bounded, so
not continuous, on the whole of the Hilbert space. The most common kinds of
unbounded operators are operators defined on a dense subset and closed operators.

\paragraph{General definitions\newline}

An unbounded operator is a linear map $X\in L\left(  D\left(  X\right)
;H\right)  $ where D(X) is a vector subspace of the Hilbert space (H,g).

\begin{definition}
The spectrum of a linear map $X\in L(D(X);H),$ where H is a Hilbert space and
D(X) a vector subspace of H is the set of scalar $\lambda\in%
\mathbb{C}
$ such that $\lambda I-X$ is injective and surjective on D(X) and has a
bounded left-inverse
\end{definition}

X is said to be regular out of its spectrum

\begin{definition}
The adjoint of a linear map $X\in L(D(X);H),$ where (H,g) is a Hilbert space
and D(X) a vector subspace of H is a map $X^{\ast}\in L\left(  D\left(
X^{\ast}\right)  ;H\right)  $ such that : $\forall u\in D(X),v\in D\left(
X^{\ast}\right)  :g\left(  Xu,v\right)  =g\left(  u,X^{\ast}v\right)  $
\end{definition}

The adjoint does not necessarily exist or be unique.

\begin{definition}
X is self-adjoint if X=X*$,$ it is normal if XX*=X*X
\end{definition}

\begin{theorem}
(von Neumann) X*X and XX* are self-adjoint
\end{theorem}

\begin{definition}
A \textbf{symmetric map} on a Hilbert space (H,g) is a linear map $X\in
L(D\left(  X\right)  ;H)$ , where D(X) is a vector subspace of H, such that
$\forall u,v\in D\left(  X\right)  ,:g\left(  u,Xv\right)  =g(Xu,v)$
\end{definition}

\begin{theorem}
(Hellinger--Toeplitz theorem) (Taylor 1 p.512) A symmetric map $X\in L(H;H)$
on a Hilbert space H is continuous and self adjoint.
\end{theorem}

The key condition is here that X is defined over the whole of H.

\begin{definition}
The \textbf{extension} X of a linear map $Y\in L(D(Y);H),$ where H is a
Hilbert space and D(Y) a vector subspace of H is a linear map $X\in L\left(
D\left(  X\right)  ;H\right)  $ where $D(Y)\subset D\left(  X\right)  $ and
X=Y on D(Y)
\end{definition}

It is usually denoted $Y\subset X$

If X is symmetric, then $X\subset X^{\ast}$ and X can be extended on D(X*) but
the extension is not necessarily unique.\ 

\begin{definition}
A symmetric operator which has a unique extension which is self adjoint is
said to be \textbf{essentially self-adjoint}.
\end{definition}

\begin{definition}
Two linear operators $X\in%
\mathcal{L}%
\left(  D\left(  X\right)  ;H\right)  ,Y\in%
\mathcal{L}%
\left(  D\left(  Y\right)  ;H\right)  $ on the Hilbert space H commute if :

i) D(X) is invariant by Y : $YD(X)\subset D(X)$

ii) $YX\subset XY$
\end{definition}

The set of maps \textit{in
$\mathcal{L}$%
(H;H)} commuting with X is still called the commutant of X and denoted X'

\paragraph{Densely defined linear maps\newline}

\begin{definition}
A \textbf{densely defined operator} is a linear map X defined on a dense
subspace D(X) of a Hilbert space
\end{definition}

\begin{theorem}
(Thill p.238, 242) A densely defined operator X has an adjoint X* which is a
closed map.

If X is self-adjoint then it is closed, X* is symmetric and has no symmetric extension.
\end{theorem}

\begin{theorem}
(Thill p.238, 242) If X,Y are densely defined operators then :

i) $X\subset Y\Rightarrow Y^{\ast}\subset X^{\ast}$

ii) if XY is \ continuous on a dense domain then Y*X* \ is continuous on a
dense domain and $Y^{\ast}X^{\ast}\subset(XY)^{\ast}$
\end{theorem}

\begin{theorem}
(Thill p.240,241) The spectrum of a self-adjoint,densely defined operator is a
closed, locally compact subset of $%
\mathbb{R}
.$
\end{theorem}

\begin{theorem}
(Thill p.240, 246) The Cayley transform $Y=\left(  X-iI\right)  \left(
X+iI\right)  ^{-1}$ of the densely defined operator X is an unitary operator
and 1 is not an eigen value. If $\lambda\in Sp(X)$ then $\left(
\lambda-i\right)  \left(  \lambda+i\right)  ^{-1}\in Sp(Y).$ Furthermore the
commutants are such that Y'=X'. If X is self-adjoint then : $X=i\left(
I+Y\right)  (1-Y)^{-1}.$ Two self adjoint densely defined operators commute
iff their Cayley transform commute.
\end{theorem}

If X is closed and densely defined, then X*X is\ self adjoint and I+X*X has a
bounded inverse.

\paragraph{Closed linear maps\newline}

\begin{definition}
A linear map $X\in L(D(X);H),$ where H is a Hilbert space and D(X) a vector
subspace of H is \textbf{closed} if its graph is closed in H$\times$H.
\end{definition}

\begin{definition}
A linear map $X\in L(D(X);H)$ is \textbf{closable} if X has a closed extension
denoted $\widetilde{X}$. Not all operators are closable.
\end{definition}

\begin{theorem}
A densely defined operator X is closable iff X* is densely defined. In this
case $\widetilde{X}=\left(  X^{\ast}\right)  ^{\ast}$ and $\left(
\widetilde{X}\right)  ^{\ast}=X^{\ast}$
\end{theorem}

\begin{theorem}
A linear map $X\in L(D(X);H)$ where D(X) is a vector subspace of an Hilbert
space H, is closed if for every sequence $\left(  u_{n}\right)  \in D(X)^{%
\mathbb{N}
}$ which converges in H to u and such that $Xu_{n}\rightarrow v\in H$ then :
$u\in D\left(  X\right)  $ and v = Xu
\end{theorem}

\begin{theorem}
(Closed graph theorem) (Taylor 1 p.511) Any closed linear operator defined on
the whole Hilbert space H is bounded thus continuous.
\end{theorem}

\begin{theorem}
The kernel of a closed linear map $X\in L(D(X);H)$ is a closed subspace of the
Hilbert space H
\end{theorem}

\begin{theorem}
If the map X is closed and injective, then its inverse $X^{-1}$ is also closed;
\end{theorem}

\begin{theorem}
If the map X is closed then $X-\lambda I$ is closed where $\lambda$ is a
scalar and I is the identity function
\end{theorem}

\begin{theorem}
An operator X is closed and densely defined iff X** = X
\end{theorem}

\subsubsection{Von Neumann algebra}

\begin{definition}
A \textbf{von Neumann algebra} W denoted W*-algebra is a *-subalgebra of
$\mathcal{L}$%
(H;H) for a Hilbert space H, such that W=W"\ 
\end{definition}

\begin{theorem}
For every Hilbert space,
$\mathcal{L}$%
(H;H), its commutant
$\mathcal{L}$%
(H;H)' and $%
\mathbb{C}
I$ are W*-algebras.
\end{theorem}

\begin{theorem}
(Thill p.203) A C*-subalgebra A of
$\mathcal{L}$%
(H;H) is a W*-algebra iff A"=A
\end{theorem}

\begin{theorem}
(Thill p.204) If W is a von Neumann algebra then W' is a von Neumann algebra
\end{theorem}

\begin{theorem}
Sakai (Bratelli 1 p.76) A C*-algebra is isomorphic to a von Neumann algebra
iff it is the dual of a Banach space.
\end{theorem}

\begin{theorem}
(Thill p.206) For any Hilbert space H and any subset S of
$\mathcal{L}$%
(H;H) the smallest W*-algebra which contains S is $W(S)=\left(  S\cup S^{\ast
}\right)  ".$ If $\forall X,Y\in S:X^{\ast}\in S,XY=YX$ then W(S) is commutative.
\end{theorem}

\begin{theorem}
von Neumann density theorem (Bratelli 1 p.74) If B is a *subalgebra of
$\mathcal{L}$%
(H;H) for a Hilbert space H, such that the orthogonal projection on the
closure of the linear span Span$\left\{  Xu,X\in B,u\in H\right\}  $\ is H,
then B is dense in B"
\end{theorem}

\begin{theorem}
(Bratelli 1 p.76) For any Hilbert space H, a state $\varphi$ of a von Neumann
algebra W in
$\mathcal{L}$%
(H;H) is normal iff there is a positive, trace class operator $\rho$ in
$\mathcal{L}$%
(H;H)\ such that : $Tr(\rho)=1,\forall X\in W:\varphi\left(  X\right)
=Tr\left(  \rho X\right)  .$
\end{theorem}

$\rho$ is called a density operator.

\begin{theorem}
(Neeb p.152) Every von Neumann algebra A is equal to the bicommutant P" of the
set P of projections belonging to A : $P=\left\{  p\in A:p=p^{2}=p^{\ast
}\right\}  $
\end{theorem}

\begin{theorem}
(Thill p.207) A von Neuman algebra is the closure of the linear span of its projections.
\end{theorem}

\bigskip

\subsection{Reproducing Kernel}

The most usual Hilbert spaces are spaces of functions.\ They can be
characterized by a single, bivariable and hermitian function, the positive
kernel, which in turn can be used to build similar Hilbert spaces.

\subsubsection{Definitions}

\begin{definition}
For any set E and field $K=%
\mathbb{R}
,%
\mathbb{C}
$, a function $N:E\times E\rightarrow K$ is a \textbf{definite positive
kernel} of E if :

i) it is definite positive : for any finite set $\left(  x_{1},...,x_{n}%
\right)  ,x_{k}\in E,$ the matrix $\left[  N\left(  x_{i},x_{j}\right)
\right]  _{n\times n}\subset K\left(  n\right)  $ is semi definite positive :
$\left[  X\right]  ^{\ast}\left[  N\left(  x_{i},x_{j}\right)  \right]
\left[  X\right]  \geq0$ with $\left[  X\right]  =\left[  x_{i}\right]
_{n\times1}.$

ii) it is either \textbf{symmetric} (if $K=%
\mathbb{R}
):N\left(  x,y\right)  ^{\ast}=N(y,x)=N(x,y)$,or \textbf{hermitian} (if $K=%
\mathbb{C}
):N\left(  x,y\right)  ^{\ast}=\overline{N(y,x)}=N(x,y)$
\end{definition}

Then $\left\vert N(x,y)\right\vert ^{2}\leq\left\vert N\left(  x,x\right)
\right\vert \left\vert N\left(  y,y\right)  \right\vert $

A Hilbert space of functions defines a reproducing kernel:

\begin{theorem}
(Neeb p.55) For any Hilbert space (H,g) of functions $H:E\rightarrow K$ , if
the evaluation maps : $x\in E:\widehat{x}:H\rightarrow K::\widehat{x}\left(
f\right)  =f\left(  x\right)  $ are continuous, then :

i) for any $x\in E$ there is $N_{x}\in H$ such that : $\forall f\in H:g\left(
N_{x},f\right)  =\widehat{x}\left(  f\right)  =f\left(  x\right)  $

ii) The set $\left(  N_{x}\right)  _{x\in E}$ spans a dense subspace of H

iii) the function $N:E\times E\rightarrow K::N\left(  x,y\right)
=N_{y}\left(  x\right)  $ called the \textbf{reproducing kernel} of H is a
definite positive kernel of E.

iv) For any Hilbert basis $\left(  e_{i}\right)  _{i\in I}$ of H: $N\left(
x,y\right)  =\sum_{i\in I}\overline{e_{i}\left(  x\right)  }e_{i}\left(
y\right)  $
\end{theorem}

Conversely, a \ reproducing kernel defines a Hilbert space:

\begin{theorem}
(Neeb p.55) If $N:E\times E\rightarrow K$ is a positive definite kernel of E,
then :

i) $H_{0}=Span\left\{  N\left(  x,.\right)  ,x\in E\right\}  $ carries a
unique positive definite hermitian form g such that :

$\forall x,y\in E:g\left(  N_{x},N_{y}\right)  =N\left(  x,y\right)  $

ii) the completion H of $H_{0}$ with injection : $\imath:H_{0}\rightarrow H$
carries a Hilbert space structure H consistent with this scalar product, and
whose reproducing kernel is N.

iii) this Hilbert space is unique and called the reproducing kernel Hilbert
space defined by N, denoted $H_{N}$

iv) If E is a topological space, N continuous, then the map : $\gamma
:E\rightarrow H_{N}::\gamma\left(  x\right)  =N_{x}$ is continuous and the
functions of $H_{N}$ are continuous
\end{theorem}

Which is summed up by :

\begin{theorem}
(Neeb p.55) A function $N:E\times E\rightarrow K$ is positive definite iff it
is the reproducing kernel of some Hilbert space $H\subset C\left(  E;K\right)
$
\end{theorem}

Notice that usually no topology is required on E.

\subsubsection{Defining other positive kernels}

From a given positive kernel N of E one can build other positive kernels, and
to each construct is attached a Hilbert space.

\begin{theorem}
(Neeb p.57,67) The set N(E) of positive definite kernels of a topological
space E is a convex cone in $K^{E\times E}$ which is closed under pointwise
convergence and pointwise multiplication :
\end{theorem}

i) $\forall P,Q\in N(E),\lambda\in%
\mathbb{R}
_{+}:P+Q\in N\left(  E\right)  ,\lambda P\in N\left(  E\right)  ,$

$PQ\in N(E)$ with $PQ::\left(  PQ\right)  (x,y)=P(x,y)Q(x,y)$

ii) If $K=%
\mathbb{C}
:P\in N\left(  E\right)  \Rightarrow\operatorname{Re}P\in N\left(  E\right)
,\overline{P}\in N\left(  E\right)  ,\left\vert P\right\vert \in N\left(
E\right)  $

iii) If $\left(  P_{i}\right)  _{i\in I}\in N\left(  E\right)  ^{I}$ and
$\forall x,y\in E,\exists\lim\sum_{i\in I}P_{i}\left(  x,y\right)  $ then
$\exists P\in N\left(  E\right)  :P=\lim\sum_{i\in I}P_{i}$

iv) If $P,Q\in N\left(  E\right)  :H_{P+Q}\simeq H_{P}\oplus H_{Q}/\left(
H_{P}\cap H_{Q}\right)  $

\begin{theorem}
(Neeb p.59) If N is a positive kernel of E, $f\in F\rightarrow E$ then
$P:F\times F\rightarrow K::P\left(  x,y\right)  =N\left(  f\left(  x\right)
,f\left(  y\right)  \right)  $ is a positive definite kernel of F
\end{theorem}

\begin{theorem}
(Neeb p.57) Let $\left(  T,S,\mu\right)  $ a measured space, $\left(
P_{t}\right)  _{t\in T}$ a family of positive definite kernels of E, such that
$\forall x,y\in E$\ the maps : $t\rightarrow P_{t}\left(  x,y\right)  $ are
measurable and the maps : $t\rightarrow P_{t}\left(  x,x\right)  $ are
integrable, then : $P\left(  x,y\right)  =\int_{T}P_{t}\left(  x,y\right)
\mu\left(  t\right)  $ is a positive definite kernel of E.
\end{theorem}

\begin{theorem}
(Neeb p.59) If the series : $f\left(  z\right)  =\sum_{n=0}^{\infty}a_{n}%
z^{n}$ over K is convergent for $\left\vert z\right\vert <r,$ if N is a
positive definite kernel of E and $\forall x,y\in E:\left\vert
N(x,y)\right\vert <r$ then : $f\left(  N\right)  \left(  x,y\right)
=\sum_{n=0}^{\infty}a_{n}N(x,y)^{n}$ is a positive definite kernel of E.
\end{theorem}

\begin{theorem}
(Neeb p.64) If N is a definite positive kernel on E, $H_{N}$ a reproducing
kernel Hilbert space for N, H is the Hilbert sum $H=\oplus_{i\in I}H_{i},$
then there are positive kernels $\left(  N_{i}\right)  _{i\in I}$ on E such
that $N=\sum_{i\in i}N_{i}$ and $H_{i}=H_{N_{i}}$
\end{theorem}

\paragraph{Extension to measures\newline}

(Neeb p.62)

Let (E,S) a measurable space with a positive measurable kernel N and canonical
$H_{N}$

For a measure $\mu:c_{N}=\int_{E}\sqrt{N\left(  x,x\right)  }\mu\left(
x\right)  <\infty$ : $\forall f\in H_{N}:f\in L^{1}\left(  E,S,K,\mu\right)  $
and $\left\Vert f\right\Vert _{L^{1}}\leq c_{N}\left\Vert f\right\Vert
_{H_{N}}$ so $\exists N_{\mu}\in H_{N}:\int_{E}f\left(  x\right)  \mu\left(
x\right)  =\left\langle N_{\mu},f\right\rangle _{H_{N}}$ .\ Then : $N_{\mu
}\left(  x\right)  =\int_{E}N\left(  x,y\right)  \mu\left(  y\right)  $ and
$\left\langle N_{\mu},N_{\mu}\right\rangle =\int_{E}N\left(  x,y\right)
\mu\left(  y\right)  \mu\left(  x\right)  $

\paragraph{Realization triple\newline}

\begin{theorem}
(Neeb p.60) For any set E, Hilbert space (H,g) on K, map $\gamma:E\rightarrow
H$ such that $\gamma\left(  E\right)  $ spans a dense subset of H, then
$N:E\times E\rightarrow K::N\left(  x,y\right)  =g\left(  \gamma\left(
x\right)  ,\gamma\left(  y\right)  \right)  $ is a positive definite kernel of E.

Conversely for any positive definite kernel N of a set E, there are : a
Hilbert space (H,g), a map : $\gamma:E\rightarrow H$ such that $\gamma(E)$
spans a dense subset of H and $N\left(  x,y\right)  =g\left(  \gamma\left(
x\right)  ,\gamma\left(  y\right)  \right)  .$

The triple $(E,\gamma,H)$ is called a realization triple of N. For any other
triple $(E,H^{\prime},\gamma^{\prime})$ there is a unique isometry :
$\varphi:H\rightarrow H^{\prime}$ such that $\gamma^{\prime}$=$\varphi
\circ\gamma$
\end{theorem}

The realization done as above with $N_{x}=N\left(  x,.\right)  $ is called the
canonical realization. So to any positive kernel corresponds a family of
isometric Hilbert spaces.

\begin{theorem}
(Neeb p.59) If $\left(  E,\gamma,H\right)  $ is a realisation triple of N,
then $P(x,y)=\overline{\gamma(x)}N(x,y)\gamma(y)$ is a positive definite
kernel of E
\end{theorem}

\begin{theorem}
(Neeb p.61) If (H,g) is a Hilbert space on K, then :

$\left(  H,Id,H\right)  $ is a realization triple of $N\left(  x,y\right)
=g\left(  x,y\right)  $

$\left(  H,\gamma,H^{\prime}\right)  $ is a realization triple of $N\left(
x,y\right)  =g\left(  y,x\right)  $ with $\gamma:H\rightarrow H^{\prime}$

If $\left(  e_{i}\right)  _{i\in I}$ is a Hilbert basis of H, then $\left(
I,\gamma,\ell^{2}\left(  I,K\right)  \right)  $ is a realization triple of
$N\left(  i,j\right)  =\delta_{ij}$ with $\gamma:I\rightarrow H::\gamma\left(
i\right)  =e_{i}$

$\left(  H,\exp,%
\mathcal{F}%
_{+}\left(  H\right)  \right)  $ is a realization triple of $N\left(
x,y\right)  =\exp g\left(  x,y\right)  $ with $%
\mathcal{F}%
_{+}\left(  H\right)  $ the symmetric Fock space of H (see below)
\end{theorem}

\begin{theorem}
(Neeb p.62) If $\left(  E,S,\mu\right)  $ is a measured space, then $\left(
E,\gamma,L^{2}\left(  E,S,K,\mu\right)  \right)  $ is a realization triple of
the kernel : $N:S\times S\rightarrow K::N\left(  \varpi,\varpi^{\prime
}\right)  =\mu\left(  \varpi\cap\varpi^{\prime}\right)  $ with $\gamma
:S\rightarrow L^{2}\left(  E,S,K,\mu\right)  ::\gamma\left(  \varpi\right)
=1_{\varpi}$
\end{theorem}

\paragraph{Inclusions of reproducing kernels\newline}

\begin{definition}
For any set E with definite positive kernel N, associated canonical Hilbert
space $H_{N},$ the \textbf{symbol} of the operator $A\in%
\mathcal{L}%
\left(  H_{N};H_{N}\right)  $ is the function : $N_{A}:E\times E\rightarrow
K::N_{A}\left(  x,y\right)  =\left\langle x,Ay\right\rangle _{H_{N}}$
\end{definition}

A is uniquely defined by its symbol.

$\left(  N_{A}\right)  ^{\ast}=N_{A^{\ast}}$ and is hermitian iff A is hermitian

$N_{A}$ is a positive definite kernel iff A is positive

\begin{theorem}
(Neeb p.68) If P,Q are definite positive kernels on the set E, then the
following are equivalent :

i) $\exists k\in%
\mathbb{R}
_{+}:kQ-P\geq0$

ii) $H_{P}\subset H_{Q}$

iii) $\exists A\in%
\mathcal{L}%
\left(  H_{Q};H_{Q}\right)  ::Q_{A}=P,$ A positive
\end{theorem}

\begin{theorem}
(Neeb p.68) For any definite positive kernel P on the set E the following are
equivalent :

i) $\dim H_{P}=1$

ii) $%
\mathbb{R}
_{+}P$ is an extremal ray of N(E)

iii) There is a non zero function f on E such that : $P\left(  x,y\right)
=\overline{f\left(  x\right)  }f\left(  y\right)  $
\end{theorem}

\paragraph{Holomorphic kernels\newline}

\begin{theorem}
(Neeb p.200) If E is a locally convex complex vector space endowed with a real
structure, then one can define the conjugate $\overline{E}$ of E. If
$O\times\overline{O}$ is an open subset of $E\times\overline{E}$ and N a
positive kernel which is holomorphic on $O\times\overline{O}$, then the
functions of $H_{N}$ are holomorphic on $O\times\overline{O}$.
\end{theorem}

If (H,g) is a Hilbert space and k
$>$
0 then $N(x,y)=\exp(kg\left(  x,y\right)  )$ is a holomorphic positive kernel
on H.

\bigskip

\subsection{Tensor product of Hilbert spaces}

The definitions and properties seen in Algebra extend to Hilbert spaces.

\subsubsection{Tensorial product}

\begin{theorem}
(Neeb p.87) If $\left(  H,g_{H}\right)  $ is a Hilbert space with hilbertian
basis $\left(  e_{i}\right)  _{i\in I}$ , $\left(  F,g_{F}\right)  $ a Hilbert
space with hilbertian basis $\left(  f_{j}\right)  _{j\in J}$ then
$\sum_{\left(  i,j\right)  \in I\times J}e_{i}\otimes f_{j}$ is a Hilbert
basis of $H\otimes F$

The scalar product is defined as : $\left\langle u_{1}\otimes v_{1}%
,u_{2}\otimes v_{2}\right\rangle =g_{H}\left(  u_{1},u_{2}\right)
g_{F}\left(  v_{1},v_{2}\right)  $

The reproducing Kernel is : $N_{H\otimes F}\left(  u_{1}\otimes v_{1}%
,u_{2}\otimes v_{2}\right)  =g_{H}\left(  u_{1},u_{2}\right)  g_{F}\left(
v_{1},v_{2}\right)  $
\end{theorem}

The subspaces $\left(  e_{i}\otimes F\right)  _{i\in I}$ are pairwise
orthogonal and span a dense subspace.

$\left\langle e_{i}\otimes u,e_{j}\otimes v\right\rangle =\left\langle
u,v\right\rangle $

$H\otimes F\simeq\ell^{2}\left(  I\times J\right)  $

The tensor product of finitely many Hilbert spaces is defined similarly and is associative.

$\otimes^{m}H$ is a Hilbert space with Hilbert basis $\left(  e_{i_{1}}%
\otimes...\otimes e_{i_{m}}\right)  _{i_{1},...i_{m}\in I^{m}}$ and scalar
product :

$\left\langle \sum T^{i_{1}..i_{m}}e_{i_{1}}\otimes...\otimes e_{i_{m}},\sum
T^{\prime j_{1}..j_{m}}e_{j_{1}}\otimes...\otimes e_{j_{m}}\right\rangle
=\sum_{i_{1},...i_{m}\in I^{m}}T^{i_{1}..i_{m}}T^{\prime i_{1}..i_{m}}$

\begin{definition}
The \textbf{Fock space} $%
\mathcal{F}%
\left(  H\right)  $\ \ of a Hilbert space is the tensorial algebra
$\oplus_{n=0}^{\infty}\otimes^{n}H$
\end{definition}

So it includes the scalars. We denote : $%
\mathcal{F}%
_{m}\left(  H\right)  =\left\{  \psi^{m},\psi^{m}\in\otimes^{m}H\right\}  $

\subsubsection{Symmetric and antisymmetric tensorial powers}

The symmetrizer is the map :

$s_{m}:%
{\displaystyle\prod\limits_{i=1}^{m}}
H\rightarrow\otimes^{m}H::s_{m}\left(  u_{1},...,u_{m}\right)  =\sum
_{\sigma\in\mathfrak{S}\left(  m\right)  }u_{\sigma\left(  1\right)  }%
\otimes...\otimes u_{\sigma\left(  m\right)  }$

which is the symmetric tensorial product of vectors : $u_{1}\odot...\odot
u_{m}=\sum_{\sigma\in\mathfrak{S}\left(  m\right)  }u_{\sigma\left(  1\right)
}\otimes...\otimes u_{\sigma\left(  m\right)  }$

The antisymmetrizer is the map :

$a_{m}:%
{\displaystyle\prod\limits_{i=1}^{m}}
H\rightarrow\otimes^{m}H::a_{m}\left(  u_{1},...,u_{m}\right)  =\sum
_{\sigma\in\mathfrak{S}\left(  m\right)  }\epsilon\left(  \sigma\right)
u_{\sigma\left(  1\right)  }\otimes...\otimes u_{\sigma\left(  m\right)  }$

which is the antisymmetric tensorial product of vectors : $u_{1}%
\wedge...\wedge u_{m}=\sum_{\sigma\in\mathfrak{S}\left(  m\right)  }%
\epsilon\left(  \sigma\right)  u_{\sigma\left(  1\right)  }\otimes...\otimes
u_{\sigma\left(  m\right)  }$

The maps $s_{m},a_{m}$ are linear, thus, with the injection : \i:$%
{\displaystyle\prod\limits_{i=1}^{m}}
H\rightarrow\otimes^{m}H$\ there are unique linear maps such that :

$S_{m}:\otimes^{m}H\rightarrow\otimes^{m}H::s_{m}=S_{m}\circ\imath$

$A_{m}:\otimes^{m}H\rightarrow\otimes^{m}H::a_{m}=A_{m}\circ\imath$

The set of symmetric tensors $\odot^{m}H\subset\otimes^{m}H$ is the subset of
$\otimes^{m}H$ such that : $S_{m}\left(  X\right)  =m!X$

The set of antisymmetric tensors $\wedge^{m}H\subset\otimes^{m}H$ is the
subset of $\otimes^{m}H$ such that : $A_{m}\left(  X\right)  =m!X$

They are closed vector subspaces of $\otimes^{m}H.$ Thus they are Hilbert spaces

It reads :

$\odot^{m}H=\left\{  \sum_{\left(  i_{1}...i_{m}\right)  \in I^{m}}%
T^{i_{1}..i_{m}}e_{i_{1}}\otimes...\otimes e_{i_{m}},T^{i_{1}..i_{m}%
}=T^{i_{\sigma\left(  1\right)  }..i_{\sigma\left(  m\right)  }},\sigma
\in\mathfrak{S}\left(  m\right)  \right\}  $

$\wedge^{m}H=\left\{  \sum_{\left(  i_{1}...i_{m}\right)  \in I^{m}}%
T^{i_{1}..i_{m}}e_{i_{1}}\otimes...\otimes e_{i_{m}},T^{i_{1}..i_{m}}%
=\epsilon\left(  \sigma\right)  T^{i_{\sigma\left(  1\right)  }..i_{\sigma
\left(  m\right)  }},\sigma\in\mathfrak{S}\left(  m\right)  \right\}  $

which is equivalent to :

$\odot^{m}H=\left\{  \sum_{\left[  i_{1}...i_{m}\right]  \in I^{m}}%
T^{i_{1}..i_{m}}e_{i_{1}}\odot...\odot e_{i_{m}},i_{1}\leq...\leq
i_{m}\right\}  $

$\wedge^{m}H=\left\{  \sum_{\left\{  i_{1}...i_{m}\right\}  \in I^{m}}%
T^{i_{1}..i_{m}}e_{i_{1}}\wedge...\wedge e_{i_{m}},i_{1}<...<i_{m}\right\}  $

The scalar product of the vectors of the basis is :

$\left\langle e_{i_{1}}\odot...\odot e_{i_{m}},e_{j_{1}}\odot...\odot
e_{j_{m}}\right\rangle $

$=\left\langle \sum_{\sigma\in\mathfrak{S}\left(  m\right)  }e_{i_{\sigma
\left(  1\right)  }}\otimes...\otimes e_{i_{\sigma\left(  m\right)  }}%
,\sum_{\tau\in\mathfrak{S}\left(  m\right)  }e_{j_{\tau\left(  1\right)  }%
}\otimes...\otimes e_{j_{\tau\left(  m\right)  }}\right\rangle $

$=\sum_{\sigma\in\mathfrak{S}\left(  m\right)  }\sum_{\tau\in\mathfrak{S}%
\left(  m\right)  }\delta\left(  \sigma\left(  i_{1}\right)  ,\tau\left(
j_{1}\right)  \right)  ...\delta\left(  \sigma\left(  i_{m}\right)
,\tau\left(  j_{m}\right)  \right)  $

$=m!\sum_{\theta\in\mathfrak{S}\left(  m\right)  }\delta\left(  i_{1}%
,\theta\left(  j_{1}\right)  \right)  ...\delta\left(  i_{m},\theta\left(
j_{m}\right)  \right)  $

If $\left[  i_{1}...i_{m}\right]  \neq\left[  j_{1}...j_{m}\right]
:\left\langle e_{i_{1}}\odot...\odot e_{i_{m}},e_{j_{1}}\odot...\odot
e_{j_{m}}\right\rangle =0$

If $\left[  i_{1}...i_{m}\right]  =\left[  j_{1}...j_{m}\right]  :\left\langle
e_{i_{1}}\odot...\odot e_{i_{m}},e_{j_{1}}\odot...\odot e_{j_{m}}\right\rangle
=m!$

$\left\langle e_{i_{1}}\wedge...\wedge e_{i_{m}},e_{j_{1}}\wedge...\wedge
e_{j_{m}}\right\rangle $

$=\left\langle \sum_{\sigma\in\mathfrak{S}\left(  m\right)  }\epsilon\left(
\sigma\right)  e_{i_{\sigma\left(  1\right)  }}\otimes...\otimes
e_{i_{\sigma\left(  m\right)  }},\sum_{\tau\in\mathfrak{S}\left(  m\right)
}\epsilon\left(  \tau\right)  e_{j_{\tau\left(  1\right)  }}\otimes...\otimes
e_{j_{\tau\left(  m\right)  }}\right\rangle $

$=\sum_{\sigma\in\mathfrak{S}\left(  m\right)  }\sum_{\tau\in\mathfrak{S}%
\left(  m\right)  }\epsilon\left(  \sigma\right)  \epsilon\left(  \tau\right)
\delta\left(  \sigma\left(  i_{1}\right)  ,\tau\left(  j_{1}\right)  \right)
...\delta\left(  \sigma\left(  i_{m}\right)  ,\tau\left(  j_{m}\right)
\right)  $

If $\left\{  i_{1}...i_{m}\right\}  \neq\left\{  j_{1}...j_{m}\right\}
:\left\langle e_{i_{1}}\wedge...\wedge e_{i_{m}},e_{j_{1}}\wedge...\wedge
e_{j_{m}}\right\rangle =0$

If $\left\{  i_{1}...i_{m}\right\}  =\left\{  j_{1}...j_{m}\right\}
:\left\langle e_{i_{1}}\wedge...\wedge e_{i_{m}},e_{j_{1}}\wedge...\wedge
e_{j_{m}}\right\rangle =m!$

Thus $\left(  \frac{1}{\sqrt{m!}}e_{i_{1}}\odot...\odot e_{i_{m}}\right)
_{\left[  i_{1}...i_{m}\right]  \in I^{m}},\left(  \frac{1}{\sqrt{m!}}%
e_{i_{1}}\wedge...\wedge e_{i_{m}}\right)  _{\left\{  i_{1}...i_{m}\right\}
\in I^{m}}$

are hilbertian basis of $\odot^{m}H,\wedge^{m}H$

The symmetric and antisymmetric products extend to symmetric and antisymmetric
tensors :

define for vectors :

$\left(  u_{1}\odot...\odot u_{m}\right)  \odot\left(  u_{m+1}\odot...\odot
u_{m+n}\right)  =u_{1}\odot...\odot u_{m+n}$

$=\sum_{\sigma\in\mathfrak{S}\left(  m+n\right)  }u_{\sigma\left(  1\right)
}\otimes...\otimes u_{\sigma\left(  m+n\right)  }$

$\left(  u_{1}\wedge...\wedge u_{m}\right)  \wedge\left(  u_{m+1}%
\wedge...\wedge u_{m+n}\right)  =u_{1}\wedge...\wedge u_{m+n}$

$=\sum_{\sigma\in\mathfrak{S}\left(  m+n\right)  }\epsilon\left(
\sigma\right)  u_{\sigma\left(  1\right)  }\otimes...\otimes u_{\sigma\left(
m+n\right)  }$

then it extends to tensors by their expression in a basis.

$\odot:\odot^{m}H\times\odot^{n}H\rightarrow\odot^{m+n}H$

$\wedge:\wedge^{m}H\times\wedge^{n}H\rightarrow\wedge^{m+n}H$

\begin{definition}
The \textbf{Bose-Fock space}\ of a Hilbert space is $%
\mathcal{F}%
_{+}\left(  H\right)  =\oplus_{n=0}^{\infty}\odot^{n}H\subset%
\mathcal{F}%
\left(  H\right)  $

The \textbf{Fermi-Fock space}\ of a Hilbert space is $%
\mathcal{F}%
_{-}\left(  H\right)  =\oplus_{n=0}^{\infty}\wedge^{n}H\subset%
\mathcal{F}%
\left(  H\right)  $
\end{definition}

They are closed vector subspaces of $%
\mathcal{F}%
\left(  H\right)  .$

If $H=\oplus_{i=1}^{n}H_{i}$ then
$\mathcal{F}$%
$_{+}\left(  H\right)  =\oplus_{i=1}^{n}$%
$\mathcal{F}$%
$_{+}\left(  H_{i}\right)  $

\subsubsection{Operators on Fock spaces}

\paragraph{Definitions\newline}

\begin{theorem}
(Neeb p.102) If E,F are Hilbert spaces, $X\in%
\mathcal{L}%
\left(  E;E\right)  ,Y\in%
\mathcal{L}%
\left(  F;F\right)  $ then there is a unique linear operator denoted $X\otimes
Y$ such that :

$\forall u\in H,v\in F:X\otimes Y\left(  u\otimes v\right)  =X\left(
u\right)  \otimes Y\left(  v\right)  .$

Moreover $X\otimes Y\in%
\mathcal{L}%
\left(  E\otimes F;E\otimes F\right)  $ and $\left\Vert X\otimes Y\right\Vert
\leq\left\Vert X\right\Vert \left\Vert Y\right\Vert ,\left(  X\otimes
Y\right)  ^{\ast}=X^{\ast}\otimes Y^{\ast}$
\end{theorem}

\begin{definition}
The \textbf{number operator} is $N\in L\left(
\mathcal{F}%
\left(  H\right)  ;%
\mathcal{F}%
\left(  H\right)  \right)  .$ Its domain is $D\left(  N\right)  =\left\{
\psi^{m},\psi^{m}\in\otimes^{m}H:\sum_{m\geq0}m^{2}\left\Vert \psi
^{m}\right\Vert ^{2}<\infty\right\}  $ and

$N\psi=\left\{  m\psi^{m},\psi^{m}\in\otimes^{m}H\right\}  .$ It is self adjoint.
\end{definition}

The \textbf{annihilation operator} $a$ is defined by extending by linearity :

$a_{0}\in%
\mathcal{L}%
\left(
\mathbb{C}
;%
\mathbb{C}
\right)  :a_{0}u=0$

$a_{n}:H\rightarrow%
\mathcal{L}%
\left(  \otimes^{n}H;\otimes^{n-1}H\right)  ::a_{n}\left(  v\right)  \left(
u_{1}\otimes...\otimes u_{n}\right)  =\frac{1}{\sqrt{n}}\left\langle
v,u_{1}\right\rangle \left(  u_{2}\otimes...\otimes u_{n}\right)  $

$\left\Vert a_{n}\left(  v\right)  \left(  \psi^{n}\right)  \right\Vert
\leq\sqrt{n}\left\Vert v\right\Vert \left\Vert \psi^{n}\right\Vert $

Its extension a(v) is a densely defined operator on the domain $D(N^{1/2})$
of
$\mathcal{F}$%
(H) and :

$\forall\psi\in D\left(  N^{1/2}\right)  \subset%
\mathcal{F}%
\left(  H\right)  :\left\Vert a\left(  v\right)  \left(  \psi\right)
\right\Vert \leq\left\Vert v\right\Vert \left\Vert \left(  N+1\right)
^{1/2}\psi\right\Vert $

The \textbf{creation operator} $a^{\ast}$ is defined by extending by linearity :

$a_{0}\in%
\mathcal{L}%
\left(
\mathbb{C}
;%
\mathbb{C}
\right)  :a_{0}u=u$

$a_{n}:H\rightarrow%
\mathcal{L}%
\left(  \otimes^{n}H;\otimes^{n+1}H\right)  ::a_{n}^{\ast}\left(  v\right)
\left(  u_{1}\otimes...\otimes u_{n}\right)  =\sqrt{n+1}v\otimes u_{1}%
\otimes...\otimes u_{n}$

$\left\Vert a_{n}^{\ast}\left(  v\right)  \left(  \psi^{n}\right)  \right\Vert
\leq\sqrt{n+1}\left\Vert v\right\Vert \left\Vert \psi^{n}\right\Vert $

Its extension a*(v) is a densely defined operator on the domain $D(N^{1/2})$
of
$\mathcal{F}$%
(H) and :

$\forall\psi\in D\left(  N^{1/2}\right)  \subset%
\mathcal{F}%
\left(  H\right)  :\left\Vert a^{\ast}\left(  v\right)  \left(  \psi\right)
\right\Vert \leq\left\Vert v\right\Vert \left\Vert \left(  N+1\right)
^{1/2}\psi\right\Vert $

Moreover a*(v) is the adjoint of a(v) :

$\left\langle a^{\ast}\left(  v\right)  \psi,\psi^{\prime}\right\rangle
=\left\langle \psi,a\left(  v\right)  \psi^{\prime}\right\rangle $

\paragraph{Operators on the Fermi and Bose spaces\newline}

(Bratelli 2 p.8)

The orthogonal projections are :

$P_{+m}:\otimes^{m}H\rightarrow\odot^{m}H::P_{+m}=\frac{1}{m!}S_{m}$

$P_{-m}:\otimes^{m}H\rightarrow\wedge^{m}H::P_{-m}=\frac{1}{m!}A_{m}$

thus : $\psi\in\odot^{m}H\Rightarrow P_{+m}\left(  \psi\right)  =\psi,$
$\psi\in\wedge^{m}H\Rightarrow P_{-m}\left(  \psi\right)  =\psi$

They extend to the orthogonal projections $P_{+},P_{-}$of $%
\mathcal{F}%
\left(  H\right)  $ on $%
\mathcal{F}%
_{+}\left(  H\right)  ,%
\mathcal{F}%
_{-}\left(  H\right)  :$

$P_{\epsilon}\in%
\mathcal{L}%
\left(
\mathcal{F}%
\left(  H\right)  ;%
\mathcal{F}%
_{\epsilon}\left(  H\right)  \right)  :P_{\epsilon}^{2}=P_{\epsilon
}=P_{\epsilon}^{\ast};P_{+}P_{-}=P_{-}P_{+}=0$

They are continuous operators with norm 1.

The operator $\exp itN$ where N is the number operator, leaves invariant the
spaces $%
\mathcal{F}%
_{\pm}\left(  H\right)  .$

\bigskip

If A is a self adjoint operator on some domain D(A) of H, then one defines the
operators $A_{n}:$ $%
\mathcal{F}%
_{n\epsilon}\left(  H\right)  $ by linear extension of :

$\forall u_{p}\in D\left(  A\right)  :A_{n}\left(  P_{\epsilon}\left(
u_{1}\otimes...\otimes u_{n}\right)  \right)  =P_{\epsilon n}\left(
\sum_{p=1}^{n}u_{1}\otimes...\otimes Au_{p}\otimes...u_{n}\right)  $

which are symmetric and closable. The direct sum $\oplus_{n}A_{n}$ is
essentially self-adjoint, and its self-adjoint closure $d\Gamma\left(
A\right)  =\overline{\oplus_{n}A_{n}}\in%
\mathcal{L}%
\left(
\mathcal{F}%
_{\epsilon}\left(  H\right)  ;%
\mathcal{F}%
_{\epsilon}\left(  H\right)  \right)  $ is called the \textbf{second
quantification} of A.

For A=Id : $d\Gamma\left(  Id\right)  =N$ the number operator.

\bigskip

If U is a unitary operator on H, then one defines the operators $U_{n}:$ $%
\mathcal{F}%
_{n\epsilon}\left(  H\right)  $ by linear extension of :

$\forall u_{p}\in D\left(  A\right)  :$ $\ \ U_{n}\left(  P_{\epsilon}\left(
u_{1}\otimes...\otimes u_{n}\right)  \right)  =P_{\epsilon n}\left(
\sum_{p=1}^{m}Uu_{1}\otimes...\otimes Uu_{p}\otimes...\otimes Uu_{n}\right)  $

which are unitary. The direct sum $\Gamma\left(  U\right)  =\oplus_{n}U_{n}\in%
\mathcal{L}%
\left(
\mathcal{F}%
_{\epsilon}\left(  H\right)  ;%
\mathcal{F}%
_{\epsilon}\left(  H\right)  \right)  $ is unitary and called the second
quantification of U.

If U(t)=exp(itA) is a strongly continuous one parameter group of unitary
operators on H, then $\Gamma\left(  U\left(  t\right)  \right)  =\exp
itd\Gamma\left(  A\right)  $

\bigskip

The creation and annihilation operators on the Fock Fermi and Bose spaces are :

$a_{\epsilon}\left(  v\right)  =P_{\epsilon}a\left(  v\right)  P_{\epsilon}\in%
\mathcal{L}%
\left(
\mathcal{F}%
_{\epsilon}\left(  H\right)  ;%
\mathcal{F}%
_{\epsilon}\left(  H\right)  \right)  $

$a_{\epsilon}^{\ast}\left(  v\right)  =P_{\epsilon}a^{\ast}\left(  v\right)
P_{\epsilon}\in%
\mathcal{L}%
\left(
\mathcal{F}%
_{\epsilon}\left(  H\right)  ;%
\mathcal{F}%
_{\epsilon}\left(  H\right)  \right)  $

with the relations :

$\left\langle a_{\epsilon}^{\ast}\left(  v\right)  \psi,\psi^{\prime
}\right\rangle =\left\langle \psi,a_{\epsilon}\left(  v\right)  \psi^{\prime
}\right\rangle $

$\left\Vert a_{\epsilon}\left(  v\right)  \left(  \psi\right)  \right\Vert
\leq\left\Vert v\right\Vert \left\Vert \left(  N+1\right)  ^{1/2}%
\psi\right\Vert ,\left\Vert a_{\epsilon}^{\ast}\left(  v\right)  \left(
\psi\right)  \right\Vert \leq\left\Vert v\right\Vert \left\Vert \left(
N+1\right)  ^{1/2}\psi\right\Vert $

$a_{\epsilon}\left(  v\right)  =a\left(  v\right)  P_{\epsilon}$

$a_{\epsilon}^{\ast}\left(  v\right)  =P_{\epsilon}a^{\ast}\left(  v\right)  $

The map $a_{\epsilon}:H\rightarrow%
\mathcal{L}%
\left(
\mathcal{F}%
_{\epsilon}\left(  H\right)  ;%
\mathcal{F}%
_{\epsilon}\left(  H\right)  \right)  $ is antilinear

The map $a_{\epsilon}^{\ast}:H\rightarrow%
\mathcal{L}%
\left(
\mathcal{F}%
_{\epsilon}\left(  H\right)  ;%
\mathcal{F}%
_{\epsilon}\left(  H\right)  \right)  $ is linear

$P_{\epsilon}\left(  u_{1}\otimes...\otimes u_{n}\right)  =\frac{1}{\sqrt{n!}%
}a_{\epsilon}^{\ast}\left(  u_{1}\right)  \circ...\circ a_{\epsilon}^{\ast
}\left(  u_{n}\right)  \Omega$ where $\Omega=\left(  1,0,....0...\right)  $
represents the "vacuum".

$P_{-}\left(  u_{1}\otimes...\otimes u_{n}\right)  =0$ whenever $u_{i}=u_{j}$
which reads also : $a_{-}^{\ast}\left(  v\right)  \circ a_{-}^{\ast}\left(
v\right)  =0$

\textbf{Canonical commutation relations} (CCR) :

$\left[  a_{+}\left(  u\right)  ,a_{+}\left(  v\right)  \right]  =\left[
a_{+}^{\ast}\left(  u\right)  ,a_{+}^{\ast}\left(  v\right)  \right]  =0$

$\left[  a_{+}\left(  u\right)  ,a_{+}^{\ast}\left(  v\right)  \right]
=\left\langle u,v\right\rangle 1$

\textbf{Canonical anticommutation relations} (CAR) :

$\left\{  a_{-}\left(  u\right)  ,a_{-}\left(  v\right)  \right\}  =\left\{
a_{-}^{\ast}\left(  u\right)  ,a_{-}^{\ast}\left(  v\right)  \right\}  =0$

$\left\{  a_{-}\left(  u\right)  ,a_{-}^{\ast}\left(  v\right)  \right\}
=\left\langle u,v\right\rangle 1$

where $\left\{  X,Y\right\}  =X\circ Y+Y\circ X$

These relations entail different features for the operators

\begin{theorem}
(Bratelli 2 p.11) If H is a complex, separable, Hilbert space then:

i) $\forall v\in H$ $a_{-}\left(  v\right)  ,a_{-}^{\ast}\left(  v\right)  $
have a bounded extension on $%
\mathcal{F}%
_{-}\left(  H\right)  $ and $\left\Vert a_{-}\left(  v\right)  \right\Vert
=\left\Vert a_{-}^{\ast}\left(  v\right)  \right\Vert =\left\Vert v\right\Vert
$

ii) if $\Omega=\left(  1,0,....0...\right)  $ and $\left(  e_{i}\right)
_{i\in I}$ is a Hilbert basis of H, then $a_{-}^{\ast}\left(  e_{i_{1}%
}\right)  \circ...\circ a_{-}^{\ast}\left(  e_{i_{n}}\right)  \Omega$ where
$\left\{  i_{1},...,i_{n}\right\}  $ runs over the ordered finite subsets of
I, is an Hilbert basis of $%
\mathcal{F}%
_{-}\left(  H\right)  $

iii) the set of operators $\left\{  a_{-}\left(  v\right)  ,a_{-}^{\ast
}\left(  v\right)  ,v\in H\right\}  $ is irreducible on $%
\mathcal{F}%
_{-}\left(  H\right)  $ (its commutant is $%
\mathbb{C}
\times Id)$
\end{theorem}

There are no similar results for $a_{+}\left(  v\right)  ,a_{+}^{\ast}\left(
v\right)  $ which are not bounded.

\paragraph{Exponential\newline}

On $%
\mathcal{F}%
_{+}\left(  H\right)  $ we have the followings (Applebaum p.5):

\begin{definition}
The exponential is defined as:

$\exp:H\rightarrow%
\mathcal{F}%
_{+}\left(  H\right)  ::\exp\psi=\sum_{n=0}^{\infty}\frac{1}{\sqrt{n!}}%
\otimes^{n}\psi=\sum_{n=0}^{\infty}\frac{1}{\left(  n!\right)  ^{3/2}}%
s_{n}\left(  \psi,...,\psi\right)  $

with $\otimes^{0}\psi=\left\{  1,0,...\right\}  $
\end{definition}

The mapping is analytic

If $\left(  e_{i}\right)  _{i\in I}$ is a basis of H, then $\left(  \exp
e_{i}\right)  _{i\in I}$ is a Hilbertian basis of $%
\mathcal{F}%
_{+}\left(  H\right)  $ with the first vector $\Omega=\left(  1,0,...\right)
$

$\left\langle \exp\psi,\exp\psi^{\prime}\right\rangle =\exp\left\langle
\psi,\psi^{\prime}\right\rangle $

$\overline{Span\left(  \exp H\right)  }=$%
$\mathcal{F}$%
$_{+}\left(  H\right)  $

If D is dense in H, then $\left\{  \exp\psi,\psi\in D\right\}  $ is dense in
$\mathcal{F}$%
$_{+}\left(  H\right)  $

\bigskip

The creation and annihilation operators read:

$a_{+}\left(  v\right)  \exp\left(  u\right)  =\left\langle v,u\right\rangle
\exp\left(  u\right)  $

$a_{+}^{\ast}\left(  v\right)  \exp\left(  u\right)  =\frac{d}{dt}\exp\left(
u+tv\right)  |_{t=0}$

the exponential annihilation operator : $U\left(  v\right)  \exp\left(
u\right)  =\exp\left\langle v,u\right\rangle \exp\left(  u\right)  $

the exponential creation operator : $U^{\ast}\left(  v\right)  \exp\left(
u\right)  =\exp\left(  u+v\right)  $

are all closable operators with $a^{\ast}\left(  v\right)  \subset a\left(
v\right)  ^{\ast},U^{\ast}\left(  v\right)  \subset U\left(  v\right)  ^{\ast
}$

If T is a contraction in H, then $\Gamma\left(  T\right)  $ is a contraction
in $%
\mathcal{F}%
_{+}\left(  H\right)  $ and $\Gamma\left(  T\right)  \exp u=\exp Tu$

If U is unitary in H, then $\Gamma\left(  T\right)  $ is unitary in $%
\mathcal{F}%
_{+}\left(  H\right)  $ and $\Gamma\left(  U\right)  \exp u=\exp Uu$

\bigskip

The universal creation and annihilation operators $\nabla^{+}$,$\nabla^{-}$
are defined as follows :

i) For $t\in%
\mathbb{R}
$\ , V(t),V*(t) are defined by extension of :

$V\left(  t\right)  :%
\mathcal{F}%
_{+}\left(  H\right)  \rightarrow%
\mathcal{F}%
_{+}\left(  H\right)  \otimes%
\mathcal{F}%
_{+}\left(  H\right)  :V(t)\exp u=\exp u\otimes\exp tu$

$V^{\ast}\left(  t\right)  :%
\mathcal{F}%
_{+}\left(  H\right)  \otimes%
\mathcal{F}%
_{+}\left(  H\right)  \rightarrow%
\mathcal{F}%
_{+}\left(  H\right)  :V(t)\left(  \exp u\otimes\exp v\right)  =\exp\left(
u+tv\right)  $

they are closable and $V^{\ast}\left(  t\right)  \subset V\left(  t\right)
^{\ast}$

ii) $\nabla^{+}$,$\nabla^{-}$ are defined by extension of :

$\nabla^{-}:%
\mathcal{F}%
_{+}\left(  H\right)  \rightarrow%
\mathcal{F}%
_{+}\left(  H\right)  \otimes H:\nabla^{-}\exp u=\frac{d}{dt}V\left(
t\right)  \exp u|_{t=0}=\left(  \exp u\right)  \otimes u$

$\nabla^{+}:%
\mathcal{F}%
_{+}\left(  H\right)  \otimes H\rightarrow%
\mathcal{F}%
_{+}\left(  H\right)  :\nabla^{+}\left(  \left(  \exp u\right)  \otimes
v\right)  =\frac{d}{dt}V^{\ast}\left(  t\right)  \exp u\otimes\exp
v|_{t=0}=a^{\ast}\left(  v\right)  \exp u$

they are closable and $\nabla^{+}\left(  t\right)  \subset\left(  \nabla
^{-}\right)  ^{\ast}$

iii) On the domain of N : $N=\nabla^{+}\nabla^{-}$

\paragraph{Bargmann's Space\newline}

(see Stochel)

Let $\ell^{2}\left(  I,K\right)  $ be the set of families $\left(
x_{i}\right)  _{i\in I}\in K^{I}$ such that $\sum_{i\in I}\left\vert
x_{i}\right\vert ^{2}<\infty$ with $K=%
\mathbb{R}
,%
\mathbb{C}
.$

Let $%
\mathbb{N}
_{0}\in%
\mathbb{N}
^{%
\mathbb{N}
}$ the set of sequences in $%
\mathbb{N}
$ such that only a finite number of elements are non null\ : $a\in%
\mathbb{N}
_{0}:a=\left\{  a_{1},...a_{n},0...\right\}  $

For $a\in%
\mathbb{N}
_{0}$\ let $e_{a}:\ell^{2}\left(
\mathbb{N}
,%
\mathbb{C}
\right)  \rightarrow%
\mathbb{C}
::e_{a}\left(  z\right)  =%
{\displaystyle\prod\limits_{k\in\mathbb{N} }}
\frac{z_{k}^{a_{k}}}{\sqrt{a_{k}!}}$ and for any $\left(  f_{a}\right)  _{a\in%
\mathbb{N}
_{0}}\in\ell^{2}\left(
\mathbb{N}
_{0},%
\mathbb{C}
\right)  :f:\ell^{2}\left(
\mathbb{N}
,%
\mathbb{C}
\right)  \rightarrow%
\mathbb{C}
::\sum_{a\in%
\mathbb{N}
_{0}}f_{a}e_{a}\left(  z\right)  $

The Bargmann's space is the set \ : $B_{\infty}=\left\{  f,\left(
f_{a}\right)  _{a\in%
\mathbb{N}
_{0}}\right\}  $ which is a Hilbert space isometric to $\ell^{2}\left(
\mathbb{N}
_{0},%
\mathbb{C}
\right)  $ with the scalar product :

$\left\langle f,g\right\rangle =\left\langle \left(  f_{a}\right)  ,\left(
g_{a}\right)  \right\rangle _{\ell^{2}}.$Moreover $e_{a}\left(  z\right)  $ is
a Hilbert basis, and all the functions in $B_{\infty}$ are entire. Its
reproducing kernel is : $N\left(  z,z^{\prime}\right)  =\exp\left\langle
z,z^{\prime}\right\rangle _{\ell^{2}\left(
\mathbb{N}
,%
\mathbb{C}
\right)  }$ and $f\left(  z\right)  =\left\langle f,N\left(  z,.\right)
\right\rangle $

There is a unitary isomorphism between $B_{\infty}$ and
$\mathcal{F}$%
$_{+}\left(  H\right)  $ for any complex, separable, infinite dimensional
Hilbert space H.

However there are entire functions which do not belong to $B_{\infty},$ the
product is not necessarily closed.

\paragraph{Free Fock space\newline}

(see Attal)

A probability algebra is an unital *-algebra A on $%
\mathbb{C}
$\ endowed with a faithful positive linear form $\varphi$ such that
$\varphi\left(  I\right)  =1$\ 

Elements of A are random variables.\ A commutative algebra gives back a
classical probability space.

A family $\left(  A_{i}\right)  _{i\in I}$ of unital subalgebras of A is free
if $\varphi\left(  a_{1}\cdot a_{2}...\cdot a_{n}\right)  $ for any finite
collection of elements $a_{p}\in A_{J\left(  p\right)  }$ with $\varphi\left(
a_{p}\right)  =0$ such that $J\left(  1\right)  \neq J\left(  2\right)
...\neq J\left(  n\right)  $

If H is a separable complex Hilbert space, $\Omega$ any vector of the full
Fock space
$\mathcal{F}$%
$\left(  H\right)  ,$ $X\in%
\mathcal{L}%
\left(
\mathcal{F}%
\left(  H\right)  ;%
\mathcal{F}%
\left(  H\right)  \right)  $ the functional $\tau$ is defined by : $\tau:%
\mathcal{L}%
\left(
\mathcal{F}%
\left(  H\right)  ;%
\mathcal{F}%
\left(  H\right)  \right)  \rightarrow%
\mathbb{C}
::\tau\left(  X\right)  =\left\langle \Omega,X\Omega\right\rangle $ which
makes of $%
\mathcal{L}%
\left(
\mathcal{F}%
\left(  H\right)  ;%
\mathcal{F}%
\left(  H\right)  \right)  $ a probability algebra.

If $X\in%
\mathcal{L}%
\left(  H;H\right)  $ then $\ \Lambda\left(  X\right)  \in%
\mathcal{L}%
\left(
\mathcal{F}%
\left(  H\right)  ;%
\mathcal{F}%
\left(  H\right)  \right)  $ is defined by : $\Lambda\left(  X\right)
\Omega=0,\Lambda\left(  X\right)  e_{1}\otimes...\otimes e_{n}=X\left(
e_{1}\right)  \otimes e_{2}...\otimes e_{n}$ and $\left\Vert \Lambda\left(
X\right)  \right\Vert =\left\Vert X\right\Vert $

\bigskip

\subsection{Representation of an algebra}

\label{Representation of an algebra}

The C*-algebras have been modelled on the set of continuous linear maps on a
Hilbert space, so it is natural to look for representations of C*algebras on
Hilbert spaces.\ In many ways this topic looks like the representation of Lie
groups.\ One of the most useful outcome of this endeavour is the spectral
theory which enables to resume the action of an operator as an integral with
measures which are projections on eigen spaces. On this subject we follow
mainly Thill. See also Bratelli.

\subsubsection{Definitions}

\begin{definition}
A linear representation of an algebra $\left(  A,\cdot\right)  $ over the
field K is a pair $(H,\rho)$ of a vector space H over the field K and the
algebra morphism $\rho:A\rightarrow L(H;H):$
\end{definition}

$\forall X,Y\in A,k,k^{\prime}\in K:$

$\rho\left(  kX+k^{\prime}Y\right)  =k\rho\left(  X\right)  +k^{\prime}%
\rho\left(  Y\right)  $

$\rho\left(  X\cdot Y\right)  =\rho\left(  X\right)  \circ\rho\left(
Y\right)  $

$\rho\left(  I\right)  =Id\Rightarrow$ if $X\in G\left(  A\right)
:\rho\left(  X\right)  ^{-1}=\rho\left(  X^{-1}\right)  $

\begin{definition}
A linear representation of a *-algebra $\left(  A,\cdot\right)  $ over the
field K is a linear representation $(H,\rho)$ of A, such that H is endowed
with an involution and : $\forall X\in A:\rho\left(  X^{\ast}\right)
=\rho\left(  X\right)  ^{\ast}$
\end{definition}

\textbf{In the following we will consider representation }$\left(
H,\rho\right)  $\textbf{\ of a Banach *-algebra A on a Hilbert space (H,g)}.

\begin{definition}
A Hilbertian representation of a Banach *-algebra A is a linear representation
$(H,\rho)$ of A, where H is a Hilbert space, and $\rho$ is a continuous
*-morphism $\rho:A\rightarrow%
\mathcal{L}%
\left(  H;H\right)  $.
\end{definition}

So: $\forall u\in H,X\in A:g\left(  \rho\left(  X\right)  u,v\right)
=g\left(  u,\rho\left(  X^{\ast}\right)  v\right)  $ with the adjoint X* of X.

The adjoint map $\rho\left(  X\right)  ^{\ast}$ is well defined if
$\rho\left(  X\right)  $ is continuous on H or at least on a dense subset of H

$\rho\in%
\mathcal{L}%
\left(  A;%
\mathcal{L}%
\left(  H;H\right)  \right)  $ and we have the norm :

$\left\Vert \rho\right\Vert =\sup_{\left\Vert X\right\Vert _{A}=1}\left\Vert
\rho\left(  X\right)  \right\Vert _{%
\mathcal{L}%
\left(  H;H\right)  }<\infty$

\bigskip

We have the usual definitions of representation theory for any linear
representation $\left(  H,\rho\right)  $ of A:

i) the representation is faithful if $\rho$\ \ is injective

ii) a vector subspace F of H is invariant if $\forall u\in F,\forall X\in
A:\rho\left(  X\right)  u\in F$

iii) $\left(  H,\rho\right)  $ is irreducible if there is no other invariant
vector space than 0,H.

iv) If $\left(  H_{k},\rho_{k}\right)  _{k\in I}$ is a family of Hilbertian
representations of A and $\forall X\in A:\left\Vert \rho_{k}\left(  X\right)
\right\Vert <\infty$ the Hilbert sum of representations $\left(  \oplus
_{i}H_{i},\oplus_{i}\rho_{i}\right)  $ is defined with : $\left(  \oplus
_{i}\rho_{i}\right)  \left(  X\right)  \left(  \oplus_{i}u_{i}\right)
=\oplus_{i}\left(  \rho_{i}\left(  X\right)  u_{i}\right)  $ and norm
$\left\Vert \oplus_{i}\rho_{i}\right\Vert =\sup_{i\in I}\left\Vert \rho
_{i}\right\Vert $

v) An operator $f\in%
\mathcal{L}%
\left(  H_{1};H_{2}\right)  $ is an interwiner between two representations
$\left(  H_{k},\rho_{k}\right)  _{k=1,2}$ if :

$\forall X\in A:f\circ\rho_{1}\left(  X\right)  =\rho_{2}\left(  X\right)
\circ f$

vi) Two representations are equivalent if there is an interwiner which an isomorphism

vii) A representation $\left(  H,\rho\right)  $ is contractive if $\left\Vert
\rho\right\Vert \leq1$

viii) A representation $\left(  H,\rho\right)  $ of the algebra A is isometric
if $\forall X\in A:\left\Vert \rho\left(  X\right)  \right\Vert _{%
\mathcal{L}%
\left(  H;H\right)  }=\left\Vert X\right\Vert _{A}$

\bigskip

Special definitions :

\begin{definition}
The \textbf{commutant} $\rho^{\prime}$\ of the linear representation $\left(
H,\rho\right)  $\ of a algebra A is the set $\left\{  \pi\in%
\mathcal{L}%
\left(  H;H\right)  :\forall X\in A:\pi\circ\rho\left(  X\right)  =\rho\left(
X\right)  \circ\pi\right\}  $
\end{definition}

\begin{definition}
A vector $u\in H$ is \textbf{cyclic} for the linear representation $\left(
H,\rho\right)  $\ of an algebra A if the set $\left\{  \rho\left(  X\right)
u,X\in A\right\}  $ is dense in H.\ $\left(  H,\rho\right)  $ is said cyclic
if there is a cyclic vector $u_{c}$\ and is denoted $\left(  H,\rho
,u_{c}\right)  $
\end{definition}

\begin{definition}
Two linear representations $\left(  H_{1},\rho_{1}\right)  ,\left(  H_{2}%
,\rho_{2}\right)  $ of the algebra A are \textbf{spatially equivalent} if
there is a unitary interwiner

$U:U\circ\rho_{1}\left(  X\right)  =\rho_{2}\left(  X\right)  U$
\end{definition}

\subsubsection{General theorems}

\begin{theorem}
(Thill p.125) If the vector subspace $F\subset H$ is invariant in the linear
representation $\left(  H,\rho\right)  $\ of A, then the orthogonal complement
$F^{\bot}$\ \ is also invariant and $\left(  F,\rho\right)  $\ is a subrepresentation
\end{theorem}

\begin{theorem}
(Thill p.125) A closed vector subspace $F\subset H$ is invariant in the linear
representation $\left(  H,\rho\right)  $\ of A iff $\forall X\in A:\pi
_{F}\circ\rho\left(  X\right)  =\rho\left(  X\right)  \circ\pi_{F}$ where
$\pi_{F}:H\rightarrow F$ is the projection on F
\end{theorem}

\begin{theorem}
If $\left(  H,\rho\right)  $ is a linear representation of A, then for every
unitary map $U\in%
\mathcal{L}%
\left(  H;H\right)  ,$ $\left(  H,U\rho U^{\ast}\right)  $ is an equivalent representation.
\end{theorem}

\begin{theorem}
(Thill p.122) Every linear representation of a Banach *-algebra with isometric
involution on a pre Hilbert space is contractive
\end{theorem}

\begin{theorem}
(Thill p.122) Every linear representation of a C*algebra on a pre Hilbert
space is contractive
\end{theorem}

\begin{theorem}
(Thill p.122) Every faithful linear representation of a C *algebra on a
Hilbert space is isometric
\end{theorem}

\begin{theorem}
If $\left(  H,\rho\right)  $ is a linear representation of a *-algebra then
the commutant $\rho^{\prime}$ is a W*-algebra.
\end{theorem}

\begin{theorem}
(Thill p.123) For every linear representation $\left(  H,\rho\right)  $\ of a
C*algebra A the sets $A/\ker\rho,\rho\left(  A\right)  $ are C*-algebras and
the representation factors to : $A/\ker\rho\rightarrow\rho\left(  A\right)  $
\end{theorem}

\begin{theorem}
(Thill p.127) For every linear representation $\left(  H,\rho\right)  $\ of a
Banach *-algebra A, and any non null vector $u\in H$, the closure of the
linear span of $F=\left\{  \rho\left(  X\right)  u,X\in A\right\}  $ is
invariant and $\left(  F,\rho,u\right)  $ is cyclic
\end{theorem}

\begin{theorem}
(Thill p.129) If $\left(  H_{1},\rho_{1},u_{1}\right)  ,\left(  H_{2},\rho
_{2},u_{2}\right)  $ are two cyclic linear representations of a Banach
*-algebra A and if $\forall X\in A:g_{1}\left(  \rho_{1}\left(  X\right)
u_{1},u_{1}\right)  =g_{2}\left(  \rho_{2}\left(  X\right)  u_{2}%
,u_{2}\right)  $ then the representations are equivalent and there is a
unitary operator U: $U\circ\rho_{1}\left(  X\right)  \circ U^{\ast}=\rho
_{2}\left(  X\right)  $
\end{theorem}

\begin{theorem}
(Thill p.136) For every linear representation $\left(  H,\rho\right)  $\ of a
C*algebra A and vector u in H such that: $\left\Vert u\right\Vert =1$ ,the map :

$\varphi:A\rightarrow%
\mathbb{C}
::\varphi\left(  X\right)  =g\left(  \rho\left(  X\right)  u,u\right)  $ is a state
\end{theorem}

\subsubsection{Representation GNS}

A Lie group can be represented on its Lie algebra through the adjoint
representation.\ Similarly an algebra has a linear representation on
itself.\ Roughly $\rho\left(  X\right)  $ is the translation operator
$\rho\left(  X\right)  Y=XY.$ A Hilbert space structure on A is built through
a linear functional.

\begin{theorem}
(Thill p.139, 141) For any linear positive functional $\varphi,$ a Banach
*-algebra has a Hilbertian representation, called \textbf{GNS} (for Gel'fand,
Naimark, Segal) and denoted $\left(  H_{\varphi},\rho_{\varphi}\right)  $
which is continuous and contractive.
\end{theorem}

The construct is the following:

i) Any linear positive functional $\varphi$ on A define the sesquilinear form
: $\left\langle X,Y\right\rangle =\varphi\left(  Y^{\ast}X\right)  $ called a
Hilbert form

ii) It can be null for non null X,Y. Let J=$\left\{  X\in A:\forall Y\in
A:\left\langle X,Y\right\rangle =0\right\}  $ .\ It is a left ideal of A and
we can pass to the quotient A/J: Define the equivalence relation : $X\sim
Y\Leftrightarrow X-Y\in J.$ A class x in A/J is comprised of elements of the
kind : X + J

iii) Define on A/J the sesquilinear form : $\left\langle x,y\right\rangle
_{A/J}=\left\langle X,Y\right\rangle _{A}.$ So A/J becomes a pre Hilbert space
which can be completed to get a Hilbert space $H_{\varphi}$.

iv) For each x in A/J define the operator on A/J : T(x)y=xy . If T is bounded
it can be extended to the Hilbert space $H_{\varphi}$ and we get a
representation of A.

v) There is a vector $u_{\varphi}\in H_{\varphi}$ such that : $\forall X\in
A:\varphi\left(  X\right)  =\left\langle x,u_{\varphi}\right\rangle ,v\left(
\varphi\right)  =\left\langle u_{\varphi},u_{\varphi}\right\rangle .$
$u_{\varphi}$ can be taken as the class of equivalence of I.

If $\varphi$\ is a state then\ the representation is cyclic with cyclic vector
$u_{\varphi}$ such that $\varphi\left(  X\right)  =\left\langle T\left(
X\right)  u_{\varphi},u_{\varphi}\right\rangle ,\upsilon\left(  \varphi
\right)  =\left\langle u_{\varphi},u_{\varphi}\right\rangle =1$

Conversely:

\begin{theorem}
(Thill p.140) If $\left(  H,\rho,u_{v}\right)  $ is a cyclic linear
representation of the Banach *-algebra A, then each cyclic vector $u_{c}$ of
norm 1 defines a state $\varphi\left(  X\right)  =g\left(  \rho\left(
X\right)  u_{c},u_{c}\right)  $ such that the associated representation
$\left(  H_{\varphi},\rho_{\varphi},u_{\varphi}\right)  $ is equivalent to
$\left(  H,\rho\right)  $ and $\rho_{\varphi}=U\circ\rho\circ U^{\ast}$ for an
unitary operator. The cyclic vectors are related by U : $u_{\varphi}=Uu_{c}$
\end{theorem}

So each cyclic representation of A on a Hilbert space can be labelled by the
equivalent GNS representation, meaning labelled by a state. Up to equivalence
the GNS representation $\left(  H_{\varphi},\rho_{\varphi}\right)  $
associated to a state $\varphi$\ is defined by the condition : $\varphi\left(
X\right)  =\left\langle \rho_{\varphi}\left(  X\right)  u_{\varphi}%
,u_{\varphi}\right\rangle $. Any other cyclic representation $\left(
H,\rho,u_{c}\right)  $ such that : $\varphi\left(  X\right)  =\left\langle
\rho\left(  X\right)  u_{c},u_{c}\right\rangle $ is equivalent to $\left(
H_{\varphi},\rho_{\varphi}\right)  $

\subsubsection{Universal representation}

The universal representation is similar to the sum of finite dimensional
representations of a compact Lie group : it contains all the classes of
equivalent representations. As any representation is sum of cyclic
representations, and that any cyclic representation is equivalent to a GNS
representation, we get all the representations with the sum of GNS representations.

\begin{theorem}
(Thill p.152) The \textbf{universal representation} of the Banach *-algebra A
is the sum :$\left(  \oplus_{\varphi\in S\left(  A\right)  }H_{\varphi}%
;\oplus_{\varphi\in S\left(  A\right)  }\rho_{\varphi}\right)  =\left(
H_{u},\rho_{u}\right)  $ where $\left(  H_{\varphi},\rho_{\varphi}\right)  $
is the GNS representation $\left(  H_{\varphi},\rho_{\varphi}\right)  $
associated to the state $\varphi$\ and S(A) is the set of states on A. It is a
$\sigma-$contractive Hilbertian representation and $\left\Vert \rho_{u}\left(
X\right)  \right\Vert \leq p\left(  X\right)  $ where p is the semi-norm :
$p\left(  X\right)  =\sup_{\varphi\in S\left(  A\right)  }\left(
\varphi\left(  X^{\ast}X\right)  \right)  ^{1/2}.$
\end{theorem}

This semi-norm is well defined as : $\forall X\in A,\varphi\in S\left(
A\right)  :\varphi\left(  X\right)  \leq p\left(  X\right)  \leq r_{\sigma
}\left(  X\right)  \leq\left\Vert X\right\Vert $ and is required to sum the
GNS representations.

The subset rad(A) of A\ such that p(X)=0 is a two-sided ideal, * stable and
closed, called the \textbf{radical}.

The quotient set A/rad(A) with the norm p(X)\ is a pre C*-algebra whose
completion is a C*-algebra denoted C*(A) called the envelopping C*algebra of
A. The map : $j:A\rightarrow C^{\ast}\left(  A\right)  $ is a *-algebra
morphism, continuous and j(C*(A)) is dense in C*(A).

To a representation $\left(  H,\rho_{\ast}\right)  $ of C*(A) one associates a
unique representation $\left(  H,\rho\right)  $ of A by : $\rho=\rho_{\ast
}\circ j.$

A is said \textbf{semi-simple} if rad(A)=0. Then A with the norm p is a
pre-C*-algebra whose completion if C*(A).

If A has a faithful representation then A is semi-simple.

\begin{theorem}
(Gelfand-Na\"{\i}mark) (Thill p.159) If A is a C*-algebra : $\left\Vert
\rho_{u}\left(  X\right)  \right\Vert =p\left(  X\right)  =r_{\sigma}\left(
X\right)  $ .The universal representation is a C* isomorphism between A and
the set $%
\mathcal{L}%
\left(  H;H\right)  $\ of a Hilbert space, thus C*(A) can be assimilated to A
\end{theorem}

If A is commutative and the set of its multiplicative linear functionals
$\Delta\left(  A\right)  \neq\varnothing,$ then C*(A) is isomorphic as a
C*-algebra to the set $C_{0v}\left(  \Delta\left(  A\right)  ;%
\mathbb{C}
\right)  $ of continuous functions vanishing at infinity.

\subsubsection{Irreducible representations}

\begin{theorem}
(Thill p.169) For every Hilbertian representation $\left(  H,\rho\right)
$\ of a Banach *-algebra the following are equivalent :

i) $\left(  H,\rho\right)  $ is irreducible

ii) any non null vector is cyclic

iii) the commutant $\rho^{\prime}$ of $\rho$ is the set $zI,z\in%
\mathbb{C}
$
\end{theorem}

\begin{theorem}
(Thill p.166) If the Hilbertian representation $\left(  H,\rho\right)  $\ of a
Banach *-algebra A is irreducible then, for any vectors u,v of H such that

$\forall X\in A:g\left(  \rho\left(  X\right)  u,u\right)  =g\left(
\rho\left(  X\right)  v,v\right)  \Rightarrow\exists z\in%
\mathbb{C}
,\left\vert z\right\vert =1:v=zu$
\end{theorem}

\begin{theorem}
(Thill p.171) For every Hilbertian representation $\left(  H,\rho\right)
$\ of a Banach *-algebra the following are equivalent :

i) $\varphi$ is a pure state

ii) $\varphi$ is indecomposable

iii) $\left(  H_{\varphi},\rho_{\varphi}\right)  $ is irreducble
\end{theorem}

Thus the pure states label the irreducible representations of A up to equivalence

\begin{theorem}
(Thill p.166) A Hilbertian representation of a commutative algebra is
irreducible iff it is unidimensional
\end{theorem}

\bigskip

\subsection{Spectral theory}

\label{Spectral theory}

Spectral theory is a general method to replace a linear map on an infinite
dimensional vector space by an integral. It is based on the following idea.
Let $X\in L(E;E)$ be a\ diagonalizable operator on a finite dimensional vector
space. On each of its eigen space $E_{\lambda}$ it acts by $u\rightarrow
\lambda u$ thus X can be written as : $X=\sum_{\lambda}\lambda\pi_{\lambda}$
where $\pi_{\lambda}$ is the projection on $E_{\lambda}$ (which can be
uniquely defined if we have a bilinear symmetric form). If E is infinite
dimensional then we can hope to replace $\sum$ by an integral.\ For an
operator on a Hilbert space the same idea involves the spectrum of X and an
integral. The interest lies in the fact that many properties of X can be
studied through the spectrum, meaning a set of complex numbers.

\subsubsection{Spectral measure}

\begin{definition}
A \textbf{spectral measure} defined on a mesurable space (E,S) and acting on a
Hilbert space (H,g) is a map $P:S\rightarrow%
\mathcal{L}%
\left(  H;H\right)  $ such that:

i) $P\left(  \varpi\right)  $ is a projection : $\forall\varpi\in S:P\left(
\varpi\right)  =P\left(  \varpi\right)  ^{\ast}=P\left(  \varpi\right)  ^{2}$

ii) $P(E)=I$

iii) $\forall u\in H$ the map : $\varpi\rightarrow g\left(  P\left(
\varpi\right)  u,u\right)  =\left\Vert P\left(  \varpi\right)  u\right\Vert
^{2}\in%
\mathbb{R}
_{+}$ is a finite measure on (E,S).
\end{definition}

Thus if g(u,u) = 1 then $\left\Vert P\left(  \varpi\right)  u\right\Vert ^{2}$
is a probability

For u,v in H we define a bounded complex measure by :

$\left\langle Pu,v\right\rangle \left(  \varpi\right)  =\frac{1}{4}\sum
_{k=1}^{4}i^{k}g\left(  P\left(  \varpi\right)  \left(  u+i^{k}v\right)
,\left(  u+i^{k}v\right)  \right)  $

$\Rightarrow\left\langle Pu,v\right\rangle \left(  \varpi\right)
=\left\langle P\left(  \varpi\right)  v,u\right\rangle $

The support of P is the complement in E of the largest open subset on which P=0

\bigskip

\begin{theorem}
(Thill p.184, 191) A spectral measure P has the following properties :

i) P is \textit{finitely} additive : for any finite disjointed family

$\left(  \varpi_{i}\right)  _{i\in I}\in S^{I},\forall i\neq j:\varpi_{i}%
\cap\varpi_{j}=\varnothing:P\left(  \cup_{i}\varpi_{i}\right)  =\sum
_{i}P\left(  \varpi_{i}\right)  $

ii) $\forall\varpi_{1},\varpi_{2}\in S:\varpi_{1}\cap\varpi_{2}=\varnothing
:P\left(  \varpi_{1}\right)  \circ P\left(  \varpi_{2}\right)  =0$

iii) $\forall\varpi_{1},\varpi_{2}\in S:P\left(  \varpi_{1}\right)  \circ
P\left(  \varpi_{2}\right)  =P\left(  \varpi_{1}\cap\varpi_{2}\right)  $

iv) $\forall\varpi_{1},\varpi_{2}\in S:P\left(  \varpi_{1}\right)  \circ
P\left(  \varpi_{2}\right)  =P\left(  \varpi_{2}\right)  \circ P\left(
\varpi_{1}\right)  $

v) If the sequence $\left(  \varpi_{n}\right)  _{n\in%
\mathbb{N}
}$ in S is disjointed or increasing then

$\forall u\in H:P\left(  \cup_{n\in%
\mathbb{N}
}\varpi_{n}\right)  u=\sum_{n\in%
\mathbb{N}
}P\left(  \varpi_{n}\right)  u$

vi) $\overline{Span\left(  P\left(  \varpi\right)  \right)  _{\varpi\in S}}$
$\ $is a commutative C*-subalgebra of
$\mathcal{L}$%
(H,H)
\end{theorem}

Warning ! P is not a measure on (E,S), $P\left(  \varpi\right)  \in%
\mathcal{L}%
\left(  H;H\right)  $

A property is said to hold P almost everywhere in E if $\forall u\in H$ if
holds almost everywhere in E for $g\left(  P\left(  \varpi\right)  u,u\right)
$

Image of a spectral measure : let (F,S') another measurable space, and
$\varphi:E\rightarrow F$ a mesurable map, then P defines a spectral measure on
(F,S') by : $\varphi^{\ast}P\left(  \varpi^{\prime}\right)  =P\left(
\varphi^{-1}\left(  \varpi^{\prime}\right)  \right)  $

\paragraph{Examples\newline}

(Neeb p.145)

1. Let (E,S,$\mu)$ be a measured space.\ Then the set $L^{2}\left(  E,S,\mu,%
\mathbb{C}
\right)  $ is a Hilbert space. The map : $\varpi\in S:P(\varpi)\varphi
=\chi_{\varpi}\varphi$ where $\chi_{\varpi}$ is the characteristic function of
$\varpi$\ , is a spectral measure on $L^{2}\left(  E,S,\mu,%
\mathbb{C}
\right)  $

2. Let $H=\oplus_{i\in I}H_{i}$ be a Hilbert sum, define $P\left(  J\right)  $
as the orthogonal projection on the closure : $\overline{\left(  \oplus_{i\in
J}H_{i}\right)  }.$ This is a spectral measure

3. If we have a family $\left(  P_{i}\right)  _{i\in I}$ of spectral measures
on some space (E,S), each valued in $%
\mathcal{L}%
\left(  H_{i};H_{i}\right)  ,$ then : $P\left(  \varpi\right)  u=\sum_{i\in
I}P_{i}\left(  \varpi\right)  u_{i}$ is a spectral measure on $H=\oplus_{i\in
I}H_{i}$.

\paragraph{Characterization of spectral measures\newline}

This theorem is new.

\begin{theorem}
For any measurable space $\left(  E,S\right)  $ with $\sigma-$algebra S, there
is a bijective correspondance between the spectral measures P on the separable
Hilbert space H and the maps : $f:S\rightarrow H$ with the following
properties :

i) f(s) is a closed vector subspace of H

ii) f(E)=H

iii) $\forall s,s^{\prime}\in S:s\cap s^{\prime}=\varnothing\Rightarrow
f\left(  s\right)  \cap f\left(  s^{\prime}\right)  =\left\{  0\right\}  $
\end{theorem}

\begin{proof}
i) Remind a theorem (see Hilbert space) : there is a bijective correspondance
between the projections on a Hilbert space H and the closed vector subspaces
$H_{P}$ of H.\ And P is the orthogonal projection on $H_{P}$

ii) With a map f define $P\left(  s\right)  $ as the unique orthogonal
projection on f(s). Let us show that the map
$\mu$
is countably additive. Take a countable family $\left(  s_{\alpha}\right)
_{\alpha\in A}$ of disjointed elements of S. Then $\left(  f\left(  s_{\alpha
}\right)  \right)  _{\alpha\in A}$ is a countable family of Hilbert vector
subspaces of H. The Hilbert sum $\oplus_{\alpha\in A}f\left(  s_{\alpha
}\right)  $ is a Hilbert space $H_{A}$, vector subspace of H, which can be
identified to $f\left(  \cup_{\alpha\in A}s_{\alpha}\right)  $\ and the
subspaces $f\left(  s_{\alpha}\right)  $ are orthogonal. Take any Hilbert
basis $\left(  \varepsilon_{\alpha i}\right)  _{i\in I_{\alpha}}$ of $f\left(
s_{\alpha}\right)  $ then its union is a Hilbert basis of $H_{A}$ and

$\forall\psi\in H_{A}:\sum_{\alpha\in A}\sum_{i\in I_{\alpha}}\left\vert
\psi^{a\alpha}\right\vert ^{2}=\sum_{\alpha\in A}\left\Vert P\left(
s_{\alpha}\right)  \psi\right\Vert ^{2}=\left\Vert P\left(  \cup_{\alpha\in
A}s_{\alpha}\right)  \psi\right\Vert ^{2}<\infty$

iii) Conversely if P is a spectral measure, using the previous lemna for each
$s\in S$ the projection P(s) defines a unique closed vector space $H_{s} $ of
H and P(s) is the orthogonal projection on $H_{s}.$

For $\psi$ fixed, because $\mu\left(  s\right)  =\left\Vert P\left(  s\right)
\psi\right\Vert ^{2}$ is a measure on E, it is countably additive. Take
$s,s^{\prime}\in S:s\cap s^{\prime}=\varnothing$ then

$\left\Vert P\left(  s\cup s^{\prime}\right)  \psi\right\Vert ^{2}=\left\Vert
P\left(  s\right)  \psi\right\Vert ^{2}+\left\Vert P\left(  s^{\prime}\right)
\psi\right\Vert ^{2}$

For any $\psi\in H_{s\cup s^{\prime}}:\left\Vert P\left(  s\cup s^{\prime
}\right)  \psi\right\Vert ^{2}=\left\Vert \psi\right\Vert ^{2}=\left\Vert
P\left(  s\right)  \psi\right\Vert ^{2}+\left\Vert P\left(  s^{\prime}\right)
\psi\right\Vert ^{2}$

With any Hilbert basis $\left(  \varepsilon_{i}\right)  _{i\in I}$ of
$H_{s},\left(  \varepsilon_{i}^{\prime}\right)  _{i\in I^{\prime}}$of
$H_{s^{\prime}},\psi\in H_{s\cup s^{\prime}}:\left\Vert \psi\right\Vert
^{2}=\sum_{i\in I}\left\vert \psi^{i}\right\vert ^{2}+\sum_{j\in I^{\prime}%
}\left\vert \psi^{\prime j}\right\vert ^{2}$ so $\left(  \varepsilon
_{i}\right)  _{i\in I}\oplus\left(  \varepsilon_{i}^{\prime}\right)  _{i\in
I^{\prime}}$ is a Hilbert basis of $H_{s\cup s^{\prime}}$ and $H_{s\cup
s^{\prime}}=H_{s}\oplus H_{s^{\prime}}$
\end{proof}

\subsubsection{Spectral integral}

This is the extension of the integral of real valued function on a measured
space $\int_{E}f\left(  \varpi\right)  \mu\left(  \varpi\right)  $ , which
gives a scalar, to the integral of a function on a space endowed with a
spectral measure : the result is a map $\int_{E}f\left(  \varpi\right)
P\left(  \varpi\right)  \in%
\mathcal{L}%
\left(  H;H\right)  $.

\begin{theorem}
If P is a spectral measure on the space (E,S), acting on the Hilbert space
(H,g), a complex valued measurable bounded function on E is
\textbf{P-integrable} if there is $X\in%
\mathcal{L}%
\left(  H;H\right)  $ such that :

$\forall u,v\in H:$ $g\left(  Xu,v\right)  =\int_{E}f\left(  \varpi\right)
g\left(  P\left(  \varpi\right)  u,v\right)  $

If so, X is unique and called the \textbf{spectral integral} of f :
$X=\int_{E}fP$
\end{theorem}

The contruct is the following (Thill p.185).

1. A step function is given by a finite set I, a partition $\left(  \varpi
_{i}\right)  _{i\in I}\subset S^{I}$ of E\ and a family of complex scalars
$\left(  \alpha_{i}\right)  _{i\in I}\in\ell^{2}\left(  I\right)  ,$ by :
$f=\sum_{i\in I}\alpha_{i}1_{\varpi_{i}},$ where $1_{\varpi_{i}}$ is the
characteristic function of $\varpi_{i}$

The set $C_{b}\left(  E;%
\mathbb{C}
\right)  $ of complex valued measurable bounded functions in E, endowed with
the norm: $\left\Vert f\right\Vert =\sup\left\vert f\right\vert $ is a
commutative C*-algebra with the involution : $f^{\ast}=\overline{f}.$

The set $C_{S}\left(  E;%
\mathbb{C}
\right)  $ of complex valued step functions on (E,S) is a C*-subalgebra of
$C_{b}\left(  E;%
\mathbb{C}
\right)  $

2. For $h\in C_{S}\left(  E;%
\mathbb{C}
\right)  $ define the integral

$\rho_{S}\left(  h\right)  =\int_{E}h\left(  \varpi\right)  P\left(
\varpi\right)  =\sum_{i\in I}\alpha_{i}h\left(  \varpi_{i}\right)  P\left(
\varpi_{i}\right)  \in%
\mathcal{L}%
\left(  H;H\right)  $

$\left(  H,\rho_{S}\right)  $ is a representation of $C_{S}\left(  E;%
\mathbb{C}
\right)  $\ with $\rho_{S}:C_{S}\left(  E;%
\mathbb{C}
\right)  \rightarrow%
\mathcal{L}%
\left(  H;H\right)  $

and : $\forall u\in H:$ $g\left(  \left(  \int_{E}h\left(  \varpi\right)
P\left(  \varpi\right)  \right)  u,u\right)  =\int_{E}h\left(  \varpi\right)
g\left(  P\left(  \varpi\right)  u,u\right)  $

3. We say that $f\in C_{b}\left(  E;%
\mathbb{C}
\right)  $ is P integrable (in norm) if there is :

$X\in%
\mathcal{L}%
\left(  H;H\right)  :\forall h\in C_{S}\left(  E;%
\mathbb{C}
\right)  :\left\Vert X-\int_{E}h\left(  \varpi\right)  P\left(  \varpi\right)
\right\Vert _{%
\mathcal{L}%
\left(  H;H\right)  }\leq\left\Vert f-h\right\Vert _{C_{b}\left(  E;%
\mathbb{C}
\right)  }$

We say that $f\in C_{b}\left(  E;%
\mathbb{C}
\right)  $ is P integrable (weakly) if there is $Y\in%
\mathcal{L}%
\left(  H;H\right)  $ such that : $\forall u\in H:$ $g\left(  Yu,u\right)
=\int_{E}f\left(  \varpi\right)  g\left(  P\left(  \varpi\right)  u,u\right)
$

4. f P integrable (in norm) $\Rightarrow$\ f P integrable (weakly) and there
is a unique $X=Y=\rho_{b}\left(  f\right)  =\int_{E}fP\in%
\mathcal{L}%
\left(  H;H\right)  $

5. conversely f P integrable (weakly) $\Rightarrow$\ f P integrable (in norm)

Remark : the norm on a C*-algebra of functions is necessarily equivalent to :
$\left\Vert f\right\Vert =\sup_{x\in E}\left\vert f\left(  x\right)
\right\vert $ (see Functional analysis). So the theorem holds for any
C*-algebra of functions on E.

\begin{theorem}
(Thill p.188) For every P integrable function f :

i) $\left\Vert \left(  \int_{E}fP\right)  u\right\Vert _{H}=\sqrt{\int
_{E}\left\vert f\right\vert ^{2}g\left(  P\left(  \varpi\right)  u,u\right)
}$

ii) $\int_{E}fP=0\Leftrightarrow f=0$ P almost everywhere

iii) $\int_{E}fP\geq0\Leftrightarrow f\geq0$ P almost everywhere
\end{theorem}

Notice that the two last results are inusual.

\begin{theorem}
(Thill p.188) For a spectral measure P on the space (E,S), acting on the
Hilbert space (H,g), H and the map : $\rho_{b}:C_{b}\left(  E;%
\mathbb{C}
\right)  \rightarrow%
\mathcal{L}%
\left(  H;H\right)  ::\rho_{b}\left(  f\right)  =\int_{E}fP$ is a
representation of the C*-algebra $C_{b}\left(  E;%
\mathbb{C}
\right)  $. $\rho_{b}\left(  C_{b}\left(  E;%
\mathbb{C}
\right)  \right)  =\overline{Span\left(  P\left(  \varpi\right)  \right)
_{\varpi\in S}}$ $\ $is the C*-subalgebra of
$\mathcal{L}$%
(H,H) generated by P and the commutants : $\rho^{\prime}=Span\left(  P\left(
\varpi\right)  \right)  _{\varpi\in S}^{\prime}.$

Every projection in $\rho_{b}\left(  C_{b}\left(  E;%
\mathbb{C}
\right)  \right)  $ is of the form : $P\left(  s\right)  $ for some $s\in S.$
\end{theorem}

\begin{theorem}
Monotone convergence theorem (Thill p.190) If P is a spectral measure P on the
space (E,S), acting on the Hilbert space (H,g), $\left(  f_{n}\right)  _{n\in%
\mathbb{N}
}$ an increasing bounded sequence of real valued mesurable functions on E,
bounded P almost everywhere, then $f=\lim f_{n}\in C_{b}\left(  E;%
\mathbb{R}
\right)  $ and $\int fP=\lim\int f_{n}P$ , $\int fP$ is self adjoint and
$\forall u\in H:g\left(  \left(  \int_{E}fP\right)  u,u\right)  =\lim\int
_{E}f_{n}\left(  \varpi\right)  g\left(  P\left(  \varpi\right)  u,u\right)  $
\end{theorem}

\begin{theorem}
Dominated convergence theorem (Thill p.190) If P is a spectral measure P on
the space (E,S), acting on the Hilbert space (H,g),$\left(  f_{n}\right)
_{n\in%
\mathbb{N}
}$ a norm bounded sequence of functions in $C_{b}\left(  E;%
\mathbb{C}
\right)  $ which converges pointwise to f, then : $\forall u\in H:\left(  \int
fP\right)  u=\lim\left(  \int f_{n}P\right)  u$
\end{theorem}

\paragraph{Extension to unbounded operators\newline}

\begin{theorem}
(Thill p.233) If P is a spectral measure on the space (E,S), acting on the
Hilbert space (H,g), for each complex valued measurable function f on (E,S)
there is a linear map $X=\int fP$ called the \textbf{spectral integral} of f,
defined on a subspace D(X) of H such that :$\ $

$\forall u\in D\left(  X\right)  :$ $g\left(  Xu,u\right)  =\int_{E}f\left(
\varpi\right)  g\left(  P\left(  \varpi\right)  u,u\right)  $

$D(\int fP)$ = $\left\{  u\in H:\int_{E}\left\vert g(u,fu)P\right\vert
^{2}<\infty\right\}  $ is dense in H
\end{theorem}

Comments:

1) the conditions on f are very weak : almost any function is integrable

2) the difference with the previous spectral integral is that $\int fP$ is
neither necessarily defined over the whole of H, nor continuous

The construct is the following :

i) For each complex valued measurable function f on (E,S)

$D(f)=\left\{  u\in H:\int_{E}\left\vert g(u,fu)P\right\vert ^{2}%
<\infty\right\}  $ is dense in H

ii) If $\exists X\in L\left(  D\left(  X\right)  ;H\right)  :D(X)=D(f)$ one
says that f is

weakly integrable if :

$\forall u\in H:$ $g\left(  Xu,u\right)  =\int_{E}f\left(  \varpi\right)
g\left(  P\left(  \varpi\right)  u,u\right)  $

pointwise integrable if : $\forall h\in C_{b}\left(  E;%
\mathbb{C}
\right)  ,\forall u\in H:$

$\left\Vert \left(  X-\int_{E}hP\right)  u\right\Vert ^{2}=\sqrt{\int
_{E}\left\Vert f\left(  \varpi\right)  -h\left(  \varpi\right)  \right\Vert
^{2}g\left(  P\left(  \varpi\right)  u,u\right)  }$

iii) f is weakly integrable $\Rightarrow$ f is pointwise integrable and X is unique.

For any complex valued measurable function f on (E,S) there exists a unique
$X=\Psi_{P}\left(  f\right)  $ such that $X=\int_{E}fP$ pointwise

f is pointwise integrable $\Rightarrow$ f is weakly integrable

\begin{theorem}
(Thill p.236, 237, 240) If P is a spectral measure on the space (E,S), acting
on the Hilbert space (H,g), and $f,f_{1},f_{2}$ are complex valued measurable
functions on (E,S) :

i) $\forall u\in D\left(  f\right)  :\left\Vert \left(  \int_{E}f\left(
\varpi\right)  P\left(  \varpi\right)  \right)  u\right\Vert _{H}=\sqrt
{\int_{E}\left\vert f\right\vert ^{2}g\left(  P\left(  \varpi\right)
u,u\right)  }$

ii) $D\left(  \left\vert f_{1}\right\vert +\left\vert f_{2}\right\vert
\right)  =D\left(  \int_{E}f_{1}P+\int_{E}f_{2}P\right)  $

$D\left(  \left(  \int_{E}f_{1}P\right)  \circ\left(  \int_{E}\left(
f_{2}\right)  P\right)  \right)  =D\left(  f_{1}\circ f_{2}\right)  \cap
D\left(  f_{2}\right)  $

which reads with the meaning of extension of operators (see Hilbert spaces)

$\int_{E}f_{1}P+\int_{E}f_{2}P\subset\int_{E}\left(  f_{1}+f_{2}\right)  P$

$\left(  \int_{E}f_{1}P\right)  \circ\left(  \int_{E}\left(  f_{2}\right)
P\right)  \subset\int_{E}\left(  f_{1}f_{2}\right)  P$

iii) $\left(  \int_{E}fP\right)  ^{\ast}=\int_{E}\overline{f}P$ so if f is a
measurable real valued function on E then $\int_{E}fP$ is self-adjoint

$\int_{E}fP$ is a closed map

$\left(  \int_{E}fP\right)  ^{\ast}\circ\left(  \int_{E}fP\right)  =\left(
\int_{E}fP\right)  \circ\left(  \int_{E}fP\right)  ^{\ast}=\int_{E}\left\vert
f\right\vert ^{2}P$
\end{theorem}

\begin{theorem}
Image of a spectral measure (Thill p.192, 236) : If P is a spectral measure on
the space (E,S), acting on the Hilbert space (H,g), (F,S') a measurable space,
and $\varphi:E\rightarrow F$ a mesurable map then for any complex valued
measurable function f on (F,S') : $\int_{F}f\varphi^{\ast}P=\int_{E}\left(
f\circ\varphi\right)  P$
\end{theorem}

This theorem holds for both cases.

\subsubsection{Spectral resolution}

The purpose is now, conversely, starting from an operator X, find f and a
spectral measure P such that X=$\int_{E}f\left(  \varpi\right)  P\left(
\varpi\right)  $

\paragraph{Spectral theorem\newline}

\begin{definition}
A \textbf{resolution of identity} is a spectral measure on a measurable
Hausdorff space (E,S) acting on a Hilbert space (H,g) such that for any $u\in
H$, $g(u,u)=1:$ $g\left(  P\left(  \varpi\right)  u,u\right)  $ is an inner
regular measure on (E,S).
\end{definition}

\begin{theorem}
Spectral theorem (Thill p.197) For any continuous normal operator X on a
Hilbert space H there is a unique resolution of identity : $P:Sp(X)\rightarrow%
\mathcal{L}%
(H;H)$ called the \textbf{spectral resolution} of X\ such that :
$X=\int_{Sp\left(  X\right)  }zP$ where Sp(X) is the spectrum of X
\end{theorem}

\begin{theorem}
(Spectral theorem for unbounded operators) (Thill p.243) For every densely
defined, linear, \textit{self-adjoint} operator X in the Hilbert space H,
there is a unique resolution of identity $P:Sp(X)\rightarrow L(H;H)$\ called
the \textbf{spectral resolution} of X, such that : $X=\int_{Sp(X)}\lambda P$
where Sp(X) is the spectrum of X.
\end{theorem}

X \textit{normal} : X*X=XX*

so the function f is here the identity map : $Id:Sp(X)\rightarrow Sp\left(
X\right)  $

We have a sometimes more convenient formulation of these theorems

\begin{theorem}
(Taylor 2 p.72) Let X be a self adjoint operator on a separable Hilbert space
H, then there is a Borel measure $\mu$ on $%
\mathbb{R}
$\ , a unitary map $W:L^{2}\left(
\mathbb{R}
,\mu,%
\mathbb{C}
\right)  \rightarrow H,$ a real valued function $a\in L^{2}\left(
\mathbb{R}
,\mu,%
\mathbb{R}
\right)  $ such that :

$\forall\varphi\in L^{2}\left(
\mathbb{R}
,\mu,%
\mathbb{C}
\right)  :W^{-1}XW\varphi\left(  x\right)  =a\left(  x\right)  \varphi\left(
x\right)  $
\end{theorem}

\begin{theorem}
(Taylor 2 p.79) If X is a self adjoint operator, defined on a dense subset
D(X) of a separable Hilbert space H, then there is a measured space ($E,\mu)$,
a unitary map $W:L^{2}\left(  E,\mu,%
\mathbb{C}
\right)  \rightarrow H,$ a real valued function $a\in L^{2}\left(  E,\mu,%
\mathbb{R}
\right)  $ such that :

$\forall\varphi\in L^{2}\left(  E,\mu,%
\mathbb{C}
\right)  :W^{-1}XW\varphi\left(  x\right)  =a\left(  x\right)  \varphi\left(
x\right)  $

$W\varphi\in D\left(  X\right)  $ iff $\varphi\in L^{2}\left(  E,\mu,%
\mathbb{C}
\right)  $
\end{theorem}

If f is a bounded measurable function on E, then : $W^{-1}f\left(  X\right)
W\varphi\left(  x\right)  =f\left(  a\left(  x\right)  \right)  \varphi\left(
x\right)  $ defines a bounded operator f(X) on $L^{2}\left(  E,\mu,%
\mathbb{C}
\right)  $

With $f(x)=e^{ia(x)}$ we get the get the strongly continuous one parameter
group $e^{iXt}=U(t)$ with generator iX.

\begin{theorem}
(Taylor 2 p.79) If $A_{k},k=1..n$ are commuting, self adjoint continuous
operators on a Hilbert space H, there are a measured space $(E,\mu),$ a
unitary map : $W:L^{2}\left(  E,\mu,%
\mathbb{C}
\right)  \rightarrow H,$ functions $a_{k}\in L^{\infty}\left(  E,\mu,%
\mathbb{R}
\right)  $ such that :

$\forall f\in L^{2}\left(  E,\mu,%
\mathbb{C}
\right)  :W^{-1}A_{k}W\left(  f\right)  \left(  x\right)  =a_{k}\left(
x\right)  f\left(  x\right)  $
\end{theorem}

\begin{theorem}
Whenever it is defined, if P is the spectral resolution of X\ :

i) Support of P = all of Sp(X)

ii) Commutants : $X\prime=Span\left(  P\left(  z\right)  \right)  _{z\in
Sp(X)}^{\prime}$
\end{theorem}

\begin{theorem}
(Thill p.198) If P is the spectral resolution of the continuous normal
operator X on a Hilbert space H, $\lambda\in Sp(X)$ is an eigen value of X iff
$P\left(  \left\{  \lambda\right\}  \right)  \neq0$ . Then the range of
$P\left(  \lambda\right)  $ is the eigen space relative to $\lambda$
\end{theorem}

So the eigen values of X are the isolated points of its spectrum.

\begin{theorem}
(Thill p.246) If P is the spectral resolution of a densely self adjoint
operator X on the Hilbert space H, $f:Sp(X)\rightarrow%
\mathbb{C}
$ a Borel measurable function, then $\int_{E}fP$ is well defined on $D\left(
\int_{E}fP\right)  $\ and denoted f(X)
\end{theorem}

\paragraph{Commutative algebras\newline}

For any algebra (see Normed algebras) :

$\Delta\left(  A\right)  \in%
\mathcal{L}%
\left(  A;%
\mathbb{C}
\right)  $ is the set of multiplicative linear functionals on A

$\widehat{X}:\Delta\left(  A\right)  \rightarrow%
\mathbb{C}
::\widehat{X}\left(  \varphi\right)  =\varphi\left(  X\right)  $ is the
Gel'fand transform of X

\begin{theorem}
(Thill p.201) For every Hilbertian representation $\left(  H,\rho\right)  $ of
a commutative *-algebra A, there is a unique resolution of identity P sur
$Sp\left(  \rho\right)  $ acting on H such that :

$\forall X\in A:\rho\left(  X\right)  =\int_{Sp\left(  X\right)  }\widehat
{X}|_{Sp\left(  X\right)  }P$ and $Supp(P)=Sp\left(  \rho\right)  =\cup_{X\in
A}Sp\left(  \rho\left(  X\right)  \right)  $
\end{theorem}

\begin{theorem}
(Neeb p.152) For any Banach commutative *-algebra A :

i) If P is a spectral measure on $\Delta\left(  A\right)  $ then $\rho\left(
X\right)  =P\left(  \widehat{X}\right)  $ defines a spectral measure on A

ii) If $(H,\rho)$ is a non degenerate Hilbertian representation of A, then
there is a unique spectral measure P on $\Delta\left(  A\right)  $ such that
$\rho\left(  X\right)  =P\left(  \widehat{X}\right)  $
\end{theorem}

\begin{theorem}
(Thill p.194) \ For every Hilbert space H, commutative C*-subalgebra A of
$\mathcal{L}$%
(H;H), there is a unique resolution of identity $P:\Delta\left(  A\right)
\rightarrow%
\mathcal{L}%
(H;H)$ such that : $\forall X\in A:X=\int_{\Delta\left(  A\right)  }%
\widehat{X}P$
\end{theorem}

\subsubsection{Application to one parameter unitary groups}

For general one parameters groups see Banach Spaces.

\begin{theorem}
(Thill p.247) A map : $U:%
\mathbb{R}
\rightarrow%
\mathcal{L}%
\left(  H;H\right)  $ such that :

i) U(t) is unitary

ii) U(t+s)=U(t)U(s)=U(s)U(t)

defines a one parameter unitary group on a Hilbert space H.

If $\forall u\in H$ the map : $%
\mathbb{R}
\rightarrow H::U(t)u$ is continuous then U is differentiable, and there is an
infinitesimal generator $S\in L\left(  D(S),H\right)  $ such that :

$\forall u\in D\left(  S\right)  :-\frac{1}{i}\frac{d}{dt}U(t)|_{t=0}u=Su$
which reads $U\left(  t\right)  =\exp\left(  itS\right)  $
\end{theorem}

\begin{theorem}
(Taylor 2 p.76) If $U:%
\mathbb{R}
\rightarrow%
\mathcal{L}%
\left(  H;H\right)  $ is an uniformly continuous one parameter group, having a
cyclic vector, and H a Hilbert space, then there exists a positive Borel
measure $\mu$\ on $%
\mathbb{R}
$\ and a unitary map : $W:L^{2}\left(
\mathbb{R}
,\mu,%
\mathbb{C}
\right)  \rightarrow H$ such that :

$\forall\varphi\in L^{2}\left(
\mathbb{R}
,\mu,%
\mathbb{C}
\right)  :W^{-1}U\left(  t\right)  W\varphi\left(  x\right)  =e^{itx}%
\varphi\left(  x\right)  $\ 
\end{theorem}

The measure $\mu=\widehat{\zeta}\left(  t\right)  dt$ where $\zeta\left(
t\right)  =\left\langle v,U\left(  t\right)  v\right\rangle $ is a tempered distribution.

Conversely :

\begin{theorem}
(Thill p.247) For every self adjoint operator S\ defined on a dense domain
D(X) of a Hilbert space H, the map :

$U:%
\mathbb{R}
\rightarrow%
\mathcal{L}%
\left(  H;H\right)  ::U(t)=\exp\left(  -itS\right)  =\int_{Sp(S)}\left(
-it\lambda\right)  P\left(  \lambda\right)  $

\ defines a one parameter unitary group on H with infinitesimal generator S.

U is differentiable and $-\frac{1}{i}\frac{d}{ds}U(s)|_{s=t}u=SU(t)u$
\end{theorem}

So U is the solution to the problem : $-\frac{1}{i}\frac{d}{ds}U(s)|_{s=t}%
=SU(t)$ with the initial value solution U(0)=S

Remark : U(t) is the Fourier transform of S

\newpage

\part{\textbf{DIFFERENTIAL\ GEOMETRY}}

\bigskip

Differential geometry is the extension of elementary geometry and deals with
manifolds. Nowodays it is customary to address many issues of differential
geometry with the fiber bundle\ formalism.\ However a more traditional
approach is sometimes useful, and enables to start working with the main
concepts without the hurdle of getting acquainted with a new theory.\ So we
will deal with fiber bundles later, after the review of Lie groups. Many
concepts and theorems about manifolds can be seen as extensions from the study
of derivatives in affine normed spaces. So we will start with a comprehensive
review of derivatives in this context.

\bigskip

\section{DERIVATIVE}

\bigskip

\subsection{Differentiable maps}

\label{Differentiable maps}

\subsubsection{Definitions}

In elementary analysis the derivative of a function f(x) at a point a is
introduced as $f^{\prime}(x)|_{x=0}=\lim_{h\rightarrow0}\frac{1}{h}\left(
f(a+h)-f(a)\right)  .$ This idea can be generalized once we have normed linear
spaces. As the derivative is taken at a point, the right structure is an
affine space (of course a vector space is an affine space and the results can
be fully implemented in this case) .

\paragraph{Differentiable at a point\newline}

\begin{definition}
A map $f:\Omega\rightarrow F$ defined on an open subset $\Omega$ of the normed
affine space $\left(  E,\overrightarrow{E}\right)  $ and valued in the normed
affine space $\left(  F,\overrightarrow{F}\right)  ,$ both on the same field
K, is \textbf{differentiable} at $a\in\Omega$ if there is a linear, continuous
map $L\in%
\mathcal{L}%
\left(  \overrightarrow{E};\overrightarrow{F}\right)  $ such that :%

\begin{equation}
\exists r>0,\forall h\in\overrightarrow{E},\left\Vert \overrightarrow
{h}\right\Vert _{E}<r:a+\overrightarrow{h}\in\Omega:f(a+\overrightarrow
{h})-f(a)=L\overrightarrow{h}+\varepsilon\left(  h\right)  \left\Vert
\overrightarrow{h}\right\Vert _{F}%
\end{equation}

where $\varepsilon\left(  h\right)  \in\overrightarrow{F}$ is such that
$\lim_{h\rightarrow0}\varepsilon\left(  h\right)  =0$

L is called the \textbf{derivative} of f in $a$.
\end{definition}

Speaking plainly : f can be approximated by an affine map in the neighborhood
of a: $f(a+\overrightarrow{h})\simeq f(a)+L\overrightarrow{h}$

\begin{theorem}
If the derivative exists at $a$, it is unique and f is continuous in $a$.
\end{theorem}

This derivative is often called Fr\'{e}chet's derivative. If we take E=F=$%
\mathbb{R}
$ we get back the usual definition of a derivative.

Notice that $f(a+\overrightarrow{h})-f(a)\in\overrightarrow{F}$ and that no
assumption is made about the dimension of E,F or the field K, but E and F must
be on the same field (because a linear map must be between vector spaces on
the same field).\ This remark will be important when K=$%
\mathbb{C}
.$

The domain $\Omega$\ must be open. If $\Omega$ is a closed subset and
$a\in\partial\Omega$ then we must have $a+\overrightarrow{h}\in\overset{\circ
}{\Omega}$ and L may not be defined over $\overrightarrow{E}.$ If E=$\left[
a,b\right]  \subset%
\mathbb{R}
$ one can define right derivative at a and left derivative at b because L is a scalar.

\begin{theorem}
(Schwartz II p.83) A map $f:\Omega\rightarrow F$ defined on an open subset
$\Omega$ of the normed affine space $\left(  E,\overrightarrow{E}\right)  $
and valued in the normed affine space $\left(  F,\overrightarrow{F}\right)  =%
{\textstyle\prod\limits_{i=1}^{r}}
\left(  F_{i},\overrightarrow{F}_{i}\right)  ,$ both on the same field K, is
differentiable at $a\in\Omega$ iff each of its components $f_{k}:E\rightarrow
F_{k}$ is differentiable at $a$ and its derivative $f^{\prime}(a)$ is the
linear map in $%
\mathcal{L}%
\left(  \overrightarrow{E};%
{\textstyle\prod\limits_{i=1}^{r}}
\overrightarrow{F}_{i}\right)  $ defined by $f_{k}^{\prime}\left(  a\right)
.$
\end{theorem}

\paragraph{Continuously differentiable in an open subset\newline}

\begin{definition}
A map $f:\Omega\rightarrow F$ defined on an open subset $\Omega$\ of the
normed affine space $\left(  E,\overrightarrow{E}\right)  $ and valued in the
normed affine space $\left(  F,\overrightarrow{F}\right)  $ both on the same
field K, is differentiable in $\Omega$ if it is differentiable at each point
of $\Omega.$ Then the map : $f^{\prime}:\Omega\rightarrow%
\mathcal{L}%
\left(  \overrightarrow{E};\overrightarrow{F}\right)  $ is the derivative map,
or more simply derivative, of f in $\Omega.$ If f' is continuous f is said to
be \textbf{continuously differentiable} or of class 1 (or $C_{1}).$
\end{definition}

\begin{notation}
$f^{\prime}$ is the derivative of f : $f^{\prime}:\Omega\rightarrow%
\mathcal{L}%
\left(  \overrightarrow{E};\overrightarrow{F}\right)  $
\end{notation}

\begin{notation}
$f^{\prime}(a)=f^{\prime}\left(  x\right)  |_{x=a}$ is the value of the
derivative in a . So $f^{\prime}(a)\in%
\mathcal{L}%
\left(  \overrightarrow{E};\overrightarrow{F}\right)  $
\end{notation}

\begin{notation}
$C_{1}\left(  \Omega;F\right)  $ is the set of continuously differentiable
maps $f:\Omega\rightarrow F.$
\end{notation}

If E,F are vector spaces then $C_{1}\left(  \Omega;F\right)  $ is a vector
space and the map which associates to each map $f:\Omega\rightarrow F$ its
derivative is a linear map on the space $C_{1}\left(  \Omega;F\right)  .$

\begin{theorem}
(Schwartz II p.87) If the map $f:\Omega\rightarrow F$ defined on an open
subset $\Omega$\ of the normed affine space $\left(  E,\overrightarrow
{E}\right)  $ and valued in the normed affine space $\left(  F,\overrightarrow
{F}\right)  $ both on the same field K, is continuously differentiable in
$\Omega$ then the map : $\Omega\times\overrightarrow{E}\rightarrow
\overrightarrow{F}::f^{\prime}(x)\overrightarrow{u}$ is continuous.
\end{theorem}

\paragraph{Differentiable along a vector\newline}

\begin{definition}
A map $f:\Omega\rightarrow F$ defined on an open subset $\Omega$\ of the
normed affine space $\left(  E,\overrightarrow{E}\right)  $ and valued in the
normed affine space $\left(  F,\overrightarrow{F}\right)  $ on the same field
K, is \textbf{differentiable }at\textbf{\ }$a\in\Omega$\textbf{\ along the
vector} $\overrightarrow{u}\in\overrightarrow{E}$ if there is $\overrightarrow
{v}\in\overrightarrow{F}$\ such that : $\lim_{z\rightarrow0}\left(  \frac
{1}{z}\left(  f(a+z\overrightarrow{u})-f(a)\right)  \right)  =\overrightarrow
{v}.$ And $\overrightarrow{v}$ is the derivative of f in $a$ with respect to
the vector $\overrightarrow{u}$
\end{definition}

\begin{notation}
$D_{u}f\left(  a\right)  \in\overrightarrow{F}$ is the derivative of f in $a$
with respect to the vector $\overrightarrow{u}$
\end{notation}

\begin{definition}
A map $f:\Omega\rightarrow F$ defined on an open subset $\Omega$\ of the
normed affine space $\left(  E,\overrightarrow{E}\right)  $ and valued in the
normed affine space $\left(  F,\overrightarrow{F}\right)  $ on the same field
K, is \textbf{G\^{a}teaux differentiable }at\textbf{\ }$a\in\Omega$\textbf{\ }
if there is $L\in L\left(  \overrightarrow{E};\overrightarrow{F}\right)
$\ such that : $\forall\overrightarrow{u}\in\overrightarrow{E}:$
$\lim_{z\rightarrow0}\left(  \frac{1}{z}\left(  f(a+z\overrightarrow
{u})-f(a)\right)  \right)  =L\overrightarrow{u}.$
\end{definition}

\begin{theorem}
If f is differentiable at a, then it is G\^{a}teaux \textbf{differentiable}
and $D_{u}f=f^{\prime}(a)\overrightarrow{u}.$\ 
\end{theorem}

But \textit{the converse is not true} : there are maps which are G\^{a}teaux
differentiable and not even continuous ! But if

$\forall\varepsilon>0,\exists r>0,\forall\overrightarrow{u}\in\overrightarrow
{E}:\left\Vert \overrightarrow{u}\right\Vert _{E}<r:\left\Vert \varphi\left(
z\right)  -\overrightarrow{v}\right\Vert <\varepsilon$

then f is differentiable in a.

\paragraph{Partial derivatives\newline}

\begin{definition}
A map $f:\Omega\rightarrow F$ defined on an open subset $\Omega$ of the normed
affine space $\left(  E,\overrightarrow{E}\right)  =%
{\textstyle\prod\limits_{i=1}^{r}}
\left(  E_{i},\overrightarrow{E}_{i}\right)  $ and valued in the normed affine
space $\left(  F,\overrightarrow{F}\right)  ,$ all on the same field K, has a
\textbf{partial derivative} at $a=\left(  a_{1},.a_{r}\right)  \in\Omega$ with
respect to the variable k if the map:

$f_{k}:\Omega_{k}=\pi_{k}\left(  \Omega\right)  \rightarrow F::$ $f_{k}\left(
x_{k}\right)  =f\left(  a_{1},...a_{k-1},x_{k},a_{k+1},..a_{r}\right)  $,

where $\pi_{k}$\ is the canonical projection $\pi_{k}:E\rightarrow E_{k}$, is
differentiable at $a$
\end{definition}

\begin{notation}
$\frac{\partial f}{\partial x_{k}}\left(  a\right)  =f_{x_{k}}^{\prime}\left(
a\right)  \in%
\mathcal{L}%
\left(  \overrightarrow{E}_{k};\overrightarrow{F}\right)  $ denotes the value
of the partial derivative at a with respect to the variable $x_{k}$
\end{notation}

\begin{definition}
If f has a \textbf{partial derivative} with respect to the variable $x_{k}$ at
each point $a\in\Omega,$and if the map : $a_{k}\rightarrow\frac{\partial
f}{\partial x_{k}}\left(  a\right)  $ is continuous, then f is said to be
continuously differentiable with respect to the variable $x_{k}$ in $\Omega$
\end{definition}

Notice that a partial derivative does not necessarily refer to a basis.

If f is differentiable at a then it has a partial derivative with respect to
each of its variable and $f^{\prime}(a)\left(  \overrightarrow{u}%
_{1},...\overrightarrow{u}_{r}\right)  =\sum_{i=1}^{r}f_{x_{i}}^{\prime
}\left(  a\right)  \left(  \overrightarrow{u}_{i}\right)  $

But the \textit{converse is not true}. We have the following :

\begin{theorem}
(Schwartz II p.118) A map $f:\Omega\rightarrow F$ defined on an open subset
$\Omega$ of the normed affine space $\left(  E,\overrightarrow{E}\right)  =%
{\textstyle\prod\limits_{i=1}^{r}}
\left(  E_{i},\overrightarrow{E}_{i}\right)  $ and valued in the normed affine
space $\left(  F,\overrightarrow{F}\right)  ,$ all on the same field K, which
is continuously differentiable in $\Omega$\ with respect to each of its
variable is continuously differentiable in \ $\Omega$
\end{theorem}

So f continuously differentiable in $\Omega$ $\Leftrightarrow$\ f has
continuous partial derivatives in $\Omega$

but f has partial derivatives in a $\nRightarrow$ f is differentiable in a

Notice that the $E_{i}$ and F can be infinite dimensional. We just need a
finite product of normed vector spaces.

\paragraph{Coordinates expressions\newline}

Let f be a map $f:\Omega\rightarrow F$ defined on an open subset $\Omega$\ of
the normed affine space $\left(  E,\overrightarrow{E}\right)  $ and valued in
the normed affine space $\left(  F,\overrightarrow{F}\right)  $ on the same
field K.

1. If E is a m dimensional affine space, it can be seen as the product of n
one dimensional affine spaces and, with a basis $\left(  \overrightarrow
{e}_{i}\right)  _{i=1}^{m}$ of $\overrightarrow{E}$ we have :

The value of f'(a) along the basis vector $\overrightarrow{e}_{i}$ is
$D_{\overrightarrow{e}_{i}}f(a)=f^{\prime}(a)\left(  \overrightarrow{e}%
_{i}\right)  \in\overrightarrow{F}$

The partial derivative with respect to $x_{i}$ is :

$\frac{\partial f}{\partial x_{i}}\left(  a\right)  $ and :
$D_{\overrightarrow{e}_{i}}f(a)=\frac{\partial f}{\partial x_{i}}\left(
a\right)  \left(  \overrightarrow{e}_{i}\right)  $

The value of f'(a) along the vector $\overrightarrow{u}=\sum_{i=1}^{m}%
u_{i}\overrightarrow{e}_{i}$ is

$D_{\overrightarrow{u}}f(a)=f^{\prime}(a)\left(  \overrightarrow{u}\right)
=\sum_{i=1}^{m}u_{i}D_{\overrightarrow{e}_{i}}f(a)=\sum_{i=1}^{m}u_{i}%
\frac{\partial f}{\partial x_{i}}\left(  a\right)  \left(  \overrightarrow
{e}_{i}\right)  \in\overrightarrow{F}$

2. If F is a n dimensional affine space, with a basis $\left(  \overrightarrow
{f}_{i}\right)  _{i=1}^{n}$ we have :

$f\left(  x\right)  =\sum_{k=1}^{n}f_{k}\left(  x\right)  $ where
$f_{k}\left(  x\right)  $ are the coordinates of f(x) in a frame $\left(
O,\left(  \overrightarrow{f}_{i}\right)  _{i=1}^{n}\right)  .$

$f^{\prime}(a)=\sum_{k=1}^{n}f_{k}^{\prime}\left(  a\right)  \overrightarrow
{f}_{k}$ where $f_{k}^{\prime}\left(  a\right)  \in K$

3. If E is m dimensional and F n dimensional, the map f'(a) is represented by
a matrix J with n rows and m columns, each column being the matrix of a
partial derivative, called the \textbf{jacobian} of f:

\bigskip

$\left[  f^{\prime}\right]  =J=\left\{  \overset{m}{\overbrace{\left[
\dfrac{\partial f_{j}}{\partial x_{i}}\right]  }}\right\}  n=%
\begin{bmatrix}
\frac{\partial f_{1}}{\partial x_{1}} & .. & \frac{\partial f_{1}}{\partial
x_{m}}\\
... & .. & ...\\
\frac{\partial f_{n}}{\partial x_{1}} & .. & \frac{\partial f_{n}}{\partial
x_{m}}%
\end{bmatrix}
$

\bigskip

If $E=F$ the determinant of J is the determinant of the linear map f'(a), thus
it does not depend on the basis.

\subsubsection{Properties of the derivative}

\paragraph{Derivative of linear maps\newline}

\begin{theorem}
A continuous affine map $f:\Omega\rightarrow F$ defined on an open subset
$\Omega$ of the normed affine space $\left(  E,\overrightarrow{E}\right)  $
and valued in the normed affine space $\left(  F,\overrightarrow{F}\right)  ,$
both on the same field K, is continuously differentiable in $\Omega$ and f' is
the linear map $\overrightarrow{f}\in%
\mathcal{L}%
\left(  \overrightarrow{E};\overrightarrow{F}\right)  $\ associated to f.
\end{theorem}

So if f=constant then f'=0

\begin{theorem}
(Schwartz II p.86) A continuous r multilinear map

$f\in%
\mathcal{L}%
^{r}\left(  \overrightarrow{E}_{1},...\overrightarrow{E}_{r};\overrightarrow
{F}\right)  $ defined on the normed vector space $%
{\textstyle\prod\limits_{i=1}^{r}}
\overrightarrow{E}_{r}$ and valued in the normed vector space $\overrightarrow
{F}$, all on the same field K, is continuously differentiable and its
derivative at $\overrightarrow{u}=\left(  \overrightarrow{u}_{1}%
,...,\overrightarrow{u}_{r}\right)  $\ is :

$f^{\prime}\left(  \overrightarrow{u}\right)  \left(  \overrightarrow{v}%
_{1},...,\overrightarrow{v}_{r}\right)  =\sum_{i=1}^{r}f\left(
\overrightarrow{u}_{1},.,\overrightarrow{u}_{i-1},\overrightarrow{v}%
_{i},\overrightarrow{u}_{i+1}..,\overrightarrow{u}_{r}\right)  $
\end{theorem}

\paragraph{Chain rule\newline}

\begin{theorem}
(Schwartz II p.93)\ Let $\left(  E,\overrightarrow{E}\right)  ,\left(
F,\overrightarrow{F}\right)  ,\left(  G,\overrightarrow{G}\right)  $ be affine
normed spaces on the same field K, $\Omega$ an open subset of E. If the map
$f:\Omega\rightarrow F$ is differentiable at $a\in E,$ and the map :
$g:F\rightarrow G$ is differentiable at $b=f(a)$, then the map $g\circ
f:\Omega\rightarrow G$ is differentiable at $a$ and :

$\left(  g\circ f\right)  ^{\prime}\left(  a\right)  =g^{\prime}\left(
b\right)  \circ f^{\prime}\left(  a\right)  \in%
\mathcal{L}%
\left(  \overrightarrow{E};\overrightarrow{G}\right)  $
\end{theorem}

Let us write : $y=f(x),z=g(y).$ Then $g^{\prime}\left(  b\right)  $ is the
derivative of g with respect to y, computed in b=f(a), and f'(a) is the
derivative of f with respect to x, computed in x=a.

If the spaces are finite dimensional then the jacobian of $g\circ f$ is the
product of the jacobians.

If E is an affine normed space and $f\in%
\mathcal{L}%
\left(  E;E\right)  $ is continuously differentiable. Consider the iterate
$F_{n}=\left(  f\right)  ^{n}=\left(  f\circ f\circ...f\right)  =F_{n-1}\circ
f$. By recursion : $F_{n}^{\prime}\left(  a\right)  =\left(  f^{\prime}\left(
a\right)  \right)  ^{n}$ the n iterate of the linear map f'(a)

\paragraph{Derivatives on the spaces of linear maps\newline}

\begin{theorem}
If E is a normed vector space, then the set $%
\mathcal{L}%
\left(  E;E\right)  $ of continuous endomorphisms is a normed vector space and
the composition : $M:%
\mathcal{L}%
\left(  E;E\right)  \times%
\mathcal{L}%
\left(  E;E\right)  \rightarrow%
\mathcal{L}%
\left(  E;E\right)  ::M(f,g)=f\circ g$ is a bilinear, continuous map $M\in%
\mathcal{L}%
^{2}\left(
\mathcal{L}%
\left(  E;E\right)  ;%
\mathcal{L}%
\left(  E;E\right)  \right)  $ thus it is differentiable and the derivative of
M at (f,g) is: $M^{\prime}\left(  f,g\right)  \left(  \delta f,\delta
g\right)  =\delta f\circ g+f\circ\delta g$
\end{theorem}

This is the application of the previous theorem.

\begin{theorem}
(Schwartz II p.181) Let E,F be Banach \textit{vector} spaces, $G%
\mathcal{L}%
\left(  E;F\right)  $\ the subset of invertible elements of
$\mathcal{L}$%
(E;F), $G%
\mathcal{L}%
\left(  F;E\right)  $ the subset of invertible elements of
$\mathcal{L}$%
(F;E), then :

i) $G%
\mathcal{L}%
\left(  E;F\right)  ,G%
\mathcal{L}%
\left(  F;E\right)  $ are open subsets

ii) the map $\Im:G%
\mathcal{L}%
\left(  E;F\right)  \rightarrow G%
\mathcal{L}%
\left(  F;E\right)  ::\Im(f)=f^{-1}$ is a $C_{\infty}-$diffeomorphism
(bijective, continuously differentiable at any order as its inverse). Its
derivative at f is : $\delta f\in G%
\mathcal{L}%
\left(  E;F\right)  :\left(  \Im(f)\right)  ^{\prime}\left(  \delta f\right)
=-f^{-1}\circ\left(  \delta f\right)  \circ f^{-1}$
\end{theorem}

\begin{theorem}
The set $G%
\mathcal{L}%
\left(  E;E\right)  $\ of continuous automorphisms of a Banach vector space E
is an open subset of $%
\mathcal{L}%
\left(  E;E\right)  .$

i) the composition law : $M:%
\mathcal{L}%
\left(  E;E\right)  \times%
\mathcal{L}%
\left(  E;E\right)  \rightarrow%
\mathcal{L}%
\left(  E;E\right)  ::M(f,g)=f\circ g $ is differentiable and

$M^{\prime}\left(  f,g\right)  \left(  \delta f,\delta g\right)  =\delta
f\circ g+f\circ\delta g$

ii) the map : $\Im:G%
\mathcal{L}%
\left(  E;E\right)  \rightarrow G%
\mathcal{L}%
\left(  E;E\right)  $ is differentiable and

$\left(  \Im(f)\right)  ^{\prime}\left(  \delta f\right)  =-f^{-1}\circ\delta
f\circ f^{-1}$
\end{theorem}

\paragraph{Diffeomorphism\newline}

\begin{definition}
A map $f:\Omega\rightarrow\Omega^{\prime}$\ between open subsets of the affine
normed spaces on the same field K, is a \textbf{diffeomorphism} if f is
bijective, continuously differentiable in $\Omega$, and $f^{-1}$\ is
continuously differentiable in $\Omega^{\prime}$.
\end{definition}

\begin{definition}
A map $f:\Omega\rightarrow\Omega^{\prime}$\ between open subsets of the affine
normed spaces on the same field K, is a \textbf{local} \textbf{diffeomorphism}
if for any $a\in\Omega$ there are a neighborhood $n(a)$ of $a$ and $n(b)$ of
$b=f(a)$ such that f is a diffeomorphism from $n(a)$ to $n(b)$
\end{definition}

A diffeomorphism is a homeomorphism, thus if E,F are finite dimensional we
have necessarily $\dim E=\dim F$. Then the jacobian of $f^{-1}$\ is the
inverse of the jacobian of f and $\det\left(  f^{\prime}(a)\right)  \neq0.$

\begin{theorem}
(Schwartz II p.96) If $f:\Omega\rightarrow\Omega^{\prime}$\ between open
subsets of the affine normed spaces on the same field K, is a diffeomorphism
then $\forall a\in\Omega,b=f\left(  a\right)  :\left(  f^{\prime}\left(
a\right)  \right)  ^{-1}=\left(  f^{-1}\right)  ^{\prime}(b)$
\end{theorem}

\begin{theorem}
(Schwartz II p.190) If the map $f:\Omega\rightarrow F$ from the open subset
$\Omega$ of the Banach affine space $\left(  E,\overrightarrow{E}\right)  $ to
the Banach affine space $\left(  F,\overrightarrow{F}\right)  $ is
continuously differentiable in $\Omega$ then :

i) if for $a\in\Omega$ the derivative $f^{\prime}(a)$ is invertible in $%
\mathcal{L}%
\left(  \overrightarrow{E};\overrightarrow{F}\right)  $ then there are A open
in E, B open in F, $a\in A,b=f(a)\in B,$ such that f is a diffeormorphism from
A to B and $\left(  f^{\prime}\left(  a\right)  \right)  ^{-1}=\left(
f^{-1}\right)  ^{\prime}(b)$

ii) If, for any $a\in\Omega$, $f^{\prime}(a)$ is invertible in $%
\mathcal{L}%
\left(  \overrightarrow{E};\overrightarrow{F}\right)  $ then f is an open map
and a local diffeomorphism in $\Omega$.

iii) If f is injective and for any $a\in\Omega$ $\ f^{\prime}(a)$ is
invertible then f is a diffeomorphism from $\Omega$ to $f\left(
\Omega\right)  $
\end{theorem}

\begin{theorem}
(Schwartz II p.192) If the map $f:\Omega\rightarrow F$ from the open subset
$\Omega$ of the Banach affine space $\left(  E,\overrightarrow{E}\right)  $ to
the normed affine space $\left(  F,\overrightarrow{F}\right)  $ is
continuously differentiable in $\Omega$ and $\forall x\in\Omega$ \ $f^{\prime
}(x)$ is invertible then f is a local homeomorphism on $\Omega.$ As a
consequence f is an open map and $f\left(  \Omega\right)  $ is open.
\end{theorem}

\paragraph{Immersion, submersion\newline}

\begin{definition}
A continuously differentiable map $f:\Omega\rightarrow F$\ between an open
subset of the affine normed space E to the affine normed space F, both on the
same field K, is

an \textbf{immersion} at $a\in\Omega$ if f'(a) is injective.

a \textbf{submersion} at $a\in\Omega$ if f'(a) is surjective.

a submersion (resp.immersion) on $\Omega$ is a submersion (resp.immersion) at
every point of $\Omega$
\end{definition}

\begin{theorem}
(Schwartz II p.193) If $f:\Omega\rightarrow F$\ between an open subset of the
affine Banach E to the affine Banach F is a submersion at $a\in\Omega$ then
the image of a neighborhood of a is a neighborhood of f(a). If f is a
submersion on $\Omega$\ then it is an open map.
\end{theorem}

\begin{theorem}
(Lang p.18) If the continuously differentiable map $f:\Omega\rightarrow
F$\ between an open subset of E to F, both Banach vector spaces on the same
field K, is such that f'(p) is an isomorphism, continuous as its inverse, from
E to a closed subspace $F_{1}$ of F and $F=F_{1}\oplus F_{2},$ then there is a
neighborhood n(p) such that $\pi_{1}\circ f$ is a diffeomorphism from n(p) to
an open subset of $F_{1}$, with $\pi_{1}$ the projection of F to $F_{1}.$
\end{theorem}

\begin{theorem}
(Lang p.19) If the continuously differentiable map $f:\Omega\rightarrow
F$\ between an open subset of $E=E_{1}\oplus E_{2}$ to F, both Banach vector
spaces on the same field K, is such that the partial derivative $\partial
_{x_{1}}f\left(  p\right)  $ is an isomorphism, continuous as its inverse,
from $E_{1}$ to F$,$ then there is a neighborhood n(p) where $f=$ $f\circ
\pi_{1}$ with $\pi_{1}$ the projection of E to $E_{1}.$
\end{theorem}

\begin{theorem}
(Lang p.19) If the continuously differentiable map $f:\Omega\rightarrow
F$\ between an open subset of E to F, both Banach vector spaces on the same
field K, is such that f'(p) is surjective and $E=E_{1}\oplus\ker f^{\prime
}(p)$, then there is a neighborhood n(p) where $f=$ $f\circ\pi_{1}$ with
$\pi_{1}$ the projection of E to $E_{1}.$
\end{theorem}

\paragraph{Rank of a map\newline}

\begin{definition}
The\textbf{\ rank} of a differentiable map is the rank of its derivative.
\end{definition}

$rank(f)|_{a}=rank(f\prime(a))=$ $\dim f^{\prime}(a)\overrightarrow{E}\leq
\min\left(  \dim\overrightarrow{E},\dim\overrightarrow{F}\right)  $

If E,F are finite dimensional the rank of f in a is the rank of the jacobian.

\begin{theorem}
Constant rank (Schwartz II p.196) Let f be a continuously differentiable map
$f:\Omega\rightarrow F$\ between an open subset of the affine normed space
$\left(  E,\overrightarrow{E}\right)  $ to the affine normed space $\left(
F,\overrightarrow{F}\right)  $, both finite dimensional on the same field K.\ Then:

i) If f has rank r at $a\in\Omega$, there is a neighborhood $n(a)$ such that f
has rank $\geq r$ in $n(a)$

ii) if f is an immersion or a submersion at $a\in\Omega$ then f has a constant
rank in a neighborhood n(a)

iii) if f has constant rank r in $\Omega$\ then there are a bases in
$\overrightarrow{E}$ and $\overrightarrow{F}$ such that f can be expressed as :

$F\left(  x_{1},...,x_{m}\right)  =\left(  x_{1},..,x_{r},0,...0\right)  $
\end{theorem}

\paragraph{Derivative of a map defined by a sequence\newline}

\begin{theorem}
(Schwartz II p.122) If the sequence $\left(  f_{n}\right)  _{n\in%
\mathbb{N}
}$ of differentiable (resp.continuously differentiable) maps : $f_{n}%
:\Omega\rightarrow F$ from an open subset $\Omega$ of the normed affine space
E, to the normed affine space F, both on the same field K, converges to f and
if for each $a\in\Omega$\ there is a neighborhood where the sequence
$f_{n}^{\prime}$ converges uniformly to g, then f is differentiable
(resp.continuously differentiable) in $\Omega$ and f'=g
\end{theorem}

\begin{theorem}
(Schwartz II p.122) If the sequence $\left(  f_{n}\right)  _{n\in%
\mathbb{N}
}$ of differentiable (resp.continuously differentiable) maps : $f_{n}%
:\Omega\rightarrow F$ from an open connected subset $\Omega$ of the normed
affine space E, to the Banach affine space F, both on the same field K,
converges to f(a) at least at a point $b\in\Omega$ , and if for each
$a\in\Omega$\ there is a neighborhood where the sequence $f_{n}^{\prime}$
converges uniformly to g, then $f_{n}$ converges locally uniformly to f in
$\Omega,$ f is differentiable (resp.continuously differentiable) in $\Omega$
and f'=g
\end{theorem}

\begin{theorem}
Logarithmic derivative (Schwartz II p.130) If the sequence $\left(
f_{n}\right)  _{n\in%
\mathbb{N}
}$ of continuously differentiable maps : $f_{n}:\Omega\rightarrow%
\mathbb{C}
$ on an open connected subset $\Omega$ of the normed affine space E are never
null on $\Omega,$ and for each $a\in\Omega$\ there is a neighborhood where the
sequence $\left(  f_{n}^{\prime}\left(  a\right)  /f_{n}\left(  a\right)
\right)  $ converges uniformly to g, if there is $b\in\Omega$ such that
$\left(  f_{n}\left(  b\right)  \right)  _{n\in%
\mathbb{N}
}$ converges to a non zero limit, then $\left(  f_{n}\right)  _{n\in%
\mathbb{N}
}$ converges to a function f which is continuously differentiable over
$\Omega$ , never null and g=f'/f
\end{theorem}

f'/f is called the \textbf{logarithmic derivative}

\paragraph{Derivative of a function defined by an integral\newline}

\begin{theorem}
(Schwartz IV p.107) Let E be an affine normed space, $\mu$ a Radon measure on
a topological space T, $f\in C\left(  E\times T;F\right)  $ with F a banach
vector space. If $f(.,t)$ is $x$ differentiable for almost every t, if for
almost every $a$ in T $\frac{\partial f}{\partial x}\left(  a,t\right)  $ is
$\mu-$measurable and there is a neighborhood $n(a)$ in E such that $\left\Vert
\frac{\partial f}{\partial x}\left(  x,t\right)  \right\Vert \leq k(t)$ in
$n(a)$ where $k(t)\geq0$ is integrable on T, then the map : $u\left(
x\right)  =\int_{T}f\left(  x,t\right)  \mu\left(  t\right)  $ is
differentiable in E and its derivative is : $\frac{du}{dx}\left(  a\right)
=\int_{T}\frac{\partial f}{\partial x}\left(  x,t\right)  \mu\left(  t\right)
. $ If $f(.,t)$ is continuously $x$ differentiable then $u$ is continuously differentiable.
\end{theorem}

\begin{theorem}
(Schwartz IV p.109) Let E be an affine normed space, $\mu$ a Radon measure on
a topological space T, f a continuous map from E$\times$T in a Banach vector
space F. If f has a continuous partial derivative with respect to $x$, if for
almost every $a$ in T there is a compact neighborhood $K(a)$ in E such that
the support of $\frac{\partial f}{\partial x}\left(  x,t\right)  $ is in
$K(a)$, then the function : $u\left(  x\right)  =\int_{T}f\left(  x,t\right)
\mu\left(  t\right)  $ is continuously differentiable in E and its derivative
is : $\frac{du}{dx}\left(  a\right)  =\int_{T}\frac{\partial f}{\partial
x}\left(  x,t\right)  \mu\left(  t\right)  .$
\end{theorem}

\paragraph{Gradient\newline}

If $f\in C_{1}\left(  \Omega;K\right)  $ with $\Omega$\ an open subset of the
affine normed space E on the field K, then $f^{\prime}(a)\in\overrightarrow
{E}^{\prime}$ the topological dual of $\overrightarrow{E}.$ If E is finite
dimensional and there is, either a bilinear symmetric or an hermitian form g,
non degenerate on $\overrightarrow{E}$, then there is an isomorphism between
$\overrightarrow{E}$ and $\overrightarrow{E}^{\prime}.$ To f'(a) we can
associate a vector, called \textbf{gradient} and denoted $grad_{a}f$ such that :%

\begin{equation}
\forall\overrightarrow{u}\in\overrightarrow{E}:f^{\prime}(a)\overrightarrow
{u}=g\left(  grad_{a}f,\overrightarrow{u}\right)
\end{equation}

If f is continuously differentiable then the map : $grad:\Omega\rightarrow
\overrightarrow{E}$ defines a vector field on E.

\bigskip

\subsection{Higher order derivatives}

\label{Higher order derivatives}

\subsubsection{Definitions}

\paragraph{Definition\newline}

\begin{theorem}
(Schwartz II p.136) If the map $f:\Omega\rightarrow F$ from the open subset
$\Omega$ of the normed affine space $\left(  E,\overrightarrow{E}\right)  $ to
the normed affine space $\left(  F,\overrightarrow{F}\right)  $ is
continuously differentiable in $\Omega$ and its derivative map f' is
differentiable in $a\in\Omega$ then f"(a) is a \textbf{continuous symmetric}
\textbf{bilinear} map in $%
\mathcal{L}%
^{2}(\overrightarrow{E};\overrightarrow{F})$
\end{theorem}

We have the map $f^{\prime}:\Omega\rightarrow%
\mathcal{L}%
\left(  \overrightarrow{E};\overrightarrow{E}\right)  $ and its derivative in
$a$ : $f"(a)=(f^{\prime}(x))|_{x=a}$ is a continuous linear map : $f"(a):$
$\overrightarrow{E}\rightarrow%
\mathcal{L}%
\left(  \overrightarrow{E};\overrightarrow{F}\right)  .$ Such a map is
equivalent to a continuous bilinear map in $%
\mathcal{L}%
^{2}(\overrightarrow{E};\overrightarrow{F})$. So we usually consider the map
f"(a) as a bilinear map valued in $\overrightarrow{F}.$ This bilinear map is
symmetric : $f"(a)(\overrightarrow{u},\overrightarrow{v}%
)=f"(a)(\overrightarrow{v},\overrightarrow{u})$

This definition can be extended by recursion to the derivative of order r.

\begin{definition}
The map $f:\Omega\rightarrow F$ from the open subset $\Omega$ of the normed
affine space $\left(  E,\overrightarrow{E}\right)  $ to the normed affine
space $\left(  F,\overrightarrow{F}\right)  $ is \textbf{r continuousy
differentiable} in $\Omega$ if it is continuously differentiable and its
derivative map f' is r-1 differentiable in $\Omega$ . Then its r order
derivative $f^{\left(  r\right)  }\left(  a\right)  $ in $a\in\Omega$\ is a
continuous symmetric r\textbf{\ }linear map in $%
\mathcal{L}%
^{r}(\overrightarrow{E};\overrightarrow{F})$.
\end{definition}

If f is r-continuously differentiable, whatever r, it is said to be
\textbf{smooth}

\begin{notation}
$C_{r}\left(  \Omega;F\right)  $ is the set of continuously r-differentiable
maps $f:\Omega\rightarrow F$.
\end{notation}

\begin{notation}
$C_{\infty}\left(  \Omega;F\right)  $ is the set of smooth maps $f:\Omega
\rightarrow F$
\end{notation}

\begin{notation}
$f^{\left(  r\right)  }$ is the r order derivative of f : $f^{\left(
r\right)  }:\Omega\rightarrow%
\mathcal{L}%
_{S}^{r}(\overrightarrow{E};\overrightarrow{F})$
\end{notation}

\begin{notation}
$f"$ is the 2nd order derivative of f : $f":\Omega\rightarrow%
\mathcal{L}%
_{S}^{2}(\overrightarrow{E};\overrightarrow{F})$
\end{notation}

\begin{notation}
$f^{\left(  r\right)  }\left(  a\right)  $ is the value at a of the r order
derivative of f : $f^{\left(  r\right)  }\left(  a\right)  \in%
\mathcal{L}%
_{S}^{r}(\overrightarrow{E};\overrightarrow{F})$
\end{notation}

\paragraph{Partial derivatives\newline}

\begin{definition}
A map $f:\Omega\rightarrow F$ defined on an open subset $\Omega$ of the normed
affine space $\left(  E,\overrightarrow{E}\right)  =%
{\textstyle\prod\limits_{i=1}^{r}}
\left(  E_{i},\overrightarrow{E}_{i}\right)  $ and valued in the normed affine
space $\left(  F,\overrightarrow{F}\right)  ,$ all on the same field K, has a
\textbf{partial derivative} \textbf{of order 2} in $\Omega$ with respect to
the variables $x_{k}=\pi_{k}\left(  x\right)  ,x_{l}=\pi_{l}\left(  x\right)
$ where $\pi_{k}:E\rightarrow E_{k}$ is the canonical projection, if f has a
partial derivative with respect to the variable $x_{k}$ \ in $\Omega$ and\ the
map $f_{x_{k}}^{\prime}$ has a partial derivative with respect to the variable
$x_{l}.$
\end{definition}

The partial derivatives must be understood as follows :

1. Let $E=E_{1}\times E_{2}$ and $\Omega=\Omega_{1}\times\Omega_{2}$\ . We
consider the map $f:\Omega\rightarrow F$ as a two variables map $f(x_{1}%
,x_{2}).$

For the first derivative we proceed as above. Let us fix $x_{1}=a_{1}$ so we
have a map : $f\left(  a_{1},x_{2}\right)  :\Omega_{2}\rightarrow F$ for
$\Omega_{2}=\left\{  x_{2}\in E_{2}:\left(  a_{1},x_{2}\right)  \in
\Omega\right\}  $. Its partial derivative with respect to $x_{2}$ at $a_{2}$
is the map $f_{x_{2}}^{\prime}\left(  a_{1},a_{2}\right)  \in%
\mathcal{L}%
\left(  \overrightarrow{E}_{2};\overrightarrow{F}\right)  $

Now allow $x_{1}=a_{1}$ to move in $E_{1}$ (but keep $a_{2}$ fixed)$.$ So we
have a map : $f_{x_{2}}^{\prime}\left(  x_{1},a_{2}\right)  :\Omega
_{1}\rightarrow%
\mathcal{L}%
\left(  \overrightarrow{E}_{2};\overrightarrow{F}\right)  $ for $\Omega
_{1}=\left\{  x_{1}\in E_{1}:\left(  x_{1},a_{2}\right)  \in\Omega\right\}  .$
Its partial derivative with respect to $x_{1}$ is a map : $f"_{x_{1}x_{2}%
}\left(  a_{1},a_{2}\right)  :\overrightarrow{E}_{1}\rightarrow%
\mathcal{L}%
\left(  \overrightarrow{E}_{2};\overrightarrow{F}\right)  $ that we assimilate
to a map $f"_{x_{1}x_{2}}\left(  a_{1},a_{2}\right)  \in%
\mathcal{L}%
^{2}\left(  \overrightarrow{E}_{1},\overrightarrow{E}_{2};\overrightarrow
{F}\right)  $

If f is 2 times differentiable in $\left(  a_{1},a_{2}\right)  $ the result
does not depend on the order for the derivations : $f"_{x_{1}x_{2}}\left(
a_{1},a_{2}\right)  =f"_{x_{2}x_{1}}\left(  a_{1},a_{2}\right)  $

We can proceed also to $f"_{x_{1}x_{1}}\left(  a_{1},a_{2}\right)
,f"_{x_{2}x_{2}}\left(  a_{1},a_{2}\right)  $ so we have 3 distinct partial
derivatives with respect to all the combinations of variables.

2. The partial derivatives are symmetric bilinear maps which act on different
vector spaces:

$f"_{x_{1}x_{2}}\left(  a_{1},a_{2}\right)  \in%
\mathcal{L}%
^{2}\left(  \overrightarrow{E}_{1},\overrightarrow{E}_{2};\overrightarrow
{F}\right)  $

$f"_{x_{1}x_{1}}\left(  a_{1},a_{2}\right)  \in%
\mathcal{L}%
^{2}\left(  \overrightarrow{E}_{1},\overrightarrow{E}_{1};\overrightarrow
{F}\right)  $

$f"_{x_{2}x_{2}}\left(  a_{1},a_{2}\right)  \in%
\mathcal{L}%
^{2}\left(  \overrightarrow{E}_{2},\overrightarrow{E}_{2};\overrightarrow
{F}\right)  $

A vector in $\overrightarrow{E}=\overrightarrow{E}_{1}\times\overrightarrow
{E}_{2}$ can be written as : $\overrightarrow{u}=\left(  \overrightarrow
{u}_{1},\overrightarrow{u}_{2}\right)  $

The action of the first derivative map $f^{\prime}(a_{1},a_{2})$ is just :
$f^{\prime}(a_{1},a_{2})\left(  \overrightarrow{u}_{1},\overrightarrow{u}%
_{2}\right)  =f_{x_{1}}^{\prime}(a_{1},a_{2})\overrightarrow{u}_{1}+f_{x_{2}%
}^{\prime}(a_{1},a_{2})\overrightarrow{u}_{2}$

The action of the second derivative map $f"(a_{1},a_{2})$ is now on the two
vectors $\overrightarrow{u}=\left(  \overrightarrow{u}_{1},\overrightarrow
{u}_{2}\right)  ,$ $\overrightarrow{v}=\left(  \overrightarrow{v}%
_{1},\overrightarrow{v}_{2}\right)  $

$f"(a_{1},a_{2})\left(  \left(  \overrightarrow{u}_{1},\overrightarrow{u}%
_{2}\right)  ,\left(  \overrightarrow{v}_{1},\overrightarrow{v}_{2}\right)
\right)  $

$=f"_{x_{1}x_{1}}\left(  a_{1},a_{2}\right)  \left(  \overrightarrow{u}%
_{1},\overrightarrow{v}_{1}\right)  +f"_{x_{1}x_{2}}\left(  a_{1}%
,a_{2}\right)  \left(  \overrightarrow{u}_{1},\overrightarrow{v}_{2}\right)
+f"_{x_{2}x_{1}}\left(  a_{1},a_{2}\right)  \left(  \overrightarrow{u}%
_{2},\overrightarrow{v}_{1}\right)  +f"_{x_{2}x_{2}}\left(  a_{1}%
,a_{2}\right)  \left(  \overrightarrow{u}_{2},\overrightarrow{v}_{2}\right)  $

\begin{notation}
$f_{x_{i_{1}}...x_{i_{r}}}^{\left(  r\right)  }=\frac{\partial^{r}f}{\partial
x_{i_{1}}...\partial x_{i_{r}}}=D_{i_{1}...i_{r}}$ is the r order partial
derivative map : $\Omega\rightarrow%
\mathcal{L}%
^{r}\left(  \overrightarrow{E}_{i_{1}},..\overrightarrow{E}_{i_{r}%
};\overrightarrow{F}\right)  $ with respect to $x_{i_{1}},...x_{i_{r}}$
\end{notation}

\begin{notation}
$f_{x_{i_{1}}...x_{i_{r}}}^{\left(  r\right)  }\left(  a\right)
=\frac{\partial^{r}f}{\partial x_{i_{1}}...\partial x_{i_{r}}}\left(
a\right)  =D_{i_{1}...i_{r}}\left(  a\right)  $ is the value of the partial
derivative map at $a\in\Omega$
\end{notation}

\paragraph{Condition for r-differentiability\newline}

\begin{theorem}
(Schwartz II p.142) A map $f:\Omega\rightarrow F$ defined on an open subset
$\Omega$ of the normed affine space $\left(  E,\overrightarrow{E}\right)  =%
{\textstyle\prod\limits_{i=1}^{r}}
\left(  E_{i},\overrightarrow{E}_{i}\right)  $ and valued in the normed affine
space $\left(  F,\overrightarrow{F}\right)  ,$ all on the same field K, is
continuously r differentiable in $\Omega$\ iff it has continuous partial
derivatives of order r with respect to every combination of r variables.in
$\Omega.$
\end{theorem}

\paragraph{Coordinates expression\newline}

If E is m dimensional and F n dimensional, the map $f_{x_{i_{1}}...x_{i_{r}}%
}^{\left(  r\right)  }\left(  a\right)  $ for r
$>$
1 is no longer represented by a matrix. This is a r covariant and 1
contravariant tensor in $\otimes_{r}\overrightarrow{E}^{\ast}\otimes F.$

With a basis $\left(  e^{i}\right)  _{i=1}^{m}$ of $\overrightarrow{E}^{\ast}$
and $\ \left(  f_{j}\right)  _{i=1}^{n}$ of $\overrightarrow{F}$ $:$%

\begin{equation}
f_{x_{i_{1}}...x_{i_{r}}}^{\left(  r\right)  }\left(  a\right)  =\sum
_{i_{1}...i_{r}=1}^{m}\sum_{j=1}^{n}T_{i_{1}...i_{r}}^{j}\left(  a\right)
e^{i_{1}}\otimes...\otimes e^{i_{r}}\otimes f_{j}%
\end{equation}

and $T_{i_{1}...i_{r}}^{j}$ is symmetric in all the lower indices.

\subsubsection{Properties of higher derivatives}

\paragraph{Polynomial\newline}

\begin{theorem}
If f is a continuous affine map $f:E\rightarrow F$ with associated linear map
$\overrightarrow{f}\in%
\mathcal{L}%
\left(  \overrightarrow{E};\overrightarrow{F}\right)  $ then f is smooth and
$f^{\prime}$=$\overrightarrow{f},f^{\left(  r\right)  }=0,r>1$
\end{theorem}

\begin{theorem}
A polynomial P of degree p in n variables over the field K, defined in an open
subset $\Omega\subset K^{n}$ is smooth. $f^{\left(  r\right)  }\equiv0$ if p
$<$
r
\end{theorem}

\begin{theorem}
(Schwartz II p.164) A map $f:\Omega\rightarrow K$ from an open connected
subset in $K^{n}$ has a r order derivative $f^{\left(  r\right)  }\equiv0$ in
$\Omega$ iff it is a polynomial of order
$<$
r.
\end{theorem}

\paragraph{Leibniz's formula\newline}

\begin{theorem}
(Schwartz II p.144) Let $E,E_{1},E_{2},F$ be normed vector spaces, $\Omega$ an
open subset of E, $B\in%
\mathcal{L}%
^{2}(E_{1},E_{2};F),$ $U_{1}\in C_{r}\left(  \Omega;E_{1}\right)  ,U_{2}\in
C_{r}\left(  \Omega;E_{2}\right)  $, then the map : $B\left(  U_{1}%
,U_{2}\right)  :\Omega\rightarrow F::B\left(  U_{1}\left(  x\right)
,U_{2}\left(  x\right)  \right)  $ is r-continuously differentiable in
$\Omega$ .
\end{theorem}

If E is n-dimensional,.with the notation above it reads :

$D_{i_{1}...i_{r}}B\left(  U_{1},U_{2}\right)  =\sum_{J\sqsubseteq\left(
i_{1}...i_{r}\right)  }B\left(  D_{J}U_{1},D_{\left(  i_{1}...i_{r}\right)
\backslash J}U_{2}\right)  $

the sum is extended to all combinations J of indices in $I=\left(
i_{1}...i_{r}\right)  $

\paragraph{Differential operator\newline}

(see Functional analysis for more)

\begin{definition}
If $\Omega$ is an open subset of a normed affine space E, $\overrightarrow{F}$
a normed \textit{vector} space, a differential operator of order $m\leq r$ is
a map : $P\in%
\mathcal{L}%
\left(  C_{r}\left(  \Omega;\overrightarrow{F}\right)  ;C_{r}\left(
\Omega;\overrightarrow{F}\right)  \right)  ::P\left(  f\right)  =\sum
_{I,}a_{I}D_{I}f$
\end{definition}

the sum is taken over any set I of m indices in (1,2,...n), the coefficients
are scalar functions $a_{I}:\Omega\rightarrow K$

Example : laplacian : $P\left(  f\right)  =\sum_{i=1}^{n}\frac{\partial^{2}%
f}{\partial x_{i}^{2}}$

If the coefficients are constant, scalars differential operators
($\overrightarrow{F}=K$) can be composed: $\left(  P\circ Q\right)  \left(
f\right)  =P\left(  Q\left(  f\right)  \right)  $ as long as the resulting
maps are differentiable, and the composition is commutative : $\left(  P\circ
Q\right)  \left(  f\right)  =\left(  Q\circ P\right)  \left(  f\right)  $

\paragraph{Taylor's formulas\newline}

\begin{theorem}
(Schwartz II p.155) If $\Omega$ is an open of an affine space E,
$\overrightarrow{F}$ a normed vector space, both on the same field K, $f\in
C_{r-1}\left(  \Omega;\overrightarrow{F}\right)  $ and has a derivative
$f^{\left(  r\right)  }$\ in $a\in\Omega,$ then, for $h\in\overrightarrow{E}%
$\ such that the segment $\left[  a,a+h\right]  \subset\Omega:$

i). $f\left(  a+h\right)  =f\left(  a\right)  +\sum_{k=1}^{r-1}\frac{1}%
{k!}f^{\left(  k\right)  }(a)h^{k}+\frac{1}{r!}f^{\left(  r\right)  }\left(
a+\theta h\right)  h^{r}$ with $\theta\in\left[  0,1\right]  $

ii) $f\left(  a+h\right)  =f\left(  a\right)  +\sum_{k=1}^{r}\frac{1}%
{k!}f^{\left(  k\right)  }(a)h^{k}+\frac{1}{r!}\varepsilon\left(  h\right)
\left\Vert h\right\Vert ^{r}$

with $\varepsilon\left(  h\right)  \in\overrightarrow{F},\varepsilon\left(
h\right)  _{h\rightarrow0}\rightarrow0$

iii) If $\forall x\in]a,a+h[:\exists f^{\left(  r\right)  }\left(  x\right)
$,$\left\Vert f^{\left(  r\right)  }\left(  x\right)  \right\Vert \leq M$ then :

$\left\Vert f\left(  a+h\right)  -\sum_{k=0}^{r-1}\frac{1}{k!}f^{\left(
k\right)  }\left(  a\right)  h^{k}\right\Vert \leq M\frac{1}{r!}\left\Vert
h\right\Vert ^{r}$
\end{theorem}

with the notation : $f^{\left(  k\right)  }\left(  a\right)  h^{k}=f^{\left(
k\right)  }\left(  a\right)  \left(  h,...h\right)  $ k times

If E is m dimensional, in a basis :

$\sum_{k=0}^{r}\frac{1}{k!}f^{\left(  k\right)  }(a)h^{k}=\sum_{\left(
\alpha_{1}...\alpha_{m}\right)  }\frac{1}{\alpha_{1}!...\alpha_{m}!}\left(
\frac{\partial}{\partial x_{1}}\right)  ^{\alpha_{1}}...\left(  \frac
{\partial}{\partial x_{m}}\right)  ^{\alpha_{m}}f\left(  a\right)
h_{1}^{\alpha_{1}}..h_{m}^{\alpha_{m}}$

where the sum is extended to all combinations of integers such that
$\sum_{k=1}^{m}\alpha_{k}\leq r$

\paragraph{Chain rule\newline}

If $f,g\in C_{r}\left(
\mathbb{R}
;%
\mathbb{R}
\right)  :$

$\left(  g\circ f\right)  ^{\left(  r\right)  }\left(  a\right)  =\sum_{I_{r}%
}\frac{r!}{i_{1}!i_{2}!...i_{r}!}g^{\left(  r\right)  }\left(  f\left(
a\right)  \right)  \left(  f^{\prime}\left(  a\right)  \right)  ^{i_{1}%
}...\left(  f^{\left(  r\right)  }\left(  a\right)  \right)  ^{i_{r}}$

where $I_{r}=\left(  i_{1},..i_{r}\right)  :i_{1}+i_{2}+..+i_{r}=r$,
$f^{\left(  p\right)  }\in%
\mathcal{L}%
^{p}\left(  \overrightarrow{%
\mathbb{R}
};%
\mathbb{R}
\right)  $

\paragraph{Convex functions\newline}

\begin{theorem}
(Berge p.209) A 2 times differentiable function $f:C\rightarrow%
\mathbb{R}
$ on a convex subset of $%
\mathbb{R}
^{m}$ is convex iff $\forall a\in C:f"\left(  a\right)  $ is a positive
bilinear map.
\end{theorem}

\bigskip

\subsection{Extremum of a function}

\label{Extremum of a function}

\subsubsection{Definitions}

E set, $\Omega$ subset of E, $f:\Omega\rightarrow%
\mathbb{R}
$

f has a \textbf{maximum} in $a\in\Omega$ if $\forall x\in\Omega:f(x)\leq
f\left(  a\right)  $

f has a \textbf{minimum} in $a\in\Omega$ if $\forall x\in\Omega:f(x)\geq
f\left(  a\right)  $

f has an \textbf{extremum} in $a\in\Omega$ if it has either a maximum of a
minimum in $a$

The extremum is :

\textbf{local} if it is an extremum in a neighborhood of $a$.\ 

\textbf{global} if it an extremum in the whole of $\Omega$

\subsubsection{General theorems}

\paragraph{Continuous functions\newline}

\begin{theorem}
A continuous real valued function $f:C\rightarrow%
\mathbb{R}
$ on a compact subset C of a topological space E has both a global maximum and
a global minimum.
\end{theorem}

\begin{proof}
$f\left(  \Omega\right)  $\ is compact in $%
\mathbb{R}
$, thus bounded and closed, so it has both an upper bound and a lower bound,
and on $%
\mathbb{R}
$\ this entails that is has a greatest lower bound and a least upper bound,
which must belong to $f\left(  \Omega\right)  $\ because it is closed.
\end{proof}

If f is continuous and C connected then f(C)\ is connected, thus is an
interval
$\vert$%
a,b%
$\vert$
with a,b possibly infinite. But it is possible that a or b are not met by f.

\paragraph{Convex functions\newline}

There are many theorems about extrema of functions involving convexity
properties. This is the basis of linear programming. See for example Berge for
a comprehensive review of the problem.

\begin{theorem}
If $f:C\rightarrow%
\mathbb{R}
$ is a stricly convex function defined on a convex subset of a real affine
space E, then a maximum of f is an extreme point of C$.$
\end{theorem}

\begin{proof}
C,f stricly convex :

$\forall M,P\in C,t\in\left[  0,1\right]  :f\left(  tM+\left(  1-t\right)
P\right)  <tf(M)+\left(  1-t\right)  f\left(  P\right)  $

If a is not an extreme point of C :

$\exists M,P\in C,t\in]0,1[:tM+\left(  1-t\right)  P=a\Rightarrow f\left(
a\right)  <tf(M)+\left(  1-t\right)  f\left(  P\right)  $

If f is a maximum : $\forall M,P:f\left(  a\right)  \geq f\left(  M\right)
,f\left(  a\right)  \geq f\left(  P\right)  $

$t\in]0,1[:tf\left(  a\right)  \geq tf\left(  M\right)  ,\left(  1-t\right)
f\left(  a\right)  \geq\left(  1-t\right)  f\left(  P\right)  $

$\Rightarrow f\left(  a\right)  \geq tf(M)+\left(  1-t\right)  f\left(
P\right)  $
\end{proof}

This theorem shows that for many functions the extrema do not lie in the
interior of the domain but at its border. So this limits seriously the
interest of the following theorems, based upon differentiability, which assume
that the domain is an open subset.

Another classic theorem (which has many versions) :

\begin{theorem}
Minimax (Berge p.220) If f is a continuous functions :$f:\Omega\times
\Omega^{\prime}\rightarrow%
\mathbb{R}
$ where $\Omega,\Omega^{\prime}$ are convex compact subsets of $%
\mathbb{R}
^{p},%
\mathbb{R}
^{q},$and f is concave in x and convex in y, then :

$\exists\left(  a,b\right)  \in\Omega\times\Omega^{\prime}:f(a,b)=\max
_{x\in\Omega}f(x,b)=\min_{y\in\Omega^{\prime}}f(a,y)$
\end{theorem}

\subsubsection{Differentiable functions}

\begin{theorem}
If a function $f:\Omega\rightarrow%
\mathbb{R}
$ ,differentiable in the open subset $\Omega$\ of a normed affine space, has a
local extremum in $a\in\Omega$ then f'(a)=0.
\end{theorem}

The proof is immediate with the Taylor's formula.

The converse is not true. It is common to say that f is \textbf{stationary}
(or that a is a \textbf{critical point}) in a if f'(a)=0, but this does not
entail that f has an extremum in a (but if $f^{\prime}(a)\neq0$ it is
\textit{not} an extremum). The condition open on $\Omega$\ is mandatory.

With the Taylor's formula the result can be precised :

\begin{theorem}
If a function $f:\Omega\rightarrow%
\mathbb{R}
$ ,r differentiable in the open subset $\Omega$\ of a normed affine space, has
a local extremum in $a\in\Omega$ and $f^{\left(  p\right)  }\left(  a\right)
=0,1\leq p<s\leq r,f^{\left(  s\right)  }\left(  a\right)  \neq0,$ then :

if $a$ is a local maximum, then s is even and $\forall h\in\overrightarrow
{E}:f^{\left(  s\right)  }\left(  a\right)  h^{s}\leq0$

if $a$ is a local minimum, then s is even and $\forall h\in\overrightarrow
{E}:f^{\left(  s\right)  }\left(  a\right)  h^{s}\geq0$
\end{theorem}

The condition is necessary, not sufficient. If s is odd then $a$ cannot be an extremum.

\subsubsection{Maximum under constraints}

They are the problems, common in engineering, to find the extremum of a map
belonging to some set defined through relations, which may be or not strict.

\paragraph{Extremum with strict constraints\newline}

\begin{theorem}
(Schwartz II p.285) Let $\Omega\ $be an open subset\ of a real affine normed
space,\ $f,L_{1},L_{2},...L_{m}$ real differentiable functions in $\Omega,$ A
the subset $A=\left\{  x\in\Omega:L_{k}\left(  x\right)  =0,k=1...m\right\}  $
. If $a\in A$ is a local extremum of f in A and the maps $L_{k}^{\prime
}\left(  a\right)  \in\overrightarrow{E}^{\prime}$ are linearly independant,
then there is a unique family of scalars $\left(  \lambda_{k}\right)
_{k=1}^{m}$ such that : $f^{\prime}(a)=\sum_{k=1}^{m}\lambda_{k}L_{k}^{\prime
}\left(  a\right)  $

if $\Omega$ is a convex set and the map $f\left(  x\right)  +\sum_{k=1}%
^{m}\lambda_{k}L_{k}\left(  x\right)  $ is concave then the condition is sufficient.
\end{theorem}

The $\lambda_{k}$ are the \textbf{Lagrange multipliers}. In physics they can
be interpreted as forces, and in economics as prices.

Notice that E can be infinite dimensional. This theorem can be restated as
follows :

\begin{theorem}
Let $\Omega\ $be an open subset $\Omega$\ of a real affine normed space
E,\ $f:\Omega\rightarrow%
\mathbb{R}
$,$L:\Omega\rightarrow F$ real differentiable functions in $\Omega,$ F a m
dimensional real vector space, A the set $A=\left\{  x\in\Omega:L\left(
x\right)  =0\right\}  .$ If $a\in A$ is a local extremum of f in A and if the
map $L^{\prime}\left(  a\right)  $ is surjective, then : $\exists\lambda\in
F^{\ast}$ such that : $f^{\prime}(a)=\lambda\circ L^{\prime}\left(  a\right)
$
\end{theorem}

\paragraph{Kuhn and Tucker theorem\newline}

\begin{theorem}
(Berge p.236) Let $\Omega\ $be an open subset of $%
\mathbb{R}
^{n}$,\ $f,L_{1},L_{2},...L_{m}$ real differentiable functions in $\Omega,$ A
the subset $A=\left\{  x\in\Omega:L_{k}\left(  x\right)  \leq
0,k=1...m\right\}  $ . If $a\in A$ is a local extremum of f in A and the maps
$L_{k}^{\prime}\left(  a\right)  \in\overrightarrow{E}^{\prime}$ are linearly
independant, then there is a family of scalars $\left(  \lambda_{k}\right)
_{k=1}^{m}$ such that :

k=1...m : $L_{k}\left(  a\right)  \leq0,\lambda_{k}\geq0,\lambda_{k}%
L_{k}\left(  a\right)  =0$

$f^{\prime}(a)+\sum_{k=1}^{m}\lambda_{k}L_{k}^{\prime}\left(  a\right)  =0$
\end{theorem}

If f is linear and L are affine functions this is the linear programming
problem :

Problem : find $a\in%
\mathbb{R}
^{n}$ extremum of $\left[  C\right]  ^{t}\left[  x\right]  $ with $\left[
A\right]  \left[  x\right]  \leq\left[  B\right]  ,\left[  A\right]  $ mxn
matrix, $\left[  B\right]  $\ mx1 matrix, $\left[  x\right]  \geq0$

An extremum point is necessarily on the border, and there are many computer
programs for solving the problem (simplex method).

\bigskip

\subsection{Implicit maps}

\label{Implicit map}

One classical problem in mathematics is to solve the equation f(x,y)=0 : find
x with respect to y.\ If there is a function g such that f(x,g(x))=0 then
y=g(x) is called the implicit function defined by f(x,y)=0. The fixed point
theorem in a Banach space is a key ingredient to resolve the problem. These
theorems are the basis of many other results in Differential Geometry and
Funcitonal Analysis. One important feature of the theorems below is that they
apply on infinite dimensional vector spaces (when they are Banach).

\paragraph{In a neighborhood of a solution\newline}

The first theorems apply when a specific solution of the equation f(a,b)=c is known.

\begin{theorem}
(Schwartz II p.176) Let E be a topological space, $\left(  F,\overrightarrow
{F}\right)  $ an affine Banach space, $\left(  G,\overrightarrow{G}\right)  $
an affine normed space, $\Omega$ an open in E$\times$F, f a continuous map
$f:\Omega\rightarrow G,$ $\left(  a,b\right)  \in E\times F,c\in G$ such that
c=f(a,b), if :

i) $\forall\left(  x,y\right)  \in\Omega$ f has a partial derivative map
$f\prime_{y}\left(  x,y\right)  \in%
\mathcal{L}%
\left(  \overrightarrow{F};\overrightarrow{G}\right)  $ and $\left(
x,y\right)  \rightarrow f_{y}^{\prime}\left(  x,y\right)  $ is continuous in
$\Omega$

ii) $Q=f_{y}^{\prime}\left(  a,b\right)  $ is invertible in $%
\mathcal{L}%
\left(  \overrightarrow{F};\overrightarrow{G}\right)  $

then there are neighborhoods $n(a)\subset E,n(b)\subset F$ of a,b such that :

for any $x\in n(a)$ there is a unique $y=g(x)\in n\left(  b\right)  $ such
that f(x,y)=c and g is continuous in n(a).
\end{theorem}

\begin{theorem}
(Schwartz II p.180) Let $\left(  E,\overrightarrow{E}\right)  ,\left(
F,\overrightarrow{F}\right)  ,\left(  G,\overrightarrow{G}\right)  $ be affine
normed spaces, $\Omega$ an open in E$\times$F, f a continuous map
$f:\Omega\rightarrow G,$ $\left(  a,b\right)  \in E\times F,c\in G$ such that
c=f(a,b), and the neighborhoods $n(a)\subset E,n(b)\subset F$ of a,b, if

i) there is a map $g:n(a)\rightarrow n(b)$ continuous at $a$ and such that
$\forall x\in n(a):f(x,g(x))=c$

ii) f is differentiable at $(a,b)$ and $f_{y}^{\prime}\left(  a,b\right)  $ invertible

then g is differentiable at $a$, and its derivative is : $g^{\prime}\left(
a\right)  =-\left(  f_{y}^{\prime}\left(  a,b\right)  \right)  ^{-1}%
\circ\left(  f_{x}^{\prime}\left(  a,b\right)  \right)  $
\end{theorem}

\paragraph{Implicit map theorem\newline}

\begin{theorem}
(Schwartz II p.185) Let $\left(  E,\overrightarrow{E}\right)  ,\left(
F,\overrightarrow{F}\right)  ,\left(  G,\overrightarrow{G}\right)  $ be affine
normed spaces, $\Omega$ an open in E$\times$F, $f:\Omega\rightarrow G$ a
continuously differentiable map in $\Omega,$

i) If there are A open in E, B open in F such that $A\times B\sqsubseteq
\Omega$ and $g:A\rightarrow B$ such that f(x,g(x))=c in A,

if $\forall x\in A:f_{y}^{\prime}\left(  x,g(x)\right)  $ is invertible in $%
\mathcal{L}%
\left(  \overrightarrow{F};\overrightarrow{G}\right)  $ then g is continuously
differentiable in A

if f is r-continuously differentiable then g is r-continuously differentiable

ii) If there are $\left(  a,b\right)  \in E\times F,c\in G$ such that
c=f(a,b), F is complete and $f_{y}^{\prime}\left(  a,b\right)  $ is invertible
in $%
\mathcal{L}%
\left(  \overrightarrow{F};\overrightarrow{G}\right)  ,$ then there are
neighborhoods $n(a)\subset A,n(b)\subset B$ of a,b such that $n(a)\times
n(b)\subset\Omega$ and for any $x\in n(a)$ there is a unique $y=g(x)\in
n\left(  b\right)  $ such that f(x,y)=c. g is continuously differentiable in
n(a) and its derivative is :$g^{\prime}\left(  x\right)  =-\left(
f_{y}^{\prime}\left(  x,y\right)  \right)  ^{-1}\circ\left(  f_{x}^{\prime
}\left(  x,y\right)  \right)  .$ If f is r-continuously differentiable then g
is r-continuously differentiable
\end{theorem}

\bigskip

\subsection{Holomorphic maps}

\label{Holomorphic maps}

In algebra we have imposed for any linear map $f\in L\left(  E;F\right)  $
that E and F shall be vector spaces over the \textit{same field} K.\ Indeed
this is the condition for the definition of linearity $f\left(  ku\right)
=kf\left(  u\right)  $ to be consistent. Everything that has been said
previously (when K was not explicitly $%
\mathbb{R}
)$ holds for complex vector spaces. But differentiable maps over complex
affine spaces have surprising properties.

\subsubsection{Differentiability}

\paragraph{Definitions\newline}

1. Let E,F be two \textit{complex} normed affine spaces with underlying vector
spaces $\overrightarrow{E},\overrightarrow{F}$ , $\Omega$ an open subset in E.

i) If f is differentiable in $a\in\Omega$ then f is said to be
\textbf{C-differentiable}, and f'(a) is a C-linear map $\in%
\mathcal{L}%
\left(  \overrightarrow{E};\overrightarrow{F}\right)  $ so :

$\forall\overrightarrow{u}\in\overrightarrow{E}:f^{\prime}(a)i\overrightarrow
{u}=if^{\prime}(a)\overrightarrow{u}$

ii) If there is a R-linear map $L:\overrightarrow{E}\rightarrow\overrightarrow
{F}$ such that :

$\exists r>0,\forall h\in\overrightarrow{E},\left\Vert \overrightarrow
{h}\right\Vert _{E}<r:f(a+\overrightarrow{h})-f(a)=L\overrightarrow
{h}+\varepsilon\left(  h\right)  \left\Vert \overrightarrow{h}\right\Vert
_{F}$

where $\varepsilon\left(  h\right)  \in\overrightarrow{F}$ is such that
$\lim_{h\rightarrow0}\varepsilon\left(  h\right)  =0$

then f is said to be \textbf{R-differentiable} in a. So the only difference is
that L is R-linear. A R-linear map is such that f(kv)=kf(v) for any real
scalar k

iii) If E is a real affine space, and F a complex affine space, one cannot
(without additional structure on E such as complexification) speak of
C-differentiability of a map $f:E\rightarrow F$ but it is still fully
legitimate to speak of R-differentiability.\ This is a way to introduce
derivatives for maps with real domain valued in a complex codomain.

2. A C-differentiable map is R-differentiable, but a R-differentiable map is
C-differentiable iff $\forall\overrightarrow{u}\in\overrightarrow{E}%
:f^{\prime}(a)\left(  i\overrightarrow{u}\right)  =if^{\prime}(a)\left(
\overrightarrow{u}\right)  $

3. Example : take a real structure on a complex vector space $\overrightarrow
{E}$.\ This is an antilinear map $\sigma:\overrightarrow{E}\rightarrow
\overrightarrow{E}.$ Apply the criterium for differentiability :
$\sigma\left(  \overrightarrow{u}+\overrightarrow{h}\right)  -\sigma\left(
\overrightarrow{u}\right)  =\sigma\left(  \overrightarrow{h}\right)  $ so the
derivative $\sigma^{\prime}$ would be $\sigma$ but this map is R-linear and
not C-linear. It is the same for the maps : $\operatorname{Re}:\overrightarrow
{E}\rightarrow\overrightarrow{E}::\operatorname{Re}\overrightarrow{u}=\frac
{1}{2}\left(  \overrightarrow{u}+\sigma\left(  \overrightarrow{u}\right)
\right)  $ and $\operatorname{Im}:\overrightarrow{E}\rightarrow\overrightarrow
{E}::\operatorname{Im}\overrightarrow{u}=\frac{1}{2i}\left(  \overrightarrow
{u}-\sigma\left(  \overrightarrow{u}\right)  \right)  .$ Thus it is not
legitimate to use the chain rule to C-differentiate a map such that $f\left(
\operatorname{Re}\overrightarrow{u}\right)  .$

4. The extension to differentiable and continuously differentiable maps over
an open subset are obvious.

\begin{definition}
A \textbf{holomorphic} map is a map $f:\Omega\rightarrow F$ continuously
differentiable in $\Omega$, where $\Omega$\ is an open subset of E, and E,F
are complex normed\ affine spaces.
\end{definition}

\paragraph{Cauchy-Riemann equations\newline}

\begin{theorem}
Let f be a map : $f:\Omega\rightarrow F$, where $\Omega$\ is an open subset of
E, and $\left(  E,\overrightarrow{E}\right)  $,$\left(  F,\overrightarrow
{F}\right)  $ are complex normed\ affine spaces. For any real structure on E,
f can be written as a map $\widetilde{f}\left(  x,y\right)  $\ on the product
$E_{%
\mathbb{R}
}\times iE_{%
\mathbb{R}
}$ of two real affine spaces. Then f is holomorphic iff :%

\begin{equation}
\widetilde{f}_{y}^{\prime}=i\widetilde{f}_{x}^{\prime}%
\end{equation}

\end{theorem}

\begin{proof}
1) Complex affine spaces can be considered as real affine spaces (see Affine
spaces) by using a real structure on the underlying complex vector space. Then
a point in E is identified by a couple of points in two real affine spaces :
it sums up to distinguish the real and the imaginary part of the coordinates.
The operation is always possible but the real structures are not unique. With
real structures on E and F, f can be written as a map :

$f\left(  \operatorname{Re}z+i\operatorname{Im}z\right)  =P\left(
\operatorname{Re}z,\operatorname{Im}z\right)  +iQ\left(  \operatorname{Re}%
z,\operatorname{Im}z\right)  $

$\widetilde{f}:\Omega_{%
\mathbb{R}
}\times i\Omega_{%
\mathbb{R}
}\rightarrow F_{%
\mathbb{R}
}\times iF_{%
\mathbb{R}
}:\widetilde{f}\left(  x,y\right)  =P\left(  x,y\right)  +iQ\left(
x,y\right)  $

where $\Omega_{%
\mathbb{R}
}\times i\Omega_{%
\mathbb{R}
}\ $is the embedding of $\Omega\ $\ in $E_{%
\mathbb{R}
}\times iE_{%
\mathbb{R}
},$ P,Q are maps valued in $F_{%
\mathbb{R}
}$

2) If f is holomorphic in $\Omega$\ then at any point $a\in\Omega$ the
derivative f'(a) is a linear map between two complex vector spaces endowed
with real structures. So for any vector $u\in\overrightarrow{E}$\ it can be
written :

$f^{\prime}(a)u=\widetilde{P}_{x}\left(  a\right)  \left(  \operatorname{Re}%
u\right)  +\widetilde{P}_{y}\left(  a\right)  (\operatorname{Im}u)+i\left(
\widetilde{Q}_{x}\left(  a\right)  \left(  \operatorname{Re}u\right)
+\widetilde{Q}_{y}\left(  a\right)  (\operatorname{Im}u)\right)  $

where $\widetilde{P}_{x}\left(  a\right)  ,\widetilde{P}_{y}\left(  a\right)
,\widetilde{Q}_{x}\left(  a\right)  ,\widetilde{Q}_{y}\left(  a\right)  $ are
real linear maps between the real kernels $E_{%
\mathbb{R}
},F_{%
\mathbb{R}
}$ which satifsfy the identities :

$\widetilde{P}_{y}\left(  a\right)  =-\widetilde{Q}_{x}\left(  a\right)
;\widetilde{Q}_{y}\left(  a\right)  =\widetilde{P}_{x}\left(  a\right)  $ (see
Complex vector spaces).

On the other hand f'(a)u reads :

$f^{\prime}(a)u=\widetilde{f}(x_{a},y_{a})^{\prime}\left(  \operatorname{Re}%
u,\operatorname{Im}u\right)  =\widetilde{f}_{x}^{\prime}(x_{a},y_{a}%
)\operatorname{Re}u+\widetilde{f}_{y}^{\prime}(x_{a},y_{a})\operatorname{Im}u$

$=P_{x}^{\prime}(x_{a},y_{a})\operatorname{Re}u+P_{y}^{\prime}(x_{a}%
,y_{a})\operatorname{Im}u+i\left(  Q_{x}^{\prime}(x_{a},y_{a}%
)\operatorname{Re}u+Q_{y}^{\prime}(x_{a},y_{a})\operatorname{Im}u\right)  $

$P_{y}^{\prime}(x_{a},y_{a})=-Q_{x}^{\prime}(x_{a},y_{a});Q_{y}^{\prime}%
(x_{a},y_{a})=P_{x}^{\prime}(x_{a},y_{a})$

Which reads : $\widetilde{f}_{x}^{\prime}=P_{x}^{\prime}+iQ_{x}^{\prime
};\widetilde{f}_{y}^{\prime}=P_{y}^{\prime}+iQ_{y}^{\prime}=-Q_{x}^{\prime
}+iP_{x}^{\prime}=i\widetilde{f}_{x}^{\prime}$

3) Conversely if there are partial derivatives $P_{x}^{\prime},P_{y}^{\prime
},Q_{x}^{\prime},Q_{y}^{\prime}\ $ continuous on $\Omega_{%
\mathbb{R}
}\times i\Omega_{%
\mathbb{R}
}$ then the map $\left(  P,Q\right)  $ is R-differentiable. It will be
C-differentiable if $f^{\prime}\left(  a\right)  i\overrightarrow
{u}=if^{\prime}\left(  a\right)  \overrightarrow{u}$ and that is just the
Cauchy-Rieman equations.\ The result stands for a given real structure, but we
have seen that there is always such a structure, thus if C-differentiable for
a real structure it will be C-differentiable in any real structure.
\end{proof}

The equations $f_{y}^{\prime}=if_{x}^{\prime}$ are the \textbf{Cauchy-Rieman
equations}.

Remarks :

i) The partial derivatives depend on the choice of a real structure $\sigma
$.\ If one starts with a basis $\left(  e_{i}\right)  _{i\in I}$\ the simplest
way is to define $\sigma\left(  e_{j}\right)  =e_{j},\sigma\left(
ie_{j}\right)  =-ie_{j}$ \ so $\overrightarrow{E}_{%
\mathbb{R}
}$\ is generated by $\left(  e_{i}\right)  _{i\in I}$\ with real components.
In a frame of reference $\left(  O,\left(  e_{j},ie_{j}\right)  _{j\in
I}\right)  $ the coordinates are expressed by two real set of scalars $\left(
x_{j},y_{j}\right)  $. Thus the Cauchy-Rieman equations reads ;%

\begin{equation}
\frac{\partial f}{\partial y_{j}}=i\frac{\partial f}{\partial x_{j}}%
\end{equation}

It is how they are usually written but we have proven that the equations hold
for E infinite dimensional.

ii) We could have thought to use $f\left(  z\right)  =f\left(  x+iy\right)  $
and the chain rule but the maps : $z\rightarrow\operatorname{Re}%
z,z\rightarrow\operatorname{Im}z$ are not differentiable.

iii) If $F=\overrightarrow{F}$\ Banach then the condition f has continuous
R-partial derivatives can be replaced by $\left\Vert f\right\Vert ^{2}$
locally integrable.

\paragraph{Differential\newline}

The notations are the same as above, E and F are assumed to be complex Banach
finite dimensional affine spaces, endowed with real structures.

Take a fixed origin O' for a frame in F. f reads :

$f(x+iy)=O^{\prime}+\overrightarrow{P}\left(  x,y\right)  +i\overrightarrow
{Q}\left(  x,y\right)  $ with $\left(  \overrightarrow{P},\overrightarrow
{Q}\right)  :E_{%
\mathbb{R}
}\times iE_{%
\mathbb{R}
}\rightarrow\overrightarrow{F}_{%
\mathbb{R}
}\times i\overrightarrow{F}_{%
\mathbb{R}
}$

1. As an affine Banach space, E is a manifold, and the open subset $\Omega
$\ is still a manifold, modelled on $\overrightarrow{E}.$ A frame of reference
$\left(  O,\left(  \overrightarrow{e}_{i}\right)  _{i\in I}\right)  $ of E
gives a map on E, and a holonomic basis on the tangent space, which is
$\overrightarrow{E}_{%
\mathbb{R}
}\times\overrightarrow{E}_{%
\mathbb{R}
}$ , and a 1-form $\left(  dx,dy\right)  $ which for any vector $\left(
\overrightarrow{u},\overrightarrow{v}\right)  \in\overrightarrow{E}_{%
\mathbb{R}
}\times\overrightarrow{E}_{%
\mathbb{R}
}$ gives the componants in the basis : $\left(  dx,dy\right)  \left(
\overrightarrow{u},\overrightarrow{v}\right)  =\left(  u^{j},v^{k}\right)
_{j,k\in I}$.

2. $\overrightarrow{f}=\overrightarrow{P}+i\overrightarrow{Q}$ can be
considered as a 0-form defined on a manifold and valued in a fixed vector
space. It is R-differentiable, so one can define the exterior derivatives :

$\varpi=d\overrightarrow{f}=\sum_{j\in I}\left(  f_{x_{j}}^{\prime}%
dx^{j}+f_{y_{j}}^{\prime}dy^{j}\right)  \in\Lambda_{1}\left(  \Omega^{\prime
};\overrightarrow{F}\right)  $

From (dx,dy) one can define the 1-forms valued in $\overrightarrow{F}:$

$dz^{j}=dx^{j}+idy^{j},$ $d\overline{z}^{j}=dx^{j}-idy^{j}$

thus : $dx^{j}=\frac{1}{2}\left(  dz^{j}+d\overline{z}^{j}\right)
,dy^{j}=\frac{1}{2i}\left(  dz^{j}-d\overline{z}^{j}\right)  $

$\varpi=\sum_{j\in I}\left(  f_{x_{j}}^{\prime}\frac{1}{2}\left(
dz^{j}+d\overline{z}^{j}\right)  +f_{y_{j}}^{\prime}\frac{1}{2i}\left(
dz^{j}-d\overline{z}^{j}\right)  \right)  $

$=\sum_{j\in I}\frac{1}{2}\left(  f_{x_{j}}^{\prime}+\frac{1}{i}f_{y_{j}%
}^{\prime}\right)  dz^{j}+\frac{1}{2}\left(  f_{x_{j}}^{\prime}-\frac{1}%
{i}f_{y_{j}}^{\prime}\right)  d\overline{z}^{j}$

It is customary to denote :%

\begin{equation}
\frac{\partial f}{\partial z^{j}}=\frac{1}{2}\left(  \frac{\partial
f}{\partial\operatorname{Re}z^{j}}+\frac{1}{i}\frac{\partial f}{\partial
\operatorname{Im}z^{j}}\right)  =\frac{1}{2}\left(  f_{x_{j}}^{\prime}%
+\frac{1}{i}f_{y_{j}}^{\prime}\right)
\end{equation}

\begin{equation}
\frac{\partial f}{\partial\overline{z}^{j}}=\frac{1}{2}\left(  \frac{\partial
f}{\partial\operatorname{Re}z^{j}}-\frac{1}{i}\frac{\partial f}{\partial
\operatorname{Im}z^{j}}\right)  =\frac{1}{2}\left(  f_{x_{j}}^{\prime}%
-\frac{1}{i}f_{y_{j}}^{\prime}\right)
\end{equation}

Then :%

\begin{equation}
d\overrightarrow{f}=\sum_{j\in I}\frac{\partial f}{\partial z^{j}}dz^{j}%
+\frac{\partial f}{\partial\overline{z}^{j}}d\overline{z}^{j}%
\end{equation}

3. f is holomorphic iff : $f_{y_{j}}^{\prime}=if_{x_{j}}^{\prime}$ that is iff
$\frac{\partial f}{\partial\overline{z}^{j}}=\frac{1}{2}\left(  f_{x_{j}%
}^{\prime}-\frac{1}{i}f_{y_{j}}^{\prime}\right)  =0$ .Then : $d\overrightarrow
{f}=\sum_{j\in I}\frac{\partial f}{\partial z^{j}}dz^{j}$

\begin{theorem}
A map is continuously C-differentiable iff it does not depend explicitly on
the conjugates $\overline{z}^{j}.$
\end{theorem}

If so the differential of f reads : df=f'(z)dz

\paragraph{Derivatives of higher order\newline}

\begin{theorem}
A holomorphic map $f:\Omega\rightarrow F$ from an open subset of an affine
normed space to an affine normed space F\ has C-derivatives of any order.
\end{theorem}

If f' exists then $f^{\left(  r\right)  }$ exists $\forall r$. So this is in
stark contrast with maps in real affine spaces. The proof is done by
differentiating the Cauchy-Rieman equations

\paragraph{Extremums\newline}

A non constant holomorphic map cannot have an extremum.

\begin{theorem}
(Schwartz III p.302, 307, 314) If $f:\Omega\rightarrow F$ is a holomorphic map
on the open $\Omega$\ of the normed affine space E, valued in the normed
affine space F, then:

i) $\left\Vert f\right\Vert $ has no strict local maximum in $\Omega$

ii) If $\Omega$ is bounded in E, f continuous on the closure of $\Omega,$ then:

$\sup_{x\in\Omega}\left\Vert f\left(  x\right)  \right\Vert =\sup
_{x\in\overset{\circ}{\Omega}}\left\Vert f\left(  x\right)  \right\Vert
=\sup_{x\in\overline{\Omega}}\left\Vert f\left(  x\right)  \right\Vert $

iii) If E is finite dimensional $\left\Vert f\right\Vert $ has a maximum on
$\partial\left(  \overline{\Omega}\right)  $.

iv) if $\Omega$ is connected and $\exists a\in\Omega:f(a)=0,$ $\forall
n:f^{\left(  n\right)  }\left(  a\right)  =0$ then f=0 in $\Omega$

v) if $\Omega$ is connected and f is constant in an open in $\Omega$ then f is
constant in $\Omega$
\end{theorem}

If f is never zero take $1/\left\Vert f\right\Vert $ and we get the same
result for a minimum.

One consequence is that any holomorphic map on a compact holomorphic manifold
is constant (Schwartz III p.307)

\begin{theorem}
(Schwartz III p.275, 312) If $f:\Omega\rightarrow%
\mathbb{C}
$ is a holomorphic function on the connected open $\Omega$\ of the normed
affine space E, then:

i) if $\operatorname{Re}f$ is constant then f is constant

ii) if $\left\vert f\right\vert $ is constant then f is constant

iii) If there is $a\in\Omega$ local extremum of $\operatorname{Re}f$ or
$\operatorname{Im}f $\ then f is constant

iv) If there is $a\in\Omega$ local maximum of $\left\vert f\right\vert $ then
f is constant

v) If there is $a\in\Omega$ local minimum of $\left\vert f\right\vert $ then f
is constant or f(a)=0
\end{theorem}

\paragraph{Sequence of holomorphic maps\newline}

\begin{theorem}
Weierstrass (Schwartz III p.326) Let $\Omega$ be an open bounded in an affine
normed space, F a Banach vector space, if the sequence $\left(  f_{n}\right)
_{n\in%
\mathbb{N}
}$ of maps : $f_{n}:\Omega\rightarrow F$, holomorphic in $\Omega$\ and
continuous on the closure of $\Omega,$ converges uniformly on $\partial\Omega$
$\ $it converges uniformly on $\overline{\Omega}.$ Its limit f is holomorphic
in $\Omega$ and continuous on the closure of $\Omega$ , and the higher
derivatives $f_{n}^{\left(  r\right)  }$ converges locally uniformly in
$\Omega$ to $f^{\left(  r\right)  }.$
\end{theorem}

\begin{theorem}
(Schwartz III p.327) Let $\Omega$ be an open in an affine normed space, F a
Banach vector space, if the sequence $\left(  f_{n}\right)  _{n\in%
\mathbb{N}
}$ of maps : $f_{n}:\Omega\rightarrow F$ ,holomorphic in $\Omega$\ and
continuous on the closure of $\Omega,$ converges locally uniformly in
$\Omega,$ then$\ $it converges locally uniformly on $\Omega,$ its limit is
holomorphic and the higher derivatives $f_{n}^{\left(  r\right)  }$ converges
locally uniformly in $\Omega$ to $f^{\left(  r\right)  }.$
\end{theorem}

\subsubsection{Maps defined on $%
\mathbb{C}
$}

The most interesting results are met when f is defined in an open of $%
\mathbb{C}
.$ But most of them cannot be extended to higher dimensions.

In this subsection : $\Omega$ is an open subset of $%
\mathbb{C}
$ and F a Banach \textit{vector} space (in the following we will drop the
arrow but F is a vector space and not an affine space). And f is a map :
$f:\Omega\rightarrow F$

\paragraph{Cauchy differentiation formula\newline}

\begin{theorem}
The map $f:\Omega\rightarrow F,\Omega$ from an open in $%
\mathbb{C}
$ to a Banach vector space F, continuously R-differentiable, is holomorphic
iff the 1-form $\lambda=f^{\prime}(z)dz$\ is closed : $d\lambda=0$
\end{theorem}

\begin{proof}
This is a direct consequence of the previous subsection. Here the real
structure of $E=%
\mathbb{C}
$ is obvious : take the "real axis" and the "imaginary axis" of the plane $%
\mathbb{R}
^{2}.$ $%
\mathbb{R}
^{2}$ as $%
\mathbb{R}
^{n},\forall n,$ is a manifold and the open subset $\Omega$\ is itself a
manifold (with canonical maps). We can define the differential $\lambda
=f^{\prime}(z)dx+if^{\prime}(z)dy=f^{\prime}(z)dz$
\end{proof}

\begin{theorem}
Morera (Schwartz III p.282): Let $\Omega$ be an open in $%
\mathbb{C}
$\ and $f:\Omega\rightarrow F$ be a continuous map valued in the Banach
\textit{vector} space F. If for any smooth compact manifold X with boundary in
$\Omega$\ we have $\int_{\partial X}f(z)dz=0$ then f is holomorphic
\end{theorem}

\begin{theorem}
(Schwartz III p.281,289,294) Let $\Omega$ be a simply connected open subset in
$%
\mathbb{C}
$\ and $f:\Omega\rightarrow F$ be a holomorphic map valued in the Banach
\textit{vector} space F. Then

i) f has indefinite integrals which are holomorphic maps $\varphi\in
H(\Omega;F):\varphi^{\prime}(z)=f(z)$ defined up to a constant\ \ 

ii) for any class 1 manifold with boundary X in $\Omega$\ :%

\begin{equation}
\int_{\partial X}f(z)dz=0
\end{equation}

if $a\notin X:\int_{\partial X}\frac{f\left(  z\right)  }{z-a}=0$ and if
$a\in\overset{\circ}{X}:$%

\begin{equation}
\int_{\partial X}\frac{f\left(  z\right)  }{z-a}dz=2i\pi f(a)
\end{equation}

If X is compact and if $a\in\overset{\circ}{X}:$%

\begin{equation}
f^{\left(  n\right)  }\left(  a\right)  =\frac{n!}{2i\pi}\int_{\partial
X}\frac{f\left(  z\right)  }{\left(  z-a\right)  ^{n+1}}dz
\end{equation}

\end{theorem}

The proofs are a direct consequence of the Stockes theorem applied to
$\Omega.$

So we have : $\int_{a}^{b}f(z)dz=\varphi\left(  b\right)  -\varphi\left(
a\right)  $ the integral being computed on any continuous curve from a to b in
$\Omega$

These theorems are the key to the computation of many definite integrals
$\int_{a}^{b}f(z)dz:$

i) f being holomorphic depends only on z, and the indefinite integral (or
antiderivative ) can be computed as in elementary analysis

ii) as we can choose any curve we can take $\gamma$ such that f or the
integral is obvious on some parts of the curve

iii) if f is real we can consider some extension of f which is holomorphic

\paragraph{Taylor's series\newline}

\begin{theorem}
(Schwartz III p.303) If the map $f:\Omega\rightarrow F,\Omega$ from an open in
$%
\mathbb{C}
$ to a Banach vector space F is holomorphic, then the series :%

\begin{equation}
f(z)=f(a)+\sum_{n=1}^{\infty}\frac{\left(  z-a\right)  ^{n}}{n!}f^{\left(
n\right)  }\left(  a\right)
\end{equation}

is absolutely convergent in the largest disc B(a,R) centered in a and
contained in $\Omega$ and convergent in any disc B(a,r), r
$<$
R.
\end{theorem}

\paragraph{Algebra of holomorphic maps}

\begin{theorem}
The set of holomorphic maps from an open in $%
\mathbb{C}
$ to a Banach vector space F is a complex vector space.

The set of holomorphic functions on an open in $%
\mathbb{C}
$ is a complex algebra with pointwise multiplication.
\end{theorem}

\begin{theorem}
Any polynomial is holomorphic, the exponential is holomorphic,
\end{theorem}

The complex logarithm is defined as the indefinite integral of $\int\frac
{dz}{z}.$ We have $\int_{R}\frac{dz}{z}=2i\pi$ where R is any circle centered
in 0. Thus complex logarithms are defined up to $2i\pi n$

\begin{theorem}
(Schwartz III p.298) If the function $f:\Omega\rightarrow%
\mathbb{C}
$ holomorphic on the simply connected open $\Omega$ is such that $\forall
z\in\Omega:f(z)\neq0$ then there is a holomorphic function $g:\Omega
\rightarrow%
\mathbb{C}
$ such that $f=\exp g$
\end{theorem}

\paragraph{Meromorphic maps\newline}

\begin{theorem}
If f is a non null holomorphic function $f:\Omega\rightarrow%
\mathbb{C}
$ on an open subset of $%
\mathbb{C}
$, then all zeros of f are isolated points.
\end{theorem}

\begin{definition}
The map $f:\Omega\rightarrow F$ from an open in $%
\mathbb{C}
$ to a Banach vector space F is \textbf{meromorphic} if it is holomorphic
except at a set of isolated points, which are called the \textbf{poles} of f.
A point a is a pole of order r
$>$
0 if there is some constant C such that $f\left(  z\right)  \simeq
C/(z-a)^{r}$ when $z\rightarrow a.$ If a is a pole and there is no such r then
a is an \textbf{essential pole}.
\end{definition}

Warning ! the poles must be isolated, thus $\sin\frac{1}{z},\ln z,..$ are not meromorphic

If $F=%
\mathbb{C}
$ then a meromorphic function can be written as the ratio u/v of two
holomorphic functions.

\begin{theorem}
(Schwartz III p.330) If f is a non null holomorphic function $f:\Omega
\rightarrow%
\mathbb{C}
$ on an open subset of $%
\mathbb{C}
$ : $\Omega=R_{1}<\left\vert z-a\right\vert <R_{2}$ then there is a family of
complex scalars $\left(  c_{n}\right)  _{n=-\infty}^{+\infty}$ such that :
$f(z)=\sum_{n=-\infty}^{n=+\infty}c_{n}\left(  z-a\right)  ^{n}$. The
coefficients are uniquely defined by : $c_{n}=\frac{1}{2i\pi}\int_{\gamma
}\frac{f(z)}{\left(  z-a\right)  ^{n+1}}dz$ where $\gamma\subset\Omega$ is a
loop which wraps only once around a.
\end{theorem}

\begin{theorem}
Weierstrass (Schwartz III p.337) : If $f:\Omega\rightarrow%
\mathbb{C}
$ is holomorphic in $\Omega=\left\{  0<\left\vert z-a\right\vert <R\right\}  $
and a is an essential pole for f, then the image by f of any subset $\left\{
0<\left\vert z-a\right\vert <r<R\right\}  $\ is dense in $%
\mathbb{C}
.$
\end{theorem}

It means that f(z) can be arbitrarily close to any complex number.

\subsubsection{Analytic maps}

Harmonic maps are treated in the Functional Analysis - Laplacian part.

\begin{definition}
A map $f:\Omega\rightarrow F$ from an open of a normed affine space and valued
in a normed affine space F, both on the field K, is \textbf{K-analytic} if it
is K-differentiable at any order and \ 

$\forall a\in\Omega,\exists n\left(  a\right)  :\forall x\in n\left(
a\right)  :f(x)-f(a)=\sum_{n=1}^{\infty}\frac{1}{n!}f^{\left(  n\right)
}\left(  a\right)  \left(  x-a\right)  ^{n}$
\end{definition}

Warning ! a K-analytic function is smooth (indefinitely K-differentiable) but
the converse is not true in general.

\begin{theorem}
(Schwartz III p.307) For a K-analytic map $f:\Omega\rightarrow F$ from a
connected open of a normed affine space and valued in a normed affine space F,
both on the field K, the following are equivalent :

i) f is constant in $\Omega$

ii) $\exists a\in\Omega:$ $\forall n\geq1:f^{\left(  n\right)  }\left(
a\right)  =0$

iii) f is constant in an open in $\Omega$
\end{theorem}

\begin{theorem}
Liouville (Schwartz III p.322): For a K-analytic map $f:E\rightarrow F$ from a
normed affine space and valued in a normed affine space F, both on the field K:

i) if f, $\operatorname{Re}f$ or $\operatorname{Im}f$ is bounded then f is constant

ii) if $\exists a\in E,n\in%
\mathbb{N}
,n>0,C>0:\left\Vert f\left(  x\right)  \right\Vert \leq C\left\Vert
x-a\right\Vert ^{n}$ then f is a polynomial of order $\leq n$
\end{theorem}

\begin{theorem}
(Schwartz III p.305) A holomorphic map $f:\Omega\rightarrow F$ on an open of a
normed affine space to a Banach vector space F is $%
\mathbb{C}
-$analytic
\end{theorem}

\begin{theorem}
(Schwartz III p.322) If $f\in C_{\infty}\left(  \Omega;%
\mathbb{R}
\right)  ,\Omega$ open in $%
\mathbb{R}
,$ then the following are equivalent

i) f is $%
\mathbb{R}
$-analytic

ii) there is a holomorphic (complex analytic) extension of f in $D\subset%
\mathbb{C}
$ such that $\Omega\subset D$

iii) for every compact set $C\subset\Omega$ there exists a constant M such
that for every $a\in C$ and every $n\in%
\mathbb{N}
:$

$\left\vert f^{\left(  n\right)  }\left(  a\right)  \right\vert \leq
M^{n+1}n!$
\end{theorem}

\newpage

\section{MANIFOLDS}

\bigskip

\subsection{Manifolds}

\label{Manifolds}

\bigskip

A manifold can be viewed as a "surface" of elementary geometry, and it is
customary to introduce manifolds as some kind of sets embedded in affine
spaces.\ However useful it can be, it is a bit misleading (notably when we
consider Lie groups) and it is good to start looking at the key ingredient of
the theory, which is the concept of charts.\ Indeed charts are really what the
name calls for : a way to locate a point through a set of figures. The beauty
of the concept is that we do not need to explicitly give all the procedure for
going to the place so designated by the coordinates : all we need to know is
that it is possible (and indeed if we have the address of a building we can go
there).\ Thus to be mathematically useful we add a condition : the procedures
from coordinates to points must be consistent with each other.\ If we have two
charts, giving different coordinates for the same point, there must be some
way to pass from one set of coordinates to the other, and this time we deal
with figures, that is mathematical objects which can be precisely dealt with.
So this "interoperability" of charts becomes the major feature of manifolds,
and enables us to forget most of the time the definition of the charts.

\subsubsection{Definitions}

\paragraph{Definition of a manifold\newline}

The most general definition of a manifold is the following (Maliavin and Lang)

\begin{definition}
An \textbf{atlas} A$\left(  E,\left(  O_{i},\varphi_{i}\right)  _{i\in
I}\right)  $ of class r on a set M is comprised of:

i) a Banach vector space E on a field K

ii) a cover $\left(  O_{i}\right)  _{i\in I}$ of M (each $O_{i}$ is a subset
of M and $\cup_{i\in I}O_{i}=M)$

iii) a family $\left(  \varphi_{i}\right)  _{i\in I},$called \textbf{charts,}
of bijective maps :

$\varphi_{i}:O_{i}\rightarrow U_{i}$ where $U_{i}$ is an open subset of E

iii) $\forall i,j\in I,O_{i}\cap O_{j}\neq\varnothing:$ $\varphi_{i}\left(
O_{i}\cap O_{j}\right)  ,\varphi_{j}\left(  O_{i}\cap O_{j}\right)  $ are open
subsets of E, and there is a r\ continuously differentiable diffeomorphism:
$\varphi_{ij}:\varphi_{i}\left(  O_{i}\cap O_{j}\right)  \rightarrow
\varphi_{j}\left(  O_{i}\cap O_{j}\right)  $ called a \textbf{transition map}
\end{definition}

\begin{definition}
Two atlas A$\left(  E,\left(  O_{i},\varphi_{i}\right)  _{i\in I}\right)
,$A'$\left(  E,\left(  O_{j}^{\prime},\varphi_{j}^{\prime}\right)  _{i\in
J}\right)  $ of class r on a set M are \textbf{compatible} if their union is
still an atlas of M
\end{definition}

which implies that whenever $O_{i}\cap O_{j}^{\prime}\neq\varnothing$ there is
a r diffeomorphism: $\varphi_{ij}:\varphi_{i}\left(  O_{i}\cap O_{j}^{\prime
}\right)  \rightarrow\varphi_{j}^{\prime}\left(  O_{i}\cap O_{j}^{\prime
}\right)  $

\begin{definition}
The relation : A,A' are compatible atlas of M is a relation of equivalence. A
\textbf{structure of manifold} on the set M is a class of compatible atlas on M
\end{definition}

\begin{notation}
$M\left(  E,\left(  O_{i},\varphi_{i}\right)  _{i\in I}\right)  $ is a set M
endowed with the manifold structure defined by the atlas $\left(  E,\left(
O_{i},\varphi_{i}\right)  _{i\in I}\right)  $
\end{notation}

\paragraph{Comments\newline}

1. M is said to be modeled on E. If E' is another Banach such that there is a
continuous bijective map between E and E', then this map is a smooth
diffeomorphism and E' defines the same manifold structure. So it is simpler to
assume that E is always the same. If E is over the field K, M is a
K-manifold.\ We will specify real or complex manifold when necessary. If E is
a Hilbert space M is a Hilbert manifold. There is also a concept of manifold
modelled on Fr\'{e}chet spaces (for example the infinite jet prolongation of a
bundle). Not all Banach spaces are alike, and the properties of E are crucial
for those of M.

2. The dimension of E is the \textbf{dimension} of the manifold (possibly
infinite). If E is finite n dimensional it is always possible to take
$E=K^{n}$.

3. There can be a unique chart in an atlas.

4. r is the \textbf{class of the manifold.} If $r=\infty$ the manifold is said
to be smooth, if r=0 (the transition maps are continuous only) M is said to be
a topological manifold. If the transition charts are K-analytic the manifold
is said to be K-analytic.

5. To a point $p\in M$ a chart associes a vector $u=\varphi_{i}\left(
p\right)  $ in E and, through a basis in E, a set of numbers $\left(
x_{j}\right)  _{j\in J}$ in K which are the \textbf{coordinates} of p in the
chart. If the manifold is finite dimensional the canonical basis of $K^{n}$ is
used and the coordinates are given by : $\alpha=1...n:$ $\left[  x_{\alpha
}\right]  =\left[  \varphi_{i}\left(  p\right)  \right]  $ matrices nx1. In
another chart : $\left[  y_{\alpha}\right]  =\left[  \varphi_{j}\left(
p\right)  \right]  =\varphi_{ij}\left(  x_{1}...x_{n}\right)  $

6. There can be different manifold structures on a given set. For $%
\mathbb{R}
^{n}$ $n\neq4$ all the smooth structures are equivalent (diffeomorphic), but
on $%
\mathbb{R}
^{4}$ there are uncountably many non equivalent smooth manifold structures
(exotic !).

7. From the definition it is clear that any open subset of a manifold is
itself a manifold.

8. Notice that no assumption is done about a topological structure on M. This
important point is addressed below.

\subsubsection{Examples}

1. Any Banach vector space, any Banach affine space, and any open subsets of
these spaces have a canonical structure of smooth differential manifold (with
an atlas comprised of a unique chart), and we will always refer to this
structure when required.

2. A finite n dimensional subspace of a topological vector space or affine
space has a manifold structure (homeomorphic to $K^{n}).$

3. An atlas for the sphere $S_{n},$ n dimensional manifold defined by
$\sum_{i=1}^{n+1}x_{i}^{2}=1$ in $%
\mathbb{R}
^{n+1}$ is the stereographic projection.\ Choose a south pole $a\in S_{n}$ and
a north pole $-a\in S_{n}.$ The atlas is comprised of 2 charts :

$O_{1}=S_{n}\backslash\left\{  a\right\}  ,\varphi_{1}\left(  p\right)
=\frac{p-\left\langle p,a\right\rangle a}{1-\left\langle p,a\right\rangle }$

$O_{2}=S_{n}\backslash\left\{  -a\right\}  ,\varphi_{2}\left(  p\right)
=\frac{p-\left\langle p,a\right\rangle a}{1+\left\langle p,a\right\rangle }$

with the scalar product : $\left\langle p,a\right\rangle =\sum_{i=1}%
^{n+1}p_{i}a_{i}$

4. For a manifold embedded in $%
\mathbb{R}
^{n},$ passing from cartesian coordinates to curvilinear coordinates (such
that polar, spheric, cylindrical coordinates) is just a change of chart on the
same manifold (see in the section Tensor bundle below).

\paragraph{Grassmanian\newline}

\begin{definition}
The \textbf{Grassmanian} denoted Gr(E;r) of a n dimensional vector space E
over a field K is the set of all r dimensional vector subspaces of E.
\end{definition}

\begin{theorem}
(Schwartz II p.236).The Grassmanian Gr(E;r) has a structure of smooth manifold
of dimension r(n-r), isomorphic to Gr(E;n-r) and homeomorphic to the set of
matrices M in K(n) such that : M%
${{}^2}$%
=M ; M*=M ; Trace(M)=r
\end{theorem}

The Grassmanian for r=1 is the projective space P(E) associated to E. It is a
n-1 smooth manifold, which is compact if $K=%
\mathbb{R}
.$.

\subsubsection{Topology}

\paragraph{The topology associated to an atlas\newline}

\begin{theorem}
An atlas A$\left(  E,\left(  O_{i},\varphi_{i}\right)  _{i\in I}\right)  $ on
a manifold M defines a topology on M for which the sets $O_{i}$\ are open in M
and the charts are continuous. This topology is the same for all compatible atlas.
\end{theorem}

\begin{theorem}
(Malliavin p.20) The topology induced by a manifold structure through an atlas
A$\left(  E,\left(  O_{i},\varphi_{i}\right)  _{i\in I}\right)  $ on a
topological space $\left(  M,\Omega\right)  $ coincides with this latter
topology iff $\forall i\in I,O_{i}\in\Omega$ and $\varphi_{i}$ is an
homeomorphism on $O_{i}.$
\end{theorem}

\begin{theorem}
If a manifold M is modelled on a Banach vector space E, then every point of M
has a neighbourhood which is homeomorphic to an open of E. Conversely a
topological space M such that every point of M has a neighbourhood which is
homeomorphic to an open of E can be endowed with a structure of manifold
modelled on E.
\end{theorem}

\begin{proof}
i) Take an atlas A=$\left(  E,\left(  O_{i},\varphi_{i}\right)  _{i\in
I}\right)  .$ Let $p\in M$ so $\exists i\in I:p\in O_{i}.$ with the topology
defined by A, $O_{i}$ is a neighborhood of p, which is homeomorphic to
$U_{i},$ which is a neighborhood of $\varphi_{i}\left(  p\right)  \in E.$

ii) Conversely let $(M,\Omega)$ be a topological space, such that for each
$p\in M$ there is a neighborhood n(p), an open subset $\mu\left(  p\right)  $
of E, and a homeomorphism $\varphi_{p}$ between n(p) and $\mu.$ The family
$\left(  n\left(  p\right)  ,\varphi_{p}\right)  _{p\in M}$ is an atlas for M :

$\forall p\in M:\varphi_{p}\left(  n\left(  p\right)  \right)  =\mu\left(
p\right)  $ is open in E

$\varphi_{p}\left(  n\left(  p\right)  \cap n\left(  q\right)  \right)
=\mu\left(  p\right)  \cap\mu\left(  q\right)  $ is an open in E (possibly empty).

$\varphi_{p}\circ\varphi_{q}^{-1}$ is the compose of two homeomorphisms, so a homeomorphism
\end{proof}

Warning ! usually there is no global homeomorphism between M and E

To sum up :

- if M has no prior topology, it gets one, uniquely defined by a class of
atlas, and it is locally homeomorphic to E.

- if M is a topological space, its topology defines the conditions which must
be met by an atlas so that it can define a manifold structure compatible with
M. So a set, endowed with a given topology, may not accept some structures of
manifolds (this is the case with structures involving scalar products).

\paragraph{Locally compact manifold\newline}

\begin{theorem}
A manifold is locally compact iff it is finite dimensional. It is then a Baire space.
\end{theorem}

\begin{proof}
i) If a manifold M modelled on a Banach E is locally compact, then E is
locally compact, and is necessarily finite dimensional, and so is M.

ii) If E is finite dimensional, it is locally compact. Take p in M, and a
chart $\varphi_{i}\left(  p\right)  =x\in E.$ x has a compact neighborhood
n(x), its image by the continous map $\varphi_{i}^{-1}$ is a compact
neighborhood of p.
\end{proof}

It implies that a compact manifold is \textit{never} infinite dimensional.

\paragraph{Paracompactness, metrizability\newline}

\begin{theorem}
A second countable, regular manifold is metrizable.
\end{theorem}

\begin{proof}
It is semi-metrizable, and metrizable if it is T1, but any manifold is T1
\end{proof}

\begin{theorem}
A regular, Hausdorff manifold with a $\sigma$-locally finite base is metrizable
\end{theorem}

\begin{theorem}
A metrizable manifold is paracompact.
\end{theorem}

\begin{theorem}
(Kobayashi 1 p.116, 166) The vector bundle of a finite dimensional paracompact
manifold M can be endowed with an inner product (a definite positive, either
symmetric bilinear or hermitian sequilinear form) and M is metrizable.
\end{theorem}

\begin{theorem}
For a finite dimensional manifold M the following properties are equivalent:

i) M is paracompact

ii) M is metrizable

iii) M admits an inner product on its vector bundle
\end{theorem}

\begin{theorem}
For a finite dimensional, paracompact manifold M it is always possible to
choose an atlas A$\left(  E,\left(  O_{i},\varphi_{i}\right)  _{i\in
I}\right)  $ such that the cover is relatively compact ($\overline{O}_{i}$ is
compact) and locally finite (each points of M meets only a finite number
of\ $O_{i}$)
\end{theorem}

If M is a Hausdorf m dimensional class 1 real manifold then we can also have
an open cover such that any non empty finite intersection of $O_{i}$\ is
diffeomorphic with an open of $%
\mathbb{R}
^{m}$ (Kobayashi p.167).

\begin{proof}
\begin{theorem}
(Lang p.35) For every open covering $\left(  \Omega_{j}\right)  _{j\in J}$ of
a locally compact, Hausdorff, second countable manifold M modelled on a Banach
E, there is an atlas $\left(  O_{i},\varphi_{i}\right)  _{i\in I}$ of M such
that $\left(  O_{i}\right)  _{i\in I}$ is a locally finite refinement of
$\left(  \Omega_{j}\right)  _{j\in J},$ $\varphi_{i}\left(  O_{i}\right)  $ is
an open ball $B(x_{i},3)\subset E$ and the open sets $\varphi_{i}^{-1}\left(
B\left(  x_{i},1\right)  \right)  $ covers M.
\end{theorem}
\end{proof}

\begin{theorem}
A finite dimensional, Hausdorff, second countable manifold is paracompact,
metrizable and can be endowed with an inner product.
\end{theorem}

\paragraph{Countable base\newline}

\begin{theorem}
A metrizable manifold is first countable.
\end{theorem}

\begin{theorem}
For a semi-metrizable manifold, separable is equivalent to second countable.\ 
\end{theorem}

\begin{theorem}
A semi-metrizable manifold has a $\sigma-$locally finite base.
\end{theorem}

\begin{theorem}
A connected, finite dimensional, metrizable, manifold is separable and second countable.
\end{theorem}

\begin{proof}
It is locally compact so the result follows the Kobayashi p.269 theorem (see
General topology)
\end{proof}

\paragraph{Separability\newline}

\begin{theorem}
A manifold is a T1 space
\end{theorem}

\begin{proof}
A Banach is a T1 space, so each point is a closed subset, and its preimage by
a chart is closed
\end{proof}

\begin{theorem}
A metrizable manifold is a Hausdorff, normal, regular topological space
\end{theorem}

\begin{theorem}
A semi-metrizable manifold is normal and regular.
\end{theorem}

\begin{theorem}
A paracompact manifold is normal
\end{theorem}

\begin{theorem}
A finite dimensional manifold is regular.
\end{theorem}

\begin{proof}
because it is locally compact
\end{proof}

\begin{theorem}
(Kobayashi 1 p.271) For a finite dimensional, connected, Hausdorff manifold M
the following are equivalent :

i) M is paracompact

ii) M is metrizable

iii) M admits an inner product

iv) M is second countable
\end{theorem}

A finite dimensional class 1 manifold has an equivalent smooth structure
(Kolar p.4) thus one can usually assume that a finite dimensional manifold is smooth

\paragraph{Infinite dimensional manifolds\newline}

\begin{theorem}
(Henderson) A separable metric manifold modelled on a separable infinite
dimensional Fr\'{e}chet space can be embedded as an open subset of an infinite
dimensional, separable Hilbert space defined uniquely up to linear isomorphism.
\end{theorem}

Of course the theorem applies to a manifold modeled on Banach space E, which
is a Fr\'{e}chet space. E is separable iff it is second countable, because
this is a metric space. Then M is second countable if it has an atlas with a
finite number of charts. If so it is also separable. It is metrizable if it is
regular (because it is T1). Then it is necessarily Hausdorff.

\begin{theorem}
A regular manifold modeled on a second countable infinite dimensional Banach
vector space, with an atlas comprised of a finite number of charts, can be
embedded as an open subset of an infinite dimensional, separable Hilbert
space, defined uniquely up to linear isomorphism.
\end{theorem}

\bigskip

\subsection{Differentiable maps}

\label{Differentiable maps manifolds}

Manifolds are the only structures, other than affine spaces, upon which
differentiable maps are defined.

\subsubsection{Definitions}

\begin{definition}
A map $f:M\rightarrow N$ between the manifolds M,N is said to be continuously
differerentiable at the order r if, for any point p in M, there are charts
$\left(  O_{i},\varphi_{i}\right)  $\ in M, and $\left(  Q_{j},\psi
_{j}\right)  $ in N, such that $p\in O_{i},f\left(  p\right)  \in Q_{j}$\ and
$\psi_{j}\circ f\circ\varphi_{i}^{-1}$ is r continuously differentiable in
$\varphi_{i}\left(  O_{i}\cap f^{-1}\left(  Q_{j}\right)  \right)  .$
\end{definition}

If so then $\psi_{j}\circ f\circ\varphi_{l}^{-1}$ is r continuously
differentiable with any other charts meeting the same conditions.

Obviously r is less or equal to the class of both M and N. If the manifolds
are smooth and f is of class r for any r then f is said to be smooth. In the
following we will assume that the classes of the manifolds and of the maps
match together.

\begin{definition}
A \textbf{r-diffeomorphism} between two manifolds is a bijective map,
r-differentiable and with a r-differentiable inverse.
\end{definition}

\begin{definition}
A \textbf{local diffeomorphism} between two manifolds M,N is a map such that
for each $p\in M$ there are neighbourhoods n(p) and n(f(p)) such that the
restriction $\widehat{f}:n(p)\rightarrow f(n(p))$ is a diffeomorphism
\end{definition}

If there is a diffeomorphism between two manifolds they are said to be
\textbf{diffeomorphic}.\ They have necessarily same dimension (possibly infinite).

The maps of charts $\left(  O_{i},\varphi_{i}\right)  $ of a class r manifold
are r-diffeomorphism : $\varphi_{i}\in C_{r}\left(  O_{i};\varphi_{i}\left(
O_{i}\right)  \right)  $:

$p\in O_{i}\cap O_{j}:y=\varphi_{j}\left(  p\right)  =\varphi_{ij}\left(
x\right)  =\varphi_{ij}\circ\varphi_{i}\left(  p\right)  \Rightarrow
\varphi_{ij}=\varphi_{j}\circ\varphi_{i}^{-1}\in C_{r}\left(  E;E\right)  $

If a manifold is an open of an affine space then its maps are smooth.

Let $M\left(  E,\left(  O_{i},\varphi_{i}\right)  \right)  ,N\left(  G,\left(
Q_{j},\psi_{j}\right)  \right)  $\ be two manifolds.\ To any map
$f:M\rightarrow N$ is associated maps between coordinates : if $x=\varphi
_{i}\left(  p\right)  $ then $y=\psi_{j}\left(  f\left(  p\right)  \right)  .$
They read :

$F:O_{i}\rightarrow Q_{j}::y=F(x)$ with $F=\psi_{j}\circ f\circ\varphi
_{i}^{-1}$

\bigskip%

\begin{tabular}
[c]{cccccc}%
M &  & $f$ &  & N & \\
$O_{i}$ & $\rightarrow$ & $\rightarrow$ & $\rightarrow$ & $Q_{i}$ & \\
$\downarrow$ &  &  &  & $\downarrow$ & \\
$\downarrow$ & $\varphi_{i}$ &  &  & $\downarrow$ & $\psi_{j}$\\
$\downarrow$ &  & F &  & $\downarrow$ & \\
$U_{i}$ & $\rightarrow$ & $\rightarrow$ & $\rightarrow$ & $V_{j}$ &
\end{tabular}

\bigskip

Then $F^{\prime}\left(  a\right)  =\left(  \psi_{j}\circ f\circ\varphi
_{i}^{-1}\right)  ^{\prime}\left(  a\right)  $ is a continuous linear map $\in%
\mathcal{L}%
\left(  E;G\right)  $ .

If f is a diffeomorphism $F=\psi_{j}\circ f\circ\varphi_{i}^{-1}$ is a
diffeomorphism between Banach vector spaces, thus :

i) $F^{\prime}(a)$ is invertible and $\left(  F^{-1}\left(  b\right)  \right)
^{\prime}=\left(  F^{\prime}(a)\right)  ^{-1}\in%
\mathcal{L}%
\left(  G;E\right)  $

ii) F is an open map (it maps open subsets to open subsets)

\begin{definition}
The \textbf{jacobian} of a differentiable map between two finite dimensional
manifolds is the matrix $F^{\prime}\left(  a\right)  =\left(  \psi_{j}\circ
f\circ\varphi_{i}^{-1}\left(  a\right)  \right)  ^{\prime}$
\end{definition}

If M is m dimensional defined over $K^{m}$,N is n dimensional defined over
$K^{n}$, then $F\left(  a\right)  =\psi_{j}\circ f\circ\varphi_{i}^{-1}\left(
a\right)  $ can be written in the canonical bases of $K^{m},K^{n}:$\ j=1...n :
$y^{j}=F^{j}\left(  x^{1},...x^{m}\right)  $ using tensorial notation for the indexes

$F^{\prime}\left(  a\right)  =\left(  \psi_{j}\circ f\circ\varphi_{i}%
^{-1}\left(  a\right)  \right)  ^{\prime}$ is expressed in bases as a
n$\times$m matrix (over K)

\bigskip

$J=\left[  F^{\prime}(a)\right]  =\left\{  \overset{m}{\overbrace{\left[
\dfrac{\partial F^{\alpha}}{\partial x^{\beta}}\right]  }}\right\}  n$

\bigskip

If f is a diffeomorphism the jacobian of $F^{-1}$\ is the inverse of the
jacobian of F.

\subsubsection{General properties}

\paragraph{Set of r differentiable maps\newline}

\begin{notation}
$C_{r}\left(  M;N\right)  $ is the set of class r\ maps from the manifold M to
the manifold N (both on the same field K)
\end{notation}

\begin{theorem}
$C_{r}\left(  M;F\right)  $ is a vector space.$C_{r}\left(  M;K\right)  $ is a
vector space and an algebra with pointwise multiplication.
\end{theorem}

\paragraph{Categories of differentiable maps\newline}

\begin{theorem}
(Schwartz II p.224) If $f\in C_{r}\left(  M;N\right)  ,g\in C_{r}\left(
N;P\right)  $ then $g\circ f\in C_{r}\left(  M.P\right)  $ (if the manifolds
have the required class)
\end{theorem}

\begin{theorem}
The class r manifolds and the class r differentiable maps (on the same field
K) constitute a category. The smooth manifolds and the smooth differentiable
maps constitute a subcategory.
\end{theorem}

There is more than the obvious : functors will transform manifolds into fiber bundles.

\paragraph{Product of manifolds\newline}

\begin{theorem}
The product M$\times$N of two class r manifolds on the same field K is a
manifold with dimension = dim(M)+dim(N) and the projections $\pi_{M}:M\times
N\rightarrow M,$ $\pi_{N}:M\times N\rightarrow N$ are of class r.
\end{theorem}

For any class r maps : $f:P\rightarrow M,g:P\rightarrow N$ between manifolds,
the mapping :

$\left(  f,g\right)  :P\rightarrow M\times N::\left(  f,g\right)  \left(
p\right)  =\left(  f\left(  p\right)  ,g\left(  p\right)  \right)  $

is the unique class r mapping with the property :

$\pi_{M}\left(  \left(  f,g\right)  \left(  p\right)  \right)  =f\left(
p\right)  ,\pi_{N}\left(  \left(  f,g\right)  \left(  p\right)  \right)
=g\left(  p\right)  $

\paragraph{Space
$\mathcal{L}$%
(E;E) for a Banach vector space\newline}

\begin{theorem}
The set $%
\mathcal{L}%
\left(  E;E\right)  $\ of continuous linear maps over a Banach vector space E
is a Banach vector space, so this is a manifold. The subset $G%
\mathcal{L}%
\left(  E;E\right)  $ of inversible map is an open subset of $%
\mathcal{L}%
(E;E)$, so this is also a manifold.

The composition law and the inverse are differentiable maps :
\end{theorem}

i) the composition law :

$M:%
\mathcal{L}%
\left(  E;E\right)  \times%
\mathcal{L}%
\left(  E;E\right)  \rightarrow%
\mathcal{L}%
\left(  E;E\right)  ::M(f,g)=f\circ g$ is differentiable and

$M^{\prime}\left(  f,g\right)  \left(  \delta f,\delta g\right)  =\delta
f\circ g+f\circ\delta g$

ii) the map : $\Im:G%
\mathcal{L}%
\left(  E;E\right)  \rightarrow G%
\mathcal{L}%
\left(  E;E\right)  $ is differentiable and

$\left(  \Im(f)\right)  ^{\prime}\left(  \delta f\right)  =-f^{-1}\circ\delta
f\circ f^{-1}$

\subsubsection{Partition of unity}

Partition of unity is a powerful tool to extend local properties to global
ones.\ They exist for paracompact Hausdorff spaces, and so for any Hausdorff
finite dimensional manifold.\ However we will need maps which are are not only
continuous but also differentiable. Furthermore difficulties arise with
infinite dimensional manifolds.

\paragraph{Definition\newline}

\begin{definition}
A \textbf{partition of unity} of class r subordinate to an open covering
$\left(  \Omega_{i}\right)  _{i\in I}$ of a manifold M is a family $\left(
f_{i}\right)  _{i\in I}$ of maps $f_{i}\in C_{r}\left(  M;%
\mathbb{R}
_{+}\right)  ,$ such that the support of $f_{i}$ is contained in $\Omega_{i}$
and :

$\forall p\in M:f_{i}\left(  p\right)  \neq0$ for at most finitely many i

$\forall p\in M:\sum_{i\in I}f_{i}\left(  p\right)  =1$
\end{definition}

As a consequence the family $(Supp\left(  f_{i}\right)  )_{i\in I}$ of the
supports of the functions is locally finite

If the support of each function is compact then the partition is compactly supported

\begin{definition}
A manifold is said to admit partitions of unity if it has a partition of unity
subordinate to any open cover.
\end{definition}

\paragraph{Conditions for the existence of a partition of unity\newline}

From the theorems of general topology:

\begin{theorem}
A paracompact Hausdorff manifold admits continuous partitions of unity
\end{theorem}

\begin{theorem}
(Lang p.37, Bourbaki) For any paracompact Hausdorff manifold and locally
finite open cover $\left(  \Omega_{i}\right)  _{i\in I}$ of M there is a
localy finite open cover $\left(  U_{i}\right)  _{i\in I}$ such that
$\overline{U}_{i}\subset\Omega_{i}$
\end{theorem}

\begin{theorem}
(Kobayashi I p.272) For any paracompact, finite dimensional manifold, and
locally finite open cover $\left(  \Omega_{i}\right)  _{i\in I}$ of M such
that each $\Omega_{i}$ has compact closure, there is a partition of unity
subodinate to $\left(  \Omega_{i}\right)  _{i\in I}.$
\end{theorem}

\begin{theorem}
(Lang p.38) A class r paracompact manifold modeled on a separable Hilbert
space admits class r partitions of unity subordinate to any locally finite
open covering.
\end{theorem}

\begin{theorem}
(Schwartz II p.242) For any class r finite dimensional second countable real
manifold M , open cover $\left(  \Omega_{i}\right)  _{i\in I}$ of M there is a
family $\left(  f_{i}\right)  _{i\in I}$ of functions $f_{i}\in C_{r}\left(
M;%
\mathbb{R}
_{+}\right)  $ with support in $\Omega_{i}$ , such that : $\forall p\in
K:\sum_{i\in I}f_{i}\left(  p\right)  =1,$ and $\forall p\in M$ there is a
neighborhood n(p) on which only a finite number of $f_{i}$ are not null.
\end{theorem}

\begin{theorem}
(Schwartz II p.240) For any class r finite dimensional real manifold M,
$\Omega$\ open in M, $p\in\Omega,$ there is a r continuously differentiable
real function f with compact support included in $\Omega$ such that :

f(p)
$>$
0 and $\forall m\in\Omega:0\leq f\left(  m\right)  \leq1 $
\end{theorem}

\begin{theorem}
(Schwartz II p.242) For any class r finite dimensional real manifold M, open
cover $\left(  \Omega_{i}\right)  _{i\in I}$ of M, compact K in M, there is a
family $\left(  f_{i}\right)  _{i\in I}$ of functions $f_{i}\in C_{r}\left(
M;%
\mathbb{R}
_{+}\right)  $ with compact support in $\Omega_{i}$ such that :

$\forall p\in M:f_{i}\left(  p\right)  \neq0$ for at most finitely many i and
$\forall p\in K:\sum_{i\in I}f_{i}\left(  p\right)  >0$
\end{theorem}

\paragraph{Prolongation of a map\newline}

\begin{theorem}
(Schwartz II p.243) Let M be a class r finite dimensional second countable
real manifold, C a closed subset of M, F real Banach vector space, then a map
$f\in C_{r}\left(  C;F\right)  $ can be extended to a map $\widehat{f}\in
C_{r}\left(  M;F\right)  :\forall p\in C:\widehat{f}\left(  p\right)  =f(p)$
\end{theorem}

Remark : the definition of a class r map on a closed set is understood in the
Whitney sense : there is a class r map g\ on M such that the derivatives for
any order $s\leq r$ of g are equal on C to the approximates of f by the
Taylor's expansion.

\bigskip

\subsection{The tangent bundle}

\label{Tangent bundle}

\subsubsection{Tangent vector space}

\begin{theorem}
A each point p on a class 1 differentiable manifold M modelled on a Banach E
on the field K, there is a set, called the \textbf{tangent space} to M at p,
which has the structure of a vector space over K, isomorphic to E
\end{theorem}

There are several ways to construct the tangent vector space.\ The simplest is
the following:

\begin{proof}
i) two differentiable functions $f,g\in C_{1}\left(  M;K\right)  $ are said to
be equivalent if for a chart $\varphi$ covering p, for every differentiable
path : $c:K\rightarrow E$ such that $\varphi^{-1}\circ c\left(  0\right)  =p$
we\ have : $\left(  f\circ\varphi^{-1}\circ c\right)  ^{\prime}|_{t=0}=\left(
g\circ\varphi^{-1}\circ c\right)  ^{\prime}|_{t=0}.$ The derivative is well
defined because this is a map : $K\rightarrow K.$ This is an equivalence
relation $\sim.$ Two maps equivalent with a chart are still equivalent with a
chart of a compatible atlas.

ii) The value of the derivative $\left(  f\circ\varphi^{-1}\right)  ^{\prime
}|_{x}$ for $\varphi\left(  p\right)  =x$\ is a continuous map from E to K, so
this is a form in E'.

The set of these values $\widetilde{T}^{\ast}\left(  p\right)  $ is a vector
space over K, because $C_{1}\left(  M;K\right)  $ is a vector space over K. If
we take the sets $T^{\ast}\left(  p\right)  $ of these values for each class
of equivalence we still have a vector space.

iii) $T^{\ast}\left(  p\right)  $ is isomorphic to E'.

The map : $T^{\ast}\left(  p\right)  \rightarrow E^{\prime}$ is injective :

If $\left(  f\circ\varphi^{-1}\right)  ^{\prime}|_{x}=\left(  g\circ
\varphi^{-1}\right)  ^{\prime}|_{x}$ then $\left(  f\circ\varphi^{-1}\circ
c\right)  ^{\prime}|_{t=0}=\left(  f\circ\varphi^{-1}\right)  ^{\prime}%
|_{x}\circ c^{\prime}|_{t=0}=\left(  g\circ\varphi^{-1}\right)  ^{\prime}%
|_{x}\circ c^{\prime}|_{t=0}\Rightarrow f\sim g$

It is surjective : for any $\lambda\in E^{\prime}$ take $f\left(  q\right)
=\lambda\left(  \varphi\left(  y\right)  \right)  $

iv) The tangent space is the topological dual of $T^{\ast}\left(  p\right)  .$
If E is reflexive then (E')' is isomorphic to E.
\end{proof}

Remarks :

i) It is crucial to notice that the tangent spaces at two different points
have no relation with each other. To set up a relation we need special tools,
such as connections.

ii) T*(p) is the 1-jet of the functions on M at p.

iii) the demonstration fails if E is not reflexive, but there are ways around
this issue by taking the weak dual.

\begin{notation}
$T_{p}M$ is the tangent vector space at p to the manifold M
\end{notation}

\begin{theorem}
The charts $\varphi_{i}^{\prime}\left(  p\right)  $ of an atlas $\left(
E,\left(  O_{i},\varphi_{i}\right)  _{i\in I}\right)  $\ are continuous
inversible linear map $\varphi_{i}^{\prime}\left(  p\right)  \in G%
\mathcal{L}%
\left(  T_{p}M;E\right)  $
\end{theorem}

A Banach vector space E has a smooth manifold structure.\ The tangent space at
any point is just E itself.

A Banach affine space $\left(  E,\overrightarrow{E}\right)  $ has a smooth
manifold structure.\ The tangent space at any point p is the affine subspace
$\left(  p,\overrightarrow{E}\right)  $.

\begin{theorem}
The tangent space at a point of a manifold modelled on a Banach space has the
structure of a Banach vector space. Different compatible charts give
equivalent norm.
\end{theorem}

\begin{proof}
Let $\left(  E,\left(  O_{i},\varphi_{i}\right)  _{i\in I}\right)  $ be an
atlas of M, and $p\in O_{i}$

The map : $\tau_{i}:T_{p}M\rightarrow E::\tau_{i}\left(  u\right)
=\varphi_{i}^{\prime}\left(  p\right)  u=v$ is a continuous isomorphism.

Define : $\left\Vert u\right\Vert _{i}=\left\Vert \varphi_{i}^{\prime}\left(
p\right)  u\right\Vert _{E}=\left\Vert v\right\Vert _{E}$

With another map :

$\left\Vert u\right\Vert _{j}=\left\Vert \varphi_{j}^{\prime}\left(  p\right)
u\right\Vert _{E}=\left\Vert \varphi_{ij}^{\prime}\circ\varphi_{i}^{\prime
}\left(  p\right)  u\right\Vert _{E}$

$\leq\left\Vert \varphi_{ij}^{\prime}\left(  \varphi_{i}\left(  p\right)
\right)  \right\Vert _{%
\mathcal{L}%
(E;E)}\left\Vert v\right\Vert _{E}=\left\Vert \varphi_{ij}^{\prime}\left(
\varphi_{i}\left(  p\right)  \right)  \right\Vert _{%
\mathcal{L}%
(E;E)}\left\Vert u\right\Vert _{i}$

and similarly : $\left\Vert u\right\Vert _{i}\leq\left\Vert \varphi
_{ji}^{\prime}\left(  \varphi_{j}\left(  p\right)  \right)  \right\Vert _{%
\mathcal{L}%
(E;E)}\left\Vert u\right\Vert _{j}$

So the two norms are equivalent (but not identical), they define the same
topology. Moreover with these norms the tangent vector space is a Banach
vector space.
\end{proof}

\subsubsection{Holonomic basis}

\begin{definition}
A \textbf{holonomic basis} of a manifold M$\left(  E,\left(  O_{i},\varphi
_{i}\right)  _{i\in I}\right)  $ is the image of a basis of E by $\varphi
_{i}^{\prime}{}^{-1}$. A each point p it is a basis of the tangent space
$T_{p}M.$
\end{definition}

\begin{notation}
$\left(  \partial x_{\alpha}\right)  _{\alpha\in A}$ is the holonomic basis at
$p=\varphi_{i}^{-1}\left(  x\right)  $ associated to the basis $\left(
e_{\alpha}\right)  _{\alpha\in A}$\ by the chart $\left(  O_{i},\varphi
_{i}\right)  $
\end{notation}

At the transition between charts :

$p\in O_{i}\cap O_{j}:y=\varphi_{j}\left(  p\right)  ,x=\varphi_{i}\left(
p\right)  \Rightarrow y=\varphi_{ij}\left(  x\right)  =\varphi_{ij}%
\circ\varphi_{i}\left(  p\right)  \Rightarrow\varphi_{ij}=\varphi_{j}%
\circ\varphi_{i}^{-1}$

$\partial x_{\alpha}=\left(  \varphi_{i}\left(  p\right)  ^{\prime}\right)
^{-1}e_{\alpha}\in T_{p}M\Leftrightarrow\varphi_{i}^{\prime}\left(  p\right)
\partial x_{\alpha}=e_{\alpha}\in E$

$\partial y_{\alpha}=\left(  \varphi_{j}\left(  p\right)  ^{\prime}\right)
^{-1}e_{\alpha}\in T_{p}M\Leftrightarrow\varphi_{j}^{\prime}\left(  p\right)
\partial y_{\alpha}=e_{\alpha}\in E$

$\partial y_{\alpha}=\left(  \varphi_{j}^{\prime}\left(  p\right)  \right)
^{-1}e_{\alpha}=\left(  \varphi_{j}^{\prime}\left(  p\right)  \right)
^{-1}\circ\varphi_{i}^{\prime}\left(  p\right)  \partial x_{\alpha}$

So a vector $u_{p}\in T_{p}M$ can be written : $u_{p}=\sum_{\alpha\in
A}u^{\alpha}\partial x_{\alpha}$ and at most finitely many $u^{\alpha}$ are
non zero. Its image by the maps $\varphi_{i}^{\prime}\left(  p\right)  $ is a
vector of E : $\varphi_{i}^{\prime}\left(  p\right)  u_{p}=\sum_{\alpha\in
A}u^{\alpha}e_{\alpha}$ which has the same components in the basis of E.

The holonomic bases are not the only bases on the tangent space. Any other
basis $\left(  \delta_{\alpha}\right)  _{\alpha\in A}$ can be defined from a
holonomic basis. They are called \textbf{non holonomic bases} . An example is
the orthonormal basis if M is endowed wih a metric. But for a non holonomic
basis the vector $u_{p}$ has not the same components in $\delta_{\alpha}$ as
its image $\varphi_{i}^{\prime}\left(  p\right)  u_{p}$ in E.

\subsubsection{Derivative of a map}

\paragraph{Definition\newline}

\begin{definition}
For a map $f\in C_{r}\left(  M;N\right)  $ between two manifolds M$\left(
E,\left(  O_{i},\varphi_{i}\right)  _{i\in I}\right)  ,$

$N\left(  G,\left(  Q_{j},\psi_{j}\right)  _{j\in J}\right)  $ there is, at
each point $p\in M$ a unique continuous linear map $f^{\prime}\left(
p\right)  \in%
\mathcal{L}%
\left(  T_{p}M;T_{f\left(  p\right)  }N\right)  ,$ called the derivative of f
at p, such that :

$\psi_{j}^{\prime}\circ f^{\prime}\left(  p\right)  \circ\left(  \varphi
_{i}^{-1}\right)  ^{\prime}\left(  x\right)  =\left(  \psi_{j}\circ
f\circ\varphi_{i}^{-1}\right)  ^{\prime}\left(  x\right)  $ with
$x=\varphi_{i}\left(  p\right)  $
\end{definition}

f'(p) is defined in a holonomic basis by :

$f^{\prime}(p)\partial x_{\alpha}=f^{\prime}(p)\circ\varphi_{i}^{\prime
}\left(  x\right)  ^{-1}e_{\alpha}\in T_{f\left(  p\right)  }N$

We have the following commuting diagrams :

\bigskip%

\begin{tabular}
[c]{cccccc}
&  & $f$ &  &  & \\
$M$ & $\rightarrow$ & $\rightarrow$ & $\rightarrow$ & $N$ & \\
$\downarrow$ &  &  &  & $\downarrow$ & \\
$\downarrow$ & $\varphi_{i}$ &  &  & $\downarrow$ & $\psi_{j}$\\
$\downarrow$ &  & $F$ &  & $\downarrow$ & \\
$E$ & $\rightarrow$ & $\rightarrow$ & $\rightarrow$ & $G$ &
\end{tabular}
\ \ \ \ \
\begin{tabular}
[c]{cccccc}
&  & $f^{\prime}\left(  p\right)  $ &  &  & \\
$T_{p}M$ & $\rightarrow$ & $\rightarrow$ & $\rightarrow$ & $T_{f\left(
p\right)  }N$ & \\
$\downarrow$ &  &  &  & $\downarrow$ & \\
$\downarrow$ & $\varphi_{i}^{\prime}\left(  p\right)  $ &  &  & $\downarrow$ &
$\psi_{j}^{\prime}\left(  q\right)  $\\
$\downarrow$ &  & $F^{\prime}\left(  x\right)  $ &  & $\downarrow$ & \\
$E$ & $\rightarrow$ & $\rightarrow$ & $\rightarrow$ & $G$ &
\end{tabular}

\bigskip

$\psi_{j}^{\prime}\circ f^{\prime}(p)\partial x_{\alpha}=\psi_{j}^{\prime
}\circ f^{\prime}(p)\circ\varphi_{i}^{\prime}\left(  a\right)  ^{-1}e_{\alpha
}\in G$

If M is m dimensional with coordinates x, N n dimensional with coordinates y
the jacobian is just the matrix of f'(p) in the holonomic bases in
$T_{p}M,T_{f\left(  p\right)  }N:$

$y=\psi_{j}\circ f\circ\varphi_{i}^{-1}\left(  x\right)  \Leftrightarrow
\alpha=1...n:y^{\alpha}=F^{\alpha}\left(  x^{1},...x^{m}\right)  $

$\left[  f^{\prime}(p)\right]  =\left[  F^{\prime}(x)\right]  =\left\{
\overset{m}{\overbrace{\left[  \dfrac{\partial F^{\alpha}}{\partial x^{\beta}%
}\right]  }}\right\}  n=\left[  \dfrac{\partial y^{\alpha}}{\partial x^{\beta
}}\right]  _{n\times m}$

Whatever the choice of the charts in M,N there is always a derivative map
f'(p), but its expression depends on the coordinates (as for any linear map).
The rules when in a change of charts are given in the Tensorial bundle subsection.

Remark : usually the use of f'(p) is the most convenient.\ But for some
demonstrations it is simpler to come back to maps between fixed vector spaces
by using $\left(  \psi_{j}\circ f\circ\varphi_{i}^{-1}\right)  ^{\prime
}\left(  x\right)  .$

\begin{definition}
The \textbf{rank of a differentiable map} $f:M\rightarrow N$ between manifolds
at a point p is the rank of its derivative f'(p). It does not depend on the
choice of the charts in M,N and is necessarily $\leq\min(\dim M,\dim N)$
\end{definition}

\begin{theorem}
Composition of maps : If $f\in C_{1}\left(  M;N\right)  ,g\in C_{1}\left(
N;P\right)  $ then

$\left(  g\circ f\right)  ^{\prime}(p)=g^{\prime}(f(p))\circ f^{\prime}(p)\in%
\mathcal{L}%
\left(  T_{p}M;T_{g\circ f(p)}P\right)  $
\end{theorem}

\subparagraph{Diffeomorphisms are very special maps :\newline}

i) This is a bijective map $f\in C_{r}\left(  M;N\right)  $ such that
$f^{-1}\in C_{r}\left(  M;N\right)  $

ii) $f^{\prime}(p)$ is invertible and $\left(  f^{-1}\left(  q\right)
\right)  ^{\prime}=\left(  f^{\prime}(p)\right)  ^{-1}\in%
\mathcal{L}%
\left(  T_{f(p)}N;T_{p}M\right)  :$ this is a continuous linear isomorphism
between the tangent spaces

iii) f is an open map (it maps open subsets to open subsets)

\subparagraph{Higher order derivatives : \newline}

With maps on affine spaces the derivative f'(a) is a linear map depending on
a, but it is still a map on fixed affine spaces, so we can consider
f"(a).\ This is no longer possible with maps on manifolds : if f is of class r
then this is the map $F(a)=\psi_{j}\circ f\circ\varphi_{i}^{-1}\left(
a\right)  \in C_{r}\left(  E;G\right)  $ which is r differentiable, and thus
for higher derivatives we have to account for $\psi_{j},\varphi_{i}^{-1}.$ In
other words f'(p) is a linear map between vector spaces which themselves
depend on p, so there is no easy way to compare f'(p) to f'(q).\ Thus we need
other tools, such as connections, to go further (see Higher tangent bundle for more).

\paragraph{Partial derivatives\newline}

The partial derivatives $\frac{\partial f}{\partial x^{\alpha}}\left(
p\right)  =f_{\alpha}^{\prime}\left(  p\right)  $ wih respect to the
coordinate $x^{\alpha}$\ is the maps
$\mathcal{L}$%
$\left(  \mathfrak{E}_{\alpha};T_{f\left(  p\right)  }N\right)  $ where
$\mathfrak{E}_{\alpha}$ is the one dimensional vector subspace in $T_{p}%
M$\ generated by $\partial x_{\alpha}$

To be consistent with the notations for affine spaces :

\begin{notation}
$f^{\prime}(p)$ $\in%
\mathcal{L}%
\left(  T_{p}M;T_{f\left(  p\right)  }N\right)  $ is the derivative
\end{notation}

\begin{notation}
$\frac{\partial f}{\partial x^{\alpha}}\left(  p\right)  =f_{\alpha}^{\prime
}\left(  p\right)  \in%
\mathcal{L}%
\left(  \mathfrak{E}_{\alpha};T_{f\left(  p\right)  }N\right)  $ are the
partial derivative with respect to the coordinate $x^{\alpha}$
\end{notation}

\paragraph{Cotangent space\newline}

\begin{definition}
The \textbf{cotangent space} to a manifold M at a point p is the topological
dual of the tangent space $T_{p}M$
\end{definition}

To follow on long custom we will not use the prime notation in this case:

\begin{notation}
$T_{p}M^{\ast}$ is the cotangent space to the manifold M at p
\end{notation}

\begin{definition}
The transpose of the derivative of $f\in C_{r}\left(  M;N\right)  $ at p is
the map : $f^{\prime}(p)^{t}\in%
\mathcal{L}%
\left(  T_{f\left(  p\right)  }N^{\ast};T_{p}N^{\ast}\right)  $
\end{definition}

The transpose of the derivative $\varphi_{i}^{\prime}\left(  p\right)  \in%
\mathcal{L}%
\left(  T_{p}M;E\right)  $ of\ a chart is :

$\varphi_{i}^{\prime}\left(  p\right)  ^{t}\in%
\mathcal{L}%
\left(  E^{^{\prime}};\left(  T_{p}M\right)  ^{\ast}\right)  $

If $e^{\alpha}$\ is a basis of E' such that $e^{\alpha}\left(  e_{\beta
}\right)  =\delta_{\beta}^{\alpha}$ (it is not uniquely defined by $e_{\alpha
}$\ if E is infinite dimensional) then $\varphi_{i}^{\prime}\left(  p\right)
^{t}\left(  e^{\alpha}\right)  $ is a (holonomic) basis of $T_{p}M^{\ast}.$

\begin{notation}
$dx^{\alpha}=\varphi_{i}^{\prime}\left(  p\right)  ^{\ast}\left(  e^{\alpha
}\right)  $ is the holonomic basis of $T_{p}M^{\ast}$ associated to the basis
$\left(  e^{\alpha}\right)  _{\alpha\in A}$\ of E' by the atlas $\left(
E,\left(  O_{i},\varphi_{i}\right)  _{i\in I}\right)  $
\end{notation}

So : $dx^{\alpha}\left(  \partial x_{\beta}\right)  =\delta_{\beta}^{\alpha}$

For a function $f\in C_{1}\left(  M;K\right)  $ : $f^{\prime}(a)\in
T_{p}M^{\ast}$ so $f^{\prime}(a)=\sum_{\alpha\in A}\varpi_{\alpha}dx^{\alpha}$

The partial derivatives $f_{\alpha}^{\prime}\left(  p\right)  \in$%
$\mathcal{L}$%
$\left(  \mathfrak{E}_{\alpha};K\right)  $ are scalars functions so :

$f^{\prime}(a)=\sum_{\alpha\in A}f_{\alpha}^{\prime}\left(  p\right)
dx^{\alpha}$

The action of f'(a) on a vector $u\in T_{p}M$ is $f^{\prime}(a)u=\sum
_{\alpha\in A}f_{\alpha}^{\prime}\left(  p\right)  u^{\alpha}$

The exterior differential of f is just $df=\sum_{\alpha\in A}f_{\alpha
}^{\prime}\left(  p\right)  dx^{\alpha}$ which is consistent with the usual
notation (and justifies the notation $dx^{\alpha})$

\paragraph{Extremum of a function\newline}

The theorem for affine spaces can be generalized .

\begin{theorem}
If a function $f\in C_{1}\left(  M;%
\mathbb{R}
\right)  $ on a class 1 real manifold has a local extremum in $p\in M$ then f'(p)=0
\end{theorem}

\begin{proof}
Take an atlas $\left(  E,\left(  O_{i},\varphi_{i}\right)  _{i\in I}\right)
$\ of M.\ If p is a local extremum on $M$ it is a local extremum on any
$O_{i}\ni p.$ Consider the map with domain an open subset of E : $F^{\prime
}:\varphi_{i}\left(  O_{i}\right)  \rightarrow%
\mathbb{R}
::F^{\prime}\left(  a\right)  =f^{\prime}\circ\varphi_{i}^{\prime-1}.$ If
$p=\varphi_{i}\left(  a\right)  $ is a local extremum on $O_{i}$ then
$a\in\varphi_{i}\left(  O_{i}\right)  $ is a local extremum for $f\circ
\varphi_{i}$ so$\ F^{\prime}\left(  a\right)  =0\Rightarrow f^{\prime}%
(\varphi_{i}\left(  a)\right)  =0.$
\end{proof}

\paragraph{Morse's theory\newline}

A real function $f:M\rightarrow%
\mathbb{R}
$ on a manifold can be seen as a map giving the heigth of some hills drawn
above M. If this map is sliced for different elevations figures (in two
dimensions) appear, highlighting characteristic parts of the landscape (such
that peaks or lakes). Morse's theory studies the topology of a manifold M
through real functions on M (corresponding to "elevation"), using the special
points where the elevation "vanishes".

\begin{definition}
For a differentiable map $f:M\rightarrow N$ a point p is \textbf{critical} is
f'(p)=0 and regular otherwise.
\end{definition}

\begin{theorem}
(Lafontaine p.77) For any smooth maps $f\in C_{\infty}\left(  M;N\right)  $\ ,
M finite dimensional manifold, union of countably many compacts, N finite
dimensional, the set of critical points is negligible.
\end{theorem}

A subset X is neligible means that, if M is modelled on a Banach E, $\forall
p\in M$ there is a chart $\left(  O,\varphi\right)  $ such that $p\in M$ and
$\varphi\left(  O\cap X\right)  $ has a null Lebesgue measure in E.

In particular :

\begin{theorem}
Sard Lemna : the set of critical values of a function defined on an open set
of $%
\mathbb{R}
^{m}$ has a null Lebesgue measure
\end{theorem}

\begin{theorem}
Reeb : For any real function f defined on a compact real manifold M:

i) if f is continuous and has exactly two critical points then M is
homeomorphic to a sphere

ii) if M is smooth then the set of non critical points is open and dense in M
\end{theorem}

\bigskip

For a class 2 real function on an open subset of $%
\mathbb{R}
^{m}$ the \textbf{Hessian} of f is the matrix of f"(p) which is a bilinear
symmetric form. A critical point is \textbf{degenerate} if f"(a) is degenerate
(then $\det\left[  F"(a)\right]  =0$)

\begin{theorem}
Morse's lemna: If $a$ is a critical non degenerate point of the function f on
an open subset M of $%
\mathbb{R}
^{m}$, then in a neighborhood of $a$ there is a chart of M such that :
$f\left(  x\right)  =f\left(  a\right)  -\sum_{\alpha=1}^{p}x_{\alpha}%
^{2}+\sum_{\alpha=p+1}^{m}x_{\alpha}^{2}$
\end{theorem}

The integer p is the \textbf{index} of $a$ (for f). It does not depend on the
chart, and is the dimension of the largest tangent vector subspace over which
f"(a) is definite negative.

A \textbf{Morse function} is a smooth real function with no critical
degenerate point.\ The set of Morse functions is dense in $C_{\infty}\left(
M;%
\mathbb{R}
\right)  .$

One extension of this theory is "catastroph theory", which studies how real
valued functions on $%
\mathbb{R}
^{n}$\ behave around a point.\ Ren\'{e} Thom has proven that there are no more
than 14 kinds of behaviour (described as polynomials around the point).

\subsubsection{The tangent bundle}

\paragraph{Definitions\newline}

\begin{definition}
The tangent bundle over a class 1 manifold M is the set : $TM=\cup_{p\in
M}\left\{  T_{p}M\right\}  $
\end{definition}

So an element of TM is comprised of a point p of M and a vector u of $T_{p}M$

\begin{theorem}
The tangent bundle over a class r manifold M$\left(  E,\left(  O_{i}%
,\varphi_{i}\right)  _{i\in I}\right)  $ is a class r-1 manifold
\end{theorem}

The cover of TM is defined by : $O_{i}^{\prime}=\cup_{p\in O_{I}}\left\{
T_{p}M\right\}  $

The maps : $O_{i}^{\prime}\rightarrow U_{i}\times E::\left(  \varphi
_{i}\left(  p\right)  ,\varphi_{i}^{\prime}\left(  p\right)  u_{p}\right)  $
define a chart of TM

If M is finite dimensional, TM is a 2$\times$dimM dimensional manifold.

\begin{theorem}
The tangent bundle over a manifold M$\left(  E,\left(  O_{i},\varphi
_{i}\right)  _{i\in I}\right)  $ is a fiber bundle $TM(M,E,\pi)$
\end{theorem}

TM is a manifold

Define the projection : $\pi:TM\rightarrow M::\pi\left(  u_{p}\right)  =p$.
This is a smooth surjective map and $\pi^{-1}\left(  p\right)  =T_{p}M$

Define (called a trivialization) : $\Phi_{i}:O_{i}\times E\rightarrow
TM::\Phi_{i}\left(  p,u\right)  =\varphi_{i}^{\prime}\left(  p\right)
^{-1}u\in T_{p}M$

If $p\in O_{i}\cap O_{j}$ then $\varphi_{j}^{\prime}\left(  p\right)  ^{-1}u$
and $\varphi_{i}^{\prime}\left(  p\right)  ^{-1}u$ define the same vector of
$T_{p}M$

All these conditions define the structure of a vector bundle with base M,
modelled on E (see Fiber bundles).

A vector $u_{p}$ in TM can be seen as the image of a couple $\left(
p,u\right)  \in M\times E$ through the maps $\Phi_{i}$ defined on the open
cover given by an atlas.

\begin{theorem}
The tangent bundle of a Banach vector space $\overrightarrow{E}$ is the set
$T\overrightarrow{E}=\cup_{p\in\overrightarrow{E}}\left\{  u_{p}\right\}  .$
As the tangent space at any point p is $\overrightarrow{E}$ then
$T\overrightarrow{E}=\overrightarrow{E}\times\overrightarrow{E}$
\end{theorem}

Similarly the tangent bundle of a Banach affine space $\left(
E,\overrightarrow{E}\right)  $\ is $E\times\overrightarrow{E}$ and can be
considered as E iself.

\paragraph{Differentiable map\newline}

Let M$\left(  E,\left(  O_{i},\varphi_{i}\right)  _{i\in I}\right)
,$N$\left(  G,\left(  Q_{j},\psi_{j}\right)  _{j\in J}\right)  $ be two class
1 manifolds on the same field, and $f\in C_{1}\left(  M;N\right)  $ then
$\forall p\in M:f^{\prime}(p)\in%
\mathcal{L}%
\left(  T_{p}M;T_{f\left(  p\right)  }N\right)  $ so there is a map :
$TM\rightarrow TN$

\begin{notation}
$Tf=\left(  f,f^{\prime}\right)  \in C\left(  TM;TN\right)  ::Tf\left(
p,v_{p}\right)  =\left(  f\left(  p\right)  ,f^{\prime}\left(  p\right)
v_{p}\right)  $
\end{notation}

\bigskip

$TM\rightarrow\rightarrow\rightarrow\rightarrow\rightarrow\rightarrow
Tf\rightarrow\rightarrow\rightarrow\rightarrow\rightarrow TN$

$\downarrow\varphi_{i}^{\prime}%
\;\;\;\;\;\;\;\;\;\ \ \ \ \ \ \ \ \ \ \ \ \;\;\;\;\ \ \ \ \ \ \ \ \ \ \ \ \ \ \ \downarrow
\psi_{j}^{\prime}$

$\varphi_{i}\left(  O_{i}\right)  \times E\rightarrow\rightarrow F^{\prime
}\rightarrow\rightarrow\rightarrow\rightarrow\rightarrow\psi_{j}\left(
Q_{j}\right)  \times G$

\bigskip

$F^{\prime}\left(  x,u\right)  =\left(  \psi_{j}\left(  f(p)\right)  ,\psi
_{j}^{\prime}\circ f^{\prime}\left(  p\right)  \circ\left(  \varphi_{i}%
^{-1}\right)  ^{\prime}u\right)  $

\paragraph{Product of manifolds\newline}

\begin{theorem}
The product M$\times$N of two class r manifolds\ has the structure of manifold
of class r with the projections $\pi_{M}:M\times N\rightarrow M,$ $\pi
_{N}:M\times N\rightarrow N$ and the tangent bundle of MxN is T(M$\times
$N)=TM$\times$TN,

$\pi_{M}^{\prime}:T(M\times N)\rightarrow TM,\pi_{N}^{\prime}:T(M\times
N)\rightarrow TN$
\end{theorem}

Similarly the \textbf{cotangent bundle} TM* is defined with $\pi^{-1}\left(
p\right)  =T_{p}M^{\ast}$

\begin{notation}
TM is the tangent bundle over the manifold M
\end{notation}

\begin{notation}
TM* is the cotangent bundle over the manifold M
\end{notation}

\paragraph{Vector fields\newline}

\begin{definition}
A \textbf{vector field} over the manifold M is a map

$V:M\rightarrow TM::V\left(  p\right)  =v_{p}$ which associes to each point p
of M a vector of the tangent space $T_{p}M$ at p
\end{definition}

In fiber bundle parlance this is a section of the vector bundle.

Warning ! With an atlas : $\left(  E,\left(  O_{i},\varphi_{i}\right)  _{i\in
I}\right)  $ of M a holonomic basis is defined as the preimage of fixed
vectors of a basis in E. So this not the same vector at the intersections :
$\partial x_{\alpha}=\varphi_{i}^{\prime}\left(  x\right)  ^{-1}\left(
e_{\alpha}\right)  \neq\partial y_{\alpha}=\varphi_{j}^{\prime}\left(
x\right)  ^{-1}\left(  e_{\alpha}\right)  $

$\partial y_{\alpha}=\left(  \varphi_{j}^{\prime}\left(  p\right)  \right)
^{-1}e_{\alpha}=\left(  \varphi_{j}^{\prime}\left(  p\right)  \right)
^{-1}\circ\varphi_{i}^{\prime}\left(  p\right)  \partial x_{\alpha}$

But \textit{a vector field V is always the same}, whatever the open $O_{i}$.
So it must be defined by a collection of maps :

$V_{i}:O_{i}\rightarrow K::V\left(  p\right)  =\sum_{\alpha\in A}V_{i}%
^{\alpha}\left(  p\right)  \partial x_{\alpha}$

If $p\in O_{i}\cap O_{j}:V\left(  p\right)  =\sum_{\alpha\in A}V_{i}^{\alpha
}\left(  p\right)  \partial x_{\alpha}=\sum_{\alpha\in A}V_{j}^{\alpha}\left(
p\right)  \partial y_{\alpha}$

In a finite dimensional manifold $\left(  \varphi_{i}^{\prime}\left(
p\right)  \right)  ^{-1}\circ\varphi_{j}^{\prime}\left(  p\right)  $ is
represented (in the holonomic bases) by a matrix : $\left[  J_{ij}\right]  $
and $\partial x_{\alpha}=\left[  J_{ij}\right]  _{\alpha}^{\beta}\left(
\partial y_{\beta}\right)  $ so : $V_{j}^{\alpha}\left(  p\right)
=\sum_{\beta\in A}V_{i}^{\beta}\left(  p\right)  \left[  J_{ij}\right]
_{\beta}^{\alpha}$

If M is a class r manifold, TM is a class r-1 manifold, so vector fields can
be defined by class r-1 maps.

\begin{notation}
$\mathfrak{X}_{r}\left(  TM\right)  $ is the set of class r vector fields on
M.\ If r is omitted it will mean smooth vector fields
\end{notation}

With the structure of vector space on $T_{p}M$ the usual operations : V+W, kV
are well defined, so the set of vector fields on M has a vector space
structure. It is infinite dimensional : the components at each p are functions
(and not constant scalars) in K.

\begin{theorem}
If $V\in\mathfrak{X}\left(  TM\right)  ,W\in\mathfrak{X}\left(  TN\right)  $ then

$X\in\mathfrak{X}\left(  TM\right)  \times\mathfrak{X}\left(  TN\right)  :$
$X\left(  p\right)  =\left(  V(p),W\left(  q\right)  \right)  \in
\mathfrak{X}\left(  T\left(  M\times N\right)  \right)  $
\end{theorem}

\begin{theorem}
(Kolar p. 16) For any manifold M modelled on E, and family $\left(
p_{j},u_{j}\right)  _{j\in J}$\ of isolated points of M and vectors of E there
is always a vector field V such that $V\left(  p_{j}\right)  =\Phi_{i}\left(
p_{j},u_{j}\right)  $
\end{theorem}

\begin{definition}
The \textbf{support} of a vector field $V\in\mathfrak{X}\left(  TM\right)  $
is the support of the map : $V:M\rightarrow TM.$
\end{definition}

It is the closure of the set : $\left\{  p\in M:V(p)\neq0\right\}  $

\begin{definition}
A \textbf{critical point} of a vector field V is a point p where V(p)=0
\end{definition}

\bigskip

Topology : if M is finite dimensional, the spaces of vector fields over M can
be endowed with the topology of a Banach or Fr\'{e}chet space (see Functional
analysis). But there is no such topology available if M is infinite
dimensional, even for the vector fields with compact support (as there is no
compact if M is infinite dimensional).

\paragraph{Commutator of vector fields\newline}

\begin{theorem}
The set of of class $r\geq1$\ functions $C_{r}\left(  M;K\right)  $ over a
manifold on the field K is a commutative algebra with pointwise multiplication
as internal operation : $f\cdot g\left(  p\right)  =f\left(  p\right)
g\left(  p\right)  .$
\end{theorem}

\begin{theorem}
(Kolar p.16) The space of vector fields $\mathfrak{X}_{r}\left(  TM\right)  $
over a manifold on the field K coincides with the set of derivations on the
algebra $C_{r}\left(  M;K\right)  $
\end{theorem}

i) A derivation over this algebra (cf Algebra) is a linear map :

$D\in L(C_{r}\left(  M;K\right)  ;C_{r}\left(  M;K\right)  )$ such that

$\forall f,g\in C_{r}\left(  M;K\right)  :D(fg)=(Df)g+f(Dg)$

ii) Take a function $f\in C_{1}\left(  M;K\right)  $ we have $f^{\prime
}(p)=\sum_{\alpha\in A}f_{\alpha}^{\prime}\left(  p\right)  dx^{\alpha}\in
T_{p}M^{\ast}$

A vector field can be seen as a differential operator DV acting on f :

$DV(f)=f^{\prime}(p)V=\sum_{\alpha\in A}f_{\alpha}^{\prime}\left(  p\right)
V^{\alpha}=\sum_{\alpha\in A}V^{\alpha}\frac{\partial}{\partial x^{\alpha}}f$

DV is a derivation on $C_{r}\left(  M;K\right)  $

\begin{theorem}
The vector space of vector fields over a manifold is a Lie algebra with the
bracket, called \textbf{commutator} of vector fields :

$\forall f\in C_{r}\left(  M;K\right)  :\left[  V,W\right]  \left(  f\right)
=DV\left(  DW(f))\right)  -DW\left(  DV\left(  f\right)  \right)  $
\end{theorem}

\begin{proof}
If r%
$>$%
1, take : $DV\left(  DW(f))\right)  -DW\left(  DV\left(  f\right)  \right)  $
it is still a derivation, thus there is a vector field denoted $\left[
V,W\right]  $ such that :

$\forall f\in C_{r}\left(  M;K\right)  :\left[  V,W\right]  \left(  f\right)
=DV\left(  DW(f))\right)  -DW\left(  DV\left(  f\right)  \right)  $

The operation : $\left[  \hspace{0in}\right]  :VM\rightarrow VM$ is bilinear
and antisymmetric, and :

$\left[  V,\left[  W,X\right]  \right]  +\left[  W,\left[  X,V\right]
\right]  +\left[  X,\left[  V,W\right]  \right]  =0$

With this operation the vector space $\mathfrak{X}_{r}\left(  M\right)  $ of
vector fields becomes a Lie algebra (of infinite dimension).
\end{proof}

The bracket $\left[  {}\right]  $ is often called "Lie bracket", but as this
is a generic name we will use the -also common - name commutator.

The components of the commutator (which is a vector field) in a holonomic
basis are given by :%

\begin{equation}
\left[  V,W\right]  ^{\alpha}\mathbf{=}\sum_{\beta\in A}\left(  V^{\beta
}\partial_{\beta}W^{\prime\alpha}-W^{\beta}\partial_{\beta}V^{\prime\alpha
}\right)
\end{equation}

By the symmetry of the second derivative of the $\varphi_{i}$ for
\textit{holonomic} bases : $\forall\alpha,\beta\in A:\left[  \partial
x_{\alpha},\partial x_{\beta}\right]  =0$

\subparagraph{Commutator of vectors fields on a Banach:\newline}

Let M be an open subset of a Banach vector space E. A vector field is a map :
$V:M\rightarrow E:V(u)$ with derivative : $V^{\prime}(u):E\rightarrow E\in%
\mathcal{L}%
\left(  E;E\right)  $

With $f\in C_{r}\left(  M;K\right)  :$

$f^{\prime}(u)\in%
\mathcal{L}%
\left(  E;K\right)  ,DV:C_{r}(M;K)\rightarrow K::DV\left(  f\right)
=f^{\prime}\left(  u\right)  \left(  V\left(  u\right)  \right)  $

$\left(  DV\left(  DW(f))\right)  -DW\left(  DV\left(  f\right)  \right)
\right)  \left(  u\right)  $

$=\left(  \frac{d}{du}\left(  f^{\prime}\left(  u\right)  \left(  W\left(
u\right)  \right)  \right)  \right)  V\left(  u\right)  -\left(  \frac{d}%
{du}\left(  f^{\prime}\left(  u\right)  \left(  V\left(  u\right)  \right)
\right)  \right)  W\left(  u\right)  $

$=f"(u)\left(  W\left(  u\right)  ,V\left(  u\right)  \right)  +f^{\prime
}(u)\left(  W^{\prime}(u)\left(  V(u)\right)  \right)  -f"(u)\left(  V\left(
u\right)  ,W\left(  u\right)  \right)  -f^{\prime}(u)\left(  V^{\prime
}(u)\left(  W\left(  u\right)  \right)  \right)  $

$=f^{\prime}(u)\left(  W^{\prime}(u)\left(  V(u)\right)  -V^{\prime}(u)\left(
W\left(  u\right)  \right)  \right)  $

that we can write :

$\left[  V,W\right]  \left(  u\right)  =W^{\prime}(u)\left(  V(u)\right)
-V^{\prime}(u)\left(  W\left(  u\right)  \right)  =\left(  W^{\prime}\circ
V-V^{\prime}\circ W\right)  \left(  u\right)  $

Let now M be either the set
$\mathcal{L}$%
(E;E) of continuous maps, or its subset of invertible maps G%
$\mathcal{L}$%
(E;E), which are both manifolds, with vector bundle the set
$\mathcal{L}$%
(E;E). A vector field is a differentiable map : $V:M\rightarrow%
\mathcal{L}%
\left(  E;E\right)  $ and

$f\in M:V^{\prime}(f):%
\mathcal{L}%
\left(  E;E\right)  \rightarrow%
\mathcal{L}%
\left(  E;E\right)  \in%
\mathcal{L}%
\left(
\mathcal{L}%
\left(  E;E\right)  ;%
\mathcal{L}%
\left(  E;E\right)  \right)  $

$\left[  V,W\right]  \left(  f\right)  =W^{\prime}(f)\left(  V(f)\right)
-V^{\prime}(f)\left(  W\left(  f\right)  \right)  =\left(  W^{\prime}\circ
V-V^{\prime}\circ W\right)  \left(  f\right)  $

\paragraph{f related vector fields\newline}

\begin{definition}
The \textbf{push forward} of vector fields by a differentiable map $f\in
C_{1}\left(  M;N\right)  $ is the linear map :%

\begin{equation}
f_{\ast}:\mathfrak{X}\left(  TM\right)  \rightarrow\mathfrak{X}\left(
TN\right)  ::f_{\ast}\left(  V\right)  \left(  f\left(  p\right)  \right)
=f^{\prime}\left(  p\right)  V\left(  p\right)
\end{equation}

\end{definition}

We have the following diagram :

\bigskip

$TM\rightarrow f^{\prime}\rightarrow TN$

$\downarrow\pi_{M}\;\;\;\;\;\;\;\ \;\downarrow\pi_{N}$

$M\rightarrow\rightarrow f\rightarrow N$

\bigskip

which reads : $f_{\ast}V=f^{\prime}V$

In components :

$V\left(  p\right)  =\sum_{\alpha}v^{\alpha}\left(  p\right)  \partial
x_{\alpha}\left(  p\right)  $

$f^{\prime}(p)V(p)=\sum_{\alpha\beta}\left[  J\left(  p\right)  \right]
_{\beta}^{\alpha}v^{\beta}\left(  p\right)  \partial y_{\alpha}\left(
f\left(  p\right)  \right)  $ with $\left[  J\left(  p\right)  \right]
_{\beta}^{\alpha}=\frac{\partial y^{\alpha}}{\partial x^{\beta}}$

The vector fields $v_{i}$ can be seen as : $v_{i}=\left(  \varphi_{i}\right)
_{\ast}V::v_{i}\left(  \varphi_{i}\left(  p\right)  \right)  =\varphi
_{i}^{\prime}\left(  p\right)  V\left(  p\right)  $ and $\varphi_{i\ast
}\partial x_{\alpha}=e_{\alpha}$

\begin{theorem}
(Kolar p.20)The map $f_{\ast}:TM\rightarrow TN$ has the following properties :

i) it is a linear map : $\forall a,b\in K:$ $f_{\ast}\left(  aV_{1}%
+bV_{2}\right)  =af_{\ast}V_{1}+bf_{\ast}V_{2}$

ii) it preserves the commutator : $\left[  f_{\ast}V_{1},f_{\ast}V_{2}\right]
=f_{\ast}\left[  V_{1},V_{2}\right]  $

iii) if f is a diffeomorphism then $f_{\ast}$ is a Lie algebra morphism
between the Lie algebras $\mathfrak{X}_{r}\left(  M\right)  $ and
$\mathfrak{X}_{r}\left(  N\right)  $.
\end{theorem}

\begin{definition}
Two vector fields $V\in\mathfrak{X}_{r}\left(  M\right)  ,W\in\mathfrak{X}%
_{r}\left(  N\right)  $ are said to be \textbf{f related} if :$W\left(
f(p)\right)  =f_{\ast}V\left(  p\right)  $
\end{definition}

\begin{theorem}
(Kolar p.19) If $V\in\mathfrak{X}\left(  TM\right)  ,W\in\mathfrak{X}\left(
TN\right)  $ , $X\in\mathfrak{X}\left(  T\left(  M\times N\right)  \right)  :$
$X\left(  p\right)  =\left(  V(p),W\left(  q\right)  \right)  $ then X and V
are $\pi_{M}$ related, X and W are $\pi_{N}$ related with the projections
$\pi_{M}:M\times N\rightarrow M,$ $\pi_{N}:M\times N\rightarrow N$ .
\end{theorem}

\begin{definition}
The \textbf{pull back} of vector fields by a diffeomorphism $f\in C_{1}\left(
M;N\right)  $ is the linear map :%

\begin{equation}
f^{\ast}:\mathfrak{X}\left(  TN\right)  \rightarrow\mathfrak{X}\left(
TM\right)  ::f^{\ast}\left(  W\right)  \left(  p\right)  =\left(  f^{\prime
}\left(  p\right)  \right)  ^{-1}W\left(  f\left(  p\right)  \right)
\end{equation}

\end{definition}

So: $f^{\ast}=\left(  f^{-1}\right)  _{\ast},\varphi_{i}^{\ast}e_{\alpha
}=\partial x_{\alpha}$

\paragraph{Frames\newline}

1. A non holonomic basis in the tangent bundle is defined by : $\delta
_{\alpha}=\sum_{\beta\in A}F_{\alpha}^{\beta}\partial x_{\beta}$ where
$F_{\alpha}^{\beta}\in K$ depends on p, and as usual if the dimension is
infinite at most a finite number of them are non zero. This is equivalent to
define vector fields $\left(  \delta_{\alpha}\right)  _{\alpha\in A}$ which at
each point represent a basis of the tangent space. Such a set of vector fields
is a (non holonomic) \textbf{frame}. One can impose some conditions to these
vectors, such as being orthonormal. But of course we need to give the
$F_{\alpha}^{\beta}$ and we cannot rely upon a chart : we need additional information.

2. If this operation is always possible locally (roughly in the domain of a
chart - which can be large), it is usually impossible to have a unique frame
of vector fields covering the whole manifold (even in finite dimensions). When
this is possible the manifold is said to be \textbf{parallelizable} . For
instance the only parallelizable spheres are $S_{1},S_{3},S_{7}.$ The tangent
bundle of a parallelizable manifold is trivial, in that it can be written as
the product M$\times$E. For the others, TM is in fact made of parts of
M$\times$E glued together in some complicated manner.

\subsubsection{Flow of a vector field}

\paragraph{Integral curve\newline}

\begin{theorem}
(Kolar p.17) For any manifold M, point $p\in M$ and vector field
$V\in\mathfrak{X}_{1}\left(  M\right)  $ there is a map : $c:J\rightarrow M$
where\ J is some interval of $%
\mathbb{R}
$ such that : c(0)=p and c'(t)=V(c(t)) for $t\in J.$ The set $\left\{
c(t),t\in J\right\}  $ is an \textbf{integral curve} of V.
\end{theorem}

With an atlas $\left(  E,\left(  O_{i},\varphi_{i}\right)  _{i\in I}\right)  $
of M, and in the domain $O_{i}$, c is the solution of the differential
equation :

Find $x:%
\mathbb{R}
\rightarrow U_{i}=\varphi_{i}\left(  O_{i}\right)  \subset E$ such that :

$\frac{dx}{dt}=v\left(  x\left(  t\right)  \right)  =\varphi_{i}^{\prime
}\left(  c\left(  t\right)  \right)  V(c(t))$ and $x(0)=\varphi_{i}\left(
p\right)  $

The map v(x) is locally Lipschitz on $U_{i}$\ : it is continuously
differentiable and:

$v\left(  x+h\right)  -v\left(  x\right)  =v^{\prime}(x)h+\varepsilon\left(
h\right)  \left\Vert h\right\Vert $ and $\left\Vert v^{\prime}(x)h\right\Vert
\leq\left\Vert v^{\prime}\left(  x\right)  \right\Vert \left\Vert h\right\Vert
$

$\varepsilon\left(  h\right)  \rightarrow0\Rightarrow\forall\delta>0,\exists
r:\left\Vert h\right\Vert \leq r\Rightarrow\left\Vert \varepsilon\left(
h\right)  \right\Vert <\delta$

$\left\Vert v\left(  x+h\right)  -v\left(  x\right)  \right\Vert \leq\left(
\left\Vert v^{\prime}\left(  x\right)  \right\Vert +\left\Vert \varepsilon
\left(  h\right)  \right\Vert \right)  \left\Vert h\right\Vert \leq\left(
\left\Vert v^{\prime}\left(  x\right)  \right\Vert +\delta\right)  \left\Vert
h\right\Vert $

So the equation has a unique solution in a neighborhood of p.

The interval J can be finite, and the curve may not be defined on a the whole
of M.

\begin{theorem}
(Lang p.94) If for a class 1 vector field V on the manifold V, and V(p)=0 for
some point p, then any integral curve of V going through p is constant,
meaning that $\forall t\in%
\mathbb{R}
:c(t)=p.$
\end{theorem}

\paragraph{Flow of a vector field\newline}

\subparagraph{1. Definition:}

\begin{theorem}
(Kolar p.18) For any class 1 vector field V on a manifold M and $p\in M$ there
is a maximal interval $J_{p}\subset%
\mathbb{R}
$ such that there is an integral curve $c:J_{p}\rightarrow M$ passing at p for
t=0. The map : $\Phi_{V}:D\left(  V\right)  \times M\rightarrow M$ , called
the \textbf{flow of the vector field}, is smooth, $D(V)=\cup_{p\in M}%
J_{p}\times\left\{  p\right\}  $ is an open neighborhood of $\left\{
0\right\}  \times M,$ and%

\begin{equation}
\Phi_{V}\left(  s+t,p\right)  =\Phi_{V}\left(  s,\Phi_{V}\left(  p,t\right)
\right)
\end{equation}

\end{theorem}

The last equality has the following meaning: if the right hand side exists,
then the left hand side exists and they are equal, if s,t are both $\geq0$ or
$\leq0$\ and if the left hand side exists, then the right hand side exists and
they are equal.

\begin{notation}
$\Phi_{V}\left(  t,p\right)  $ is the flow of the vector field V, defined for
$t\in J$ and $p\in M$
\end{notation}

The theorem from Kolar can be extended to infinite dimensional manifolds (Lang p.89)

As $\Phi_{V}\left(  0,p\right)  =p$ always exist, whenever $t,-t\in J_{p}$ then%

\begin{equation}
\Phi_{V}\left(  t,\Phi_{V}\left(  -t,p\right)  \right)  =p
\end{equation}

$\Phi_{V}$ is differentiable \textit{with respect to t} and :%

\begin{equation}
\frac{\partial}{\partial t}\Phi_{V}\left(  t,p\right)  |_{t=0}=V(p);\frac
{\partial}{\partial t}(\Phi_{V}\left(  t,p\right)  |_{t=\theta}=V(\Phi
_{V}\left(  \theta,p\right)  )
\end{equation}

Warning ! the partial derivative of $\Phi_{V}\left(  t,p\right)  $ with
respect to p is more complicated (see below)

\begin{theorem}
For t fixed $\Phi_{V}\left(  t,p\right)  $ is a class r local diffeomorphism :
there is a neighborhood n(p) such that $\Phi_{V}\left(  t,p\right)  $ is a
diffeomorphism from n(p) to its image.
\end{theorem}

\bigskip

\subparagraph{2. Examples on M=$%
\mathbb{R}
^{n}$\newline}

i) V(p) = V is a constant vector field.\ Then the integral curves are straigth
lines parallel to V and passing by a given point. Take the point $A=\left(
a_{1},...a_{n}\right)  .$ Then $\Phi_{V}\left(  a,t\right)  =\left(
y_{1},...y_{n}\right)  $ such that : $\frac{\partial y_{i}}{\partial t}%
=V_{i},y_{i}\left(  a,0\right)  =a_{i}\Leftrightarrow y_{i}=tV_{i}+a_{i}$ so
the flow of V is the affine map : $\Phi_{V}\left(  a,t\right)  =Vt+a$

ii) if V(p)=Ap $\Leftrightarrow$ $V_{i}\left(  x_{1},...x_{n}\right)
=\sum_{,j=1}^{n}A_{i}^{j}x_{j}$ where A is a constant matrix.\ Then $\Phi
_{V}\left(  a,t\right)  =\left(  y_{1},...y_{n}\right)  $ such that :

$\frac{\partial y_{i}}{\partial t}|_{t=\theta}=\sum_{,j=1}^{n}A_{i}^{j}%
y_{j}\left(  \theta\right)  \Rightarrow y\left(  a,t\right)  =\left(  \exp
tA\right)  a$

iii) in the previous example, if $A=rI$ then $y\left(  a,t\right)  =\left(
\exp tr\right)  a$ and we have a radial flow

\bigskip

\subparagraph{3. Complete flow:}

\begin{definition}
The flow of a vector field is said to be \textbf{complete} if it is defined on
the whole of $%
\mathbb{R}
\times M.$ Then $\forall t$ $\Phi_{V}\left(  t,\cdot\right)  $ is a
diffeomorphism on M.
\end{definition}

\begin{theorem}
(Kolar p.19) Every vector field with compact support is complete.
\end{theorem}

So on compact manifold every vector field is complete.

There is an extension of this theorem :

\begin{theorem}
(Lang p.92) For any class 1 vector field V on a manifold M$\left(  E,\left(
O_{i},\varphi_{i}\right)  \right)  $ $v_{i}=\left(  \varphi_{i}\right)
_{\ast}V$, if :

$\forall p\in M,\exists i\in I,\exists k,r\in%
\mathbb{R}
:$

$p\in O_{i},\max\left(  \left\Vert v_{i}\right\Vert ,\left\Vert \frac{\partial
v_{i}}{\partial x}\right\Vert \right)  \leq k,B\left(  \varphi_{i}\left(
p\right)  ,r\right)  \subset\varphi_{i}\left(  O_{i}\right)  $

then the flow of V is complete.
\end{theorem}

\bigskip

\subparagraph{4. Properties of the flow:}

\begin{theorem}
(Kolar p.20,21) For any class 1 vector fields V,W on a manifold M:

$\frac{\partial}{\partial t}\left(  \Phi_{V}(t,p)_{\ast}W\right)  |_{t=0}%
=\Phi_{V}(t,p)_{\ast}\left[  V,W\right]  $

$\frac{\partial}{\partial t}\Phi_{W}\left(  -t,\Phi_{V}\left(  -t,\Phi
_{W}\left(  t,\Phi_{V}\left(  t,p\right)  \right)  \right)  \right)
|_{t=0}=0$

$\frac{1}{2}\frac{\partial^{2}}{\partial t^{2}}\Phi_{W}\left(  -t,\Phi
_{V}\left(  -t,\Phi_{W}\left(  t,\Phi_{V}\left(  t,p\right)  \right)  \right)
\right)  |_{t=0}=\left[  V,W\right]  $

The following are equivalent :

i) $\left[  V,W\right]  =0$

ii) $\left(  \Phi_{V}\right)  ^{\ast}W=W$ whenever defined

iii) $\Phi_{V}\left(  t,\Phi_{W}\left(  s,p\right)  \right)  =\Phi_{W}\left(
s,\Phi_{V}\left(  t,p\right)  \right)  $ whenever defined
\end{theorem}

\begin{theorem}
(Kolar p.20) For a differentiable map $f\in C_{1}\left(  M;N\right)  $ between
the manifolds M,N, and any vector field $V\in\mathfrak{X}_{1}\left(
TM\right)  :$ $f\circ\Phi_{V}=\Phi_{f_{\ast}V}\circ f$ whenever both sides are
defined. If f is a diffeomorphism then similarly for $W\in\mathfrak{X}%
_{1}\left(  TN\right)  :$ $f\circ\Phi_{f^{\ast}W}=\Phi_{W}\circ f$
\end{theorem}

\begin{theorem}
(Kolar p.24) For any vector fields $V_{k}\in\mathfrak{X}_{1}\left(  TM\right)
,k=1...n$\ on a real n-dimensional manifold M such that :

i) $\forall k,l:\left[  V_{k},V_{l}\right]  =0$

i) $V_{k}\left(  p\right)  $ are linearly independent at p

there is a chart centered at p such that $V_{k}=\partial x_{k}$
\end{theorem}

\bigskip

\subparagraph{5. Remarks:\newline}

i) $\frac{\partial}{\partial t}\Phi_{V}\left(  t,p\right)  |_{t=0}%
=V(p)\Rightarrow\frac{\partial}{\partial t}(\Phi_{V}\left(  t,p\right)
|_{t=\theta}=V(\Phi_{V}\left(  \theta,p\right)  )$

\begin{proof}
Let be $T=t+\theta,\theta$ fixed

$\Phi_{V}\left(  T,p\right)  =\Phi_{V}\left(  t,\Phi_{V}\left(  \theta
,p\right)  \right)  $

$\frac{\partial}{\partial t}\Phi_{V}\left(  T,p\right)  |_{t=0}=\frac
{\partial}{\partial t}(\Phi_{V}\left(  t,p\right)  |_{t=\theta}=\frac
{\partial}{\partial t}\Phi_{V}\left(  t,\Phi_{V}\left(  \theta,p\right)
\right)  |_{t=0}=V(\Phi_{V}\left(  \theta,p\right)  )$
\end{proof}

So the flow is fully defined by the equation : $\frac{\partial}{\partial
t}(\Phi_{V}\left(  t,p\right)  |_{t=0}=V(p)$

ii) If we proceed to the change of parameter : $s\rightarrow t=f\left(
s\right)  $ with $f:J\rightarrow J$ some function such that f(0)=0,f'(s)$\neq
0$

$\Phi_{V}\left(  t,p\right)  =\Phi_{V}\left(  f(s),p\right)  =\widehat{\Phi
}_{V}\left(  s,p\right)  $

$\frac{\partial}{\partial s}(\widehat{\Phi}_{V}\left(  s,p\right)
|_{s=0}=\frac{\partial}{\partial t}(\Phi_{V}\left(  t,p\right)  |_{t=f\left(
0\right)  }\frac{df}{ds}|_{s=0}$

$=V(\Phi_{V}\left(  f\left(  0\right)  ,p\right)  )\frac{df}{ds}%
|_{s=0}=V(p)\frac{df}{ds}|_{s=0}$

So it sums up to replace the vector field V by $\widehat{V}\left(  p\right)
=$ $V(p)\frac{df}{ds}|_{s=0}$

iii) the Lie derivative (see next sections)

$\pounds _{V}W=\left[  V,W\right]  =\frac{\partial}{\partial t}\left(
\frac{\partial}{\partial p}\Phi_{V}\left(  -t,p\right)  \circ W\circ
\frac{\partial}{\partial p}\Phi_{V}\left(  t,p\right)  \right)  |_{t=0}$

\paragraph{One parameter group of diffeomorphisms\newline}

\begin{definition}
A \textbf{one parameter group of diffeomorphims} on a manifold M is a map :
$F:%
\mathbb{R}
\times M\rightarrow M$ such that for each t fixed $F\left(  t,.\right)  $ is a
diffeomorphism on M and%

\begin{equation}
\forall t,s\in%
\mathbb{R}
,p\in M:F\left(  t+s,p\right)  =F\left(  t,F\left(  s,p\right)  \right)
=F\left(  s,F\left(  t,p\right)  \right)  ;F(0,p)=p
\end{equation}

\end{definition}

$%
\mathbb{R}
\times M$ has a manifold structure so F has partial derivatives.\ 

For p fixed $F(.,p):%
\mathbb{R}
\rightarrow M$ and $F_{t}^{\prime}\left(  t,p\right)  \in T_{F(t,p)}M$ so
$F_{t}^{\prime}\left(  t,p\right)  |_{t=0}\in T_{p}M$ and there is a vector field

$V\left(  p\right)  =\Psi_{i}\left(  p,v(p)\right)  $ with $v\left(  p\right)
=\varphi_{i}^{\prime}\left(  p\right)  \left(  F_{t}^{\prime}\left(
t,p\right)  |_{t=0}\right)  $

So V is the \textbf{infinitesimal generator} of F :$F\left(  t,p\right)
=\Phi_{V}\left(  t,p\right)  $

Warning ! If M has the atlas $\left(  E,\left(  O_{i},\varphi_{i}\right)
_{i\in I}\right)  $ the partial derivative with respect to p : $F_{p}^{\prime
}\left(  t,p\right)  \in%
\mathcal{L}%
\left(  T_{p}M;T_{F(t,p)}M\right)  $ and

$U\left(  t,p\right)  =\varphi_{i}^{\prime}\circ F_{p}^{\prime}\circ
\varphi_{i}^{\prime}{}^{-1}\left(  a\right)  \in%
\mathcal{L}%
\left(  E;E\right)  $

$U\left(  t+s,p\right)  =\varphi_{i}^{\prime}\circ F_{p}^{\prime}\left(
t+s,p\right)  \circ\varphi_{i}^{\prime}{}^{-1}\left(  a\right)  $

$=\varphi_{i}^{\prime}\circ F_{p}^{\prime}\left(  t,F\left(  s,p\right)
\right)  \circ F_{p}^{\prime}\left(  s,p\right)  \circ\varphi_{i}^{\prime}%
{}^{-1}\left(  a\right)  $

$=\varphi_{i}^{\prime}\circ F_{p}^{\prime}\left(  t,F\left(  s,p\right)
\right)  \circ\varphi_{i}^{\prime}{}^{-1}\circ\varphi_{i}^{\prime}{}\circ
F_{p}^{\prime}\left(  s,p\right)  \circ\varphi_{i}^{\prime}{}^{-1}\left(
a\right)  =U(t,F\left(  s,p\right)  )\circ U(s,p)$

So we \textit{do not} have a one parameter group on the Banach E which would
require : U(t+s,p)=U(t,p)oU(s,p).

\bigskip

\subsection{Submanifolds}

\label{Submanifolds}

A submanifold is a part of a manifold that is itself a manifold, meaning that
there is an atlas to define its structure. This can be conceived in several
ways.\ The choice that has been made is that the structure of a submanifold
must come from its "mother". Practically this calls for a specific map which
injects the submanifold structure into the manifold : an embedding. But there
are other ways to relate two manifolds, via immersion and submersion. The
definitions vary according to the authors. We have chosen the definitions
which are the most illuminating and practical, without loss of generality. The
theorems cited have been adjusted to account for these differences.

The key point is that most of the relations between the manifolds M,N stem
from the derivative of the map $f:M\rightarrow N$ which is linear and falls
into one of 3 cases : injective, surjective or bijective.

For finite dimensional manifolds the results sum up in the following :

\begin{theorem}
(Kobayashi I p.8) For a differentiable map f from the m dimensional manifold M
to the n dimensional manifold N, at any point p in M:

i) if f'(p) is bijective there is a neighborhood n(p) such that f is a
diffeomorphism from n(p) to f(n(p))

ii) if f'(p) is injective from a neighborhood n(p) to n(f(p)), f is a
homeomorphism from n(p) to f(n(p)) and there are charts $\varphi$ of M, $\psi$
of N such that $F=\psi\circ f\circ\varphi^{-1}$ reads : i=1..m : $y^{i}\left(
f\left(  p\right)  \right)  =x^{i}\left(  p\right)  $

iii) if f'(p) is surjective from a neighborhood n(p) to n(f(p)),
$f:n(p)\rightarrow N$ is open, and there are charts $\varphi$ of M, $\psi$ of
N such that $F=\psi\circ f\circ\varphi^{-1}$ reads : i=1..n : $y^{i}\left(
f\left(  p\right)  \right)  =x^{i}\left(  p\right)  $
\end{theorem}

\subsubsection{Submanifolds}

\paragraph{Submanifolds\newline}

\begin{definition}
A subset M of a manifold N is a \textbf{submanifold} of N if :

i) $G=G_{1}\oplus G_{2}$ where $G_{1},G_{2}$ are vector subspaces of G

ii) there is an atlas $\left(  G,\left(  Q_{i},\psi_{j}\right)  _{j\in
J}\right)  $\ of N such that M is a manifold with atlas $\left(  G_{1},\left(
M\cap Q_{i},\psi_{j}|_{M\cap Q_{i}}\right)  _{i\in I}\right)  $
\end{definition}

So a condition is that $\dim M\leq\dim N$

The key point is that the manifold structure of M is defined through the
structure of manifold of N. M has no manifold structure of its own. But it is
clear that not any subset can be a submanifold.

Topologically M can be any subset, so it can be closed in N and so we have the
concept of closed manifold.

\begin{theorem}
For any point p of the submanifold M in N, the tangent space $T_{p}M$ is a
subspace of $T_{p}N$
\end{theorem}

\begin{proof}
$\forall q\in N,$ $\psi_{j}\left(  q\right)  $\ can be uniquely written as :

$\psi_{j}\left(  q\right)  =\sum_{\alpha\in B_{1}}x^{\alpha}e_{\alpha}%
+\sum_{\beta\in B_{2}}x^{\beta}e_{\beta}$ with $\left(  e_{\alpha}\right)
_{\alpha\in B_{1}},\left(  e_{\beta}\right)  _{\beta\in B_{2}}$ bases of
$G_{1},G_{2}$

$q\in M\Leftrightarrow\forall\beta\in B_{2}:x^{\beta}=0$

For any vector $u_{q}\in T_{q}N:u_{q}=\sum_{\alpha\in B}u_{q}^{\alpha}\partial
x_{\alpha}$

$\psi_{j}^{\prime}\left(  q\right)  u_{q}=\sum_{\alpha\in B_{1}}u_{q}^{\alpha
}\partial x_{\alpha}+\sum_{\beta\in B_{2}}u_{q}^{\beta}\partial x_{\beta}$

and $u_{q}\in T_{q}M\Leftrightarrow\forall\beta\in B_{2}:u_{q}^{\beta}=0$

So : $\forall p\in M:T_{p}M\subset T_{p}N$ and a vector tangent to N at p can
be written uniquely :

$u_{p}=u_{1}+u_{2}:u_{1}\in T_{p}M$ with $u_{p}\in T_{p}M\Leftrightarrow
u_{2}=0$
\end{proof}

The vector $u_{2}$ is said to be \textbf{transversal} to M at p

\begin{definition}
If N is a n finite dimensional manifold and M is a submanifold of dimension n
- 1 then M is called an \textbf{hypersurface}.
\end{definition}

\begin{theorem}
Extension of a map (Schwartz II p.442) A map $f\in C_{r}\left(  M;E\right)
,$r$\geq1$\ from a m dimensional class r submanifold M of a real manifold N
which is the union of countably many compacts, to a Banach vector space E can
be extended to $\widehat{f}\in C_{r}\left(  N;E\right)  $
\end{theorem}

\paragraph{Conditions for a subset to be a manifold\newline}

\begin{theorem}
An open subset of a manifold is a submanifold with the same dimension.
\end{theorem}

\begin{theorem}
A connected component of a manifold is a a submanifold with the same dimension.
\end{theorem}

\begin{theorem}
(Schwartz II p.261) For a subset M of a n dimensional class r manifold
N$\left(  E,\left(  Q_{i},\psi_{i}\right)  _{i\in I}\right)  $ of a field K,
if, $\forall p\in M,$ there is, in a neighborhood of p, a chart $\left(
Q_{i},\psi_{i}\right)  $ of N such that :

i) either $\psi_{i}\left(  M\cap Q_{i}\right)  =\left\{  x\in K^{n}%
:x_{m+1}=..=x_{n}=0\right\}  $ and M is closed

i) or $\psi_{i}\left(  M\cap Q_{i}\right)  =\psi_{i}\left(  Q_{i}\right)  \cap
K^{m}$

then M is a m dimensional class r submanifold of N
\end{theorem}

\begin{theorem}
Smooth retract (Kolar p.9): If M is a class r connected finite dimensional
manifold, $f\in C_{r}\left(  M;M\right)  $ such that $f\circ f=f$ then f(M) is
a submanifold of M
\end{theorem}

\paragraph{Embedding\newline}

The previous definition is not practical in many cases. It is more convenient
to use a map, as it is done in a parametrized representation of a submanifold
in $%
\mathbb{R}
^{n}.$\ There are different definitions of an embedding.\ The simplest if the following.

\begin{definition}
An \textbf{embedding} is a map $f:C_{r}\left(  M;N\right)  $ between two
manifolds M,N such that:

i) f is a diffeomorphism from M to f(M)

ii) f(M) is a submanifold of N
\end{definition}

M is the origin of the parameters, f(M) is the submanifold. So M must be a
manifod, and we must know that f(M) is a submanifold. To be the image by a
diffeomorphism is not sufficient. The next subsection deals with this issue.

$\dim M=\dim f(M)\leq\dim N$

If M,N are finite dimensional, in coordinates f can be written in a
neighborhood of $q\in f\left(  M\right)  $ and adaptated charts :

$\beta=1...m:y^{\beta}=F^{\beta}\left(  x^{1},...x^{m}\right)  $

$\beta=m+1...n:y^{\beta}=0$

The image of a vector $u_{p}\in M$ is $f^{\prime}(p)u_{p}=v_{1}+v_{2}:v_{1}\in
T_{p}f\left(  M\right)  $ and $v_{2}=0$

The jacobian $\left[  f^{\prime}(p)\right]  _{m}^{n}$ is of rank m.

If M is a m dimensional embedded submanifold of N then it is said that M has
codimension n-m.

Example :

\begin{theorem}
Let $c:J\rightarrow N$ be a path in the manifold N with J an interval in $%
\mathbb{R}
$. The curve C=$\left\{  c(t),t\in J\right\}  \subset N$ is a connected 1
dimensional submanifold iff c is class 1 and c'(t) is never zero. If J is
closed then C is compact.
\end{theorem}

\begin{proof}
$c^{\prime}(t)\neq0:$ then c is injective and a homeomorphism in N

$\psi_{j}^{\prime}\circ c^{\prime}(t)$ is a vector in the Banach G and there
is an isomorphism between $%
\mathbb{R}
$ as a vector space and the 1 dimensional vector space generated by $\psi
_{j}^{\prime}\circ c^{\prime}(t)$ in G
\end{proof}

\paragraph{Submanifolds defined by embedding\newline}

The following important theorems deal with the pending issue : is f(M) a
submanifold of N ?

\begin{theorem}
Theorem of constant rank (Schwartz II .263) : If the map $f\in C_{1}\left(
M;N\right)  $\ on a m dimensional manifold M to a manifold N has a constant
rank s\ in M then :

i) $\forall p\in M,$ there is a neighborhood n(p) such that f(n(p)) is a s
dimensional submanifold of N.\ For any $\mu\in n\left(  p\right)  $ we have :
$T_{f(\mu)}f(n\left(  p\right)  )=f^{\prime}(\mu)T_{\mu}M$ .

ii) $\forall q\in f(M),$ the set $f^{-1}(q)$ is a closed m-s submanifold of M and

$\forall p\in f^{-1}(q):T_{p}f^{-1}(q)=\ker f^{\prime}(p)$
\end{theorem}

\begin{theorem}
(Schwartz II p.263) If the map $f\in C_{1}\left(  M;N\right)  $\ on a m
dimensional manifold M is such that f is injective and $\forall p\in M$ f'(p)
is injective

i) if M is compact then f(M) is a submanifold of N and f is an embedding.

ii) if f is an homeomorphism of M to f(M) then f(M) is a submanifold of N and
f is an embedding.
\end{theorem}

\begin{theorem}
(Schwartz II p.264) If, for the map $f\in C_{1}\left(  M;N\right)  $\ on a m
dimensional manifold M, f'(p) is injective at some point p, there is a
neighborhood n(p) such that f(n(p)) is a submanifold of N and f an embedding
of n(p) into f(n(p)).
\end{theorem}

Remark : L.Schwartz used a slightly different definition of an embedding.\ His
theorems are adjusted to our definition.

\begin{theorem}
(Kolar p.10) A smooth n dimensional real manifold can be embedded in $%
\mathbb{R}
^{2n+1}$ and $%
\mathbb{R}
^{2n}$
\end{theorem}

\paragraph{Immersion\newline}

\begin{definition}
A map $f\in C_{1}\left(  M;N\right)  $ from the manifold M to the manifold N
is an \textbf{immersion} at p is f'(p) is injective. It is an immersion of M
into N if it is an immersion at each point of M.
\end{definition}

In an immersion $\dim M\leq\dim N$\ (f(M) is "smaller" than N so it is
immersed in N)

\begin{theorem}
(Kolar p.11) If the map $f\in C_{1}\left(  M;N\right)  $ from the manifold M
to the manifold N is an immersion on M, both finite dimensional, then for any
p in M there is a neighborhood n(p) such that f(n(p)) is a submanifold of N
and f an embedding from n(p) to f(n(p)).
\end{theorem}

\begin{theorem}
(Kolar p.12) If the map $f\in C_{1}\left(  M;N\right)  $ from the manifold M
to the manifold N both finite dimensional, is an immersion, if f is injective
and a homeomorphism on f(M), then f(M) is a submanifold of N.
\end{theorem}

\begin{theorem}
(Kobayashi I p.178) If the map $f\in C_{1}\left(  M;N\right)  $ from the
manifold M to the manifold N, both connected and of the same dimension, is an
immersion, if M is compact then N is compact and a covering space for M and f
is a projection.
\end{theorem}

\paragraph{Real submanifold of a complex manifold\newline}

We always assume that M,N are defined, as manifolds or other structure, on the
same field K. However it happens that a subset of a complex manifold has the
structure of a real manifold.\ For instance the matrix group U(n) is a real
manifold comprised of complex matrices and a subgroup of GL($%
\mathbb{C}
,n).$ To deal with such situations we define the following :

\begin{definition}
A real manifold M$\left(  E,\left(  O_{i},\varphi_{i}\right)  _{i\in
I}\right)  $ is an immersed submanifold of the complex manifold N$\left(
G,\left(  Q_{i},\psi_{i}\right)  _{i\in I^{\prime}}\right)  $ if there is a
map : $f:M\rightarrow N$ such that the map : $F=\psi_{j}\circ f\circ
\varphi_{i}^{-1}$, whenever defined, is R-differentiable and its derivative is injective.
\end{definition}

The usual case is f=Identity.

\paragraph{Submersions\newline}

Submersions are the converse of immersions.\ Here M is "larger" than N so it
is submersed by M. They are mainly projections of M on N and used in fiber bundles.

\begin{definition}
A map $f\in C_{1}\left(  M;N\right)  $ from the manifold M to the manifold N
is a \textbf{submersion} at p is f'(p) is surjective. It is an submersion of M
into N if it is an submersion at each point of M.
\end{definition}

In an submersion $\dim N\leq\dim M$

\begin{theorem}
(Kolar p.11) A submersion on finite dimensional manifolds is an open map
\end{theorem}

A fibered manifold $M\left(  N,\pi\right)  $ is a triple of two manifolds M, N
and a map $\pi:M\rightarrow N$ which is both surjective and a submersion. It
has the universal property : if f is a map $f\in C_{r}\left(  N;P\right)  $ in
another manifold P then $f\circ\pi$ is class r iff f is class r (all the
manifolds are assumed to be of class r).

\paragraph{Independant maps\newline}

This an aplication of the previous theorems to the following problem : let
$f\in C_{1}\left(  \Omega;K^{n}\right)  ,\Omega$ open in $K^{m}.$ We want to
tell when the n scalar maps $f_{i}$ are "independant".

We can give the following meaning to this concept. f is a map between two
manifolds.\ If $f\left(  \Omega\right)  $ is a p$\leq n$ dimensional
submanifold of $K^{n}$ , any point q in $f\left(  \Omega\right)  $\ can be
coordinated by p scalars y. If p%
$<$%
m we could replace the m variables x by y and get a new map which can meet the
same values with fewer variables.

1) Let m$\geq n$ .\ If f'(x) has a constant rank p then the maps are independant

2) If f'(x) has a constant rank r%
$<$%
m then locally $f\left(  \Omega\right)  $ is a r dimensional submanifold of
$K^{n}$\ and we have n-r independent maps.

\subsubsection{Distributions}

Given a vector field, it is possible to define an integral curve such that its
tangent at any point coincides with the vector. \ A distribution is a
generalization of this idea : taking several vector fields, they define at
each point a vector space and we look for a submanifold which admits this
vector space as tangent space.

Distributions of Differential Geometry are not related in any way to the
distributions of Functional Analysis.

\paragraph{Definitions\newline}

\subparagraph{1. Distribution:}

\begin{definition}
A r dimensional \textbf{distribution} on the manifold M is a map :
$W:M\rightarrow\left(  TM\right)  ^{r}$ such that W(p) is a r dimensional
vector subspace of $T_{p}M$
\end{definition}

If M is an open in K$^{m}$ a r dimensional distribution is a map between M and
the grassmanian Gr(K$^{m}$;r) which is a (m-r)r dimensional manifold.

The definition can be generalized : W(p) can be allowed to have different
dimensions at different points, and even be infinite dimensional.\ We will
limit ourselves to more usual conditions.

\begin{definition}
A family $\left(  V_{j}\right)  _{j\in J}$ of vector fields on a manifold M
generates a distribution W if for any point p in M the vector subspace spanned
by the family is equal to W(p) : $\forall p\in M:W(p)=Span\left(  V_{j}\left(
p\right)  \right)  $
\end{definition}

So two families are equivalent with respect to a distribution if they generate
the same distribution. To generate a m dimensional distribution the family
must be comprised at least of m pointwise linearly independent vector fields.

\subparagraph{2. Integral manifold:}

\begin{definition}
A connected submanifold L of M is an \textbf{integral manifold} for the
distribution W on M if $\forall p\in L:$ $T_{p}L=W\left(  p\right)  $
\end{definition}

So dimL=dimW. A distribution is not always integrable, and the submanifolds
are usually different at each point.

An integral manifold is said to be maximal if it is not strictly contained in
another integral manifold. If there is an integral manifold, there is always a
unique maximal integral manifold. Thus we will assume in the following that
the integral manifolds are maximal.

\begin{definition}
A distribution W on M is \textbf{integrable} if there is a family $\left(
L_{\lambda}\right)  _{\lambda\in\Lambda}$\ of maximal integral manifolds of W
such that : $\forall p\in M:\exists\lambda:p\in L_{\lambda}.$ This family
defines a partition of M, called a \textbf{folliation}, and each $L_{\lambda}$
is called a \textbf{leaf} of the folliation.
\end{definition}

Notice that the condition is about points of M.

$p\sim q\Leftrightarrow\left(  p\in L_{\lambda}\right)  \&\left(  q\in
L_{\lambda}\right)  $ is a relation of equivalence for points in M which
defines the partition of M.

Example : take a single vector field.\ An integral curve is an integral
manifold. If there is an integral curve passing through each point then the
distribution given by the vector field is integrable, but we have usually many
integral submanifolds. We have a folliation, whose leaves are the curves.

\subparagraph{3. Stability of a distribution:}

\begin{definition}
A distribution W on a manifold M is \textbf{stable} by a map $f\in
C_{1}\left(  M;M\right)  $ if : $\forall p\in M:f^{\prime}(p)W\left(
p\right)  \subset W\left(  f\left(  p\right)  \right)  $
\end{definition}

\begin{definition}
A vector field V on a manifold M is said to be an \textbf{infinitesimal
automorphism of the distribution} W on M if W is stable by the flow of V
\end{definition}

meaning that : $\frac{\partial}{\partial p}\Phi_{V}\left(  t,p\right)  \left(
W(p)\right)  \subset W\left(  \Phi_{V}\left(  t,p\right)  \right)  $ whenever
the flow is defined.

The set Aut(W) of vector fields which are infinitesimal generators of W is stable.

\subparagraph{4. Family of involutive vector fields:}

\begin{definition}
A subset $V\subset\mathfrak{X}_{1}\left(  TM\right)  $ is \textbf{involutive }if

$\forall V_{1},V_{2}\in V,\exists V_{3}\in V:\left[  V_{1},V_{2}\right]
=V_{3} $
\end{definition}

\paragraph{Conditions for integrability of a distribution\newline}

There are two main formulations, one purely geometric, the other relying on forms.

\subparagraph{1. Geometric formulation:\newline}

\begin{theorem}
(Maliavin p.123) A distribution W on a finite dimensional manifold M is
integrable iff there is an atlas $\left(  E,\left(  O_{i},\varphi_{i}\right)
_{i\in I}\right)  $ of M such that, for any point p in M and neighborhood n(p) :

$\forall q\in n(p)\cap O_{i}:\varphi_{i}^{\prime}\left(  q\right)  W\left(
q\right)  =E_{i1}$ where $E=E_{i1}\oplus E_{i2}$
\end{theorem}

\begin{theorem}
(Kolar p.26) For a distribution W on a finite dimensional manifold M the
following conditions are equivalent:

i) the distribution W is integrable

ii) the subset $V_{W}=\left\{  V\in\mathfrak{X}\left(  TM\right)  :\forall
p\in M:V\left(  p\right)  \in W\left(  p\right)  \right\}  $ is stable:

$\forall V_{1},V_{2}\in V_{W},\exists X\in V_{W}:\frac{\partial}{\partial
p}\Phi_{V_{1}}\left(  t,p\right)  \left(  V_{2}\left(  p\right)  )\right)
=X\left(  \Phi_{V_{1}}\left(  t,p\right)  \right)  $ whenever the flow is defined.

iii) The set $Aut(W)\cap V_{W}$\ spans W

iv) There is an involutive family $\left(  V_{j}\right)  _{j\in J}$ which
generates W
\end{theorem}

\bigskip

\subparagraph{2. Formulation using forms :}

\begin{theorem}
(Malliavin p.133) A class 2 form $\varpi\in\Lambda_{1}\left(  M;V\right)  $ on
a class 2 finite dimensional manifold M valued in a finite dimensional vector
space V such that $\ker\varpi\left(  p\right)  $ has a constant finite
dimension on M defines a distribution on M : $W\left(  p\right)  =\ker
\varpi\left(  p\right)  .$ This distribution is integrable iff :

$\forall u,v\in W(p):\varpi\left(  p\right)  u=0,\varpi\left(  p\right)
v=0\Rightarrow d\varpi\left(  u,v\right)  =0$
\end{theorem}

\begin{corollary}
A function $f\in C_{2}\left(  M;%
\mathbb{R}
\right)  $ on a m dimensional manifold M such that $\dim\ker f^{\prime}(p)=Ct$
defines an integrable distribution, whose folliation is given by $f\left(
p\right)  =Ct$
\end{corollary}

\begin{proof}
The derivative $f^{\prime}(p)$ defines a 1-form df on N. Its kernel has
dimension m-1 at most. d(df)=0 thus we have always $d\varpi\left(  u,v\right)
=0.$
\end{proof}

$W\left(  p\right)  =\ker\varpi\left(  p\right)  $ is represented by a system
of partial differential equations called a Pfaff system.

\subsubsection{Manifold with boundary}

In physics usually manifolds enclose a system.\ The walls are of paramount
importance as it is where some conditions determining the evolution of the
system are defined.\ Such manifolds are manifolds with boundary. They are the
geometrical objects of the Stokes' theorem and are essential in partial
differential equations. We present here a new theorem which gives a stricking
definition of these objects.

\paragraph{Hypersurfaces\newline}

A hypersurface divides a manifold in two disjoint parts :

\begin{theorem}
(Schwartz IV p.305) For any n-1 dimensional class 1 submanifold M of a n
dimensional class 1 real manifold N, every point p of M has a neighborhood
n(p) in N such that :

i) n(p) is homeomorphic to an open ball

ii) $M\cap n(p)$ is closed in n(p) and there are two disjoint connected
subsets $n_{1},n_{2}$ such that :

$n(p)=\left(  M\cap n(p)\right)  \cup n_{1}\cup n_{2},$

$\forall q\in M\cap n(p):q\in\overline{n_{1}}\cap\overline{n_{2}}$

iii) there is a function $f:N\rightarrow%
\mathbb{R}
$ such that :

$n(p)=\left\{  q:f(q)=0\right\}  ,n_{1}=\left\{  q:f(q)<0\right\}
,n_{2}=\left\{  q:f(q)>0\right\}  $
\end{theorem}

\begin{theorem}
Lebesgue (Schwartz IV p.305) : Any closed class 1 hypersurface M of a finite
dimensional real affine space E parts E in at least 2 regions, and exactly two
if M is connected.
\end{theorem}

\paragraph{Definition\newline}

There are several ways to define a manifold with boundary, always in finite
dimensions. We will use only the following, which is the most general and
useful (Schwartz\ IV p.343) :

\begin{definition}
A \textbf{manifold with boundary} is a set M :

i) which is a subset of a n dimensional \textit{real} manifold N

ii) identical to the closure of its interior : $M=\overline{\left(
\overset{\circ}{M}\right)  }$

iii) whose border $\partial M$ called its boundary is a hypersurface in N
\end{definition}

Remarks :

i) M inherits the topology of N so the interior $\overset{\circ}{M},$ the
border $\partial M$ are well defined (see topology).\ The condition ii)
prevents "spikes" or "barbed" areas protuding from M. So M is exactly the
disjointed union of its interior and its boundary:

$M=\overline{M}=\overset{\circ}{M}\cup\partial M=\left(  \overset{\circ
}{\left(  M^{c}\right)  }\right)  ^{c}$

$\overset{\circ}{M}\cap\partial M=\varnothing$

$\partial M=M\cap\overline{\left(  M^{c}\right)  }=\partial\left(
M^{c}\right)  $

ii) M is closed in N, so usually \textit{it is not a manifold}

iii) we will always assume that $\partial M\neq\varnothing$

iv) N must be a real manifold as the sign of the coordinates plays a key role

\paragraph{Properties\newline}

\begin{theorem}
(Schwartz IV p.343) If M is a manifold with boundary in N, N and $\partial M$
both connected then :

i) $\partial M$ splits N in two disjoint regions : $\overset{\circ}{M}$ and
$M^{c}$

ii) If O is an open in N and $M\cap O\neq\varnothing$\ \ then $M\cap O$ is
still a manifod with boundary : $\partial M\cap O$

iii) any point p of $\partial M$ is adherent to M, $\overset{\circ}{M}$ and
$M^{c}$
\end{theorem}

\begin{theorem}
(Lafontaine p.209) If M is a manifold with boundary in N, then there is an
atlas $\left(  O_{i},\varphi_{i}\right)  _{i\in I}$ of N such that :

$\varphi_{i}\left(  O_{i}\cap\overset{\circ}{M}\right)  =\{x\in\varphi
_{i}(O_{i}):x_{1}<0\}$

$\varphi_{i}\left(  O_{i}\cap\partial M\right)  =\{x\in\varphi_{i}%
(O_{i}):x_{1}=0\}$
\end{theorem}

\begin{theorem}
(Taylor 1 p.97) If M is a compact manifold with boundary in an oriented
manifold N then there is no continuous retraction from M to $\partial M.$
\end{theorem}

\paragraph{Transversal vectors\newline}

The tangent spaces $T_{p}\partial M$\ to the boundary are hypersurfaces of the
tangent space $T_{p}N.$ The vectors of $T_{p}N$ which are not in
$T_{p}\partial M$ are said to be \textbf{transversal}.

If N and $\partial M$ are both connected then any class 1 path c(t) :
$c:\left[  a,b\right]  \rightarrow N$ such that $c(a)\in\overset{\circ}{M}$
and $c(b)\in M^{c}$ meets $\partial M$ at a unique point (see topology). For
any transversal vector : $u\in T_{p}N,p\in\partial M,$ if there is such a path
with $c^{\prime}(t)=ku,k>0$ then u is said to be \textbf{outward oriented},
and inward oriented if $c^{\prime}(t)=ku,k<0.$ Notice that we do not need to
define an orientation on N.

Equivalently if V is a vector field such that its flow is defined from
$p\in\overset{\circ}{M}$ to $q\in M^{c}$ then V is outward oriented if
$\exists t>0:q=\Phi_{V}\left(  t,p\right)  .$

\paragraph{Fundamental theorems\newline}

Manifolds with boundary have a unique characteristic : they can be defined by
a function : $f:N\rightarrow%
\mathbb{R}
.$

It seems that the following theorems are original, so we give a full proof.

\bigskip

\begin{theorem}
Let N be a n dimensional smooth Hausdorff real manifold.

i) Let $f\in C_{1}\left(  N;%
\mathbb{R}
\right)  $ and $P=f^{-1}\left(  0\right)  \neq\varnothing,$ if f'(p)$\neq0$ on
P then the set $M=\left\{  p\in N:f(p)\leq0\right\}  $ is a manifold with
boundary in N, with boundary $\partial M=P.$ And : $\forall p\in\partial
M,\forall u\in T_{p}\partial M:f^{\prime}(p)u=0$

ii) Conversely if M is a manifold with boundary in N there is a function :
$f\in C_{1}\left(  N;%
\mathbb{R}
\right)  $ such that :

$\overset{\circ}{M}=\left\{  p\in N:f(p)<0\right\}  ,\partial M=\left\{  p\in
N:f(p)=0\right\}  $

$\forall p\in\partial M:f^{\prime}(p)\neq0$ and : $\forall u\in T_{p}\partial
M:f^{\prime}(p)u=0,$

for any transversal vector v : $f^{\prime}(p)v\neq0$

If M and $\partial M$ are connected then for any transversal outward oriented
vector v : $f^{\prime}(p)v>0$

iii) for any riemannian metric on N the vector $gradf$ defines a vector field
outward oriented normal to the boundary
\end{theorem}

N is smooth finite dimensional Hausdorff, thus paracompact and admits a
Riemanian metric

Proof of i)

\begin{proof}
f is continuous thus P is closed in N and M'=$\left\{  p\in N:f(p)<0\right\}
$ is open.

The closure of M' is the set of points which are limit of sequences in M' :
$\overline{M^{\prime}}=\left\{  \lim q_{n},q_{n}\in M^{\prime}\right\}
=\left\{  p\in N:f(p)\leq0\right\}  =M$

f has constant rank 1 on P, thus the set P is a closed n-1 submanifold of N
and $\forall p\in P:T_{p}P=\ker f^{\prime}(p)$ thus $\forall u\in
T_{p}\partial M:f^{\prime}(p)u=0.$
\end{proof}

Proof of ii)

\begin{proof}
1) there is an atlas $\left(  O_{i},\varphi_{i}\right)  _{i\in I}$ of N such
that :

$\varphi_{i}\left(  O_{i}\cap\overset{\circ}{M}\right)  =\{x\in\varphi
_{i}(O_{i}):x_{1}<0\}$

$\varphi_{i}\left(  O_{i}\cap\partial M\right)  =\{x\in\varphi_{i}%
(O_{i}):x_{1}=0$

Denote : $\varphi_{i}^{1}\left(  p\right)  =x_{1}$ thus $\forall p\in
M:\varphi_{i}^{1}\left(  p\right)  \leq0$

N admits a smooth partition of unity subordinated to O$_{i}:$

$\chi_{i}\in C_{\infty}\left(  N;%
\mathbb{R}
_{+}\right)  :\forall p\in O_{i}^{c}:\chi_{i}\left(  p\right)  =0;\forall p\in
N:\sum_{i}\chi_{i}\left(  p\right)  =1$

Define : $f\left(  p\right)  =\sum_{i}\chi_{i}\left(  p\right)  \varphi
_{i}^{1}\left(  p\right)  $

Thus :

$\forall p\in\overset{\circ}{M}:f\left(  p\right)  =\sum_{i}\chi_{i}\left(
p\right)  \varphi_{i}^{1}\left(  p\right)  <0$

$\forall p\in\partial M:f\left(  p\right)  =\sum_{i}\chi_{i}\left(  p\right)
\varphi_{i}^{1}\left(  p\right)  =0$

Conversely :

$\sum_{i}\chi_{i}\left(  p\right)  =1\Rightarrow J=\left\{  i\in I:\chi
_{i}\left(  p\right)  \neq0\right\}  \neq\varnothing$

let be : $L=\left\{  i\in I:p\in O_{i}\right\}  \neq\varnothing$ \ so $\forall
i\notin L:\chi_{i}\left(  p\right)  =0$

Thus $J\cap L\neq\varnothing$ and $f\left(  p\right)  =\sum_{i\in J\cap L}%
\chi_{i}\left(  p\right)  \varphi_{i}^{1}\left(  p\right)  $

let $p\in N:f\left(  p\right)  <0$ : there is at least one $j\in J\cap L$ such
that $\varphi_{i}^{1}\left(  p\right)  <0\Rightarrow p\in\overset{\circ}{M}$

let $p\in N:f\left(  p\right)  =0$ : $\sum_{i\in J\cap L}\chi_{i}\left(
p\right)  \varphi_{i}^{1}\left(  p\right)  =0,\varphi_{i}^{1}\left(  p\right)
\leq0\Rightarrow\varphi_{i}^{1}\left(  p\right)  =0$

2) Take a path on the boundary : $c:\left[  a,b\right]  \rightarrow\partial M
$

$c\left(  t\right)  \in\partial M\Rightarrow\varphi_{i}^{1}(c(t))=0\Rightarrow
\left(  \varphi_{i}^{1}\right)  ^{\prime}\left(  c(t)\right)  c^{\prime}(t)=0$

$\Rightarrow\forall p\in\partial M,\forall u\in T_{p}\partial M:\left(
\varphi_{i}^{1}\left(  p\right)  \right)  ^{\prime}u=0$

$f^{\prime}(p)u=\sum_{i}\left(  \chi_{i}^{\prime}(p)\varphi_{i}^{1}%
(p)u+\chi_{i}(p)\left(  \varphi_{i}^{1}\right)  ^{\prime}\left(  p\right)
u_{x}\right)  $

$p\in\partial M\Rightarrow\varphi_{i}^{1}(p)=0\Rightarrow f^{\prime}%
(p)u=\sum_{i}\chi_{i}(p)\left(  \varphi_{i}^{1}\right)  ^{\prime}\left(
p\right)  u=0$

3) Let $p\in\partial M$ and $v_{1}$ transversal vector. We can take a basis of
$T_{p}N$ comprised of $v_{1}$ and n-1 vectors $\left(  v_{\alpha}\right)
_{\alpha=2}^{n}$ of $T_{p}\partial M$

$\forall u\in T_{p}N:u=\sum_{\alpha=1}^{n}u_{\alpha}v_{\alpha}$

$f^{\prime}(p)u=\sum_{\alpha=1}^{n}u_{\alpha}f^{\prime}(p)v_{\alpha}%
=u_{1}f^{\prime}(p)v_{1}$

As $f^{\prime}(p)\neq0$\ thus for any transversal vector we have $f^{\prime
}(p)u\neq0$

4) Take a vector field V such that its flow is defined from $p\in
\overset{\circ}{M}$ to $q\in M^{c}$ and $V\left(  p\right)  =v_{1}$

$v_{1}$ is outward oriented if $\exists t>0:q=\Phi_{V}\left(  t,p\right)  .$
Then :

$t\leq0\Rightarrow\Phi_{V}\left(  t,p\right)  \in\overset{\circ}{M}\Rightarrow
f\left(  \Phi_{V}\left(  t,p\right)  \right)  \leq0$

$t=0\Rightarrow f\left(  \Phi_{V}\left(  t,p\right)  \right)  =0$

$\frac{d}{dt}\Phi_{V}\left(  t,p\right)  |_{t=0}=V(p)=v_{1}$

$\frac{d}{dt}f\left(  \Phi_{V}\left(  t,p\right)  \right)  |_{t=0}=f^{\prime
}(p)v_{1}=\lim_{t\rightarrow0^{-}}\frac{1}{t}f\left(  \Phi_{V}\left(
t,p\right)  \right)  \geq0$

5) Let g be a riemannian form on N. So we can associate to the 1-form df a
vector field : $V^{\alpha}=g^{\alpha\beta}\partial_{\beta}f$ and $f^{\prime
}(p)V=g^{\alpha\beta}\partial_{\beta}f\partial_{\alpha}f\geq0$ is zero only if
f'(p)=0.\ So we can define a vector field outward oriented.
\end{proof}

Proof of iii)

\begin{proof}
V is normal (for the metric g) to the boundary :

$u\in\partial M:g_{\alpha\beta}u^{\alpha}V^{\beta}=g_{\alpha\beta}u^{\alpha
}g^{\beta\gamma}\partial_{\gamma}f=u^{\gamma}\partial_{\gamma}f=f^{\prime
}(p)u=0$
\end{proof}

\bigskip

\begin{theorem}
Let M be a m dimensional smooth Hausdorff real manifold, $f\in C_{1}\left(  M;%
\mathbb{R}
\right)  $ such that f'(p)$\neq0$ on M

i) Then $M_{t}=\left\{  p\in M:f(p)\leq t\right\}  $ , for any $t\in f(M)$\ is
a family of manifolds with boundary $\partial M_{t}=\left\{  p\in
M:f(p)=t\right\}  $

ii) f defines a folliation of M with leaves $\partial M_{t}$

iii) if M is connected compact then f(M)=[a,b] and there is a transversal
vector field \ V whose flow is a diffeomorphism for the boundaries

$\partial M_{t}=\Phi_{V}\left(  \partial M_{a},t\right)  $
\end{theorem}

\begin{proof}
M is smooth finite dimensional Hausdorff, thus paracompact and admits a
Riemanian metric

i) f'(p)$\neq0.$ Thus f'(p) has constant rank m-1.

The theorem of constant rank tells us that for any t in f(M)$\subset%
\mathbb{R}
$ the set $f^{-1}(t)$ is a closed m-1 submanifold of M and $\forall p\in
f^{-1}(t):T_{p}f^{-1}(t)=\ker f^{\prime}(p)$

We have a family of manifolds with boundary :

$M_{t}=\left\{  p\in N:f(p)\leq t\right\}  $ for $t\in f(M)$

ii) Frobenius theorem tells us that f defines a foliation of $M,$ with leaves
the boundary $\partial M_{t}=\left\{  p\in N:f(p)=t\right\}  $

And we have $\forall p\in M_{t},\ker f^{\prime}(p)=T_{p}\partial M_{t}$

$\Rightarrow\forall u\in T_{p}\partial M_{t}:f^{\prime}(p)u=0,\forall u\in
T_{p}M,u\notin T_{p}\partial M_{t}:f^{\prime}(p)u\neq0$

iii) If M is connected then f(M)=$\left\vert a,b\right\vert $ an interval in $%
\mathbb{R}
.$ If M is compact then f has a maximum and a minimum :

$a\leq f\left(  p\right)  \leq b$

There is a Riemannian structure on M, let g be the bilinear form and define
the vector field :

$V=\frac{gradf}{\left\Vert gradf\right\Vert ^{2}}::\forall p\in M:V\left(
p\right)  =\frac{1}{\lambda}\left(  g\left(  p\right)  ^{\alpha\beta}%
\partial_{\beta}f\left(  p\right)  \right)  \partial_{\alpha}$

with $\lambda=\sum_{\alpha\beta}g^{\alpha\beta}\left(  \partial_{\alpha
}f\right)  \left(  \partial_{\beta}f\right)  >0$

$f^{\prime}(p)V=\frac{1}{\lambda}g^{\alpha\beta}\partial_{\beta}%
f\partial_{\alpha}f=1$ So V is a vector field everywhere transversal and
outward oriented. Take $p_{a}\in\partial M_{a}$

The flow $\Phi_{V}\left(  p_{a},s\right)  $ of V is such that : $\forall
s\geq0:\exists\theta\in\left[  a,b\right]  :\Phi_{V}\left(  p_{a},s\right)
\in\partial M_{\theta}$ whenever defined.

Define : $h:%
\mathbb{R}
\rightarrow\left[  a,b\right]  :h(s)=f\left(  \Phi_{V}\left(  p_{a},s\right)
\right)  $

$\frac{\partial}{\partial s}\Phi_{V}\left(  p,s\right)  |_{s=\theta}%
=V(\Phi_{V}\left(  p,\theta\right)  )$

$\frac{d}{ds}h(s)|_{s=\theta}=f^{\prime}(\Phi_{V}\left(  p,\theta\right)
)V(\Phi_{V}\left(  p,\theta\right)  )=1\Rightarrow h\left(  s\right)  =s$

and we have : $\Phi_{V}\left(  p_{a},s\right)  \in\partial M_{s}$
\end{proof}

\bigskip

An application of these theorems is the propagation of waves. Let us take $M=%
\mathbb{R}
^{4}$ endowed with the Lorentz metric, that is the space of special
relativity. Take a constant vector field V of components $\left(  v_{1}%
,v_{2},v_{3},c\right)  $ with $\sum_{\alpha=1}^{3}\left(  v_{\alpha}\right)
^{2}=c^{2}.$ This is a field of rays of ligth. Take $f\left(  p\right)
=\left\langle p,V\right\rangle =p_{1}v_{1}+p_{2}v_{2}+p_{3}v_{3}-cp_{4}$

The folliation is the family of hyperplanes orthogonal to V. A wave is
represented by a map : $F:M\rightarrow E$ with E some vector space, such that
: $F\left(  p\right)  =\chi\circ f\left(  p\right)  $ where $\chi:%
\mathbb{R}
\rightarrow E$ . So the wave has the same value on any point on the front
wave, meaning the hyperplanes f(p)=s. f(p) is the phase of the wave.

For any component $F_{i}\left(  p\right)  $ we have the following derivatives :

$\alpha=1,2,3:\frac{\partial}{\partial p_{\alpha}}F_{i}=F_{i}^{\prime}\left(
-v_{\alpha}\right)  \rightarrow\frac{\partial^{2}}{\partial p_{\alpha}^{2}%
}F_{i}=F_{i}^{"}\left(  v_{\alpha}^{2}\right)  $

$\frac{\partial}{\partial p_{4}}F_{i}=F_{i}^{\prime}\left(  -c\right)
\rightarrow\frac{\partial^{2}}{\partial p_{\alpha}^{2}}F_{i}=F_{i}^{"}\left(
c^{2}\right)  $

so : $\frac{\partial^{2}}{\partial p_{1}^{2}}F_{i}+\frac{\partial^{2}%
}{\partial p_{2}^{2}}F_{i}+\frac{\partial^{2}}{\partial p_{3}^{2}}F_{i}%
-\frac{\partial^{2}}{\partial p_{4}^{2}}F_{i}=0=\square F_{i}$

F follows the wave equation. We have plane waves with wave vector V.

We would have spherical waves with $f(p)=\left\langle p,p\right\rangle $

Another example is the surfaces of constant energy in symplectic manifolds.

\subsubsection{Homology on manifolds}

This is the generalization of the concepts in the Affine Spaces, exposed in
the Algebra part. On this subject see Nakahara p.230.

A r-simplex $S^{r}$ on $%
\mathbb{R}
^{n}$\ is the convex hull of the r dimensional subspaces defined by r+1
\ independants points $\left(  A_{i}\right)  _{i1}^{k}$ :

$S^{r}=\left\langle A_{0},...A_{r}\right\rangle =\{P\in%
\mathbb{R}
^{n}:P=\sum_{i=0}^{r}t_{i}A_{i};0\leq t_{i}\leq1,\sum_{i=0}^{r}t_{i}=1\}$

A simplex is not a differentiable manifold, but is a topological (class 0)
manifold with boundary. It can be oriented.

\begin{definition}
A \textbf{r-simplex on a manifold} M modeled on $%
\mathbb{R}
^{n}$ is the image of a r-simplex $S^{r}$ on $%
\mathbb{R}
^{n}$\ by a smooth map : $f:$ $%
\mathbb{R}
^{n}\rightarrow M$
\end{definition}

It is denoted : $M^{r}=\left\langle p_{0},p_{1},...p_{r}\right\rangle
=\left\langle f\left(  A_{0}\right)  ,...f\left(  A_{r}\right)  \right\rangle
$

\begin{definition}
A \textbf{r-chain on a manifold} M is the formal sum : $\sum_{i}k_{i}M_{i}%
^{r}$ where $M_{i}^{r}$ is any r-simplex on M counted positively with its
orientation, and $k_{i}\in%
\mathbb{R}
$
\end{definition}

Notice two differences with the affine case :

i) here the coefficients $k_{i}\in%
\mathbb{R}
$ (in the linear case the coefficients are in $%
\mathbb{Z}
).$

ii) we do not precise a simplicial complex C : any r simplex on M is suitable

The set of r-chains on M is denoted $G_{r}\left(  M\right)  .$ It is a group
with formal addition.

\begin{definition}
The \textbf{border of the simplex} $\left\langle p_{0},p_{1},...p_{r}%
\right\rangle $ on the manifold M is the r-1-chain :

$\partial\left\langle p_{0},p_{1},...p_{r}\right\rangle =\sum_{k=0}^{r}\left(
-1\right)  ^{k}\left\langle p_{0},p_{1},.,\widehat{p}_{k},...p_{r}%
\right\rangle $

where the point $p_{k}$ has been removed. Conventionnaly : $\partial
\left\langle p_{0}\right\rangle =0$
\end{definition}

$M^{r}=f\left(  S^{r}\right)  \Rightarrow\partial M^{r}=f\left(  \partial
S^{r}\right)  $

$\partial^{2}=0$

A r-chain such that $\partial M^{r}=0$ is a \textbf{r-cycle}. The set
$Z_{r}\left(  M\right)  =\ker\left(  \partial\right)  $ is the r-cycle
subgroup of $G_{r}\left(  M\right)  $ and $Z_{0}(M)=G_{0}(M)$

Conversely if there is $M^{r+1}\in G_{r+1}\left(  M\right)  $ such that
$N=\partial M\in G_{r}\left(  C\right)  $ then N is called a \textbf{r-border}%
.\ The set of r-borders is a subgroup $B_{r}\left(  M\right)  $ of
$G_{r}\left(  M\right)  $ and $B_{n}(M)=0$. One has : $B_{r}(M)\subset
Z_{r}(M)\subset G_{r}(M)$

The \textbf{r-homology group} of M is the quotient set : $H_{r}\left(
M\right)  =Z_{r}(M)/B_{r}\left(  M\right)  $

The rth Betti number of M is $b_{r}\left(  M\right)  =\dim H_{r}\left(
M\right)  $

\newpage

\section{TENSORIAL\ BUNDLE}

\subsection{Tensor fields}

\label{Tensor fields}

\subsubsection{Tensors in the tangent space}

1. The tensorial product of copies of the vectorial space tangent and its
topological dual at every point of \ a manifold is well defined as for any
other vector space (see Algebra).\ So contravariant and covariant tensors, and
mixed tensors of any type (r,s) are defined in the usual way at every point of
a manifold.

2. All operations valid on tensors apply fully on the tangent space at one
point p of a manifold M : $\otimes_{s}^{r}T_{p}M$ is a vector space over the
field K (the same as M), product or contraction of tensors are legitimate
operations. The space $\otimes T_{p}M$\ of tensors of all types is an algebra
over K.

3. With an atlas $\left(  E,\left(  O_{i},\varphi_{i}\right)  _{i\in
I}\right)  $\ of the manifold M, at any point p the maps : $\varphi
_{i}^{\prime}\left(  p\right)  :T_{p}M\rightarrow E$ are vector space
isomorphisms, so there is a unique extension to an isomorphism of algebras in
$L(\otimes T_{p}M;\otimes E)$ which preserves the type of tensors and commutes
with contraction (see Tensors). So any chart $\left(  O_{i},\varphi
_{i}\right)  $ can be uniquely extended to a chart $\left(  O_{i}%
,\varphi_{i,r,s}\right)  :$

$\varphi_{i,r,s}\left(  p\right)  :\otimes_{s}^{r}T_{p}M\rightarrow\otimes
_{s}^{r}E$

$\forall S_{p},T_{p}\in\otimes_{s}^{r}T_{p}M,k,k^{\prime}\in K:$

$\varphi_{i,r,s}\left(  p\right)  \left(  kS_{p}+k^{\prime}T_{p}\right)
=k\varphi_{i,r,s}\left(  p\right)  S_{p}+k^{\prime}\varphi_{i,r,s}\left(
p\right)  \left(  T_{p}\right)  $

$\varphi_{i,r,s}\left(  p\right)  \left(  S_{p}\otimes T_{p}\right)
=\varphi_{i,r,s}\left(  p\right)  \left(  S_{p}\right)  \otimes\varphi
_{i,r,s}\left(  p\right)  \left(  T_{p}\right)  $

$\varphi_{i,r,s}\left(  p\right)  \left(  Trace\left(  S_{p}\right)  \right)
=Trace\left(  \varphi_{i,r,s}\left(  p\right)  \left(  \left(  S_{p}\right)
\right)  \right)  $

with the property :

$\left(  \varphi_{i}^{\prime}\left(  p\right)  \otimes\varphi_{i}^{\prime
}\left(  p\right)  \right)  \left(  u_{p}\otimes v_{p}\right)  =\varphi
_{i}^{\prime}\left(  p\right)  \left(  u_{p}\right)  \otimes\varphi
_{i}^{\prime}\left(  p\right)  \left(  v_{p}\right)  $

$\varphi_{i}^{\prime}\left(  p\right)  \otimes\left(  \varphi_{i}^{\prime
}\left(  p\right)  ^{t}\right)  ^{-1}\left(  u_{p}\otimes\mu_{p}\right)
=\varphi_{i}^{\prime}\left(  p\right)  \left(  u_{p}\right)  \otimes\left(
\varphi_{i}^{\prime}\left(  p\right)  ^{t}\right)  ^{-1}\left(  \mu
_{p}\right)  ,...$

4. Tensors on $T_{p}M$ can be expressed locally in any basis of $T_{p}M.$ The
natural bases are the bases induced by a chart, with vectors $\left(  \partial
x_{\alpha}\right)  _{\alpha\in A}$ and covectors $\left(  dx^{\alpha}\right)
_{\alpha\in A}$ with : $\partial x_{\alpha}=\varphi_{i}^{\prime}\left(
p\right)  ^{-1}e_{\alpha},dx^{\alpha}=\varphi_{i}^{\prime}\left(  p\right)
^{t}e^{\alpha}$ where $\left(  e_{\alpha}\right)  _{\alpha\in A}$ is a basis
of E and $\left(  e^{\alpha}\right)  _{\alpha\in A}$ a basis of E'.

The components of a tensor $S_{p}$ in $\otimes_{s}^{r}T_{p}M$
\textit{expressed in a holonomic basis} are :%

\begin{equation}
S_{p}=\sum_{\alpha_{1}...\alpha_{r}}\sum_{\beta_{1}....\beta_{s}}S_{\beta
_{1}...\beta_{q}}^{\alpha_{1}...\alpha_{r}}\partial x_{\alpha_{1}}%
\otimes..\otimes\partial x_{\alpha_{r}}\otimes dx^{\beta_{1}}\otimes...\otimes
dx^{\beta_{s}}%
\end{equation}

$\varphi_{i,r,s}\left(  p\right)  \left(  \partial x_{\alpha_{1}}%
\otimes..\otimes\partial x_{\alpha_{r}}\otimes dx^{\beta_{1}}\otimes...\otimes
dx^{\beta_{s}}\right)  =e_{\alpha_{1}}\otimes..\otimes e_{\alpha_{r}}\otimes
e^{\beta_{1}}\otimes...\otimes e^{\beta_{s}}$

The image of $S_{p}$ by the previous map $\varphi_{i,r,s}\left(  p\right)  $
is a tensor S in $\otimes_{s}^{r}E$ which has the same components in the basis
of $\otimes_{s}^{r}E:$%

\begin{equation}
\varphi_{i,r,s}\left(  p\right)  S_{p}=\sum_{\alpha_{1}...\alpha_{r}}%
\sum_{\beta_{1}....\beta_{s}}S_{\beta_{1}...\beta_{q}}^{\alpha_{1}%
...\alpha_{r}}e_{\alpha_{1}}\otimes..\otimes e_{\alpha_{r}}\otimes
e^{\beta_{1}}\otimes...\otimes e^{\beta_{s}}%
\end{equation}

\subsubsection{Change of charts}

1. In a change of basis in the tangent space the usual rules apply (see
Algebra).\ When the change of bases is induced by a change of chart the matrix
giving the new basis with respect to the old one is given by the jacobian.

2. If the old chart is $\left(  O_{i},\varphi_{i}\right)  $ and the new chart
: $\left(  O_{i},\psi_{i}\right)  $ (we can assume that the domains are the
same, this issue does not matter here).

Coordinates in the old chart : $x=\varphi_{i}\left(  p\right)  $

Coordinates in the new chart : $y=\psi_{i}\left(  p\right)  $

Old holonomic basis :

$\partial x_{\alpha}=\varphi_{i}^{\prime}\left(  p\right)  ^{-1}e_{\alpha}, $

$dx^{\alpha}=\varphi_{i}^{\prime}\left(  x\right)  ^{t}e^{\alpha}$ with
$dx^{\alpha}\left(  \partial x_{\beta}\right)  =\delta_{\beta}^{\alpha}$

New holonomic basis :

$\partial y_{\alpha}=\psi_{i}^{\prime}\left(  p\right)  ^{-1}e_{\alpha},$

$dy^{\alpha}=\psi_{i}^{\prime}\left(  y\right)  ^{\ast}e^{\alpha}$ with
$dy^{\alpha}\left(  \partial y_{\beta}\right)  =\delta_{\beta}^{\alpha}$

In a n-dimensional manifold the new coordinates $\left(  y^{i}\right)
_{i=1}^{n}$ are expressed with respect to the old coordinates by :

$\alpha=1..n:y^{\alpha}=F^{\alpha}\left(  x^{1},...x^{n}\right)
\Leftrightarrow\psi_{i}\left(  p\right)  =F\circ\varphi_{i}\left(  p\right)
\Leftrightarrow F\left(  x\right)  =\psi_{i}\circ\varphi_{i}^{-1}\left(
x\right)  $

\bigskip

The jacobian is : $J=\left[  F^{\prime}(x)\right]  =\left[  J_{\beta}^{\alpha
}\right]  =\left[  \dfrac{\partial F^{\alpha}}{\partial x^{\beta}}\right]
_{n\times n}\simeq\left[  \frac{\partial y^{\alpha}}{\partial x^{\beta}%
}\right]  $

$\partial y_{\alpha}=\sum_{\beta}\left[  J^{-1}\right]  _{\alpha}^{\beta
}\partial x_{\beta}\simeq\frac{\partial}{\partial y^{\alpha}}=\sum_{\beta
}\frac{\partial x^{\beta}}{\partial y^{\alpha}}\frac{\partial}{\partial
x^{\beta}}\Leftrightarrow\partial y_{\alpha}=\psi_{i}^{\prime}{}^{-1}%
\circ\varphi_{i}^{\prime}\left(  p\right)  \partial x_{\alpha}$

$dy^{\alpha}=\sum_{\beta}\left[  J\right]  _{\alpha}^{\beta}dx_{\beta}\simeq
dy^{\alpha}=\sum_{\beta}\frac{\partial y^{\alpha}}{\partial x^{\beta}%
}dx^{\beta}\Leftrightarrow dy^{\alpha}=\psi_{i}^{\prime}{}^{t}\circ\left(
\varphi_{i}^{\prime}\left(  x\right)  ^{t}\right)  ^{-1}dx_{\alpha}$

The components of vectors :

$u_{p}=\sum_{\alpha}u_{p}^{\alpha}\partial x_{\alpha}=\sum_{\alpha}\widehat
{u}_{p}^{\alpha}\partial y_{\alpha}$ \ with $\widehat{u}_{p}^{\alpha}%
=\sum_{\beta}J_{\beta}^{\alpha}u_{p}^{\beta}\simeq\sum_{\beta}\frac{\partial
y^{\alpha}}{\partial x^{\beta}}u_{p}^{\beta}$

The components of covectors :

$\mu_{p}=\sum_{\alpha}\mu_{p\alpha}dx^{\alpha}=\sum_{\alpha}\widehat{\mu
}_{p\alpha}dy^{\alpha}$ \ with $\widehat{\mu}_{p\alpha}=\sum_{\beta}\left[
J^{-1}\right]  _{\alpha}^{\beta}\mu_{p\beta}\simeq\sum_{\beta}\frac{\partial
x^{\beta}}{\partial y^{\alpha}}\mu_{p\beta}$

For a type (r,s) tensor :

$T=\sum_{\alpha_{1}...\alpha_{r}}\sum_{\beta_{1}....\beta_{s}}t_{\beta
_{1}...\beta_{q}}^{\alpha_{1}...\alpha_{r}}\partial x_{\alpha_{1}}%
\otimes..\otimes\partial x_{\alpha_{r}}\otimes dx^{\beta_{1}}\otimes...\otimes
dx^{\beta_{s}}$

$T=\sum_{\alpha_{1}...\alpha_{r}}\sum_{\beta_{1}....\beta_{s}}\widehat
{t}_{\beta_{1}...\beta_{q}}^{\alpha_{1}...\alpha_{r}}\partial y_{\alpha_{1}%
}\otimes..\otimes\partial y_{\alpha_{r}}\otimes dy^{\beta_{1}}\otimes
...\otimes dy^{\beta_{s}}$

with :

$\widehat{t}_{\beta_{1}...\beta_{q}}^{\alpha_{1}...\alpha_{r}}=\sum
_{\lambda_{1}...\lambda_{r}}\sum_{\mu_{1}....\mu_{s}}t_{\mu_{1}...\mu_{s}%
}^{\lambda_{1}...\lambda_{r}}\left[  J\right]  _{\lambda_{1}}^{\alpha_{1}%
}..\left[  J\right]  _{\lambda_{r}}^{\alpha_{r}}\left[  J^{-1}\right]
_{\beta_{1}}^{\mu_{1}}..\left[  J^{-1}\right]  _{\beta_{s}}^{\mu_{s}}$%

\begin{equation}
\widehat{t}_{\beta_{1}...\beta_{q}}^{\alpha_{1}...\alpha_{r}}\left(  q\right)
=\sum_{\lambda_{1}...\lambda_{r}}\sum_{\mu_{1}....\mu_{s}}t_{\mu_{1}...\mu
_{s}}^{\lambda_{1}...\lambda_{r}}\frac{\partial y^{\alpha_{1}}}{\partial
x^{\lambda_{1}}}...\frac{\partial y^{\alpha_{r}}}{\partial x^{\lambda_{r}}%
}\frac{\partial x^{\mu_{1}}}{\partial y^{\beta_{1}}}...\frac{\partial
x^{\mu_{s}}}{\partial y^{\beta_{s}}}%
\end{equation}

For a r-form :

$\varpi=\sum_{\left(  \alpha_{1}...\alpha_{r}\right)  }\varpi_{\alpha
_{1}...\alpha_{r}}dx^{\alpha_{1}}\otimes dx^{\alpha_{2}}\otimes...\otimes
dx^{\alpha_{r}}$

$=\sum_{\left\{  \alpha_{1}...\alpha_{r}\right\}  }\varpi_{\alpha_{1}%
...\alpha_{r}}dx^{\alpha_{1}}\wedge dx^{\alpha_{2}}\wedge...\wedge
dx^{\alpha_{r}}$

$\varpi=\sum_{\left(  \alpha_{1}...\alpha_{r}\right)  }\widehat{\varpi
}_{\alpha_{1}...\alpha_{r}}dy^{\alpha_{1}}\otimes dy^{\alpha_{2}}%
\otimes...\otimes dy^{\alpha_{r}}$

$=\sum_{\left\{  \alpha_{1}...\alpha_{r}\right\}  }\widehat{\varpi}%
_{\alpha_{1}...\alpha_{r}}dy^{\alpha_{1}}\wedge dy^{\alpha_{2}}\wedge...\wedge
dy^{\alpha_{r}}$

with%

\begin{equation}
\widehat{\varpi}_{\alpha_{1}...\alpha_{r}}=\sum_{\left\{  \beta_{1}%
....\beta_{r}\right\}  }\varpi_{\beta_{1}...\beta_{r}}\det\left[
J^{-1}\right]  _{\alpha_{1}...\alpha_{r}}^{\beta_{1}...\beta_{r}}%
\end{equation}

where $\det\left[  J^{-1}\right]  _{\alpha_{1}...\alpha_{r}}^{\beta
_{1}...\beta_{r}}$ is the determinant of the matrix $\left[  J^{-1}\right]  $
with elements row $\beta_{k}$\ column $\alpha_{l}$

\subsubsection{Tensor bundle}

The tensor bundle is defined in a similar way as the vector bundle.

\begin{definition}
The (r,s) tensor bundle is the set $\otimes_{s}^{r}TM=\cup_{p\in M}\otimes
_{s}^{r}T_{p}M$
\end{definition}

\begin{theorem}
$\otimes_{s}^{r}TM$ has the structure of a class r-1 manifold, with dimension
(rs+1)$\times$dimM
\end{theorem}

The open cover of $\otimes_{s}^{r}TM$ is defined by : $O_{i}^{\prime}%
=\cup_{p\in O_{I}}\left\{  \otimes_{s}^{r}T_{p}M\right\}  $

The maps : $O_{i}^{\prime}\rightarrow U_{i}\times\otimes_{s}^{r}E::\left(
\varphi_{i}\left(  p\right)  ,\varphi_{i,r,s}\left(  p\right)  T_{p}\right)  $
define an atlas of $\otimes_{s}^{r}TM$

The dimension of $\otimes_{s}^{r}TM$ is (rs+1)$\times$dimM. Indeed we need m
coordinates for p and m$\times$r$\times$s components for $T_{p}$.

\begin{theorem}
$\otimes_{s}^{r}TM$ has the structure of vector bundle over M, modeled on
$\otimes_{s}^{r}E$
\end{theorem}

$\otimes_{s}^{r}TM$ is a manifold

Define the projection : $\pi_{r,s}:\otimes_{s}^{r}TM\rightarrow M::\pi
_{r,s}\left(  T_{p}\right)  =p$. This is a smooth surjective map and
$\pi_{r,s}^{-1}\left(  p\right)  =\otimes_{s}^{r}T_{p}M$

Define the trivialization :

$\Phi_{i,r,s}:O_{i}\times\otimes_{s}^{r}E\rightarrow\otimes_{s}^{r}%
TM::\Phi_{i,r,s}\left(  p,t\right)  =\varphi_{i,r,s}^{-1}\left(  \varphi
_{i}\left(  p\right)  \right)  t\in\otimes_{s}^{r}T_{p}M.$

This is a class c-1 map if the manifold is of class c.

If $p\in O_{i}\cap O_{j}$ then $\varphi_{i,r,s}^{-1}\circ\varphi
_{j,r,s}\left(  p\right)  t$ and $\varphi_{j,r,s}^{-1}\circ\varphi
_{i,r,s}\left(  p\right)  t$ define the same tensor of $\otimes_{s}^{r}T_{p}M$

\begin{theorem}
$\otimes_{s}^{r}TM$ has a structure of a vector space with pointwise operations.
\end{theorem}

\subsubsection{Tensor fields}

\paragraph{Definition\newline}

\begin{definition}
A \textbf{tensor field} of type (r,s) is a map : $T:M\rightarrow$ $\otimes
_{s}^{r}TM$ which associates at each point p of M a tensor T(p)
\end{definition}

A tensor field of type (r,s) over the open $U_{i}\subset E$ is a map :
$t_{i}:U_{i}\rightarrow\otimes_{s}^{r}E$

A tensor field is a collection of maps :

$T_{i}:O_{i}\times\otimes_{s}^{r}E\rightarrow\otimes_{s}^{r}TM::T(p)=\Phi
_{i,r,s}\left(  p,t_{i}\left(  \varphi_{i}(p)\right)  \right)  $ with $t_{i}$
a tensor field on E.

This reads :

$T\left(  p\right)  =\sum_{\alpha_{1}...\alpha_{r}}\sum_{\beta_{1}%
....\beta_{s}}t_{\beta_{1}...\beta_{q}}^{\alpha_{1}...\alpha_{r}}%
\partial_{\alpha_{1}}\otimes..\otimes\partial_{\alpha_{r}}\otimes
dx^{\beta_{1}}\otimes...\otimes dx^{\beta_{s}}$

$\varphi_{i,r,s}\left(  p\right)  \left(  T\left(  p\right)  \right)  $

$=\sum_{\alpha_{1}...\alpha_{r}}\sum_{\beta_{1}....\beta_{s}}t_{\beta
_{1}...\beta_{q}}^{\alpha_{1}...\alpha_{r}}\left(  \varphi_{i}\left(
p\right)  \right)  e_{\alpha_{1}}\otimes..\otimes e_{\alpha_{r}}\otimes
e^{\beta_{1}}\otimes...\otimes e^{\beta_{s}}$

The tensor field if of class c if all the functions $t_{i}:U_{i}%
\rightarrow\otimes_{s}^{r}E$ are of class c.

Warning! As with vector fields, the components of a given tensor fields vary
through the domains of an atlas.

\begin{notation}
$\mathfrak{X}_{c}\left(  \otimes_{s}^{r}TM\right)  $ is the set of fields of
class c type (r,s) tensors on the manifold M
\end{notation}

\begin{notation}
$\mathfrak{X}_{c}\left(  \Lambda_{s}TM\right)  $ is the set of fields of class
c antisymmetric type (0,s) tensors on the manifold M
\end{notation}

A vector field can be seen as a (1,0) type contravariant tensor field

$\mathfrak{X}\left(  \otimes_{0}^{1}TM\right)  \simeq\mathfrak{X}\left(
TM\right)  $

A vector field on the cotangent bundle is a (0,1) type covariant tensor field
$\mathfrak{X}\left(  \otimes_{1}^{0}TM\right)  \simeq\mathfrak{X}\left(
TM^{\ast}\right)  $

Scalars can be seen a (0,0) tensors.\ Similarly a map : $T:M\rightarrow K$ is
just a scalar \textit{function}. So the 0-covariant tensor fields are scalar maps:

$\mathfrak{X}\left(  \otimes_{0}^{0}TM\right)  =\mathfrak{X}\left(  \wedge
_{0}TM\right)  \simeq C\left(  M;K\right)  $

\paragraph{Operations on tensor fields\newline}

1. All usual operations with tensors are available with tensor fields when
they are implemented at the same point of M.

With the tensor product (pointwise) the set of tensor fields over a manifold
is an algebra denoted $\mathfrak{X}\left(  \otimes TM\right)  =\oplus
_{r,s}\mathfrak{X}\left(  \otimes_{s}^{r}TM\right)  $\ .

If the manifold is of class c, $\otimes_{s}^{r}TM$ is a class c-1 manifold,
the tensor field is of class c-1 if the map : $t:U_{i}\rightarrow\otimes
_{s}^{r}E$ is of class c-1. So the maps : $t_{\beta_{1}...\beta_{q}}%
^{\alpha_{1}...\alpha_{r}}:M\rightarrow%
\mathbb{R}
$ giving the components of the tensor field in a holonomic basis are class c-1
scalar functions. And this property does not depend of the choice of an atlas
of class c.

2. The trace operator (see the Algebra part) is the unique linear map :

$Tr:\mathfrak{X}\left(  \otimes_{1}^{1}TM\right)  \rightarrow C\left(
M;K\right)  $ such that $Tr\left(  \varpi\otimes V\right)  =\varpi\left(
V\right)  $

From the trace operator one can define the contraction on tensors as a linear
map : $\mathfrak{X}\left(  \otimes_{s}^{r}TM\right)  \rightarrow
\mathfrak{X}\left(  \otimes_{s-1}^{r-1}TM\right)  $ which depends on the
choice of the indices to be contracted.

3. It is common to meet complicated operators over vector fields, including
derivatives, and to wonder if they have some tensorial significance.\ A useful
criterium is the following (Kolar p.61):

If the multilinear (with scalars) map on vector fields

$F\in%
\mathcal{L}%
^{s}\left(  \mathfrak{X}\left(  TM\right)  ^{s};\mathfrak{X}\left(
\otimes^{r}TM\right)  \right)  $ is still linear for any function, meaning $:$

$\forall f_{k}\in C_{\infty}\left(  M;K\right)  ,\forall\left(  V_{k}\right)
_{k=1}^{s},F\left(  f_{1}V_{1},...f_{s}V_{s}\right)  =f_{1}f_{2}%
...f_{s}F\left(  V_{1},...V_{s}\right)  $

then $\exists T\in\mathfrak{X}\left(  \otimes_{s}^{r}TM\right)  ::\forall
\left(  V_{k}\right)  _{k=1}^{s},F\left(  V_{1},...V_{s}\right)  =T\left(
V_{1},...V_{s}\right)  $

\subsubsection{Pull back, push forward}

The push-forward and the pull back of a vector field by a map can be
generalized but work differently according to the type of tensors. For some
transformations we need only a differentiable map, for others we need a
diffeomorphism, and then the two operations - push forward and pull back - are
the opposite of the other.

\paragraph{Definitions\newline}

1. For any differentiable map f between the manifolds M,N (on the same field):

Push-forward for vector fields :

$f_{\ast}:\mathfrak{X}\left(  TM\right)  \rightarrow\mathfrak{X}\left(
TN\right)  ::f_{\ast}V=f^{\prime}V\Leftrightarrow f_{\ast}V\left(  f\left(
p\right)  \right)  =f^{\prime}(p)V\left(  p\right)  $

Pull-back for 0-forms (functions) :

$f^{\ast}:\mathfrak{X}\left(  \Lambda_{0}TN^{\ast}\right)  \rightarrow
\mathfrak{X}\left(  \Lambda_{0}TM^{\ast}\right)  ::f^{\ast}h=h\circ
f\Leftrightarrow f^{\ast}h\left(  p\right)  =h\left(  f\left(  p\right)
\right)  $

Pull-back for 1-forms :

$f^{\ast}:\mathfrak{X}\left(  \Lambda_{1}TN^{\ast}\right)  \rightarrow
\mathfrak{X}\left(  \Lambda_{1}TM^{\ast}\right)  ::f^{\ast}\mu=\mu\circ
f^{\prime}\Leftrightarrow f^{\ast}\mu\left(  p\right)  =\mu\left(  f\left(
p\right)  \right)  \circ f^{\prime}\left(  p\right)  $

Notice that the operations above do not need a diffeormorphism, so M,N do not
need to have the same dimension.

2. For any diffeomorphism f between the manifolds M,N (which implies that they
must have the same dimension) we have the inverse operations :

Pull-back for vector fields :

$f^{\ast}:\mathfrak{X}\left(  TN\right)  \rightarrow\mathfrak{X}\left(
TM\right)  ::f^{\ast}W=\left(  f^{-1}\right)  ^{\prime}V\Leftrightarrow
f^{\ast}W\left(  p\right)  =\left(  f^{-1}\right)  ^{\prime}(f(p))W\left(
f(p)\right)  $

Push-forward for 0-forms (functions) :

$f_{\ast}:\mathfrak{X}\left(  \Lambda_{0}TM^{\ast}\right)  \rightarrow
\mathfrak{X}\left(  \Lambda_{0}TN^{\ast}\right)  ::f_{\ast}g=g\circ
f^{-1}\Leftrightarrow f_{\ast}g\left(  q\right)  =g\left(  f^{-1}\left(
q\right)  \right)  $

Push-forward for 1-forms :

$f_{\ast}:\mathfrak{X}\left(  \Lambda_{1}TM^{\ast}\right)  \rightarrow
\mathfrak{X}\left(  \Lambda_{1}TN^{\ast}\right)  ::f_{\ast}\lambda=\varpi
\circ\left(  f^{-1}\right)  ^{\prime}$

$\Leftrightarrow f_{\ast}\lambda\left(  q\right)  =\lambda\left(
f^{-1}\left(  q\right)  \right)  \circ\left(  f^{-1}\right)  ^{\prime}\left(
q\right)  $

3. For any mix (r,s) type tensor, on finite dimensional manifolds M,N with the
same dimension, and any diffeomorphism $f:M\rightarrow N$

Push-forward of a tensor :%

\begin{equation}
f_{\ast}:\mathfrak{X}\left(  \otimes_{s}^{r}TM\right)  \rightarrow
\mathfrak{X}\left(  \otimes_{s}^{r}TN\right)  ::\left(  f_{\ast}S_{p}\right)
\left(  f\left(  p\right)  \right)  =f_{r,s}^{\prime}\left(  p\right)  S_{p}%
\end{equation}

Pull-back of a tensor :%

\begin{equation}
f^{\ast}:\mathfrak{X}\left(  \otimes_{s}^{r}TM\right)  \rightarrow
\mathfrak{X}\left(  \otimes_{s}^{r}TN\right)  ::\left(  f^{\ast}S_{q}\right)
\left(  f^{-1}\left(  q\right)  \right)  =\left(  f_{r,s}^{\prime}\right)
^{-1}\left(  q\right)  S_{q}%
\end{equation}

where $f_{r,s}^{\prime}\left(  p\right)  :\otimes_{s}^{r}T_{p}M\rightarrow
\otimes_{s}^{r}T_{f\left(  p\right)  }N$ is the extension to the algebras of
the isomorphism : $f^{\prime}(p):T_{p}M\rightarrow T_{f\left(  p\right)  }N$

\paragraph{Properties\newline}

\begin{theorem}
(Kolar p.62) Whenever they are defined, the push forward $f_{\ast}$ and pull
back $f^{\ast}$ of tensors are linear operators (with scalars) :

$f^{\ast}\in%
\mathcal{L}%
\left(  \mathfrak{X}\left(  \otimes_{s}^{r}TM\right)  ;\mathfrak{X}\left(
\otimes_{s}^{r}TN\right)  \right)  $

$f_{\ast}\in%
\mathcal{L}%
\left(  \mathfrak{X}\left(  \otimes_{s}^{r}TM\right)  ;\mathfrak{X}\left(
\otimes_{s}^{r}TN\right)  \right)  $

which are the inverse map of the other :

$f^{\ast}=\left(  f^{-1}\right)  _{\ast}$

$f_{\ast}=\left(  f^{-1}\right)  ^{\ast}$

They preserve the commutator of vector fields:

$\left[  f_{\ast}V_{1},f_{\ast}V_{2}\right]  =f_{\ast}\left[  V_{1}%
,V_{2}\right]  $

$\left[  f^{\ast}V_{1},f^{\ast}V_{2}\right]  =f^{\ast}\left[  V_{1}%
,V_{2}\right]  $

and the exterior product of r-forms :

$f^{\ast}\left(  \varpi\wedge\pi\right)  =f^{\ast}\varpi\wedge f^{\ast}\pi$

$f_{\ast}\left(  \varpi\wedge\pi\right)  =f_{\ast}\varpi\wedge f_{\ast}\pi$

They can be composed with the rules :

$\left(  f\circ g\right)  ^{\ast}=g^{\ast}\circ f^{\ast}$

$\left(  f\circ g\right)  _{\ast}=f_{\ast}\circ g_{\ast}$

They commute with the exterior differential (if f is of class 2) :

$d(f^{\ast}\varpi)=f^{\ast}(d\varpi)$

$d(f_{\ast}\varpi)=f_{\ast}(d\varpi)$
\end{theorem}

\paragraph{Components expressions\newline}

For a diffeomorphism f between the n dimensional manifolds M$\left(
K^{n},\left(  O_{i},\varphi_{i}\right)  _{i\in I}\right)  $ and the manifold
N$\left(  K^{n},\left(  Q_{j},\psi_{j}\right)  _{j\in J}\right)  $ the
formulas are

\bigskip

Push forward : $f_{\ast}:\mathfrak{X}\left(  \otimes_{s}^{r}TM\right)
\rightarrow\mathfrak{X}\left(  \otimes_{s}^{r}TN\right)  $

$S\left(  p\right)  =\sum_{\alpha_{1}...\alpha_{r}}\sum_{\beta_{1}%
....\beta_{s}}S_{\beta_{1}...\beta_{q}}^{\alpha_{1}...\alpha_{r}}\left(
p\right)  \partial x_{\alpha_{1}}\otimes..\otimes\partial x_{\alpha_{r}%
}\otimes dx^{\beta_{1}}\otimes...\otimes dx^{\beta_{s}}$

$\left(  f_{\ast}S\right)  \left(  q\right)  =\sum_{\alpha_{1}...\alpha_{r}%
}\sum_{\beta_{1}....\beta_{s}}\widehat{S}_{\beta_{1}...\beta_{q}}^{\alpha
_{1}...\alpha_{r}}\left(  q\right)  \partial y_{\alpha_{1}}\otimes
..\otimes\partial y_{\alpha_{r}}\otimes dy^{\beta_{1}}\otimes...\otimes
dy^{\beta_{s}}$

with :

$\widehat{S}_{\beta_{1}...\beta_{q}}^{\alpha_{1}...\alpha_{r}}\left(
q\right)  =\sum_{\lambda_{1}...\lambda_{r}}\sum_{\mu_{1}....\mu_{s}}S_{\mu
_{1}...\mu_{s}}^{\lambda_{1}...\lambda_{r}}\left(  f^{-1}\left(  q\right)
\right)  \left[  J\right]  _{\lambda_{1}}^{\alpha_{1}}..\left[  J\right]
_{\lambda_{r}}^{\alpha_{r}}\left[  J^{-1}\right]  _{\beta_{1}}^{\mu_{1}%
}..\left[  J^{-1}\right]  _{\beta_{s}}^{\mu_{s}}$

$\widehat{S}_{\beta_{1}...\beta_{q}}^{\alpha_{1}...\alpha_{r}}\left(
q\right)  =\sum_{\lambda_{1}...\lambda_{r}}\sum_{\mu_{1}....\mu_{s}}S_{\mu
_{1}...\mu_{s}}^{\lambda_{1}...\lambda_{r}}\left(  f^{-1}\left(  q\right)
\right)  \frac{\partial y^{\alpha_{1}}}{\partial x^{\lambda_{1}}}%
...\frac{\partial y^{\alpha_{r}}}{\partial x^{\lambda_{r}}}\frac{\partial
x^{\mu_{1}}}{\partial y^{\beta_{1}}}...\frac{\partial x^{\mu_{s}}}{\partial
y^{\beta_{s}}}$

\bigskip

Pull-back : $f^{\ast}:\mathfrak{X}\left(  \otimes_{s}^{r}TN\right)
\rightarrow\mathfrak{X}\left(  \otimes_{s}^{r}TM\right)  $

$S\left(  q\right)  =\sum_{\alpha_{1}...\alpha_{r}}\sum_{\beta_{1}%
....\beta_{s}}S_{\beta_{1}...\beta_{q}}^{\alpha_{1}...\alpha_{r}}\left(
q\right)  \partial y_{\alpha_{1}}\otimes..\otimes\partial y_{\alpha_{r}%
}\otimes dy^{\beta_{1}}\otimes...\otimes dy^{\beta_{s}}$

$f^{\ast}S\left(  p\right)  =\sum_{\alpha_{1}...\alpha_{r}}\sum_{\beta
_{1}....\beta_{s}}\widehat{S}_{\beta_{1}...\beta_{q}}^{\alpha_{1}...\alpha
_{r}}\left(  p\right)  \partial x_{\alpha_{1}}\otimes..\otimes\partial
x_{\alpha_{r}}\otimes dx^{\beta_{1}}\otimes...\otimes dx^{\beta_{s}}$

with :

$\widehat{S}_{\beta_{1}...\beta_{q}}^{\alpha_{1}...\alpha_{r}}\left(
p\right)  =\sum_{\lambda_{1}...\lambda_{r}}\sum_{\mu_{1}....\mu_{s}}S_{\mu
_{1}...\mu_{s}}^{\lambda_{1}...\lambda_{r}}\left(  f\left(  p\right)  \right)
\left[  J^{-1}\right]  _{\lambda_{1}}^{\alpha_{1}}..\left[  J^{-1}\right]
_{\lambda_{r}}^{\alpha_{r}}\left[  J\right]  _{\beta_{1}}^{\mu_{1}}..\left[
J\right]  _{\beta_{s}}^{\mu_{s}}$

$\widehat{S}_{\beta_{1}...\beta_{q}}^{\alpha_{1}...\alpha_{r}}\left(
q\right)  =\sum_{\lambda_{1}...\lambda_{r}}\sum_{\mu_{1}....\mu_{s}}S_{\mu
_{1}...\mu_{s}}^{\lambda_{1}...\lambda_{r}}\left(  f\left(  p\right)  \right)
\frac{\partial x^{\alpha_{1}}}{\partial y^{\lambda_{1}}}...\frac{\partial
x^{\alpha_{r}}}{\partial y^{\lambda_{r}}}\frac{\partial y^{\mu_{1}}}{\partial
x^{\beta_{1}}}...\frac{\partial y^{\mu_{s}}}{\partial x^{\beta_{s}}}$

where x are the coordinates on M, y the coordinates on N, and J is the
jacobian :

$\left[  J\right]  =\left[  F^{\prime}(x)\right]  =\left[  \frac{\partial
y^{\alpha}}{\partial x^{\beta}}\right]  ;\left[  F^{\prime}(x)\right]
^{-1}=\left[  J\right]  ^{-1}=\left[  \frac{\partial x^{\alpha}}{\partial
y^{\beta}}\right]  $

F is the transition map : $F:\varphi_{i}\left(  O_{i}\right)  \rightarrow
\psi_{j}\left(  Q_{j}\right)  ::$ $y=\psi_{j}\circ f\circ\varphi_{i}%
^{-1}\left(  x\right)  =F\left(  x\right)  $

\bigskip

For a r-form these formulas simplify :

Push forward :

$\varpi=\sum_{\left(  \alpha_{1}...\alpha_{r}\right)  }\varpi_{\alpha
_{1}...\alpha_{r}}dx^{\alpha_{1}}\otimes dx^{\alpha_{2}}\otimes...\otimes
dx^{\alpha_{r}}$

$=\sum_{\left\{  \alpha_{1}...\alpha_{r}\right\}  }\varpi_{\alpha_{1}%
...\alpha_{r}}dx^{\alpha_{1}}\wedge dx^{\alpha_{2}}\wedge...\wedge
dx^{\alpha_{r}}$

$\left(  f_{\ast}\varpi\right)  \left(  q\right)  =\sum_{\left(  \alpha
_{1}...\alpha_{r}\right)  }\widehat{\varpi}_{\alpha_{1}...\alpha_{r}}\left(
q\right)  dy^{\alpha_{1}}\otimes dy^{\alpha_{2}}\otimes...\otimes
dy^{\alpha_{r}}$

$=\sum_{\left\{  \alpha_{1}...\alpha_{r}\right\}  }\widehat{\varpi}%
_{\alpha_{1}...\alpha_{r}}\left(  q\right)  dy^{\alpha_{1}}\wedge
dy^{\alpha_{2}}\wedge...\wedge dy^{\alpha_{r}}$

with :

$\widehat{\varpi}_{\alpha_{1}...\alpha_{r}}\left(  q\right)  =\sum_{\left\{
\beta_{1}....\beta_{r}\right\}  }\varpi_{\beta_{1}...\beta_{r}}\left(
f^{-1}\left(  q\right)  \right)  \det\left[  J^{-1}\right]  _{\alpha
_{1}...\alpha_{r}}^{\beta_{1}...\beta_{r}}$

$=\sum_{\mu_{1}....\mu_{s}}\varpi_{\mu_{1}...\mu_{s}}\left(  f^{-1}\left(
q\right)  \right)  \left[  J^{-1}\right]  _{\alpha_{1}}^{\mu_{1}}..\left[
J^{-1}\right]  _{\alpha_{r}}^{\mu_{r}}$

Pull-back :

$\varpi\left(  q\right)  =\sum_{\left(  \alpha_{1}...\alpha_{r}\right)
}\varpi_{\alpha_{1}...\alpha_{r}}\left(  q\right)  dy^{\alpha_{1}}\otimes
dy^{\alpha_{2}}\otimes...\otimes dy^{\alpha_{r}}$

$=\sum_{\left\{  \alpha_{1}...\alpha_{r}\right\}  }\varpi_{\alpha_{1}%
...\alpha_{r}}\left(  q\right)  dy^{\alpha_{1}}\wedge dy^{\alpha_{2}}%
\wedge...\wedge dy^{\alpha_{r}}$

$f^{\ast}\varpi\left(  p\right)  =\sum_{\left(  \alpha_{1}...\alpha
_{r}\right)  }\widehat{\varpi}_{\alpha_{1}...\alpha_{r}}\left(  p\right)
dx^{\alpha_{1}}\otimes dx^{\alpha_{2}}\otimes...\otimes dx^{\alpha_{r}}$

$=\sum_{\left\{  \alpha_{1}...\alpha_{r}\right\}  }\widehat{\varpi}%
_{\alpha_{1}...\alpha_{r}}\left(  p\right)  dx^{\alpha_{1}}\wedge
dx^{\alpha_{2}}\wedge...\wedge dx^{\alpha_{r}}$

with :

$\widehat{\varpi}_{\alpha_{1}...\alpha_{r}}\left(  p\right)  =\sum_{\left\{
\beta_{1}....\beta_{r}\right\}  }\varpi_{\beta_{1}...\beta_{r}}\left(
f\left(  p\right)  \right)  \det\left[  J\right]  _{\alpha_{1}...\alpha_{r}%
}^{\beta_{1}...\beta_{r}}$

$=\sum_{\mu_{1}....\mu_{s}}\varpi_{\mu_{1}...\mu_{s}}\left(  f\left(
p\right)  \right)  \left[  J\right]  _{\alpha_{1}}^{\mu_{1}}..\left[
J\right]  _{\alpha_{r}}^{\mu_{r}}$

where $\det\left[  J^{-1}\right]  _{\alpha_{1}...\alpha_{r}}^{\beta
_{1}...\beta_{r}}$ is the determinant of the matrix $\left[  J^{-1}\right]  $
with r column $\left(  \alpha_{1},..\alpha_{r}\right)  $ comprised each of the
components $\left\{  \beta_{1}...\beta_{r}\right\}  $

\bigskip

Remark :

A change of chart can also be formalized as a push-forward :

$\varphi_{i}:O_{i}\rightarrow U_{i}::x=\varphi_{i}\left(  p\right)  $

$\psi_{i}:O_{i}\rightarrow V_{i}::y=\psi_{i}\left(  p\right)  $

$\psi_{i}\circ\varphi_{i}^{-1}:O_{i}\rightarrow O_{i}::y=\psi_{i}\circ
\varphi_{i}^{-1}\left(  x\right)  $

The change of coordinates of a tensor is the push forward : $\widehat{t}%
_{i}=\left(  \psi_{i}\circ\varphi_{i}^{-1}\right)  _{\ast}t_{i}.$ As the
components in the holonomic basis are the same as in E, we have the same
relations between S and $\widehat{S}$

\bigskip

\subsection{Lie derivative}

\label{Lie derivative}

\subsubsection{Invariance, transport and derivation}

\paragraph{Covariance\newline}

1. Let be two observers doing some experiments about the same
phenomenon.\ They use models which are described in the tensor bundle of the
same manifold M modelled on a Banach E, but using different charts.

Observer 1 : charts $\left(  O_{i},\varphi_{i}\right)  _{i\in I},\varphi
_{i}\left(  O_{i}\right)  =U_{i}\subset E$ with coordinates x

Observer 2 : charts $\left(  O_{i},\psi_{i}\right)  _{i\in I},\psi_{i}\left(
O_{i}\right)  =V_{i}\subset E$ with coordinates y

We assume that the cover $O_{i}$ is the same (it does not matter here).

The physical phenomenon is represented in the models by a tensor $T\in
\otimes_{s}^{r}TM.$ This is a geometrical quantity : it does not depend on the
charts used. The measures are done at the same point p.

Observer 1 mesures the components of $T:$

$T\left(  p\right)  =\sum_{\alpha_{1}...\alpha_{r}}\sum_{\beta_{1}%
....\beta_{s}}t_{\beta_{1}...\beta_{q}}^{\alpha_{1}...\alpha_{r}}\left(
p\right)  \partial x_{\alpha_{1}}\otimes..\otimes\partial x_{\alpha_{r}%
}\otimes dx^{\beta_{1}}\otimes...\otimes dx^{\beta_{s}}$

Observer 2 mesures the components of $T:$

$T\left(  p\right)  =\sum_{\alpha_{1}...\alpha_{r}}\sum_{\beta_{1}%
....\beta_{s}}s_{\beta_{1}...\beta_{q}}^{\alpha_{1}...\alpha_{r}}\left(
q\right)  \partial y_{\alpha_{1}}\otimes..\otimes\partial y_{\alpha_{r}%
}\otimes dy^{\beta_{1}}\otimes...\otimes dy^{\beta_{s}}$

So in their respective charts the measures are :

$t=\left(  \varphi_{i}\right)  _{\ast}T$

$s=\left(  \psi_{i}\right)  _{\ast}T$

Passing from one set of measures to the other is a change of charts :

$s=\left(  \psi_{i}\circ\varphi_{i}^{-1}\right)  _{\ast}t=\left(  \psi
_{i}\right)  _{\ast}\circ\left(  \varphi_{i}^{-1}\right)  _{\ast}t$

So the measures are related : they are \textbf{covariant}. They change
according to the rules of the charts.

2. This is just the same rule as in affine space : when we use different
frames, we need to adjust the mesures according to the proper rules in order
to be able to make any sensible comparison. The big difference here is that
these rules should apply for any point p, and any set of transition maps
$\psi_{i}\circ\varphi_{i}^{-1}.$ So we have stronger conditions for the
specification of the functions $t_{\beta_{1}...\beta_{q}}^{\alpha_{1}%
...\alpha_{r}}\left(  p\right)  .$

\paragraph{Invariance\newline}

1. If both observers find the \textit{same} numerical results the tensor is
indeed special : $t=\left(  \psi_{i}\right)  _{\ast}\circ\left(  \varphi
_{i}^{-1}\right)  _{\ast}t$ \ .\ It is \textbf{invariant} by some specific
diffeomorphism $\left(  \psi_{i}\circ\varphi_{i}^{-1}\right)  $ and the
physical phenomenon has a \textbf{symmetry} which is usually described by the
action of a group. Among these groups the one parameter groups of
diffeomorphisms have a special interest because they are easily related to
physical systems and can be characterized by an infinitesimal generator which
is a vector field (they are the axes of the symmetry).

2. Invariance can also occur when one single operator does measurements of the
same phenomenon at two different points. If he uses the same chart $\left(
O_{i},\varphi_{i}\right)  _{i\in I}$ with coordinates x as above :

Observation 1 at point p :

$T\left(  p\right)  =\sum_{\alpha_{1}...\alpha_{r}}\sum_{\beta_{1}%
....\beta_{s}}t_{\beta_{1}...\beta_{q}}^{\alpha_{1}...\alpha_{r}}\left(
p\right)  \partial x_{\alpha_{1}}\otimes..\otimes\partial x_{\alpha_{r}%
}\otimes dx^{\beta_{1}}\otimes...\otimes dx^{\beta_{s}}$

Observation 2 at point q :

$T\left(  q\right)  =\sum_{\alpha_{1}...\alpha_{r}}\sum_{\beta_{1}%
....\beta_{s}}t_{\beta_{1}...\beta_{q}}^{\alpha_{1}...\alpha_{r}}\left(
q\right)  \partial x_{\alpha_{1}}\otimes..\otimes\partial x_{\alpha_{r}%
}\otimes dx^{\beta_{1}}\otimes...\otimes dx^{\beta_{s}}$

Here we have a big difference with affine spaces, where we can always use a
common basis $\left(  e_{\alpha}\right)  _{\alpha\in A}$. Even if the chart is
the same, the tangent spaces are not the same, and we cannot tell much without
some tool to compare the holonomic bases at p and q. Let us assume that we
have such a tool.\ So we can "transport" T(p) at q and express it in the
holonomic frame at q. If we find the same figures we can say that T is
invariant when we go from p to q. More generally if we have such a procedure
we can give a precise meaning to the variation of the tensor field between p
and q.

In differential geometry we have several tools to transport tensors on tensor
bundles : the "push-forward", which is quite general, and derivations.

\paragraph{Transport by push forward\newline}

If there is a diffeomorphism : $f:M\rightarrow M$ then with the push-forward
$\widehat{T}=f_{\ast}T$ reads :

$\widehat{T}\left(  f\left(  p\right)  \right)  =f^{\ast}T_{j}\left(
p\right)  =\Phi_{i,r,s}\left(  p,t_{j}\left(  \varphi_{j}\circ f\left(
p\right)  \right)  \right)  =\Phi_{i,r,s}\left(  p,t_{j}\left(  \varphi
_{i}\left(  p\right)  \right)  \right)  $

The components of the tensor $\widehat{T}$, expressed in the holonomic basis
are :

$\widehat{T}_{\beta_{1}...\beta_{q}}^{\alpha_{1}...\alpha_{r}}\left(
f(p)\right)  =\sum_{\lambda_{1}...\lambda_{r}}\sum_{\mu_{1}....\mu_{s}}%
T_{\mu_{1}...\mu_{s}}^{\lambda_{1}...\lambda_{r}}\left(  p\right)  \left[
J\right]  _{\lambda_{1}}^{\alpha_{1}}..\left[  J\right]  _{\lambda_{r}%
}^{\alpha_{r}}\left[  J^{-1}\right]  _{\beta_{1}}^{\mu_{1}}..\left[
J^{-1}\right]  _{\beta_{s}}^{\mu_{s}}$

where $\left[  J\right]  =\left[  \frac{\partial y^{\alpha}}{\partial
x^{\beta}}\right]  $ is the matrix of f'(p)

So they are a linear (possibly complicated) combination of the components of T.

\begin{definition}
A tensor T is said to be \textbf{invariant} by a diffeomorphism f on the
manifold M if : $T=f^{\ast}T\Leftrightarrow T=f_{\ast}T$
\end{definition}

If T is invariant then the components of the tensor at p and f(p) must be
linearly dependent.

If there is a one parameter group of diffeomorphisms, it has an infinitesimal
generator which is a vector field V. If a tensor T is invariant by such a one
parameter group the Lie derivative $\pounds _{V}T=0.$

\paragraph{Derivation\newline}

1. Not all physical phenomenons are invariant, and of course we want some tool
to measure how a tensor changes when we go from p to q. This is just what we
do with the derivative : $T\left(  a+h\right)  =T\left(  a\right)  +T^{\prime
}(a)h+\epsilon\left(  h\right)  h$ .So we need a derivative for tensor fields.
Manifolds are not isotropic : all directions on the tangent spaces are not
equivalent. Thus it is clear that a derivation depends on the direction
u\ along which we differentiate, meaning something like the derivative $D_{u}%
$\ along a vector, and the direction u will vary at each point.\ There are two
ways to do it : either u is the tangent c'(t) to some curve p=c(t), or u=V(p)
with V a vector field. For now on let us assume that u is given by some vector
field V (we would have the same results with c'(t)).

So we shall look for a map : $D_{V}:\mathfrak{X}\left(  \otimes_{s}%
^{r}TM\right)  \rightarrow\mathfrak{X}\left(  \otimes_{s}^{r}TM\right)  $ with
$V\in\mathfrak{X}\left(  TM\right)  $ which preserves the type of the tensor field.

2. We wish also that this derivation D has some nice useful properties, as
classical derivatives :

i) it should be linear in V :

$\forall V,V^{\prime}\in\mathfrak{X}\left(  TM\right)  ,k,k^{\prime}\in
K:D_{kV+k^{\prime}V^{\prime}}T=kD_{V}T+k^{\prime}D_{V^{\prime}}T$

so that we can compute easily the derivative along the vectors of a basis.
This condition, joined with that $D_{V}T$ should be a tensor of the same type
as T leads to say that :

$D:\mathfrak{X}\left(  \otimes_{s}^{r}TM\right)  \rightarrow\mathfrak{X}%
\left(  \otimes_{s+1}^{r}TM\right)  $

For a (0,0) type tensor, meaning a function on M, the result is a 1-form.

ii) $D$ should be a linear operator on the tensor fields :

$\forall S,T\in\mathfrak{X}\left(  \otimes_{s}^{r}TM\right)  ,k,k^{\prime}\in
K:D\left(  kS+k^{\prime}T\right)  =kDS+k^{\prime}DT$

iii) $D$ should obey the Leibnitz rule with respect to the tensorial product :

$D\left(  S\otimes T\right)  =\left(  DS\right)  \otimes T+S\otimes\left(
DT\right)  $

The tensor fields have a structure of algebra $\mathfrak{X}\left(  \otimes
TM\right)  $ with the tensor product. These conditions make D a derivation on
$\mathfrak{X}\left(  \otimes TM\right)  $ (see the Algebra part).

iv) In addition we wish some kind of relation between the operation on TM and
TM*. Without a bilinear form the only general relation which is available is
the trace operator, well defined and the unique linear map : $Tr:\mathfrak{X}%
\left(  \otimes_{1}^{1}TM\right)  \rightarrow C\left(  M;K\right)  $ such that
$\ \forall\varpi\in\mathfrak{X}\left(  \otimes_{1}^{0}TM\right)
,V\in\mathfrak{X}\left(  \otimes_{0}^{1}TM\right)  \ :Tr\left(  \varpi\otimes
V\right)  =\varpi\left(  V\right)  $

So we impose that $D$ commutes with the trace operator. Then it commutes with
the contraction of tensors.

3. There is a general theorem (Kobayashi p.30) which tells that any derivation
can be written as a linear combination of a Lie derivative and a covariant
derivative, which are seen in the next subsections. So the tool that we are
looking for is a linear combination of Lie derivative and covariant derivative.

4. The parallel transport of a tensor T by a derivation D along a vector field
is done by defining the "transported tensor" $\widehat{T}$ as the solution of
a differential equation $D_{V}\widehat{T}=0$ and the initial condition
$\widehat{T}\left(  p\right)  =T\left(  p\right)  .$ Similarly a tensor is
invariant if $D_{V}T=0.$

5. Conversely with a derivative we can look for the curves such that a given
tensor is invariant. We can see these curves as integral curves for both the
transport and the tensor. Of special interest are the curves such that their
tangent are themselves invariant by parallel transport. They are the
geodesics. If the covariant derivative comes from a metric these curves are
integral curves of the length.

\subsubsection{Lie derivative}

The idea is to use the flow of a vector field to transport a tensor : at each
point along a curve we use the diffeomorphism to push forward the tensor along
the curve and we compute a derivative at this point. It is clear that the
result depends on the vector field : in some way the Lie derivative is a
generalization of the derivative along a vector. This is a very general tool,
in that it does not require any other ingredient than the vector field V.

\paragraph{Definition\newline}

Let T be a tensor field $T\in\mathfrak{X}\left(  \otimes_{s}^{r}TM\right)  $
and V a vector field $V\in\mathfrak{X}\left(  TM\right)  .$ The flow $\Phi
_{V}$ is defined in a domain which is an open neighborhood of 0xM and in this
domain it is a diffeomorphism $M\rightarrow M$. For t small the tensor at
$\Phi_{V}\left(  -t,p\right)  $ is pushed forward at p by $\Phi_{V}\left(
t,.\right)  :$

$\left(  \Phi_{V}\left(  t,.\right)  _{\ast}T\right)  \left(  p\right)
=\left(  \Phi_{V}\left(  t,.\right)  \right)  _{r,s}\left(  p\right)  T\left(
\Phi_{V}\left(  -t,p\right)  \right)  $

The two tensors are now in the same tangent space at p, and it is possible to
compute for any p in M :

$\pounds _{V}T\left(  p\right)  =\lim_{t\rightarrow0}\frac{1}{t}(\left(
\Phi_{V}\left(  t,.\right)  _{\ast}T\right)  \left(  p\right)  -T(p))$

$=\lim_{t\rightarrow0}\frac{1}{t}(\left(  \Phi_{V}\left(  t,.\right)  _{\ast
}T\right)  (p)-\left(  \Phi_{V}\left(  0,.\right)  _{\ast}T\right)  \left(
p\right)  )$

The limit exists as the components and the jacobian J are differentiable and :

\begin{definition}
The \textbf{Lie derivative of a tensor} field $T\in\mathfrak{X}\left(
\otimes_{s}^{r}TM\right)  $ along the vector field $V\in\mathfrak{X}\left(
TM\right)  $ is :%

\begin{equation}
\pounds _{V}T\left(  p\right)  =\frac{d}{dt}\left(  \left(  \Phi_{V}\left(
t,.\right)  _{\ast}T\right)  \left(  p\right)  \right)  |_{t=0}%
\end{equation}

\end{definition}

In components :

$\left(  \Phi_{V}\left(  t,.\right)  _{\ast}T\right)  _{\beta_{1}...\beta_{q}%
}^{\alpha_{1}...\alpha_{r}}\left(  p\right)  $

$=\sum_{\lambda_{1}...\lambda_{r}}\sum_{\mu_{1}....\mu_{s}}T_{\mu_{1}%
...\mu_{s}}^{\lambda_{1}...\lambda_{r}}\left(  \Phi_{V}\left(  -t,p\right)
\right)  \left[  J\right]  _{\lambda_{1}}^{\alpha_{1}}..\left[  J\right]
_{\lambda_{r}}^{\alpha_{r}}\left[  J^{-1}\right]  _{\beta_{1}}^{\mu_{1}%
}..\left[  J^{-1}\right]  _{\beta_{s}}^{\mu_{s}}$

with : $F:U_{i}\rightarrow U_{i}::$ $y=\varphi_{i}\circ\Phi_{V}\left(
t,.\right)  \circ\varphi_{i}^{-1}\left(  x\right)  =F\left(  x\right)  $

$\left[  F^{\prime}(x)\right]  =\left[  J\right]  =\left[  \frac{\partial
y^{\alpha}}{\partial x^{\beta}}\right]  $

so the derivatives of $\Phi_{V}\left(  t,p\right)  $ with respect to p are involved

\paragraph{Properties of the Lie derivative\newline}

\begin{theorem}
(Kolar p.63) The Lie derivative along a vector field $V\in\mathfrak{X}\left(
TM\right)  $ on a manifold M is a derivation on the algebra $\mathfrak{X}%
\left(  \otimes TM\right)  :$

i) it is a linear operator : $\pounds _{V}\in%
\mathcal{L}%
\left(  \mathfrak{X}\left(  \otimes_{s}^{r}TM\right)  ;\mathfrak{X}\left(
\otimes_{s}^{r}TM\right)  \right)  $

ii) it is linear with respect to the vector field V

iii) it follows the Leibnitz rule with respect to the tensorial product

Moreover:

iv) it commutes with any contraction between tensors

v) antisymmetric tensors go to antisymmetric tensors
\end{theorem}

So $\forall V,W\in\mathfrak{X}\left(  TM\right)  ,\forall k,k^{\prime}\in
K,\forall S,T\in\mathfrak{X}\left(  \otimes TM\right)  $

$\pounds _{V+W}=\pounds _{V}+\pounds _{W}$

$\pounds _{V}\left(  kS+k^{\prime}T\right)  =k\pounds _{V}S+k^{\prime
}\pounds _{V}T$

$\pounds _{V}\left(  S\otimes T\right)  =\left(  \pounds _{V}S\right)  \otimes
T+S\otimes\left(  \pounds _{V}T\right)  $

which gives with $f\in C\left(  M;K\right)  :\pounds _{V}\left(  f\times
T\right)  =\left(  \pounds _{V}f\right)  \times T+f\times\left(
\pounds _{V}T\right)  $ (pointwise multiplication)

\begin{theorem}
(Kobayashi I p.32) For any vector field $V\in\mathfrak{X}\left(  TM\right)  $
and tensor field $T\in\mathfrak{X}\left(  \otimes TM\right)  :$

$\Phi_{V}\left(  -t,.\right)  ^{\ast}\pounds _{V}T=-\frac{d}{dt}\left(
\Phi_{V}\left(  -t,.\right)  ^{\ast}T\right)  |_{t=0}$
\end{theorem}

\begin{theorem}
The Lie derivative of a vector field is the commutator of the vectors fields :
\end{theorem}

\begin{equation}
\forall V,W\in\mathfrak{X}\left(  TM\right)  :\pounds _{V}W=-\pounds _{W}%
V=\left[  V,W\right]
\end{equation}

$f\in C\left(  M;K\right)  :\pounds _{V}f=i_{V}f=V\left(  f\right)
=\sum_{\alpha}V^{\alpha}\partial_{\alpha}f=f^{\prime}\left(  V\right)  $

Remark : $V\left(  f\right)  $ is the differential operator associated to V
acting on the function $f$

\begin{theorem}
Exterior product:%

\begin{equation}
\forall\lambda,\mu\in\mathfrak{X}\left(  \Lambda TM^{\ast}\right)
:\pounds _{V}\left(  \lambda\wedge\mu\right)  =\left(  \pounds _{V}%
\lambda\right)  \wedge\mu+\lambda\wedge\left(  \pounds _{V}\mu\right)
\end{equation}

\end{theorem}

\begin{theorem}
Action of a form on a vector:

$\forall\lambda\in\mathfrak{X}\left(  \Lambda_{1}TM^{\ast}\right)
,W\in\mathfrak{X}\left(  TM\right)  :\pounds _{V}\left(  \lambda\left(
W\right)  \right)  =\left(  \pounds _{V}\lambda\right)  \left(  W\right)
+\lambda\left(  \pounds _{V}W\right)  $

$\forall\lambda\in\mathfrak{X}\left(  \Lambda_{r}TM^{\ast}\right)
,W_{1},..W_{r}\in\mathfrak{X}\left(  TM\right)  :$

$\left(  \pounds _{V}\lambda\right)  \left(  W_{1},...W_{r}\right)  =V\left(
\lambda\left(  W_{1},...W_{r}\right)  \right)  -\sum_{k=1}^{r}\lambda\left(
W_{1},...\left[  V,W_{k}\right]  ..W_{r}\right)  $
\end{theorem}

Remark : $V\left(  \lambda\left(  W_{1},...W_{r}\right)  \right)  $ is the
differential operator associated to V acting on the function $\lambda\left(
W_{1},...W_{r}\right)  $

\begin{theorem}
Interior product of a r form and a vector field :

$\forall\lambda\in\mathfrak{X}\left(  \Lambda_{r}TM^{\ast}\right)
,V,W\in\mathfrak{X}\left(  TM\right)  :\pounds _{V}(i_{W}\lambda
)=i_{\pounds _{V}W}\left(  \lambda\right)  +i_{W}\left(  \pounds _{V}%
\lambda\right)  $
\end{theorem}

Remind that : $\left(  i_{W}\lambda\right)  \left(  W_{1},...W_{r-1}\right)
=\lambda\left(  W,W_{1},...W_{r-1}\right)  $

\begin{theorem}
The \textbf{bracket of the Lie derivative operators} $\pounds _{V}%
,\pounds _{W}$ for the vector fields V,W is : $\left[  \pounds _{V}%
,\pounds _{W}\right]  =\pounds _{V}\circ\pounds _{W}-\pounds _{W}%
\circ\pounds _{V}$ and we have :$\left[  \pounds _{V},\pounds _{W}\right]
=\pounds _{\left[  V,W\right]  }$
\end{theorem}

\paragraph{Parallel transport}

The Lie derivative along a curve is defined only if this is the integral curve
of a tensor field V. The transport is then equivalent to the push forward by
the flow of the vector field.

\begin{theorem}
(Kobayashi I p.33) A tensor field T is invariant by the flow of a vector field
V iff $\pounds _{V}T=0$
\end{theorem}

This result holds for any one parameter group of diffeomorphism, with V = its
infinitesimal generator.

In the next subsections are studied several one parameter group of
diffeomorphisms which preserve some tensor T (the metric of a pseudo
riemannian manifold, the 2 form of a symplectic manifold).\ These groups have
an infinitesimal generator V and $\pounds _{V}T=0.$

\bigskip

\subsection{Exterior algebra}

\label{Exterior algebra manifold}

\subsubsection{Definitions}

For any manifold M a r-form in $T_{p}M^{\ast}$ is an antisymmetric r covariant
tensor in the tangent space at p. A field of r-form is a field of
antisymmetric r covariant tensor in the tangent bundle TM. All the operations
on the exterior algebra of $T_{p}M$ are available, and similarly for the
fields of r-forms, whenever they are implemented pointwise (for a fixed p). So
the exterior product of two r forms fields can be computed.

\begin{notation}
$\mathfrak{X}\left(  \Lambda TM^{\ast}\right)  =\oplus_{r=0}^{\dim
M}\mathfrak{X}\left(  \Lambda_{r}TM^{\ast}\right)  $ is the \textbf{exterior
algebra} of the manifold M.
\end{notation}

This is an algebra over the same field K as M with pointwise operations.

$\sigma\in\mathfrak{S}\left(  r\right)  :\varpi_{\sigma\left(  \alpha
_{1}...\alpha_{r}\right)  }=\epsilon\left(  \sigma\right)  \varpi_{\alpha
_{1}...\alpha_{r}}$

In a holonomic basis a field of r forms reads :

i) $\varpi\left(  p\right)  =\sum_{\left\{  \alpha_{1}...\alpha_{r}\right\}
}\varpi_{\alpha_{1}...\alpha_{r}}\left(  p\right)  dx^{\alpha_{1}}\wedge
dx^{\alpha_{2}}\wedge...\wedge dx^{\alpha_{r}}$ with ordered indices

ii) $\varpi\left(  p\right)  =\frac{1}{r!}\sum_{\left(  \alpha_{1}%
...\alpha_{r}\right)  }\varpi_{\alpha_{1}...\alpha_{r}}\left(  p\right)
dx^{\alpha_{1}}\wedge dx^{\alpha_{2}}\wedge...\wedge dx^{\alpha_{r}}$ with non
ordered indices

iii) $\varpi\left(  p\right)  =\sum_{\left(  \alpha_{1}...\alpha_{r}\right)
}\varpi_{\alpha_{1}...\alpha_{r}}\left(  p\right)  dx^{\alpha_{1}}\otimes
dx^{\alpha_{2}}\otimes...\otimes dx^{\alpha_{r}}$ with non ordered indices

$\varpi_{\alpha_{1}...\alpha_{r}}:M\rightarrow K$ \ the form is of class c if
the functions are of class c.

To each r form is associated a r multilinear antisymmetric map, valued in the
field K :

$\forall\varpi\in\mathfrak{X}\left(  \wedge_{r}TM^{\ast}\right)
,V_{1},...,V_{r}\in\mathfrak{X}\left(  TM\right)  :$

$\varpi\left(  V_{1},...,V_{r}\right)  =\sum_{\left(  \alpha_{1}...\alpha
_{r}\right)  }\varpi_{\alpha_{1}...\alpha_{r}}v_{1}^{\alpha_{1}}v_{2}%
^{\alpha_{2}}...v_{r}^{\alpha_{r}}$

Similarly a r-form on M can be valued in a \textit{fixed} Banach vector space
F. It reads :

$\varpi=\sum_{\left\{  \alpha_{1}...\alpha_{r}\right\}  }\sum_{i=1}^{q}%
\varpi_{\alpha_{1}...\alpha_{r}}^{i}f_{i}\otimes dx^{\alpha_{1}}\wedge
dx^{\alpha_{2}}\wedge...\wedge dx^{\alpha_{r}}$

where $\left(  f_{i}\right)  _{i=1}^{q}$ is a basis of F.

All the results for r-forms valued in K can be extended to these forms.

\begin{notation}
$\Lambda_{r}\left(  M;F\right)  $ is the space of fields of r-forms on the
manifold M valued in the fixed vector space F
\end{notation}

So $\mathfrak{X}\left(  \Lambda_{r}TM^{\ast}\right)  =\Lambda_{r}\left(
M;K\right)  .$

\begin{definition}
The \textbf{canonical form} on the manifold M modeled on E is the field of 1
form valued in E : $\Theta=\sum_{\alpha\in A}dx^{\alpha}\otimes e_{\alpha}$
\end{definition}

So : $\Theta\left(  p\right)  \left(  u_{p}\right)  =\sum_{\alpha\in A}%
u_{p}^{\alpha}e_{\alpha}\in E$

It is also possible to consider r-forms valued in TM. They read :

$\varpi=\sum_{\beta\left\{  \alpha_{1}...\alpha_{r}\right\}  }\sum_{\beta
}\varpi_{\alpha_{1}...\alpha_{r}}^{\beta}\partial x_{\beta}\otimes
dx^{\alpha_{1}}\wedge dx^{\alpha_{2}}\wedge...\wedge dx^{\alpha_{r}}%
\in\mathfrak{X}\left(  \Lambda_{r}TM^{\ast}\otimes TM\right)  $

So this is a field of mixed tensors $\otimes_{r}^{1}TM$ which is antisymmetric
in the lower indices. To keep it short we use the :

\begin{notation}
$\Lambda_{r}\left(  M;TM\right)  $ is the space of fields of r-forms on the
manifold M valued in the tangent bundle
\end{notation}

Their theory involves the derivatives on graded algebras and leads to the
Fr\"{o}licher-Nijenhuis bracket (see Kolar p.67). We will see more about them
in the Fiber bundle part.

\subsubsection{Interior product}

The interior product $i_{V}\varpi$ of a r-form $\varpi$\ and a vector V is an
operation which, when implemented pointwise, can be extended to fields of r
forms and vectors on a manifold M, with the same properties. In a holonomic
basis of M:

$\forall\varpi\in\mathfrak{X}\left(  \wedge_{r}TM^{\ast}\right)  ,\pi
\in\mathfrak{X}\left(  \wedge_{s}TM^{\ast}\right)  ,V,W\in\mathfrak{X}\left(
TM\right)  ,f\in C\left(  M;K\right)  ,k\in K:$%

\begin{equation}
i_{V}\varpi=\sum_{k=1}^{r}(-1)^{k-1}\sum_{\left\{  \alpha_{1}...\alpha
_{r}\right\}  }V^{\alpha_{k}}\varpi_{\alpha_{1}...\alpha_{r}}dx^{\alpha_{1}%
}\Lambda...\Lambda\widehat{dx^{\alpha_{k}}}...dx^{\alpha_{r}}%
\end{equation}

where\ \symbol{94} is for a variable that shall be omitted.

$i_{V}\left(  \varpi\wedge\pi\right)  =\left(  i_{V}\varpi\right)  \wedge
\pi+\left(  -1\right)  ^{\deg\varpi}\varpi\wedge\left(  i_{V}\pi\right)  $

$i_{V}\circ i_{V}=0$

$i_{fV}=fi_{V}$

$i_{\left[  V,W\right]  }\varpi=\left(  i_{W}\varpi\right)  V-\left(
i_{V}\varpi\right)  W$

$i_{V}\varpi(kV)=0$

$\varpi\in\mathfrak{X}\left(  \wedge_{2}TM^{\ast}\right)  :\left(  i_{V}%
\varpi\right)  W=\varpi(V,W)=-\varpi(W,V)=-\left(  i_{W}\varpi\right)  V$

\subsubsection{Exterior differential}

The exterior differential is an operation which is specific both to
differential geometry and r-forms. But, as functions are 0 forms, it extends
to functions on a manifold.

\begin{definition}
On a m dimensional manifold M the \textbf{exterior differential} is the
operator : $d:\mathfrak{X}_{1}\left(  \wedge_{r}TM^{\ast}\right)
\rightarrow\mathfrak{X}_{0}\left(  \wedge_{r+1}TM^{\ast}\right)  $ defined in
a holonomic basis by :%

\begin{equation}
d\left(  \sum_{\left\{  \alpha_{1}...\alpha_{r}\right\}  }\varpi_{\alpha
_{1}...\alpha_{r}}dx^{\alpha_{1}}\wedge...dx^{\alpha_{r}}\right)
=\sum_{\left\{  \alpha_{1}...\alpha_{r}\right\}  }\sum_{\beta=1}^{m}%
\partial_{\beta}\varpi_{\alpha_{1}...\alpha_{r}}dx^{\beta}\wedge
dx^{\alpha_{1}}\wedge...dx^{\alpha_{r}}%
\end{equation}

\end{definition}

Even if this definition is based on components one can show that d is the
unique "natural" operator $\Lambda_{r}TM^{\ast}\rightarrow\Lambda
_{r+1}TM^{\ast}.$ So the result does not depend on the choice of a chart.

$d\varpi=\sum_{\left\{  \alpha_{1}...\alpha_{r+1}\right\}  }\left(  \sum
_{k=1}^{r+1}(-1)^{k-1}\partial_{\alpha_{k}}\varpi_{\alpha_{1}.\widehat
{\alpha_{k}}...\alpha_{r+1}}\right)  dx^{\alpha_{1}}\wedge dx^{\alpha_{2}%
}\wedge...\wedge dx^{\alpha_{r+1}}$

For :

$f\in C_{2}\left(  M;K\right)  :df=\sum_{\alpha\in A}\left(  \partial_{\alpha
}f\right)  dx^{\alpha}$ so df(p) = f'(p)$\in%
\mathcal{L}%
\left(  T_{p}M;K\right)  $

$\varpi\in\Lambda_{1}TM^{\ast}:$

$d(\sum_{\alpha\in A}\varpi_{\alpha}dx^{\alpha})=\sum_{\alpha<\beta}%
(\partial_{\beta}\varpi_{\alpha}-\partial_{\alpha}\varpi_{\beta})(dx^{\beta
}\Lambda dx^{\alpha})$

$=\sum_{\alpha<\beta}(\partial_{\beta}\varpi_{\alpha})(dx^{\beta}\otimes
dx^{\alpha}-dx^{\alpha}\otimes dx^{\beta})$

\begin{theorem}
(Kolar p.65) On a m dimensional manifold M the exterior differential is a
linear operator : $d\in%
\mathcal{L}%
\left(  \mathfrak{X}_{1}\left(  \wedge_{r}TM^{\ast}\right)  ;\mathfrak{X}%
_{0}\left(  \wedge_{r+1}TM^{\ast}\right)  \right)  $ which has the following
properties :

i) it is nilpotent : d%
${{}^2}$%
=0

ii) it commutes with the pull back by any differential map and push forward by diffeomorphisms

iii) it commutes with the Lie derivative $\pounds _{V}$ for any vector field V
\end{theorem}

So :

$\forall\lambda,\mu\in\mathfrak{X}\left(  \Lambda_{r}TM^{\ast}\right)  ,\pi
\in\mathfrak{X}\left(  \Lambda_{s}TM^{\ast}\right)  ,\forall k;k^{\prime}\in
K,V\in\mathfrak{X}\left(  TM\right)  ,f\in C_{2}\left(  M;K\right)  $

$d\left(  k\lambda+k^{\prime}\mu\right)  =kd\lambda+k^{\prime}d\mu$

$d\left(  d\varpi\right)  =0$

$f^{\ast}\circ d=d\circ f^{\ast}$

$f_{\ast}\circ d=d\circ f_{\ast}$

$\pounds _{V}\circ d=d\circ\pounds _{V}$

$\forall\varpi\in\mathfrak{X}\left(  \Lambda_{\dim M}TM^{\ast}\right)
:d\varpi=0$

\begin{theorem}
On a m dimensional manifold M the exterior differential d, the Lie derivative
along a vector field V and the interior product are linked in the formula :%

\begin{equation}
\forall\varpi\in\mathfrak{X}\left(  \Lambda_{r}TM^{\ast}\right)
,V\in\mathfrak{X}\left(  TM\right)  :\pounds _{V}\varpi=i_{V}d\varpi+d\circ
i_{V}\varpi
\end{equation}

\end{theorem}

This is an alternate definition of the exterior differential.

\begin{theorem}
$\forall\lambda\in\mathfrak{X}\left(  \Lambda_{r}TM^{\ast}\right)  ,\mu
\in\mathfrak{X}\left(  \Lambda_{s}TM^{\ast}\right)  :$%

\begin{equation}
d\left(  \lambda\wedge\mu\right)  =\left(  d\lambda\right)  \wedge\mu+\left(
-1\right)  ^{\deg\lambda}\lambda\wedge\left(  d\mu\right)
\end{equation}

\end{theorem}

so for $f\in C_{2}\left(  M;K\right)  :d\left(  f\varpi\right)  =\left(
df\right)  \wedge\varpi+fd\varpi$

\begin{theorem}
Value for vector fields :

$\forall\varpi\in\mathfrak{X}\left(  \Lambda_{r}TM^{\ast}\right)
,V_{1},...,V_{r+1}\in\mathfrak{X}\left(  TM\right)  :$

$d\varpi(V_{1},V_{2},...V_{r+1})$

$=\sum_{i=1}^{r+1}(-1)^{i}V_{i}\left(  \varpi(V_{1},...\widehat{V_{i}%
}...V_{r+1})\right)  +\sum_{\{i,j\}}(-1)^{i+j}\varpi([V_{i},V_{j}%
],V_{1},...\widehat{V_{i}},...\widehat{V_{j}}...V_{r+1})$
\end{theorem}

where $V_{i}$ is the differential operator linked to V$_{i}$ acting on the
function $\varpi(V_{1},...\widehat{V_{i}}...V_{r+1})$

Which gives : $d\varpi(V,W)=\left(  i_{W}\varpi\right)  V-\left(  i_{V}%
\varpi\right)  W-i_{\left[  V,W\right]  }\varpi$

and if\emph{\ }$\varpi\in X\left(  \Lambda_{1}TM^{\ast}\right)  :d\varpi
(V,W)=\pounds _{V}\left(  i_{V}\varpi\right)  -\pounds _{W}\left(  i_{V}%
\varpi\right)  -i_{\left[  V,W\right]  }\varpi$

If $\varpi$ is a r-form valued in a fixed vector space, the exterior
differential is computed by :

$\varpi=\sum_{\left\{  \alpha_{1}...\alpha_{r}\right\}  }\sum_{i}%
\varpi_{\alpha_{1}...\alpha_{r}}^{i}e_{i}\otimes dx^{\alpha_{1}}\wedge
dx^{\alpha_{2}}\wedge...\wedge dx^{\alpha_{r}}$

$\rightarrow d\varpi=\sum_{\left\{  \alpha_{1}...\alpha_{r}\right\}  }%
\sum_{\beta=1}^{m}\sum_{i}\partial_{\beta}\varpi_{\alpha_{1}...\alpha_{r}}%
^{i}e_{i}\otimes dx^{\beta}\wedge dx^{\alpha_{1}}\wedge dx^{\alpha_{2}}%
\wedge...\wedge dx^{\alpha_{r}}$

\subsubsection{Poincar\'{e}'s lemna}

\begin{definition}
On a manifold M :

a \textbf{closed form} is a field of r-form $\varpi\in X\left(  \Lambda
_{r}TM^{\ast}\right)  $ such that $d\varpi=0$

an \textbf{exact form} is a field of r-form $\varpi\in X\left(  \Lambda
_{r}TM^{\ast}\right)  $ such that there is $\lambda\in X\left(  \Lambda
_{r-1}TM^{\ast}\right)  $ with $\varpi=d\lambda$
\end{definition}

An exact form is closed, the lemna of Poincar\'{e} gives a converse.

\begin{theorem}
Poincar\'{e}'s lemna : A closed differential form is locally exact.
\end{theorem}

Which means that : If $\varpi\in X\left(  \Lambda_{r}TM^{\ast}\right)  $ such
that $d\varpi=0$ then, for any $p\in M,$ there is a neighborhood n(p) and
$\lambda\in X\left(  \Lambda_{r-1}TM^{\ast}\right)  $ such that $\varpi
=d\lambda$ in n(p).

\textit{The solution is not unique} : $\lambda+d\mu$ is still a solution,
whatever $\mu.$ The study of the subsets of closed forms which differ only by
an exact form is the main topic of cohomology (see below).

If M is an open simply connected subset of a real finite dimensional affine
space, $\varpi\in\Lambda_{1}TM^{\ast}$ of class q, such that $d\varpi=0, $
then there is a function $f\in C_{q+1}(M;%
\mathbb{R}
)$ such that $df=\varpi$

If $M=%
\mathbb{R}
^{n}$:$\varpi=\sum_{\alpha=1}^{n}a_{\alpha}(x)dx^{\alpha},d\varpi=0$

$\Rightarrow\lambda(x)=\sum_{\alpha=1}^{n}x^{\alpha}\int_{0}^{1}a_{\alpha
}(tx)dt$ and $d\lambda=\varpi$

\bigskip

\subsection{Covariant derivative}

\label{Covariant derivative}

The general theory of connections is seen in the Fiber bundle part. We will
limit here to the theory of covariant derivation, which is part of the story,
but simpler and very useful for many practical purposes. A covariant
derivative is a derivative for tensor fields, which meets the requirements for
the transportation of tensors (see Lie Derivatives)

In this section the manifold M is a m dimensional smooth real manifold with
atlas $\left(  O_{i},\varphi_{i}\right)  _{i\in I}$

The theory of affine connection and covariant derivative can be extended to
Banach manifolds of infinite dimension (see Lewis).

\subsubsection{Covariant derivative}

\paragraph{Definition\newline}

\begin{definition}
A \textbf{covariant derivative} on a manifold M is a linear operator
$\nabla\in%
\mathcal{L}%
\left(  \mathfrak{X}\left(  TM\right)  ;D\right)  $ from the space of vector
fields to the space D of derivations on the tensorial bundle of M, such that
for every $V\in\mathfrak{X}\left(  TM\right)  :$

i) $\nabla_{V}\in%
\mathcal{L}%
\left(  \mathfrak{X}\left(  \otimes_{s}^{r}TM\right)  ;\mathfrak{X}\left(
\otimes_{s+1}^{r}TM\right)  \right)  $

ii) $\nabla_{V}$ follows the Leibnitz rule with respect to the tensorial product

iii) $\nabla_{V}$ commutes with the trace operator
\end{definition}

\begin{definition}
An \textbf{affine connection} on a manifold M over the field K is a bilinear
operator $\nabla\in%
\mathcal{L}%
^{2}\left(  \mathfrak{X}\left(  TM\right)  ;\mathfrak{X}\left(  TM\right)
\right)  $ such that :

$\forall f\in C_{1}\left(  M;K\right)  :\nabla_{fX}Y=f\nabla_{X}Y$

$\nabla_{X}(fY)=f\nabla_{X}Y+(i_{X}df)Y$
\end{definition}

In a holonomic basis of M the coefficients of $\nabla$ are the
\textbf{Christoffel symbols} of the connection : $\Gamma_{\beta\eta}^{\alpha
}\left(  p\right)  $

\begin{theorem}
An affine connection defines uniquely a covariant derivative and conversely a
covariant derivative defines an affine connection.
\end{theorem}

\begin{proof}
i) According to the rules above, a covariant derivative is defined if we have
the derivatives of $\partial_{\alpha},dx^{\alpha}$ which are tensor fields. So
let us denote :

$\left(  \nabla\partial_{\alpha}\right)  \left(  p\right)  =\sum_{\gamma
,\eta=1}^{m}X_{\eta\alpha}^{\gamma}\left(  p\right)  dx^{\eta}\otimes
\partial_{\gamma}$

$\nabla dx^{\alpha}=\sum_{\gamma,\eta=1}^{m}Y_{\eta\gamma}^{\alpha}\left(
p\right)  dx^{\eta}\otimes dx^{\gamma}$

By definition : $\left(  dx^{\alpha}\left(  \partial_{\beta}\right)  \right)
=\delta_{\beta}^{\alpha}$

$\Rightarrow\nabla\left(  Tr\left(  dx^{\alpha}\left(  \partial_{\beta
}\right)  \right)  \right)  =Tr\left(  \left(  \nabla dx^{\alpha}\right)
\otimes\partial_{\beta}\right)  +Tr\left(  dx^{\alpha}\otimes\nabla
\partial_{\beta}\right)  =0$

$Tr\left(  \sum_{\eta,\gamma=1}^{m}Y_{\eta\gamma}^{\alpha}dx^{\eta}\otimes
dx^{\gamma}\otimes\partial_{\beta}\right)  =-Tr\left(  dx^{\alpha}\otimes
\sum_{\eta,\gamma=1}^{m}X_{\eta\beta}^{\gamma}dx^{\eta}\otimes\partial
_{\gamma}\right)  $

$\sum_{\eta,\gamma=1}^{m}Y_{\eta\gamma}^{\alpha}dx^{\eta}\left(  dx^{\gamma
}\left(  \partial_{\beta}\right)  \right)  =-\sum_{\eta,\gamma=1}^{m}%
X_{\eta\beta}^{\gamma}dx^{\eta}\left(  dx^{\alpha}\left(  \partial_{\gamma
}\right)  \right)  $

$\sum_{\eta,\gamma=1}^{m}Y_{\eta\beta}^{\alpha}dx^{\eta}=-\sum_{\eta,\gamma
=1}^{m}X_{\eta\beta}^{\alpha}dx^{\eta}\Leftrightarrow Y_{\eta\beta}^{\alpha
}=-X_{\eta\beta}^{\alpha}$

So the derivation is fully defined by the value of the Christofell
coefficients $\Gamma_{\beta\eta}^{\alpha}\left(  p\right)  $ scalar functions
for a holonomic basis and we have:

$\nabla\partial_{\alpha}=\sum_{\beta,\gamma=1}^{m}\Gamma_{\beta\alpha}%
^{\gamma}dx^{\beta}\otimes\partial_{\gamma}$

$\nabla dx^{\alpha}=-\sum_{\beta,\gamma=1}^{m}\Gamma_{\beta\gamma}^{\alpha
}dx^{\beta}\otimes dx^{\gamma}$

ii) Conversely an affine connection with Christofell coefficients
$\Gamma_{\beta\eta}^{\alpha}\left(  p\right)  $ defines a unique covariant
connection (Kobayashi I p.143).
\end{proof}

\paragraph{Christoffel symbols in a change of charts\newline}

A covariant derivative is not unique : it depends on the coefficients $\Gamma$
which have been computed in a chart. However a given covariant derivative
$\nabla$\ is a geometric object, which is independant of a the choice of a
basis. In a change of charts \textit{the Christoffel coefficients are not
tensors}, and change according to specific rules.

\begin{theorem}
The Christoffel symbols in the new basis are :

$\widehat{\Gamma}_{\beta\gamma}^{\alpha}=\sum_{\lambda\mu}\left[
J^{-1}\right]  _{\beta}^{\mu}\left[  J^{-1}\right]  _{\gamma}^{\lambda}\left(
\Gamma_{\mu\lambda}^{\nu}\left[  J\right]  _{\nu}^{\alpha}-\partial_{\mu
}\left[  J\right]  _{\lambda}^{\alpha}\right)  $ with theJacobian $J=\left[
\frac{\partial y^{\alpha}}{\partial x^{\beta}}\right]  $
\end{theorem}

\begin{proof}
Coordinates in the old chart : $x=\varphi_{i}\left(  p\right)  $

Coordinates in the new chart : $y=\psi_{i}\left(  p\right)  $

Old holonomic basis :

$\partial x_{\alpha}=\varphi_{i}^{\prime}\left(  p\right)  ^{-1}e_{\alpha}, $

$dx^{\alpha}=\varphi_{i}^{\prime}\left(  x\right)  ^{t}e^{\alpha}$ with
$dx^{\alpha}\left(  \partial x_{\beta}\right)  =\delta_{\beta}^{\alpha}$

New holonomic basis :

$\partial y_{\alpha}=\psi_{i}^{\prime}\left(  p\right)  ^{-1}e_{\alpha}%
=\sum_{\beta}\left[  J^{-1}\right]  _{\alpha}^{\beta}\partial x_{\beta}$

$dy^{\alpha}=\psi_{i}^{\prime}\left(  y\right)  ^{\ast}e^{\alpha}=\sum_{\beta
}\left[  J\right]  _{\alpha}^{\beta}dx_{\beta}$ with $dy^{\alpha}\left(
\partial y_{\beta}\right)  =\delta_{\beta}^{\alpha}$

Transition map:

$\alpha=1..n:y^{\alpha}=F^{\alpha}\left(  x^{1},...x^{n}\right)
\Leftrightarrow F\left(  x\right)  =\psi_{i}\circ\varphi_{i}^{-1}\left(
x\right)  $

Jacobian : $J=\left[  F^{\prime}(x)\right]  =\left[  J_{\beta}^{\alpha
}\right]  =\left[  \dfrac{\partial F^{\alpha}}{\partial x^{\beta}}\right]
_{n\times n}\simeq\left[  \frac{\partial y^{\alpha}}{\partial x^{\beta}%
}\right]  $

$V=\sum_{\alpha}V^{\alpha}\partial x_{\alpha}=\sum_{\alpha}\widehat{V}%
^{\alpha}\partial y_{\alpha}$ \ with $\widehat{V}^{\alpha}=\sum_{\beta
}J_{\beta}^{\alpha}V^{\beta}\simeq\sum_{\beta}\frac{\partial y^{\alpha}%
}{\partial x^{\beta}}V^{\beta}$

If we want the same derivative with both charts, we need for any vector field :

$\nabla V=\sum_{\alpha,\beta=1}^{m}\left(  \frac{\partial}{\partial x^{\beta}%
}V^{\alpha}+\Gamma_{\beta\gamma}^{\alpha}V^{\gamma}\right)  dx^{\beta}%
\otimes\partial x_{\alpha}$

$=\sum_{\alpha,\beta=1}^{m}\left(  \frac{\partial}{\partial y^{\beta}}%
\widehat{V}^{\alpha}+\widehat{\Gamma}_{\beta\gamma}^{\alpha}\widehat
{V}^{\gamma}\right)  dy^{\beta}\otimes\partial y_{\alpha}$

$\nabla V$ is a (1,1) tensor, whose components change according to :

$T=\sum_{\alpha\beta}T_{\beta}^{\alpha}\partial x_{\alpha}\otimes dx^{\beta
}=\sum_{\alpha\beta}\widehat{T}_{\beta}^{\alpha}\partial y_{\alpha}\otimes
dy^{\beta}$ with $\widehat{T}_{\beta}^{\alpha}=\sum_{\lambda\mu}T_{\mu
}^{\lambda}\left[  J\right]  _{\lambda}^{\alpha}\left[  J^{-1}\right]
_{\beta}^{\mu}$

So : $\frac{\partial}{\partial y^{\beta}}\widehat{V}^{\alpha}+\widehat{\Gamma
}_{\beta\gamma}^{\alpha}\widehat{V}^{\gamma}=\sum_{\lambda\mu}\left(
\frac{\partial}{\partial x^{\mu}}V^{\lambda}+\Gamma_{\mu\nu}^{\lambda}V^{\nu
}\right)  \left[  J\right]  _{\lambda}^{\alpha}\left[  J^{-1}\right]  _{\beta
}^{\mu}$

which gives : $\ \widehat{\Gamma}_{\beta\gamma}^{\alpha}=\left[
J^{-1}\right]  _{\beta}^{\mu}\left[  J^{-1}\right]  _{\gamma}^{\lambda}\left(
\Gamma_{\mu\lambda}^{\nu}\left[  J\right]  _{\nu}^{\alpha}-\partial_{\mu
}\left[  J\right]  _{\lambda}^{\alpha}\right)  $
\end{proof}

\paragraph{Properties\newline}

$\forall V,W\in\mathfrak{X}\left(  TM\right)  ,\forall S,T\in\mathfrak{X}%
\left(  \otimes_{s}^{r}TM\right)  ,k,k^{\prime}\in K,\forall f\in C_{1}\left(
M;K\right)  $

$\nabla_{V}\in%
\mathcal{L}%
\left(  \mathfrak{X}\left(  \otimes_{s}^{r}TM\right)  ;\mathfrak{X}\left(
\otimes_{s+1}^{r}TM\right)  \right)  $

$\nabla_{V}\left(  kS+k^{\prime}T\right)  =k\nabla_{V}S+k^{\prime}\nabla_{V}T$

$\nabla_{V}\left(  S\otimes T\right)  =\left(  \nabla_{V}S\right)  \otimes
T+S\otimes\left(  \nabla_{V}T\right)  $

$\nabla_{fV}W=f\nabla_{V}W$

$\nabla_{V}(fW)=f\nabla_{V}W+(i_{V}df)W$

$\nabla f=df\in\mathfrak{X}\left(  \otimes_{1}^{0}TM\right)  $

$\nabla_{V}(Tr\left(  T\right)  )=Tr\left(  \nabla_{V}T\right)  $

Coordinate expressions in a holonomic basis:

for a vector field : $V=\sum_{\alpha=1}^{m}V^{\alpha}\partial_{\alpha}:$

$\nabla V=\sum_{\alpha,\beta=1}^{m}\left(  \partial_{\beta}V^{\alpha}%
+\Gamma_{\beta\gamma}^{\alpha}V^{\gamma}\right)  dx^{\beta}\otimes
\partial_{\alpha}$

for a 1-form : $\varpi=\sum_{\alpha=1}^{m}\varpi_{\alpha}dx^{\alpha}:$

$\nabla\varpi=\sum_{\alpha,\beta=1}^{m}\left(  \partial_{\beta}\varpi_{\alpha
}-\Gamma_{\beta\alpha}^{\gamma}\varpi_{\gamma}\right)  dx^{\beta}\otimes
dx^{\alpha}$

for a mix tensor :

$T\left(  p\right)  =\sum_{\alpha_{1}...\alpha_{r}}\sum_{\beta_{1}%
....\beta_{s}}T_{\beta_{1}...\beta_{s}}^{\alpha_{1}...\alpha_{r}}\left(
p\right)  \partial x_{\alpha_{1}}\otimes..\otimes\partial x_{\alpha_{r}%
}\otimes dx^{\beta_{1}}\otimes...\otimes dx^{\beta_{s}}$

$\nabla T\left(  p\right)  $

$=\sum_{\alpha_{1}...\alpha_{r}}\sum_{\beta_{1}....\beta_{s}}\sum_{\gamma
}\widehat{T}_{\gamma\beta_{1}...\beta_{s}}^{\alpha_{1}...\alpha_{r}}\partial
x_{\alpha_{1}}\otimes..\otimes\partial x_{\alpha_{r}}\otimes dx^{\gamma
}\otimes dx^{\beta_{1}}\otimes...\otimes dx^{\beta_{s}}$%

\begin{equation}
\widehat{T}_{\gamma\beta_{1}...\beta_{s}}^{\alpha_{1}...\alpha_{r}}%
=\partial_{\gamma}T_{\beta_{1}...\beta_{s}}^{\alpha_{1}...\alpha_{r}}%
+\sum_{k=1}^{r}\Gamma_{\gamma\eta}^{\alpha_{k}}T_{\beta_{1}...\beta_{s}%
}^{\alpha_{1}.\alpha_{k-1}\eta\alpha_{k+1}..\alpha_{r}}-\sum_{k=1}^{s}%
\Gamma_{\gamma\beta_{k}}^{\eta}T_{\beta_{1}..\beta_{k-1}\eta\beta_{k+1}%
..\beta_{s}}^{\alpha_{1}...\alpha_{r}}%
\end{equation}

\subsubsection{Exterior covariant derivative}

The covariant derivative of a r-form is not an antisymmetric tensor.\ In order
to get an operator working on r-forms, one defines the exterior covariant
derivative which applies to r-forms on M, \textit{valued in the tangent
bundle}. Such a form reads :

$\varpi=\sum_{\beta,\left\{  \alpha_{1}...\alpha_{r}\right\}  }\varpi
_{\alpha_{1}...\alpha_{r}}^{\beta}dx^{\alpha_{1}}\wedge dx^{\alpha_{2}}%
\wedge...\wedge dx^{\alpha_{r}}\otimes\partial x_{\beta}\in\Lambda_{r}\left(
M;TM\right)  $

\begin{definition}
The \textbf{exterior covariant derivative} associated to the covariant
derivative $\nabla$, is the linear map :

$\nabla_{e}\in%
\mathcal{L}%
\left(  \Lambda_{r}\left(  M;TM\right)  ;\Lambda_{r+1}\left(  M;TM\right)
\right)  $

with the condition : $\forall X_{0},X_{1},..X_{r}\in\mathfrak{X}\left(
TM\right)  ,\varpi\in\mathfrak{X}\left(  \Lambda_{r}TM^{\ast}\right)  $

$\left(  \nabla_{e}\varpi\right)  (X_{0},X_{1},...X_{r})$

$=\sum_{i=0}^{r}(-1)^{i}\nabla_{X_{i}}\varpi(X_{0},X_{1},...\widehat{X_{i}%
}...X_{r})+\sum_{\{i,j\}}(-1)^{i+j}\varpi([X_{i},X_{j}],X_{0},X_{1}%
,...\widehat{X_{i}},...\widehat{X_{j}}...X_{r})$
\end{definition}

This formula is similar to the one for the exterior differential ($\nabla$
replacing \pounds ). Which leads to the formula :%

\begin{equation}
\nabla_{e}\varpi=\sum_{\beta}\left(  d\varpi^{\beta}+\left(  \sum_{\gamma
}\left(  \sum_{\alpha}\Gamma_{\alpha\gamma}^{\beta}dx^{\alpha}\right)
\wedge\varpi^{\gamma}\right)  \right)  \partial x_{\beta}%
\end{equation}

\begin{proof}
Let us denote : $\sum_{\beta}\varpi^{\beta}(X_{0},...\widehat{X_{i}}%
...X_{r})\partial x_{\beta}=\sum_{\beta}\Omega_{i}^{\beta}\partial x_{\beta}$

From the exterior differential formulas , $\beta$ fixed:

$\left(  d\varpi^{\beta}\right)  (X_{0},X_{1},...X_{r})$

$=\sum_{i=0}^{r}(-1)^{i}X_{i}^{\alpha}\partial_{\alpha}\left(  \varpi^{\beta
}(X_{0},...\widehat{X_{i}}...X_{r})\right)  $

$+\sum_{\{i,j\}}(-1)^{i+j}\varpi^{\beta}([X_{i},X_{j}],X_{0},...\widehat
{X_{i}},...\widehat{X_{j}}...X_{r})$

$\nabla_{e}\varpi(X_{0},X_{1},...X_{r})$

$=\sum_{\beta}\left(  d\varpi^{\beta}\right)  (X_{0},X_{1},...X_{r})\partial
x_{\beta}$

$+\sum_{i=0}^{r}(-1)^{i}\left(  \nabla_{X_{i}}\left(  \sum_{\beta}\Omega
_{i}^{\beta}\partial x_{\beta}\right)  -\sum_{\alpha\beta}X_{i}^{\alpha
}\partial_{\alpha}\left(  \Omega_{i}^{\beta}\right)  \partial x_{\beta
}\right)  $

$=\sum_{\beta}\left(  d\varpi^{\beta}\right)  (X_{0},X_{1},...X_{r})\partial
x_{\beta}$

$+\sum_{i=0}^{r}(-1)^{i}\left(  \sum_{\alpha\beta\gamma}\left(  \partial
_{\alpha}\Omega_{i}^{\beta}+\Gamma_{\alpha\gamma}^{\beta}\Omega_{i}^{\gamma
}\right)  X_{i}^{\alpha}\partial x_{\beta}-\sum_{\alpha\beta}X_{i}^{\alpha
}\partial_{\alpha}\left(  \Omega_{i}^{\beta}\right)  \partial x_{\beta
}\right)  $

$=\sum_{\beta}\left(  d\varpi^{\beta}\right)  (X_{0},X_{1},...X_{r})\partial
x_{\beta}+\sum_{i=0}^{r}(-1)^{i}\left(  \sum_{\alpha\beta\gamma}\Gamma
_{\alpha\gamma}^{\beta}\Omega_{i}^{\gamma}X_{i}^{\alpha}\partial x_{\beta
}\right)  $

$\Omega_{i}^{\gamma}=\sum_{\left\{  \lambda_{0}...\widehat{\lambda_{i}%
}..\lambda_{r-1}\right\}  }\varpi_{\left\{  \lambda_{0}...\widehat{\lambda
_{i}}..\lambda_{r-1}\right\}  }^{\gamma}X_{0}^{\lambda_{0}}X_{1}^{\lambda_{1}%
}...\widehat{X_{i}^{\lambda_{i}}}...X_{r}^{\lambda_{r}}$

$\sum_{\alpha\gamma}\Gamma_{\alpha\gamma}^{\beta}\sum_{i=0}^{r}(-1)^{i}%
\Omega_{i}^{\gamma}X_{i}^{\alpha}$

$=\sum_{\gamma}\sum_{i=0}^{r}\sum_{\left\{  \lambda_{0}....\lambda
_{r-1}\right\}  }\Gamma_{\lambda_{i}\gamma}^{\beta}\varpi_{\left\{
\lambda_{0}....\lambda_{r-1}\right\}  }^{\gamma}X_{0}^{\lambda_{0}}%
X_{1}^{\lambda_{1}}...X_{i}^{\lambda_{i}}...X_{r}^{\lambda_{r}}$

$=\left(  \sum_{\gamma}\left(  \sum_{\alpha}\Gamma_{\alpha\gamma}^{\beta
}dx^{\alpha}\right)  \wedge\varpi^{\gamma}\right)  (X_{0},...X_{r})$

$\nabla_{e}\varpi(X_{0},X_{1},...X_{r})=\sum_{\beta}\left(  d\varpi^{\beta
}+\left(  \sum_{\gamma}\left(  \sum_{\alpha}\Gamma_{\alpha\gamma}^{\beta
}dx^{\alpha}\right)  \wedge\varpi^{\gamma}\right)  \right)  (X_{0}%
,...X_{r})\partial x_{\beta}$
\end{proof}

A vector field can be considered as a 0-form valued in TM, and $\forall
X\in\mathfrak{X}\left(  TM\right)  :\nabla_{e}X=\nabla X$ (we have the usual
covariant derivative of a vector field on M)

\begin{theorem}
Exterior product:

$\forall\varpi_{r}\in\Lambda_{r}\left(  M;TM\right)  ,\varpi_{s}\in\Lambda
_{s}\left(  M;TM\right)  :$

$\nabla_{e}\left(  \varpi_{r}\wedge\varpi_{s}\right)  =\left(  \nabla
_{e}\varpi_{r}\right)  \wedge\varpi_{s}+\left(  -1\right)  ^{r}\varpi
_{r}\wedge\nabla_{e}\varpi_{s}$
\end{theorem}

So the formula is the same as for the exterior differential d.

\begin{theorem}
Pull-back, push forward (Kolar p.112) The exterior covariant derivative
commutes with the pull back of forms :

$\forall f\in C_{2}\left(  N;M\right)  ,\varpi\in\mathfrak{X}\left(
\Lambda_{r}TN^{\ast}\right)  :\nabla_{e}\left(  f^{\ast}\varpi\right)
=f^{\ast}\left(  \nabla_{e}\varpi\right)  $
\end{theorem}

\subsubsection{Curvature}

\begin{definition}
The \textbf{Riemann} \textbf{curvature} of a covariant connection $\nabla$\ is
the multilinear map :%

\begin{equation}
R:\left(  \mathfrak{X}\left(  TM\right)  \right)  ^{3}\rightarrow
\mathfrak{X}\left(  M\right)  ::R(X,Y,Z)=\nabla_{X}\nabla_{Y}Z-\nabla
_{Y}\nabla_{X}Z-\nabla_{\left[  X,Y\right]  }Z
\end{equation}

\end{definition}

It is also called the Riemann tensor or curvature tensor. As there are many
objects called curvature we opt for Riemann curvature.

The name curvature comes from the following : for a vector field V :

$R(\partial_{\alpha},\partial_{\beta},V)=\nabla_{\partial_{\alpha}}%
\nabla_{\partial_{\beta}}V-\nabla_{\partial_{\beta}}\nabla_{\partial_{\alpha}%
}V-\nabla_{\left[  \partial_{\alpha},\partial_{\beta}\right]  }V=\left(
\nabla_{\partial_{\alpha}}\nabla_{\partial_{\beta}}-\nabla_{\partial_{\beta}%
}\nabla_{\partial_{\alpha}}\right)  V$ because $\left[  \partial_{\alpha
},\partial_{\beta}\right]  =0.$ So R is a measure of the obstruction of the
covariant derivative to be commutative : $\nabla_{\partial_{\alpha}}%
\nabla_{\partial_{\beta}}-\nabla_{\partial_{\beta}}\nabla_{\partial_{\alpha}%
}\neq0$

\begin{theorem}
The Riemann curvature is a tensor valued in the tangent bundle :
$R\in\mathfrak{X}\left(  \Lambda_{2}TM^{\ast}\overset{1}{\underset{1}{\otimes
}}TM\right)  $

$R=\sum_{\left\{  \gamma\eta\right\}  }\sum_{\alpha\beta}R_{\gamma\eta\beta
}^{\alpha}dx^{\gamma}\wedge dx^{\eta}\otimes dx^{\beta}\otimes\partial
x_{\alpha}$ with%

\begin{equation}
R_{\alpha\beta\gamma}^{\varepsilon}=\partial_{\alpha}\Gamma_{\beta\gamma
}^{\varepsilon}-\partial_{\beta}\Gamma_{\alpha\gamma}^{\varepsilon}%
+\Gamma_{\alpha\eta}^{\varepsilon}\Gamma_{\beta\gamma}^{\eta}-\Gamma
_{\beta\eta}^{\varepsilon}\Gamma_{\alpha\gamma}^{\eta}%
\end{equation}

\end{theorem}

\begin{proof}
$R(X,Y,Z)=\nabla_{X}\nabla_{Y}Z-\nabla_{Y}\nabla_{X}Z-\nabla_{\left[
X,Y\right]  }Z$

$=\nabla_{X}\left(  \left(  \partial_{\alpha}Z^{\varepsilon}+\Gamma
_{\alpha\gamma}^{\varepsilon}Z^{\gamma}\right)  Y^{\alpha}\partial
_{\varepsilon}\right)  -\nabla_{Y}\left(  \left(  \partial_{\alpha
}Z^{\varepsilon}+\Gamma_{\alpha\gamma}^{\varepsilon}Z^{\gamma}\right)
X^{\alpha}\partial_{\varepsilon}\right)  $

$-\left(  \partial_{\alpha}Z^{\varepsilon}+\Gamma_{\alpha\gamma}^{\varepsilon
}Z^{\gamma}\right)  \left(  (X^{\eta}\partial_{\eta}Y^{\alpha}-Y^{\eta
}\partial_{\eta}X^{\alpha})\right)  $

$=\left(  \partial_{\beta}\left(  \left(  \partial_{\alpha}Z^{\varepsilon
}+\Gamma_{\alpha\gamma}^{\varepsilon}Z^{\gamma}\right)  Y^{\alpha}\right)
+\Gamma_{\beta\eta}^{\varepsilon}\left(  \left(  \partial_{\alpha}Z^{\eta
}+\Gamma_{\alpha\gamma}^{\eta}Z^{\gamma}\right)  Y^{\alpha}\right)  \right)
X^{\beta}\partial_{\varepsilon}$

$-\left(  \partial_{\beta}\left(  \left(  \partial_{\alpha}Z^{\varepsilon
}+\Gamma_{\alpha\gamma}^{\varepsilon}Z^{\gamma}\right)  X^{\alpha}\right)
-\Gamma_{\beta\eta}^{\varepsilon}\nabla_{Y}\left(  \left(  \partial_{\alpha
}Z^{\eta}+\Gamma_{\alpha\gamma}^{\eta}Z^{\gamma}\right)  X^{\alpha}\right)
\right)  Y^{\beta}\partial_{\varepsilon}$

$-\left(  \left(  \partial_{\alpha}Z^{\varepsilon}\right)  \left(  (X^{\eta
}\partial_{\eta}Y^{\alpha}-Y^{\eta}\partial_{\eta}X^{\alpha})\right)
+\Gamma_{\alpha\gamma}^{\varepsilon}Z^{\gamma}\left(  (X^{\eta}\partial_{\eta
}Y^{\alpha}-Y^{\eta}\partial_{\eta}X^{\alpha})\right)  \right)  \partial
_{\varepsilon}$

The component of $\partial_{\varepsilon}$ is:

$=\left(  \partial_{\beta}\left(  \partial_{\alpha}Z^{\varepsilon}%
+\Gamma_{\alpha\gamma}^{\varepsilon}Z^{\gamma}\right)  \right)  X^{\beta
}Y^{\alpha}+\left(  \partial_{\alpha}Z^{\varepsilon}+\Gamma_{\alpha\gamma
}^{\varepsilon}Z^{\gamma}\right)  X^{\beta}\partial_{\beta}Y^{\alpha}%
+\Gamma_{\beta\eta}^{\varepsilon}\left(  \partial_{\alpha}Z^{\eta}\right)
Y^{\alpha}X^{\beta}+\Gamma_{\beta\eta}^{\varepsilon}\Gamma_{\alpha\gamma
}^{\eta}Z^{\gamma}X^{\beta}Y^{\alpha}$

$-\left(  \partial_{\beta}\left(  \partial_{\alpha}Z^{\varepsilon}%
+\Gamma_{\alpha\gamma}^{\varepsilon}Z^{\gamma}\right)  \right)  X^{\alpha
}Y^{\beta}-\left(  \partial_{\alpha}Z^{\varepsilon}+\Gamma_{\alpha\gamma
}^{\varepsilon}Z^{\gamma}\right)  Y^{\beta}\partial_{\beta}X^{\alpha}%
-\Gamma_{\beta\eta}^{\varepsilon}\partial_{\alpha}Z^{\eta}X^{\alpha}Y^{\beta
}-\Gamma_{\beta\eta}^{\varepsilon}\Gamma_{\alpha\gamma}^{\eta}Z^{\gamma
}X^{\alpha}Y^{\beta}$

$-\left(  \partial_{\alpha}Z^{\varepsilon}\right)  X^{\eta}\left(
\partial_{\eta}Y^{\alpha}\right)  +\left(  \partial_{\alpha}Z^{\varepsilon
}\right)  Y^{\eta}\left(  \partial_{\eta}X^{\alpha}\right)  -\Gamma
_{\alpha\gamma}^{\varepsilon}Z^{\gamma}X^{\eta}\left(  \partial_{\eta
}Y^{\alpha}\right)  +\Gamma_{\alpha\gamma}^{\varepsilon}Z^{\gamma}Y^{\eta
}\left(  \partial_{\eta}X^{\alpha}\right)  $

$=\left(  \partial_{\beta}\partial_{\alpha}Z^{\varepsilon}\right)  X^{\beta
}Y^{\alpha}+\left(  \partial_{\beta}\Gamma_{\alpha\gamma}^{\varepsilon
}\right)  X^{\beta}Z^{\gamma}Y^{\alpha}+\Gamma_{\alpha\gamma}^{\varepsilon
}\left(  \partial_{\beta}Z^{\gamma}\right)  X^{\beta}Y^{\alpha}+\left(
\partial_{\alpha}Z^{\varepsilon}\right)  \left(  \partial_{\beta}Y^{\alpha
}\right)  X^{\beta}$

$+\Gamma_{\alpha\gamma}^{\varepsilon}Z^{\gamma}\left(  \partial_{\beta
}Y^{\alpha}\right)  X^{\beta}+\Gamma_{\beta\eta}^{\varepsilon}\left(
\partial_{\alpha}Z^{\eta}\right)  Y^{\alpha}X^{\beta}+\Gamma_{\beta\eta
}^{\varepsilon}\Gamma_{\alpha\gamma}^{\eta}Z^{\gamma}Y^{\alpha}X^{\beta}$

$-\left(  \partial_{\beta}\partial_{\alpha}Z^{\varepsilon}\right)  X^{\alpha
}Y^{\beta}-\left(  \partial_{\beta}\Gamma_{\alpha\gamma}^{\varepsilon}\right)
Z^{\gamma}X^{\alpha}Y^{\beta}-\Gamma_{\alpha\gamma}^{\varepsilon}\left(
\partial_{\beta}Z^{\gamma}\right)  X^{\alpha}Y^{\beta}-\left(  \partial
_{\alpha}Z^{\varepsilon}\right)  \left(  \partial_{\beta}X^{\alpha}\right)
Y^{\beta}$

$+\Gamma_{\alpha\gamma}^{\varepsilon}Z^{\gamma}\left(  \partial_{\beta
}X^{\alpha}\right)  Y^{\beta}-\Gamma_{\beta\eta}^{\varepsilon}\left(
\partial_{\alpha}Z^{\eta}\right)  X^{\alpha}Y^{\beta}-\Gamma_{\beta\eta
}^{\varepsilon}\Gamma_{\alpha\gamma}^{\eta}Z^{\gamma}X^{\alpha}Y^{\beta}$

$-\left(  \partial_{\alpha}Z^{\varepsilon}\right)  X^{\eta}\partial_{\eta
}Y^{\alpha}+\left(  \partial_{\alpha}Z^{\varepsilon}\right)  Y^{\eta}\left(
\partial_{\eta}X^{\alpha}\right)  -\Gamma_{\alpha\gamma}^{\varepsilon
}Z^{\gamma}X^{\eta}\left(  \partial_{\eta}Y^{\alpha}\right)  +\Gamma
_{\alpha\gamma}^{\varepsilon}Z^{\gamma}Y^{\eta}\left(  \partial_{\eta
}X^{\alpha}\right)  $

$=\left(  \partial_{\beta}\partial_{\alpha}Z^{\varepsilon}\right)  X^{\beta
}Y^{\alpha}-\left(  \partial_{\beta}\partial_{\alpha}Z^{\varepsilon}\right)
X^{\alpha}Y^{\beta}$

$+\left(  \partial_{\alpha}Z^{\varepsilon}\right)  \left(  \partial_{\beta
}Y^{\alpha}\right)  X^{\beta}-\left(  \partial_{\alpha}Z^{\varepsilon}\right)
\left(  \partial_{\beta}X^{\alpha}\right)  Y^{\beta}-\left(  \partial_{\alpha
}Z^{\varepsilon}\right)  X^{\eta}\left(  \partial_{\eta}Y^{\alpha}\right)
+\left(  \partial_{\alpha}Z^{\varepsilon}\right)  Y^{\eta}\left(
\partial_{\eta}X^{\alpha}\right)  $

$+\Gamma_{\alpha\gamma}^{\varepsilon}\left(  \partial_{\beta}Z^{\gamma
}\right)  X^{\beta}Y^{\alpha}+\Gamma_{\beta\eta}^{\varepsilon}\left(
\partial_{\alpha}Z^{\eta}\right)  Y^{\alpha}X^{\beta}-\Gamma_{\beta\eta
}^{\varepsilon}\left(  \partial_{\alpha}Z^{\eta}\right)  X^{\alpha}Y^{\beta
}-\Gamma_{\alpha\gamma}^{\varepsilon}\left(  \partial_{\beta}Z^{\gamma
}\right)  X^{\alpha}Y^{\beta}$

$+\Gamma_{\alpha\gamma}^{\varepsilon}Z^{\gamma}\left(  \partial_{\beta
}Y^{\alpha}\right)  X^{\beta}-\Gamma_{\alpha\gamma}^{\varepsilon}Z^{\gamma
}X^{\eta}\left(  \partial_{\eta}Y^{\alpha}\right)  +\Gamma_{\alpha\gamma
}^{\varepsilon}Z^{\gamma}\left(  \partial_{\beta}X^{\alpha}\right)  Y^{\beta
}+\Gamma_{\alpha\gamma}^{\varepsilon}Z^{\gamma}Y^{\eta}\left(  \partial_{\eta
}X^{\alpha}\right)  $

$+\left(  \partial_{\beta}\Gamma_{\alpha\gamma}^{\varepsilon}\right)
X^{\beta}Z^{\gamma}Y^{\alpha}+\Gamma_{\beta\eta}^{\varepsilon}\Gamma
_{\alpha\gamma}^{\eta}Z^{\gamma}Y^{\alpha}X^{\beta}-\left(  \partial_{\beta
}\Gamma_{\alpha\gamma}^{\varepsilon}\right)  Z^{\gamma}X^{\alpha}Y^{\beta
}-\Gamma_{\beta\eta}^{\varepsilon}\Gamma_{\alpha\gamma}^{\eta}Z^{\gamma
}X^{\alpha}Y^{\beta}$

$=\left(  \partial_{\alpha}\partial_{\beta}Z^{\varepsilon}\right)  X^{\alpha
}Y^{\beta}-\left(  \partial_{\beta}\partial_{\alpha}Z^{\varepsilon}\right)
X^{\alpha}Y^{\beta}$

$+\left(  \partial_{\alpha}Z^{\varepsilon}\right)  \left(  \left(
\partial_{\beta}Y^{\alpha}\right)  X^{\beta}-X^{\beta}\left(  \partial_{\beta
}Y^{\alpha}\right)  +Y^{\beta}\left(  \partial_{\beta}X^{\alpha}\right)
-\left(  \partial_{\beta}X^{\alpha}\right)  Y^{\beta}\right)  $

$+\Gamma_{\beta\eta}^{\varepsilon}X^{\alpha}Y^{\beta}\left(  \partial_{\alpha
}Z^{\eta}\right)  -\Gamma_{\beta\eta}^{\varepsilon}X^{\alpha}Y^{\beta}\left(
\partial_{\alpha}Z^{\eta}\right)  +\Gamma_{\alpha\eta}^{\varepsilon}X^{\alpha
}Y^{\beta}\left(  \partial_{\beta}Z^{\eta}\right)  -\Gamma_{\alpha\eta
}^{\varepsilon}X^{\alpha}Y^{\beta}\left(  \partial_{\beta}Z^{\eta}\right)  $

$+\Gamma_{\alpha\gamma}^{\varepsilon}Z^{\gamma}\left(  \left(  \partial
_{\beta}Y^{\alpha}\right)  X^{\beta}-X^{\beta}\left(  \partial_{\beta
}Y^{\alpha}\right)  \right)  +\Gamma_{\alpha\gamma}^{\varepsilon}Z^{\gamma
}\left(  \left(  \partial_{\beta}X^{\alpha}\right)  Y^{\beta}+Y^{\beta}\left(
\partial_{\beta}X^{\alpha}\right)  \right)  $

$+\left(  \partial_{\beta}\Gamma_{\alpha\gamma}^{\varepsilon}\right)
X^{\beta}Z^{\gamma}Y^{\alpha}-\left(  \partial_{\beta}\Gamma_{\alpha\gamma
}^{\varepsilon}\right)  Z^{\gamma}X^{\alpha}Y^{\beta}+\Gamma_{\beta\eta
}^{\varepsilon}\Gamma_{\alpha\gamma}^{\eta}Z^{\gamma}Y^{\alpha}X^{\beta
}-\Gamma_{\beta\eta}^{\varepsilon}\Gamma_{\alpha\gamma}^{\eta}Z^{\gamma
}X^{\alpha}Y^{\beta}$

$=\left(  \partial_{\beta}\Gamma_{\alpha\gamma}^{\varepsilon}\right)
X^{\beta}Y^{\alpha}Z^{\gamma}-\left(  \partial_{\beta}\Gamma_{\alpha\gamma
}^{\varepsilon}\right)  X^{\alpha}Y^{\beta}Z^{\gamma}+\Gamma_{\beta\eta
}^{\varepsilon}\Gamma_{\alpha\gamma}^{\eta}X^{\beta}Y^{\alpha}Z^{\gamma
}-\Gamma_{\beta\eta}^{\varepsilon}\Gamma_{\alpha\gamma}^{\eta}X^{\alpha
}Y^{\beta}Z^{\gamma}$

$R(X,Y,Z)$

$=\left(  \left(  \partial_{\alpha}\Gamma_{\beta\gamma}^{\varepsilon}\right)
X^{\alpha}Y^{\beta}Z^{\gamma}-\left(  \partial_{\beta}\Gamma_{\alpha\gamma
}^{\varepsilon}\right)  X^{\alpha}Y^{\beta}Z^{\gamma}+\Gamma_{\alpha\eta
}^{\varepsilon}\Gamma_{\beta\gamma}^{\eta}X^{\alpha}Y^{\beta}Z^{\gamma}%
-\Gamma_{\beta\eta}^{\varepsilon}\Gamma_{\alpha\gamma}^{\eta}X^{\alpha
}Y^{\beta}Z^{\gamma}\right)  \partial_{\varepsilon}$

$R(X,Y,Z)=R_{\alpha\beta\gamma}^{\varepsilon}X^{\alpha}Y^{\beta}Z^{\gamma
}\partial_{\varepsilon}$

With : $R_{\alpha\beta\gamma}^{\varepsilon}=\partial_{\alpha}\Gamma
_{\beta\gamma}^{\varepsilon}-\partial_{\beta}\Gamma_{\alpha\gamma
}^{\varepsilon}+\Gamma_{\alpha\eta}^{\varepsilon}\Gamma_{\beta\gamma}^{\eta
}-\Gamma_{\beta\eta}^{\varepsilon}\Gamma_{\alpha\gamma}^{\eta}$

Clearly : $R_{\alpha\beta\gamma}^{\varepsilon}=-R_{\beta\alpha\gamma
}^{\varepsilon}$ so : $R=\sum_{\left\{  \alpha\beta\right\}  \gamma
\varepsilon}R_{\alpha\beta\gamma}^{\varepsilon}dx^{\alpha}\wedge dx^{\beta
}\otimes dx^{\gamma}\otimes\partial_{\varepsilon}$
\end{proof}

\begin{theorem}
For any covariant derivative $\nabla,$ its exterior covariant derivative
$\nabla_{e},$ and its Riemann curvature:%

\begin{equation}
\forall\varpi\in\Lambda_{r}\left(  M;TM\right)  :\nabla_{e}\left(  \nabla
_{e}\varpi\right)  =R\wedge\varpi
\end{equation}

\end{theorem}

More precisely in a holonomic basis :

$\nabla_{e}\left(  \nabla_{e}\varpi\right)  =\sum_{\alpha\beta}\left(
\sum_{\left\{  \gamma\eta\right\}  }R_{\gamma\eta\beta}^{\alpha}dx^{\gamma
}\wedge dx^{\eta}\right)  \wedge\varpi^{\beta}\otimes\partial x_{\alpha}$

\begin{proof}
$\nabla_{e}\varpi=\sum_{\alpha}\left(  d\varpi^{\alpha}+\left(  \sum_{\beta
}\left(  \sum_{\alpha}\Gamma_{\gamma\beta}^{\alpha}dx^{\gamma}\right)
\wedge\varpi^{\beta}\right)  \right)  \partial x_{\alpha}$

$=\sum_{\alpha}\left(  d\varpi^{\alpha}+\sum_{\beta}\Omega_{\beta}^{\alpha
}\wedge\varpi^{\beta}\right)  \otimes\partial x_{\alpha}$

with $\Omega_{\beta}^{\alpha}=\sum_{\gamma}\Gamma_{\gamma\beta}^{\alpha
}dx^{\gamma}$

$\nabla_{e}\left(  \nabla_{e}\varpi\right)  =\sum_{\alpha}\left(  d\left(
\nabla_{e}\varpi\right)  ^{\alpha}+\sum_{\beta}\Omega_{\beta}^{\alpha}%
\wedge\left(  \nabla_{e}\varpi\right)  ^{\beta}\right)  \otimes\partial
x_{\alpha}$

$=\sum_{\alpha}\left(  d\left(  d\varpi^{\alpha}+\sum_{\beta}\Omega_{\beta
}^{\alpha}\wedge\varpi^{\beta}\right)  +\sum_{\beta}\Omega_{\beta}^{\alpha
}\wedge\left(  d\varpi^{\beta}+\sum_{\gamma}\Omega_{\gamma}^{\beta}%
\wedge\varpi^{\gamma}\right)  \right)  \otimes\partial x_{\alpha}$

$=\sum_{\alpha\beta}\left(  d\Omega_{\beta}^{\alpha}\wedge\varpi^{\beta
}-\Omega_{\beta}^{\alpha}\wedge d\varpi^{\beta}+\Omega_{\beta}^{\alpha}\wedge
d\varpi^{\beta}+\Omega_{\beta}^{\alpha}\wedge\sum_{\gamma}\Omega_{\gamma
}^{\beta}\wedge\varpi^{\gamma}\right)  \otimes\partial x_{\alpha}$

$=\sum_{\alpha\beta}\left(  d\Omega_{\beta}^{\alpha}\wedge\varpi^{\beta}%
+\sum_{\gamma}\Omega_{\gamma}^{\alpha}\wedge\Omega_{\beta}^{\gamma}%
\wedge\varpi^{\beta}\right)  \otimes\partial x_{\alpha}$

$\nabla_{e}\left(  \nabla_{e}\varpi\right)  =\sum_{\alpha\beta}\left(
d\Omega_{\beta}^{\alpha}+\sum_{\gamma}\Omega_{\gamma}^{\alpha}\wedge
\Omega_{\beta}^{\gamma}\right)  \wedge\varpi^{\beta}\otimes\partial x_{\alpha
}$

$d\Omega_{\beta}^{\alpha}+\sum_{\gamma}\Omega_{\gamma}^{\alpha}\wedge
\Omega_{\beta}^{\gamma}=d\left(  \sum_{\eta}\Gamma_{\eta\beta}^{\alpha
}dx^{\eta}\right)  +\sum_{\gamma}\left(  \sum_{\varepsilon}\Gamma
_{\varepsilon\gamma}^{\alpha}dx^{\varepsilon}\right)  \wedge\left(  \sum
_{\eta}\Gamma_{\eta\beta}^{\gamma}dx^{\eta}\right)  $

$=\sum_{\eta\gamma}\partial_{\gamma}\Gamma_{\eta\beta}^{\alpha}dx^{\gamma
}\wedge dx^{\eta}+\sum_{\eta\varepsilon\gamma}\Gamma_{\varepsilon\gamma
}^{\alpha}\Gamma_{\eta\beta}^{\gamma}dx^{\varepsilon}\wedge dx^{\eta}$

$=\sum_{\eta\gamma}\left(  \partial_{\gamma}\Gamma_{\eta\beta}^{\alpha}%
+\sum_{\varepsilon}\Gamma_{\gamma\varepsilon}^{\beta}\Gamma_{\eta\beta
}^{\gamma}\right)  dx^{\gamma}\wedge dx^{\eta}$
\end{proof}

\begin{definition}
The \textbf{Ricci tensor }is the contraction of R with respect to the indexes
$\varepsilon,\beta:$%

\begin{equation}
Ric=\sum_{\alpha\gamma}Ric_{\alpha\gamma}dx^{\alpha}\otimes dx^{\gamma}\text{
with }Ric_{\alpha\gamma}=\sum_{\beta}R_{\alpha\beta\gamma}^{\beta}%
\end{equation}

\end{definition}

It is a symmetric tensor if R comes from the Levi-Civita connection.

Remarks :

i) The curvature tensor can be defined for any covariant derivative : there is
no need of a Riemannian metric or a symmetric connection.

ii) The formula above is written in many ways in the litterature, depending on
the convention used to write $\Gamma_{\beta\gamma}^{\alpha}.$ This is why I
found useful to give the complete calculations.

iii) R is always antisymmetric in the indexes $\alpha,\beta$

\subsubsection{Torsion}

\begin{definition}
The \textbf{torsion} of an affine connection $\nabla$ is the map:%

\begin{equation}
T:\mathfrak{X}\left(  TM\right)  \times\mathfrak{X}\left(  TM\right)
\rightarrow\mathfrak{X}\left(  TM\right)  ::T(X,Y)=\nabla_{X}Y-\nabla
_{Y}X-\left[  X,Y\right]
\end{equation}

\end{definition}

It is a tensor field : $T=\sum_{\alpha,\beta,\gamma}T_{\alpha\beta}^{\gamma
}dx^{\alpha}\otimes dx^{\beta}\otimes\partial x_{\gamma}\in\mathfrak{X}\left(
\otimes_{2}^{1}TM\right)  $ with $\ T_{\alpha\beta}^{\gamma}=-T_{\alpha\beta
}^{\gamma}=\Gamma_{\alpha\beta}^{\gamma}-\Gamma_{\beta\alpha}^{\gamma}$ so
this is a 2 form valued in the tangent bundle : $T=\sum_{\left\{  \alpha
,\beta\right\}  }\sum_{\gamma}\left(  \Gamma_{\alpha\beta}^{\gamma}%
-\Gamma_{\beta\alpha}^{\gamma}\right)  dx^{\alpha}\Lambda dx^{\beta}%
\otimes\partial x_{\gamma}\in\Lambda_{2}\left(  M;TM\right)  $

\begin{definition}
An affine connection is \textbf{torsion free} if its torsion vanishes.\ 
\end{definition}

\begin{theorem}
An affine connection is torsion free iff the covariant derivative is
\textbf{symmetric} : $T=0\Leftrightarrow\Gamma_{\alpha\beta}^{\gamma}%
=\Gamma_{\beta\alpha}^{\gamma}$
\end{theorem}

\begin{theorem}
(Kobayashi I p.149) If the covariant connection $\nabla$ is torsion free then :

$\forall\varpi\in\mathfrak{X}\left(  \Lambda_{r}TM^{\ast}\right)
:d\varpi=\frac{1}{r!}\sum_{\sigma\in\mathfrak{S}\left(  r\right)  }%
\epsilon\left(  \sigma\right)  \nabla\varpi$
\end{theorem}

\begin{definition}
A covariant connection on a manifold whose curvature and torsion vanish is
said to be \textbf{flat} (or locally affine).
\end{definition}

\subsubsection{Parallel transport by a covariant connection}

\paragraph{Parallel transport of a tensor\newline}

\begin{definition}
A tensor field $T\in\mathfrak{X}\left(  \otimes_{s}^{r}TM\right)  $ on a
manifold is invariant by a covariant connection along a path $c:\left[
a,b\right]  \rightarrow M$ on M if its covariant derivative along the tangent,
evaluated at each point of the path, is null : $\nabla_{c^{\prime}(t)}T\left(
c\left(  t\right)  \right)  =0$
\end{definition}

\begin{definition}
The transported tensor $\widehat{T}$\ of a tensor field $T\in\mathfrak{X}%
\left(  \otimes_{s}^{r}TM\right)  $ along a path $c:\left[  a,b\right]
\rightarrow M$ on the manifold M is defined as a solution of the differential
equation : $\nabla_{c^{\prime}(t)}\widehat{T}\left(  c\left(  t\right)
\right)  =0$ with initial condition : $\widehat{T}\left(  c\left(  a\right)
\right)  =T\left(  c\left(  a\right)  \right)  $
\end{definition}

If T in a holonomic basis reads :

$T\left(  p\right)  $

$=\sum_{\alpha_{1}...\alpha_{r}}\sum_{\beta_{1}....\beta_{s}}\sum_{\gamma
}T\left(  p\right)  _{\beta_{1}...\beta_{s}}^{\alpha_{1}...\alpha_{r}}\partial
x_{\alpha_{1}}\otimes..\otimes\partial x_{\alpha_{r}}\otimes dx^{\gamma
}\otimes dx^{\beta_{1}}\otimes...\otimes dx^{\beta_{s}}$

$\nabla_{c^{\prime}(t)}\widehat{T}\left(  t\right)  =0\Leftrightarrow
\sum_{\gamma}\widehat{T}_{\gamma\beta_{1}...\beta_{s}}^{\alpha_{1}%
...\alpha_{r}}v^{\gamma}=0$ with c'(t)=$\sum_{\gamma}v^{\gamma}\partial
x_{\gamma}$

The tensor field $\widehat{T}$ is defined by the first order linear
differential equations :

$\sum_{\gamma}v^{\gamma}\partial_{\gamma}\widehat{T}_{\beta_{1}...\beta_{s}%
}^{\alpha_{1}...\alpha_{r}}$

$=-\sum_{k=1}^{r}v^{\gamma}\widehat{\Gamma}_{\gamma\eta}^{\alpha_{k}}%
T_{\beta_{1}...\beta_{s}}^{\alpha_{1}.\alpha_{k-1}\eta\alpha_{k+1}..\alpha
_{r}}+\sum_{k=1}^{s}v^{\gamma}\widehat{\Gamma}_{\gamma\beta_{k}}^{\eta
}T_{\beta_{1}..\beta_{k-1}\eta\beta_{k+1}..\beta_{s}}^{\alpha_{1}...\alpha
_{r}}$

$\widehat{T}\left(  c\left(  a\right)  \right)  =T\left(  c\left(  a\right)
\right)  $

where $\Gamma,$ c(t) and v are assumed to be known.

They define a map : $Pt_{c}:\left[  a,b\right]  \times\mathfrak{X}\left(
\otimes_{s}^{r}TM\right)  \rightarrow\mathfrak{X}\left(  \otimes_{s}%
^{r}TM\right)  $

If $S,T\in\mathfrak{X}\left(  \otimes_{s}^{r}TM\right)  ,k,k^{\prime}\in K$
then :

$Pt_{c}\left(  t,kS+k^{\prime}T\right)  =kPt_{c}\left(  t,S\right)
+k^{\prime}Pt_{c}\left(  t,T\right)  $

But the components of $\widehat{T}$ do not depend linearly of the components
of T$.$

The map : $Pt_{c}\left(  .,T\right)  :\left[  a,b\right]  \rightarrow
\mathfrak{X}\left(  \otimes_{s}^{r}TM\right)  ::Pt_{c}\left(  t,T\right)  $ is
a path in the tensor bundle.\ So it is common to say that one "lifts" the
curve c on M to a curve in the tensor bundle.

Given a vector field V, a point p in M, the set of vectors $u_{p}\in T_{p}M$
such that $\nabla_{V}u_{p}=0\Leftrightarrow\sum_{\alpha}\left(  \partial
_{\alpha}V^{\gamma}+\Gamma_{\alpha\beta}^{\gamma}V^{\beta}\right)
u_{p}^{\alpha}=0$ is a vector subspace of $T_{p}M$, called the horizontal
vector subspace at p (depending on V). So parallel transported vectors are
horizontal vectors.

Notice the difference with the transports previously studied :

i) transport by "push-forward" : it can be done everywhere, but the components
of the transported tensor depend linearly of the components of T$_{0}$

ii) transport by the Lie derivative : it is the transport by push forward with
the flow of a vector field, with similar constraints

\paragraph{Holonomy\newline}

If the path c is a loop : $c:\left[  a,b\right]  \rightarrow M::c(a)=c(b)=p$
the parallel transport goes back to the same tangent space at p. In the vector
space $T_{p}M,$ which is isomorphic to $K^{n},$ the parallel transport for a
given loop is a linear map on $T_{p}M,$ which has an inverse (take the
opposite loop with the reversed path) and the set of all such linear maps at p
has a group structure : this is the \textbf{holonomy group }$H\left(
M,p\right)  $ at p $.$If the loops are restricted to loops which are homotopic
to a point this is the restricted holonomy group $H_{0}\left(  M,p\right)  .$
The holonomy group is a finite dimensional Lie group (Kobayashi I p.72).

\paragraph{Geodesic\newline}

\subparagraph{1. Definitions:\newline}

\begin{definition}
A path $c\in C_{1}\left(  \left[  a,b\right]  ;M\right)  $ in a manifold M
endowed with a covariant derivative $\nabla$ describes a \textbf{geodesic}
c$\left(  \left[  a,b\right]  \right)  $ if the tangent to the curve c$\left(
\left[  a,b\right]  \right)  $\ is parallel transported.
\end{definition}

So c describes a geodesic if :

$\nabla_{c^{\prime}(t)}c^{\prime}(t)=0\Leftrightarrow\sum\left(
\frac{dV^{\beta}}{dt}+\Gamma_{\alpha\gamma}^{\beta}\left(  c\left(  t\right)
\right)  V^{\alpha}V^{\gamma}\right)  =0$ with $V\left(  t\right)  =c^{\prime
}(t)$

A curve C, that is a 1 dimensional submanifold in M, can be described by
different paths.

If C is a geodesic for some parameter t, then it is still a geodesic for a
parameter $\tau=h(t)$ iff $\tau=kt+k^{\prime}$ meaning iff h is an affine map.

Conversely a given curve C is a geodesic iff there is a parametrization $c\in
C_{1}\left(  \left[  a,b\right]  ;M\right)  $ such that $c\left(  \left[
a,b\right]  \right)  =C$ and for which $\nabla_{c^{\prime}(t)}c^{\prime}(t)=0$
. If it exists, such parametrization is called an \textbf{affine parameter}.
They are all linked to each other by an affine map.

If a geodesic is a class 1 path and the covariant derivative is smooth (the
coefficients $\Gamma$ are smooth maps), then c is smooth.

If we define $\widehat{\Gamma}_{\beta\gamma}^{\alpha}=k\Gamma_{\beta\gamma
}^{\alpha}+\left(  1-k\right)  \Gamma_{\gamma\beta}^{\alpha}$ with a fixed
scalar k, we still have a covariant derivative, which has the same geodesics.
In particular with k=1/2 this covariant derivative is torsion free.

\subparagraph{2. Fundamental theorem:\newline}

\begin{theorem}
(Kobayashi I p.139, 147) For any point p and any vector v in $T_{p}M$ of a
finite dimensional real manifold M endowed with a covariant connection, there
is a unique geodesic $c\in C_{1}\left(  J_{p};M\right)  $ such that
$c(0)=p,c^{\prime}(0)=v$ where $J_{p}$ is an open interval in $%
\mathbb{R}
,$ including 0, depending both of p and $v_{p}$.

For each p there is a neighborhood N(p) of $\left(  p,\overrightarrow
{0}\right)  \times0$ in TM$\times$R in which the \textbf{exponential map} :
$\exp:TM\times%
\mathbb{R}
\rightarrow M::\exp tv_{p}=c(t)$ is defined. The point c(t) is the point on
the geodesic located at the affine parameter t from p.This map is
differentiable and smooth if the covariant derivative is smooth. It is a
diffeomorphism from N(p) to a neighborhood n(p) of p in M.
\end{theorem}

Warning ! this map exp is not the flow of a vector field, even if its
construct is similar. $\frac{d}{dt}\left(  \exp tv_{p}\right)  |_{t=\theta}$
is the vector $v_{p}$ parallel transported along the geodesic.

\begin{theorem}
In a finite dimensional real manifold M endowed with a covariant connection,
if there is a geodesic passing through p$\neq$q in M, it is unique. \ A
geodesic is never a loop.
\end{theorem}

This is a direct consequence of the previous theorem.

\subparagraph{3. Normal coordinates:\newline}

\begin{definition}
In a m dimensional real manifold M endowed with a covariant connection, a
system of \textbf{normal coordinates} is a local chart defined in a
neighborhood n(p) of a point p, with m independant vectors $\left(
\varepsilon_{i}\right)  _{i=1}^{m}$ in $T_{p}M,$ by which to a point $q\in
n\left(  p\right)  $ is associated the coordinates $\left(  y_{1}%
,...,y_{m}\right)  $ such that : $q=\exp v$ with $v=\sum_{i=1}^{m}%
y^{i}\varepsilon_{i}.$
\end{definition}

In this coordinate system the geodesics are expressed as straigth lines :
$c_{i}(t)\simeq tv$ and the Christoffel coefficients are such that \textit{at
p} : $\forall i,j,k:\widehat{\Gamma}_{jk}^{i}\left(  p\right)  +\widehat
{\Gamma}_{kj}^{i}\left(  p\right)  =0$ so they vanish if the connection is
torsion free. Then the covariant derivative of any tensor coincides with the
derivative of its components.

\begin{theorem}
(Kobayashi I p.149) Any point p of a finite dimensional real manifold M
endowed with a covariant connection has a convex neighborhood n(p) : two
points in n(p) can be joigned by a geodesic which lies in n(p). So there is a
system of normal coordinates centered at any point.
\end{theorem}

n(p) is defined by a ball centered at p with radius given in a normal
coordinate system.

\paragraph{Affine transformation\newline}

\begin{definition}
A map $f\in C_{1}\left(  M;N\right)  $ between the manifolds M,N endowed with
the covariant derivatives $\nabla,\widehat{\nabla}$, is an \textbf{affine
transformation} if it maps a parallel transported vector along a curve c in M
into a parallel transported vector along the curve f(c) in N.
\end{definition}

\begin{theorem}
(Kobayashi I p.225) An affine transformation f between the manifolds M,N
endowed with the covariant derivatives $\nabla,\widehat{\nabla}$, and the
corresponding torsions and curvature tensors $T,\widehat{T},R,\widehat{R}$

i) maps geodesics into geodesics

ii) commutes with the exponential : $\exp t\left(  f^{\prime}(p)v_{p}\right)
=f(\exp tv_{p})$

iii) for $X,Y,Z\in\mathfrak{X}\left(  TM\right)  :$

$f^{\ast}\left(  \nabla_{X}Y\right)  =\widehat{\nabla}_{f^{\ast}X}f^{\ast}Y$

$f^{\ast}\left(  T\left(  X,Y\right)  \right)  =\widehat{T}\left(  f^{\ast
}X,f^{\ast}Y\right)  $

$f^{\ast}R\left(  X,Y,Z\right)  =\widehat{R}\left(  f^{\ast}X,f^{\ast
}Y,f^{\ast}Z\right)  $

iv) is smooth if the connections have smooth Christoffel symbols
\end{theorem}

\begin{definition}
A vector field V on a manifold M endowed with the covariant derivatives
$\nabla$ is an infinitesimal generator of affine transformations if

$f_{t}:M\rightarrow M::f_{t}\left(  p\right)  =\exp Vt\left(  p\right)  $ is
an affine transformation on M.
\end{definition}

$V\in\mathfrak{X}\left(  TM\right)  $ is an infinitesimal generator of affine
transformationsis on M iff :

$\forall X\in\mathfrak{X}\left(  TM\right)  :\nabla_{X}\left(  \pounds _{V}%
-\nabla_{V}\right)  =R\left(  V,X\right)  $

The set of affine transformations\ on a manifold M is a group.\ If M has a
finite number of connected components it is a Lie group with the open compact
topology. The set of vector fields which are infinitesimal generators of
affine transformations is a Lie subalgebra of $\mathfrak{X}\left(  TM\right)
$, with dimension at most m%
${{}^2}$%
+m. If its dimension is m%
${{}^2}$%
+m then the torsion and the riemann tensors vanish.

\paragraph{Jacobi field\newline}

\begin{definition}
Let a family of geodesics in a manifold M endowed with a covariant derivatives
$\nabla$ be defined by a smooth map : $C:\left[  0,1\right]  \times\left[
-a,+a\right]  \rightarrow M$ ,$a\in%
\mathbb{R}
$ such that$\ \forall s\in\left[  -a,+a\right]  :C\left(  .,s\right)
\rightarrow M$ is a geodesic on M. The \textbf{deviation vector} of the family
of geodesics is defined as : $J_{t}=\frac{\partial C}{\partial s}|_{s=0}\in
T_{C(t,0)}M$
\end{definition}

It measures the variation of the family of geodesics along a transversal
vector $J_{t}$

\begin{theorem}
(Kobayashi II p.63) The deviation vector J of a family of geodesics satisfies
the equation :

$\nabla_{v_{t}}^{2}J_{t}+\nabla_{v_{t}}\left(  T\left(  J_{t},v_{t}\right)
\right)  +R\left(  J_{t},v_{t},v_{t}\right)  =0$ with $v_{t}=\frac{\partial
C}{\partial t}|_{s=0}$

It is fully defined by the values of $J_{t},\nabla_{v_{t}}J_{t}$ at a point t.

Conversely a vector field $J\in\mathfrak{X}\left(  TM\right)  \ $is said to be
a \textbf{Jacobi field} if there is a geodesic c(t) in M such that :

$\forall t:\nabla_{v_{t}}^{2}J\left(  c\left(  t\right)  \right)
+\nabla_{v_{t}}\left(  T\left(  J\left(  c\left(  t\right)  \right)
,v_{t}\right)  \right)  +R\left(  J\left(  c\left(  t\right)  \right)
,v_{t},v_{t}\right)  =0$ with $v_{t}=\frac{dc}{\partial t}$

It is then the deviation vector for a family of geodesics built from c(t).

Jacobi fields are the infinitesimal generators of affine transformations.
\end{theorem}

\begin{definition}
Two points p,q on a geodesic are said to be conjugate if there is a Jacobi
field which vanishes both at p and q.
\end{definition}

\subsubsection{Submanifolds}

If M is a submanifold in N, a covariant derivative $\nabla$\ defined on N does
not necessarily induces a covariant derivative $\widehat{\nabla}$ on M :
indeed even if X,Y are in $\mathfrak{X}\left(  TM\right)  $, $\nabla_{X}Y$ is
not always in $\mathfrak{X}\left(  TM\right)  $.

\begin{definition}
A submanifold M of a manifold N endowed with a covariant derivatives $\nabla$
is \textbf{autoparallel} if for each curve in M, the parallel transport of a
vector $v_{p}\in T_{p}M$\ stays in M, or equivalently if

$\forall X,Y\in\mathfrak{X}\left(  TM\right)  ,\nabla_{X}Y\in\mathfrak{X}%
\left(  TM\right)  .$
\end{definition}

\begin{theorem}
(Kobayashi II p.54) If a submanifold M of a manifold N endowed with a
covariant derivatives $\nabla$ is \textbf{autoparallel} then $\nabla$ induces
a covariant derivative $\widehat{\nabla}$ on M and :$\forall X,Y\in
\mathfrak{X}\left(  TM\right)  :\nabla_{X}Y=\widehat{\nabla}_{X}Y.$

Moreover the curvature and the torsion are related by :

$\forall X,Y,Z\in\mathfrak{X}\left(  TM\right)  :R\left(  X,Y,Z\right)
=\widehat{R}(X,Y,Z),T\left(  X,Y\right)  =\widehat{T}(X,Y)$
\end{theorem}

M is said to be \textbf{totally geodesic} at p if $\forall v_{p}\in T_{p}M$
the geodesic of N defined by (p,$v_{p})$ lies in M for small values of the
parameter t.\ A submanifold is totally geodesic if it is totally geodesic at
each of its point.

An autoparallel submanifold is totally geodesic. But the converse is true only
if the covariant derivative on N is torsion free.

\newpage

\section{INTEGRAL}

Orientation of a manifold and therefore integral are meaningful only for
finite dimensional manifolds.\ So in this subsection we will limit ourselves
to this case.

\subsection{Orientation of a manifold}

\label{Orientation of a manifold}

\subsubsection{Orientation function}

\begin{definition}
Let M be a class 1 finite dimensional manifold with atlas $\left(  E,\left(
O_{i},\varphi_{i}\right)  _{i\in I}\right)  ,$ where an orientation has been
chosen on E. An \textbf{orientation function }is the\textbf{\ }map :
$\theta_{i}:O_{i}\rightarrow\left\{  +1,-1\right\}  $ with $\theta\left(
p\right)  =+1$ if the holonomic basis defined by $\varphi_{i}$ at p has the
same orientation as the basis of E and $\theta\left(  p\right)  =-1$ if not.
\end{definition}

If there is an atlas of M such that it is possible to define a continuous
orientation function over M then it is possible to define continuously an
orientation in the tangent bundle.

This leads to the definition :

\subsubsection{Orientable manifolds}

\begin{definition}
A manifold M is \textbf{orientable} if there is a continuous system of
orientation functions.\ It is then oriented if an orientation function has
been chosen.
\end{definition}

\begin{theorem}
A class 1 finite dimensional real manifold M is \textbf{orientable} iff there
is an atlas atlas $\left(  E,\left(  O_{i},\varphi_{i}\right)  _{i\in
I}\right)  $ such that $\forall i,j\in I:\det\left(  \varphi_{j}\circ
\varphi_{i}^{-1}\right)  ^{\prime}>0$
\end{theorem}

\begin{proof}
We endow the set $\Theta=\left\{  +1,-1\right\}  $ with the discrete topology
: $\left\{  +1\right\}  $ and $\left\{  -1\right\}  $ are both open and closed
subsets, so we can define continuity for $\theta_{i}.$ If $\theta_{i}$ is
continuous on $O_{i}$ then the subsets $\theta_{i}^{-1}\left(  +1\right)
=O_{i}^{+},\theta_{i}^{-1}\left(  -1\right)  =O_{i}^{-}$ are both open and
closed in $O_{i}$\ .\ If $O_{i}$ is connected then we have either $O_{i}%
^{+}=O_{i},$ or $O_{i}^{-}=O_{i}.$ More generally $\theta_{i}$ has the same
value over each of the connected components of $O_{i}.$

Let be another chart j such that $p\in O_{i}\cap O_{j}.$ We have now two maps
: $\theta_{k}:O_{k}\rightarrow\left\{  +1,-1\right\}  $ for k=i,j. We go from
one holonomic basis to the other by the transition map :

$e_{\alpha}=\varphi_{i}^{\prime}\left(  p\right)  \partial x_{\alpha}%
=\varphi_{j}^{\prime}\left(  p\right)  \partial y_{\alpha}\Rightarrow\partial
y_{\alpha}=\varphi_{j}^{\prime}\left(  p\right)  ^{-1}\circ\varphi_{i}%
^{\prime}\left(  p\right)  \partial x_{\alpha}$

The bases $\partial x_{\alpha},\partial y_{\alpha}$ have the same orientation
iff $\det\varphi_{j}^{\prime}\left(  p\right)  ^{-1}\circ\varphi_{i}^{\prime
}\left(  p\right)  >0.$ As the maps are class 1 diffeomorphisms, the
determinant does not vanish and thus keep a constant sign in the neighborhood
of p. So in the neighborhood of each point p the functions $\theta_{i}%
,\theta_{j}$ will keep the same value (which can be different), and so all
over the connected components of $O_{i},O_{j}.$
\end{proof}

There are manifolds which are not orientable.\ The most well known examples
are the M\"{o}bius strip and the Klein bottle.

If M is disconnected it can be orientable but the orientation is distinct on
each connected component.

By convention a set of disconnected points $M=\cup_{i\in M}\left\{
p_{i}\right\}  $ is a 0 dimensional orientable manifold and its orientation is
given by a function $\theta\left(  p_{i}\right)  =\pm1.$

\begin{theorem}
A finite dimensional complex manifold is orientable
\end{theorem}

\begin{proof}
At any point p there is a canonical orientation of the tangent space, which
does not depend of the choice of a real basis or a chart.
\end{proof}

\begin{theorem}
An open subset of an orientable manifold is orientable.
\end{theorem}

\begin{proof}
Its atlas is a restriction of the atlas of the manifold.
\end{proof}

An open subset of $%
\mathbb{R}
^{m}$ is an orientable m dimensional manifold.

A curve on a manifold M defined by a path : $c:J\rightarrow M::c(t)$ is a
submanifold if c'(t) is never zero. Then it is orientable (take as direct
basis the vectors such that c'(t)u%
$>$%
0).

If $\left(  V_{i}\right)  _{i=1}^{m}$are m linearly independant continuous
vector fields over M then the orientation of the basis given by them is
continuous in a neighborhood of each point. But it does not usually defines an
orientation on M, because if M is not parallelizable there is not such vector fields.

A diffeomorphism $f:M\rightarrow N$ between two finite dimensional real
manifolds preserves (resp.reverses) the orientation if in two atlas:
$\det\left(  \psi_{j}\circ f\circ\varphi_{i}^{-1}\right)  ^{\prime}>0$ (resp.%
$<$%
0). As $\det\left(  \psi_{j}\circ f\circ\varphi_{i}^{-1}\right)  ^{\prime}$ is
never zero and continuous it has a constant sign : If two manifolds M,N are
diffeomorphic, if M is orientable then N is orientable. Notice that M,N must
have the same dimension.

\subsubsection{Volume form}

\begin{definition}
A \textbf{volume form} on a m dimensional manifold M is a m-form $\varpi
\in\mathfrak{X}\left(  \Lambda_{m}TM^{\ast}\right)  $ which is never zero on M.
\end{definition}

Any m form $\mu$ on M can then be written $\mu=f\varpi$ with $f\in C\left(  M;%
\mathbb{R}
\right)  .$

Warning ! the symbol "$dx^{1}\wedge...\wedge dx^{m}"$ is not a volume form,
except if M is an open of $%
\mathbb{R}
^{m}$. Indeed it is the coordinate expression of a m form in some chart
$\varphi_{i}$ :$\varpi_{i}\left(  p\right)  =1$ $\forall p\in O_{i}.$ At a
transition $p\in O_{i}\cap O_{j}$ we have, for the same form : $\varpi
_{j}=\det\left[  J^{-1}\right]  \neq0$ so we still have a volume form, but it
is defined only on the part of $O_{j}$ which intersects $O_{i}.$ We cannot say
anything outside $O_{i}.$ And of course put $\varpi_{j}\left(  q\right)  =1$
would not define the same form. More generally $f\left(  p\right)
dx^{1}\wedge...\wedge dx^{m}$ were f is a function on M, meaning that its
value is the same in any chart, does not define a volume form, not even a m
form. In a pseudo-riemannian manifold the volume form is $\sqrt{\left\vert
\det g\right\vert }dx^{1}\wedge...\wedge dx^{m}$ where the value of
$\left\vert \det g\right\vert $ is well defined but depends of the charts and
changes according to the usual rules in a change of basis.

\begin{theorem}
(Lafontaine p.201) A class 1 m dimensional manifold M which is the union of
countably many compact sets is orientable iff there is a volume form.
\end{theorem}

As a consequence a m dimensional submanifold of M is itself orientable (take
the restriction of the volume form). It is not true for a n%
$<$%
m submanifold.

A riemannian, pseudo-riemannian or symplectic manifold has such a form, thus
is orientable if it is the union of countably many compact sets.

\subsubsection{Orientation of an hypersurface}

\begin{definition}
Let M be a hypersurface of a class 1 n dimensional manifold N. A vector
$u_{p}\in T_{p}N,p\in M$ is \textbf{transversal} if $u_{p}\notin T_{p}M$
\end{definition}

At any point we can have a basis comprised of $\left(  u_{p},\varepsilon
_{2},...\varepsilon_{n}\right)  $ where $\left(  \varepsilon_{\beta}\right)
_{\beta=2}^{n}$ is a local basis of $T_{p}M$ . Thus we can define a
transversal orientation function by the orientation of this basis : say that
$\theta\left(  u_{p}\right)  =+1$ if $\left(  u_{p},\varepsilon_{2}%
,...\varepsilon_{n}\right)  $ is direct and $\theta\left(  u_{p}\right)  =-1$
if not.

M is transversally orientable if there is a continuous map $\theta.$

\begin{theorem}
The boundary of a manifold with boundary is transversally orientable
\end{theorem}

See manifold with boundary.\ It does not require N to be orientable.

\begin{theorem}
A manifold M with boundary $\partial M$ in an orientable class 1 manifold N is orientable.
\end{theorem}

\begin{proof}
The interior of M is an open subset of N, so is orientable. There is an
outward going vector field n on $\partial M$ , so we can define a direct basis
$\left(  e_{\alpha}\right)  $ on $\partial M$\ as a basis such that $\left(
n,e_{1},...,e_{m-1}\right)  $ is direct in N and $\partial M$ is an orientable manifold
\end{proof}

\bigskip

\subsection{Integral}

\label{Integral on a manifold}

In the Analysis part measures and integral are defined on any set .\ A m
dimensional real manifold M is locally homeomorphic to $%
\mathbb{R}
^{m},$\ thus it implies some constraints on the Borel measures on M, whose
absolutely continuous part must be related to the Lebesgue measure. Conversely
any m form on a m dimensional manifold defines an absolutely continuous
measure, called a Lebesgue measure on the manifold, and we can define the
integral of a m form.

\subsubsection{Definitions}

\paragraph{Principle\newline}

1. Let M be a Hausdorff, m dimensional real manifold with atlas

$\left(
\mathbb{R}
^{m},\left(  O_{i},\varphi_{i}\right)  _{i\in I}\right)  ,U_{i}=\varphi
_{i}\left(  O_{i}\right)  $ and $\mu$ a positive, locally finite Borel measure
on M. It is also a Radon measure.

i) On $%
\mathbb{R}
^{m}$\ is defined the Lebesgue measure $d\xi$ which can be seen as the
tensorial product of the measures $d\xi^{k},k=1...m$ and reads : $d\xi
=d\xi^{1}\otimes...\otimes d\xi^{n}$ or more simply : $d\xi=d\xi^{1}..d\xi
^{m}$

ii) The charts define push forward positive Radon measures $\nu_{i}%
=\varphi_{i\ast}\mu$ on $U_{i}\subset%
\mathbb{R}
^{m}$ such that $\forall B\subset U_{i}:\ \varphi_{i\ast}\mu\left(  B\right)
=\mu\left(  \varphi_{i}^{-1}\left(  B\right)  \right)  $

Each of the measures $\nu_{i}\ $can be uniquely decomposed in a singular part
$\lambda_{i}$ and an absolute part $\widehat{\nu}_{i}$, which itself can be
written as the integral of some positive function $g_{i}\in C\left(  U_{i};%
\mathbb{R}
\right)  $ with respect to the Lebesgue measure on $%
\mathbb{R}
^{m}$

Thus for each chart there is a couple $\left(  g_{i},\lambda_{i}\right)  $
such that : $\nu_{i}=\varphi_{i\ast}\mu=\widehat{\nu}_{i}+\lambda_{i},$
$\widehat{\nu}_{i}=g_{i}\left(  \xi\right)  d\xi$

If a measurable subset A belongs to the intersection of the domains $O_{i}\cap
O_{j}$ and for any i,j :

$\varphi_{i\ast}\mu\left(  \varphi_{i}\left(  A\right)  \right)  =\mu\left(
A\right)  =\varphi_{j\ast}\mu\left(  \varphi_{j}\left(  A\right)  \right)  $

Thus there is a unique Radon measure $\nu$ on $U=\cup_{i}U_{i}\subset%
\mathbb{R}
^{m}$ such that : $\nu=\nu_{i}$ on each $U_{i}.$ $\nu$ can be seen as the push
forward on $%
\mathbb{R}
^{m}$ of the measure $\mu$ on M by the atlas. This measure can be decomposed
as above :

$\nu=\widehat{\nu}+\lambda,$ $\widehat{\nu}=g\left(  \xi\right)  d\xi$

iii) Conversely the pull back $\varphi_{i}^{\ast}\nu$ of $\nu$ by each chart
on each open $O_{i}$ gives a Radon measure $\mu_{i}$\ on $O_{i}$ and $\mu$ is
the unique Radon measure on M such that $\mu|_{O_{i}}=\varphi_{i}^{\ast}\nu$
on each $O_{i}.$

iv) Pull back and push forward are linear operators, they apply to the
singular and the absolutely continuous parts of the measures. So the
absolutely continuous part of $\mu$\ denoted $\widehat{\mu}$\ is the pull back
of the product of g with the Lebesgue measure :

$\widehat{\mu}|_{O_{i}}=\varphi_{i}^{\ast}\left(  \widehat{\nu}|_{U_{i}%
}\right)  =\varphi_{i}^{\ast}\widehat{\nu}_{i}=\varphi_{i}^{\ast}\left(
g_{i}\left(  \xi\right)  d\xi\right)  $

$\widehat{\nu}|_{U_{i}}=\varphi_{i\ast}\left(  \widehat{\mu}|_{O_{i}}\right)
=\varphi_{i\ast}\widehat{\mu}_{i}=g_{i}\left(  \xi\right)  d\xi$

2. On the intersections $U_{i}\cap U_{j}$ the maps : $\varphi_{ij}=\varphi
_{j}\circ\varphi_{i}^{-1}:U_{i}\rightarrow U_{j}$ are class r diffeomorphisms,
the push forward of $\nu_{i}=\varphi_{i\ast}\mu$ by $\varphi_{ij}$ is
:$\left(  \varphi_{ij}\right)  _{\ast}\varphi_{i\ast}\mu=\left(  \varphi
_{j}\circ\varphi_{i}^{-1}\right)  _{\ast}\varphi_{i\ast}\mu=\varphi_{j\ast}%
\mu$

$\widehat{\nu}_{j}=\varphi_{j\ast}\widehat{\mu}$ being the image of
$\widehat{\nu}_{i}=\varphi_{i\ast}\widehat{\mu}$ by the diffeomorphism
$\varphi_{ij}$ reads :

$\widehat{\nu}_{j}=\left(  \varphi_{ij}\right)  _{\ast}\widehat{\nu}%
_{i}=\left\vert \det\left[  \varphi_{ij}^{\prime}\right]  \right\vert
\widehat{\nu}_{i}$

which resumes to : $g_{j}=\left\vert \det\left[  \varphi_{ij}^{\prime}\right]
\right\vert g_{i}$

So, even if there is a function g such that $\nu$ is the Radon integral of g,
g itself is defined as a family $\left(  g_{i}\right)  _{i\in I}$\ of
functions changing according to the above formula through the open cover of M.

3. On the other hand a m form on M reads $\varpi=\varpi\left(  p\right)
dx^{1}\wedge dx^{2}...\wedge dx^{m}$ in the holonomic basis. Its components
are a family $\left(  \varpi_{i}\right)  _{i\in I}$ of functions $\varpi
_{i}:O_{i}\rightarrow%
\mathbb{R}
$ such that : $\varpi_{j}=\det\left[  \varphi_{ij}^{\prime}\right]
^{-1}\varpi_{i}$ on the intersection $O_{i}\cap O_{j}.$

The push forward of $\varpi$ by a chart gives a m form on $%
\mathbb{R}
^{m}$\ :

$\left(  \varphi_{i\ast}\varpi_{i}\right)  \left(  \xi\right)  =\varpi
_{i}\left(  \varphi_{i}^{-1}\left(  \xi\right)  \right)  e^{1}\wedge...\Lambda
e^{m}$ in the corresponding basis $\left(  e^{k}\right)  _{k=1}^{m} $ of
$\left(
\mathbb{R}
^{m}\right)  ^{\ast}$

and on $O_{i}\cap O_{j}:$

$\left(  \varphi_{j\ast}\varpi_{j}\right)  =\left(  \varphi_{ij}\right)
_{\ast}\varphi_{i\ast}\varpi_{i}=\det\left[  \varphi_{ij}^{\prime}\right]
^{-1}\varpi_{i}\left(  \varphi_{i}^{-1}\left(  \xi\right)  \right)
e^{1}\wedge...\Lambda e^{m}$

So the rules for the transformations of the component of a m-form, and the
functions $g_{i}$ are similar (but not identical). Which leads to the
following definitions.

\paragraph{Integral of a m form on a manifold\newline}

\begin{theorem}
On a m dimensional oriented Hausdorff class 1 real manifold M, any continuous
m form $\varpi$ defines a unique, absolutely continuous, Radon measure on M,
called the \textbf{Lebesgue measure} associated to $\varpi.$
\end{theorem}

\begin{proof}
Let $\left(
\mathbb{R}
^{m},\left(  O_{i},\varphi_{i}\right)  _{i\in I}\right)  ,U_{i}=\varphi
_{i}\left(  O_{i}\right)  $\ be an atlas of M as above. As M is oriented the
atlas can be chosen such that $\det\left[  \varphi_{ij}^{\prime}\right]  >0$ .
Take a continuous m form $\varpi$ on M .On each open $U_{i}=\varphi_{i}\left(
O_{i}\right)  $ we define the Radon measure : $\nu_{i}=\varphi_{i\ast}\left(
\varpi_{i}\right)  d\xi.$ It is locally finite and finite if $\int_{U_{i}%
}\left\vert \left(  \varphi_{i\ast}\varpi_{i}\right)  \right\vert d\xi<\infty$
.Then on the subsets $U_{i}\cap U_{j}\neq\varnothing$ : $\nu_{i}=\nu_{j}$
.Thus the family $\left(  \nu_{i}\right)  _{i\in I}$ defines a unique Radon
measure, absolutely continuous, on $U=\cup_{i}U_{i}\subset%
\mathbb{R}
^{m}$ .The pull back, on each chart, of the $\nu_{i}$ give a family $\left(
\mu_{i}\right)  _{i\in I}$ of Radon measures on each $O_{i}$ and from there a
locally compact, absolutely continuous, Radon measure on M.

It can be shown (Schwartz IV p.319) that the measure does not depend on the
atlas with the same orientation on M.
\end{proof}

\begin{definition}
The \textbf{Lebesgue integral} of a m form $\varpi$ on M is $\int_{M}%
\mu_{\varpi}$ where $\mu_{\varpi}$ is the Lebesgue measure on M which is
defined by $\varpi.$
\end{definition}

It is denoted $\int_{M}\varpi$

An open subset $\Omega$ of an orientable manifold is an orientable manifold of
the same dimension, so the integral of a m-form on any open of M is given by
restriction of the measure $\mu$ : $\int_{\Omega}\varpi$

\paragraph{Remaks\newline}

i) The measure is linked to the Lebesgue measure but, from the definition,
whenever we have an absolutely continuous Radon measure $\mu$ on M, there is a
m form such that $\mu$ is the Lebesgue measure for some form. However there
are singular measures on M which are not linked to the Lebesgue measure.

ii) Without orientation on each domain there are two measures, different by
the sign, but there is no guarantee that one can define a unique measure on
the whole of M. Such "measures" are called densities.

iii) On $%
\mathbb{R}
^{m}$ we have the canonical volume form : $dx=dx^{1}\wedge...\wedge dx^{m},$
which naturally induces the Lebesgue measure, also denoted dx=$dx^{1}%
\otimes...\otimes dx^{m}=dx^{1}dx^{2}...dx^{m}$

iv) The product of the Lebesgue form $\varpi_{\mu}$ by a function
$f:M\rightarrow%
\mathbb{R}
$\ gives another measure and : $f\varpi_{\mu}=\varpi_{f\mu}.$Thus, given a m
form $\varpi$, the integral of any continuous function on M can be defined,
but its value depends on the choice of $\varpi.$

If there is a volume form $\varpi_{0}$, then for any function $f:M\rightarrow%
\mathbb{R}
$ the linear functional $f\rightarrow\int_{M}f\varpi_{0}$ can be defined.

Warning ! the quantity $\int_{M}fdx^{1}\wedge...\wedge dx^{m}$ where f is a
function is not defined (except if M is an open in $%
\mathbb{R}
^{m}$\ ) because $fdx^{1}\wedge...\wedge dx^{m}$ is not a m form.

v) If M is a set of a finite number of points $M=\left\{  p_{i}\right\}
_{i\in I}$ then this is a 0-dimensional manifold, a 0-form on M is just a map
: $f:M\rightarrow%
\mathbb{R}
$ and the integral is defined as :$\int_{M}f=\sum_{i\in I}f(p_{i})$

vi) For manifolds M with compact boundary in $%
\mathbb{R}
^{m}$ the integral $\int_{M}dx$ is proportionnal to the usual euclidean
"volume" delimited by M.

\paragraph{Integrals on a r-simplex\newline}

It is useful for practical purposes to be able to compute integrals on subsets
of a manifold M which are not submanifolds, for instance subsets delimited
regularly by a finite number of points of M. The r-simplices on a manifold
meet this purpose (see Homology on manifolds).

\begin{definition}
The integral of a r form $\varpi\in\mathfrak{X}\left(  \Lambda_{r}TM^{\ast
}\right)  $ on a r-simplex $M^{r}=f\left(  S^{r}\right)  $ of a m dimensional
oriented Hausdorff class 1 real manifold M is given by : $\int_{M^{r}}%
=\int_{S^{r}}f^{\ast}\varpi dx$
\end{definition}

$f\in C_{\infty}\left(
\mathbb{R}
^{m};M\right)  $ and

$S^{r}=\left\langle A_{0},...A_{r}\right\rangle =\{P\in%
\mathbb{R}
^{m}:P=\sum_{i=0}^{r}t_{i}A_{i};0\leq t_{i}\leq1,\sum_{i=0}^{r}t_{i}=1\}$ is a
r-simplex on $%
\mathbb{R}
^{m}.$

$f^{\ast}\varpi\in\mathfrak{X}\left(  \Lambda_{r}%
\mathbb{R}
^{m}\right)  $ and the integral $\int_{S^{r}}f^{\ast}\varpi dx$ is computed in
the classical way. Indeed $f^{\ast}\varpi=\sum\pi_{\alpha_{1}...\alpha_{r}%
}dx^{\alpha_{1}}\wedge...\wedge dx^{\alpha_{r}}$ so the integrals are of the
kind : $\int_{s^{r}}\pi_{\alpha_{1}...\alpha_{r}}dx^{\alpha_{1}}%
...dx^{\alpha_{r}}$ on domains $s^{r}$ which are the convex hull of the r
dimensional subspaces generated by r+1 points, there are r variables and a r
dimensional domain of integration.

Notice that here a m form (meaning a form of the same order as the dimension
of the manifold) is not needed. But the condition is to have a r-simplex and a
r form.

For a r-chain $C^{r}=\sum_{i}k_{i}M_{i}^{r}$ on M then :

$\int_{C^{r}}\varpi=\sum_{i}k_{i}\int_{M^{r}}\varpi=\sum_{i}k_{i}\int
_{S_{i}^{r}}f^{\ast}\varpi dx.$ and : $\int_{C^{r}+D^{r}}\varpi=\int_{C^{r}%
}\varpi+\int_{D^{r}}\varpi$

\subsubsection{Properties of the integral}

\begin{theorem}
$\int_{M}$ is a linear operator : $\mathfrak{X}\left(  \Lambda_{m}TM^{\ast
}\right)  \rightarrow%
\mathbb{R}
$
\end{theorem}

$\forall k,k^{\prime}\in%
\mathbb{R}
,\varpi,\pi\in\Lambda_{m}TM^{\ast}:\int_{M}\left(  k\varpi+k^{\prime}%
\pi\right)  \mu=k\int_{M}\varpi\mu+k^{\prime}\int_{M}\pi\mu$

\begin{theorem}
If the orientation on M is reversed, $\int_{M}\varpi\mu\rightarrow-\int
_{M}\varpi\mu$
\end{theorem}

\begin{theorem}
If a manifold is endowed with a continuous volume form $\varpi_{0}$ the
induced Lebesgue measure $\mu_{0}$ on M can be chosen such that it is
positive, locally compact, and M is $\sigma-$additive with respect to $\mu
_{0}$.
\end{theorem}

\begin{proof}
If the component of $\varpi_{0}$ is never null and continuous it keeps its
sign over M and we can choose $\varpi_{0}$ such it is positive. The rest comes
from the measure theory.
\end{proof}

\begin{theorem}
(Schwartz IV p.332) If $f\in C_{1}\left(  M;N\right)  $ is a diffeomorphism
between two oriented manifolds, which preserves \ the orientation, then :

$\forall\varpi\in\mathfrak{X}_{1}\left(  \Lambda_{m}TM^{\ast}\right)  $%
:$\int_{M}\varpi=\int_{N}\left(  f_{\ast}\varpi\right)  $
\end{theorem}

This result is not surprising : the integrals can be seen as the same integral
computed in different charts.

Conversely :

\begin{theorem}
Moser's theorem (Lang p.406) Let M be a compact, real, finite dimensional
manifold with volume forms $\varpi,\pi$ such that : $\int_{M}\varpi=\int
_{M}\pi$ then there is a diffeomorphism $f:M\rightarrow M$ such that
$\pi=f^{\ast}\varpi$
\end{theorem}

If M is a m dimensional submanifold of the n%
$>$%
m manifold N, both oriented, f an embedding of M into N, then the integral on
M of a m form in N can be defined by :

$\forall\varpi\in\mathfrak{X}_{1}\left(  \Lambda_{m}TN^{\ast}\right)
:\int_{M}\varpi=\int_{f(M)}\left(  f_{\ast}\varpi\right)  $

because f is a diffeormophism of M to f(M) and f(M) an open subset of N.

Example : a curve $c:J\rightarrow N::c(t)$ on the manifold N is a orientable
submanifold if c'(t)$\neq0.$ For any 1-form over N : $\varpi\left(  p\right)
=\sum_{\alpha}\varpi_{\alpha}\left(  p\right)  dx^{\alpha}.$ \ So $c_{\ast
}\varpi=\varpi\left(  c(t)\right)  c^{\prime}(t)dt$ and $\int_{c}\varpi
=\int_{J}\varpi\left(  c(t)\right)  c^{\prime}(t)dt$

\subsubsection{Stokes theorem}

1. For the physicists it is the most important theorem of differential
geometry. It can be written :

\begin{theorem}
Stokes theorem : For any manifold with boundary M in a n dimensional real
orientable manifold N and any n--1 form :%

\begin{equation}
\forall\varpi\in\mathfrak{X}_{1}\left(  \wedge_{n-1}TN^{\ast}\right)
:\int_{M}d\varpi=\int_{\partial M}\varpi
\end{equation}

\end{theorem}

This theorem requires some comments and conditions .

2. Comments :

i) the exterior differential $d\varpi$ is a n-form, so its integral in N makes
sense, and the integration over M, which is a closed subset of N, must be read
as : $\int_{\overset{\circ}{M}}d\varpi,$ meaning the integral over the open
subset $\overset{\circ}{M}$ of N (which is a n-dimensional submanifold of N).

ii) the boundary is a n-1 orientable submanifold in N, so the integral of a
the n-1 form $\varpi$ makes sense. Notice that the Lebesgue measures are not
the same : on M is is induced by $d\varpi$\ , on $\partial M$\ it is induced
by the restriction $\varpi|_{\partial M}$\ of $\varpi$ on $\partial M$

iii) the n-1 form $\varpi$ does not need to be defined over the whole of N :
only the domain included in M (with boundary) matters, but as we have not
defined forms over manifold with boundary it is simpler to look at it this
way. And of course it must be at least of class 1 to compute its exterior derivative.

3. Conditions :

There are several alternate conditions. The theorem stands if one of the
following condition is met:

i) the simplest : M is compact

ii) $\varpi$ is compactly supported : the support Supp($\varpi)$\ is the
closure of the set : $\left\{  p\in M:\varpi(p)\neq0\right\}  $

iii) Supp($\varpi)\cap M$ is compact

Others more complicated conditions exist.

4. If C is a r-chain on M, then both the integral $\int_{C}\varpi$ and the
border $\partial C$ ot the r chain are defined. And the equivalent of the
Stokes theorem reads :

If C is a r-chain on M, $\varpi\in\mathfrak{X}_{1}\left(  \wedge_{r-1}%
TM^{\ast}\right)  $ $\ $then $\int_{C}d\varpi=\int_{\partial C}\varpi$

\begin{theorem}
Integral on a curve (Schwartz IV p.339) Let E be a finite dimensional real
normed affine space. A continuous curve C generated by a path $c:\left[
a,b\right]  \rightarrow E$\ on E is rectifiable if $\ell\left(  c\right)
<\infty$ with $\ell\left(  c\right)  =\sup\sum_{k=1}^{n}d(p\left(
t_{k+1}\right)  ),p(t_{k}))$ for any increasing sequence $\left(
t_{n}\right)  _{n\in%
\mathbb{N}
}$ in [a,b] and d the metric induced by the norm. The curve is oriented in the
natural way (t increasing).

For any function $f\in C_{1}\left(  E;%
\mathbb{R}
\right)  $ : $\int_{C}df=f(c\left(  b)\right)  -f\left(  c\left(  a\right)
\right)  $
\end{theorem}

\subsubsection{Divergence}

\paragraph{Definition\newline}

\begin{theorem}
For any vector field $V\in\mathfrak{X}\left(  TM\right)  $ on a manifold
endowed with a volume form $\varpi_{0}$\ there is a function div(V) on M,
called the \textbf{divergence} of the vector field, such that%

\begin{equation}
\pounds _{V}\varpi_{0}=\left(  divV\right)  \varpi_{0}%
\end{equation}

\end{theorem}

\begin{proof}
If M is m dimensional, $\varpi_{0},\ \pounds _{V}\varpi_{0}\in\mathfrak{X}%
\left(  \wedge_{m}TM^{\ast}\right)  .$ All m forms are proportional on M and
$\varpi_{0}$ is never null, then : $\forall p\in M,\exists k\in K:\pounds _{V}%
\varpi_{0}\left(  p\right)  =k\varpi_{0}\left(  p\right)  $
\end{proof}

\paragraph{Expression in a holonomic basis\newline}

$\forall V\in\mathfrak{X}\left(  TM\right)  :\pounds _{V}\varpi_{0}%
=i_{V}d\varpi_{0}+d\circ i_{V}\varpi_{0}$ and $d\varpi_{0}=0$ so
$\pounds _{V}\varpi_{0}=d\left(  i_{V}\varpi_{0}\right)  $

$\varpi_{0}=\varpi_{0}\left(  p\right)  dx^{1}\Lambda...\Lambda dx^{m}%
:\pounds _{V}\omega_{0}=d\left(  \sum_{\alpha}V^{\alpha}\left(  -1\right)
^{\alpha-1}\varpi_{0}dx^{1}\Lambda..\widehat{dx^{\alpha}}.\Lambda
dx^{m}\right)  $

$=\sum_{\beta}\partial_{\beta}\left(  V^{\alpha}\left(  -1\right)  ^{\alpha
-1}\varpi_{0}\right)  dx^{\beta}\wedge dx^{1}\Lambda..\widehat{dx^{\alpha}%
}.\Lambda dx^{m}=\left(  \sum_{\alpha}\partial_{\alpha}\left(  V^{\alpha
}\varpi_{0}\right)  \right)  dx^{1}\Lambda..\Lambda dx^{m}$%

\begin{equation}
divV=\frac{1}{\varpi_{0}}\sum_{\alpha}\partial_{\alpha}\left(  V^{\alpha
}\varpi_{0}\right)
\end{equation}

\paragraph{Properties\newline}

For any $f\in C_{1}\left(  M;%
\mathbb{R}
\right)  ,V\in\mathfrak{X}\left(  M\right)  :$ $fV\in\mathfrak{X}\left(
M\right)  $ and

$div\left(  fV\right)  \varpi_{0}=d\left(  i_{fV}\varpi_{0}\right)  =d\left(
fi_{V}\varpi_{0}\right)  =df\wedge i_{V}\varpi_{0}+fd\left(  i_{V}\varpi
_{0}\right)  $

$=df\wedge i_{V}\varpi_{0}+fdiv(V)\varpi_{0}$

$df\wedge i_{V}\varpi_{0}=\left(  \sum_{\alpha}\partial_{\alpha}fdx^{\alpha
}\right)  \wedge\left(  \sum_{\beta}\left(  -1\right)  ^{\beta}V^{\beta}%
\varpi_{0}dx^{1}\wedge...\widehat{dx^{\beta}}...\wedge dx^{m}\right)  $

$=\left(  \sum_{\alpha}V^{\alpha}\partial_{\alpha}f\right)  \varpi
_{0}=f^{\prime}\left(  V\right)  \varpi_{0}$

So : $div\left(  fV\right)  =f^{\prime}(V)+fdiv(V)$

\paragraph{Divergence theorem\newline}

\begin{theorem}
For any vector field $V\in\mathfrak{X}_{1}\left(  TM\right)  $ on a manifold N
endowed with a volume form $\varpi_{0}$\ , and manifold with boundary M in N:%

\begin{equation}
\int_{M}\left(  divV\right)  \varpi_{0}=\int_{\partial M}i_{V}\varpi_{0}%
\end{equation}

\end{theorem}

\begin{proof}
$\pounds _{V}\varpi_{0}=\left(  divV\right)  \varpi_{0}=d\left(  i_{V}%
\varpi_{0}\right)  $

In conditions where the Stockes theorem holds :

$\int_{M}d\left(  i_{V}\varpi_{0}\right)  =\int_{M}\left(  divV\right)
\varpi_{0}=\int_{\partial M}i_{V}\varpi_{0}$
\end{proof}

$\varpi_{0}$ defines a volume form on N, and the interior of M (which is an
open subset of N).\ So any class 1 vector field on N defines a Lebesgue
measure on $\partial M$ by $i_{V}\varpi_{0}.$

If M is endowed with a Riemannian metric there is an outgoing unitary vector n
on $\partial M$ (see next section) which defines a measure $\varpi_{1}$ on
$\partial M$ and : $i_{V}\varpi_{0}=\left\langle V,n\right\rangle \varpi
_{1}=i_{V}\varpi_{0}$ so $\int_{M}\left(  divV\right)  \varpi_{0}%
=\int_{\partial M}\left\langle V,n\right\rangle \varpi_{1}$

\subsubsection{Integral on domains depending on a parameter}

\paragraph{Layer integral:\newline}

\begin{theorem}
(Schwartz 4 p.99) Let M be a m dimensional class 1 real riemannian manifold
with the volume form $\varpi_{0}$, then for any function $f\in C_{1}\left(  M;%
\mathbb{R}
\right)  $ $\varpi_{0}-$integrable on M such that $f^{\prime}(x)\neq0$ on M,
for almost every value of t, the function $g(x)=\frac{f\left(  x\right)
}{\left\Vert gradf\right\Vert }$ is integrable on the hypersurface $\partial
N\left(  t\right)  =\left\{  x\in M:f(x)=t\right\}  $ and we have :%

\begin{equation}
\int_{M}f\varpi_{0}=\int_{0}^{\infty}\left(  \int_{\partial N\left(  t\right)
}\frac{f\left(  x\right)  }{\left\Vert gradf\right\Vert }\sigma\left(
t\right)  \right)  dt
\end{equation}

where $\sigma\left(  t\right)  $ is the volume form induced on $\partial N(t)$
by $\varpi_{0}$
\end{theorem}

(the Schwartz's demontration for an affine space is easily extended to a real manifold)

$\sigma\left(  t\right)  =i_{n\left(  t\right)  }\varpi_{0}$ where
$n=\frac{gradf}{\left\Vert gradf\right\Vert }$ (see Pseudo riemannian
manifolds )\ so

$\int_{M}f\varpi_{0}=\int_{0}^{\infty}\left(  \int_{\partial N\left(
t\right)  }\frac{f\left(  x\right)  }{\left\Vert gradf\right\Vert ^{2}%
}i_{gradf}\varpi_{0}\right)  dt$

The meaning of this theorem is the following : f defines a family of manifolds
with boundary $N(t)=\left\{  x\in M:p(x)\leq t\right\}  $ in M, which are
diffeomorphic by the flow of the gradiant grad(f). Using an atlas of M there
is a folliation in $%
\mathbb{R}
^{m}$\ and using the Fubini theorem the integral can be computed by summing
first over the hypersurface defined by N(t) then by taking the integral over t.

The previous theorem does not use the fact that M is riemannian, and the
formula is valid whenever g is not degenerate on N(t), but we need both
$f^{\prime}\neq0,\left\Vert gradf\right\Vert \neq0$ which cannot be guaranted
without a positive definite metric.

\paragraph{Integral on a domain depending on a parameter :\newline}

\begin{theorem}
Reynold's theorem: Let (M,g) be a class 1 m dimensional real riemannian
manifold with the volume form $\varpi_{0}$, f a function $f\in C_{1}\left(
\mathbb{R}
\times M;%
\mathbb{R}
\right)  ,$ N(t) a family of manifolds with boundary in M, then :%

\begin{equation}
\frac{d}{dt}\int_{N\left(  t\right)  }f\left(  t,x\right)  \varpi_{0}\left(
x\right)  =\int_{N\left(  t\right)  }\frac{\partial f}{\partial t}\left(
t,x\right)  \varpi_{0}\left(  x\right)  +\int_{\partial N\left(  t\right)
}f\left(  x,t\right)  \left\langle v,n\right\rangle \sigma\left(  t\right)
\end{equation}

where $v\left(  q\left(  t\right)  \right)  =\frac{dq}{dt}$ for $q(t)\in
N\left(  t\right)  $
\end{theorem}

This assumes that there is some map :

$\phi:%
\mathbb{R}
\times M\rightarrow M::\phi\left(  t,q\left(  s\right)  \right)  =q\left(
t+s\right)  \in N\left(  t+s\right)  $

If N(t) is defined by a function p : $N\left(  t\right)  =\left\{  x\in
M:p(x)\leq t\right\}  $ then :

$\frac{d}{dt}\int_{N\left(  t\right)  }f\left(  t,x\right)  \varpi_{0}\left(
x\right)  =\int_{N\left(  t\right)  }\frac{\partial f}{\partial t}\left(
t,x\right)  \varpi_{0}\left(  x\right)  +\int_{\partial N\left(  t\right)
}\frac{f\left(  x,t\right)  }{\left\Vert gradp\right\Vert }\sigma\left(
t\right)  $

\begin{proof}
the boundaries are diffeomorphic by the flow of the vector field (see
Manifolds with boundary) :

$V=\frac{gradp}{\left\Vert gradp\right\Vert ^{2}}::\forall q_{t}\in\partial
N\left(  t\right)  :\Phi_{V}\left(  q_{t},s\right)  \in\partial N_{t+s}$

So : $v\left(  q\left(  t\right)  \right)  =\frac{\partial}{\partial t}%
\Phi_{V}\left(  q_{t},s\right)  |_{t=s}=V\left(  q\left(  t\right)  \right)
=\frac{gradp}{\left\Vert gradp\right\Vert ^{2}}|_{q\left(  t\right)  }$

On the other hand : $n=\frac{gradp}{\left\Vert gradp\right\Vert }$

$\left\langle v,n\right\rangle =\frac{\left\Vert gradp\right\Vert ^{2}%
}{\left\Vert gradp\right\Vert ^{3}}=\frac{1}{\left\Vert gradp\right\Vert }$
\end{proof}

Formula which is consistent with the previous one if f does not depend on t.

\paragraph{m forms depending on a parameter:\newline}

$\mu$ is a family $\mu\left(  t\right)  $ of m form on M such that : $\mu:%
\mathbb{R}
\rightarrow\mathfrak{X}\left(  \Lambda_{m}TM^{\ast}\right)  $ is a class 1 map
and one considers the integral : $\int_{N\left(  t\right)  }\mu$ where N(t) is
a manifold with boundary defined by $N\left(  t\right)  =\left\{  x\in
M:p(x)\leq t\right\}  $

M is extended to $%
\mathbb{R}
\times M$ with the riemannian metric

$G=dt\otimes dt+\sum g_{\alpha\beta}dx^{\alpha}\otimes dx^{\beta}$

With $\lambda=dt\wedge\mu\left(  t\right)  :D\mu=\frac{\partial\mu}{\partial
t}dt\wedge\mu+d\mu\wedge\mu=\frac{\partial\mu}{\partial t}dt\wedge\mu$

With the previous theorem : $\int_{M\times I\left(  t\right)  }D\mu
=\int_{N(t)}\mu\varpi_{0}$ where $I\left(  t\right)  =\left[  0,t\right]  $

$\frac{d}{dt}\int_{N\left(  t\right)  }\mu=\int_{N\left(  t\right)  }%
i_{v}\left(  d_{x}\varpi\right)  +\int_{N\left(  t\right)  }\frac{\partial\mu
}{\partial t}+\int_{\partial N\left(  t\right)  }i_{V}\mu$

where $d_{x}\varpi$ is the usual exterior derivative with respect to x, and
$v=gradp$

\bigskip

\subsection{Cohomology}

\label{cohomology}

Also called de Rahm cohomology (there are other concepts of cohomology). It is
a branch of algebraic topology adapted to manifolds, which gives a
classification of manifolds and is related to the homology on manifolds.

\subsubsection{Spaces of cohomology}

\paragraph{Definition\newline}

Let M be a smooth manifold modelled over the Banach E on the field K.

The \textbf{de Rahm complex} is the sequence :

$0\rightarrow\mathfrak{X}\left(  \Lambda_{0}TM^{\ast}\right)  \overset
{d}{\rightarrow}\mathfrak{X}\left(  \Lambda_{1}TM^{\ast}\right)  \overset
{d}{\rightarrow}\mathfrak{X}\left(  \Lambda_{2}TM^{\ast}\right)  \overset
{d}{\rightarrow}...$

In the categories parlance this is a sequence because the image of the
operator d is just the kernel for the next operation :

if $\varpi\in\mathfrak{X}\left(  \Lambda_{r}TM^{\ast}\right)  $ then
$d\varpi\in\mathfrak{X}\left(  \Lambda_{r+1}TM^{\ast}\right)  $ and
$d^{2}\varpi=0$

An exact form is a closed form, the Poincar\'{e} lemna tells that the converse
is locally true, and cohomology studies this fact.

Denote the sets of :

closed r-forms : $F^{r}\left(  M\right)  =\left\{  \varpi\in\mathfrak{X}%
\left(  \Lambda_{r}TM^{\ast}\right)  :d\varpi=0\right\}  $ sometimes called
the set of cocycles with $F^{0}\left(  M\right)  $ the set of locally constant functions.

exact r--1 forms : $G^{r-1}\left(  M\right)  =\left\{  \varpi\in
\mathfrak{X}\left(  \Lambda_{r}TM^{\ast}\right)  :\exists\pi\in\mathfrak{X}%
\left(  \Lambda_{r-1}TM^{\ast}\right)  :\varpi=d\pi\right\}  $ sometimes
called the set of coboundary.

\begin{definition}
The rth \textbf{space of cohomology} of a manifold M is the quotient space :
$H^{r}\left(  M\right)  =F^{r}\left(  M\right)  /G^{r-1}\left(  M\right)  $
\end{definition}

The definition makes sense : $F^{r}\left(  M\right)  ,G^{r-1}\left(  M\right)
$ are vector spaces over K and $G^{r-1}\left(  M\right)  $ is a vector
subspace of $F^{r}\left(  M\right)  .$ Two closed forms in a class of
equivalence denoted $\left[  {}\right]  $\ differ by an exact form :

$\varpi_{1}\sim\varpi_{2}\Leftrightarrow\exists\pi\in\mathfrak{X}\left(
\Lambda_{r-1}TM^{\ast}\right)  :\varpi_{2}=\varpi_{1}+d\pi$

The exterior product extends to $H^{r}\left(  M\right)  $

$\left[  \varpi\right]  \in H^{p}(V),\left[  \pi\right]  \in H^{q}(V):\left[
\varpi\right]  \wedge\left[  \pi\right]  =\left[  \varpi\wedge\pi\right]  \in
H^{p+q}(V)$

$\oplus_{r=0}^{\dim M}H^{r}\left(  M\right)  =H^{\ast}\left(  M\right)  $ has
the structure of an algebra over the field K

\paragraph{Properties\newline}

\begin{definition}
The r \textbf{Betti number }$b_{r}\left(  M\right)  $ of the manifold M
is\textbf{\ }the dimension of $H^{r}\left(  M\right)  $. The \textbf{Euler
characteristic} of the manifold M is :

$\chi\left(  M\right)  =\sum_{r=1}^{\dim M}\left(  -1\right)  ^{r}b_{r}\left(
M\right)  $
\end{definition}

They are topological invariant : two diffeomorphic manifolds have the same
Betti numbers and Euler characteristic.

Betti numbers count the number of "holes" of dimension r in the manifold.

$\chi\left(  M\right)  =0$ if dimM is odd.

\begin{definition}
The \textbf{Poincar\'{e} polynomial} on the field K is :

$P\left(  M\right)  :K\rightarrow K:P\left(  M\right)  \left(  z\right)
=\sum_{r}b_{r}\left(  M\right)  z^{r}$
\end{definition}

For two manifolds M,N : $P\left(  M\times N\right)  =P\left(  M\right)  \times
P\left(  N\right)  $

The Poincar\'{e} polynomials can be computed for Lie groups (see Wikipedia,
Betti numbers).

If M has n connected components then : $H^{0}\left(  M\right)  \simeq%
\mathbb{R}
^{n}.$ This follows from the fact that any smooth function on M with zero
derivative (i.e. locally constant) is constant on each of the connected
components of M. So $b_{0}\left(  M\right)  $ is the number of connected
components of M,

If M is a simply connected manifold then $H^{1}\left(  M\right)  $ is trivial
(it has a unique class of equivalence which is $\left[  0\right]  $) and
$b_{1}\left(  M\right)  =0$.

\begin{theorem}
If M,N are two real smooth manifolds and $f:M\rightarrow N$ then :

i) the pull back $f^{\ast}\varpi$ of a closed (resp.exact) form $\varpi$ is a
closed (resp.exact) form so : $f^{\ast}\left[  \varpi\right]  =\left[
f^{\ast}\varpi\right]  \in H^{r}\left(  M\right)  $

ii) if $f,g\in C_{\infty}\left(  M;N\right)  $ are homotopic then
$\forall\varpi\in H^{r}\left(  N\right)  :f^{\ast}\left[  \varpi\right]
=g^{\ast}\left[  \varpi\right]  $
\end{theorem}

\begin{theorem}
\textbf{K\"{u}nneth formula} : Let $M_{1},M_{2}$ smooth finite dimensional
real manifolds :

$H^{r}(M_{1}\times M_{2})=\underset{p+q=r}{\oplus}[H^{p}(M_{1})\otimes
H^{q}(M_{2})]$

$H^{\ast}(M_{1}\times M_{2})=H^{\ast}(M_{1})\times H^{\ast}(M_{2})$

$b_{r}(M_{1}\times M_{2})=\sum_{q+p=r}b_{p}(M_{1})b_{q}(M_{2})$

$\chi(M_{1}\times M_{2})=\chi(M_{1})\chi(M_{2})$
\end{theorem}

\subsubsection{de Rahm theorem}

Let M be a real smooth manifold. The sets $C^{r}\left(  M\right)  $ of
r-chains on M and $\mathfrak{X}\left(  \Lambda_{r}TM^{\ast}\right)  $ of
r-forms on M are real vector spaces. The map :

$\left\langle {}\right\rangle :C^{r}\left(  M\right)  \times\mathfrak{X}%
\left(  \Lambda_{r}TM^{\ast}\right)  \rightarrow%
\mathbb{R}
::\left\langle C,\varpi\right\rangle =\int_{C}\varpi$

is bilinear. And the Stokes theorem reads : $\left\langle C,d\varpi
\right\rangle =\left\langle \partial C,\varpi\right\rangle $

This map stands with the quotient spaces $H^{r}\left(  M\right)  $ of
homologous r-chains and $H_{r}\left(  M\right)  $ of cohomologous r-forms:

$\left\langle {}\right\rangle :H^{r}\left(  M\right)  \times H_{r}\left(
M\right)  \rightarrow%
\mathbb{R}
::\left\langle \left[  C\right]  ,\left[  \varpi\right]  \right\rangle
=\int_{\left[  C\right]  }\left[  \varpi\right]  $

These two vector spaces can be seen as "dual" from each other.

\begin{theorem}
de Rahm : If M is a real, m dimensional, compact manifold, then :

i) the vector spaces $H^{r}\left(  M\right)  ,H_{r}\left(  M\right)  $ have
the same finite dimension equal to the rth Betti number $b_{r}\left(
M\right)  $

$b_{r}\left(  M\right)  =0$ if r
$>$
dimM, $b_{r}\left(  M\right)  =b_{m-r}\left(  M\right)  $

ii) the map $\left\langle {}\right\rangle :H^{r}\left(  M\right)  \times
H_{r}\left(  M\right)  \rightarrow%
\mathbb{R}
$\ is non degenerate

iii)\ $H^{r}\left(  M\right)  =H_{r}\left(  M\right)  ^{\ast}$

iv) $H^{r}\left(  M\right)  \simeq H^{m-r}\left(  M\right)  $

v) Let $M_{i}^{r}\in C^{r}\left(  M\right)  ,i=1...b_{r}\left(  M\right)
:\forall i\neq j:\left[  M_{i}^{r}\right]  \neq\left[  M_{j}^{r}\right]  $
then :

a closed r-form $\varpi$ is exact iff $\forall i=1...b_{r}:\int_{M_{i}^{r}%
}\varpi=0$

$\forall k_{i}\in%
\mathbb{R}
,i=1...b_{r},\exists\varpi\in\Lambda_{r}TM^{\ast}:d\varpi=0,\int_{M_{i}^{r}%
}\varpi=k_{i}$
\end{theorem}

\begin{theorem}
(Lafontaine p.233) Let M be a smooth real m dimensional, compact, connected
manifold, then:

i) a m form $\varpi$ is exact iff $\int_{M}\varpi=0$

ii) $H^{m}\left(  M\right)  $ is isomorphic to $%
\mathbb{R}
$
\end{theorem}

Notice that these theorems require \textit{compact} manifolds.

\subsubsection{Degree of a map}

\begin{theorem}
(Lafontaine p.235) Let M,N be smooth real m dimensional, compact, oriented
manifolds, $f\in C_{\infty}\left(  M;N\right)  $ then there is a signed
integer k(f) called the \textbf{degree of the map} such that :

$\exists k\left(  f\right)  \in%
\mathbb{Z}
:\forall\varpi\in\Lambda_{m}TM^{\ast}:\int_{M}f^{\ast}\varpi=k\left(
f\right)  \int_{N}\varpi$
\end{theorem}

If f is not surjective then k(f)=0

If f,g are homotopic then k(f)=k(g)

\begin{theorem}
(Taylor 1 p.101) Let M be a compact manifold with boundary, N a smooth compact
oriented real manifold, $f\in C_{1}\left(  M;N\right)  $ then :

$Degree\left(  f|_{\partial M}\right)  =0$
\end{theorem}

\newpage

\section{COMPLEX\ MANIFOLDS}

\label{Complex manifolds}

Everything which has been said before for manifolds holds for complex
manifolds, if not stated otherwise.\ However complex manifolds have specific
properties linked on one hand to the properties of holomorphic maps and on the
other hand to the relationship between real and complex structures. The most
usual constructs, which involve only the tangent bundle, not the manifold
structure itself, are simple extensions of the previous theorems. Complex
manifolds, meaning manifolds whose manifold structure is complex, are a
different story.

It is useful to refer to the Algebra part about complex vector spaces.

\subsection{Complex manifolds}

\subsubsection{General properties}

1. Complex manifolds are manifolds modelled on a Banach vector space E over $%
\mathbb{C}
.$ The transition maps : $\varphi_{j}\circ\varphi_{i}^{-1}$ are
C-differentiable maps between Banach vector spaces, so they are holomorphic
maps, and smooth. Thus a differentiable complex manifold is smooth.

2. The tangent vector spaces are complex vector spaces (their introduction
above does not require the field to be $%
\mathbb{R}
).$ So \textit{on the tangent space} of complex manifolds real structures can
be defined (see below).

3. A map $f\in C_{r}\left(  M;N\right)  $ between complex manifolds M,N
modeled on E,G is $%
\mathbb{R}
$-differentiable iff the map $F=\psi_{j}\circ f\circ\varphi_{i}^{-1}%
:E\rightarrow G$\ \ is $%
\mathbb{R}
$- differentiable. If F is 1-$%
\mathbb{C}
$-differentiable, it is holomorphic, thus smooth and f itself is said to be holomorphic.

F is $%
\mathbb{C}
$-differentiable iff it is R-differentiable and meets the Cauchy-Riemann
conditions on partial derivatives $F_{y}^{\prime}=iF_{x}^{\prime}$ where y,x
refer to any real structure on E.

4. A complex manifold of (complex) dimension 1 is called a Riemann manifold.
The compactified (as in topology) of $%
\mathbb{C}
$ is the \textbf{Riemann sphere}. Important properties of holomorphic
functions stand only when the domain is an open of $%
\mathbb{C}
.$ So many of these results (but not all of them) are still valid for maps
(such as functions or forms) defined on Riemann manifolds, but not on general
complex manifolds. We will not review them as they are in fact very specific
(see Schwartz).

\subsubsection{Maps on complex manifolds}

In the previous parts or sections several theorems address specifically
complex vector spaces and holomorphic maps.\ We give their obvious extensions
on manifolds.

\begin{theorem}
A holomorphic map $f:M\rightarrow F$ from a finite dimensional connected
complex manifold M to a normed vector space F is constant if one of the
following conditions is met:

i) f is constant in an open of M

ii) if $F=%
\mathbb{C}
$ and $\operatorname{Re}f$ \ or $\operatorname{Im}f$\ has an extremum or
$\left\vert f\right\vert $ has a maximum

iii) if M is compact
\end{theorem}

\begin{theorem}
If M,N are finite dimensional connected complex manifolds, f a holomorphic map
$f:M\rightarrow N$ ,if f is constant in an open of M then it is constant in M
\end{theorem}

A compact connected finite dimensional complex manifold cannot be an affine
submanifold of $%
\mathbb{C}
^{n}$ because its charts would be constant.

\subsubsection{Real structure on the tangent bundle}

A real structure on a complex manifold involves only the tangent bundle, not
the manifold itself, which keeps its genuine complex manifold structure.

\begin{theorem}
The tangent bundle of any manifold modeled on a complex space E admits real
structures, defined by a real continuous real stucture on E.
\end{theorem}

\begin{proof}
If M is a complex manifold with atlas $\left(  E,\left(  O_{k},\varphi
_{k}\right)  _{k\in K}\right)  $\ it is always possible to define real
structures on E : antilinear maps $\sigma:E\rightarrow E$ such that
$\sigma^{2}=Id_{E}$ and then define a real kernel $E_{%
\mathbb{R}
}$ and split any vector u of E in a real and an imaginary part both belonging
to the kernel, such that : $u=\operatorname{Re}u+i\operatorname{Im}u$. If E is
infinite dimensional we will require $\sigma$ to be continuous

At any point $p\in O_{k}$ of M the real structure $S_{k}(p)$ on $T_{p}M$ is
defined by :

$S_{k}\left(  p\right)  =\varphi_{k}^{\prime-1}\circ\sigma\circ\varphi
_{k}^{\prime}:T_{p}M\rightarrow T_{p}M$

This is an antilinear map and $S^{2}\left(  p\right)  =Id_{T_{p}M}.$

The real kernel of $T_{p}M$ is

$\left(  T_{p}M\right)  _{%
\mathbb{R}
}=\left\{  u_{p}\in T_{p}M:S_{k}\left(  p\right)  u_{p}=u_{p}\right\}
=\varphi_{k}^{\prime-1}\left(  x\right)  \left(  E_{%
\mathbb{R}
}\right)  $

Indeed : $u\in E_{%
\mathbb{R}
}:\sigma\left(  u\right)  =u\rightarrow u_{p}=\varphi_{k}^{\prime-1}\left(
x\right)  u$

$S_{k}\left(  p\right)  u_{p}=\varphi_{k}^{\prime-1}\circ\sigma\circ
\varphi_{k}^{\prime}\circ\varphi_{k}^{\prime-1}\left(  x\right)  \left(
u\right)  =\varphi_{k}^{\prime-1}\left(  x\right)  \left(  u\right)  =u_{p}$

At the transitions :

$\sigma=\varphi_{k}^{\prime}\circ S_{k}\left(  p\right)  \circ\varphi
_{k}^{\prime-1}=\varphi_{j}^{\prime}\circ S_{j}\left(  p\right)  \circ
\varphi_{j}^{\prime-1}$

$S_{j}\left(  p\right)  =\left(  \varphi_{k}^{\prime-1}\circ\varphi
_{j}^{\prime}\right)  ^{-1}\circ S_{k}\left(  p\right)  \circ\left(
\varphi_{k}^{\prime-1}\circ\varphi_{j}^{\prime}\right)  $

From the definition of the tangent space $S_{j}\left(  p\right)  ,S_{k}\left(
p\right)  $ give the same map so this definition is intrinsic and we have a
map :$S:M\rightarrow C\left(  TM;TM\right)  $ such that S(p) is a real
structure on $T_{p}M.$
\end{proof}

The tangent bundle splits in a real and an imaginary part :

$TM=\operatorname{Re}TM\oplus i\operatorname{Im}TM$

We can define tensors on the product of vector spaces

$\left(  \left(  T_{p}M\right)  _{%
\mathbb{R}
}\times\left(  T_{p}M\right)  _{%
\mathbb{R}
}\right)  ^{r}\otimes\left(  \left(  T_{p}M\right)  _{%
\mathbb{R}
}\times\left(  T_{p}M\right)  _{%
\mathbb{R}
}\right)  ^{\ast s}$

We can always choose a basis $\left(  e_{a}\right)  _{a\in A}$ of E such that
: $\sigma\left(  e_{a}\right)  =e_{a},\sigma\left(  ie_{a}\right)  =-e_{a}$ so
that the holonomic basis of the real vector space $E_{\sigma}=E_{%
\mathbb{R}
}\oplus E_{%
\mathbb{R}
}$ reads $\left(  e_{a},ie_{a}\right)  _{a\in A}$.

\bigskip

\subsection{Complex structures on real manifolds}

There are two ways to build a complex structure \textit{on the tangent bundle}
of a real manifold : the easy way by complexification, and the complicated way
by a special map.

\subsubsection{Complexified tangent bundle}

This is the implementation, in the manifold context, of the general procedure
for vector spaces (see Algebra). The tangent vector space at each point p of a
real manifold M can be complexified : $T_{p}M_{%
\mathbb{C}
}=T_{p}M\oplus iT_{p}M.$ If M is modeled on the real Banach E, then $T_{p}M_{%
\mathbb{C}
}$ is isomorphic to the complexified of E, by taking the complexified of the
derivatives of the charts. This procedure does not change anything to the
manifold structure of M, it is similar to the tensorial product : the
complexified tangent bundle is $TM_{%
\mathbb{C}
}=TM\otimes%
\mathbb{C}
.$

A holonomic basis of M is still a holonomic basis in $TM_{%
\mathbb{C}
},$ the vectors may have complex components.

On $TM_{%
\mathbb{C}
}$ we can define a complexified tangent bundle, and r forms valued in $%
\mathbb{C}
:\mathfrak{X}\left(  \wedge_{r}TM_{%
\mathbb{C}
}^{\ast}\right)  =\wedge_{r}\left(  M;%
\mathbb{C}
\right)  .$

All the operations in complex vector space are available at each point p of M.
The complexified structure is fully dependent on the tangent bundle, so there
is no specific rule for a change of charts. This construct is strictly
independant of the manifold structure itself.

However there is another way to define a complex structure on a real vector
space, by using a complex structure.

\subsubsection{Almost complex structures}

\begin{definition}
An \textbf{almost complex structure} on a real manifold M is a tensor field
$J\in\mathfrak{X}\left(  \otimes_{1}^{1}TM\right)  $ such that $\forall u\in
T_{p}M:J^{2}\left(  p\right)  \left(  u\right)  =-u$
\end{definition}

\begin{theorem}
A complex structure on a real manifold M defines a structure of complex vector
space on each tangent space, and on the tangent bundle. A necessary condition
for the existence of a complex structure on a manifold M is that the dimension
of M is infinite or even.
\end{theorem}

A complex structure defines in each tangent space a map : $J\left(  p\right)
\in%
\mathcal{L}%
\left(  T_{p}M;T_{p}M\right)  $ such that $J^{2}\left(  p\right)  \left(
u\right)  =-u.$ Such a map is a complex structure on $T_{p}M,$ it cannot exist
if M is finite dimensional with an odd dimension, and otherwise defines,
continuously, a structure of complex vector space on each tangent space by :
$iu=J\left(  u\right)  $ (see Algebra).

A complex vector space has a canonical orientation. So a manifold endowed with
a complex structure is orientable, and one can deduce that there are
obstructions to the existence of almost complex structures on a manifold.

A complex manifold has an almost complex structure : J(u)=iu but a real
manifold endowed with an almost complex structure does not necessarily admits
the structure of a complex manifold. There are several criteria for this purpose.

\subsubsection{K\"{a}hler manifolds}

\begin{definition}
An \textbf{almost K\"{a}hler manifold} is a real manifold M endowed with a non
degenerate bilinear symmetric form g, an almost complex structure J, and such
its fundamental 2-form is closed. If M is also a complex manifold then it is
a\textbf{\ K\"{a}hler manifold}.
\end{definition}

i) It is always possible to assume that J preserves g by defining :
$\widehat{g}\left(  p\right)  \left(  u_{p},v_{p}\right)  =\frac{1}{2}\left(
g\left(  p\right)  \left(  u_{p},v_{p}\right)  +g\left(  p\right)  \left(
Ju_{p},Jv_{p}\right)  \right)  $ and so assume that : $g\left(  p\right)
\left(  u_{p},v_{p}\right)  =g\left(  p\right)  \left(  Ju_{p},Jv_{p}\right)
$

ii) The \textbf{fundamental 2-form} is then defined as :

$\varpi\left(  p\right)  \left(  u_{p},v_{p}\right)  =g\left(  p\right)
\left(  u_{p},Jv_{p}\right)  $

This is a 2-form, which is invariant by J and non degenerate if g is non
degenerate. It defines a structure of symplectic manifold over M.

\newpage

\section{PSEUDO-RIEMANNIAN MANIFOLDS}

So far we have not defined a metric on manifolds. The way to define a metric
on the topological space M is to define a differentiable norm on the tangent
bundle.\ If M is a real manifold and the norm comes from a bilinear positive
definite form we have a Riemanian manifold, which is the equivalent of an
euclidean vector space (indeed M is then modelled on an euclidean vector
space). Riemannian manifolds have been the topic of many studies and in the
litterature most of the results are given in this context.\ Unfortunately for
the physicists the Universe of General Relativity is not riemannian but
modelled on a Minkovski space.\ Most, but not all, the results stand if there
is a non degenerate, but non positive definite, metric on M. So we will strive
to stay in this more general context.

One of the key point of pseudo-riemanian manifolds is the isomorphism with the
dual, which requires finite dimensional manifolds. So in this section we will
assume that the manifolds are finite dimensional. For infinite dimensional
manifold the natural extension is Hilbert structure.

\subsection{General properties}

\label{General pseudo riemannian}

\subsubsection{Definitions}

\begin{definition}
A \textbf{pseudo-riemannian manifold} (M,g) is a real finite dimensional
manifold M endowed with a (0,2) symmetric tensor which induces a bilinear
symmetric non degenerate form g on TM.
\end{definition}

Thus g has a signature (+p,-q) with p+q=dimM, and we will say that M is a
pseudo-riemannian manifold of signature (p,q).

\begin{definition}
A \textbf{riemannian manifold} (M,g) is a real finite dimensional manifold M
endowed with a (0,2) symmetric tensor which induces a bilinear symmetric
definite positive form g on TM.
\end{definition}

Thus a riemannian manifold is a pseudo riemannian manifold of signature (m,0).

The manifold and g will be assumed to be at least of class 1. In the following
if not otherwise specified M is a pseudo-riemannian manifold. It will be
specified when a theorem holds for riemannian manifold only.

Any real finite dimensional Hausdorff manifold which is either paracompact or
second countable admits a riemannian metric.

Any open subset M of a pseudo-riemannian manifold (N,g) is a a
pseudo-riemannian manifold (M,g%
$\vert$%
$_{M}).$

The bilinear form is called a \textbf{scalar product}, and an \textbf{inner
product} if it is definite positive. It is also usually called the
\textbf{metric} (even if it is not a metric in the topological meaning)

The coordinate expressions are in holonomic bases:%

\begin{equation}
g\in\otimes_{2}^{0}TM:g\left(  p\right)  =\sum_{\alpha\beta}g_{\alpha\beta
}\left(  p\right)  dx^{\alpha}\otimes dx^{\beta}%
\end{equation}

$g_{\alpha\beta}=g_{\beta\alpha}$

$u_{p}\in T_{p}M:\forall v_{p}\in T_{p}M:g\left(  p\right)  \left(
u_{p},v_{p}\right)  =0\Rightarrow u_{p}=0\Leftrightarrow\det\left[  g\left(
p\right)  \right]  \neq0$

\paragraph{Isomorphism between the tangent and the cotangent bundle\newline}

\begin{theorem}
A scalar product g on a finite dimensional real manifold M defines an
isomorphim between the tangent space $T_{p}M$\ and the cotangent space
$T_{p}M^{\ast}$ at any point, and then an isomorphism \j\ between the tangent
bundle TM and the cotangent bundle TM*.
\end{theorem}

$j:TM\rightarrow TM^{\ast}::u_{p}\in T_{p}M\rightarrow\mu_{p}=\jmath\left(
u_{p}\right)  \in T_{p}M^{\ast}::$

$\forall v_{p}\in T_{p}M:g\left(  p\right)  \left(  u_{p},v_{p}\right)
=\mu_{p}\left(  v_{p}\right)  $

g induces a scalar product g* on the cotangent bundle,

$g^{\ast}\left(  p\right)  \left(  \mu_{p},\lambda_{p}\right)  =\sum
_{\alpha\beta}\mu_{p\alpha}\lambda_{p\beta}g^{\alpha\beta}\left(  p\right)  $

which is defined by the (2,0) symmetric tensor on M:%

\begin{equation}
g^{\ast}=\sum_{\alpha\beta}g^{\alpha\beta}\left(  p\right)  \partial
x_{\alpha}\otimes\partial x_{\beta}%
\end{equation}

with : $\sum_{\beta}g^{\alpha\beta}\left(  p\right)  g_{\beta\gamma}\left(
p\right)  =\delta_{\gamma}^{\alpha}$ so the matrices of g and g* are inverse
from each other : $\left[  g^{\ast}\right]  =\left[  g\right]  ^{-1} $

For any vector : $u_{p}=\sum_{\alpha}u_{p}^{\alpha}\partial x_{\alpha}\in
T_{p}M:\mu_{p}=\jmath\left(  u_{p}\right)  =\sum_{\alpha\beta}g_{\alpha\beta
}u_{p}^{\beta}dx^{\alpha}$

and conversely : $\mu_{p}=\sum_{\alpha}\mu_{p\alpha}dx^{\alpha}\in
T_{p}M^{\ast}\rightarrow\jmath^{-1}\left(  \mu_{p}\right)  =u_{p}=\sum
_{\alpha\beta}g^{\alpha\beta}\mu_{\beta}\partial x_{\alpha} $

The operation can be done with any mix tensor. Say that one "lifts" or
"lowers" the indices with g.

If $f\in C_{1}\left(  M;%
\mathbb{R}
\right)  $ the gradient of f is the vector field grad(f) such that :

$\forall u\in\mathfrak{X}\left(  TM\right)  :g\left(  p\right)  \left(
gradf,u\right)  =f^{\prime}(p)u\Leftrightarrow\left(  gradf\right)  ^{\alpha
}=\sum_{\beta}g^{\alpha\beta}\partial_{\beta}f$

\paragraph{Orthonormal basis}

\begin{theorem}
A pseudo-riemannian manifold admits an \textbf{orthonormal basis} at each
point :
\end{theorem}

$\forall p:\exists\left(  e_{i}\right)  _{i=1}^{m},e_{i}\in T_{p}M:g\left(
p\right)  \left(  e_{i},e_{j}\right)  =\eta_{ij}=\pm\delta_{ij}$

The coefficients $\eta_{ij}$\ define the signature of the metric, they do not
depend on the choice of the orthonormal basis or p. We will denote by $\left[
\eta\right]  $ the matrix $\eta_{ij}$ so that for any orthonormal basis :
$\left[  E\right]  =\left[  e_{i}^{\alpha}\right]  ::\left[  E\right]
^{t}\left[  g\right]  \left[  E\right]  =\left[  \eta\right]  $

Warning ! Even if one can find an orthonomal basis at each point, usually
there is no chart such that the holonomic basis is orthonormal at each point.
And there is no distribution of m vector fields which are orthonormal at each
point if M is not parallelizable.

\paragraph{Volume form\newline}

At each point p a volume form is a m-form $\varpi_{p}$ such that $\varpi
_{p}\left(  e_{1},...,e_{m}\right)  =+1$ for any orthonormal basis
(cf.Algebra). Such a form is given by :%

\begin{equation}
\varpi_{0}\left(  p\right)  =\sqrt{\left\vert \det g\left(  p\right)
\right\vert }dx^{1}\wedge dx^{2}...\wedge dx^{m}%
\end{equation}

As it never vanishes, this is a \textbf{volume form} (with the meaning used
for integrals) on M, and a pseudo-riemanian manifold is orientable if it is
the union of countably many compact sets.

\paragraph{Divergence\newline}

\begin{theorem}
The \textbf{divergence} of a vector field V is the function $div(V)\in
C\left(  M;%
\mathbb{R}
\right)  $ such that : $\pounds _{V}\varpi_{0}=\left(  divV\right)  \varpi_{0}
$ and

$divV=\sum_{\alpha}\left(  \partial_{\alpha}V^{\alpha}+V^{\alpha}\frac{1}%
{2}\sum_{\beta\gamma}g^{\gamma\beta}\left(  \partial_{\alpha}g_{\beta\gamma
}\right)  \right)  $
\end{theorem}

\begin{proof}
$divV=\frac{1}{\varpi_{0}}\sum_{\alpha}\partial_{\alpha}\left(  V^{\alpha
}\varpi_{0}\right)  $ (see Integral)

So : $divV=\frac{1}{\sqrt{\left\vert \det g\left(  p\right)  \right\vert }%
}\sum_{\alpha}\partial_{\alpha}\left(  V^{\alpha}\sqrt{\left\vert \det
g\left(  p\right)  \right\vert }\right)  $

$=\sum_{\alpha}\partial_{\alpha}V^{\alpha}+V^{\alpha}\frac{1}{\sqrt{\left\vert
\det g\left(  p\right)  \right\vert }}\partial_{\alpha}\sqrt{\left\vert \det
g\left(  p\right)  \right\vert }$

$=\sum_{\alpha}\partial_{\alpha}V^{\alpha}+V^{\alpha}\frac{1}{\sqrt{\left\vert
\det g\left(  p\right)  \right\vert }}\frac{1}{2}\frac{1}{\sqrt{\left\vert
\det g\left(  p\right)  \right\vert }}\left(  -1\right)  ^{p}\left(  \det
g\right)  Tr\left(  \left(  \frac{\partial}{\partial x^{\alpha}}\left[
g\right]  \right)  \left[  g\right]  ^{-1}\right)  $

$=\sum_{\alpha}\partial_{\alpha}V^{\alpha}+V^{\alpha}\frac{1}{2}Tr\left(
\left[  \sum_{\gamma}\left(  \partial_{\alpha}g_{\beta\gamma}\right)
g^{\gamma\eta}\right]  \right)  $

$=\sum_{\alpha}\left(  \partial_{\alpha}V^{\alpha}+V^{\alpha}\frac{1}{2}%
\sum_{\beta\gamma}g^{\gamma\beta}\left(  \partial_{\alpha}g_{\beta\gamma
}\right)  \right)  $
\end{proof}

\paragraph{Complexification\newline}

It is always possible to define a complex structure on the tangent bundle of a
real manifold by complexification. The structure of the manifold stays the
same, only the tangent bundle is involved.

If (M,g) is pseudo-riemannian then, pointwise, g(p) can be extended to a
hermitian, sequilinear, non degenerate form $\gamma\left(  p\right)  $ (see
Algebra) $:$

$\forall u,v\in T_{p}M:$

$\gamma\left(  p\right)  \left(  u,v\right)  =g\left(  p\right)  \left(
u,v\right)  ;\gamma\left(  p\right)  \left(  iu,v\right)  =-ig\left(
p\right)  \left(  u,v\right)  ;\gamma\left(  p\right)  \left(  u,iv\right)
=ig\left(  p\right)  \left(  u,v\right)  $

$\gamma$ defines a tensor field on the complexified tangent bundle
$\mathfrak{X}\left(  \otimes_{2}TM_{%
\mathbb{C}
}^{\ast}\right)  .$ The holonomic basis stays the same (with complex
components) and $\gamma$ has same components as g.

Most of the operations on the complex bundle can be extended, as long as they
do not involve the manifold structure itself (such as derivation). We will use
it in this section only for the Hodge duality, because the properties will be
useful in Functional Analysis. Of course if M is also a complex manifold the
extension is straightforward.

\subsubsection{Hodge duality}

Here we use the extension of a symmetric bilinear form g to a hermitian,
sequilinear, non degenerate form that we still denote g$.$ The field K is $%
\mathbb{R}
$ or $%
\mathbb{C}
.$

\paragraph{Scalar product of r-forms\newline}

This is the direct application of the definitions and results of the Algebra part.

\begin{theorem}
On a finite dimensional manifold (M,g) endowed with a scalar product the map :

$G_{r}:\mathfrak{X}\left(  \Lambda_{r}TM^{\ast}\right)  \times\mathfrak{X}%
\left(  \Lambda_{r}TM^{\ast}\right)  \rightarrow K::$%

\begin{equation}
G_{r}\left(  \lambda,\mu\right)  =\sum_{\left\{  \alpha_{1}..\alpha
_{r}\right\}  \left\{  \beta_{1}..\beta_{r}\right\}  }\overline{\lambda
}_{\alpha_{1}..\alpha_{r}}\mu_{\beta_{1}...\beta_{r}}\det\left[
g^{-1}\right]  ^{\left\{  \alpha_{1}..\alpha_{r}\right\}  ,\left\{  \beta
_{1}..\beta_{r}\right\}  }%
\end{equation}

is a non degenerate hermitian form and defines a scalar product which does not
depend on the basis.

It is definite positive if g is definite positive
\end{theorem}

In the matrix $\left[  g^{-1}\right]  $\ one takes the elements $g^{\alpha
_{k}\beta_{l}}$ with $\alpha_{k}\in\left\{  \alpha_{1}..\alpha_{r}\right\}
,\beta_{l}\in\left\{  \beta_{1}..\beta_{r}\right\}  $

$G_{r}\left(  \lambda,\mu\right)  =\sum_{\left\{  \alpha_{1}...\alpha
_{r}\right\}  }\overline{\lambda}_{\{\alpha_{1}...\alpha_{r}\}}\sum_{\beta
_{1}...\beta_{r}}g^{\alpha_{1}\beta_{1}}...g^{\alpha_{r}\beta_{r}}\mu
_{\beta_{1}...\beta_{r}}$

$=\sum_{\left\{  \alpha_{1}...\alpha_{r}\right\}  }\overline{\lambda
}_{\{\alpha_{1}...\alpha_{r}\}}\mu^{\left\{  \beta_{1}\beta_{2}...\beta
_{r}\right\}  }$

where the indexes are lifted and lowered with g.

The result does not depend on the basis.

\begin{proof}
In a change of charts for a r-form :

$\lambda=\sum_{\left\{  \alpha_{1}...\alpha_{r}\right\}  }\lambda_{\alpha
_{1}...\alpha_{r}}dx^{\alpha_{1}}\wedge dx^{\alpha_{2}}\wedge...\wedge
dx^{\alpha_{r}}$

$=\sum_{\left\{  \alpha_{1}...\alpha_{r}\right\}  }\widehat{\lambda}%
_{\alpha_{1}...\alpha_{r}}dy^{\alpha_{1}}\wedge dy^{\alpha_{2}}\wedge...\wedge
dy^{\alpha_{r}}$

with $\widehat{\lambda}_{\alpha_{1}...\alpha_{r}}=\sum_{\left\{  \beta
_{1}....\beta_{r}\right\}  }\lambda_{\beta_{1}...\beta_{r}}\det\left[
J^{-1}\right]  _{\alpha_{1}...\alpha_{r}}^{\beta_{1}...\beta_{r}}$

$G_{r}(\lambda,\mu)=\sum_{\left\{  \alpha_{1}..\alpha_{r}\right\}  \left\{
\beta_{1}..\beta_{r}\right\}  }\overline{\widehat{\lambda}}_{\alpha
_{1}..\alpha_{r}}\widehat{\mu}_{\beta_{1}...\beta_{r}}\det\left[  \widehat
{g}^{-1}\right]  ^{\left\{  \alpha_{1}..\alpha_{r}\right\}  ,\left\{
\beta_{1}..\beta_{r}\right\}  }$

$=\sum_{\left\{  \alpha_{1}..\alpha_{r}\right\}  \left\{  \beta_{1}..\beta
_{r}\right\}  }\sum_{\left\{  \gamma_{1}....\gamma_{r}\right\}  }%
\overline{\lambda}_{\gamma_{1}...\gamma_{r}}\det\left[  J^{-1}\right]
_{\alpha_{1}...\alpha_{r}}^{\gamma_{1}...\gamma_{r}}$

$\times\sum_{\left\{  \eta_{1}....\eta_{r}\right\}  }\widehat{\mu}_{\eta
_{1}...\eta_{r}}\det\left[  J^{-1}\right]  _{\beta_{1}...\beta_{r}}^{\eta
_{1}...\eta_{r}}\det\left[  \widehat{g}^{-1}\right]  ^{\left\{  \alpha
_{1}..\alpha_{r}\right\}  ,\left\{  \beta_{1}..\beta_{r}\right\}  }$

$=\sum_{\left\{  \gamma_{1}....\gamma_{r}\right\}  \left\{  \eta_{1}%
....\eta_{r}\right\}  }\overline{\lambda}_{\gamma_{1}...\gamma_{r}}%
\widehat{\mu}_{\eta_{1}...\eta_{r}}$

$\times\sum_{\left\{  \alpha_{1}..\alpha_{r}\right\}  \left\{  \beta
_{1}..\beta_{r}\right\}  }\det\left[  J^{-1}\right]  _{\alpha_{1}...\alpha
_{r}}^{\gamma_{1}...\gamma_{r}}\det\left[  J^{-1}\right]  _{\beta_{1}%
...\beta_{r}}^{\eta_{1}...\eta_{r}}\det\left[  \widehat{g}^{-1}\right]
^{\left\{  \alpha_{1}..\alpha_{r}\right\}  ,\left\{  \beta_{1}..\beta
_{r}\right\}  }$

$\widehat{g}_{\alpha\beta}=\left[  J^{-1}\right]  _{\alpha}^{\gamma}\left[
J^{-1}\right]  _{\beta}^{\eta}g_{\gamma\eta}$

$\det\left[  \widehat{g}^{-1}\right]  ^{\left\{  \alpha_{1}..\alpha
_{r}\right\}  ,\left\{  \beta_{1}..\beta_{r}\right\}  }=\det\left[
g^{-1}\right]  ^{\left\{  \gamma_{1}..\gamma_{r}\right\}  ,\left\{  \eta
_{1}..\eta_{r}\right\}  }\det\left[  J\right]  _{\gamma_{1}...\gamma_{r}%
}^{\alpha_{1}...\alpha_{r}}\det\left[  J\right]  _{\eta_{1}...\eta_{r}}%
^{\beta_{1}...\beta_{r}}$
\end{proof}

In an orthonormal basis :

$G_{r}\left(  \lambda,\mu\right)  =\sum_{\left\{  \alpha_{1}..\alpha
_{r}\right\}  \left\{  \beta_{1}..\beta_{r}\right\}  }\overline{\lambda
}_{\alpha_{1}..\alpha_{r}}\mu_{\beta_{1}...\beta_{r}}\eta^{\alpha_{1}\beta
_{1}}...\eta^{\alpha_{r}\beta_{r}}$

For $r=1$ one gets the usual bilinear symmetric form over $\mathfrak{X}\left(
\otimes_{1}^{0}TM\right)  :$

$G_{1}\left(  \lambda,\mu\right)  =\sum_{\alpha\beta}\overline{\lambda
}_{\alpha}\mu_{\beta}g^{\alpha\beta}$

For r=m : $G_{m}\left(  \lambda,\mu\right)  =\overline{\lambda}\mu\left(  \det
g\right)  ^{-1}$

\begin{theorem}
For a 1 form $\pi$ fixed in $\mathfrak{X}\left(  \Lambda_{1}TM^{\ast}\right)
$, the map :

$\lambda\left(  \pi\right)  :\mathfrak{X}\left(  \Lambda_{r}TM^{\ast}\right)
\rightarrow\mathfrak{X}\left(  \Lambda_{r+1}TM^{\ast}\right)  ::\lambda\left(
\pi\right)  \mu=\pi\wedge\mu$

has an adjoint with respect to the scalar product of forms :

$G_{r+1}\left(  \lambda\left(  \pi\right)  \mu,\mu^{\prime}\right)
=G_{r}\left(  \mu,\lambda^{\ast}\left(  \pi\right)  \mu^{\prime}\right)  $
which is

$\lambda^{\ast}\left(  \pi\right)  :\mathfrak{X}\left(  \Lambda_{r}TM^{\ast
}\right)  \rightarrow\mathfrak{X}\left(  \Lambda_{r-1}TM^{\ast}\right)
::\lambda^{\ast}\left(  \pi\right)  \mu=i_{grad\pi}\mu$
\end{theorem}

It suffices to compute the two quantities.

\paragraph{Hodge duality\newline}

\begin{theorem}
On a m dimensional manifold (M,g) endowed with a scalar product, with the
volume form $\varpi_{0}$ the map :

$\ast:\mathfrak{X}\left(  \Lambda_{r}TM^{\ast}\right)  \rightarrow
\mathfrak{X}\left(  \Lambda_{m-r}TM^{\ast}\right)  $ defined by the condition%

\begin{equation}
\forall\mu\in\mathfrak{X}\left(  \Lambda_{r}TM^{\ast}\right)  :\ast\lambda
_{r}\wedge\mu=G_{r}\left(  \lambda,\mu\right)  \varpi_{0}%
\end{equation}

is an anti-isomorphism
\end{theorem}

A direct computation gives the value of the \textbf{Hodge dual} $\ast\lambda$
in a holonomic basis :

$\ast\left(  \sum_{\left\{  \alpha_{1}...\alpha_{r}\right\}  }\lambda
_{\{\alpha_{1}...\alpha_{r}\}}dx^{\alpha_{1}}\wedge..\wedge dx^{\alpha_{r}%
}\right)  =$

$\sum_{\left\{  \alpha_{1}..\alpha_{n-r}\right\}  \left\{  \beta_{1}%
..\beta_{r}\right\}  }\epsilon\left(  \beta_{1}..\beta_{r},\alpha
_{1},...\alpha_{m-r}\right)  \overline{\lambda}^{\beta_{1}...\beta_{r}}%
\sqrt{\left\vert \det g\right\vert }dx^{\alpha_{1}}\wedge dx^{\alpha_{2}%
}...\wedge dx^{\alpha_{m-r}}$

( $\det g$\ is always real)

For r=0:

$\ast\lambda=\overline{\lambda}\varpi_{0}$

For r=1 :

$\ast\left(  \sum_{\alpha}\lambda_{\alpha}dx^{\alpha}\right)  =\sum_{\beta
=1}^{m}\left(  -1\right)  ^{\beta+1}g^{\alpha\beta}\overline{\lambda}_{\beta
}\sqrt{\left\vert \det g\right\vert }dx^{1}\wedge..\widehat{dx^{\beta}}%
\wedge...\wedge dx^{m}$

For r=m-1:

$\ast\left(  \sum_{\alpha=1}^{m}\lambda_{1..\widehat{\alpha}...m}dx^{1}%
\wedge..\widehat{dx^{\alpha}}\wedge...\wedge dx^{m}\right)  =\sum_{\alpha
=1}^{m}\left(  -1\right)  ^{\alpha-1}\overline{\lambda}^{1..\widehat{\alpha
}...n}\sqrt{\left\vert \det g\right\vert }dx^{\alpha}$

For r=m:

$\ast\left(  \lambda dx^{1}\wedge....\wedge dx^{m}\right)  =\epsilon\frac
{1}{\sqrt{\left\vert \det g\right\vert }}\overline{\lambda}$ With
$\epsilon=sign\det\left[  g\right]  $

\bigskip

\begin{theorem}
The inverse of the map * is :

$\ast^{-1}:\mathfrak{X}\left(  \Lambda_{r}TM^{\ast}\right)  \rightarrow
\mathfrak{X}\left(  \Lambda_{m-r}TM^{\ast}\right)  ::$

$\ast^{-1}\lambda_{r}=\epsilon(-1)^{r\left(  n-r\right)  }\ast\lambda
_{r}\Leftrightarrow\ast\ast\lambda_{r}=\epsilon(-1)^{r\left(  n-r\right)
}\lambda_{r}$

$G_{q}(\lambda,\ast\mu)=G_{n-q}(\ast\lambda,\mu)$

$G_{n-q}(\ast\lambda,\ast\mu)=G_{q}(\lambda,\mu)$
\end{theorem}

\paragraph{Codifferential\newline}

\begin{definition}
On a m dimensional manifold (M,g) endowed with a scalar product, with the
volume form $\varpi_{0}$, the \textbf{codifferential} is the operator :

$\delta:\mathfrak{X}\left(  \Lambda_{r+1}TM^{\ast}\right)  \rightarrow
\mathfrak{X}\left(  \Lambda_{r}TM^{\ast}\right)  ::\delta\lambda
=\epsilon(-1)^{r(m-r)+r}\ast d\ast\lambda=\left(  -1\right)  ^{r}\ast
d\ast^{-1}\lambda$

where $\epsilon=\left(  -1\right)  ^{p}$ with p the number of - in the
signature of g
\end{definition}

It has the following properties :

$\delta^{2}=0$

For $f\in C\left(  M;%
\mathbb{R}
\right)  :\delta f=0$

For $\lambda_{r}\in\mathfrak{X}\left(  \Lambda_{r+1}TM^{\ast}\right)
:\ast\delta\lambda=\left(  -1\right)  ^{m-r-1}d\ast\lambda$

$\left(  \delta\lambda\right)  _{\left\{  \gamma_{1}..\gamma_{r-1}\right\}  }
$

$=\epsilon(-1)^{r\left(  m-r\right)  }\sqrt{\left\vert \det g\right\vert }%
\sum_{\left\{  \eta_{1}..\eta_{m-r+1}\right\}  }\epsilon\left(  \eta_{1}%
..\eta_{m-r+1},\gamma_{1},...\gamma_{r-1}\right)  $

$\times\sum g^{\eta_{1}\beta_{1}}...g^{\eta_{m-r+1}\beta_{m-r+1}}$

$\times\sum_{k=1}^{r+1}(-1)^{k-1}\sum_{\left\{  \alpha_{1}..\alpha
_{r}\right\}  }\epsilon\left(  \alpha_{1}..\alpha_{r},\beta_{1},.\widehat
{\beta}_{k}..\beta_{m-r+1}\right)  \partial_{\beta_{k}}\left(  \lambda
^{\alpha_{1}...\alpha_{r}}\sqrt{\left\vert \det g\right\vert }\right)  $

For r=1 : $\delta\left(  \sum_{i}\lambda_{\alpha}dx^{\alpha}\right)
=(-1)^{m}\frac{1}{\sqrt{\left\vert \det g\right\vert }}\sum_{\alpha,\beta
=1}^{m}\partial_{\alpha}\left(  g^{\alpha\beta}\lambda_{\beta}\sqrt{\left\vert
\det g\right\vert }\right)  $

\begin{proof}
$\delta\left(  \sum_{i}\lambda_{\alpha}dx^{\alpha}\right)  =\epsilon
(-1)^{m}\ast d\ast\left(  \sum_{i}\lambda_{\alpha}dx^{\alpha}\right)  $

$=\epsilon(-1)^{m}\ast d\left(  \sum_{\alpha=1}^{m}\left(  -1\right)
^{\alpha+1}g^{\alpha\beta}\overline{\lambda_{\beta}}\sqrt{\left\vert \det
g\right\vert }dx^{1}\wedge..\widehat{dx^{\alpha}}\wedge...\wedge
dx^{m}\right)  $

$=\epsilon(-1)^{m}\ast\sum_{\alpha,\beta,\gamma=1}^{m}\left(  -1\right)
^{\alpha+1}\partial_{\beta}\left(  g^{\alpha\gamma}\overline{\lambda_{\gamma}%
}\sqrt{\left\vert \det g\right\vert }\right)  dx^{\beta}\wedge dx^{1}%
\wedge..\widehat{dx^{\alpha}}\wedge...\wedge dx^{m}$

$=\epsilon(-1)^{m}\ast\sum_{\alpha,\beta=1}^{m}\partial_{\alpha}\left(
g^{\alpha\beta}\overline{\lambda_{\beta}}\sqrt{\left\vert \det g\right\vert
}\right)  dx^{1}\wedge....\wedge dx^{m}$

$=(-1)^{m}\frac{1}{\sqrt{\left\vert \det g\right\vert }}\sum_{\alpha,\beta
=1}^{m}\partial_{\alpha}\left(  g^{\alpha\beta}\lambda_{\beta}\sqrt{\left\vert
\det g\right\vert }\right)  $
\end{proof}

The codifferential is the adjoint of the exterior derivative with respect to
the interior product $G_{r}$ (see Functional analysis).

\paragraph{Laplacian\newline}

\begin{definition}
On a m dimensional manifold (M,g) endowed with a scalar product the
\textbf{Laplace-de Rahm} operator is :

$\Delta:$ $\mathfrak{X}\left(  \Lambda_{r}TM^{\ast}\right)  \rightarrow
\mathfrak{X}\left(  \Lambda_{r}TM^{\ast}\right)  ::\Delta=-\left(  \delta
d+d\delta\right)  =-\left(  d+\delta\right)  ^{2}$
\end{definition}

Remark : one finds also the definition $\Delta=\left(  \delta d+d\delta
\right)  .$

Properties : see Functional analysis

\subsubsection{Isometries}

\begin{definition}
A class 1 map $f:M\rightarrow N$ between the pseudo-riemannian manifolds
(M,g),(N,h) is \textbf{isometric} at $p\in M$ if :

$\forall u_{p},v_{p}\in T_{p}M:h\left(  f\left(  p\right)  \right)  \left(
f^{\prime}(p)u_{p},f^{\prime}(p)v_{p}\right)  =g\left(  p\right)  \left(
u_{p},v_{p}\right)  $
\end{definition}

Then f'(p) is injective. If f is isometric on M this is an immersion, and if
it is bijective, this an embedding.

\begin{definition}
An \textbf{isometry} is a class 1 bijective map on the pseudo-riemannian
manifolds (M,g)\ which is isometric for all p in M
\end{definition}

The isometries play a specific role in that they define the symmetries of the manifold.

\begin{theorem}
(Kobayashi I p.162) An isometry maps geodesics to geodesics, and orthonormal
bases to orthonormal bases.
\end{theorem}

\paragraph{Killing vector fields\newline}

\begin{definition}
A \textbf{Killing vector field} is a vector field on a pseudo-riemannian
manifold which is the generator of a one parameter group of isometries.
\end{definition}

For any t in its domain of definition, $\Phi_{V}\left(  t,p\right)  $ is an
isometry on M.

A Killing vector field is said to be complete if its flow is complete (defined
over all $%
\mathbb{R}
).$

\begin{theorem}
(Kobayashi I p.237) For a vector field V on a pseudo-riemannian manifold (M,g)
the followings are equivalent :

i) V is a Killing vector field

ii) $\pounds _{V}g=0$

iii) $\forall Y,Z\in\mathfrak{X}\left(  TM\right)  :g\left(  \left(
\pounds _{V}-\nabla_{V}\right)  Y,Z\right)  =-g\left(  \left(  \pounds _{V}%
-\nabla_{V}\right)  Z,Y\right)  $ where $\nabla$ is the Levy-Civita connection

iv) $\forall\alpha,\beta:\sum_{\gamma}\left(  g_{\gamma\beta}\partial_{\alpha
}V^{\gamma}+g_{\alpha\gamma}\partial_{\beta}V^{\gamma}+V^{\gamma}%
\partial_{\gamma}g_{\alpha\beta}\right)  =0$
\end{theorem}

\begin{theorem}
(Wald p.442) If V is a Killing vector field and $c^{\prime}(t)$\ the tangent
vector to a geodesic then $g\left(  c\left(  t\right)  \right)  \left(
c^{\prime}(t),V\right)  =Ct$
\end{theorem}

\paragraph{Group of isometries\newline}

\begin{theorem}
(Kobayashi I p.238) The set of vector fields $\mathfrak{X}\left(  M\right)  $
over a m dimensional real pseudo-riemannian manifold (M,g) has a structure of
Lie algebra (infinite dimensional) with the commutator as bracket. The set of
Killing vector fields is a subalgebra of dimension at most equal to m(m+1)/2.
If it is equal to m(m+1)/2 then M is a space of constant curvature.

The set I(M) of isometries over M, endowed with the compact-open topology (see
Topology), is a Lie group whose Lie algebra is naturally isomorphic to the Lie
algebra of all complete Killing vector fields (and of same dimension). The
isotropy subgroup at any point is compact. If M is compact then the group I(M)
is compact.
\end{theorem}

\bigskip

\subsection{L\'{e}vi-Civita connection}

\label{Levy-Civita connection}

\subsubsection{Definitions}

\paragraph{Metric connection\newline}

\begin{definition}
A covariant derivative $\nabla$\ on a pseudo-riemannian manifold (M,g) is said
to be \textbf{metric} if $\nabla g=0$
\end{definition}

Then we have similarly for the metric on the cotangent bundle : $\nabla
g^{\ast}=0$

So : $\forall\alpha,\beta,\gamma:\nabla_{\gamma}g_{\alpha\beta}=\nabla
_{\gamma}g^{\alpha\beta}=0$

By simple computation we have the theorem :

\begin{theorem}
For a covariant derivative $\nabla$\ on a pseudo-riemannian manifold (M,g) the
following are equivalent :

i) the covariant derivative is metric

ii) $\forall\alpha,\beta,\gamma:\partial_{\gamma}g_{\alpha\beta}=\sum_{\eta
}\left(  g_{\alpha\eta}\Gamma_{\gamma\beta}^{\eta}+g_{\beta\eta}\Gamma
_{\gamma\alpha}^{\eta}\right)  $

iii) the covariant derivative preserves the scalar product of transported vectors

iv) the riemann tensor is such that :

$\forall X,Y,Z\in\mathfrak{X}\left(  TM\right)  :R(X,Y)Z+R(Y,Z)X+R(Z,X)Y=0$
\end{theorem}

\paragraph{L\'{e}vi-Civita connection\newline}

\begin{theorem}
On a pseudo-riemannian manifold (M,g) there is a unique affine connection,
called the \textbf{L\'{e}vi-Civita connection}, which is both torsion free and
metric. It is fully defined by the metric, through the relations :%

\begin{equation}
\Gamma_{\beta\gamma}^{\alpha}=\dfrac{1}{2}\sum_{\eta}(g^{\alpha\eta}\left(
\partial_{\beta}g_{\gamma\eta}+\partial_{\gamma}g_{\beta\eta}-\partial_{\eta
}g_{\beta\gamma}\right)
\end{equation}

$=-\dfrac{1}{2}\sum_{\eta}(g_{\gamma\eta}\partial_{\beta}g^{\alpha\eta
}+g_{\beta\eta}\partial_{\gamma}g^{\alpha\eta}+g^{\alpha\eta}\partial_{\eta
}g_{\beta\gamma})$
\end{theorem}

The demonstration is a straightforward computation from the property ii) above.

Warning ! these are the most common definitions for $\Gamma$, but they can
vary (mainly in older works) according to the definitions of the Christoffel
symbols. Those above are fully consistent with all other definitions used in
this book.

As it is proven in the Fiber bundle part, the only genuine feature of the
L\'{e}vy-Civita connection is that it is torsion free.

\subsubsection{Curvature}

With the L\'{e}vi-Civita connection many formulas take a simple form :

\begin{theorem}
$\Gamma_{\alpha\beta\gamma}=\sum_{\eta}g_{\alpha\eta}\Gamma_{\beta\gamma
}^{\eta}=\dfrac{1}{2}(\partial_{\beta}g_{\gamma\alpha}+\partial_{\gamma
}g_{\beta\alpha}-\partial_{\alpha}g_{\beta\gamma})$

$\partial_{\alpha}g_{\beta\gamma}=\sum_{\eta}g_{\gamma\eta}\Gamma_{\alpha
\beta}^{\eta}+g_{\beta\eta}\Gamma_{\alpha\gamma}^{\eta}$

$\partial_{\alpha}g^{\beta\gamma}=-\sum_{\eta}g^{\beta\eta}\Gamma_{\alpha\eta
}^{\gamma}+g^{\gamma\eta}\Gamma_{\alpha\eta}^{\beta}$

$\sum_{\gamma}\Gamma_{\gamma\alpha}^{\gamma}=\frac{1}{2}\frac{\partial
_{\alpha}\left\vert \det g\right\vert }{\left\vert \det g\right\vert }%
=\frac{\partial_{\alpha}\left(  \sqrt{\left\vert \det g\right\vert }\right)
}{\sqrt{\left\vert \det g\right\vert }}$
\end{theorem}

\begin{proof}
$\frac{d}{d\xi^{\alpha}}\det g=\left(  \frac{d}{dg_{\lambda\mu}}\det g\right)
\frac{d}{d\xi^{\alpha}}g_{\lambda\mu}=g^{\mu\lambda}\left(  \det g\right)
\frac{d}{d\xi^{\alpha}}g_{\lambda\mu}=\left(  \det g\right)  g^{\mu\lambda
}\partial_{\alpha}g_{\lambda\mu}$

$g^{\lambda\mu}\left(  \partial_{\alpha}g_{\lambda\mu}\right)  =g^{\lambda\mu
}\left(  g_{\mu l}\Gamma_{\alpha\lambda}^{l}+g_{\lambda l}\Gamma_{\alpha\mu
}^{l}\right)  =\left(  g^{\lambda\mu}g_{\mu l}\Gamma_{\alpha\lambda}%
^{l}+g^{\lambda\mu}g_{\lambda l}\Gamma_{\alpha\mu}^{l}\right)  =\left(
\Gamma_{\alpha\lambda}^{\lambda}+\Gamma_{\alpha\mu}^{\mu}\right)
=2\Gamma_{\gamma\alpha}^{\gamma}$

$\partial_{\alpha}\det g=2\left(  \det g\right)  \Gamma_{\gamma\alpha}%
^{\gamma}$

$\frac{d}{d\xi^{\alpha}}\sqrt{\left\vert \det g\right\vert }=\frac{d}%
{d\xi^{\alpha}}\sqrt{\left\vert \det g\right\vert }=\frac{1}{2\sqrt{\left\vert
\det g\right\vert }}\frac{d}{d\xi^{\alpha}}\left\vert \det g\right\vert
=\frac{1}{2\sqrt{\left\vert \det g\right\vert }}\left(  2\left(  -1\right)
^{p}\left(  \det g\right)  \Gamma_{\gamma\alpha}^{\gamma}\right)
=\sqrt{\left\vert \det g\right\vert }\Gamma_{\gamma\alpha}^{\gamma}$
\end{proof}

\begin{theorem}
$div(V)=\sum_{\alpha}\nabla_{\alpha}V^{\alpha}$
\end{theorem}

\begin{proof}
$div(V)=\sum_{\alpha}\partial_{\alpha}V^{\alpha}+V^{\alpha}\frac{1}%
{\sqrt{\left\vert \det g\left(  p\right)  \right\vert }}\partial_{\alpha}%
\sqrt{\left\vert \det g\left(  p\right)  \right\vert }=\sum_{\alpha}%
\partial_{\alpha}V^{\alpha}+V^{\alpha}\sum_{\beta}\Gamma_{\beta\alpha}^{\beta
}=\sum_{\alpha}\nabla_{\alpha}V^{\alpha}$
\end{proof}

\begin{theorem}
For the volume form $\varpi_{0}=\sqrt{\left\vert \det g\right\vert }$ :

$\nabla\varpi_{0}=0$

$\partial_{\beta}\varpi_{0}=\frac{1}{2}\frac{\partial_{\beta}\left\vert \det
g\right\vert }{\varpi_{0}}$

$\sum_{\gamma}\Gamma_{\gamma\alpha}^{\gamma}=\frac{\partial_{\alpha}\varpi
_{0}}{\varpi_{0}}$
\end{theorem}

\begin{proof}
$\left(  \nabla\varpi_{12...n}\right)  _{\alpha}=\dfrac{\partial
\varpi_{12...n}}{\partial x^{\alpha}}-\sum_{\beta=1}^{n}\sum_{l=1}^{n}%
\Gamma_{\alpha l}^{\beta}\varpi_{12..l-1,\beta,l+1...n}$

$=\partial_{\alpha}\sqrt{\left\vert \det g\right\vert }-\sum_{l=1}^{n}%
\Gamma_{\alpha l}^{l}\varpi_{12...n}=\partial_{\alpha}\sqrt{\left\vert \det
g\right\vert }-\varpi_{12...n}\frac{1}{2}\frac{1}{\det g}\partial_{\alpha}\det
g$

$=\partial_{\alpha}\sqrt{\left\vert \det g\right\vert }-\frac{1}{2}%
\sqrt{\left\vert \det g\right\vert }\frac{1}{\det g}\partial_{\alpha}(\det g)$

$\left\vert \det g\right\vert =\left(  -1\right)  ^{p}\det g$

$\left(  \nabla\varpi\right)  _{\alpha}=\left(  \frac{1}{2}\frac{\left(
-1\right)  ^{p}\partial_{\alpha}\det g}{\sqrt{\left\vert \det g\right\vert }%
}-\frac{1}{2}\sqrt{\left\vert \det g\right\vert }\frac{1}{\det g}%
\partial_{\alpha}\det g\right)  $

$=\frac{1}{2}\left(  \partial_{\alpha}\det g\right)  \left(  \frac{\left(
-1\right)  ^{p}}{\sqrt{\left\vert \det g\right\vert }}-\frac{(-1)^{p}}%
{\sqrt{\left\vert \det g\right\vert }}\right)  =0$
\end{proof}

For the Levi-Civitta connection the Riemann tensor, is :

$R_{\alpha\beta\gamma}^{\eta}=\dfrac{1}{2}\sum_{\lambda}\{g^{\eta\lambda
}\left(  \partial_{\alpha}\partial_{\gamma}g_{\beta\lambda}-\partial_{\alpha
}\partial_{\lambda}g_{\beta\gamma}-\partial_{\beta}\partial_{\gamma}%
g_{\alpha\lambda}+\partial_{\beta}\partial_{\lambda}g_{\alpha\gamma}\right)  $

$+\left(  \partial_{\alpha}g^{\eta\lambda}+\dfrac{1}{2}g^{\eta\varepsilon
}g^{s\lambda}\left(  \partial_{\alpha}g_{s\varepsilon}+\partial_{s}%
g_{\alpha\varepsilon}-\partial_{\varepsilon}g_{\alpha s}\right)  \right)
\left(  \partial_{\beta}g_{\gamma\lambda}+\partial_{\gamma}g_{\beta\lambda
}-\partial_{\lambda}g_{\beta\gamma}\right)  $

$-\left(  \partial_{\beta}g^{\eta\lambda}-\dfrac{1}{2}g^{\eta\varepsilon
}g^{s\lambda}\left(  \partial_{\beta}g_{s\varepsilon}+\partial_{s}%
g_{\beta\varepsilon}-\partial_{\varepsilon}g_{\beta s}\right)  \right)
\left(  \partial_{\alpha}g_{\gamma\lambda}+\partial_{\gamma}g_{\alpha\lambda
}-\partial_{\lambda}g_{\alpha\gamma}\right)  \}$

It has the properties :

\begin{theorem}
(Wald p.39)

$R_{\alpha\beta\gamma}^{\eta}+R_{\beta\gamma\alpha}^{\eta}+R_{\gamma
\alpha\beta}^{\eta}=0$

Bianchi's identity : $\nabla_{\alpha}R_{\beta\gamma\delta}^{\eta}%
+\nabla_{\gamma}R_{\alpha\beta\delta}^{\eta}+\nabla_{\beta}R_{\alpha
\gamma\delta}^{\eta}=0$

$R_{\alpha\beta\gamma\eta}=\sum_{\lambda}g_{\alpha\lambda}R_{\beta\gamma
\delta}^{\lambda}=-R_{\alpha\gamma\beta\eta}$

$R_{\alpha\beta\gamma\eta}+R_{\alpha\gamma\eta\beta}+R_{\alpha\eta\beta\gamma
}$
\end{theorem}

The Weyl's tensor is C such that :

$R_{\alpha\beta\gamma\delta}=C_{\alpha\beta\gamma\delta}+\frac{2}{n-2}\left(
g_{\alpha\lbrack\gamma}R_{\delta]\beta}-g_{\beta\lbrack\gamma}R_{\delta
]\alpha}\right)  -\frac{2}{\left(  n-1\right)  \left(  n-2\right)  }%
Rg_{\alpha\lbrack\gamma}g_{\delta]\beta}$

The Ricci tensor is :

Ric = $\sum_{\alpha\gamma}Ric_{\alpha\gamma}dx^{\alpha}\otimes dx^{\gamma}$

$Ric_{\alpha\gamma}=\sum_{\beta}R_{\alpha\beta\gamma}^{\beta}=\sum_{\eta
}\left(  \partial_{\alpha}\Gamma_{\eta\gamma}^{\eta}-\partial_{\eta}%
\Gamma_{\alpha\gamma}^{\eta}+\sum_{\varepsilon}\left(  \Gamma_{\alpha
\varepsilon}^{\eta}\Gamma_{\eta\gamma}^{\varepsilon}-\Gamma_{\eta\varepsilon
}^{\eta}\Gamma_{\alpha\gamma}^{\varepsilon}\right)  \right)  $

So it is symmetric : $Ric_{\alpha\gamma}=Ric_{\gamma\alpha}$

\begin{definition}
The \textbf{scalar curvature} is : $R=\sum_{\alpha\beta}g^{\alpha\beta
}Ric_{\alpha\beta}\in C\left(  M;%
\mathbb{R}
\right)  $
\end{definition}

\begin{definition}
A Riemannian manifold whose Levi-Civita connection is flat (the torsion and
the riemann tensor vanish on M) is said to be locally euclidean.
\end{definition}

\begin{definition}
An Einstein manifold is a pseudo-riemannian manifold whose Ricci tensor is
such that : $R_{\alpha\beta}\left(  p\right)  =\lambda\left(  p\right)
g_{\alpha\beta}$
\end{definition}

Then $R=Cte;\lambda=Cte$

\subsubsection{Sectional curvature}

(Kobayashi I p.200)

\begin{definition}
On a pseudo-riemannian manifold (M,g) the \textbf{sectional curvature} K(p) at
$p\in M$ is the scalar : $K(p)=g\left(  p\right)  \left(  R\left(  u_{1}%
,u_{2},u_{2}\right)  ,u_{1}\right)  $ where $u_{1},u_{2}$ are two orthonormal
vectors in $T_{p}M.$
\end{definition}

K(p) depends only on the plane P spanned by $u_{1},u_{2}.$

\begin{definition}
If M is connected, with dimension
$>$%
2, and K(p) is constant for all planes P in p, and for all p in M, then M is
called a \textbf{space of constant curvature}.
\end{definition}

So there are positive (resp.negative) curvature according to the sign of K(p).

Then :

$R(X,Y,Z)=K\left(  g\left(  Z,Y\right)  X-g\left(  Z,X\right)  Y\right)  $

$R_{\beta\gamma\eta}^{\alpha}=K\left(  \delta_{\gamma}^{\alpha}g_{\beta\eta
}-\delta_{\eta}^{\alpha}g_{\beta\gamma}\right)  $

\subsubsection{Geodesics}

Geodesics can be defined by different ways :

- by a connection, as the curves whose tangent is paralllel transported

- by a metric on a topological space

- by a scalar product on a pseudo-riemannian manifold

The three concepts are close, but not identical. In particular geodesics in
pseudo-riemannian manifolds have special properties used in General Relativity.

\paragraph{Length of a curve\newline}

A curve C is a 1 dimensional submanifold of a manifold M defined by a class 1
path : $c:\left[  a,b\right]  \rightarrow M$ such that c'(t)$\neq0.$

The volume form on C induced by a scalar product on (M,g) is

$\lambda\left(  t\right)  =\sqrt{\left\vert g\left(  c\left(  t\right)
\right)  \left(  c^{\prime}(t),c^{\prime}(t)\right)  \right\vert }dt=c_{\ast
}\varpi_{0}.$

So the "volume" of C is the length of the curve :

$\ell\left(  a,b\right)  =\int_{\left[  a,b\right]  }\varpi_{0}=\int
_{C}c_{\ast}\varpi_{0}=\int_{a}^{b}\sqrt{\left\vert g\left(  c\left(
t\right)  \right)  \left(  c^{\prime}(t),c^{\prime}(t)\right)  \right\vert
}dt$ , which is always finite because the function is the image of a compact
by a continuous function, thus bounded. And it does not depend on the parameter.

The sign of $g\left(  c\left(  t\right)  \right)  \left(  c^{\prime
}(t),c^{\prime}(t)\right)  $ does not depend of the parameter, it is
$>$
0 if M is Riemannian, but its sign can change on c if not. If, for
$t\in\left[  a,b\right]  $ :

$g\left(  c\left(  t\right)  \right)  \left(  c^{\prime}(t),c^{\prime
}(t)\right)  <0$ we say that the path is time like

$g\left(  c\left(  t\right)  \right)  \left(  c^{\prime}(t),c^{\prime
}(t)\right)  >0$ we say that the path is space like

$g\left(  c\left(  t\right)  \right)  \left(  c^{\prime}(t),c^{\prime
}(t)\right)  =0$ we say that the path is light like

Let us denote the set of paths $P=\left\{  c\in C_{1}\left(  \left[
a,b\right]  ;M\right)  ,c^{\prime}(t)\neq0\right\}  $ where [a,b] is any
compact interval in $%
\mathbb{R}
$ . For any c in P the image c([a,b]) is a compact, connected curve of finite
length which does not cross itself. P can be endowed with the open-compact
topology with base the maps $\varphi\in P$ such that the image of any compact
K of $%
\mathbb{R}
$ is included in an open subset of M. With this topology P is first countable
and Hausdorff (each c is itself an open).

The subsets :

$P^{+}$ such that $g\left(  c\left(  t\right)  \right)  \left(  c^{\prime
}(t),c^{\prime}(t)\right)  >0$

$P^{-}$ such that $g\left(  c\left(  t\right)  \right)  \left(  c^{\prime
}(t),c^{\prime}(t)\right)  <0$ \ 

are open in P.

\begin{theorem}
In a pseudo-riemannian manifold (M,g), the curve of extermal length, among all
the class 1 path from p to q in M of the same type, is a geodesic for the
L\'{e}vy-Civita connection.
\end{theorem}

\begin{proof}
So we restrict ourselves to the subset of P such that c(a)=p,c(b)=q. At first
a,b are fixed.

To find an extremal curve is a problem of variational calculus.

The map c in $P^{+/-}$ for which the functionnal :

$\ell\left(  a,b\right)  =\int_{a}^{b}\sqrt{\epsilon g\left(  c\left(
t\right)  \right)  \left(  c^{\prime}(t),c^{\prime}(t)\right)  }dt$ is
extremum is such that the derivative vanishes.

The Euler-Lagrange equations give with c'(t)=$\sum_{\alpha}u^{\alpha}\left(
t\right)  \partial x_{\alpha}:$

For $\alpha:\frac{1}{2}\frac{\epsilon\left(  \partial_{\alpha}g_{\beta\gamma
}\right)  u^{\gamma}u^{\beta}}{\sqrt{\epsilon g_{\lambda\mu}u^{\lambda}u^{\mu
}}}-\frac{d}{dt}\left(  \frac{\epsilon g_{\alpha\beta}u^{\beta}}%
{\sqrt{\epsilon g_{\lambda\mu}u^{\lambda}u^{\mu}}}\right)  =0$

Moreover the function : $L=\sqrt{\epsilon g\left(  c\left(  t\right)  \right)
\left(  c^{\prime}(t),c^{\prime}(t)\right)  }$ is homogeneous of degree 1 so
we have the integral $\sqrt{\epsilon g\left(  c\left(  t\right)  \right)
\left(  c^{\prime}(t),c^{\prime}(t)\right)  }=\theta=Ct$

The equations become :

$\frac{1}{2}\left(  \partial_{\alpha}g_{\beta\gamma}\right)  u^{\gamma
}u^{\beta}=\frac{d}{dt}\left(  g_{\alpha\beta}u^{\beta}\right)  =\left(
\frac{d}{dt}g_{\alpha\beta}\right)  u^{\beta}+g_{\alpha\beta}\frac{d}%
{dt}u^{\beta}$

using : $\partial_{\alpha}g_{\beta\gamma}=g_{\eta\gamma}\Gamma_{\alpha\beta
}^{\eta}+g_{\beta\eta}\Gamma_{\alpha\gamma}^{\eta}$ and $\frac{du^{\beta}}%
{dt}=u^{\gamma}\partial_{\gamma}u^{\beta},\frac{dg_{\alpha\beta}}{dt}=\left(
\partial_{\gamma}g_{\alpha\beta}\right)  u^{\gamma}=\left(  g_{\eta\beta
}\Gamma_{\gamma\alpha}^{\eta}+g_{\alpha\eta}\Gamma_{\gamma\beta}^{\eta
}\right)  u^{\gamma}$

$\frac{1}{2}\left(  g_{\eta\gamma}\Gamma_{\alpha\beta}^{\eta}+g_{\beta\eta
}\Gamma_{\alpha\gamma}^{\eta}\right)  u^{\gamma}u^{\beta}=\left(  g_{\eta
\beta}\Gamma_{\gamma\alpha}^{\eta}+g_{\alpha\eta}\Gamma_{\gamma\beta}^{\eta
}\right)  u^{\gamma}u^{\beta}+g_{\alpha\beta}u^{\gamma}\partial_{\gamma
}u^{\beta}$

$g_{\eta\gamma}\left(  \nabla_{\alpha}u^{\eta}-\partial_{\alpha}u^{\eta
}\right)  u^{\gamma}+g_{\beta\eta}\left(  \nabla_{\alpha}u^{\eta}%
-\partial_{\alpha}u^{\eta}\right)  u^{\beta}$

$=2\left(  g_{\eta\beta}\left(  \nabla_{\alpha}u^{\eta}-\partial_{\alpha
}u^{\eta}\right)  +g_{\alpha\eta}\left(  \nabla_{\beta}u^{\eta}-\partial
_{\beta}u^{\eta}\right)  \right)  u^{\beta}+2g_{\alpha\beta}u^{\gamma}%
\partial_{\gamma}u^{\beta}-2g_{\alpha\gamma}u^{\beta}\nabla_{\beta}u^{\gamma
}=0$

$\nabla_{u}u=0$

Thus a curve with an extremal length must be a geodesic, with an affine parameter.

If this an extremal for [a,b] it will be an extremal for any other affine parameter.
\end{proof}

\begin{theorem}
The quantity $g\left(  c\left(  t\right)  \right)  \left(  c^{\prime
}(t),c^{\prime}(t)\right)  $ is constant along a geodesic with an affine parameter.
\end{theorem}

\begin{proof}
Let us denote : $\theta(t)=g_{\alpha\beta}u^{\alpha}u^{\beta}$

$\dfrac{d\theta}{dt}=\left(  \partial_{\gamma}\theta\right)  u^{\gamma
}=\left(  \nabla_{\gamma}\theta\right)  u^{\gamma}=u^{\alpha}u^{\beta
}u^{\gamma}\left(  \nabla_{\gamma}g_{\alpha\beta}\right)  +g_{\alpha\beta
}(\nabla_{\gamma}u^{\alpha})u^{\beta}u^{\gamma}+g_{\alpha\beta}(\nabla
_{\gamma}u^{\beta})u^{\alpha}u^{\gamma}=2g_{\alpha\beta}(\nabla_{\gamma
}u^{\alpha})u^{\beta}u^{\gamma}=0$
\end{proof}

So a geodesic is of constant type, which is defined by its tangent at any
point. If there is a geodesic joining two points it is unique, and its type is
fixed by its tangent vector at p. Moreover at any point p has a convex
neighborhood n(p). To sum up :

\begin{theorem}
On a pseudo-riemannian manifold M, any point p has a convex neighborhood n(p)
in which the points q can be sorted according to the fact that they can be
reached by a geodesic which is either time like, space like or light like.
This geodesic is unique and is a curve of extremal length among the curves for
which $g\left(  c\left(  t\right)  \right)  \left(  c^{\prime}(t),c^{\prime
}(t)\right)  $ has a constant sign.
\end{theorem}

Remarks :

i) there can be curves of extremal length such that $g\left(  c\left(
t\right)  \right)  \left(  c^{\prime}(t),c^{\prime}(t)\right)  $ has not a
constant sign.

ii) one cannot say if the length is a maximum or a minimum

iii) as a loop cannot be a geodesic the relation p,q are joined by a geodesic
is not an equivalence relation, and therefore does not define a partition of M
(we cannot have $p\sim p$\ ).

iv) the definition of a geodesic depends on the connection. These results hold
for the L\'{e}vi-Civita connection.

\paragraph{General relativity context\newline}

In the context of general relativity (meaning g is of Lorentz type and M is
four dimensional) the set of time like vectors at any point is disconnected,
so it is possible to distinguish future oriented and past oriented time like
vectors.\ The manifold is said to be \textbf{time orientable} if it is
possible to make this distinction in a continuous manner all over M.

The future of a given point p is the set I(p) of all points q in p which can
be reached from p by a curve whose tangent is time like, future oriented.

There is a theorem saying that, in a convex neighborhood n(p) of p, I(p)
consists of all the points which can be reached by future oriented geodesic
staying in n(p).

In the previous result we could not exclude that a point q reached by a space
like geodesic could nevertheless be reached by a time like curve (which cannot
be a geodesic). See Wald p.191 for more on the subject.

\paragraph{Riemannian manifold\newline}

\begin{theorem}
(Kobayashi I p. 168) For a riemannian manifold (M,g) the map : $d:M\times
M\rightarrow%
\mathbb{R}
_{+}::d(p,q)=\min_{c}\ell\left(  p,q\right)  $ for all the piecewise class 1
paths from p to q is a metric on M, which defines a topology equivalent to the
topology of M. If the length of a curve between p and q is equal to d, this is
a geodesic. Any maping f in M which preserves d is an isometry.
\end{theorem}

\begin{theorem}
(Kobayashi I p. 172) For a connected Riemannian manifold M, the following are
equivalent :

i) the geodesics are complete (defined for $t\in%
\mathbb{R}
)$

ii) M is a complete topological space with regard to the metric d

iii) every bounded (with d) subset of M is relatively compact

iv) any two points can be joined by a geodesic (of minimal length)

v) any geodesic is infinitely extendable
\end{theorem}

As a consequence:

- a compact riemannian manifold is complete

- the affine parameter of a geodesic on a riemannian manifold is the arc
length $\ell.$

\bigskip

\subsection{Submanifolds}

\label{Submanifolds riemannian}

On a pseudo-riemannian manifold (N,g) g induces a bilinear symmetric form in
the tangent space of any submanifold M, but this form can be degenerate if it
is not definite positive. Similarly the L\'{e}vy-Civita connection does not
always induces an affine connection on a submanifold, even if N is riemmanian.

\subsubsection{Induced scalar product}

\begin{theorem}
If (N,g) is a real, finite dimensional pseudo-riemannian manifold, any
submanifold M, embedded into N by f, is endowed with a bilinear symmetric form
which is non degenerate at $p\in f\left(  M\right)  $ iff $\det\left[
f^{\prime}(p)\right]  ^{t}\left[  g\left(  p\right)  \right]  \left[
f^{\prime}(p)\right]  \neq0.$ It is non degenerate on f(M) if N is riemannian.
\end{theorem}

\begin{proof}
Let us denote f(M)=$\widehat{M}$ as a subset of N.

Because $\widehat{M}$ is a submanifold any vector $v_{q}\in T_{q}%
N,q\in\widehat{M}$ has a unique decomposition : $v_{q}=v_{q}^{\prime}+w_{q}$
with $v_{q}^{\prime}\in T_{q}\widehat{W}$

Because f is an embedding, thus a diffeomorphism, for any vector
$v_{q}^{\prime}\in T_{q}\widehat{W}$ there is $u_{p}\in T_{p}M,q=f(p):v_{q}%
^{\prime}=f^{\prime}\left(  p\right)  u_{p}$

So $v_{q}=f^{\prime}\left(  p\right)  u_{p}+w_{q}$

g reads for the vectors of $T_{q}\widehat{W}:g\left(  f\left(  p\right)
\right)  \left(  f^{\prime}\left(  p\right)  u_{p},f^{\prime}\left(  p\right)
u_{p}^{\prime}\right)  =f_{\ast}g\left(  p\right)  \left(  u_{p},u_{p}%
^{\prime}\right)  $

$h=f_{\ast}g$ is a bilinear symmetric form on M. And $T_{q}\widehat{W}$ is
endowed with the bilinear symmetric form which has the same matrix : $\left[
h\left(  p\right)  \right]  =\left[  f^{\prime}(p)\right]  ^{t}\left[
g\left(  p\right)  \right]  \left[  f^{\prime}(p)\right]  $ in an adaptated chart.

If g is positive definite : $\forall u_{p}^{\prime}:g\left(  f\left(
p\right)  \right)  \left(  f^{\prime}\left(  p\right)  u_{p},f^{\prime}\left(
p\right)  u_{p}^{\prime}\right)  =0\Rightarrow f^{\prime}\left(  p\right)
u_{p}=0\Leftrightarrow u_{p}\in\ker f^{\prime}(p)\Rightarrow u_{p}=0$
\end{proof}

If N is n dimensional, M m%
$<$%
n dimensional, there is a chart in which $\left[  f^{\prime}(p)\right]
_{n\times m}$ and $h_{\lambda\mu}=\sum_{\alpha\beta=1}^{n}g_{\alpha\beta
}\left(  f\left(  p\right)  \right)  \left[  f^{\prime}\left(  p\right)
\right]  _{\lambda}^{\alpha}\left[  f^{\prime}\left(  p\right)  \right]
_{\mu}^{\beta}$

If g is riemannian there is an orthogonal complement to each vector space
tangent at M.

\subsubsection{Covariant derivative}

1. With the notations of the previous subsection, h is a symmetric bilinear
form on M, so, whenever it is not degenerate (meaning $\left[  h\right]  $
invertible) it defines a Levi-Civita connection $\widehat{\nabla}$ on M:

$\widehat{\Gamma}_{\mu\nu}^{\lambda}=\dfrac{1}{2}\sum_{\rho}(h^{\lambda\rho
}\left(  \partial_{\mu}h_{\nu\rho}+\partial_{\nu}h_{\mu\rho}-\partial_{\rho
}h_{\mu\nu}\right)  $

A lenghty but straightforward computation gives the formula :

$\widehat{\Gamma}_{\mu\nu}^{\lambda}=\sum_{\alpha\beta\gamma}G_{\alpha
}^{\lambda}\left(  \partial_{\nu}F_{\mu}^{\alpha}+\Gamma_{\beta\gamma}%
^{\alpha}F_{\mu}^{\beta}F_{\nu}^{\gamma}\right)  $

with $\left[  F\right]  =\left[  f^{\prime}\left(  p\right)  \right]  $ and

$\left[  G\left(  p\right)  \right]  _{m\times n}=\left[  h\right]  _{m\times
m}^{-1}\left[  F\right]  _{m\times n}^{t}\left[  g\right]  _{n\times
n}\Rightarrow\left[  G\right]  \left[  F\right]  =\left[  h\right]
^{-1}\left[  F\right]  ^{t}\left[  g\right]  \left[  F\right]  =I_{m}$

This connection $\widehat{\nabla}$ is symmetric and metric, with respect to h.

2. A vector field U$\in\mathfrak{X}\left(  TM\right)  $ gives by push-forward
a vector field on $T\widehat{M}$ and every vector field on $T\widehat{M}$ is
of this kind.

The covariant derivative of such a vector field on N gives :

$\nabla_{f^{\ast}V}\left(  f^{\ast}U\right)  =X+Y\in\mathfrak{X}\left(
TN\right)  $ with $X=f^{\ast}U^{\prime}$ for some U'$\in\mathfrak{X}\left(
TM\right)  $

On the other hand the push forward of the covariant derivative on M gives :
$f^{\ast}\left(  \widehat{\nabla}_{V}U\right)  \in\mathfrak{X}\left(
T\widehat{M}\right)  $

It can be schown (Lovelock p.269)\ that $\nabla_{f^{\ast}V}\left(  f^{\ast
}U\right)  =f^{\ast}\left(  \widehat{\nabla}_{V}U\right)  +S\left(  f^{\ast
}U,f^{\ast}V\right)  $

S is a bilinear map called the \textbf{second fundamental form. }If g is
riemannian $S:\mathfrak{X}\left(  T\widehat{M}\right)  \times\mathfrak{X}%
\left(  T\widehat{M}\right)  \rightarrow\mathfrak{X}\left(  T\widehat{M}%
^{\bot}\right)  $ sends $u_{q},v_{q}\in T_{q}\widehat{M}$ to the orthogonal
complement $T_{q}\widehat{M}^{\bot}$\ of $T_{q}\widehat{M}$ (Kobayashi II p.11).

So usually the induced connection is not a connection\ on $T\widehat{M}$ , as
$\nabla_{f^{\ast}V}\left(  f^{\ast}U\right)  $ has a component out of
$T\widehat{M}.$

\subsubsection{Vectors normal to a hypersurface}

\begin{theorem}
If (N,g) is a real, finite dimensional pseudo-riemannian manifold, M a
hypersurface embedded into N by f, then the symmetric bilinear form h induced
in M is not degenerate at $q\in f(M)$ iff there is a normal $\nu$\ at f(M)
such that $g(q)(\nu,\nu)\neq0$
\end{theorem}

\begin{proof}
1. If f(M) is a hypersurface then $\left[  F\right]  $ is a nx(n-1) matrix of
rank n-1, the system of n-1 linear equations : $\left[  \mu\right]  _{1\times
n}\left[  F\right]  _{n\times n-1}=0$ has a non null solution, unique up to a
multiplication by a scalar. If we take : $\left[  \nu\right]  =\left[
g\right]  ^{-1}\left[  \mu\right]  ^{t}$ we have the components of a vector
orthogonal to $T_{q}\widehat{M}:\left[  F\right]  ^{t}\left[  g\right]
\left[  \nu\right]  =0\Rightarrow\forall u_{p}\in T_{p}M:g\left(  f\left(
p\right)  \right)  \left(  f^{\prime}(p)u_{p},\nu_{q}\right)  =0.$ Thus we
have a non null normal, unique up to a scalar. Consider the matrix
$\widehat{F}=\left[  F,F_{\lambda}\right]  _{n\times n}$ were the last column
is any column of F. It is of rank n-1 and by development along the last column
we gets :

$\det\widehat{F}=0=\sum_{\alpha}\left(  -1\right)  ^{\alpha+n}\widehat
{F}_{\lambda}^{\alpha}\det\left[  \widehat{F}\right]  _{\left(
1...n\backslash n\right)  }^{\left(  1...n`\backslash\alpha\right)  }%
=\sum_{\alpha}\left(  -1\right)  ^{\alpha+n}F_{\lambda}^{\alpha}\det\left[
F\right]  ^{\left(  1...n`\backslash\alpha\right)  }$

And the component expression of vectors normal to f(M) is :

$\nu=\sum_{\alpha\beta}\left(  -1\right)  ^{\alpha}g^{\alpha\beta}\det\left[
F\right]  ^{\left(  1...n`\backslash\beta\right)  }\partial y_{\alpha} $

2. If h is not degenerate at p, then there is an orthonormal basis $\left(
e_{i}\right)  _{i=1}^{n-1}$ at p in M, and $\left[  h\right]  =I_{n-1}=\left[
F\right]  ^{t}\left[  g\right]  \left[  F\right]  $

If $\nu\in T_{q}\widehat{M}$ we would have $\left[  \nu\right]  =\left[
F\right]  \left[  u\right]  $ for some vector u$_{p}\in T_{p}M$ and $\left[
F\right]  ^{t}\left[  g\right]  \left[  \nu\right]  =\left[  F\right]
^{t}\left[  g\right]  \left[  F\right]  \left[  u\right]  =\left[  u\right]
=0$

So $\nu\notin T_{q}\widehat{M}.$ If $g\left(  q\right)  \left(  \nu
,\nu\right)  =0$ then, because $\left(  \left(  f^{\prime}(p)e_{i}%
,i=1...n-1\right)  ,\nu\right)  $ are linearly independant, they constitute a
basis of $T_{q}N$. And we would have $\forall u\in T_{q}N:\exists v_{q}\in
T_{q}\widehat{M},y\in%
\mathbb{R}
:u=v_{p}+y\nu$ and $g\left(  q\right)  \left(  \nu,v_{p}+y\nu\right)  =0$ so g
would be degenerate.

3. Conversely let $g\left(  q\right)  \left(  \nu,\nu\right)  \neq0$ . If
$\nu\in T_{q}\widehat{M}:\exists u\in T_{p}M:\nu=f^{\prime}(p)u.$ As $\nu$ is
orthogonal to all vectors of $T_{q}\widehat{M}$ we would have : $g\left(
q\right)  \left(  f^{\prime}(p)u,f^{\prime}(p)u\right)  =0$ .So $\nu\notin
T_{q}\widehat{M}$ and the nxn matrix : $\widehat{F}=\left[  F,\nu\right]
_{n\times n}$ is the matrix of coordinates of n independant vectors.

$\left[  \widehat{F}\right]  ^{t}\left[  g\right]  \left[  \widehat{F}\right]
=%
\begin{bmatrix}
\left[  F\right]  ^{t}\left[  g\right]  \left[  F\right]  & \left[  F\right]
^{t}\left[  g\right]  \left[  \nu\right] \\
\left[  \nu\right]  ^{t}\left[  g\right]  \left[  F\right]  & \left[
\nu\right]  ^{t}\left[  g\right]  \left[  \nu\right]
\end{bmatrix}
=%
\begin{bmatrix}
\left[  F\right]  ^{t}\left[  g\right]  \left[  F\right]  & 0\\
0 & \left[  \nu\right]  ^{t}\left[  g\right]  \left[  \nu\right]
\end{bmatrix}
$

$\det\left[  \widehat{F}\right]  ^{t}\left[  g\right]  \left[  \widehat
{F}\right]  =\det\left(  \left[  F\right]  ^{t}\left[  g\right]  \left[
F\right]  \right)  \det\left(  \left[  \nu\right]  ^{t}\left[  g\right]
\left[  \nu\right]  \right)  =g\left(  p\right)  \left(  v,\nu\right)  \left(
\det\left[  h\right]  \right)  =\left(  \det\left[  \widehat{F}\right]
\right)  ^{2}\det\left[  g\right]  \neq0$

$\Rightarrow\det\left[  h\right]  \neq0$
\end{proof}

\bigskip

\begin{theorem}
If M is a connected manifold with boundary $\partial M$ in a real, finite
dimensional pseudo-riemannian manifold (N,g), given by a function $f\in
C_{1}\left(  M;%
\mathbb{R}
\right)  ,$ the symmetric bilinear form h induced in M by g is not degenerate
at $q\in\partial M$ iff $g\left(  gradf,gradf\right)  \neq0$\ at q, and then
the unitary, outward oriented normal vector to $\partial M$\ is : $\nu
=\frac{gradf}{\left\vert g\left(  gradf,gradf\right)  \right\vert }$
\end{theorem}

\begin{proof}
On the tangent space at p to $\partial M:\forall u_{p}\in T_{p}\partial
M:f^{\prime}(p)u_{p}=0$

$\forall u\in T_{p}N:g\left(  gradf,u\right)  =f^{\prime}(p)u$ so $f^{\prime
}(p)u_{p}=0\Leftrightarrow g\left(  gradf,u_{p}\right)  =0$

So the normal n is proportional to gradf and the metric is non degenerate iff
$g\left(  gradf,gradf\right)  \neq0$

If M and $\partial M$ are connected then for any transversal outward oriented
vector v : $f^{\prime}(p)v>0$ so a normal outward oriented n is such that :
$f^{\prime}(p)n=g\left(  gradf,n\right)  >0$ with n=kgradf : $g\left(
gradf,gradf\right)  k>0$
\end{proof}

A common notation in this case is with the normal $\nu=\frac{gradf}{\left\vert
g\left(  gradf,gradf\right)  \right\vert }:\forall f\in C_{1}\left(  M;%
\mathbb{R}
\right)  :$

$\frac{\partial f}{\partial\nu}=g\left(  grad\varphi,\nu\right)  =\sum
_{\alpha\beta}g_{\alpha\beta}g^{\alpha\gamma}\left(  \partial_{\gamma
}f\right)  \nu^{\beta}=\sum_{\alpha}\left(  \partial_{\alpha}f\right)
\nu^{\alpha}=f^{\prime}(p)\nu$

If N is compact and we consider the family of manifolds with boundary :
$M_{t}=\left\{  p\in N:f(p)\leq t\right\}  $ then the flow of the vector $\nu$
is a diffeomorphism between the boundaries : $\partial M_{t}=\Phi_{\nu}\left(
t-a,\partial M_{a}\right)  $ (see Manifold with boundary).

\subsubsection{Volume form}

\begin{theorem}
If (N,g) is a real, finite dimensional pseudo-riemannian manifold with its
volume form $\varpi_{0}$, M a hypersurface embedded into N by f, and the
symmetric bilinear form h induced in f(M) is not degenerate at $q\in f\left(
M\right)  $, then the volume form $\varpi_{1}$ induced by h on f(M) is such
that $\varpi_{1}=i_{\nu}\varpi_{0},$ where $\nu$ is the outgoing unitary
normal to f(M). Conversely $\varpi_{0}=\nu^{\ast}\wedge\varpi_{1}$ where $\nu
$* is the 1-form $\nu_{\alpha}^{\ast}=\sum_{\beta}g_{\alpha\beta}\nu^{\beta}$
\end{theorem}

\begin{proof}
if the metric h is not degenerate at q, there is an orthonormal basis $\left(
\varepsilon_{i}\right)  _{i=1}^{n-1}$ in $T_{q}f(M)$ and a normal unitary
vector $\nu$, so we can choose the orientation of $\nu$ such that $\left(
\nu,\varepsilon_{1},...,\varepsilon_{n-1}\right)  $ is direct in TN, and :

$\varpi_{0}\left(  \nu,\varepsilon_{1},...,\varepsilon_{n-1}\right)
=1=i_{\nu}\varpi_{0}$

Let us denote $\nu^{\ast}\in T_{q}N^{\ast}$ the 1-form such that $\nu_{\alpha
}^{\ast}=\sum_{\beta}g_{\alpha\beta}\nu^{\beta}$

$\nu^{\ast}\wedge\varpi_{1}\in\Lambda_{n}TN^{\ast}$ and all n-forms are
proportional so : $\nu^{\ast}\wedge\varpi_{1}=k\varpi_{0}$ as $\varpi_{0}$ is
never null.

$\left(  \nu^{\ast}\wedge\varpi_{1}\right)  \left(  n,\varepsilon
_{1},...,\varepsilon_{n-1}\right)  =k\varpi_{0}\left(  n,\varepsilon
_{1},...,\varepsilon_{n-1}\right)  =k$

$=\frac{1}{\left(  n-1\right)  !}\sum_{\sigma\in\mathfrak{S}_{n-1}}%
\epsilon\left(  \sigma\right)  \nu^{\ast}\left(  n\right)  \varpi_{1}\left(
\varepsilon_{\sigma\left(  1\right)  },...\varepsilon_{\sigma\left(  q\right)
}\right)  =1$

So $\nu^{\ast}\wedge\varpi_{1}=\varpi_{0}$
\end{proof}

Notice that this result needs only that the induced metric be not degenerate
on M (the Levy Civita connection has nothing to do in this matter).

The volume form on N is : $\varpi_{0}=\sqrt{\left\vert \det g\right\vert
}dy^{1}\wedge...dy^{n}$

The volume form on $\widehat{M}=f\left(  M\right)  $ is $\varpi_{1}%
=\sqrt{\left\vert \det h\right\vert }du^{1}\wedge...\wedge du^{n-1}$

\subsubsection{Stockes theorem}

\begin{theorem}
If M is a connected manifold with boundary $\partial M$ in a real,
pseudo-riemannian manifold (N,g), given by a function $f\in C_{1}\left(  M;%
\mathbb{R}
\right)  ,$

if $g\left(  gradf,gradf\right)  \neq0$\ on $\partial M$ then for any vector
field on N :%

\begin{equation}
\int_{M}\left(  divV\right)  \varpi_{0}=\int_{\partial M}g\left(
V,\nu\right)  \varpi_{1}%
\end{equation}

where $\varpi_{1}$ the volume form induced by g on $\partial M$ and $\nu$ the
unitary, outward oriented normal vector to $\partial M$
\end{theorem}

\begin{proof}
The boundary is an hypersurface embedded in N and given by f(p)=0.

In the conditions where the Stockes theorem holds, for a vector field V on M
and $\varpi_{0}\in\mathfrak{X}\left(  \Lambda_{n}TN^{\ast}\right)  $ the
volume form induced by g in TN :

$\int_{M}\left(  divV\right)  \varpi_{0}=\int_{\partial M}i_{V}\varpi_{0}$

$i_{V}\varpi_{0}=i_{V}\left(  \nu^{\ast}\wedge\varpi_{1}\right)  =\left(
i_{V}\nu^{\ast}\right)  \wedge\varpi_{1}+\left(  -1\right)  ^{\deg\nu^{\ast}%
}\nu^{\ast}\wedge\left(  i_{V}\varpi_{1}\right)  $

$=g\left(  V,\nu\right)  \varpi_{1}-\nu^{\ast}\wedge\left(  i_{V}\varpi
_{1}\right)  $

On $\partial M:\nu^{\ast}\wedge\left(  i_{V}\varpi_{1}\right)  =0$

$\int_{M}\left(  divV\right)  \varpi_{0}=\int_{\partial M}g\left(
V,\nu\right)  \varpi_{1}$
\end{proof}

The unitary, outward oriented normal vector to $\partial M$\ is : $\nu
=\frac{gradf}{\left\vert g\left(  gradf,gradf\right)  \right\vert }$

$g\left(  V,\nu\right)  =\sum_{\alpha\beta}g_{\alpha\beta}V^{\alpha}\nu
^{\beta}=\frac{1}{\left\Vert gradf\right\Vert }\sum_{\alpha\beta}%
g_{\alpha\beta}V^{\alpha}\sum_{\beta}g^{\beta\gamma}\partial_{\gamma}f$

$=\frac{1}{\left\Vert gradf\right\Vert }\left(  \sum_{\alpha}V^{\alpha
}\partial_{\alpha}f\right)  =\frac{1}{\left\Vert gradf\right\Vert }f^{\prime
}(p)V$

$\int_{M}\left(  divV\right)  \varpi_{0}=\int_{\partial M}\frac{1}{\left\Vert
gradf\right\Vert }f^{\prime}(p)V\varpi_{1}$

If V is a transversal outgoing vector field : $f^{\prime}(p)V>0$ and $\int
_{M}\left(  divV\right)  \varpi_{0}>0$

Notice that this result needs only that the induced metric be not degenerate
on the boundary.\ If N is a riemannian manifold then the condition is always met.

Let $\varphi\in C_{1}\left(  M;%
\mathbb{R}
\right)  ,V\in\mathfrak{X}\left(  M\right)  $ then for any volume form (see
Lie derivative) :

$div\left(  \varphi V\right)  =\varphi^{\prime}(V)+\varphi div\left(
V\right)  $

but $\varphi$'(V) reads : $\varphi^{\prime}(V)=g\left(  grad\varphi,V\right)
$

With a manifold with boundary it gives the usual formula :

$\int_{M}div\left(  \varphi V\right)  \varpi_{0}=\int_{\partial M}g\left(
\varphi V,n\right)  \varpi_{1}=\int_{\partial M}\varphi g\left(  V,n\right)
\varpi_{1}$

$=\int_{M}g\left(  grad\varphi,V\right)  \varpi_{0}+\int_{M}\varphi div\left(
V\right)  \varpi_{0}$

If $\varphi$ or V has a support in the interior of M then $\int_{\partial
M}\varphi g\left(  V,n\right)  \varpi_{1}=0$ and

$\int_{M}g\left(  grad\varphi,V\right)  \varpi_{0}=-\int_{M}\varphi div\left(
V\right)  \varpi_{0}$

\newpage

\section{SYMPLECTIC\ MANIFOLDS}

\label{Symplectic manifolds}

Symplectic manifolds,used in mechanics, are a powerful tool to study
lagrangian models. We consider only finite dimensional manifolds as one of the
key point of symplectic manifold (as of pseudo-riemanian manifolds) is the
isomorphism with the dual.

We will follow mainly Hofer.

\bigskip

\subsection{Symplectic manifold}

\begin{definition}
A \textbf{symplectic manifold} M is a finite dimensional real manifold endowed
with a 2-form $\varpi$ closed non degenerate, called the \textbf{symplectic
form}:
\end{definition}

$\varpi\in\mathfrak{X}\left(  \Lambda_{2}TM^{\ast}\right)  $ is such that :

it is non degenerate : $\forall u_{p}\in TM:\forall v_{p}\in T_{p}%
M:\varpi\left(  p\right)  \left(  u_{p},v_{p}\right)  =0\Rightarrow u_{p}=0$

it is closed :$d\varpi=0$

As each tangent vector space is a symplectic vector space a \textit{necessary
condition is that the dimension of M be even}, say 2m.

M must be at least of class 2.

Any open subset M of a symplectic manifold (N,$\varpi$) is a symplectic
manifold (M,$\varpi$%
$\vert$%
$_{M}).$

\begin{definition}
The \textbf{Liouville form} on a 2m dimensional symplectic manifold $\left(
M,\varpi\right)  $ is $\Omega=\left(  \wedge\varpi\right)  ^{m}.$ This is a
volume form on M.
\end{definition}

So if M is the union of countably many compact sets it is orientable

\begin{theorem}
The product $M=M_{1}\times M_{2}$\ of symplectic manifolds $\left(
M_{1},\varpi_{1}\right)  ,\left(  M_{2},\varpi_{2}\right)  $ is a symplectic manifold:
\end{theorem}

$V_{1}\in VM_{1},V_{2}\in VM_{2},V=\left(  V_{1},V_{2}\right)  \in
VM_{1}\times VM_{2}$

$\varpi\left(  V,W\right)  =\varpi_{1}\left(  V_{1},W_{1}\right)  +\varpi
_{2}\left(  V_{2},W_{2}\right)  $

\subsubsection{Symplectic maps}

\begin{definition}
A map $f\in C_{1}\left(  M_{1};M_{2}\right)  $ between two symplectic
manifolds $\left(  M_{1},\varpi_{1}\right)  ,\left(  M_{2},\varpi_{2}\right)
$ is \textbf{symplectic} if it preserves the symplectic forms : $f^{\ast
}\varpi_{2}=\varpi_{1}$
\end{definition}

f'(p) is a symplectic linear map, thus it is injective and we must have :
$\dim M_{1}\leq\dim M_{2}$

f preserves the volume form : $f^{\ast}\Omega_{2}=\Omega_{1}$ so we must have :

$\int_{M_{1}}f^{\ast}\Omega_{2}=\int_{M_{2}}\Omega_{2}=\int_{M_{1}}\Omega_{1}$

Total volumes measures play a special role in symplectic manifolds, with a
special kind of invariant called capacity (see Hofer).

Symplectic maps are the morphisms of the category of symplectic manifolds.

A \textbf{symplectomorphism} is a symplectic diffeomorphism.

\begin{theorem}
There cannot be a symplectic map between compact smooth symplectic manifolds
of different dimension.
\end{theorem}

\begin{proof}
If the manifolds are compact, smooth and oriented :

$\exists k\left(  f\right)  \in%
\mathbb{Z}
:\int_{M_{1}}f^{\ast}\Omega_{2}=k\left(  f\right)  \int_{M_{2}}\Omega_{2}%
=\int_{M_{1}}\Omega_{1}$

where k(f) is the degree of the map. If f is not surjective then k(f)=0. Thus
if $\dim M_{1}<\dim M_{2}$ then $\int_{M_{1}}\Omega_{1}=0$ which is not
possible for a volume form.

if $\dim M_{1}=\dim M_{2}$ then f is a local diffeomorphism
\end{proof}

Conversely: from the Moser's theorem, if M is a compact, real, oriented,
finite dimensional manifold with volume forms $\varpi,\pi$ such that :
$\int_{M}\varpi=\int_{M}\pi$ then there is a diffeomorphism $f:M\rightarrow M$
such that $\pi=f^{\ast}\varpi.$ So f is a symplectomorphism.

\subsubsection{Canonical symplectic structure}

The symplectic structure on $%
\mathbb{R}
^{2m}$ is given by $\varpi_{0}=\sum_{k=1}^{m}e^{k}\wedge f^{k}$ with matrix
$J_{m}=%
\begin{bmatrix}
0 & I_{m}\\
-I_{m} & 0
\end{bmatrix}
_{2m\times2m}$ and any basis $\left(  e_{k},f_{k}\right)  _{k=1}^{m}$ and its
dual $\left(  e^{k},f^{k}\right)  _{k=1}^{m}$ (see Algebra)

\begin{theorem}
Darboux's theorem (Hofer p.10) For any symplectic manifold $\left(
M,\varpi\right)  $ there is an atlas $\left(
\mathbb{R}
^{2m},\left(  O_{i},\varphi_{i}\right)  _{i\in I}\right)  $ such that, if $%
\mathbb{R}
^{2m}$ is endowed with its canonical symplectic structure with the symplectic
form $\varpi_{0}$, the transitions maps $\varphi_{j}^{-1}\circ\varphi_{i}$ are
symplectomorphisms on $%
\mathbb{R}
^{2m}$
\end{theorem}

They keep invariant $\varpi_{0}:\left(  \varphi_{j}^{-1}\circ\varphi
_{i}\right)  ^{\ast}\varpi_{0}=\varpi_{0}$

The maps $\varphi_{i}$ are symplectic diffeomorphisms.

Then there is a family $\left(  \varpi_{i}\right)  _{i\in I}\in\mathfrak{X}%
\left(  \Lambda_{2}TM^{\ast}\right)  ^{I}$ such that $\forall p\in
O_{i}:\varpi\left(  p\right)  =\varpi_{i}\left(  p\right)  $ and $\varpi
_{i}=\varphi_{i}^{\ast}\varpi_{0}$

We will denote $\left(  x_{\alpha},y_{\alpha}\right)  _{\alpha=1}^{m}$ the
coordinates in $%
\mathbb{R}
^{2m}$ associated to the maps $\varphi_{i}$ :and the canonical symplectic
basis. Thus there is a holonomic basis of M which is also a canonical
symplectic basis : $dx^{\alpha}=\varphi_{i}^{\ast}\left(  e^{\alpha}\right)
,dy^{\alpha}=\varphi_{i}^{\ast}\left(  f^{\alpha}\right)  $ and : $\varpi
=\sum_{\alpha=1}^{m}dx^{\alpha}\wedge dy^{\alpha}$

The Liouville forms reads : $\Omega=\left(  \wedge\varpi\right)  ^{m}%
=dx^{1}\wedge...\wedge dx^{m}\wedge dy^{1}\wedge..\wedge dy^{m}$.

So locally all symplectic manifolds of the same dimension look alike.

Not all manifolds can be endowed with a symplectic structure (the spheres
$S_{n}$ for n%
$>$%
1 have no symplectic structure).

\subsubsection{Generating functions}

(Hofer p.273)

In $\left(
\mathbb{R}
^{2m},\varpi_{0}\right)  $ symplectic maps can be represented in terms of a
single function, called generating function.

Let $f:%
\mathbb{R}
^{2m}\rightarrow%
\mathbb{R}
^{2m}::f\left(  \xi,\eta\right)  =\left(  x,y\right)  $ be a symplectic map
with x=$\left(  x^{i}\right)  _{i=1}^{m},$...

$x=X\left(  \xi,\eta\right)  $

$y=Y\left(  \xi,\eta\right)  $

If $\det\left[  \frac{\partial X}{\partial\xi}\right]  \neq0$ we can change
the variables and express f as :

$\xi=A\left(  x,\eta\right)  $

$y=B\left(  x,\eta\right)  $

Then f is symplectic if there is a function $W:%
\mathbb{R}
^{2m}\rightarrow%
\mathbb{R}
^{2m}::W\left(  x,\eta\right)  $ such that :

$\xi=A\left(  x,\eta\right)  =\frac{\partial W}{\partial\eta}$

$y=B\left(  x,\eta\right)  =\frac{\partial W}{\partial x}$

\bigskip

\subsection{Hamiltonian vector fields}

\subsubsection{Isomorphism between the tangent and the cotangent bundle}

\begin{theorem}
On a symplectic manifold $\left(  M,\varpi\right)  $, there is a canonical
isomorphism at any point p between the tangent space $T_{p}M$\ and the
cotangent space $T_{p}M^{\ast}$, and\ between the tangent bundle TM and the
cotangent bundle TM*.
\end{theorem}

$j:TM\rightarrow TM^{\ast}::u_{p}\in T_{p}M\rightarrow\mu_{p}=\jmath\left(
u_{p}\right)  \in T_{p}M^{\ast}::$

$\forall v_{p}\in T_{p}M:\varpi\left(  p\right)  \left(  u_{p},v_{p}\right)
=\mu_{p}\left(  v_{p}\right)  $

\subsubsection{Hamilonian vector fields}

As a particular case, if f is a function then its differential is a 1-form.

\begin{definition}
The \textbf{Hamiltonian vector field} $V_{f}$ associated to a function $f\in
C_{1}\left(  M;%
\mathbb{R}
\right)  $ on a symplectic manifold $\left(  M,\varpi\right)  $ is the unique
vector field such that :%

\begin{equation}
i_{V_{f}}\varpi=-df\Leftrightarrow\forall W\in\mathfrak{X}\left(  TM\right)
:\varpi\left(  V_{f},W\right)  =-df\left(  W\right)
\end{equation}

\end{definition}

The vector $V_{f}$ is usually denoted grad(f) : $\varpi(gradf,u)=-df\left(
u\right)  $

\begin{theorem}
The flow of a Hamiltonian vector field preserves the symplectic form and is a
one parameter group of symplectic diffeomorphisms. Conversely Hamiltonian
vector fields are the infinitesimal generators of one parameter group of
symplectic diffeomorphisms.
\end{theorem}

\begin{proof}
We have $\pounds _{V_{f}}\varpi=i_{V_{f}}d\varpi+d\circ$ $i_{V_{f}}\varpi=0$
so the flow of a Hamiltonian vector field preserves the symplectic form.

$\forall t:\Phi_{V_{f}}\left(  t,.\right)  ^{\ast}\varpi=\varpi$ : the flow is
a one parameter group of symplectic diffeomorphisms.
\end{proof}

So symplectic structures show a very nice occurence : the infinitesimal
generators of one parameter groups of diffeomorphisms which preserve the
structure (the form $\varpi)$ are directly related to functions f on M.

\begin{theorem}
The divergence of Hamiltonian vector fields is null.
\end{theorem}

\begin{proof}
Hamiltonian vector field preserves the Liouville form : $\pounds _{Vf}%
\Omega=0$ . So we have $\pounds _{Vf}\Omega=\left(  divV_{f}\right)
\Omega\Rightarrow divV_{f}=0$
\end{proof}

\begin{theorem}
Symplectic maps S maps hamiltonian vector fields to hamiltonian vector fields
\end{theorem}

$V_{f\circ S}=S^{\ast}V_{f}:\Phi_{V_{f}}\left(  t,.\right)  \circ S=S\circ
\Phi_{V_{f\circ S}}\left(  t,.\right)  $

\subsubsection{Poisson brackets}

\begin{definition}
The \textbf{Poisson bracket} of two functions $f,h\in C_{1}\left(  M;%
\mathbb{R}
\right)  $ on a symplectic manifold $\left(  M,\varpi\right)  $ is the
function :%

\begin{equation}
(f,h)=\varpi(grad(f),grad(h))=\varpi(JV_{f},JV_{h})=\varpi(V_{f},V_{h})\in
C\left(  M;%
\mathbb{R}
\right)
\end{equation}

\end{definition}

\begin{theorem}
With the Poisson bracket the vector space $C_{\infty}\left(  M;%
\mathbb{R}
\right)  $ is a Lie algebra (infinite dimensional)
\end{theorem}

i) The Poisson bracket is a an antisymmetric, bilinear map$.$:

$\forall f_{1},f_{2}\in C_{1}\left(  M;%
\mathbb{R}
\right)  ,k,k^{\prime}\in%
\mathbb{R}
:$

$(f_{1},f_{2})=-(f_{2},f_{1})$

$\left(  kf_{1}+k^{\prime}f_{2},h\right)  =k\left(  f_{1},h\right)
+k^{\prime}\left(  f_{2},h\right)  $

$\left(  f_{1}f_{2},h\right)  =(f_{1},h)f_{2}+\left(  f_{2},h\right)  f_{1}$

ii) $\left(  f_{1},\left(  f_{2},f_{3}\right)  \right)  +(f_{2},\left(
f_{3},f_{1}\right)  +\left(  f_{3},(f_{1},f_{2}\right)  )=0$

Furthermore the Poisson bracket has the properties:

i) for any function : $\phi\in C_{1}\left(
\mathbb{R}
;%
\mathbb{R}
\right)  :\left(  f,\phi\circ h\right)  =\phi^{\prime}(h)(f,h)$

ii) If f is constant then : $\forall h\in C_{1}\left(  M;%
\mathbb{R}
\right)  :\left(  f,h\right)  =0$

iii) $\left(  f,h\right)  =\pounds _{grad(h)}(f)=-\pounds _{grad(f)}(h)$

\subsubsection{Complex structure}

\begin{theorem}
(Hofer p.14) On a symplectic manifold $\left(  M,\varpi\right)  $ there is an
almost complex structure J and a Riemannian metric g such that :

$\forall u,p\in T_{p}M:\varpi\left(  p\right)  \left(  u,Jv\right)  =g\left(
p\right)  \left(  u,v\right)  $
\end{theorem}

$\varpi\left(  p\right)  \left(  Ju,Jv\right)  =\varpi\left(  p\right)
\left(  u,v\right)  $ so J(p) is a symplectic map in the tangent space

$J^{2}=-Id,J^{\ast}=J^{-1}$ where J* is the adjoint of J by g : $g\left(
Ju,v\right)  =g\left(  u,J^{\ast}v\right)  $

Finite dimensional real manifolds generally admit Riemannian metrics (see
Manifolds), but g is usually not this metric.\ The solutions J,g are not unique.

So a symplectic manifold has an almost complex K\"{a}hler manifold structure.

In the holonomic symplectic chart :

$\left[  J^{\ast}\right]  =\left[  g\right]  ^{-1}\left[  J\right]  \left[
g\right]  $

$\varpi\left(  p\right)  \left(  u,Jv\right)  =\left[  u\right]  ^{t}\left[
J\right]  ^{t}\left[  J_{m}\right]  \left[  v\right]  =\left[  u\right]
^{t}\left[  g\right]  \left[  v\right]  \Leftrightarrow\left[  g\right]
=\left[  J\right]  ^{t}\left[  J_{m}\right]  $

J is not necessarily $J_{m}$

$\det g=\det J\det J_{m}=1$ because $\left[  J\right]  ^{2}=\left[
J_{m}\right]  ^{2}=-I_{2m}$

So the volume form for the Riemannian metric is identical to the Liouville
form $\Omega.$

If $V_{f}$ is a Hamiltonian vector field : $\varpi\left(  V_{f},W\right)
=-df\left(  W\right)  =\varpi\left(  W,-J^{2}V_{f}\right)  =g\left(
W,-JV_{f}\right)  $ so : $JV_{f}=grad\left(  f\right)  $.

\bigskip

\subsection{Symplectic structure on the cotangent bundle}

\subsubsection{Definition}

\begin{theorem}
The cotangent bundle TM* of a m dimensional real manifold can be endowed with
the structure of a symplectic manifold
\end{theorem}

The symplectic form is $\varpi=\sum_{\alpha}dy^{\alpha}\wedge dx^{\alpha} $
where $\left(  x^{\alpha},y^{\beta}\right)  _{\alpha,\beta=1}^{m}$ are the
coordinates in TM*

\begin{proof}
Let $\left(
\mathbb{R}
^{m},\left(  O_{i},\varphi_{i}\right)  _{i\in I}\right)  $ be an atlas of M,
with coordinates $\left(  x^{\alpha}\right)  _{\alpha=1}^{m}$\ 

The cotangent bundle TM* has the structure of a 2m dimensional real manifold
with atlas $\left(
\mathbb{R}
^{m}\times%
\mathbb{R}
^{m\ast},\left(  O_{i}\times\cup_{p\in O_{i}}T_{p}M^{\ast},\left(  \varphi
_{i},\left(  \varphi_{i}^{\prime}\right)  ^{t}\right)  \right)  \right)  $\ 

A point $\mu_{p}\in TM^{\ast}$\ can be coordinated by the pair\ $\left(
x^{\alpha},y^{\beta}\right)  _{\alpha,\beta=1}^{m}$ where x stands for p and y
for the components of $\mu_{p}$ in the holonomic basis. A vector $Y\in
T_{\mu_{p}}TM^{\ast}$ has components $\left(  u^{\alpha},\sigma_{\alpha
}\right)  _{\alpha\in A}$ expressed in the holonomic basis $\left(  \partial
x_{\alpha},\partial y_{\alpha}\right)  $

Let be the projections :

$\pi_{1}:TM^{\ast}\rightarrow M::\pi_{1}\left(  \mu_{p}\right)  =p$

$\pi_{2}:T\left(  TM^{\ast}\right)  \rightarrow TM^{\ast}::\pi_{2}\left(
Y\right)  =\mu_{p}$

So : $\pi_{1}^{\prime}\left(  \mu_{p}\right)  :T_{\mu_{p}}T_{p}M^{\ast
}\rightarrow T_{p}M::\pi_{1}^{\prime}\left(  \mu_{p}\right)  Y=u\in T_{p}M$

Define the 1-form over $TM^{\ast}:\lambda\left(  \mu_{p}\right)  \in L\left(
T_{\mu_{p}}T_{p}M^{\ast};%
\mathbb{R}
\right)  $

$\lambda\left(  \mu_{p}\right)  \left(  Y\right)  =\pi_{2}\left(  Y\right)
\left(  \pi^{\prime}\left(  \mu_{p}\right)  \left(  Y\right)  \right)
=\mu_{p}\left(  u\right)  \in%
\mathbb{R}
$

It is a well defined form. Its components in the holonomic basis associated to
the coordinates $\left(  x^{\alpha},y_{\alpha}\right)  _{\alpha\in A}$ are :

$\lambda\left(  \mu_{p}\right)  =\sum_{\alpha}y_{\alpha}dy^{\alpha}$

The components in $dy^{\alpha}$ are zero.

The exterior differential of $\lambda$ is $\varpi=d\lambda=\sum_{\alpha
}dy^{\alpha}\wedge dx^{\alpha}$ so $\varpi$ is closed, and it is not degenerate.
\end{proof}

If $X,Y\in T_{\mu_{p}}TM^{\ast}:X=\sum_{\alpha}\left(  u^{\alpha}\partial
x_{\alpha}+\sigma^{\alpha}\partial y^{\alpha}\right)  ;Y=\sum_{\alpha}\left(
v^{\alpha}\partial x_{\alpha}+\theta^{\alpha}\partial y^{\alpha}\right)  $

$\varpi\left(  \mu_{p}\right)  \left(  X,Y\right)  =\sum_{\alpha}\left(
\sigma^{\alpha}u^{\alpha}-\theta^{\alpha}v^{\alpha}\right)  $

\subsubsection{Application to analytical mechanics}

In analytic mechanics the state of the system is described as a point q in
some m dimensional manifold M coordinated by m positional variables $\left(
q^{i}\right)  _{i=1}^{m}$ which are coordinates in M (to account for the
constraints of the system modelled as liaisons between variables) called the
configuration space. Its evolution is some path $%
\mathbb{R}
\rightarrow M:q(t)$\ The quantities $\left(  q^{i},\frac{dq^{i}}{dt}%
=\overset{\cdot}{q}^{i}\right)  $ belong to the tangent bundle TM.

The dynamic of the system is given by the principle of least action with a
Lagrangian : $L\in C_{2}\left(  M\times%
\mathbb{R}
^{m};%
\mathbb{R}
\right)  :L\left(  q_{i},u_{i}\right)  $

q(t) is such that : $\int_{t_{0}}^{t_{1}}L(q\left(  t\right)  ,\overset{\cdot
}{q}\left(  t\right)  )dt$ is extremal.

The Euler-Lagrange equations give for the solution :

$\frac{\partial L}{\partial q^{i}}=\frac{d}{dt}\frac{\partial L}{\partial
u^{i}}=\frac{\partial}{\partial u^{i}}\frac{d}{dt}L$

If the Hessian $\left[  \frac{\partial^{2}L}{\partial u_{i}\partial u_{j}%
}\right]  $ has a determinant which is non zero it is possible to implement
the change of variables :

$\left(  q^{i},u^{i}\right)  \rightarrow\left(  q^{i},p_{i}\right)
:p_{i}=\frac{\partial L}{\partial u^{i}}$

and the equations become : $q^{i\prime}=\frac{\partial H}{\partial p_{i}%
};p_{i}^{\prime}=-\frac{\partial H}{\partial q^{i}}$ with the Hamiltonian :
$H(q,p)=\sum_{i=1}^{n}p_{i}u^{i}-L(q,u)$

The new variables $p_{i}\in TM^{\ast}$ are called the moments of the system
and the cotangent bundle TM* is the phase space. The evolution of the system
is a path

$C:%
\mathbb{R}
\rightarrow TM^{\ast}::$ $\left(  q^{i}\left(  t\right)  ,p_{i}\left(
t\right)  \right)  $ and $C^{\prime}(t)=\left(  \overset{\cdot}{q}%
^{i},\overset{\cdot}{p}_{i}\right)  $

The Hamiltonian vector field $V_{H}$ associated with H has the components :
$V_{H}=(\frac{\partial H}{\partial p_{i}}\partial q_{i},-\frac{\partial
H}{\partial q^{i}}\partial p_{i})_{i=1..m}$

So the solutions C(t) are just given by the flow of $V_{H}.$

\subsubsection{Surfaces of constant energy}

As infinitesimal generators of the one parameter group of diffeomorphisms
(Hamiltonian vector fields) are related to functions, the submanifolds where
this function is constant play a specific role.\ They are manifolds with
boundary. So manifolds with boundary are in some way the integral submanifolds
of symplectic structures.

In physics usually a function, the energy H, plays a special role, and one
looks for the evolutions of a system such that this energy is constant.

\paragraph{Principles\newline}

If a system is modelled by a symplectic manifold (M,$\varpi)$ (such as above)
with a function $H\in C_{1}\left(  M;%
\mathbb{R}
\right)  :$

1. The value of H is constant along the integral curves of its Hamiltonian
vector field $V_{H}$ of H

The Hamiltonian vector field and its flow $\Phi_{V_{H}}$ are such that :

$\varpi\left(  V_{H},\frac{\partial}{\partial t}\Phi_{V_{H}}\left(
t,.\right)  |_{t=\theta}\right)  =-dH\frac{\partial}{\partial t}\Phi_{V_{H}%
}\left(  t,.\right)  |_{t=\theta}=0=\frac{d}{dt}H\left(  \Phi_{V_{H}}\left(
t,p\right)  \right)  |_{t=\theta}$

So : $\forall t,\forall p:H\left(  \Phi_{V_{H}}\left(  t,p\right)  \right)
=H\left(  p\right)  \Leftrightarrow\Phi_{V_{H}}\left(  t,.\right)  _{\ast}H=H$ .

2. The divergence of $V_{H}$\ is null.

3. H defines a folliation of M with leaves the surfaces of constant energy
$\partial S_{c}=\left\{  p\in M:H\left(  p\right)  =c\right\}  $

If H'(p)=0 then $V_{H}=0$ because $\forall W:\varpi\left(  V_{H},W\right)
=-dH\left(  W\right)  =0$

If H'(p)$\neq$0 the sets : $S_{c}=\left\{  p\in M:H\left(  p\right)  \leq
c\right\}  $ are a family of manifolds with boundary the hypersurfaces
$\partial S.$ The vector $V_{H}$ belongs to the tangent space to $\partial
S_{c}$ The hypersurfaces $\partial S_{c}$\ are preserved by the flow of
$V_{H}.$

4. If there is a Riemannian metric g (as above) on M then the unitary normal
outward oriented\ $\nu$ to $\partial S_{c}$ is : $\nu=\frac{JV_{H}}%
{\varpi\left(  V_{H},JV_{H}\right)  }$

$\nu=\frac{gradH}{\left\vert g\left(  gradH,gradH\right)  \right\vert }$ with
$gradH=JV_{H}\Rightarrow g\left(  gradH,gradH\right)  =g\left(  JV_{H}%
,JV_{H}\right)  =g\left(  V_{H},V_{H}\right)  =\varpi\left(  V_{H}%
,JV_{H}\right)  >0$

The volume form on $\partial S_{t}$ is $\Omega_{1}=i_{\nu}\Omega
\Leftrightarrow\Omega=\Omega_{1}\wedge\nu$

If M is compact then H(M)=[a,b] and the flow of the vector field\ $\nu$ is a
diffeomorphism for the boundaries $\partial S_{c}=\Phi_{\nu}\left(  \partial
S_{a},c\right)  .$ (see Manifolds with boundary).

\paragraph{Periodic solutions\newline}

If t is a time parameter and the energy is constant, then the system is
described by some curve c(t) in M, staying on the boundary $\partial S_{c}.$
There is a great deal of studies about the kind of curve that can be found,
depending of the surfaces S. They must meet the equations :

$\forall u\in T_{c\left(  t\right)  }M:\varpi\left(  V_{H}\left(  c\left(
t\right)  \right)  ,u\right)  =-dH\left(  c\left(  t\right)  \right)  u$

A T \textbf{periodic solution} is such that if : $c^{\prime}(t)=V_{H}\left(
t\right)  $ on M, then c(T)=c(0). The system comes back to the initial state
after T.

There are many results about the existence of periodic solutions and about
ergodicity.\ The only general theorem is :

\begin{theorem}
Poincar\'{e}'s recurrence theorem (Hofer p.20): If M is a Hausdorff, second
countable, symplectic $\left(  M,\varpi\right)  $ manifold and $H\in
C_{1}\left(  M;%
\mathbb{R}
\right)  $\ such that H'(p) is not null on M, then for almost every point p of
$\partial S_{c}=\left\{  p\in M:H\left(  p\right)  =c\right\}  $ there is an
increasing sequence $\left(  t_{n}\right)  _{n\in%
\mathbb{N}
}$ such that $\lim_{n\rightarrow\infty}\Phi_{V_{H}}\left(  t_{n},p\right)  =p$
\end{theorem}

the null measure is with respect to the measure $\Omega_{1}.$ The proof in
Hofer can easily be extended to the conditions above.

One says that p is a recurring point : if we follow an integral curve of the
hamiltonian vector, then we will come back infinitely often close to any point.

\newpage

\part{\textbf{LIE\ GROUPS}}

\bigskip

The general properties and definitions about groups are seen in the Algebra
part.\ Groups with countably many elements have been classified.\ When the
number of elements is uncountable the logical way is to endow the set with a
topological structure : when the operations (product and inversion) are
continuous we get the continuous groups. Further if we endow the set with a
manifold structure compatible with the group structure we get Lie groups. The
combination of the group and manifold structures gives some stricking
properties. First the manifold is smooth, and even analytic.\ Second, the
tangent space at the unity element of the group in many ways summarize the
group itself, through a Lie algebra.\ Third, most (but not all) Lie groups are
isomorphic to a linear group, meaning a group of matrices, that we get through
a linear representation. So the study of Lie groups is closely related to the
study of Lie algebras and linear representations of groups.

In this part we start with Lie algebras.\ As such a Lie algebra is a vector
space endowed with an additional internal operation (the bracket). The most
common example of Lie algebra is the set of vector fields over a manifold
equipped with the commutator. In the finite dimensional case we have more
general results, and indeed all finite dimensional Lie algebras are isomorphic
to some algebra of matrices, and have been classified. Their study involves
almost uniquely algebraic methods. Thus the study of finite dimensional Lie
groups stems mostly from their Lie algebra.

The theory of linear representation of Lie groups is of paramount importance
in physics. There is a lot of litterature on the subject, but unfortunately it
is rather confusing. This is a fairly technical subject, with many traditional
notations and conventions which are not very helpful.\ I will strive to put
some ligth on the topic, with the main purpose to give to the reader the most
useful and practical grasp on these questions.

\newpage

\section{LIE\ ALGEBRAS}

\subsection{Definitions}

\label{Lie algebra defintions}

\subsubsection{Lie algebra}

\begin{definition}
A \textbf{Lie algebra} over a field K is a vector space A over K endowed with
a bilinear map (\textbf{bracket}) :$\left[  {}\right]  :A\times A\rightarrow
A$

$\forall X,Y,Z\in A,\forall\lambda,\mu\in K:\left[  \lambda X+\mu Y,Z\right]
=\lambda\left[  X,Z\right]  +\mu\left[  Y,Z\right]  $ such that :

$\left[  X,Y\right]  =-\left[  Y,X\right]  $

$\left[  X,\left[  Y,Z\right]  \right]  +\left[  Y,\left[  Z,X\right]
\right]  +\left[  Z,\left[  X,Y\right]  \right]  =0$ (Jacobi identities)
\end{definition}

Notice that a Lie algebra is not an algebra because the bracket is not
associative.\ But any algebra becomes a Lie algebra with the bracket :
$\left[  X,Y\right]  =X\cdot Y-Y\cdot X.$

The \textbf{dimension of the Lie algebra} is the dimension of the vector
space.\ In the following A can be infinite dimensional if not otherwise
specified. The field K is either $%
\mathbb{R}
$ or $%
\mathbb{C}
$.

A Lia algebra is said to be \textbf{abelian} if it is commutative, then the
bracket is null : $\forall X,Y:\left[  X,Y\right]  =0$

\begin{notation}
\textbf{ad} is the linear map $ad:A\rightarrow L\left(  A;A\right)
::ad(X)(Y)=\left[  X,Y\right]  $ induced by the bracket
\end{notation}

\paragraph{Classical examples\newline}

The set of linear endomorphisms over a vector space L(E;E) with the bracket ;
$\left[  f,g\right]  =f\circ g-g\circ f$.\ 

The set K(r) of square rxr matrices endowed with the bracket :

$\left[  X,Y\right]  =\left[  X\right]  \left[  Y\right]  -\left[  Y\right]
\left[  X\right]  .$

The set of vector fields $\mathfrak{X}\left(  TM\right)  $ over a manifold
endowed with the commutator : $\left[  V,W\right]  $

Any vector space with the trivial bracket : $\left[  X,Y\right]  =0.$ Indeed
any commutative Lie algebra has this bracket.

\paragraph{Structure coefficients\newline}

As any bilinear map, the bracket of a Lie algebra can be expressed in any
basis $\left(  e_{i}\right)  _{i\in I}$ by its value $\left[  e_{i}%
,e_{j}\right]  $ for each pair of vectors of the basis, which reads :%

\begin{equation}
\forall i,j:\left[  e_{i},e_{j}\right]  =\sum_{k\in I}C_{ij}^{k}e_{k}%
\end{equation}

The scalars $\left(  C_{ij}^{k}\right)  _{k\in I}$ are called the
\textbf{structure coefficients} of the algebra. They depend on the basis as
any set of components, and at most finitely many of them are non zero.

Because of the antisymmetry of the bracket and the Jacobi identities the
structure coefficients are not independant.\ They must satisfy the identities :

$C_{ij}^{k}=-C_{ji}^{k}$

$\forall i,j,k,m:\sum_{l\in I}\left(  C_{jk}^{l}C_{il}^{m}+C_{ki}^{l}%
C_{jl}^{m}+C_{ij}^{l}C_{kl}^{m}\right)  =0$

Conversely a family of structure coefficients meeting these relations
define,\ in any basis of a vector space, a bracket and a unique Lie algebra
structure. Because of the particularities of this system of equations there
are only, for finite dimensional algebras, a finite number of kinds of
solutions.\ This is the starting point to the classification of Lie algebras.

\subsubsection{Morphisms of Lie algebras}

\begin{definition}
A \textbf{Lie algebra morphism} (also called a homomorphism) is a linear map f
between Lie algebras $\left(  A,\left[  {}\right]  _{A}\right)  $,$\left(
B,\left[  {}\right]  _{B}\right)  $\ which preserves the bracket
\end{definition}

$f\in L\left(  A;B\right)  :\forall X,Y\in A:f\left(  \left[  X,Y\right]
_{A}\right)  =\left[  f\left(  X\right)  ,f\left(  Y\right)  \right]  _{B}$

They are the morphisms of the category of Lie algebras over the same field.

A Lie algebra isomorphism is an \textbf{isomorphism} of vector spaces which is
also a Lie algebra morphism.

Two Lie algebras A,B are \textbf{isomorphic} if there is an isomorphism of Lie
algebra : $A\rightarrow B$

\begin{theorem}
The set L(A;A) of linear maps over a Lie algebra A is a Lie algebra with the
composition law and the map $ad:A\rightarrow L\left(  A;A\right)  $ is a Lie
algebra morphism :
\end{theorem}

\begin{proof}
i) $\forall X,Y\in A:f\left(  \left[  X,Y\right]  \right)  =\left[  f\left(
X\right)  ,f\left(  Y\right)  \right]  ,g\left(  \left[  X,Y\right]  \right)
=\left[  g\left(  X\right)  ,g\left(  Y\right)  \right]  $

$h\left(  \left[  X,Y\right]  \right)  =f\circ g\left(  \left[  X,Y\right]
\right)  -g\circ f\left(  \left[  X,Y\right]  \right)  =f\left(  \left[
g\left(  X\right)  ,g\left(  Y\right)  \right]  \right)  -g\left(  \left[
f\left(  X\right)  ,f\left(  Y\right)  \right]  \right)  =\left[  f\circ
g\left(  X\right)  ,f\circ g\left(  Y\right)  \right]  -\left[  g\circ
f\left(  X\right)  ,g\circ f\left(  Y\right)  \right]  $

So : $\forall f,g\in\hom\left(  A;A\right)  :h=\left[  f,g\right]  =f\circ
g-g\circ f\in\hom\left(  A;A\right)  $

ii) Take$\ U\in A:$

$ad\left(  X\right)  \circ ad\left(  Y\right)  \left(  U\right)  -ad\left(
Y\right)  \circ ad\left(  X\right)  \left(  U\right)  =\left[  X,\left[
Y,U\right]  \right]  -\left[  Y,\left[  X,U\right]  \right]  $

$=\left[  X,\left[  Y,U\right]  \right]  +\left[  Y,\left[  U,X\right]
\right]  =-\left[  U,\left[  X,Y\right]  \right]  =\left[  \left[  X,Y\right]
,U\right]  =ad\left(  \left[  X,Y\right]  \right)  \left(  U\right)  $

So : $ad\in\hom\left(  A,L\left(  A;A\right)  \right)  $ :

$\left[  ad\left(  X\right)  ,ad\left(  Y\right)  \right]  _{L\left(
A;A\right)  }=ad\left(  X\right)  \circ ad\left(  Y\right)  -ad\left(
Y\right)  \circ ad\left(  X\right)  =ad\left(  \left[  X,Y\right]
_{A}\right)  $
\end{proof}

\begin{definition}
An \textbf{automorphism} over a Lie algebra $\left(  A,\left[  {}\right]
\right)  $ is a linear automorphism of vector space (thus it must be
invertible) which preserves the bracket
\end{definition}

$f\in GL(A):\forall X,Y\in A:f\left(  \left[  X,Y\right]  \right)  =\left[
f\left(  X\right)  ,f\left(  Y\right)  \right]  $ .

Then $f^{-1}$ is a Lie algebra automorphism.

The set GL(A) of automorphisms over A is a group with the composition law.

\begin{definition}
A \textbf{derivation} over a Lie algebra $\left(  A,\left[  {}\right]
\right)  $ is an endomorphism D such that :

$D\in L\left(  A;A\right)  :\forall X,Y\in A:D\left(  \left[  X,Y\right]
\right)  =\left[  D\left(  X\right)  ,Y\right]  +\left[  X,D\left(  Y\right)
\right]  $
\end{definition}

For any X the map ad(X) is a derivation.

\begin{theorem}
(Knapp p.38) \ The set of derivations over a Lie algebra $\left(  A,\left[
{}\right]  \right)  $ , denoted Der(A), has the structure of a Lie algebra
with the composition law : $D,D^{\prime}\in Der\left(  A\right)  :D\circ
D^{\prime}-D^{\prime}\circ D\in Der(A)$

The map : $ad:A\rightarrow Der\left(  A\right)  $ is a Lie algebra morphism
\end{theorem}

\subsubsection{Killing form}

\begin{definition}
The Killing form over a finite dimensional Lie algebra $\left(  A,\left[
{}\right]  \right)  $ is the map B :%

\begin{equation}
B\in L^{2}(A,A;K):A\times A\rightarrow K::B(X,Y)=Trace(ad(X)\circ ad(Y))
\end{equation}

\end{definition}

\begin{theorem}
The Killing form is a bilinear symmetric form
\end{theorem}

\begin{proof}
In a basis $\left(  e_{i}\right)  _{i\in I}$ of A and its dual $\left(
e^{i}\right)  _{i\in I}$ the map ad can be read as the tensor :

$ad(X)=\sum_{i,j,k\in I}C_{kj}^{i}x^{k}e_{i}\otimes e^{j}$

$ad(Y)=\sum_{i,j,k\in I}C_{kj}^{i}y^{k}e_{i}\otimes e^{j}$

$ad(X)\circ ad(Y)\left(  Z\right)  =\sum_{i,j,k\in I}ad(X)\left(  C_{kj}%
^{i}y^{k}z^{j}e_{i}\right)  =\sum_{i,j,k\in I}C_{kj}^{i}y^{k}z^{j}ad(X)\left(
e_{i}\right)  =\sum_{i,j,k\in I}C_{kj}^{i}y^{k}z^{j}\sum_{l,m\in I}C_{mi}%
^{l}x^{m}e_{l}$

$ad(X)\circ ad(Y)=\sum_{i,j,k,l,m\in I}C_{kj}^{i}C_{mi}^{l}y^{k}x^{m}%
e^{j}\otimes e_{l}$

$Trace(ad(X)\circ ad(Y))=\sum_{i,j,k,m\in I}C_{kj}^{i}C_{mi}^{j}y^{k}x^{m}$

So $Trace(ad(X)\circ ad(Y))=Trace(ad(Y)\circ ad(X))$
\end{proof}

The Killing form reads in a basis of A : $B=\sum_{i,j,k,m\in I}C_{kj}%
^{i}C_{mi}^{j}e^{k}\otimes e^{m}$

\begin{theorem}
The Killing form B is such that :

$\forall X,Y,Z\in A:B\left(  \left[  X,Y\right]  ,Z\right)  =B\left(
X,\left[  Y,Z\right]  \right)  $
\end{theorem}

It comes from the Jacobi identities

\begin{theorem}
(Knapp p.100) Any automorphism of a Lie algebra $\left(  A,\left[  {}\right]
\right)  $ preserves the Killing form
\end{theorem}

$f\left(  \left[  X,Y\right]  \right)  =\left[  f\left(  X\right)  ,f\left(
Y\right)  \right]  \Rightarrow B\left(  f\left(  X\right)  ,f\left(  Y\right)
\right)  =B\left(  X,Y\right)  $

\subsubsection{Subsets of a Lie algebra}

\paragraph{Subalgebra\newline}

\begin{definition}
A \textbf{Lie subalgebra} (say also subalgebra) of a Lie algebra $\left(
A,\left[  {}\right]  \right)  $ is a vector subspace B of A which is closed
under the bracket operation : $\forall X,Y\in B:\left[  X,Y\right]  \in B$.
\end{definition}

\begin{notation}
$\left[  B,C\right]  $ denotes the vector space $Span\left\{  \left[
X,Y\right]  ,X\in B,Y\in C\right\}  $ generated by all the brackets of
elements of B,C in A
\end{notation}

\begin{theorem}
If B is a subalgebra and f an automorphism then f(B) is a subalgebra.
\end{theorem}

\begin{definition}
The \textbf{normalizer} N(B) of a subalgeba B of the Lie algebra $\left(
A,\left[  {}\right]  \right)  $ is the set of vectors : $N\left(  B\right)
=\left\{  X\in A:\forall Y\in B:\left[  X,Y\right]  \in B\right\}  $
\end{definition}

\paragraph{Ideal\newline}

\begin{definition}
An \textbf{ideal} of of the Lie algebra $\left(  A,\left[  {}\right]  \right)
$ is a vector subspace B of A such that : $\left[  A,B\right]  \sqsubseteq B$
\end{definition}

An ideal is a subalgebra (the converse is not true)

If B,C are ideals then the sets $B+C,\left[  B,C\right]  ,B\cap C$ are ideal

If A,B are Lie algebras and $f\in\hom(A,B)$ then $\ker f$ is an ideal of A.

If B is an ideal of A the quotient set A/B : $X\sim Y\Leftrightarrow X-Y\in B
$ is a Lie algebra with the bracket : $\left[  \left[  X\right]  ,\left[
Y\right]  \right]  =\left[  X,Y\right]  $ because $\forall U,V\in B,\exists
W\in B:\left[  X+U,Y+V\right]  =\left[  X,Y\right]  +W$ .Then the map :
$A\rightarrow A/B$ is a Lie algebra morphism

\paragraph{Center\newline}

\begin{definition}
The \textbf{centralizer} Z(B) of a subset B of the Lie algebra $\left(
A,\left[  {}\right]  \right)  $ is the set : $Z\left(  B\right)  =\left\{
X\in A:\forall Y\in B:\left[  X,Y\right]  =0\right\}  $

The \textbf{center} Z(A) of the Lie algebra $\left(  A,\left[  {}\right]
\right)  $ \ is the centralizer of A itself
\end{definition}

So : $Z\left(  A\right)  =\left\{  X\in A:\forall Y\in A:\left[  X,Y\right]
=0\right\}  $ is the set of vectors which commute with any vector of A. Z(A)
is an ideal.

\subsubsection{Complex and real Lie algebra}

\paragraph{Complexified\newline}

There are two ways to define a complex vector space structure on a real vector
space and so for a real Lie algbra (see Complex vector spaces in the Algebra part).

\subparagraph{1. Complexification:\newline}

\begin{theorem}
Any real Lie algebra $\left(  A,\left[  {}\right]  \right)  $ can be endowed
with the structure of a complex Lie algebra, called its \textbf{complexified}
$\left(  A_{%
\mathbb{C}
},\left[  {}\right]  _{%
\mathbb{C}
}\right)  $ which has same basis and structure coefficients as A.
\end{theorem}

\begin{proof}
i) It is always possible to define the complexified vector space $A_{%
\mathbb{C}
}=A\oplus iA$ over A

ii) define the bracket by :

$\left[  X+iY,X^{\prime}+iY^{\prime}\right]  _{%
\mathbb{C}
}=\left[  X,X^{\prime}\right]  -\left[  Y,Y^{\prime}\right]  +i\left(  \left[
X,Y^{\prime}\right]  +\left[  X^{\prime},Y\right]  \right)  $
\end{proof}

\begin{definition}
A real Lie algebra $\left(  A_{0},\left[  {}\right]  \right)  $ is a
\textbf{real form} of the complex Lie algebra $\left(  A,\left[  {}\right]
\right)  $ if its complexified is equal to A : $A\equiv A_{0}\oplus iA_{0}$
\end{definition}

\subparagraph{2. Complex structure:\newline}

\begin{theorem}
A complex structure J on a real Lie algebra $\left(  A,\left[  {}\right]
\right)  $ defines a structure of complex Lie algebra on the set A iff $J\circ
ad=ad\circ J$
\end{theorem}

\begin{proof}
J is a linear map $J\in L\left(  E;E\right)  $ such that $J^{2}=-Id_{E}$ ,
then for any $X\in A:iX$ is defined as J(X).

The complex vector space structure $A_{J}$ is defined by
$X=x+iy\Leftrightarrow X=x+J(y)$ then the bracket

$\left[  X,X^{\prime}\right]  _{A_{J}}=\left[  x,x^{\prime}\right]  +\left[
J\left(  y\right)  ,J\left(  y^{\prime}\right)  \right]  +\left[  x,J\left(
y^{\prime}\right)  \right]  +\left[  x^{\prime},J\left(  y\right)  \right]
=\left[  x,x^{\prime}\right]  -\left[  Jy,y^{\prime}\right]  +i\left(  \left[
x,y^{\prime}\right]  +\left[  x^{\prime},y\right]  \right)  $

if $\left[  x,J\left(  y^{\prime}\right)  \right]  =-J\left(  \left[
x,y^{\prime}\right]  \right)  \Leftrightarrow\forall x\in A_{1}:J\left(
ad\left(  x\right)  \right)  \circ ad\left(  J\left(  x\right)  \right)
\Leftrightarrow J\circ ad=ad\circ J$
\end{proof}

If A is finite dimensional a necessary condition is that its dimension is even.

\paragraph{Real structure\newline}

\begin{theorem}
Any real structure on a complex Lie algebra $\left(  A,\left[  {}\right]
\right)  $ defines a structure of real Lie algebra with same bracket on the
real kernel.
\end{theorem}

\begin{proof}
There are always a real structure, an antilinear map $\sigma$ such that
$\sigma^{2}=Id_{A}$ and any vector can be written as : $X=\operatorname{Re}%
X+i\operatorname{Im}X$ where $\operatorname{Re}X,\operatorname{Im}X\in A_{%
\mathbb{R}
}.$ The real kernel $A_{%
\mathbb{R}
}$ of A is a real vector space, subset of A, defined by $\sigma\left(
X\right)  =X$. It is simple to check that the bracket is closed in $A_{%
\mathbb{R}
}.$
\end{proof}

Notice that there are two real vector spaces and Lie algebras $A_{%
\mathbb{R}
},iA_{%
\mathbb{R}
}$ which are isomorphic (in A) by multiplication with i. The \textbf{real
form} of the Lie algebra is : $A_{\sigma}=A_{%
\mathbb{R}
}\times iA_{%
\mathbb{R}
}$ which can be seen either as the direct product of two real algebras, or a
real algebras of two times the complex dimension of $A_{%
\mathbb{R}
}$.

If $\sigma$\ is a real structure of A then $A_{%
\mathbb{R}
}$ is a real form of A.

\bigskip

\subsection{Sum and product of Lie algebras}

\label{Sum and product Lie algebras}

\subsubsection{Free Lie algebra}

\begin{definition}
A \textbf{free Lie algebra} over any set X is a pair $\left(  L,\jmath\right)
$ of a Lie algebra L and a map : $\jmath:X\rightarrow L$ with the universal
property : whatever the Lie algebra A and the map : $f:X\rightarrow A$ there
is a unique Lie algebra morphism $F:L\rightarrow A$ such that : $f=F\circ j$
\end{definition}

\bigskip%

\begin{tabular}
[c]{ccccc}
&  & \j &  & \\
X & $\rightarrow$ & $\rightarrow$ & $L$ & \\
& $\searrow$ &  & $\downarrow$ & \\
& f & $\searrow$ & $\downarrow$ & F\\
&  &  & A &
\end{tabular}

\bigskip

\begin{theorem}
(Knapp p.188) For any non empty set X there is a free Lie algebra over X and
the image \j(X) generates L. Two free Lie algebras over X are isomorphic.
\end{theorem}

\subsubsection{Sum of Lie algebras}

\begin{definition}
The sum of two Lie algebras $\left(  A,\left[  {}\right]  _{A}\right)
,\left(  B,\left[  {}\right]  _{B}\right)  $ is the vector space $A\oplus B$
with the bracket :

$X,Y\in A,X^{\prime},Y^{\prime}\in B:\left[  X+X^{\prime},Y+Y^{\prime}\right]
_{A\oplus B}=\left[  X,Y\right]  _{A}+\left[  X^{\prime},Y^{\prime}\right]
_{B}$
\end{definition}

then A'=(A,0), B'=(0,B) are ideals of $A\oplus B$

\begin{definition}
A Lie algebra A is said to be \textbf{reductive} if for any ideal B there is
an ideal C such that $A=B\oplus C$
\end{definition}

A real Lie algebra of matrices over the fields $%
\mathbb{R}
,%
\mathbb{C}
,H$ which is closed under the operation conjugate / transpose is reductive.

\subsubsection{Product}

\begin{definition}
The \textbf{product} of two Lie algebras $\left(  A,\left[  {}\right]
_{A}\right)  ,\left(  B,\left[  {}\right]  _{B}\right)  $ is the direct
product of the vector spaces $A\times B$ with the bracket:

$\left[  \left(  X,Y\right)  ,\left(  X^{\prime},Y^{\prime}\right)  \right]
_{A\times B}=\left(  \left[  X,X^{\prime}\right]  _{A},\left[  Y,Y^{\prime
}\right]  _{B}\right)  $
\end{definition}

\begin{theorem}
(Knapp p.38) If $\left(  A,\left[  {}\right]  _{A}\right)  ,\left(  B,\left[
{}\right]  _{B}\right)  $ are two Lie algebras over the same field, F a Lie
algebra morphism $F:A\rightarrow Der(B)$ where Der(B) is the set of
derivations over B, then there is a unique Lie algebra structure over $A\oplus
B$ called \textbf{semi-direct product} of A,B, denoted $C=A\oplus_{F}B$ such
that :

$\forall X,Y\in A:\left[  X,Y\right]  _{C}=\left[  X,Y\right]  _{A}$

$\forall X,Y\in B:\left[  X,Y\right]  _{C}=\left[  X,Y\right]  _{B}$

$\forall X\in A,Y\in B:\left[  X,Y\right]  _{C}=F\left(  X\right)  \left(
Y\right)  $
\end{theorem}

Then A is a subalgebra of C, and B is an ideal of C. The direct sum is a
special case of semi-direct product with F=0.

\subsubsection{Universal envelopping algebra}

A Lie algebra is not an algebra because the bracket is not associative.\ It
entails that the computations in a Lie algebras, when they involve many
brackets, become quickly unmanageable. This is the case with the linear
representations $\left(  F,\rho\right)  $ of Lie algebras where it is natural
to deal with products of the kind $\rho\left(  X_{1}\right)  \rho\left(
X_{2}\right)  ...\rho\left(  X_{p}\right)  $ which are product of matrices,
image of tensorial products $X_{1}\otimes...\otimes X_{p}.$ Moreover it is
useful to be able to use some of the many theorems about "true" algebras. But
to build an algebra over a Lie algebra A requires to use many copies of A.

\paragraph{Definition\newline}

\begin{definition}
The universal envelopping algebra $U_{r}\left(  A\right)  $ of order r of a
Lie algebra $\left(  A,\left[  {}\right]  \right)  $ over the field K is the
quotient space: $\left(  T\left(  A\right)  \right)  ^{r}$ of the tensors of
order r over A, by the two sided ideal :$J=\left\{  X\otimes Y-Y\otimes
X-[X,Y],X,Y\in T^{1}(A)\right\}  $

The \textbf{universal envelopping algebra} $U\left(  A\right)  $ is the direct
sum :

$U\left(  A\right)  =\oplus_{r=0}^{\infty}U_{r}\left(  A\right)  $
\end{definition}

\begin{theorem}
(Knapp p.214) \ With the tensor product U(A) is a unital algebra over the
field K
\end{theorem}

The scalars K belong to U(A) and the unity element is 1. The product in U(A)
is denoted XY without other symbol.

The map : $\imath:A\rightarrow U(A)$ is one-one with the founding identity :

$\imath\left[  X,Y\right]  =\imath(X)\imath(Y)-\imath(Y)\imath(X)$ so all
elements of the kind :

$X\otimes Y-Y\otimes X-[X,Y]\sim0$

The subset $U_{r}\left(  A\right)  $ of the elements of U(A) which can be
written as products of exactly r elements of A is a vector subspace of U(A).

U(A) \textit{is not a Lie algebra}. Notice that A can be infinite dimensional.

\paragraph{Properties\newline}

\begin{theorem}
(Knapp p.215) The universal envelopping algebra of a Lie algebra $\left(
A,\left[  {}\right]  \right)  $ over the field K has the universal property
that, whenever L is a unital algebra on the field K and $\rho:A\rightarrow L$
a map such that

$\rho\left(  X\right)  \rho\left(  Y\right)  -\rho(Y)\rho(X)=\rho\left(
\left[  X,Y\right]  \right)  $ ,

there is a unique algebra morphism $\widetilde{\rho}$\ such that :
$\widetilde{\rho}:U(A)\rightarrow L:\rho=\widetilde{\rho}\circ\imath$
\end{theorem}

\begin{theorem}
Poincar\'{e}-Birkhoff-Witt (Knapp p.217): If A is a Lie algebra with basis
$\left(  e_{i}\right)  _{i\in I}$ where the set I has some total ordering,
then the set of monomials : $\left(  \text{\i}\left(  e_{i_{1}}\right)
\right)  ^{n_{1}}$ $\left(  \text{\i}\left(  e_{i_{2}}\right)  \right)
^{n_{2}}....\left(  \text{\i}\left(  e_{i_{p}}\right)  \right)  ^{n_{p}}%
,i_{1}<i_{2}...<i_{p}\in I,n_{1},...n_{p}\in%
\mathbb{N}
$ is a basis of its universal envelopping algebra U(A).
\end{theorem}

\begin{theorem}
(Knapp p.216) \textbf{Transpose} is the unique automorphism

$^{t}:U(A)\rightarrow U(A)$ on the universal envelopping algebra U(A) of a Lie
algebra : such that : $\forall X\in A:\imath(X)^{t}=-\imath(X)$
\end{theorem}

\begin{theorem}
(Knapp p.492)\ If $\left(  A,\left[  {}\right]  \right)  $ is a finite
dimensional Lie algebra , then the following are equivalent for any element U
of its universal envelopping algebra U(A):

i) U is in the center of U(A)

ii) $\forall X\in A:XU=UX$

iii) $\forall X\in A:\exp\left(  ad\left(  X\right)  \right)  \left(
U\right)  =U $
\end{theorem}

\begin{theorem}
(Knapp p.221) If B is a Lie subalgebra of A, then the associative subalgebra
of U(A) generated by 1 and B is canonically isomorphic to U(B).
\end{theorem}

If $A=A_{1}\oplus A_{2}$ then :

$U(A)\simeq U\left(  A_{1}\right)  \otimes_{K}U\left(  A_{2}\right)  $

$U(A)\simeq S_{\dim A_{1}}\left(  A_{1}\right)  S_{\dim A_{2}}\left(
A_{2}\right)  $ with the symmetrization operator on U(A) :

$S_{r}\left(  U\right)  =\sum_{\left(  i_{1}...i_{r}\right)  }U^{i_{1}%
...i_{r}}\sum_{\sigma\in\mathfrak{s}_{r}}e_{\sigma\left(  1\right)
}..e_{\sigma\left(  r\right)  }$

\begin{theorem}
(Knapp p.230) If A is the Lie algebra of a Lie group G then U(A) can be
identified with the left invariant differential operators on G, A can be
identified with the first order operators .
\end{theorem}

\begin{theorem}
If the Lie algebra A is also a Banach algebra (possibly infinite dimensional),
then U(A) is a Banach algebra and a C*-algebra with the involution :
$\ast:U\left(  A\right)  \rightarrow U\left(  A\right)  :U^{\ast}=U^{t}$
\end{theorem}

\begin{proof}
the tensorial product is a Banach vector space and the map \i\ is continuous.
\end{proof}

\paragraph{Casimir elements\newline}

\begin{definition}
(Knapp p.293) The Casimir element of the finite dimensional Lie algebra
$\left(  A,\left[  {}\right]  \right)  $ , is :

$\Omega=\sum_{i,j=1}^{n}B\left(  e_{i},e_{j}\right)  \imath\left(
E_{i}\right)  \imath\left(  E_{j}\right)  \in U\left(  A\right)  $

where B is the Killing form B on A and $E_{i}\in A$ is such that $B\left(
E_{i},e_{j}\right)  =\delta_{ij}$ with $\left(  e_{i}\right)  _{i=1}^{n}$ is a
basis of A
\end{definition}

Warning ! the basis $\left(  E_{i}\right)  _{i=1}^{n}$ is a basis of A and not
a basis of its dual A*, so $\Omega$ is an element of U(A) and not a bilinear form.

The matrix of the components of $\left(  E_{i}\right)  _{i=1}^{n}$ is just
$\left[  E\right]  =\left[  B\right]  ^{-1}$ where $\left[  B\right]  $ is the
matrix of B in the basis.\ So the vectors $\left(  E_{i}\right)  _{i=1}^{n}$
are another basis of A

In sl(2,$%
\mathbb{C}
)$ : $\Omega=\frac{1}{4}\sum_{j=1}^{3}\sigma_{j}^{2}$ with the Pauli matrices
in the standard representation.

\begin{theorem}
(Knapp p.293) In a semi-simple complex Lie algebra the Casimir element does
not depend on the basis and belongs to the center of A, so it commutes with
any element of A
\end{theorem}

Casimir elements are extended in representations (see representation).

\bigskip

\subsection{Classification of Lie algebras}

\label{Classification of Lie algebras}

\subsubsection{Fundamental theorems}

\begin{theorem}
Third Lie-Cartan theorem (Knapp p.663) : Every finite dimensional real Lie
algebra is isomorphic to the Lie algebra of an analytic real Lie group.
\end{theorem}

\begin{theorem}
Ado's Theorem (Knapp p.663) : Let A be a finite dimensional real Lie algebra,
N its unique largest nilpotent ideal.\ Then there is a one-one finite
dimensional representation (E,f) of A on a complex vector space E such that
f(X) is nilpotent for every X in N.\ If A is complex then f can be taken
complex linear.
\end{theorem}

Together these theorems show that any finite dimensional Lie algebra on a
field K is the Lie algebra of some group on the field K and can be represented
as a Lie algebra of matrices. So the classification of the finite dimensional
Lie algebras sums up to a classification of the Lie algebras of matrices. This
is not true for topological groups which are a much more diversified breed.

The first step is the identification of the elementary bricks from which other
Lie algebras are built : the simple Lie algebras. This is done by the study of
the subsets generated by the brackets. Then the classification of the simple
Lie algebras relies on their generators : a set of vectors such that, by
linear combination or brackets, generates any element of the Lie algebra.
There are only 9 types of generators, which are represented by diagrams.

The classification of Lie algebras is also the starting point to the linear
representation of both Lie algebras and Lie groups and in fact the
classification is based upon a representation of the algebra on itself through
the operator ad.

\subsubsection{Solvable and nilpotent algebras}

\begin{definition}
For any Lie algebra $\left(  A,\left[  {}\right]  \right)  $ we define the
sequences :

$A^{0}=A\sqsupseteq A^{1}=\left[  A^{0},A^{0}\right]  \supseteq...A^{k+1}%
=\left[  A^{k},A^{k}\right]  ..$

$A_{0}=A\supseteq A_{1}=\left[  A,A_{0}\right]  ...\supseteq A_{k+1}=\left[
A,A_{k}\right]  ,...$
\end{definition}

\begin{theorem}
(Knapp p.42) For any Lie algebra $\left(  A,\left[  {}\right]  \right)  $ :

$A^{k}\sqsubseteq A_{k}$

Each $A^{k},A_{k}$ is an ideal of A.
\end{theorem}

\paragraph{Solvable algebra\newline}

\begin{definition}
A Lie algebra is said \textbf{solvable} if $\exists k:A^{k}=0.$ Then $A^{k-1}
$ is abelian.
\end{definition}

\begin{theorem}
A solvable finite dimensional Lie algebra can be represented as a set of
triangular matrices.
\end{theorem}

\begin{theorem}
Any 1 or 2 dimensional Lie algebra is solvable
\end{theorem}

\begin{theorem}
If\ B is a solvable ideal and A/B is solvable, then A is solvable
\end{theorem}

\begin{theorem}
The image of a solvable algebra by a Lie algebra morphism is solvable
\end{theorem}

\begin{theorem}
(Knapp p.40) A n dimensional Lie algebra is solvable iff there is a decreasing
sequence of subalgebras $B_{k}:$

$B_{0}=B\supseteq B_{1}...\supseteq B_{k+1}..\supseteq B_{n}=0$

such that $B_{k+1}$ is an ideal of $B_{k}$ and dim$\left(  B_{k}%
/B_{k+1}\right)  =1$
\end{theorem}

\begin{theorem}
Cartan (Knapp p.50) : A finite dimensional Lie algebra is solvable iff its
Killing form B is such that : $\forall X\in A,Y\in\lbrack A,A]:B(X,Y)=0$
\end{theorem}

\begin{theorem}
Lie (Knapp p.40) : If A is a solvable Lie algebra on a field K and $\left(
E,f\right)  $ a finite dimensional representation of A, then there is a non
null vector u in E which is a simulaneous eigen vector for $f\left(  X\right)
,X\in A$ if all the eigen values are in the field K.
\end{theorem}

\begin{theorem}
(Knapp p.32) If A is finite dimensional Lie algebra, there is a unique
solvable ideal, called the \textbf{radical} of A which contains all the
solvable ideals.
\end{theorem}

\paragraph{Nilpotent algebra\newline}

\begin{definition}
A Lie algebra A is said \textbf{nilpotent} if $\exists k:A_{k}=0.$
\end{definition}

\begin{theorem}
A nilpotent algebra is solvable, has a non null center Z(A) and $A_{k-1}%
\subseteq Z\left(  A\right)  $.
\end{theorem}

\begin{theorem}
The image of a nilpotent algebra by a Lie algebra morphism is nilpotent
\end{theorem}

\begin{theorem}
(Knapp p.46) A Lie algebra A is nilpotent iff $\forall X\in A:ad\left(
X\right)  $ is a nilpotent linear map (meaning that $\exists k:\left(
adX\right)  ^{k}=0).$

\begin{theorem}
(Knapp p.49) Any finite dimensional Lie algebra has a unique largest nilpotent
ideal n, which is contained in the radical rad(A) of A and
$[A,rad(A)]\sqsubseteq n.$ Any derivation D is such that $D(rad(A))\sqsubseteq
n.$
\end{theorem}
\end{theorem}

\begin{theorem}
(Knapp p.48, 49) A finite dimensional solvable Lie algebra A :

i) has a unique largest nilpotent ideal n, namely the set of elements X for
which ad(X) is nilpotent. For any derivation D : $D(A)\sqsubseteq n$

ii) [A,A] is nilpotent
\end{theorem}

\begin{theorem}
Engel (Knapp p.46) If E is a finite dimensional vector space, A a sub Lie
algebra of L(E;E) of nilpotent endomorphisms, then A is nilpotent and there is
a non null vector u of E such that f(u)=0 for any f in A. There is a basis of
E such that the matrix of any f in A is triangular with 0 on the diagonal.
\end{theorem}

\subsubsection{Simple and semi-simple Lie algebras}

\begin{definition}
A Lie algebra is :

\textbf{simple} if it is non abelian and has no non zero ideal.

\textbf{semi-simple} if it has no non zero solvable ideal.
\end{definition}

A simple algebra is semi-simple, the converse is not true.

There is no complex semi-simple Lie algebra of dimension 4,5 or 7.

\begin{theorem}
(Knapp p.33) If A is simple then $\left[  A,A\right]  =A$
\end{theorem}

\begin{theorem}
(Knapp p.33) A semi-simple Lie algebra has 0 center: Z(A)=0.
\end{theorem}

\begin{theorem}
(Knapp p.32,50,54) For a finite dimensional Lie algebra A the following are equivalent:

i) A is semi-simple

ii) its Killing form B is non degenerate

iii) rad(A)=0

iv) $A=A_{1}\oplus A_{2}...\oplus A_{k}$ where the $A_{i}$ are ideal and
simple Lie subalgebras.\ Then the decomposition is unique and the only ideals
of A are sum of some $A_{i}.$
\end{theorem}

\begin{theorem}
(Knapp p.54) If $A_{0}$ is the real form of a complex Lie algebra A, then
$A_{0}$ is a semi simple real Lie algebra iff A is a semi simple complex Lie algebra.\ 
\end{theorem}

Thus the way to classify finite dimensional Lie algebras is the following :

i) for any algebra H=A/rad(A) is semi-simple.

ii) rad(A) is a solvable Lie algebra, and can be represented as a set of
triangular matrices

iii) H is the sum of simple Lie algebras

iv) A is the semi-direct product $H\rtimes rad\left(  A\right)  $

If we have a classification of simple Lie algebras we have a classification of
all finite dimensional Lie algebras.

The classification is based upon the concept of abstract roots system, which
is...abstract and technical (it is of little use elwewhere in mathematics),
but is detailed here because it is frequently cited about the representation
of groups. We follow Knapp (II p.124).

\subsubsection{Abstract roots system}

\paragraph{Abstract roots system\newline}

\subparagraph{1. Definition:\newline}

\begin{definition}
An \textbf{abstract roots system} is a finite set $\Delta$\ of non null
vectors of a finite dimensional real vector space $\left(  V,\left\langle
{}\right\rangle \right)  $ endowed with an inner product (definite positive),
such that :

i) $\Delta$ spans V

ii) $\forall\alpha,\beta\in\Delta:2\frac{\left\langle \alpha,\beta
\right\rangle }{\left\langle \alpha,\alpha\right\rangle }\in%
\mathbb{N}
$

iii) the set W of linear maps on V defined by $s_{\alpha}\left(  \beta\right)
=\beta-2\frac{\left\langle \alpha,\beta\right\rangle }{\left\langle
\alpha,\alpha\right\rangle }\alpha$ for $\alpha\in\Delta$ carries $\Delta$ on itself
\end{definition}

W is a subgroup of the orthonormal group of V, comprised of reflexions, called
the \textbf{Weyl's group}.

The vectors of $\Delta$ are linearly dependant, and the identities above are
very special, and indeed they can be deduced from only 9 possible
configurations. It is easy to see that any integer multiple of the vectors
still meet the identities.\ So to find solutions we can impose additional conditions.

\subparagraph{2. Reduced system:\newline}

\begin{definition}
An abstract roots sytem is :

\textbf{reduced} if $\alpha\in\Delta\Rightarrow2\alpha\notin\Delta$

\textbf{reducible} if it is the direct sum of two sets, which are themselves
abstract roots systems, and are orthogonal.

\textbf{irreducible} if it is not reducible.
\end{definition}

An irreducible system is necessarily reduced, but the converse is not true.

The previous conditions give rise to a great number of identities. The main
result is that $\alpha\in\Delta\Rightarrow-\alpha\in\Delta$ even if the system
is reduced.

\subparagraph{3. Simple system:\newline}

\begin{definition}
An \textbf{ordering} on a finite dimensional real vector space is an order
relation where :

a set of positive vectors is identified

if u is in V, either u or -u is positive

if u and v are positive then u+v is positive.\ 
\end{definition}

There are many ways to achieve that, the simplest is the lexicographic
ordering.\ Take a basis $\left(  e_{i}\right)  _{i=1}^{l}$ of V and say that u%
$>$%
0 if there is a k such that $\left\langle u,e_{i}\right\rangle =0$ for
i=1...k-1 and $\left\langle u,e_{k}\right\rangle >0.$

\begin{definition}
A root $\alpha\in\Delta$ \ of an ordered abstract root system is
\textbf{simple} if $\alpha>0$ and there is no $\beta,\beta^{\prime}>0$ such
that $\alpha=\beta+\beta^{\prime}.$
\end{definition}

\begin{theorem}
For any abstract roots system $\Delta$ there is a set $\Pi=\left(  \alpha
_{1},...,\alpha_{l}\right)  $\ with l=dim(V) of linearly independant simple
roots, called a \textbf{simple system}, which fully defines the system:

i) for any root $\beta\in\Delta:$

$\exists\left(  n_{i}\right)  _{i=1}^{l},n_{i}\in%
\mathbb{N}
:$ either $\beta=\sum_{i=1}^{l}n_{i}\alpha_{i}$ or $\beta=-\sum_{i=1}^{l}%
n_{i}\alpha_{i}$

ii) W is generated by the $s_{\alpha_{i}},i=1...l$

iii) $\forall\alpha\in\Delta,\exists w\in W,\alpha_{i}\in\Pi:\alpha
=w\alpha_{i}$

iv) if $\Pi,\Pi^{\prime}$ are simple systems then $\exists w\in W,$ w unique
$:\Pi^{\prime}=w\Pi$
\end{theorem}

The set of positive roots is denoted $\Delta^{+}=\left\{  \alpha\in
\Delta:\alpha>0\right\}  .$

Remarks :

i) $\Pi\sqsubset\Delta^{+}\sqsubset\Delta$ but $\Delta^{+}$ is usually larger
than $\Pi$

ii) any $\pm\left(  \sum_{i=1}^{l}n_{i}\alpha_{i}\right)  $ does not
necessarily belong to $\Delta$

\begin{definition}
A vector $\lambda\in V$ is said to be \textbf{dominant} if : $\forall\alpha
\in\Delta^{+}:\left\langle \lambda,\alpha\right\rangle \geq0.$
\end{definition}

For any vector $\lambda$ of V there is always a simple system $\Pi$ for which
it is dominant, and there is always $w\in W$ such that $w\lambda$ is dominant.

\paragraph{Abstract Cartan matrix\newline}

The next tool is a special kind of matrix, adjusted to abstract roots systems.

1. For an abstract root system represented by a simple system $\Pi=\left(
\alpha_{1},...,\alpha_{l}\right)  $\ the matrix :

$\left[  A\right]  _{ij}=2\frac{\left\langle \alpha_{i},\alpha_{j}%
\right\rangle }{\left\langle \alpha_{i},\alpha_{i}\right\rangle }$

has the properties :

i) $\left[  A\right]  _{ij}\in%
\mathbb{Z}
$

ii) $\left[  A\right]  _{ii}=2;\forall i\neq j:\left[  A\right]  _{ij}\leq0$

iii) $\left[  A\right]  _{ij}=0\Leftrightarrow\left[  A\right]  _{ji}=0$

iv) there is a diagonal matrix D with positive elements such that $DAD^{-1}$
is symmetric definite positive.

v) does not depend ot the choice of positive ordering, up to a permutation of
the indices (meaning up to conjugation by permutation matrices).

2. A matrix meeting the conditions i) through iv) is called a \textbf{Cartan
matrix}. To any Cartan matrix one can associate a unique simple system, thus a
reduced abstract root system, unique up to isomorphism.

3. By a permutation of rows and columns it is always possible to bring a
Cartan matrix in the block diagonal form : a triangular matrix which is the
assembling of triangular matrices above the main diagonal. The matrix has a
unique block iff the associated abstract root system is irreducible and then
is also said to be irreducible.

The diagonal matrix D above is unique up a multiplicative scalar on each
block, thus D is unique up to a multiplicative scalar if A is irreducible.

\paragraph{Dynkin's diagram\newline}

Dynkin's diagram are a way to represent abstract root systems.\ They are also
used in the representation of Lie algebras.

1. The \textbf{Dymkin's diagram} of a simple system of a reduced abstract
roots $\Pi$ is built as follows :

i) to each simple root $\alpha_{i}$ we associate a vertex of a graph

ii) to each vertex we associate a weigth $w_{i}=k\left\langle \alpha
_{i},\alpha_{i}\right\rangle $ where k is some fixed scalar

iii) two vertices i, j are connected by $\left[  A\right]  _{ij}\times\left[
A\right]  _{ji}=4\frac{\left\langle \alpha_{i},\alpha_{j}\right\rangle ^{2}%
}{\left\langle \alpha_{i},\alpha_{i}\right\rangle \left\langle \alpha
_{j},\alpha_{j}\right\rangle }$ edges

The graph is connected iff the system is irreducible.

2. Conversely given a Dynkin's diagram the matrix A is defined up to a
multiplicative scalar for each connected component, thus it defines a unique
reduced abstract root system up to isomorphism.

3. To understand the usual representation of Dynkin's diagram :

i) the abstract roots system is a set of vectors of a finite dimensional real
vector space V.\ This is $V=%
\mathbb{R}
^{m}$ or exceptionnaly a vector subspace of $%
\mathbb{R}
^{m}$

ii) the vectors $\left(  e_{i}\right)  _{i=1}^{m}$ are the usual basis of $%
\mathbb{R}
^{m}$ with the usual euclidian inner product.

iii) a simple roots system $\Delta$ is then defined as a special linear
combination of the $\left(  e_{i}\right)  _{i=1}^{m}$

4. There are only 9 types of connected Dynkin's diagrams, which define all the
irreducible abstract roots systems. They are often represented in books about
Lie groups (see Knapp p.182). They are the following :

a) $A_{n}:n\geq1:V=\sum_{k=1}^{n+1}x_{k}e_{k},\sum_{k=1}^{n+1}x_{k}=0$

$\Delta=e_{i}-e_{j},i\neq j$

$\Pi=\left\{  e_{1}-e_{2},e_{2}-e_{3},..e_{n}-e_{n+1}\right\}  $

b) $B_{n}:n\geq2:V=%
\mathbb{R}
^{n}$

$\Delta=\left\{  \pm e_{i}\pm e_{j},i<j\right\}  \cup\left\{  \pm
e_{k}\right\}  $

$\Pi=\left\{  e_{1}-e_{2},e_{2}-e_{3},..e_{n-1}-e_{n},e_{n}\right\}  $

c) $C_{n}:n\geq3:V=%
\mathbb{R}
^{n}$

$\Delta=\left\{  \pm e_{i}\pm e_{j},i<j\right\}  \cup\left\{  \pm
2e_{k}\right\}  $

$\Pi=\left\{  e_{1}-e_{2},e_{2}-e_{3},..e_{n-1}-e_{n},2e_{n}\right\}  $

d) $D_{n}:n\geq4:V=%
\mathbb{R}
^{n}$

$\Delta=\left\{  \pm e_{i}\pm e_{j},i<j\right\}  $

$\Pi=\left\{  e_{1}-e_{2},e_{2}-e_{3},..e_{n-1}-e_{n},e_{n-1}+e_{n}\right\}  $

e) 5 exceptional types, in $%
\mathbb{R}
^{n}$ : $E_{6},E_{7},E_{8},F_{4},G_{2}$

\subsubsection{Classification of semi-simple complex Lie algebras}

The procedure is to exhibit for any complex semi-simple Lie algebra an
abstract roots system, which gives also a set of generators of the algebra.
And conversely to prove that a Lie algebra can be associated to any abstract
roots system.

In the following A is a complex semi-simple finite dimensional Lie algebra
with dimension n.

\paragraph{Cartan subalgebra\newline}

\begin{definition}
A \textbf{Cartan subalgebra} of a complex Lie algebra $\left(  A,\left[
{}\right]  \right)  $ is an abelian subalgebra h such that there is a set of
linearly independant eigen vectors $X_{k}$ of A such that $A=Span(X_{k})$ and
$\forall H\in h:ad_{H}X_{k}=\lambda\left(  H\right)  X_{k}$ with an eigen
value $\lambda\left(  H\right)  $ which can depend on H
\end{definition}

We assume that h is a maximal Cartan algebra : it does not contain any other
subset with these properties.

\begin{theorem}
(Knapp p.134) Any complex semi-simple finite dimensional Lie algebra A has a
Cartan subalgebra.\ All Cartan subalgebras of A have the same dimension,
called the \textbf{rank} of A
\end{theorem}

Cartan subalgebras are not necessarily unique.\ If h,h' are Cartan subalgebras
of A then there is some automorphism $a\in Int(A):h^{\prime}=ah$

\begin{definition}
A Cartan subalgebra h of a real Lie algebra $\left(  A,\left[  {}\right]
\right)  $ is a subalgebra of h such that the complexified of h is a Cartan
subalgebra of the complexified of A.
\end{definition}

\begin{theorem}
(Knapp p.384) Any real semi simple finite dimensional Lie algebra has a Cartan
algebra . All Cartan subalgebras have the same dimension.
\end{theorem}

\paragraph{Root-space decomposition\newline}

1. Let h be a Cartan subalgebra, then we have the following properties :

i) h itself is an eigen space of $ad_{H}$ with eigen value 0 : indeed $\forall
H,H^{\prime}\in h:\left[  H,H^{\prime}\right]  =ad_{H}H^{\prime}=0$

ii) the eigenspaces for the non zero eigen values are unidimensional

iii) the non zero eigen values are linear functions $\alpha\left(  H\right)  $
of the vectors H

2. Thus there are n linearly independant vectors denoted $X_{k}$ such that :

for k=1...l \ $\left(  X_{k}\right)  _{k=1}^{l}$ is a basis of h and $\forall
H\in h:ad_{H}X_{k}=\left[  H,X_{k}\right]  =0$

for k=l+1,...n\ : $\forall H\in h:ad_{H}X_{k}=\left[  H,X_{k}\right]
=\alpha_{k}\left(  H\right)  X_{k}$ where : $\alpha_{k}:h\rightarrow%
\mathbb{C}
$ is linear, meaning that \ $\alpha_{k}\in h^{\ast}$

These functions do not depend on the choice of $X_{k}$ in the eigenspace
because they are unidimensional. Thus it is customary to label the eigenspaces
by the function itself : indeed they are no more than vectors of the dual, and
we have exactly n-l of them. And one writes :

$\Delta\left(  h\right)  =\left\{  \alpha_{k}\in h^{\ast}\right\}  $

$A_{\alpha}=\left\{  X\in A:\forall H\in h:ad_{H}X=\alpha\left(  H\right)
X\right\}  $

$A=h\oplus_{\alpha\in\Delta}A_{\alpha}$

The functionals $\alpha\in\Delta$ are called the \textbf{roots}, the vectors
of each $A_{\alpha}$ are the \textbf{root vectors}, and the equation above is
the root-space decomposition of the algebra.

3. Let B be the Killing form on A. Because A is semi-simple B is non
degenerate thus it can be used to implement the duality between A and A*, and
h and h*, both as vector spaces on $%
\mathbb{C}
.$

Then : $\forall H,H^{\prime}\in h:B\left(  H,H^{\prime}\right)  =\sum
_{\alpha\in\Delta}\alpha\left(  H\right)  \alpha\left(  H^{\prime}\right)  $

Define :

i) V the linear real span of $\Delta$\ \ in h*: $V=\left\{  \sum_{\alpha
\in\Delta}x_{\alpha}\alpha;x_{\alpha}\in%
\mathbb{R}
\right\}  $

ii) the n-l B-dual vectors $H_{\alpha}$ of $\alpha$ in h :

$H_{\alpha}\in h:\forall H\in h:B\left(  H,H_{\alpha}\right)  =\alpha\left(
H\right)  $

iii) the bilinear symmetric form in V :

$\left\langle H_{\alpha},H_{\beta}\right\rangle =B\left(  H_{\alpha},H_{\beta
}\right)  =\sum_{{}}\sum_{\gamma\in\Delta}\gamma\left(  H_{\alpha}\right)
\gamma\left(  H_{\beta}\right)  $

$\left\langle u,v\right\rangle =\sum_{\alpha,\beta\in\Delta}x_{\alpha}%
y_{\beta}B\left(  H_{\alpha},H_{\beta}\right)  $

iv) $h_{0}$ the real linear span of the $H_{\alpha}:h_{0}=\left\{
\sum_{\alpha\in\Delta}x_{\alpha}H_{\alpha};x_{\alpha}\in%
\mathbb{R}
\right\}  $

Then :

i) V is a real form of h*: $h^{\ast}=V\oplus iV$

ii) $h_{0}$ is a real form of h: $h=h_{0}\oplus ih_{0}$ and V is exactly the
set of covectors such that $\forall H\in h_{0}:u\left(  H\right)  \in%
\mathbb{R}
$ and V is real isomorphic to $h_{0}^{\ast}$

iii) $\left\langle {}\right\rangle $ is a definite positive form, that is an
inner product, on V

iv) the set $\Delta$ is an abstract roots system on V, with $\left\langle
{}\right\rangle $

v) up to isomorphism this abstract roots system does not depend on a choice of
a Cartan algebra

vi) the abstract root system is irreducible iff the Lie algebra is simple

4. Thus, using the results of the previous subsection there is a simple system
of roots $\Pi=\left(  \alpha_{1},...\alpha_{l}\right)  $ with l roots, because

$V=span_{%
\mathbb{R}
}(\Delta),\dim_{%
\mathbb{R}
}V=\dim_{%
\mathbb{C}
}h^{\ast}=\dim_{%
\mathbb{C}
}h=l$

Define :

$h_{i}=\frac{2}{\left\langle \alpha_{i},\alpha_{i}\right\rangle }H_{\alpha
_{i}}$

$E_{i}\neq0\in A_{\alpha_{i}}:\forall H\in h:ad_{H}E_{i}=\alpha_{i}\left(
H\right)  E_{i}$

$F_{i}\neq0\in A_{-\alpha_{i}}:\forall H\in h:ad_{H}F_{i}=-\alpha_{i}\left(
H\right)  F_{i}$

Then the set $\left\{  h_{i},E_{i},F_{i}\right\}  _{i=1}^{l}$ generates A as a
Lie algebra (by linear combination and bracket operations).

As the irreducible abstract roots systems have been classified, so are the
simple complex Lie algebras.

Notice that any semi-simple complex Lie algebra is associated to a diagram,
but it is deconnected.

5. So \textit{a semi-simple complex Lie algebra has a set of at most 3
}$\times$\textit{\ rank generators}. But notice that it can have fewer
generators : indeed if dim(A)=n%
$<$%
3l.

This set of generators follows some specific identities, called Serre's
relations, expressed with a Cartan matrix C, which can be useful (see Knapp p.187):

$\left[  h_{i},h_{j}\right]  =0$

$\left[  E_{i},F_{j}\right]  =\delta_{ij}h_{i}$

$\left[  h_{i},E_{j}\right]  =C_{ij}E_{j}$

$\left[  h_{i},F_{j}\right]  =-C_{ij}F_{j}$

$\left(  adE_{i}\right)  ^{-C_{ij}+1}E_{j}=0$ when $i\neq j$

$\left(  adF_{i}\right)  ^{-A_{ij}+1}F_{j}=0$ when $i\neq j$

6. Conversely if we start with an abstract roots system it can be proven :

i) Given an abstract Cartan matrix C there is a complex semi-simple Lie
algebra whose roots system has C as Cartan matrix

ii) and that this Lie algebra is unique, up to isomorphism. More precisely :

let A,A' be complex semi-simple algebras with Cartan subalgebras h, h', and
roots systems $\Delta,\Delta^{\prime}.$ Assume that there is a vector space
isomorphism :$\varphi:h\rightarrow h^{\prime}$ such that its dual
$\varphi^{\ast}:h^{\prime\ast}\rightarrow h^{\ast}::\varphi^{\ast}\left(
\Delta^{\prime}\right)  =\Delta.$ For $\alpha\in\Delta$ define $\alpha
^{\prime}=\varphi^{\ast-1}\left(  \alpha\right)  .$ Take a simple system
$\Pi\subset\Delta,$ root vectors $E_{\alpha},E_{\alpha^{\prime}} $ then there
is one unique Lie algebra isomorphism $\Phi:A\rightarrow A^{\prime}$ such that
$\Phi|_{h}=\varphi$ and $\Phi\left(  E_{\alpha}\right)  =E_{\alpha^{\prime}}$

\paragraph{List of simple complex Lie algebras\newline}

The classification of simple complex Lie algebras follows the classification
of irreducible abstract roots systems (Knapp p.683).\ The Lie algebras are
expressed as matrices algebras\ in their standard linear representation (other
Lie algebras are isomorphic to the ones listed here, see below the list of
classical algebras):

$A_{n},n\geq1:sl(n+1,C)$ \ \ $\dim A_{n}=n\left(  n+2\right)  $

$B_{n},n\geq2:so(2n+1,C)$ \ \ $\dim B_{n}=n\left(  2n+1\right)  $

$C_{n},n\geq3:sp(n,C)$ \ \ $\dim C_{n}=n\left(  2n+1\right)  $

$D_{n},n\geq4:so(2n,C)$ \ \ $\dim D_{n}=n\left(  2n-1\right)  $

The exceptional systems give rise to 5 exceptional Lie algebras (their
dimension is in the brackets) :

$E_{6}\left[  78\right]  ,E_{7}\left[  133\right]  ,E_{8}\left[  248\right]
,F_{4}\left[  52\right]  ,G_{2}\left[  14\right]  $

Practically : \textit{usually there is no need for all the material above}. We
know that any finite dimensional simple complex Lie algebra belongs to one the
9 types above, and semi-simple Lie algebras can be decomposed in the sum of
simple complex Lie algebras. So we can proceed directly with them.

The structure and classification of real Lie algebras are a bit more
complicated than complex ones.\ However the outcome is very similar and based
on the list above.

\subsubsection{Compact algebras}

This the only topic for which we use analysis concepts in Lie algebra study.

\paragraph{Definition of Int(A)\newline}

Let A be a Lie algebra such that A is also a Banach vector space (it will be
the case if A is a finite dimensional vector space). Then :

i) For any continuous map $f\in%
\mathcal{L}%
(A;A)$ the map : $\exp f=\sum_{n=0}^{\infty}\frac{1}{n!}f^{n}\in%
\mathcal{L}%
\left(  A;A\right)  $ ($f^{n}$ is the n iterate of f) is well defined and has
an inverse.

ii) If f is a continuous morphism then $f^{n}\left(  \left[  X,Y\right]
\right)  =\left[  f^{n}\left(  X\right)  ,f^{n}\left(  Y\right)  \right]  $
and $\exp(f)$ is a continuous automorphism of A

iii) If for any X the map ad(X) is continuous then : $\exp ad(X)$ is a
continuous automorphism of A

iv) The subset of continuous (then smooth) automorphisms of A is a Lie group
whose connected component of the identity is denoted \textbf{Int(A)}.

\paragraph{Compact Lie algebra\newline}

\begin{definition}
A Banach Lie algebra A is said to be \textbf{compact} if Int(A) is compact
with the topology of $%
\mathcal{L}%
(A;A).$
\end{definition}

Int(A) is a manifold, if it is compact, it must be locally compact, so it
cannot be infinite dimensional.\ Therefore there is no compact infinite
dimensional Lie algebra. Moreover :

\begin{theorem}
(Duistermaat p.151) A compact complex Lie algebra is abelian.
\end{theorem}

And their Lie algebra is trivial. So the story of compact Lie algebras is
limited to real finite dimensional Lie algebras.

\begin{theorem}
(Duistermaat p.149) For a real finite dimensional Lie algebra A the following
are equivalent :

i) A is compact

ii) its Killing form is negative semi-definite and its kernel is the center of A

iii) A is the Lie algebra of a compact group
\end{theorem}

\begin{theorem}
(Duistermaat p.151) For a real finite dimensional Lie algebra A the following
are equivalent :

i) A is compact and semi-simple

ii) A is compact and has zero center

iii) its Killing form is negative definite

iv) Every Lie group with Lie algebra Lie isomorphic to A is compact
\end{theorem}

\subsubsection{Semi-simple real Lie algebra}

\paragraph{Compact real forms\newline}

\begin{theorem}
(Knapp p.434) The isomorphisms classes of :

finite dimensional, compact, semi simple real Lie algebras $A_{0}$ on one hand,

and of finite dimensional, semi simple, complex Lie algebras A on the other hand,

are in one-one correspondance : A is the complexification of $A_{0}$ and
$A_{0}$ is a compact real form of A. Under this correspondance simple Lie
algebras correspond to simple Lie algebras.
\end{theorem}

So \textit{any finite dimensional complex semi-simple Lie algebra A\ has a
compact real form} $u_{0}.$ A can be written \ : $A=u_{0}\oplus iu_{0}$ where
$u_{0}$ is a real compact Lie algebra. Any two compact real forms are
conjugate via Int(A).

Given an abstract Cartan matrix C there is a unique, up to isomorphism,
compact real semi simple algebra such that its complexified has C as Cartan matrix.

\paragraph{Cartan involution\newline}

\begin{definition}
A \textbf{Cartan involution} on a real semi-simple Lie algebra A is an
automorphism on A, such that $\theta^{2}=Id$ and $B_{\theta}:B_{\theta}\left(
X,Y\right)  =-B\left(  X,\theta Y\right)  $\ is positive definite.
\end{definition}

Let A be a semi-simple complex Lie algebra, $u_{0}$ its compact real
form.\ Then $\forall Z\in A,\exists x,y\in u_{0}:Z=x+iy.$

Define :$\theta:A\rightarrow A::\theta\left(  Z\right)  =x-iy$ this is a
Cartan involution on $A_{%
\mathbb{R}
}=\left(  u_{0},u_{0}\right)  $ and all the Cartan involutions are of this kind.

\begin{theorem}
(Knapp p.445) Any real semi-simple finite dimensional Lie algebra $A_{0}$ has
a Cartan involution. And any two Cartan involutions are conjugate via
$Int(A_{0})$.
\end{theorem}

\paragraph{Cartan decomposition:}

\begin{definition}
A Cartan decomposition of a real finite dimensional Lie algebra $\left(
A,\left[  {}\right]  \right)  $ is a pair of vector subspaces $l_{0},p_{0}$ of
A such that :

i) $A=l_{0}\oplus p_{0}$

ii) $l_{0}$ is a subalgebra of A

iii) $\left[  l_{0},l_{0}\right]  \sqsubseteq l_{0},\left[  l_{0}%
,p_{0}\right]  \sqsubseteq p_{0},,\left[  p_{0},p_{0}\right]  \sqsubseteq
l_{0},$

iv) the Killing form B of A is negative definite on $l_{0}$ and positive
definite on $p_{.0}$

v) $l_{0},p_{0}$ are orthogonal under B and $B_{\theta}$
\end{definition}

\begin{theorem}
Any real semi-simple finite dimensional Lie algebra A has a Cartan decomposition
\end{theorem}

\begin{proof}
Any real semi-simple finite dimensional Lie algebra A has a Cartan involution
$\theta,$ which has two eigenvalues : $\pm1.$ Taking the eigenspaces
decomposition of A with respect to $\theta$ :$\theta\left(  l_{0}\right)
=l_{0},\theta\left(  p_{0}\right)  =-p_{0}$ we have a Cartan decomposition.
\end{proof}

Moreover $l_{0},p_{0}$ are orthogonal under $B_{\theta}.$

Conversely a Cartan decomposition gives a Cartan involution with the
definition $\theta=+Id$ on $l_{0}$ and $\theta=-Id$ on $p_{0}$

If $A=l_{0}\oplus p_{0}$ is a Cartan decomposition of A, then the real Lie
algebra $A=l_{0}\oplus ip_{0}$ is a compact real form of the complexified
$\left(  A_{0}\right)  _{%
\mathbb{C}
}.$

\begin{theorem}
(Knapp p.368) A finite dimensional real semi simple Lie algebra is isomorphic
to a Lie algebra of real matrices that is closed under transpose. The
isomorphism can be specified so that a Cartan involution is carried to
negative transpose.
\end{theorem}

\paragraph{Classification of simple real Lie algebras\newline}

The procedure is to go from complex semi simple Lie algebras to real forms by
Cartan involutions. It uses Vogan diagrams, which are Dynkin diagrams with
additional information about the way to get the real forms.

The results are the following (Knapp p.421) :

Up to isomorphism, any \textit{simple} real finite dimensional Lie algebra
belongs to one the following types :

i) the compact real forms of the complex simple Lie algebras (they are compact
real simple Lie algebras) :

A$_{n},n\geq1:su(n+1,%
\mathbb{C}
)$

B$_{n},n\geq2:so(2n+1,%
\mathbb{R}
)$

C$_{n},n\geq3:sp(n,%
\mathbb{C}
)\cap u(2n)$

D$_{n},n\geq4:so(2n,%
\mathbb{R}
)$

and similarly for the exceptional Lie algebras (Knapp p.413).

ii) the real structures of the complex simple Lie algebras.\ Considered as
real Lie algebras they are couples (A,B) of matrices:

A$_{n},n\geq1:su(n+1,%
\mathbb{C}
)\oplus isu(n+1,%
\mathbb{C}
)$

B$_{n},n\geq2:so(2n+1,%
\mathbb{R}
)\oplus iso(2n+1,%
\mathbb{R}
)$

C$_{n},n\geq3:\left(  sp(n,%
\mathbb{C}
)\cap u(2n)\right)  \oplus i\left(  sp(n,%
\mathbb{C}
)\cap u(2n)\right)  $

D$_{n},n\geq4:so(2n,%
\mathbb{R}
)\oplus iso(2n,%
\mathbb{R}
)$

and similarly for \ $E_{6},E_{7},E_{8},F_{4},G_{2}$

iii) the linear matrix algebras (they are non compact) :

$su(p,q,%
\mathbb{C}
):p\geq q>0,p+q>1$

$so(p,q,%
\mathbb{R}
):p>q>0,p+q\;odd\;and>4\ or\;p\geq q>0,p+q\;even\;p+q>7$

$sp(p,q,H):p\geq q,p+q>2$

$sp(n,R):n>2$

$so^{\ast}(2n,%
\mathbb{C}
):n>3$

$sl(n,%
\mathbb{R}
):n>2$

$sl(n,H):n>1$

iv) 12 non complex, non compact exceptional Lie algebras (p 416)

Semi-simple Lie algebras are direct sum of simple Lie algebras.

\newpage

\section{LIE GROUPS}

\subsection{General definitions and results}

\label{Lie groups general definitions and results}

\subsubsection{Definition of Lie groups}

Whereas Lie algebras involve essentially only algebraic tools, at the group
level analysis is of paramount importance.\ And there are two ways to deal
with this : with simple topological structure and we have the topological
groups, or with manifold structure and we have the Lie groups.

We will denote :

the operation $G\times G\rightarrow G::xy=z$

the inverse : $G\rightarrow G::x\rightarrow x^{-1}$

the unity : 1

\paragraph{Topological group\newline}

\begin{definition}
A \textbf{topological grou}p is a Hausdorff topological space, endowed with an
algebraic structure such that the operations product and inverse are continuous.
\end{definition}

With such structure we can handle all the classic concepts of general topology
: convergence, integration, continuity of maps over a group,...What we will
miss is what is related to derivatives. Lie algebras can be related to Lie
groups only.

\begin{definition}
A discrete group is a group endowed with the discrete topology.
\end{definition}

Any set endowed with an algebraic group structure can be made a topological
group with the discrete topology. A discrete group which is second-countable
has necessarily countably many elements. A discrete group is compact iff it is
finite. A finite topological group is necessarily discrete.

\begin{theorem}
(Wilansky p.240) Any product of topological groups is a topological group
\end{theorem}

\begin{theorem}
(Wilansky p.243) A topological group is a regular topological space
\end{theorem}

\begin{theorem}
(Wilansky p.250) A locally compact topological group is paracompact and normal
\end{theorem}

\paragraph{Lie group\newline}

\begin{definition}
A \textbf{Lie group} is a class r manifold G, modeled on a Banach space E over
a field K, endowed with a group structure such that the product and the
inverse are class r maps.
\end{definition}

Moreover we will assume that G is a normal, Hausdorff, second countable
topological space, which is equivalent to say that G is a metrizable,
separable manifold (see Manifolds).

The manifold structure (and thus the differentiability of the operations) are
defined with respect to the field K. As E is a Banach we need the field K to
be complete (practically K=$%
\mathbb{R}
$ or $%
\mathbb{C}
).$ While a topological group is not linked to any field, a Lie group is
defined over a field K, through its manifold structure that is necessary
whenever we use derivative on G.

The \textbf{dimension of the Lie group} is the dimension of the manifold.
Notice that we do not assume that the manifold is finite dimensional : we will
precise this point when it is necessary. Thus if G is infinite dimensional,
following the Henderson theorem, it can be embedded as an open subset of an
infinite dimensional, separable, Hilbert space.

For the generality of some theorems we take the convention that finite groups
with the discrete topology are Lie groups of dimension zero.

A Lie group is locally compact iff it is finite dimensional.

If G has a complex manifold structure then it is a smooth manifold, and the
operations being C-differentiable are holomorphic.

\begin{theorem}
Montgomery and Zippin (Kolar p.43) If G is a separable, locally compact
topological group, with a neighbourhood of 1 which does not contain a proper
subgroup, then G is a Lie group .
\end{theorem}

\begin{theorem}
Gleason, Montgomery and Zippin (Knapp p.99 for the real case) : For a real
finite dimensional Lie group G there is exactly one analytic manifold
structure on G which is consistent with the Lie group structure
\end{theorem}

As the main advantage of the manifold structure (vs the topological structure)
is the use of derivatives, in the following we will always assume that a Lie
group has the structure of a smooth (real or complex) manifold, with smooth
group operations.

\paragraph{Connectedness\newline}

Topological groups are locally connected, but usually not connected.

The \textbf{connected component of the identity}, denoted usually $G_{0}$ is
of a particular importance. This is a group and a for a Lie group this is a
Lie subgroup.

The direct product of connected groups, with the product topology, is
connected. So $SU(3)\times SU(2)\times U\left(  1\right)  $ is connected.

Example : $GL(%
\mathbb{R}
,1)=\left(
\mathbb{R}
,\times\right)  $ has two connected components, $GL_{0}(%
\mathbb{R}
,1)=\left\{  x,x>0\right\}  $

\paragraph{Examples of Lie groups\newline}

1. The group GL(K,n) of square nxn inversible matrices over a field K : it is
a vector subspace of $K^{n^{2}}$ which is open as the preimage of $\det
X\neq0$ so it is a manifold, and the operations are smooth.

2. A Banach vector space is an abelian Lie group with addition

\subsubsection{Translations}

\paragraph{Basic operations\newline}

The \textbf{translations} over a group are just the rigth (R) and left (L)
products (same definition as for any group - see Algebra). They are smooth
diffeomorphisms.\ There is a commonly used notation for them :

\begin{notation}
$R_{a}$ is the right multiplication by a : $R_{a}:G\rightarrow G:R_{a}x=xa$
\end{notation}

\begin{notation}
$L_{a}$ is the left multiplication by a : $L_{a}:G\rightarrow G::L_{a}x=ax$
\end{notation}

and $R_{a}x=xa=L_{x}a$

These operations commute : $L_{a}\circ R_{b}=R_{b}\circ L_{a}$

Because the product is associative we have the identities :

$abc=R_{c}ab=R_{c}\left(  R_{a}b\right)  =L_{a}bc=L_{a}\left(  L_{b}c\right)
$

$L_{ab}=L_{a}\circ L_{b};R_{a}\circ R_{b}=R_{ab};$

$L_{a^{-1}}=\left(  L_{a}\right)  ^{-1};R_{a^{-1}}=\left(  R_{a}\right)
^{-1};$

$L_{a^{-1}}(a)=1;R_{a^{-1}}(a)=1$

$L_{1}=R_{1}=Id$

The \textbf{conjugation} with respect to a is the map :

$Conj_{a}:G\rightarrow G::Conj_{a}x=axa^{-1}$

\begin{notation}
$Conj_{a}x=L_{a}\circ R_{a^{-1}}(x)=R_{a^{-1}}\circ L_{a}(x)$
\end{notation}

If the group is commutative then $Conj_{a}x=x$

Conjugation is an inversible map.

\paragraph{Derivatives\newline}

If G is a Lie group all these operations are smooth diffeomorphisms over G so
we have the linear bijective maps :

\begin{notation}
$L_{a}^{\prime}x$ is the derivative of $L_{a}\left(  g\right)  $ with respect
to g, at g=x; $L_{a}^{\prime}x\in G%
\mathcal{L}%
\left(  T_{x}G;T_{ax}G\right)  $
\end{notation}

\begin{notation}
$R_{a}^{\prime}x$ is the derivative of $R_{a}\left(  g\right)  $ with respect
to g, at g=x; $R_{a}^{\prime}x\in G%
\mathcal{L}%
\left(  T_{x}G;T_{xa}G\right)  $
\end{notation}

The product can be seen as a two variables map :

$M:G\times G\rightarrow G:M(x,y)=xy$ with partial derivatives :

$u\in T_{x}G,v\in T_{y}G:M^{\prime}(x,y)(u,v)=R_{y}^{\prime}(x)u+L_{x}%
^{\prime}(y)v\in T_{xy}G$

$\frac{\partial}{\partial x}\left(  xy\right)  =\frac{\partial}{\partial
z}\left(  R_{y}\left(  z\right)  \right)  |_{z=x}=R_{y}^{\prime}(x)$

$\frac{\partial}{\partial y}\left(  xy\right)  =\frac{\partial}{\partial
z}\left(  L_{x}\left(  z\right)  \right)  |_{z=y}=L_{x}^{\prime}(y)$

Let g,h be differentiable maps $G\rightarrow G$ and :

$f:G\rightarrow G::f(x)=g(x)h(x)=M\left(  g(x),h(x)\right)  $%

\begin{equation}
f^{\prime}(x)=\frac{d}{dx}\left(  g\left(  x\right)  h\left(  x\right)
\right)  =R_{h(x)}^{\prime}(g(x))\circ g^{\prime}(x)+L_{g(x)}^{\prime
}(h(x))\circ h^{\prime}(x)
\end{equation}

Similarly for the inverse map :$\Im:G\rightarrow G::\Im\left(  x\right)
=x^{-1}$%

\begin{equation}
\frac{d}{dx}\Im(x)|_{x=a}=\Im^{\prime}(a)=-R_{a^{-1}}^{\prime}(1)\circ
L_{a^{-1}}^{\prime}(a)=-L_{a^{-1}}^{\prime}(1)\circ R_{a^{-1}}^{\prime}(a)
\end{equation}

and for the map : $f:G\rightarrow G::f(x)=g(x)^{-1}=\Im\circ g\left(
x\right)  $

$\frac{d}{dx}\left(  g(x)^{-1}\right)  =f^{\prime}(x)$

$=-R_{g(x)^{-1}}^{\prime}(1)\circ L_{g(x)^{-1}}^{\prime}(g(x))\circ g^{\prime
}(x)=-L_{g(x)^{-1}}^{\prime}(1)\circ R_{g(x)^{-1}}^{\prime}(g(x))\circ
g^{\prime}(x)$

From the relations above we get the useful identities :%

\begin{equation}
\left(  L_{g}^{\prime}1\right)  ^{-1}=L_{g^{-1}}^{\prime}(g);\left(
R_{g}^{\prime}1\right)  ^{-1}=R_{g^{-1}}^{\prime}(g)
\end{equation}

\begin{equation}
\left(  L_{g}^{\prime}h\right)  ^{-1}=L_{g^{-1}}^{\prime}(gh);\left(
R_{g}^{\prime}h\right)  ^{-1}=R_{g^{-1}}^{\prime}(hg)
\end{equation}

\begin{equation}
L_{g}^{\prime}(h)=L_{gh}^{\prime}(1)L_{h^{-1}}^{\prime}(h);R_{g}^{\prime
}(h)=R_{hg}^{\prime}(1)R_{h^{-1}}^{\prime}(h)
\end{equation}

\begin{equation}
L_{gh}^{\prime}(1)=L_{g}^{\prime}(h)L_{h}^{\prime}\left(  1\right)
;R_{hg}^{\prime}(1)=R_{g}^{\prime}(h)R_{h}^{\prime}(1)
\end{equation}

\begin{equation}
\left(  L_{g}^{\prime}\left(  h\right)  \right)  ^{-1}=L_{h}^{\prime}\left(
1\right)  L_{\left(  gh\right)  ^{-1}}^{\prime}(gh)
\end{equation}

\paragraph{Group of invertible endomorphisms of a Banach\newline}

Let E be a Banach vector space, then the set
$\mathcal{L}$%
$\left(  E;E\right)  $\ of continuous linear map is a Banach vector space,
thus a manifold. The set $G%
\mathcal{L}%
\left(  E;E\right)  $ of continuous automorphisms over E is an open subset of
$\mathcal{L}$%
$\left(  E;E\right)  $\ thus a manifold. It is a group with the composition
law and the operations are differentiable . So $G%
\mathcal{L}%
\left(  E;E\right)  $ is a Lie group (but not a Banach algebra).

The derivative of the composition law and the inverse are (see derivatives) :

$M:G%
\mathcal{L}%
\left(  E;E\right)  \times G%
\mathcal{L}%
\left(  E;E\right)  \rightarrow G%
\mathcal{L}%
\left(  E;E\right)  ::M\left(  f,g\right)  =f\circ g$

$M^{\prime}\left(  f,g\right)  \left(  \delta f,\delta g\right)  =\delta
f\circ g+f\circ\delta g$

$\Im:G%
\mathcal{L}%
\left(  E;E\right)  \rightarrow G%
\mathcal{L}%
\left(  E;E\right)  ::\Im\left(  f\right)  =f^{-1}$

$\left(  \Im(f)\right)  ^{\prime}\left(  \delta f\right)  =-f^{-1}\circ\delta
f\circ f^{-1}$

Thus we have :

$M^{\prime}\left(  f,g\right)  \left(  \delta f,\delta g\right)
=R_{g}^{\prime}(f)\delta f+L_{f}^{\prime}(g)\delta g$

$R_{g}^{\prime}(f)\delta f=\delta f\circ g=R_{g}\left(  \delta f\right)  ,$

$L_{f}^{\prime}(g)\delta g=f\circ\delta g=L_{f}\left(  \delta g\right)  $

\paragraph{Tangent bundle of a Lie group\newline}

The tangent bundle $TG=\cup_{x\in G}T_{x}G$ of a Lie group is a manifold
G$\times$E, on the same field K with dimTG=2$\times$dimG.\ We can define a
multiplication on TG as follows :

$M:TG\times TG\rightarrow TG::M(U_{x},V_{y})=R_{y}^{\prime}(x)U_{x}%
+L_{x}^{\prime}(y)V_{y}\in T_{xy}G$

$\Im:TG\rightarrow TG::\Im(V_{x})=-R_{x^{-1}}^{\prime}(1)\circ L_{x^{-1}%
}^{\prime}(x)V_{x}=-L_{x^{-1}}^{\prime}(x)\circ R_{x^{-1}}^{\prime}(x)V_{x}\in
T_{x^{-1}}G$

Identity : $U_{x}=0_{1}\in T_{1}G$

Notice that the operations are between vectors on the tangent bundle TG, and
not vector fields (the set of vector fields is denoted $\mathfrak{X}\left(
TG\right)  $).

The operations are well defined and smooth. So TG has a Lie group structure.
It is isomorphic to the semi direct group product : $TG\simeq\left(
T_{1}G,+\right)  \varpropto_{Ad}G$ with the map $Ad:G\times T_{1}G\rightarrow
T_{1}G$\ (Kolar p.98).

\bigskip

\subsection{Lie algebra of a Lie group}

\subsubsection{Subalgebras of invariant vector fields}

\begin{theorem}
The subspace of the vector fields on a Lie group G over a field K, which are
invariant by the left translation have a structure of Lie subalgebra over K
with the commutator of vector fields as bracket. Similarly for the vector
fields invariant by the right translation
\end{theorem}

\begin{proof}
i) Left translation, right translation are diffeomorphisms, so the push
forward of vector fields is well defined.

A left invariant vector field is such that:

$X\in\mathfrak{X}\left(  TG\right)  :\forall g\in G:L_{g\ast}%
X=X\Leftrightarrow L_{g}^{\prime}\left(  x\right)  X(x)=X\left(  gx\right)  $

so with x=1 : $\forall g\in G:L_{g}^{\prime}\left(  1\right)  X(1)=X\left(
g\right)  $

The set of left smooth invariant vector fields is

LVG=$\left\{  X\in\mathfrak{X}_{\infty}\left(  TG\right)  :X\left(  g\right)
=L_{g}^{\prime}\left(  1\right)  u,u\in T_{1}G\right\}  $

Similarly for the smooth right invariant vector fields :

RVG=$\left\{  X\in\mathfrak{X}_{\infty}\left(  TG\right)  :X\left(  g\right)
=R_{g}^{\prime}\left(  1\right)  u,u\in T_{1}G\right\}  $

ii) The set $\mathfrak{X}_{\infty}\left(  TG\right)  $ of smooth vector fields
over G has an infinite dimensional Lie algebra structure over the field K,
with the commutator as bracket. The push forward of a vector field over G
preserves the commutator (see Manifolds).

$X,Y\in LVG:\left[  X,Y\right]  =Z\in\mathfrak{X}_{\infty}\left(  TG\right)  $

$L_{g\ast}\left(  Z\right)  =L_{g\ast}\left(  \left[  X,Y\right]  \right)
=\left[  L_{g\ast}X,L_{g\ast}Y\right]  =\left[  X,Y\right]  =Z\Rightarrow
\left[  X,Y\right]  \in LVG$

So the sets LVG of left invariant and RVG right invariant vector fields are
both Lie subalgebras of $\mathfrak{X}_{\infty}\left(  TG\right)  $.
\end{proof}

\subsubsection{Lie algebra structure of $T_{1}G$}

\begin{theorem}
The derivative $L_{g}^{\prime}(1)$ of the left translation at x=1 is an
isomorphism to the set of left invariant vector fields. The tangent space
$T_{1}G$\ becomes a Lie algebra over K with the bracket

$\left[  u,v\right]  _{T_{1}G}=L_{g^{-1}}^{\prime}\left(  g\right)  \left[
L_{g}^{\prime}\left(  1\right)  u,L_{g}^{\prime}\left(  1\right)  v\right]
_{LVG}$

So for any two left invariant vector fields X,Y :%

\begin{equation}
\left[  X,Y\right]  \left(  g\right)  =L_{g}^{\prime}\left(  1\right)  \left(
\left[  X\left(  1\right)  ,Y\left(  1\right)  \right]  _{T_{1}G}\right)
\end{equation}

\end{theorem}

\begin{proof}
The map : $\lambda:T_{1}G\rightarrow LVG::\lambda\left(  u\right)
(g)=L_{g}^{\prime}\left(  1\right)  u$ is an injective linear map. It has an inverse:

$\forall X\in LVG,\exists u\in T_{1}G:X\left(  g\right)  =L_{g}^{\prime
}\left(  1\right)  u$

$\lambda^{-1}:LVG\rightarrow T_{1}G::u=L_{g^{-1}}^{\prime}\left(  g\right)
X\left(  g\right)  $ which is linear.

So we map : $\left[  {}\right]  _{T_{1}G}:T_{1}G\times T_{1}G\rightarrow
T_{1}G::\left[  u,v\right]  _{T_{1}G}=\lambda^{-1}\left(  \left[
\lambda\left(  u\right)  ,\lambda\left(  v\right)  \right]  _{LVG}\right)  $
is well defined and it is easy to check that it defines a Lie bracket.
\end{proof}

Remarks :

i) if M is finite dimensional, there are several complete topologies available
on $\mathfrak{X}_{\infty}\left(  TG\right)  $ so the continuity of the map
$\lambda$\ is well assured.\ There is no such thing if G is infinite
dimensional, however $\lambda$\ and the bracket are well defined algebraically.

ii) some authors (Duistermaat) define the Lie bracket through rigth invariant
vector fields.

With this bracket the tangent space $T_{1}G$ becomes a Lie algebra, the
\textbf{Lie algebra of the Lie group G }on the field K, with the same
dimension as G as manifold, which is Lie isomorphic to the Lie algebra LVG.

\begin{notation}
$T_{1}G$ is the Lie algebra of the Lie group G
\end{notation}

\subsubsection{Right invariant vector fields}

The right invariant vector fields define also a Lie algebra structure on
$T_{1}G,$ which is Lie isomorphic to the previous one and the bracket have
opposite signs (Kolar p.34).

\begin{theorem}
If X,Y are two smooth right invariant vector fields on the Lie group G, then

$\left[  X,Y\right]  \left(  g\right)  =-R_{g}^{\prime}(1)\left[  X\left(
1\right)  ,X\left(  1\right)  \right]  _{T_{1}G}$
\end{theorem}

\begin{proof}
The derivative of the inverse map (see above) is :

$\frac{d}{dx}\Im(x)|_{x=g}=\Im^{\prime}(g)=-R_{g^{-1}}^{\prime}(1)\circ
L_{g^{-1}}^{\prime}(g)\Rightarrow\Im^{\prime}(g^{-1})=-R_{g}^{\prime}(1)\circ
L_{g}(g^{-1})$

$R_{g}^{\prime}(1)=-\Im^{\prime}(g^{-1})\circ L_{g^{-1}}^{\prime}\left(
1\right)  $

So if X is a right invariant vector field : $X\left(  g\right)  =R_{g}%
^{\prime}\left(  1\right)  X(1)$ then $X\left(  g\right)  =-\Im^{\prime
}(g^{-1})X_{L}$ with X$_{L}=L_{g^{-1}}^{\prime}\left(  1\right)  X\left(
1\right)  $ a left invariant vector field.

For two right invariant vector fields :

$\left[  X,Y\right]  =\left[  -\Im^{\prime}(g^{-1})X_{L},-\Im^{\prime}%
(g^{-1})Y_{L}\right]  =\Im^{\prime}(g^{-1})\left[  X_{L},Y_{L}\right]  $

$=\Im^{\prime}(g^{-1})\left[  L_{g^{-1}}^{\prime}\left(  1\right)  X\left(
1\right)  ,L_{g^{-1}}^{\prime}\left(  1\right)  X\left(  1\right)  \right]  $

$=\Im^{\prime}(g^{-1})L_{g^{-1}}^{\prime}\left(  1\right)  \left[  X\left(
1\right)  ,X\left(  1\right)  \right]  _{T_{1}G}=-R_{g}^{\prime}(1)\left[
X\left(  1\right)  ,X\left(  1\right)  \right]  _{T_{1}G}$
\end{proof}

\begin{theorem}
(Kolar p.34) If X,Y are two smooth vector fields on the Lie group G,
respectively right invariant and left invariant, then $\left[  X,Y\right]  =0$
\end{theorem}

\subsubsection{Lie algebra of the group of automorphisms of a Banach space}

\begin{theorem}
For a Banach vector space E, the Lie algebra of $G%
\mathcal{L}%
\left(  E;E\right)  $\ is $%
\mathcal{L}%
\left(  E;E\right)  $ . It is a Banach Lie algebra with bracket $\left[
u,v\right]  =u\circ v-v\circ u$
\end{theorem}

\begin{proof}
If E is a Banach vector space then the set $G%
\mathcal{L}%
\left(  E;E\right)  $ of continuous automorphisms over E with the composition
law is an open of the Banach vector space
$\mathcal{L}$%
$\left(  E;E\right)  .$ Thus the tangent space at any point is
$\mathcal{L}$%
$\left(  E;E\right)  $ .

A left invariant vector field is :

$f\in G%
\mathcal{L}%
\left(  E;E\right)  ,u\in%
\mathcal{L}%
\left(  E;E\right)  :X_{L}\left(  f\right)  =L_{f}^{\prime}\left(  1\right)
u=f\circ u=L_{f}\left(  1\right)  u$

The commutator of two vector fields V,W:$G%
\mathcal{L}%
\left(  E;E\right)  \rightarrow%
\mathcal{L}%
\left(  E;E\right)  $ is (see Differential geometry) :

$\left[  V,W\right]  \left(  f\right)  =\left(  \frac{d}{df}W\right)  \left(
V(f)\right)  -\left(  \frac{d}{df}V\right)  \left(  W\left(  f\right)
\right)  $

with the derivative : $V^{\prime}(f):%
\mathcal{L}%
\left(  E;E\right)  \rightarrow%
\mathcal{L}%
\left(  E;E\right)  $ and here :

$X_{L}\left(  f\right)  =f\circ u=R_{u}\left(  f\right)  \Rightarrow\left(
X_{L}\right)  ^{\prime}\left(  f\right)  =\frac{d}{df}\left(  R_{u}\left(
f\right)  \right)  =R_{u}$

$\left[  L_{f}^{\prime}\left(  1\right)  u,L_{f}^{\prime}\left(  1\right)
v\right]  =R_{v}\left(  f\circ u\right)  -R_{u}\left(  f\circ v\right)
=f\circ\left(  u\circ v-v\circ u\right)  =L_{f}\left(  1\right)  \left[
u,v\right]  _{%
\mathcal{L}%
\left(  E;E\right)  }=f\circ\left[  u,v\right]  _{%
\mathcal{L}%
\left(  E;E\right)  }$

So the Lie bracket on the Lie algebra
$\mathcal{L}$%
(E;E) is : $u,v\in%
\mathcal{L}%
\left(  E;E\right)  :\left[  u,v\right]  =u\circ v-v\circ u.$ This is a
continuous bilinear map because the composition of maps is itself a continuous operation.
\end{proof}

\subsubsection{The group of automorphisms of the Lie algebra}

Following the conditions imposed to the manifold structure of G, the tangent
space $T_{x}G$ at any point can be endowed with the structure of a Banach
space (see Differential geometry), which is diffeomorphic to E itself. So this
is the case for $T_{1}G$\ which becomes a Banach Lie algebra, linear
diffeomorphic to E.\ 

The set
$\mathcal{L}$%
$\left(  T_{1}G;T_{1}G\right)  $ of continuous maps over $T_{1}G$ has a Banach
algebra structure with composition law and the subset $G%
\mathcal{L}%
\left(  T_{1}G;T_{1}G\right)  $\ of continuous automorphisms of $T_{1}G$ is a
Lie group. Its component of the identity Int($T_{1}G)$ is also a Lie group.
Its Lie algebra is $%
\mathcal{L}%
\left(  T_{1}G;T_{1}G\right)  $ endowed with the bracket : $f,g\in%
\mathcal{L}%
\left(  T_{1}G;T_{1}G\right)  ::\left[  f,g\right]  =f\circ g-g\circ f$

One consequence of these results is :

\begin{theorem}
A Lie group is a parallellizable manifold
\end{theorem}

\begin{proof}
take any basis $\left(  e_{\alpha}\right)  _{\alpha\in A}$\ of the Lie algebra
and transport the basis in any point x by left invariant vector fields :
$\left(  L_{x}^{\prime}\left(  1\right)  e_{\alpha}\right)  _{\alpha\in A}$ is
a basis of $T_{x}G.$
\end{proof}

\begin{theorem}
On the Lie algebra $T_{1}G$ of a Lie group G, endowed with a symmetric scalar
product $\left\langle {}\right\rangle _{T_{1}G}$ which is preserved by the
adjoint map :
\end{theorem}

\begin{proposition}
$\forall X,Y,Z\in T_{1}G:\left\langle X,\left[  Y,Z\right]  \right\rangle
=\left\langle \left[  X,Y\right]  ,Z\right\rangle $
\end{proposition}

\begin{proof}
$\forall g\in G:\left\langle Ad_{g}X,Ad_{g}Y\right\rangle =\left\langle
X,Y\right\rangle $

take the derivative with respect to g at g = 1 for $Z\in T_{1}G:$

$\left(  Ad_{g}X\right)  ^{\prime}\left(  Z\right)  =ad\left(  Z\right)
\left(  X\right)  =\left[  Z,X\right]  $

$\left\langle \left[  Z,X\right]  ,Y\right\rangle +\left\langle X,\left[
Z,Y\right]  \right\rangle =0\Leftrightarrow\left\langle X,\left[  Y,Z\right]
\right\rangle =\left\langle \left[  Z,X\right]  ,Y\right\rangle $

exchange X,Z:

$\Rightarrow\left\langle Z,\left[  Y,X\right]  \right\rangle =\left\langle
\left[  X,Z\right]  ,Y\right\rangle =-\left\langle \left[  Z,X\right]
,Y\right\rangle =-\left\langle X,\left[  Y,Z\right]  \right\rangle
=-\left\langle Z,\left[  X,Y\right]  \right\rangle $
\end{proof}

\subsubsection{Exponential map}

The exponential map $\exp:%
\mathcal{L}%
\left(  E;E\right)  \rightarrow G%
\mathcal{L}%
\left(  E;E\right)  $ is well defined on the set of continuous linear maps on
a\ Banach space E . It is \ related to one paramer groups, meaning the
differential equation $\frac{dU}{dt}=SU(t)$ between operators on E, where S is
the infinitesimal generator and U(t)=exptS.\ On manifolds the flow of vector
fields provides another concept of one parameter group of diffeomorphisms.

\paragraph{One parameter subgroup\newline}

\begin{theorem}
(Kolar p.36) On a Lie group G, left and right invariant vector fields are the
infinitesimal generators of one parameter groups of diffeomorphism. The flow
of these vector fields is complete.
\end{theorem}

\begin{proof}
i) Let $\phi:%
\mathbb{R}
\rightarrow G$ be a smooth map such that : $\forall s,t,s+t\in%
\mathbb{R}
:\phi\left(  s+t\right)  =\phi\left(  s\right)  \phi\left(  t\right)  $ so
$\phi\left(  0\right)  =1$

Then : $F_{R}:%
\mathbb{R}
\times G\rightarrow G::F_{R}\left(  t,x\right)  =\phi\left(  t\right)
x=R_{x}\phi\left(  t\right)  $ is a one parameter group of diffeomorphism on
the manifold G, as defined previously (see Differential geometry). Similarly
with $F_{L}\left(  t,x\right)  =\phi\left(  t\right)  x=L_{x}\phi\left(
t\right)  $

So $F_{R},F_{L}$ have an infinitesimal generator, which is given by the vector
field :

$X_{L}\left(  x\right)  =L_{x}^{\prime}\left(  1\right)  \left(  \frac{d\phi
}{dt}|_{t=0}\right)  =L_{x}^{\prime}\left(  1\right)  u$

$X_{R}\left(  x\right)  =R_{x}^{\prime}\left(  1\right)  \left(  \frac{d\phi
}{dt}|_{t=0}\right)  =R_{x}^{\prime}\left(  1\right)  u$

with $u=\frac{d\phi}{dt}|_{t=0}\in T_{1}G$

2. And conversely any left (right) invariant vector field gives rise to the
flow $\Phi_{X_{L}}\left(  t,x\right)  $ (or $\Phi_{X_{R}}\left(  t,x\right)
)$ which is defined on some domain $D\left(  \Phi_{X_{L}}\right)  =\cup_{x\in
G}\left\{  J_{x}\times\left\{  x\right\}  \right\}  \subset%
\mathbb{R}
\times G$ which is an open neighborhood of 0xG

Define : $\phi\left(  t\right)  =\Phi_{X_{L}}\left(  t,1\right)
\Rightarrow\phi\left(  s+t\right)  =\Phi_{X_{L}}\left(  s+t,1\right)
=\Phi_{X_{L}}\left(  s,\Phi_{X_{L}}\left(  t,1\right)  \right)  =\Phi_{X_{L}%
}\left(  s,\phi\left(  t\right)  \right)  $ thus $\phi$ is defined over $%
\mathbb{R}
$

Define $F_{L}\left(  t,x\right)  =L_{x}\phi\left(  t\right)  $ then :
$\frac{\partial}{\partial t}F_{L}\left(  t,x\right)  |_{t=0}=L_{x}^{\prime
}\left(  1\right)  \frac{d\phi}{dt}|_{t=0}=X_{L}\left(  x\right)  $ so this is
the flow of $X_{L}$ and

$F_{L}\left(  t+s,x\right)  =F_{L}\left(  t,F_{L}\left(  s,x\right)  \right)
=L_{F_{L}\left(  s,x\right)  }\phi\left(  t\right)  =L_{x\phi\left(  s\right)
}\phi\left(  t\right)  =x\phi\left(  s\right)  \phi\left(  t\right)
\Rightarrow\phi\left(  s+t\right)  =\phi\left(  s\right)  \phi\left(
t\right)  $

Thus the flow of left and rigth invariant vector fields are complete.
\end{proof}

\paragraph{The exponential map\newline}

\begin{definition}
The \textbf{exponential map} on a Lie group is the map :%

\begin{equation}
\exp:T_{1}G\rightarrow G::\exp u=\Phi_{X_{L}}\left(  1,1\right)  =\Phi_{X_{R}%
}\left(  1,1\right)  \text{ with }X_{L}\left(  x\right)  =L_{x}^{\prime
}\left(  1\right)  u,X_{R}(x)=R_{x}^{\prime}\left(  1\right)  u,u\in T_{1}G
\end{equation}

\end{definition}

From the definition and the properties of the flow :

\begin{theorem}
On a Lie group G over the field K the exponential has the following properties:

i) $\exp\left(  0\right)  =1,\left(  \exp u\right)  ^{\prime}|_{u=0}%
=Id_{T_{1}G}$

ii) $\exp\left(  (s+t)u\right)  =\left(  \exp su\right)  \left(  \exp
tu\right)  ;\exp(-u)=\left(  \exp u\right)  ^{-1}$

iii) $\frac{\partial}{\partial t}\exp tu|_{t=\theta}=L_{\exp\theta u}^{\prime
}(1)u=R_{\exp\theta u}^{\prime}\left(  1\right)  u$

iv) $\forall x\in G,u\in T_{1}G:\exp\left(  Ad_{x}u\right)  =x\left(  \exp
u\right)  x^{-1}=Conj_{x}\left(  \exp u\right)  $

v) For any left $X_{L}$ and right $X_{R}$ invariant vector fields :
$\ \Phi_{X_{L}}\left(  x,t\right)  =x\exp tu;$ $\Phi_{X_{R}}\left(
x,t\right)  =\left(  \exp tu\right)  x$
\end{theorem}

\begin{theorem}
(Kolar p.36) On a finite dimensional Lie group G the exponential has the
following properties:

i) it is a smooth map from the vector space $T_{1}G$\ \ to G,

ii) it is a diffeomorphism of a neighbourhood n(0) of 0 in $T_{1}G$ to a
neighborhood of 1 in G. The image expn(0) generates the connected component of
the identity.
\end{theorem}

Remark : the theorem still holds for infinite dimensional Lie groups, if the
Lie algebra is a Banach algebra, with a continuous bracket (Duistermaat p.35).
This is not usually the case, except for the automorphisms of a Banach space.

\begin{theorem}
(Knapp p.91) On a finite dimensional Lie group G for any vectors $u,v\in
T_{1}G:$

$\left[  u,v\right]  _{T_{1}G}=0\Leftrightarrow\forall s,t\in%
\mathbb{R}
:\exp su\circ\exp tv=\exp tv\circ\exp su$
\end{theorem}

Warning !

i) we do not have exp(u+v)=(expu)(expv) and the exponential do not
commute.\ See below the formula.

ii) usually exp is not surjective : there can be elements of G which cannot be
written as $g=\exp X.$ But the subgroup generated (through the operation of
the group) by the elements $\left\{  \exp v,v\in n(0)\right\}  $ is the
component of the identity $G_{0}$.\ See coordinates of the second kind below.

iii) the derivative of expu with respect to u is \textit{not} expu (see below
logarithmic derivatives)

iv) this exponential map is not related to the exponential map deduced from
geodesics on a manifold with connection.

\begin{theorem}
\textbf{Campbell-Baker-Hausdorff formula} (Kolar p.40) : In the Lie algebra
$T_{1}G$\ of a finite dimensional group G there is a heighborhood of 0 such
that $\forall u,v\in n\left(  0\right)  :\exp u\exp v=\exp w$ where

$w=u+v+\frac{1}{2}\left[  u,v\right]  +\sum_{n=2}^{\infty}\frac{\left(
-1\right)  ^{n}}{n+1}\int_{0}^{1}\left(  \sum_{k,l\geq0;n\geq k+l\geq1}%
\frac{t^{k}}{k!l!}\left(  adu\right)  ^{k}\left(  adv\right)  ^{l}\right)
^{n}\left(  u\right)  dt$
\end{theorem}

\paragraph{Group of automorphisms of a Banach space\newline}

\begin{theorem}
The exponential map on the group of continuous automorphisms $G%
\mathcal{L}%
\left(  E;E\right)  $ of a Banach vector space E is the map :

$\exp:%
\mathbb{R}
\times%
\mathcal{L}%
\left(  E;E\right)  \rightarrow G%
\mathcal{L}%
\left(  E;E\right)  ::\exp tu=\sum_{n=0}^{\infty}\frac{t^{n}}{n!}u^{n}$

and $\left\Vert \exp tu\right\Vert \leq\exp t\left\Vert u\right\Vert $

if E is finite dimensional : $\det(\exp u)=\exp(Tr\left(  u\right)  )$
\end{theorem}

where the power is understood as the n iterate of u.

\begin{proof}
$G%
\mathcal{L}%
\left(  E;E\right)  $ is a Lie group, with Banach Lie algebra
$\mathcal{L}$%
$\left(  E;E\right)  $ and bracket : $u,v\in%
\mathcal{L}%
\left(  E;E\right)  ::\left[  u,v\right]  =u\circ v-v\circ u$

For $u\in%
\mathcal{L}%
\left(  E;E\right)  ,X_{L}=L_{f}^{\prime}\left(  1\right)  u=f\circ u$ fixed,
the map : $\phi:%
\mathbb{R}
\times%
\mathcal{L}%
(E;E)\rightarrow%
\mathcal{L}%
(E;E)::\phi\left(  t\right)  =f^{-1}\Phi_{X_{L}}\left(  f,t\right)  =\exp tu$
with the relations : $\phi\left(  0\right)  =Id,\phi\left(  s+t\right)
=\phi\left(  s\right)  \circ\phi\left(  t\right)  $ is a one parameter group
over
$\mathcal{L}$%
(E;E) (see Banach spaces). It is uniformly continuous : $\lim_{t\rightarrow
0}\left\Vert \phi(t)-Id\right\Vert =\lim_{t\rightarrow0}\left\Vert \exp
tu-Id\right\Vert =0$ because exp is smooth (%
$\mathcal{L}$%
$\left(  E;E\right)  $ is a Banach algebra). So there is an infinitesimal
generator : $S\in%
\mathcal{L}%
\left(  E;E\right)  :\phi(t)=\exp tS$ with the exponential defined as the
series : $\exp tS=\sum_{n=0}^{\infty}\frac{t^{n}}{n!}S^{n}.$ Thus we can
identified the two exponential maps. The exponential map has all the
properties seen in Banach spaces :$\left\Vert \exp tu\right\Vert \leq\exp
t\left\Vert u\right\Vert $ and if E is finite dimensional :$\det(\exp
u)=\exp(Tr\left(  u\right)  )$
\end{proof}

\paragraph{\textbf{Logarithmic derivatives}\newline}

\begin{definition}
(Kolar p.38) : For a map $f\in C_{\infty}\left(  M;G\right)  $ from a manifold
M to a Lie group G, on the same field,

the \textbf{right logarithmic derivative} of f is the map : $\delta
_{R}f:TM\rightarrow T_{1}G::\delta_{R}f\left(  u_{p}\right)  =R_{f(p)^{-1}%
}^{\prime}\left(  f\left(  p\right)  \right)  f^{\prime}(p)u_{p}$

the \textbf{left logarithmic derivative} of f is the map : $\delta
_{L}f:TM\rightarrow T_{1}G::\delta_{L}f\left(  u_{p}\right)  =L_{f(p)^{-1}%
}^{\prime}\left(  f\left(  p\right)  \right)  f^{\prime}(p)u_{p}$
\end{definition}

$\delta_{R},\delta_{L}\in\Lambda_{1}\left(  M;T_{1}G\right)  :$ they are
1-form on M valued in the Lie algebra of G

If $f=Id_{G}$ then $\delta_{L}\left(  f\right)  \left(  x\right)
=L_{x}^{\prime}\left(  x^{-1}\right)  \in\Lambda_{1}\left(  G;T_{1}G\right)  $
is the \textbf{Maurer-Cartan form} of G

If $f,g\in C_{\infty}\left(  M;G\right)  :$

$\delta_{R}\left(  fg\right)  \left(  p\right)  =\delta_{R}f\left(  p\right)
+Ad_{f\left(  p\right)  }\delta_{R}g\left(  p\right)  $

$\delta_{L}\left(  fg\right)  \left(  p\right)  =\delta_{L}g\left(  p\right)
+Ad_{g\left(  p\right)  ^{-1}}\delta_{L}f\left(  p\right)  $

\begin{theorem}
(Kolar p.39,Duistermaat p.23) If $T_{1}G$ is a Banach Lie algebra of a Lie
group G, then

i)\ The derivative of the exponential is given by :

$\left(  \exp u\right)  ^{\prime}=R_{\exp u}^{\prime}(1)\circ\int_{0}%
^{1}e^{sad(u)}ds=L_{\exp u}^{\prime}(1)\circ\int_{0}^{1}e^{-sad(u)}ds\in%
\mathcal{L}%
\left(  T_{1}G;T_{\exp u}G\right)  $

with :

$\int_{0}^{1}e^{sad(u)}ds=(ad(u))^{-1}\circ(e^{ad(u)}-I)=\sum_{n=0}^{\infty
}\dfrac{(ad(u))^{n}}{(n+1)!}\in%
\mathcal{L}%
\left(  T_{1}G;T_{1}G\right)  $

$\int_{0}^{1}e^{-sad(u)}ds=(ad(u))^{-1}\circ(I-e^{-ad(z)})=\sum_{n=0}^{\infty
}\dfrac{(-ad(u))^{n}}{(n+1)!}\in%
\mathcal{L}%
\left(  T_{1}G;T_{1}G\right)  $

the series, where the power is understood as the n iterate of ad(u), being
convergent if ad(u) is invertible

ii) we have :

$\delta_{R}\left(  \exp\right)  \left(  u\right)  =\int_{0}^{1}e^{sad(u)}ds$

$\delta_{L}\left(  \exp\right)  \left(  u\right)  =\int_{0}^{1}e^{-sad(u)}ds$

iii) The eigen values of $\int_{0}^{1}e^{sad(u)}ds$ are $\frac{e^{z}-1}{z}$
where z is eigen value of ad(u)

iv) The map $\frac{d}{dv}\left(  \exp v\right)  |_{v=u}:T_{1}G\rightarrow
T_{1}G$ is bijective except for the u which are eigen vectors of ad(u) with
eigenvalue of the form $\pm i2k\pi$ with $k\in%
\mathbb{Z}
/0$
\end{theorem}

\paragraph{Coordinates of the second kind\newline}

\begin{definition}
On a n dimensional Lie group on a field K, there is a neighborhood n(0) of 0
in $K^{n}$ such that\ the map to the connected component of the identity
$G_{0}:\phi:n\left(  0\right)  \rightarrow G_{0}::\phi\left(  t_{1}%
,..t_{n}\right)  =\exp t_{1}e_{1}\times\exp t_{2}e_{2}...\times\exp t_{n}%
e_{n}$ is a diffeomorphism. The map $\phi^{-1}$ is a \textbf{coordinate system
of the second kind} on G.
\end{definition}

Warning ! The product is not commutative.

\subsubsection{Adjoint map}

\begin{definition}
The \textbf{adjoint map} over a Lie group G is the derivative of the
conjugation taken at x=1
\end{definition}

\begin{notation}
Ad is the adjoint map : $Ad:G\rightarrow%
\mathcal{L}%
\left(  T_{1}G;T_{1}G\right)  ::$%

\begin{equation}
Ad_{g}=(Conj_{g}(x))^{\prime}|_{x=1}=L_{g}^{\prime}(g^{-1})\circ R_{g^{-1}%
}^{\prime}(1)=R_{g^{-1}}^{\prime}(g)\circ L_{g}^{\prime}(1)
\end{equation}

\end{notation}

\begin{theorem}
(Knapp p.79) The \textbf{adjoint map} over a Lie group G is a bijective,
smooth morphism from G to the group of continuous automorphism $G%
\mathcal{L}%
\left(  T_{1}G;T_{1}G\right)  .$ Moreover $\forall x\in G:Ad_{x}$ belongs to
the connected component $Int\left(  T_{1}G\right)  $ of the automorphisms of
the Lie algebra.
\end{theorem}

\begin{theorem}
On a Lie group G the adjoint map Ad is the exponential of the map ad in the
following meaning :%

\begin{equation}
\forall u\in T_{1}G:Ad_{\exp_{G}u}=\exp_{G%
\mathcal{L}%
\left(  T_{1}G;T_{1}G\right)  }ad(u)
\end{equation}

\end{theorem}

\begin{proof}
$\forall x\in G,Ad_{x}\in G%
\mathcal{L}%
\left(  T_{1}G;T_{1}G\right)  :$ so Ad is a Lie a Lie group morphism :
$Ad:G\rightarrow G%
\mathcal{L}%
\left(  T_{1}G;T_{1}G\right)  $ and we have :

$\forall u\in T_{1}G:Ad_{\exp_{G}u}=\exp_{G%
\mathcal{L}%
\left(  T_{1}G;T_{1}G\right)  }\left(  Ad_{x}\right)  _{x=1}^{\prime}u=\exp_{G%
\mathcal{L}%
\left(  T_{1}G;T_{1}G\right)  }ad(u)$
\end{proof}

The exponential over the Lie group $G%
\mathcal{L}%
\left(  T_{1}G;T_{1}G\right)  $ is computed as for any group of automorphisms
over a Banach vector space :

$\exp_{G%
\mathcal{L}%
\left(  T_{1}G;T_{1}G\right)  }ad(u)=\sum_{n=0}^{\infty}\frac{1}{n!}\left(
ad\left(  u\right)  \right)  ^{n}$ where the power is understood as the n
iterate of ad(u).

And we have :

$\det(\exp ad\left(  u\right)  )=\exp(Tr\left(  ad\left(  u\right)  \right)
)=\det Ad_{\exp u}$

It is easy to check that :

$Ad_{xy}=Ad_{x}\circ Ad_{y}$

$Ad_{1}=Id$

$\left(  Ad_{x}\right)  ^{-1}=Ad_{x^{-1}}$

$\forall u,v\in T_{1}G,x\in G:Ad_{x}\left[  u,v\right]  =\left[
Ad_{x}u,Ad_{x}v\right]  $

Conjugation being differentiable at any order, we can compute the derivative
of $Ad_{x}$ with respect to x :

$\frac{d}{dx}Ad_{x}|_{x=1}\in%
\mathcal{L}%
\left(  T_{1}G;%
\mathcal{L}%
\left(  T_{1}G;T_{1}G\right)  \right)  =%
\mathcal{L}%
^{2}\left(  T_{1}G;T_{1}G\right)  $

$\left(  \frac{d}{dx}Ad_{x}u\right)  |_{x=1}\left(  v\right)  =\left[
u,v\right]  _{T_{1}G}=ad(u)(v)$

If M is a manifold on the same field K as G, $f:M\rightarrow G$ a smooth map
then for a fixed $u\in T_{1}G$\ \ let us define :

$\phi_{u}:M\rightarrow T_{1}G::\phi_{u}\left(  p\right)  =Ad_{f\left(
p\right)  }u$

The value of its derivative for $v_{p}\in T_{p}M$ is $:$%

\begin{equation}
\left(  Ad_{f\left(  p\right)  }u\right)  \mathbf{\prime}_{p}\left(
v_{p}\right)  =Ad_{f\left(  p\right)  }\left[  L_{f(p)^{-1}}^{\prime}(f\left(
p\right)  )f^{\prime}(p)v_{p},u\right]  _{T_{1}G}%
\end{equation}

If G is the set $G%
\mathcal{L}%
\left(  E;E\right)  $ of automorphims of a Banach vector space or a subset of
a matrices group, then the derivatives of the translations are the
translations. Thus :

$G%
\mathcal{L}%
\left(  E;E\right)  :Ad_{x}u=x\circ u\circ x^{-1}$

Matrices : $Ad_{x}\left[  u\right]  =\left[  x\right]  \left[  u\right]
\left[  x\right]  ^{-1}$

\subsubsection{Morphisms}

\paragraph{Definitions\newline}

\begin{definition}
A \textbf{group morphism} is a map f between two groups G,H such that :
$\forall x,y\in G:f\left(  xy\right)  =f\left(  x\right)  f\left(  y\right)
,f\left(  x^{-1}\right)  =f\left(  x\right)  ^{-1}$
\end{definition}

\begin{definition}
A morphism between topological groups is a group morphism which is also continuous
\end{definition}

\begin{definition}
A class s \textbf{Lie group morphism} between class r Lie groups over the same
field K is a group morphism which is a class s differentiable map between the
manifolds underlying the groups.
\end{definition}

If not precised otherwise the Lie groups and the Lie morphisms are assumed to
be smooth.

A morphism is usually called also a homomorphism.

Thus the categories of :

i) topological groups, comprises topological groups and continuous morphisms

ii) Lie groups comprises Lie groups on the same field K as objects, and smooth
Lie groups morphisms as morphisms.

The set of continuous (resp. Lie) groups morphisms between topological
(resp.Lie) groups G,H is denoted $\hom\left(  G;H\right)  .$

If a continuous (resp.Lie) group morphism is bijective and its inverse is also
a continuous (resp.Lie) group morphism then it is a continuous (resp.Lie)
group \textbf{isomorphism}. An isomorphism over the same set is an automorphism.

If there is a continuous (resp.Lie) group isomorphism between two topological
(resp.Lie) groups they are said to be isomorphic.

Continuous Lie group morphisms are continuous, and their derivative at the
unity is a Lie algebra morphism.\ The converse is true with a condition.

\paragraph{Lie group morphisms\newline}

\begin{theorem}
(Kolar p.37, Duistermaat p.49, 58) A continuous group morphism between the Lie
groups G,H on the same field K:

i) is a smooth Lie group morphism

ii) if it is bijective and H has only many countably connected components,
then it is a smooth diffeomorphism and a Lie group isomorphism.

iii) if : at least G or H has finitely many connected components, $f_{1}%
\in\hom(G;H),f_{2}\in\hom(H,G)$ are continuous injective group
morphisms.\ Then $f_{1}\left(  G\right)  =H$,$f_{2}\left(  H\right)  =G$ and
$f_{1},f_{2}$ are Lie group isomorphisms.
\end{theorem}

\begin{theorem}
(Kolar p.36) If f is a smooth Lie group morphism $f\in\hom(G,H)$\ then its
derivative at the unity f'(1) is a Lie algebra morphism $f^{\prime}(1)\in
\hom\left(  T_{1}G,T_{1}H\right)  .$
\end{theorem}

The following diagram commutes :

\bigskip%

\begin{tabular}
[c]{ccccc}%
T$_{1}$G & $\rightarrow$ & f'(1) & $\rightarrow$ & T$_{1}$H\\
$\downarrow$ &  &  &  & $\downarrow$\\
exp$_{G}$ &  &  &  & exp$_{H}$\\
$\downarrow$ &  &  &  & $\downarrow$\\
G & $\rightarrow$ & f & $\rightarrow$ & H
\end{tabular}

\bigskip%

\begin{equation}
\forall u\in T_{1}G:f\left(  \exp_{G}u\right)  =\exp_{H}f^{\prime}(1)u
\end{equation}

\bigskip

and conversely:

\begin{theorem}
(Kolar p.42) If $f:T_{1}G\rightarrow T_{1}H$ is Lie algebra morphism between
the Lie algebras of the finite dimensional Lie groups G,H, there is a Lie
group morphism F locally defined in a neighborhood of $1_{G}$ such that
$F^{\prime}\left(  1_{G}\right)  =f.$ If G is simply connected then there is a
globally defined morphism of Lie group with this property.
\end{theorem}

\begin{theorem}
(Knapp p.90) Any two \textit{simply connected} Lie groups whose Lie algebras
are Lie isomorphic are Lie isomorphic.
\end{theorem}

Notice that in the converse there is a condition : G must be simply connected.

Warning ! two Lie groups with isomorphic Lie algebras are not Lie isomorphic
in general, so even if they have the same universal cover they are not
necessarily Lie isomorphic.

\bigskip

\subsection{Action of a group on a set}

\subsubsection{Definitions}

These definitions are mainly an adaptation of those given in Algebra (groups).

\begin{definition}
Let G be a topological group, E a topological space.

A \textbf{left-action} of G on E is a continuous map : $\lambda:G\times
E\rightarrow E$ such that :

$\forall p\in E,\forall g,g^{\prime}\in G:\lambda\left(  g,\lambda\left(
g^{\prime},p\right)  \right)  =\lambda\left(  g\cdot g^{\prime},p\right)
;\lambda\left(  1,p\right)  =p$

A \textbf{right-action} of G on E is a continuous map : $\rho:E\times
G\rightarrow E$ such that :

$\forall p\in E,\forall g,g^{\prime}\in G:\rho\left(  \rho\left(  p,g^{\prime
}),g\right)  \right)  =\rho\left(  p,g^{\prime}\cdot g\right)  ;\rho\left(
p,1\right)  =p$
\end{definition}

For g fixed, the maps $\lambda\left(  g,.\right)  :E\rightarrow E,\rho\left(
.,g\right)  :E\rightarrow E$ are bijective.

In the following every definition holds for a right action.

If G is a Lie group and E is a manifold M on the same field then we can define
class r actions. It is assumed to be smooth if not specified otherwise.

A manifold endowed with a right or left action is called a G-space.

The orbit of the action through $p\in E$ is the subset of E : $Gp=\left\{
\lambda\left(  g,p\right)  ,g\in G\right\}  .$ The relation $q\in Gp$ is an
equivalence relation between p,q denoted $R_{\lambda},$\ the classes of
equivalence form a partition of G called the orbits of the action.

A subset F of E is \textbf{invariant} by the action if : $\forall p\in
F,\forall g\in G:\lambda\left(  g,p\right)  \in F.$ F is invariant iff it is
the union of a collection of orbits. The minimal non empty invariant sets are
the orbits.

The action is said to be :

transitive if $\forall p,q\in E,\exists g\in G:q=\lambda\left(  g,p\right)  .$
: there is only one orbit.

free if : $\lambda\left(  g,p\right)  =p\Rightarrow g=1.$ Then each orbit is
in bijective correspondance with G and the map : $\lambda\left(  .,p\right)
:G\rightarrow\lambda\left(  G,p\right)  $ is bijective.

effective if : $\forall p:\lambda(g,p)=\lambda(h,p)=>g=h$

free $\Leftrightarrow$ effective

\begin{theorem}
(Kolar p.44) If $\lambda:G\times M\rightarrow M$\ is a continuous effective
left action from a locally compact topological group G on a smooth manifold M,
then G is a Lie group and the action is smooth
\end{theorem}

\begin{theorem}
(Duistermaat p.94) If $\lambda:G\times M\rightarrow M$\ is a left action from
a Lie group G on a smooth manifold M, then for any $p\in E$ the set $A\left(
p\right)  =\left\{  g\in G:\lambda\left(  g,p\right)  =p\right\}  $ is a
closed Lie subgroup of G called the \textbf{isotropy subgroup} of p.\ The map
$\lambda\left(  .,p\right)  :G\rightarrow M$ factors over the projection :
$\pi:G\rightarrow G/A\left(  p\right)  $ to an injective immersion :
$\imath:G/G/A\left(  p\right)  \rightarrow M$ which is G equivariant :
$\lambda\left(  g,\imath\left(  \left[  x\right]  \right)  \right)
=\imath\left(  \left[  \lambda\left(  g,p\right)  \right]  \right)  .$ The
image of \i\ is the orbit through p.
\end{theorem}

\subsubsection{Proper actions}

\begin{definition}
A left action $\lambda:G\times M\rightarrow M$\ of a Lie group G on a manifold
M is \textbf{proper} if the preimage of a compact of M is a compact of
G$\times$M
\end{definition}

If G and M are compact and Hausdorff, and $\lambda$ continuous then it is
proper (see topology)

\begin{theorem}
(Duistermaat p.98) A left action $\lambda:G\times M\rightarrow M$\ of a Lie
group G on a manifold M is proper if for any convergent sequences
$p_{n}\rightarrow p,g_{m}\rightarrow g$\ there is a subsequence $\left(
g_{m},p_{n}\right)  $\ such that $\lambda(g_{m},p_{n})\rightarrow\lambda(g,p)$
\end{theorem}

\begin{theorem}
(Duistermaat p.53) If the left action $\lambda:G\times M\rightarrow M$\ of a
Lie group G on a manifold M is proper and continuous then the quotient set
$M/R_{\lambda}$ whose elements are the orbits of the action, is Hausdorff with
the quotient topology.

If moreover M,G are finite dimensional and of class r, $\lambda$ is free and
of class r, then the quotient set $M/R_{\lambda}$ has a unique structure of
class r real manifold of dimension = dim M - dim G.\ M has a principal fiber
bundle structure with group G.
\end{theorem}

That means the following :

$R_{\lambda}$ is the relation of equivalence $q\in Gp$ between p,q

The projection $\pi:M\rightarrow M/R_{\lambda}$ is a class r map;

$\forall p\in M/R_{\lambda}$ there is a neighborhood n(p) and a diffeomorphism

$\tau:\pi^{-1}\left(  n\left(  p\right)  \right)  \rightarrow G\times n\left(
p\right)  ::\tau\left(  m\right)  =\left(  \tau_{1}\left(  m\right)  ,\tau
_{2}\left(  m\right)  \right)  $ such that

$\forall g\in G,m\in\pi^{-1}\left(  n\left(  p\right)  \right)  :\tau\left(
\lambda\left(  g,p\right)  \right)  =\left(  \lambda\left(  g,\tau_{1}\left(
m\right)  \right)  ,\pi\left(  m\right)  \right)  $

\subsubsection{Identities}

From the definition of an action of a group over a manifold one can deduce
some identities which are useful.

1. As a consequence of the definition :

$\lambda\left(  g^{-1},p\right)  =\lambda\left(  g,p\right)  ^{-1};\rho\left(
p,g^{-1}\right)  =\rho\left(  p,g\right)  ^{-1}$

2. By taking the derivative of $\lambda(h,\lambda(g,p))=\lambda(hg,p)$ and
putting successively $g=1,h=1,h=g^{-1}$

$\lambda_{p}^{\prime}(1,p)=Id_{TM}$%

\begin{equation}
\lambda_{g}^{\prime}(g,p)=\lambda_{g}^{\prime}(1,\lambda(g,p))R_{g^{-1}%
}^{\prime}(g)=\lambda_{p}^{\prime}(g,p)\lambda_{g}^{\prime}(1,p)L_{g^{-1}%
}^{\prime}(g)
\end{equation}

\begin{equation}
\left(  \lambda_{p}^{\prime}(g,p)\right)  ^{-1}=\lambda_{p}^{\prime}%
(g^{-1},\lambda(g,p))
\end{equation}

Notice that $\lambda_{g}^{\prime}(1,p)\in%
\mathcal{L}%
\left(  T_{1}G;T_{p}M\right)  $\ is not necessarily invertible.

3. Similarly :

$\rho_{p}^{\prime}(p,1)=Id_{TM}$%

\begin{equation}
\rho_{g}^{\prime}(p,g)=\rho_{g}^{\prime}(\rho(p,g),1)L_{g^{-1}}^{\prime
}(g)=\rho_{p}^{\prime}(p,g)\rho_{g}^{\prime}(p,1)R_{g^{-1}}^{\prime}(g)
\end{equation}

\begin{equation}
\left(  \rho_{p}^{\prime}(p,g)\right)  ^{-1}=\rho_{p}^{\prime}(\rho
(p,g),g^{-1})
\end{equation}

\subsubsection{Fundamental vector fields}

\begin{definition}
For a differentiable left action $\lambda:G\times M\rightarrow M$\ of a Lie
group G on a manifold M, the \textbf{fundamental vector fields }$\zeta_{L}$
are the vectors fields on M generated by a vector of the Lie algebra of G:%

\begin{equation}
\zeta_{L}:T_{1}G\rightarrow TM::\zeta_{L}\left(  u\right)  \left(  p\right)
=\lambda_{g}^{\prime}\left(  1,p\right)  u
\end{equation}

\end{definition}

We have similarly for a right action :

$\zeta_{R}:T_{1}G\rightarrow TM::\zeta_{R}\left(  u\right)  \left(  p\right)
=\rho_{g}^{\prime}\left(  p,1\right)  u$

\begin{theorem}
(Kolar p.46) For a differentiable action of a Lie group G on a manifold M, the
\textbf{fundamental vector fields} have the following properties :

i) the maps $\zeta_{L},\zeta_{R}$ are linear

ii) $\left[  \zeta_{L}\left(  u\right)  ,\zeta_{L}\left(  v\right)  \right]
_{\mathfrak{X}\left(  TM\right)  }=-\zeta_{L}\left(  \left[  u,v\right]
_{T_{1}G}\right)  $

$\left[  \zeta_{R}\left(  u\right)  ,\zeta_{R}\left(  v\right)  \right]
_{\mathfrak{X}\left(  TM\right)  }=\zeta_{R}\left(  \left[  u,v\right]
_{T_{1}G}\right)  $

iii) $\lambda_{p}^{\prime}\left(  x,q\right)  |_{p=q}\zeta_{L}\left(
u\right)  \left(  q\right)  =\zeta_{L}\left(  Ad_{x}u\right)  \left(
\lambda\left(  x,q\right)  \right)  $

$\rho_{p}^{\prime}\left(  q,x\right)  |_{p=q}\zeta_{R}\left(  u\right)
\left(  q\right)  =\zeta_{R}\left(  Ad_{x^{-1}}u\right)  \left(  \rho\left(
q,x\right)  \right)  $

iv) $\zeta_{L}\left(  u\right)  =\lambda_{\ast}\left(  R_{x}^{\prime}\left(
1\right)  u,0\right)  ,\zeta_{R}\left(  u\right)  =\rho_{\ast}\left(
L_{x}^{\prime}\left(  1\right)  u,0\right)  $

v) the fundamental vector fields span an integrable distribution over M, whose
leaves are the connected components of the orbits.
\end{theorem}

\begin{theorem}
The flow of the fundamental vector fields is :

$\Phi_{\zeta_{L}\left(  u\right)  }\left(  t,p\right)  =\lambda\left(  \exp
tu,p\right)  $

$\Phi_{\zeta_{L}\left(  u\right)  }\left(  t,p\right)  =\rho\left(  p,\exp
tu\right)  $
\end{theorem}

\begin{proof}
use the relation : $f\circ\Phi_{V}=\Phi_{f_{\ast}V}\circ f$ with
$\lambda\left(  \Phi_{X_{R}\left(  u\right)  }\left(  t,x\right)  ,p\right)
=\Phi_{\zeta_{L}\left(  u\right)  }\left(  t,\lambda\left(  x,p\right)
\right)  $ and x=1
\end{proof}

\subsubsection{Equivariant mapping}

\begin{definition}
A map $f:M\rightarrow N$ between the manifolds M,N is \textbf{equivariant} by
the left actions of a Lie group G, $\lambda_{1}$ on M, $\lambda_{2}$ on N, if :

$\forall p\in M,\forall g\in G:f\left(  \lambda_{1}\left(  g,p\right)
\right)  =\lambda_{2}\left(  g,f\left(  p\right)  \right)  $
\end{definition}

\begin{theorem}
(Kolar p.47) If G is connected then f is equivariant iff the fundamental
vector fields $\zeta_{L1},\zeta_{L2}$ are f related :

$f^{\prime}\left(  p\right)  \left(  \zeta_{L1}\left(  u\right)  \right)
=\zeta_{L2}\left(  u\right)  \left(  f\left(  p\right)  \right)
\Leftrightarrow f_{\ast}\zeta_{L1}\left(  u\right)  =\zeta_{L2}\left(
u\right)  $
\end{theorem}

A special case is of bilinear symmetric maps, which are invariant under the
action of a map. This includes the isometries.

\begin{theorem}
(Duistermaat p.105) If there is a class r
$>$
0 proper action of a finite dimensional Lie group G on a smooth finite
dimensional Riemannian manifold M, then M has a G-invariant class r-1
Riemannian structure.

Conversely if M is a smooth finite dimensional riemannian manifold (M,g) with
finitely many connected components, and if g is a class k
$>$%
1 map, then the group of isometries of M is equal to the group of
automorphisms of (M;g), it is a finite dimensional Lie group, with finitely
many connected components. Its action is proper and of class k+1.
\end{theorem}

\bigskip

\subsection{Structure of Lie groups}

\label{Structure of Lie groups}

\subsubsection{Subgroups}

\paragraph{Topological groups\newline}

\begin{definition}
A subset H of a topological group G is a \textbf{subgroup} of G if:

i) H is an algebraic subgroup of G

ii) H has itself the structure of a topologic group

iii) the injection map : $\imath:H\rightarrow G$ is continuous.
\end{definition}

Explanation : Let $\Omega$ be the set of open subsets of G.\ Then H inherits
the relative topology given by $\Omega\cap H.$ But an open in H is not
necessarily open in G. So we take another $\Omega_{H}$ and the continuity of
the map : $\imath:H\rightarrow G$ is checked with respect to $\left(
G,\Omega\right)  ,\left(  H,\Omega_{H}\right)  .$

\begin{theorem}
If H is an algebraic subgroup of G and is a closed subset of a topological
group G then it is a topological subgroup of G.
\end{theorem}

But a topological subgroup of G is not necessarily closed.

\begin{theorem}
(Knapp p.84) For a topological group G, with a separable, metric topology :

i) any open subgroup H is closed and G/H has the discrete topology

ii) the identity component $G_{0}$ is open if G is locally connected

iii) any discrete subgroup (meaning whose relative topology is the discrete
topology) is closed

iv) if G is connected then any discrete normal subgroup lies in the center of G.
\end{theorem}

\paragraph{Lie groups}

\begin{definition}
A subset H of a Lie group is a \textbf{Lie subgroup} of G if :

i) H is an algebraic subgroup of G

ii) H is itself a Lie group

iii) the inclusion $\imath:H\rightarrow G$ is smooth. Then it is an immersion
and a smooth morphism of Lie group $\imath\in\hom\left(  H;G\right)  $.
\end{definition}

So to be a Lie subgroup requires more than to be an algebraic subgroup. Notice
that one can endows any algebraic subgroup with a Lie group structure, but it
can be non separable (Kolar p.43), thus the restriction of iii).

The most useful theorem is the following (the demonstration is still valid for
G infinite dimensional) :

\begin{theorem}
(Kolar p.42) An algebraic subgroup H of a lie group G which is topologicaly
closed in G is a Lie subgroup of G.
\end{theorem}

But the converse is not true : a Lie subgroup is not necessarily closed.

As a corollary :

\begin{theorem}
If G is a closed subgroup of matrices in GL(K,n), then it is a Lie subgroup
(and a\ Lie group).
\end{theorem}

For instance if M is some Lie group of matrices in GL(K,n), the subset of M
such that detg=1 is closed, thus it is a Lie subgroup of M.

\begin{theorem}
(Kolar p.41) If H is a Lie subgroup of the Lie group G, then the Lie algebra
$T_{1}H$ is a Lie subalgebra of $T_{1}G.$
\end{theorem}

Conversely :

\begin{theorem}
If h is a Lie subalgebra of the Lie algebra of the finite dimensional Lie
group G there is a unique connected Lie subgroup H of G which has h as Lie
algebra. H is generated by exp(h) (that is the product of elements of exp(h)).
\end{theorem}

(Duistermaat p.42) The theorem is still true if G is infinite dimensional and
h is a closed linear subspace of $T_{1}G.$

\begin{theorem}
Yamabe (Kolar p.43) An arc wise connected algebraic subgroup of a Lie group is
a connected Lie subgroup
\end{theorem}

\subsubsection{Centralizer}

Reminder of algebra (see Groups):

The centralizer of a subset A of a group G is the set of elements of G which
commute with the elements of A

The center of a group G is the subset of the elements which commute with all
other elements.

\begin{theorem}
The center of a topological group is a topological subgroup.
\end{theorem}

\begin{theorem}
(Kolar p.44) For a Lie group G and any subset A of G:

i) the centralizer $Z_{A}$ of A is a subgroup of G.

ii) If G is connected then the Lie algebra of $Z_{A}$ is the subset $:$

$T_{1}Z_{A}=\left\{  u\in T_{1}G:\forall a\in Z_{A}:Ad_{a}u=u\right\}  $

If A and G are connected then $T_{1}Z_{A}=\left\{  u\in T_{1}G:\forall v\in
T_{1}Z_{A}:\left[  u,v\right]  =0\right\}  $

iii) the center $Z_{G}$ of G is a Lie subgroup of G and its algebra is the
center of $T_{1}G.$
\end{theorem}

\begin{theorem}
(Knapp p.90) A connected Lie subgroup H of a connected Lie group G is
contained in the center of G iff $T_{1}H$ is contained in the center of
$T_{1}G$.
\end{theorem}

\subsubsection{Quotient spaces}

\paragraph{Reminder of Algebra (Groups)\newline}

If H is a subgroup of the group G :

The quotient set $G/H$ is the set $G/\sim$ of classes of equivalence $:$
$x\sim y\Leftrightarrow\exists h\in H:x=y\cdot h$

The quotient set $H\backslash G$ is the set $G/\sim$ of classes of equivalence
$:$ $x\sim y\Leftrightarrow\exists h\in H:x=h\cdot y$

Usually they are not groups.

The projections give the classes of equivalences denoted $\left[  x\right]  $ :

$\pi_{L}:G\rightarrow G/H:\pi_{L}\left(  x\right)  =\left[  x\right]
_{L}=\left\{  y\in G:\exists h\in H:x=y\cdot h\right\}  =x\cdot H$

$\pi_{R}:G\rightarrow H\backslash G:\pi_{R}\left(  x\right)  =\left[
x\right]  _{R}=\left\{  y\in G:\exists h\in H:x=h\cdot y\right\}  =H\cdot x$

$x\in H\Rightarrow\pi_{L}\left(  x\right)  =\pi_{R}\left(  x\right)  =\left[
x\right]  =1$

By choosing one element in each class, we have two maps :

For $G/H:\lambda:G/H\rightarrow G:x\neq y\Leftrightarrow\lambda\left(
x\right)  \neq\lambda\left(  y\right)  $

For $H\backslash G:\rho:H\backslash G\rightarrow G:x\neq y\Leftrightarrow
\rho\left(  x\right)  \neq\rho\left(  y\right)  $

any $x\in G$ can be written as :$x=\lambda\left(  x\right)  \cdot h$ or
$x=h^{\prime}\cdot\rho\left(  x\right)  $ for unique h,h'$\in H$

G/H=H%
$\backslash$%
G iff H is a normal subgroup.\ If so then G/H=H%
$\backslash$%
G is a group. Then $\pi_{L}$ is a morphism with kernel H.

\paragraph{Topological groups\newline}

\begin{theorem}
(Knapp p.83) If H is a closed subgroup of the separable, metrisable,
topological group G, then :

i) the projections : $\pi_{L},\pi_{R}$ are open maps

ii) G/H is a separable metrisable space

iii) if H and G/H (or H%
$\backslash$%
G) are connected then G is connected

iv) if H and G/H (or H%
$\backslash$%
G) are compact then G is compact
\end{theorem}

\paragraph{Lie groups\newline}

\begin{theorem}
(Kolar p.45, 88, Duistermaat p.56) If H is a closed Lie subgroup of the Lie
group G then :

i) the maps :

$\lambda:H\times G\rightarrow G::\lambda\left(  h,g\right)  =L_{h}g=hg$

$\rho:G\times H\rightarrow G:\rho\left(  g,h\right)  =R_{h}g=gh$

are left (rigth) actions of H on G, which are smooth, proper and free.

ii) There is a unique smooth manifold structure on G/H,H%
$\backslash$%
G, called \textbf{homogeneous space}s of G.

If G is finite dimensional then dimG/H=dimG - dimH.

iii) The projections $\pi_{L},\pi_{R}$ are submersions, so they are open maps
and $\pi_{L}^{\prime}\left(  g\right)  ,\pi_{L}^{\prime}\left(  g\right)  $
are surjective

iv) G is a principal fiber bundle G$\left(  G/H,H,\pi_{L}\right)  ,G\left(
H\backslash G,H,\pi_{R}\right)  $

v) The translation induces a smooth transitive right (left) action of G on H%
$\backslash$%
G (G/H):

$\Lambda:G\times G/H\rightarrow G/H::\Lambda\left(  g,x\right)  =\pi
_{L}\left(  g\lambda\left(  x\right)  \right)  $

$P:H\backslash G\times G\rightarrow H\backslash G:P\left(  x,g\right)
=\pi_{R}\left(  \rho\left(  x\right)  g\right)  $

vi) If H is a normal Lie subgroup then G/H=H%
$\backslash$%
G=N is a Lie group (possibly finite) and the projection $G\rightarrow N$ is a
Lie group morphism with kernel H.
\end{theorem}

The action is free so each orbit, that is each coset $\left[  x\right]  $, is
in bijective correspondance with H

If H is not closed and G/H is provided with a topology so that the projection
is continuous then G/H is not Hausdorff.

\begin{theorem}
(Duistermaat p.58) For any Lie group morphism $f\in\hom\left(  G,H\right)  $:

i) $K=\ker f=\left\{  x\in G:f(x)=1_{H}\right\}  $ is a\ normal Lie subgroup
of G with Lie algebra $\ker f^{\prime}(1)$

ii) if $\pi:G\rightarrow G/K$ is the canonical projection, then the unique
homomorphism $\phi:G/K\rightarrow H$ such that $f=\phi\circ\pi$ is a smooth
immersion making f(G)=$\phi\left(  G/K\right)  $ into a Lie subgroup of H with
Lie algebra $f\prime(1)T_{1}G$

iii) with this structure on f(G), G is a principal fiber bundle with base f(G)
and group K.

iv) If G has only many countably components, and f is surjective then G is a
principal fiber bundle with base H and group K.
\end{theorem}

\paragraph{Normal subgroups\newline}

A subgroup is normal if for all g in G, $gH=Hg\Leftrightarrow\forall x\in
G:x\cdot H\cdot x^{-1}\in H.$

\begin{theorem}
(Knapp p.84) The identity component of a topological group is a closed normal
subgroup .
\end{theorem}

\begin{theorem}
(Kolar p.44, Duistermaat p.57) A connected Lie subgroup H of a connected Lie
group is normal iff its Lie algebra $T_{1}H$ is an ideal in $T_{1}G.$
Conversely : If h is an ideal of the Lie algebra of a Lie group G then the
group H generated by exp(h) is a connected Lie subgroup of G, normal in the
connected component $G_{0}$ of the identity and has h as Lie algebra.
\end{theorem}

\begin{theorem}
(Duistermaat p.57) For a closed Lie subgroup H of Lie group G, and their
connected component of the identity $G_{0},H_{0}$ the following are equivalent :

i) $H_{0}$ is normal in $G_{0}$

ii) $\forall x\in G_{0},u\in T_{1}H:Ad_{x}u\in T_{1}H$

iii) $T_{1}H$ is an ideal in $T_{1}G$

If H is normal in G then $H_{0}$ is normal in $G_{0}$
\end{theorem}

\begin{theorem}
(Duistermaat p.58) If f is a Lie group morphism between the Lie groups G,H
then $\ K=\ker f=\left\{  x\in G:f(x)=1_{H}\right\}  $ is a\ normal Lie
subgroup of G with Lie algebra $\ker f^{\prime}(1)$
\end{theorem}

\begin{theorem}
(Kolar p.44) For any closed subset A of a Lie group G, the normalizer
$N_{A}=\{x\in G:Conj_{x}(A)=A\}$ is a Lie subgroup of G. If A is a Lie
subgroup of G, A and G connected, then $N_{A}=\{x\in G:\forall u\in T_{1}%
N_{a}:Ad_{x}u\in T_{1}N_{A}\}$ and $T_{1}N_{A}=\left\{  u\in T_{1}G:\forall
v\in T_{1}N_{a}ad(u)v\in T_{1}N_{A}\right\}  $
\end{theorem}

\subsubsection{Connected component of the identity}

\begin{theorem}
The connected component of the identity $G_{0}$ in a Lie group G:

i) is a normal Lie subgoup of G, both closed and open in G.\ It is the only
open connected subgroup of G.

ii) is arcwise connected

iii) is contained in any open algebraic subgroup of G

iv) is generated by $\left\{  \exp u,u\in T_{1}G\right\}  $

v) $G/G_{0}$ is a discrete group
\end{theorem}

The connected components of G are generated by $xG_{0}$\ or $G_{0}x$ : so it
suffices to know one element of each of the other connected components to
generate G.

\subsubsection{Product of Lie groups}

\begin{definition}
The \textbf{direct product} of the topological groups G,H is the algebraic
product $G\times H$ endowed with the product topology, and is a topological group.
\end{definition}

\begin{definition}
The \textbf{direct product} of the Lie groups G,H is the algebraic product
$G\times H$ endowed with the manifold structure of product of manifolds and is
a Lie group.
\end{definition}

\begin{theorem}
(Kolar p.47) If G,T are Lie groups, and $f:G\times T\rightarrow T$ is such that:

i) f is a left action of G on T

ii) for every $g\in G$ the map $f\left(  g,.\right)  $ is a Lie group
automorphism on T

then the set $G\times T$ endowed with the operations :

$\left(  g,t\right)  \times\left(  g^{\prime},t^{\prime}\right)  =\left(
gg^{\prime},f\left(  g,t^{\prime}\right)  \cdot t\right)  $

$\left(  g,t\right)  ^{-1}=\left(  g^{-1},f\left(  g^{-1},t^{-1}\right)
\right)  $

is a Lie group, called the \textbf{semi-direct product} of G and T denoted
$G\ltimes_{f}T$

Then :

i) the projection : $\pi:G\ltimes_{f}T\rightarrow G$ is a smooth morphism with
kernel $1_{G}\times T$

ii) the insertion : $\imath:G\rightarrow G\ltimes_{f}T::\imath\left(
g\right)  =\left(  g,1_{T}\right)  $ is a smooth Lie group morphism with
$\pi\circ\imath=Id_{G}$
\end{theorem}

The conditions on f read :

$\forall g,g^{\prime}\in G,t,t^{\prime}\in T:$

$f\left(  g,t\cdot t^{\prime}\right)  =f\left(  g,t\right)  \cdot f\left(
g,t^{\prime}\right)  ,f\left(  g,1_{T}\right)  =1_{T},f\left(  g,t\right)
^{-1}=f\left(  g,t^{-1}\right)  $

$f\left(  gg^{\prime},t\right)  =f\left(  g,f(g^{\prime},t)\right)  ,f\left(
1_{G},t\right)  =t$

\begin{theorem}
(Neeb p.55) If $G\ltimes_{f}T$ is a semi-direct product of the Lie groups G,T,
and the map : $F:T\rightarrow G$\ is such that :

$F\left(  t\cdot t^{\prime}\right)  =f\left(  F\left(  t\right)  ,t^{\prime
}\right)  $ then the map $\left(  F,Id_{T}\right)  :T\rightarrow G\ltimes
_{f}T$ is a group morphism and conversely every group morphism is of this form.
\end{theorem}

such a map F is called a 1-cocycle.

\paragraph{Example : Group of displacements\newline}

Let $\left(  F,\rho\right)  $ be a representation of the group G on the vector
space F. The group T of translations on F can be identified with F itself with
the addition of vectors. The semi-direct product $G\ltimes_{+}F$ is the set
$G\times F$ with the operations :

$\left(  g,v\right)  \times\left(  g^{\prime},v^{\prime}\right)  =\left(
g\cdot g^{\prime},\rho\left(  g\right)  v^{\prime}+v\right)  $

$\left(  g,v\right)  ^{-1}=\left(  g^{-1},-\rho\left(  g^{-1}\right)
v\right)  $

$\left(  F,\rho^{\prime}\left(  1\right)  \right)  $ is a representation of
the Lie algebra $T_{1}G,$ and F is itself a Lie algebra with the null bracket.

For any $\kappa\in T_{1}G$ the map $\rho^{\prime}\left(  1\right)  \kappa\in
L\left(  F;F\right)  $ is a derivation.

The set $T_{1}G\times F$ is the semi-direct product $T_{1}G\times
_{\rho^{\prime}\left(  1\right)  }F$ of Lie algebras with the bracket :

$\left[  \left(  \kappa_{1},u_{1}\right)  ,\left(  \kappa_{2},u_{1}\right)
\right]  _{T_{1}G\ltimes F}=\left(  \left[  \kappa_{1},\kappa_{2}\right]
_{T_{1}G},\rho^{\prime}\left(  1\right)  \left(  \kappa_{1}\right)  u_{2}%
-\rho^{\prime}\left(  1\right)  \left(  \kappa_{2}\right)  u_{1}\right)  $

\subsubsection{Third Lie's theorem}

A Lie group has a Lie algebra, the third Lie's theorem adresses the converse :
given a Lie algebra, can we build a Lie group ?

\begin{theorem}
(Kolar p.42, Duistermaat p.79) Let g be a finite dimensional real Lie algebra,
then there is a simply connected Lie group with Lie algebra g. The restriction
of the exponential mapping to the center Z of g induces an isomorphism from
(Z,+) to the identity component of the center of G (the center Z of g and of G
are abelian).
\end{theorem}

Notice that that the group is not necessarily a group of matrices : there are
finite dimensional Lie groups which are not isomorphic to a matrices group
(meanwhile a real finite dimensional Lie algebra is isomorphic to a matrices algebra).

This theorem does not hold if g is infinite dimensional.

\begin{theorem}
Two simply connected Lie groups with isomorphic Lie algebras are Lie isomorphic.
\end{theorem}

But this is generally untrue if they are not simply connected. However if we
have a simply connected Lie group, we can deduce all the other Lie groups,
simply connected or not, sharing the same Lie algebra, as quotient groups.
This is the purpose of the next topic.

\subsubsection{Covering group}

\begin{theorem}
(Knapp p.89) Let G be a connected Lie group, there is a unique connected,
simply connected Lie group $\widetilde{G}$ and a smooth Lie group morphism :
$\pi:\widetilde{G}\rightarrow G$ such that $\left(  \widetilde{G},\pi\right)
$ is a universal covering of G. $\widetilde{G}$ and G have the same dimension,
and same Lie algebra. G is Lie isomorphic to $\widetilde{G}/H$ where H is some
discrete subgroup in the center of G. Any connected Lie group G' with the same
Lie algebra as G is isomorphic to $\widetilde{G}/D$ for some discrete subgroup
D in the center of G.
\end{theorem}

See topology for the definition of covering spaces.

So for any connected Lie group G, there is a unique simply connected Lie group
$\widetilde{G}$ which has the same Lie algebra. And $\widetilde{G}$ is the
direct product of G and some finite groups. The other theorems give results
useful with topological groups.

\begin{theorem}
(Knapp p.85) Let G be a connected, locally pathwise connected, separable
topological metric group, H be a closed locally pathwise connected subgroup,
$H_{0}$\ the identity component of H.Then

i) the quotient G/H is connected and pathwise connected

ii) if G/H is simply connected then H is connected

iii) the map $G/H_{0}\rightarrow G/H$ is a covering map

iv) if H is discrete, then the quotient map $G\rightarrow G/H$ is a covering map

v) if H is connected ,G simply connected, G/H locally simply connected, then
G/H is simply connected
\end{theorem}

\begin{theorem}
(Knapp p.88) Let G be a locally connected, pathwise connected, locally simply
connected, separable topological metric group, $\left(  \widetilde{G}%
,\pi:\widetilde{G}\rightarrow G\right)  $ a simply connected covering of G
with $1_{\widetilde{G}}=\pi^{-1}\left(  1_{G}\right)  .$ Then there is a
unique multiplication in $\widetilde{G}$ such that it is a topological group
and $\pi$\ is a continuous group morphism. $\widetilde{G}$ with this structure
is called the \textbf{universal covering group} of G. It is unique up to isomorphism.
\end{theorem}

\begin{theorem}
(Knapp p.88) Let G be a connected, locally pathwise connected, locally simply
connected, separable topological metric group, H a closed subgroup, locally
pathwise connected, locally simply connected.\ If G/H is simply connected then
the fundamental group $\pi_{1}\left(  G,1\right)  $ is isomorphic to a
quotient group of $\pi_{1}\left(  H,1\right)  $
\end{theorem}

\subsubsection{Complex structures}

The nature of the field K matters only for Lie groups, when the manifold
structure is involved. All the previous results are valid for K=$%
\mathbb{R}
,%
\mathbb{C}
$ whenever it is not stated otherwise. So if G is a complex manifold its Lie
algebra is a complex algebra and the exponential is a holomorphic map.

The converse (how a real Lie group can be made a complex Lie group) is less
obvious, as usual.\ The group structure is not involved, so the problem is to
define a complex manifold structure. The way to do it is through the Lie algebra.

\begin{definition}
A complex Lie group $G_{%
\mathbb{C}
}$ is the \textbf{complexification} of a real Lie group G if G is a Lie
subgroup of $G_{%
\mathbb{C}
}$ and if the Lie algebra of $G_{%
\mathbb{C}
}$ is the complexification of the Lie algebra of G.
\end{definition}

There are two ways to "complexify" G.

1. By a complex structure J on $T_{1}G$. Then $T_{1}G$\ \ and the group G stay
the same as set. But there are compatibility conditions (the dimension of G
must be even and J compatible with the bracket), moreover the exponential must
be holomorphic (Knapp p.96) with this structure : $\frac{d}{dv}\exp J\left(
v\right)  |_{v=u}=J(\left(  \frac{d}{dv}\exp v\right)  |_{v=u}). $ We have a
partial answer to this problem :

\begin{theorem}
(Knapp p.435) A semi simple real Lie group G whose Lie algebra has a complex
structure admits uniquely the structure of a complex Lie group such that the
exponential is holomorphic.
\end{theorem}

2. By complexification of the Lie algebra. This is always possible, but the
sets do not stay the same. The new complex algebra $g_{%
\mathbb{C}
}$ can be the Lie algebra of some complex Lie group $G_{%
\mathbb{C}
}$ with complex dimension equal to the real dimension of G. But the third's
Lie theorem does not apply, and more restrictive conditions are imposed to G.
If there is a complex Lie group $G_{%
\mathbb{C}
}$ such that : its Lie algebra is $\left(  T_{1}G\right)  _{%
\mathbb{C}
}$ and G is a subgroup of $G_{%
\mathbb{C}
}$ then one says that $G_{%
\mathbb{C}
}$\ is the complexified of G. Complexified of a Lie group do not always exist,
and they are usually not unique. Anyway then $G_{%
\mathbb{C}
}\neq G.$

If G is a real semi simple finite dimensional Lie group, its Lie algebra is
semi-simple and its complexified is still semi-simple, thus $G_{%
\mathbb{C}
}$ must be a complex semi simple group, isomorphic to a Lie group of matrices,
and so for G.

\begin{theorem}
(Knapp p.537) A compact finite dimensional real Lie group admits a unique
complexification (up to isomorphism)
\end{theorem}

\subsubsection{Solvable, nilpotent Lie groups}

\begin{theorem}
(Kolar p.130) The commutator of two elements of a group G is the operation :
$K:G\times G\rightarrow G::K\left(  g,h\right)  =ghg^{-1}h^{-1}$

If G is a Lie group the map is continuous.\ If $G_{1},G_{2}$ are two closed
subgroup, then the set $K\left[  G_{1},G_{2}\right]  $ generated by all the
commutators $K\left(  g_{1},g_{2}\right)  $ with $g_{1}\in G_{1},g_{2}\in
G_{2}$ is a closed subgroup of G, thus a Lie group.
\end{theorem}

From there one can build sequences similar to the sequences of brackets of Lie
algebra :

$G^{0}=G=G_{0},G^{n}=K\left[  G^{n-1},G^{n-1}\right]  ,G_{n}=K\left[
G,G_{n-1}\right]  ,G_{n}\subset G^{n}$

A Lie group is said to be solvable if $\exists n\in%
\mathbb{N}
:G^{n}=1$

A Lie group is said to be nilpotent if $\exists n\in%
\mathbb{N}
:G_{n}=1$

But the usual and most efficient way is to proceed through the Lie algebra.

\begin{theorem}
A Lie group is solvable (resp.nilpotent) if its Lie algebra is solvable (resp.nilpotent).
\end{theorem}

\begin{theorem}
(Knapp p.106) If A is a finite dimensional, solvable, real, Lie algebra, then
there is a simply connected Lie group G with Lie algebra A, and G is
diffeomorphic to an euclidean space with coordinates of the second kind.

If $\left(  e_{i}\right)  _{i=1}^{n}$ is a basis of A, then :

$\forall g\in G,\exists t_{1},..t_{n}\in%
\mathbb{R}
:g=\exp t_{1}e_{1}\times\exp t_{2}e_{2}...\times\exp t_{n}e_{n}$

There is a sequence $\left(  G_{p}\right)  $ of closed simply connected Lie
subgroups of G such that :

$G=G_{0}\supseteq G_{1}...\supseteq G_{n}=\left\{  1\right\}  $

$G_{p}=%
\mathbb{R}
^{p}\varpropto G_{p+1}$

$G_{p+1}$ normal in $G_{p}$
\end{theorem}

\begin{theorem}
(Knapp p.107) On a simply connected finite dimensional nilpotent real Lie
group G the exponential map is a diffeomorphism from $T_{1}G$ to G (it is
surjective). Moreover any Lie subgroup of G is simply connected and closed.
\end{theorem}

\subsubsection{Abelian groups}

\paragraph{Main result\newline}

\begin{theorem}
(Duistermaat p.59) A connected Lie group G is abelian (=commutative) iff its
Lie algebra is abelian . Then the exponential map is onto and its kernel is a
discrete (closed, zero dimensional) subgroup of $\left(  T_{1}G,+\right)  $.
The exponential induces an isomorphism of Lie groups : $T_{1}G/\ker
\exp\rightarrow G$
\end{theorem}

That means that there are $0<p\leq\dim T_{1}G$ linearly independant vectors
$V_{k}$ of $T_{1}G$ such that :

$\ker\exp=\sum_{k=1}^{p}z_{k}V_{k},z_{k}\in%
\mathbb{Z}
$

Such a subset is called a p dimensional integral lattice.

Any n dimensional abelian Lie group over the field K is isomorphic to the
group (with addition) : $\left(  K/%
\mathbb{Z}
\right)  ^{p}\times K^{n-p}$ with $p=\dim span\ker\exp$

\begin{definition}
A \textbf{torus} is a compact abelian topological group
\end{definition}

\begin{theorem}
Any torus which is a finite n dimensional Lie group on a field K is isomorphic
to $\left(  \left(  K/%
\mathbb{Z}
\right)  ^{n},+\right)  $
\end{theorem}

A subgroup of $\left(
\mathbb{R}
,+\right)  $ is of the form $G=\left\{  ka,k\in%
\mathbb{Z}
\right\}  $ or is dense in $%
\mathbb{R}
$

Examples :

the nxn diagonal matrices diag$\left(  \lambda_{1},...\lambda_{n}\right)
,\lambda_{k}\neq0\in K$ is a commutative n dimensional Lie group isomorphic to
$K^{n}.$

the nxn diagonal matrices diag$\left(  \exp\left(  i\lambda_{1}\right)
,...\exp\left(  i\lambda_{n}\right)  \right)  ,\lambda_{k}\neq0\in%
\mathbb{R}
$ is a commutative n dimensional Lie group which is a torus.

\paragraph{Pontryagin duality\newline}

\begin{definition}
The "\textbf{Pontryagin dual"} $\widehat{G}$ of an \textit{abelian}
topological group G is the set of continuous morphisms, called
\textbf{characters, }$\chi:G\rightarrow T$ where T is the set $T=\left(
\left\{  z\in%
\mathbb{C}
:\left\vert z\right\vert =1\right\}  ,\times\right)  $ of complex number of
module 1 endowed with the product as internal operation .\ Endowed with the
compact-open topology and the pointwise product as internal operation
$\widehat{G}$ is a topological abelian group.
\end{definition}

$\chi\in\widehat{G},g,h\in G:\chi\left(  g+h\right)  =\chi\left(  g\right)
\chi\left(  h\right)  ,\chi\left(  -g\right)  =\chi\left(  g\right)
^{-1},\chi\left(  1\right)  =1$

$\left(  \chi_{1}\chi_{2}\right)  \left(  g\right)  =\chi_{1}\left(  g\right)
\chi_{2}\left(  g\right)  $

The "double-dual" of G : $\widehat{\left(  \widehat{G}\right)  }%
:\theta:\widehat{G}\rightarrow T$

The map : $\tau:G\times\widehat{G}\rightarrow T::\tau\left(  g,\chi\right)
=\chi\left(  g\right)  $ is well defined and depends only on G.

For any $g\in G$\ the map : $\tau_{g}:\widehat{G}\rightarrow T::\tau
_{g}\left(  \chi\right)  =\tau\left(  g,\chi\right)  =\chi\left(  g\right)  $
is continuous and $\tau_{g}\in\widehat{\left(  \widehat{G}\right)  }$

The map, called Gel'fand transformation : $\widehat{}:G\rightarrow
\widehat{\left(  \widehat{G}\right)  }::\widehat{g}=\tau_{g}$ has the defining
property : $\forall\chi\in\widehat{G}:\widehat{g}\left(  \chi\right)
=\chi\left(  g\right)  $

\begin{theorem}
Pontryagin-van Kampen theorem: If G is an abelian, locally compact topological
group, then G is continuously isomorphic to its bidual $\widehat{\left(
\widehat{G}\right)  }$ through the Gel'fand transformation. Then if G is
compact, its Pontryagin dual $\widehat{G}$ is discrete and is isomorphic to a
closed subgroup of $T^{\widehat{G}}$. Conversely if $\widehat{G}$ is discrete,
then G is compact. If G is finite then $\widehat{G}$ is finite.
\end{theorem}

A subset E of $\widehat{G}$ is said to \textbf{separate} G if :

$\forall g,h\in G,g\neq h,\exists\chi\in E:\chi\left(  g\right)  \neq
\chi\left(  h\right)  $

Any subset E which separates G is dense in $\widehat{G}.$

\begin{theorem}
Peter-Weyl: If G is an abelian, compact topological group, then its
topological dual $\widehat{G}$ separates G:

$\forall g,h\in G,g\neq h,\exists\chi\in\widehat{G}:\chi\left(  g\right)
\neq\chi\left(  h\right)  $
\end{theorem}

\subsubsection{Compact groups}

Compact groups have many specific properties that we will find again in
representation theory.

\begin{definition}
A topological or Lie group is compact if it is compact with its topology.
\end{definition}

\begin{theorem}
The Lie algebra of a compact Lie group is compact.

Any closed algebraic subgroup of a compact Lie group is a compact Lie
subgroup. Its Lie algebra is compact.
\end{theorem}

\begin{theorem}
A compact Lie group is necessarily

i) finite dimensional

ii) a torus if it is a connected complex Lie group

iii) a torus if it is abelian
\end{theorem}

\begin{proof}
i) a compact manifold is locally compact, thus it cannot be infinite dimensional

ii) the Lie algebra of a compact complex Lie group is a complex compact Lie
algebra, thus an abelian algebra, and the Lie group is an abelian Lie group.
The only abelian Lie groups are the product of torus and euclidean spaces, so
a complex compact Lie group must be a torus.
\end{proof}

\begin{theorem}
(Duistermaat p.149) A real finite dimensional Lie group G :

i) is compact iff the Killing form of its Lie algebra $T_{1}G$ is negative
semi-definite and its kernel is the center of $T_{1}G$

ii) is compact, semi-simple, iff the Killing form of its Lie algebra $T_{1}G$
is negative definite (so it has zero center).
\end{theorem}

\begin{theorem}
(Knapp p.259) For any connected compact Lie group the exponential map is
onto.\ Thus : $\exp:T_{1}G\rightarrow G$ is a diffeomorphism
\end{theorem}

\begin{theorem}
Weyl's theorem (Knapp p.268): If G is a compact semi-simple real Lie group,
then its fundamental group is finite, and its universal covering group is compact.
\end{theorem}

\paragraph{Structure of compact real Lie groups\newline}

The study of the internal structure of a compact group proceeds along lines
similar to the complex simple Lie algebras, the tori replacing the Cartan
algebras. It mixes analysis at the algebra and group levels (Knapp IV.5 for more).

Let G be a compact, connected, real Lie group.

1. Torus:

A torus of G is an abelian Lie subgroup.\ It is said to be maximal if it is
not contained in another torus. Maximal tori are conjugate from each others$.$
Each element of G lies in some maximal torus and is conjugate to an element of
any maximal torus. The center of G lies in all maximal tori. So let T be a
maximal torus, then : $\forall g\in G:\exists t\in T,x\in G:g=xtx^{-1}.$ The
relation : $x\sim y\Leftrightarrow\exists z:y=zxz^{-1}$ is an equivalence
relation, thus we have a partition of G in classes of conjugacy, T is one
class, pick up $\left(  x_{i}\right)  _{i\in I}$ in the other classes and
$G=\left\{  x_{i}tx_{i}^{-1},i\in I,t\in T\right\}  .$

2. Root space decomposition:

Let t be the Lie algebra of T$.$ If we take the complexified $\left(
T_{1}G\right)  _{C}$ of the Lie algebra of G, and $t_{C}$ of t, then $t_{C}$
is a Cartan subalgebra of $\left(  T_{1}G\right)  _{C}$ and we have a
root-space decomposition similar to a semi simple complex Lie algebra :

$\left(  T_{1}G\right)  _{C}=t_{C}\oplus_{\alpha}g_{\alpha}$

where the root vectors $g_{\alpha}=\left\{  X\in\left(  T_{1}G\right)
_{C}:\forall H\in t_{C}:\left[  H,X\right]  =\alpha\left(  H\right)
X\right\}  $ are the unidimensional eigen spaces of ad over $t_{C},$ with
eigen values $\alpha\left(  H\right)  ,$ which are the roots of $\left(
T_{1}G\right)  _{C}$ with respect to t.

The set of roots $\Delta\left(  \left(  T_{1}G\right)  _{C},t_{C}\right)  $
has the properties of a roots system except that we do not have $t_{C}^{\ast
}=span\Delta.$

For any $H\in t:\alpha\left(  H\right)  \in i%
\mathbb{R}
:$ the roots are purely imaginary.

3. Any $\lambda\in t_{c}^{\ast}$ is said to be analytically integral if it
meets one of the following properties :

i) $\forall H\in t:\exp H=1\Rightarrow\exists k\in%
\mathbb{Z}
:\lambda\left(  H\right)  =2i\pi k$

ii) there is a continuous homomorphism $\xi$\ from T to the complex numbers of
modulus 1 (called a multiplicative character) such that : $\forall H\in
t:\exp\lambda\left(  H\right)  =\xi\left(  \exp H\right)  $

Then $\lambda$ is real valued on t. All roots have these properties.

remark : $\lambda\in t_{c}^{\ast}$ is said to be algebraically integral if
$\frac{2\left\langle \lambda,\alpha\right\rangle }{\left\langle \alpha
,\alpha\right\rangle }\in%
\mathbb{Z}
$ with some inner product on the Lie algebra as above.

\subsubsection{Semi simple Lie groups}

\begin{definition}
A Lie group is :

\textbf{simple} if the only normal subgroups are 1 and G itself.

\textbf{semi-simple} if its Lie algebra is semi-simple (it has no non zero
solvable ideal).
\end{definition}

The simplest criterium is that the Killing form of the Lie algebra of a
semi-simple Lie group is non degenerate. The center of a connected semi-simple
Lie group is just 1.

Any real semi-simple finite dimensional Lie algebra A has a Cartan
decomposition, that is a pair of subvector spaces $l_{0},p_{0}$ of A such that
: $A=l_{0}\oplus p_{0},$ $l_{0}$ is a subalgebra of A, and an involution
$\theta\left(  l_{0}\right)  =l_{0},\theta\left(  p_{0}\right)  =-p_{0}$. We
have something similar at the group level, which is both powerful and useful
because semi-simple Lie groups are common.

\begin{theorem}
(Knapp p.362) For any real, finite dimensional, semi-simple Lie group G, with
the subgroup L corresponding to $l_{0}\in T_{1}G$\ :

i) There is a Lie group automorphism $\Theta$ on G such that $\Theta^{\prime
}\left(  g\right)  |_{g=1}=\theta$

ii) L is invariant by $\Theta$

iii) the maps : $L\times p_{0}\rightarrow G::g=l\exp p$ and $p_{0}\times
L\rightarrow G::g=\left(  \exp p\right)  l$\ are diffeomorphisms onto.\ 

iv) L is closed

v) L contains the center Z of G

vi) L is compact iff Z is finite

vii) when Z is finite then L is a maximal compact subgroup of G.
\end{theorem}

So any element of G can be written as : $g=l\exp p$ or equivalently as
$g=(\exp p)l.$ Moreover if L is compact the exponential is onto :
$l=\exp\lambda,\lambda\in l_{0}$

$\Theta$ is called the \textbf{global Cartan involution}

The decomposition $g=\left(  \exp p\right)  l$ is the \textbf{global Cartan
decomposition}.

Warning ! usually the set $\left\{  \exp p,p\in p_{0}\right\}  $ is not a group.

As an application :

\begin{theorem}
(Knapp p.436) For a complex semi simple finite dimensional Lie group G:

i) its algebra is complex semi simple, and has a real form $u_{0}$ which is a
compact semi simple real Lie algebra and the Lie algebra can be written as the
real vector space $T_{1}G=u_{0}\oplus iu_{0}$\ 

ii) G has necessarily a finite center.

iii) G is Lie complex isomorphic to a complex Lie group of matrices. And the
same is true for its universal covering group (which has the same algebra).
\end{theorem}

Remark : while semi simple Lie algebras can be realized as matrices algebras,
semi simple \textit{real} Lie groups need not to be realizable as group of
matrices : there are examples of such groups which have no linear faithful
representation (ex : the universal covering group of $SL(2,%
\mathbb{R}
)).$

\subsubsection{Classification of Lie groups}

The isomorphisms classes of finite dimensional :

i) simply connected compact semi simple real Lie groups

ii) complex semi simple Lie algebras

iii) compact semi simple real Lie algebras

iv) reduced abstract roots system

v) abstract Cartan matrices and their associated Dynkin diagrams

are in one one correspondance, by passage from a Lie group to its Lie algebra,
then to its complexification and eventually to the roots system.

So the list of all simply connected compact semi simple real Lie groups is
deduced from the list of Dynkin diagrams given in the Lie algebra section, and
we go from the Lie algebra to the Lie group by the exponential.

\bigskip

\subsection{Integration on a group}

\label{Integration on a group}

The integral can be defined on any measured set, and so on topological groups,
and we start with this case which is the most general.\ The properties of
integration on Lie groups are similar, even if they proceed from a different approach.

\subsubsection{Integration on a topological group}

\paragraph{Haar Radon measure\newline}

The integration on a topological group is based upon Radon measure on a
topological group. A Radon measure is a Borel, locally finite, regular, signed
measure on a topological Hausdorff locally compact space (see Measure). So if
the group is also a Lie group it must be finite dimensional.

\begin{definition}
(Neeb p.46) A left (right) Haar Radon measure on a locally compact topological
group G is a positive Radon measure $\mu$ such that : $\forall f\in
C_{0c}\left(  G;%
\mathbb{C}
\right)  ,\forall g\in G:$

left invariant : $\ell\left(  f\right)  =\int_{G}f\left(  gx\right)
\mu\left(  x\right)  =\int_{G}f\left(  x\right)  \mu\left(  x\right)  $\ 

right invariant : $\ell\left(  f\right)  =\int_{G}f\left(  xg\right)
\mu\left(  x\right)  =\int_{G}f\left(  x\right)  \mu\left(  x\right)  $\ 
\end{definition}

\begin{theorem}
(Neeb p.46) Any locally compact topological group has Haar Radon measures and
they are proportional.
\end{theorem}

The Lebesgue measure is a Haar measure on $\left(
\mathbb{R}
^{m},+\right)  $ so any Haar measure on $\left(
\mathbb{R}
^{m},+\right)  $ is proportional to the Lebesgue measure.

If G is a discrete group a Haar Radon measure is just a map :

$\int_{G}f\mu=\sum_{g\in G}f\left(  g\right)  \mu\left(  g\right)  ,\mu\left(
g\right)  \in%
\mathbb{R}
_{+}$

On the circle group T=$\left\{  \exp it,t\in%
\mathbb{R}
\right\}  :\ell\left(  f\right)  =\frac{1}{2\pi}\int_{0}^{2\pi}f\left(  \exp
it\right)  dt$\ 

\begin{theorem}
All connected, locally compact groups G are $\sigma$-finite under Haar measure.
\end{theorem}

\paragraph{Modular function\newline}

\begin{theorem}
For any left Haar Radon measure $\mu_{L}$ on the group G there is a continuous
homomorphism, called the \textbf{modular function} $\Delta:G\rightarrow%
\mathbb{R}
_{+}$ such that : $\forall a\in G:R_{a}^{\ast}\mu_{L}=\Delta\left(  a\right)
^{-1}\mu_{L}$ . It does not depend on the choice of $\mu_{L}.$
\end{theorem}

\begin{definition}
If the group G is such that $\Delta\left(  a\right)  =1$ then G is said to be
\textbf{unimodular} and then any left invariant Haar Radon measure is also
right invariant, and called a Haar Radon measure.
\end{definition}

Are unimodular the topological, locally compact, goups which are either :
compact, abelian, or for which the commutator group (G,G) is dense.

Remark : usually affine groups are not unimodular.

\paragraph{Spaces of functions\newline}

1. If G is a topological space endowed with a Haar Radon measure (or even a
left invariant or right invariant measure) $\mu$, then one can implement the
classical definitions of spaces of integrable functions on G (Neeb p.32). See
the part Functional Analysis for the definitions.

$L^{p}\left(  G,S,\mu,%
\mathbb{C}
\right)  $ is a Banach vector space with the norm: $\left\Vert f\right\Vert
_{p}=\left(  \int_{E}\left\vert f\right\vert ^{p}\mu\right)  ^{1/p}$

$L^{2}\left(  G,S,\mu,%
\mathbb{C}
\right)  $ is a Hilbert vector space with the scalar product : $\left\langle
f,g\right\rangle =\int_{E}\overline{f}g\mu$

$%
\mathcal{L}%
^{\infty}\left(  G,S,\mu,%
\mathbb{C}
\right)  =\left\{  f:G\rightarrow%
\mathbb{C}
:\exists C\in%
\mathbb{R}
:\left\vert f(x)\right\vert <C\right\}  $

$L^{\infty}\left(  G,S,\mu,%
\mathbb{C}
\right)  $ is a C*-algebra (with pointwise multiplication and the norm
$\left\Vert f\right\Vert _{\infty}=\inf\left(  C\in%
\mathbb{R}
:\mu\left(  \left\{  \left\vert \mu\left(  f\right)  \right\vert >C\right\}
\right)  =0\right)  $

2. If H is a separable Hilbert space the definition can be extended to maps
$\varphi:G\rightarrow H$ valued in H (Knapp p.567)..

One uses the fact that :

if $\left(  e_{i}\right)  _{\in I}$ is a Hilbert basis, then for any
measurable maps $G\rightarrow H.$\ 

$\left(  \varphi\left(  g\right)  ,\psi\left(  g\right)  \right)  =\sum_{i\in
I}\left\langle \varphi\left(  g\right)  e_{i},e_{i}\right\rangle \left\langle
e_{i},\psi\left(  g\right)  e_{i}\right\rangle $ is a measurable map :
$G\rightarrow%
\mathbb{C}
$

The scalar product is defined as : $\left(  \varphi,\psi\right)  =\int
_{G}\left(  \varphi\left(  g\right)  ,\psi\left(  g\right)  \right)  \mu.$

Then we can define the spaces $L^{p}\left(  G,\mu,H\right)  $ as above and
$L^{2}\left(  G,\mu,H\right)  $ is a separable Hilbert space

\paragraph{Convolution\newline}

\begin{definition}
(Neeb p.134) Let $\mu_{L}$\ be a left Haar measure on the locally compact
topological group G. The \textbf{convolution} on $L^{1}\left(  G,S,\mu_{L},%
\mathbb{C}
\right)  $ is defined as the map :

$\ast:L^{1}\left(  G,S,\mu_{L},%
\mathbb{C}
\right)  \times L^{1}\left(  G,S,\mu_{L},%
\mathbb{C}
\right)  \rightarrow L^{1}\left(  G,S,\mu_{L},%
\mathbb{C}
\right)  ::$%

\begin{equation}
\varphi\ast\psi\left(  g\right)  =\int_{G}\varphi\left(  x\right)  \psi\left(
x^{-1}g\right)  \mu_{L}\left(  x\right)  =\int_{G}\varphi\left(  gx\right)
\psi\left(  x^{-1}\right)  \mu\left(  x\right)
\end{equation}

\end{definition}

\begin{definition}
(Neeb p.134) Let $\mu_{L}$\ be a left Haar measure on the locally compact
topological group G. On the space $L^{1}\left(  G,S,\mu_{L},%
\mathbb{C}
\right)  $ \ the involution is defined as :%

\begin{equation}
\varphi^{\ast}\left(  g\right)  =\left(  \Delta\left(  g\right)  \right)
^{-1}\overline{\varphi\left(  g^{-1}\right)  }%
\end{equation}

so it will be $\varphi^{\ast}\left(  g\right)  =\overline{\varphi\left(
g^{-1}\right)  }$ if G is unimodular.
\end{definition}

\bigskip

Supp$\left(  \varphi\ast\psi\right)  \subset Supp(\varphi)Supp\left(
\psi\right)  $

convolution is associative

$\left\Vert \varphi\ast\psi\right\Vert _{1}\leq\left\Vert \varphi\right\Vert
_{1}\left\Vert \psi\right\Vert _{1}$

$\left\Vert \varphi^{\ast}\right\Vert _{1}=\left\Vert \varphi\right\Vert _{1}$

$\left(  \varphi\ast\psi\right)  ^{\ast}=\psi^{\ast}\ast\varphi^{\ast} $

If G is abelian the convolution is commutative (Neeb p.161)

With the left and right actions of G on $L^{1}\left(  G,S,\mu_{L},%
\mathbb{C}
\right)  :$

$\Lambda\left(  g\right)  \varphi\left(  x\right)  =\varphi\left(
g^{-1}x\right)  $

$P\left(  g\right)  \varphi\left(  x\right)  =\varphi\left(  xg\right)  $

then :

$\Lambda\left(  g\right)  \left(  \varphi\ast\psi\right)  =\left(
\Lambda\left(  g\right)  \varphi\right)  \ast\psi$

$P\left(  g\right)  \left(  \varphi\ast\psi\right)  =\varphi\ast\left(
P\left(  g\right)  \psi\right)  $

$\left(  P\left(  g\right)  \varphi\right)  \ast\psi=\varphi\ast\left(
\Delta\left(  g\right)  \right)  ^{-1}\Lambda\left(  g^{-1}\right)  \psi$

$\left(  \Lambda\left(  g\right)  \varphi\right)  ^{\ast}=\Delta\left(
g\right)  \left(  P\left(  g\right)  \varphi^{\ast}\right)  $

$\left(  P\left(  g\right)  \varphi\right)  ^{\ast}=\left(  \Delta\left(
g\right)  \right)  ^{-1}\left(  \Lambda\left(  g\right)  \varphi^{\ast
}\right)  $

$\left\Vert \Lambda\left(  g\right)  \varphi\right\Vert _{1}=\left\Vert
\varphi\right\Vert _{1}$

$\left\Vert P\left(  g\right)  \varphi\right\Vert _{1}=\left(  \Delta\left(
g\right)  \right)  ^{-1}\left\Vert \varphi\right\Vert _{1}$

\begin{theorem}
With convolution as internal operation $L^{1}\left(  G,S,\mu_{L},%
\mathbb{C}
\right)  $ is a complex Banach *-algebra and $\Lambda\left(  g\right)
,\Delta\left(  g\right)  P\left(  g\right)  $ are isometries.
\end{theorem}

\subsubsection{Integration on a Lie group}

The definition of a Haar measure on a Lie group proceeds differently since it
is the integral of a n-form.

\paragraph{Haar measure\newline}

\begin{definition}
A \textbf{left Haar measure} on a real n dimensional Lie group is a n-form
$\varpi$\ on TG such that : $\forall a\in G:L_{a}^{\ast}\varpi=\varpi$

A \textbf{right Haar measure} on a real n dimensional Lie group is a n-form
$\varpi$\ on TG such that : $\forall a\in G:R_{a}^{\ast}\varpi=\varpi$
\end{definition}

that is : $\forall u_{1},..u_{r}\in T_{x}G:\varpi\left(  ax\right)  \left(
L_{a}^{\prime}\left(  x\right)  u_{1},...,L_{a}^{\prime}\left(  x\right)
u_{r}\right)  =\varpi\left(  x\right)  \left(  u_{1},...u_{r}\right)  $

\begin{theorem}
Any real finite dimensional Lie group has left and right Haar measures, which
are volume forms on TG
\end{theorem}

\begin{proof}
Take the dual basis $\left(  e^{i}\right)  _{i=1}^{n}$ of $T_{1}G$\ and its
pull back in x :

$e^{i}\left(  x\right)  \left(  u_{x}\right)  =e^{i}\left(  L_{x^{-1}}%
^{\prime}\left(  x\right)  \right)  \Rightarrow e^{i}\left(  x\right)  \left(
e_{j}\left(  x\right)  \right)  =e^{i}\left(  L_{x^{-1}}^{\prime}\left(
x\right)  e_{j}\left(  x\right)  \right)  =\delta_{j}^{i}$

Then $\varpi_{r}\left(  x\right)  =e^{1}\left(  x\right)  \wedge...\wedge
e^{r}\left(  x\right)  $ is left invariant :

$\varpi\left(  ax\right)  \left(  L_{a}^{\prime}\left(  x\right)
u_{1},...,L_{a}^{\prime}\left(  x\right)  u_{r}\right)  =\sum_{\left(
i_{1}...i_{r}\right)  }\epsilon\left(  i_{1},..i_{r}\right)  e^{i_{1}}\left(
ax\right)  \left(  L_{a}^{\prime}\left(  x\right)  u_{1}\right)  ...e^{i_{r}%
}\left(  ax\right)  \left(  L_{a}^{\prime}\left(  x\right)  u_{r}\right)  $

$=\sum_{\left(  i_{1}...i_{r}\right)  }\epsilon\left(  i_{1},..i_{r}\right)
e^{i_{1}}\left(  L_{\left(  ax\right)  ^{-1}}^{\prime}\left(  ax\right)
L_{a}^{\prime}\left(  x\right)  u_{1}\right)  ...e^{i_{r}}\left(  L_{\left(
ax\right)  ^{-1}}^{\prime}\left(  ax\right)  L_{a}^{\prime}\left(  x\right)
u_{r}\right)  $

$L_{\left(  ax\right)  ^{-1}}^{\prime}\left(  ax\right)  =L_{x^{-1}}^{\prime
}\left(  x\right)  \left(  L_{a}^{\prime}\left(  x\right)  \right)
^{-1}\Rightarrow e^{i_{1}}\left(  L_{\left(  ax\right)  ^{-1}}^{\prime}\left(
ax\right)  L_{a}^{\prime}\left(  x\right)  u_{1}\right)  =e^{i_{1}}\left(
L_{x^{-1}}^{\prime}\left(  x\right)  u_{1}\right)  =e^{i_{1}}\left(  x\right)
\left(  u_{1}\right)  $

Such a n form is never null, so $\varpi_{n}\left(  x\right)  =e^{1}\left(
x\right)  \wedge...\wedge e^{n}\left(  x\right)  $ defines a left invariant
volume form on G. And G is orientable.
\end{proof}

All left (rigth) Haar measures are proportionnal. A particularity of Haar
measures is that any open non empty subset of G has a non null measure :
indeed if there was such a subset by translation we could always cover any
compact, which would have measure 0.

Remarks :

i) Haar measure is a bit a misnomer.\ Indeed it is a volume form, the measure
itself is defined through charts on G (see Integral on manifolds). The use of
notations such that $d_{L},d_{R}$ for Haar measures is just confusing.

ii) A Haar measure on a Lie group is a volume form so it is a Lebesgue measure
on the manifold G, and is necessarily absolutely continuous. A Haar measure on
a topological Group is a Radon measure, without any reference to charts, and
can have a discrete part.

\paragraph{Modular function\newline}

\begin{theorem}
For any left Haar measure $\varpi_{L}$\ on a finite dimensional Lie group G
there is some non null function $\Delta\left(  a\right)  $ , called a
\textbf{modular function}, such that : $R_{a}^{\ast}\varpi_{L}=\Delta\left(
a\right)  ^{-1}\varpi_{L}$
\end{theorem}

\begin{proof}
$R_{a}^{\ast}\varpi_{L}=R_{a}^{\ast}\left(  L_{b}^{\ast}\varpi_{L}\right)
=\left(  L_{b}R_{a}\right)  ^{\ast}\varpi_{L}=\left(  R_{a}L_{b}\right)
^{\ast}\varpi_{L}=L_{b}^{\ast}R_{a}^{\ast}\varpi_{L}=L_{b}^{\ast}\left(
R_{a}^{\ast}\varpi_{L}\right)  $ thus $R_{a}^{\ast}\varpi_{L}$ is still a left
invariant measure, and because all left invariant measure are proportionnal
there is some non null function $\Delta\left(  a\right)  $ such that :
$R_{a}^{\ast}\varpi_{L}=\Delta\left(  a\right)  ^{-1}\varpi_{L}$
\end{proof}

\begin{theorem}
(Knapp p.532) The modular function on a finite dimensional Lie group G, with
$\varpi_{L},\varpi_{R}\ $left, rigth Haar measure, has the following
properties :

i) its value is given by : $\Delta\left(  a\right)  =\left\vert \det
Ad_{a}\right\vert $

ii) $\Delta:G\rightarrow%
\mathbb{R}
_{+}$ is a smooth group homomorphism

iii) if $a\in H$ and H is a compact or semi-simple Lie subgroup of G then
$\Delta\left(  a\right)  =1$

iv) $\Im^{\ast}\varpi_{L}=\Delta\left(  x\right)  \varpi_{L}$ are rigth Haar
measures (with $\Im$ = inverse map)

v) $\Im^{\ast}\varpi_{R}=\Delta\left(  x\right)  ^{-1}\varpi_{R}$ are left
Haar measures

vi) $L_{a}\varpi_{R}=\Delta\left(  a\right)  \varpi_{R}$
\end{theorem}

\begin{definition}
A Lie group is said to be \textbf{unimodular} if any left Haar measure is a
right measure (and vice versa). Then we say that any right or left invariant
volume form is a \textbf{Haar measure}.
\end{definition}

A Lie group is unimodular iff $\forall a\in g:\Delta\left(  a\right)  =1.$

Are unimodular the following Lie groups : abelian, compact, semi simple,
nilpotent, reductive.

\paragraph{Decomposition of Haar measure\newline}

We have something similar to the Fubini theorem for Haar measures.

\begin{theorem}
(Knapp p.535) If S,T are closed Lie subgroups of the finite dimensional real
Lie group G, such that $S\cap T$ is compact, then the multiplication
$M:S\times T\rightarrow G$ is an open map, the products ST exhausts the whole
of G except for a null subset. Let $\Delta_{S},\Delta_{T}$ be the modular
functions on S,T.\ Then any left Haar measure $\varpi_{L}$ on S, T and G can
be normalized so that :

$\forall f\in C\left(  G;%
\mathbb{R}
\right)  :$

$\int_{G}f\varpi_{L}=\int_{S\times T}M^{\ast}\left(  f\frac{\Delta_{T}}%
{\Delta_{S}}\left(  \varpi_{L}\right)  _{S}\otimes\left(  \varpi_{L}\right)
_{T}\right)  =\int_{S}\varpi_{L}\left(  s\right)  \int_{T}\frac{\Delta
_{T}\left(  t\right)  }{\Delta_{S}\left(  s\right)  }f\left(  st\right)
\varpi_{L}\left(  t\right)  $
\end{theorem}

\begin{theorem}
(Knapp p.538) If H is a closed Lie subgroup of a finite dimensional real Lie
group G, $\Delta_{G},\Delta_{H}$ are the modular functions on G,H,\ there is a
volume form $\mu$\ on G/H invariant with respect to the right action if and
only if the restriction of $\Delta_{G}$ to H is equal to $\Delta_{H}$.\ Then
it is unique up to a scalar and can be normalized such that :

$\forall f\in C_{0c}\left(  G;%
\mathbb{C}
\right)  :\int_{G}f\varpi_{L}=\int_{G/H}\mu\left(  x\right)  \int_{H}f\left(
xh\right)  \varpi_{L}\left(  h\right)  $

with $\pi_{L}:G\rightarrow G/H::\pi_{L}\left(  g\right)  h=g\Leftrightarrow
g^{-1}\pi_{L}\left(  g\right)  \in H$
\end{theorem}

G/H is not a group (if H is not normal) but can be endowed with a manifold
structure, and the left action of G on G/H is continuous.

\newpage

\section{REPRESENTATION\ THEORY}

\subsection{Definitions and general results}

\label{Representation Definitions and general results}

Let E be a vector space on a field K. Then the set L(E;E) of linear
endomorphisms of E is : an algebra, a Lie algebra and its subset GL(E;E) of
inversible elements is a Lie group. Thus it is possible to consider morphisms
between algebra, Lie algebra or groups and L(E;E) or GL(E;E). Moreover if E is
a Banach vector space then we can consider continuous, differentiable or
smooth morphisms.

\subsubsection{Definitions}

\begin{definition}
A linear \textbf{representation of the Banach algebra} (A,$\cdot)$ over a
field K is a couple (E,f) of a Banach vector space E over K and a smooth map
$f:A\rightarrow%
\mathcal{L}%
\left(  E;E\right)  $ which is an algebra morphism :
\end{definition}

$\forall X,Y\in A,k,k^{\prime}\in K:f\left(  kX+k^{\prime}Y\right)
=kf(X)+k^{\prime}f(Y)$

$\forall X,Y\in A,:f\left(  X\cdot Y\right)  =f(X)\circ f(Y)$

$f\in%
\mathcal{L}%
\left(  A;%
\mathcal{L}%
\left(  E;E\right)  \right)  $

If A is unital (there is a unit element I) then we require that f(I)=Id$_{E}$

Notice that f must be K linear.

The representation is over an algebra of linear maps, so this is a geometrical
representation (usually called linear representation). A Clifford algebra is
an algebra, so it enters the present topic, but morphisms of Clifford algebras
have some specificities which are addressed in the Algebra part.

The representation of Banach algebras has been addressed in the Analysis part.
So here we just recall some basic definitions.

\begin{definition}
A linear \textbf{representation of the Banach Lie algebra} (A,$\left[
{}\right]  )$ over a field K is a couple (E,f) of a Banach vector space E over
K and a smooth map $f:A\rightarrow%
\mathcal{L}%
\left(  E;E\right)  $ which is a Lie algebra morphism:
\end{definition}

$\forall X,Y\in A,k,k^{\prime}\in K:f\left(  kX+k^{\prime}Y\right)
=kf(X)+k^{\prime}f(Y)$

$\forall X,Y\in A,:f\left(  \left[  X,Y\right]  \right)  =f(X)\circ
f(Y)-f\left(  Y\right)  \circ f\left(  X\right)  =\left[  f\left(  X\right)
,f\left(  Y\right)  \right]  _{%
\mathcal{L}%
\left(  E;E\right)  }$

$f\in%
\mathcal{L}%
\left(  A;%
\mathcal{L}%
\left(  E;E\right)  \right)  $

Notice that f must be K linear.

If E is a Hilbert space H then $%
\mathcal{L}%
\left(  H;H\right)  $ is a C*-algebra.

\begin{definition}
A linear \textbf{representation of the topological group G} is a couple (E,f)
of a topological vector space E and a continuous map $f:G\rightarrow G%
\mathcal{L}%
\left(  E;E\right)  $ which is a continuous group morphism :
\end{definition}

$\forall g,h\in G:f\left(  gh\right)  =f\left(  g\right)  \circ f\left(
h\right)  ,f\left(  g^{-1}\right)  =\left(  f\left(  g\right)  \right)
^{-1};f\left(  1\right)  =Id_{E}$

$f\in C_{0}\left(  G;G%
\mathcal{L}%
\left(  E;E\right)  \right)  $

f is usually not linear but f(g) must be invertible. E can be over any field.

\begin{definition}
A linear \textbf{representation of the Lie group G} over a field K is a couple
(E,f) of a Banach vector space E \ over the field K and a differentiable class
$r$ $\geq1$ map $f:G\rightarrow G%
\mathcal{L}%
\left(  E;E\right)  $ which is a Lie group morphism :
\end{definition}

$\forall g,h\in G:f\left(  gh\right)  =f\left(  g\right)  \circ f\left(
h\right)  ,f\left(  g^{-1}\right)  =\left(  f\left(  g\right)  \right)
^{-1};f\left(  1\right)  =Id_{E}$

$f\in C_{r}\left(  G;G%
\mathcal{L}%
\left(  E;E\right)  \right)  $

A continuous Lie group morphism is necessarily smooth

In all the cases above the vector space E is a \textbf{module} over the set
$\left\{  f\left(  g\right)  ,g\in G\right\}  $.

\bigskip

\begin{definition}
The \textbf{trivial representation} (E,f) of an algebra or a Lie algebra is
$f\left(  X\right)  =0$ with any E.

The trivial representation (E,f) of a group is $f\left(  g\right)  =Id$ with
any E.
\end{definition}

\begin{definition}
The \textbf{standard representation} of a Lie group of matrices in GL(K,n) is
$\left(  K^{n},\imath\right)  $ where \i(g) is just the linear map in
$GL\left(  K^{n};K^{n}\right)  $ whose matrix is g in the canonical basis .
\end{definition}

\textbf{Matrix representation }: any representation (E,f) on a finite
dimensional vector space E becomes a representation on a set of matrices by
choosing a basis. But a representation is not necessarily faithful, and the
algebra or the group may not to be isomorphic to a set of matrices. The matrix
representation depends on the choice of a basis, which can be specific
(usually orthonormal).

\bigskip

\begin{definition}
A representation (E,f) is \textbf{faithful} if the map f is bijective.
\end{definition}

Then we have an isomorphism, and conversely if we have an isomorphism with
some subset of linear maps over a vector space we have a representation.

\begin{definition}
An \textbf{interwiner} between two representations $(E_{1},f_{1}),\left(
E_{2},f_{2}\right)  $ of a set X is a morphism : $\phi\in%
\mathcal{L}%
\left(  E_{1};E_{2}\right)  $ such that :

$\forall x\in X:\phi\circ f_{1}\left(  x\right)  =f_{2}\left(  x\right)
\circ\phi$

If $\phi$ is an isomorphism then the two representations are said to be
\textbf{equivalent}.
\end{definition}

Conversely, if there is an isomorphism $\phi\in G%
\mathcal{L}%
\left(  E_{1};E_{2}\right)  $ between two vector spaces $E_{1},E_{2}$ and if
$(E_{1},f_{1})$ is a representation, then with $f_{2}\left(  x\right)
=\phi\circ f_{1}\left(  x\right)  \circ\phi^{-1},\ \left(  E_{2},f_{2}\right)
$ is an equivalent representation$.$

Notice that $f_{1},f_{2}$ are morphisms, of algebras or groups, so that $\phi$
must also meet this requirement.

\bigskip

An \textbf{invariant vector space} in a representation (E,f) is a vector
subspace F of E such that: $\forall x\in X,\forall u\in F:f\left(  x\right)
\left(  u\right)  \in F$

\begin{definition}
A representation (E,f) is \textbf{irreducible} if the only \textit{closed}
invariant vector subspaces are 0 and E.
\end{definition}

Irreducible representations are useful because many representations can be
built from simpler irreducible representations.

\begin{definition}
If H is a Lie subgroup of G, (E,f) a representation of G, then $(E,f_{H})$
where $f_{H}$ is the restriction of f to H, is a representation of H, called a
\textbf{subrepresentation}.
\end{definition}

Similarly If B is a Lie subalgebra of A, (E,f) a representation of A, then
$(E,f_{B})$ where $f_{B}$ is the restriction of f to B, is a representation of B.

\subsubsection{Complex and real representations}

These results seem obvious but are very useful, as many classical
representations can be derived from each other by using simultaneously complex
and real representations.

1. For any complex representation (E,f), with E a complex vector space and f a
$%
\mathbb{C}
$-differentiable map, $\left(  E_{%
\mathbb{R}
}\oplus iE_{%
\mathbb{R}
},f\right)  $ is a real representation with any real structure on E.

So if (E,f) is a complex representation of a complex Lie algebra or Lie group
we have easily a representation of any of their real forms.

2. There is a bijective correspondance between\ the real representations of a
real Lie algebra A and the complex representations of its complexified $A_{%
\mathbb{C}
}=A\oplus iA$. And one representation is irreducible iff the other is irreducible.

A real representation (E,f) of a Lie algebra A can be extended uniquely in a
complex representation $\left(  E_{%
\mathbb{C}
},f_{%
\mathbb{C}
}\right)  $ of $A_{%
\mathbb{C}
}$ by $f_{%
\mathbb{C}
}\left(  X+iY\right)  =f\left(  X\right)  +if\left(  Y\right)  $

Conversely if A is a real form of the complex Lie algebra $B=A\oplus iA$, any
complex representation (E,f) of B gives a representation of A by taking the
restriction of f to the vector subspace A.

3. If (E,f) is a complex representation of a Lie group G and $\sigma$\ a real
structure on E, there is always a conjugate complex vector space $\overline
{E}$ and a bijective antilinear map : $\sigma:E\rightarrow\overline{E}.$ To
any $f\left(  g\right)  \in L\left(  E;E\right)  $ we can associate a unique
conjugate map :

$\overline{f}(g)=\sigma\circ f\left(  g\right)  \circ\sigma^{-1}\in L\left(
E;E\right)  ::\overline{f}(g)u=\overline{f(g)\overline{u}}$

If f is a real map then $\overline{f}(g)=f\left(  g\right)  .$ If not they are
different maps, and we have the \textbf{conjugate} \textbf{representation
(}also called contragredient\textbf{)} : $\left(  E;\overline{f}\right)  $
which is not equivalent to (E,f).

\subsubsection{Sum and product of representations}

Given a representation (E,f) we can define infinitely many other
representations, and in physics finding the right representation is often a
big issue (example in the standard model).

\paragraph{Lie algebras - Sum of representations\newline}

\begin{definition}
The \textbf{sum} of the representations $(E_{i},f_{i})_{i=1}^{r}$ of the Lie
algebra $\left(  A,\left[  {}\right]  \right)  $, is the representation
$\left(  \oplus_{i=1}^{r}E_{i},\oplus_{i=1}^{r}f_{i}\right)  $
\end{definition}

For two representations :%

\begin{equation}
f_{1}\oplus f_{2}:A\rightarrow%
\mathcal{L}%
\left(  E_{1}\oplus E_{2};E_{1}\oplus E_{2}\right)  ::\left(  f_{1}\oplus
f_{2}\right)  \left(  X\right)  =f_{1}\left(  X\right)  +f_{2}\left(
X\right)
\end{equation}

So : $\left(  f_{1}\oplus f_{2}\right)  \left(  X\right)  \left(  u_{1}\oplus
u_{2}\right)  =f_{1}\left(  X\right)  u_{1}+f_{2}\left(  X\right)  u_{2}$

The bracket on $E_{1}\oplus E_{2}$ is $\left[  u_{1}+u_{2},v_{1}+v_{2}\right]
=\left[  u_{1},v_{1}\right]  +\left[  u_{2},v_{2}\right]  $

The bracket on $%
\mathcal{L}%
\left(  E_{1}\oplus E_{2};E_{1}\oplus E_{2}\right)  $ is $\left[  \varphi
_{1}\oplus\varphi_{2},\psi_{1}\oplus\psi_{2}\right]  =\left[  \varphi_{1}%
,\psi_{1}\right]  +\left[  \varphi_{2},\psi_{2}\right]  =\varphi_{1}\circ
\psi_{1}-\psi_{1}\circ\varphi_{1}+\varphi_{2}\circ\psi_{2}-\psi_{2}%
\circ\varphi_{2}$

The direct sum of representations is not irreducible. Conversely a
representation is said to \textbf{completely reducible} if it can be expressed
as the direct sum of irreducible representations.

\paragraph{Lie algebras - tensorial product of representations\newline}

\begin{definition}
The \textbf{tensorial product of the representations} $(E_{i},f_{i})_{i=1}%
^{r}$ of the Lie algebra $\left(  A,\left[  {}\right]  \right)  $ is a
representation $\left(  E=E_{1}\otimes E_{2}...\otimes E_{r},D\right)  $ with
the morphism $D$ defined as follows : for any $X\in A,$ $D\left(  X\right)  $
is the unique extension of

$\phi\left(  X\right)  =\sum_{k=1}^{r}Id\otimes...\otimes f\left(  X\right)
\otimes...\otimes Id\in L^{r}\left(  E_{1},..E_{r};E\right)  $

to a map $L\left(  E;E\right)  $ such that : $\phi\left(  X\right)  =D\left(
X\right)  \circ\imath$ with the canonical map : $\imath:%
{\textstyle\prod\limits_{k=1}^{r}}
E^{r}\rightarrow E$
\end{definition}

As $\phi\left(  X\right)  $ is a multilinear map such an extension always exist.

So for two representations :%

\begin{equation}
D\left(  X\right)  \left(  u_{1}\otimes u_{2}\right)  =\left(  f_{1}\left(
X\right)  u_{1}\right)  \otimes u_{2}+u_{1}\otimes\left(  f_{2}\left(
X\right)  u_{2}\right)
\end{equation}

The bracket on $L\left(  E_{1}\otimes E_{2};E_{1}\otimes E_{2}\right)  $ is

$\left[  f_{1}\left(  X\right)  \otimes Id_{2}+Id_{1}\otimes f_{2}\left(
X\right)  ,f_{1}\left(  Y\right)  \otimes Id_{2}+Id_{1}\otimes f_{2}\left(
Y\right)  \right]  $

$=\left(  f_{1}\left(  X\right)  \otimes Id_{2}+Id_{1}\otimes f_{2}\left(
X\right)  \right)  \circ\left(  f_{1}\left(  Y\right)  \otimes Id_{2}%
+Id_{1}\otimes f_{2}\left(  Y\right)  \right)  -\left(  f_{1}\left(  Y\right)
\otimes Id_{2}+Id_{1}\otimes f_{2}\left(  Y\right)  \right)  \circ\left(
f_{1}\left(  X\right)  \otimes Id_{2}+Id_{1}\otimes f_{2}\left(  X\right)
\right)  $

$=\left[  f_{1}\left(  X\right)  ,f_{1}\left(  Y\right)  \right]  \otimes
Id_{2}+Id_{1}\otimes\left[  f_{2}\left(  X\right)  ,f_{2}\left(  Y\right)
\right]  =\left(  f_{1}\times f_{2}\right)  \left(  \left[  X,Y\right]
\right)  $

If all representations are irreducible, then their tensorial product is irreducible.

If (E,f) is a representation the procedure gives a representation
($\otimes^{r}E,D^{r}f)$

\begin{definition}
If (E,f) is a representation of the Lie algebra $\left(  A,\left[  {}\right]
\right)  $ the representation ($\wedge^{r}E;,D_{A}^{r}f)$ is defined by
extending the antisymmetric map : $\phi_{A}\left(  X\right)  \in L^{r}\left(
E^{r};\wedge^{r}E\right)  ::\phi_{A}\left(  X\right)  =\sum_{k=1}^{r}%
Id\wedge...\wedge f\left(  X\right)  \wedge...\wedge Id$
\end{definition}

\begin{definition}
If (E,f) is a representation of the Lie algebra $\left(  A,\left[  {}\right]
\right)  $ the representation ($\odot^{r}E;,D_{S}^{r}f)$ is defined by
extending the symmetric map $\phi_{S}\left(  X\right)  \in L^{r}\left(
E^{r};S^{r}\left(  E\right)  \right)  ::\phi_{S}\left(  X\right)  =\sum
_{k=1}^{r}Id\odot...\odot f\left(  X\right)  \odot...\odot Id$
\end{definition}

Remarks :

i) $\odot^{r}E,\wedge^{r}E\subset\otimes^{r}E$ as vector subspaces.

ii) If a vector subspace F of E is invariant by f, then $\otimes^{r}F$\ is
invariant by $D^{r}f,$ as $\odot^{r}E,\wedge^{r}F$ for $D_{s}^{r}f,D_{a}%
^{r}f.$

iii) If all representations are irreducible, then their tensorial product is irreducible.

\paragraph{Groups - sum of representations\newline}

\begin{definition}
The \textbf{sum of the representations} $(E_{i},f_{i})_{i=1}^{r}$ of the group
G, is a representation $\left(  \oplus_{i=1}^{r}E_{i},\oplus_{i=1}^{r}%
f_{i}\right)  $
\end{definition}

For two representations :%

\begin{equation}
\left(  f_{1}\oplus f_{2}\right)  \left(  g\right)  \left(  u_{1}\oplus
u_{2}\right)  =f_{1}\left(  g\right)  u_{1}+f_{2}\left(  g\right)  u_{2}%
\end{equation}

The direct sum of representations is not irreducible. Conversely a
representation is said to \textbf{completely reducible} if it can be expressed
as the direct sum of irreducible representations.

\paragraph{Groups - tensorial product of representations\newline}

\begin{definition}
The \textbf{tensorial product of the representations} $(E_{i},f_{i})_{i=1}%
^{r}$ of the group G is a representation

$\left(  E=E_{1}\otimes E_{2}...\otimes E_{r},D\right)  $ with the morphism
$D$ defined as follows : for any $g\in G,$ $D\left(  g\right)  $ is the unique
extension of

$\phi\left(  g\right)  \left(  u_{1},..,u_{r}\right)  =f_{1}\left(  g\right)
u_{1}\otimes...\otimes f_{r}\left(  g\right)  u_{r}\in%
\mathcal{L}%
^{r}\left(  E_{1},..,E_{r};E\right)  $

to $D\left(  g\right)  \in%
\mathcal{L}%
\left(  E;E\right)  $ such that $\phi\left(  g\right)  =D\left(  g\right)
\circ\imath$ with the canonical map : $\imath:%
{\textstyle\prod\limits_{k=1}^{r}}
E^{r}\rightarrow E$
\end{definition}

As $\phi\left(  g\right)  $ is a multilinear map such an extension always exist.

If (E,f) is a representation the procedure gives a representation
($\otimes^{r}E,D^{r}f)$

For two representations :%

\begin{equation}
\left(  f_{1}\otimes f_{2}\right)  \left(  g\right)  \left(  u_{1}\otimes
u_{2}\right)  =f_{1}\left(  g\right)  u_{1}\otimes f_{2}\left(  g\right)
u_{2}%
\end{equation}

\begin{definition}
If (E,f) is a representation of the group G the representation ($\wedge
^{r}E;,D_{A}^{r}f)$ is defined by extending the antisymmetric map : $\phi
_{A}\left(  g\right)  \in L^{r}\left(  E^{r};\wedge^{r}E\right)  ::\phi
_{A}\left(  g\right)  =\sum_{k=1}^{r}f(g)\wedge...\wedge f\left(  g\right)
\wedge...\wedge f(g)$
\end{definition}

\begin{definition}
If (E,f) is a representation of the group G the representation ($\odot
^{r}E;,D_{S}^{r}f)$ is defined by extending the symmetric map $\phi_{S}\left(
g\right)  \in L^{r}\left(  E^{r};S^{r}\left(  E\right)  \right)  ::\phi
_{S}\left(  g\right)  =\sum_{k=1}^{r}f(g)\odot...\odot f\left(  g\right)
\odot...\odot f(g)$
\end{definition}

Remarks :

i) $\odot^{r}E,\wedge^{r}E\subset\otimes^{r}E$ as vector subspaces.

ii) If a vector subspace F of E is invariant by f, then $\otimes^{r}F$\ is
invariant by $D^{r}f,$ as $\odot^{r}E,\wedge^{r}F$ for $D_{s}^{r}f,D_{a}%
^{r}f.$

iii) If all representations are irreducible, then their tensorial product is irreducible.

\subsubsection{Representation of a Lie group and its Lie algebra}

\paragraph{From the Lie group to the Lie algebra\newline}

\begin{theorem}
If (E,f) is a representation of the Lie group G then (E,f'(1)) is a
representation of $T_{1}G$ and
\begin{equation}
\forall u\in T_{1}G:f\left(  \exp_{G}u\right)  =\exp_{G%
\mathcal{L}%
\left(  E;E\right)  }f^{\prime}(1)u
\end{equation}

\end{theorem}

\begin{proof}
f is a smooth morphism $f\in C_{\infty}\left(  G;G%
\mathcal{L}%
\left(  E;E\right)  \right)  $ and its derivative f'(1) is a morphism :
$f^{\prime}(1)\in%
\mathcal{L}%
\left(  T_{1}G;%
\mathcal{L}%
\left(  E;E\right)  \right)  $
\end{proof}

The exponential on the right side is computed by the usual series.

\begin{theorem}
If $(E_{1},f_{1}),\left(  E_{2},f_{2}\right)  $ are two equivalent
representations of the Lie group G, then $(E_{1},f_{1}^{\prime}(1)),\left(
E_{2},f_{2}^{\prime}\left(  1\right)  \right)  $ are two equivalent
representations of $T_{1}G$
\end{theorem}

\begin{theorem}
If the closed vector subspace F is invariant in the representation (E,f) of
the Lie group G, then F is invariant in the representation (E,f'(1)) of
$T_{1}G$ .
\end{theorem}

\begin{proof}
(F,f) is a representation is a representation of G, so is (F,f'(1))
\end{proof}

\begin{theorem}
If the Lie group G is connected, the representation (E,f) of G is irreducible
(resp.completely reducible) iff the representation (E,f'(1)) of its Lie
algebra is irreducible (resp.completely reducible)
\end{theorem}

Remark :

If $(E_{1},f_{1}),\left(  E_{2},f_{2}\right)  $ are two representations of the
Lie group G, the derivative of the product of the representations is :

$\left(  f_{1}\otimes f_{2}\right)  ^{\prime}(1):T_{1}G\rightarrow%
\mathcal{L}%
\left(  E_{1}\otimes E_{2};E_{1}\otimes E_{2}\right)  ::\left(  f_{1}\otimes
f_{2}\right)  ^{\prime}\left(  1\right)  \left(  u_{1}\otimes u_{2}\right)
=\left(  f_{1}^{\prime}\left(  1\right)  u_{1}\right)  \otimes u_{2}%
+u_{1}\otimes\left(  f_{2}^{\prime}\left(  1\right)  u_{2}\right)  $

that is the product $(E_{1}\otimes E_{2},f_{1}^{\prime}\left(  1\right)
\times f_{2}^{\prime}\left(  1\right)  )$ of the representations $(E_{1}%
,f_{1}^{\prime}\left(  1\right)  ),\left(  E_{2},f_{2}^{\prime}\left(
1\right)  \right)  $ of $T_{1}G$

We have similar results for the sum of representations.

\paragraph{From the Lie algebra to the Lie group\newline}

The converse is more restrictive.

\begin{theorem}
If (E,f) is a representation of the Lie algebra $T_{1}G$ of a connected finite
dimensional Lie group, $\widetilde{G}$ a universal covering group of G with
the smooth morphism : $\pi:\widetilde{G}\rightarrow G$ , there is a smooth Lie
group morphism $F\in C_{\infty}\left(  \widetilde{G};G%
\mathcal{L}%
\left(  E;E\right)  \right)  $ such that F'(1)=f and (E,F) is a representation
of $\widetilde{G},$ $(E,F\circ\pi)$ is a representation of G.
\end{theorem}

\begin{proof}
G and $\widetilde{G}$\ have the same Lie algebra, f is a Lie algebra morphism
$T_{1}\widetilde{G}\rightarrow%
\mathcal{L}%
\left(  E;E\right)  $ which can be extended globally to $\widetilde{G}$
because it is simply connected. As a product of Lie group morphisms $F\circ
\pi$ is still a smooth morphism in $C_{\infty}\left(  G;G%
\mathcal{L}%
\left(  E;E\right)  \right)  $
\end{proof}

F can be computed from the exponential : $\forall u\in T_{1}G:\exp f^{\prime
}(1)u=F(\exp_{\widetilde{G}}u)=F\left(  \widetilde{g}\right)  .$

\paragraph{Weyl's unitary trick\newline}

(Knapp p.444)

It allows to go from different representations involving a semi simple Lie
group. The context is the following :

Let G be a semi simple, finite dimensional, real Lie group.\ This is the case
if G is a group of matrices closed under negative conjugate transpose.There is
a Cartan decomposition : $T_{1}G=l_{0}\oplus p_{0}$. If $l_{0}\cap ip_{0}=0$
the real Lie algebra $u_{0}=l_{0}\oplus ip_{0}$ is a compact real form of the
complexified $\left(  T_{1}G\right)  _{%
\mathbb{C}
}.$ So there is a compact, simply connected real Lie group U with Lie algebra
$u_{0}$.

Assume that there is a complex Lie group $G_{%
\mathbb{C}
}$ with Lie algebra $\left(  T_{1}G\right)  _{%
\mathbb{C}
}$ which is the complexified of G. Then $G_{%
\mathbb{C}
}$ is simply connected, semi simple and G,U are Lie subgroup of $G_{%
\mathbb{C}
}.$

We have the identities : $\left(  T_{1}G\right)  _{%
\mathbb{C}
}=\left(  T_{1}G\right)  \oplus i\left(  T_{1}G\right)  =u_{0}\oplus iu_{0}$

Then we have the trick :

1. If (E,f) is a complex representation of $G_{%
\mathbb{C}
}$ we get real representations (E,f) of G,U by restriction to the subgroups.

2. If (E,f) is a representation of G we have a the representation (E,f'(1)) of
$T_{1}G$ or $u_{0}.$

3. If (E,f) is a representation of U we have a the representation (E,f'(1)) of
$T_{1}G$ or $u_{0}$

4. If (E,f) is a representation of $T_{1}G$ or $u_{0}$ we have a the
representation (E,f'(1)) of $\left(  T_{1}G\right)  _{%
\mathbb{C}
}$

5. A representation (E,f) of $\left(  T_{1}G\right)  _{%
\mathbb{C}
}$ lifts to a representation of $G_{%
\mathbb{C}
}$

6. Moreover in all these steps the invariant subspaces and the equivalences of
representations are preserved.

\subsubsection{Universal envelopping algebra}

\paragraph{Principle\newline}

\begin{theorem}
(Knapp p.216) : The representations (E,f) of the Lie algebra A are in
bijective correspondance with the representations (E,F) of its universal
envelopping algebra U(A) by : $f=F\circ\imath$ where $\imath:A\rightarrow
U(A)$ is the canonical injection.
\end{theorem}

If V is an invariant closed vector subspace in the representation (E,f) of the
Lie algebra A, then V is invariant for (E,F). So if (E,F) is irreducible iff
(E,f) is irreducible.

\paragraph{Components expressions\newline}

If $\left(  e_{i}\right)  _{i\in I}$ is a basis of A then a basis of U(A) is
given by monomials :

$\left(  \text{\i}\left(  e_{i_{1}}\right)  \right)  ^{n_{1}}$ $\left(
\text{\i}\left(  e_{i_{2}}\right)  \right)  ^{n_{2}}....\left(  \text{\i
}\left(  e_{i_{p}}\right)  \right)  ^{n_{p}},i_{1}<i_{2}...<i_{p}\in
I,n_{1},...n_{p}\in%
\mathbb{N}
$

and F reads :

$F\left(  \left(  \left(  e_{i_{1}}\right)  \right)  ^{n_{1}}\left(  \left(
e_{i_{2}}\right)  \right)  ^{n_{2}}....\left(  \left(  e_{i_{p}}\right)
\right)  ^{n_{p}}\right)  =\left(  f\left(  \left(  e_{i_{1}}\right)  \right)
\right)  ^{n_{1}}\circ\left(  f\left(  \left(  e_{i_{2}}\right)  \right)
\right)  ^{n_{2}}....\circ\left(  f\left(  \left(  e_{i_{p}}\right)  \right)
\right)  ^{n_{p}}$

On the right hand side the powers are for the iterates of f.

$F(1_{K})=Id_{E}\Rightarrow\forall k\in K:F(k)\left(  U\right)  =kU$

If the representation (E,f) is given by matrices $\left[  f\left(  X\right)
\right]  $ then F reads as a product of matrices :

$F\left(  \left(  \text{\i}\left(  e_{i_{1}}\right)  \right)  ^{n_{1}}\left(
\text{\i}\left(  e_{i_{2}}\right)  \right)  ^{n_{2}}....\left(  \text{\i
}\left(  e_{i_{p}}\right)  \right)  ^{n_{p}}\right)  =\left[  f\left(
\text{\i}\left(  e_{i_{1}}\right)  \right)  \right]  ^{n_{1}}\left[  f\left(
\text{\i}\left(  e_{i_{2}}\right)  \right)  \right]  ^{n_{2}}....\left[
f\left(  \text{\i}\left(  e_{i_{p}}\right)  \right)  \right]  ^{n_{p}}$

\paragraph{Casimir elements\newline}

\begin{theorem}
(Knapp p.295,299 ) If A is a semi-simple, complex, finite dimensional Lie
algebra, then in any irreducible representation (E,f) of A the image $F\left(
\Omega\right)  $ of the Casimir element $\Omega$ acts by a non zero scalar :
$F\left(  \Omega\right)  =kId$
\end{theorem}

The Casimir elements are then extended to any order $r\in%
\mathbb{N}
$ by the formula

$\Omega_{r}=\sum_{\left(  i_{1}...i_{r}\right)  }Tr\left(  F\left(
e_{i_{1}...}e_{i_{r}}\right)  \right)  \imath\left(  E_{i_{1}}\right)
...\imath\left(  E_{i_{r}}\right)  \in U\left(  A\right)  $

$\Omega_{r}$ does not depend on the choice of a basis, belongs to the center
of U(A), commutes with any element of A, and its image $F\left(  \Omega
_{r}\right)  $ acts by scalar.

\paragraph{Infinitesimal character\newline}

If (E,F) is an irreducible representation of U(A) there is a function $\chi$,
called the infinitesimal character of the representation, such that :
$\chi:Z\left(  U\left(  A\right)  \right)  \rightarrow K::F\left(  U\right)
=\chi\left(  U\right)  Id_{E}$ where Z(U(A)) is the center of U(A).

U is in the center of U(A) iff $\forall X\in A:XU=UX$ or $\exp\left(
ad\left(  X\right)  \right)  \left(  U\right)  =U.$

\paragraph{Hilbertian representations\newline}

U(A) is a Banach C*-algebra with the involution : $U^{\ast}=U^{t}$ such that :
$\imath\left(  X\right)  ^{\ast}=-\imath\left(  X\right)  $

If (H,f) is a representation of the Lie algebra over a Hilbert space H, then
$\mathcal{L}$%
(H;H) is a C*-algebra.

(H,f) is a representation of the Banach C*-algebra U(A) if $\forall U\in
U\left(  A\right)  :F\left(  U^{\ast}\right)  =F\left(  U\right)  ^{\ast}$ and
this condition is met if : $f\left(  X\right)  ^{\ast}=-f\left(  X\right)  $ :
the representation of A must be anti-hermitian.

\subsubsection{Adjoint representations}

\paragraph{Lie algebras\newline}

\begin{theorem}
For any Lie algebra A, $\left(  A,ad\right)  $ is a representation of A on itself.
\end{theorem}

This representation is extended to representations $\left(  U_{n}\left(
A\right)  ,f_{n}\right)  $ of A on its universal envelopping algebra U(A):

$U_{n}\left(  A\right)  $ is the subspace of homogeneous elements of U(A) of
order n

$f_{n}:A\rightarrow L(U_{n}\left(  A\right)  ;U_{n}\left(  A\right)
)::f_{n}\left(  X\right)  u=Xu-uX$ is a Lie algebra morphism.

\begin{theorem}
(Knapp p.291) If A is a Banach algebra there is a representation (U(A),f) of
the component of identity Int(A) of G%
$\mathcal{L}$%
(A;A)
\end{theorem}

\begin{proof}
If A is a Banach algebra, then G%
$\mathcal{L}$%
(A;A) is a Lie group with Lie algebra
$\mathcal{L}$%
(A;A), and it is the same for its component of the identity Int(A). With any
automorphism $g\in Int(A)$ and the canonical map : $\imath:A\rightarrow U(A)$
the map : $\imath\circ g:A\rightarrow U(A)$ is such that : $i\circ g\left(
X\right)  \imath\circ g\left(  Y\right)  -i\circ g\left(  Y\right)
\imath\circ g\left(  X\right)  =\imath\circ g\left[  X,Y\right]  $ and
$\imath\circ g$ can be extended uniquely to an algebra morphism $f\left(
g\right)  $\ such that : $f\left(  g\right)  :U(A)\rightarrow U\left(
A\right)  :\imath\circ g=f\left(  g\right)  \circ\imath$ . Each $f(g)\in G%
\mathcal{L}%
\left(  U\left(  A\right)  ;U\left(  A\right)  \right)  $ is an algebra
automorphism of U(A) and each $U_{n}\left(  A\right)  $ is invariant.

The map : $f:Int(A)\rightarrow%
\mathcal{L}%
\left(  U\left(  A\right)  ;U\left(  A\right)  \right)  $ is smooth and we
have : $f\left(  g\right)  \circ f\left(  h\right)  =f\left(  g\circ h\right)
$ so (U(A),f) is a representation of Int(A).
\end{proof}

\paragraph{Lie groups\newline}

\begin{theorem}
For any Lie group G the \textbf{adjoint representation} is the representation
$\left(  T_{1}G,Ad\right)  $ of G on its Lie algebra
\end{theorem}

The map : $Ad:G\rightarrow G%
\mathcal{L}%
\left(  T_{1}G;T_{1}G\right)  $ is a smooth Lie group homorphism

This representation is not necessarily faithful. It is irreducible iff G has
no normal subgroup other than 1. The adjoint representation is faithful for
simple Lie groups but not for semi-simple Lie groups.

It can be extended to a representation on the universal envelopping algebra
$U(T_{1}G).$ There is a representation (U(A),f) of the component of identity
Int(A) of G%
$\mathcal{L}$%
(A;A)). $Ad_{g}\in Int(T_{1}G)$ so it gives a family of representations
$\left(  U_{n}\left(  T_{1}G\right)  ,Ad\right)  $ of G on the universal
envelopping algebra.

\subsubsection{Unitary and orthogonal representations}

\paragraph{Definition\newline}

Unitary or orthogonal representations are considered when there is some scalar
product on E. So we will assume that H is a complex Hilbert space (the
definitions and results are easily adjusted for the real case) with scalar
product $\left\langle {}\right\rangle ,$ antilinear in the first variable.

Each operator X in
$\mathcal{L}$%
(H;H) (or at least defined on a dense domain of H) has an adjoint X* in
$\mathcal{L}$%
(H;H) such that :

$\left\langle Xu,v\right\rangle =\left\langle u,X^{\ast}v\right\rangle $

The map $\ast:%
\mathcal{L}%
(H;H)\rightarrow%
\mathcal{L}%
(H;H)$ is an involution, antilinear, bijective, continuous, isometric and if X
is invertible, then X* is invertible and $\left(  X^{-1}\right)  ^{\ast
}=\left(  X^{\ast}\right)  ^{-1}$ .With this involution
$\mathcal{L}$%
(H;H) is a C*-algebra.

\begin{definition}
A \textbf{unitary representation} (H,f) of \textit{a group G}\ is a
representation on a Hilbert space H such that%

\begin{equation}
\forall g\in G:f\left(  g\right)  ^{\ast}f\left(  g\right)  =f\left(
g\right)  f\left(  g\right)  ^{\ast}=I\Leftrightarrow\forall g\in G,u,v\in
H:\left\langle f\left(  g\right)  u,f\left(  g\right)  v\right\rangle
=\left\langle u,v\right\rangle
\end{equation}

\end{definition}

If H is finite dimensional then f(g) is represented \textit{in a Hilbert
basis} by a unitary matrix. In the real case it is represented by an
orthogonal matrix.

If there is a dense subsace E of H such that :$\forall u,v\in E$ the map
$G\rightarrow K::\left\langle u,f\left(  g\right)  v\right\rangle $ is
continuous then f is continuous.

\paragraph{Sum of unitary representations of a group\newline}

\begin{theorem}
(Neeb p.24) The Hilbert sum of the unitary representations $\left(
H_{i},f_{i}\right)  _{i\in I}$ is a unitary representation $\left(
H,f\right)  $ where :

$H=\oplus_{i\in I}H_{i}$ the Hilbert sum of the spaces

$f:G\rightarrow%
\mathcal{L}%
\left(  H;H\right)  ::f\left(  \sum_{i\in I}u_{i}\right)  =\sum_{i\in I}%
f_{i}\left(  u_{i}\right)  $
\end{theorem}

This definition generalizes the sum for any set I for a Hilbert space.

\paragraph{Representation of the Lie algebra\newline}

\begin{theorem}
If (H,f) is a unitary representation of the Lie group G, then (H,f'(1)) is an
anti-hermitian representation of $T_{1}G$
\end{theorem}

\begin{proof}
(H,f'(1)) is a representation of $T_{1}G.$ The scalar product is a continuous
form so it is differentiable and :

$\forall X\in T_{1}G,u,v\in H:\left\langle f^{\prime}\left(  1\right)  \left(
X\right)  u,v\right\rangle +\left\langle u,f^{\prime}\left(  1\right)  \left(
X\right)  v\right\rangle =0\Leftrightarrow\left(  f^{\prime}\left(  1\right)
\left(  X\right)  \right)  ^{\ast}=-f^{\prime}\left(  1\right)  \left(
X\right)  $
\end{proof}

(H,f'(1)) is a representation of the C*-algebra U(A).

\paragraph{Dual representation\newline}

\begin{theorem}
If (H,f) is a unitary representation of the Lie group G, then there is a
unitary representation $(H^{\prime},\widetilde{f})$ of G
\end{theorem}

\begin{proof}
The dual H' of H is also Hilbert. There is a continuous anti-isomorphism
$\tau:H^{\prime}\rightarrow H$\ such that :

$\forall\lambda\in H^{\prime},\forall u\in H:\left\langle \tau\left(
\varphi\right)  ,u\right\rangle =\varphi\left(  u\right)  $

$\widetilde{f}$ is defined by : $\widetilde{f}\left(  g\right)  \varphi
=\tau^{-1}\left(  f\left(  g\right)  \tau\left(  \varphi\right)  \right)
\Leftrightarrow\widetilde{f}\left(  g\right)  =\tau^{-1}\circ f\left(
g\right)  \circ\tau$

Which is $%
\mathbb{C}
$ linear. If (H,f) is unitary then $\left(  H^{\prime},\widetilde{f}\right)  $
is unitary:

$\left\langle \widetilde{f}\left(  g\right)  \varphi,\widetilde{f}\left(
g\right)  \psi\right\rangle _{H^{\prime}}=\left\langle \tau\circ\widetilde
{f}\left(  g\right)  \varphi,\tau\circ\widetilde{f}\left(  g\right)
\psi\right\rangle _{H}=\left\langle f\left(  g\right)  \circ\tau
\varphi,f\left(  g\right)  \circ\tau\psi\right\rangle _{H}=\left\langle
\tau\varphi,\tau\psi\right\rangle _{H}=\left\langle \varphi,\psi\right\rangle
_{H^{\ast}}$
\end{proof}

The dual representation is also called the contragredient representation.

\bigskip

\subsection{Representation of Lie groups}

\label{Representation of Lie groups}

\subsubsection{Action of the group}

A representation (E,f) of G can be seen as a left (or right) action of G on
E$:$

$\rho:E\times G\rightarrow E::\rho\left(  u,g\right)  =f(g)u$

$\lambda:G\times E\rightarrow E::\lambda\left(  g,u\right)  =f(g)u$

\begin{theorem}
The action is smooth and proper
\end{theorem}

\begin{proof}
As $f:G\rightarrow%
\mathcal{L}%
\left(  E;E\right)  $ is assumed to be continuous, the map $\phi:%
\mathcal{L}%
\left(  E;E\right)  \times E\rightarrow E$ is bilinear, continuous with norm
1, so $\lambda\left(  g,u\right)  =\phi\left(  f\left(  g\right)  ,u\right)  $
is continuous.

The set GxE has a trivial manifold structure, and group structure.\ This is a
Lie group.\ The maps $\lambda$ is a continuous Lie group morphism, so it is
smooth and a diffeomorphism. The inverse if continuous, and $\lambda$ is proper.
\end{proof}

An invariant vector space is the union of orbits.

The representation is irreducible iff the action is transitive.

The representation is faithful iff the action is effective.

The map : $%
\mathbb{R}
\rightarrow G%
\mathcal{L}%
\left(  E;E\right)  ::f(\exp tX)$ is a diffeomorphism in a neighborhood of 0,
thus f'(1) is inversible.

We have the identities :

$\forall g\in G,X\in T_{1}G:$%

\begin{equation}
f^{\prime}(g)=f(g)\circ f^{\prime}(1)\circ L_{g^{-1}}^{\prime}g
\end{equation}

\begin{equation}
f^{\prime}(g)\left(  R_{g}^{\prime}1\right)  X=f^{\prime}(1)\left(  X\right)
\circ f(g)
\end{equation}

\begin{equation}
Ad_{f\left(  g\right)  }f^{\prime}\left(  1\right)  =f^{\prime}\left(
1\right)  Ad_{g}%
\end{equation}

The fundamental vector fields are :

$\zeta_{L}:T_{1}G\rightarrow%
\mathcal{L}%
\left(  E;E\right)  ::\zeta_{L}\left(  X\right)  =f^{\prime}(1)X$

\subsubsection{Functional representations}

A functional representation (E,f) is a representation where E is a space of
functions or maps and the action of the group is on the argument of the
map.\ Functional representations are the paradigm of infinite dimensional
representations of a group. They exist for any group, and there are "standard"
functional representations which have nice properties.

\paragraph{Right and left representations\newline}

\begin{definition}
The \textbf{left representation} of a topological group G on a Banach vector
space of maps $H\subset C\left(  E;F\right)  $ is defined, with a continuous
left action $\lambda$ of G on the topological space E by :%

\begin{equation}
\Lambda:G\rightarrow%
\mathcal{L}%
\left(  H;H\right)  ::\Lambda\left(  g\right)  \varphi\left(  x\right)
=\varphi\left(  \lambda\left(  g^{-1},x\right)  \right)
\end{equation}

\end{definition}

G acts on the variable inside $\varphi$ and $\Lambda\left(  g\right)
\varphi=\lambda_{g^{-1}}^{\ast}\varphi$ with $\lambda_{g^{-1}}=\lambda\left(
g^{-1},.\right)  $

\begin{proof}
$\Lambda$ is a morphism:

For g fixed in G consider the map : $H\rightarrow H::\varphi\left(  x\right)
\rightarrow\varphi\left(  \lambda\left(  g^{-1},x\right)  \right)  $

$\Lambda\left(  gh\right)  \varphi=\lambda_{\left(  gh\right)  ^{-1}}^{\ast
}\varphi=\left(  \lambda_{h^{-1}}\circ\lambda_{g^{-1}}\right)  ^{\ast}%
\varphi=\left(  \lambda_{g^{-1}}^{\ast}\circ\lambda_{h^{-1}}^{\ast}\right)
\varphi=\left(  \Lambda\left(  g\right)  \circ\Lambda\left(  h\right)
\right)  \varphi$

$\Lambda\left(  1\right)  \varphi=\varphi$

$\Lambda\left(  g^{-1}\right)  \varphi=\lambda_{g}^{\ast}\varphi=\left(
\lambda_{g^{-1}}^{\ast}\right)  ^{-1}\varphi$
\end{proof}

We have similarly the \textbf{right representation} with a right action :

$H\rightarrow H::\varphi\left(  x\right)  \rightarrow\varphi\left(
\rho\left(  x,g\right)  \right)  $%

\begin{equation}
P:G\rightarrow%
\mathcal{L}%
\left(  H;H\right)  ::P\left(  g\right)  \varphi\left(  x\right)
=\varphi\left(  \rho\left(  x,g\right)  \right)
\end{equation}

$P\left(  g\right)  \varphi=\rho_{g}^{\ast}\varphi$

Remark : some authors call right the left representation and vice versa.

\begin{theorem}
If there is a finite Haar Radon measure $\mu$\ on the topological group G any
left representation on a Hilbert space H is unitary
\end{theorem}

\begin{proof}
as H is a Hilbert space there is a scalar product denoted $\left\langle
\varphi,\psi\right\rangle $

$\forall g\in G,\varphi\in H:\Lambda\left(  g\right)  \varphi\in H$ so
$\left\langle \Lambda\left(  g\right)  \varphi,\Lambda\left(  g\right)
\psi\right\rangle $ is well defined

$\left(  \varphi,\psi\right)  =\int_{G}\left\langle \Lambda\left(  g\right)
\varphi,\Lambda\left(  g\right)  \psi\right\rangle \mu$ is well defined and
$<\infty.$ This is a scalar product (it has all the properties of
$\left\langle {}\right\rangle )$ over H

It is invariant by the action of G, thus with this scalar product the
representation is unitary.
\end{proof}

\begin{theorem}
A left representation $\left(  H,\Lambda\right)  $ of a Lie group G on a
Banach vector space of differentiables maps $H\subset C_{1}\left(  M;F\right)
$, with a differentiable left action $\lambda$ of G on the manifold M, induces
a representation $\left(  H,\Lambda^{\prime}\left(  1\right)  \right)  $\ of
the Lie algebra $T_{1}G$ where $T_{1}G$ acts by differential operators.
\end{theorem}

\begin{proof}
$\left(  H,\Lambda^{\prime}\left(  1\right)  \right)  $ is a representation of
$T_{1}G$

By the differentiation of : $\Lambda\left(  g\right)  \varphi\left(  x\right)
=\varphi\left(  \lambda\left(  g^{-1},x\right)  \right)  $

$\Lambda^{\prime}\left(  g\right)  \varphi\left(  x\right)  |_{g=1}%
=\varphi^{\prime}\left(  \lambda\left(  g^{-1},x\right)  \right)
|_{g=1}\lambda_{g}^{\prime}\left(  g^{-1},x\right)  |_{g=1}\left(  -R_{g^{-1}%
}^{\prime}(1)\circ L_{g^{-1}}^{\prime}(g)\right)  |_{g=1}$

$\Lambda^{\prime}\left(  1\right)  \varphi\left(  x\right)  =-\varphi^{\prime
}\left(  x\right)  \lambda_{g}^{\prime}\left(  1,x\right)  $

$X\in T_{1}G:\Lambda^{\prime}\left(  1\right)  \varphi\left(  x\right)
X=-\varphi^{\prime}\left(  x\right)  \lambda_{g}^{\prime}\left(  1,x\right)
X$

$\Lambda^{\prime}\left(  1\right)  \varphi\left(  x\right)  X$ is a local
differential operator (x does not change)
\end{proof}

Similarly : $P^{\prime}\left(  1\right)  \varphi\left(  x\right)
=\varphi^{\prime}\left(  x\right)  \rho_{g}^{\prime}\left(  x,1\right)  $

It is usual to write these operators as :

$\Lambda^{\prime}\left(  1\right)  \varphi\left(  x\right)  X=\frac{d}%
{dt}\varphi\left(  \lambda\left(  \left(  \exp\left(  -tX\right)  \right)
,x\right)  \right)  |_{t=0}$

$P^{\prime}\left(  1\right)  \varphi\left(  x\right)  X=\frac{d}{dt}%
\varphi\left(  \rho\left(  x,\exp\left(  tX\right)  \right)  \right)  |_{t=0}$

These representations can be extended to representations of the universal
envelopping algebra $U\left(  T_{1}G\right)  .$ We have differential operators
on H of any order. These operators have an algebra structure, isomorphic to
$U\left(  T_{1}G\right)  .$

\paragraph{Polynomial representations\newline}

If G is a set of matrices in K(n) and $\lambda$ the action of G on $E=K^{n}%
$\ associated to the standard representation of G, then for any Banach space H
of functions of n variables on K we have the left representation $\left(
H,\Lambda\right)  :$

$\Lambda:G\rightarrow%
\mathcal{L}%
\left(  H;H\right)  ::\Lambda\left(  g\right)  \varphi\left(  x_{1}%
,x_{2},..x_{n}\right)  =\varphi\left(  y_{1},..y_{n}\right)  $ with $\left[
Y\right]  =\left[  g\right]  ^{-1}\left[  X\right]  $

The set $K_{p}\left[  x_{1},...x_{n}\right]  $\ of polynomials of degree p
with n variables over a field K has the structure of a finite dimensional
vector space, which is a Hilbert vector space with a norm on $K^{p+1}.$ Thus
with $H=K_{p}\left[  x_{1},...x_{n}\right]  $ we have a finite dimensional
left representation of G.

The tensorial product of two polynomial representations :

$\left(  K_{p}\left[  x_{1},...x_{p}\right]  ,\Lambda_{p}\right)  ,\left(
K_{q}\left[  y_{1},...y_{q}\right]  ,\Lambda_{q}\right)  $

is given by :

- the tensorial product of the vector spaces, which is : $K_{p+q}\left[
x_{1},...x_{p},y_{1},...y_{q}\right]  $ represented in the canonical basis as
the product of the polynomials

- the morphism : $\left(  \Lambda_{p}\otimes\Lambda_{q}\right)  \left(
g\right)  \left(  \varphi_{p}\left(  X\right)  \otimes\varphi_{q}\left(
Y\right)  \right)  =\varphi_{p}\left(  \left[  g\right]  ^{-1}\left[
X\right]  \right)  \varphi_{q}\left(  \left[  g\right]  ^{-1}\left[  Y\right]
\right)  $

\paragraph{Representations on $L^{2}\left(  E,\mu,%
\mathbb{C}
\right)  $\newline}

See Functional analysis for the properties of these spaces.

\begin{theorem}
(Neeb p.45) If G is a topological group, E a topological locally compact
space, $\lambda:G\times E\rightarrow E$ a continuous left action of G on E,
$\mu$ a G invariant Radon measure on E, then the left representation $\left(
L^{2}\left(  E,\mu,%
\mathbb{C}
\right)  ,f\right)  $ with $\left(  g\right)  \varphi\left(  x\right)
=\varphi\left(  \lambda\left(  g^{-1},x\right)  \right)  $ is an unitary
representation of G.
\end{theorem}

\paragraph{Representations given by kernels\newline}

(Neeb p.97)

Let $\left(  H,\Lambda\right)  $ be a left representation of the topological
group G, with H a Hilbert space of functions $H\subset C\left(  E;%
\mathbb{C}
\right)  $ on a topological space E, valued in a field K and a left action
$\lambda:G\times E\rightarrow E$

$H$ can be defined uniquely by a definite positive kernel $N:E\times
E\rightarrow K.$ (see Hilbert spaces).

So let J be a map (which is a cocycle) :$J:G\times E\rightarrow K^{E}$ such
that :

$J\left(  gh,x\right)  =J\left(  g,x\right)  J\left(  h,\lambda\left(
g^{-1},x\right)  \right)  $

Then $\left(  H,f\right)  $ with the morphism : $f\left(  g\right)  \left(
\varphi\right)  \left(  x\right)  =J\left(  g,x\right)  \varphi\left(
\lambda\left(  g^{-1},x\right)  \right)  $ is a unitary representation of G
iff :

$N\left(  \lambda\left(  g,x\right)  ,\lambda\left(  g,y\right)  \right)
=J\left(  g,\lambda\left(  g,x\right)  \right)  N\left(  x,y\right)
\overline{J\left(  g,\lambda\left(  g,y\right)  \right)  }$

If J, N, $\lambda$ are continuous, then the representation is continuous.

Any G invariant closed subspace $A\sqsubseteq H_{N}$ has for reproducing
kernel P which satisfies :

$P\left(  \lambda\left(  g,x\right)  ,\lambda\left(  g,y\right)  \right)
=J\left(  g,\lambda\left(  g,x\right)  \right)  P\left(  x,y\right)
\overline{J\left(  g,\lambda\left(  g,y\right)  \right)  }$

Remarks :

i) if N is G invariant then take J=1 and we get back the left representation.

ii) if N(x,x)$\neq0$ by normalization $Q\left(  x,y\right)  =\frac{N\left(
x,y\right)  }{\sqrt{\left\vert N\left(  x,x\right)  \right\vert \left\vert
N\left(  y,y\right)  \right\vert }},J_{Q}\left(  g,x\right)  =\frac{J\left(
g,x\right)  }{\left\vert J\left(  g,x\right)  \right\vert },$ we have an
equivalent representation where all the maps of $H_{Q}$ are valued in the
circle T.

3. Example : the Heisenberg group Heis(H) has the continuous unitary
representation on the Fock space given by :

$f\left(  t,v\right)  \varphi\left(  u\right)  =\exp\left(  it+\left\langle
u,v\right\rangle -\frac{1}{2}\left\langle v,v\right\rangle \right)
\varphi\left(  u-v\right)  $

\paragraph{Regular representations\newline}

The \textbf{regular representations} are functional representations on spaces
of maps \textit{defined on G itself} $H\subset C\left(  G;E\right)  .$ The
rigth and left actions are then the translations on G. The right and left
actions commute, and we have a representation of G$\times$G on H by :

$\Phi:\left(  G\times G\right)  \times H\rightarrow H::\Phi\left(
g,g^{\prime}\right)  \left(  \varphi\right)  \left(  x\right)  =\Lambda\left(
g\right)  \circ P\left(  g^{\prime}\right)  \left(  \varphi\right)  \left(
x\right)  =\varphi\left(  g^{-1}xg^{\prime}\right)  $

\begin{theorem}
(Neeb p.49) For any locally compact, topological group G$,$ left invariant
Haar Radon measure $\mu_{L}$ on G

the \textbf{left regular representation} $\left(  L^{2}\left(  G,\mu_{L},%
\mathbb{C}
\right)  ,\Lambda\right)  $ with : $\Lambda\left(  g\right)  \varphi\left(
x\right)  =\varphi\left(  g^{-1}x\right)  $

the \textbf{right regular representation} $\left(  L^{2}\left(  G,\mu_{R},%
\mathbb{C}
\right)  ,P\right)  $ with : $P\left(  g\right)  \varphi\left(  x\right)
=\sqrt{\Delta\left(  g\right)  }\varphi\left(  xg\right)  $

are both unitary.
\end{theorem}

This left regular representation is injective.

So any locally compact, topological group has a least one faithful unitary
representation (usually infinite dimensional).

\paragraph{Averaging\newline}

$L^{1}\left(  G,S,\mu,%
\mathbb{C}
\right)  $ is a Banach *-algebra with convolution as internal product (see
Integral on Lie groups) :

$\varphi\ast\psi\left(  g\right)  =\int_{G}\varphi\left(  x\right)
\psi\left(  x^{-1}g\right)  \mu_{L}\left(  x\right)  =\int_{G}\varphi\left(
gx\right)  \psi\left(  x^{-1}\right)  \mu\left(  x\right)  $

\begin{theorem}
(Knapp p.557, Neeb p.134,143) \ If (H,f) is a unitary representation of a
locally compact topological group G endowed with a finite Radon Haar measure
$\mu,$ and H a Hilbert space, then the map :

$F:L^{1}\left(  G,S,\mu,%
\mathbb{C}
\right)  \rightarrow%
\mathcal{L}%
\left(  H;H\right)  ::F\left(  \varphi\right)  =\int_{G}\varphi\left(
g\right)  f\left(  g\right)  \mu\left(  g\right)  $

gives a representation (H,F) of the Banach *-algebra $L^{1}\left(  G,S,\mu,%
\mathbb{C}
\right)  $ with convolution as internal product. The representations
(H,f),(H,F) have the same invariant subspaces, (H,F) is irreducible\ iff (H,f)
is irreducible.

Conversely for each non degenerate Banach *-algebra representation (H,F) of
$L^{1}\left(  G,S,\mu,%
\mathbb{C}
\right)  $ there is a unique unitary continuous representation (H,f) of G such
that : $f\left(  g\right)  F\left(  \varphi\right)  =F\left(  \Lambda\left(
g\right)  \varphi\right)  $ where $\Lambda$\ is the left regular action :
$\Lambda\left(  g\right)  \varphi\left(  x\right)  =\varphi\left(
g^{-1}x\right)  $.
\end{theorem}

F is defined as follows : for any $\varphi\in L^{1}\left(  G,S,\mu,%
\mathbb{C}
\right)  $ fixed, the map : $B:H\times H\rightarrow%
\mathbb{C}
::B\left(  u,v\right)  =\int_{G}\left\langle u,\varphi\left(  g\right)
f\left(  g\right)  v\right\rangle \mu$ is sesquilinear and bounded because
$\left\vert B\left(  u,v\right)  \right\vert \leq\left\Vert f\left(  g\right)
\right\Vert \left\Vert u\right\Vert \left\Vert v\right\Vert \int_{G}\left\vert
\varphi\left(  g\right)  \right\vert \mu$ and there is a unique map : $A\in%
\mathcal{L}%
\left(  H;H\right)  :\forall u,v\in H:B\left(  u,v\right)  =\left\langle
u,Av\right\rangle .$ We put $A=F\left(  \varphi\right)  .$ It is linear
continuous and $\left\Vert F\left(  \varphi\right)  \right\Vert \leq\left\Vert
\varphi\right\Vert $

$F\left(  \varphi\right)  \in%
\mathcal{L}%
\left(  H;H\right)  $ and can be seen as the integral of f(g) "averaged" by
$\varphi.$

F has the following properties :

$F\left(  \varphi\right)  ^{\ast}=F\left(  \varphi^{\ast}\right)  $ with
$\varphi^{\ast}\left(  g\right)  =\overline{\varphi\left(  g^{-1}\right)  }$

$F\left(  \varphi\ast\psi\right)  =F\left(  \varphi\right)  \circ F\left(
\psi\right)  $

$\left\Vert F\left(  \varphi\right)  \right\Vert _{%
\mathcal{L}%
\left(  H;H\right)  }\leq\left\Vert \varphi\right\Vert _{L^{1}}$

$F\left(  g\right)  F\left(  \varphi\right)  \left(  x\right)  =F\left(
f\left(  gx\right)  \right)  $

$F\left(  \varphi\right)  F\left(  g\right)  \left(  x\right)  =\Delta
_{G}\left(  g\right)  F\left(  f\left(  xg\right)  \right)  $

For the commutants : $\left(  F\left(  L^{1}\left(  G,S,\mu,%
\mathbb{C}
\right)  \right)  \right)  ^{\prime}=\left(  f\left(  G\right)  \right)
^{\prime} $

\paragraph{Induced representations\newline}

Induced representations are representations of a subgroup S of the Lie group G
which are extended to G.

Let :

S be a closed subgroup of a Lie group G,

(E,f) a representation of S: $f:S\rightarrow%
\mathcal{L}%
\left(  E;E\right)  $

H a space of continuous maps $H::G\rightarrow E$

The left and rigth actions of G on H :

$\Lambda:G\times H\rightarrow H::\Lambda\left(  g,\varphi\right)  \left(
x\right)  =\varphi\left(  g^{-1}x\right)  $

$P:H\times G\rightarrow H::P\left(  \varphi,g\right)  \left(  x\right)
=\varphi\left(  xg\right)  $

are well defined and commute.

$\left(  H,f\circ P\right)  $ is a representation of S : $f\circ P\left(
s\right)  \left(  \varphi\right)  \left(  x\right)  =f\left(  s\right)
\varphi\left(  xs\right)  .$ The subspace $H_{S}$ of H of maps which are
invariant in this representation are such that : $\forall s\in S,x\in
G:f\left(  s\right)  \varphi\left(  xs\right)  =\varphi\left(  x\right)
\Leftrightarrow\varphi\left(  xs\right)  =\left(  f\left(  s\right)  \right)
^{-1}\varphi\left(  x\right)  .$

$H_{S}$ is invariant under the left action $\Lambda:\forall\varphi\in
H_{0},\Lambda\left(  g,\varphi\right)  \in H_{0}$

So $\left(  H_{S},\Lambda\right)  $ is a representation of G, called the
\textbf{representation induced} by (E,f) and usually denoted Ind$_{S}^{G}f$

G is a principal fiber bundle $G\left(  G/S,S,\pi_{L}\right)  $ and $G\left[
E,f\right]  $ an associated vector bundle, with the equivalence relation :
$\left(  g,u\right)  \sim\left(  gs^{-1},f\left(  s\right)  u\right)  .$ A
section $X\in\mathfrak{X}\left(  G\left[  E,f\right]  \right)  $ is a map :
$X:G/H\rightarrow G\left[  E,f\right]  ::\left(  y,u\right)  \sim\left(
ys,f\left(  s^{-1}\right)  u\right)  $ which assigns to each class of
equivalence $y=\left\{  ys,s\in S\right\}  $ equivalent vectors.\ So that
$H_{S}$ can be assimilated to the space of continuous sections of $G\left[
E,f\right]  $ belonging to H.

\begin{theorem}
Frobenius reciprocity (Duistermatt p.242) Let S a closed subgroup of a compact
Lie group G, (E,f) a finite dimensional representation of S, (H,F) a finite
dimensional representation of G, then :

i) the linear spaces of interwiners between (H,F) and Ind$_{S}^{G}f$ on one
hand, and between (H,F)$_{|S}$ and (E,f) on the other hand, are isomorphic

ii) if (E,f), (H,F) are irreducible, then the number of occurences of H in
Ind$_{S}^{G}f$ is equal to the number of occurences of (E,f) in (H,F)$_{|S}$
\end{theorem}

\begin{theorem}
(Knapp p.564) A unitary representation (H,f) of a closed Lie subgroup S of a
Lie group G endowed with a left invariant Haar measure $\varpi_{L}$ can be
extended to a unitary representation $(W,\Lambda_{W})$ of G where W is a
subset of $L^{2}\left(  G;\varpi_{L};H\right)  $ and $\Lambda_{W}$ the left
regular representation on W.
\end{theorem}

The set $L^{2}\left(  G;\varpi_{L};H\right)  $ is a Hilbert space.

W is the set of continuous maps $\varphi$\ in $L^{2}\left(  G;\varpi
_{L};H\right)  $ such that :

$W=\left\{  \varphi\in L^{2}\left(  G;\varpi_{L};H\right)  \cap C_{0}\left(
G;H\right)  :\forall s\in S,g\in G:\varphi\left(  gs\right)  =f\left(
s^{-1}\right)  \varphi\left(  g\right)  \right\}  $

$\Lambda_{W}:G\rightarrow L\left(  W;W\right)  ::\Lambda_{W}\left(  g\right)
\left(  \varphi\right)  \left(  g^{\prime}\right)  =\varphi\left(
g^{-1}g^{\prime}\right)  $

\subsubsection{Irreducible representations}

\paragraph{General theorems\newline}

\begin{theorem}
Schur's lemna : An interwiner $\phi\in%
\mathcal{L}%
\left(  E_{1};E_{2}\right)  $ of two irreducible representations$\ (E_{1}%
,f_{1}),\left(  E_{2},f_{2}\right)  $ of a group G is either 0 or an isomorphism.
\end{theorem}

\begin{proof}
From the theorems below:

$\ker\phi$ is either 0, and then $\phi$ is injective, or $E_{1}$ and then
$\phi=0$

$\operatorname{Im}\phi$ is either 0, and then $\phi=0,$\ or $E_{2}$ and then
$\phi$ is surjective

Thus $\phi$\ is either 0 or bijective, and then the representations are
isomorphic :

$\forall g\in G:f_{1}\left(  g\right)  =\phi^{-1}\circ f_{2}\left(  g\right)
\circ\phi$
\end{proof}

Therefore for any two irreducible representations either they are not
equivalent, or they are isomorphic, and we can define \textbf{classes of
irreducible representations}. If a representation(E,f) is reducible, we can
define the number of occurences of a given class j of irreducible
representation, which is called the \textbf{multiplicity} $d_{j}$ of the class
of representations j in (E,f).

\begin{theorem}
If $(E,f_{1}),\left(  E,f_{2}\right)  $ are two irreducible equivalent
representations of a Lie group G on the same complex space then $\exists
\lambda\in%
\mathbb{C}
$ and an interwiner $\phi=\lambda Id$
\end{theorem}

\begin{proof}
There is a bijective interwiner $\phi$\ because the representations are
equivalent. The spectrum of $\phi\in G%
\mathcal{L}%
\left(  E;E\right)  $ is a compact subset of $%
\mathbb{C}
$\ with at least a non zero element $\lambda,$ thus $\phi-\lambda Id$ is not
injective in
$\mathcal{L}$%
(E;E) but continuous, it is an interwiner of $(E,f_{1}),\left(  E,f_{2}%
\right)  ,$ thus it must be zero.
\end{proof}

\begin{theorem}
(Kolar p.131) If F is an invariant vector subspace in the finite dimensional
representation (E,f) of a group G, then any tensorial product of (E,f) is
completely reducible.
\end{theorem}

\begin{theorem}
If (E,f) is a representation of the group G and F an invariant subspace, then :

$\forall u\in F,\exists g\in G,v\in F:u=f(g)v$

$\left(  E/F,\widehat{f}\right)  $ is a representation of G, with :
$\widehat{f}:G\rightarrow G%
\mathcal{L}%
\left(  E/F;E/F\right)  ::\widehat{f}\left(  g\right)  \left(  \left[
u\right]  \right)  =\left[  f\left(  g\right)  u\right]  $
\end{theorem}

\begin{proof}
$\forall v\in F,\forall g\in G:f(g)v\in F\Rightarrow v=f(g)\left(
f(g^{-1})v\right)  =f(g)w$ with $w=\left(  f(g^{-1})v\right)  $

u$\sim v\Leftrightarrow u-v=w\in F\Rightarrow f\left(  g\right)  u-f\left(
g\right)  v=f\left(  g\right)  w\in F$

and if F is a closed vector subspace of E, and E a Banach, then E/F is still a
Banach space.
\end{proof}

\begin{theorem}
If $\phi\in%
\mathcal{L}%
\left(  E_{1};E_{2}\right)  $ is an interwiner of the representations$\ (E_{1}%
,f_{1})$,$\left(  E_{2},f_{2}\right)  $ of a group G then : $\ker
\phi,\operatorname{Im}\phi$ are invariant subspaces of $E_{1},E_{2}$ respectively
\end{theorem}

\begin{proof}
$\forall g\in G:\phi\circ f_{1}\left(  g\right)  =f_{2}\left(  g\right)
\circ\phi$

$u\in\ker\phi\Rightarrow\phi\circ f_{1}\left(  g\right)  u=f_{2}\left(
g\right)  \circ\phi u=0\Rightarrow f_{1}\left(  g\right)  u\in\ker
\phi\Leftrightarrow\ker\phi$ is invariant for $f_{1}$

$v\in\operatorname{Im}\phi\Rightarrow\exists u\in E_{1}:v=\phi u\Rightarrow
\phi\left(  f_{1}\left(  g\right)  u\right)  =f_{2}\left(  g\right)  \circ\phi
u=f_{2}\left(  g\right)  v\Rightarrow f_{2}\left(  g\right)  v\in
\operatorname{Im}\phi\Leftrightarrow\operatorname{Im}\phi$ is invariant for
$f_{2}$
\end{proof}

\paragraph{Theorems for unitary representations\newline}

\begin{theorem}
(Neeb p.77) If (H,f) is a unitary representation of the topological group G,
$H_{d}$ the closed vector subspace generated by all the irreducible
subrepresentations in (H,f), then :

i) $H_{d}$ is invariant by G, and $(H_{d},f)$ is a unitary representation of G
which is the direct sum of irreducible representations

ii) the orthogonal complement $H_{d}^{\perp}$ does not contain any irreducible representation.
\end{theorem}

So a unitary representation of a topological group can be written as the
direct sum (possibly infinite) of subrepresentations :

$H=\left(  \oplus_{j}d_{j}H_{j}\right)  \oplus H_{c}$

each $H_{j}$ is a class of irreducible representation, and $d_{j}$ their
multiplicity in the representation (H,f)

$H_{c}$ does not contain any irreducible representation.

The components are mutually orthogonal : $H_{j}\perp H_{k}$ for $j\neq
k,H_{j}\perp H_{c}$

The representation (H,f) is completely reducible iff $H_{c}=0$

Are completely reducible in this manner :

- the continuous unitary representations of a topological finite or compact group;

- the continuous unitary finite dimensional representations of a topological group

Moreover we have the important result for compact groups :

\begin{theorem}
(Knapp p.559) Any irreducible unitary representation of a compact group is
finite dimensional. Any compact Lie group has a faithful finite dimensional
representation, and thus is isomorphic to a closed group of matrices.
\end{theorem}

Thus for a compact group any continuous unitary representation is completely
reducible in the direct sum of orthogonal finite dimensional irreducible
unitary representations.

The tensorial product of irreducible representations is not necessarily
irreducible.\ But we have the following result :

\begin{theorem}
(Neeb p.91) If $\left(  H_{1},f_{1}\right)  ,\left(  H_{2},f_{2}\right)  $ are
two irreducible \textit{unitary} \textit{infinite dimensional} representations
of G, then $\left(  H_{1}\otimes H_{2},f_{1}\otimes f_{2}\right)  $ is an
irreducible representation of G.
\end{theorem}

This is untrue if the representations are not unitary or infinite dimensional.

\begin{theorem}
(Neeb p.76) A unitary representation (H,f) of the topological group G on a
Hilbert space over the field K is irreducible if the commutant S' of the
subset $S=\left\{  f\left(  g\right)  ,g\in G\right\}  $ of
$\mathcal{L}$%
(H;H) is trivial : S'=K$\times$Id
\end{theorem}

\begin{theorem}
If E is an invariant vector subspace in a unitary representation (H,f) of the
topological group G, then its orthogonal complement $E^{\intercal}$is still a
closed invariant vector subspace.
\end{theorem}

\begin{proof}
The orthogonal complement $E^{\intercal}$ is a closed vector subspace, and
also a Hilbert space and $H=E\oplus E^{\intercal}$

Let be $u\in E,v\in E^{\intercal},$ then $\left\langle u,v\right\rangle
=0,\forall g\in G:f\left(  g\right)  u\in E$

$\left\langle f\left(  g\right)  u,v\right\rangle =0=\left\langle u,f\left(
g\right)  ^{\ast}v\right\rangle =\left\langle u,f\left(  g\right)
^{-1}v\right\rangle =\left\langle u,f\left(  g^{-1}\right)  v\right\rangle
\Rightarrow\forall g\in G:f\left(  g\right)  u\in E^{\bot}$
\end{proof}

\begin{definition}
A unitary representation (H,f) of the topological group G is \textbf{cyclic}
if there is a vector u in H such that $F(u)=\left\{  f(g)u,g\in G\right\}  $
is dense in H.
\end{definition}

\begin{theorem}
(Neeb p.117) If (H,f,u),(H',f',u') are two continuous unitary cyclic
representations of the topological group G there is a unitary interwining
operator F with u'=F(u) iff $\forall g:$ $\left\langle u,f\left(  g\right)
u\right\rangle _{H}=\left\langle u^{\prime},f^{\prime}\left(  g\right)
u^{\prime}\right\rangle _{H^{\prime}}$
\end{theorem}

\begin{theorem}
If F is a closed invariant vector subspace in the unitary representation (H,f)
of the topological group G, then each vector of F is cyclic in F, meaning that
$\forall u\neq0\in F:F(u)=\left\{  f(g)u,g\in G\right\}  $ is dense in F
\end{theorem}

\begin{proof}
Let S=$\left\{  f(g),g\in G\right\}  \sqsubset G%
\mathcal{L}%
\left(  H;H\right)  .$ We have S=S* because f(g) is unitary, so $f(g)^{\ast
}=f(g^{-1})\in S.$

F is a closed vector subspace in H, thus a Hilbert space, and is invariant by
S.\ Thus (see Hilbert spaces) :

$\forall u\neq0\in F:F(u)=\left\{  f(g)u,g\in G\right\}  $ is dense in F and
the orthogonal complement F'(u)\ of F(u) in F is 0.
\end{proof}

\begin{theorem}
(Neeb p.77) If $\left(  H_{1},f\right)  ,(H_{2},f)$ are two inequivalent
irreducible subrepresentations of the unitary representation (H,f) of the
topological group G, then $H_{1}\perp H_{2}.$
\end{theorem}

\begin{theorem}
(Neeb p.24) Any unitary representation (H,f) of a topological group G is
equivalent to the Hilbert sum of mutually orthogonal cyclic
subrepresentations: $\left(  H,f\right)  =\oplus_{i\in I}\left(
H_{i},f|_{H_{i}}\right)  $
\end{theorem}

\subsubsection{Character}

\begin{definition}
The \textbf{character} of a \textit{finite dimensional} representation (E,f)
of the topological group G is the function :%

\begin{equation}
\chi_{f}:G\rightarrow K::\chi_{f}(g)=Tr(f(g))
\end{equation}

\end{definition}

The trace of any endomorphism always exists if E is finite dimensional. If E
is an infinite dimensional Hilbert space H there is another definition, but a
unitary operator is never trace class, so the definition does not hold any more.

The character reads in any orthonormal basis : $\chi_{f}(g)=\sum_{i\in
I}\left\langle e_{i},f\left(  g\right)  e_{i}\right\rangle $

\paragraph{Properties for a unitary representation\newline}

\begin{theorem}
(Knapp p.242) The character $\chi$ of the unitary finite dimensional
representation (H,f) of the group G has the following properties :

$\chi_{f}\left(  1\right)  =\dim E$

$\forall g,h\in G:\chi_{f}(ghg^{-1})=\chi_{f}(h)$

$\chi_{f^{\ast}}\left(  g\right)  =\chi_{f}\left(  g^{-1}\right)  $
\end{theorem}

\begin{theorem}
(Knapp p.243) If $(H_{1},f_{1}),\left(  H_{2},f_{2}\right)  $ are unitary
finite dimensional representations of the group G :

For the sum $\left(  E_{1}\oplus E_{2},f=f_{1}\oplus f_{2}\right)  $ of the
representations : $\chi_{f}=\chi_{f_{1}}+\chi_{f_{2}}$

For the tensorial product $\left(  E_{1}\otimes E_{2},f=f_{1}\otimes
f_{2}\right)  $ of the representations : $\chi_{f_{1}\otimes f_{2}}%
=\chi_{f_{1}}\chi_{f_{2}}$

If the two representations $(E_{1},f_{1}),\left(  E_{2},f_{2}\right)  $ are
equivalent then : $\chi_{f_{1}}=\chi_{f_{2}}$
\end{theorem}

So if (H,f) is the direct sum of $(H_{j},f_{j})_{j=1}^{p}:\chi_{f}=\sum
_{q=1}^{r}d_{q}\chi_{f_{q}}$ where $\chi_{f_{q}}$ is for a class of equivalent
representations, and $d_{q}$ is the number of representations in the family
$(H_{j},f_{j})_{j=1}^{p}$ which are equivalent to $(H_{q},f_{q}).$

If G is a compact connected Lie group, then there is a maximal torus T and any
element of G is conjugate to an element of T : $\forall g\in G,\exists x\in
G,t\in T:g=xtx^{-1}$ thus : $\chi_{f}\left(  g\right)  =\chi_{f}\left(
t\right)  .$ So all the characters of the representation can be obtained by
taking the characters of a maximal torus.

\paragraph{Compact Lie groups\newline}

\begin{theorem}
Schur's orthogonality relations (Knapp p.239) : Let G be a compact Lie group,
endowed with a Radon Haar measure $\mu.$

i) If the unitary finite dimensional representation (H,f) is irreducible $:$

$\forall u_{1},v_{1},u_{2},v_{2}\in H:\int_{G}\left\langle u_{1},f\left(
g\right)  v_{1}\right\rangle \overline{\left\langle u_{2},f\left(  g\right)
v_{2}\right\rangle }\mu=\frac{1}{\dim H}\left\langle u_{1},v_{1}\right\rangle
\left\langle u_{2},v_{2}\right\rangle $

$\chi_{f}\in%
\mathcal{L}%
^{2}\left(  G,\mu,%
\mathbb{C}
\right)  $ and $\left\Vert \chi_{f}\right\Vert =1$

ii) If $(H_{1},f_{1}),\left(  H_{2},f_{2}\right)  $ are two
\textit{inequivalent} irreducible unitary finite dimensional representations
of G :

$\forall u_{1},v_{1}\in H_{1},u_{2},v_{2}\in H_{2}:\int_{G}\left\langle
u_{1},f\left(  g\right)  v_{1}\right\rangle \overline{\left\langle
u_{2},f\left(  g\right)  v_{2}\right\rangle }\mu=0$

$\int_{G}\chi_{f_{1}}\overline{\chi_{f_{2}}}\mu=0$

$\chi_{f_{1}}\ast\chi_{f_{2}}=0$ with the definition of convolution above.

iii) If $(H_{1},f_{1}),\left(  H_{2},f_{2}\right)  $ are two
\textit{equivalent} irreducible unitary finite dimensional representations of
G : $\chi_{f_{1}}\ast\chi_{f_{2}}=d_{f_{1}}^{-1}\chi_{f_{1}}$ where $d_{f_{1}%
}$ is the multiplicity of the class of representations of both $(H_{1}%
,f_{1}),\left(  H_{2},f_{2}\right)  .$
\end{theorem}

\begin{theorem}
Peter-Weyl theorem (Knapp p.245) : For a compact Lie group G the linear span
of all matrix coefficients for all finite dimensional \ irreducible unitary
representation of G is dense in $L^{2}\left(  G,S,\mu,%
\mathbb{C}
\right)  $
\end{theorem}

If (H,f) is a unitary representation of G, $u,v\in H,$ a matrix coefficient is
a map : $G\rightarrow C\left(  G;%
\mathbb{C}
\right)  :m\left(  g\right)  =\left\langle u,f\left(  g\right)  v\right\rangle
$

Thus if $(H_{j},f_{j})_{j=1}^{r}$ is a family of mutually orthogonal and
inequivalent unitary, finite dimensional representations, take an orthonormal
basis $\left(  \varepsilon_{\alpha}\right)  $\ in each $H_{j}$ and define :
$\varphi_{j\alpha\beta}\left(  g\right)  =\left\langle \varepsilon_{\alpha
},f_{j}\left(  g\right)  \varepsilon_{\beta}\right\rangle ,$ then the family
of functions $\left(  \sqrt{d_{j}}\varphi_{j\alpha\beta}\right)
_{j,\alpha,\beta}$ is an orthonormal basis of $L^{2}\left(  G,S,\mu,%
\mathbb{C}
\right)  .$

\subsubsection{Abelian groups}

\begin{theorem}
(Neeb p.76) Any irreducible unitary representation of a topological abelian
group is one dimensional.
\end{theorem}

\begin{theorem}
(Duistermaat p.213) Any finite dimensional irreducible representation of an
abelian group is one dimensional.
\end{theorem}

\begin{theorem}
Any finite dimensional irreducible unitary representation of an
\textit{abelian} topological group G is of the form : $\left(  T,\chi
\in\widehat{G}\right)  $ where $\widehat{G}$ is the Pontryagin dual of G, T
the set of complexscalars of module 1. The map $\chi\left(  g\right)  =\exp
i\theta\left(  g\right)  $ is a character.
\end{theorem}

\begin{theorem}
(Neeb p.150-163) There is a bijective correspondance between the continuous
unitary representations (H,f) of G in a Hilbert space H and the regular
spectral measure P on $\widehat{G}$\ 

If P is a regular spectral measure on $\widehat{G}$\ , valued in
$\mathcal{L}$%
(H;H) for a Hilbert space H, (H,f) is a unitary representation of G with :
$f\left(  g\right)  =P\circ\widehat{g}=\int_{\widehat{G}}\chi\left(  g\right)
P\left(  \chi\right)  $ where $\widehat{g}:G\rightarrow\widehat{G}%
::\widehat{g}\left(  \chi\right)  =\chi\left(  g\right)  $

Conversely, for any unitary continuous representation (H,f) of G there is a
unique regular spectral measure P such that :

$P:S\rightarrow%
\mathcal{L}%
\left(  H;H\right)  ::P\left(  \chi\right)  =P^{2}\left(  \chi\right)
=P^{\ast}\left(  \chi\right)  $

$f\left(  g\right)  =P\left(  \widehat{g}\right)  $ where $\widehat
{g}:G\rightarrow\widehat{G}::\widehat{g}\left(  \chi\right)  =\chi\left(
g\right)  $

Moreover any operator in
$\mathcal{L}$%
(H;H) commutes with f iff it commutes with each $P\left(  \chi\right)  .$
\end{theorem}

The support of P is the smallest closed subset A of $\widehat{G}$ such that P(A)=1

Any unitary representation takes the form :

$\left(  H=\left(  \oplus_{\chi\in\widehat{G}}d_{\chi}e_{\chi}\right)  \oplus
H_{c},f=\left(  \oplus_{\chi\in\widehat{G}}d_{\chi}\chi e_{\chi}\right)
\oplus f_{c}\right)  $

where the vectors $e_{j}$ are orthonormal, and orthogonal to $H_{c}.$ The
irreducible representations are given with their multiplicity by the $e_{\chi
},$ indexed by the characters, and $H_{c}$ does not contain any irreducible
representation. It can happen that $H_{c}$ is non trivial : a unitary
representation of an abelian group is not necessarily completely reducible.

\begin{theorem}
A unitary representation (H,f) of an abelian topological group G, isomorphic
to a m dimensional vector space E reads :%

\begin{equation}
f\left(  g\right)  =\int_{E^{\ast}}\left(  \exp ip(g)\right)  P\left(
p\right)
\end{equation}

where P is a spectral measure on the dual E*.
\end{theorem}

\begin{proof}
There is an isomorphism between the dual E* of E and $\widehat{G}:$

$\Phi:E^{\ast}\rightarrow\widehat{G}::\chi\left(  g\right)  =\exp ip\left(
g\right)  $

So any unitary representation (H,f) of G can be written :

$\varphi\left(  g\right)  =\int_{E^{\ast}}\left(  \exp ip(g)\right)  P\left(
p\right)  $

where $P:\sigma\left(  E^{\ast}\right)  \rightarrow%
\mathcal{L}%
\left(  H;H\right)  $ is a spectral measure and $\sigma\left(  E^{\ast
}\right)  $ the Borel $\sigma-$algebra of the dual E*,
\end{proof}

If H is finite dimensional then the representation can be decomposed in a sum
of orthogonal irreducible one dimensional representations and we have with a
basis $\left(  e_{k}\right)  _{k=1}^{n}$ of H : $P\left(  p\right)
=\sum_{k=1}^{n}\pi_{k}\left(  p\right)  e_{k}$ where $\pi_{k}$ is a measure on
E* and $\varphi\left(  g\right)  =\sum_{k=1}^{n}\left(  \int_{p}\left(  \exp
ip\left(  g)\right)  \right)  \pi_{k}\left(  p\right)  \right)  e_{k}.$

\bigskip

\subsection{Representation of Lie algebras}

\label{Representaition of Lie algebra}

Lie algebras are classified, therefore it is possible to exhibit almost all
their representations, and this is the base for the classification of the
representation of groups.

\subsubsection{Irreducible representations}

The results are similar to the results for groups.

\begin{theorem}
(Schur's lemna) Any interwiner $\phi\in%
\mathcal{L}%
\left(  E_{1};E_{2}\right)  $ between the irreducible representations
$(E_{1},f_{1}),\left(  E_{2},f_{2}\right)  $ of the Lie algebra A are either 0
or an isomorphism.
\end{theorem}

\begin{proof}
with the use of the theorem below

$\ker\phi$ is either 0,and then $\phi$ is injective, or $E_{1}$ and then
$\phi=0$

$\operatorname{Im}\phi$ is either 0, and then $\phi=0,$\ or $E_{2}$ and then
$\phi$ is surjective

Thus $\phi$\ is either 0 or bijective, and then the representations are
isomorphic :

$\forall X\in X:f_{1}\left(  X\right)  =\phi^{-1}\circ f_{2}\left(  X\right)
\circ\phi$
\end{proof}

\begin{theorem}
If $(E,f_{1}),\left(  E,f_{2}\right)  $ are two irreducible equivalent
representations of a Lie algebra A on the same complex space then
$\exists\lambda\in%
\mathbb{C}
$ and an interwiner $\phi=\lambda Id$
\end{theorem}

\begin{proof}
The spectrum of $\phi\in G%
\mathcal{L}%
\left(  E;E\right)  $ is a compact subset of $%
\mathbb{C}
$\ with at least a non zero element $\lambda,$ thus $\phi-\lambda Id$ is not
injective in
$\mathcal{L}$%
(E;E) but continuous, it is an interwiner of $(E,f_{1}),\left(  E,f_{2}%
\right)  ,$ thus it must be zero.
\end{proof}

Therefore for any two irreducible representations either they are not
equivalent, or they are isomorphic, and we can define \textbf{classes of
irreducible representations}. If a representation (E,f) is reducible, we can
define the number of occurences of a given class j of irreducible
representation, which is called the \textbf{multiplicity} $d_{j}$ of the class
of representations j in (E,f).

\begin{theorem}
(Knapp p.296) Any finite dimensional representation (E,f) of a complex
semi-simple finite dimensional Lie algebra A is completely reducible:
\end{theorem}

$E=\oplus_{k=1}^{p}E_{k},\left(  E_{k},f|_{E_{k}}\right)  $ is an irreducible
representation of A

\begin{theorem}
If $\phi\in%
\mathcal{L}%
\left(  E_{1};E_{2}\right)  $ is an interwiner between the representations
$(E_{1},f_{1}),\left(  E_{2},f_{2}\right)  $ of a Lie algebra A, then :
$\ker\phi,\operatorname{Im}\phi$ are invariant subspaces of $E_{1},E_{2}$ respectively
\end{theorem}

\begin{proof}
$u\in\ker\phi\Rightarrow\phi\circ f_{1}\left(  X\right)  u=f_{2}\left(
X\right)  \circ\phi u=0\Rightarrow f_{1}\left(  X\right)  u\in\ker
\phi\Leftrightarrow\ker\phi$ is invariant for $f_{1}$

$v\in\operatorname{Im}\phi\Rightarrow\exists u\in E_{1}:v=\phi u\Rightarrow
\phi\left(  f_{1}\left(  X\right)  u\right)  =f_{2}\left(  X\right)  \circ\phi
u=f_{2}\left(  X\right)  v\Rightarrow f_{2}\left(  X\right)  v\in
\operatorname{Im}\phi\Leftrightarrow\operatorname{Im}\phi$ is invariant for
$f_{2}$
\end{proof}

\begin{theorem}
(Knapp p.250) Any 1 dimensional representation of a semi simple Lie algebra is
trivial (=0). Any 1-dimensional representation of a connected semi simple Lie
group is trivial (=1).
\end{theorem}

\subsubsection{Classification of representations}

Any finite dimensional Lie algebra has a representation as a matrix space over
a finite dimensional vector space. All finite dimensional Lie algebras are
classified, according to the abstract roots system. In a similar way one can
build any irreducible representation from such a system. The theory is very
technical (see Knapp) and the construction is by itself of little practical
use, because all common Lie algebras are classified and well documented with
their representations.\ However it is useful as a way to classify the
representations, and decompose reducible representations into irreducible
representations. The procedure starts with semi-simple complex Lie algebras,
into which any finite dimensional Lie algebra can be decomposed.

\paragraph{Representations of a complex semi-simple Lie algebra\newline}

(Knapp p.278, Fulton p.202)

Let A be a complex semi-simple n dimensional Lie algebra., B its Killing form
and B* the form on the dual A*. There is a Cartan subalgebra h of dimension r,
the rank of A. Let h* be its dual.

The key point is that in any representation (E,f) of A, for any element H of
h, f(H) acts diagonally with eigen values which are linear functional of the
$H\in h.$ As the root-space decomposition of the algebra is just the
representation (A,ad) of A on itself, we have many similarities.

i) So there are forms $\lambda\in h^{\ast},$ called \textbf{weights}, and
eigenspaces, called \textbf{weigth spaces}, denoted $E_{\lambda}$ such that :

$E_{\lambda}=\left\{  u\in E:\forall H\in h:f\left(  H\right)  u=\lambda
\left(  H\right)  u\right\}  $

we have similarly :

$A_{\alpha}=\left\{  X\in A:\forall H\in h:ad\left(  H\right)  X=\alpha\left(
H\right)  X\right\}  $

The set of weights is denoted $\Delta\left(  \lambda\right)  \in h^{\ast}$ as
the set of roots $\Delta\left(  \alpha\right)  $

Whereas the $A_{\alpha}$ are one dimensional, the $E_{\lambda}$\ can have any
dimension $\leq n$ called the multiplicity of the weight

ii) E is the direct sum of all the weight spaces :

$E=\oplus_{\lambda}E_{\lambda}$

we have on the other hand : $A=h\oplus_{\lambda}A_{\lambda}$ because 0 is a
common eigen value for h.

iii) every weight $\lambda$ is real valued on $h_{0}$\ and algebraically
integral in the meaning that :

$\forall\alpha\in\Delta:2\frac{B^{\ast}\left(  \lambda,\alpha\right)
}{B^{\ast}\left(  \alpha,\alpha\right)  }\in%
\mathbb{Z}
$

where $h_{0}=\sum_{\alpha\in\Delta}k^{\alpha}H_{\alpha}$ is the real vector
space generated by $H_{\alpha}$ the vectors of A, dual of each root $\alpha
$\ with respect to the Killing form : $\forall H\in h:B\left(  H,H_{\alpha
}\right)  =\alpha\left(  H\right)  $

iv) for any weigth $\lambda:\ \forall\alpha\in\Delta:f\left(  H_{\alpha
}\right)  E_{\lambda}\sqsubseteq E_{\lambda+\alpha}$

As seen previously it is possible to introduce an ordering \textit{of the
roots} and compute a simple system of roots : $\Pi\left(  \alpha\right)
=\Pi\left(  \alpha_{1},...\alpha_{l}\right)  $ and distinguish positive roots
$\Delta^{+}\left(  \alpha\right)  $ and negative roots $\Delta^{-}\left(
\alpha\right)  $

\begin{theorem}
\textbf{Theorem of the highest weigth }: If (E,f) is an irreducible finite
dimensional representation of a complex semi-simple n dimensional Lie algebra
A then there is a unique vector $V\in E$, called the highest weigth vector,
such that :

i) $V\in E_{\mu}$ for some $\mu\in\Delta\left(  \lambda\right)  $ called the
highest weigth

ii) $E_{\mu}$ is one dimensional

iii) up to a scalar, V is the only vector such that : $\forall\alpha\in
\Delta^{+}\left(  \alpha\right)  ,\forall H\in A_{\alpha}:f\left(  H\right)
V=0$
\end{theorem}

Then :

i) successive applications of $\forall\beta\in\Delta^{-}\left(  \alpha\right)
$ to V generates E:

$E=Span\left(  f\left(  H_{\beta_{1}}\right)  f\left(  H_{\beta_{2}}\right)
...f\left(  H_{\beta_{p}}\right)  V,\beta_{k}\in\Delta^{-}\left(
\alpha\right)  \right)  $

ii) all the weights $\lambda$ of the representation are of the form :
$\lambda=\mu-\sum_{k=1}^{l}n_{k}\alpha_{k}$ with $n_{k}\in%
\mathbb{N}
$ and $\left\vert \lambda\right\vert \leq\left\vert \mu\right\vert $

iii) $\mu$ depends on the simple system $\Pi\left(  \alpha\right)  $\ and not
the ordering

Conversely :

Let A be a finite dimensional complex semi simple Lie algebra.\ We can use the
root-space decomposition to get $\Delta\left(  \alpha\right)  .$ We know that
a weight for any representation is real valued on $h_{0}$\ and algebraically
integral, that is : $2\frac{B^{\ast}\left(  \lambda,\alpha\right)  }{B^{\ast
}\left(  \alpha,\alpha\right)  }\in%
\mathbb{Z}
.$

So choose an ordering on the roots, and on a simple system $\Pi\left(
\alpha_{1},...\alpha_{l}\right)  $ define the \textbf{fundamental weights} :
$\left(  w_{i}\right)  _{i=1}^{l}$ by : $2\frac{B^{\ast}\left(  w_{i}%
,\alpha_{j}\right)  }{B^{\ast}\left(  \alpha_{i},\alpha_{i}\right)  }%
=\delta_{ij}.$ Then any highest weight will be of the form : $\mu=\sum
_{k=1}^{l}n_{k}w_{k}$ with $n_{k}\in%
\mathbb{N}
$

The converse of the previous theorem is that, for any choice of such highest
weight, there is a unique irreducible representation, up to isomorphism.

The irreducible representation $\left(  E_{i},f_{i}\right)  $ related to a
fundamental weight $w_{i}$ is called a \textbf{fundamental representation}.
The dimension $p_{i}$ of $E_{i}$ is not a parameter : it is fixed by the
choice of $w_{i}.$

To build this representation the procedure, which is complicated, is, starting
with any vector V which will be the highest weight vector, compute
successively other vectors by successive applications of $f\left(
\beta\right)  $\ and prove that we get a set of independant vectors which
consequently generates E. The dimension p of E is not fixed before the
process, it is a result. As we have noticed, the choice of the vector space
itself is irrelevant with regard to the representation problem, what matters
is the matrix of f in some basis of E.

From fundamental representations one can build other irreducible
representations, using the following result

\begin{theorem}
(Knapp p.341) If $(E_{1},f_{1}),(E_{2},f_{2})$ are irreducible representations
of the same algebra A, associated to the highest weights $\mu_{1},\mu_{2}.$
then the tensorial product of the representations, $\left(  E_{1}\otimes
E_{2},f_{1}\times f_{2}\right)  $ is an irreducible representation of A,
associated to the highest weight $\mu_{1}+\mu_{2}.$
\end{theorem}

\begin{notation}
$\left(  E_{i},f_{i}\right)  $ denotes in the following the fundamental
representation corresponding to the fundamental weight $w_{i}$
\end{notation}

\paragraph{Practically\newline}

Any finite dimensional semi simple Lie algebra belongs to one the 4 general
families, or one of the 5 exceptional algebras. And, for each of them, the
fundamental weigths $\left(  w_{i}\right)  _{i=1}^{l}$ (expressed as linear
combinations of the roots) and the corresponding fundamental representations
$(E_{i},f_{i})$, have been computed and are documented. They are given in the
next subsection with all the necessary comments.

Any finite dimensional irreducible representation of a complex semi simple Lie
algebra of rank l can be labelled by l integers $\left(  n_{i}\right)
_{i=1}^{l}$ : $\Gamma_{n_{1}...n_{l}}$\ identifies the representation given by
the highest weight : $w=\sum_{k=1}^{l}n_{k}w_{k}.$ It is given by the
tensorial product of fundamental representations. As the vector spaces $E_{i}
$ are distinct, we have the isomorphism $E_{1}\otimes E_{2}\simeq E_{2}\otimes
E_{1}$ and we can collect together the tensorial products related to the same
vector space. So the irreducible representation labelled by $w=\sum_{k=1}%
^{l}n_{k}w_{k}$ is :

$\Gamma_{n_{1}...n_{l}}=\left(  \otimes_{i=1}^{l}\left(  \otimes_{k=1}^{n_{i}%
}E_{i}\right)  ,\times_{i=1}^{l}\left(  \times_{k=1}^{n_{i}}f_{i}\right)
\right)  $

And any irreducible representation is of this kind.

Each $E_{i}$ has its specific dimension, thus if we want an irreducible
representation on a vector space with a given dimension n, we have usually to
patch together several representations through tensorial products.

Any representation is a combination of irreducible representations, however
the decomposition is not unique. When we have a direct sum of such irreducible
representations, it is possible to find equivalent representations, with a
different sum of irreducible representations. The coefficients involved in
these decompositions are called the \textbf{Clebsh-Jordan} coefficients. There
are softwares which manage most of the operations.

\paragraph{Representation of compact Lie algebras\newline}

Compact complex Lie algebras are abelian, so only real compact Lie algebras
are concerned. A root-space decomposition on a compact real Lie algebra A can
be done (see Compact Lie groups) by : choosing a maximal Cartan subalgebra t
(for a Lie group it comes from a torus, which is abelian, so its Lie
subalgebra is also abelian), taking the complexified $A_{%
\mathbb{C}
},t_{%
\mathbb{C}
}$\ of A and t , and the roots $\alpha$ are elements of $t_{%
\mathbb{C}
}^{\ast}$ such that :

$A_{\alpha}=\left\{  X\in A_{C}:\forall H\in t_{C}:ad(H)X=\alpha\left(
H\right)  X\right\}  $

$A_{C}=t_{C}\oplus_{\alpha}A_{\alpha}$

For any $H\in t:\alpha\left(  H\right)  \in i%
\mathbb{R}
:$ the roots are purely imaginary.

If we have a finite dimensional representation (E,f) of A, then we have
weights with the same properties as above (except that t replaces h): there
are forms $\lambda\in t_{C}^{\ast},$ called weights, and eigenspaces, called
weigth spaces, denoted $E_{\lambda}$ such that :

$E_{\lambda}=\left\{  u\in E:\forall H\in t_{C}:f\left(  H\right)
u=\lambda\left(  H\right)  u\right\}  $

The set of weights is denoted $\Delta\left(  \lambda\right)  \in t_{C}^{\ast}$.

The theorem of the highest weight extends in the same terms.\ In addition the
result stands for the irreducible representations of compact connected Lie
group, which are in bijective correspondance with the representations of the
highest weight of their algebra.

\bigskip

\subsection{Summary of results on the representation theory}

- any irreducible representation of an abelian group is unidimensional (see
the dedicated subsection)

- any continuous unitary representation of a compact or a finite group is
completely reducible in the direct sum of orthogonal finite dimensional
irreducible unitary representations.

- any continuous unitary finite dimensional representation of a topological
group is completely reducible

- any 1 dimensional representation of a semi simple Lie algebra is trivial
(=0). Any 1-dimensional representation of a connected semi simple Lie group is
trivial (=1).

- any topological, locally compact, group has a least one faithful unitary
representation (usually infinite dimensional) : the left (right) regular
representations on the spaces $L^{2}\left(  G,\mu_{L},%
\mathbb{C}
\right)  $.

- any Lie group has the adjoint representations over its Lie algebra and its
universal envelopping algebra

- any group of matrices in K(n) has the standard representation over $K^{n}%
$\ where the matrices act by multiplication the usual way.

- there is a bijective correspondance between representations of real Lie
algebras (resp real groups) and its complexified. And one representation is
irreducible iff the other is irreducible.

- any Lie algebra has the adjoint representations over itself and its
universal envelopping algebra

- any finite dimensional Lie algebra has a representation as a matrix group
over a finite dimensional vector space.

- the finite dimensional representations of finite dimensional semi-simple
complex Lie algebras are computed from the fundamental representations, which
are documented

- the finite dimensional representations of finite dimensional real compact
Lie algebras are computed from the fundamental representations, which are documented.

- Whenever we have a representation $(E_{1},f_{1})$ and $\phi:E_{1}\rightarrow
E_{2}$\ is an isomorphism we have an equivalent representation $\left(
E_{2},f_{2}\right)  $ with $f_{2}\left(  g\right)  =\phi\circ f_{1}\left(
g\right)  \circ\phi^{-1}.$ So for finite dimensional representations we can
take $K^{n}=E.$

\newpage

\section{CLASSICAL\ LIE\ GROUPS\ AND\ ALGEBRAS}

\label{Classical Lie groups and algebras}

There are a finite number of types of finite dimensional Lie groups and Lie
algebras, and most of them are isomorphic to algebras or groups of
matrices.\ These are well documented and are the workhorses of studies on Lie
groups, so they deserve some attention, for all practical purpose. We will
start with some terminology and general properties.

\bigskip

\subsection{General properties of linear groups and algebras}

\subsubsection{Definitions}

\begin{definition}
A \textbf{linear Lie algebra} on the field K' is a Lie algebra of square
$n\times n$ matrices on a field K, with the bracket : $\left[  M,N\right]
=MN-NM$

A \textbf{linear Lie group} on the field K' is a Lie group of square $n\times
n$ matrices on a field K
\end{definition}

Notice that K,K' can be different, but $K^{\prime}\subset K.$ There are
\textit{real Lie algebras or groups of complex matrices.}

The most general classes are :

L(K,n) is the linear Lie algebra of square nxn matrices on a field K with the
bracket : $\left[  X,Y\right]  =\left[  X\right]  \left[  Y\right]  -\left[
Y\right]  \left[  X\right]  $

GL(K,n) the linear Lie group of inversible square nxn matrices on a field K,

SL(K,n) the linear Lie group of inversible square nxn matrices on a field K
with determinant=1

The identity matrix is $I_{n}=diag(1,...,1)\in GL(K,n)$

\paragraph{Matrices on the quaternions ring}

It is common to consider matrices on the division ring of quaternions denoted
H.\ This is not a field (it is not commutative), thus there are some
difficulties (and so these matrices should be avoided).

A quaternion reads (see the Algebra part) : : $x=a+bi+cj+dk$ with $a,b,c,d\in%
\mathbb{R}
$ and

$i^{2}=j^{2}=k^{2}=-1,ij=k=-ji,jk=i=-kj,ki=j=-ik$

So we can write : $x=z_{1}+j\left(  c-id\right)  =z_{1}+jz_{2},z_{1},z_{2}\in%
\mathbb{C}
$

$xx^{\prime}=z_{1}z_{1}^{\prime}+j^{2}z_{2}z_{2}^{\prime}+j\left(  z_{1}%
z_{2}^{\prime}+z_{1}^{\prime}z_{2}\right)  =z_{1}z_{1}^{\prime}-z_{2}%
z_{2}^{\prime}+j\left(  z_{1}z_{2}^{\prime}+z_{1}^{\prime}z_{2}\right)
=x^{\prime}x$

and a matrix on L(H,n) reads : $M=M_{1}+jM_{2},M_{1},M_{2}\in L\left(
\mathbb{C}
,n\right)  $ so it can be considered as a couple of complex matrices with the
multiplication rule :

$\left[  MM^{\prime}\right]  _{q}^{p}=\sum_{r}\left[  M_{1}\right]  _{r}%
^{p}\left[  M_{1}^{\prime}\right]  _{q}^{r}-\left[  M_{2}\right]  _{r}%
^{p}\left[  M_{2}^{\prime}\right]  _{q}^{r}+j\sum_{r}\left[  M_{1}\right]
_{r}^{p}\left[  M_{2}^{\prime}\right]  _{q}^{r}+\left[  M_{1}^{\prime}\right]
_{r}^{p}\left[  M_{2}\right]  _{q}^{r}$

that is : $MM^{\prime}=M_{1}M_{1}^{\prime}-M_{2}M_{2}^{\prime}+j\left(
M_{1}M_{2}^{\prime}+M_{1}^{\prime}M_{2}\right)  $

The identity is : $\left(  I_{n},0\right)  $

\subsubsection{Lie algebras}

1. A Lie algebra of matrices on the field K is a subspace L of L(K,n) for some
n, such that :

$\forall X,Y\in L,\forall r,r^{\prime}\in K:r\left[  X\right]  +r^{\prime
}\left[  Y\right]  \in L,\left[  X\right]  \left[  Y\right]  -\left[
Y\right]  \left[  X\right]  \in L.$

The dimension m of L is usually different from n. A basis of L is a set of
matrices $\left[  e_{i}\right]  _{i=1}^{m}$ which is a basis of L as vector
space, and so the structure coefficients are given by :

$C_{jk}^{i}\left[  e_{i}\right]  =\left[  e_{j}\right]  \left[  e_{k}\right]
-\left[  e_{k}\right]  \left[  e_{j}\right]  $

with the Jacobi identies : $\forall i,j,k,p:\sum_{l=1}^{m}\left(  C_{jk}%
^{l}C_{il}^{p}+C_{ki}^{l}C_{jl}^{p}+C_{ij}^{l}C_{kl}^{p}\right)  =0$

2. L(K,n) is a Banach finite dimensional vector space with the norm :
$\left\Vert M\right\Vert =Tr\left(  MM^{\ast}\right)  =Tr\left(  M^{\ast
}M\right)  $ where M* is the transpose conjugate.

3. If L is a real linear Lie algebra, its complexified is just the set :
$\left\{  X+iY,X,Y\in L\right\}  \subset L\left(
\mathbb{C}
,n\right)  $ which is a complex Lie algebra with Lie bracket :

$\left[  X+iY,X^{\prime}+iY^{\prime}\right]  _{L_{%
\mathbb{C}
}}=\left[  X,X^{\prime}\right]  _{L}-\left[  Y,Y^{\prime}\right]
_{L}+i\left(  \left[  X,Y^{\prime}\right]  _{L}+\left[  X^{\prime},Y\right]
_{L}\right)  $

If L is a complex Lie algebra, the obvious real structure is just : $\left[
X\right]  =\left[  x\right]  +i\left[  y\right]  ,\left[  x\right]  ,\left[
y\right]  \in L\left(
\mathbb{R}
,n\right)  .$ We have a real Lie algebra $L_{%
\mathbb{R}
}=L_{0.}\oplus iL_{0}$, comprised of \textit{couples} of real matrices, with
the bracket above, with two isomorphic real subalgebras $L_{0.},iL_{0}.$
$L_{0}$ is a real form of L.

If L is a even dimensional real Lie algebra endowed with a complex structure,
meaning a linear map $J\in L\left(  L;L\right)  $ such that $J^{2}=-Id_{L}$
and $J\circ ad=ad\circ J$ then take a basis of L :$\left(  e_{j}\right)
_{j=1}^{2m}$ with p=1...m : $J\left(  e_{j}\right)  =e_{j+m},J\left(
e_{j+m}\right)  =-e_{j}$

The complex Lie algebra reads :

$L_{%
\mathbb{C}
}=\sum_{p=1}^{m}\left(  x^{p}+iy^{p}\right)  \left[  e_{p}\right]  =\sum
_{p=1}^{m}\left(  x^{p}\left[  e_{p}\right]  +y^{p}\left[  e_{p+m}\right]
\right)  $

\subsubsection{Linear Lie groups}

\paragraph{Definition\newline}

Let G be an algebraic subgroup G of GL(K,n) for some n. The group operations
are always smooth and GL(K,n) is a Lie group. So G is a Lie subgroup if it is
a submanifold of L(K,n).

There are several criteria to make a Lie group :

i) if G is closed in GL(K,n) it is a Lie group

ii) if G is a finite group (it is then open and closed) it is a Lie group

It is common to have a group of matrices defined as solutions of some
equation.\ Let $F:L\left(  K,n\right)  \rightarrow L\left(  K,n\right)  $ be a
differentiable map on the manifold L(K,n) and define G as a group whose
matrices are solutions of F(M)=0. Then following the theorem of constant rank
(see Manifolds) if F' has a constant rank r in L(K,n) the set $F^{-1}\left(
0\right)  $ is a closed n%
${{}^2}$%
-r submanifold of L(K,n) and thus a Lie subgroup. The map F can involve the
matrix M and possibly its transpose and its conjugate. We know that a map on a
complex Banach is not C-differentiable if its involve the conjugate.\ So
usually a group of complex matrices defined by an equation involving the
conjugate is not a complex Lie group, but can be a real Lie group (example :
U(n) see below).

Remark : if F is continuous then the set $F^{-1}\left(  0\right)  $ is closed
in L(K,n), but this is not sufficient : it should be closed in the Lie group
GL(K,n), which is an open subset of L(K,n).

If G,H are Lie group of matrices, then $G\cap H$ is a Lie group of matrices
with Lie algebra $T_{1}G\cap T_{1}H$

\paragraph{Connectedness\newline}

The connected component of the identity $G_{0}$ is a normal Lie subgroup of G,
with the same dimension as G. The quotient set $G/G_{0}$ is a finite group.
The other components of G can be written as : $g=g_{L}x=yg_{R}$ where x,y are
in one of the other connected components, and $g_{R},g_{L}$ run over $G_{0}.$

If G is connected there is always a universal covering group $\widetilde{G}%
$\ which is a Lie group.\ It is a compact group if G is compact. It is a group
of matrices if G is a a complex semi simple Lie group, but otherwise
$\widetilde{G}$ is not necessarily a group of matrices (ex : $GL(%
\mathbb{R}
,n)).$

If G is not connected we can consider the covering group of the connected
component of the identity $G_{0}.$

\paragraph{Translations\newline}

The translations are $L_{A}\left(  M\right)  =\left[  A\right]  \left[
M\right]  ,R_{A}\left(  M\right)  =\left[  M\right]  \left[  A\right]  $

The conjugation is : $Conj_{A}M=\left[  A\right]  \left[  M\right]  \left[
A\right]  ^{-1}$ and the derivatives :

$\forall X\in L\left(  K,n\right)  :$

$L_{A}^{\prime}\left(  M\right)  \left(  X\right)  =\left[  A\right]  \left[
X\right]  ,R_{A}^{\prime}\left(  M\right)  =\left[  X\right]  \left[
A\right]  ,\left(  \Im(M)\right)  ^{\prime}\left(  X\right)  =-M^{-1}XM^{-1}$

So : $Ad_{M}X=Conj_{M}X=\left[  M\right]  \left[  X\right]  \left[  M\right]
^{-1}$

\paragraph{Lie algebra of a linear Lie group\newline}

\begin{theorem}
The Lie algebra of a linear Lie group of matrices in K(n) is a linear Lie
subalgebra of matrices in L(K,n)
\end{theorem}

If the Lie group G is defined through a matrix equation involving $M,M^{\ast
},M^{t},\overline{M}:$

$P\left(  M,M^{\ast},M^{t},\overline{M}\right)  =0$

Take a path : $M:%
\mathbb{R}
\rightarrow G$ such that M(0)=I. Then $X=M^{\prime}(0)\in T_{1}G$ satisfies
the polynomial equation :

$\left(  \frac{\partial P}{\partial M}X+\frac{\partial P}{\partial M^{\ast}%
}X^{\ast}+\frac{\partial P}{\partial M^{t}}X^{t}+\frac{\partial P}%
{\partial\overline{M}}\overline{X}\right)  |_{M=I}=0$

Then a left invariant vector field is $X_{L}\left(  M\right)  =MX$. Its flow
is given by the equation : $\frac{d}{dt}\Phi_{X_{L}}\left(  t,g\right)
|_{t=\theta}=X_{L}\left(  \Phi_{X_{L}}\left(  \theta,g\right)  \right)
=\Phi_{X_{L}}\left(  \theta,g\right)  \times X$ whose solution is :
$\Phi_{X_{L}}\left(  t,g\right)  =g\exp tX$ where the exponential is computed as

$\exp tu=\sum_{p=0}^{\infty}\frac{t^{p}}{p!}\left[  X\right]  ^{p}$.

\paragraph{Complex structures\newline}

\begin{theorem}
(Knapp p.442) For any real compact connected Lie group of matrices G there is
a unique (up to isomorphism) closed complex Lie group of matrices whose Lie
algebra is the complexified $T_{1}G\oplus iT_{1}G$ of $T_{1}G.$
\end{theorem}

\paragraph{Cartan decomposition\newline}

\begin{theorem}
(Knapp p.445) Let be the maps :

$\Theta:GL(K,n)\rightarrow GL(K,n)::\Theta\left(  M\right)  =\left(
M^{-1}\right)  ^{\ast},$

$\theta:L(K,n)\rightarrow L\left(  K,n\right)  ::\theta\left(  X\right)
=-X^{\ast}$

If G is a connected closed semi simple Lie group of matrices in GL(K,n),
invariant under $\Theta$\ , then :

i) its Lie algebra is invariant under $\theta,$

ii) $T_{1}G=l_{0}\oplus p_{0}$ where $l_{0},p_{0}$\ are the eigenspaces
corresponding to the eigen values +1,-1 of $\theta$

iii) the map : $K\times p_{0}\rightarrow G::g=k\exp X$ where $K=\left\{  x\in
G:\Theta x=x\right\}  $ is a diffeomorphism onto.
\end{theorem}

\paragraph{Groups of tori\newline}

A group of tori is defined through a family $\left[  e_{k}\right]  _{k=1}^{m}$
of commuting matrices in L(K,n) which is a basis of the abelian algebra.\ Then
the group G is generated by : $\left[  g\right]  _{k}=\exp t\left[
e_{k}\right]  =\sum_{p=0}^{\infty}\frac{t^{p}}{p!}\left[  e_{k}\right]
^{p},t\in K$

A group of diagonal matrices is a group of tori, but they are not the only ones.

The only compact complex Lie group of matrices are groups of tori.

\subsubsection{Representations}

1. Any linear Lie group G on the field K in GL(K,n) has the \textbf{standard
representation} $(K^{n},\imath)$ where the matrices act the usual way on
column matrices.

By the choice of a fixed basis it is isomorphic to the representation (E,f) in
any n dimensional vector space on the field K, where f is the endomorphism
which is represented by a matrix of g. And two representations (E,f),(E,f')
are isomorphic if the matrix for passing from one basis to the other belongs
to G.

2. From the standard representation $(K^{n},\imath)$ of a Lie group G one
deduces the standard representation $(K^{n},\imath^{\prime}\left(  1\right)
)$ of its Lie algebra $T_{1}G$ as a subalgebra of L(K,n).

3. From the standard representation one can deduces other representations by
the usual operations (sum, product,...).

4. Conversely, a n dimensional vector space E and a subset L of endomorphisms
of GL(E;E) define, by taking the matrices of L in a fixed basis, a subset G of
matrices which are a linear group in GL(K,n).\ But the representation may not
be faithful and G may be a direct sum of smaller matrices (take $E=E_{1}\oplus
E_{2}$ then G is a set of couples of matrices).

\bigskip

\subsection{Finite groups}

The case of finite groups, meaning groups with a finite number of elements
(which are not usual Lie groups) has not been studied so far. We denote \#G
the number of its elements (its cardinality).

\paragraph{Standard representation\newline}

\begin{definition}
The standard representation (E,f) of the finite group G is :

E is any \#G dimensional vector space (such as $K^{\#G})$ on any field K

$f:G\rightarrow L\left(  E;E\right)  ::f(g)e_{h}=e_{gh}$ with any basis of E :
$\left(  e_{g}\right)  _{g\in G}$
\end{definition}

$f(g)\left(  \sum_{h\in G}x_{h}e_{h}\right)  =\sum_{h\in G}x_{h}e_{gh}$

$f(1)=I,$

$f\left(  gh\right)  u=\sum_{k\in G}x_{k}e_{ghk}=\sum_{k\in G}x_{k}f\left(
g\right)  \circ f\left(  h\right)  e_{k}=f\left(  g\right)  \circ f\left(
h\right)  \left(  u\right)  $

\paragraph{Unitary representation\newline}

\begin{theorem}
For any representation (E,f) of the finite group G, and any hermitian
sequilinear form $\left\langle {}\right\rangle $ on E, the representation
(E,f) is unitary with the scalar product : $\left(  u,v\right)  =\frac{1}%
{\#G}\sum_{g\in G}\left\langle f\left(  g\right)  u,f\left(  g\right)
v\right\rangle $
\end{theorem}

Endowed with the discrete topology G is a compact group. So we can use the
general theorem :

\begin{theorem}
Any representation of the finite group G is completely reducible in the direct
sum of orthogonal finite dimensional irreducible unitary representations.
\end{theorem}

\begin{theorem}
(Kosmann p.35) The number N of irreducible representations of the finite group
G is equal to the number of conjugacy classes of G
\end{theorem}

So there is a family $\left(  E_{i},f_{i}\right)  _{i=1}^{N}$ of irreducible
representations from which is deduced any other representation of G and
conversely a given representation can be reduced to a sum and tensorial
products of thse irreducible representations.

\paragraph{Irreducible representations\newline}

The irreducible representations $\left(  E_{i},f_{i}\right)  _{i=1}^{N}$ are
deduced from the standard representation, in some ways. A class of conjugacy
is a subset $G_{k}$ of G such that all elements of $G_{k}$\ commute with each
other, and the $G_{k}$\ form a partition of G. Any irreducible representation
$\left(  E_{i},f_{i}\right)  $ of G gives an irreducible subrepresentation of
each $G_{k}$ which is necessarily one dimensional because $G_{k}$ is abelian.
A representation $\left(  E_{i},f_{i}\right)  $ is built by patching together
these one dimensional representations. There are many examples of these
irreducible representations for the permutations group (Kosmann, Fulton).

\paragraph{Characters\newline}

The characters $\chi_{f}:G\rightarrow%
\mathbb{C}
::\chi_{f}(g)=Tr(f(g))$ are well defined for any representation. They are
represented as a set of \#G scalars.

They are the same for equivalent representations. Moreover for any two
irreducible representations $\left(  E_{p},f_{p}\right)  ,\left(  E_{q}%
,f_{q}\right)  :\frac{1}{\#G}\sum_{g\in G}\overline{\chi_{f_{p}}\left(
g\right)  }\chi_{f_{q}}\left(  g\right)  =\delta_{pq}$

The table of characters of G is built as the matrix : $\left[  \chi_{f_{p}%
}\left(  g_{q}\right)  \right]  _{q=1..N}^{p=1...N}$ where $g_{q}$ is a
representative of each class of conjugacy of G. It is an orthonormal system
with the previous relations. So it is helpful in patching together the one
dimensional representations of the class of conjugacy.

For any representation (E,f) which is the direct sum of $\left(  E_{i}%
,f_{i}\right)  _{i=1}^{N},$ each with a multiplicity $d_{j}$. : $\chi_{f}%
=\sum_{q\in I}d_{q}\chi_{f_{q}},\chi_{f}\left(  1\right)  =\dim E$

\paragraph{Functionnal representations\newline}

\begin{theorem}
The set $%
\mathbb{C}
^{G}$ of functions $\varphi:G\rightarrow%
\mathbb{C}
$ on the finite group G can be identified with the vector space $%
\mathbb{C}
^{\#G}.$
\end{theorem}

\begin{proof}
Any map is fully defined by \#G complex numbers : $\left\{  \varphi\left(
g\right)  =a_{g},g\in G\right\}  $ so it is a vector space on $%
\mathbb{C}
$ with dimension is \#G
\end{proof}

The natural basis of $%
\mathbb{C}
^{G}$\ is :$\left(  e_{g}\right)  _{g\in G}::e_{g}\left(  h\right)
=\delta_{gh} $

It is is orthonormal with the scalar product : $\left\langle \varphi
,\psi\right\rangle =\sum_{g\in G}\overline{\varphi\left(  h\right)  }%
\psi\left(  h\right)  \mu=\sum_{g\in G}\frac{1}{\#G}\overline{\varphi\left(
g\right)  }\psi\left(  g\right)  $

The Haar measure over G has for $\sigma-$algebra the set $2^{G}$\ of subsets
of G and for values:

$\varpi\in2^{G}:\mu\left(  \varpi\right)  =\frac{1}{\#G}\delta\left(
g\right)  ::\mu\left(  \varpi\right)  =0$ if $g\notin\varpi,:\mu\left(
\varpi\right)  =\frac{1}{\#G}$ if $g\in\varpi$

The left regular representation $\left(
\mathbb{C}
^{G},\Lambda\right)  :\varphi\left(  x\right)  \rightarrow\Lambda\left(
g\right)  \left(  \varphi\right)  \left(  x\right)  =\varphi\left(
g^{-1}x\right)  $ is unitary , finite dimensional, and $\Lambda\left(
g\right)  \left(  e_{h}\right)  =e_{gh}.$ The characters in this
representation are : $\chi_{\Lambda}\left(  g\right)  =Tr\left(
\Lambda\left(  g\right)  \right)  =\left(  \#G\right)  \delta_{1g}$

This representation is reducible : it is the sum of all the irreducible
representations $\left(  E_{p},f_{p}\right)  _{p=1}^{N}$\ of G, each with a
multiplicty equal to its dimension : $\left(
\mathbb{C}
^{G},\Lambda\right)  =\sum_{p=1}^{N}\left(  \otimes^{\dim E_{p}}\left(
\mathbb{C}
^{G}\right)  ,\otimes^{\dim E_{p}}f_{p}\right)  $

\bigskip

\subsection{GL(K,n) and L(K,n)}

$K=%
\mathbb{R}
,%
\mathbb{C}
$

If n=1 we have the trivial group $G=\left\{  1\right\}  $ so we assume that n%
$>$%
1

\subsubsection{General properties}

GL(K,n) is comprised of all square nxn matrices on K which are inversible. It
is a Lie group of dimension n%
${{}^2}$
over K.

Its Lie algebra L(K,n) is comprised of all square nxn matrices on K.\ It is a
Lie algebra on K with dimension n%
${{}^2}$
over K.

The center of GL(K,n) is comprised of scalar matrices $k\left[  I\right]  $

GL(K,n) is not semi simple, not compact, not connected.

$GL(%
\mathbb{C}
,n)$ is the complexified of $GL(%
\mathbb{R}
,n)$

\subsubsection{Representations of GL(K,n)}

\begin{theorem}
All finite dimensional irreducible representations of $GL$($K,n)$ are
alternate\ tensorial product of $\left(  \wedge^{k}K^{n},D_{A}^{k}%
\imath\right)  $ of the standard representation.$\left(
\mathbb{C}
^{n},\imath\right)  .$
\end{theorem}

$\dim\wedge^{k}K^{n}=C_{n}^{k}$.

For k=n we have the one dimensional representation : $\left(  K,\det\right)
.$

The infinite dimensional representations are functional representations

$GL$($%
\mathbb{C}
,n),SL(%
\mathbb{C}
,n)$ are respectively the complexified of $GL$($%
\mathbb{R}
,n),SL(%
\mathbb{R}
,n)$ so the representations of the latter are restrictions of the
representations of the former. For more details about the representations of
$SL$($%
\mathbb{R}
,n)$ see Knapp 1986.

\bigskip

\subsection{SL(K,n) and sl(K,n)}

\subsubsection{General properties}

SL(K,n) is the Lie subgroup of GL(K,n) comprised of all square nxn matrices on
K such that detM=1. It has the dimension n%
${{}^2}$%
-1 over K.

Its Lie algebra sl(K,n) is the set comprised of all square nxn matrices on K
with null trace $sl(K,n)=\{X\in L(K,n):Trace(X)=0\}$.

SL(K,n) is a connected, semi-simple, not compact group.

$SL(%
\mathbb{C}
,n)$ is simply connected, and simple for n%
$>$%
1

$SL(%
\mathbb{R}
,n)$ is not simply connected. For n%
$>$%
1 the universal covering group of $SL(%
\mathbb{R}
,n)$ is not a group of matrices.

The complexified of $sl(%
\mathbb{R}
,n)$ is $sl(%
\mathbb{C}
,n),$ and $SL(%
\mathbb{C}
,n)$ is the complexified of $SL\left(
\mathbb{R}
,n\right)  $

The Cartan algebra of $sl(%
\mathbb{C}
,n)$ is the subset of diagonal matrices.

The simple root system of $sl(%
\mathbb{C}
,n)$ is A$_{n-1},n\geq2:$

$V=\sum_{k=1}^{n}x_{k}e_{k},\sum_{k=1}^{n}x_{k}=0$

$\Delta=e_{i}-e_{j},i\neq j$

$\Pi=\left\{  e_{1}-e_{2},e_{2}-e_{3},..e_{n-1}-e_{n}\right\}  $

The fundamental weights are : $w_{l}=\sum_{k=1}^{l}e_{k},1\leq l\leq n-1$

\begin{theorem}
(Knapp p.340, Fulton p.221) If n
$>$
2 the fundamental representation $\left(  E_{l},f_{l}\right)  $ of SL$\left(
\mathbb{C}
,n\right)  $\noindent for the fundamental weight $w_{l}$ is the alternate
tensorial product $\left(  \Lambda^{l}%
\mathbb{C}
^{n},D_{A}^{l}\imath\right)  $\ of the standard representation $\left(
\mathbb{C}
^{n},\imath\right)  $. The $%
\mathbb{C}
-$dimension of $\Lambda^{l}%
\mathbb{C}
^{n}$ is C$_{n}^{l}$
\end{theorem}

The irreducible finite dimensional representations are then tensorial products
of the fundamental representations.

\subsubsection{Group SL($%
\mathbb{C}
,2)$}

The group SL($%
\mathbb{C}
,2)$ is of special importance, as it is the base of many morphisms with other
groups, and its representations are in many ways the "mother of all representations".

\paragraph{Algebra sl$\left(
\mathbb{C}
,2\right)  $\newline}

1. The Lie algebra $sl(%
\mathbb{C}
,2)$ is the algebra of 2x2 complex matrices with null trace. There are several
possible basis.

2. The most convenient is the following :

$i\sigma_{1}=%
\begin{bmatrix}
0 & i\\
i & 0
\end{bmatrix}
;i\sigma_{2}=%
\begin{bmatrix}
0 & 1\\
-1 & 0
\end{bmatrix}
;i\sigma_{3}=%
\begin{bmatrix}
i & 0\\
0 & -i
\end{bmatrix}
$

with the Pauli's matrices :

$\sigma_{1}=%
\begin{bmatrix}
0 & 1\\
1 & 0
\end{bmatrix}
;\sigma_{2}=%
\begin{bmatrix}
0 & -i\\
i & 0
\end{bmatrix}
;\sigma_{3}=%
\begin{bmatrix}
1 & 0\\
0 & -1
\end{bmatrix}
$

Then any matrix of $sl\left(
\mathbb{C}
,2\right)  $ can be written as :

$\sum_{j=1}^{3}z_{j}i\sigma_{j}=%
\begin{bmatrix}
iz_{3} & z_{2}+iz_{1}\\
-z_{2}+iz_{1} & -iz_{3}%
\end{bmatrix}
\in sl(2,C),$

that we denote $N=f\left(  z\right)  $ and the components are given by a
vector $z\in%
\mathbb{C}
^{3}$

We have the identities :

$\det f\left(  z\right)  =z_{3}^{2}+z_{2}^{2}+z_{1}^{2}$

$f\left(  z\right)  f\left(  z\right)  =-\det f\left(  z\right)  I$

$\left[  f\left(  z\right)  ,f\left(  z^{\prime}\right)  \right]  =2f\left(
\left(  -z_{2}z_{3}^{\prime}+z_{3}z_{2}^{\prime}\right)  ,z_{1}z_{3}^{\prime
}-z_{3}z_{1}^{\prime},\left(  -z_{1}z_{2}^{\prime}+z_{2}z_{1}^{\prime}\right)
\right)  =-2f\left(  j\left(  z\right)  z^{\prime}\right)  $

where j is the map (for any field K) :

$j:K^{3}\rightarrow L\left(  K,3\right)  ::j\left(  z\right)  =%
\begin{bmatrix}
0 & -z_{3} & z_{2}\\
z_{3} & 0 & -z_{1}\\
-z_{2} & z_{1} & 0
\end{bmatrix}
$

Using the properties above it is easy to prove that :

$\exp f\left(  z\right)  =\sum_{n=0}^{\infty}\frac{1}{\left(  2n\right)
!}\left(  -\det f\left(  z\right)  \right)  ^{n}I+\sum_{n=0}^{\infty}\frac
{1}{\left(  2n+1\right)  !}\left(  -\det f\left(  z\right)  \right)
^{n}f\left(  z\right)  $

By denoting $D\in%
\mathbb{C}
$ one of the solution of $:D^{2}=-\det f\left(  z\right)  $

$\exp f\left(  z\right)  =I\cosh D+\frac{\sinh D}{D}f\left(  z\right)  $

with the convention : $z^{t}z=0\Rightarrow\exp f\left(  z\right)  =I+f\left(
z\right)  $

One can see that $\det\left(  \pm\left(  I\cosh D+\frac{\sinh D}{D}f\left(
z\right)  \right)  \right)  =1$ so the exponential is not surjective, and :

$\forall g\in SL\left(
\mathbb{C}
,2\right)  ,\exists z\in%
\mathbb{C}
^{3}::g=\exp f\left(  z\right)  $ or $g=-\exp f\left(  z\right)  $

$\left(  \exp f\left(  z\right)  \right)  ^{-1}=\exp\left(  -f\left(
z\right)  \right)  =\exp f\left(  -z\right)  $

3. An usual basis in physics comprises the 3 matrices :

$J_{3}=\frac{1}{2}%
\begin{bmatrix}
-1 & 0\\
0 & 1
\end{bmatrix}
,J_{+}=J_{1}+iJ_{2}=%
\begin{bmatrix}
0 & 0\\
-1 & 0
\end{bmatrix}
,J_{-}=J_{1}-iJ_{2}=%
\begin{bmatrix}
0 & -1\\
0 & 0
\end{bmatrix}
;$

and one denotes $J_{\epsilon}=J_{+},J_{-}$ with $\epsilon=\pm1$

So the bracket has the value : $\left[  J_{+},J_{-}\right]  =2J_{3},\left[
J_{3},J_{\epsilon}\right]  =\epsilon J_{\varepsilon}$

SL(C,2) has several unitary representations which are not finite dimensional

\paragraph{Finite dimensional representation of sl($%
\mathbb{C}
$,2)\newline}

\begin{theorem}
For any n%
$>$%
0, $sl(%
\mathbb{C}
,2)$ has a unique (up to isomorphism) irreducible representation (E,f) on a
complex n dimensional vector space E.
\end{theorem}

1. There are specific notational conventions for these representations, which
are used in physics.

i) The representations of sl(C,2) are labelled by a scalar j which is an
integer or half an integer : j = 1/2,1,3/2,...

ii) The representation (E,f) is then denoted $\left(  E^{j},d^{j}\right)  $
and the dimension of the vector space $E^{j}$ is 2j

The vectors of a basis of $E^{j}$ are denoted
$\vert$%
j,m%
$>$
with m integer or half integer, and varies by increment of 1,$-j\leq m\leq j$

n even, j integer : m = -j, -j+1,...,-1, +1,...j-1,j

n odd, j half integer : m = -j, -j+1,...,-3/2,0, +3/2,...j-1,j

iii) f is defined by computing its matrix on a basis of E. The morphism
$d^{j}$ is defined by the action of the vector of a basis of sl(C,2) on the
vectors of $E^{j}$

With the basis $\left(  i\sigma_{j}\right)  :$

$d^{j}\left(  i\sigma_{1}\right)  =-i\left(  \sqrt{\left(  j-m\right)  \left(
j+m-1\right)  }|j,m+1>+\sqrt{\left(  j+m\right)  \left(  j-m+1\right)
}|j,m-1>\right)  $

$d^{j}\left(  i\sigma_{2}\right)  =\sqrt{\left(  j-m\right)  \left(
j+m-1\right)  }|j,m+1>-\sqrt{\left(  j+m\right)  \left(  j-m+1\right)
}|j,m-1>$

$d^{j}\left(  i\sigma_{3}\right)  =-2m|j,m>$

With the basis $J_{3},J_{+},J_{-}$:

$d^{j}\left(  J_{3}\right)  \left(  |j,m>\right)  =m|j,m>$

$d^{j}\left(  J_{\epsilon}\right)  \left(  |j,m>\right)  =\sqrt{j\left(
j+1\right)  -m\left(  m+\epsilon\right)  }\left(  |j,m+\epsilon>\right)  $

As one can check the matrix deduced from this action is not antihermitian so
the \textit{representation at the group level is not unitary}.

2. Casimir elements: the elements of the representation of the universal
envelopping algebra are expressed as products of matrices. The Casimir element
$d^{j}\left(  \Omega\right)  $ is represented by the matrix $J^{2}=J_{1}%
^{2}+J_{2}^{2}+J_{3}^{2}=J_{+}J_{-}+J_{3}\left(  J_{3}-I\right)  =J_{-}%
J_{+}+J_{3}\left(  J_{3}+I\right)  .$ It acts by scalar :

$d^{j}\left(  \Omega_{2}\right)  |j,m>=j\left(  j+1\right)
|j,m>\Leftrightarrow\left[  d^{j}\left(  \Omega_{2}\right)  \right]  =j\left(
j+1\right)  \left[  I\right]  $

3. A sesquilinear form over $E^{j}$ is defined by taking $|j,m>$ as an
orthonormal basis : $<m\prime,j||j,m>=\delta_{mm^{\prime}}:$

$\left\langle \sum_{m}x^{m}|j,m>,\sum_{m}y^{m}|j,m>\right\rangle =\sum
_{m}\overline{x}^{m}y^{m}$

$E^{j}$ becomes a Hilbert space, and the adjoint of an operator has for matrix
in this basis the adjoint of the matrix of the operator : $\left[
d^{j}\left(  X\right)  ^{\ast}\right]  =\left[  d^{j}\left(  X\right)
\right]  ^{\ast}$ and we have : $\left[  d^{j}\left(  J_{\epsilon}\right)
^{\ast}\right]  =\left[  d^{j}\left(  J_{\epsilon}\right)  \right]  ^{\ast
}=\left[  d^{j}\left(  J_{-\epsilon}\right)  \right]  ,\left[  d^{j}\left(
J_{3}\right)  ^{\ast}\right]  =\left[  d^{j}\left(  J_{3}\right)  \right]
,\left[  d^{j}\left(  \Omega_{2}\right)  ^{\ast}\right]  =\left[  d^{j}\left(
\Omega_{r}\right)  \right]  $ so $J_{3},\Omega$ are hermitian operators. But
notice that $d^{j}$ itself is not antihermitian.

\paragraph{Finite dimensional representations of SL($%
\mathbb{C}
$,2)\newline}

\begin{theorem}
Any finite dimensional representation of the Lie algebra sl(C,2) lifts to a
representation of the group SL(C,2) and can be computed by the exponential of matrices.
\end{theorem}

\begin{proof}
Any $g\in SL\left(
\mathbb{C}
,2\right)  $ can be written as : $g=\epsilon\exp X$ for a unique $X\in sl(%
\mathbb{C}
,2),\epsilon=\pm1$

Take : $\Phi\left(  g\left(  t\right)  \right)  =\epsilon\exp\left(
tD^{j}\left(  X\right)  \right)  $

$\frac{\partial g}{\partial t}|_{t=0}=\Phi^{\prime}\left(  g\right)
|_{g=1}=\epsilon D^{j}\left(  X\right)  $

so, as SL(C,2) is simply connected $\left(  E,\epsilon\exp\left(  D^{j}\left(
X\right)  \right)  \right)  $ is a representation of SL(C,2).

As the vector spaces are finite dimensional, the exponential of the morphisms
can be computed as exponential of matrices.
\end{proof}

As the computation of these exponential is complicated, the finite dimensional
representations of SL(C,2) are obtained more easily as functional
representations on spaces of polynomials (see below).

\paragraph{Infinite dimensional representations of SL($%
\mathbb{C}
$,2)\newline}

1. The only unitary representations of $SL(%
\mathbb{C}
,2)$ are infinite dimensional.

2. Functional representations can be defined over a Banach vector space of
functions through a left or a right action. They have all be classified.

\begin{theorem}
(Knapp 1986 p. 31) The only irreducible representations (other than the
trivial one) (H,f) of SL(C,2) are the following :
\end{theorem}

i) The principal unitary series :

on the Hilbert space H of functions $\varphi:%
\mathbb{R}
^{2}\rightarrow%
\mathbb{C}
$ such that : $\int_{%
\mathbb{R}
^{2}}\left\vert \varphi\left(  x+iy\right)  \right\vert ^{2}dxdy<\infty$ with
the scalar product : $\left\langle \varphi,\psi\right\rangle =\int_{%
\mathbb{R}
}\overline{\varphi\left(  x+iy\right)  }\psi\left(  x+iy\right)  dxdy$

the morphisms f are parametrized by two scalars $\left(  k,v\right)
\in\left(
\mathbb{Z}
,%
\mathbb{R}
\right)  $ and z=x+iy

$f_{k,v}\left(
\begin{bmatrix}
a & b\\
c & d
\end{bmatrix}
\right)  \varphi(z)=-\left\vert -bz+d\right\vert ^{-2-iv}\left(  \frac
{-bz+d}{\left\vert -bz+d\right\vert }\right)  ^{-k}\varphi\left(  \frac
{az-c}{-bz+d}\right)  $

We have : $\left(  H,f_{k,v}\right)  \sim\left(  H,f_{-k,-v}\right)  $

ii) The non unitary principal series :

on the Hilbert space H of functions $\varphi:%
\mathbb{R}
^{2}\rightarrow%
\mathbb{C}
$ such that :

$\int_{%
\mathbb{R}
^{2}}\left\vert \varphi\left(  x+iy\right)  \right\vert ^{2}\left(
1+\left\vert x+iy\right\vert ^{2}\right)  ^{\operatorname{Re}w}dxdy<\infty$
with the scalar product :

$\left\langle \varphi,\psi\right\rangle =\int_{%
\mathbb{R}
^{2}}\overline{\varphi\left(  x+iy\right)  }\psi\left(  x+iy\right)  \left(
1+\left\vert x+iy\right\vert ^{2}\right)  ^{\operatorname{Re}w}dxdy$

the morphisms f are parametrized by two scalars $\left(  k,v\right)
\in\left(
\mathbb{Z}
,%
\mathbb{C}
\right)  $

$f_{k,w}\left(
\begin{bmatrix}
a & b\\
c & d
\end{bmatrix}
\right)  \varphi(z)=-\left\vert -bz+d\right\vert ^{-2-w}\left(  \frac
{-bz+d}{\left\vert -bz+d\right\vert }\right)  ^{-k}\varphi\left(  \frac
{az-c}{-bz+d}\right)  $

If w if purely imaginary we get back the previous series.

the non unitary principal series contain all the finite dimensional
irreducible representations, by taking : H= polynomial of degree m in z and of
degree n in $\overline{z}$

iii) the complementary unitary series :

on the Hilbert space H of functions $\varphi:%
\mathbb{C}
\rightarrow%
\mathbb{C}
$ with the scalar product : $\left\langle \varphi,\psi\right\rangle =\int_{%
\mathbb{C}
\times%
\mathbb{C}
}\frac{\overline{\varphi(z_{1})}\psi\left(  z_{2}\right)  }{\left\vert
z_{1}-z_{2}\right\vert ^{-2-w}}dz_{1}dz_{2}=\int_{%
\mathbb{C}
\times%
\mathbb{C}
}\overline{\varphi(z_{1})}\psi\left(  z_{2}\right)  \nu$

the morphisms f are parametrized by two scalars $\left(  k,w\right)
\in\left(
\mathbb{Z}
,]0,2[\subset%
\mathbb{R}
\right)  $

$f_{k,w}\left(
\begin{bmatrix}
a & b\\
c & d
\end{bmatrix}
\right)  f(z)=-\left\vert -bz+d\right\vert ^{-2-w}\left(  \frac{-bz+d}%
{\left\vert -bz+d\right\vert }\right)  ^{-k}f\left(  \frac{az-c}%
{-bz+d}\right)  $

\bigskip

\subsection{Orthogonal groups}

\subsubsection{Definitions and general properties}

1. On a vector space of dimension n on the field K, endowed with a
\textit{bilinear symmetric} form g, the orthogonal group O(E,g) is the group
of endomorphisms f which preserve the scalar product : $\forall u,v\in
E:g\left(  f\left(  u\right)  ,f\left(  v\right)  \right)  =g\left(
u,v\right)  .$

SO(E,g) is the subgroup of O(E,g) of endomorphisms with det(f) =1.

2. If $K=$ $%
\mathbb{C}
$ a bilinear symmetric form g can always be written : $g=\sum_{jk}\delta
_{jk}e^{j}\otimes e^{k}$ and

O(E,g) is isomorphic to $O\left(
\mathbb{C}
,n\right)  ,$ the linear group of matrices of $GL\left(
\mathbb{C}
,n\right)  $ such that $M^{t}M=I_{n}.$

SO(E,g) is isomorphic to $SO\left(
\mathbb{C}
,n\right)  ,$ the linear group of matrices of $GL\left(
\mathbb{C}
,n\right)  $ such that $M^{t}M=I_{n}$ detM = 1.

3. If $K=$ $%
\mathbb{R}
$ the group depends on the signature (p,q) with p + q = n of the bilinear
symmetric form g

O(E,g) is isomorphic to $O\left(
\mathbb{R}
,p,q\right)  ,$ the linear group of matrices of $GL\left(
\mathbb{R}
,n\right)  $ such that $M^{t}\left[  \mathbf{\eta}_{p,q}\right]  M=\left[
\mathbf{\eta}_{p,q}\right]  .$

SO(E,g) is isomorphic to $SO\left(
\mathbb{R}
,p,q\right)  ,$ the linear group of matrices of $GL\left(
\mathbb{R}
,n\right)  $ such that $M^{t}\left[  \mathbf{\eta}_{p,q}\right]  M=\left[
\mathbf{\eta}_{p,q}\right]  $ detM = 1.

with $\left[  \mathbf{\eta}_{p,q}\right]  =\left[
\begin{array}
[c]{cc}%
I_{p} & 0\\
0 & -I_{q}%
\end{array}
\right]  $

If p or q =0 the groups are simply denoted $O\left(
\mathbb{R}
,n\right)  ,SO\left(
\mathbb{R}
,n\right)  $

4. In all cases the standard representation $\left(  K^{n},\imath\right)  $ is
on $K^{n}$ endowed with the bilinear form g, with the proper signature.\ So
the canonical basis is orthonormal.

\subsubsection{Groups O(K,n), SO(K,n)}

\paragraph{General properties\newline}

$K=%
\mathbb{R}
,%
\mathbb{C}
$

Their dimension is n(n-1)/2 over K, and their Lie algebra is: $o(K,n)=\{X\in
L(K,n):X+X^{t}=0\}$.

O(K,n), SO(K,n) are semi-simple for n%
$>$%
2

O(K,n), SO(K,n) are compact.

O(K,n) has two connected components, with detM=+1 and detM=-1. The connected
components are not simply connected.

SO(K,n) is the connected component of the identity of O(K,n)

SO(K,n) is not simply connected. The universal covering group of SO(K,n) is
the Spin group Spin(K,n) (see below) which is a double cover.

$so(%
\mathbb{C}
,n)$ is the complexified of $so(%
\mathbb{R}
,n)$, $SO(%
\mathbb{C}
,n)$ is the complexified of $SO(%
\mathbb{R}
,n)$

Roots system of $so(%
\mathbb{C}
,n):$ it depends upon the parity of n

For $so(%
\mathbb{C}
,2n+1),n\geq1:B_{n}$ system:

$V=%
\mathbb{R}
^{n}$

$\Delta=\left\{  \pm e_{i}\pm e_{j},i<j\right\}  \cup\left\{  \pm
e_{k}\right\}  $

$\Pi=\left\{  e_{1}-e_{2},e_{2}-e_{3},..e_{n-1}-e_{n},e_{n}\right\}  $

For $so(%
\mathbb{C}
,2n),n\geq2:D_{n}$ system:

$V=%
\mathbb{R}
^{n}$

$\Delta=\left\{  \pm e_{i}\pm e_{j},i<j\right\}  $

$\Pi=\left\{  e_{1}-e_{2},e_{2}-e_{3},..e_{n-1}-e_{n},e_{n-1}+e_{n}\right\}  $

\paragraph{Group SO($%
\mathbb{R}
,3)$\newline}

This is the group of rotations in the euclidean space.\ As it is used very
often in physics it is good to give more details and some useful computational results.

1. The algebra $o(%
\mathbb{R}
;3)$ is comprised of 3x3 skewsymmetric matrices.

Take as basis for $o(%
\mathbb{R}
;3)$ the matrices :

\bigskip

$\kappa_{1}=%
\begin{bmatrix}
0 & 0 & 0\\
0 & 0 & -1\\
0 & 1 & 0
\end{bmatrix}
;\kappa_{2}=%
\begin{bmatrix}
0 & 0 & 1\\
0 & 0 & 0\\
-1 & 0 & 0
\end{bmatrix}
;\kappa_{3}=%
\begin{bmatrix}
0 & -1 & 0\\
1 & 0 & 0\\
0 & 0 & 0
\end{bmatrix}
$

\bigskip

then a matrix of o(3) reads with the operator :

\bigskip

$j:%
\mathbb{R}
\left(  3,1\right)  \rightarrow o(%
\mathbb{R}
;3)::j\left(
\begin{bmatrix}
r_{1}\\
r_{2}\\
r_{3}%
\end{bmatrix}
\right)  =%
\begin{bmatrix}
0 & -r_{3} & r_{2}\\
r_{3} & 0 & -r_{1}\\
-r_{2} & r_{1} & 0
\end{bmatrix}
$

\bigskip

which has some nice properties :

$j(r)^{t}=-j(r)=j(-r)$

$j(x)y=-j(y)x=x\times y$

(this is just the "vector product $\times$" of elementary geometry)

$y^{t}j(x)=-x^{t}j(y)$

$j(x)y=0\Leftrightarrow\exists k\in R:y=kx$

$j(x)j(y)=yx^{t}-\left(  y^{t}x\right)  I$

$j(j(x)y)=yx^{t}-xy^{t}=j(x)j(y)-j(y)j(x)$

$j(x)j(y)j(x)=-\left(  y^{t}x\right)  j(x)$

$M\in L(%
\mathbb{R}
,3):M^{t}j(Mx)M=\left(  \det M\right)  j(x)$

$M\in O(%
\mathbb{R}
,3):j(Mx)My=Mj(x)y\Leftrightarrow Mx\times My=M\left(  x\times y\right)  $

$k>0:j(r)^{2k}=\left(  -r^{t}r\right)  ^{k-1}j(r)j(r)$

$k\geq0:J(r)^{2k+1}=\left(  -r^{t}r\right)  ^{k}j(r)$

2. The group $SO$($%
\mathbb{R}
,3)$ is compact, thus the exponential is onto and any matrix can be written as :

$\exp\left(  j(r)\right)  =I_{3}+j(r)\frac{\sin\sqrt{r^{t}r}}{\sqrt{r^{t}r}%
}+j(r)j(r)\frac{1-\cos\sqrt{r^{t}r}}{r^{r}r}$

The eigen values of g=expj(r) are $\left(  1;\exp i\sqrt{r_{1}^{2}+r_{2}%
^{2}+r_{3}^{2}};\exp\left(  -i\sqrt{r_{1}^{2}+r_{2}^{2}+r_{3}^{2}}\right)
\right)  $

3. The axis of rotation is by definition the unique eigen vector of expj(r)
with eigen value 1. It is easy to see that its components are proportional to
$\left[  r\right]  =\left[
\begin{array}
[c]{c}%
r^{1}\\
r^{2}\\
r^{3}%
\end{array}
\right]  $ .

For any vector u of norm 1 : $\left\langle u,gu\right\rangle =\cos\theta$
where $\theta$ is an angle which depends on u.

With the formula above, it is easy to see that

$\left\langle u,\exp\left(  j(r)\right)  u\right\rangle =1+\left(
\left\langle u,r\right\rangle ^{2}-\left\langle r,r\right\rangle \right)
\frac{1-\cos\sqrt{\left\langle r,r\right\rangle }}{\left\langle
r,r\right\rangle }$

which is minimum for $\left\langle u,r\right\rangle =0$ that is for the
vectors orthogonal to the axis, and : $\cos\theta=\cos\sqrt{\left\langle
r,r\right\rangle }$. So we can define the angle of rotation by the scalar
$\sqrt{\left\langle r,r\right\rangle }.$

4. The universal covering group of $SO\left(
\mathbb{R}
,3\right)  $ is Spin($%
\mathbb{R}
,3)$ isomorphic to SU(2) which is a subgroup of $SL(%
\mathbb{C}
,2)$.

su(2) and so(3) are isomorphic, with $r\in%
\mathbb{R}
,$ by :

$\sum_{i=1}^{3}ir_{i}\sigma_{i}\in su(2)\rightarrow j\left(  r\right)  \in
so(3)$

SU(2) is isomorphic to Spin($%
\mathbb{R}
,3)$ by

$\exp\left(  \sum_{i=1}^{3}ir_{i}\sigma_{i}\right)  \in SU(2)\rightarrow
\pm\exp j\left(  r\right)  $

We go from so(3) to SO(3) by :

$\left(  \sum_{i=1}^{3}r_{i}\kappa_{i}\right)  \in so\left(  3\right)
\rightarrow\exp j\left(  r\right)  $

\paragraph{Representations of SO(K,n)\newline}

\begin{theorem}
Any irreducible representation of SO(K,n) is finite dimensional. The
representations of $SO(%
\mathbb{R}
,n)$ can be deduced by restriction from the representations of $SO(%
\mathbb{C}
,n)$.
\end{theorem}

\begin{proof}
SO(K,n) is semi-simple for n%
$>$%
2, compact, connected, not simply connected.

$so(%
\mathbb{C}
,n)$ is the complexified of $so(%
\mathbb{R}
,n)$, $SO(%
\mathbb{C}
,n)$ is the complexified of $SO(%
\mathbb{R}
,n)$

As the groups are compact, all unitary irreducible representations are finite dimensional.
\end{proof}

\begin{theorem}
(Knapp p.344) so(C,m) has two sets of fundamental weights and fundamental
representations according to the parity of n:

i) m odd : m = 2n + 1 : $so(C,2n+1),n\geq2$ belongs to the $B_{n}$ family

Fundamental weights : $w_{l}=\sum_{k=1}^{l}e_{k},l\leq n-1$ and $w_{n}%
=\frac{1}{2}\sum_{k=1}^{n}e_{k}$

The fundamental representation for l
$<$
n \ is the tensorial product $\left(  \Lambda^{l}%
\mathbb{C}
^{2n+1},D_{A}^{l}\imath\right)  $ of the standard representation $(%
\mathbb{C}
^{2n+1},\imath)$ on orthonormal bases. The $%
\mathbb{C}
-$dimension of $\Lambda^{l}%
\mathbb{C}
^{2n+1}$ is $C_{2n+1}^{l}$

The fundamental representation for $w_{n}$ is the spin representation
$\Gamma_{2n+1}$

ii) m even : m = 2n : $so(2n,C),n\geq2$ belongs to the $D_{n}$ family

Fundamental weights :

for $l\leq n-2:\ w_{l}=\sum_{k=1}^{l}e_{k},l<n$ and $w_{n-1}=\left(  \frac
{1}{2}\sum_{k=1}^{n-1}e_{k}\right)  -e_{n}$ and $w_{n}=\left(  \frac{1}{2}%
\sum_{k=1}^{n-1}e_{k}\right)  +e_{n}$

The fundamental representation for l
$<$
n \ is the tensorial product $\left(  \Lambda^{l}%
\mathbb{C}
^{2n},D_{A}^{l}f\right)  $ of the standard representation $(%
\mathbb{C}
^{2n+1},\imath)$ on orthonormal bases. The $%
\mathbb{C}
-$dimension of the representation is $C_{2n}^{l}$

The fundamental representation for $w_{n}$ is the spin representation
$\Gamma_{2n}$
\end{theorem}

For l
$<$
n the representations of the group are alternate\ tensorial product of the
standard representation, but for l = n the "spin representations" are
different and come from the Spin groups. Some comments.

i) Spin(K,n) and SO(K,n) have the same Lie algebra, isomorphic to so(K,n). So
they share the same representations for their Lie algebras. These
representations are computed by the usual method of roots, and the "spin
representations" above correspond to specific value of the roots.

ii) At the group level the picture is different.\ The representations of the
Lie algebras lift to a representation of the double cover, which is the Spin
group. The representations of the Spin groups are deduced from the
representations of the Clifford algebra. For small size of the parameters
isomorphims open the possibility to compute representations of the spin group
by more direct methods.

iii) There is a Lie group morphism : $\pi:Spin(K,n)\rightarrow SO(K,n)::\pi
\left(  \pm s\right)  =g$

If (E,f)\ is a representation of SO(K,n), then $(E,f\circ\pi)$ is a
representation of Spin(K,n). Conversely a representation $(E,\widehat{f})$ of
Spin(K,n) is a representation of SP(K,n) iff it meets the condition :
$f\circ\pi\left(  -s\right)  =\widehat{f}\left(  -s\right)  =f\circ\pi\left(
s\right)  =\widehat{f}\left(  s\right)  $

$\widehat{f}$ must be such that $\widehat{f}\left(  -s\right)  =\widehat
{f}\left(  s\right)  ,$ and we can find all the related representations of
SO(K,n) among the representations of Spin(K,n) which have this symmetry.

The following example - one of the most important - explains the procedure.

\paragraph{Representations of SO($%
\mathbb{R}
,$3)\newline}

1. Irreducible representations:

\begin{theorem}
(Kosmann) The only irreducible representations of $SO(%
\mathbb{R}
,3)$ are equivalent to the $\left(  P^{j},D^{j}\right)  $ of SU(2) with $j\in%
\mathbb{N}
$. The spin representation for j=1 is equivalent to the standard representation.
\end{theorem}

\begin{proof}
To find all the representations of $SO(%
\mathbb{R}
,3)$ we explore the representations of $Spin(%
\mathbb{R}
,3)\simeq SU(2)$ . So we have to look among the representations $\left(
P^{j},D^{j}\right)  $ and find which ones meet the condition above.\ It is
easy to check that j must be an integer (so the half integer representations
are excluded).
\end{proof}

2. Representations by harmonic functions:

It is more convenient to relate the representations to functions in $%
\mathbb{R}
^{3}$\ than to polynomials of complex variables as it comes from SU(2).

The representation $\left(  P^{j},D^{j}\right)  $\ is equivalent to the
representation :

i) Vector space : the homogeneous polynomials $\varphi\left(  x_{1}%
,x_{2},x_{3}\right)  $ on $%
\mathbb{R}
^{3}$ (meaning three real variables) with degree j with complex coefficients,
which are harmonic, that is : $\Delta\varphi=0$ where $\Delta$\ is the
laplacian $\Delta=\sum_{k=1}^{3}\frac{\partial^{2}}{\partial x_{k}^{2}}$\ \ .
This is a 2j+1 vector space, as $P^{j}$

ii) Morphism : the left regular representation $\Lambda$, meaning $f\left(
g\right)  \varphi\left(  x\right)  =\varphi\left(  \left[  g\right]
_{3\times3}^{-1}\left[  x\right]  _{3\times1}\right)  $

So $f\left(  \exp j(r)\right)  \varphi\left(  x\right)  =\varphi\left(
\left(  I_{3}-j(r)\frac{\sin\sqrt{r^{t}r}}{\sqrt{r^{t}r}}+j(r)j(r)\frac
{1-\cos\sqrt{r^{t}r}}{r^{r}r}\right)  \left[  x\right]  \right)  $

3. Representation by spherical harmonics:

In order to have a unitary representation we need a Hilbert space. Harmonic
functions are fully defined by their value on any sphere centered at 0. Let
$L^{2}\left(  S^{2},\sigma,%
\mathbb{C}
\right)  $ the space of function defined over the shere S%
${{}^2}$
of $%
\mathbb{R}
^{3}$ and valued in $%
\mathbb{C}
,$ endowed with the sesquilinear form : $\left\langle \varphi,\psi
\right\rangle =\int_{S^{2}}\overline{\varphi\left(  x\right)  }\psi\left(
x\right)  \sigma\left(  x\right)  $ where $\sigma$ is the measure induced on S%
${{}^2}$
by the Lebesgue measure of $%
\mathbb{R}
^{3}.$

This is a Hilbert space, and the set of harmonic homogeneous polynomials of
degree j on $%
\mathbb{R}
^{3}$ is a closed vector subspace, so a Hilbert space denoted $H^{j}$. Thus
$\left(  H^{j},\Lambda\right)  $ is a unitary irreducible representation of
SO(R,3) equivalent to $\left(  P^{j},D^{j}\right)  $\ .

$L^{2}\left(  S^{2},\sigma,%
\mathbb{C}
\right)  $ is the Hilbert sum of the $H^{j}:$ $L^{2}\left(  S^{2},\sigma,%
\mathbb{C}
\right)  =\oplus_{j\in%
\mathbb{N}
}H^{j}$

By derivation we get a representations of the Lie algebra so(R,3) on the same
Hilbert spaces $H^{j}$ and the elements of so(R,3) are represented by
antihermitian differential operators on $H^{j}:$

$D\Lambda\left(  j\left(  r\right)  \right)  =j\left(  r\right)  \left[
\frac{\partial}{\partial x_{i}}\right]  $

The Casimir operator $D\Lambda\left(  \Omega\right)  =-\Delta_{S^{2}}$ where
$\Delta_{S^{2}}$ is the spherical laplacian defined on S%
${{}^2}$%
. Its spectrum is discrete, with values j(j+1).

As we stay on S%
${{}^2}$
it is convenient to use spherical coordinates :

$x_{1}=\rho\sin\theta\cos\phi,x,x_{2}=\rho\sin\theta\sin\phi,x_{3}=\rho
\cos\phi$

and on S%
${{}^2}$
: $\rho=1$ so $\varphi\in H^{j}:\varphi\left(  \theta,\phi\right)  $

A Hilbert basis of $L^{2}\left(  S^{2},\sigma,%
\mathbb{C}
\right)  $ (thus orthonormal) comprises the vectors, called the
\textbf{spherical harmonics} :

$j\in%
\mathbb{N}
,-j\leq m\leq+j:|j,m>=Y_{m}^{j}\left(  \theta,\phi\right)  $

$m\geq0:Y_{m}^{j}\left(  \theta,\phi\right)  =C_{m}^{j}Z_{m}^{j}\left(
\theta\right)  e^{im\phi},Z_{m}^{j}\left(  \theta\right)  =\left(  \sin
^{m}\theta\right)  Q_{m}^{j}\left(  \cos\theta\right)  ;Q_{m}^{j}\left(
z\right)  =\frac{d^{j+m}}{dz^{j+m}}\left(  1-z^{2}\right)  ^{j};C_{m}%
^{j}=\frac{\left(  -1\right)  ^{j+m}}{2^{l}l!}\sqrt{\frac{2j+1}{4\pi}}%
\sqrt{\frac{\left(  j-m\right)  !}{\left(  j+m\right)  !}}$

$m<0:Y_{m}^{j}=\left(  -1\right)  ^{m}\overline{Y_{-m}^{j}}$

which are eigen vectors of $D\Lambda\left(  \Omega\right)  $ with the eigen
value j(j+1)

\subsubsection{Special orthogonal groups O($%
\mathbb{R}
$,p,q),SO($%
\mathbb{R}
$,p,q)}

\paragraph{General properties\newline}

p
$>$
0,q
$>$
0, p+q= n
$>$%
1

1. $O(%
\mathbb{R}
,p,q)$, $SO(%
\mathbb{R}
,p,q)$ have dimension n(n-1)/2, and their Lie algebra is comprised of the real
matrices :

$o(%
\mathbb{R}
,p,q)=\{X\in L(%
\mathbb{R}
,n):\left[  \mathbf{\eta}_{p,q}\right]  X+X^{t}\left[  \mathbf{\eta}%
_{p,q}\right]  =0\}$.

$O(%
\mathbb{R}
,p,q),O\left(
\mathbb{R}
,q,p\right)  $ are identical : indeed $\left[  \mathbf{\eta}_{p,q}\right]
=-\left[  \mathbf{\eta}_{q,p}\right]  $

2. $O(%
\mathbb{R}
,p,q)$ has four connected components, and each component is not simply connected.

$SO(%
\mathbb{R}
,p,q)$ is not connected, and has two connected components. Usually one
considers the connected component of the identity $SO_{0}(%
\mathbb{R}
,p,q)$. The universal covering group of $SO_{0}(%
\mathbb{R}
,p,q)$ is $Spin(%
\mathbb{R}
,p,q)$.

3. $O(%
\mathbb{R}
,p,q)$, $SO(%
\mathbb{R}
,p,q)$ are semi-simple for n%
$>$%
2

4. $O(%
\mathbb{R}
,p,q)$ is not compact.\ The maximal compact subgroup is $O(%
\mathbb{R}
,p)\times O(%
\mathbb{R}
,q)$.

$SO(%
\mathbb{R}
,p,q)$ is not compact.\ The maximal compact subgroup is $SO(%
\mathbb{R}
,p)\times SO(%
\mathbb{R}
,q)$.

5. $O(%
\mathbb{C}
,p+q)$ is the complexified of $O(%
\mathbb{R}
,p,q)$.

$SO(%
\mathbb{C}
,p+q)$ is the complexified of $SO(%
\mathbb{R}
,p,q)$.

The group $SO(%
\mathbb{C}
,p,q)$ is isomorphic to $SO(%
\mathbb{C}
,p+q)$

\paragraph{Cartan decomposition\newline}

$O(%
\mathbb{R}
,p,q)$ is invariant by transpose, and admits a Cartan decomposition :

$o(%
\mathbb{R}
,p,q)=l_{0}\oplus p_{0}$ with : $l_{0}=\left\{  l=%
\begin{bmatrix}
M_{p\times p} & 0\\
0 & N_{q\times q}%
\end{bmatrix}
\right\}  ,p_{0}=\left\{  p=%
\begin{bmatrix}
0 & P_{p\times q}\\
P_{q\times p}^{t} & 0
\end{bmatrix}
\right\}  $

$\left[  l_{0},l_{0}\right]  \subset l_{0},\left[  l_{0},p_{0}\right]  \subset
p_{0},\left[  p_{0},p_{0}\right]  \subset l_{0}$

So the maps :

$\lambda:l_{0}\times p_{0}\rightarrow SO(%
\mathbb{R}
,p,q)::\lambda\left(  l,p\right)  =\left(  \exp l\right)  \left(  \exp
p\right)  ; $

$\rho:p_{0}\times l_{0}\rightarrow SO(%
\mathbb{R}
,p,q)::\rho\left(  p,l\right)  =\left(  \exp p\right)  \left(  \exp l\right)
;$

are diffeomorphisms;

It can be proven (see Algebra - Matrices) that :

i) the Killing form is $B\left(  X,Y\right)  =\frac{n}{2}Tr\left(  XY\right)
$

ii) $\exp p=%
\begin{bmatrix}
I_{p}+H\left(  \cosh D-I_{q}\right)  H^{t} & H(\sinh D)U^{t}\\
U(\sinh D)H^{t} & U(\cosh D)U^{t}%
\end{bmatrix}
$ with $H_{p\times q}$ such that : $H^{t}H=I_{q},P=HDU^{t}$ where D is a real
diagonal qxq matrix and U is a qxq real orthogonal matrix. The powers of
exp(p) can be easily deduced.

\paragraph{Group SO($%
\mathbb{R}
,3,1)$\newline}

This is the group of rotations in the Minkovski space (one considers also $SO(%
\mathbb{R}
,1,3)$ which is the same).

1. If we take as basis of the algebra the matrices :

\bigskip

$l_{0}:\kappa_{1}=%
\begin{bmatrix}
0 & 0 & 0 & 0\\
0 & 0 & -1 & 0\\
0 & 1 & 0 & 0\\
0 & 0 & 0 & 0
\end{bmatrix}
;\kappa_{2}=%
\begin{bmatrix}
0 & 0 & 1 & 0\\
0 & 0 & 0 & 0\\
-1 & 0 & 0 & 0\\
0 & 0 & 0 & 0
\end{bmatrix}
;\kappa_{3}=%
\begin{bmatrix}
0 & -1 & 0 & 0\\
1 & 0 & 0 & 0\\
0 & 0 & 0 & 0\\
0 & 0 & 0 & 0
\end{bmatrix}
$

$p_{0}:\kappa_{4}=%
\begin{bmatrix}
0 & 0 & 0 & 1\\
0 & 0 & 0 & 0\\
0 & 0 & 0 & 0\\
1 & 0 & 0 & 0
\end{bmatrix}
;\kappa_{5}=%
\begin{bmatrix}
0 & 0 & 0 & 0\\
0 & 0 & 0 & 1\\
0 & 0 & 0 & 0\\
0 & 1 & 0 & 0
\end{bmatrix}
;\kappa_{6}=%
\begin{bmatrix}
0 & 0 & 0 & 0\\
0 & 0 & 0 & 0\\
0 & 0 & 0 & 1\\
0 & 0 & 1 & 0
\end{bmatrix}
$

\bigskip

It is easy to show that the map j of SO(3) extends to a map :

\bigskip

$J:%
\mathbb{R}
\left(  3,1\right)  \rightarrow o(%
\mathbb{R}
;3,1)::J\left(
\begin{bmatrix}
r_{1}\\
r_{2}\\
r_{3}%
\end{bmatrix}
\right)  =%
\begin{bmatrix}
0 & -r_{3} & r_{2} & 0\\
r_{3} & 0 & -r_{1} & 0\\
-r_{2} & r_{1} & 0 & 0\\
0 & 0 & 0 & 0
\end{bmatrix}
$

with the same identities as above with j.

We have similarly :

\bigskip

$K:%
\mathbb{R}
\left(  3,1\right)  \rightarrow o(%
\mathbb{R}
;3,1)::K\left(
\begin{bmatrix}
v_{1}\\
v_{2}\\
v_{3}%
\end{bmatrix}
\right)  =%
\begin{bmatrix}
0 & 0 & 0 & v_{1}\\
0 & 0 & 0 & v_{2}\\
0 & 0 & 0 & v_{3}\\
v_{1} & v_{2} & v_{3} & 0
\end{bmatrix}
$

\bigskip

And $\forall X\in O\left(
\mathbb{R}
,3,1\right)  :\exists r,v\in%
\mathbb{R}
\left(  3,1\right)  :X=J(r)+K(v)$

The identities above read:

\bigskip

$\exp K(v)=%
\begin{bmatrix}
I_{3}+\left(  \cosh\sqrt{v^{t}v}-1\right)  \frac{vv^{t}}{v^{t}v} & \frac
{v}{\sqrt{v^{t}v}}\left(  \sinh\sqrt{v^{t}v}\right) \\
\left(  \sinh\sqrt{v^{t}v}\right)  \frac{v^{t}}{\sqrt{v^{t}v}} & \cosh
\sqrt{v^{t}v}%
\end{bmatrix}
$

\bigskip

that is :

$\exp K(v)=I_{4}+\frac{\sinh\sqrt{v^{t}v}}{\sqrt{v^{t}v}}K(v)+\frac{\cosh
\sqrt{v^{t}v}-1}{v^{t}v}K(v)K(v)$

Similarly :

$\exp J(r)=I_{4}+\frac{\sin\sqrt{r^{t}r}}{\sqrt{r^{t}r}}J(r)+\frac{1-\cos
\sqrt{r^{t}r}}{r^{r}r}J(r)J(r)$

\paragraph{Isomorphism SL($%
\mathbb{C}
,2)$ with Spin($%
\mathbb{R}
,3,1)$\newline}

\begin{theorem}
There are Lie algebra isomorphisms :

$\phi:so(%
\mathbb{R}
,3,1)\rightarrow sl(%
\mathbb{C}
,2)::\phi\left(  J\left(  r\right)  +K\left(  w\right)  \right)  =-\frac{1}%
{2}f\left(  r+\epsilon iw\right)  $ with $\epsilon=\pm1$

There are Lie group isomorphisms :

$\Phi:Spin(%
\mathbb{R}
,3,1)\rightarrow SL(%
\mathbb{C}
,2)::$

$\Phi\left(  \epsilon^{\prime}\exp\upsilon\left(  \left[  J(r)+K(w)\right]
\right)  \right)  =\epsilon^{\prime}\left(  I\cosh D-\frac{1}{2}\frac{\sinh
D}{D}f\left(  r+\epsilon iw\right)  \right)  $ with $D^{2}=\frac{1}{4}\left(
r+\epsilon iw\right)  ^{t}\left(  r+\epsilon iw\right)  $
\end{theorem}

\begin{proof}
i) We define a morphism $\phi:so(%
\mathbb{R}
,3,1)\rightarrow sl(%
\mathbb{C}
,2)$ through the map

$f:%
\mathbb{C}
^{3}\rightarrow sl(2,C):f\left(  z\right)  =\sum_{j=1}^{3}z_{j}i\sigma_{j}=%
\begin{bmatrix}
iz_{3} & z_{2}+iz_{1}\\
-z_{2}+iz_{1} & -iz_{3}%
\end{bmatrix}
$

by associating to any matrix N = J(r) + K(w) of $so$($%
\mathbb{R}
,3,1)$ the matrix of $sl(%
\mathbb{C}
,2):\phi\left(  J(r)+K(w)\right)  =f\left(  \alpha w+\beta r\right)  $ where
$\alpha,\beta$ are fixed complex scalars. In order to have a morphism of Lie
algebras :

$\left[  J(r)+K(w),J(r^{\prime})+K(w^{\prime})\right]  =J(j(r)r^{\prime
}-j(w)w^{\prime})+K(j(r)w^{\prime}+j(w)r^{\prime})$

it is necessary and sufficient that :

$\alpha(j(r)w^{\prime}+j(w)r^{\prime})+\beta\left(  j(r)r^{\prime
}-j(w)w^{\prime}\right)  =-2j\left(  \alpha w+\beta r\right)  \left(  \alpha
w^{\prime}+\beta r^{\prime}\right)  $

that is $\beta=-\frac{1}{2};\alpha=\pm i\frac{1}{2};$

So we have two possible, conjugate, isomorphisms :

$\phi\left(  J\left(  r\right)  +K\left(  w\right)  \right)  =-\frac{1}%
{2}f\left(  r+\epsilon iw\right)  $ with $\epsilon=\pm1$

ii) There is an isomorphism of Lie algebras : $\upsilon:so(%
\mathbb{R}
,3,1)\rightarrow T_{1}Spin(%
\mathbb{R}
,3,1)$ which reads :

$\upsilon\left(  J(r)+K(w)\right)  =\frac{1}{4}\sum_{ij}$ $\left(  \left[
J(r)+K(w)\right]  \left[  \eta\right]  \right)  _{j}^{i}\varepsilon_{i}%
\cdot\varepsilon_{j}$

iii) $\phi$ lifts to a group isomorphism through the exponential, from the
universal covering group $Spin\left(
\mathbb{R}
,3,1\right)  $ of $SO(%
\mathbb{R}
,3,1)$ to $SL(%
\mathbb{C}
,2)$, which are both simply connected, but the exponential over the Lie
algebras are not surjective :

$\Phi:Spin\left(
\mathbb{R}
,3,1\right)  \simeq Spin\left(
\mathbb{R}
,1,3\right)  \rightarrow SL\left(
\mathbb{C}
,2\right)  ::$

$\Phi\left(  \pm\exp\upsilon\left(  \left[  J(r)+K(w)\right]  \right)
\right)  =\pm\exp\left(  -\frac{1}{2}f\left(  r+\epsilon iw\right)  \right)  $

But as $\Phi\left(  \exp0\right)  =\Phi\left(  1\right)  =I\cosh0$ and $\Phi$
is a morphism : $\Phi\left(  -1\right)  =-I$

$\Phi\left(  \epsilon^{\prime}\exp\upsilon\left(  \left[  J(r)+K(w)\right]
\right)  \right)  =\epsilon^{\prime}\left(  I\cosh D-\frac{1}{2}\frac{\sinh
D}{D}f\left(  r+\epsilon iw\right)  \right)  $

with $D^{2}=\frac{1}{4}\left(  r+\epsilon iw\right)  ^{t}\left(  r+\epsilon
iw\right)  $ is a group isomorphism.
\end{proof}

To $\exp\left(  \left[  J(r)+K(w)\right]  \right)  \in SO_{0}\left(
\mathbb{R}
,3,1\right)  $ one can associate :

$I\cosh D-\frac{1}{2}\frac{\sinh D}{D}f\left(  r+\epsilon iw\right)  \in
SL\left(
\mathbb{C}
,2\right)  $

\paragraph{Representations of SO($%
\mathbb{R}
$,p,q)\newline}

$SO(%
\mathbb{C}
,p+q)$ is the complexified of $SO(%
\mathbb{R}
,p,q)$. So if we take the complexified we are back to representations of
SO(C,p+q), with its tensorial product of the standard representations and the
Spin representations. The Spin representations are deduced as restrictions of
representations of the Clifford algebra Cl(p,q). The representations of the
connected components $SO_{0}(%
\mathbb{R}
,p,q)$ are then selected through the double cover in a way similar to $SO(%
\mathbb{C}
,n):$ the representations of Spin(R,p,q) must be such that $\forall s\in Spin(%
\mathbb{R}
,p,q):\widehat{f}\left(  -s\right)  =\widehat{f}\left(  s\right)  $

Another common way to find representations of SO(R,3,1) is to notice that
so(R,3,1) is isomorphic to the direct product su(2)xsu(2) and from there the
finite dimensional representations of SO(R,3,1) are the tensorial product of
two irreducible representations of SU(2): $\left(  P^{j_{1}}\otimes P^{j_{2}%
},D^{j_{1}}\otimes D^{j_{2}}\right)  $, which is then reducible to a sum of
irreducible representations $\left(  P^{k},D^{k}\right)  $ with the
Clebsch-Jordan coefficients. This way we can use the well known tricks of the
SU(2) representations, but the link with the generators of so(R,3,1) is less obvious.

\bigskip

\subsection{Unitary groups U(n)}

\subsubsection{Definition and general properties}

1. On a \textit{complex} vector space of dimension n, endowed with an
\textit{hermitian form} g, the unitary group U(E,g) is the group of
endomorphisms f which preserve the scalar product : $\forall u,v\in E:g\left(
f\left(  u\right)  ,f\left(  v\right)  \right)  =g\left(  u,v\right)  .$

SU(E,g) is the subgroup of U(E,g) of endomorphisms with det(f) =1.

2. U(E,g) is isomorphic to $U\left(  n\right)  ,$ the linear group of complex
matrices of $GL\left(
\mathbb{C}
,n\right)  $ such that $M^{\ast}M=I_{n}.$

SU(E,g) is isomorphic to $SU\left(  n\right)  ,$ the linear group of complex
matrices of $GL\left(
\mathbb{C}
,n\right)  $ such that $M^{\ast}M=I_{n}$ detM = 1.

3. The matrices of $GL(%
\mathbb{C}
,n)$ such that $M^{\ast}\left[  \mathbf{\eta}_{p,q}\right]  M=\left[
\mathbf{\eta}_{p,q}\right]  $\ are a \textit{real} Lie subgroup of $GL(%
\mathbb{C}
,n)$ denoted U(p,q) with p + q = n.

$U(p,q),U\left(  q,p\right)  $ are identical.

The matrices of U(p,q) such that detM=1\ are a \textit{real} Lie subgroup of
U(p,q) denoted SU(p,q).

4. All are real Lie groups, and not complex Lie groups.

5. The Lie algebra of :

U(n) is the set of matrices $u(n)=\{X\in L(%
\mathbb{C}
,n):X+X^{\ast}=0\}$

SU(n) is the set of matrices $su(n)=\{X\in L(%
\mathbb{C}
,n):X+X^{\ast}=0,TrX=0\}$

U(p,q), SU(p,q) is the set of matrices $u(p,q)=\{X\in L(%
\mathbb{C}
,n):\left[  \mathbf{\eta}_{p,q}\right]  X+X^{\ast}\left[  \mathbf{\eta}%
_{p,q}\right]  =0\}$

They are real Lie algebras of complex matrices.

The complexified of the Lie algebra $sl\left(
\mathbb{C}
,n\right)  _{%
\mathbb{C}
}=sl\left(
\mathbb{C}
,n\right)  $ and the complexified $SU\left(  n\right)  _{%
\mathbb{C}
}=SL\left(
\mathbb{C}
,n\right)  .$

6. U(n) is connected but not simply connected. Its universal covering group is
$T\times SU(n)=\left\{  e^{it},t\in%
\mathbb{R}
\right\}  \times SU(n)$ with $\pi:T\rightarrow U(n)::\pi\left(  \left(
e^{it}\times\left[  g\right]  \right)  \right)  =it\left[  g\right]  $ so for
n=1 the universal cover is $\left(
\mathbb{R}
,+\right)  .$

The matrices of $U(n)\cap GL(%
\mathbb{R}
,n)$ comprised of real elements are just $O(%
\mathbb{R}
,n).$

SU(n) is connected and simply connected.

U(p,q) has two connected components.

SU(p,q) is connected and is the connected component of the identity of U(p,q).

7. U(n) is not semi simple.\ The center of U(n) are the purely imaginary
scalar matrices $kiI_{n}$

SU(n) is semi simple for n
$>$
1.

U(p,q), SU(p,q) are semi-simple

8. U(n), SU(n) are compact.

U(p,q), SU(p,q) are not compact.

9. In all cases the standard representation $\left(
\mathbb{C}
^{n},\imath\right)  $ is on $%
\mathbb{C}
^{n}$ endowed with the hermitian form g, with the proper signature.\ So the
canonical basis is orthonormal.

\subsubsection{Representations of SU(n)}

As SU(n) is compact, any unitary, any irreducible representation is finite dimensional.

The complexified of the Lie algebra $su\left(  n\right)  _{%
\mathbb{C}
}=sl\left(
\mathbb{C}
,n\right)  $ and the complexified $SU\left(  n\right)  _{%
\mathbb{C}
}=SL\left(
\mathbb{C}
,n\right)  .$ So the representations (E,f) of SU(n) are in bijective
correspondance with the representations of SL(C,n), by restriction of f to the
subgroup SU(n) of SL(C,n). And the irreducible representations of SU(n) are
the restrictions of the irreducible representations of SL(C,n). The same
applies to the representations of the algebra su(n). So one finds the
representations of SU(2) in the non unitary principal series of SL(C,2).

\paragraph{Representations of SU(2)\newline}

(from Kosmann)

1. Basis of su(2):

We must restrict the actions of the elements of sl(C,2) to elements of su(2).
The previous basis (J) of sl(C,2) is not a basis of su(2), so it is more
convenient to take the matrices :

$K_{1}=%
\begin{bmatrix}
0 & i\\
i & 0
\end{bmatrix}
=i\sigma_{1};K_{2}=%
\begin{bmatrix}
0 & -1\\
1 & 0
\end{bmatrix}
=-i\sigma_{2};K_{3}=%
\begin{bmatrix}
i & 0\\
0 & -i
\end{bmatrix}
=i\sigma_{2}$

where $\sigma_{i}$\ are the Pauli's matrices. We add for convenience :
$K_{0}=I_{2\times2}$

$K_{i}^{2}=-I,\left[  K_{1},K_{2}\right]  =2K_{3},\left[  K_{2},K_{3}\right]
=2K_{1};\left[  K_{3},K_{1}\right]  =2K_{2}$

The exponential is surjective on SU(2) so : $\forall g\in SU(2),\exists X\in
su(2):g=\exp X$

$\forall g\in SU(2):\left[  g\right]  =\sum_{k=0}^{3}a_{k}\left[
K_{k}\right]  $ with $a_{k}\in%
\mathbb{R}
,\sum_{k=0}^{3}\left\vert a_{k}\right\vert ^{2}=1$

2. Finite dimensional representations :

All the finite dimensional representations of SL(C,2) stand in representations
over polynomials. After adjusting to SU(2) (basically that $\left[  g\right]
^{-1}=\left[  g\right]  ^{\ast})$ we have the following :

\begin{theorem}
The finite dimensional representations $\left(  P^{j},D^{j}\right)  $\ of
SU(2) are the left regular representations over the degree 2j homogeneous
polynomials with two complex variables $z_{1},z_{2}.$
\end{theorem}

$P^{j}$ is a 2j+1 complex dimensional vector space with basis : $|j,m>=z_{1}%
^{j+m}z_{2}^{j-m},-j\leq m\leq j$

$D^{j}\left(  g\right)  P\left(
\begin{bmatrix}
z_{1}\\
z_{2}%
\end{bmatrix}
\right)  =P\left(  \left[  g\right]  ^{-1}%
\begin{bmatrix}
z_{1}\\
z_{2}%
\end{bmatrix}
\right)  =P\left(  \left[  g\right]  ^{\ast}%
\begin{bmatrix}
z_{1}\\
z_{2}%
\end{bmatrix}
\right)  $

The representation is unitary with the scalar product over $P^{j}$ defined by
taking $|j,m>$ as orthonormal basis :

$\left\langle \sum_{m}x^{m}|j,m>,\sum_{m}y^{m}|j,m>\right\rangle =\sum
_{m}\overline{x}^{m}y^{m}$

Any irreducible representation of SU(2) is equivalent to one of the $\left(
P^{j},D^{j}\right)  $ for some $j\in\frac{1}{2}%
\mathbb{N}
.$

3. Characters:

The characters of the representations can be obtained by taking the characters
for a maximal torus which are of the kind:

$T\left(  t\right)  =%
\begin{bmatrix}
e^{it} & 0\\
0 & e^{-it}%
\end{bmatrix}
\in SU\left(  2\right)  $

and we have :

$t\in]0,\pi\lbrack:\chi_{D^{j}}\left(  T\left(  t\right)  \right)  =\frac
{\sin\left(  2j+1\right)  t}{\sin t}$

$\chi_{D^{j}}\left(  T\left(  0\right)  \right)  =2j+1=\dim P^{j}$

$\chi_{D^{j}}\left(  T\left(  \pi\right)  \right)  =\left(  -1\right)
^{2j}\left(  2j+1\right)  $

So if we take the tensorial product of two representations : $\left(
P^{j_{1}}\otimes P^{j_{2}},D^{j_{1}}\otimes D^{j_{2}}\right)  $ we have :

$\chi_{j_{1}\otimes j_{2}}\left(  t\right)  =\chi_{j_{1}}\left(  t\right)
\chi_{j_{2}}\left(  t\right)  =\chi_{\left\vert j_{2}-j_{1}\right\vert
}\left(  t\right)  +\chi_{\left\vert j_{2}-j_{1}\right\vert +1}\left(
t\right)  +...\chi_{j_{2}+j_{1}}\left(  t\right)  $

And the tensorial product is reducible in the sum of representations according
to the \textbf{Clebsch-Jordan formula} :

$D^{j_{1}\otimes j_{2}}=D^{\left\vert j_{2}-j_{1}\right\vert }\oplus
D^{\left\vert j_{2}-j_{1}\right\vert +1}\oplus...\oplus D^{j_{2}+j_{1}}$

The basis of $\left(  P^{j_{1}}\otimes P^{j_{2}},D^{j_{1}}\otimes D^{j_{2}%
}\right)  $ is comprised of the vectors :

$|j_{1},m_{1}>\otimes|j_{2},m_{2}>$ with $,-j_{1}\leq m_{1}\leq j_{1}%
,,-j_{2}\leq m_{2}\leq j_{2}$

it can be decomposed in a sum of bases on each of these vector spaces :

$|J,M>$ with $\left\vert j_{1}-j_{2}\right\vert \leq J\leq j_{1}+j_{2},-J\leq
M\leq J$

So we have a matrix to go from one basis to the other :

$|J,M>=\sum_{m_{1},m_{2}}C\left(  J,M,j_{1},j_{2},m_{1},m_{2}\right)
|j_{1},m_{1}>\otimes|j_{2},m_{2}>$

whose coefficients are the \textbf{Clebsch-Jordan coefficients. }They are
tabulated\textbf{.}

\bigskip

\subsection{Pin and Spin groups}

\subsubsection{Definition and general properties}

1. These groups are defined starting from Clifford algebra over a finite
dimensional vector space F on a field K, endowed with a bilinear symmetric
form g (valued in K) (see the Algebra part). The scalars +1, - 1 belongs to
the groups, so they are not linear groups (but they have matricial representations).

2. All Clifford algebras on vector spaces with the same dimension on the same
field, with bilinear form of the same signature are isomorphic.\ So we can
speak of Pin(K,p,q), Spin(K,p,q).

There are groups morphism :

$\mathbf{Ad:}Pin(K,p,q)\rightarrow O\left(  K,p,q\right)  $

$\mathbf{Ad:}Spin(K,p,q)\rightarrow SO\left(  K,p,q\right)  $

with a double cover (see below) : for any $g\in O\left(  K,p,q\right)  $ (or
SO(K,p,q) there are two elements $\pm w$ of Pin(K,p,q) (or Spin(K,p,q) such
that : $\mathbf{Ad}_{w}=h.$

They have the same Lie algebra : o(K,p,q) is the Lie algebra of Pin(K,p,q) and
so(K,p,q) is the Lie algebra of Spin(K,p,q).

3. The situation with respect to the cover is a bit complicated. We have
always two elements of the Pin or Spin group for one element of the orthogonal
group, and they are a cover as a manifold, but not necessarily the universal
cover as Lie group, which has been defined only for connected Lie groups. When
they are connected, they have a unique universal cover as a topological space,
which has a unique Lie group structure $\widetilde{G}$ which is a group of
matrices, which can be identified to Pin(K,p,q) or Spin(K,p,q) respectively.
When they are disconnected, the same result is valid for their connected
component,\ which is a Lie subgroup.

4. If $K=%
\mathbb{C}
$ :

$Pin(%
\mathbb{C}
,p,q)\simeq Pin(%
\mathbb{C}
,p+q)$

$Spin(%
\mathbb{C}
,p,q)\simeq Spin(%
\mathbb{C}
,p+q)$

$Spin(%
\mathbb{C}
,n)$ is connected, simply connected.

$SO(%
\mathbb{C}
,n)\simeq SO\left(
\mathbb{C}
,p,q\right)  $ is a semi simple, complex Lie group, thus its universal
covering group is a group of matrices which can be identified with
$Spin\left(
\mathbb{C}
,n\right)  .$

$Spin\left(
\mathbb{C}
,n\right)  $ and $SO(%
\mathbb{C}
,n)$ have the same Lie algebra which is compact, thus $Spin\left(
\mathbb{C}
,n\right)  $\ is compact.

We have the isomorphisms :

$SO(%
\mathbb{C}
,n,)\simeq Spin(%
\mathbb{C}
,n)/U(1)$

$Spin(%
\mathbb{C}
,2)\simeq%
\mathbb{C}
$

$Spin(%
\mathbb{C}
,3)\simeq SL(%
\mathbb{C}
,2)$

$Spin(%
\mathbb{C}
,4)\simeq SL(%
\mathbb{C}
,2)\times SL(%
\mathbb{C}
,2)$

$Spin(%
\mathbb{C}
,5)\simeq Sp\left(
\mathbb{C}
,4\right)  $

$Spin(%
\mathbb{C}
,6)\simeq SL(%
\mathbb{C}
,4)$

5. If $K=%
\mathbb{R}
$

$Pin(%
\mathbb{R}
,p,q),Pin(%
\mathbb{R}
,q,p)$ are not isomorphic if $p\neq q$\ 

$Pin(%
\mathbb{R}
,p,q)$ is not connected, it maps to $O(%
\mathbb{R}
,p,q)$ but the map is not surjective and it is not a cover of $O(%
\mathbb{R}
,p,q)$

$Spin(%
\mathbb{R}
,p,q)$ and $Spin(%
\mathbb{R}
,q,p)$ are isomorphic, and simply connected if p+q%
$>$%
2

$Spin(%
\mathbb{R}
,0,n)$ and $Spin(%
\mathbb{R}
,n,0)$ are equal to $Spin(%
\mathbb{R}
,n)$

For n%
$>$%
2 $Spin(%
\mathbb{R}
,n)$ is connected, simply connected and is the universal cover of $SO(%
\mathbb{R}
,n)$ and has the same Lie algebra, so it is compact.

If p+q%
$>$%
2 $Spin(%
\mathbb{R}
,p,q)$ is connected, simply connected and is a double cover of $SO_{0}(%
\mathbb{R}
,p,q)$ and has the same Lie algebra, so it is not compact.

We have the isomorphisms :

Spin($%
\mathbb{R}
,$1) $\simeq$ O($%
\mathbb{R}
,$1)

Spin($%
\mathbb{R}
,$2) $\simeq$ U(1) $\simeq$ SO($%
\mathbb{R}
,$2)

Spin($%
\mathbb{R}
,$3) $\simeq$ Sp(1) $\simeq$ SU(2)

Spin($%
\mathbb{R}
,$4) $\simeq$ Sp(1) x\ Sp(1)

Spin($%
\mathbb{R}
,$5) $\simeq$ Sp(2)

Spin($%
\mathbb{R}
,$6) $\simeq$ SU(4)

Spin($%
\mathbb{R}
,$1,1) $\simeq%
\mathbb{R}
$

Spin($%
\mathbb{R}
,$2,1) = SL(2,R)

Spin($%
\mathbb{R}
,$3,1) = SL($%
\mathbb{C}
,$2)

Spin($%
\mathbb{R}
,$2,2) = SL($%
\mathbb{R}
,$2) x\ SL($%
\mathbb{R}
,$2)

Spin($%
\mathbb{R}
,$4,1) = Sp(1,1)

Spin($%
\mathbb{R}
,$3,2) = Sp(4)

Spin($%
\mathbb{R}
,$4,2) = SU(2,2)

\subsubsection{Representations of the Spin groups Spin(K,n)}

The Pin and Spin groups are subsets of their respective Clifford algebras.
Every Clifford algebra is isomorphic to an algebra of matrices over a field
K', so a Clifford algebra has a faithful irreducible representation on a
vector space over the field K', with the adequate dimension and orthonormal
basis such that the representative of a Clifford member is one of these
matrices. Then we get a representation of the Pin or Spin group by restriction
of the representation of the Clifford algebra. In the Clifford algebra section
the method to build the matrices algebra is given.

Notice that the dimension of the representation is given. Other
representations can be deduced from there by tensorial products, but they are
not necessarily irreducible.

$Spin(%
\mathbb{C}
,n),Spin\left(
\mathbb{R}
,n\right)  $ are double covers of $SO(%
\mathbb{C}
,n),SO\left(
\mathbb{R}
,n\right)  ,$ thus they have the same Lie algebra, which is compact. All their
irreducible unitary representations are finite dimensional.

\bigskip

\subsection{Symplectic groups Sp(K,n)}

1. On a symplectic vector space (E,h), that is a real vector space E endowed
with a non degenerate antisymmetric 2-form h, the endomorphisms which preserve
the form h, called the symplectomorphisms constitute a group Sp(E,h). The
groups Sp(E,h) are all isomorphic for the same dimension 2n of E (which is
necessarily even).

2. The symplectic group denoted Sp(K,n), $K=%
\mathbb{R}
,%
\mathbb{C}
,$ is the linear group of matrices of GL(K,2n) such that $M^{t}J_{n}M=J_{n}%
$\ where $J_{n}$ is the 2nx2n matrix : $J_{n}=%
\begin{bmatrix}
0 & I_{n}\\
-I_{n} & 0
\end{bmatrix}
$

The groups Sp(E,h) for E real and dimension n are isomorphic to $Sp\left(
\mathbb{R}
,n\right)  .$

Notice that we have either real or complex Lie groups.

3. Sp(K,n) is a Lie subgroup over K of GL(K,2n). Its Lie algebra comprises the
matrices $sp(K,p,q)=\{X\in L(K,n):JX+X^{t}J=0\}$.

4. Sp(K,n) is a semi-simple, connected, non compact group.

5. Their algebra belongs to the $C_{n}$ family with the fundamental weights :
$w_{i}=\sum_{k=1}^{i}e_{k}$

Root system for $sp\left(
\mathbb{C}
,n\right)  ,n\geq3:C_{n}$

$V=%
\mathbb{R}
^{n}$

$\Delta=\left\{  \pm e_{i}\pm e_{j},i<j\right\}  \cup\left\{  \pm
2e_{k}\right\}  $

$\Pi=\left\{  e_{1}-e_{2},e_{2}-e_{3},..e_{n-1}-e_{n},2e_{n}\right\}  $

\bigskip

\subsection{Heisenberg group}

\subsubsection{Finite dimensional Heisenberg group}

\paragraph{Definition\newline}

Let E be a symplectic vector space, meaning a real finite n-dimensional vector
space endowed with a non degenerate 2-form $h\in\Lambda_{2}E$. So n must be
even : n=2m

Take the set $E\times%
\mathbb{R}
,$ endowed with its natural structure of vector space $E\oplus%
\mathbb{R}
$ and the internal product $\cdot:$

$\forall u,v\in E,x,y\in%
\mathbb{R}
:\left(  u,x\right)  \cdot\left(  v,y\right)  =\left(  u+v,x+y+\frac{1}%
{2}h\left(  u,v\right)  \right)  $

The product is associative. The identity element is (0,0) and each element has
an inverse :

$\left(  u,x\right)  ^{-1}=\left(  -u,-x\right)  $

So it has a group structure.\ $E\times%
\mathbb{R}
$ with this structure is called the \textbf{Heisenberg group} H(E,h).

As all symplectic vector spaces with the same dimension are isomorphic, all
Heisenberg group for dimension n are isomorphic and the common structure is
denoted H(n).

\paragraph{Properties\newline}

The Heisenberg group is a connected, simply-connected Lie group. It is
isomorphic (as a group) to the matrices of $GL$($%
\mathbb{R}
,n+1)$ which read :

\bigskip

$%
\begin{bmatrix}
1 & \left[  p\right]  _{1\times n} & \left[  c\right]  _{1\times1}\\
0 & I_{n} & \left[  q\right]  _{n\times1}\\
0 & 0 & 1
\end{bmatrix}
$

\bigskip

H(E,h) is a vector space and a Lie group.\ Its Lie algebra denoted also H(E,h)
is the set $E\times%
\mathbb{R}
$ itself with the bracket:

$\left[  \left(  u,x\right)  ,\left(  v,y\right)  \right]  =\left(
u,x\right)  \cdot\left(  v,y\right)  -\left(  v,y\right)  \cdot\left(
u,x\right)  =\left(  0,h\left(  u,v\right)  \right)  $

Take a canonical basis of E : $\left(  e_{i},f_{i}\right)  _{i=1}^{m}$ then
the structure coefficients of the Lie algebra H(n) are :

$\left[  \left(  e_{i},1\right)  ,\left(  f_{j},1\right)  \right]  =\left(
0,\delta_{ij}\right)  $ all the others are null

It is isomorphic (as Lie algebra) to the matrices of $L(%
\mathbb{R}
,n+1):$

\bigskip

$%
\begin{bmatrix}
0 & \left[  p\right]  _{1\times n} & \left[  c\right]  _{1\times1}\\
0 & \left[  0\right]  _{n\times n} & \left[  q\right]  _{n\times1}\\
0 & 0 & 0
\end{bmatrix}
$

\bigskip

There is a complex structure on E defined from a canonical basis $\left(
e_{j},f_{j}\right)  _{j=1}^{m}$\ by taking a complex basis $\left(
e_{j},if_{j}\right)  _{j=1}^{m}$ with complex components. Define the new
complex basis :

$a_{k}=\frac{1}{\sqrt{2}}\left(  e_{k}-if_{k}\right)  ,a_{k}^{\dag}=\frac
{1}{\sqrt{2}}\left(  e_{k}+if_{k}\right)  $

and the commutation relations becomes : $\left[  a_{j},a_{k}\right]  =\left[
a_{j}^{\dag},a_{k}^{\dag}\right]  =0;\left[  a_{j},a_{k}^{\dag}\right]
=\delta_{jk}$ called CAR

\paragraph{Representations\newline}

All irreducible finite dimensional linear representation of H(n) are
1-dimensional.\ \ The character is :

$\chi_{ab}\left(  x,y,t\right)  =e^{-2i\pi\left(  ax+by\right)  }$ where
$\left(  a,b\right)  \in%
\mathbb{R}
^{n}\times%
\mathbb{R}
^{n}$

The only unitary representations are infinite dimensional over $F=L^{2}\left(
%
\mathbb{R}
^{n}\right)  $

$\lambda\neq0\in r,f\in L^{2}\left(
\mathbb{R}
^{n}\right)  ,\left(  x,y,t\right)  \in H\left(  n\right)  $

$\rightarrow\rho_{\lambda}\left(  x,y,t\right)  f\left(  s\right)  =\left(
\exp\left(  -2i\pi\lambda t-i\pi\lambda\left\langle x,y\right\rangle
+2i\pi\lambda\left\langle s,y\right\rangle \right)  \right)  f(s-x)$

Two representations $\rho_{\lambda},\rho_{\mu}$\ are equivalent iff
$\lambda=\mu.$ Then they are unitary equivalent.

\subsubsection{Heisenberg group on Hilbert space}

There is a generalization of the previous definition for complex\ Hilbert
spaces H (possibly infinite dimensional)

If H is a complex Hilbert space, then $h:H\times H\rightarrow%
\mathbb{C}
::h\left(  u,v\right)  =-\operatorname{Im}\left\langle u,v\right\rangle $ is
an antisymmetric real 2-form, non degenerate. Thus one can consider the
Heisenberg group $Heis(H)=H\times%
\mathbb{R}
$ with product :

$\left(  u,s\right)  \times\left(  v,t\right)  =\left(  u+v,s+t+\frac{1}%
{2}h\left(  u,v\right)  \right)  $

If $\sigma:H\rightarrow H$ is an antilinear involution, then it defines a real
form $H_{\sigma}$ of H, and $H_{\sigma}\times\left\{  0\right\}  $ is a
subgroup of Heis(H).

\begin{theorem}
(Neeb p.102) The Heisenberg group Heis(H) is a topological group. The action :
$\lambda:Heis(H)\times H\rightarrow H::\lambda\left(  u,s\right)  v=u+v$ is continuous.

$\left(
\mathcal{F}%
_{+}\left(  H\right)  ,\rho\right)  $ with :$\
\mathcal{F}%
_{+}\left(  H\right)  $ the symmetric Fock space of H and$\ \left(
\rho\left(  u,s\right)  \exp\right)  \left(  v\right)  =\exp\left(
is+\left\langle v,u\right\rangle -\frac{1}{2}\left\langle u,u\right\rangle
\right)  \exp\left(  v-u\right)  $ is a continuous unitary representation of Heis(H).
\end{theorem}

\begin{theorem}
(Neeb p.103) Let H be a complex Hilbert space, $%
\mathcal{F}%
_{+}\left(  H\right)  \subset C\left(  H;%
\mathbb{C}
\right)  $ be the Hilbert space associated to the positive kernel $N\left(
u,v\right)  =\exp\left\langle u,v\right\rangle ,%
\mathcal{F}%
_{m+}\left(  H\right)  $ the subspace of m homogeneous functions $f\left(
\lambda u\right)  =\lambda^{m}f\left(  u\right)  ,$ then the following
assertions hold :

i) $\left(
\mathcal{F}%
_{m+}\left(  H\right)  ,\rho\right)  $ is a unitary continuous representation
of the group of unitary operators on H, with $\rho\left(  U\right)  \left(
f\right)  \left(  u\right)  =f\left(  U^{-1}u\right)  .$ The closed subspaces
$\mathcal{F}$%
$_{m+}\left(  H\right)  $ are invariant and their positive kernel is
$N_{m}\left(  u,v\right)  =\exp\left\langle u,v\right\rangle ^{m}\frac{1}{m!}$

ii) If $\left(  e_{i}\right)  _{i\in I}$ is a Hilbertian basis of H, then
$p_{M}\left(  u\right)  =%
{\displaystyle\prod\limits_{i\in I}}
\left\langle e_{i},u\right\rangle ^{m_{i}}$ with $M=\left(  m_{i}\right)
_{i\in I}\in\left(
\mathbb{N}
-0\right)  ^{I}$ is a hilbertian basis of $%
\mathcal{F}%
_{+}\left(  H\right)  $ and $\left\Vert p_{M}\right\Vert ^{2}=%
{\displaystyle\prod\limits_{i\in I}}
m_{i}$
\end{theorem}

\bigskip

\subsection{Groups of displacements}

1. An affine map d over an affine space E is the combination of a linear map g
and a translation t. If g belongs to some linear group G, with the operations :

$\left(  g,t\right)  \times\left(  g^{\prime},t^{\prime}\right)  =\left(
gg^{\prime},\lambda\left(  g,t^{\prime}\right)  +t\right)  ,\left(
g,t\right)  ^{-1}=\left(  g^{-1},-\lambda\left(  g^{-1},t\right)  \right)  $

we have a group D, called a group of displacements, which is the semi product
of G and the abelian group $T\simeq\left(  \overrightarrow{E},+\right)  $ of
translations : :$D=G\varpropto_{\lambda}T,$ $\lambda$ being the action of G on
the vectors of $\overrightarrow{E}:\lambda:G\times\overrightarrow
{E}\rightarrow\overrightarrow{E}$.

Over finite dimensional space, in a basis, an affine group is characterized by
a couple (A,B) of a square inversible matrix A belonging to a group of
matrices G and a column matrix B, corresponding to the translation.

2. An affine group in a n dimensional affine space over the field K has a
standard representation by (n+1)x(n+1) matrices over K as follows :

$D=%
\begin{bmatrix}
a_{11} & .. & a_{1n} & b_{1}\\
.. & .. & .. & ..\\
a_{n1} & .. & a_{nn} & b_{n}\\
0 & 0 & 0 & 1
\end{bmatrix}
=%
\begin{bmatrix}
A & B\\
0 & 1
\end{bmatrix}
$

and we can check that the composition of two affine maps is given by the
product of the matrices. The inverse is :

$D^{-1}=%
\begin{bmatrix}
A^{-1} & -A^{-1}B\\
0 & 1
\end{bmatrix}
$

From there finding representations of groups come back to find representations
of a group of matrices. Of special interest are the affine maps such that A
belongs to a group of orthonormal matrices SO(K,n) or, in the Minkovski space,
to SO(R,3,1) (this is the Poincar\'{e} group).

\newpage

\part{FIBER\ BUNDLES AND\ JETS}

\bigskip

The theory of fiber bundles is fairly recent - for mathematics - but in 40 or
50 years it has become an essential tool in many fields of research.\ ln
mathematics it gives a unified and strong foundation for the study of whatever
object one puts over a manifold, and thus it is the natural prolongation of
differential geometry, notably whith the language of categories. In physics it
is the starting point of all gauge theories. As usual we address first the
general definitions, for a good understanding of the key principles, before
the study of vector bundles, principal bundles and associated bundles. A fiber
bundle E can be seen, at least locally and let us say in the practical
calculii, as the product of two manifolds MxV, associated with a projection
from E onto M. If V is a vector space we have a vector bundle, if V is a Lie
group we have a principal bundle, and an associated fiber bundle is just a
vector bundle where a preferred frame of reference has been identified. All
that has been said previously about tensor bundle, vector and tensor fields
can be generalized very simply to vector bundles using functors from the
category of manifolds to the category of fibered manifolds.

However constructions above manifolds involve quite often differential
operators, which link in some manner the base manifold and the objects upon
it, for example connection on a tangent bundle. Objects involving partial
derivatives on a manifold can be defined in the terms of jets : jets are just
maps which have same derivatives at some order. Using these concepts we have a
comprehensive body of tools to classify almost all non barbaric differential
operators on manifolds, and using once more functors on manifolds one can show
that they turn out to be combinations of well known operators.

One of these is connection, which is studied here in the more general context
of fiber bundle, but with many similarities to the object defined over the
tangent bundle in the previous part of this book.

In differential geometry we usually strive to get "intrinsic" results, meaning
which are expressed by the same formulas whatever the charts. With jets,
coordinates become the central focus, giving a precise and general study of
all objects with derivatives content. So the combination of fiber bundle and
jets enables to give a consistent definition of connections, matching the pure
geometrical definition and the usual definition through Christoffel symbols.

\newpage

\section{FIBER\ BUNDLES}

\bigskip

\subsection{General fiber bundles}

\label{Fiber bundles definition and principal results}

There are several generally accepted definitions of fiber bundles, with more
or less strict requirements. We will use two definitions which encompass all
current usages, while offering the useful ingredients. The definition of fiber
bundle is new, but more in line with that for manifolds, and make simpler the
subtle issue of equivalent trivializations.

\subsubsection{Fiber bundle}

\paragraph{Fibered manifold\newline}

\begin{definition}
A \textbf{fibered manifold} $E(M,\pi)$ is a triple of two Hausdorff manifolds
E,M and a surjective submersion $\pi:E\rightarrow M$ . M is called the
\textbf{base space} and $\pi$ projects E on M. The fibered manifold is of
class r if E,M and $\pi$ are of class r.
\end{definition}

\begin{theorem}
(Giachetta p.7) E is a fibered manifold iff E admits an atlas $\left(
O_{a},\varphi_{a}\right)  _{a\in A}$ such that :

$\varphi_{a}:O_{a}\rightarrow B_{M}\times B_{V}$ are two Banach vector spaces

$\varphi_{a}\left(  x,.\right)  :\pi\left(  O_{a}\right)  \rightarrow B_{M}$
is a chart $\psi_{a}$ of M

and the transitions functions on $O_{a}\cap O_{b}$ are $\varphi_{ba}\left(
\xi_{a},\eta_{a}\right)  =\left(  \psi_{ba}\left(  \xi_{a}\right)
,\phi\left(  \xi_{a},\eta_{a}\right)  \right)  =\left(  \xi_{b},\eta
_{b}\right)  $
\end{theorem}

Such a chart of E is said to be a \textbf{chart adaptated} to the fibered
structure. The coordinates on E are a pair $\left(  \xi_{a},\eta_{a}\right)
,$ $\xi$ corresponds to x.

\begin{definition}
A fibered manifold $E^{\prime}(N,\pi^{\prime})$ is a \textbf{subbundle} of
$E(M,\pi)$ if N is a submanifold of M and the restriction $\pi_{E^{\prime}%
}=\pi^{\prime}$
\end{definition}

\begin{theorem}
A fibered manifold $E(M,\pi)$ has the universal property : if f is a map $f\in
C_{r}\left(  M;P\right)  $ in another manifold P then $f\circ\pi$ is class r
iff f is class r (all the manifolds are assumed to be of class r).
\end{theorem}

\paragraph{Fiber bundle\newline}

\begin{definition}
On a fibered manifold $E\left(  M,\pi\right)  $\ an \textbf{atlas of fiber
bundle} is a set $\left(  V,\left(  O_{a},\varphi_{a}\right)  _{a\in
A}\right)  $ where :

V is a Hausdorff manifold, called the \textbf{fiber}

$\left(  O\right)  _{a\in A}$ is an open cover of M

$\left(  \varphi_{a}\right)  _{a\in A}$ is a family of diffeomorphisms, called
\textbf{trivialization} :

$\varphi_{a}:O_{a}\times V\subset M\times V\rightarrow\pi^{-1}\left(
O_{a}\right)  \subset E$ $::p=\varphi_{a}\left(  x,u\right)  $

and there is a family of maps $\left(  \varphi_{ab}\right)  _{\left(
a,b\right)  \in A\times A}$,\ called the \textbf{transition maps}, defined on
$O_{a}\cap O_{b}$ whenever $O_{a}\cap O_{b}\neq\varnothing$\ , such that
$\varphi_{ab}\left(  x\right)  $ is a diffeomorphism on V and

$\forall p\in\pi^{-1}\left(  O_{a}\cap O_{b}\right)  ,p=\varphi_{a}\left(
x,u_{a}\right)  =\varphi_{b}\left(  x,u_{b}\right)  \Rightarrow u_{b}%
=\varphi_{ba}\left(  x\right)  \left(  u_{a}\right)  $

meeting the \textbf{cocycle conditions}, whenever they are defined:

$\forall a,b,c\in A:\varphi_{aa}\left(  x\right)  =Id;\varphi_{ab}\left(
x\right)  \circ\varphi_{bc}\left(  x\right)  =\varphi_{ac}\left(  x\right)  $

The atlas is of class r if V and the maps are of class r.
\end{definition}

\bigskip%

\begin{tabular}
[c]{ccc}%
$E$ &  & \\
$\pi^{-1}\left(  O_{a}\right)  $ &  & \\
$\pi\downarrow$ & $\nwarrow\varphi_{a}$ & \\
$O_{a}$ & $\rightarrow$ & $O_{a}\times V$\\
$M$ &  & $M\times V$%
\end{tabular}

\bigskip

\begin{definition}
Two atlas $\left(  V,\left(  O_{a},\varphi_{a}\right)  _{a\in A}\right)
,\left(  V,\left(  Q_{i},\psi_{i}\right)  _{i\in I}\right)  $ over the same
fibered manifold are \textbf{compatible} if their union is still an atlas.
\end{definition}

Which means that, whenever $O_{a}\cap Q_{i}\neq\varnothing$ there is a
diffeomorphism : $\chi_{ai}\left(  x\right)  :V\rightarrow V$ such that :

$\varphi_{a}\left(  x,u\right)  =\psi_{i}\left(  x,v\right)  \Rightarrow
v=\chi_{ai}\left(  x\right)  \left(  u\right)  $

and the family $\left(  \chi_{ai}\right)  _{\left(  a,i\right)  \in A\times
I}$ meets the cocycle conditions

\begin{definition}
To be compatible is an equivalence relation between atlas of fiber bundles.\ A
\textbf{fiber bundle} structure is a class of equivalence of compatible atlas.
\end{definition}

\begin{notation}
$E(M,V,\pi)$ is a fiber bundle with total space E, base M, fiber V, projection
$\pi$
\end{notation}

\paragraph{The key points\newline}

1. A fiber bundle is a fibered manifold, and, as the theorems below show,
conversely a fibered manifold is usually a fiber bundle. The added value of
the fiber bundle structure is the identification of a manifold V, which can be
given an algebraic structure (vector space or group). Fiber bundles are also
called "locally trivial fibered manifolds with fiber V" (Hosemoller). We opt
for simplicity.

2. For any p in E there is a \textit{unique} x in M, but \textit{every} u in V
provides a \textit{different} p in E (for a given open $O_{a})$. The set
$\pi^{-1}\left(  x\right)  =E\left(  x\right)  \subset E$ is called the
\textbf{fiber over x}.

3.The projection $\pi$\ is a surjective submersion so : $\pi^{\prime}\left(
p\right)  $ is surjective, rank $\pi^{\prime}\left(  p\right)  =\dim M,$ $\pi$
is an open map. For any x in M, the fiber $\pi^{-1}\left(  x\right)  $\ is a
submanifold of E.

4. Any product M$\times$V=E is a fiber bundle, called a \textbf{trivial fiber
bundle}.\ Then all the transition maps are the identity. But the converse is
not true : usually E is built from pieces of M$\times$V patched together
through the trivializations. Anyway locally a fiber bundle is isomorphic to
the product M$\times$V, so it is useful to keep this model in mind.

5. If the manifolds are finite dimensional we have : dimE = dimM + dimV and
$\dim\pi^{-1}\left(  x\right)  =\dim V.$

So If dimE=dimM then V is a manifold of dim0, meaning a discrete set. We will
always assume that this is not the case.

6. The couples $(O_{a},\varphi_{a})_{a\in A}$ and the set $\left(
\varphi_{ab}\right)  _{a,b\in A}$\ play a role similar to the charts and
transition maps for manifolds, but here they are defined between manifolds,
without a Banach space. The same fiber bundle can be defined by different
atlas. Conversely on the same manifold E different structures of fiber bundle
can be defined.

7. For a general fiber bundle no algebraic feature is assumed on V or the
transition maps. If an algebraic structure exists on V then we will require
that the transition maps are also morphisms.

8. It is common to define the trivializations as maps : $\varphi_{a}:\pi
^{-1}\left(  O_{a}\right)  \rightarrow O_{a}\times V::\varphi_{a}\left(
p\right)  =\left(  x,u\right)  $. As the maps $\varphi_{a}$\ are
diffeomorphism they give the same result. But from my personal experience the
convention adopted here is more convenient.

9. The indexes a,b in the transition maps : $p=\varphi_{a}\left(
x,u_{a}\right)  =\varphi_{b}\left(  x,u_{b}\right)  \Rightarrow u_{b}%
=\varphi_{ba}\left(  x\right)  \left(  u_{a}\right)  $ are a constant
hassle.\ Careful to stick as often as possible to some simple rules about the
meaning of $\varphi_{ba}:$ the order matters !

\paragraph{Example : the tangent bundle\newline}

Let M be a manifold with atlas $\left(  B,\left(  \Omega_{i},\psi_{i}\right)
_{i\in I}\right)  $ then $TM=\cup_{x\in M}\left\{  u_{x}\in T_{x}M\right\}  $
is a fiber bundle $TM\left(  M,B,\pi\right)  $ with base M, the canonical
projection : $\pi:TM\rightarrow M::\pi\left(  u_{x}\right)  =x$ and the
trivializations : $\left(  \Omega_{i},\varphi_{i}\right)  _{i\in I}%
:\varphi_{i}:O_{i}\times B\rightarrow TM::\left(  \psi_{i}^{\prime}\left(
x\right)  \right)  ^{-1}u=u_{x}$

The transition maps : $\varphi_{ji}\left(  x\right)  =\psi_{j}^{\prime}\left(
x\right)  \circ\left(  \psi_{i}^{\prime}\left(  x\right)  \right)  ^{-1}$ are
linear and do not depend on u. The tangent bundle is trivial iff it is parallelizable.

\paragraph{General theorems\newline}

\begin{theorem}
(Hosemoller p.15) A fibered manifold $E(M,\pi)$ is a fiber bundle $E(M,V,\pi)$
iff each fiber $\pi^{-1}\left(  x\right)  $\ is diffeomorphic to V.
\end{theorem}

\begin{theorem}
(Kolar p.77) If $\pi:E\rightarrow M$ is a proper surjective submersion from a
manifold E to a connected manifold M, both real finite dimensional, then there
is a structure of fiber bundle $E(M,V,\pi)$ for some V.
\end{theorem}

\begin{theorem}
For any Hausdorff manifolds M,V, M connected, if there are an open cover
$\left(  O_{a}\right)  _{a\in A}$ of M , a set of diffeomorphisms $x\in
O_{a}\cap O_{b}$,\ $\varphi_{ab}\left(  x\right)  \in C_{r}\left(  V;V\right)
, $ there is a fiber bundle structure $E(M,V,\pi).$
\end{theorem}

This shows that the transitions maps are an essential ingredient in the
definition of fiber bundles.

\begin{proof}
Let X be the set : $X=\cup_{a\in A}O_{a}\times V$ and the equivalence relation :

$\Re:(x,u)\sim(x^{\prime},u^{\prime}),x\in O_{a},x^{\prime}\in O_{b}%
\Leftrightarrow x=x^{\prime},u^{\prime}=\varphi_{ba}\left(  x\right)  \left(
u\right)  $

Then the fiber bundle is $E=X/\Re$

The projection is the map : $\pi:E\rightarrow M::\pi\left(  \left[
x,u\right]  \right)  =x$

The trivializations are : $\varphi_{a}:O_{a}\times V\rightarrow E::\varphi
_{a}\left(  x,u\right)  =\left[  x,u\right]  $

and the cocycle conditions are met.
\end{proof}

\begin{theorem}
(Giachetta p.10) If M is reducible to a point, then any fiber bundle with
basis M is trivial (but this is untrue for fibered manifold).
\end{theorem}

\begin{theorem}
(Giachetta p.9) A fibered manifold whose fibers are diffemorphic either to a
compact manifold or $%
\mathbb{R}
^{r}$ is a fiber bundle.
\end{theorem}

\begin{theorem}
(Giachetta p.10) Any finite dimensional fiber bundle admits a
\textit{countable} open cover whose each element has a compact closure.
\end{theorem}

\subsubsection{Change of trivialization}

As manifolds, a fiber bundle can be defined by different atlas. This topic is
a bit more complicated for fiber bundles, all the more so because there are
different ways to procede and many denominations (change of trivialization, of
gauge,...). The definition above gives the conditions for two atlas to be
compatible. But it is convenient to have a general procedure to build another
compatible atlas from a given atlas. The fiber bundle structure and each
element of E stay the same. Usually the open cover is not a big issue, what
matters is the couple (x,u). Because of the projection $\pi$\ the element x is
always the same.\ So the change can come only from $u\in V.$

\begin{definition}
A \textbf{change of trivialization} on a fiber bundle $E\left(  M,V,\pi
\right)  $ with atlas $\left(  O_{a},\varphi_{a}\right)  _{a\in A}$\ is the
definition of a new, compatible atlas $\left(  O_{a},\widetilde{\varphi}%
_{a}\right)  _{a\in A}$ , the trivialization $\widetilde{\varphi}_{a}$\ is
defined by : $p=\varphi_{a}\left(  x,u_{a}\right)  =\widetilde{\varphi}%
_{a}\left(  x,\widetilde{u}_{a}\right)  \Leftrightarrow\widetilde{u}_{a}%
=\chi_{a}\left(  x\right)  \left(  u_{a}\right)  $ where $\left(  \chi
_{a}\left(  x\right)  \right)  _{a\in A}$ is a family of diffeomorphisms on V.
The new transition maps are : $\widetilde{\varphi}_{ba}\left(  x\right)
=\chi_{b}\left(  x\right)  \circ\varphi_{ba}\left(  x\right)  \circ\chi
_{a}\left(  x\right)  ^{-1}$
\end{definition}

\begin{proof}
For $x\in O_{a}$ the transition map between the two atlas is $\widetilde
{u}_{a}=\chi_{a}\left(  x\right)  \left(  u_{a}\right)  $

For $x\in O_{a}\cap O_{b}:$

$p=\varphi_{a}\left(  x,u_{a}\right)  =\varphi_{b}\left(  x,u_{b}\right)
\Leftrightarrow u_{b}=\varphi_{ba}\left(  x\right)  \left(  u_{a}\right)  $

$p=\widetilde{\varphi}_{a}\left(  x,\widetilde{u}_{a}\right)  =\widetilde
{\varphi}_{b}\left(  x,\widetilde{u}_{b}\right)  $

$u_{b}=\varphi_{ba}\left(  x\right)  \left(  u_{a}\right)  $

$\widetilde{u}_{b}=\chi_{b}\left(  x\right)  u_{b}$

$\widetilde{u}_{a}=\chi_{a}\left(  x\right)  u_{a}$

$\widetilde{u}_{b}=\chi_{b}\left(  x\right)  \circ\varphi_{ba}\left(
x\right)  \circ\chi_{a}\left(  x\right)  ^{-1}\left(  \widetilde{u}%
_{a}\right)  $

So the new transition maps are :

$\widetilde{\varphi}_{ba}\left(  x\right)  =\chi_{b}\left(  x\right)
\circ\varphi_{ba}\left(  x\right)  \circ\chi_{a}\left(  x\right)  ^{-1}$

The cocycle conditions are met.\ They read : $:$

$\widetilde{\varphi}_{aa}\left(  x\right)  =\chi_{a}\left(  x\right)
\circ\varphi_{aa}\left(  x\right)  \circ\chi_{a}\left(  x\right)  ^{-1}=Id$

$\widetilde{\varphi}_{ab}\left(  x\right)  \circ\widetilde{\varphi}%
_{bc}\left(  x\right)  =\chi_{a}\left(  x\right)  \circ\varphi_{ab}\left(
x\right)  \circ\chi_{b}\left(  x\right)  ^{-1}\circ\chi_{b}\left(  x\right)
\circ\varphi_{bc}\left(  x\right)  \circ\chi_{c}\left(  x\right)  ^{-1}%
=\chi_{a}\left(  x\right)  \circ\varphi_{ac}\left(  x\right)  \circ\chi
_{c}\left(  x\right)  ^{-1}$
\end{proof}

$\chi_{a}$ is defined in $O_{a}$\ and valued in the set of diffeomorphisms
over V : we have a \textbf{local change of trivialization}.\ When $\chi_{a}$
is constant in $O_{a}$ (this is the same diffeomorphism whatever x) this is a
\textbf{global change of trivialization}. When V has an algebraic structure
additional conditions are required from $\chi.$

The interest of change of trivializations, notably for physicists, is the following.

A property on a fiber bundle is \textbf{intrinsic} (purely geometric) if it
does not depend on the trivialization. This is similar to what we say for
manifolds or vector spaces : a property is intrinsic if it does not depend on
the charts or the basis in which it is expressed. If a quantity is tensorial
it must change according to precise rules in a change of basis. In physics to
say that a quantity is intrinsic is the implementation of the general
principle that the laws of physics should not depend on the observer. If an
observer uses a trivialization, and another one uses another trivialization,
then their measures (the $u\in V)$ of the same phenomenon should be related by
a precise mathematical transformation : they are \textbf{equivariant}.

The observer has the "freedom of gauge". So, when a physical law is expressed
in coordinates, its specification shall be equivariant in \textit{any} change
of trivialization : we are free to choose the family $\left(  \chi_{a}\left(
x\right)  \right)  _{a\in A}$ to verify this principle. This is a powerful
tool to help in the specification of the laws. It is the amplification of the
well known "rule of dimensionality" with respect to the units of measures.

A change of trivialization is a change of atlas, both atlas being compatible.
One can see that the conditions to be met are the same as the transition
conditions (up to a change of notation). So we have the simple rule (warning !
as usual the order of the indices a,b matters):

\begin{theorem}
Whenever a theorem is proven with the usual transition conditions, it is
proven for any change of trivialization. The formulas for a change of
trivialization read as the formulas for the transitions by taking
$\mathbf{\varphi}_{ba}\left(  x\right)  \mathbf{=\chi}_{a}\left(  x\right)  .$
\end{theorem}

$p=\varphi_{a}\left(  x,u_{a}\right)  =\varphi_{b}\left(  x,u_{b}\right)
=\varphi_{b}\left(  x,\varphi_{ba}\left(  x\right)  \left(  u_{a}\right)
\right)  \leftrightarrow p=\varphi_{a}\left(  x,u_{a}\right)  =\widetilde
{\varphi}_{a}\left(  x,\widetilde{u}_{a}\right)  =\widetilde{\varphi}%
_{a}\left(  x,\chi_{a}\left(  x\right)  \left(  u_{a}\right)  \right)  $

\begin{definition}
A \textbf{one parameter group of change of trivialization} on a fiber bundle
$E\left(  M,V,\pi\right)  $ with atlas $\left(  O_{a},\varphi_{a}\right)
_{a\in A}$\ is the family of diffeomorphisms on V defined by a vector field
$W\in\mathfrak{X}\left(  TV\right)  $ with complete flow.
\end{definition}

The flow $\Phi_{W}\left(  u,t\right)  $ is a diffeomorphism on $%
\mathbb{R}
\times V$. If sums up to a change of trivialization $\varphi_{a}%
\rightarrow\widetilde{\varphi}_{at}$ with the diffeormorphisms $\left(
\Phi_{Wt}\right)  _{t\in%
\mathbb{R}
}:\Phi_{Wt}=\Phi_{W}\left(  .,t\right)  $ : $p=\widetilde{\varphi}_{at}\left(
x,\widetilde{u}_{a}\left(  t\right)  \right)  =\varphi_{a}\left(  x,\Phi
_{W}\left(  u_{a},t\right)  \right)  $

\textit{For given t and W} we have a compatible atlas with trivializations
$\widetilde{\varphi}_{at}$ and transition maps : $\widetilde{\varphi}%
_{bat}\left(  x\right)  =\Phi_{Wt}\circ\varphi_{ba}\left(  x\right)  \circ
\Phi_{W-t}.$

They are of special interest because they can be easily labeled (they are
parametrized by W and $t\in%
\mathbb{R}
)$ and their properties are simpler.

\subsubsection{Sections}

\paragraph{Fibered manifolds\newline}

\begin{definition}
A \textbf{section} on a fibered manifold $E(M,\pi)$ is a map $S:M\rightarrow
E$ such that $\pi\circ S=Id$
\end{definition}

\paragraph{Fiber bundles\newline}

A \textit{local section} on a fiber bundle $E(M,V,\pi)$ is a map :
$S:O\rightarrow E$ with domain $O\subset M$

A \textit{section} on a fiber bundle is defined by a map : $\sigma
:M\rightarrow V$ such that $S=\varphi\left(  x,\sigma\left(  x\right)
\right)  .$ However we must solve the transition problem between two opens
$O_{a},O_{b}.$ To be consistent with the rule chosen for tensors over a
manifold, a section is defined by a family of maps $\sigma_{a}\left(
x\right)  $\ with the condition that they define the \textit{same element}
S(x) at the transitions.

\begin{definition}
A class r section S on a fiber bundle $E(M,V,\pi)$ with trivialization
$(O_{a},\varphi_{a})_{a\in A}$ is a family of maps $\left(  \sigma_{a}\right)
_{a\in A},\sigma_{a}\in C_{r}\left(  O_{a};V\right)  $ such that:
\end{definition}

$\forall a\in A,x\in O_{a}:S\left(  x\right)  =\varphi_{a}\left(  x,\sigma
_{a}\left(  x\right)  \right)  $

$\forall a,b\in A,O_{a}\cap O_{b}\neq\varnothing,\forall x\in O_{a}\cap
O_{b}:\sigma_{b}\left(  x\right)  =\varphi_{ba}\left(  x,\sigma_{a}\left(
x\right)  \right)  $

\begin{notation}
$\mathfrak{X}_{r}\left(  E\right)  $ is the set of class r sections of the
fiber bundle E
\end{notation}

Example : On the tangent bundle $TM\left(  M,B,\pi\right)  $ over a manifold
the sections are vector fields V(p) and the set of sections is $\mathfrak{X}%
\left(  TM\right)  $

A \textbf{global section }on a fiber bundle $E(M,V,\pi)$ is a section defined
by a single map : $\sigma\in C_{r}\left(  M;V\right)  $. It implies : $\forall
x\in O_{a}\cap O_{b},\sigma\left(  x\right)  =\varphi_{ba}\left(
x,\sigma\left(  x\right)  \right)  .$ On a fiber bundle we always have
sections but not all fiber bundles have a global section.

\subsubsection{Morphisms}

Morphisms have a general definition for fibered manifolds.\ For fiber bundles
it depends on the structure of V, and it is addressed in the following sections.

\begin{definition}
A fibered manifold morphism or \textbf{fibered map} between two fibered
manifolds $E_{1}(M_{1},\pi_{1}),E_{2}(M_{2},\pi_{2})$ is a couple (F,f) of
maps :

$F:E_{1}\rightarrow E_{2},f:M_{1}\rightarrow M_{2}$ such that : $\pi_{2}\circ
F=f\circ\pi_{1}$.
\end{definition}

The following diagram commutes :

\bigskip

$%
\begin{tabular}
[c]{cccccc}%
E$_{1}$ & $\rightarrow$ & $\rightarrow$ & $\rightarrow$ & E$_{2}$ & \\
$\downarrow$ &  & F &  & $\downarrow$ & \\
$\downarrow$ & $\pi_{1}$ &  &  & $\downarrow$ & $\pi_{2}$\\
$\downarrow$ &  &  &  & $\downarrow$ & \\
M$_{1}$ & $\rightarrow$ & $\rightarrow$ & $\rightarrow$ & M$_{2}$ & \\
&  & f &  &  &
\end{tabular}
$

\bigskip

\begin{definition}
A \textbf{base preserving morphism} is a\ morphism $F:E_{1}\rightarrow E_{2}$
between two fibered manifolds $E_{1}(M,\pi_{1}),E_{2}(M,\pi_{2})$ over the
same base M such that $\pi_{2}\circ F=\pi_{1}$ , which means that f is the identity.
\end{definition}

\begin{definition}
A morphism (F,f) of fibered manifolds is injective if F,f are both injective,
surjective if F,f are both surjective and is an isomorphism if F,f are both
diffeomorphisms. Two fibered manifolds are \textbf{isomorphic} if there is a
fibered map (F,f) such that F and f are diffeomorphisms.
\end{definition}

\begin{theorem}
If (F,f) is a fibered manifold morphism $E_{1}(M_{1},\pi_{1})\rightarrow
E_{2}(M_{2},\pi_{2})$ and f is a local diffeomorphim, then if $S_{1}$\ is a
section on $E_{1}$ then $S_{2}=F\circ S_{1}\circ f^{-1}$ is a section on
$E_{2}:$
\end{theorem}

\begin{proof}
$\pi_{2}\circ S_{2}\left(  x_{2}\right)  =\pi_{2}\circ F\circ S_{1}\left(
x_{1}\right)  =f\circ\pi_{1}\circ S_{1}\left(  x_{1}\right)  =f\left(
x_{1}\right)  =x_{2}$
\end{proof}

\subsubsection{Product and sum of fibered bundles}

\begin{definition}
The \textbf{product} of two fibered manifolds $E_{1}(M_{1},\pi_{1}%
),E_{2}(M_{2},\pi_{2})$ is the fibered manifold $\left(  E_{1}\times
E_{2}\right)  \left(  M_{1}\times M_{2},\pi_{1}\times\pi_{2}\right)  $

The product of two fiber bundle $E_{1}(M_{1},V_{1},\pi_{1}),E_{2}(M_{2}%
,V_{2},\pi_{2})$ is the fiber bundle $\left(  E_{1}\times E_{2}\right)
\left(  M_{1}\times M_{2},V_{1}\times V_{2},\pi_{1}\times\pi_{2}\right)  $
\end{definition}

\begin{definition}
The \textbf{Whitney sum} of two fibered manifolds $E_{1}(M,\pi_{1}%
),E_{2}(M,\pi_{2})$ is the fibered manifold denoted $E_{1}\oplus E_{2}$ where
: M is the base, the total space is : $E_{1}\oplus E_{2}=\left\{  \left(
p_{1},p_{2}\right)  :\pi_{1}\left(  p_{1}\right)  =\pi_{2}\left(
p_{2}\right)  \right\}  ,$ the projection$\ \pi\left(  p_{1},p_{2}\right)
=\pi_{1}\left(  p_{1}\right)  =\pi_{2}\left(  p_{2}\right)  $

The Whitney sum of two fiber bundles $E_{1}(M,V_{1},\pi_{1}),E_{2}(M,V_{2}%
,\pi_{2})$ is the fiber bundle denoted $E_{1}\oplus E_{2} $ where : M is the
base, the total space is : $E_{1}\oplus E_{2}=\left\{  \left(  p_{1}%
,p_{2}\right)  :\pi_{1}\left(  p_{1}\right)  =\pi_{2}\left(  p_{2}\right)
\right\}  ,$ the projection$\ \pi\left(  p_{1},p_{2}\right)  =\pi_{1}\left(
p_{1}\right)  =\pi_{2}\left(  p_{2}\right)  ,$ the fiber is $V=V_{1}\times
V_{2}$ and the trivializations : $\varphi\left(  x,\left(  u_{1},u_{2}\right)
\right)  =\left(  p_{1},p_{2}\right)  =\left(  \varphi_{1}\left(
x,u_{1}\right)  ,\varphi_{2}\left(  x,u_{2}\right)  \right)  $
\end{definition}

If the fibers bundles are trivial then their Whitney sum is trivial.

\subsubsection{Pull back}

Also called induced bundle

\begin{definition}
The \textbf{pull back} $f^{\ast}E$ of a fibered manifold $E\left(
M,\pi\right)  $\ on a manifold N by a continuous map $f:N\rightarrow M$ is the
fibered manifold $f^{\ast}E\left(  N,\widetilde{\pi}\right)  $ with total
space : $f^{\ast}E=\left\{  \left(  y,p\right)  \in N\times E:f\left(
y\right)  =\pi\left(  p\right)  \right\}  ,$projection : $\widetilde{\pi
}:f^{\ast}E\rightarrow N::\widetilde{\pi}\left(  y,p\right)  =y$
\end{definition}

\begin{definition}
The pull back $f^{\ast}E$ of a fiber bundle $E\left(  M,V,\pi\right)  $\ with
atlas $(O_{a},\varphi_{a})_{a\in A}$\ on a manifold N by a continuous map
$f:N\rightarrow M$ is the fiber bundle $f^{\ast}E\left(  N,V,\widetilde{\pi
}\right)  $ with :

total space : $f^{\ast}E=\left\{  \left(  y,p\right)  \in N\times E:f\left(
y\right)  =\pi\left(  p\right)  \right\}  ,$

projection : $\widetilde{\pi}:f^{\ast}E\rightarrow N::\widetilde{\pi}\left(
y,p\right)  =y,$

open cover : $f^{-1}\left(  O_{a}\right)  ,$

trivializations : $\widetilde{\varphi}_{a}:f^{-1}\left(  O_{a}\right)  \times
V\rightarrow\widetilde{\pi}^{-1}\circ f^{-1}\left(  O_{a}\right)
::\widetilde{\varphi}_{a}(y,u)=\left(  y,\varphi_{a}(f(y),u)\right)  $
\end{definition}

\bigskip

$\ \ \ \ f^{\ast}E\rightarrow\overset{\pi^{\ast}f}{\rightarrow}\rightarrow
\rightarrow E$

\ \ \ \ \ \ $\downarrow$ $\ \ \ \ \ \ \ \ \ \;\ \ \ \ \ \downarrow$

$f^{\ast}\pi\downarrow$ $\ \ \ \ \ \ \ \ \ \ \ \;\ \ \ \downarrow\pi$

\ \ \ \ \ \ $\downarrow$ $\ \ \ \ \ \ f$ $\ \ \ \ \ \ \downarrow$

$\ \ \ \ \ \ N\rightarrow\rightarrow\rightarrow\rightarrow\rightarrow M$

\bigskip

The projection $\widetilde{\pi}$ is an open map.

For any section $S:M\rightarrow E$ we have the pull back

$f^{\ast}S:N\rightarrow E::f^{\ast}S\left(  y\right)  =\left(  f\left(
y\right)  ,S\left(  f\left(  y\right)  \right)  \right)  $

\subsubsection{Tangent space over a fiber bundle}

The key point is that the tangent space $T_{p}E$ of a fiber bundle is
isomorphic to $T_{x}M\times T_{u}V$

\paragraph{Local charts\newline}

\begin{theorem}
The total space E of a fiber bundle $E\left(  M,V,\pi\right)  $ with atlas
$\left(  O_{a},\varphi_{a}\right)  _{a\in A}$\ admits, as a manifold, an atlas
$\left(  B_{M}\times B_{V},\left(  \varphi_{a}\left(  O_{a},V_{i}\right)
,\Phi_{ai}\right)  _{\left(  a,i\right)  \in A\times I}\right)  $\ where

$\left(  B_{M},\left(  O_{a},\psi_{a}\right)  _{a\in A}\right)  $ is an atlas
of the manifold M

$\left(  B_{V},\left(  U_{i},\phi_{i}\right)  _{i\in I}\right)  $ is an atlas
of the manifold V

$\Phi_{ai}\left(  p\right)  =\left(  \psi_{a}\circ\pi\left(  p\right)
,\phi_{i}\tau_{a}\left(  p\right)  \right)  =\left(  \xi_{a},\eta_{a}\right)
$
\end{theorem}

\begin{proof}
By taking a refinement of the open covers we can assume that $\left(
B_{M},\left(  O_{a},\psi_{a}\right)  _{a\in A}\right)  $ is an atlas of M

As the $\varphi_{a}$ are diffeomorphisms the following maps are well defined :

$\tau_{a}:\pi^{-1}\left(  O_{a}\right)  \rightarrow V::\varphi_{a}\left(
x,\tau_{a}\left(  p\right)  \right)  =p=\varphi_{a}\left(  \pi\left(
p\right)  ,\tau_{a}\left(  p\right)  \right)  $

$\forall p\in\pi^{-1}\left(  O_{a}\right)  \cap\pi^{-1}\left(  O_{b}\right)
:\tau_{b}\left(  p\right)  =\varphi_{ba}\left(  \pi\left(  p\right)  ,\tau
_{a}\left(  p\right)  \right)  $

and the $\varphi_{a}$ are open maps so $\varphi_{a}\left(  O_{a},V_{i}\right)
=\Omega_{ai}$ is an open cover of E.

The map :

$\Phi_{ai}:\varphi_{a}\left(  O_{a},V_{i}\right)  \rightarrow B_{M}\times
B_{V}::\Phi_{ai}\left(  p\right)  =\left(  \psi_{a}\circ\pi\left(  p\right)
,\phi_{i}\tau_{a}\left(  p\right)  \right)  =\left(  \xi_{a},\eta_{a}\right)
$

is bijective and differentiable.

If $p\in\Omega_{ai}\cap\Omega_{bj}:\Phi_{bj}\left(  p\right)  =\left(
\psi_{b}\circ\pi\left(  p\right)  ,\phi_{j}\circ\tau_{b}\left(  p\right)
\right)  =\left(  \xi_{b},\eta_{bj}\right)  $

$\xi_{b}=\psi_{b}\circ\pi\circ\pi^{-1}\circ\psi_{a}^{-1}\left(  \xi
_{a}\right)  =\psi_{b}\circ\psi_{a}^{-1}\left(  \xi_{a}\right)  =\psi
_{ba}\left(  \xi_{a}\right)  $

$\eta_{bj}=\phi_{j}\circ\tau_{b}\circ\tau_{a}^{-1}\circ\phi_{i}^{-1}\left(
\eta_{ai}\right)  =\phi_{j}\circ\varphi_{ba}\left(  \psi_{a}^{-1}\left(
\xi_{a}\right)  ,\phi_{i}^{-1}\left(  \eta_{ai}\right)  \right)  $
\end{proof}

\paragraph{Tangent space\newline}

\begin{theorem}
Any vector $v_{p}\in T_{p}E$ of the fiber bundle $E\left(  M,V,\pi\right)  $
has a unique decomposition : $v_{p}=\varphi_{a}^{\prime}\left(  x,u_{a}%
\right)  \left(  v_{x},v_{au}\right)  $ where : $v_{x}=\pi^{\prime}\left(
p\right)  v_{p}\in T_{\pi\left(  p\right)  }M$ does not depend on the
trivialization and $v_{au}=\tau_{a}^{\prime}\left(  p\right)  v_{p}$
\end{theorem}

\begin{proof}
The differentiation of :

$p=\varphi_{a}\left(  x,u_{a}\right)  \rightarrow v_{p}=\varphi_{ax}^{\prime
}\left(  x,u_{a}\right)  v_{ax}+\varphi_{au}^{\prime}\left(  x,u_{a}\right)
v_{au}$

$x=\pi\left(  p\right)  \rightarrow v_{ax}=\pi^{\prime}\left(  p\right)
v_{p}\Rightarrow v_{ax}$ does not depend on the trivialization

$p=\varphi_{a}\left(  \pi\left(  p\right)  ,\tau_{a}\left(  p\right)  \right)
\rightarrow v_{p}=\varphi_{ax}^{\prime}\left(  \pi\left(  p\right)  ,\tau
_{a}\left(  p\right)  \right)  \pi^{\prime}\left(  p\right)  v_{p}%
+\varphi_{au}^{\prime}\left(  \pi\left(  p\right)  ,\tau_{a}\left(  p\right)
\right)  \tau_{a}^{\prime}\left(  p\right)  v_{p}$

$\varphi_{au}^{\prime}\left(  x,u_{a}\right)  v_{au}=\varphi_{au}^{\prime
}\left(  \pi\left(  p\right)  ,\tau_{a}\left(  p\right)  \right)  \tau
_{a}^{\prime}\left(  p\right)  v_{p}\Rightarrow v_{au}=\tau_{a}^{\prime
}\left(  p\right)  v_{p}$
\end{proof}

\begin{theorem}
Any vector of $T_{p}E$\ can be uniquely written$:$%

\begin{equation}
v_{p}=\sum_{\alpha}v_{x}^{\alpha}\partial x_{\alpha}+\sum_{i}v_{au}%
^{i}\partial u_{i}%
\end{equation}

with the basis, called a \textbf{holonomic} basis,%

\begin{equation}
\partial x_{\alpha}=\varphi_{ax}^{\prime}\left(  x,u\right)  \partial
\xi_{\alpha},\partial u_{i}=\varphi_{au}^{\prime}\left(  x,u\right)
\partial\eta_{i}%
\end{equation}

where $\partial\xi_{\alpha},\partial\eta_{i}$ are holonomic bases of
$T_{x}M,T_{u}V$
\end{theorem}

\begin{proof}
$v_{x}=\sum_{\alpha}v_{x}^{\alpha}\partial\xi_{\alpha},$ $v_{au}=\sum
_{i}v_{au}^{i}\partial\eta_{i}$

$v_{p}=\varphi_{ax}^{\prime}\left(  x,u_{a}\right)  v_{x}+\varphi_{au}%
^{\prime}\left(  x,u_{a}\right)  v_{au}=\sum_{\alpha}v_{x}^{\alpha}%
\varphi_{ax}^{\prime}\left(  x,u\right)  \partial\xi_{\alpha}+\sum_{i}%
v_{au}^{i}\varphi_{au}^{\prime}\left(  x,u\right)  \partial\eta_{i}$
\end{proof}

In the following in this book:

\begin{notation}
$\partial x_{\alpha}$ (latine letter, greek indices) is the part of the basis
on TE induced by M
\end{notation}

\begin{notation}
$\partial u_{i}$ (latine letter, latine indices) is the part of the basis on
TE induced by V.
\end{notation}

\begin{notation}
$\partial\xi_{\alpha}$ (greek letter, greek indices) is a holonomic basis on TM.
\end{notation}

\begin{notation}
$\partial\eta_{i}$ (greek letter, latine indices) is a holonomic basis on TV.
\end{notation}

\begin{notation}
$v_{p}=\sum_{\alpha}v_{x}^{\alpha}\partial x_{\alpha}+\sum_{i}v_{u}%
^{i}\partial u_{i}$ is any vector $v_{p}\in T_{p}E$
\end{notation}

With this notation it is clear that a holonomic basis on TE splits in a part
related to M (the base) and V (the standard fiber).\ Notice that the
decomposition is unique for a given trivialization, but depends on the trivialization.

\paragraph{Transition\newline}

A transition does not involve the atlas of the manifolds M, V. Only the way
the atlas of E is defined. Thus the vectors $v_{p}\in T_{p}E,v_{x}\in
T_{\pi\left(  p\right)  }M$ do not change. For clarity we write $\varphi
_{ba}\left(  x\right)  \left(  u_{a}\right)  =\varphi_{ba}\left(
x,u_{a}\right)  $

\begin{theorem}
At the transitions between charts : $v_{p}=\varphi_{a}^{\prime}\left(
x,u_{a}\right)  \left(  v_{x},v_{au}\right)  =\varphi_{b}^{\prime}\left(
x,u_{b}\right)  \left(  v_{x},v_{bu}\right)  $ we have the identities :

$\varphi_{ax}^{\prime}\left(  x,u_{a}\right)  =\varphi_{bx}^{\prime}\left(
x,u_{b}\right)  +\varphi_{bu}^{\prime}\left(  x,u_{b}\right)  \varphi
_{bax}^{\prime}\left(  x,u_{a}\right)  $

$\varphi_{au}^{\prime}\left(  x,u_{a}\right)  =\varphi_{bu}^{\prime}\left(
x,u_{b}\right)  \varphi_{bau}^{\prime}\left(  x,u_{a}\right)  $

$v_{bu}=\left(  \varphi_{ba}\left(  x,u_{a}\right)  \right)  ^{\prime}\left(
v_{x},v_{au}\right)  $
\end{theorem}

\begin{proof}
The differentiation of :$\varphi_{a}\left(  x,u_{a}\right)  =\varphi
_{b}\left(  x,\varphi_{ba}\left(  x,u_{a}\right)  \right)  $ with respect to
$u_{a}$ gives :

$\varphi_{au}^{\prime}\left(  x,u_{a}\right)  v_{au}=\varphi_{bu}^{\prime
}\left(  x,u_{b}\right)  \varphi_{bau}^{\prime}\left(  x,u_{a}\right)  v_{au}$

The differentiation of :$\varphi_{a}\left(  x,u_{a}\right)  =\varphi
_{b}\left(  x,\varphi_{ba}\left(  x,u_{a}\right)  \right)  $ with respect to x
gives :

$\varphi_{ax}^{\prime}\left(  x,u_{a}\right)  v_{x}=\left(  \varphi
_{bx}^{\prime}\left(  x,u_{b}\right)  +\varphi_{bu}^{\prime}\left(
x,u_{b}\right)  \varphi_{bax}^{\prime}\left(  x,u_{a}\right)  \right)  v_{x}$

$v_{p}=\varphi_{ax}^{\prime}\left(  x,u_{a}\right)  v_{x}+\varphi_{au}%
^{\prime}\left(  x,u_{a}\right)  v_{au}=\varphi_{bx}^{\prime}\left(
x,u_{b}\right)  v_{x}+\varphi_{bu}^{\prime}\left(  x,u_{b}\right)  v_{bu}$

$\left(  \varphi_{bx}^{\prime}\left(  x,u_{b}\right)  +\varphi_{bu}^{\prime
}\left(  x,u_{b}\right)  \varphi_{bax}^{\prime}\left(  x,u_{a}\right)
\right)  v_{x}+\varphi_{bu}^{\prime}\left(  x,u_{b}\right)  \varphi
_{bau}^{\prime}\left(  x,u_{a}\right)  v_{au}$

$=\varphi_{bx}^{\prime}\left(  x,u_{b}\right)  v_{x}+\varphi_{bu}^{\prime
}\left(  x,u_{b}\right)  v_{bu}$

$\varphi_{bu}^{\prime}\left(  x,u_{b}\right)  \varphi_{bax}^{\prime}\left(
x,u_{a}\right)  v_{x}+\varphi_{bu}^{\prime}\left(  x,u_{b}\right)
\varphi_{bau}^{\prime}\left(  x,u_{a}\right)  v_{au}=\varphi_{bu}^{\prime
}\left(  x,u_{b}\right)  v_{bu}$

$v_{bu}=\varphi_{bax}^{\prime}\left(  x,u_{a}\right)  v_{x}+\varphi
_{bau}^{\prime}\left(  x,u_{a}\right)  v_{au}$
\end{proof}

The holonomic bases are at p:

$\partial_{a}x_{\alpha}=\varphi_{ax}^{\prime}\left(  x,u_{a}\right)
\partial\xi_{\alpha}=\varphi_{bx}^{\prime}\left(  x,u_{b}\right)  \partial
\xi_{\alpha}+\varphi_{bu}^{\prime}\left(  x,u_{b}\right)  \varphi
_{bax}^{\prime}\left(  x,u_{a}\right)  \partial\xi_{\alpha}=\partial
_{b}x_{\alpha}+\sum_{j}\left[  \varphi_{bax}^{\prime}\left(  x,u_{a}\right)
\right]  _{\alpha}^{j}\partial_{b}u_{j}$

$\partial_{a}u_{i}=\varphi_{a}^{\prime}\left(  x,u_{a}\right)  \partial
\eta_{i}=\varphi_{bu}^{\prime}\left(  x,u_{b}\right)  \varphi_{bau}^{\prime
}\left(  x,u_{a}\right)  \partial\eta_{i}=\sum_{j}\left[  \varphi
_{bau}^{\prime}\left(  x,u_{a}\right)  \right]  _{i}^{j}\partial_{b}u_{j}$

\begin{theorem}
If S is a section on E, then $S^{\prime}\left(  x\right)  :T_{x}M\rightarrow
T_{S\left(  x\right)  }E$ is such that : $\pi^{\prime}\left(  S\left(
x\right)  \right)  S^{\prime}\left(  x\right)  =Id$
\end{theorem}

\begin{proof}
$\pi\left(  S\left(  x\right)  \right)  =x\Rightarrow\pi^{\prime}\left(
S\left(  x\right)  \right)  S^{\prime}\left(  x\right)  v_{x}=v_{x}$
\end{proof}

\paragraph{Vertical space}

\begin{definition}
The \textbf{vertical space} at p to the fiber bundle $E(M,V,\pi)$ is :
$V_{p}E=\ker\pi^{\prime}\left(  p\right)  .$ It is isomorphic to $T_{u}V$
\end{definition}

This is a vector subspace of $T_{p}E$ which does not depend\ on the trivialization

As : $\pi\left(  p\right)  =x\Rightarrow\pi\left(  p\right)  ^{\prime}%
\varphi_{a}^{\prime}\left(  x,u\right)  \left(  v_{x},v_{u}\right)  =v_{x}$ so

$V_{p}E=\left\{  \varphi_{a}^{\prime}\left(  x,u\right)  \left(  v_{x}%
,v_{u}\right)  ,v_{x}=0\right\}  =\left\{  \varphi_{au}^{\prime}\left(
x,u\right)  v_{u},v_{u}\in T_{u}V\right\}  $

It is isomorphic to $T_{u}V$ thus to $B_{V}$

\paragraph{Fiber bundle structures}

\begin{theorem}
The tangent bundle TE\ of a fiber bundle $E\left(  M,V,\pi\right)  $ is the
vector bundle $TE\left(  TM,TV,T\pi\right)  $
\end{theorem}

Total space : $TE=\cup_{p\in E}\left\{  v_{p}\in T_{p}E\right\}  $

Base : $TM=\cup_{x\in M}\left\{  x,v_{x}\in T_{x}M\right\}  $

Projection : $T\pi\left(  p,v_{p}\right)  =\left(  x,v_{x}\right)  $

Open cover : $\{TM,x\in O_{a}\}$

Trivializations :

$\left(  x,v_{x}\right)  \times\left(  u,v_{u}\right)  \in TM\times
TV\rightarrow(\varphi_{a}(x,u),\varphi_{a}^{\prime}(x,u)(v_{x},v_{u})\}\in TE$

Transitions :

$\left(  \left(  x,v_{bx}\right)  ,\left(  u_{b},v_{bu}\right)  \right)
=\left(  \left(  x,v_{ax}\right)  ,\left(  \varphi_{ba}\left(  x,u_{a}\right)
,\left(  \varphi_{ba}\left(  x,u_{a}\right)  \right)  ^{\prime}\left(
v_{ax},v_{au}\right)  \right)  \right)  $

Basis : $v_{p}=\sum_{\alpha}v_{x}^{\alpha}\partial x_{\alpha}+\sum_{i}%
v_{u}^{i}\partial u_{i},\partial x_{\alpha}=\varphi_{ax}^{\prime}\left(
x,u_{a}\right)  \partial\xi_{\alpha},\partial u_{i}=\varphi_{au}^{\prime
}\left(  x,u_{a}\right)  \partial\eta_{i}$

The coordinates of $v_{p}$ in this atlas are : $\left(  \xi^{\alpha},\eta
^{i},v_{x}^{\alpha},v_{u}^{i}\right)  $

\begin{theorem}
The \textbf{vertical bundle }\ of a fiber bundle $E\left(  M,V,\pi\right)  $
is the vector bundle : $VE\left(  M,TV,\pi\right)  $
\end{theorem}

total space : $VE=\cup_{p\in E}\left\{  v_{p}\in V_{p}E\right\}  $

base : M

projection : $\pi\left(  v_{p}\right)  =x$

trivializations : $M\times TV\rightarrow VE::\varphi_{au}^{\prime}%
(x,u)v_{u}\in VE$

open cover : $\{O_{a}\}$

transitions : $\left(  u_{b},v_{bu}\right)  =\left(  \varphi_{ba}\left(
x,u_{a}\right)  ,\varphi_{bau}^{\prime}\left(  x,u_{a}\right)  v_{au}\right)
$

\subsubsection{Vector fields on a fiber bundle}

\begin{definition}
A vector field on the tangent bundle TE of the fiber bundle $E(M,V,\pi)$ is a
map : $W:E\rightarrow TE.$ In an atlas $\left(  O_{a},\varphi_{a}\right)
_{a\in A}$ of E it is defined by a family $\left(  W_{ax},W_{au}\right)
_{a\in A}$ with

$W_{ax}:\pi^{-1}\left(  O_{a}\right)  \rightarrow TM,W_{au}:\pi^{-1}\left(
O_{a}\right)  \rightarrow TV$

such that for $x\in O_{a}\cap O_{b},p=\varphi_{a}\left(  x,u_{a}\right)  $

$\left(  W_{bx}\left(  p\right)  ,W_{bu}\left(  p\right)  \right)  =\left(
W_{bx}\left(  p\right)  ,\left(  \varphi_{ba}\left(  x,u_{a}\right)  \right)
^{\prime}\left(  W_{ax}\left(  p\right)  ,W_{au}\left(  p\right)  \right)
\right)  $
\end{definition}

$W_{a}\left(  \varphi_{a}\left(  x,u_{a}\right)  \right)  =\varphi_{a}%
^{\prime}\left(  x,u_{a}\right)  \left(  W_{ax}\left(  p\right)
,W_{au}\left(  p\right)  \right)  =\sum_{\alpha\in A}W_{ax}^{\alpha}\partial
x_{\alpha}+\sum_{i\in I}W_{au}^{i}\partial u_{ai}$

Notice that the components $W_{ax}^{\alpha},W_{au}^{i}$ \textit{depend on p,
that is both on x and u}.

\begin{definition}
A \textbf{vertical vector field} on the tangent bundle TE of the fiber bundle
$E(M,V,\pi)$ is a vector field such that : $\pi^{\prime}\left(  p\right)
W\left(  p\right)  =0.$
\end{definition}

$W_{a}\left(  \varphi_{a}\left(  x,u_{a}\right)  \right)  =\sum_{i\in I}%
W_{au}^{i}\partial u_{ai},$ the components $W_{au}^{i}$ depend on p, that is
both on x and u.

\begin{definition}
The \textbf{commutator} of the vector fields $X,Y\in\mathfrak{X}\left(
TE\right)  $ on the tangent bundle TE of the fiber bundle $E(M,V,\pi)$ is the
vector field :

$\left[  X,Y\right]  _{TE}\left(  \varphi_{a}\left(  x,u_{a}\right)  \right)
=\varphi_{a}^{\prime}\left(  x,u_{a}\right)  \left(  \left[  X_{ax}%
,Y_{ax}\right]  _{TM}\left(  p\right)  ,\left[  X_{au},Y_{au}\right]
_{TV}\left(  p\right)  \right)  $
\end{definition}

$\left[  X,Y\right]  _{TE}=\sum_{\alpha\beta}\left(  X_{x}^{\beta}%
\partial_{\beta}Y_{x}^{\alpha}-Y_{x}^{\beta}\partial_{\beta}X_{x}^{\alpha
}\right)  \partial x_{\alpha}+\sum_{ij}\left(  X_{u}^{j}\partial_{j}Y_{u}%
^{i}-Y_{u}^{j}\partial_{j}X_{u}^{i}\right)  \partial u_{i}$

\subsubsection{Forms defined on a fiber bundle}

To be consistent with previous notations (E is a manifold) for a fiber bundle
$E\left(  M,V,\pi\right)  :$

\begin{notation}
$\Lambda_{r}\left(  E\right)  =\Lambda_{r}\left(  E;K\right)  $ is the usual
space of r forms on E valued in the field K.
\end{notation}

$\varpi=\sum_{\left\{  \alpha_{1}...\alpha_{r}\right\}  }\varpi_{\alpha
_{1}...\alpha_{r}}\left(  x\right)  dx^{\alpha_{1}}\wedge..\wedge dx^{r}$

\begin{notation}
$\Lambda_{r}\left(  E;H\right)  $ is the space of r forms on E valued in a
fixed vector space H.
\end{notation}

$\varpi=\sum_{\left\{  \alpha_{1}...\alpha_{r}\right\}  }\sum_{i}%
\varpi_{\alpha_{1}...\alpha_{r}}^{i}\left(  x\right)  dx^{\alpha_{1}}%
\wedge..\wedge dx^{r}\otimes e_{i}$

\begin{notation}
$\Lambda_{r}\left(  E;TE\right)  $ is the space of r forms on E valued in the
tangent bundle TE of E,
\end{notation}

$\varpi=\sum_{\left\{  \alpha_{1}...\alpha_{r}\right\}  }\sum_{\beta}%
\varpi_{\alpha_{1}...\alpha_{r}}^{\beta i}\left(  p\right)  dx^{\alpha_{1}%
}\wedge..\wedge dx^{r}\otimes\partial x_{\beta}\otimes\partial u_{i}$

\begin{notation}
$\Lambda_{r}\left(  E;VE\right)  $ is the space of r forms on E valued in the
vertical bundle VE of E,
\end{notation}

$\varpi=\sum_{\left\{  \alpha_{1}...\alpha_{r}\right\}  }\sum_{\beta}%
\varpi_{\alpha_{1}...\alpha_{r}}^{i}\left(  p\right)  dx^{\alpha_{1}}%
\wedge..\wedge dx^{r}\otimes\partial u_{i}$

Such a form is said to be \textbf{horizontal} : it is null whenever one of the
vector is vertical.

A horizontal 1-form on E, valued in the vertical bundle, called a soldering
form, reads :

$\theta=\sum_{\alpha,i}\theta_{\alpha}^{i}\left(  p\right)  dx^{\alpha}%
\otimes\partial u_{i}$

\begin{notation}
$\Lambda_{r}\left(  M;TE\right)  $ is the space of r forms on M valued in the
tangent bundle TE of E,
\end{notation}

$\varpi=\sum_{\left\{  \alpha_{1}...\alpha_{r}\right\}  }\sum_{\beta}%
\varpi_{\alpha_{1}...\alpha_{r}}^{\beta i}\left(  p\right)  d\xi^{\alpha_{1}%
}\wedge..\wedge d\xi^{r}\otimes\partial x_{\beta}\otimes\partial u_{i}$

\begin{notation}
$\Lambda_{r}\left(  M;VE\right)  $ is the space of r forms on M valued in the
vertical bundle VE of E,
\end{notation}

$\varpi=\sum_{\left\{  \alpha_{1}...\alpha_{r}\right\}  }\sum_{\beta}%
\varpi_{\alpha_{1}...\alpha_{r}}^{i}\left(  p\right)  d\xi^{\alpha_{1}}%
\wedge..\wedge d\xi^{r}\otimes\partial u_{i}$

The sections over E are considered as 0 forms on M valued in E

\subsubsection{Lie derivative}

\paragraph{Projectable vector field\newline}

For any vector field $W\in\mathfrak{X}\left(  TE\right)  $\ : $T\pi\left(
W\left(  p\right)  \right)  \in TM.$ But this projection is not necessarily a
vector field over M.

\begin{definition}
A vector field $W\in\mathfrak{X}\left(  TE\right)  $ on a fiber bundle
$E\left(  M,V,\pi\right)  $ is \textbf{projectable} if $T\pi\left(  W\right)
\in\mathfrak{X}\left(  TM\right)  $
\end{definition}

$\forall p\in E:T\pi\left(  W\left(  p\right)  \right)  =\left(  \pi\left(
p\right)  ,Y\left(  \pi\left(  p\right)  \right)  \right)  ,Y\in
\mathfrak{X}\left(  M\right)  $

That we can write : $\pi_{\ast}W\left(  \pi\left(  p\right)  \right)
=\pi^{\prime}\left(  p\right)  W\left(  p\right)  \Leftrightarrow Y=\pi_{\ast
}W$

For any vector field : $\pi^{\prime}\left(  p\right)  W\left(  p\right)
=\pi^{\prime}\left(  p\right)  \left(  \varphi_{ax}^{\prime}\left(
x,u_{a}\right)  \left(  v_{x}\left(  p\right)  ,v_{au}\left(  p\right)
\right)  \right)  =v_{x}\left(  p\right)  $ so W is projectable iff
$v_{x}\left(  p\right)  $ does not depend on u. In particular a vertical
vector field is projectable (on 0).

\begin{theorem}
The flow of a projectable vector field on a fiber bundle E is a fibered
manifold morphism
\end{theorem}

\begin{proof}
The flow of a vector field W on E is a local diffeomorphism on E with :
$\frac{\partial}{\partial t}\Phi_{W}\left(  p,t\right)  |_{t=\theta}=W\left(
\Phi_{W}\left(  p,\theta\right)  \right)  $

It is a fibered manifold morphism if there is : $f:M\rightarrow M$ such that :
$\pi\left(  \Phi_{W}\left(  p,t\right)  \right)  =f\left(  \pi\left(
p\right)  ,t\right)  $

If W is a projectable vector field :

$\frac{\partial}{\partial t}\pi\left(  \Phi_{W}\left(  p,t\right)  \right)
|_{t=\theta}=\pi^{\prime}\left(  \Phi_{W}\left(  p,\theta\right)  \right)
\frac{\partial}{\partial t}\Phi_{W}\left(  p,t\right)  |_{t=\theta}$

$=\pi^{\prime}\left(  \Phi_{W}\left(  p,\theta\right)  \right)  W\left(
\Phi_{W}\left(  p,\theta\right)  \right)  =Y\left(  \pi\left(  \Phi_{W}\left(
p,t\right)  \right)  \right)  $

So we have : $\pi\left(  \Phi_{W}\left(  p,t\right)  \right)  =\Phi_{Y}\left(
\pi\left(  p\right)  ,t\right)  $
\end{proof}

If W is vertical, then Y=0 : $\pi\left(  \Phi_{W}\left(  p,t\right)  \right)
=\pi\left(  p\right)  $ the morphism is base preserving

\begin{theorem}
A projectable vector field defines a section on E
\end{theorem}

\begin{proof}
take p in E, in a neighborhood n(x) of $x=\pi\left(  p\right)  $ the flow
$\Phi_{Y}\left(  x,t\right)  $ is defined for some interval J, and $\forall
x^{\prime}\in n(x),\exists t:x^{\prime}=\Phi_{Y}\left(  x,t\right)  $

take $S(x^{\prime})=\Phi_{W}\left(  p,t\right)  \Rightarrow\pi\left(  S\left(
x^{\prime}\right)  \right)  =\Phi_{Y}\left(  \pi\left(  p\right)  ,t\right)
=x^{\prime}$
\end{proof}

\paragraph{Lie derivative of a section\newline}

Lie derivatives can be extended to sections on fiber bundles, but with some
adjustments (from Kolar p.377). As usual they are a way to compare sections on
a fiber bundle at different points, without the need for a covariant derivative.

\begin{theorem}
The Lie derivative of a section S of a fiber bundle $E(M,V,\pi)$\ along a
projectable vector field W is the section of the vertical bundle :%

\begin{equation}
\pounds _{W}S=\frac{\partial}{\partial t}\Phi_{W}\left(  S\left(  \Phi
_{Y}\left(  x,-t\right)  \right)  ,t\right)  |_{t=0}\in\mathfrak{X}\left(
VE\right)
\end{equation}

with $\pi^{\prime}\left(  p\right)  W\left(  p\right)  =Y\left(  \pi\left(
p\right)  \right)  $
\end{theorem}

\begin{proof}
The flow $\Phi_{W}\left(  p,t\right)  $\ is a fibered manifold morphism and a
local diffeomorphism on E

On a neighborhood of (x,0) in $M\times%
\mathbb{R}
$ the map :

$F_{W}\left(  x,t\right)  =\Phi_{W}\left(  S\left(  \Phi_{Y}\left(
x,-t\right)  \right)  ,t\right)  :E\rightarrow E$

defines the transport of a section S on E : $S\rightarrow\widetilde
{S}:\widetilde{S}\left(  x\right)  =F_{W}\left(  x,t\right)  $

We stay in the same fiber because W is projectable and S is a section :

$\pi\left(  \Phi_{W}\left(  S\left(  \Phi_{Y}\left(  x,-t\right)  \right)
,t\right)  \right)  =\Phi_{Y}\left(  \pi\left(  S\left(  \Phi_{Y}\left(
x,-t\right)  \right)  \right)  ,t\right)  =\Phi_{Y}\left(  \Phi_{Y}\left(
x,-t\right)  ,t\right)  =x$

Therefore if we differentiate in t=0 we have a vector in x, which is vertical :

$\frac{\partial}{\partial t}\Phi_{W}\left(  S\left(  \Phi_{Y}\left(
x,-t\right)  \right)  ,t\right)  |_{t=0}$

$=\frac{\partial}{\partial t}\Phi_{W}\left(  p\left(  t\right)  ,t\right)
|_{t=0}=\frac{\partial}{\partial p}\Phi_{W}\left(  p\left(  t\right)
,t\right)  |_{t=0}\frac{\partial p}{\partial t}|_{t=0}+W\left(  p\left(
t\right)  \right)  |_{t=0}$

$=\frac{\partial}{\partial p}\Phi_{W}\left(  p\left(  t\right)  ,t\right)
|_{t=0}\frac{\partial}{\partial t}S\left(  \Phi_{Y}\left(  x,-t\right)
\right)  |_{t=0}+W\left(  S\left(  \Phi_{Y}\left(  x,-t\right)  \right)
\right)  |_{t=0}$

$=\frac{\partial}{\partial p}\Phi_{W}\left(  p\left(  t\right)  ,t\right)
|_{t=0}\frac{\partial}{\partial x}S\left(  y\left(  t\right)  \right)
|_{t=0}\frac{\partial}{\partial t}\Phi_{Y}\left(  x,-t\right)  |_{t=0}%
+W\left(  S\left(  x\right)  \right)  $

$=-\frac{\partial}{\partial p}\Phi_{W}\left(  S\left(  x\right)  \right)
\frac{\partial}{\partial x}S\left(  x\right)  Y(x)+W\left(  S\left(  x\right)
\right)  $

$\pounds _{W}S=-\frac{\partial\Phi_{W}}{\partial p}\left(  S\left(  x\right)
\right)  S^{\prime}\left(  x\right)  Y(x)+W\left(  S\left(  x\right)  \right)
\in V_{p}E$
\end{proof}

So the Lie derivative is a map : $\pounds _{W}S:M\rightarrow VE$

$\Phi_{W}\left(  x,t\right)  \circ\Phi_{W}\left(  x,s\right)  =\Phi_{W}\left(
x,s+t\right)  $ whenever the flows are defined

so : $\frac{\partial}{\partial s}\Phi_{W}\left(  x,s+t\right)  |_{s=0}%
=\Phi_{W}\left(  x,t\right)  \circ\frac{\partial}{\partial s}\Phi_{W}\left(
x,s\right)  |_{s=0}=\Phi_{W}\left(  x,t\right)  \circ\pounds _{W}%
S=\pounds _{W}\left(  \Phi_{W}\left(  x,t\right)  \right)  $

\bigskip

In coordinates :

$S\left(  x\right)  =\varphi\left(  x,\sigma\left(  x\right)  \right)
\rightarrow S^{\prime}(x)=\sum_{\alpha}\partial x_{\alpha}\otimes d\xi
^{\alpha}+\left(  \sum_{i\alpha}\left(  \partial_{\alpha}\sigma^{i}\right)
\partial u_{i}\right)  \otimes d\xi^{\alpha}$

$S^{\prime}\left(  x\right)  Y\left(  x\right)  =\sum_{\alpha}Y^{\alpha
}\left(  x\right)  \partial x_{\alpha}+\sum_{i\alpha}Y^{\alpha}\left(
x\right)  \left(  \partial_{\alpha}\sigma^{i}\right)  \partial u_{i}$

$\pounds _{W}S=-\frac{\partial\Phi_{W}}{\partial p}\left(  S\left(  x\right)
\right)  \left(  \sum_{\alpha}Y^{\alpha}\left(  x\right)  \partial x_{\alpha
}+\sum_{i}Y^{\alpha}\left(  x\right)  \left(  \partial_{\alpha}\sigma
^{i}\right)  \partial u_{i}\right)  +W\left(  S\left(  x\right)  \right)  $

$\frac{\partial\Phi_{W}}{\partial p}=\frac{\partial\Phi_{W}^{\beta}}{\partial
x^{\alpha}}dx^{\alpha}\otimes\partial x_{\beta}+\frac{\partial\Phi_{W}^{i}%
}{\partial x^{\alpha}}dx^{\alpha}\otimes\partial u_{i}+\frac{\partial\Phi
_{W}^{\alpha}}{\partial u^{i}}du^{i}\otimes\partial x_{\alpha}+\frac
{\partial\Phi_{W}^{j}}{\partial u^{i}}du^{i}\otimes\partial u_{j}$

$\pounds _{W}S=\left(  W^{\alpha}-\frac{\partial\Phi_{W}^{\alpha}}{\partial
x^{\beta}}Y^{\beta}-\frac{\partial\Phi_{W}^{\alpha}}{\partial u^{i}}Y^{\beta
}\left(  \partial_{\beta}\sigma^{i}\right)  \right)  \partial x_{\alpha
}+\left(  W^{i}-\frac{\partial\Phi_{W}^{i}}{\partial x^{\alpha}}Y^{\alpha
}-\frac{\partial\Phi_{W}^{i}}{\partial u^{j}}Y^{\alpha}\left(  \partial
_{\alpha}\sigma^{j}\right)  \right)  \partial u_{i}$

$W^{\alpha}=\frac{\partial\Phi_{W}^{\alpha}}{\partial x^{\beta}}Y^{\beta
}+\frac{\partial\Phi_{W}^{\alpha}}{\partial u^{i}}Y^{\beta}\left(
\partial_{\beta}\sigma^{i}\right)  $ because $\pounds _{W}S$ is vertical

$\pounds _{W}S=\sum_{i}\left(  W^{i}-\frac{\partial\Phi_{W}^{i}}{\partial
x^{\alpha}}Y^{\alpha}-\frac{\partial\Phi_{W}^{i}}{\partial u^{j}}Y^{\alpha
}\left(  \partial_{\alpha}\sigma^{j}\right)  \right)  \partial u_{i}$

Notice that this expression involves the derivative of the flow with respect
to x and u.

\begin{theorem}
(Kolar p.57) The Lie derivative of a section S of a fiber bundle E along a
projectable vector field W has the following properties :

i) A section is invariant by the flow of a projectable vector field iff its
lie derivative is null.

ii) If W,Z are two projectable vector fields on E then $\left[  W,Z\right]  $
is a projectable vector field and we have : $\pounds _{\left[  W,Z\right]
}S=\pounds _{W}\circ\pounds _{Z}S-\pounds _{Z}\circ\pounds _{W}S$

iii) If W is a vertical vector field then $\pounds _{W}S\left(  x\right)
=W\left(  S\left(  x\right)  \right)  $
\end{theorem}

For ii) $\pi_{\ast}\left(  \left[  W,Z\right]  \right)  \left(  \pi\left(
p\right)  \right)  =\left[  \pi_{\ast}W,\pi_{\ast}Z\right]  _{M}=\left[
W_{x},Z_{x}\right]  \left(  \pi\left(  p\right)  \right)  $

The result holds because the functor which makes TE a vector bundle is natural
(Kolar p.391).

For iii) : If W is a vertical vector field it is projectable and Y=0 so :

$\pounds _{W}S=\frac{\partial}{\partial t}\Phi_{W}\left(  S\left(  x\right)
,t\right)  |_{t=0}=\sum_{i}W^{i}\partial u_{i}=W\left(  S\left(  x\right)
\right)  $

\paragraph{Lie derivative of a morphism\newline}

The definition can be extended as follows (Kolar p.378):

\begin{definition}
The Lie derivative of the base preserving morphism $F:E_{1}\rightarrow E_{2}$
between two fibered manifolds over the same base : $E_{1}\left(  M,\pi
_{1}\right)  ,E_{2}\left(  M,\pi_{2}\right)  $ , with respect to the vector
fields $W_{1}\in\mathfrak{X}\left(  TE_{1}\right)  ,W_{2}\in\mathfrak{X}%
\left(  TE_{2}\right)  $ projectable on the same vector field $Y\in
\mathfrak{X}\left(  TM\right)  $ is :

$\pounds _{\left(  W_{1},W_{2}\right)  }F\left(  p\right)  =\frac{\partial
}{\partial t}\Phi_{W2}\left(  F\left(  \Phi_{W_{1}}\left(  p,-t\right)
\right)  ,t\right)  |_{t=0}\in\mathfrak{X}\left(  VE_{2}\right)  $
\end{definition}

\begin{proof}
$p\in E_{1}:\Phi_{W_{1}}\left(  p,-t\right)  ::\pi_{1}\left(  \Phi_{W_{1}%
}\left(  p,-t\right)  \right)  =\pi_{1}\left(  p\right)  $ because $W_{1}$ is projectable

$F\left(  \Phi_{W_{1}}\left(  p,-t\right)  \right)  \in E_{2}::\pi_{2}\left(
F\left(  \Phi_{W_{1}}\left(  p,-t\right)  \right)  \right)  =\pi_{1}\left(
p\right)  $ because $F$ is base preserving

$\Phi_{W2}\left(  F\left(  \Phi_{W_{1}}\left(  p,-t\right)  \right)
,t\right)  \in E_{2}::\pi_{2}\left(  \Phi_{W2}\left(  F\left(  \Phi_{W_{1}%
}\left(  p,-t\right)  \right)  ,t\right)  \right)  =\pi_{1}\left(  p\right)  $
because $W_{2}$ is projectable

$\frac{\partial}{\partial t}\pi_{2}\left(  \Phi_{W2}\left(  F\left(
\Phi_{W_{1}}\left(  p,-t\right)  \right)  ,t\right)  \right)  |_{t=0}=0$

So $\pounds _{\left(  W_{1},W_{2}\right)  }F\left(  p\right)  $ is a vertical
vector field on $E_{2}$
\end{proof}

\bigskip

\subsection{Vector bundles}

\label{Vector bundle}

This is the generalization of the vector bundle over a manifold. With a vector
bundle we can use vector spaces located at each point over a manifold and
their tensorial products, for any dimension. The drawback is that each vector
bundle must be defined through specific atlas, whereas the common vector
bundle comes from the manifold structure itself. Many theorems are just the
implementation of results for general fiber bundles.

\subsubsection{Definitions}

\begin{definition}
A \textbf{vector bundle} $E(M,V,\pi)$ is a fiber bundle whose standard fiber V
is a Banach vector space and transitions maps $\ \varphi_{ab}\left(  x\right)
:V\rightarrow V$ are continuous linear invertible maps : $\varphi_{ab}\left(
x\right)  \in G%
\mathcal{L}%
\left(  V;V\right)  $
\end{definition}

So : with an atlas $\left(  O_{a},\varphi_{a}\right)  _{a\in A}$ of E :

$\forall a,b:O_{a}\cap O_{b}\neq\varnothing:\exists\varphi_{ba}:O_{a}\cap
O_{b}\rightarrow G%
\mathcal{L}%
\left(  V;V\right)  ::$

$\forall p\in\pi^{-1}\left(  O_{a}\cap O_{b}\right)  ,p=\varphi_{a}\left(
x,u_{a}\right)  =\varphi_{b}\left(  x,u_{b}\right)  \Rightarrow u_{b}%
=\varphi_{ba}\left(  x\right)  u_{a}$

The transitions maps must meet the cocycle conditions :

$\forall a,b,c\in A:\varphi_{aa}\left(  x\right)  =1;\varphi_{ab}\left(
x\right)  \varphi_{bc}\left(  x\right)  =\varphi_{ac}\left(  x\right)  $

The set of transition maps is not fixed : if it is required that they belong
to some subgroup of $G%
\mathcal{L}%
\left(  V;V\right)  $\ we have a G-bundle (see associated bundles).

Examples :

the tangent bundle $TM\left(  M,B,\pi\right)  $ of a manifold M modelled on
the Banach B.

the tangent bundle $TE\left(  TM,TV,T\pi\right)  $ of a fiber bundle $E\left(
M,V,\pi\right)  $

\begin{theorem}
(Kolar p.69) Any finite dimensional vector bundle admits a finite vector
bundle atlas
\end{theorem}

In the definition of fiber bundles we have required that all the manifolds are
on the same field K. However for vector bundles we can be more flexible.

\begin{definition}
A \textbf{complex vector bundle} $E(M,V,\pi)$ over a real manifold M is a
fiber bundle whose standard fiber V is a Banach complex vector space and
transitions functions at each point $\varphi_{ab}\left(  x\right)  \in G%
\mathcal{L}%
\left(  V;V\right)  $ are complex continuous linear maps.
\end{definition}

\subsubsection{Vector space structure}

The main property of a vector bundle is that each fiber has a vector space
structure, isomorphic to V : the fiber $\pi^{-1}\left(  x\right)  $\ over x is
just a copy of V located at x.

\paragraph{Definition of the operations\newline}

\begin{theorem}
The fiber over each point of a vector bundle $E\left(  M,V,\pi\right)  $ has a
canonical structure of vector space, isomorphic to V
\end{theorem}

\begin{proof}
Define the operations, pointwise with an atlas $\left(  O_{a},\varphi
_{a}\right)  _{a\in A}$ of E

$p=\varphi_{a}\left(  x,u_{a}\right)  ,q=\varphi_{a}\left(  x,v_{a}\right)
,k,k^{\prime}\in K:kp+k^{\prime}q=\varphi_{a}\left(  x,ku_{a}+k^{\prime}%
v_{a}\right)  $

then :

$p=\varphi_{b}\left(  x,u_{b}\right)  ,q=\varphi_{b}\left(  x,v_{b}\right)
,u_{b}=\varphi_{ba}\left(  x,u_{a}\right)  ,v_{b}=\varphi_{ba}\left(
x,v_{a}\right)  $

$kp+k^{\prime}q=\varphi_{b}\left(  x,ku_{b}+k^{\prime}v_{b}\right)
=\varphi_{b}\left(  x,k\varphi_{ba}\left(  x,u_{a}\right)  +k^{\prime}%
\varphi_{ba}\left(  x,v_{a}\right)  \right)  $

$=\varphi_{b}\left(  x,\varphi_{ba}\left(  x,ku_{a}+k^{\prime}v_{a}\right)
\right)  =\varphi_{a}\left(  x,ku_{a}+k^{\prime}v_{a}\right)  $
\end{proof}

With this structure of vector space on E(x) the trivializations are linear in
u : $\varphi\left(  x,.\right)  \in%
\mathcal{L}%
\left(  E\left(  x\right)  ;E\left(  x\right)  \right)  $

\paragraph{Holonomic basis\newline}

\begin{definition}
The \textbf{holonomic basis} of the vector bundle $E\left(  M,V,\pi\right)  $
associated to the atlas $\left(  O_{a},\varphi_{a}\right)  _{a\in A}$ of E and
a basis $\left(  e_{i}\right)  _{i\in I}$ of V is the basis of each fiber
defined by : $\mathbf{e}_{ia}\in C_{r}\left(  O_{a},E\right)  :\mathbf{e}%
_{ia}\left(  x\right)  =\varphi_{a}\left(  x,e_{i}\right)  .$ At the
transitions : $\mathbf{e}_{ib}\left(  x\right)  =\varphi_{ab}\left(  x\right)
\mathbf{e}_{ia}\left(  x\right)  $
\end{definition}

Warning ! we take the image of the \textit{same} vector $e_{i}$ in each open.
So at the transitions $e_{ia}\left(  x\right)  \neq e_{ib}\left(  x\right)  $.
\textit{They are not sections}. This is similar to the holonomic bases of
manifolds : $\partial x_{\alpha}=\varphi_{a}^{\prime}\left(  x\right)
^{-1}\left(  \varepsilon_{\alpha}\right)  $

\begin{proof}
At the transitions :

$\mathbf{e}_{ib}\left(  x\right)  =\varphi_{b}\left(  x,e_{i}\right)
=\varphi_{a}\left(  x,u_{a}\right)  \Rightarrow e_{i}=\varphi_{ba}\left(
x\right)  u_{a}\Leftrightarrow u_{a}=\varphi_{ab}\left(  x\right)  e_{i}$

$\mathbf{e}_{ib}\left(  x\right)  =\varphi_{a}\left(  x,\varphi_{ab}\left(
x\right)  e_{i}\right)  =\varphi_{ab}\left(  x\right)  \varphi_{a}\left(
x,e_{i}\right)  =\varphi_{ab}\left(  x\right)  \mathbf{e}_{ia}\left(
x\right)  $
\end{proof}

A vector of E does not depend on a basis : it is the same as in any other
vector space. It reads in the holonomic basis :%

\begin{equation}
U=\sum_{i\in I}u_{a}^{i}\mathbf{e}_{ai}\left(  x\right)  =\varphi_{a}\left(
x,\sum_{i\in I}u_{a}^{i}\mathbf{e}_{ai}\right)  =\varphi_{a}\left(
x,u_{a}\right)
\end{equation}

and at the transitions the components $u_{b}^{i}=\sum_{j}\varphi_{ba}\left(
x\right)  _{j}^{i}u_{a}^{j}$

The bases of a vector bundle are not limited to the holonomic basis defined by
a trivialization.\ Any other basis can be defined at any point, by an
endomorphism in the fiber.

\begin{definition}
A change of trivialization on a vector bundle $E\left(  M,V,\pi\right)  $ with
atlas $\left(  O_{a},\varphi_{a}\right)  _{a\in A}$ is the definition of a
new, compatible atlas $\left(  O_{a},\widetilde{\varphi}_{a}\right)  _{a\in
A}$ , the trivialization $\widetilde{\varphi}_{a}$\ is defined by :
$p=\varphi_{a}\left(  x,u_{a}\right)  =\widetilde{\varphi}_{a}\left(
x,\chi_{a}\left(  x\right)  \left(  u_{a}\right)  \right)  \Leftrightarrow
\widetilde{u}_{a}=\chi_{a}\left(  x\right)  \left(  u_{a}\right)  $ where
$\left(  \chi_{a}\left(  x\right)  \right)  _{a\in A}$ is a family of linear
diffeomorphisms on V.
\end{definition}

It is equivalent to the change of holonomic basis (we have still the rule :
$\chi_{a}=\varphi_{ba})$

$e_{ai}\left(  x\right)  =\varphi_{a}\left(  x,e_{i}\right)  \rightarrow
\widetilde{e}_{ia}\left(  x\right)  =\widetilde{\varphi}_{a}\left(
x,e_{i}\right)  =\varphi_{a}\left(  x,\chi_{a}\left(  x\right)  ^{-1}%
e_{i}\right)  =\chi_{a}\left(  x\right)  ^{-1}\varphi_{a}\left(
x,e_{i}\right)  =\chi_{a}\left(  x\right)  ^{-1}e_{ai}\left(  x\right)  $%

\begin{equation}
e_{ai}\left(  x\right)  =\varphi_{a}\left(  x,e_{i}\right)  \rightarrow
\widetilde{e}_{ia}\left(  x\right)  =\widetilde{\varphi}_{a}\left(
x,e_{i}\right)  =\chi_{a}\left(  x\right)  ^{-1}e_{ai}\left(  x\right)
\end{equation}

\paragraph{Sections\newline}

\begin{definition}
With an atlas $\left(  O_{a},\varphi_{a}\right)  _{a\in A}$ of E a\ section
$U\in\mathfrak{X}\left(  E\right)  $ of the vector bundle $E\left(
M,V,\pi\right)  $ is defined by a family of maps $\left(  u_{a}\right)  _{a\in
A},u_{a}:O_{a}\rightarrow V$ such that :

$x\in O_{a}:U\left(  x\right)  =\varphi_{a}\left(  x,u_{a}\left(  x\right)
\right)  =\sum_{i\in I}u_{a}^{i}\left(  x\right)  e_{ai}\left(  x\right)  $

$\forall x\in O_{a}\cap O_{b}:u_{b}\left(  x\right)  =\varphi_{ba}\left(
x\right)  u_{a}\left(  x\right)  $
\end{definition}

\begin{theorem}
The set of sections $\mathfrak{X}\left(  E\right)  $\ over a vector bundle has
the structure of a vector space with pointwise operations.
\end{theorem}

Notice that it is infinite dimensional, and thus usually not isomorphic to V.

\begin{theorem}
(Giachetta p.13) If a finite dimensional vector bundle E admits a family of
global sections which spans each fiber then the fiber bundle is trivial.
\end{theorem}

Warning ! there is no commutator of sections $\mathfrak{X}\left(  E\right)  $
on a vector bundle.\ But there is a commutator for sections $\mathfrak{X}%
\left(  TE\right)  $ of TE.

\paragraph{Complex and real structures\newline}

\begin{definition}
A \textbf{real structure on a complex vector bundle} $E(M,V,\pi)$ is a
continuous map $\sigma$ defined on M\ such that $\sigma\left(  x\right)  $\ is
antilinear on E(x) and $\sigma^{2}\left(  x\right)  =Id_{E\left(  x\right)  }
$
\end{definition}

\begin{theorem}
A real structure on a complex vector bundle $E(M,V,\pi)$ with atlas $\left(
O_{a},\varphi_{a}\right)  _{a\in A}$ and transition maps $\varphi_{ab}$ is
defined by a family of maps $\left(  \sigma_{a}\right)  _{a\in A}$ defined on
each domain $O_{a}$ such that at the transitions :$\sigma_{b}\left(  x\right)
=\varphi_{ba}\left(  x\right)  \circ\sigma_{a}\left(  x\right)  \circ
\overline{\varphi_{ab}\left(  x\right)  }$
\end{theorem}

\begin{theorem}
A real structure $\sigma$ on a complex vector space V induces a real structure
on a complex vector bundle $E(M,V,\pi)$ iff the transition maps $\varphi_{ab}$
are real maps with respect to $\sigma$
\end{theorem}

\begin{proof}
A real structure $\sigma$ on V it induces a real structure on E by :
$\sigma\left(  x\right)  \left(  u_{a}\right)  =\varphi_{a}\left(
x,\sigma\left(  u_{a}\right)  \right)  $

The definition is consistent iff : $u_{b}=\varphi_{ba}\left(  x\right)
\left(  u_{a}\right)  \Rightarrow\varphi_{a}\left(  x,\sigma\left(
u_{a}\right)  \right)  =\varphi_{b}\left(  x,\sigma\left(  u_{b}\right)
\right)  $

$\sigma\left(  \varphi_{ba}\left(  x\right)  \left(  u_{a}\right)  \right)
=\varphi_{ba}\left(  x\right)  \sigma\left(  u_{a}\right)  \Leftrightarrow
\sigma\circ\varphi_{ba}\left(  x\right)  =\varphi_{ba}\left(  x\right)
\circ\sigma$

So iff $\varphi_{ab}\left(  x\right)  $ is a real map with respect to $\sigma$
\end{proof}

\subsubsection{Tangent bundle}

\begin{theorem}
The tangent space to a vector bundle $E(M,V,\pi)$\ has the vector bundle
structure : $TE\left(  TM,V\times V,T\pi\right)  $. A vector $v_{p}$ of
$T_{p}E$\ is a couple $\left(  v_{x},v_{u}\right)  \in T_{x}M\times V$ which
reads :

$v_{p}=\sum_{\alpha\in A}v_{x}^{\alpha}\partial x_{\alpha}+\sum_{i\in I}%
v_{u}^{i}\mathbf{e}_{ia}\left(  x\right)  $
\end{theorem}

\begin{proof}
With an atlas $\left(  O_{a},\varphi_{a}\right)  _{a\in A}$ of E the tangent
space $T_{p}E$ to E at $p=\varphi_{a}\left(  x,u\right)  $ has the basis
$\partial x_{\alpha}=\varphi_{ax}^{\prime}\left(  x,u\right)  \partial
\xi_{\alpha},\partial u_{i}=\varphi_{au}^{\prime}\left(  x,u\right)
\partial\eta_{i}$ with holonomic bases $\partial\xi_{\alpha}\in T_{x}%
M,\partial\eta_{i}\in T_{u}V$

$v_{p}=\sum_{\alpha}v_{x}^{\alpha}\partial x_{\alpha}+\sum_{i}v_{u}%
^{i}\partial u_{i}$

But : $\partial\eta_{i}=e_{i}$ and $\varphi$\ is linear with respect to u, so
: $\partial u_{i}=\varphi_{au}^{\prime}\left(  x,u\right)  e_{i}=\varphi
_{a}\left(  x,e_{i}\right)  =\mathbf{e}_{ai}\left(  x\right)  $
\end{proof}

So the basis of $T_{p}E$\ is $\partial x_{\alpha}=\varphi_{ax}^{\prime}\left(
x,u\right)  \partial\xi_{\alpha},\partial u_{i}=\mathbf{e}_{ai}\left(
x\right)  $

The coordinates of $v_{p}$ in this atlas are : $\left(  \xi^{\alpha},\eta
^{i},v_{x}^{\alpha},v_{u}^{i}\right)  .$

At the transitions we have the identities : $v_{p}=\varphi_{a}^{\prime}\left(
x,u_{a}\right)  \left(  v_{x},v_{au}\right)  =\varphi_{b}^{\prime}\left(
x,u_{b}\right)  \left(  v_{x},v_{bu}\right)  $ with $v_{bu}=\left(
\varphi_{ba}\left(  x\right)  ^{\prime}v_{x}\right)  \left(  u_{a}\right)
+\varphi_{ba}\left(  x\right)  \left(  v_{au}\right)  $

\begin{theorem}
The vertical bundle $VE\left(  M,V,\pi\right)  $ is a trivial bundle
isomorphic to $E\times_{M}E:v_{p}=\sum_{i\in I}v_{u}^{i}e_{i}\left(  x\right)
$
\end{theorem}

we need both p (for $e_{i}\left(  x\right)  )$ and v$_{u}$ for a point in VE

The vertical cotangent bundle is the dual of the vertical tangent bundle, and
is not a subbundle of the cotangent bundle.

Vector fields on the tangent bundle TE are defined by a family $\left(
W_{ax},W_{au}\right)  _{a\in A}$ with $W_{ax}\in\mathfrak{X}\left(
TO_{a}\right)  ,W_{au}\in C\left(  O_{a};V\right)  $

$W_{a}\left(  \varphi_{a}\left(  x,u_{a}\right)  \right)  =\sum_{\alpha\in
A}W_{ax}^{\alpha}\left(  p\right)  \partial x_{\alpha}+\sum_{i\in I}W_{au}%
^{i}\left(  p\right)  \mathbf{e}_{ai}\left(  x\right)  $

such that for

$x\in O_{a}\cap O_{b}:W_{ax}\left(  p\right)  =W_{bx}\left(  p\right)
,W_{bu}\left(  p\right)  =\left(  \varphi_{ba}\left(  x\right)  \left(
u_{a}\right)  \right)  ^{\prime}\left(  W_{x}\left(  p\right)  ,W_{au}\left(
p\right)  \right)  $

\paragraph{Lie derivative\newline}

Projectable vector fields W on TE are such that :

$W_{a}\left(  \varphi_{a}\left(  x,u_{a}\right)  \right)  =\sum_{\alpha\in
A}Y_{x}^{\alpha}\left(  \pi\left(  p\right)  \right)  \partial x_{\alpha}%
+\sum_{i\in I}W_{au}^{i}\left(  p\right)  \mathbf{e}_{ai}\left(  x\right)  $

The Lie derivative of a section X \textit{on E}, that is a vector field
$X\left(  x\right)  =\sum X^{i}\left(  x\right)  e_{i}\left(  x\right)  $
along a projectable vector field W \textit{on TE} is :

$\pounds _{W}:\mathfrak{X}\left(  E\right)  \rightarrow\mathfrak{X}\left(
VE\right)  ::\pounds _{W}X=\frac{\partial}{\partial t}\Phi_{W}\left(  X\left(
\Phi_{Y}\left(  x,-t\right)  \right)  ,t\right)  |_{t=0}$

$\pounds _{W}X=\sum_{i}\left(  W^{i}-\frac{\partial\Phi_{W}^{i}}{\partial
x^{\alpha}}Y^{\alpha}-\frac{\partial\Phi_{W}^{i}}{\partial u^{j}}Y^{\alpha
}\left(  \partial_{\alpha}X^{j}\right)  \right)  \mathbf{e}_{i}\left(
x\right)  $

If W is a vertical vector field, then ; $\pounds _{W}X=W\left(  X\right)  $

\subsubsection{Maps between vectors bundles}

\paragraph{Morphism of vector bundles\newline}

\begin{definition}
A \textbf{morphism F between vector bundles} $E_{1}(M_{1},V_{1},\pi_{1}%
)$,$E_{2}(M_{2},V_{2},\pi_{2})$\ is a couple (F,f) of maps $F:E_{1}\rightarrow
E_{2},f:M_{1}\rightarrow M_{2}$ such that :

$\forall x\in M_{1}:F\left(  \pi_{1}^{-1}\left(  x\right)  \right)  \in%
\mathcal{L}%
\left(  \pi_{1}^{-1}\left(  x\right)  ;\pi_{2}^{-1}\left(  f\left(  x\right)
\right)  \right)  $ \ 

$f\circ\pi_{1}=\pi_{2}\circ F$
\end{definition}

so it preserves both the fiber and the vector structure of the fiber.

If the morphism is base preserving (f is the identity) it can be defined by a
single map : $\forall x\in M:F\left(  x\right)  \in%
\mathcal{L}%
\left(  E_{1}\left(  x\right)  ;E_{2}\left(  x\right)  \right)  $

\begin{definition}
A vector bundle $E_{1}(M_{1},V_{1},\pi_{1})$ is a \textbf{vector subbundle} of
$E_{2}(M_{2},V_{2},\pi_{2})$ if :

i) $E_{1}(M_{1},\pi_{1})$ is a fibered submanifold of $E_{2}(M_{2},\pi
_{2}):M_{1}$ is a submanifold of $M_{2},\pi_{2}|_{M_{1}}=\pi_{1}$

ii) there is a vector bundle morphism $F:E_{1}\rightarrow E_{2}$
\end{definition}

\paragraph{Pull back of vector bundles\newline}

\begin{theorem}
The pull back of a vector bundle is a vector bundle, with the same standard fiber.
\end{theorem}

\begin{proof}
Let $E\left(  M,V,\pi\right)  $\ a vector bundle with atlas $\left(
O_{a},\varphi_{a}\right)  _{a\in A}$, N a manifold, f a continuous map
$f:N\rightarrow M$ then the fiber in f*E over (y,p) is

$\widetilde{\pi}^{-1}\left(  y,p\right)  =\left(  y,\varphi_{a}%
(f(y),u)\right)  $ with $p=\varphi_{a}(f(y),u)$

This is a vector space with the operations (the first factor is neutralized) :

$k\left(  y,\varphi_{a}(f(y),u)\right)  +k^{\prime}\left(  y,\varphi
_{a}(f(y),v)\right)  =\left(  y,\varphi_{a}(f(y),ku+k^{\prime}v)\right)  $

For a basis $\mathbf{e}_{ia}\left(  x\right)  =\varphi_{a}\left(
x,e_{i}\right)  $ we have the pull back $f^{\ast}\mathbf{e}_{ia}:N\rightarrow
E::f^{\ast}e_{i}\left(  y\right)  =\left(  f\left(  y\right)  ,\mathbf{e}%
_{ia}\left(  f\left(  y\right)  \right)  \right)  $ so it is a basis of
$\widetilde{\pi}^{-1}\left(  y,p\right)  $ with the previous vector space structure.
\end{proof}

\paragraph{Whitney sum\newline}

\begin{theorem}
The Whitney sum $E_{1}\oplus E_{2}$ of two vector bundles $E_{1}(M,V_{1}%
,\pi_{1}),$ $E_{2}(M,V_{2},\pi_{2})$ can be identified with $E(M,V_{1}\oplus
V_{2},\pi)$ with :

$E=\left\{  p_{1}+p_{2}:\pi_{1}\left(  p_{1}\right)  =\pi_{2}\left(
p_{2}\right)  \right\}  ,$ $\pi\left(  p_{1},p_{2}\right)  =\pi_{1}\left(
p_{1}\right)  =\pi_{2}\left(  p_{2}\right)  $
\end{theorem}

\begin{theorem}
(Kolar p.69) For any finite dimensional vector bundle $E_{1}(M,V_{1},\pi_{1})$
there is a second vector bundle $E_{2}(M,V_{2},\pi_{2})$ such that the Whitney
sum $E_{1}\oplus E_{2}$ is trivial
\end{theorem}

\subsubsection{Tensorial bundles}

Tensorial bundles on a vector bundle are similar to the tensors on the tangent
bundle of a manifold. Notice that here the tensors are defined with vectors of
E. There are similarly tensors defined over TE (as for any manifold) but they
are not seen here.

\paragraph{Functor on vector bundles\newline}

\begin{theorem}
The vector bundles on a field K and their morphisms constitute a category
$\mathfrak{VM}$
\end{theorem}

The functors over the category of vector spaces on a field K : $\digamma
=\mathfrak{D,T}^{r},\mathfrak{T}_{s}:\mathfrak{V\rightarrow V}$ (see
Algebra-tensors) transform a vector space V in its dual V$^{\ast}$, its tensor
powers $\otimes^{r}V,\otimes_{s}V^{\ast}$ and linear maps between vector
spaces in morphisms between tensor spaces.

Their action can be restricted to the subcategory of Banach vector spaces and
continuous morphisms.

We define the functors $\mathfrak{\digamma M=DM,T}^{r}\mathfrak{M}%
,\mathfrak{T}_{s}\mathfrak{M}:\mathfrak{VM\rightarrow VM}$ as follows :

i) To any vector bundle $E(M,V,\pi)$ we associate the vector bundle :

- with the same base manifold

- with standard fiber $\digamma\left(  V\right)  $

- to each transition map $\varphi_{ba}\left(  x\right)  \in G%
\mathcal{L}%
\left(  V;V\right)  $ the map :

$\digamma\left(  \varphi_{ba}\left(  x\right)  \right)  \in G%
\mathcal{L}%
\left(  \digamma\left(  V\right)  ;\digamma\left(  V\right)  \right)  $

then we have a vector bundle $\mathfrak{\digamma M}\left(  E\left(
M,V,\pi\right)  \right)  =\digamma E\left(  M,\digamma V,\pi\right)  $

ii) To any morphism of vector bundle $F\in\hom_{\mathfrak{VM}}\left(
E_{1}(M_{1},V_{1},\pi_{1}),E_{2}(M_{2},V_{2},\pi_{2})\right)  $\ such that :
$\forall x\in M_{1}:F\left(  \pi_{1}^{-1}\left(  x\right)  \right)  \in%
\mathcal{L}%
\left(  \pi_{1}^{-1}\left(  x\right)  ;\pi_{2}^{-1}\left(  f\left(  x\right)
\right)  \right)  $ \ where : $f:M_{1}\rightarrow M_{2}::f\circ\pi_{1}\left(
p\right)  =\pi_{2}\circ F$ we associate the morphism

$\mathfrak{\digamma M}F\in\hom_{\mathfrak{VM}}\left(  \mathfrak{\digamma
M}E_{1}(M_{1},\digamma V_{1},\pi_{1}),\digamma E_{2}(M_{2},\digamma V_{2}%
,\pi_{2})\right)  $

with : $\mathfrak{\digamma M}f=f;\mathfrak{\digamma M}F\left(  \pi_{1}%
^{-1}\left(  x\right)  \right)  =\mathfrak{\digamma}F\left(  \pi_{1}%
^{-1}\left(  x\right)  \right)  $

The tensors over a vector bundle do not depend on their definition through a
basis, either holonomic or not.

\paragraph{Dual bundle\newline}

The application of the functor $\mathfrak{DM}$ to the vector bundle
$E(M,V,\pi)$ with atlas $\left(  O_{a},\varphi_{a}\right)  _{a\in A}$ and
transitions maps : $\varphi_{ab}\left(  x\right)  $ gives the dual vector
bundle denoted $E^{\prime}\left(  M,V^{\prime},\pi\right)  $ :

with same base M and open cover $\left(  O_{a}\right)  _{a\in A}$

with fiber the topological dual V' of V, this is a Banach vector space if V is
a Banach.

for each transition map the transpose :

$\varphi_{ab}^{t}\left(  x\right)  \in%
\mathcal{L}%
\left(  V^{\prime};V^{\prime}\right)  :\varphi_{ab}^{t}\left(  x\right)
\left(  \lambda\right)  \left(  u\right)  =\lambda\left(  \varphi_{ab}\left(
x\right)  \left(  u\right)  \right)  $

for trivialization $\left(  \varphi_{a}^{t}\right)  _{a\in A}$ the
maps\ defined through the holonomic basis : $\mathbf{e}_{a}^{i}\left(
x\right)  \left(  \mathbf{e}_{j}\left(  x\right)  \right)  =\delta_{ij}$ and
$\mathbf{e}_{a}^{i}\left(  x\right)  =\varphi_{a}^{t}\left(  x,e^{i}\right)  $
where $\left(  e^{i}\right)  _{i\in I}$ is a basis of V' such that :
$e^{i}\left(  e_{j}\right)  =\delta_{ij}$.

At the transitions we have :

$\mathbf{e}_{b}^{i}\left(  x\right)  =\varphi_{b}^{t}\left(  x,e^{i}\right)
=\varphi_{a}^{t}\left(  x,\varphi_{ba}^{t}\left(  x\right)  e^{i}\right)
=\sum_{j\in I}\left[  \varphi_{ab}^{t}\left(  x\right)  \right]  _{j}%
^{i}\mathbf{e}_{a}^{j}\left(  x\right)  $

$\mathbf{e}_{b}^{i}\left(  x\right)  \left(  \mathbf{e}_{jb}\left(  x\right)
\right)  =\left(  \sum_{k\in I}\left[  \varphi_{ab}^{t}\left(  x\right)
\right]  _{k}^{i}\mathbf{e}_{a}^{k}\left(  x\right)  \right)  \left(
\sum_{l\in I}\left[  \varphi_{ab}\left(  x\right)  \right]  _{j}^{l}%
\mathbf{e}_{la}\left(  x\right)  \right)  $

$=\sum_{k\in I}\left[  \varphi_{ab}^{t}\left(  x\right)  \right]  _{k}%
^{i}\left[  \varphi_{ab}\left(  x\right)  \right]  _{j}^{k}=\delta_{j}^{i}$

Thus : $\left[  \varphi_{ab}^{t}\left(  x\right)  \right]  =\left[
\varphi_{ab}\left(  x\right)  \right]  ^{-1}$

$\mathbf{e}_{b}^{i}\left(  x\right)  =\sum_{j\in I}\left[  \varphi_{ba}\left(
x\right)  \right]  _{j}^{i}e_{a}^{j}\left(  x\right)  $

A section $\Lambda$ of the dual bundle is defined by a family of maps $\left(
\lambda_{a}\right)  _{a\in A},\lambda_{a}:O_{a}\rightarrow V^{\prime}$ such
that :

$x\in O_{a}:\Lambda\left(  x\right)  =\varphi_{a}^{t}\left(  x,\lambda
_{a}\left(  x\right)  \right)  =\sum_{i\in I}\lambda_{ai}\left(  x\right)
\mathbf{e}_{a}^{i}\left(  x\right)  $

$\forall x\in O_{a}\cap O_{b}:\lambda_{bi}\left(  x\right)  =\sum_{j\in
I}\left[  \varphi_{ab}\left(  x\right)  \right]  _{i}^{j}\lambda_{aj}\left(
x\right)  $

So we can define pointwise the action of $\mathfrak{X}\left(  E^{\prime
}\right)  $ on $\mathfrak{X}\left(  E\right)  :$

$\mathfrak{X}\left(  E^{\prime}\right)  \times\mathfrak{X}\left(  E\right)
\rightarrow C\left(  M;K\right)  ::\Lambda\left(  x\right)  \left(  U\left(
x\right)  \right)  =\lambda_{a}\left(  x\right)  u_{a}\left(  x\right)  $

\paragraph{Tensorial product of vector bundles\newline}

As the transition maps are invertible, we can implement the functor
$\mathfrak{T}_{s}^{r}$ . The action of the functors on a vector bundle :
$E(M,V,\pi)$ with trivializations $\left(  O_{a},\varphi_{a}\right)  _{a\in A}
$ and transitions maps : $\varphi_{ab}\left(  x\right)  $\ gives :

\begin{notation}
$\otimes^{r}E$ is the vector bundle $\mathfrak{\otimes}^{r}E\left(
M,\mathfrak{\otimes}^{r}V,\pi)\right)  =\mathfrak{T}^{r}\mathfrak{M}\left(
E(M,V,\pi)\right)  $
\end{notation}

\begin{notation}
$\otimes_{s}E$ is the vector bundle $\mathfrak{\otimes}_{s}E\left(
M,\mathfrak{\otimes}_{s}V,\pi)\right)  =\mathfrak{T}_{s}\mathfrak{M}\left(
E(M,V,\pi)\right)  $
\end{notation}

\begin{notation}
$\otimes_{s}^{r}E$ is the vector bundle $\mathfrak{\otimes}_{s}^{r}E\left(
M,\mathfrak{\otimes}_{s}V^{\prime},\pi)\right)  =\mathfrak{T}_{s}%
^{r}\mathfrak{M}\left(  E(M,V,\pi)\right)  $
\end{notation}

Similarly we have the algebras of symmetric tensors and antisymmetric tensors.

\begin{notation}
$\odot^{r}E$ is the vector bundle $\odot^{r}E\left(  M,\odot^{r}V,\pi)\right)
$
\end{notation}

\begin{notation}
$\wedge_{s}E^{\prime}$ is the vector bundle $\mathfrak{\wedge}_{s}E\left(
M,\mathfrak{\wedge}_{s}V^{\prime},\pi)\right)  $
\end{notation}

Notice :

i) The tensorial bundles are not related to the tangent bundle of the vector
bundle or of the base manifold

ii) $\Lambda_{s}E^{\prime}\left(  M,\Lambda_{s}V,\pi\right)  \neq\Lambda
_{s}\left(  M;E\right)  $

These vector bundles have all the properties of the tensor bundles over a
manifold, as seen in the Differential geometry part: linear combination of
tensors of the same type, tensorial product, contraction, as long as we
consider only pointwise operations which do not involve the tangent bundle of
the base M.

The trivializations are defined on the same open cover and any map
$\varphi_{a}\left(  x,u\right)  $ can be uniquely extended to a map:

$\Phi_{ar,s}:O_{a}\times\mathfrak{\otimes}^{r}V\otimes\mathfrak{\otimes}%
_{s}V^{\prime}\rightarrow\otimes_{s}^{r}E$ such that :

$\forall i_{k},j_{l}\in I:U=\Phi_{ar,s}\left(  x,e_{i_{1}}\otimes...\otimes
e_{i_{r}}\otimes e^{j_{1}}\otimes...\otimes e^{j_{s}}\right)  $

$=\mathbf{e}_{ai_{1}}\left(  x\right)  \otimes...\otimes\mathbf{e}_{ai_{r}%
}\left(  x\right)  \otimes\mathbf{e}^{aj_{1}}\left(  x\right)  \otimes
...\otimes\mathbf{e}^{aj_{s}}\left(  x\right)  $

The sections of these vector bundles are denoted accordingly : $\mathfrak{X}%
\left(  \otimes_{s}^{r}E\right)  .$They are families of maps : $T_{a}%
:O_{a}\rightarrow\otimes_{s}^{r}E$ such that on the transitions we have :

$T\left(  x\right)  =\sum_{i_{1}...i_{r}j_{1}...j_{s}}T_{aj_{1}...j_{s}%
}^{i_{1}...i_{r}}\mathbf{e}_{ai_{1}}\left(  x\right)  \otimes...\otimes
\mathbf{e}_{ai_{r}}\left(  x\right)  \otimes\mathbf{e}^{aj_{1}}\left(
x\right)  \otimes...\otimes\mathbf{e}^{aj_{s}}\left(  x\right)  $

$T\left(  x\right)  =\sum_{i_{1}...i_{r}j_{1}...j_{s}}T_{bj_{1}...j_{s}%
}^{i_{1}...i_{r}}\mathbf{e}_{bi_{1}}\left(  x\right)  \otimes...\otimes
\mathbf{e}_{bi_{r}}\left(  x\right)  \otimes\mathbf{e}^{bj_{1}}\left(
x\right)  \otimes...\otimes\mathbf{e}^{bj_{s}}\left(  x\right)  $

$e_{ib}\left(  x\right)  =\sum_{j\in I}\left[  \varphi_{ab}\left(  x\right)
\right]  _{i}^{j}\mathbf{e}_{ja}\left(  x\right)  $

$e_{b}^{j}\left(  x\right)  =\sum_{j\in I}\left[  \varphi_{ba}\left(
x\right)  \right]  _{i}^{j}\mathbf{e}_{a}^{j}\left(  x\right)  $

So :

$T_{bj_{1}...j_{s}}^{i_{1}...i_{r}}=\sum_{k_{1}...k_{r}}\sum_{l_{1}....l_{s}%
}T_{al_{1}...l_{s}}^{k_{1}...k_{r}}\left[  J\right]  _{k_{1}}^{i_{1}}..\left[
J\right]  _{k_{r}}^{i_{r}}\left[  J^{-1}\right]  _{j_{1}}^{l_{1}}..\left[
J^{-1}\right]  _{j_{s}}^{l_{s}}$

with $\left[  J\right]  =\left[  \varphi_{ba}\left(  x\right)  \right]  ,$
$\left[  J\right]  ^{-1}=\left[  \varphi_{ab}\left(  x\right)  \right]  $

For a r-form in $\Lambda_{r}E^{\prime}$ these expressions give the formula :

$T_{bi_{1}...i_{r}}\left(  x\right)  =\sum_{\left\{  j_{1}....j_{r}\right\}
}T_{aj_{1}...j_{r}}\det\left[  J^{-1}\right]  _{i_{1}...i_{r}}^{j_{1}..._{r}}$

where $\det\left[  J^{-1}\right]  _{i_{1}...i_{r}}^{j_{1}..._{r}}$ is the
determinant of the matrix $\left[  J^{-1}\right]  $ with r columns $\left(
i_{1},..i_{r}\right)  $ comprised each of the components $\left\{
j_{1}...j_{r}\right\}  $

We still call contravariant a section of $\otimes^{r}E$ and covariant a
section of $\otimes_{s}E^{\prime}.$

\paragraph{Tensorial product of vector bundles\newline}

\begin{definition}
The tensorial product $E_{1}\otimes E_{2}$ of two vector bundles $E_{1}\left(
M,V_{1},\pi_{1}\right)  ,$ $E_{2}\left(  M,V_{2},\pi_{2}\right)  $ over the
same manifold is the set : $\cup_{x\in M}E_{1}\left(  x\right)  \otimes
E_{2}\left(  x\right)  $, which has the structure of the vector bundle
$E_{1}\otimes E_{2}\left(  M,V_{1}\otimes V_{2},\pi\right)  $
\end{definition}

\subsubsection{Scalar product on a vector bundle}

The key point is that a scalar product on V induces a scalar product on
$E(M,V,\pi)$ if it is preserved by the transition maps. M is not involved,
except if E=TM.

\paragraph{General definition\newline}

\begin{definition}
A scalar product on a vector bundle $E(M,V,\pi)$ is a map g defined on M such
that $\forall x\in M$ g(x) is a non degenerate, bilinear symmetric form if E
is real, or a sesquilinear hermitian form if E is complex.
\end{definition}

For any atlas $\left(  O_{a},\varphi_{a}\right)  _{a\in A}$ of E with
transition maps $\varphi_{ba},$ g is defined by a family $\left(
g_{a}\right)  _{a\in A}$ of forms defined on each domain $O_{a}$ :

$g\left(  x\right)  \left(  \sum_{i}u^{i}\mathbf{e}_{ai}\left(  x\right)
,\sum_{i}v^{i}\mathbf{e}_{ai}\left(  x\right)  \right)  =\sum_{ij}\overline
{u}^{i}v^{j}g_{aij}\left(  x\right)  $\ 

with : $g_{aij}\left(  x\right)  =g\left(  x\right)  \left(  \mathbf{e}%
_{a}^{i}\left(  x\right)  ,\mathbf{e}_{a}^{j}\left(  x\right)  \right)  $

At the transitions :

$x\in O_{a}\cap O_{b}:g_{bij}\left(  x\right)  =g\left(  x\right)  \left(
\mathbf{e}_{b}^{i}\left(  x\right)  ,\mathbf{e}_{b}^{j}\left(  x\right)
\right)  =g\left(  x\right)  \left(  \varphi_{ab}\left(  x\right)
\mathbf{e}_{a}^{i}\left(  x\right)  ,\varphi_{ab}\left(  x\right)
\mathbf{e}_{a}^{j}\left(  x\right)  \right)  $

So we have the condition : $\left[  g_{b}\left(  x\right)  \right]  =\left[
\varphi_{ab}\left(  x\right)  \right]  ^{\ast}\left[  g_{a}\left(  x\right)
\right]  \left[  \varphi_{ab}\left(  x\right)  \right]  $

There are always orthonormal basis. If the basis $\mathbf{e}_{ai}\left(
x\right)  $\ is orthonormal then it will be orthonormal all over E if the
transition maps are orthonormal : $\sum_{k}\left[  \varphi_{ab}\left(
x\right)  \right]  _{i}^{k}\left[  \varphi_{ab}\left(  x\right)  \right]
_{j}^{k}=\eta_{ij}$

\paragraph{Tensorial definition\newline}

1. Real case :

\begin{theorem}
A symmetric covariant tensor $g\in\mathfrak{X}\left(  \odot_{2}E\right)  $
defines a bilinear symmetric form on each fiber of a real vector bundle
$E(M,V,\pi)$ and a scalar product on E if this form is non degenerate.
\end{theorem}

Such a tensor reads in a holonomic basis of E :

$g\left(  x\right)  =\sum_{ij}g_{aij}\left(  x\right)  \mathbf{e}_{a}%
^{i}\left(  x\right)  \otimes\mathbf{e}_{a}^{j}\left(  x\right)  $ with
$g_{aij}\left(  x\right)  =g_{aji}\left(  x\right)  $

at the transitions : $g_{bij}\left(  x\right)  =\sum_{kl}\left[  \varphi
_{ab}\left(  x\right)  \right]  _{i}^{k}\left[  \varphi_{ab}\left(  x\right)
\right]  _{j}^{l}g_{akl}\left(  x\right)  \Leftrightarrow\left[  g_{b}\right]
=\left[  \varphi_{ab}\left(  x\right)  \right]  ^{t}\left[  g_{a}\right]
\left[  \varphi_{ab}\left(  x\right)  \right]  $

For : $X,Y\in\mathfrak{X}\left(  E\right)  :g\left(  x\right)  \left(
X\left(  x\right)  ,Y\left(  x\right)  \right)  =\sum_{ij}g_{aij}\left(
x\right)  X_{a}^{i}\left(  x\right)  Y_{a}^{j}\left(  x\right)  $

2. Complex case :

\begin{theorem}
A real structure $\sigma$\ and a covariant tensor $g\in\mathfrak{X}\left(
\otimes_{2}E\right)  $ such that \ $g\left(  x\right)  \left(  \sigma\left(
x\right)  \left(  u\right)  ,\sigma\left(  x\right)  \left(  v\right)
\right)  =\overline{g\left(  x\right)  \left(  v,u\right)  }$\ define a
hermitian sequilinear form, on a complex vector bundle $E(M,V,\pi)$ by
$\gamma\left(  x\right)  \left(  u,v\right)  =g\left(  x\right)  \left(
\sigma\left(  x\right)  u,v\right)  $
\end{theorem}

\begin{proof}
g defines a C-bilinear form g on E(x). This is the application of a general
theorem (see Algebra).

g is a tensor and so is defined by a family of maps in a holonomic basis of E

$g\left(  x\right)  =\sum_{ij}g_{aij}\left(  x\right)  \mathbf{e}_{a}%
^{i}\left(  x\right)  \otimes\mathbf{e}_{a}^{j}\left(  x\right)  $ at the
transitions : $\left[  g_{b}\right]  =\left[  \varphi_{ab}\left(  x\right)
\right]  ^{t}\left[  g_{a}\right]  \left[  \varphi_{ab}\left(  x\right)
\right]  $

g is not a symmetric tensor. The condition $g\left(  x\right)  \left(
\sigma\left(  x\right)  \left(  u\right)  ,\sigma\left(  x\right)  \left(
v\right)  \right)  =\overline{g\left(  x\right)  \left(  v,u\right)  }$ reads
in matrix notation :

$\overline{\left[  u_{a}\right]  }^{t}\left[  \sigma_{a}\left(  x\right)
\right]  ^{t}\left[  g_{a}\left(  x\right)  \right]  \left[  \sigma_{a}\left(
x\right)  \right]  \overline{\left[  v_{a}\right]  }=\overline{\left[
v_{a}\right]  ^{t}\left[  g_{a}\left(  x\right)  \right]  \left[
u_{a}\right]  }=\overline{\left[  v_{a}\right]  }^{t}\overline{\left[
g_{a}\left(  x\right)  \right]  }\overline{\left[  u_{a}\right]  }$

$=\overline{\left[  u_{a}\right]  }^{t}\left[  g_{a}\left(  x\right)  \right]
^{\ast}\left[  v_{a}\right]  $

$\left[  \sigma_{a}\left(  x\right)  \right]  ^{t}\left[  g_{a}\left(
x\right)  \right]  \left[  \sigma_{a}\left(  x\right)  \right]  =\left[
g_{a}\left(  x\right)  \right]  ^{\ast}$

At the transitions this property is preserved :

$\left[  \sigma_{b}\left(  x\right)  \right]  ^{t}\left[  g_{b}\left(
x\right)  \right]  \left[  \sigma_{b}\left(  x\right)  \right]  $

$=\left[  \varphi_{ab}\left(  x\right)  \right]  ^{\ast}\left[  \sigma\left(
e_{a}\left(  x\right)  \right)  \right]  ^{t}\left[  g_{a}\right]  \left[
\sigma\left(  e_{a}\left(  x\right)  \right)  \right]  \overline{\left[
\varphi_{ab}\left(  x\right)  \right]  }=\left[  \varphi_{ab}\left(  x\right)
\right]  ^{\ast}\left[  g_{a}\left(  x\right)  \right]  ^{\ast}\overline
{\left[  \varphi_{ab}\left(  x\right)  \right]  }=\left[  g_{b}\right]
^{\ast}$
\end{proof}

\paragraph{Induced scalar product\newline}

\begin{theorem}
A scalar product g on a vector space V induces a scalar product on a vector
bundle $E(M,V,\pi)$ iff the transitions maps preserve g
\end{theorem}

\begin{proof}
Let $E(M,V,\pi)$ a vector bundle on a field K=$%
\mathbb{R}
$ or $%
\mathbb{C}
,$ with trivializations $\left(  O_{a},\varphi_{a}\right)  _{a\in A}$ and
transitions maps : $\varphi_{ab}\left(  x\right)  \in%
\mathcal{L}%
\left(  V;V\right)  $\ and V endowed with a map : $\gamma:V\times V\rightarrow
K$ which is either symmetric bilinear (if K=$%
\mathbb{R}
)$ or sesquilinear (if K=$%
\mathbb{C}
)$ and non degenerate.

Define : $x\in O_{a}:g_{a}\left(  x\right)  \left(  \varphi_{a}\left(
x,u_{a}\right)  ,\varphi_{a}\left(  x,v_{a}\right)  \right)  =\gamma\left(
u_{a},v_{a}\right)  $

The definition is consistent iff :

$\forall x\in O_{a}\cap O_{b}:g_{a}\left(  x\right)  \left(  \varphi
_{a}\left(  x,u_{a}\right)  ,\varphi_{a}\left(  x,v_{a}\right)  \right)
=g_{b}\left(  x\right)  \left(  \varphi_{b}\left(  x,u_{a}\right)
,\varphi_{b}\left(  x,v_{a}\right)  \right)  $

$\Leftrightarrow\gamma\left(  u_{b},v_{b}\right)  =\gamma\left(  u_{a}%
,v_{a}\right)  =g\left(  \varphi_{ba}\left(  x\right)  u_{a},\varphi
_{ba}\left(  x\right)  v_{a}\right)  $

that is : $\forall x,\varphi_{ba}\left(  x\right)  $ preserves the scalar
product, they are orthogonal (unitary) with respect to g.
\end{proof}

Notice that a scalar product on V induces a scalar product on E, but this
scalar product is not necessarily defined by a tensor in the complex
case.\ Indeed for this we need also a real structure which is defined iff the
transitions maps are real.

\paragraph{Norm on a vector bundle\newline}

\begin{theorem}
Each fiber of a vector bundle is a normed vector space
\end{theorem}

We have assumed that V is a Banach vector space, thus a normed vector space.
With an atlas $\left(  O_{a},\varphi_{a}\right)  _{a\in A}$ of E we can define
\textit{pointwise} a norm on $E(M,V,\pi)$ by :

$\left\Vert {}\right\Vert _{E}:E\times E\rightarrow%
\mathbb{R}
_{+}::\left\Vert \varphi_{a}\left(  x,u_{a}\right)  \right\Vert _{E}%
=\left\Vert u_{a}\right\Vert _{V}$

The definition is consistent iff :

$\forall a,b\in A,O_{a}\cap O_{b}\neq\varnothing,\forall x\in O_{a}\cap
O_{b}:$

$\left\Vert \varphi_{a}\left(  x,u_{a}\right)  \right\Vert _{E}=\left\Vert
u_{a}\right\Vert _{V}=\left\Vert \varphi_{b}\left(  x,u_{b}\right)
\right\Vert _{E}=\left\Vert u_{b}\right\Vert _{V}=\left\Vert \varphi
_{ba}\left(  x\right)  u_{a}\left(  x\right)  \right\Vert $

The transitions maps : $\varphi_{ba}\left(  x\right)  \in G%
\mathcal{L}%
\left(  V;V\right)  $ are continuous so :

$\left\Vert u_{b}\right\Vert _{V}\leq\left\Vert \varphi_{ba}\left(  x\right)
\right\Vert \left\Vert u_{a}\left(  x\right)  \right\Vert _{V},\left\Vert
u_{a}\right\Vert _{V}\leq\left\Vert \varphi_{ab}\left(  x\right)  \right\Vert
\left\Vert u_{b}\left(  x\right)  \right\Vert _{V}$

and the norms defined on $O_{a},O_{b}$ are equivalent : they define the same
topology (see Normed vector spaces). So for any matter involving only the
topology (such as convergence or continuity) we have a unique definition,
however the norms are not the same.

Norms and topology on the space of sections $\mathfrak{X}\left(  E\right)  $
can be defined, but we need a way to agregate the results. This depends upon
the maps $\sigma_{a}:O_{a}\rightarrow V$ (see Functional analysis) and without
further restrictions upon the maps $M\rightarrow TE$ we have \textit{not} a
normed vector space $\mathfrak{X}\left(  E\right)  $.\ 

If the norm on V is induced by a definite positive scalar product, then V is a
Hilbert space, and each fiber is itself a Hilbert space if the transitions
maps preserve the scalar product.

\subsubsection{Affine bundles}

\begin{definition}
An \textbf{affine bundle} is a fiber bundle $E\left(  M,V,\pi\right)  $\ where
V is an affine space and the transitions maps are affine maps
\end{definition}

V is an affine space modelled on $\overrightarrow{V}$ : $A=\left(
O,\overrightarrow{u}\right)  \in V$ with an origin O

If $\left(  O_{a},\varphi_{a}\right)  $ is an atlas of E, there is a vector
bundle $\overrightarrow{E}\left(  M,\overrightarrow{V},\overrightarrow{\pi
}\right)  $ with atlas $\left(  O_{a},\overrightarrow{\varphi}_{a}\right)  $
such that :

$\forall A\in V,\overrightarrow{u}\in\overrightarrow{V}:\varphi_{a}\left(
x,A+\overrightarrow{u}\right)  =\varphi_{a}\left(  x,A\right)
+\overrightarrow{\varphi}_{a}\left(  x,\overrightarrow{u}\right)  $

At the transitions :

$\varphi_{a}\left(  x,O_{a}+\overrightarrow{u}_{a}\right)  =\varphi_{b}\left(
x,O_{b}+\overrightarrow{u}_{b}\right)  \Rightarrow$

$O_{b}=O_{a}+L_{ba}\left(  x\right)  ,L_{ba}\left(  x\right)  \in
\overrightarrow{V}$

$\overrightarrow{u}_{b}=\overrightarrow{\varphi}_{ba}\left(  x\right)  \left(
\overrightarrow{u}_{a}\right)  ,\overrightarrow{\varphi}_{ba}\left(  x\right)
\in G%
\mathcal{L}%
\left(  \overrightarrow{V};\overrightarrow{V}\right)  $

\subsubsection{Higher order tangent bundle}

\paragraph{Definition\newline}

For any manifold M the tangent bundle TM is a manifold, therefore it has a
tangent bundle, denoted T%
${{}^2}$%
M and called the bitangent bundle, which is a manifold with dimension
2x2xdimM. More generally we can define the r order tangent bundle :
$T^{r}M=T\left(  T^{r-1}M\right)  $ which is a manifold of dimension
$2^{r}\times\dim M.$ The set T%
${{}^2}$%
M can be endowed with different structures.

\begin{theorem}
If M is a manifold with atlas $\left(  E,\left(  O_{i},\varphi_{i}\right)
_{i\in I}\right)  $ then the bitangent bundle is a vector bundle
$T^{2}M\left(  TM,E\times E,\pi\times\pi^{\prime}\right)  $
\end{theorem}

This is the application of the general theorem about vector bundles.

A point in TM is some $u_{x}\in T_{x}M$ and a vector at $u_{x}$ in the tangent
space $T_{u_{x}}\left(  TM\right)  $ has for coordinates : $\left(
x,u,v,w\right)  \in O_{i}\times E^{3}$

Let us write the maps :

$\psi_{i}:U_{i}\rightarrow M::x=\psi_{i}\left(  \xi\right)  $ with
$U_{i}=\varphi_{i}\left(  O_{i}\right)  \subset E$

$\psi_{i}^{\prime}\left(  \xi\right)  :U_{i}\times E\rightarrow T_{x}%
M::\psi_{i}^{\prime}\left(  \xi\right)  u=u_{x}$

$\psi"_{i}\left(  \xi\right)  $ is a symmetric bilinear map:

$\psi"_{i}\left(  \xi,u\right)  :U_{i}\times E\times E\rightarrow T_{u_{x}}TM
$

By derivation of : $\psi_{i}^{\prime}\left(  \xi\right)  u=u_{x}$ with respect
to $\xi:$

$U_{u_{x}}=\psi"_{i}\left(  \xi\right)  \left(  u,v\right)  +\psi_{i}^{\prime
}\left(  \xi\right)  w$

The trivialization is :

$\Psi_{i}:O_{i}\times E^{3}\rightarrow T^{2}M::\Psi_{i}\left(  x,u,v,w\right)
=\psi"_{i}\left(  \xi\right)  \left(  u,v\right)  +\psi_{i}^{\prime}\left(
\xi\right)  w$

With $\partial_{\alpha\beta}\psi_{i}\left(  \xi\right)  =\partial
x_{\alpha\beta},\partial_{\alpha}\psi_{i}\left(  \xi\right)  =\partial
x_{\alpha}:U_{u_{x}}=\sum\left(  u^{\alpha}v^{\beta}\partial x_{\alpha\beta
}\left(  u,v\right)  +w^{\gamma}\partial x_{\gamma}\right)  $

The \textbf{canonical flip} is the map :

$\kappa:T^{2}M\rightarrow T^{2}M::\kappa\left(  \Psi_{i}\left(
x,u,v,w\right)  \right)  =\Psi_{i}\left(  x,v,u,w\right)  $

\paragraph{Lifts on T%
${{}^2}$%
M\newline}

1. Acceleration:

The acceleration of of a path $c:%
\mathbb{R}
\rightarrow M$ on a manifold M is the path in T%
${{}^2}$%
M :

$C_{2}:t\rightarrow T^{2}M::C_{2}\left(  t\right)  =c"(t)c^{\prime}%
(t)=\psi"_{i}\left(  x\left(  t\right)  \right)  \left(  x^{\prime
}(t),x^{\prime}(t)\right)  +\psi_{i}^{\prime}\left(  x(t)\right)  x"(t)$

2. Lift of a path:

The lift of a path $c:%
\mathbb{R}
\rightarrow M$ on a manifold M to a path in the fiber bundle TM is a path:

$C:%
\mathbb{R}
\rightarrow TM$ such that $\pi^{\prime}\left(  C\left(  t\right)  \right)
=c^{\prime}(t)\Leftrightarrow C(t)=\left(  c(t),c^{\prime}(t)\right)  \in
T_{c(t)}M$

3. Lift of a vector field:

The lift of a vector field $V\in\mathfrak{X}\left(  TM\right)  $ is a
projectable vector field $W\in\mathfrak{X}\left(  T^{2}M\right)  $ with
projection V : $\pi^{\prime}\left(  V(x)\right)  W\left(  V(x)\right)  =V(x)$

W is defined through the flow of V: it is the derivative of the lift to T%
${{}^2}$%
M of an integral curve of V

$W\left(  V\left(  x\right)  \right)  =\frac{\partial}{\partial y}\left(
V\left(  y\right)  \right)  |_{y=x}\left(  V\left(  x\right)  \right)  $

\begin{proof}
The lift of an integral curve $c(t)=\Phi_{V}\left(  x,t\right)  $ of V :

$C(t)=\left(  c(t),c^{\prime}(t)\right)  =\left(  \Phi_{V}\left(  x,t\right)
,\frac{d}{d\theta}\Phi_{V}\left(  x,\theta\right)  |_{\theta=t}\right)
=\left(  \Phi_{V}\left(  x,t\right)  ,V\left(  \Phi_{V}\left(  x,t\right)
\right)  \right)  $

Its derivative is in T%
${{}^2}$%
M with components :

$\frac{d}{dt}C\left(  t\right)  =\left(  \Phi_{V}\left(  x,t\right)  ,V\left(
\Phi_{V}\left(  x,t\right)  \right)  ,\frac{d}{dy}V\left(  \Phi_{V}\left(
x,t\right)  \right)  |_{y=x}\frac{d}{d\theta}\Phi_{V}\left(  x,\theta\right)
|_{\theta=t}\right)  $

$=\left(  \Phi_{V}\left(  x,t\right)  ,V\left(  \Phi_{V}\left(  x,t\right)
\right)  ,\frac{d}{dy}V\left(  \Phi_{V}\left(  x,t\right)  \right)
|_{y=x}V\left(  \Phi_{V}\left(  x,\theta\right)  \right)  \right)  $

with t=0 :

$W\left(  V\left(  x\right)  \right)  =\left(  x,V\left(  x\right)  ,\frac
{d}{dy}V\left(  \Phi_{V}\left(  x,0\right)  \right)  |_{y=x}V(x)\right)  $
\end{proof}

A vector field $W\in\mathfrak{X}\left(  T^{2}M\right)  $ such that :
$\pi^{\prime}\left(  u_{x}\right)  W\left(  u_{x}\right)  =u_{x}$ is called a
second order vector field (Lang p.96).

\paragraph{Family of curves\newline}

A classic problem in differential geometry is, given a curve c, to build a
family of curves which are the deformation of c (they are homotopic) and
depend on a real parameter s\ .

So let be : $c:\left[  0,1\right]  \rightarrow M$ with c(0)=A,c(1)=B the given curve.

We want a map $f:%
\mathbb{R}
\times\left[  0,1\right]  \rightarrow M$ with f(s,0)=A,f(s,1)=B

Take a compact subset P of M such that A,B are in P, and a vector field V with
compact support in P, such that V(A)=V(B)=0

The flow of V is complete so define : $f\left(  s,t\right)  =\Phi_{V}\left(
c(t),s\right)  $

$f\left(  s,0\right)  =\Phi_{V}\left(  A,s\right)  =A,f\left(  s,1\right)
=\Phi_{V}\left(  B,s\right)  =B$ because V(A)=V(B)=0, so all the paths :
$f\left(  s,.\right)  :\left[  0,1\right]  \rightarrow M$ go through A and B.
The family $\left(  f\left(  s,.\right)  \right)  _{s\in%
\mathbb{R}
}$ is what we wanted.

The lift of f(s,t) on TM gives :

$F\left(  s,.\right)  :\left[  0,1\right]  \rightarrow TM::F(s,t)=\frac
{\partial}{\partial t}F(s,t)=V\left(  f\left(  s,t\right)  \right)  $

The lift of V on T%
${{}^2}$%
M gives :

$W\left(  u_{x}\right)  =\frac{\partial}{\partial y}\left(  V\left(  y\right)
\right)  |_{y=x}\left(  u_{x}\right)  $

so for $u_{x}=V\left(  f\left(  s,t\right)  \right)  :W\left(  V\left(
f\left(  s,t\right)  \right)  \right)  =\frac{\partial}{\partial y}\left(
V\left(  y\right)  \right)  |_{y=f(s,t)}\left(  V\left(  f\left(  s,t\right)
\right)  \right)  $

But : $\frac{\partial}{\partial s}V\left(  f\left(  s,t\right)  \right)
=\frac{\partial}{\partial y}\left(  V\left(  y\right)  \right)  |_{y=f(s,t)}%
\frac{\partial}{\partial s}f\left(  s,t\right)  =\frac{\partial}{\partial
y}\left(  V\left(  y\right)  \right)  |_{y=f(s,t)}\left(  V\left(  f\left(
s,t\right)  \right)  \right)  $

So we can write : $W\left(  V\left(  f\left(  s,t\right)  \right)  \right)
=\frac{\partial}{\partial s}V\left(  f\left(  s,t\right)  \right)  $ : the
vector field $W$\ gives the transversal deviation of the family of curves.

\bigskip

\subsection{Principal fiber bundles}

\label{Principal fiber bundles}

\subsubsection{Definitions}

\paragraph{Principal bundle\newline}

\begin{definition}
A \textbf{principal (fiber) bundle} P$\left(  M,G,\pi\right)  $\ is a fiber
bundle where G is a Lie group and the transition maps act by left translation
on G
\end{definition}

So we have :

i) 3 manifolds : the total bundle P, the base M, a Lie group G, all manifolds
on the same field

ii) a class r surjective submersion : $\pi:P\rightarrow M$

iii) an atlas $\left(  O_{a},\varphi_{a}\right)  _{a\in A}$\ with an open
cover $\left(  O_{a}\right)  _{a\in A}$ of M and a set of diffeomorphisms

$\varphi_{a}:O_{a}\times G\subset M\times G\rightarrow\pi^{-1}\left(
O_{a}\right)  \subset P$ $::p=\varphi_{a}\left(  x,g\right)  $.

iv) a set of class r maps $\left(  g_{ab}\right)  _{a,b\in A}$ \ $g_{ab}%
:\left(  O_{a}\cap O_{b}\right)  \rightarrow G$ such that :

$\forall p\in\pi^{-1}\left(  O_{a}\cap O_{b}\right)  ,p=\varphi_{a}\left(
x,g_{a}\right)  =\varphi_{b}\left(  x,g_{b}\right)  \Rightarrow g_{b}%
=g_{ba}\left(  x\right)  g_{a}$

meeting the cocycle conditions :

$\forall a,b,c\in A:g_{aa}\left(  x\right)  =1;g_{ab}\left(  x\right)
g_{bc}\left(  x\right)  =g_{ac}\left(  x\right)  $

Remark : $\varphi_{ba}\left(  x,g_{a}\right)  =g_{b}=g_{ba}\left(  x\right)
g_{a}$ \ is not a morphism on G.

\paragraph{Change of trivialization\newline}

\begin{definition}
A change of trivialization on a principal bundle $P\left(  M,G,\pi\right)
$\ with atlas $\left(  O_{a},\varphi_{a}\right)  _{a\in A}$ is defined by a
family $\left(  \chi_{a}\right)  _{a\in A}$\ of maps $\chi_{a}\left(
x\right)  \in C\left(  O_{a};G\right)  $\ and the new atlas is $\left(
O_{a},\widetilde{\varphi}_{a}\right)  _{a\in A}$ with $p=\varphi_{a}\left(
x,g_{a}\right)  =\widetilde{\varphi}_{a}\left(  x,\chi_{a}\left(  x\right)
g_{a}\right)  $
\end{definition}

The new transition maps are the translation by :

$p=\varphi_{a}\left(  x,g_{a}\right)  =\widetilde{\varphi}_{b}\left(
x,\widetilde{g}_{b}\right)  $

$\widetilde{\varphi}_{ba}\left(  x\right)  =\chi_{b}\left(  x\right)
\circ\varphi_{ba}\left(  x\right)  \circ\chi_{a}\left(  x\right)  ^{-1}$

$\widetilde{g}_{b}=\chi_{b}\left(  x\right)  g_{b}=\chi_{b}\left(  x\right)
g_{ba}\left(  x\right)  g_{a}=\chi_{b}\left(  x\right)  g_{ba}\left(
x\right)  \chi_{a}\left(  x\right)  ^{-1}\widetilde{g}_{a}$%

\begin{equation}
p=\varphi_{a}\left(  x,g_{a}\right)  =\widetilde{\varphi}_{b}\left(
x,\widetilde{g}_{b}\right)  \Leftrightarrow\widetilde{g}_{b}=\chi_{b}\left(
x\right)  g_{ba}\left(  x\right)  \chi_{a}\left(  x\right)  ^{-1}\widetilde
{g}_{a}%
\end{equation}

Rule : whenever a theorem is proven with the usual transition conditions, it
is proven for any change of trivialization. The formulas for a change of
trivialization read as the formulas for the transitions by taking
$\mathbf{g}_{ba}\left(  x\right)  \mathbf{=\chi}_{a}\left(  x\right)  .$

\paragraph{Pull back\newline}

\begin{theorem}
if $f:N\rightarrow M$ is a smooth map, then the pull back $f^{\ast}P$ \ of the
principal bundle $P\left(  M,G;\pi\right)  $\ is still a principal bundle.
\end{theorem}

\paragraph{Right action\newline}

One of the main features of principal bundles is the existence of a right
action of G on P

\begin{definition}
The right action of G on a principal bundle P$\left(  M,G,\pi\right)  $ with
atlas $\left(  O_{a},\varphi_{a}\right)  _{a\in A}$ is the map :%

\begin{equation}
\rho:P\times G\rightarrow P::\rho\left(  \varphi_{a}\left(  x,g\right)
,h\right)  =\varphi_{a}\left(  x,gh\right)
\end{equation}

\end{definition}

So $\pi\left(  \rho\left(  p,g\right)  \right)  =\pi\left(  p\right)  $

This action is free and for $p\in\pi^{-1}\left(  x\right)  $ the map
$\rho\left(  p,.\right)  :G\rightarrow\pi^{-1}\left(  x\right)  $ is a
diffeomorphism, so $\rho_{g}^{\prime}(p,g)$ is invertible.

The orbits are the sets $\pi^{-1}\left(  x\right)  $.

\begin{theorem}
The right action of G on a principal bundle $P\left(  M,G,\pi\right)  $ does
not depend on the trivialization
\end{theorem}

\begin{proof}
$p=\varphi_{a}\left(  x,g_{a}\right)  =\widetilde{\varphi}_{a}\left(
x,\chi_{a}\left(  x\right)  g_{a}\right)  \Leftrightarrow p=\widetilde
{\varphi}_{a}\left(  x,\widetilde{g}_{a}\right)  =\varphi_{a}\left(
x,\chi_{a}\left(  x\right)  ^{-1}\widetilde{g}_{a}\right)  $

The images by the rigth action belong to the same fiber, thus :

$\rho\left(  p,h\right)  =\varphi_{a}\left(  x,g_{a}h\right)  $

$\widetilde{\rho}\left(  p,h\right)  =\widetilde{\varphi}_{a}\left(
x,\widetilde{g}_{a}h\right)  =\varphi_{a}\left(  x,\chi_{a}\left(  x\right)
^{-1}\widetilde{g}_{a}h\right)  =\rho\left(  \varphi_{a}\left(  x,\chi
_{a}\left(  x\right)  ^{-1}\widetilde{g}_{a}\right)  ,h\right)  =\rho\left(
p,h\right)  $
\end{proof}

The right action is an action of G on the manifold P, so we have the usual
identities with the derivatives :

$\rho_{p}^{\prime}(p,1)=\Im_{P}$%

\begin{equation}
\rho_{g}^{\prime}(p,g)=\rho_{g}^{\prime}(\rho(p,g),1)L_{g^{-1}}^{\prime
}(g)=\rho_{p}^{\prime}(p,g)\rho_{g}^{\prime}(p,1)R_{g^{-1}}^{\prime}(g)
\end{equation}

\begin{equation}
\left(  \rho_{p}^{\prime}(p,g)\right)  ^{-1}=\rho_{p}^{\prime}(\rho
(p,g),g^{-1})
\end{equation}

and more with an atlas $\left(  O_{a},\varphi_{a}\right)  _{a\in A}$ of E

$\pi\left(  \rho\left(  p,g\right)  \right)  =\pi\left(  p\right)
\Rightarrow$

$\pi^{\prime}\left(  \rho\left(  p,g\right)  \right)  \rho_{p}^{\prime
}(p,g)=\pi^{\prime}\left(  p\right)  $

$\pi^{\prime}\left(  \rho\left(  p,g\right)  \right)  \rho_{g}^{\prime
}(p,g)=0$

$\rho_{g}^{\prime}(\varphi_{a}\left(  x,h\right)  ,g)=\rho_{g}^{\prime
}(\varphi_{a}\left(  x,hg\right)  ,1)L_{g^{-1}}^{\prime}(g)=\varphi
_{ag}^{\prime}\left(  x,hg\right)  \left(  L_{hg}^{\prime}1\right)  L_{g^{-1}%
}^{\prime}(g)$%

\begin{equation}
\rho_{g}^{\prime}(p,1)=\varphi_{ag}^{\prime}\left(  x,g\right)  \left(
L_{g}^{\prime}1\right)  \in G%
\mathcal{L}%
\left(  T_{1}V;TP\right)
\end{equation}

\begin{proof}
Take : $p_{a}\left(  x\right)  =\varphi_{a}\left(  x,1\right)  \Rightarrow
\varphi_{a}\left(  x,g\right)  =\rho\left(  p_{a}\left(  x\right)  ,g\right)
$

$\Rightarrow\varphi_{ag}^{\prime}\left(  x,g\right)  =\rho_{g}^{\prime}\left(
p_{a}\left(  x\right)  ,g\right)  =\rho_{g}^{\prime}(\rho(p_{a}\left(
x\right)  ,g),1)L_{g^{-1}}^{\prime}(g)=\rho_{g}^{\prime}(p,1)L_{g^{-1}%
}^{\prime}(g)$
\end{proof}

A principal bundle can be defined equivalently by a right action on a fibered
manifold :

\begin{theorem}
(Kolar p.87) Let $P\left(  M,\pi\right)  $ a fibered manifold, and G a Lie
group which acts freely on P on the right, such that the orbits of the action
$\left\{  \rho\left(  p,g\right)  ,g\in G\right\}  =\pi^{-1}\left(  x\right)
.$ Then $P\left(  M,G;\pi\right)  $ is a principal bundle.
\end{theorem}

Remark : this is still true with P infinite dimensional, if all the manifolds
are modeled on Banach spaces as usual.

The trivializations are defined by : $O_{a}=$ the orbits of the action. Then
pick up any $p_{a}=\tau_{a}\left(  x\right)  $ in each orbit and define the
trivializations by : $\varphi_{a}\left(  x,g\right)  =\rho\left(
p_{a},g\right)  \in\pi^{-1}\left(  x\right)  $

\paragraph{Sections\newline}

\begin{definition}
A class r section S(x) on a principal bundle $P(M,G,\pi)$ with an atlas
$\left(  O_{a},\varphi_{a}\right)  _{a\in A}$ is defined by a family of maps :
$\sigma_{a}\in C_{r}\left(  O_{a};G\right)  $ such that :

$\forall a\in A,x\in O_{a}:S\left(  x\right)  =\varphi_{a}\left(  x,\sigma
_{a}\left(  x\right)  \right)  $

$\forall a,b\in A,O_{a}\cap O_{b}\neq\varnothing,\forall x\in O_{a}\cap
O_{b}:\sigma_{b}\left(  x\right)  =g_{ba}\left(  x\right)  \sigma_{a}\left(
x\right)  $
\end{definition}

\begin{theorem}
(Giachetta p.172) A principal bundle admits a global section iff it is a
trivial bundle.
\end{theorem}

In a change of trivialization :%

\begin{equation}
S\left(  x\right)  =\varphi_{a}\left(  x,\sigma_{a}\left(  x\right)  \right)
=\widetilde{\varphi}_{a}\left(  x,\widetilde{\sigma}_{a}\left(  x\right)
\right)  \Rightarrow\widetilde{\sigma}_{a}\left(  x\right)  =\chi_{a}\left(
x\right)  \sigma_{a}\left(  x\right)
\end{equation}

\subsubsection{Morphisms and Gauge}

\paragraph{Morphisms\newline}

\begin{definition}
A \textbf{principal bundle morphism} between the principal bundles
$P_{1}(M_{1},G_{1},\pi_{1}),P_{2}(M_{2},G_{2},\pi_{2})$\ is a couple
$(F,\chi)$ of maps : $F:P_{1}\rightarrow P_{2},$ and a Lie group morphism
$\chi:G_{1}\rightarrow G_{2}$ such that :

$\forall p\in P_{1},g\in G_{1}:F\left(  \rho_{1}\left(  p,g\right)  \right)
=\rho_{2}\left(  F\left(  p\right)  ,\chi\left(  g\right)  \right)  $
\end{definition}

Then we have a fibered manifold morphism by taking :

$f:M_{1}\rightarrow M_{2}::f\left(  \pi_{1}\left(  p\right)  \right)  =\pi
_{2}\left(  F\left(  p\right)  \right)  $

f is well defined because :

$f\left(  \pi_{1}\left(  \rho_{1}\left(  p,g\right)  \right)  \right)
=\pi_{2}\left(  F\left(  \rho_{1}\left(  p,g\right)  \right)  \right)
=\pi_{2}\left(  \rho_{2}\left(  F\left(  p\right)  ,\chi\left(  g\right)
\right)  \right)  =\pi_{2}\left(  F\left(  p\right)  \right)  =f\circ\pi
_{1}\left(  p\right)  $

\begin{definition}
An \textbf{automorphism} on a principal bundle P with right action $\rho$ is a
diffeomorphism $F:P\rightarrow P$ such that :$\forall g\in G:F\left(
\rho\left(  p,g\right)  \right)  =\rho\left(  F\left(  p\right)  ,g\right)  $
\end{definition}

Then $f:M\rightarrow M::f=\pi\circ F$ it is a diffeomorphism on M

\paragraph{Gauge\newline}

\begin{definition}
A \textbf{gauge} (or holonomic map\textbf{)} of a principal bundle
$P(M,G,\pi)$ with an atlas $\left(  O_{a},\varphi_{a}\right)  _{a\in A}$ is
the family of maps :%

\begin{equation}
\mathbf{p}_{a}:O_{a}\rightarrow P::\mathbf{p}_{a}\left(  x\right)
=\varphi_{a}\left(  x,1\right)
\end{equation}

At the transitions : $\mathbf{p}_{b}\left(  x\right)  =\rho\left(
\mathbf{p}_{a}\left(  x\right)  ,g_{ab}\left(  x\right)  \right)  $
\end{definition}

so it has not the same value at the intersections : \textit{this is not a
section}. This is the equivalent of the holonomic basis of a vector bundle. It
depends on the trivialization. In a change of trivialization :

$\mathbf{p}_{a}\left(  x\right)  =\varphi_{a}\left(  x,1\right)
=\widetilde{\varphi}_{a}\left(  x,\widetilde{g}_{a}\right)  \Rightarrow
\widetilde{g}_{a}=\chi_{a}\left(  x\right)  $

$\mathbf{p}_{a}\left(  x\right)  =\rho\left(  \widetilde{\varphi}_{a}\left(
x,1\right)  ,\chi_{a}\left(  x\right)  \right)  =\rho\left(  \widetilde
{\mathbf{p}}_{a}\left(  x\right)  ,\chi_{a}\left(  x\right)  \right)
\Leftrightarrow$%

\begin{equation}
\widetilde{\mathbf{p}}_{a}\left(  x\right)  =\rho\left(  \mathbf{p}_{a}\left(
x\right)  ,\chi_{a}\left(  x\right)  ^{-1}\right)
\end{equation}

A section on the principal bundle can be defined by : $S(x)=\rho\left(
\mathbf{p}_{a}\left(  x\right)  ,\sigma_{a}\left(  x\right)  \right)
=\rho\left(  \widetilde{\mathbf{p}}_{a}\left(  x\right)  ,\widetilde{\sigma
}_{a}\left(  x\right)  \right)  $

\paragraph{Gauge group\newline}

\begin{definition}
The \textbf{gauge group} of a principal bundle P is the set of fiber
preserving automorphisms F, called \textbf{vertical morphisms} $F:P\rightarrow
P$ such that:

$\pi\left(  F\left(  p\right)  \right)  =\pi\left(  p\right)  $

$\forall g\in G:F\left(  \rho\left(  p,g\right)  \right)  =\rho\left(
F\left(  p\right)  ,g\right)  $
\end{definition}

\begin{theorem}
The elements of the gauge group of a principal bundle P with atlas $\left(
O_{a},\varphi_{a}\right)  _{a\in A}$ are characterized \ by a collection of
maps : $j_{a}:O_{a}\rightarrow G$ such that :

$F\left(  \varphi_{a}\left(  x,g\right)  \right)  =\varphi_{a}\left(
x,j_{a}\left(  x\right)  g\right)  ,j_{b}\left(  x\right)  =g_{ba}\left(
x\right)  j_{a}\left(  x\right)  g_{ba}\left(  x\right)  ^{-1}$
\end{theorem}

\begin{proof}
i) define j from F

define : $\tau_{a}:\pi^{-1}\left(  O_{a}\right)  \rightarrow G::p=\varphi
_{a}\left(  \pi\left(  p\right)  ,\tau_{a}\left(  p\right)  \right)  $

define : $J_{a}\left(  p\right)  =\tau_{a}\left(  F\left(  p\right)  \right)
$

$J_{a}\left(  \rho\left(  p,g\right)  \right)  =\tau_{a}\left(  F\left(
\rho\left(  p,g\right)  \right)  \right)  =\tau_{a}\left(  \rho\left(
F\left(  p\right)  ,g\right)  \right)  =\tau_{a}\left(  F\left(  p\right)
\right)  g=J_{a}(p)g$

define : $j_{a}\left(  x\right)  =J\left(  \varphi_{a}\left(  x,1\right)
\right)  $

$J_{a}\left(  p\right)  =J_{a}\left(  \rho\left(  \varphi_{a}\left(
x,1\right)  ,\tau_{a}\left(  p\right)  \right)  \right)  =J_{a}\left(
\varphi_{a}\left(  x,1\right)  \right)  \tau_{a}\left(  p\right)
=j_{a}\left(  x\right)  \tau_{a}\left(  p\right)  $

$F\left(  p\right)  =\varphi_{a}\left(  \pi\left(  p\right)  ,J_{a}\left(
p\right)  \right)  =\varphi_{a}\left(  \pi\left(  p\right)  ,j_{a}\left(
x\right)  \tau_{a}\left(  p\right)  \right)  $

$F\left(  \varphi_{a}\left(  x,g\right)  \right)  =\varphi_{a}\left(
x,j_{a}\left(  x\right)  g\right)  $

ii) conversely define F from j :

$F\left(  \varphi_{a}\left(  x,g\right)  \right)  =\varphi_{a}\left(
x,j_{a}\left(  x\right)  g\right)  $

then : $\pi\circ F=\pi$ and

$F\left(  \rho\left(  p,g\right)  \right)  =F\left(  \rho\left(  \varphi
_{a}\left(  x,g_{a}\right)  ,g\right)  \right)  =F\left(  \varphi_{a}\left(
x,g_{a}g\right)  \right)  =F\left(  \varphi_{a}\left(  x,j_{a}\left(
x\right)  g_{a}g\right)  \right)  =\rho\left(  F\left(  p\right)  ,g\right)  $

iii) At the transitions j must be such that :\ 

$F\left(  \varphi_{a}\left(  x,g_{a}\right)  \right)  =\varphi_{a}\left(
x,j_{a}\left(  x\right)  g_{a}\right)  =F\left(  \varphi_{b}\left(
x,g_{b}\right)  \right)  =\varphi_{b}\left(  x,j_{b}\left(  x\right)
g_{b}\right)  $

$j_{b}\left(  x\right)  g_{b}=g_{ba}\left(  x\right)  j_{a}\left(  x\right)
g_{a}$

$g_{b}=g_{ba}\left(  x\right)  g_{a}$

$j_{b}\left(  x\right)  g_{ba}\left(  x\right)  g_{a}=g_{ba}\left(  x\right)
j_{a}\left(  x\right)  g_{a}\Rightarrow j_{b}\left(  x\right)  =g_{ba}\left(
x\right)  j_{a}\left(  x\right)  g_{ba}\left(  x\right)  ^{-1}$
\end{proof}

Remarks :

i) In a change of trivialization : we keep p \ and change $\varphi$

$p=\varphi_{a}\left(  x,g_{a}\right)  =\widetilde{\varphi}_{a}\left(
x,\chi_{a}\left(  x\right)  g_{a}\right)  ;$

$\mathbf{p}_{a}\left(  x\right)  =\varphi_{a}\left(  x,1\right)
\rightarrow\widetilde{\mathbf{p}}_{a}\left(  x\right)  =\widetilde{\varphi
}_{a}\left(  x,1\right)  =\varphi_{a}\left(  x,\chi_{a}\left(  x\right)
^{-1}\right)  =\rho\left(  \mathbf{p}_{a}\left(  x\right)  ,\chi_{a}\left(
x\right)  ^{-1}\right)  $

In a vertical morphism : we keep $\varphi$ and change p

$p=\varphi_{a}\left(  x,g_{a}\right)  \rightarrow F\left(  p\right)
=\varphi_{a}\left(  x,j_{a}\left(  x\right)  g_{a}\right)  $

$\mathbf{p}_{a}\left(  x\right)  =\varphi_{a}\left(  x,1\right)  \rightarrow
F\left(  \mathbf{p}_{a}\left(  x\right)  \right)  =\varphi_{a}\left(
x,j_{a}\left(  x\right)  \right)  $

If we define an element of P through the rigth action : $p=\varphi_{a}\left(
x,g_{a}\right)  =\rho\left(  \mathbf{p}_{a}\left(  x\right)  ,g_{a}\right)  $ because

$F\left(  \rho\left(  \mathbf{p}_{a}\left(  x\right)  ,g_{a}\right)  \right)
=\rho\left(  F\left(  \mathbf{p}_{a}\left(  x\right)  \right)  ,g_{a}\right)
$

a vertical morphism sums up to replace the gauge : $\mathbf{p}_{a}=\varphi
_{a}\left(  x,1\right)  $ by $F\left(  \mathbf{p}_{a}\left(  x\right)
\right)  =\varphi_{a}\left(  x,j_{a}\left(  x\right)  \right)  .$

ii) A vertical morphism is a \textit{left} action by $j_{a}\left(  x\right)
,$ a change of trivialization is a \textit{left} action by $\chi_{a}\left(
x\right)  .$

iii) $j_{a}$ depends on the trivialization

\subsubsection{Tangent space of a principal bundle}

\paragraph{Tangent bundle\newline}

The tangent bundle $TG=\cup_{g\in G}T_{g}G$ of a Lie group is still a Lie
group with the actions :

$M:TG\times TG\rightarrow TG::M(X_{g},Y_{h})=R_{h}^{\prime}(g)X_{g}%
+L_{g}^{\prime}(g)Y_{h}\in T_{gh}G$

$\Im:TG\rightarrow TG::\Im(X_{g})=-R_{g^{-1}}^{\prime}(1)\circ L_{g^{-1}%
}^{\prime}(g)X_{g}=-L_{g^{-1}}^{\prime}(g)\circ R_{g^{-1}}^{\prime}(g)X_{g}\in
T_{g^{-1}}G$

Identity : $X_{1}=0_{1}\in T_{1}G$

\begin{theorem}
(Kolar p.99) The tangent bundle TP of a principal fiber bundle $P\left(
M,G,\pi\right)  $\ is a principal bundle $TP\left(  TM,TG,T\pi\right)  .$

The vertical bundle VP is :

a trivial vector bundle over P : $VP(P,T_{1}G,\pi)\simeq P\times T_{1}G$

a principal bundle over M with group TG : $VP(M,TG,\pi)$
\end{theorem}

With an atlas $\left(  O_{a},\varphi_{a}\right)  _{a\in A}$ of P the right
action of TG on TP is :

$T\rho:TP\times TG\rightarrow TP::T\rho\left(  \left(  p,v_{p}\right)
,\left(  g,\kappa_{g}\right)  \right)  =\left(  \rho\left(  p,g\right)
,\rho^{\prime}\left(  p,g\right)  \left(  v_{p},\kappa_{g}\right)  \right)  $

$T\rho\left(  \left(  \varphi(x,h),\varphi^{\prime}\left(  x,h\right)  \left(
v_{x},v_{h}\right)  \right)  ,\left(  g,\kappa_{g}\right)  \right)  $

$=\left(  \varphi\left(  x,hg\right)  ,\varphi^{\prime}\left(  p,gh\right)
\left(  v_{x},\left(  R_{g}^{\prime}h\right)  v_{h}+\left(  L_{h}^{\prime
}g\right)  \kappa_{g}\right)  \right)  $

\paragraph{Fundamental vector fields\newline}

Fundamental vector fields are defined as for any action of a group on a
manifold, with the same properties (see Lie Groups).

\begin{theorem}
On a principal bundle $P(M,G,\pi)$ with atlas $\left(  O_{a},\varphi
_{a}\right)  _{a\in A}$ and right action $\rho:$

i) The map :%

\begin{equation}
\zeta:T_{1}G\rightarrow\mathfrak{X}\left(  VP\right)  ::\zeta\left(  X\right)
\left(  p\right)  =\rho_{g}^{\prime}\left(  p,1\right)  X=\varphi_{ag}%
^{\prime}\left(  x,g\right)  \left(  L_{g}^{\prime}1\right)  X
\end{equation}

is linear and does not depend on the trivialization,

$\zeta\left(  X\right)  =\rho_{\ast}\left(  L_{g}^{\prime}\left(  1\right)
X,0\right)  $

$\rho_{p}^{\prime}\left(  p,g\right)  \left(  \zeta\left(  X\right)  \left(
p\right)  \right)  =\zeta\left(  Ad_{g^{-1}}X\right)  \left(  \rho\left(
p,g\right)  \right)  $

ii) The \textbf{fundamental vector fields} on P are defined, for any fixed X
in $T_{1}G$ , by :

$\zeta\left(  X\right)  :M\rightarrow VP::\zeta\left(  X\right)  \left(
p\right)  =$ $\rho_{g}^{\prime}\left(  \mathbf{p},1\right)  X$

They belong to the vertical bundle, and span an integrable distribution over
P, whose leaves are the connected components of the orbits

$\forall X,Y\in T_{1}G:\left[  \zeta\left(  X\right)  ,\zeta\left(  Y\right)
\right]  _{VP}=\zeta\left(  \left[  X,Y\right]  _{T_{1}G}\right)  $

iii) The Lie derivative of a section S on P along a fundamental vector field
$\zeta\left(  X\right)  $\ is : $\pounds _{\zeta\left(  X\right)  }%
S=\zeta\left(  X\right)  \left(  S\right)  $
\end{theorem}

\paragraph{Component expressions\newline}

\begin{theorem}
A vector $v_{p}\in T_{p}P$ of the principal bundle $P(M,G,\pi)$ with atlas
$\left(  O_{a},\varphi_{a}\right)  _{a\in A}$ can be written%

\begin{equation}
v_{p}=\sum_{\alpha}v_{x}^{\alpha}\partial x_{\alpha}+\zeta\left(
v_{g}\right)  \left(  p\right)
\end{equation}

where

$v_{g}\in T_{1}G,\partial x_{\alpha}=\varphi_{x}^{\prime}\left(  x,g\right)
\partial\xi_{\alpha},\zeta\left(  v_{g}\right)  \left(  p\right)  =\varphi
_{g}^{\prime}\left(  x,g\right)  \left(  L_{g}^{\prime}1\right)  v_{g} $

At the transitions :%

\begin{equation}
v_{ax}=v_{bx}=v_{x}%
\end{equation}

\begin{equation}
v_{bg}=v_{ag}+Ad_{g_{a}^{-1}}L_{g_{ba}^{-1}}^{\prime}\left(  g_{ba}\right)
g_{ba}^{\prime}\left(  x\right)  v_{x}%
\end{equation}

\end{theorem}

\begin{proof}
With the basis $\left(  \varepsilon_{i}\right)  _{i\in I}$ of the Lie algebra
of G $\left(  L_{g}^{\prime}1\right)  \varepsilon_{i}$ is a basis of $T_{g}G.
$

A basis of $T_{p}P$ at $p=\varphi_{a}\left(  x,g\right)  $\ is :

$\partial x_{\alpha}=\varphi_{ax}^{\prime}\left(  x,g\right)  \partial
\xi_{\alpha},\partial g_{i}=\varphi_{ag}^{\prime}\left(  x,g\right)  \left(
L_{g}^{\prime}1\right)  \varepsilon_{i}=\zeta\left(  \varepsilon_{i}\right)
\left(  p\right)  $

$\left(  p,v_{p}\right)  \in T_{p}P:v_{p}=\sum_{\alpha}v_{x}^{\alpha}\partial
x^{\alpha}+\sum_{i}v_{g}^{i}\zeta\left(  \varepsilon_{i}\right)  \left(
p\right)  =\sum_{\alpha}v_{x}^{\alpha}\partial x_{\alpha}+\zeta\left(
v_{g}\right)  \left(  p\right)  $ where $v_{g}=\sum_{i}v_{g}^{i}%
\varepsilon_{i}\in T_{1}G$

At the transitions :

$v_{bg}=\left(  g_{ba}\left(  x\right)  g_{a}\right)  ^{\prime}\left(
v_{x},v_{ag}\right)  =R_{g_{a}}^{\prime}\left(  g_{ba}\left(  x\right)
\right)  g_{ba}^{\prime}\left(  x\right)  v_{x}+L_{g_{ba}\left(  x\right)
}^{\prime}\left(  g_{a}\right)  v_{ag}$

$v_{bg}=L_{g_{b}^{-1}}^{\prime}g_{b}\left(  R_{g_{a}}^{\prime}\left(
g_{ba}\right)  g_{ba}^{\prime}\left(  x\right)  v_{x}+L_{g_{ba}}^{\prime
}\left(  g_{a}\right)  L_{g_{a}}^{\prime}\left(  1\right)  v_{ag}\right)  $

$v_{bg}=L_{g_{b}^{-1}}^{\prime}\left(  g_{b}\right)  \left(  R_{g_{b}}%
^{\prime}(1)R_{g_{ba}^{-1}}^{\prime}(g_{ba})g_{ba}^{\prime}v_{x}+L_{g_{b}%
}^{\prime}\left(  1\right)  L_{g_{a}^{-1}}^{\prime}\left(  g_{a}\right)
L_{g_{a}}^{\prime}\left(  1\right)  v_{ag}\right)  $

$v_{bg}=Ad_{g_{b}^{-1}}R_{g_{ba}^{-1}}^{\prime}(g_{ba})L_{g_{ba}}^{\prime
}\left(  1\right)  L_{g_{ba}^{-1}}^{\prime}\left(  g_{ba}\right)
g_{ba}^{\prime}\left(  x\right)  v_{x}+v_{ag}$

$v_{bg}=Ad_{g_{b}^{-1}}Ad_{g_{ba}}L_{g_{ba}^{-1}}^{\prime}\left(
g_{ba}\right)  g_{ba}^{\prime}\left(  x\right)  v_{x}+v_{ag}$

$v_{bg}=Ad_{g_{a}^{-1}}L_{g_{ba}^{-1}}^{\prime}\left(  g_{ba}\right)
g_{ba}^{\prime}\left(  x\right)  v_{x}+v_{ag}$
\end{proof}

\paragraph{Component expression of the right action of TG on TP:\newline}

\begin{theorem}
The right action of TG on the tangent bundle reads:

$T\rho\left(  \left(  \varphi(x,h),\varphi^{\prime}\left(  x,h\right)
v_{x}+\zeta\left(  v_{h}\right)  p\right)  ,\left(  g,\left(  L_{g}^{\prime
}1\right)  \kappa_{g}\right)  \right)  $

$=\left(  \rho\left(  p,g\right)  ,\varphi_{x}^{\prime}\left(  x,hg\right)
v_{x}+\zeta\left(  Ad_{g^{-1}}v_{h}+\kappa_{g}\right)  \left(  \rho\left(
p,g\right)  \right)  \right)  $
\end{theorem}

\begin{proof}
$T\rho\left(  \left(  \varphi(x,h),\varphi^{\prime}\left(  x,h\right)
v_{x}+\zeta\left(  v_{h}\right)  p\right)  \left(  g,\left(  L_{g}^{\prime
}1\right)  \kappa_{g}\right)  \right)  $

$=\left(  \rho\left(  p,g\right)  ,\varphi_{x}^{\prime}\left(  x,hg\right)
v_{x}+\varphi_{g}^{\prime}\left(  x,hg\right)  \left(  \left(  R_{g}^{\prime
}h\right)  \left(  L_{h}^{\prime}1\right)  v_{h}+\left(  L_{h}^{\prime
}g\right)  \left(  L_{g}^{\prime}1\right)  \kappa_{g}\right)  \right)  $

$\varphi_{g}^{\prime}\left(  x,gh\right)  \left(  \left(  R_{g}^{\prime
}h\right)  \left(  L_{h}^{\prime}1\right)  v_{h}+\left(  L_{h}^{\prime
}g\right)  \left(  L_{g}^{\prime}1\right)  \kappa_{g}\right)  $

$=\varphi_{g}^{\prime}\left(  x,hg\right)  \left(  L_{hg}^{\prime}1\right)
L_{\left(  hg\right)  ^{-1}}^{\prime}\left(  hg\right)  \left(  \left(
R_{g}^{\prime}h\right)  \left(  L_{h}^{\prime}1\right)  v_{h}+\left(
L_{h}^{\prime}g\right)  \left(  L_{g}^{\prime}1\right)  \kappa_{g}\right)  $

$=\zeta\left(  L_{\left(  hg\right)  ^{-1}}^{\prime}\left(  hg\right)  \left(
\left(  R_{g}^{\prime}h\right)  \left(  L_{h}^{\prime}1\right)  v_{h}+\left(
L_{h}^{\prime}g\right)  \left(  L_{g}^{\prime}1\right)  \kappa_{g}\right)
\right)  \left(  \rho\left(  p,g\right)  \right)  $

$L_{\left(  hg\right)  ^{-1}}^{\prime}\left(  hg\right)  \left(  \left(
R_{g}^{\prime}h\right)  \left(  L_{h}^{\prime}1\right)  v_{h}+\left(
L_{h}^{\prime}g\right)  \left(  L_{g}^{\prime}1\right)  \kappa_{g}\right)  $

$=L_{\left(  hg\right)  ^{-1}}^{\prime}\left(  hg\right)  \left(  \left(
R_{hg}^{\prime}(1)R_{h^{-1}}^{\prime}(h)\left(  L_{h}^{\prime}1\right)
v_{h}+L_{hg}^{\prime}(1)L_{g^{-1}}^{\prime}(g)\left(  L_{g}^{\prime}1\right)
\kappa_{g}\right)  \right)  $

$=L_{\left(  hg\right)  ^{-1}}^{\prime}\left(  hg\right)  R_{hg}^{\prime
}(1)R_{h^{-1}}^{\prime}(h)\left(  L_{h}^{\prime}1\right)  v_{h}+L_{\left(
hg\right)  ^{-1}}^{\prime}\left(  hg\right)  L_{hg}^{\prime}(1)\kappa_{g}$

$=Ad_{\left(  hg\right)  ^{-1}}Ad_{h}v_{h}+\kappa_{g}=Ad_{g^{-1}}v_{h}%
+\kappa_{g}$
\end{proof}

\paragraph{One parameter group of vertical morphisms\newline}

\begin{theorem}
The infinitesimal generators of continuous one parameter groups of vertical
automorphisms on a principal bundle P are fundamental vector fields
$\zeta\left(  \kappa_{a}\left(  x\right)  \right)  \left(  p\right)  $ given
by sections $\kappa_{a}$ of the adjoint bundle $P\left[  T_{1}G,Ad\right]  .$
\end{theorem}

It reads : $\Phi_{\zeta\left(  \kappa\right)  }\left(  \varphi_{a}\left(
x,g\right)  ,t\right)  =\varphi_{a}\left(  x,\exp t\kappa_{a}\left(  x\right)
g\right)  $

\begin{proof}
Let $F:P\times%
\mathbb{R}
\rightarrow P$ be a one parameter group of vertical automorphisms, that is :

$F\left(  p,t+t^{\prime}\right)  =F\left(  F\left(  p,t\right)  ,t^{\prime
}\right)  ,F\left(  p,0\right)  =p$

$\forall t\in%
\mathbb{R}
$ F(.,t) is a vertical automorphism, so with atlas $\left(  O_{a},\varphi
_{a}\right)  _{a\in A}$ of P there is a collection $\left(  j_{at}\right)
_{a\in A}$ of maps such that : $F\left(  \varphi_{a}\left(  x,g\right)
,t\right)  =\varphi_{a}\left(  x,j_{at}\left(  x\right)  g\right)
,j_{bt}\left(  x\right)  =g_{ba}\left(  x\right)  j_{at}\left(  x\right)
g_{ba}\left(  x\right)  ^{-1}$. $j_{at}\left(  x\right)  $ is a one parameter
group of morphisms on G, so there is an infinitesimal generator $\kappa
_{a}\left(  x\right)  \in T_{1}G $ and $j_{at}\left(  x\right)  =\exp
t\kappa_{a}\left(  x\right)  .$ The condition :

$j_{bt}\left(  x\right)  =g_{ba}\left(  x\right)  j_{at}\left(  x\right)
g_{ba}\left(  x\right)  ^{-1}\Leftrightarrow\exp t\kappa_{b}\left(  x\right)
=g_{ba}\left(  x\right)  \exp t\kappa_{a}\left(  x\right)  g_{ba}\left(
x\right)  ^{-1}$

gives by derivation at t=0 : $\kappa_{b}\left(  x\right)  =Ad_{g_{ba}\left(
x\right)  }\kappa_{a}\left(  x\right)  $ which is the characteristic of a
section of the adjoint bundle $P\left[  T_{1}G,Ad\right]  $ of P (see
Associated bundles for more). On TP the infinitesimal generator of F is given
by the vector field :

$\frac{d}{dt}F\left(  p,t\right)  |_{t=0}=\varphi_{ag}^{\prime}\left(
x,g\right)  \left(  L_{g}^{\prime}1\right)  \kappa_{a}\left(  x\right)
=\zeta\left(  \kappa_{a}\left(  x\right)  \right)  \left(  p\right)  $
\end{proof}

\paragraph{One parameter group of change of trivialization\newline}

\begin{theorem}
A section $\kappa\in\mathfrak{X}\left(  P\left[  T_{1}G,Ad\right]  \right)  $
of the adjoint bundle to a principal bundle $P\left(  M,G,\pi_{P}\right)  $
with atlas $\left(  O_{a},\varphi_{a}\right)  _{a\in A}$ defines a one
parameter group of change of trivialization on P by : $p=\varphi_{a}\left(
x,g_{a}\right)  =\widetilde{\varphi}_{at}\left(  x,\exp t\kappa_{a}\left(
x\right)  \left(  g_{a}\right)  \right)  $\ with infinitesimal generator the
fundamental vector field $\zeta\left(  \kappa_{a}\left(  x\right)  \right)
\left(  p\right)  $ .\ The transition maps are unchanged : $\widetilde{g}%
_{ba}\left(  x,t\right)  =g_{ba}\left(  x\right)  $
\end{theorem}

\begin{proof}
A section $\kappa\in\mathfrak{X}\left(  P\left[  T_{1}G,Ad\right]  \right)
$\ is defined by a collection maps $\left(  \kappa_{a}\right)  _{a\in A}\in
C\left(  O_{a};T_{1}G\right)  $ with the transition maps : $\kappa
_{b}=Ad_{g_{ba}}\kappa_{a}.$

It is the infinitesimal generator of the one parameter group of
diffeomorphisms on G :

$\Phi_{\kappa_{a}\left(  x\right)  }\left(  g,t\right)  =\left(  \exp
t\kappa_{a}\left(  x\right)  \right)  g$

By taking $\chi_{at}\left(  x\right)  =\exp t\kappa_{a}\left(  x\right)  $ we
have the one parameter group of change of trivialisations on P :

$p=\varphi_{a}\left(  x,g_{a}\right)  =\widetilde{\varphi}_{at}\left(  x,\exp
t\kappa_{a}\left(  x\right)  \left(  g_{a}\right)  \right)  $

Its infinitesimal generator is : $\frac{d}{dt}\widetilde{\varphi}_{at}\left(
x,\exp t\kappa_{a}\left(  x\right)  \left(  g_{a}\right)  \right)
|_{t=0}=\varphi_{a}^{\prime}\left(  x,g_{a}\right)  L_{g_{a}}^{\prime}%
\kappa_{a}\left(  x\right)  =\zeta\left(  \kappa_{a}\left(  x\right)  \right)
\left(  p\right)  $ that is the fundamental vector at p.

Moreover the new transition maps are :

$\widetilde{g}_{ba}\left(  x,t\right)  =\left(  \exp t\kappa_{b}\left(
x\right)  \right)  g_{ba}\left(  x\right)  \exp\left(  -t\kappa_{a}\left(
x\right)  \right)  =\left(  \exp tAd_{g_{ba}}\kappa_{a}\left(  x\right)
\right)  g_{ba}\left(  x\right)  \exp\left(  -t\kappa_{a}\left(  x\right)
\right)  $

$=g_{ba}\left(  \exp t\kappa_{a}\left(  x\right)  \right)  g_{ba}^{-1}%
g_{ba}\left(  x\right)  \exp\left(  -t\kappa_{a}\left(  x\right)  \right)
=g_{ba}\left(  x\right)  $
\end{proof}

\subsubsection{Principal bundle of frames}

\paragraph{Reduction of a principal bundle by a subgroup\newline}

\begin{definition}
A morphism between the principal bundles $P(M,G,\pi_{P}),Q(M,H,\pi_{Q})$\ is a
\textbf{reduction} of P is H is a Lie subgroup of G
\end{definition}

\begin{theorem}
A principal bundle $P\left(  M,G,\pi\right)  $ admits a reduction to
$Q(M,H,\pi_{Q})$\ if it has an atlas with transition maps valued in H
\end{theorem}

\begin{proof}
If P has the atlas $\left(  O_{a},\varphi_{a},g_{ab}\right)  $ the restriction
of the trivializations to H reads: $\psi_{a}:O_{a}\times H\rightarrow
P:\psi_{a}\left(  x,h\right)  =\varphi_{a}\left(  x,h\right)  $

For the transitions : $\psi_{a}\left(  x,h_{a}\right)  =\psi_{b}\left(
x,h_{b}\right)  \Rightarrow h_{b}=g_{ba}\left(  x\right)  h_{a}$ and because H
is a subgroup : $g_{ba}\left(  x\right)  =h_{b}h_{a}^{-1}\in H$
\end{proof}

\paragraph{Principal bundle of linear frames\newline}

\begin{theorem}
The \textbf{bundle of linear frames }of\textbf{\ }a vector bundle $E\left(
M,V,\pi\right)  $ with atlas $\left(  O_{a},\varphi_{a}\right)  _{a\in A}$ is
the set of its bases. It has the structure of a principal bundle P$\left(  M,G%
\mathcal{L}%
(V;V),\pi_{P}\right)  $ with trivializations $\left(  O_{a},L_{a}\right)
_{a\in A}$ where $L_{a}\in C_{0}\left(  O_{a};G%
\mathcal{L}%
\left(  V;V\right)  \right)  $
\end{theorem}

\begin{proof}
A holonomic basis of E associated to the atlas is defined by : $\mathbf{e}%
_{ia}:O_{a}\rightarrow E::\mathbf{e}_{ai}\left(  x\right)  =\varphi_{a}\left(
x,e_{i}\right)  $ where $\left(  e_{i}\right)  _{i\in I}$ is a basis of V. At
the transitions : $\mathbf{e}_{ib}\left(  x\right)  =\varphi_{b}\left(
x,e_{i}\right)  =\varphi_{a}\left(  x,\varphi_{ab}\left(  x\right)
e_{i}\right)  $ where $\varphi_{ab}\left(  x\right)  \in G%
\mathcal{L}%
(V,V).$ Any family of maps : $\left(  L_{a}\in C_{0}\left(  O_{a};G%
\mathcal{L}%
\left(  V;V\right)  \right)  \right)  _{a\in A}$ defines a new trivialization
of the same vector bundle, with new holonomic bases $\widetilde{\mathbf{e}%
}_{ai}\left(  x\right)  =L_{a}\left(  x\right)  \mathbf{e}_{ai}\left(
x\right)  $.\ Define $P=\left\{  \mathbf{e}_{ai}\left(  x\right)  ,a\in A,i\in
I,x\in O_{a}\right\}  $ the set of all bases of E.\ This is a fibered
manifold, and the action of G on P defines a principal bundle.
\end{proof}

As a particular case the bundle of linear frames\textbf{\ }of a n dimensional
manifold on a field K\ is the bundle of frames of its tangent bundle. It has
the structure of a principal bundle $GL\left(  M,GL\left(  K,n\right)
,\pi_{GL}\right)  .$ A basis $\left(  e_{i}\right)  _{i=1}^{n}$ of $T_{x}M$ is
deduced from the holonomic basis $\left(  \partial x_{\alpha}\right)
_{\alpha=1}^{n}$ of TM associated to an atlas of M by an endomorphism of
$K^{n}$ represented by a matrix of GL(K,n). If its bundle of linear frame
admits a global section, it is trivial and M is parallelizable.

\paragraph{Principal bundle of orthonormal frames\newline}

\begin{theorem}
The \textbf{bundle of orthonormal frames }on a m dimensional real manifold
(M,g) endowed with a scalar product is the set of its orthogonal bases.\ It
has the structure of a principal bundle $O(M,O(%
\mathbb{R}
,r,s),\pi)$ if g has the signature (r,s) with r+s=m.
\end{theorem}

\begin{proof}
There is always an orthonormal basis at each point, by diagonalization of the
matrix of g. These operations are continuous and differentiable if g is
smooth. So at least locally we can define charts such that the holonomic bases
are orthonormal.\ The transitions maps belong to $O(%
\mathbb{R}
,r,s)$ because the bases are orthonormal. We have a reduction of the bundle of
linear frames.
\end{proof}

If M is orientable, it is possible to define a continuous orientation, and
then a principal bundle $SO\left(  M,SO\left(
\mathbb{R}
,r,s\right)  ,\pi\right)  $

$O\left(
\mathbb{R}
,r,s\right)  $ has four connected components, and two for $SO\left(
\mathbb{R}
,r,s\right)  $ . One can always restrict a trivialization to the connected
component $SO_{0}\left(
\mathbb{R}
,r,s\right)  $\ and then the frames for the other connected component are
given by a global gauge transformation.

\bigskip

\subsection{Associated bundles}

\label{Associated fiber bundles}

In a fiber bundle, the fibers $\pi^{-1}\left(  x\right)  $ ,which are
diffeomorphic to V, have usually some additional structure. The main example
is the vector bundle where the bases $\left(  \mathbf{e}_{i}\left(  x\right)
\right)  _{i\in I}$ are chosen to be some specific frame, such that
orthonormal. So we have to combine in the same structure a fiber bundle with
its fiber content given by V, and a principal bundle with its organizer G.
This is the basic idea about associated bundles.\ They will be defined in the
more general setting, even if the most common are the associated vector bundles.

\subsubsection{General associated bundles}

\paragraph{G-bundle\newline}

\begin{definition}
A fiber bundle $E\left(  M,V,\pi\right)  $ with atlas $\left(  O_{a}%
,\varphi_{a}\right)  _{a\in A}$ has a \textbf{G-bundle structure} if there are :

- a Lie group G (on the same field as E)

- a left action of G on the standard fiber V : $\lambda:G\times V\rightarrow
V$

- a family $\left(  g_{ab}\right)  _{a,b\in A}$\ of maps : $g_{ab}:\left(
O_{a}\cap O_{b}\right)  \rightarrow G$

such that the transition maps read : $\varphi_{ba}\left(  x\right)  \left(
u\right)  =\lambda(g_{ba}(x),u)$
\end{definition}

$p=\varphi_{a}(x,u_{a})=\varphi_{b}(x,u_{b})\Rightarrow u_{b}=\lambda
(g_{ba}(x),u_{a})$

All maps are assumed to be of the same class r.

\begin{notation}
$E=M\times_{G}V$ is a G-bundle with base M, standard fiber V and left action
of the group G
\end{notation}

Example : a vector bundle $E\left(  M,V,\pi\right)  $ has a G-bundle structure
with G = GL(V).

\paragraph{Associated bundle\newline}

\begin{definition}
An associated bundle is a structure which consists of :

i) a principal bundle $P(M,G,\pi_{P})$ with right action $\rho:P\times
G\rightarrow G$

ii) a manifold V and a left action $\lambda:G\times V\rightarrow V$ which is
effective :

$\forall g,h\in G,u\in V:\lambda\left(  g,v\right)  =\lambda\left(
h,v\right)  \Rightarrow g=h$

iii) an action of G on $P\times V$ defined as :

$\Lambda:G\times\left(  P\times V\right)  \rightarrow P\times V::\Lambda
\left(  g,\left(  p,u\right)  \right)  =\left(  \rho\left(  p,g\right)
,\lambda\left(  g^{-1},u\right)  \right)  $

iv) the equivalence relation $\sim$ on $P\times V:\forall g\in G:\left(
p,u\right)  \sim\left(  \rho\left(  p,g\right)  ,\lambda\left(  g^{-1}%
,u\right)  \right)  $

The \textbf{associated bundle} is the quotient set $E=\left(  P\times
V\right)  /\sim$

The projection $\pi_{E}:E\rightarrow M::\pi_{E}\left(  \left[  p,u\right]
_{\sim}\right)  =\pi_{P}\left(  p\right)  $
\end{definition}

\begin{notation}
$P\left[  V,\lambda\right]  $ is the associated bundle, with principal bundle
P, fiber V and action $\lambda:G\times V\rightarrow V$
\end{notation}

Remarks:

1. To each couple $\left(  p,u\right)  \in P\times V$ is associated its class
of equivalence, defined by the projection :

$\Pr:P\times V\rightarrow E=\left(  P\times V\right)  /\sim::q=\left[
p,u\right]  $

2. By construction $\Pr$ is invariant by $\Lambda:\Pr\left(  \Lambda\left(
g,\left(  p,u\right)  \right)  \right)  =\Pr\left(  p,u\right)  $

3. If P and V are finite dimensional then : $\dim E=\dim M+\dim V$

4. Because $\lambda$ is effective, $\lambda\left(  .,u\right)  $ is injective
and $\lambda_{g}^{\prime}\left(  g,u\right)  $ is invertible

5. The power -1 comes from linear frames : coordinates change according to the
inverse of the matrix to pass from one frame to the other.

We have the VERY useful identity at the transitions :

$q_{a}=\left(  \varphi_{a}\left(  x,g_{a}\right)  ,u_{a}\right)  =\left(
\varphi_{b}\left(  x,g_{ba}g_{a}\right)  ,u_{a}\right)  \sim\left(
\rho\left(  \varphi_{b}\left(  x,g_{ba}g_{a}\right)  ,g_{a}^{-1}g_{ab}\right)
,\lambda\left(  g_{ba}g_{a},u_{a}\right)  \right)  =\left(  \varphi_{b}\left(
x,1\right)  ,\lambda\left(  g_{ba}g_{a},u_{a}\right)  \right)  $

$q_{b}=\left(  \varphi_{b}\left(  x,g_{b}\right)  ,u_{b}\right)  \sim\left(
\rho\left(  \varphi_{b}\left(  x,g_{b}\right)  ,g_{b}^{-1}\right)
,\lambda\left(  g_{b},u_{b}\right)  \right)  =\left(  \varphi_{b}\left(
x,1\right)  ,\lambda\left(  g_{b},u_{b}\right)  \right)  $

$q_{a}=q_{b}\Leftrightarrow\lambda\left(  g_{ba}g_{a},u_{a}\right)
=\lambda\left(  g_{b},u_{b}\right)  \Leftrightarrow u_{b}=\lambda\left(
g_{b}^{-1}g_{ba}g_{a},u_{a}\right)  $%

\begin{equation}
\left(  \varphi_{a}\left(  x,g_{a}\right)  ,u_{a}\right)  \sim\left(
\varphi_{b}\left(  x,g_{b}\right)  ,u_{b}\right)  \Leftrightarrow
u_{b}=\lambda\left(  g_{b}^{-1}g_{ba}g_{a},u_{a}\right)
\end{equation}

if $g_{b}=g_{ba}g_{a}:u_{b}=u_{a}$

An associated bundle is a fiber bundle with a G-bundle structure, in which the
principal bundle has been identified :

\begin{theorem}
(Kolar p.91) For any fiber bundle $E\left(  M,V,\pi\right)  $ endowed with a
G-bundle structure with a left action of G on V, there is a unique principal
bundle $P(M,G,\pi)$ such that E is an associated bundle $P\left[
V,\lambda\right]  $\ 
\end{theorem}

To deal with associated bundle it is usually convenient to use a gauge on P,
which emphasizes the G-bundle structure (from Kolar p.90) :

\begin{theorem}
For any principal bundle $P\left(  M,G,\pi_{P}\right)  $ with atlas $\left(
O_{a},\varphi_{a}\right)  _{a\in A}$ , transition maps $g_{ab}\left(
x\right)  $, manifold V and effective left action : $\lambda:G\times
V\rightarrow V$ there is an associated bundle $E=P[V,\lambda].$

E is the quotient set : $P\times V/\sim$ with the equivalence relation
$\left(  p,u\right)  \sim\left(  \rho\left(  p,g\right)  ,\lambda\left(
g^{-1},u\right)  \right)  $

If P,G, V are class r manifolds, then there is a unique structure of class r
manifold on E such that the map :

$\Pr:P\times V\rightarrow E::\Pr\left(  p,u\right)  =\left[  p,u\right]
_{\sim}$is a submersion

E is a fiber bundle $E\left(  M,V,\pi_{E}\right)  $ with the \textbf{standard
trivialization} and atlas $\left(  O_{a},\psi_{a}\right)  _{a\in A}:$

projection : $\pi_{E}:E\rightarrow M::\pi_{E}\left(  \left[  p,u\right]
\right)  =\pi_{P}\left(  p\right)  $

\begin{theorem}
trivializations $\psi_{a}:O_{a}\times V\rightarrow E::\psi_{a}\left(
x,u\right)  =\Pr\left(  \left(  \varphi_{a}(x,1),u\right)  \right)  $

transitions maps : $\psi_{a}\left(  x,u_{a}\right)  =\psi_{b}\left(
x,u_{b}\right)  \Rightarrow u_{b}=\lambda\left(  g_{ba}\left(  x\right)
,u_{a}\right)  =\psi_{ba}\left(  x,u_{a}\right)  $
\end{theorem}
\end{theorem}

So : $\psi_{a}\left(  x,u_{a}\right)  =\left(  \varphi_{a}\left(  x,1\right)
,u_{a}\right)  \sim\psi_{b}\left(  x,u_{b}\right)  =\left(  \varphi_{b}\left(
x,1\right)  ,u_{b}\right)  \Leftrightarrow u_{b}=\lambda\left(  g_{ba}%
,u_{a}\right)  $

The map : $\mathbf{p}_{a}\left(  x\right)  =\varphi_{a}\left(  x,1\right)  \in
P $ is the \textbf{gauge} of the associated bundle.

\paragraph{Sections\newline}

\begin{definition}
A section of an associated bundle $P\left[  V,\lambda\right]  $ is a pair of a
section s on P and a map : $u:M\rightarrow V$
\end{definition}

The pair is not unique : any other pair s',u' such that :

$\forall x\in M:\exists g\in G:\left(  \rho\left(  s^{\prime}\left(  x\right)
,g\right)  ,\lambda\left(  g^{-1},u^{\prime}\left(  x\right)  \right)
\right)  =\left(  s\left(  x\right)  ,u\left(  x\right)  \right)  $ defines
the same section.

With the standard trivialization $\left(  O_{a},\psi_{a}\right)  _{a\in A}$
associated to an atlas of P a section S on $P\left[  V,\lambda\right]  $ is
defined by a collection of maps : $u_{a}:O_{a}\rightarrow V$ :: $S\left(
x\right)  =\psi_{a}\left(  x,u_{a}\left(  x\right)  \right)  $ such that at
the transitions : $x\in O_{a}\cap O_{b}:u_{b}\left(  x\right)  =\lambda\left(
g_{ba}\left(  x\right)  ,u_{a}\left(  x\right)  \right)  $

\paragraph{Morphisms\newline}

\begin{definition}
(Kolar p.92)\ : A morphism between the associated bundles $P_{1}\left[
V_{1},\lambda_{1}\right]  ,P_{2}\left[  V_{2},\lambda_{2}\right]  $ is a map :

$\Phi:P_{1}\rightarrow P_{2}::\Phi\left(  p,u\right)  =\left(  F\left(
p\right)  ,f\left(  u\right)  \right)  $ such that

$F:P_{1}\rightarrow P_{2}$ is a morphism of principal bundles : there is a Lie
group morphism $\chi:G_{1}\rightarrow G_{2}$ such that : $\forall p\in
P_{1},g\in G_{1}:F\left(  \rho_{1}\left(  p,g\right)  \right)  =\rho
_{2}\left(  F\left(  p\right)  ,\chi\left(  g\right)  \right)  $

$f:V_{1}\rightarrow V_{2}$is equivariant : $\forall g\in G,u\in V_{1}:f\left(
\lambda_{1}\left(  g,u\right)  \right)  =\lambda_{2}\left(  \chi\left(
g\right)  ,f\left(  u\right)  \right)  $
\end{definition}

With a morphism of principal bundle $F:P_{1}\rightarrow P_{2}$ and an
associated bundle $P_{1}\left[  V,\lambda_{1}\right]  $ it is always possible
to define an associated bundle $P_{2}\left[  V,\lambda_{2}\right]  $ by f=Id
and $\lambda_{2}\left(  \chi\left(  g\right)  ,u\right)  =\lambda_{1}\left(
g,u\right)  $

\paragraph{Change of gauge\newline}

\subparagraph{Summary of the procedures\newline}

Let be the associated bundle $E=P\left[  V,\lambda\right]  $ to the principal
bundle $P\left(  M,G,\pi_{P}\right)  $ with atlas $\left(  O_{a},\varphi
_{a}\right)  _{a\in A}$ and transition maps $g_{ab}\left(  x\right)  \in G$

i) In the standard trivialization E has the atlas $\left(  O_{a},\psi
_{a}\right)  _{a\in A}$ and the transition maps : $u_{b}=\psi_{ba}\left(
x,u_{a}\right)  =\lambda\left(  g_{ba}\left(  x\right)  ,u_{a}\right)  .$

A change of standard trivialization is defined by a family $\left(  \phi
_{a}\right)  _{a\in A}$\ of maps $\phi_{a}\left(  x\right)  \in C\left(
V;V\right)  $\ and the new atlas is $\left(  O_{a},\widetilde{\psi}%
_{a}\right)  _{a\in A}$ with $q=\psi_{a}\left(  x,u_{a}\right)  =\widetilde
{\psi}_{a}\left(  x,\phi_{a}\left(  x\right)  \left(  u_{a}\right)  \right)
.$ The new transition maps are : $\widetilde{\psi}_{ba}\left(  x\right)
=\phi_{b}\left(  x\right)  \circ\psi_{ba}\left(  x\right)  \circ\phi
_{a}\left(  x\right)  ^{-1}$

ii) A change of trivialization on the principal bundle P defined by a family
$\left(  \chi_{a}\right)  _{a\in A}$\ of maps $\chi_{a}\in C\left(
O_{a};G\right)  $\ gives the new atlas $\left(  O_{a},\widetilde{\varphi}%
_{a}\right)  _{a\in A}$ with $p=\varphi_{a}\left(  x,g_{a}\right)
=\widetilde{\varphi}_{a}\left(  x,\chi_{a}\left(  x\right)  g_{a}\right)  $
and transition maps: $\widetilde{g}_{ba}\left(  x\right)  =\chi_{b}\left(
x\right)  \circ g_{ba}\left(  x\right)  \circ\chi_{a}\left(  x\right)  ^{-1}.$
The gauge becomes :

$\mathbf{p}_{a}\left(  x\right)  =\varphi_{a}\left(  x,1\right)
\rightarrow\widetilde{\mathbf{p}}_{a}\left(  x\right)  =\widetilde{\varphi
}_{a}\left(  x,1\right)  =\varphi_{a}\left(  x,\chi_{a}\left(  x\right)
^{-1}\right)  =\rho\left(  \mathbf{p}_{a}\left(  x\right)  ,\chi_{a}\left(
x\right)  ^{-1}\right)  $

It induces on the associated bundle E the change of trivialization :
$q=\psi_{a}\left(  x,u_{a}\right)  =\widetilde{\psi}_{a}\left(  x,\lambda
\left(  \chi_{a}\left(  x\right)  ,u_{a}\right)  \right)  $

iii) A vertical morphism on P, $p=\varphi_{a}\left(  x,g\right)
\rightarrow\widetilde{p}=\varphi_{a}\left(  x,j_{a}\left(  x\right)  g\right)
$ defined by a family of maps $j_{a}\left(  x\right)  \in C\left(
O_{a};G\right)  $ changes the gauge as : $\widetilde{\mathbf{p}}_{a}\left(
x\right)  =\rho\left(  \mathbf{p}_{a}\left(  x\right)  ,j_{a}\left(  x\right)
\right)  $

It induces on the associated bundle E the change of trivialization :
$q=\psi_{a}\left(  x,u_{a}\right)  =\widetilde{\psi}_{a}\left(  x,\lambda
\left(  j_{a}\left(  x\right)  ^{-1},u_{a}\right)  \right)  $

\subparagraph{Rules\newline}

As one can see a change of trivialization by $\chi_{a}$ and a vertical
morphism by $j_{a}^{-1}$ in P have the same impact on E and are expressed as a
change of the standard trivialization.

Uusually one wants to know the consequences of a change of trivialization,
both on the principal bundle and the associated bundle. So we are in case ii).
The general rule expressed previously still holds :

\begin{theorem}
Whenever a theorem is proven with the usual transition conditions, it is
proven for any change of trivialization. The formulas for a change of
trivialization read as the formulas for the transitions by taking
$\mathbf{g}_{ba}\left(  x\right)  \mathbf{=\chi}_{a}\left(  x\right)  .$
\end{theorem}

Principal bundle : $p=\varphi_{a}\left(  x,g_{a}\right)  =\varphi_{b}\left(
x,g_{ba}\left(  x\right)  \left(  g_{a}\right)  \right)  \leftrightarrow
p=\varphi_{a}\left(  x,g_{a}\right)  =\widetilde{\varphi}_{a}\left(
x,\chi_{a}\left(  x\right)  \left(  g_{a}\right)  \right)  $

Associated bundle : $q=\psi_{a}\left(  x,u_{a}\right)  =\psi_{b}\left(
x,\lambda\left(  g_{ba}\left(  x\right)  ,u_{a}\right)  \right)
\leftrightarrow q=\psi_{a}\left(  x,u_{a}\right)  =\widetilde{\psi}_{a}\left(
x,\lambda\left(  \chi_{a}\left(  x\right)  ,u_{a}\right)  \right)  $

\subparagraph{One parameter group of change of trivialization\newline}

\begin{theorem}
A section $\kappa\in\mathfrak{X}\left(  P\left[  T_{1}G,Ad\right]  \right)  $
of the adjoint bundle to a principal bundle $P\left(  M,G,\pi_{P}\right)  $
with atlas $\left(  O_{a},\varphi_{a}\right)  _{a\in A}$ defines a one
parameter group of change of trivialization on P by : $p=\varphi_{a}\left(
x,g_{a}\right)  =\widetilde{\varphi}_{at}\left(  x,\exp t\kappa_{a}\left(
x\right)  \left(  g_{a}\right)  \right)  $ . It induces on any associated
bundle $P\left[  V,\lambda\right]  $\ a one parameter group of changes of
trivialization : $q=\psi_{a}\left(  x,u_{a}\right)  =\widetilde{\psi}%
_{at}\left(  x,\lambda\left(  \exp t\kappa_{a}\left(  x\right)  ,u_{a}\right)
\right)  $\ with infinitesimal generator the vector field on TV : $W\left(
x,u_{a}\right)  =\lambda_{g}^{\prime}\left(  1,u_{a}\right)  \kappa_{a}\left(
x\right)  \in\mathfrak{X}\left(  TV\right)  .$ The transition maps on E are
unchanged : $\widetilde{u}_{bt}=\lambda\left(  g_{ba}\left(  x\right)
,\widetilde{u}_{at}\right)  $
\end{theorem}

A section $\kappa\in\mathfrak{X}\left(  P\left[  T_{1}G,Ad\right]  \right)
$\ is defined by a collection maps $\left(  \kappa_{a}\right)  _{a\in A}\in
C\left(  O_{a};T_{1}G\right)  $ with the transition maps : $\kappa
_{b}=Ad_{g_{ba}}\kappa_{a}.$

The change of trivialization on P is : $p=\varphi_{a}\left(  x,g_{a}\right)
=\widetilde{\varphi}_{at}\left(  x,\exp t\kappa_{a}\left(  x\right)  \left(
g_{a}\right)  \right)  $ and the transition maps are unchanged.

The change of trivialization on E is : $q=\psi_{a}\left(  x,u_{a}\right)
=\widetilde{\psi}_{at}\left(  x,\lambda\left(  \exp t\kappa_{a}\left(
x\right)  ,u_{a}\right)  \right)  $

This is a one parameter group on the manifold V, with infinitesimal generator
the fundamental vector $\zeta_{L}\left(  \kappa\left(  x\right)  \right)
\left(  u\right)  $

$W\left(  x,u_{a}\right)  =\frac{d}{dt}\lambda\left(  \exp t\kappa_{a}\left(
x\right)  ,u_{a}\right)  |_{t=0}=\lambda_{g}^{\prime}\left(  1,u_{a}\right)
\kappa_{a}\left(  x\right)  \in\mathfrak{X}\left(  TV\right)  $

$W\left(  x,u_{b}\right)  =\lambda_{g}^{\prime}\left(  g_{ba},u_{a}\right)
L_{g_{ba}}^{\prime}(1)\kappa_{a}$

The new transition maps are :

$\widetilde{u}_{bt}=\lambda\left(  \exp t\kappa_{b},\lambda\left(
g_{ba},\lambda\left(  \exp\left(  -t\kappa_{a}\right)  ,\widetilde{u}%
_{at}\right)  \right)  \right)  =\lambda\left(  \left(  \exp t\kappa
_{b}\right)  g_{ba}\exp\left(  -t\kappa_{a}\right)  ,\widetilde{u}%
_{at}\right)  $

$=\lambda\left(  \left(  \exp tAd_{g_{ba}}\kappa_{a}\right)  g_{ba}\exp\left(
-t\kappa_{a}\right)  ,\widetilde{u}_{at}\right)  =\lambda\left(  g_{ba}\left(
\exp t\kappa_{a}\right)  g_{ab}g_{ba}\exp\left(  -t\kappa_{a}\right)
,\widetilde{u}_{at}\right)  $

$=\lambda\left(  g_{ba},\widetilde{u}_{at}\right)  $

\paragraph{Tangent bundle\newline}

\begin{theorem}
(Kolar p.99) The tangent bundle TE of the associated bundle $E=P\left[
V,\lambda\right]  $ is the associated bundle $TP\left[  TV,T\lambda\right]
\sim TP\times_{TG}TV$
\end{theorem}

\begin{proof}
TP is a principal bundle $TP\left(  TM,TG,T\pi\right)  $ with right action of TG

$T\rho=\left(  \rho,\rho^{\prime}\right)  :TP\times TG\rightarrow
TP::T\rho\left(  \left(  p,v_{p}\right)  ,\left(  g,\kappa_{g}\right)
\right)  =\left(  \rho\left(  p,g\right)  ,\rho^{\prime}\left(  p,g\right)
\left(  v_{p},\kappa_{g}\right)  \right)  $

TV is a manifold with the left action of TG :

$T\lambda=\left(  \lambda,\lambda^{\prime}\right)  :TG\times TV\rightarrow
TV::T\lambda\left(  \left(  g,\kappa_{g}\right)  ,\left(  u,v_{u}\right)
\right)  =\left(  \lambda\left(  g,u\right)  ,\lambda^{\prime}\left(
g,u\right)  \left(  \kappa_{g},v_{u}\right)  \right)  $

$TP\times TV=T\left(  P\times V\right)  $ is a manifold with the left action
of TG$:$

$T\Lambda=\left(  \Lambda,\Lambda^{\prime}\right)  :TG\times T\left(  P\times
V\right)  \rightarrow T\left(  P\times V\right)  ::$

$T\Lambda\left(  \left(  g,\kappa_{g}\right)  ,\left(  \left(  p,v_{p}\right)
,\left(  u,v_{u}\right)  \right)  \right)  $

$=\left(  T\rho\left(  \left(  p,v_{p}\right)  ,\left(  g,\kappa_{g}\right)
\right)  ,T\lambda\left(  \left(  g^{-1},-R_{g^{-1}}^{\prime}(1)L_{g^{-1}%
}^{\prime}(g)\kappa_{g}\right)  ,\left(  u,v_{u}\right)  \right)  \right)  $

$=\left(  \left(  \rho\left(  p,g\right)  ,\rho^{\prime}\left(  p,g\right)
\left(  v_{p},\kappa_{g}\right)  \right)  ,\left(  \lambda\left(
g^{-1},u\right)  ,\lambda^{\prime}\left(  g^{-1},u\right)  \left(  -R_{g^{-1}%
}^{\prime}(1)L_{g^{-1}}^{\prime}(g)\kappa_{g},v_{u}\right)  \right)  \right)
$

with : $\kappa_{g}\in T_{g}G\rightarrow-R_{g^{-1}}^{\prime}(1)L_{g^{-1}%
}^{\prime}(g)\kappa_{g}=-L_{g^{-1}}^{\prime}(1)R_{g^{-1}}^{\prime}%
(g)\kappa_{g}\in T_{g^{-1}}G$

The equivalence relation is :

$\forall g\in G,\kappa_{g}\in T_{g}G:\left(  \left(  p,v_{p}\right)  ,\left(
u,v_{u}\right)  \right)  \sim T\Lambda\left(  \left(  g,\kappa_{g}\right)
,\left(  \left(  p,v_{p}\right)  ,\left(  u,v_{u}\right)  \right)  \right)  $

So by the general theorem TE is an associated bundle.
\end{proof}

TE is the quotient set $TE=T\left(  P\times V\right)  /\sim$ with the
equivalence relation given by the action and the projection :

$\Pr:TP\times TV\rightarrow TE::\Pr\left(  \left(  p,v_{p}\right)  ,\left(
u,v_{u}\right)  \right)  =\left[  \left(  p,v_{p}\right)  ,\left(
u,v_{u}\right)  \right]  $ is a submersion.

The equivalence relation reads in $TP\times TV:$

$\forall g\in G,\kappa_{g}\in T_{g}G:\left(  \left(  p,v_{p}\right)  ,\left(
u,v_{u}\right)  \right)  $

$\sim\left[  \left(  \rho\left(  p,g\right)  ,\rho^{\prime}\left(  p,g\right)
\left(  v_{p},\kappa_{g}\right)  \right)  ,\left(  \lambda\left(
g^{-1},u\right)  ,\lambda^{\prime}\left(  g^{-1},u\right)  \left(  -R_{g^{-1}%
}^{\prime}(1)L_{g^{-1}}^{\prime}(g)\kappa_{g},v_{u}\right)  \right)  \right]
$

Using the standard trivialization, we have :

\begin{theorem}
TE has the structure of a fiber bundle $TE\left(  TM,TV,\pi_{TE}\right)  $
with atlas $\left(  \cup_{x\in O_{a}}T_{x}M,\Psi_{a}\right)  _{a\in A}$
\end{theorem}

This is the implementation of general theorems on the tangent bundle of
$E\left(  M,V,\pi_{E}\right)  $ with the standard trivialization :

The trivializations are $\Psi_{a}:\pi^{\prime}\left(  \cup_{x\in O_{a}}%
T_{x}M\right)  ^{-1}\times TV\rightarrow TE::$

$\Psi_{a}\left(  \left(  x,v_{x}\right)  ,\left(  u_{a},v_{au}\right)
\right)  =\Pr\left(  \left(  \mathbf{p}_{a}\left(  x\right)  ,\varphi
_{ax}^{\prime}\left(  x,1\right)  v_{x}\right)  ,\left(  u_{a},v_{au}\right)
\right)  $

with the holonomic maps : $\partial x_{\alpha}=\psi_{ax}^{\prime}\left(
x,u_{a}\right)  \partial\xi_{\alpha},\partial u_{i}=\psi_{au}^{\prime}\left(
x,u_{a}\right)  \partial\eta_{i}$

so the vectors of TE read at $q=\psi_{a}\left(  x,u_{a}\right)  $

$W_{q}=\psi_{ax}^{\prime}\left(  x,u_{a}\right)  v_{x}+\psi_{au}^{\prime
}\left(  x,u_{a}\right)  v_{au}=\sum_{\alpha}v_{x}^{\alpha}\partial x_{\alpha
}+\sum_{i}v_{u}^{i}\partial u_{i}$

The projection $\pi_{TE}:TE\rightarrow TM::\pi_{TE}\left(  q\right)
W_{q}=v_{x}$ is a submersion

The transitions maps are : $\forall x\in O_{a}\cap O_{b}:u_{b}=\lambda\left(
g_{ba}\left(  x\right)  ,u_{a}\right)  =\psi_{ba}\left(  x,u_{a}\right)  $

$\psi_{ax}^{\prime}\left(  x,u_{a}\right)  =\psi_{bx}^{\prime}\left(
x,u_{b}\right)  +\psi_{bu}^{\prime}\left(  x,u_{b}\right)  \lambda_{g}%
^{\prime}\left(  g_{ba}\left(  x\right)  ,u_{a}\right)  g_{abx}^{\prime
}\left(  x\right)  $

$\psi_{au}^{\prime}\left(  x,u_{a}\right)  =\psi_{bu}^{\prime}\left(
x,u_{b}\right)  \lambda_{u}^{\prime}\left(  g_{ba}\left(  x\right)
,u_{a}\right)  $

$\partial x_{\alpha}=\partial\widetilde{x}_{\alpha}+\psi_{bu}^{\prime}\left(
x,u_{b}\right)  \lambda_{g}^{\prime}\left(  g_{ba}\left(  x\right)
,u_{a}\right)  g_{abx}^{\prime}\left(  x\right)  \partial\xi^{\alpha}$

$\partial u_{i}=\psi_{bu}^{\prime}\left(  x,u_{b}\right)  \lambda_{u}^{\prime
}\left(  g_{ba}\left(  x\right)  ,u_{a}\right)  \partial\eta_{i}$

$v_{bu}=\left(  \lambda\left(  g_{ba}\left(  x\right)  ,u_{a}\right)  \right)
^{\prime}\left(  v_{x},v_{au}\right)  $

\begin{theorem}
(Kolar p.99) The vertical bundle of the associated bundle $P\left[
V,\lambda\right]  $ is the associated bundle $P\left[  TV,\lambda_{u}^{\prime
}\right]  $ isomorphic to the G bundle $P\times_{G}TV$
\end{theorem}

The vertical vectors are : $\Psi_{a}\left(  \left(  x,0\right)  ,\left(
u_{a},v_{au}\right)  \right)  =\Pr\left(  \left(  \mathbf{p}_{a}\left(
x\right)  ,0\right)  ,\left(  u_{a},v_{au}\right)  \right)  $

$q=\psi_{a}\left(  x,u_{a}\right)  :W_{q}=\psi_{au}^{\prime}\left(
x,u_{a}\right)  v_{au}$

With the transitions maps for TE they transform as:

$W_{q}=\psi_{au}^{\prime}\left(  x,u_{a}\right)  v_{au}=\psi_{bu}^{\prime
}\left(  x,u_{a}\right)  v_{bu}=\psi_{bu}^{\prime}\left(  x,u_{b}\right)
\lambda_{u}^{\prime}\left(  g_{ba}\left(  x\right)  ,u_{a}\right)  v_{au}$

$v_{bu}=\lambda_{u}^{\prime}\left(  g_{ba}\left(  x\right)  ,u_{a}\right)
v_{au}$

\paragraph{Fundamental vector fields\newline}

For any associated bundle $E=P\left[  V,\lambda\right]  $ with $P\left(
M,G,\pi_{P}\right)  ,$\ to the action

$\Lambda:G\times\left(  P\times V\right)  \rightarrow P\times V::\Lambda
\left(  g,\left(  p,u\right)  \right)  =\left(  \rho\left(  p,g\right)
,\lambda\left(  g^{-1},u\right)  \right)  $

correspond the fundamental vector fields $Z\left(  X\right)  \in T\left(
P\times V\right)  $, generated for any $X\in T_{1}G$ by (cf.Lie groups) :

$Z\left(  X\right)  \left(  p,u\right)  =\Lambda_{g}^{\prime}\left(  1,\left(
p,u\right)  \right)  \left(  X\right)  =\left(  \rho_{g}^{\prime}\left(
p,1\right)  X,-\lambda_{g}^{\prime}\left(  1,u\right)  X\right)  $

They have the properties :

$\left[  Z\left(  X\right)  ,Z\left(  Y\right)  \right]  _{\mathfrak{X}\left(
TM\times TV\right)  }=Z\left(  \left[  X,Y\right]  _{T_{1}G}\right)  $

$\Lambda_{(p,u)}^{\prime}\left(  g,\left(  p,u\right)  \right)  |_{\left(
p,u\right)  =\left(  p_{0},u_{0}\right)  }Z\left(  X\right)  \left(
p_{0},u_{0}\right)  =Z\left(  Ad_{g}X\right)  \left(  \Lambda\left(  g,\left(
p_{0},u_{0}\right)  \right)  \right)  $

The fundamental vector fields span an integrable distribution over $P\times V
$, whose leaves are the connected components of the orbits. The orbits are the
elements of $E=P\left[  V,\lambda\right]  ,$ thus :

\begin{theorem}
The kernel of the projection $\Pr:TP\times TV\rightarrow TE$\ is spanned by
the fundamental vector fields of $P\times V$
\end{theorem}

\begin{proof}
$Z\left(  X\right)  \left(  p,u\right)  $ reads in $TP\times TV:$

$Z\left(  X\right)  \left(  p,u\right)  =\left(  \left(  p,\zeta\left(
X\right)  \left(  p\right)  \right)  ,\left(  u,-\lambda_{g}^{\prime}\left(
1,u\right)  X\right)  \right)  $

which is equivalent, $\forall h\in G,\kappa_{g}\in T_{h}G$ to :

$\left\{  \left(  \rho\left(  p,h\right)  ,\rho^{\prime}\left(  p,h\right)
\left(  \zeta\left(  X\right)  \left(  p\right)  ,\kappa_{g}\right)  \right)
,\left(  \lambda\left(  h^{-1},u\right)  ,\lambda^{\prime}\left(
h^{-1},u\right)  \left(  -R_{h^{-1}}^{\prime}(1)L_{h^{-1}}^{\prime}%
(h)\kappa_{g},-\lambda_{g}^{\prime}\left(  1,u\right)  X\right)  \right)
\right\}  $

If $p=\rho\left(  \mathbf{p}_{a}\left(  x\right)  ,g\right)  $ take $h=g^{-1}$

$\rho_{p}^{\prime}\left(  p,g^{-1}\right)  \zeta\left(  X\right)  \left(
p\right)  =\zeta\left(  Ad_{g}X\right)  \left(  \rho\left(  p,g^{-1}\right)
\right)  =\zeta\left(  Ad_{g}X\right)  \left(  \mathbf{p}_{a}\right)  $

$\rho_{g}^{\prime}\left(  p,g^{-1}\right)  \kappa_{g}=\rho_{g}^{\prime}%
(\rho(p,g^{-1}),1)L_{g}^{\prime}(g^{-1})\kappa_{g}=\rho_{g}^{\prime
}(\mathbf{p}_{a},1)L_{g}^{\prime}(g^{-1})\kappa_{g}=\zeta\left(  L_{g}%
^{\prime}(g^{-1})\kappa_{g}\right)  \left(  \mathbf{p}_{a}\right)  $

$\rho^{\prime}\left(  p,g^{-1}\right)  \left(  \zeta\left(  X\right)  \left(
p\right)  ,\kappa_{g}\right)  =\zeta\left(  Ad_{g}X+L_{g}^{\prime}%
(g^{-1})\kappa_{g}\right)  \left(  \mathbf{p}_{a}\right)  $

Take : $L_{g}^{\prime}(g^{-1})\kappa_{g}=-Ad_{g}X\Leftrightarrow\kappa
_{g}=-R_{g^{-1}}^{\prime}\left(  1\right)  X$

$-R_{g}^{\prime}(1)L_{g}^{\prime}(g^{-1})\kappa_{g}=R_{g}^{\prime}%
(1)Ad_{g}X=R_{g}^{\prime}(1)R_{g^{-1}}^{\prime}\left(  g\right)  L_{g}%
^{\prime}\left(  1\right)  X=L_{g}^{\prime}\left(  1\right)  X$

$\lambda_{u}^{\prime}\left(  g^{-1},u\right)  \lambda_{g}^{\prime}\left(
1,u\right)  X=\lambda_{g}^{\prime}(g,u)L_{g}^{\prime}\left(  1\right)  X$

$Z\left(  X\right)  \left(  p,u\right)  \sim\left\{  \left(  \mathbf{p}%
_{a}\left(  x\right)  ,0\right)  ,\left(  \lambda\left(  g,u\right)
,0\right)  \right\}  $

and : $\Pr\left\{  \left(  \mathbf{p}_{a}\left(  x\right)  ,0\right)  ,\left(
\lambda\left(  g,u\right)  ,0\right)  \right\}  =0$
\end{proof}

\subsubsection{Associated vector bundles}

\paragraph{Definition\newline}

\begin{definition}
An \textbf{associated vector bundle} is an associated bundle $P\left[
V,r\right]  $ where (V,r) is a continuous representation of G on a Banach
vector space V on the same field as $P\left(  M,G,\pi\right)  $
\end{definition}

So we have the equivalence relation on P$\times$V :%

\begin{equation}
\left(  p,u\right)  \sim\left(  \rho\left(  p,g\right)  ,r\left(
g^{-1}\right)  u\right)
\end{equation}

\paragraph{Vector bundle structure\newline}

\begin{theorem}
An associated vector bundle $P\left[  V,r\right]  $ on a principal bundle P
with atlas $\left(  O_{a},\varphi_{a}\right)  _{a\in A}$ \ is a vector bundle
$E\left(  M,V,\pi_{E}\right)  $ with standard trivialization $\left(
O_{a},\psi_{a}\right)  _{a\in A}$, $\psi_{a}\left(  x,u_{a}\right)
=\Pr\left(  \varphi_{a}\left(  x,1\right)  ,u_{a}\right)  _{\sim}$and
transition maps : $\psi_{ba}\left(  x\right)  =r\left(  g_{ba}\left(
x\right)  \right)  $ where $g_{ba}\left(  x\right)  $ are the transition maps
on P :%

\begin{equation}
\psi_{a}\left(  x,u_{a}\right)  =\psi_{b}\left(  x,u_{b}\right)
\Leftrightarrow u_{b}=r\left(  g_{ba}\left(  x\right)  \right)  u_{a}%
\end{equation}

\end{theorem}

The "p" in the couple (p,u) represents the frame, and u acts as the
components. So the vector space operations are completed in E fiberwise and in
the same frame :

Let $U_{1}=\left(  p_{1},u_{1}\right)  ,U_{2}=\left(  p_{2},u_{2}\right)
,\pi\left(  p_{1}\right)  =\pi\left(  p_{2}\right)  $

There is some g such that: $p_{2}=\rho\left(  p_{1},g\right)  $ thus :
$U_{2}\sim\left(  \rho\left(  p_{1},g^{-1}\right)  ,r\left(  g\right)
u_{2}\right)  =\left(  p_{1},r\left(  g\right)  u_{2}\right)  $

and : $\forall k,k^{\prime}\in K:kU_{1}+k^{\prime}U_{2}=\left(  p_{1}%
,ku_{1}+k^{\prime}r(g)u_{2}\right)  $

\paragraph{Associated vector bundles of linear frames\newline}

\begin{theorem}
Any vector bundle $E\left(  M,V,\pi\right)  $ has the structure of an
associated bundle $P\left[  V,r\right]  $ where $P\left(  M,G,\pi_{P}\right)
$ is the bundle of its linear frames and r the natural action of the group G
on V.
\end{theorem}

If M is a m dimensional manifold on a field K, its principal bundle\ of linear
frames $GL\left(  M,GL\left(  K,m\right)  ,\pi\right)  $ gives, with the
standard representation $\left(  K^{m},\imath\right)  $ of $GL\left(
K,m\right)  $ the associated vector bundle $GL\left[  K^{m},\imath\right]
$\ .This is the usual definition of the tangent bundle, where one can use any
basis of the tangent space $T_{x}M.$

Similarly if M is endowed with a metric g of signature (r,s)\ its principal
bundle $O\left(  M,O\left(  K,r,s\right)  ,\pi\right)  $\ of orthonormal
frames gives, with the standard unitary representation $\left(  K^{m}%
,\jmath\right)  $ of $O\left(  K,r,s\right)  $ the associated vector bundle
$O\left[  K^{m},\jmath\right]  .$

Conversely if we have a principal bundle $P\left(  M,O\left(  K,r,s\right)
,\pi\right)  ,$ then with the standard unitary representation $\left(
K^{m},\jmath\right)  $ of $O\left(  K,r,s\right)  $ the associated vector
bundle $E=P\left[  K^{n},\jmath\right]  $ is endowed with a scalar product
corresponding to O(K,r,s).

A section $U\in\mathfrak{X}\left(  P\left[  K^{m},\jmath\right]  \right)  $ is
a vector field whose components are defined in orthonomal frames.

\paragraph{Holonomic basis\newline}

\begin{definition}
A \textbf{holonomic basis }$\mathbf{e}_{ai}\left(  x\right)  =\left(
\mathbf{p}_{a}\left(  x\right)  ,e_{i}\right)  $ on an associated vector
bundle $P\left[  V,r\right]  $ is defined by a gauge $\mathbf{p}_{a}\left(
x\right)  =\varphi_{a}\left(  x,1\right)  $\ of P and a basis $\left(
e_{i}\right)  _{i\in I}$\ of V. At the transitions : $\mathbf{e}_{bi}\left(
x\right)  =r\left(  g_{ab}\left(  x\right)  \right)  \mathbf{e}_{ai}\left(
x\right)  $
\end{definition}

This is not a section.%

\begin{equation}
\mathbf{e}_{bi}\left(  x\right)  =\left(  \mathbf{p}_{b}\left(  x\right)
,e_{i}\right)  =\left(  \rho\left(  \mathbf{p}_{a}\left(  x\right)
,g_{ab}\left(  x\right)  \right)  ,e_{i}\right)  )\sim\left(  \mathbf{p}%
_{a}\left(  x\right)  ,r\left(  g_{ab}\left(  x\right)  \right)  e_{i}\right)
\end{equation}

\begin{theorem}
A section U on an associated vector bundle $P\left[  V,r\right]  $ is defined
by a family of maps : $u_{a}:O_{a}\rightarrow K$ such that :

$U\left(  x\right)  =\psi_{a}\left(  x,u_{a}\left(  x\right)  \right)  $

$u_{b}\left(  x\right)  =r\left(  g_{ba}\left(  x\right)  \right)
u_{a}\left(  x\right)  $
\end{theorem}

$U\left(  x\right)  =\sum_{i}u_{a}^{i}\left(  x\right)  \mathbf{e}_{ai}\left(
x\right)  )=\sum_{j}u_{b}^{j}\left(  x\right)  \mathbf{e}_{bj}\left(
x\right)  )=\sum u_{b}^{j}\left(  x\right)  \left[  r\left(  g_{ab}\left(
x\right)  \right)  \right]  _{j}^{i}\mathbf{e}_{ai}\left(  x\right)
)\Leftrightarrow u_{b}^{i}\left(  x\right)  =\sum_{ij}\left[  r\left(
g_{ba}\left(  x\right)  \right)  \right]  _{j}^{i}u_{a}^{j}\left(  x\right)  $

\paragraph{Change of gauge on an associated vector bundle\newline}

\begin{theorem}
A change of trivialization on the principal bundle P with atlas $\left(
O_{a},\varphi_{a}\right)  _{a\in A}$ defined by a family $\left(  \chi
_{a}\right)  _{a\in A}$\ of maps $\chi_{a}\in C\left(  O_{a};G\right)
$\ induces on any associated vector bundle $E=P\left[  V,r\right]  $ the
change of standard trivialization : $q=\psi_{a}\left(  x,u_{a}\right)
=\widetilde{\psi}_{a}\left(  x,r\left(  \chi_{a}\left(  x\right)  \right)
u_{a}\right)  .$ The holonomic basis changes as :%

\begin{equation}
\mathbf{e}_{ai}\left(  x\right)  =\psi_{a}\left(  x,e_{i}\right)
\rightarrow\widetilde{\mathbf{e}}_{ai}\left(  x\right)  =\widetilde{\psi}%
_{a}\left(  x,e_{i}\right)  =r\left(  \chi_{a}\left(  x\right)  ^{-1}\right)
\mathbf{e}_{ai}\left(  x\right)
\end{equation}

\end{theorem}

The components of a vector on the associated bundle changes as :%

\begin{equation}
\widetilde{u}_{a}^{i}=\sum_{j\in I}\left[  r\left(  \chi_{a}\left(  x\right)
\right)  \right]  _{j}^{i}u_{a}^{j}%
\end{equation}

\begin{theorem}
A one parameter group of changes of trivialization on a principal bundle
$P\left(  M,G,\pi\right)  $, defined by the section $\kappa\in\mathfrak{X}%
\left(  P\left[  T_{1}G,Ad\right]  \right)  $ of the adjoint bundle of P,
defines a one parameter group of changes of trivialization on any bundle
$P\left[  V,r\right]  $ associated to P. An holonomic basis $\mathbf{e}%
_{ai}\left(  x\right)  $ of $P\left[  V,r\right]  $ changes as :

$\mathbf{e}_{ai}\left(  x\right)  =\left(  \mathbf{p}_{a}\left(  x\right)
,e_{i}\right)  \rightarrow\widetilde{\mathbf{e}}_{ai}\left(  x,t\right)
=\exp(-tr^{\prime}(1)\kappa_{a}\left(  x\right)  )\mathbf{e}_{ai}\left(
x\right)  $
\end{theorem}

This is the application of the theorems on General associated bundles with
$r\left(  \exp t\kappa_{a}\left(  x\right)  \right)  =\exp tr^{\prime
}(1)\kappa_{a}\left(  x\right)  $.

\paragraph{Tensorial bundle\newline}

Because E has a vector bundle structure we can import all the tensorial
bundles defined with V : $\otimes_{s}^{r}V,V^{\ast},\wedge_{r}V,...$The change
of bases can be implemented fiberwise, the rules are the usual using the
matrices of r(g) in V. They are not related to any holonomic map on M : both
structures are fully independant. Everything happens as if we had a casual
vector space, copy of V, located at some point x in M. The advantage of the
structure of associated vector bundle over that of simple vector bundle is
that we have at our disposal the mean to define any frame p at any point x
through the principal bundle structure. So the picture is fully consistent.

\paragraph{Complexification of an associated vector bundle\newline}

\begin{theorem}
The complexified of a real associated vector bundle $P\left[  V,r\right]  $ is
$P\left[  V_{%
\mathbb{C}
},r_{%
\mathbb{C}
}\right]  $
\end{theorem}

with

$V_{%
\mathbb{C}
}=V\oplus iV$

$r_{%
\mathbb{C}
}:G\rightarrow V_{%
\mathbb{C}
}::r_{%
\mathbb{C}
}\left(  g\right)  \left(  u+iv\right)  =r\left(  g\right)  u+ir\left(
g\right)  v$ so $r_{%
\mathbb{C}
}$ is complex linear

A holonomic basis of $P\left[  V,r\right]  $ is a holonomic basis of $P\left[
V_{%
\mathbb{C}
},r_{%
\mathbb{C}
}\right]  $ with complex components.

Notice that the group stays the same and the principal bundle is still P.

\paragraph{Scalar product\newline}

\begin{theorem}
On a complex vector bundle $E=P\left[  V,r\right]  $ associated to $P\left(
M,G,\pi\right)  $ with an atlas $\left(  O_{a},\varphi_{a}\right)  _{a\in A}%
$\ and transition maps $g_{ab}$\ of P a scalar product is defined by a family
$\gamma_{a}$\ of hermitian sesquilinear maps on V, such that, for a holonomic
basis at the transitions : $\gamma_{a}\left(  x\right)  \left(  e_{i}%
,e_{j}\right)  =\sum_{kl}\overline{\left[  r\left(  g_{ba}\left(  x\right)
\right)  \right]  }_{i}^{k}\left[  r\left(  g_{ba}\left(  x\right)  \right)
\right]  _{j}^{l}\gamma_{b}\left(  x\right)  \left(  e_{k},e_{l}\right)  $
\end{theorem}

\begin{proof}
$\widehat{\gamma}_{a}\left(  x\right)  \left(  e_{ia}\left(  x\right)
,e_{ja}\left(  x\right)  \right)  =\gamma_{a}\left(  x\right)  \left(
e_{i},e_{j}\right)  =\gamma_{aij}\left(  x\right)  $

For two sections U,V on E the scalar product reads : $\widehat{\gamma}\left(
x\right)  \left(  U\left(  x\right)  ,V\left(  x\right)  \right)  =\sum
_{ij}\overline{u_{a}^{i}\left(  x\right)  }v_{a}^{j}\left(  x\right)
\gamma_{aij}\left(  x\right)  $

At the transitions : $\forall x\in O_{a}\cap O_{b}:$

$\gamma\left(  x\right)  \left(  U\left(  x\right)  ,V\left(  x\right)
\right)  =\sum_{ij}\overline{u_{a}^{i}\left(  x\right)  }v_{a}^{j}\left(
x\right)  \gamma_{aij}\left(  x\right)  =\sum_{ij}\overline{u_{b}^{i}\left(
x\right)  }v_{b}^{j}\left(  x\right)  \gamma_{bij}\left(  x\right)  $

with $u_{b}^{i}\left(  x\right)  =\sum_{j}\left[  r\left(  g_{ba}\left(
x\right)  \right)  \right]  _{j}^{i}u_{a}^{j}\left(  x\right)  $

$\gamma_{aij}\left(  x\right)  =\sum_{kl}\overline{\left[  r\left(
g_{ba}\left(  x\right)  \right)  \right]  }_{i}^{k}\left[  r\left(
g_{ba}\left(  x\right)  \right)  \right]  _{j}^{l}\gamma_{bkl}\left(
x\right)  $
\end{proof}

Conversely if $\gamma$\ is a hermitian sesquilinear maps on V it induces a
scalar product on E iff : $\left[  \gamma\right]  =\left[  r\left(
g_{ab}\left(  x\right)  \right)  \right]  ^{\ast}\left[  \gamma\right]
\left[  r\left(  g_{ab}\left(  x\right)  \right)  \right]  $ that is iff (V,r)
is a unitary representation of G.

\begin{theorem}
For any vector space V endowed with a scalar product $\gamma$, and any unitary
representation (V,r) of a group G, there is a vector bundle $E=P\left[
V,r\right]  $ associated to any principal bundle P$\left(  M,G,\pi\right)  ,$
and E is endowed with a scalar product induced by $\gamma.$
\end{theorem}

\paragraph{Tangent bundle\newline}

By applications of the general theorems

\begin{theorem}
The tangent bundle of the associated bundle $P\left[  V,r\right]  $ is the G
bundle $TP\times_{G}TV$

The vertical bundle is isomorphic to $P\times_{G}TV$
\end{theorem}

\paragraph{Ajoint bundle of a principal bundle\newline}

\begin{definition}
The \textbf{adjoint bundle} of a principal bundle $P\left(  M,G,\pi\right)  $
is the associated vector bundle $P\left[  T_{1}G,Ad\right]  .$
\end{definition}

With the gauge $\mathbf{p}_{a}\left(  x\right)  =\varphi_{a}\left(
x,1\right)  $ on P and a basis $\mathbf{\varepsilon}_{i}$ of $T_{1}G.$: the
holonomic basis of $P\left[  T_{1}G,Ad\right]  $ is $\mathbf{\varepsilon}%
_{ai}\left(  x\right)  =\left(  \mathbf{p}_{a}\left(  x\right)  ,\varepsilon
_{i}\right)  $ and a section U of $P\left[  T_{1}G,Ad\right]  $ is defined by
a collection of maps : $\kappa_{a}:O_{a}\rightarrow T_{1}G$ such that :
$\kappa_{b}\left(  x\right)  =Ad_{g_{ba}\left(  x\right)  }\left(  \kappa
_{a}\left(  x\right)  \right)  $ and $\kappa\left(  x\right)  =\sum_{i\in
I}u_{a}^{i}\left(  x\right)  \mathbf{\varepsilon}_{ai}\left(  x\right)  $

\begin{theorem}
The tangent bundle of $P\left[  T_{1}G,Ad\right]  $ is the associated bundle
$TP\left[  T_{1}G\times T_{1}G,TAd\right]  $ with group TG
\end{theorem}

A vector $U_{q}$\ at $p=\left(  p,u\right)  $ reads $w_{q}=\sum_{i}v_{g}%
^{i}\mathbf{\varepsilon}_{ai}\left(  x\right)  +\sum_{\alpha}v_{x}^{\alpha
}\partial x_{\alpha}$

\begin{theorem}
The vertical bundle of the adjoint bundle is $P\times_{G}\left(  T_{1}G\times
T_{1}G\right)  $
\end{theorem}

\subsubsection{Homogeneous spaces}

Homogeneous spaces are the quotient sets of a Lie group G by a one of its
subgroup H (see Lie groups).\ They have the structure of a manifold (but not
of a Lie group except if H is normal).

\paragraph{Principal fiber structure\newline}

If H is a subgroup of the group G :

The quotient set $G/H$ is the set $G/\sim$ of classes of equivalence $:$

$x\sim y\Leftrightarrow\exists h\in H:x=y\cdot h$

The quotient set $H\backslash G$ is the set $G/\sim$ of classes of equivalence
$:$

$x\sim y\Leftrightarrow\exists h\in H:x=h\cdot y$

They are groups iff H\ is normal : gH = Hg

\begin{theorem}
For any closed subgroup H of the Lie group G, G has the structure of a
principal fiber bundle $G\left(  G/H,H,\pi_{L}\right)  $ (resp.$G\left(
H\backslash G,H,\pi_{R}\right)  $
\end{theorem}

\begin{proof}
The homogeneous space G/H (resp. H%
$\backslash$%
G) is a manifold

The projection $\pi_{L}:G\rightarrow G/H$ (resp $\pi_{R}:G\rightarrow
H\backslash G)$ is a submersion.

On any open cover $\left(  O_{a}\right)  _{a\in A}$ of G/H (resp. H%
$\backslash$%
G), by choosing one element of G in each class, we can define the smooth maps :

$\lambda_{a}:O_{a}\rightarrow\pi_{L}^{-1}\left(  O_{a}\right)  :\lambda
_{a}\left(  x\right)  \in G$

$\rho_{a}:O_{a}\rightarrow\pi_{R}^{-1}\left(  O_{a}\right)  :\rho_{a}\left(
x\right)  \in G$

the trivializations are :

$\varphi_{a}:O_{a}\times H\rightarrow\pi_{L}^{-1}\left(  O_{a}\right)
::g=\varphi_{a}\left(  x,h\right)  =\lambda_{a}\left(  x\right)  h$

$\varphi_{a}:H\times O_{a}\rightarrow\pi_{R}^{-1}\left(  O_{a}\right)
::g=\varphi_{a}\left(  x,h\right)  =h\rho_{a}\left(  x\right)  $

For $x\in O_{a}\cap O_{b}:\lambda_{a}\left(  x\right)  h_{a}=\lambda
_{b}\left(  x\right)  h_{b}\Leftrightarrow h_{b}=\lambda_{b}\left(  x\right)
^{-1}\lambda_{a}\left(  x\right)  h_{a}$

$\varphi_{ba}\left(  x\right)  =\lambda_{b}\left(  x\right)  ^{-1}\lambda
_{a}\left(  x\right)  =h_{b}h_{a}^{-1}\in H$
\end{proof}

The right actions of H on G are:

$\rho\left(  g,h^{\prime}\right)  =\varphi_{a}\left(  x,hh^{\prime}\right)
=\lambda_{a}\left(  x\right)  hh^{\prime}$

$\rho\left(  g,h^{\prime}\right)  =\varphi_{a}\left(  x,hh^{\prime}\right)
=hh^{\prime}\rho_{a}\left(  x\right)  $

The translation induces a smooth transitive right (left) action of G on H%
$\backslash$%
G (G/H) :

$\Lambda:G\times G/H\rightarrow G/H::\Lambda\left(  g,x\right)  =\pi
_{L}\left(  g\lambda_{a}\left(  x\right)  \right)  $

$P:H\backslash G\times G\rightarrow H\backslash G:P\left(  x,g\right)
=\pi_{R}\left(  \rho_{a}\left(  x\right)  g\right)  $

\begin{theorem}
(Giachetta p.174) If G is a real finite dimensional Lie group, H a maximal
compact subgroup, then the principal fiber bundle $G\left(  G/H,H,\pi
_{L}\right)  $ is trivial.
\end{theorem}

Examples :

$O(%
\mathbb{R}
,n)\left(  S^{n-1},O(%
\mathbb{R}
,n-1),\pi_{L}\right)  $

$SU(n)\left(  S^{2n-1},SU(n-1),\pi_{L}\right)  $

$U(n)\left(  S^{2n-1},U(n-1),\pi_{L}\right)  $

$Sp(%
\mathbb{R}
,n)\left(  S^{4n-1},Sp(%
\mathbb{R}
,n-1),\pi_{L}\right)  $

where $S^{n-1}\subset%
\mathbb{R}
^{n}$\ \ is the sphere

\paragraph{Tangent bundle\newline}

\begin{theorem}
(Kolar p.96) If H is a closed subgroup H of the Lie group G, the tangent
bundle T(G/H) has the structure of a G-bundle

$\left(  G\times_{H}\left(  T_{1}G/T_{1}H\right)  \right)  \left(
G/H,T_{1}G/T_{1}H,\pi\right)  $
\end{theorem}

Which implies the following :

$\forall v_{x}\in T_{x}G/H,\exists Y_{H}\in T_{1}H,Y_{G}\in T_{1}G,h\in
H:v_{x}=L_{\lambda_{a}\left(  x\right)  }^{\prime}\left(  1\right)
Ad_{h}\left(  Y_{G}-Y_{H}\right)  $

\begin{proof}
By differentiation of : $g=\lambda_{a}\left(  x\right)  h$

$v_{g}=R_{h}^{\prime}\left(  \lambda_{a}\left(  x\right)  \right)
\circ\lambda_{a}^{\prime}\left(  x\right)  v_{x}+L_{\lambda_{a}\left(
x\right)  }^{\prime}\left(  h\right)  v_{h}=L_{g}^{\prime}1Y_{G}$

$Y_{G}=L_{g^{-1}}^{\prime}\left(  g\right)  R_{h}^{\prime}\left(  \lambda
_{a}\left(  x\right)  \right)  \circ\lambda_{a}^{\prime}\left(  x\right)
v_{x}+L_{g^{-1}}^{\prime}\left(  g\right)  L_{\lambda_{a}\left(  x\right)
}^{\prime}\left(  h\right)  L_{h}^{\prime}1Y_{H}$

$L_{g^{-1}}^{\prime}\left(  g\right)  R_{h}^{\prime}\left(  \lambda_{a}\left(
x\right)  \right)  =L_{g^{-1}}^{\prime}\left(  g\right)  R_{\lambda_{a}\left(
x\right)  h}^{\prime}\left(  1\right)  R_{\lambda_{a}\left(  x\right)  ^{-1}%
}^{\prime}\left(  \lambda_{a}\left(  x\right)  \right)  $

$=L_{g^{-1}}^{\prime}\left(  g\right)  R_{g}^{\prime}\left(  1\right)
R_{\lambda_{a}\left(  x\right)  ^{-1}}^{\prime}\left(  \lambda_{a}\left(
x\right)  \right)  =Ad_{g^{-1}}\circ R_{\lambda_{a}\left(  x\right)  ^{-1}%
}^{\prime}\left(  \lambda_{a}\left(  x\right)  \right)  $

$L_{g^{-1}}^{\prime}\left(  g\right)  \left(  L_{\lambda_{a}\left(  x\right)
}^{\prime}\left(  h\right)  \right)  L_{h}^{\prime}1=L_{g^{-1}}^{\prime
}\left(  g\right)  \left(  L_{\lambda_{a}\left(  x\right)  h}^{\prime
}(1)L_{h^{-1}}^{\prime}(h)\right)  L_{h}^{\prime}1$

$=L_{g^{-1}}^{\prime}\left(  g\right)  L_{g}^{\prime}(1)=Id_{T_{1}H}$

$Y_{G}=Ad_{g^{-1}}\circ R_{g_{a}\left(  x\right)  ^{-1}}^{\prime}\left(
g_{a}\left(  x\right)  \right)  v_{x}+Y_{H}$

$v_{x}=R_{g_{a}\left(  x\right)  }^{\prime}\left(  1\right)  Ad_{g_{a}\left(
x\right)  h}\left(  Y_{G}-Y_{H}\right)  =L_{g_{a}\left(  x\right)  }^{\prime
}\left(  1\right)  Ad_{h}\left(  Y_{G}-Y_{H}\right)  $
\end{proof}

\subsubsection{Spin bundles}

\paragraph{Clifford bundles}

\begin{theorem}
On a real finite dimensional manifold M endowed with a bilinear symmetric form
g with signature (r,s) which has a bundle of orthornormal frames P, there is a
vector bundle Cl(TM), called a \textbf{Clifford bundle}, such that each fiber
is Clifford isomorphic to $Cl\left(
\mathbb{R}
,r,s\right)  $. And $O\left(
\mathbb{R}
,r,s\right)  $ has a left action on each fiber for which Cl(TM) is a G-bundle
: $P\times_{O\left(
\mathbb{R}
,r,s\right)  }Cl\left(
\mathbb{R}
,r,s\right)  $.
\end{theorem}

\begin{proof}
i) On each tangent space $\left(  T_{x}M,g(x)\right)  $ there is a structure
of Clifford algebra $Cl\left(  T_{x}M,g(x)\right)  $\ . All Clifford algebras
on vector spaces of same dimension, endowed with a bilinear symmetric form
with the same signature are Clifford isomorphic. So there are Clifford
isomorphisms : $T\left(  x\right)  :Cl\left(
\mathbb{R}
,r,s\right)  \rightarrow Cl\left(  T_{x}M,g\left(  x\right)  \right)  .$ These
isomorphisms are geometric and do not depend on a basis. However it is useful
to see how it works.

ii) Let be :

$\left(
\mathbb{R}
^{m},\gamma\right)  $ with basis $\left(  \varepsilon_{i}\right)  _{i=1}^{m}$
and bilinear symmetric form of signature (r,s)

$\left(
\mathbb{R}
^{m},\jmath\right)  $ the standard representation of $O\left(
\mathbb{R}
,r,s\right)  $

$O\left(  M,O\left(
\mathbb{R}
,r,s\right)  ,\pi\right)  $ the bundle of orthogonal frames of (M,g) with
atlas $\left(  O_{a},\varphi_{a}\right)  $ and transition maps $g_{ba}\left(
x\right)  $

$E=P\left[
\mathbb{R}
^{m},\jmath\right]  $ the associated vector bundle with holonomic basis :
$\mathbf{\varepsilon}_{ai}\left(  x\right)  =\left(  \varphi_{a}\left(
x,1\right)  ,\varepsilon_{i}\right)  $ endowed with the induced scalar product
g(x). E is just TM with orthogonal frames. So there is a structure of Clifford
algebra $Cl\left(  E\left(  x\right)  ,g\left(  x\right)  \right)  .$

On each domain $O_{a}$ the maps : $t_{a}\left(  x\right)  :%
\mathbb{R}
^{m}\rightarrow E\left(  x\right)  :t_{a}\left(  x\right)  \left(
\varepsilon_{i}\right)  =\mathbf{\varepsilon}_{ai}\left(  x\right)  $ preserve
the scalar product, using the product of vectors on both Cl$\left(
\mathbb{R}
^{m},\gamma\right)  $ and Cl$\left(  E\left(  x\right)  ,g\left(  x\right)
\right)  $ : $t_{a}\left(  x\right)  \left(  \varepsilon_{i}\cdot
\varepsilon_{j}\right)  =\mathbf{\varepsilon}_{ai}\left(  x\right)
\cdot\mathbf{\varepsilon}_{aj}\left(  x\right)  $ the map $t_{a}\left(
x\right)  $ can be extended to a map : $T_{a}\left(  x\right)  :Cl\left(
\mathbb{R}
^{m},\gamma\right)  \rightarrow Cl\left(  E\left(  x\right)  ,g\left(
x\right)  \right)  $ which is an isomorphism of Clifford algebra. The
trivializations are :

$\Phi_{a}\left(  x,w\right)  =T_{a}\left(  x\right)  \left(  w\right)  $

$T_{a}\left(  x\right)  $ is a linear map between the vector spaces $Cl\left(
%
\mathbb{R}
^{m},\gamma\right)  ,Cl\left(  E\left(  x\right)  ,g\left(  x\right)  \right)
$ which can be expressed in their bases.

The transitions are :

$x\in O_{a}\cap O_{b}:\Phi_{a}\left(  x,w_{a}\right)  =\Phi_{b}\left(
x,w_{b}\right)  \Leftrightarrow w_{b}=T_{b}\left(  x\right)  ^{-1}\circ
T_{a}\left(  x\right)  \left(  w_{a}\right)  $

so the transition maps are linear, and Cl(TM) is a vector bundle.

iii) The action of $O\left(
\mathbb{R}
,r,s\right)  $ on the Clifford algebra $Cl\left(
\mathbb{R}
,r,s\right)  $ is :

$\lambda:O\left(
\mathbb{R}
,r,s\right)  \times Cl\left(
\mathbb{R}
,r,s\right)  \rightarrow Cl\left(
\mathbb{R}
,r,s\right)  ::\lambda\left(  h,u\right)  =\alpha\left(  s\right)  \cdot
u\cdot s^{-1}$ where $s\in Pin\left(
\mathbb{R}
,r,s\right)  :\mathbf{Ad}_{w}=h$

This action is extended on Cl(TM) fiberwise :

$\Lambda:O\left(
\mathbb{R}
,r,s\right)  \times Cl(TM)\left(  x\right)  \rightarrow Cl(TM)\left(
x\right)  ::\Lambda\left(  h,W_{x}\right)  =\Phi_{a}\left(  x,\alpha\left(
s\right)  \right)  \cdot W_{x}\cdot\Phi_{a}\left(  x,s^{-1}\right)  $

For each element of $O\left(
\mathbb{R}
,r,s\right)  $ there are two elements $\pm w$ of $Pin\left(
\mathbb{R}
,r,s\right)  $ but they give the same result.

With this action $Cl\left(  M,Cl\left(
\mathbb{R}
,r,s\right)  ,\pi_{c}\right)  $ is a bundle : $P\times_{O\left(
\mathbb{R}
,r,s\right)  }Cl(%
\mathbb{R}
,r,s)$.
\end{proof}

Comments :

i) Cl(TM) can be seen as a vector bundle $Cl\left(  M,Cl\left(
\mathbb{R}
,r,s\right)  ,\pi_{c}\right)  $\ with standard fiber $Cl\left(
\mathbb{R}
,p,q\right)  ,$ or an associated vector bundle $P[Cl\left(
\mathbb{R}
,r,s\right)  ,\lambda],$ or as a G-bundle : $P\times_{O\left(
\mathbb{R}
,p,q\right)  }Cl\left(
\mathbb{R}
,r,s\right)  $. But it has additional properties, as we have all the
operations of Clifford algebras, notably the product of vectors, available fiberwise.

ii) This structure can always be built whenever we have a principal bundle
modelled over an orthogonal group. This is always possible for a riemannian
metric but there are topological obstruction for the existence of
pseudo-riemannian manifolds.

iii) With the same notations as above, if we take the restriction
$\widetilde{T}_{a}\left(  x\right)  $ of the isomorphisms $T_{a}\left(
x\right)  $ to the Pin group we have a group isomorphism : $\widetilde{T}%
_{a}\left(  x\right)  :Pin\left(
\mathbb{R}
,r,s\right)  \rightarrow Pin\left(  T_{x}M,g\left(  x\right)  \right)  $ and
the maps $\widetilde{T}_{b}\left(  x\right)  ^{-1}\circ\widetilde{T}%
_{a}\left(  x\right)  $ are group automorphims on $Pin\left(
\mathbb{R}
,r,s\right)  ,$ so we have the structure of a fiber bundle $Pin\left(
M,Pin\left(
\mathbb{R}
,r,s\right)  ,\pi_{p}\right)  .$ However there is no guarantee that this is a
principal fiber bundle, which requires $w_{b}=\left(  T_{b}\left(  x\right)
^{-1}\circ T_{a}\left(  x\right)  \right)  \left(  w_{a}\right)
=T_{ba}\left(  x\right)  \cdot w_{a}$ with $T_{ba}\left(  x\right)  \in
Pin\left(
\mathbb{R}
,r,s\right)  .$ Indeed an automorphism on a group is not necessarily a translation.

\paragraph{Spin structure\newline}

For the reasons above, it is useful to define a principal spin bundle with
respect to a principal bundle with an orthonormal group : this is a spin structure.

\begin{definition}
On a pseudo-riemannian manifold (M,g), with its principal bundle of orthogonal
frames O$\left(  M,O\left(
\mathbb{R}
,r,s\right)  ,\pi\right)  ,$ an atlas $\left(  O_{a},\varphi_{a}\right)  $ of
O, a \textbf{spin structure} is a family of maps $\left(  \chi_{a}\right)
_{a\in A}$ such that : $\chi_{a}\left(  x\right)  :Pin\left(
\mathbb{R}
,r,s\right)  \rightarrow O\left(
\mathbb{R}
,r,s\right)  $ is a group morphism.
\end{definition}

So there is a continuous map which selects, for each $g\in O\left(
\mathbb{R}
,r,s\right)  $ , one of the two elements of the Pin group $Pin\left(
\mathbb{R}
,r,s\right)  $.

\begin{theorem}
A spin structure defines a principal pin bundle Pin$\left(  M,O\left(
\mathbb{R}
,r,s\right)  ,\pi\right)  $
\end{theorem}

\begin{proof}
The trivializations are :

$\psi_{a}:O_{a}\times Pin\left(
\mathbb{R}
,r,s\right)  \rightarrow Pin\left(  M,O\left(
\mathbb{R}
,r,s\right)  ,\pi\right)  ::\psi_{a}\left(  x,s\right)  =\varphi_{a}\left(
x,\chi\left(  s\right)  \right)  $

At the transitions :

$\psi_{b}\left(  x,s_{b}\right)  =\psi_{a}\left(  x,s_{a}\right)  =\varphi
_{a}\left(  x,\chi\left(  s_{a}\right)  \right)  =\varphi_{b}\left(
x,\chi\left(  s_{b}\right)  \right)  $

$\Leftrightarrow\chi\left(  s_{b}\right)  =g_{ba}\left(  x\right)  \chi\left(
s_{a}\right)  =\chi\left(  s_{ba}\left(  x\right)  \right)  \chi\left(
s_{a}\right)  $

$s_{b}=s_{ba}\left(  x\right)  s_{a}$
\end{proof}

There are topological obstructions to the existence of spin structures on a
manifold (see Giachetta p.245).

If M is oriented we have a similar definition and result for a principal Spin bundle.

With a spin structure any associated bundle $P\left[  V,\rho\right]  $ can be
extended to an associated bundle $Sp\left[  V,\widetilde{\rho}\right]  $ with
the left action of Spin on V : $\widetilde{\rho}\left(  s,u\right)
=\rho\left(  \chi\left(  g\right)  ,u\right)  $

One can similarly build a principal manifold $Sp_{c}\left(  M,Spin_{c}\left(
\mathbb{R}
,p,q\right)  ,\pi_{s}\right)  $ with the group $Spin_{c}\left(
\mathbb{R}
,p,q\right)  $ and define complex spin structure, and complex associated bundle.

\paragraph{Spin bundle\newline}

\begin{definition}
A \textbf{spin bundle, }on a manifold M endowed with a Clifford bundle
structure Cl(TM) is a vector bundle $E\left(  M,V,\pi\right)  $\ with a map R
on M such that $\left(  E\left(  x\right)  ,R\left(  x\right)  \right)  $ are
geometric equivalent representations of $Cl(TM)\left(  x\right)  $
\end{definition}

A spin bundle is not a common vector bundle, or associated bundle. Such a
structure does not always exist and may be not unique. For a given
representation\ there are topological obstructions to their existence,
depending on M. \textbf{Spin manifolds} are manifolds such that there is a
spin bundle structure for any representation. The following theorem (which is
new) shows that any manifold with a structure of a principal bundle of Spin
group is a spin bundle. So a manifold with a spin structure is a spin manifold.

\begin{theorem}
If there is a principal bundle $Sp(M,Spin\left(
\mathbb{R}
,t,s\right)  ,\pi_{S})$ on the t+s=m dimensional real manifold M, then for any
representation (V,r) of the Clifford algebra $Cl\left(
\mathbb{R}
,t,s\right)  $ there is a spin bundle on M.
\end{theorem}

\begin{proof}
The ingredients are the following :

a principal bundle Sp(M,$Spin\left(
\mathbb{R}
,t,s\right)  $,$\pi_{S})$ with atlas $\left(  O_{a},\varphi_{a}\right)  _{a\in
A}$ transition maps $\varphi_{ba}\left(  x\right)  \in Spin\left(
\mathbb{R}
,t,s\right)  $ and right action $\rho$.

$\left(
\mathbb{R}
^{m},\gamma\right)  $ endowed with the symmetric bilinear form $\gamma$ of
signature (t,s) on $%
\mathbb{R}
^{m}$ and its basis $\left(  \varepsilon_{i}\right)  _{i=1}^{m}$

$\left(
\mathbb{R}
^{m},\mathbf{Ad}\right)  $ the representation of $Spin\left(
\mathbb{R}
,t,s\right)  $

(V,r) a representation of $Cl\left(
\mathbb{R}
,t,s\right)  $, with a basis $\left(  e_{i}\right)  _{i=1}^{n}$ of V

From which we have :

an associated vector bundle $F=Sp\left[
\mathbb{R}
^{m},\mathbf{Ad}\right]  $ with atlas $\left(  O_{a},\phi_{a}\right)  _{a\in
A} $ and holonomic basis : $\mathbf{\varepsilon}_{ai}\left(  x\right)
=\phi_{a}\left(  x,\varepsilon_{i}\right)  ,$ $\mathbf{\varepsilon}%
_{bi}\left(  x\right)  =Ad_{\varphi_{ab}\left(  x\right)  }\mathbf{\varepsilon
}_{ai}\left(  x\right)  .$ Because \textbf{Ad }preserves $\gamma$ the vector
bundle F can be endowed with a scalar product g.

an associated vector bundle E=Sp[V,r] with atlas $\left(  O_{a},\psi
_{a}\right)  _{a\in A}$ and holonomic basis : $\mathbf{e}_{ai}\left(
x\right)  =\psi_{a}\left(  x,e_{i}\right)  ,$ $\mathbf{e}_{bi}\left(
x\right)  =r\left(  \varphi_{ab}\left(  x\right)  \right)  \mathbf{e}%
_{ai}\left(  x\right)  $

$\psi_{a}:O_{a}\times V\rightarrow E::\psi_{a}\left(  x,u\right)  =\left(
\varphi_{a}\left(  x,1\right)  ,u\right)  $

For $x\in O_{a}\cap O_{b}:$

$U_{x}=\left(  \varphi_{a}\left(  x,s_{a}\right)  ,u_{a}\right)  \sim\left(
\varphi_{b}\left(  x,s_{b}\right)  ,u_{b}\right)  \Leftrightarrow
u_{b}=r\left(  s_{b}^{-1}\varphi_{ba}s_{a}\right)  u_{a}$

Each fiber (F(x),g(x)) has a Clifford algebra structure Cl(F(x),g(x))
isomorphic to Cl(TM)(x). There is a family $\left(  O_{a},T_{a}\right)  _{a\in
A}$\ of Clifford isomorphism : $T_{a}\left(  x\right)
:Cl(F(x),g(x))\rightarrow Cl\left(
\mathbb{R}
,t,s\right)  $ defined by identifying the bases : $T_{a}\left(  x\right)
\left(  \mathbf{\varepsilon}_{ai}\left(  x\right)  \right)  =\varepsilon_{i}$
and on $x\in O_{a}\cap O_{b}:T_{a}\left(  x\right)  \left(
\mathbf{\varepsilon}_{ai}\left(  x\right)  \right)  =T_{b}\left(  x\right)
\left(  \mathbf{\varepsilon}_{bi}\left(  x\right)  \right)  =\varepsilon
_{i}=Ad_{\varphi_{ab}\left(  x\right)  }T_{b}\left(  x\right)  \left(
\mathbf{\varepsilon}_{ai}\left(  x\right)  \right)  $

$\forall W_{x}\in Cl(F(x),g(x)):T_{b}\left(  x\right)  \left(  W_{x}\right)
=Ad_{\varphi_{ba}\left(  x\right)  }T_{a}\left(  x\right)  \left(
W_{x}\right)  $

The action $R(x)$ is defined by the family $\left(  O_{a},R_{a}\right)  _{a\in
A}$\ of maps:

$R\left(  x\right)  :Cl(F(x),g(x))\times E\left(  x\right)  \rightarrow
E\left(  x\right)  ::R_{a}\left(  x\right)  \left(  W_{x}\right)  \left(
\varphi_{a}\left(  x,s_{a}\right)  ,u_{a}\right)  =\left(  \varphi_{a}\left(
x,s_{a}\right)  ,r\left(  s_{a}^{-1}\cdot T_{a}\left(  x\right)  \left(
W_{x}\right)  \cdot s_{a}\right)  u_{a}\right)  $

The definition is consistent and does not depend on the trivialization:

For $x\in O_{a}\cap O_{b}:$

$R_{b}\left(  x\right)  \left(  W_{x}\right)  \left(  U_{x}\right)  =\left(
\varphi_{b}\left(  x,s_{b}\right)  ,r\left(  s_{b}^{-1}\cdot T_{b}\left(
x\right)  \left(  W_{x}\right)  \cdot s_{b}\right)  u_{b}\right)  $

$\sim\left(  \varphi_{a}\left(  x,s_{a}\right)  ,r\left(  s_{a}^{-1}%
\varphi_{ab}\left(  x\right)  s_{b}\right)  r\left(  s_{b}^{-1}\cdot
T_{b}\left(  x\right)  \left(  W_{x}\right)  \cdot s_{b}\right)  u_{b}\right)
$

$=\left(  \varphi_{a}\left(  x,s_{a}\right)  ,r\left(  s_{a}^{-1}\varphi
_{ab}\left(  x\right)  s_{b}\right)  r\left(  s_{b}^{-1}\cdot T_{b}\left(
x\right)  \left(  W_{x}\right)  \cdot s_{b}\right)  r\left(  s_{b}^{-1}%
\varphi_{ba}\left(  x\right)  s_{a}\right)  u_{a}\right)  $

$=\left(  \varphi_{a}\left(  x,s_{a}\right)  ,r\left(  s_{a}^{-1}\cdot
\varphi_{ab}\left(  x\right)  \cdot T_{b}\left(  x\right)  \left(
W_{x}\right)  \cdot\varphi_{ba}\left(  x\right)  \cdot s_{a}\right)
u_{a}\right)  $

$=\left(  \varphi_{a}\left(  x,s_{a}\right)  ,r\left(  s_{a}^{-1}\cdot
Ad_{\varphi_{ab}\left(  x\right)  }T_{b}\left(  x\right)  \left(
W_{x}\right)  \cdot s_{a}\right)  u_{a}\right)  =R_{a}\left(  x\right)
\left(  W_{x}\right)  \left(  U_{x}\right)  $
\end{proof}

\newpage

\section{JETS}

\label{Jets}

Physicists are used to say that two functions f(x),g(x) are "closed at the rth
order" in the neighborhood of a if $\left\vert f\left(  x\right)
-g(x)\right\vert <k\left\vert x-a\right\vert ^{r}$ which translates as the rth
derivatives are equal at a : $f^{k}\left(  a\right)  =g^{k}\left(  a\right)
,k\leq r.$ If we take differentiable maps between manifolds the equivalent
involves partial derivatives. This is the starting point of the jets
framework. Whereas in differential geometry one strives to get "intrinsic
formulations", without any reference to the coordinates, they play a central
role in jets. In this section the manifolds will be assumed real finite
dimensional and smooth.

\bigskip

\subsection{Jets on a manifold}

\subsubsection{Definition of a jet\newline}

\begin{definition}
(Kolar Chap.IV) Two paths $P_{1},P_{2}\in C_{\infty}\left(
\mathbb{R}
;M\right)  $ on a real finite dimensional smooth manifold M, going through
$p\in M$ at t=0, are said to have a \textbf{r-th order contact} at p if for
any smooth real function f on M the function $\left(  f\circ P_{1}-f\circ
P_{2}\right)  $ has derivatives equal to zero for all order $k\leq r$ at p :

$0\leq k\leq r:\left(  f\circ P_{1}-f\circ P_{2}\right)  \left(  0\right)
^{\left(  k\right)  }=0$
\end{definition}

\begin{definition}
Two maps $f,g\in C_{r}\left(  M;N\right)  $ between the real finite
dimensional, smooth manifolds M,N\ are said to have a r-th order contact at
$p\in M$ if for any smooth path P on M going through p then the paths $f\circ
P,g\circ P$ have a r-th order contact at p.

A \textbf{r-jet} at p is a class of equivalence in the relation "to have a
r-th order contact at p" for maps in $C_{r}\left(  M;N\right)  .$
\end{definition}

Two maps belonging to the same r-jet at p have same value f(p), and same
derivatives at all order up to r at p.

$f,g\in C_{r}\left(  M;N\right)  :f\sim g\Leftrightarrow0\leq s\leq
r:f^{\left(  s\right)  }\left(  p\right)  =g^{\left(  s\right)  }\left(
p\right)  $

The \textbf{source} is p and the \textbf{target} is f(p)=g(p)=q

\begin{notation}
$j_{p}^{r}$ is a r-jet at the source $p\in M$
\end{notation}

\begin{notation}
$j_{p}^{r}f$ is the class of equivalence of $f\in C_{r}\left(  M;N\right)  $
at p
\end{notation}

Any s order derivative of a map f is a s symmetric linear map $z_{s}\in%
\mathcal{L}%
_{S}^{s}\left(  T_{p}M;T_{q}N\right)  ,$ or equivalently a tensor $\odot
^{s}T_{p}M^{\ast}\otimes T_{q}N$ where $\odot$ denote the symmetric tensorial
product (see Algebra).

So a r jet at p in M denoted $j_{p}^{r}$ with target q can be identified with
a set :

$\left\{  q,z_{s}\in%
\mathcal{L}%
_{S}^{s}\left(  T_{p}M;T_{q}N\right)  ,s=1...r\right\}  \simeq\left\{
q,z_{s}\in%
\mathcal{L}%
_{S}^{s}\left(
\mathbb{R}
^{\dim M};%
\mathbb{R}
^{\dim N}\right)  ,s=1...r\right\}  $

and the set $J_{p}^{r}\left(  M,N\right)  _{q}$ of all r jets $j_{p}^{r}$ with
target q can be identified with the set $\left\{  \oplus_{s=1}^{r}%
\mathcal{L}%
_{S}^{s}\left(  T_{p}M;T_{q}N\right)  ,s=1...r\right\}  $

We have in particular : $J_{p}^{1}\left(  M,N\right)  _{q}=T_{p}M^{\ast
}\otimes T_{q}N.$ Indeed the 1 jet is just f'(p) and this is the set of linear
map from $T_{p}M$ to $T_{q}N.$

\begin{notation}
For two class r manifolds M,N on the same field :

$J_{p}^{r}\left(  M,N\right)  _{q}$ is the set of r order jets at $p\in M$
(the source)$,$ with value $q\in N$ (the target)

$J_{p}^{r}\left(  M,N\right)  _{q}=\left\{  \oplus_{s=1}^{r}%
\mathcal{L}%
_{S}^{s}\left(  T_{p}M;T_{q}N\right)  ,s=1...r\right\}  $

$J_{p}^{r}\left(  M,N\right)  =\cup_{q}J_{p}^{r}\left(  M,N\right)
_{q}=\left\{  p,\left\{  \oplus_{s=1}^{r}%
\mathcal{L}%
_{S}^{s}\left(  T_{p}M;T_{q}N\right)  ,s=1...r\right\}  \right\}  $

$J^{r}\left(  M,N\right)  _{q}=\cup_{p}J_{p}^{r}\left(  M,N\right)
_{q}=\left\{  q,\left\{  \oplus_{s=1}^{r}%
\mathcal{L}%
_{S}^{s}\left(  T_{p}M;T_{q}N\right)  ,s=1...r\right\}  \right\}  $

$J^{r}\left(  M,N\right)  =\left\{  p,q,\left\{  \oplus_{s=1}^{r}%
\mathcal{L}%
_{S}^{s}\left(  T_{p}M;T_{q}N\right)  ,s=1...r\right\}  \right\}  $

$J^{0}(M;N)=M\times N$ Conventionally

$L_{m,n}^{r}=J_{0}^{r}\left(
\mathbb{R}
^{m},%
\mathbb{R}
^{n}\right)  _{0}$
\end{notation}

\begin{definition}
the \textbf{r jet prolongation} of a map $f\in C_{r}\left(  M;N\right)  $ is
the map : $J^{r}f:M\rightarrow J^{r}\left(  M,N\right)  ::\left(
J^{r}f\right)  \left(  p\right)  =j_{p}^{r}f$
\end{definition}

A r-jet comprises several elements, the projections are :

\begin{notation}
$\pi_{s}^{r}:J^{r}\left(  M,N\right)  \rightarrow J^{s}\left(  M,N\right)
::\pi_{s}^{r}\left(  j^{r}\right)  =j^{s}$ where we drop the s+1...r terms

$\pi_{0}^{r}:J^{r}\left(  M,N\right)  \rightarrow M\times N::\pi_{0}%
^{r}\left(  j_{p}^{r}f\right)  =\left(  p,f\left(  p\right)  \right)  $

$\pi^{r}:J^{r}\left(  M,N\right)  \rightarrow M:\pi_{0}^{r}\left(  j_{p}%
^{r}f\right)  =p$
\end{notation}

\subsubsection{Structure of the space of r-jets}

Because a r-jet is a class of equivalence, we can take any map in the class to
represent the r-jet : we "forget" the map f to keep only the value of f(p) and
the value of the derivarives $f^{\left(  s\right)  }\left(  p\right)  .$ In
the following we follow Krupka (2000).

\paragraph{Coordinates expression\newline}

Any map $f\in C_{r}\left(  M;N\right)  $ between the manifolds M,N with atlas
$\left(
\mathbb{R}
^{m},\left(  O_{a},\varphi_{a}\right)  _{a\in A}\right)  ,$ $\left(
\mathbb{R}
^{n},\left(  Q_{b},\psi_{b}\right)  _{b\in B}\right)  $ is represented in
coordinates by the functions:

$F=\varphi_{a}\circ f\circ\psi_{b}^{-1}:%
\mathbb{R}
^{m}\rightarrow%
\mathbb{R}
^{n}::\zeta^{i}=F^{i}\left(  \xi^{1},...,\xi^{m}\right)  $

The partial derivatives of f expressed in the holonomic bases of the atlas are
the same as the partial derivatives of f, with respect to the coordinates. So :

\begin{theorem}
A r-jet $j_{p}^{r}\in J_{p}^{r}\left(  M,N\right)  _{q}$ ,in atlas of the
manifolds, is represented by a set of scalars :

$\left(  \zeta_{\alpha_{1}...\alpha_{s}}^{i},1\leq\alpha_{k}\leq
m,i=1..n,s=1..r\right)  $ with m=dimM, n=dim N where the $\zeta_{\alpha
_{1}...\alpha_{s}}^{i}$ are symmetric in the lower indices
\end{theorem}

These scalars are the components of the symmetric tensors :$\ $

$\left\{  z_{s}\in\odot^{s}T_{p}M^{\ast}\otimes T_{q}N,s=1...r\right\}  $ in
the holonomic bases.$\left(  dx^{\alpha}\right)  $ of $T_{p}M^{\ast},\left(
\partial y_{i}\right)  $ of $T_{q}N$

\paragraph{Structure with p,q fixed\newline}

The source and the target are the same for all maps in $J_{p}^{r}\left(
M,N\right)  _{q}$

$%
\mathcal{L}%
_{S}^{s}\left(  T_{p}M;T_{q}N\right)  \simeq%
\mathcal{L}%
_{S}^{s}\left(
\mathbb{R}
^{m};%
\mathbb{R}
^{n}\right)  $ with m=dimM, n=dim N

$J_{p}^{r}\left(  M,N\right)  _{q}=\left\{  \oplus_{s=1}^{r}%
\mathcal{L}%
_{S}^{s}\left(  T_{p}M;T_{q}N\right)  ,s=1...r\right\}  $

$\simeq L_{m,n}^{r}=J_{0}^{r}\left(
\mathbb{R}
^{m},%
\mathbb{R}
^{n}\right)  _{0}=\left\{  \oplus_{s=1}^{r}%
\mathcal{L}%
_{S}^{s}\left(
\mathbb{R}
^{m};%
\mathbb{R}
^{n}\right)  ,s=1...r\right\}  $

This is a vector space. A point $Z\in L_{m,n}^{r}$ has for coordinates :

$\left(  \zeta_{\alpha_{1}...\alpha_{s}}^{i},1\leq\alpha_{1}\leq\alpha
_{2}..\leq\alpha_{s}\leq m,i=1..n,s=1..r\right)  $ to account for the symmetries.

$J_{0}^{r}\left(
\mathbb{R}
^{m},%
\mathbb{R}
^{n}\right)  _{0}$ is a smooth real $J=n\left(  C_{m+r}^{m}-1\right)  $
dimensional manifold $L_{m,n}^{r}$ embedded in $%
\mathbb{R}
^{mn\frac{r\left(  r+1\right)  }{2}}$

\paragraph{Structure of $J^{r}\left(  M,N\right)  $\newline}

If we do not specify p and q : $J^{r}\left(  M,N\right)  =\left\{
p,q,J_{p}^{r}\left(  M,N\right)  _{q},p\in M,q\in N\right\}  $.

$J^{r}\left(  M,N\right)  $ has the structure of a smooth J+m+n=$nC_{m+r}%
^{m}+m $ real manifold with charts $\left(  O_{a}\times Q_{b}\times
L_{m,n}^{r},\left(  \varphi_{a}^{\left(  s\right)  },\psi_{b}^{\left(
s\right)  },s=0..r\right)  \right)  $ and coordinates :

$\left(  \xi^{\alpha},\zeta^{i},\zeta_{\alpha_{1}...\alpha_{s}}^{i}%
,1\leq\alpha_{1}\leq\alpha_{2}..\leq\alpha_{s}\leq m,i=1..n,s=1..r\right)  $

The projections $\pi_{s}^{r}:J^{r}\left(  M,N\right)  \rightarrow J^{s}\left(
M,N\right)  $\ are smooth submersions and translate in the chart by dropping
the terms s
$>$
r

For each p the set $J_{p}^{r}\left(  M,N\right)  $ is a submanifold of
$J^{r}\left(  M,N\right)  $

For each q the set $J^{r}\left(  M,N\right)  _{q}$ is a submanifold of
$J^{r}\left(  M,N\right)  $

$J^{r}\left(  M,N\right)  $ is a smooth fibered manifold $J^{r}\left(
M,N\right)  \left(  M\times N,\pi_{0}^{r}\right)  $

$J^{r}\left(  M,N\right)  $ is an affine bundle over $J^{r-1}\left(
M,N\right)  $ with the projection $\pi_{r-1}^{r}$

(Krupka p.66 ).$J^{r}\left(  M,N\right)  $ is an associated fiber bundle
$GT_{m}^{r}\left(  M\right)  \left[  T_{n}^{r}\left(  N\right)  ,\lambda
\right]  $ - see below

\paragraph{Associated polynomial map\newline}

\begin{theorem}
To any r-jet $j_{p}^{r}\in J_{p}^{r}\left(  M,N\right)  _{q}$ is associated a
symmetric polynomial with m real variables, of order r
\end{theorem}

This is the polynomial :

i=1..n : $P^{i}\left(  \xi^{1},...,\xi^{m}\right)  =\zeta_{0}^{i}+\sum
_{s=1}^{r}\sum_{\alpha_{1}...\alpha_{s}=1}^{m}\zeta_{\alpha_{1}...\alpha_{s}%
}^{i}\left(  \xi^{\alpha_{1}}-\xi_{0}^{\alpha_{1}}\right)  ...\left(
\xi^{\alpha_{s}}-\xi_{0}^{\alpha_{s}}\right)  $

with : $\left(  \xi_{0}^{1},...,\xi_{0}^{m}\right)  =\varphi_{a}\left(
p\right)  ,\left(  \zeta_{0}^{1},...,\zeta_{0}^{n}\right)  =\psi_{b}\left(
q\right)  $

At the transitions between open subsets of M,N we have local maps :

$\varphi_{aa^{\prime}}\left(  p\right)  \in%
\mathcal{L}%
\left(
\mathbb{R}
^{m};%
\mathbb{R}
^{m}\right)  ,\psi_{bb^{\prime}}\left(  q\right)  \in%
\mathcal{L}%
\left(
\mathbb{R}
^{n};%
\mathbb{R}
^{n}\right)  $

$\left(  \zeta-\zeta_{0}\right)  =\psi_{bb^{\prime}}\left(  q\right)  \left(
\zeta^{\prime}-\zeta_{0}^{\prime}\right)  $

$\left(  \xi-\xi_{0}\right)  =\varphi_{aa^{\prime}}\left(  p\right)  \left(
\xi^{\prime}-\xi_{0}^{\prime}\right)  $

$\left(  \zeta^{\prime}-\zeta_{0}^{\prime}\right)  $

$=\sum_{s=1}^{r}\sum_{\alpha_{1}...\alpha_{s}=1}^{m}P_{\alpha_{1}...\alpha
_{s}}^{i}\sum_{\beta_{1}..\beta_{s}=1}^{m}\left[  \varphi_{aa^{\prime}}\left(
p\right)  \right]  _{\beta_{1}}^{\alpha_{1}}\left(  \xi^{\prime\beta_{1}}%
-\xi_{0}^{\prime\beta_{1}}\right)  ...\left[  \varphi_{aa^{\prime}}\left(
p\right)  \right]  _{\beta_{s}}^{\alpha_{s}}\left(  \xi^{\prime\alpha_{s}}%
-\xi_{0}^{\prime\alpha_{s}}\right)  $

$P_{\alpha_{1}...\alpha_{s}}^{\prime i}=\left[  \psi_{bb^{\prime}}\left(
q\right)  ^{-1}\right]  _{j}^{i}P_{\alpha_{1}...\alpha_{s}}^{j}\sum_{\beta
_{1}..\beta_{s}=1}^{m}\left[  \varphi_{aa^{\prime}}\left(  p\right)  \right]
_{\beta_{1}}^{\alpha_{1}}...\left[  \varphi_{aa^{\prime}}\left(  p\right)
\right]  _{\beta_{s}}^{\alpha_{s}}$

They transform according to the rules for tensors in $\odot^{s}T_{p}M^{\ast
}\otimes T_{q}N$

It is good to keep in mind this representation of a r jet as a polynomial :
they possess all the information related to a r-jet and so can be used to
answer some subtle questions about r-jet, which are sometimes very abstract objects.

\subsubsection{Jets groups}

The composition of maps and the chain rule give the possibility to define the
product of r-jets, and then invertible elements and a group structure.

\paragraph{Composition of jets :\newline}

\begin{definition}
The composition of two r jets is given by the rule :

$\circ:J^{r}\left(  M;N\right)  \times J^{r}\left(  N;P\right)  \rightarrow
J^{r}\left(  M;P\right)  ::j_{x}^{r}\left(  g\circ f\right)  =\left(
j_{f(x)}^{r}g\right)  \circ\left(  j_{x}^{r}f\right)  $
\end{definition}

The definition makes sense because :

For $f_{1},f_{2}\in C_{r}\left(  M;N\right)  ,g_{1},g_{2}\in C_{r}\left(
N;P\right)  $,

$j_{x}^{r}f_{1}=j_{x}^{r}f_{2},j_{y}^{r}g_{1}=j_{y}^{r}g_{2},y=f_{1}%
(x)=f_{2}\left(  x\right)  \Rightarrow j_{x}^{r}\left(  g_{1}\circ
f_{1}\right)  =j_{x}^{r}\left(  g_{2}\circ f_{2}\right)  $

\begin{theorem}
The composition of r jets is associative and smooth
\end{theorem}

In coordinates the map $L_{m,n}^{r}\times L_{n,p}^{r}\rightarrow L_{m,p}^{r}$
can be obtained by the product of the polynomials $P^{i},Q^{i}$ and discarding
all terms of degree
$>$
r.

$\left(  \sum_{s=1}^{r}\sum_{\alpha_{1},...,\alpha_{s}}a_{\alpha_{1}%
...\alpha_{s}}^{i}t_{\alpha_{1}}...t_{\alpha_{s}}\right)  \times\left(
\sum_{s=1}^{r}\sum_{\alpha_{1},...,\alpha_{s}}b_{\alpha_{1}...\alpha_{s}}%
^{i}t_{\alpha_{1}}...t_{\alpha_{s}}\right)  $

$=\left(  \sum_{s=1}^{2r}\sum_{\alpha_{1},...,\alpha_{s}}c_{\alpha
_{1}...\alpha_{s}}^{i}t_{\alpha_{1}}...t_{\alpha_{s}}\right)  $

$c_{\alpha_{1}...\alpha_{s}}^{j}=\sum_{k=1}^{s}\sum a_{\beta_{1}...\beta_{k}%
}^{j}b_{I_{1}}^{\beta_{1}}...b_{I_{k}}^{\beta_{k}}$ where $\left(  I_{1}%
,I_{2},..I_{k}\right)  =$ any partition of $\left(  \alpha_{1},...\alpha
_{s}\right)  $

For r= 2 : $c_{\alpha}^{i}=a_{\beta}^{i}b_{\beta}^{\beta};c_{\alpha\beta}%
^{i}=a_{\lambda\mu}^{i}b_{\alpha}^{\lambda}b_{\beta}^{\mu}+a_{\gamma}%
^{i}b_{\alpha\beta}^{\gamma}$ and the coefficients b for the inverse are given
by : $\delta_{\alpha}^{i}=a_{\beta}^{i}b_{\beta}^{\beta};a_{\alpha\beta}%
^{i}=-b_{\lambda\mu}^{j}a_{\alpha}^{\lambda}a_{\beta}^{\mu}a_{j}^{i}$

(see Krupka for more values)

\paragraph{Invertible r jets :\newline}

\begin{definition}
If dimM=dimN=n \ a r-jet $X\in J_{p}^{r}\left(  M;N\right)  _{q}$ is said
\textbf{invertible} (for the composition law)\ if :

$\exists Y\in J_{q}^{r}\left(  N;M\right)  _{p}:X\circ Y=Id_{M},Y\circ
X=Id_{N} $
\end{definition}

and then it will be denoted $X^{-1}.$

X is inversible iff $\pi_{1}^{r}X$ is invertible.

The set of invertible elements of $J_{p}^{r}\left(  M,N\right)  _{q}$ is
denoted $GJ_{p}^{r}\left(  M,N\right)  _{q}$

\begin{definition}
The \textbf{rth differential group} (or r jet group) is the set, denoted
$GL^{r}(%
\mathbb{R}
,n),$\ of invertible elements of $L_{n,n}^{r}=J_{0}^{r}(%
\mathbb{R}
^{n},%
\mathbb{R}
^{n})_{0}$
\end{definition}

\begin{theorem}
(Kolar p.129) The set $GL^{r}(%
\mathbb{R}
,n)$ is a Lie group, with Lie algebra $L^{r}(%
\mathbb{R}
,n)$ given by the vector space $\left\{  j_{0}^{r}X,X\in C_{r}\left(
\mathbb{R}
^{n};%
\mathbb{R}
^{n}\right)  ,X\left(  0\right)  =0\right\}  $ and bracket : $\left[
j_{0}^{r}X,j_{0}^{r}Y\right]  =-j_{0}^{r}\left(  \left[  X,Y\right]  \right)
.$

The exponential mapping is : $\exp j_{0}^{r}X=j_{0}^{r}\Phi_{X}$
\end{theorem}

For r=1 we have $GL^{1}(%
\mathbb{R}
,n)=GL(%
\mathbb{R}
,n)$.

The canonical coordinates of $G\in GL^{r}(%
\mathbb{R}
,n)$ are $\left\{  g_{\alpha_{1}...\alpha_{s}}^{i},i,\alpha_{j}%
=1...n,s=1...r\right\}  $

The group $GL^{r}(%
\mathbb{R}
,n)$ has many special properties (see Kolar IV.13) which can be extended to representations.

\paragraph{Velocities\newline}

\begin{definition}
The space of \textbf{k velocities} to a manifold M is the set

$T_{k}^{r}(M)=J_{0}^{r}(%
\mathbb{R}
^{k},M)$
\end{definition}

$T_{k}^{r}(M)=\left\{  q,z_{s}\in\odot^{s}%
\mathbb{R}
^{k\ast}\otimes TM,s=1...r\right\}  $

$=\left\{  q,z_{\alpha_{1}...\alpha_{s}}^{i}e^{\alpha_{1}}\otimes..\otimes
e^{\alpha_{k}}\otimes\partial x_{i},\alpha_{j}=1...k,s=1...r,i=1..m\right\}  $

where the $z_{\alpha_{1}...\alpha_{s}}^{i}$ are symmetric in the lower
indices. Notice that it includes the target in M.

For k=1,r=1 we have just the tangent bundle $T_{1}^{1}(M)=TM=\left\{
q,z^{i}\partial x_{i}\right\}  $

$GL^{r}(%
\mathbb{R}
,k)$ acts on the right on $T_{k}^{r}(M):$

$\rho:T_{k}^{r}(M)\times GL^{r}(%
\mathbb{R}
,k)\rightarrow T_{k}^{r}(M)::\rho\left(  j_{0}^{r}f,j_{0}^{r}\varphi\right)
=j_{0}^{r}\left(  f\circ\varphi\right)  $ \ 

with $\varphi\in Diff_{r}\left(
\mathbb{R}
^{k};%
\mathbb{R}
^{k}\right)  $

\begin{theorem}
(Kolar p.120) $T_{k}^{r}(M)$ is :

a smooth $mC_{k+r}^{k}$ dimensional manifold,

a smooth fibered manifold $T_{k}^{r}(M)\rightarrow M$

a smooth fiber bundle $T_{k}^{r}(M)\left(  M,L_{km}^{r},\pi^{r}\right)  $

an associated fiber bundle $GT_{m}^{r}(M)\left[  L_{mn}^{r},\lambda\right]  $
with $\lambda=$ the left action of $GL^{r}(%
\mathbb{R}
,m)$ on $L_{mn}^{r}$
\end{theorem}

\begin{theorem}
If G is a Lie group then $T_{k}^{r}(G)$ is a Lie group with multiplication :
$\left(  j_{0}^{r}f\right)  \cdot\left(  j_{0}^{r}g\right)  =j_{0}^{r}\left(
f\cdot g\right)  ,f,g\in C_{r}\left(
\mathbb{R}
^{k};G\right)  $
\end{theorem}

\begin{definition}
(Kolar p.122) The \textbf{bundle of r-frames} over a m dimensional manifold M
is the set $GT_{m}^{r}(M)$ = the jets of r differentiable frames on M, or
equivalently the set of all invertible jets of $T_{m}^{r}(M)$. This is a
principal fiber bundle over M $:GT_{m}^{r}(M)\left(  M,GL^{r}(%
\mathbb{R}
,m),\pi^{r}\right)  $
\end{definition}

For r=1 we have the usual linear frame bundle.

For any local r diffeormorphism $f:M\rightarrow N$ (with dimM=dimN=m) the map
$GT_{m}^{r}f:GT_{m}^{r}\left(  M\right)  \rightarrow GT_{m}^{r}\left(
N\right)  ::GT_{m}^{r}f\left(  j_{0}^{r}\varphi\right)  =j_{0}^{r}\left(
f\circ\varphi\right)  $ is a morphism of principal bundles.

\paragraph{Covelocities\newline}

\begin{definition}
The space of \textbf{k covelocities} from a manifold M is the set :

$T_{k}^{r\ast}(M)=J^{r}(M,%
\mathbb{R}
^{k})_{0}$
\end{definition}

$T_{k}^{r\ast}(M)=\left\{  p,z_{\alpha_{1}...\alpha_{s}}^{i}dx^{\alpha_{1}%
}\otimes...\otimes dx^{\alpha_{s}}\otimes e_{i},s=1...r,\alpha_{j}%
=1...m,i=1..k\right\}  $ where the $z_{\alpha_{1}...\alpha_{s}}^{i}$ are
symmetric in the lower indices.

$T_{k}^{r\ast}(M)$ is a vector bundle :

for $f,g\in C_{r}\left(
\mathbb{R}
^{k};M\right)  :j_{0}^{r}\left(  \lambda f+\mu g\right)  =\lambda j_{0}%
^{r}f+\mu j_{0}^{r}g$

For k=r=1 we have the cotangent bundle : $T_{1}^{1\ast}(M)=TM^{\ast}$

The projection : $\pi_{r-1}^{r}:T_{k}^{r\ast}(M)\rightarrow T_{k}^{r-1\ast
}(M)$ is a linear morphism of vector bundles

If k=1 : $T_{1}^{r\ast}(M)$ is an algebra with multiplication of maps :

$j_{0}^{r}\left(  f\times g\right)  =\left(  j_{0}^{r}f\right)  \times\left(
j_{0}^{r}g\right)  $

There is a canonical bijection between $J_{p}^{r}\left(  M,N\right)  _{q}$ and
the set of algebras morphisms $\hom\left(  J_{p}^{r}(M,%
\mathbb{R}
)_{0},J_{q}^{r}(N,%
\mathbb{R}
)_{0}\right)  $

\bigskip

\subsection{Jets on fiber bundles}

\subsubsection{r-jet prolongation of fiber bundles}

\paragraph{Fibered manifold\newline}

\begin{definition}
The \textbf{r-jet prolongation} of a fibered manifold E, denoted $J^{r}E,$ is
the space of r-jets of sections over E.
\end{definition}

A section on a fibered manifold E is a local map S from an open O of M to E
which has the additional feature that $\pi\left(  S\left(  x\right)  \right)
=x.$ So the space of r-jets over E is a subset of $J^{r}\left(  M,E\right)  $.
More precisely :

\begin{theorem}
(Kolar p.124) The r jet prolongation $J^{r}E$ of a fibered manifold $E\left(
M,\pi\right)  $ is a closed submanifold of $J^{r}\left(  M,E\right)  $ and a
fibered submanifold of $J^{r}\left(  M,E\right)  \left(  M\times E,\pi_{0}%
^{r}\right)  $
\end{theorem}

The dimension of $J^{r}E$ is $nC_{m+r}^{m}+m$ with dim(E)=m+n

Warning ! $J\left(  J^{r-1}E\right)  \neq J^{r}E$ and similalry an element of
$J^{r}E$ is not the simple r derivative of $\varphi\left(  x,u\right)  $

\begin{notation}
The projections are defined as :

$\pi^{r}:J^{r}E\rightarrow M:\pi^{r}\left(  j_{x}^{r}S\right)  =x$

$\pi_{0}^{r}:J^{r}E\rightarrow E:\pi_{0}^{r}\left(  j_{x}^{r}S\right)
=S\left(  x\right)  $

$\pi_{s}^{r}:J^{r}E\rightarrow J^{s}E:\pi_{s}^{r}\left(  j_{x}^{r}S\right)
=j_{x}^{s}$
\end{notation}

\paragraph{General fiber bundles\newline}

\begin{theorem}
(Krupka 2000 p.75) The r jet prolongation $\ J^{r}E$\ of the smooth fiber
bundle $E(M,V,\pi)$ is a fiber bundle $J^{r}E\left(  M,J_{0}^{r}\left(
\mathbb{R}
^{\dim M},V\right)  ,\pi^{r}\right)  $ and a vector bundle $J^{r}E\left(
E,J_{0}^{r}\left(
\mathbb{R}
^{\dim M},V\right)  _{0},\pi_{0}^{r}\right)  $
\end{theorem}

Let $\left(  O_{a},\varphi_{a}\right)  _{a\in A}$ be an atlas of E, $\left(
\mathbb{R}
^{m},\left(  O_{a},\psi_{a}\right)  _{a\in A}\right)  $ be an atlas of M,
$\left(
\mathbb{R}
^{n},\left(  U_{i},\phi_{i}\right)  _{i\in I}\right)  $ be an atlas of the
manifold V.

Define : $\tau_{a}:\pi^{-1}\left(  O_{a}\right)  \rightarrow V::\varphi
_{a}\left(  x,\tau_{a}\left(  p\right)  \right)  =p=\varphi_{a}\left(
\pi\left(  p\right)  ,\tau_{a}\left(  p\right)  \right)  $

The map :

$\Phi_{ai}:\varphi_{a}\left(  O_{a},V_{i}\right)  \rightarrow%
\mathbb{R}
^{m}\times%
\mathbb{R}
^{n}::\Phi_{ai}\left(  p\right)  =\left(  \psi_{a}\circ\pi\left(  p\right)
,\phi_{i}\tau_{a}\left(  p\right)  \right)  =\left(  \xi_{a},\eta_{a}\right)
$

is bijective and differentiable. Then $\left(
\mathbb{R}
^{m}\times%
\mathbb{R}
^{n},\left(  \varphi_{a}\left(  O_{a},V_{i}\right)  ,\Phi_{ai}\right)
_{\left(  a,i\right)  \in A\times I}\right)  $\ is an atlas of E as manifold.

A section $S\in\mathfrak{X}_{r}\left(  E\right)  $ on E\ is defined by a
family $\left(  \sigma_{a}\right)  _{a\in A}$ of maps $\sigma_{a}\in
C_{r}\left(  O_{a};V\right)  $ such that : $S\left(  x\right)  =\varphi
_{a}\left(  x,\sigma_{a}\left(  x\right)  \right)  $ . The definition of
$J^{r}E $ is purely local, so the transition conditions are not involved. Two
sections are in the same r-jet\ at x if the derivatives $\sigma_{a}^{\left(
s\right)  }\left(  x\right)  ,0\leq s\leq r$ have same value.

$\sigma_{a}^{\left(  s\right)  }\left(  x\right)  $ is a s symmetric linear
map $z_{s}\in%
\mathcal{L}%
_{S}^{s}\left(  T_{x}M;T_{\tau_{a}\left(  S\left(  x\right)  \right)
}V\right)  ,$ which is represented in the canonical bases of $%
\mathbb{R}
^{m}\times%
\mathbb{R}
^{n}$ by a set of scalars: $\sigma_{a}^{\left(  s\right)  }\left(  x\right)
=\left\{  \eta_{\alpha_{1}..\alpha_{s}}^{i},i=1..n,s=0...r,\alpha
_{k}=1...m\right\}  $ which are symmetrical in the indices $\alpha
_{1},...\alpha_{s}$ with $\eta_{\alpha_{1}..\alpha_{s}}^{i}=\frac{\partial
\phi_{i}\left(  \sigma_{a}\right)  }{\partial\xi^{\alpha_{1}}...\partial
\xi^{\alpha_{s}}}|_{x}$ , $\eta^{i}=\phi_{i}\left(  \sigma_{a}\right)  $\ and
conversely any r-jet at x is identified by the same set.

So the map :

$\Phi_{a}^{r}:\left(  \pi^{r}\right)  ^{-1}\left(  O_{a}\right)  \rightarrow%
\mathbb{R}
^{m}\times%
\mathbb{R}
^{n}\times J_{0}^{r}\left(
\mathbb{R}
^{m},%
\mathbb{R}
^{n}\right)  _{0}::$

$\Phi^{r}_{a}\left(  j^{r}Z\right)  =\left(  \xi^{\alpha},\eta^{i}%
,\eta_{\alpha_{1}..\alpha_{s}}^{i},s=1...r,1\leq\alpha_{k}\leq\alpha_{k+1}\leq
m,i=1..n\right)  $

is a chart of $J^{r}E$ \textit{as manifold} and

$\left(  \xi^{\alpha},\eta^{i},\eta_{\alpha_{1}..\alpha_{s}}^{i}%
,s=1...r,1\leq\alpha_{k}\leq\alpha_{k+1}\leq m,i=1..n\right)  $ are the
coordinates of $j^{r}Z$

So the map :

$\psi_{ai}:O_{a}\times J_{0}^{r}\left(
\mathbb{R}
^{\dim M},V\right)  \rightarrow J^{r}E::$

$\psi_{ai}\left(  x,\sigma,\eta_{\alpha_{1}..\alpha_{s}}^{i}%
,i=1..n,s=1...r,\alpha_{k}=1...m\right)  $

$=\left\{  p=\varphi_{a}\left(  x,\sigma\right)  ,\sigma_{a}^{\left(
s\right)  }=\sum\eta_{\alpha_{1}..\alpha_{s}}^{i}d\xi^{\alpha_{1}}%
\otimes..d\xi^{\alpha_{s}}\otimes\partial u_{i},s=0...r\right\}  $

is a trivialization of $J^{r}E\left(  M,J_{0}^{r}\left(
\mathbb{R}
^{\dim M},V\right)  ,\pi^{r}\right)  $

So the map :

$\widehat{\psi}_{ai}:\pi^{-1}\left(  O_{a}\right)  \times J_{0}^{r}\left(
\mathbb{R}
^{\dim M},V\right)  _{0}\rightarrow J^{r}E::$

$\psi_{ai}\left(  p,\eta_{\alpha_{1}..\alpha_{s}}^{i},i=1..n,s=1...r,\alpha
_{k}=1...m\right)  $

$=\left\{  p,\sigma_{a}^{\left(  s\right)  }=\sum\eta_{\alpha_{1}..\alpha_{s}%
}^{i}d\xi^{\alpha_{1}}\otimes..d\xi^{\alpha_{s}}\otimes\partial u_{i}%
,s=0...r\right\}  $

is a trivialization of $J^{r}E\left(  E,J_{0}^{r}\left(
\mathbb{R}
^{\dim M},V\right)  _{0},\pi^{r}\right)  $

Remark : as we can see the bases on TE are not involved : only the bases
$d\xi^{\alpha}\in%
\mathbb{R}
^{m\ast}$,$\partial u_{i}\in TV$ show. E appears only in the first component
(by p). However in the different cases of vector bundle, principal bundle,
associated bundle, additional structure appears, but they are complicated and
the vector bundle structure $J^{r}E\left(  E,J_{0}^{r}\left(
\mathbb{R}
^{\dim M},V\right)  _{0},\pi_{0}^{r}\right)  $ is usually sufficient.

\begin{theorem}
The fiber bundle defined by the projection :

$\pi_{r-1}^{r}:J^{r}E\rightarrow J^{r-1}E$\ \ 

is an affine bundle modelled on the vector bundle $J^{r-1}E\left(  E,J_{0}%
^{r}\left(
\mathbb{R}
^{\dim M},V\right)  _{0},\pi_{0}^{r}\right)  $
\end{theorem}

$J^{1}E$ is an affine bundle over E, modelled on the vector bundle

$TM^{\ast}\otimes VE\rightarrow E$

\paragraph{Prolongation of a vector bundle\newline}

\begin{theorem}
The r jet prolongation $J^{r}E$\ of a vector bundle $E\left(  M,V,\pi\right)
$ is the vector bundle $J^{r}E\left(  M,J_{0}^{r}\left(
\mathbb{R}
^{\dim M},V\right)  ,\pi^{r}\right)  $
\end{theorem}

In an atlas $\left(  O_{a},\varphi_{a}\right)  $\ of E, a basis $\left(
e_{i}\right)  _{i=1}^{n}$\ of V, a vector of $J^{r}E$ reads :%

\begin{equation}
Z\left(  x\right)  =Z_{0}^{i}\left(  x\right)  \mathbf{e}_{i}\left(  x\right)
+\sum_{s=1}^{r}\sum_{\alpha_{1}..\alpha_{s}=1}^{m}\sum_{i=1}^{n}Z_{\alpha
_{1}...\alpha_{s}}^{i}\mathbf{e}_{i}^{\alpha_{1}...\alpha_{s}}\left(
x\right)
\end{equation}

where $Z_{\alpha_{1}...\alpha_{s}}^{i}$ are symmetric in the subscripts

$\mathbf{e}_{i}^{\alpha_{1}...\alpha_{s}}\left(  x\right)  =j_{x}^{s}%
\varphi\left(  x,e_{i}\right)  $ is the r-derivative of $\varphi\left(
x,e_{i}\right)  $ with respect to x

In a change of basis in V $\mathbf{e}_{i}^{\alpha_{1}...\alpha_{s}}\in
\odot_{s}TM^{\ast}\otimes V$ changes as a vector of V.

\paragraph{Prolongation of a principal bundle\newline}

\begin{theorem}
(Kolar p.150) If P$\left(  M,G,\pi\right)  $ is a principal fiber bundle with
M m dimensional, then its r jet prolongation is the principal bundle
$W^{r}P(M,W_{m}^{r}G,\pi^{r})$ with $W^{r}P=GT_{m}^{r}(M)\times_{M}%
J^{r}P,W_{m}^{r}G=GL^{r}(%
\mathbb{R}
,m)\rtimes T_{m}^{r}\left(  G\right)  $
\end{theorem}

The trivialization : $S\left(  x\right)  =\varphi\left(  x,\sigma\left(
x\right)  \right)  $ is r-differentiated with respect to x. So the ingredients
are derivatives $\frac{\partial\sigma}{\partial\xi^{\alpha_{1}}..\partial
\xi^{\alpha_{s}}}\in T_{m}^{r}\left(  G\right)  $ , the set $W_{m}^{r}G$ is a
group with the product :

$\left(  A_{1},\theta_{1}\right)  ,\left(  A_{2},\theta_{2}\right)  \in
GL^{r}(%
\mathbb{R}
,m)\times T_{m}^{r}\left(  G\right)  :\left(  A_{1},\theta_{1}\right)
\times\left(  A_{2},\theta_{2}\right)  =\left(  A_{1}\times A_{2},\left(
\theta_{1}\times A_{2}\right)  \cdot\theta_{2}\right)  $

\paragraph{Prolongation of an associated bundle\newline}

\begin{theorem}
(Kolar p.152) If P$\left(  M,G,\pi\right)  $ is a principal fiber bundle with
G n dimensional, $P\left[  V,\lambda\right]  $ an associated bundle, then the
r-jet prolongation of $P\left[  V,\lambda\right]  $ is $W^{r}P\left[
T_{n}^{r}V,\Lambda\right]  $ with the equivalence :

$j_{x}^{r}\left(  p,u\right)  \sim j_{x}^{r}\left(  \rho\left(  p,g\right)
,\lambda\left(  g^{-1},u\right)  \right)  $
\end{theorem}

The action :

$\Lambda:G\times\left(  P\times V\right)  \rightarrow P\times V::\Lambda
\left(  g\right)  \left(  \mathbf{p}_{a}\left(  x\right)  ,u\left(  x\right)
\right)  =\left(  \rho\left(  \mathbf{p}_{a}\left(  x\right)  ,g\left(
x\right)  \right)  ,\lambda\left(  g\left(  x\right)  ^{-1},u\left(  x\right)
\right)  \right)  $

is r-differentiated with respect to x, g and u are maps depending on x. So the
ingredients are derivatives $\frac{\partial g}{\partial\xi^{\alpha_{1}%
}..\partial\xi^{\alpha_{s}}},\frac{\partial u}{\partial\xi^{\alpha_{1}%
}..\partial\xi^{\alpha_{s}}}$

\paragraph{Sections of $J^{r}E$\newline}

\begin{notation}
$\mathfrak{X}\left(  J^{r}E\right)  $ is the set of sections of the
prolongation $J^{r}E$\ of the fibered manifold (or fiber bundle) E
\end{notation}

\begin{theorem}
A class r section $S\in\mathfrak{X}_{r}\left(  E\right)  $ induces a\ section
called its \textbf{r jet prolongation}\ denoted $J^{r}S\in\mathfrak{X}\left(
J^{r}E\right)  $ by : $J^{r}S\left(  x\right)  =j_{x}^{r}S$
\end{theorem}

So we have two cases :

i) On $J^{r}E$ a section $Z\in\mathfrak{X}\left(  J^{r}E\right)  $ is defined
by a set of coordinates :%

\begin{equation}
Z\left(  x\right)  =\left(  \xi^{\alpha}\left(  x\right)  ,\eta^{i}\left(
x\right)  ,\eta_{\alpha_{1}..\alpha_{s}}^{i}\left(  x\right)  ,s=1...r,1\leq
\alpha_{k}\leq\alpha_{k+1}\leq m,i=1..n\right)  \in\mathfrak{X}\left(
J^{r}E\right)
\end{equation}

which depend on $x\in M.$ But $\eta_{\beta\alpha_{1}..\alpha_{s}}^{i}\left(
x\right)  $ is \textit{not} the derivative $\frac{\partial}{\partial\xi
^{\beta}}\eta_{\beta\alpha_{1}..\alpha_{s}}^{i}\left(  x\right)  .$

ii) Any section $S\in\mathfrak{X}_{r}\left(  E\right)  $ gives by derivation a
section $J^{r}S\in\mathfrak{X}\left(  J^{r}E\right)  $ represented by
coordinates as above, depending on x, and

$\eta_{\alpha_{1}..\alpha_{s}}^{i}\left(  x\right)  =\frac{\partial\sigma^{i}%
}{\partial\xi^{\alpha_{1}}...\partial\xi^{\alpha_{s}}}|_{x}.$

In a change a chart $\left(  \xi^{\alpha},\eta^{i}\right)  \rightarrow\left(
\widetilde{\xi}^{\alpha},\widetilde{\eta}^{i}\right)  $ in the manifold E, the
coordinates of the prolongation $J^{r}S\in\mathfrak{X}\left(  J^{r}E\right)  $
of a section of E change as (Giachetta p.46):

$\widetilde{\eta}_{\beta\alpha_{1}...\alpha_{s}}^{i}=\sum_{\gamma}%
\frac{\partial\xi^{\gamma}}{\partial\widetilde{\xi}^{\beta}}d_{\gamma}%
\eta_{\alpha_{1}...\alpha_{s}}^{i}$ where $d_{\gamma}=\frac{\partial}%
{\partial\xi^{\gamma}}+\sum_{s=1}^{r}\sum_{\beta_{1\leq}...\leq\beta_{s}}%
\eta_{\gamma\beta_{1}...\beta_{s}}^{i}\frac{\partial}{\partial y_{\beta
_{1}...\beta_{s}}^{i}}$ is the total differential (see below)

The map : $J^{r}:\mathfrak{X}_{r}\left(  E\right)  \rightarrow\mathfrak{X}%
\left(  J^{r}E\right)  $ is neither injective or surjective. So the image of M
by $J^{r}S$ is a subset of $J^{r}E.$

If E is a single manifold M then $J^{r}M\equiv M.$ So the r jet prolongation
$J^{r}S$ of a section S is a morphism of fibered manifolds : $M\rightarrow
J^{r}M$

\paragraph{Tangent space to the prolongation of a fiber bundle\newline}

$J^{r}E$ is a manifold, with coordinates in trivializations and charts (same
notation as above) :

$\left(  \xi^{\alpha},\eta^{i},\eta_{\alpha_{1}..\alpha_{s}}^{i}%
,s=1...r,1\leq\alpha_{k}\leq\alpha_{k+1}\leq m,i=1..n\right)  $

A vector on the tangent space $T_{Z}J^{r}E$ of $J^{r}E$ reads :%

\begin{equation}
W_{Z}=\sum_{\alpha=1}^{m}w^{\alpha}\partial\xi_{\alpha}+\sum_{s=0}^{r}%
\sum_{1\leq\alpha_{1}\leq..\leq\alpha_{s}\leq m}w_{\alpha_{1}...\alpha
_{\alpha_{s}}}^{i}\partial\eta_{i}^{\alpha_{1}...\alpha_{\alpha_{s}}}%
\end{equation}

with the honolomic basis :

$\left(  \partial\xi_{\alpha},\partial\eta_{i},\partial\eta_{i}^{\alpha
_{1}...\alpha_{\alpha_{s}}},s=1..r,i=1...n,1\leq\alpha_{1}\leq\alpha_{2}%
..\leq\alpha_{m}\leq m\right)  $

We need to take an ordered set of indices in order to account for the symmetry
of the coordinates.

A vector field on $TJ^{r}E$ is a section $W\in\mathfrak{X}\left(
TJ^{r}E\right)  $ and the components $\left\{  w^{\alpha},w_{\alpha
_{1}...\alpha_{\alpha_{s}}}^{i}\right\}  $ \textit{depend on Z }(and not only x).

We have the dual basis of $TJ^{r}E^{\ast}$

$\left(  dx^{\alpha},du^{i},du_{\alpha_{1}...\alpha_{\alpha_{s}}}%
^{i},s=1..r,i=1...n,1\leq\alpha_{1}\leq\alpha_{2}..\leq\alpha_{m}\leq
m\right)  $

\subsubsection{Prolongation of morphisms}

\paragraph{Definition\newline}

\begin{definition}
The \textbf{r jet prolongation of a fibered manifold morphism} (F,f) :
$E_{1}(M_{1},\pi_{1})\rightarrow E_{2}(M_{2},\pi_{2})$ where f is a smooth
local diffeomorphism, is the morphism\ of fibered manifolds $\left(
J^{r}F,J^{r}f\right)  :J^{r}E_{1}\rightarrow J^{r}E_{2}$
\end{definition}

(F,f) is such that :

$F:E_{1}\rightarrow E_{2},f:M_{1}\rightarrow M_{2}$ and : $\pi_{2}\circ
F=f\circ\pi_{1}$

$\left(  J^{r}F,J^{r}f\right)  $ is defined as :

$j^{r}f:J^{r}M_{1}\rightarrow J^{r}M_{2}::j^{r}f\left(  x_{1}\right)
=j_{x_{1}}^{r}f$

$J^{r}F:J^{r}E_{1}\rightarrow J^{r}E_{2}::\forall S\in\mathfrak{X}\left(
E_{1}\right)  :J^{r}F\left(  j^{r}S\left(  x\right)  \right)  =j_{f\left(
x\right)  }^{r}\left(  F\circ S\circ f^{-1}\right)  $

$\pi_{s}^{r}\left(  J^{r}F\right)  =J^{s}\left(  \pi_{s}^{r}\right)  ,\pi
^{r}(J^{r}F)=f(\pi^{r})$

If the morphism is between fibered manifolds $E_{1}(M,\pi_{1})\rightarrow
E_{2}(M,\pi_{2})$ on the same base and is base preserving, then f = Id and
$\pi_{2}\circ F=\pi_{1}.$

$\forall S\in\mathfrak{X}\left(  E_{1}\right)  :J^{r}F\left(  j^{r}S\left(
x\right)  \right)  =j_{x}^{r}\left(  F\circ S\right)  =j_{S\left(  x\right)
}^{r}F\circ j_{x}^{r}S$

$\pi_{s}^{r}\left(  J^{r}F\right)  =J^{s}\left(  \pi_{s}^{r}\right)  ,\pi
^{r}(J^{r}F)=\pi^{r}$

\begin{theorem}
The r jet prolongation of an injective (resp.surjective) fibered manifold
morphism is injective (resp.surjective)
\end{theorem}

\begin{theorem}
The r jet prolongation of a morphism between vector bundles is a morphism
between the r-jet prolongations of the vector bundles. The r jet prolongation
of a morphism between principal bundles is a morphism between the r-jet
prolongations of the principal bundles.
\end{theorem}

\paragraph{Total differential\newline}

\begin{definition}
(Kolar p.388) The \textbf{total differential of a base preserving morphism}
$F:J^{r}E\rightarrow\Lambda_{p}TM^{\ast}$ \ from a fibered manifold $E\left(
M,\pi\right)  $ is the map: $\mathfrak{D}F:J^{r+1}E\rightarrow\Lambda
_{p+1}TM^{\ast}$ defined as : $\mathfrak{D}F\left(  j_{x}^{r+1}S\right)
=d\left(  F\circ j^{r}S\right)  \left(  x\right)  $ for any local section S on E
\end{definition}

The total differential is also called formal differential.

F reads : $F=\sum_{\left\{  \beta_{1}...\beta_{p}\right\}  }\varpi_{\left\{
\beta_{1}...\beta_{k}\right\}  }\left(  \xi^{\alpha},\eta^{i},\eta_{\alpha
}^{i},...\eta_{\alpha_{1}...\alpha_{r}}^{i}\right)  d\xi^{\beta_{1}}%
\wedge...\wedge d\xi^{\beta_{p}}$

then : $\mathfrak{D}F=\sum_{\alpha=1}^{m}\sum_{\left\{  \beta_{1}...\beta
_{p}\right\}  }\left(  d_{\alpha}F\right)  d\xi^{\alpha}\wedge d\xi^{\beta
_{1}}\wedge...\wedge d\xi^{\beta p}$

with : $d_{\alpha}F=\frac{\partial F}{\partial\xi^{\alpha}}+\sum_{s=0}^{r}%
\sum_{i=1}^{n}\sum_{\beta_{1}\leq..\leq\beta_{s}}\frac{\partial F}%
{\partial\eta_{\beta_{1}...\beta_{s}}^{i}}\eta_{\alpha\beta_{1}...\beta_{s}%
}^{i}$

Its definition comes from :

$d_{\alpha}F=\frac{\partial F}{\partial\xi^{\alpha}}+\sum_{s=0}^{r}\sum
_{i=1}^{n}\sum_{\beta_{1}\leq..\leq\beta_{s}}\frac{\partial F}{\partial
\eta_{\beta_{1}...\beta_{s}}^{i}}\frac{d\eta_{\beta_{1}...\beta_{s}}^{i}}%
{d\xi^{\alpha}}$

$=\frac{\partial F}{\partial\xi^{\alpha}}+\sum_{s=0}^{r}\sum_{i=1}^{n}%
\sum_{\beta_{1}\leq..\leq\beta_{s}}\frac{\partial F}{\partial\eta_{\beta
_{1}...\beta_{s}}^{i}}\eta_{\alpha\beta_{1}...\beta_{s}}^{i}$

\begin{theorem}
The total differential of morphisms has the properties :

$\mathfrak{D}\circ d=d\circ\mathfrak{D}$

$\mathfrak{D}\left(  \varpi\wedge\mu\right)  =\mathfrak{D}\varpi\wedge
\mu+\varpi\wedge\mathfrak{D}\mu,$

$\mathfrak{D}\left(  dx^{\alpha}\right)  =0$

$d_{\gamma}\left(  du_{\alpha_{1}...\alpha_{s}}^{i}\right)  =du_{\gamma
\alpha_{1}...\alpha_{s}}^{i}$
\end{theorem}

\begin{definition}
The \textbf{total differential of a function} $f:J^{r}E\rightarrow%
\mathbb{R}
$ is a map : $\mathfrak{D:}J^{r}E\rightarrow\Lambda_{1}TM^{\ast}%
::\mathfrak{D}f=\sum_{\alpha}\left(  d_{\alpha}f\right)  d\xi^{\alpha}%
\in\Lambda_{1}TM$ with :%

\begin{equation}
d_{\alpha}f=\frac{\partial f}{\partial\xi^{\alpha}}+\sum_{s=0}^{r}\sum
_{i=1}^{n}\sum_{\beta_{1}\leq...\leq\beta_{s}}\frac{\partial f}{\partial
\eta_{\beta_{1}...\beta_{s}}^{i}}\eta_{\alpha\beta_{1}...\beta_{s}}^{i}%
\end{equation}

\end{definition}

\paragraph{Prolongation of a projectable vector field\newline}

\begin{definition}
The \textbf{r-jet prolongation of the one parameter group }$\Phi_{W}$
associated to a projectable vector field W on a fibered manifold E is the one
parameter group of base preserving morphisms :%

\begin{equation}
J^{r}\Phi_{W}:J^{r}E\rightarrow J^{r}E::J^{r}\Phi_{W}\left(  j^{r}S\left(
x\right)  ,t\right)  =j_{\Phi_{Y}\left(  x,t\right)  }^{r}\left(  \Phi
_{W}\left(  S(\Phi_{Y}\left(  x,-t\right)  ),t\right)  \right)
\end{equation}

with any section S on E and Y the projection of W on TM

The \textbf{r-jet prolongation of the vector field} W is the vector field
$J^{r}W\in\mathfrak{X}\left(  TJ^{r}E\right)  $ :%

\begin{equation}
Z\in\mathfrak{X}\left(  J^{r}E\right)  :J^{r}W\left(  Z\right)  =\frac
{\partial}{\partial t}J^{r}\Phi_{W}\left(  Z,t\right)  |_{t=0}%
\end{equation}

\end{definition}

A vector field on the fiber bundle $E(M,V,\pi)$ is a section $W\in
\mathfrak{X}\left(  TE\right)  .$ In an atlas $\left(  O_{a},\varphi
_{a}\right)  _{a\in A}$ of E it is defined by a family $\left(  W_{ax}%
,W_{au}\right)  _{a\in A}$ : $W_{ax}:\pi^{-1}\left(  O_{a}\right)  \rightarrow
TM,W_{au}:\pi^{-1}\left(  O_{a}\right)  \rightarrow TV$ depending both on p,
that is x and u. It reads :

$W_{a}\left(  \varphi_{a}\left(  x,u_{a}\right)  \right)  =\varphi_{a}%
^{\prime}\left(  x,u_{a}\right)  \left(  W_{ax}\left(  p\right)
,W_{au}\left(  p\right)  \right)  =\sum_{\alpha\in A}W_{ax}^{\alpha}\partial
x_{\alpha}+\sum_{i\in I}W_{au}^{i}\partial u_{ai}$

It is projectable if $\exists Y\in\mathfrak{X}\left(  TM\right)  :\forall p\in
E:\pi^{\prime}\left(  p\right)  W\left(  p\right)  =Y\left(  \pi\left(
p\right)  \right)  .$ Then its components $W_{ax}^{\alpha}$ do not depend on
u. It is vertical if $W_{ax}^{\alpha}=0.$ A projectable vector field defines
with any section S on E a one parameter group of base preserving morphisms on
E through its flow :

$U\left(  t\right)  :\mathfrak{X}\left(  E\right)  \rightarrow\mathfrak{X}%
\left(  E\right)  ::U\left(  t\right)  S\left(  x\right)  =\Phi_{W}\left(
S(\Phi_{Y}\left(  x,-t\right)  ),t\right)  $

The r-jet prolongation $J^{r}U\left(  t\right)  $ of U(t) is the
diffeomorphism :

$J^{r}U\left(  t\right)  :J^{r}E\rightarrow J^{r}E::J^{r}U\left(  t\right)
\left(  j_{x}^{r}S\left(  x\right)  \right)  =j_{U\left(  t\right)  S\left(
x\right)  }^{r}\left(  U\left(  t\right)  S\left(  x\right)  \right)  $

for any section S.

The value of $J^{r}U\left(  t\right)  $\ at $Z=j_{x}^{r}S\left(  x\right)
$\ is computed by taking the r derivative of $U\left(  t\right)  S\left(
x\right)  $ with respect to x for any section S on E.

This is a one parameter group of base preserving morphisms on $J^{r}E$ :

$U\left(  t+t^{\prime}\right)  =U\left(  t\right)  \circ U\left(  t^{\prime
}\right)  $

$j_{U\left(  t+t^{\prime}\right)  S\left(  x\right)  }^{r}\left(  U\left(
t+t^{\prime}\right)  S\left(  x\right)  \right)  =j_{U(t)\circ U\left(
t^{\prime}\right)  S\left(  x\right)  }^{r}\left(  U\left(  t\right)  \right)
\circ j_{U\left(  t^{\prime}\right)  S\left(  x\right)  }^{r}\left(
U(t^{\prime})S\left(  x\right)  \right)  =J^{r}U\left(  t\right)  \circ
J^{r}U\left(  t^{\prime}\right)  \left(  S\left(  x\right)  \right)  $

So it has an infinitesimal generator which is a vector field on the tangent
bundle $TJ^{r}E$ defined at $Z\in J^{r}E$\ by :

$J^{r}W\left(  Z\right)  =\frac{\partial}{\partial t}J^{r}\Phi_{W}\left(
Z,t\right)  |_{t=0}$

So by construction :

\begin{theorem}
The r jet prolongation of the one parameter group of morphisms induced by the
projectable vector field W on E is the one parameter group of morphisms
induced by the r jet prolongation $J^{r}W$ on $J^{r}E$ $:$%

\begin{equation}
J^{r}\Phi_{W}=\Phi_{J^{r}W}%
\end{equation}

\end{theorem}

Because $J^{r}W\in\mathfrak{X}\left(  TJ^{r}E\right)  $ it reads in an
holonomic basis of $TJ^{r}E:$

$J^{r}W\left(  Z\right)  =\sum_{\alpha=1}^{m}X^{\alpha}\partial\xi_{\alpha
}+\sum_{s=0}^{r}\sum_{1\leq\alpha_{1}\leq..\leq\alpha_{s}\leq m}X_{\alpha
_{1}...\alpha_{\alpha_{s}}}^{i}\partial\eta_{i}^{\alpha_{1}...\alpha
_{\alpha_{s}}}$

where $\left\{  X^{\alpha},X_{\alpha_{1}...\alpha_{\alpha_{s}}}^{i}\right\}  $
\textit{depend on Z }(and not only x).

The projections give :

$\left(  \pi^{r}\right)  ^{\prime}\left(  Z\right)  J^{r}W=Y\left(  \pi
^{r}\left(  Z\right)  \right)  $

$\left(  \pi_{0}^{r}\right)  ^{\prime}\left(  Z\right)  J^{r}W=W\left(
p\right)  $

thus : $X^{\alpha}=Y^{\alpha},X^{i}=W^{i}$

The components of $J^{r}W$ at $Z=\left(  \xi^{\alpha},\eta^{i},\eta
_{\alpha_{1}...\alpha_{s}}^{i},s=1...r,1\leq\alpha_{k}\leq m\right)  $ where
$\eta_{\alpha_{1}...\alpha_{s}}^{i}\left(  x\right)  $ are (Giachetta p.49,
Kolar p.360):

$J^{r}W\left(  \xi^{\alpha},\eta^{i},\eta_{\alpha_{1}...\alpha_{s}}%
^{i},s=1...r,1\leq\alpha_{k}\leq m\right)  $

$=\sum_{\alpha}Y^{\alpha}\partial\xi_{\alpha}+\sum_{i=1}^{n}W^{i}\partial
\eta_{i}$

$+\sum_{s=1}^{r}\sum_{1\leq\alpha_{1}\leq..\leq\alpha_{s}\leq m}\sum_{\beta
=1}^{m}\left(  d_{\alpha_{1}}...d_{\alpha_{s}}\left(  W^{i}-\eta_{\beta}%
^{i}W^{\beta}\right)  +\eta_{\beta\alpha_{1}...\alpha_{s}}^{i}W^{\beta
}\right)  \partial\eta_{i}^{\alpha_{1}...\alpha_{\alpha_{s}}}$%

\begin{equation}
J^{1}W\left(  \xi^{\alpha},\eta^{i},\eta_{\alpha}^{i}\right)  =\sum_{\alpha
}Y^{\alpha}\partial x_{\alpha}+\sum_{i}W^{i}\partial\eta_{i}+\sum_{i\alpha
}\left(  \frac{\partial W^{i}}{\partial\xi^{\alpha}}+\sum_{j=1}^{n}%
\eta_{\alpha}^{j}\frac{\partial W^{i}}{\partial\eta^{j}}-\sum_{\beta=1}%
^{m}\eta_{\beta}^{i}\frac{\partial Y^{\beta}}{\partial\xi^{\alpha}}\right)
\partial\eta_{i}^{\alpha}%
\end{equation}

The r jet prolongation of a vertical vector field ($W^{\alpha}=0)$ is a
vertical vector field:

$J^{r}W\left(  \xi^{\alpha},\eta^{i},\eta_{\alpha_{1}...\alpha_{s}}%
^{i},s=1...r,1\leq\alpha_{k}\leq m\right)  $

$=\sum_{i=1}^{n}W^{i}\partial\eta_{i}+\sum_{s=1}^{r}\sum_{1\leq\alpha_{1}%
\leq..\leq\alpha_{s}\leq m}d_{\alpha_{1}}...d_{\alpha_{s}}\left(
W^{i}\right)  \partial\eta_{i}^{\alpha_{1}...\alpha_{\alpha_{s}}}$

The components $Y_{\alpha_{1}...\alpha_{s}}^{i}\partial\eta_{i}^{\alpha
_{1}...\alpha_{\alpha_{s}}}=d_{\alpha_{1}}...d_{\alpha_{s}}\left(
W^{i}\right)  \partial\eta_{i}^{\alpha_{1}...\alpha_{\alpha_{s}}}$ can be
computed by recursion :

$Y_{\beta\alpha_{1}...\alpha_{s}}^{i}=d_{\beta}d_{\alpha_{1}}...d_{\alpha_{s}%
}\left(  W^{i}\right)  =d_{\beta}Y_{\alpha_{1}...\alpha_{s}}^{i}$

$=\frac{\partial Y_{\alpha_{1}...\alpha_{s}}^{i}}{\partial\xi^{\beta}}%
+\sum_{s=0}^{r}\sum_{i=1}^{n}\sum_{\beta_{1}\leq...\leq\beta_{s}}%
\frac{\partial Y_{\alpha_{1}...\alpha_{s}}^{i}}{\partial\eta_{\gamma
_{1}...\gamma_{s}}^{i}}\eta_{\alpha\gamma_{1}...\gamma_{s}}^{i}$

If $W^{i}$ does not depend on $\eta:$

$J^{r}W\left(  \xi^{\alpha},\eta^{i},\eta_{\alpha_{1}...\alpha_{s}}%
^{i},s=1...r,1\leq\alpha_{k}\leq m\right)  $

$=\sum_{i=1}^{n}W^{i}\partial\eta_{i}+\sum_{s=1}^{r}\sum_{1\leq\alpha_{1}%
\leq..\leq\alpha_{s}\leq m}\left(  D_{\alpha_{1}}.._{\alpha_{s}}W^{i}\right)
\partial\eta_{i}^{\alpha_{1}...\alpha_{\alpha_{s}}}$

\begin{theorem}
The Lie derivative of a section $Z\in\mathfrak{X}\left(  J^{r}E\right)  $
along the r-jet prolongation of a projectable vector field W is the section of
the vertical bundle:%

\begin{equation}
\pounds _{J^{r}W}Z=\frac{\partial}{\partial t}\Phi_{J^{r}W}\left(  \pi_{0}%
^{r}\left(  Z\left(  \Phi_{Y}\left(  x,-t\right)  \right)  \right)  ,t\right)
|_{t=0}%
\end{equation}

If W is a vertical vector field then $\pounds _{J^{r}W}Z=J^{r}W\left(
Z\right)  $
\end{theorem}

$J^{r}E$ is a fiber bundle with base M. $J^{r}W$ is a projectable vector
field. The theorem follows from the definition of the Lie derivative.

Moreover for s
$<$
r : $\pi_{s}^{r}\left(  Z\right)  \pounds _{J^{r}W}Z=\pounds _{J^{s}W}\pi
_{s}^{r}\left(  Z\right)  $

\paragraph{Change of trivialization on the r-jet prolongation of a fiber
bundle\newline}

1. A section $\kappa\in\mathfrak{X}\left(  P\left[  T_{1}G,Ad\right]  \right)
$ of the adjoint bundle to a principal bundle $P\left(  M,G,\pi\right)  $ with
atlas $\left(  O_{a},\varphi_{a}\right)  _{a\in A}$\ induces a change of
trivialization on P : $p=\varphi_{a}\left(  x,g\right)  =\widetilde{\varphi
}_{a}\left(  x,\exp t\kappa_{a}\left(  x\right)  g\right)  $ so for a section
$S=\varphi_{a}\left(  x,\sigma_{a}\left(  x\right)  \right)  $\ $\widetilde
{\sigma}_{a}\left(  x,t\right)  =\left(  \exp t\kappa_{a}\left(  x\right)
\right)  \sigma_{a}\left(  x\right)  .$ It induces on any associated fiber
bundle $E=P\left[  V,\lambda\right]  $ a change of trivialisation for a
section $U=\psi_{a}\left(  x,u_{a}\left(  x\right)  \right)  $ $\widetilde
{u}_{a}\left(  x,t\right)  =\lambda\left(  \exp t\kappa_{a}\left(  x\right)
,u_{a}\left(  x\right)  \right)  $

The r-jet prolongation $J^{r}U$ has for coordinates :

$\left(  U\left(  x\right)  ,\frac{\partial u^{i}}{\partial\xi^{\alpha_{1}%
}...\partial\xi^{\alpha_{s}}},s=1...r,1\leq\alpha_{k}\leq\alpha_{k+1}\leq\dim
M,i=1..\dim G\right)  $

2. The new components $\widetilde{u}_{a}$ are given by the flow of the
vertical vector field $W=\lambda^{\prime}\left(  1,u_{a}\left(  x\right)
\right)  \kappa_{a}\left(  x\right)  $ on TV, which has a r-jet prolongation
and \ $J\widetilde{u}_{a}\left(  x,t\right)  =\Phi_{J^{r}W}\left(  J^{r}%
u_{a},t\right)  $ and :

$\frac{d}{dt}J\widetilde{u}_{a}\left(  x,t\right)  |_{t=0}=J^{r}W\left(
J^{r}u_{a}\right)  $

$J^{r}W\left(  \xi^{\alpha},\eta^{i},\eta_{\alpha_{1}..\alpha_{s}}%
^{i}...\right)  =\sum_{\alpha=1}^{m}Y^{\alpha}\partial\xi_{\alpha}+\sum
_{s=0}^{r}\sum_{1\leq\alpha_{1}\leq..\leq\alpha_{s}\leq m}Y_{\alpha
_{1}...\alpha_{\alpha_{s}}}^{i}\partial\eta_{i}^{\alpha_{1}...\alpha
_{\alpha_{s}}}$

The components $X_{\alpha_{1}...\alpha_{\alpha_{s}}}^{i}$ are given by the
formula above. Because the vector is vertical :

$Y_{\beta\alpha_{1}...\alpha_{s}}^{i}=d_{\beta}d_{\alpha_{1}}...d_{\alpha_{s}%
}\left(  W^{i}\right)  =d_{\beta}Y_{\alpha_{1}...\alpha_{s}}^{i}$

A direct computation is more illuminating.

We start with : $\widetilde{u}\left(  x,t\right)  =\lambda\left(  \exp\left(
t\kappa\left(  x\right)  \right)  ,u\left(  x\right)  \right)  $

$\frac{d}{dt}\widetilde{u}\left(  x,t\right)  |_{t=0}=\lambda_{g}^{\prime
}\left(  1,u\right)  \kappa$

$\frac{\partial}{\partial\xi^{\alpha}}\widetilde{u}\left(  x,t\right)
=t\lambda_{g}^{\prime}\left(  \exp t\kappa,u\right)  \left(  \exp
t\kappa\right)  ^{\prime}\frac{\partial\kappa}{\partial\xi^{\alpha}}%
+\lambda_{u}^{\prime}\left(  \exp t\kappa,u\right)  \frac{\partial u}%
{\partial\xi^{\alpha}}$

$\frac{d}{dt}\frac{\partial\widetilde{u}\left(  x,t\right)  }{\partial
\xi^{\alpha}}=\lambda_{g}^{\prime}\left(  \exp t\kappa,u\right)  \left(  \exp
t\kappa\right)  ^{\prime}\frac{\partial\kappa}{\partial\xi^{\alpha}}+t\frac
{d}{dt}\left(  \lambda_{g}^{\prime}\left(  \exp t\kappa,u\right)  \left(  \exp
t\kappa\right)  ^{\prime}\frac{\partial\kappa}{\partial\xi^{\alpha}}\right)
+\lambda"_{ug}\left(  \exp t\kappa,u\right)  \left(  L_{\exp t\kappa}^{\prime
}\left(  1\right)  \kappa,\frac{\partial u}{\partial\xi^{\alpha}}\right)  $

$\frac{d}{dt}\frac{\partial\widetilde{u}\left(  x,t\right)  }{\partial
\xi^{\alpha}}|_{t=0}=\lambda_{g}^{\prime}\left(  1,u\right)  \frac
{\partial\kappa}{\partial\xi^{\alpha}}+\lambda"_{ug}\left(  1,u\right)
\left(  \kappa,\frac{\partial u}{\partial\xi^{\alpha}}\right)  $

Similarly :

$\frac{\partial^{2}\widetilde{u}\left(  x,t\right)  }{\partial\xi^{\alpha
}\partial\xi^{\beta}}|_{t=0}=\lambda_{g}^{\prime}\left(  1,u\right)
\frac{\partial^{2}\kappa}{\partial\xi^{\alpha}\partial\xi^{\beta}}%
+\lambda"_{gu}\left(  1,u\right)  \left(  \frac{\partial u}{\partial\xi
^{\beta}},\frac{\partial\kappa}{\partial\xi^{\alpha}}\right)  +\lambda
"_{ug}\left(  1,u\right)  \left(  \frac{\partial\kappa}{\partial\xi^{\beta}%
},\frac{\partial u}{\partial\xi^{\alpha}}\right)  +\lambda_{gu^{2}}^{\left(
3\right)  }\left(  1,u\right)  \left(  \kappa,\frac{\partial u}{\partial
\xi^{\beta}},\frac{\partial u}{\partial\xi^{\alpha}}\right)  $

One can check that : $\frac{d}{dt}\frac{\partial\widetilde{u}\left(
x,t\right)  }{\partial\xi^{\beta}\partial\xi^{\alpha_{1}}...\partial
\xi^{\alpha_{s}}}|_{t=0}=\frac{\partial}{\partial\xi^{\beta}}\left(
\frac{\partial\widetilde{u}\left(  x,t\right)  }{\partial\xi^{\alpha_{1}%
}...\partial\xi^{\alpha_{s}}}|_{t=0}\right)  $

3. These operations are useful in gauge theories. Assume that we have a
function : $L:J^{r}E\rightarrow%
\mathbb{R}
$ (say a lagrangian) which is invariant by a change of trivialization. Then we
must have : $L\left(  J\widetilde{U}\left(  t\right)  \right)  =L\left(
J^{r}U\right)  $ for any t and change of gauge $\kappa.$ By differentiating
with respect to t at t=0 we get :

$\forall\kappa:\sum\frac{\partial L}{\partial u_{\alpha_{1}...\alpha_{s}}^{i}%
}\frac{d}{dt}\frac{\partial\widetilde{u}^{i}}{\partial\xi^{\alpha_{1}%
}...\partial\xi^{\alpha_{s}}}=0$

that is a set of identities between $J^{r}U$ and the partial derivatives
$\frac{\partial L}{\partial u_{\alpha_{1}...\alpha_{s}}^{i}}$ of L.

\subsubsection{Infinite order jet}

(Kolar p.125)

We have the inverse sequence :

$E\overset{\pi_{0}^{1}}{\leftarrow}J^{1}E\overset{\pi_{1}^{2}}{\leftarrow
}J^{2}E\overset{\pi_{2}^{3}}{\leftarrow}....$

If the base M and the fiber V are smooth the infinite prolongation $J^{\infty
}E$ is defined as the minimal set such that the projections :

$\pi_{r}^{\infty}:J^{\infty}E\rightarrow J^{r}E,\pi^{\infty}:J^{\infty
}E\rightarrow M,\pi_{0}^{\infty}:J^{\infty}E\rightarrow E$ are submersions and
follow the relations : $\pi_{r}^{\infty}=\pi_{r}^{s}\circ\pi_{s}^{\infty}$

This set exists and, provided with the inductive limit topology, all the
projections are continuous and $J^{\infty}E$ is a paracompact Fr\'{e}chet
space.\ It is not a manifold according to our definition : it is modelled on a
Fr\'{e}chet space.

\newpage

\section{CONNECTIONS}

\bigskip

With fiber bundles the scope of mathematical objects defined over a manifold
can be extended. When dealing with them the frames are changing with the
location making the comparisons and their derivation more
complicated.\ Connections are the tool for this job : they "connect" the
frames at different point. So this is an of extension of the parallel
transport. To do so we need to distinguish between transport in the base
manifold, which becomes "horizontal transport", and transport along the fiber
manifold, which becomes "vertical transport". Indeed vertical transports are
basically changes in the frame without changing the location. So we can split
a variation in an object between what can be attributed to a change of
location, and what comes from a change of the frame.

General connections on general fiber bundles are presented first, with their
general properties, and the many related objects: covariant derivative,
curvature, exterior covariant derivative,... Then, for each of the 3 classes
of fiber bundles : vector bundles, principal bundles, associated bundles,
there are connections which takes advantage of the special feature of the
respective bundle. For the most part it is an adaptation of the general framework.

\bigskip

\subsection{General connections}

\label{General connections}

Connections on a fiber bundle can be defined in a purely geometrical way,
without any reference to coordinates, or through jets. Both involve
Christoffel symbols. The first is valid in a general context (whatever the
dimension and the field of the manifolds), the second is restricted to the
common case of finite dimensional real fiber bundles. They give the same
results and the choice is mainly a matter of personnal preference.\ We will
follow the geometrical way, as it is less abstract and more intuitive.

\subsubsection{Definitions}

\paragraph{Geometrical definition\newline}

The tangent space $T_{p}E$ at any point p to a fiber bundle $E\left(
M,V,\pi\right)  $ has a preferred subspace : the vertical space corresponding
to the kernel of $\pi^{\prime}\left(  p\right)  ,$ which does not depend on
the trivialization. And any vector $v_{p}$\ can be decomposed between a part
$\varphi_{x}^{\prime}\left(  x,u\right)  v_{x}$\ related to $T_{x}M$ and
another part $\varphi_{u}^{\prime}\left(  x,u\right)  v_{u}$\ related to
$T_{u}V$\ . If this decomposition is unique for a given trivialization, it
depends on the trivialization. A connection is a geometric decomposition,
independant from the trivialization.

\bigskip

\begin{definition}
A \textbf{connection} on a fiber bundle $E\left(  M,V,\pi\right)  $\ is a
1-form $\Phi$ on E valued in the vertical bundle, which is a projection :

$\Phi\in\Lambda_{1}\left(  E;VE\right)  :$ $TE\rightarrow VE::\Phi\circ
\Phi=\Phi,\Phi\left(  TE\right)  =VE$
\end{definition}

So $\Phi$ acts on vectors of the tangent bundle TE, and the result is in the
vertical bundle.

$\Phi$ has constant rank, and $\ker\Phi$ is a vector subbundle of TE.

Remark : this is a first order connection, with differential operators one can
define r-order connections.

\begin{definition}
The \textbf{horizontal bundle} of the tangent bundle TE is the vector
subbundle $HE=\ker\Phi$
\end{definition}

The tangent bundle TE is the direct sum of two vector bundles :

$TE=HE\oplus VE$ :

$\forall p\in E:T_{p}E=H_{p}E\oplus V_{p}E$

$V_{p}E=\ker\pi^{\prime}(p)$

$H_{p}E=\ker\Phi(p)$

The horizontal bundle can be seen as "displacements along the base M(x)" and
the vertical bundle as "displacements along V(u)". The key point here is that
the decomposition does not depend on the trivialization : it is purely geometric.

\paragraph{Christoffel form\newline}

\begin{theorem}
A connection $\Phi$ on a fiber bundle $E\left(  M,V,\pi\right)  $ with atlas
$\left(  O_{a},\varphi_{a}\right)  _{a\in A}$\ is uniquely defined by a family
of maps $\left(  \Gamma_{a}\right)  _{a\in A}$\ called the \textbf{Christoffel
forms} of the connection.%

\begin{equation}
\Gamma_{a}\in C\left(  \pi^{-1}\left(  O_{a}\right)  ;TM^{\ast}\otimes
TV\right)  ::\Phi(p)v_{p}=\varphi_{a}^{\prime}(x,u)\left(  0,v_{u}+\Gamma
_{a}\left(  p\right)  v_{x}\right)
\end{equation}

\end{theorem}

\begin{proof}
$\varphi_{a}:O_{a}\times V\rightarrow\pi^{-1}\left(  O_{a}\right)  $

$\Rightarrow\varphi_{a}^{\prime}:T_{x}O_{a}\times T_{u}V\rightarrow
T_{p}E::v_{p}=\varphi_{a}^{\prime}(x,u)(v_{x},v_{u})$

and $\varphi_{a}^{\prime}\left(  x,u_{a}\right)  $ is invertible.

$\Phi\left(  p\right)  v_{p}\in V_{p}E\Rightarrow\exists w_{u}\in T_{u}%
V:\Phi(p)v_{p}=\varphi_{a}^{\prime}(x,u)(0,w_{u})$

$\Phi(p)v_{p}=\Phi(p)\varphi_{a}^{\prime}(x,u)(0,v_{u})+\Phi(p)\varphi
^{\prime}(x,u)(v_{x},0)=\varphi_{a}^{\prime}(x,u)(0,w_{u})$

$\varphi_{a}^{\prime}(x,u)(0,v_{u})\in V_{p}E\Rightarrow\Phi(p)\varphi
_{a}^{\prime}(x,u)(0,v_{u})=\varphi_{a}^{\prime}(x,u)(0,v_{u})$

So $\Phi(p)\varphi^{\prime}(x,u)(v_{x},0)=\varphi_{a}^{\prime}(x,u)(0,w_{u}%
-v_{u})$ depends linearly on $w_{u}-v_{u}$

Let us define : $\Gamma_{a}:\pi^{-1}\left(  O_{a}\right)  \rightarrow%
\mathcal{L}%
\left(  T_{x}M;T_{u}V\right)  ::\Gamma_{a}\left(  p\right)  v_{x}=w_{u}-v_{u}$

So : $\Phi(p)v_{p}=\varphi_{a}^{\prime}(x,u)\left(  0,v_{u}+\Gamma_{a}\left(
p\right)  v_{x}\right)  $
\end{proof}

$\Gamma$ is a map, defined on E (it depends on p) and valued in the tensorial
product $TM^{\ast}\otimes TV$ :%

\begin{equation}
\Gamma\left(  p\right)  =\sum_{i=1}^{n}\sum_{\alpha=1}^{m}\Gamma\left(
p\right)  _{\alpha}^{i}d\xi^{\alpha}\otimes\partial u_{i}%
\end{equation}

with a dual holonomic basis $d\xi^{\alpha}$\ of TM* \footnote{I will keep the
notations $\partial x_{\alpha},dx^{\alpha}$ for the part of the basis on TE
related to M, and denote $\partial\xi_{\alpha},d\xi^{\alpha}$ for holonomic
bases in TM,TM*}. This is a 1-form acting on vectors of $T_{\pi\left(
p\right)  }M$ and valued in $T_{u}V$

Remark : there are different conventions regarding the sign in the above
expression.\ I have taken a sign which is consistent with the common
definition of affine connections on the tangent bundle of manifolds, as they
are the same objects.

\bigskip

\begin{theorem}
On a fiber bundle $E\left(  M,V,\pi\right)  $\ \ a family of maps

$\Gamma_{a}\in C\left(  \pi^{-1}\left(  O_{a}\right)  ;TM^{\ast}\otimes
TV\right)  $ defines a connection on E iff it satisfies the transition
conditions :

$\Gamma_{b}\left(  p\right)  =\varphi_{ba}^{\prime}\left(  x,u_{a}\right)
\circ\left(  -Id_{TM},\Gamma_{a}\left(  p\right)  \right)  $ in an atlas
$\left(  O_{a},\varphi_{a}\right)  _{a\in A}$ of E.
\end{theorem}

\begin{proof}
At the transitions between charts (see Tangent space in Fiber bundles):

$x\in O_{a}\cap O_{b}:v_{p}=\varphi_{a}^{\prime}\left(  x,u_{a}\right)
v_{x}+\varphi_{au}^{\prime}\left(  x,u_{a}\right)  v_{au}=\varphi_{b}^{\prime
}\left(  x,u_{b}\right)  v_{x}+\varphi_{bu}^{\prime}\left(  x,u_{b}\right)
v_{bu}$

we have the identities :

$\varphi_{ax}^{\prime}\left(  x,u_{a}\right)  =\varphi_{bx}^{\prime}\left(
x,u_{b}\right)  +\varphi_{bu}^{\prime}\left(  x,u_{b}\right)  \varphi
_{bax}^{\prime}\left(  x,u_{a}\right)  $

$\varphi_{au}^{\prime}\left(  x,u_{a}\right)  =\varphi_{bu}^{\prime}\left(
x,u_{b}\right)  \varphi_{bau}^{\prime}\left(  x,u_{a}\right)  $

$v_{bu}=\varphi_{ba}^{\prime}\left(  x,u_{a}\right)  \left(  v_{x}%
,v_{au}\right)  $

i) If there is a connection :

$\Phi(p)v_{p}=\varphi_{a}^{\prime}(x,u_{a})\left(  0,v_{au}+\Gamma_{a}\left(
p\right)  v_{x}\right)  =\varphi_{b}^{\prime}(x,u_{b})\left(  0,v_{bu}%
+\Gamma_{b}\left(  p\right)  v_{x}\right)  $

$v_{bu}+\Gamma_{b}\left(  p\right)  v_{x}=\varphi_{ba}^{\prime}\left(
x,u_{a}\right)  \left(  0,v_{au}+\Gamma_{a}\left(  p\right)  v_{x}\right)
=\varphi_{bau}^{\prime}\left(  x,u_{a}\right)  v_{au}+\varphi_{bau}^{\prime
}\left(  x,u_{a}\right)  \Gamma_{a}\left(  p\right)  v_{x}$

$=\varphi_{bax}^{\prime}\left(  x,u_{a}\right)  v_{x}+\varphi_{bau}^{\prime
}\left(  x,u_{a}\right)  v_{au}+\Gamma_{b}\left(  p\right)  v_{x}$

$\Gamma_{b}\left(  p\right)  v_{x}=\varphi_{bau}^{\prime}\left(
x,u_{a}\right)  \Gamma_{a}\left(  p\right)  v_{x}-\varphi_{bax}^{\prime
}\left(  x,u_{a}\right)  v_{x}=\varphi_{ba}^{\prime}\left(  x,u_{a}\right)
\left(  -v_{x},\Gamma_{a}\left(  p\right)  v_{x}\right)  $

ii) Conversely let be a set of maps $\left(  \Gamma_{a}\right)  _{a\in
A},\Gamma_{a}\in C\left(  O_{a};TM^{\ast}\otimes TV\right)  $ such that
$\Gamma_{b}\left(  p\right)  =\varphi_{ba}^{\prime}\left(  x,u_{a}\right)
\circ\left(  -Id_{TM},\Gamma_{a}\left(  p\right)  \right)  $

define $\Phi_{a}(p)v_{p}=\varphi_{a}^{\prime}(x,u_{a})\left(  0,v_{au}%
+\Gamma_{a}\left(  p\right)  v_{x}\right)  $

let us show that $\Phi_{b}(p)v_{p}=\varphi_{b}^{\prime}(x,u_{b})\left(
0,v_{bu}+\Gamma_{b}\left(  p\right)  v_{x}\right)  =\Phi_{a}(p)v_{p}$

$v_{bu}=\varphi_{ba}^{\prime}\left(  x,u_{a}\right)  \left(  v_{x}%
,v_{au}\right)  ;\Gamma_{b}\left(  p\right)  v_{x}=-\varphi_{bax}^{\prime
}\left(  x,u_{a}\right)  v_{x}+\varphi_{bax}^{\prime}\left(  x,u_{a}\right)
\Gamma_{a}\left(  p\right)  v_{x}$

$\varphi_{b}^{\prime}(x,u_{b})=\varphi_{bu}^{\prime}\left(  x,\varphi
_{bau}\left(  x,u_{a}\right)  \right)  $

$\Phi_{b}(p)v_{p}=\varphi_{bu}^{\prime}(x,u_{b})\varphi_{bax}^{\prime}\left(
x,u_{a}\right)  v_{x}+\varphi_{bu}^{\prime}(x,u_{b})\varphi_{bau}^{\prime
}\left(  x,u_{a}\right)  v_{au}$

$-\varphi_{bu}^{\prime}(x,u_{b})\varphi_{bax}^{\prime}\left(  x,u_{a}\right)
v_{x}+\varphi_{bu}^{\prime}(x,u_{b})\varphi_{bax}^{\prime}\left(
x,u_{a}\right)  \Gamma_{a}\left(  p\right)  v_{x}$

$=\varphi_{bu}^{\prime}(x,u_{b})\varphi_{bau}^{\prime}\left(  x,u_{a}\right)
\left(  v_{au}+\Gamma_{a}\left(  p\right)  v_{x}\right)  =\varphi_{au}%
^{\prime}(x,u_{a})\left(  v_{au}+\Gamma_{a}\left(  p\right)  v_{x}\right)  $
\end{proof}

In a change of trivialization on E :

$p=\varphi_{a}\left(  x,u_{a}\right)  =\widetilde{\varphi}_{a}\left(
x,\chi_{a}\left(  x\right)  \left(  u_{a}\right)  \right)  \Leftrightarrow
\widetilde{u}_{a}=\chi_{a}\left(  x,u_{a}\right)  $%

\begin{equation}
\Gamma_{a}\left(  p\right)  \rightarrow\widetilde{\Gamma}_{a}\left(  p\right)
=\chi_{a}^{\prime}\left(  x,u_{a}\right)  \circ\left(  -Id_{TM},\Gamma
_{a}\left(  p\right)  \right)  =-\chi_{ax}^{\prime}\left(  x,u_{a}\right)
+\chi_{au}^{\prime}\left(  x,u_{a}\right)  \Gamma_{a}\left(  p\right)
\end{equation}

\paragraph{Jet definition\newline}

The 1-jet prolongation of E is an affine bundle $J^{1}E$ over E, modelled on
the vector bundle $TM^{\ast}\otimes VE\rightarrow E$ . So a section of this
bundle reads :

$j_{\alpha}^{i}\left(  p\right)  d\xi^{\alpha}\otimes\partial u^{i}$

On the other hand $\Gamma\left(  p\right)  \in%
\mathcal{L}%
\left(  T_{x}M;T_{u}V\right)  $ has the coordinates in charts : $\Gamma\left(
p\right)  _{\alpha}^{i}d\xi^{\alpha}\otimes\partial u_{i}$ and we can define a
connection through a section of the 1-jet prolongation $J^{1}E$ of E.

The geometric definition is focused on $\Phi$ and the jet definition is
focused on $\Gamma\left(  p\right)  .$

\paragraph{Pull back of a connection\newline}

\begin{theorem}
For any connection $\Phi$ on a fiber bundle $E\left(  M,V,\pi\right)  $\ , N
smooth manifold and $f:N\rightarrow M$ smooth map, the pull back $f^{\ast}%
\Phi$ of $\Phi$\ is a connection on $f^{\ast}E:\left(  f^{\ast}\Phi\right)
\left(  y,p\right)  \left(  v_{y},v_{p}\right)  =\left(  0,\Phi\left(
p\right)  v_{p}\right)  $
\end{theorem}

\begin{proof}
If $f:N\rightarrow M$ is a smooth map, then the pull back of the fiber bundle
is a fiber bundle $f^{\ast}E\left(  N,V,f^{\ast}\pi\right)  $ such that :

Base : N, Standard fiber : V

Total space : $f^{\ast}E=\left\{  \left(  y,p\right)  \in N\times E:f\left(
y\right)  =\pi\left(  p\right)  \right\}  $

Projection : $\widetilde{\pi}:f^{\ast}E\rightarrow N::\widetilde{\pi}\left(
y,p\right)  =y$

So $q=\left(  y,p\right)  \in f^{\ast}E\rightarrow v_{q}=\left(  v_{y}%
,v_{p}\right)  \in T_{q}f^{\ast}E$

$\widetilde{\pi}^{\prime}\left(  q\right)  v_{q}=\widetilde{\pi}^{\prime
}\left(  q\right)  \left(  v_{y},v_{p}\right)  =\left(  v_{y},0\right)  $

Take : $\left(  f^{\ast}\Phi\right)  \left(  q\right)  v_{q}=\left(
0,\Phi\left(  p\right)  v_{p}\right)  $

$\widetilde{\pi}^{\prime}\left(  q\right)  \left(  0,\Phi\left(  p\right)
v_{p}\right)  =0\Leftrightarrow\left(  0,\Phi\left(  p\right)  v_{p}\right)
\in V_{q}f^{\ast}E$
\end{proof}

\subsubsection{Covariant derivative}

The common covariant derivative is a map which transforms vector fields,
$\otimes^{1}TM$ tensors, into $\otimes_{1}^{1}TM$ . So it acts on sections of
the tangent bundle.\ Similarly the covariant derivative associated to a
connection acts on sections of E.

\begin{definition}
The covariant derivative $\nabla$\ associated to a connection $\Phi$ on a
fiber bundle $E\left(  M,V,\pi\right)  $ is the map :%

\begin{equation}
\nabla:\mathfrak{X}\left(  E\right)  \rightarrow\Lambda_{1}\left(
M;VE\right)  ::\nabla S=S^{\ast}\Phi
\end{equation}

\end{definition}

So the covariant derivative along a vector field X on M is :

$\nabla_{X}S(x)=\Phi(S(x))(S^{\prime}(x)X)\in\mathfrak{X}\left(  VE\right)  $

If $S\left(  x\right)  =\varphi_{a}\left(  x,\sigma_{a}\left(  x\right)
\right)  :\nabla_{X}S(x)=\sum_{\alpha i}(\partial_{\alpha}\sigma_{a}%
^{i}+\Gamma_{a}(S\left(  x\right)  )_{\alpha}^{i})X_{a}^{\alpha}\partial
u_{i}$

Notice the difference :

the connection $\Phi$ acts on sections of TE,

the covariant derivative $\nabla$ acts on sections of E.

The covariant derivative is also called the absolute differential.

$\nabla S$ is linear with respect to the vector field X :

$\nabla_{X+Y}S(x)=\nabla_{X}S(x)+\nabla_{Y}S(x)$,$\nabla_{kX}S(x)=k\nabla
_{X}S(x)$

but we cannot say anything about linearity with respect to S for a general
fiber bundle.

\begin{definition}
A section S is said to be an integral of a connection on a fiber bundle with
covariant derivative $\nabla$\ if $\nabla S=0.$
\end{definition}

\begin{theorem}
(Giachetta p.33) For any global section S on a fiber bundle $E\left(
M,V,\pi\right)  $ there is always a connection such that S is an integral
\end{theorem}

\subsubsection{Lift to a fiber bundle}

A connection is a projection on the vertical bundle.\ Similarly we can define
a projection on the horizontal bundle.

\paragraph{Horizontal form\newline}

\begin{definition}
The \textbf{horizontal form} of a connection $\Phi$ on a fiber bundle
$E\left(  M,V,\pi\right)  $\ is the 1 form

$\chi\in\Lambda_{1}\left(  E;HE\right)  :\chi\left(  p\right)  =Id_{TE}%
-\Phi\left(  p\right)  $
\end{definition}

A connection can be equivalently defined by its horizontal form $\chi\in
\wedge_{1}\left(  E;HE\right)  $ and we have :

$\chi\circ\chi=\chi;$

$\chi(\Phi)=0;$

$VE=\ker\chi;$

$\chi\left(  p\right)  v_{p}=\varphi_{a}^{\prime}(x,u)\left(  v_{x}%
,-\Gamma\left(  p\right)  v_{x}\right)  \in H_{p}E$

The horizontal form is directly related to TM, as we can see in the formula
above which involves only $v_{x}$\ . So we can "lift" any object defined on TM
onto TE by "injecting" $v_{x}\in T_{x}M$ in the formula.

\paragraph{Horizontal lift of a vector field\newline}

\begin{definition}
The \textbf{horizontal lift }of a vector field on M by a connection $\Phi$ on
a fiber bundle $E\left(  M,V,\pi\right)  $\ with trivialization $\varphi$\ is
the map :%

\begin{equation}
\chi_{L}:\mathfrak{X}\left(  TM\right)  \rightarrow\mathfrak{X}\left(
HE\right)  ::\chi_{L}\left(  p\right)  \left(  X\right)  =\varphi^{\prime
}(x,u)\left(  X\left(  x\right)  ,-\Gamma\left(  p\right)  X\left(  x\right)
\right)
\end{equation}

\end{definition}

$\chi_{L}\left(  p\right)  \left(  X\right)  $ is a horizontal vector field on
TE, which is projectable on TM as X.

\begin{theorem}
(Kolar p.378) For any connection $\Phi$\ on a fiber bundle $E\left(
M,V,\pi\right)  $ with covariant derivative $\nabla$\ ,\ the horizontal lift
$\chi_{L}\left(  X\right)  $\ of a vector field X on M is a projectable vector
field on E and for any section $S\in$\ $\mathfrak{X}\left(  E\right)  $:%

\begin{equation}
\nabla_{X}S=\pounds _{\chi_{L}\left(  X\right)  }S
\end{equation}

\end{theorem}

\paragraph{Horizontal lift of a curve\newline}

By lifting the tangent to a curve we can lift the curve itself.

\begin{theorem}
(Kolar p. 80) For any connection $\Phi$ on a fiber bundle $E\left(
M,V,\pi\right)  $\ and path $c:\left[  a,b\right]  \rightarrow M$ in M, with
$0\in\left[  a,b\right]  ,c\left(  0\right)  =x,A\in\pi^{-1}\left(  x\right)
$ there is a neighborhood n(x) and a unique smooth map : $P:n\left(  x\right)
\rightarrow E$ such that :

i) $\pi\left(  P\left(  c\left(  t\right)  \right)  \right)  =c\left(
t\right)  ,$\ $P(x)=A$

ii) $\Phi\left(  \frac{dP}{dt}\right)  =0$ when defined

The curve is inchanged in a smooth change of parameter.\ If c depends smoothly
on other parameters, then P depends smoothly on those parameters
\end{theorem}

P is defined through the equation : $\Phi\left(  \frac{dP}{dt}\right)  =0$
that is equivalent to :

$\Phi(P\left(  c\left(  t\right)  \right)  (\frac{dP}{dt})=\Phi(P\left(
c\left(  t\right)  \right)  (P^{\prime}\frac{dc}{dt})=\nabla_{c^{\prime}%
(t)}P\left(  c\left(  t\right)  \right)  =0$

\begin{theorem}
(Kolar p.81) A connection is said to be \textbf{complete} if the lift is
defined along any path on M. Each fiber bundle admits complete connections.
\end{theorem}

A complete connection is sometimes called an Ehresmann connection.

Remarks : the lift is sometime called "parallel transport" (Kolar), but there
are significant differences with what is defined usually on a simple manifold.

i) the lift transports curves on the base manifold to curves on the fiber
bundle, whereas the parallel transport transforms curves in the same manifold.

ii ) a vector field on a manifold can be parallel transported, but there is no
"lift" of a section on a vector bundle. But a section can be parallel
transported by the flow of a projectable vector field.

iii) there is no concept of geodesic on a fiber bundle.\ It would be a curve
such that its tangent is parallel transported along the curve, which does not
apply here.\ Meanwhile on a fiber bundle there are horizontal curves :
$C:\left[  a,b\right]  \rightarrow E$\ such that $\Phi\left(  C\left(
t\right)  \right)  \left(  \frac{dC}{dt}\right)  =0$\ so its tangent is a
horizontal vector. Given any curve c(t) on M there is always a horizontal
curve which projects on c(t), this is just the lift of c.

\paragraph{Holonomy group\newline}

If the path c is a loop in M : $c:\left[  a,b\right]  \rightarrow
M::c(a)=c(b)=x,$ the lift goes from a point $A\in E\left(  x\right)  $\ to a
point B in the same fiber E(x) over x, so we have a map in V : $A=\varphi
\left(  x,u_{A}\right)  \rightarrow B=$ $\varphi\left(  x,u_{B}\right)
::u_{B}=\phi\left(  u_{A}\right)  .$This map has an inverse (take the opposite
loop with the reversed path) and is a diffeomorphism in V. The set of all
these diffeomorphisms has a group structure : this is the \textbf{holonomy
group }$H\left(  \Phi,x\right)  $ at x $.$If we restrict the loops to loops
which are homotopic to a point we have the restricted holonomy group
$H_{0}\left(  \Phi,x\right)  .$

\bigskip

\subsubsection{Curvature}

There are several objects linked to connections which are commonly called
curvature.\ The following is the curvature of the connection $\Phi$.

\begin{definition}
The \textbf{curvature} of a connection $\Phi$ on the fiber bundle $E\left(
M,V,\pi\right)  $\ is the 2-form $\Omega\in\Lambda_{2}\left(  E;VE\right)  $
such that for any vector field X,Y on E :

$\Omega(X,Y)=\Phi([\chi X,\chi Y]_{TE})$ where $\chi$ is the horizontal form
of $\Phi$
\end{definition}

\begin{theorem}
The local components of the curvature are given by the Maurer-Cartan formula :

$\varphi_{a}^{\ast}\Omega=\sum_{i}\left(  -d_{M}\Gamma^{i}+\left[
\Gamma,\Gamma\right]  _{V}^{i}\right)  \otimes\partial u_{i}$%

\begin{equation}
\Omega=\sum_{\alpha\beta}\left(  -\partial_{\alpha}\Gamma_{\beta}^{i}%
+\Gamma_{\alpha}^{j}\partial_{j}\Gamma_{\beta}^{i}\right)  dx^{\alpha}\wedge
dx^{\beta}\otimes\partial u_{i}%
\end{equation}

\end{theorem}

the holonomic basis of TE is $\left(  \partial x_{\alpha},\partial
u_{i}\right)  $

Notice that the curvature is a 2-form on TE (and not TM), in the bracket we
have $\chi X$\ and not $\chi_{L}X$\ ,whence the notation $dx^{\alpha}\wedge
dx^{\beta},\partial u_{i}.$The bracket is well defined for vector fields on
the tangent bundle. See below the formula.

The curvature is zero if one of the vector X,Y is vertical (because then $\chi
X=0)$ so the curvature is an horizontal form, valued in the vertical bundle.

\begin{proof}
in an atlas $\left(  O_{a},\varphi_{a}\right)  $ of E:

$\chi\left(  p\right)  X_{p}=\varphi_{a}^{\prime}(x,u)\left(  v_{x}%
,-\Gamma\left(  p\right)  v_{x}\right)  =v_{x}^{\alpha}\partial x_{\alpha
}-\Gamma\left(  p\right)  _{\alpha}^{i}v_{x}^{\alpha}\partial u_{i}$

$\chi\left(  p\right)  Y_{p}=\varphi_{a}^{\prime}(x,u)\left(  w_{x}%
,-\Gamma\left(  p\right)  w_{x}\right)  =w_{x}^{\alpha}\partial x_{\alpha
}-\Gamma\left(  p\right)  _{\alpha}^{i}w_{x}^{\alpha}\partial u_{i}$

$\left[  \left(  v_{x},-\Gamma\left(  p\right)  v_{x}\right)  ,\left(
w_{x},-\Gamma\left(  p\right)  w_{x}\right)  \right]  $

$=\sum_{\alpha}\left(  \sum_{\beta}v_{x}^{\beta}\partial_{\beta}w_{x}^{\alpha
}-w_{x}^{\beta}\partial_{\beta}v_{x}^{\alpha}+\sum_{j}\left(  -\Gamma_{\beta
}^{j}v_{x}^{\beta}\partial_{j}w_{x}^{\alpha}+\Gamma_{\beta}^{j}w_{x}^{\beta
}\partial_{j}v_{x}^{\alpha}\right)  \right)  \partial x_{\alpha}$

$+\sum_{i}(\sum_{\alpha}v_{x}^{\alpha}\partial_{\alpha}\left(  -\Gamma_{\beta
}^{i}w_{x}^{\beta}\right)  -w_{x}^{\alpha}\partial_{\alpha}\left(
-\Gamma_{\beta}^{i}v_{x}^{\beta}\right)  $

$+\sum_{j}\left(  -\Gamma_{\alpha}^{j}v_{x}^{\alpha}\right)  \partial
_{j}\left(  -\Gamma_{\beta}^{i}w_{x}^{\beta}\right)  -\left(  -\Gamma_{\alpha
}^{j}w_{x}^{\alpha}\right)  \partial_{j}\left(  -\Gamma_{\beta}^{i}%
v_{x}^{\beta}\right)  \partial u_{i})$

$\Phi\left[  \left(  v_{x},-\Gamma\left(  p\right)  v_{x}\right)  ,\left(
w_{x},-\Gamma\left(  p\right)  w_{x}\right)  \right]  $

$=\sum_{i}\left(  \sum_{\alpha}-v_{x}^{\alpha}\partial_{\alpha}\left(
\Gamma_{\beta}^{i}w_{x}^{\beta}\right)  +w_{x}^{\alpha}\partial_{\alpha
}\left(  \Gamma_{\beta}^{i}v_{x}^{\beta}\right)  +\sum_{j}\Gamma_{\alpha}%
^{j}v_{x}^{\alpha}\partial_{j}\left(  \Gamma_{\beta}^{i}w_{x}^{\beta}\right)
-\Gamma_{\alpha}^{j}w_{x}^{\alpha}\partial_{j}\left(  \Gamma_{\beta}^{i}%
v_{x}^{\beta}\right)  \right)  \partial u_{i}$

$+\sum_{\alpha}\Gamma_{\alpha}^{i}\left(  \sum_{\beta}v_{x}^{\beta}%
\partial_{\beta}w_{x}^{\alpha}-w_{x}^{\beta}\partial_{\beta}v_{x}^{\alpha
}\right)  \partial u_{i}$

$=\sum\left(  v_{x}^{\alpha}w_{x}^{\beta}-v_{x}^{\beta}w_{x}^{\alpha}\right)
\left(  -\partial_{\alpha}\Gamma_{\beta}^{i}+\Gamma_{\alpha}^{j}\partial
_{j}\Gamma_{\beta}^{i}\right)  \partial u_{i}$

$+\Gamma_{\alpha}^{i}\left(  w_{x}^{\beta}\partial_{\beta}v_{x}^{\alpha}%
-v_{x}^{\beta}\partial_{\beta}w_{x}^{\alpha}\right)  +\Gamma_{\alpha}%
^{i}\left(  v_{x}^{\beta}\partial_{\beta}w_{x}^{\alpha}-w_{x}^{\beta}%
\partial_{\beta}v_{x}^{\alpha}\right)  \partial u_{i}$

$=\sum\left(  v_{x}^{\alpha}w_{x}^{\beta}-v_{x}^{\beta}w_{x}^{\alpha}\right)
\left(  -\partial_{\alpha}\Gamma_{\beta}^{i}+\Gamma_{\alpha}^{j}\partial
_{j}\Gamma_{\beta}^{i}\right)  \partial u_{i}$

$\Omega=-\partial_{\alpha}\Gamma_{\beta}^{i}dx^{\alpha}\otimes dx^{\beta
}\otimes\partial u_{i}+\partial_{\alpha}\Gamma_{\beta}^{i}dx^{\beta}\otimes
dx^{\alpha}\otimes\partial u_{i}+\Gamma_{\alpha}^{j}\partial_{j}\Gamma_{\beta
}^{i}dx^{\alpha}\otimes dx^{\beta}\otimes\partial u_{i}-\Gamma_{\alpha}%
^{j}\partial_{j}\Gamma_{\beta}^{i}dx^{\beta}\otimes dx^{\alpha}\otimes\partial
u_{i}$

$\Omega=\sum_{\alpha\beta}\left(  -\partial_{\alpha}\Gamma_{\beta}^{i}%
+\Gamma_{\alpha}^{j}\partial_{j}\Gamma_{\beta}^{i}\right)  dx^{\alpha}\wedge
dx^{\beta}\otimes\partial u_{i}$

The sign - on the first term comes from the convention in the definition of
$\Gamma.$
\end{proof}

\bigskip

\begin{theorem}
For any vector fields \ X,Y on M :

$\nabla_{X}\circ\nabla_{Y}-\nabla_{Y}\circ\nabla_{X}=\nabla_{\left[
X,Y\right]  }+\pounds _{\Omega\left(  \chi_{L}\left(  X\right)  ,\chi
_{L}\left(  Y\right)  \right)  }$

$\left[  \chi_{L}\left(  X\right)  ,\chi_{L}\left(  Y\right)  \right]  _{TE}$
is a projectable vector field :

$\pi_{\ast}\left(  \left[  \chi_{L}\left(  X\right)  ,\chi_{L}\left(
Y\right)  \right]  _{TE}\right)  \left(  \pi\left(  p\right)  \right)
=\left[  \pi_{\ast}\chi_{L}\left(  X\right)  ,\pi_{\ast}\chi_{L}\left(
Y\right)  \right]  =\left[  X,Y\right]  $

$\Omega\left(  p\right)  \left(  \chi_{L}\left(  X\right)  ,\chi_{L}\left(
Y\right)  \right)  =\left[  \chi_{L}\left(  X\right)  ,\chi_{L}\left(
Y\right)  \right]  _{TE}-\chi_{L}\left(  \left[  X,Y\right]  _{TM}\right)  $
\end{theorem}

\bigskip

\begin{proof}
$\left[  \chi_{L}\left(  X\right)  ,\chi_{L}\left(  Y\right)  \right]
=\left[  \varphi_{a}^{\prime}(x,u)\left(  X\left(  x\right)  ,-\Gamma\left(
p\right)  X\left(  x\right)  \right)  ,\varphi_{a}^{\prime}(x,u)\left(
Y\left(  x\right)  ,-\Gamma\left(  p\right)  Y\left(  x\right)  \right)
\right]  $

The same computation as above gives :

$\left[  \chi_{L}\left(  X\right)  ,\chi_{L}\left(  Y\right)  \right]
=\sum_{\alpha}\left(  \sum_{\beta}X^{\beta}\partial_{\beta}Y^{\alpha}%
-Y^{\beta}\partial_{\beta}X^{\alpha}\right)  \partial x_{\alpha}$

$+\sum_{i}(\sum_{\alpha\beta}X^{\alpha}\partial_{\alpha}\left(  -\Gamma
_{\beta}^{i}Y^{\beta}\right)  -Y^{\alpha}\partial_{\alpha}\left(
-\Gamma_{\beta}^{i}X^{\beta}\right)  $

$+\sum_{j\alpha\beta}\left(  -\Gamma_{\alpha}^{j}X^{\alpha}\right)
\partial_{j}\left(  -\Gamma_{\beta}^{i}Y^{\beta}\right)  -\left(
-\Gamma_{\alpha}^{j}Y^{\alpha}\right)  \partial_{j}\left(  -\Gamma_{\beta}%
^{i}X^{\beta}\right)  \partial u_{i})$

$=\sum_{\alpha}\left(  \sum_{\beta}X^{\beta}\partial_{\beta}Y^{\alpha
}-Y^{\beta}\partial_{\beta}X^{\alpha}\right)  \partial x_{\alpha}$

$+\sum_{i}(\sum_{\alpha\beta}-X^{\alpha}Y^{\beta}\partial_{\alpha}%
\Gamma_{\beta}^{i}-X^{\alpha}\Gamma_{\beta}^{i}\partial_{\alpha}Y^{\beta
}+Y^{\alpha}X^{\beta}\partial_{\alpha}\Gamma_{\beta}^{i}+\Gamma_{\beta}%
^{i}Y^{\alpha}\partial_{\alpha}X^{\beta}$

$+\sum_{j}\Gamma_{\alpha}^{j}X^{\alpha}Y^{\beta}\partial_{j}\Gamma_{\beta}%
^{i}-\Gamma_{\alpha}^{j}Y^{\alpha}X^{\beta}\partial_{j}\Gamma_{\beta}%
^{i}\partial u_{i})$

$=\sum_{\alpha}\left(  \sum_{\beta}X^{\beta}\partial_{\beta}Y^{\alpha
}-Y^{\beta}\partial_{\beta}X^{\alpha}\right)  \left(  \partial x_{\alpha}%
-\sum_{i}\Gamma_{\alpha}^{i}\partial u_{i}\right)  $

$+\sum_{i}\sum_{\alpha\beta}\left(  X^{\alpha}Y^{\beta}-Y^{\alpha}X^{\beta
}\right)  \left(  -\partial_{\alpha}\Gamma_{\beta}^{i}+\sum_{j}\Gamma_{\alpha
}^{j}\partial_{j}\Gamma_{\beta}^{i}\right)  \partial u_{i}$

$=\chi_{L}\left(  \left[  X,Y\right]  \right)  +\Omega\left(  \chi_{L}\left(
X\right)  ,\chi_{L}\left(  Y\right)  \right)  $

Moreover for any projectable vector fields W,U :

$\pounds _{\left[  W,U\right]  }S=\pounds _{W}\circ\pounds _{U}S-\pounds _{U}%
\circ\pounds _{W}S$

So : $\pounds _{\Omega\left(  \chi_{L}\left(  X\right)  ,\chi_{L}\left(
Y\right)  \right)  +\chi_{L}\left(  p\right)  \left(  \left[  X,Y\right]
_{TM}\right)  }$

$=\pounds _{\chi_{L}\left(  X\right)  }\circ\pounds _{\chi_{L}\left(
Y\right)  }-\pounds _{\chi_{L}\left(  X\right)  }\circ\pounds _{\chi
_{L}\left(  Y\right)  }=\pounds _{\Omega\left(  \chi_{L}\left(  X\right)
,\chi_{L}\left(  Y\right)  \right)  }+\pounds _{\chi_{L}\left(  p\right)
\left(  \left[  X,Y\right]  _{TM}\right)  }$

$\Omega\left(  \chi_{L}\left(  X\right)  ,\chi_{L}\left(  Y\right)  \right)  $
is a vertical vector field, so projectable in 0

$\nabla_{X}\circ\nabla_{Y}-\nabla_{Y}\circ\nabla_{X}=\nabla_{\left[
X,Y\right]  }+\pounds _{\Omega\left(  \chi_{L}\left(  X\right)  ,\chi
_{L}\left(  Y\right)  \right)  }$
\end{proof}

So $\Omega\circ\chi_{L}$ is a 2-form on M :

$\Omega\circ\chi_{L}:TM\times TM\rightarrow TE::$

$\Omega\left(  p\right)  \left(  \chi_{L}\left(  X\right)  ,\chi_{L}\left(
Y\right)  \right)  =\left[  \chi_{L}\left(  X\right)  ,\chi_{L}\left(
Y\right)  \right]  _{TE}-\chi_{L}\left(  p\right)  \left(  \left[  X,Y\right]
_{TM}\right)  $

which measures the obstruction against lifting the commutator of vectors.

The horizontal bundle HE is integrable if the connection has a null curvature
(Kolar p.79).

\bigskip

\subsection{Connections on vectors bundles}

\label{Connections on vector bundles}

The main feature of vector bundles is the vector space structure of E itself,
and the linearity of the transition maps.\ So the connections specific to
vector bundles are linear connections.

\subsubsection{Linear connection}

Each fiber E(x) has the structure of a vector space and in an atlas $\left(
O_{a},\varphi_{a}\right)  _{a\in A}$ of E the chart $\varphi_{a}$ is linear
with respect to u.

The tangent bundle of the vector bundle $E\left(  M,V,\pi\right)  $ is the
vector bundle $TE\left(  TM,V\times V,T\pi\right)  $. With a holonomic basis
$\left(  \mathbf{e}_{i}\left(  x\right)  \right)  _{i\in I}$ a vector
$v_{p}\in T_{p}E$ reads : $v_{p}=\sum_{\alpha\in A}v_{x}^{\alpha}\partial
x^{\alpha}+\sum_{i\in I}v_{u}^{i}\mathbf{e}_{i}\left(  x\right)  $ and the
vertical bundle is isomorphic to ExE : $VE\left(  M,V,\pi\right)  \simeq
E\times_{M}E$.

A connection $\Phi\in\Lambda_{1}\left(  E;VE\right)  $ is defined by a family
of Chrisfoffel forms $\Gamma_{a}\in C\left(  \pi^{-1}\left(  O_{a}\right)
;TM^{\ast}\otimes V\right)  :\Gamma_{a}\left(  p\right)  \in\Lambda_{1}\left(
M;V\right)  $

$\Phi$ reads in this basis : $\Phi(p)\left(  v_{x}^{\alpha}\partial x_{\alpha
}+v_{u}^{i}\mathbf{e}_{i}\left(  x\right)  \right)  =\sum_{i}\left(  v_{u}%
^{i}+\sum_{\alpha}\Gamma\left(  p\right)  _{\alpha}^{i}v_{x}^{\alpha}\right)
\mathbf{e}_{i}\left(  x\right)  $

\begin{definition}
A \textbf{linear connection} $\Phi$ on a vector bundle is a connection such
that its Christoffel forms are linear with respect to the vector space
structure of each fiber :
\end{definition}

$\forall L\in%
\mathcal{L}%
\left(  V;V\right)  ,v_{x}\in T_{x}M$ : $\Gamma_{a}\left(  \varphi_{a}\left(
x,L\left(  u_{a}\right)  \right)  \right)  v_{x}=L\circ\left(  \Gamma
_{a}\left(  \varphi\left(  x,u_{a}\right)  \right)  v_{x}\right)  $

A linear connection can then be defined by maps with domain in M.

\begin{theorem}
A linear connection $\Phi$ on a vector bundle $E\left(  M,V,\pi\right)  $ with
atlas $\left(  O_{a},\varphi_{a}\right)  _{a\in A}$\ is uniquely defined by a
family $\left(  \Gamma_{a}\right)  _{a\in A}$ of 1-form on M valued in
$\mathcal{L}$%
(V;V) : $\Gamma_{a}\in\Lambda_{1}\left(  O_{a};%
\mathcal{L}%
\left(  V;V\right)  \right)  $ such that at the transitions%

\begin{equation}
\Gamma_{b}\left(  x\right)  =-\varphi_{ba}^{\prime}\left(  x\right)
\varphi_{ab}\left(  x\right)  +\Gamma_{a}\left(  x\right)
\end{equation}

\end{theorem}

\begin{proof}
i) Let $\Phi$ be a linear connection :

$\Phi(p)\left(  v_{x}^{\alpha}\partial x_{\alpha}+v_{u}^{i}e_{i}\left(
x\right)  \right)  =\sum_{i}\left(  v_{u}^{i}+\sum_{\alpha}\widehat{\Gamma
}\left(  p\right)  _{\alpha}^{i}v_{x}^{\alpha}\right)  e_{i}\left(  x\right)
$

At the transitions :

$x\in O_{a}\cap O_{b}:p=\varphi_{b}\left(  x,u_{b}\right)  =\varphi_{a}\left(
x,u_{a}\right)  $

$\widehat{\Gamma}_{b}\left(  p\right)  =\varphi_{ba}^{\prime}\left(
x,u_{a}\right)  \circ\left(  -Id_{TM},\widehat{\Gamma}_{a}\left(  p\right)
\right)  =-\varphi_{ba}^{\prime}\left(  x\right)  u_{a}+\varphi_{ba}\left(
x\right)  \widehat{\Gamma}_{a}\left(  p\right)  $

$\widehat{\Gamma}_{b}\left(  \varphi_{b}\left(  x,u_{b}\right)  \right)
=-\varphi_{ba}^{\prime}\left(  x\right)  u_{a}+\varphi_{ba}\left(  x\right)
\widehat{\Gamma}_{a}\left(  \varphi_{a}\left(  x,u_{a}\right)  \right)  $

$=-\varphi_{ba}^{\prime}\left(  x\right)  u_{a}+\widehat{\Gamma}_{a}\left(
\varphi_{a}\left(  x,\varphi_{ba}\left(  x\right)  u_{a}\right)  \right)
=-\varphi_{ba}^{\prime}\left(  x\right)  u_{a}+\widehat{\Gamma}_{a}\left(
\varphi_{a}\left(  x,u_{b}\right)  \right)  $

Define : $\Gamma_{a}\in\Lambda_{1}\left(  O_{a};%
\mathcal{L}%
\left(  V;V\right)  \right)  ::\Gamma_{a}\left(  x\right)  \left(
v_{x}\right)  \left(  u\right)  =\widehat{\Gamma}_{a}\left(  \varphi
_{a}\left(  x,u\right)  \right)  v_{x}$

$\widehat{\Gamma}_{b}\left(  \varphi_{b}\left(  x,u_{b}\right)  \right)
=\Gamma_{b}\left(  x\right)  \left(  u_{b}\right)  =-\varphi_{ba}^{\prime
}\left(  x\right)  \varphi_{ab}\left(  x\right)  u_{b}+\Gamma_{a}\left(
x\right)  \left(  u_{b}\right)  $

$\Gamma_{b}\left(  x\right)  =-\varphi_{ba}^{\prime}\left(  x\right)
\varphi_{ab}\left(  x\right)  +\Gamma_{a}\left(  x\right)  $

ii) Conversely if there is a family

$\Gamma_{a}\in\Lambda_{1}\left(  O_{a};%
\mathcal{L}%
\left(  V;V\right)  \right)  ::\Gamma_{b}\left(  x\right)  =-\varphi
_{ba}^{\prime}\left(  x\right)  \varphi_{ab}\left(  x\right)  +\Gamma
_{a}\left(  x\right)  $

$\widehat{\Gamma}_{a}\left(  \varphi_{a}\left(  x,u\right)  \right)
v_{x}=\Gamma_{a}\left(  x\right)  \left(  v_{x}\right)  \left(  u\right)  $
defines a family of linear Christoffel forms

At the transitions :

$\widehat{\Gamma}_{b}\left(  \varphi_{b}\left(  x,u_{b}\right)  \right)
v_{x}=\Gamma_{b}\left(  x\right)  \left(  v_{x}\right)  \left(  u_{b}\right)
=-\varphi_{ba}^{\prime}\left(  x\right)  v_{x}\varphi_{ab}\left(  x\right)
u_{b}+\left(  \Gamma_{a}\left(  x\right)  v_{x}\right)  u_{b}$

$=-\varphi_{ba}^{\prime}\left(  x\right)  v_{x}u_{a}+\left(  \Gamma_{a}\left(
x\right)  v_{x}\right)  \varphi_{ba}\left(  x\right)  u_{a}=\left(
\varphi_{ba}\left(  x\right)  u_{a}\right)  ^{\prime}\left(  -v_{x}%
,\widehat{\Gamma}_{a}\left(  \varphi_{a}\left(  x,u_{a}\right)  \right)
v_{x}\right)  $
\end{proof}

In a holonomic basis $\mathbf{e}_{ai}\left(  x\right)  =\varphi_{a}\left(
x,e_{i}\right)  $ the maps $\Gamma$ still called Christoffel forms are :

$\Gamma\left(  x\right)  =\sum_{ij\alpha}\Gamma_{\alpha j}^{i}\left(
x\right)  d\xi^{\alpha}\otimes\mathbf{e}_{i}\left(  x\right)  \otimes
\mathbf{e}^{j}\left(  x\right)  $

$\leftrightarrow\widehat{\Gamma}\left(  \varphi\left(  x,u\right)  \right)
=\sum_{ij\alpha}\Gamma_{\alpha j}^{i}\left(  x\right)  u^{j}d\xi^{\alpha
}\otimes\mathbf{e}_{i}\left(  x\right)  $%

\begin{equation}
\Phi(\varphi\left(  x,\sum_{i\in I}u^{i}e_{i}\right)  )\left(  v_{x}^{\alpha
}\partial x_{\alpha}+v_{u}^{i}\mathbf{e}_{i}\left(  x\right)  \right)
=\sum_{i}\left(  v_{u}^{i}+\sum_{j\alpha}\Gamma_{\alpha j}^{i}\left(
x\right)  u^{j}v_{x}^{\alpha}\right)  \mathbf{e}_{i}\left(  x\right)
\end{equation}

In a change of trivialization on E :

$p=\varphi_{a}\left(  x,u_{a}\right)  =\widetilde{\varphi}_{a}\left(
x,\chi_{a}\left(  x\right)  \left(  u_{a}\right)  \right)  \Leftrightarrow
\widetilde{u}_{a}=\chi_{a}\left(  x\right)  u_{a}$

$\Gamma_{a}\left(  x\right)  \rightarrow\widetilde{\Gamma}_{a}\left(
x\right)  =-\chi_{a}^{\prime}\left(  x\right)  \chi_{a}\left(  x\right)
^{-1}+\Gamma_{a}\left(  x\right)  $

The \textbf{horizontal form} of the linear connection $\Phi$\ is the 1 for:

$\chi\in\Lambda_{1}\left(  E;HE\right)  :\chi\left(  p\right)  =Id_{TE}%
-\Phi\left(  p\right)  $

$\chi\left(  p\right)  (\varphi\left(  x,\sum_{i\in I}u^{i}e_{i}\right)
)\left(  v_{x}^{\alpha}\partial x_{\alpha}+v_{u}^{i}\mathbf{e}_{i}\left(
x\right)  \right)  =\sum_{i}\left(  v_{u}^{i}+\sum_{j\alpha}\Gamma_{\alpha
j}^{i}\left(  x\right)  u^{j}v_{x}^{\alpha}\right)  \mathbf{e}_{i}\left(
x\right)  $

The horizontal lift of a vector on TM is :

$\chi_{L}\left(  \varphi\left(  x,u\right)  \right)  \left(  v_{x}\right)
=\varphi^{\prime}(x,u)\left(  v_{x},-\Gamma\left(  p\right)  v_{x}\right)
=\sum_{\alpha}v_{x}^{\alpha}\partial x^{\alpha}-\sum_{i\alpha}\Gamma_{j\alpha
}^{i}\left(  x\right)  u^{j}v_{x}^{\alpha}\mathbf{e}_{i}\left(  x\right)  $

\subsubsection{Curvature}

\begin{theorem}
For any linear connection $\Phi$ defined by the Christoffel form $\Gamma$ on
the vector bundle E, there is a 2-form $\Omega\in\Lambda_{2}\left(  E;E\otimes
E^{\ast}\right)  $ such that the curvature form $\widehat{\Omega} $\ of $\Phi$ :

$\widehat{\Omega}\left(  \sum_{i}u^{i}e_{i}\left(  x\right)  \right)
=\sum_{j}u^{j}\Omega_{j}\left(  x\right)  $

in a holonomic basis of E

$\Omega\left(  x\right)  =\sum_{\alpha\beta j}\left(  -\partial_{\alpha}%
\Gamma_{j\beta}^{i}\left(  x\right)  +\sum_{k}\Gamma_{j\alpha}^{k}\left(
x\right)  \Gamma_{k\beta}^{i}\left(  x\right)  \right)  dx^{\alpha}\wedge
dx^{\beta}\otimes\mathbf{e}_{i}\left(  x\right)  \otimes\mathbf{e}^{j}\left(
x\right)  $,
\end{theorem}

\begin{proof}
The Cartan formula gives for the curvature with $\partial u_{i}=\mathbf{e}%
_{i}\left(  x\right)  $ in a holonomic basis :

$\widehat{\Omega}\left(  p\right)  =\sum_{\alpha\beta}\left(  -\partial
_{\alpha}\widehat{\Gamma}_{\beta}^{i}+\widehat{\Gamma}_{\alpha}^{j}%
\partial_{j}\widehat{\Gamma}_{\beta}^{i}\right)  dx^{\alpha}\wedge dx^{\beta
}\otimes\mathbf{e}_{i}\left(  x\right)  $

So with a linear connection :

$\widehat{\Gamma}\left(  \varphi\left(  x,\sum_{i\in I}u^{i}e_{i}\right)
\right)  =\sum_{i,j\in I}\Gamma_{\beta j}^{i}\left(  x\right)  u^{j}%
d\xi^{\beta}\otimes\mathbf{e}_{i}\left(  x\right)  $

$\partial_{\alpha}\widehat{\Gamma}_{\beta}^{i}\left(  \varphi\left(
x,\sum_{i\in I}u^{i}e_{i}\right)  \right)  =\sum_{k\in I}u^{k}\partial
_{\alpha}\Gamma_{\beta k}^{i}\left(  x\right)  $

$\partial_{j}\widehat{\Gamma}_{\beta}^{i}\left(  \varphi\left(  x,\sum_{i\in
I}u^{i}e_{i}\right)  \right)  =\Gamma_{\beta j}^{i}\left(  x\right)  $

$\widehat{\Omega}\left(  p\right)  $

$=\sum_{\alpha\beta}\left(  -\sum_{j\in I}u^{j}\partial_{\alpha}\Gamma
_{j\beta}^{i}\left(  x\right)  +\sum_{j,k\in I}u^{k}\Gamma_{k\alpha}%
^{j}\left(  x\right)  \Gamma_{j\beta}^{i}\left(  x\right)  \right)
dx^{\alpha}\wedge dx^{\beta}\otimes\mathbf{e}_{i}\left(  x\right)  $

$=\sum_{\alpha\beta}\sum_{j\in I}u^{j}\left(  -\partial_{\alpha}\Gamma
_{j\beta}^{i}\left(  x\right)  +\sum_{k\in I}\Gamma_{j\alpha}^{k}\left(
x\right)  \Gamma_{k\beta}^{i}\left(  x\right)  \right)  dx^{\alpha}\wedge
dx^{\beta}\otimes\mathbf{e}_{i}\left(  x\right)  $

$\widehat{\Omega}\left(  \varphi\left(  x,u\right)  \right)  =\sum_{j}%
u^{j}\Omega_{j}\left(  x\right)  $
\end{proof}

\subsubsection{Covariant derivative}

\paragraph{Covariant derivative of sections\newline}

The covariant derivative of a section X on a vector bundle $E\left(
M,V,\pi\right)  $ is a map : $\nabla X\in\Lambda_{1}\left(  M;E\right)  \simeq
TM^{\ast}\otimes E$ . It is independant on the trivialization. It has the
following coordinates expression for a linear connection:

\begin{theorem}
The covariant derivative $\nabla$\ associated to a linear connection $\Phi$ on
a vector bundle $E\left(  M,V,\pi\right)  $ with Christoffel form $\Gamma$\ is
the map :%

\begin{equation}
\nabla:\mathfrak{X}\left(  E\right)  \rightarrow\Lambda_{1}\left(  M;E\right)
::\nabla X(x)=\sum_{\alpha i}(\partial_{\alpha}X^{i}+X^{j}\Gamma_{j\alpha}%
^{i}(x))d\xi^{\alpha}\otimes\mathbf{e}_{i}\left(  x\right)
\end{equation}

in a holonomic basis of E
\end{theorem}

\begin{proof}
A section on E is a vector field : $X:M\rightarrow E::X\left(  x\right)
=\sum_{i}X^{i}\left(  x\right)  \mathbf{e}_{i}\left(  x\right)  $ with
$\pi\left(  X\right)  =x$ .Its derivative is with a dual holonomic basis
$d\xi^{\alpha}\in T_{x}M^{\ast}:$

$X^{\prime}\left(  x\right)  :T_{x}M\rightarrow T_{X\left(  x\right)
}E::X^{\prime}\left(  x\right)  =\sum_{i,\alpha}\left(  \partial_{\alpha}%
X^{i}\left(  x\right)  \right)  \mathbf{e}_{i}\left(  x\right)  \otimes
d\xi^{\alpha}\in T_{x}M^{\ast}\otimes E\left(  x\right)  $
\end{proof}

The covariant derivative of a section \textit{on E} along a vector field
\textit{on M} is a \textit{section on E} which reads :

$\nabla_{Y}X(x)=\sum_{\alpha i}(\partial_{\alpha}X^{i}+X^{j}\Gamma_{j\alpha
}^{i}(x))Y^{\alpha}\mathbf{e}_{i}\left(  x\right)  \in\mathfrak{X}\left(
E\right)  $

With the tangent bundle TM of a manifold we get back the usual definition of
the covariant derivative (which justifies the choice of the sign before
$\Gamma$\ ).

\paragraph{Covariant derivative of tensor fields\newline}

Tensorial functors $\mathfrak{T}_{s}^{r}$ can extend a vector bundle $E\left(
M,V,\pi\right)  $ to a tensor bundle $\mathfrak{T}_{s}^{r}E\left(
M,\mathfrak{F}_{s}^{r}V,\pi\right)  .$ There are connections defined on these
vector bundles as well, but we can extend a linear connection from E to
$\mathfrak{T}_{s}^{r}E.$

If we look for a derivation D on the algebra $\mathfrak{T}_{s}^{r}E\left(
x\right)  $ at some fixed point x, with the required properties listed in
Differential geometry (covariant derivative), we can see, by the same
reasonning, that this derivation is determined by its value over a basis:

$\nabla\mathbf{e}_{i}\left(  x\right)  =\sum_{i,j,\alpha}\Gamma_{i\alpha}%
^{j}\left(  x\right)  d\xi^{\alpha}\otimes\mathbf{e}_{j}\left(  x\right)  \in
T_{x}M^{\ast}\otimes E\left(  x\right)  $

$\nabla\mathbf{e}^{i}\left(  x\right)  =-\sum_{i,j,\alpha}\Gamma_{j\alpha}%
^{i}d\xi^{\alpha}\otimes\mathbf{e}^{j}\left(  x\right)  \in T_{x}M^{\ast
}\otimes E^{\ast}\left(  x\right)  $

with the dual bases $\mathbf{e}^{i}\left(  x\right)  ,d\xi^{\alpha}$ of
$\mathbf{e}_{i}\left(  x\right)  ,\partial\xi_{\alpha}$ in the fiber E(x) and
the tangent space $T_{x}M$.

Notice that we have $d\xi^{\alpha}\in T_{x}M^{\ast}$ because the covariant
derivative acts on vector fields on M. So the covariant derivative is a map :
$\nabla:\mathfrak{T}_{s}^{r}E\rightarrow TM^{\ast}\otimes\mathfrak{T}_{s}%
^{r}E$

The other properties are the same as the usual covariant derivative on the
tensorial bundle $\otimes_{s}^{r}TM$

Linearity :

$\forall S,T\in\mathfrak{T}_{s}^{r}E,k,k^{\prime}\in K,Y,Z\in\mathfrak{X}%
\left(  TM\right)  $

$\nabla\left(  kS+k^{\prime}T\right)  =k\nabla S+k^{\prime}\nabla T$

$\nabla_{kY+k^{\prime}Z}S=k\nabla_{Y}S+k^{\prime}\nabla_{Z}S$

Leibnitz rule with respect to the tensorial product (because it is a derivation):

$\nabla\left(  S\otimes T\right)  =\left(  \nabla S\right)  \otimes
T+S\otimes\left(  \nabla T\right)  $

If $f\in C_{1}\left(  M;%
\mathbb{R}
\right)  ,X\in\mathfrak{X}\left(  TM\right)  ,Y\in\mathfrak{X}\left(
E\right)  :\nabla_{X}fY=df\left(  X\right)  Y+f\nabla_{X}Y$

Commutative with the trace operator :

$\nabla(Tr\left(  T\right)  )=Tr\left(  \nabla T\right)  $

The formulas of the covariant derivative are :

for a section on E :

$X=\sum_{i\in I}X^{i}\left(  x\right)  \mathbf{e}_{i}(x)\rightarrow\nabla
_{Y}X(x)=\sum_{\alpha i}(\partial_{\alpha}X^{i}+X^{j}\Gamma_{j\alpha}%
^{i}(x))Y^{\alpha}\mathbf{e}_{i}\left(  x\right)  $

for a section of E* :

$\varpi=\sum_{i\in I}\varpi_{i}\left(  x\right)  \mathbf{e}^{i}\left(
x\right)  \rightarrow\nabla_{Y}\varpi=\sum_{\alpha i}\left(  \partial_{\alpha
}\varpi_{i}-\Gamma_{i\alpha}^{j}\varpi_{j}\right)  Y^{\alpha}\otimes
\mathbf{e}^{i}\left(  x\right)  $

for a mix tensor, section of $\mathfrak{T}_{s}^{r}E$:

$T\left(  x\right)  =\sum_{i_{1}...i_{r}}\sum_{j_{1}....j_{s}}T_{j_{1}%
...j_{s}}^{i_{1}...i_{r}}\left(  x\right)  \mathbf{e}_{i_{1}}\left(  x\right)
\otimes..\otimes\mathbf{e}_{i_{r}}\left(  x\right)  \otimes\mathbf{e}^{j_{1}%
}\left(  x\right)  \otimes...\otimes\mathbf{e}^{j_{s}}\left(  x\right)  $

$\nabla T=\sum_{i_{1}...i_{r}}\sum_{j_{1}....j_{s}}\sum_{\alpha}\widehat
{T}_{\alpha j_{1}...j_{s}}^{i_{1}...i_{r}}d\xi^{\alpha}\otimes\mathbf{e}%
_{i_{1}}\left(  x\right)  \otimes..\otimes\mathbf{e}_{i_{r}}\left(  x\right)
\otimes\mathbf{e}^{j_{1}}\left(  x\right)  \otimes...\otimes\mathbf{e}^{j_{s}%
}\left(  x\right)  $

$\widehat{T}_{\alpha j_{1}...j_{s}}^{i_{1}...i_{r}}=\partial_{\alpha}%
T_{j_{1}...j_{s}}^{i_{1}...i_{r}}+\sum_{k=1}^{r}\sum_{m}\Gamma_{m\alpha
}^{i_{k}}T_{j_{1}...j_{s}}^{i_{1}.i_{k-1}mi_{k+1}..i_{r}}-\sum_{k=1}^{s}%
\sum_{m}\Gamma_{j_{k}\alpha}^{m}T_{j_{1}..j_{k-1}mj_{k+1}..j_{s}}%
^{i_{1}...i_{r}}$

\paragraph{Horizontal lift of a vector field\newline}

The horizontal lift of a vector field $X\in\mathfrak{X}\left(  TM\right)  $ on
a vector bundle $E\left(  M,V,\pi\right)  $ by a linear connection on HE is
the linear map :

$\chi_{L}:\mathfrak{X}\left(  TM\right)  \rightarrow\mathfrak{X}\left(
HE\right)  ::\chi_{L}\left(  \mathbf{e}_{i}\left(  x\right)  \right)  \left(
X\right)  =\sum_{i\alpha}\left(  \partial x_{\alpha}-\Gamma_{\alpha i}%
^{j}\left(  x\right)  \mathbf{e}_{j}\left(  x\right)  \right)  X^{\alpha
}\left(  x\right)  $ in a holonomic basis of TE

This is a projectable vector field $\pi^{\prime}\left(  \mathbf{e}_{i}\left(
x\right)  \right)  \left(  \chi_{L}\left(  \mathbf{e}_{i}\left(  x\right)
\right)  \left(  X\right)  \right)  =X$ and for any $S\in\mathfrak{X}\left(
E\right)  :\nabla_{X}S=\pounds _{\chi_{L}\left(  X\right)  }S$

Notice that this is a lift to HE and not E. Indeed a lift of a vector field X
in TM on E is given by the covariant derivative $\nabla_{X}S$ of a section S
on E along X.

\paragraph{Lift of a curve\newline}

\begin{theorem}
For any path $c\in C_{1}\left(  \left[  a;b\right]  ;M\right)  $ with
$0\in\left[  a;b\right]  $\ there is a unique path $C\in C_{1}\left(  \left[
a;b\right]  ;E\right)  $\ with $C(0)=\pi^{-1}\left(  c\left(  0\right)
\right)  $\ lifted on the vector bundle $E\left(  M,V,\pi\right)  $ by the
linear connection with Christoffel form $\Gamma$ such that :

$\nabla_{c^{\prime}(t)}C=0,\pi\left(  C\left(  t\right)  \right)  =c\left(
t\right)  $
\end{theorem}

\begin{proof}
C is defined in a holonomic basis by :

$C\left(  t\right)  =\sum_{i}C^{i}\left(  t\right)  \mathbf{e}_{i}\left(
c\left(  t\right)  \right)  $

$\nabla_{c^{\prime}(t)}C=\sum_{i}\left(  \frac{dC^{i}}{dt}+\sum_{ij\alpha
}\Gamma_{j\alpha}^{i}\left(  c\left(  t\right)  \right)  C^{j}\left(
t\right)  \frac{dc}{dt}\right)  \mathbf{e}_{i}\left(  c\left(  t\right)
\right)  =0$

$\forall i:\frac{dC^{i}}{dt}+\sum_{j\alpha}\Gamma_{j\alpha}^{i}\left(
c\left(  t\right)  \right)  C^{j}\left(  t\right)  \frac{dc}{dt}=0$

$C(0)=C_{0}=\sum_{i}C^{i}\left(  t\right)  \mathbf{e}_{i}\left(  c\left(
0\right)  \right)  $

This is a linear ODE which has a solution.
\end{proof}

Notice that this is a lift of a curve on M to a curve in E (and not TE).

\subsubsection{Exterior covariant derivative}

\paragraph{Exterior covariant derivative\newline}

In differential geometry one defines the exterior covariant derivative of a
r-form $\varpi$\ on M valued in the tangent bundle TM : $\varpi\in\Lambda
_{r}\left(  M;TM\right)  .$ There is an exterior covariant derivative of r
forms \textit{on M} valued \textit{in a vector bundle E} (and not TE).

A r-form $\varpi$\ \textit{on M} valued \textit{in the vector bundle E} :
$\varpi\in\Lambda_{r}\left(  M;E\right)  $ reads in an holonomic basis of M
and a basis of E:

$\varpi=\sum_{i\left\{  \alpha_{1}...\alpha_{r}\right\}  }\varpi_{\alpha
_{1}...\alpha_{r}}^{i}\left(  x\right)  d\xi^{\alpha_{1}}\wedge d\xi
^{\alpha_{2}}\wedge...\wedge d\xi^{\alpha_{r}}\otimes\mathbf{e}_{i}\left(
x\right)  $

\begin{definition}
The exterior covariant derivative $\nabla_{e}$\ of r-forms $\varpi
$\ \textit{on M} valued in the vector bundle $E\left(  M,V,\pi\right)  $\ , is
a map : $\nabla_{e}:\Lambda_{r}\left(  M;E\right)  \rightarrow\Lambda
_{r+1}\left(  M;E\right)  $ . For a linear connection with Christoffel form
$\Gamma$ \ it is given in a holonomic basis by the formula:
\begin{equation}
\nabla_{e}\varpi=\sum_{i}\left(  d\varpi^{i}+\left(  \sum_{j}\left(
\sum_{\alpha}\Gamma_{j\alpha}^{i}d\xi^{\alpha}\right)  \wedge\varpi
^{j}\right)  \right)  \otimes\mathbf{e}_{i}\left(  x\right)
\end{equation}

\end{definition}

the exterior differential $d\varpi^{i}$\ is taken on M.

\begin{theorem}
(Kolar p.112) The exterior covariant derivative $\nabla_{e}$ on a vector
bundle $E\left(  M,V,\pi\right)  $\ with linear covariant derivative $\nabla$
has the following properties :

i) if $\varpi\in\Lambda_{0}\left(  M;E\right)  :\nabla_{e}\varpi=\nabla\varpi$
(we have the usual covariant derivative of a section on E)

ii) the exterior covariant derivative is the only map :

$\nabla_{e}:\Lambda_{r}\left(  M;E\right)  \rightarrow\Lambda_{r+1}\left(
M;E\right)  $ such that :

$\forall\mu_{r}\in\Lambda_{r}\left(  M;%
\mathbb{R}
\right)  ,\forall\varpi_{s}\in\Lambda_{s}\left(  M;E\right)  :$

$\nabla_{e}\left(  \mu_{r}\wedge\varpi_{s}\right)  =\left(  d\mu_{r}\right)
\wedge\varpi_{s}+\left(  -1\right)  ^{r}\mu_{r}\wedge\nabla_{e}\varpi_{s}$

iii) if $f\in C_{\infty}\left(  N;M\right)  ,\varpi\in\Lambda_{r}\left(
N;f^{\ast}E\right)  :\nabla_{e}\left(  f^{\ast}\varpi\right)  =f^{\ast}\left(
\nabla_{e}\varpi\right)  $
\end{theorem}

Accounting for the last property, we can implement the covariant exterior
derivative for $\varpi\in\Lambda_{r}\left(  N;E\right)  ,$ meaning when the
base of E is not N.

\paragraph{Riemann curvature\newline}

\begin{definition}
The \textbf{Riemann curvature} of a linear connection $\Phi$\ on a vector
bundle $E\left(  M,V,\pi\right)  $ with the covariant derivative $\nabla$\ is
the map :%

\begin{equation}
\mathfrak{X}\left(  TM\right)  ^{2}\times\mathfrak{X}\left(  E\right)
\rightarrow\mathfrak{X}\left(  E\right)  ::R(Y_{1},Y_{2})X=\nabla_{Y_{1}%
}\nabla_{Y_{2}}X-\nabla_{Y_{2}}\nabla_{Y_{1}}X-\nabla_{\left[  Y_{1}%
,Y_{2}\right]  }X
\end{equation}

\end{definition}

The formula makes sense : $\nabla:\mathfrak{X}\left(  E\right)  \rightarrow
\Lambda_{1}\left(  M;E\right)  $ so $\nabla_{Y}X\in\mathfrak{X}\left(
E\right)  $ and $\nabla_{Y_{1}}\left(  \nabla_{Y_{2}}X\right)  \in
\mathfrak{X}\left(  E\right)  $

If $Y_{1}=\partial\xi_{\alpha},Y_{2}=\partial\xi_{\beta}$ then $\left[
\partial\xi_{\alpha},\partial\xi_{\beta}\right]  =0$ and

$R(\partial\xi_{\alpha},\partial\xi_{\beta},X)=\left(  \nabla_{\partial
\xi_{\alpha}}\nabla_{\partial\xi_{\beta}}-\nabla_{\partial\xi_{\beta}}%
\nabla_{\partial\xi_{\alpha}}\right)  X$

so R is a measure of the obstruction of the covariant derivative to be
commutative. The name is inspired by the corresponding object on manifolds.

\bigskip

\begin{theorem}
The Riemann curvature is a tensor valued in the tangent bundle : $R\in
\Lambda_{2}\left(  M;E\otimes E^{\prime}\right)  $

$R=\sum_{\left\{  \alpha\beta\right\}  }\sum_{ij}R_{j\alpha\beta}^{i}%
d\xi^{\alpha}\wedge d\xi^{\beta}\otimes\mathbf{e}^{j}\left(  x\right)
\otimes\mathbf{e}_{i}\left(  x\right)  $ and%

\begin{equation}
R_{j\alpha\beta}^{i}=\partial_{\alpha}\Gamma_{j\beta}^{i}-\partial_{\beta
}\Gamma_{j\alpha}^{i}+\sum_{k}\left(  \Gamma_{k\alpha}^{i}\Gamma_{j\beta}%
^{k}-\Gamma_{k\beta}^{i}\Gamma_{j\alpha}^{k}\right)
\end{equation}

\end{theorem}

\begin{proof}
This is a straightforward computation similar to the one given for the
curvature in Differential Geometry
\end{proof}

\bigskip

\begin{theorem}
For any r-form $\varpi$\ \textit{on M} valued in the vector bundle $E\left(
M,V,\pi\right)  $\ endowed with a linear connection and covariant derivative
$\nabla$\ :%

\begin{equation}
\nabla_{e}\left(  \nabla_{e}\varpi\right)  =R\wedge\varpi
\end{equation}

where R is the Riemann curvature tensor
\end{theorem}

More precisely in a holonomic basis :

$\nabla_{e}\left(  \nabla_{e}\varpi\right)  =\sum_{ij}\left(  \sum
_{\alpha\beta}R_{j\alpha\beta}^{i}d\xi^{\alpha}\wedge d\xi^{\beta}\right)
\wedge\varpi^{j}\otimes\mathbf{e}_{i}\left(  x\right)  $

Where $R=\sum_{\alpha\beta}\sum_{ij}R_{j\alpha\beta}^{i}d\xi^{\alpha}\wedge
d\xi^{\beta}\otimes e^{j}\left(  x\right)  \otimes\mathbf{e}_{i}\left(
x\right)  $

and $R_{j\alpha\beta}^{i}=\partial_{\alpha}\Gamma_{j\beta}^{i}+\sum_{k}%
\Gamma_{k\alpha}^{i}\Gamma_{j\beta}^{k}$

\begin{proof}
$\nabla_{e}\varpi=\sum_{i}\left(  d\varpi^{i}+\sum_{j}\Omega_{j}^{i}%
\wedge\varpi^{j}\right)  \otimes\mathbf{e}_{i}\left(  x\right)  $ with
$\Omega_{j}^{i}=\sum_{\alpha}\Gamma_{j\alpha}^{i}d\xi^{\alpha}$

$\nabla_{e}\left(  \nabla_{e}\varpi\right)  =\sum_{i}\left(  d\left(
\nabla_{e}\varpi\right)  ^{i}+\sum_{j}\Omega_{j}^{i}\wedge\left(  \nabla
_{e}\varpi\right)  ^{j}\right)  \otimes\mathbf{e}_{i}\left(  x\right)  $

$=\sum_{i}\left(  d\left(  d\varpi^{i}+\sum_{j}\Omega_{j}^{i}\wedge\varpi
^{j}\right)  +\sum_{j}\Omega_{j}^{i}\wedge\left(  d\varpi^{j}+\sum_{k}%
\Omega_{k}^{j}\wedge\varpi^{k}\right)  \right)  \otimes\mathbf{e}_{i}\left(
x\right)  $

$=\sum_{ij}\left(  d\Omega_{j}^{i}\wedge\varpi^{j}-\Omega_{j}^{i}\wedge
d\varpi^{j}+\Omega_{j}^{i}\wedge d\varpi^{j}+\Omega_{j}^{i}\wedge\sum
_{k}\Omega_{k}^{j}\wedge\varpi^{k}\right)  \otimes\mathbf{e}_{i}\left(
x\right)  $

$=\sum_{ij}\left(  d\Omega_{j}^{i}\wedge\varpi^{j}+\sum_{k}\Omega_{k}%
^{i}\wedge\Omega_{j}^{k}\wedge\varpi^{j}\right)  \otimes\mathbf{e}_{i}\left(
x\right)  $

$\nabla_{e}\left(  \nabla_{e}\varpi\right)  =\sum_{ij}\left(  d\Omega_{j}%
^{i}+\sum_{k}\Omega_{k}^{i}\wedge\Omega_{j}^{k}\right)  \wedge\varpi
^{j}\otimes\mathbf{e}_{i}\left(  x\right)  $

$d\Omega_{j}^{i}+\sum_{k}\Omega_{k}^{i}\wedge\Omega_{j}^{k}=d\left(
\sum_{\alpha}\Gamma_{j\alpha}^{i}d\xi^{\alpha}\right)  +\sum_{k}\left(
\sum_{\alpha}\Gamma_{k\alpha}^{i}d\xi^{\alpha}\right)  \wedge\left(
\sum_{\beta}\Gamma_{j\alpha}^{k}d\xi^{\beta}\right)  $

$=\sum_{\alpha\beta}\left(  \partial_{\beta}\Gamma_{j\alpha}^{i}d\xi^{\beta
}\wedge d\xi^{\alpha}+\sum_{k}\Gamma_{k\alpha}^{i}\Gamma_{j\beta}^{k}%
d\xi^{\alpha}\wedge d\xi^{\beta}\right)  $

$=\sum_{\alpha\beta}\left(  \partial_{\alpha}\Gamma_{j\beta}^{i}+\sum
_{k}\Gamma_{k\alpha}^{i}\Gamma_{j\beta}^{k}\right)  d\xi^{\alpha}\wedge
d\xi^{\beta}$
\end{proof}

\bigskip

\begin{theorem}
The commutator of vector fields on M lifts to E iff R=0.
\end{theorem}

\begin{proof}
We have for any connection with $Y_{1},Y_{2}\in\mathfrak{X}\left(  TM\right)
,X\in\mathfrak{X}\left(  E\right)  $

$\nabla_{Y_{1}}\circ\nabla_{Y_{2}}X-\nabla_{Y_{2}}\circ\nabla_{Y_{1}}%
X=\nabla_{\left[  Y_{1},Y_{2}\right]  }X+\pounds _{\Omega\left(  \chi
_{L}\left(  Y_{1}\right)  ,\chi_{L}\left(  Y_{2}\right)  \right)  }X$

So : $R(Y_{1},Y_{2})X=\pounds _{\Omega\left(  \chi_{L}\left(  X\right)
,\chi_{L}\left(  Y\right)  \right)  }X=$

$R\left(  \varphi\left(  x,\sum_{i}u^{i}e_{i}\right)  \right)  (Y_{1}%
,Y_{2})X=\sum_{i}u^{i}\pounds _{\widehat{\Omega}\left(  \chi_{L}\left(
Y_{1}\right)  ,\chi_{L}\left(  Y_{2}\right)  \right)  e_{i}\left(  x\right)
}X$

$R\left(  e_{i}\left(  x\right)  \right)  (Y_{1},Y_{2})X=\pounds _{\widehat
{\Omega}\left(  \chi_{L}\left(  Y_{1}\right)  ,\chi_{L}\left(  Y_{2}\right)
\right)  e_{i}\left(  x\right)  }X$
\end{proof}

\bigskip

\begin{theorem}
The exterior covariant derivative of $\nabla$ is :

$\nabla_{e}\left(  \nabla X\right)  =\sum_{\left\{  \alpha\beta\right\}
}R_{j\alpha\beta}^{i}X^{j}d\xi^{\alpha}\wedge d\xi^{\beta}\otimes
\mathbf{e}_{i}\left(  x\right)  $
\end{theorem}

\begin{proof}
The covariant derivative $\nabla$ is a 1 form valued in E, so we can compute
its exterior covariant derivative :

$\nabla X=\sum_{\alpha i}(\partial_{\alpha}X^{i}+X^{j}\Gamma_{j\alpha}%
^{i}(x))d\xi^{\alpha}\otimes\mathbf{e}_{i}\left(  x\right)  $

$\nabla_{e}\left(  \nabla X\right)  =\sum_{i}(d\left(  \sum_{\alpha}%
(\partial_{\alpha}X^{i}+X^{j}\Gamma_{j\alpha}^{i}(x))d\xi^{\alpha}\right)  $

$+\sum_{j}\left(  \sum_{\beta}\Gamma_{j\beta}^{i}d\xi^{\beta}\right)
\wedge\left(  \sum_{\alpha i}(\partial_{\alpha}X^{j}+X^{k}\Gamma_{k\alpha}%
^{j}(x))d\xi^{\alpha}\right)  )\otimes\mathbf{e}_{i}\left(  x\right)  $

$=\sum\left(  \partial_{\beta\alpha}^{2}X^{i}+\Gamma_{j\alpha}^{i}%
\partial_{\beta}X^{j}+X^{j}\partial_{\beta}\Gamma_{j\alpha}^{i}+\Gamma
_{j\beta}^{i}\partial_{\alpha}X^{j}+X^{k}\Gamma_{j\beta}^{i}\Gamma_{k\alpha
}^{j}\right)  d\xi^{\beta}\wedge d\xi^{\alpha}\otimes\mathbf{e}_{i}\left(
x\right)  $

$=\sum\left(  \partial_{\beta\alpha}^{2}X^{i}\right)  d\xi^{\beta}\wedge
d\xi^{\alpha}\otimes e_{i}\left(  x\right)  +\left(  \Gamma_{j\alpha}%
^{i}\partial_{\beta}X^{j}+\Gamma_{j\beta}^{i}\partial_{\alpha}X^{j}\right)
d\xi^{\beta}\wedge d\xi^{\alpha}\otimes\mathbf{e}_{i}\left(  x\right)  $

$+X^{j}\left(  \partial_{\beta}\Gamma_{j\alpha}^{i}+\Gamma_{k\beta}^{i}%
\Gamma_{j\alpha}^{k}\right)  d\xi^{\beta}\wedge d\xi^{\alpha}\otimes
\mathbf{e}_{i}\left(  x\right)  $

$=\sum X^{j}\left(  \partial_{\beta}\Gamma_{j\alpha}^{i}+\Gamma_{k\beta}%
^{i}\Gamma_{j\alpha}^{k}\right)  d\xi^{\beta}\wedge d\xi^{\alpha}%
\otimes\mathbf{e}_{i}\left(  x\right)  $

$=\sum\left(  X^{j}\partial_{\beta}\Gamma_{j\alpha}^{i}d\xi^{\beta}\wedge
d\xi^{\alpha}+X^{j}\Gamma_{k\beta}^{i}\Gamma_{j\alpha}^{k}d\xi^{\beta}\wedge
d\xi^{\alpha}\right)  \otimes\mathbf{e}_{i}\left(  x\right)  $

$=\sum X^{j}\left(  \partial_{\alpha}\Gamma_{j\beta}^{i}+\Gamma_{k\alpha}%
^{i}\Gamma_{j\beta}^{k}\right)  d\xi^{\alpha}\wedge d\xi^{\beta}%
\otimes\mathbf{e}_{i}\left(  x\right)  $

$=\sum R_{j\alpha\beta}^{i}X^{j}d\xi^{\alpha}\wedge d\xi^{\beta}%
\otimes\mathbf{e}_{i}\left(  x\right)  $
\end{proof}

\bigskip

Remarks :

The curvature of the connection reads :

$\Omega=-\sum_{\alpha\beta}\sum_{j\in I}\left(  \partial_{\alpha}%
\Gamma_{j\beta}^{i}+\sum_{k\in I}\Gamma_{k\alpha}^{i}\Gamma_{j\beta}%
^{k}\right)  dx^{\alpha}\wedge dx^{\beta}\otimes\mathbf{e}_{i}\left(
x\right)  \otimes\mathbf{e}^{j}\left(  x\right)  $

The Riemann curvature reads :

$R=\sum_{\alpha\beta}\sum_{ij}\left(  \partial_{\alpha}\Gamma_{j\beta}%
^{i}+\sum_{k}\Gamma_{k\alpha}^{i}\Gamma_{j\beta}^{k}\right)  d\xi^{\alpha
}\wedge d\xi^{\beta}\otimes\mathbf{e}_{i}\left(  x\right)  \otimes
\mathbf{e}^{j}\left(  x\right)  $

i) The curvature is defined for any fiber bundle, the Riemann curvature is
defined only for vector bundle

ii) Both are valued in $E\otimes E^{\prime}$\ but the curvature is a two
horizontal form on TE and the Riemann curvature is a two form on TM

iii) The formulas are not identical but opposite of each other

\subsubsection{Metric connection}

\begin{definition}
A linear connection with covariant derivative $\nabla$ on a vector bundle
E$\left(  M,V,\pi_{E}\right)  $ endowed with a scalar product g is said to be
\textbf{metric} if $\nabla g=0$
\end{definition}

We have the general characterization of such connexions :

\begin{theorem}
A linear connection with Christoffel forms $\Gamma$ on a complex or real
vector bundle E$\left(  M,V,\pi_{E}\right)  $ endowed with a scalar product g
is metric if :%

\begin{equation}
\forall\alpha,i,j:\partial_{\alpha}g_{ij}=\sum_{k}\left(  \Gamma_{\alpha
i}^{k}g_{kj}+\Gamma_{\alpha j}^{k}g_{ik}\right)  \Leftrightarrow\left[
\partial_{\alpha}g\right]  =\left[  \Gamma_{\alpha}\right]  ^{t}\left[
g\right]  +\left[  g\right]  \left[  \Gamma_{\alpha}\right]
\end{equation}

\end{theorem}

\begin{proof}
Real case :

the scalar product is defined by a tensor $g\in\mathfrak{X}\left(  \odot
_{2}E\right)  $

At the transitions : $g_{bij}\left(  x\right)  =\sum_{kl}\overline{\left[
\varphi_{ab}\left(  x\right)  \right]  }_{i}^{k}\left[  \varphi_{ab}\left(
x\right)  \right]  _{j}^{l}g_{akl}\left(  x\right)  $

The covariant derivative of g reads with the Christoffel forms $\Gamma\left(
x\right)  $ of the connection :

$\nabla g=\sum_{\alpha ij}\left(  \partial_{\alpha}g_{ij}-\sum_{k}\left(
\Gamma_{\alpha i}^{k}g_{kj}+\Gamma_{\alpha j}^{k}g_{ik}\right)  \right)
e_{a}^{i}\left(  x\right)  \otimes e_{a}^{j}\left(  x\right)  \otimes
d\xi^{\alpha}$

So : $\forall\alpha,i,j:\partial_{\alpha}g_{ij}=\sum_{k}\left(  \Gamma_{\alpha
i}^{k}g_{kj}+\Gamma_{\alpha j}^{k}g_{ik}\right)  $

$\left[  \partial_{\alpha}g\right]  _{j}^{i}=\sum_{k}\left(  \left[
\Gamma_{\alpha}\right]  _{i}^{k}\left[  g\right]  _{j}^{k}+\left[
\Gamma_{\alpha}\right]  _{j}^{k}\left[  g\right]  _{k}^{i}\right)
\Leftrightarrow\left[  \partial_{\alpha}g\right]  =\left[  \Gamma_{\alpha
}\right]  ^{t}\left[  g\right]  +\left[  g\right]  \left[  \Gamma_{\alpha
}\right]  $

Complex case :

the scalar product is defined by a real structure $\sigma$\ and a tensor
$g\in\mathfrak{X}\left(  \otimes_{2}E\right)  $ which is not symmetric.

The covariant derivative of g is computed as above with the same result.
\end{proof}

Remarks :

i) The scalar product of two covariant derivatives, which are 1-forms on M
valued in E, has no precise meaning, so it is necessary to go through the
tensorial definition to stay rigorous. And there is no clear equivalent of the
properties of metric connections on manifolds : preservation of the scalar
product of transported vector field, or the formula : $\forall X,Y,Z\in
\mathfrak{X}\left(  TM\right)  :R(X,Y)Z+R(Y,Z)X+R(Z,X)Y=0.$

ii) A scalar product $\gamma$ on V induces a scalar product on E iff the
transition maps preserve the scalar product : $\left[  \gamma\right]  =\left[
\varphi_{ba}\right]  ^{\ast}\left[  \gamma\right]  \left[  \varphi
_{ba}\right]  $.\ 

If E is real then this scalar product defines a tensor $g_{aij}\left(
x\right)  =g_{aij}\left(  x\right)  \left(  \mathbf{e}_{i}\left(  x\right)
,\mathbf{e}_{j}\left(  x\right)  \right)  =\gamma\left(  e_{i},e_{j}\right)  $
which is metric iff$\ \forall\alpha:\left[  \Gamma_{a\alpha}\right]
^{t}\left[  \gamma\right]  +\left[  \gamma\right]  \left[  \Gamma_{a\alpha
}\right]  =0$ and it is easy to check that if it is met for $\left[
\Gamma_{a\alpha}\right]  $ it is met for $\left[  \Gamma_{b\alpha}\right]
=\left[  \Gamma_{a\alpha}\right]  -\left[  \partial_{\alpha}\varphi
_{ba}\right]  \left[  \varphi_{ab}\right]  $ .

If E is complex there is a tensor associated to the scalar product iff there
is a real structure on E.\ A real structure $\sigma$\ on V induces a real
structure on E iff the transition maps are real : $\varphi_{ab}\left(
x\right)  \circ\sigma=\sigma\circ\varphi_{ab}\left(  x\right)  $

The connection is real iff : $\widehat{\Gamma}\left(  \varphi\left(
x,\sigma\left(  u\right)  \right)  \right)  =\sigma\left(  \widehat{\Gamma
}\left(  \varphi\left(  x,u\right)  \right)  \right)  \Leftrightarrow
\Gamma\left(  x\right)  =\overline{\Gamma\left(  x\right)  }$

Then the condition above reads :

$\left[  \Gamma_{\alpha}\right]  ^{t}\left[  \gamma\right]  +\left[
\gamma\right]  \left[  \Gamma_{\alpha}\right]  =0\Leftrightarrow\left[
\Gamma_{\alpha}\right]  ^{\ast}\left[  \gamma\right]  +\left[  \gamma\right]
\left[  \Gamma_{\alpha}\right]  =0$

\bigskip

\subsection{Connections on principal bundles}

\label{Connection on principal bundles}

The main feature of principal bundles is the right action of the group on the
bundle.\ So connections specific to principal bundle are connections which are
equivariant under this action.

\subsubsection{Principal connection}

\paragraph{Definition\newline}

The tangent bundle TP of a principal fiber bundle P$\left(  M,G,\pi\right)
$\ is a principal bundle $TP\left(  TM,TG,T\pi\right)  .$ Any vector $v_{p}\in
T_{p}P$ can be written $v_{p}=\varphi_{ax}^{\prime}(x,g)v_{x}+\zeta\left(
u_{g}\right)  \left(  p\right)  $ where $\zeta$\ is a fundamental vector field
and $u_{g}\in T_{1}G:\zeta\left(  u_{g}\right)  \left(  p\right)  =\rho
_{g}^{\prime}\left(  p,1\right)  X=\varphi_{ag}^{\prime}\left(  x,g\right)
\left(  L_{g}^{\prime}1\right)  u_{g}.$ The vertical bundle VP is a trivial
vector bundle over P : $VP(P,T_{1}G,\pi)\simeq P\times T_{1}G$.

So a connection $\Phi$ on P reads in an atlas $\left(  O_{a},\varphi
_{a}\right)  _{a\in A}$ of P:

$\Phi\left(  p\right)  \left(  \varphi_{ax}^{\prime}(x,g)v_{x}+\zeta\left(
u_{g}\right)  \left(  p\right)  \right)  =\varphi_{ag}^{\prime}(x,g)\left(
\left(  L_{g}^{\prime}1\right)  u_{g}+\Gamma_{a}\left(  p\right)
v_{x}\right)  $

$=\varphi_{ag}^{\prime}(x,g)\left(  L_{g}^{\prime}1\right)  \left(
u_{g}+L_{g^{-1}}^{\prime}\left(  g\right)  \Gamma_{a}\left(  p\right)
v_{x}\right)  =\zeta\left(  u_{g}+L_{g^{-1}}^{\prime}\left(  g\right)
\Gamma_{a}\left(  p\right)  v_{x}\right)  \left(  p\right)  $

As we can see in this formula the Lie algebra $T_{1}G$ will play a significant
role in a connection : this is a fixed vector space, and through the
fundamental vectors\ formalism any equivariant vector field on VP can be
linked to a fixed vector on $T_{1}G$ .

\begin{definition}
A connection on a principal bundle P$\left(  M,G,\pi\right)  $ is said to be
\textbf{principal} if it is equivariant under the right action of G :
\end{definition}

$\forall g\in G,p\in P:\rho\left(  p,g\right)  ^{\ast}\Phi\left(  p\right)
=\rho_{p}^{\prime}\left(  p,g\right)  \Phi(p)$

$\Leftrightarrow\Phi(\rho\left(  p,g\right)  )\rho_{p}^{\prime}\left(
p,g\right)  v_{p}=\rho_{p}^{\prime}\left(  p,g\right)  \Phi(p)v_{p}$

\begin{theorem}
(Kolar p.101) Any finite dimensional principal bundle admits principal connections
\end{theorem}

\begin{theorem}
A connection is principal iff%

\begin{equation}
\Gamma_{a}\left(  \varphi_{a}\left(  x,g\right)  \right)  =R_{g}^{\prime
}(1)\Gamma_{a}\left(  \varphi_{a}\left(  x,1\right)  \right)
\end{equation}

\end{theorem}

\begin{proof}
Take $\mathbf{p}_{a}\left(  x\right)  =\varphi_{a}\left(  x,1\right)  $

$v_{\mathbf{p}}=\varphi_{ax}^{\prime}(x,1)v_{x}+\zeta\left(  u_{g}\right)
\left(  \mathbf{p}\right)  $

$\rho_{p}^{\prime}\left(  \mathbf{p},g\right)  v_{\mathbf{p}}=\varphi
_{ax}^{\prime}(x,1)v_{x}+\rho_{p}^{\prime}\left(  p,g\right)  \zeta\left(
u_{g}\right)  \left(  \mathbf{p}\right)  =\varphi_{ax}^{\prime}(x,g)v_{x}%
+\zeta\left(  Ad_{g^{-1}}u_{g}\right)  \left(  \rho\left(  \mathbf{p}%
,g\right)  \right)  $

$\Phi(\rho\left(  \mathbf{p},g\right)  )\rho_{p}^{\prime}\left(
\mathbf{p},g\right)  v_{\mathbf{p}}=\zeta\left(  Ad_{g^{-1}}u_{g}+L_{g^{-1}%
}^{\prime}\left(  g\right)  \Gamma_{a}\left(  \rho\left(  \mathbf{p},g\right)
\right)  v_{x}\right)  \left(  \rho\left(  \mathbf{p},g\right)  \right)  $

$\rho_{p}^{\prime}\left(  \mathbf{p},g\right)  \Phi(\mathbf{p})v_{\mathbf{p}%
}=\zeta\left(  Ad_{g^{-1}}\left(  u_{g}+\Gamma_{a}\left(  \mathbf{p}\right)
v_{x}\right)  \right)  \left(  \rho\left(  \mathbf{p},g\right)  \right)  $

$Ad_{g^{-1}}u_{g}+L_{g^{-1}}^{\prime}\left(  g\right)  \Gamma_{a}\left(
\rho\left(  \mathbf{p},g\right)  \right)  v_{x}=Ad_{g^{-1}}\left(
u_{g}+\Gamma_{a}\left(  \mathbf{p}\right)  v_{x}\right)  $

$\Gamma_{a}\left(  \rho\left(  \mathbf{p},g\right)  \right)  =L_{g}^{\prime
}\left(  1\right)  Ad_{g^{-1}}\Gamma_{a}\left(  \mathbf{p}\right)
=L_{g}^{\prime}\left(  1\right)  L_{g^{-1}}^{\prime}(g)R_{g}^{\prime}%
(1)\Gamma_{a}\left(  \mathbf{p}\right)  $

So : $\Gamma_{a}\left(  \varphi_{a}\left(  x,g\right)  \right)  =R_{g}%
^{\prime}(1)\Gamma_{a}\left(  \varphi_{a}\left(  x,1\right)  \right)  $
\end{proof}

\bigskip

We will use the convenient notation and denomination, which is common in
physics :

\begin{notation}
$\grave{A}_{a}\left(  x\right)  =\Gamma_{a}\left(  \varphi_{a}\left(
x,1\right)  \right)  =\Gamma_{a}\left(  \mathbf{p}_{a}\left(  x\right)
\right)  \in\Lambda_{1}\left(  O_{a};T_{1}G\right)  $ is the
\textbf{potential} of the connection
\end{notation}

\begin{theorem}
A principal connection $\Phi$ on a principal bundle fiber bundle $P\left(
M,G,\pi\right)  $\ with atlas $\left(  O_{a},\varphi_{a}\right)  _{a\in A}$ is
uniquely defined by a family $\left(  \grave{A}_{a}\right)  _{a\in A}$ of maps
$\grave{A}_{a}\in\Lambda_{1}\left(  O_{a};T_{1}G\right)  $ such that :%

\begin{equation}
\grave{A}_{b}\left(  x\right)  =Ad_{g_{ba}}\left(  \grave{A}_{a}\left(
x\right)  -L_{g_{ba}^{-1}}^{\prime}(g_{ba})g_{ba}^{\prime}\left(  x\right)
\right)
\end{equation}

by :%

\begin{equation}
\Phi\left(  \varphi_{a}\left(  x,g\right)  \right)  \left(  \varphi
_{ax}^{\prime}(x,g)v_{x}+\zeta\left(  u_{g}\right)  \left(  p\right)  \right)
=\zeta\left(  u_{g}+Ad_{g^{-1}}\grave{A}_{a}\left(  x\right)  v_{x}\right)
\left(  p\right)
\end{equation}

\end{theorem}

\begin{proof}
i) If $\Phi$ is a principal connection :

$\Phi\left(  \varphi_{a}\left(  x,g\right)  \right)  \left(  \varphi
_{ax}^{\prime}(x,g)v_{x}+\zeta\left(  u_{g}\right)  \left(  p\right)  \right)
=\zeta\left(  u_{g}+L_{g^{-1}}^{\prime}\left(  g\right)  \Gamma_{a}\left(
\varphi_{a}\left(  x,g\right)  \right)  v_{x}\right)  \left(  p\right)
=\zeta\left(  u_{g}+L_{g^{-1}}^{\prime}\left(  g\right)  R_{g}^{\prime}\left(
1\right)  \Gamma_{a}\left(  \varphi_{a}\left(  x,1\right)  \right)
v_{x}\right)  \left(  p\right)  $

Define : $\grave{A}_{a}\left(  x\right)  =\Gamma_{a}\left(  \varphi_{a}\left(
x,1\right)  \right)  \in\Lambda_{1}\left(  O_{a};T_{1}G\right)  $

$\Phi\left(  \varphi_{a}\left(  x,g\right)  \right)  \left(  \varphi
_{ax}^{\prime}(x,g)v_{x}+\zeta\left(  u_{g}\right)  \left(  p\right)  \right)
=\zeta\left(  u_{g}+L_{g^{-1}}^{\prime}\left(  g\right)  R_{g}^{\prime}\left(
1\right)  \grave{A}_{a}\left(  x\right)  v_{x}\right)  \left(  p\right)
=\zeta\left(  u_{g}+Ad_{g^{-1}}\grave{A}_{a}\left(  x\right)  v_{x}\right)
\left(  p\right)  $

$\Gamma_{a}\left(  \varphi_{a}\left(  x,g_{a}\right)  \right)  =R_{g_{a}%
}^{\prime}(1)\grave{A}_{a}\left(  x\right)  $

In a transition $x\in O_{a}\cap O_{b}:$

$p=\varphi_{a}\left(  x,g\right)  =\varphi_{b}\left(  x,g_{b}\right)
\Rightarrow\Gamma_{b}\left(  p\right)  =\varphi_{ba}^{\prime}\left(
x,g_{a}\right)  \circ\left(  -Id_{TM},\Gamma_{a}\left(  p\right)  \right)  $

$\Gamma_{b}\left(  p\right)  =\left(  g_{ba}\left(  x\right)  \left(
g_{a}\right)  \right)  ^{\prime}\circ\left(  -Id_{TM},\Gamma_{a}\left(
p\right)  \right)  =-R_{g_{a}}^{\prime}(g_{ba}(x))g_{ba}^{\prime}\left(
x\right)  +L_{g_{ba}\left(  x\right)  }^{\prime}(g_{a})\Gamma_{a}\left(
p\right)  $

with the general formula :

$\frac{d}{dx}\left(  g\left(  x\right)  h\left(  x\right)  \right)
=R_{h(x)}^{\prime}(g(x))\circ g^{\prime}(x)+L_{g(x)}^{\prime}(h(x))\circ
h^{\prime}(x)$

$R_{g_{b}}^{\prime}(1)\grave{A}_{b}\left(  x\right)  =-R_{g_{a}}^{\prime
}(g_{ba}(x))g_{ba}^{\prime}\left(  x\right)  +L_{g_{ba}\left(  x\right)
}^{\prime}(g_{a})R_{g_{a}}^{\prime}(1)\grave{A}_{a}\left(  x\right)  $

$\grave{A}_{b}\left(  x\right)  =-R_{g_{b}^{-1}}^{\prime}(g_{b})R_{g_{a}%
}^{\prime}(g_{ba})g_{ba}^{\prime}\left(  x\right)  +R_{g_{b}^{-1}}^{\prime
}(g_{b})L_{g_{ba}}^{\prime}(g_{a})R_{g_{a}}^{\prime}(1)\grave{A}_{a}\left(
x\right)  $

$=-Ad_{g_{b}}L_{g_{b}^{-1}}^{\prime}\left(  g_{b}\right)  R_{g_{ba}g_{a}%
}^{\prime}\left(  1\right)  R_{g_{ba}^{-1}}^{\prime}\left(  g_{ba}\right)
g_{ba}^{\prime}\left(  x\right)  +R_{g_{b}^{-1}}^{\prime}(g_{b})L_{g_{ba}%
g_{a}}^{\prime}\left(  1\right)  L_{g_{a}^{-1}}^{\prime}\left(  g_{a}\right)
R_{g_{a}}^{\prime}(1)\grave{A}_{a}\left(  x\right)  $

$=-Ad_{g_{b}}L_{g_{b}^{-1}}^{\prime}\left(  g_{b}\right)  R_{g_{b}}^{\prime
}\left(  1\right)  R_{g_{ba}^{-1}}^{\prime}\left(  g_{ba}\right)
g_{ba}^{\prime}\left(  x\right)  +R_{g_{b}^{-1}}^{\prime}(g_{b})L_{g_{b}%
}^{\prime}\left(  1\right)  L_{g_{a}^{-1}}^{\prime}\left(  g_{a}\right)
R_{g_{a}}^{\prime}(1)\grave{A}_{a}\left(  x\right)  $

$=-Ad_{g_{b}}Ad_{g_{b}^{-1}}R_{g_{ba}^{-1}}^{\prime}\left(  g_{ba}\right)
g_{ba}^{\prime}\left(  x\right)  +Ad_{g_{b}}Ad_{g_{a}^{-1}}\grave{A}%
_{a}\left(  x\right)  $

$=-R_{g_{ba}^{-1}}^{\prime}\left(  g_{ba}\right)  g_{ba}^{\prime}\left(
x\right)  +Ad_{g_{ba}}\grave{A}_{a}\left(  x\right)  $

$=-Ad_{g_{ba}}L_{g_{ba}^{-1}}^{\prime}\left(  g_{ba}\right)  g_{ba}^{\prime
}\left(  x\right)  +Ad_{g_{ba}}\grave{A}_{a}\left(  x\right)  $

ii) If there is a family of maps :

$\grave{A}_{a}\in\Lambda_{1}\left(  O_{a};T_{1}G\right)  $ such that :
$\grave{A}_{b}\left(  x\right)  =Ad_{g_{ba}}\left(  \grave{A}_{a}\left(
x\right)  -L_{g_{ba}^{-1}}^{\prime}(g_{ba})g_{ba}^{\prime}\left(  x\right)
\right)  $

Define : $\Gamma_{a}\left(  \varphi_{a}\left(  x,g_{a}\right)  \right)
=R_{g_{a}}^{\prime}(1)\grave{A}_{a}\left(  x\right)  $

$\Gamma_{b}\left(  \varphi_{b}\left(  x,g_{b}\right)  \right)  =R_{g_{b}%
}^{\prime}(1)\grave{A}_{b}\left(  x\right)  =R_{g_{b}}^{\prime}(1)Ad_{g_{ba}%
}\left(  \grave{A}_{a}\left(  x\right)  -L_{g_{ba}^{-1}}^{\prime}%
(g_{ba})g_{ba}^{\prime}\left(  x\right)  \right)  $

$=R_{g_{b}}^{\prime}(1)Ad_{g_{ba}}R_{g_{a}^{-1}}^{\prime}\left(  g_{a}\right)
\Gamma_{a}\left(  \varphi_{a}\left(  x,g_{a}\right)  \right)  -R_{g_{a}%
}^{\prime}(g_{ba})R_{g_{ba}}^{\prime}\left(  1\right)  )R_{g_{ba}^{-1}%
}^{\prime}\left(  g_{ba}\right)  L_{g_{ba}}^{\prime}\left(  1\right)
L_{g_{ba}^{-1}}^{\prime}(g_{ba})g_{ba}^{\prime}\left(  x\right)  $

$=R_{g_{b}}^{\prime}(1)Ad_{g_{ba}}Ad_{g_{a}}L_{g_{a}^{-1}}^{\prime}\left(
g_{a}\right)  \Gamma_{a}\left(  p\right)  -R_{g_{a}}^{\prime}(g_{ba}%
)g_{ba}^{\prime}\left(  x\right)  $

$=R_{g_{b}}^{\prime}(1)Ad_{g_{b}}L_{g_{a}^{-1}}^{\prime}\left(  g_{a}\right)
\Gamma_{a}\left(  p\right)  -R_{g_{a}}^{\prime}(g_{ba})g_{ba}^{\prime}\left(
x\right)  $

$=L_{g_{b}}^{\prime}(1)L_{g_{a}^{-1}}^{\prime}\left(  g_{a}\right)  \Gamma
_{a}\left(  p\right)  -R_{g_{a}}^{\prime}(g_{ba})g_{ba}^{\prime}\left(
x\right)  =L_{g_{ba}}^{\prime}(g_{a})\Gamma_{a}\left(  p\right)  -R_{g_{a}%
}^{\prime}(g_{ba})g_{ba}^{\prime}\left(  x\right)  $

$\Gamma_{b}\left(  p\right)  =\left(  g_{ba}\left(  x\right)  \left(
g_{a}\right)  \right)  ^{\prime}\circ\left(  -Id_{TM},\Gamma_{a}\left(
p\right)  \right)  $

So the family defines a connection, and it is principal.
\end{proof}

So with $p=\varphi_{a}\left(  x,g\right)  :$

$\Gamma_{a}\left(  p\right)  =R_{g}^{\prime}(1)\grave{A}_{a}\left(  x\right)
$

$\Phi\left(  p\right)  \left(  \varphi_{ax}^{\prime}(x,g)v_{x}+\zeta\left(
u_{g}\right)  \left(  p\right)  \right)  =\zeta\left(  u_{g}+Ad_{g^{-1}}%
\grave{A}_{a}\left(  x\right)  v_{x}\right)  \left(  p\right)  $

In a change of trivialization on P :

$p=\varphi_{a}\left(  x,g_{a}\right)  =\widetilde{\varphi}_{a}\left(
x,\chi_{a}\left(  x\right)  g_{a}\right)  \Leftrightarrow\widetilde{g}%
_{a}=\chi_{a}\left(  x\right)  g_{a}$

the potential becomes with $g_{ba}\left(  x\right)  =\chi_{a}\left(  x\right)
$%

\begin{equation}
\grave{A}_{a}\left(  x\right)  \rightarrow\widetilde{\grave{A}}_{a}\left(
x\right)  =Ad_{\chi_{a}}\left(  \grave{A}_{a}\left(  x\right)  -L_{\chi
_{a}^{-1}}^{\prime}(\chi_{a})\chi_{a}^{\prime}\left(  x\right)  \right)
\end{equation}

\begin{theorem}
The fundamental vectors are invariant by a principal connection: $\forall X\in
T_{1}G:\Phi(p)\left(  \zeta\left(  X\right)  \left(  p\right)  \right)
=\zeta\left(  X\right)  \left(  p\right)  $
\end{theorem}

\begin{proof}
$\Phi(p)\left(  \zeta\left(  X\right)  \left(  p\right)  \right)
=\zeta\left(  X+Ad_{g^{-1}}\grave{A}_{a}\left(  x\right)  0\right)  \left(
p\right)  =\zeta\left(  X\right)  \left(  p\right)  $
\end{proof}

\paragraph{Connexion form \newline}

\begin{definition}
The \textbf{connexion form} of the connection $\Phi$\ is the form :%

\begin{equation}
\widehat{\Phi}\left(  p\right)  :TP\rightarrow T_{1}G::\Phi\left(  p\right)
\left(  v_{p}\right)  =\zeta\left(  \widehat{\Phi}\left(  p\right)  \left(
v_{p}\right)  \right)  \left(  p\right)
\end{equation}

\end{definition}

$\widehat{\Phi}\left(  \varphi_{a}\left(  x,g\right)  \right)  \left(
\varphi_{ax}^{\prime}(x,g)v_{x}+\zeta\left(  u_{g}\right)  \left(  p\right)
\right)  =u_{g}+Ad_{g^{-1}}\grave{A}_{a}\left(  x\right)  v_{x}$

It has the property for a principal connection : $\rho\left(  p,g\right)
_{\ast}\widehat{\Phi}\left(  p\right)  =Ad_{g^{-1}}\widehat{\Phi}(p)$

\begin{proof}
$\rho\left(  p,g\right)  ^{\ast}\Phi\left(  p\right)  =\rho_{p}^{\prime
}\left(  p,g\right)  \Phi(p)$

$\rho_{p}^{\prime}\left(  p,g\right)  \Phi\left(  p\right)  =\zeta\left(
Ad_{g^{-1}}\widehat{\Phi}\left(  p\right)  \right)  \left(  \rho\left(
p,g\right)  \right)  $
\end{proof}

This is a 1-form on TP valued in a fixed vector space. For any $X\in T_{1}G$
the fundamendal vector field $\zeta\left(  X\right)  \left(  p\right)
\in\mathfrak{X}\left(  VP\right)  \sim\mathfrak{X}\left(  P\times
T_{1}G\right)  .$ So it makes sens to compute the Lie derivative, in the usual
meaning, of the 1 form $\widehat{\Phi}$ along a fundamental vector field and
(Kolar p.100):

$\pounds _{\zeta\left(  X\right)  }\widehat{\Phi}=-ad\left(  X\right)  \left(
\widehat{\Phi}\right)  =\left[  \widehat{\Phi},X\right]  _{T_{1}G}$\ 

\paragraph{The bundle of principal connections\newline}

\begin{theorem}
(Kolar p.159) There is a bijective correspondance between principal
connections on a principal fiber bundle $P(M,G,\pi)$ and the equivariant
sections of the first jet prolongation $J^{1}P$ given by :

$\Gamma:P\rightarrow J^{1}P$::\ \ \ $\Gamma\left(  p\right)  =\left(
\Gamma\left(  p\right)  _{\alpha}^{i}\right)  $ \ such that $\Gamma
(\rho\left(  p,g\right)  )=\left(  R_{g}^{\prime}1\right)  \Gamma(p)$
\end{theorem}

The set of potentials $\left\{  \grave{A}\left(  x\right)  _{\alpha}%
^{i}\right\}  $ on a principal bundle has the structure of an affine bundle,
called the \textbf{bundle of principal connections}.

Its structure is defined as follows :

The adjoint bundle of P is the vector bundle $E=P\left[  T_{1}G,Ad\right]  $
associated to P.

$QP=J^{1}P$ is an affine bundle over E, modelled on the vector bundle

$TM^{\ast}\otimes VE\rightarrow E$

$A=A_{\alpha}^{i}dx^{\alpha}\otimes\varepsilon_{ai}\left(  x\right)  $

With :

$QP\left(  x\right)  \times QP\left(  x\right)  \rightarrow TM^{\ast}\otimes
VE::\left(  A,B\right)  =B-A$

At the transitions :

$\left(  A_{b},B_{b}\right)  =B_{b}-A_{b}=Ad_{g_{ba}}\left(  B_{a}%
-A_{a}\right)  $

\paragraph{Holonomy group\newline}

\begin{theorem}
(Kolar p.105) The holonomy group Hol($\Phi,p)$ on a principal fiber bundle
$P\left(  M,G,\pi\right)  $ with principal connection $\Phi$\ is a Lie
subgroup of G, and Hol$_{0}$($\Phi,p)$ is a connected Lie subgroup of G and
$Hol(\Phi,\rho\left(  p,g\right)  )=Conj_{g^{-1}}Hol(\Phi,p)$

The Lie algebra hol($\Phi,p)$ of Hol($\Phi,p)$ is a subalgebra of $T_{1}G,$
linearly generated by the vectors of the curvature form $\widehat{\Omega
}\left(  v_{p},w_{p}\right)  $

The set $P_{c}\left(  p\right)  $ of all curves on P, lifted from curves on M
and going through a point p in P, is a principal fiber bundle, with group
Hol($\Phi,p),$ subbundle of P. The pullback of $\Phi$ on this bundle is still
a principal connection. P is foliated by $P_{c}\left(  p\right)  .$If the
curvature of the connection $\Omega=0$ then $Hol(\Phi,p)=\{1\}$ and each
P$_{c}\left(  p\right)  $ is a covering of M.
\end{theorem}

\paragraph{Horizontal form\newline}

The horizontal form of the connection is :

$\chi\left(  p\right)  =Id_{TE}-\Phi\left(  p\right)  $

With : $p=\varphi\left(  x,g\right)  :$

$\chi\left(  p\right)  \left(  \varphi_{ax}^{\prime}(x,g)v_{x}+\zeta\left(
u_{g}\right)  \left(  p\right)  \right)  =\varphi_{ax}^{\prime}(x,g)v_{x}%
-\zeta\left(  Ad_{g^{-1}}\grave{A}_{a}\left(  x\right)  v_{x}\right)  \left(
p\right)  $

It is equivariant if the connection is principal :

$\rho\left(  p,g\right)  ^{\ast}\chi\left(  p\right)  =\chi(\rho\left(
p,g\right)  )\rho_{p}^{\prime}\left(  p,g\right)  =\rho_{p}^{\prime}\left(
p,g\right)  \chi(p)$

\begin{theorem}
The horizontal lift\textbf{\ }of a vector field on M by a principal connection
with potential \`{A} on a principal bundle $P\left(  M,G,\pi\right)  $\ with
trivialization $\varphi$\ is the map : $\chi_{L}:\mathfrak{X}\left(
TM\right)  \rightarrow\mathfrak{X}\left(  HP\right)  ::$

$\chi_{L}\left(  p\right)  \left(  X\right)  =\varphi_{x}^{\prime
}(x,g)X\left(  x\right)  -\zeta\left(  Ad_{g^{-1}}\left(  \grave{A}_{a}\left(
x\right)  X\left(  x\right)  \right)  \right)  \left(  p\right)  $
\end{theorem}

$\chi_{L}\left(  p\right)  \left(  X\right)  $ is a horizontal vector field on
TE, which is projectable on X.

For any section S on P we have : $S\in$\ $\mathfrak{X}\left(  E\right)  $:
$\nabla_{X}S=\pounds _{\chi_{L}\left(  X\right)  }S$

\begin{definition}
The \textbf{horizontalization} of a r-form on a principal bundle $P\left(
M,G,\pi\right)  $ endowed with a principal connection is the map :

$\chi^{\ast}:\mathfrak{X}\left(  \Lambda_{r}TP\right)  \rightarrow
\mathfrak{X}\left(  \Lambda_{r}TP\right)  ::\chi^{\ast}\varpi\left(  p\right)
\left(  v_{1},...,v_{r}\right)  =\varpi\left(  p\right)  \left(  \chi\left(
p\right)  \left(  v_{1}\right)  ,...,\chi\left(  p\right)  \left(
v_{r}\right)  \right)  $
\end{definition}

$\chi^{\ast}\varpi$ is a horizontal form : it is null whenever one of the
vector $v_{k}$ is vertical. It reads in the holonomic basis of P:

$\chi^{\ast}\varpi\left(  p\right)  =\sum_{\left\{  \alpha_{1}..\alpha
_{r}\right\}  }\mu_{\alpha_{1}..\alpha_{r}}dx^{\alpha_{1}}\wedge...\wedge
dx^{\alpha_{r}}$

\begin{theorem}
(Kolar p.103) The horizontalization of a r-form has the following properties :

$\chi^{\ast}\circ\chi^{\ast}=\chi^{\ast}$

$\forall\mu\in\mathfrak{X}\left(  \Lambda_{r}TP\right)  ,\varpi\in
\mathfrak{X}\left(  \Lambda_{s}TP\right)  :\chi^{\ast}\left(  \mu\wedge
\varpi\right)  =\chi^{\ast}\left(  \mu\right)  \wedge\chi^{\ast}\left(
\varpi\right)  ,$

$\chi^{\ast}\widehat{\Phi}=0$

$X\in T_{1}G:\chi^{\ast}\pounds \zeta\left(  X\right)  =\pounds \zeta\left(
X\right)  \circ\chi^{\ast}$
\end{theorem}

\subsubsection{Curvature}

The curvature of the connection is :

$\Omega\left(  p\right)  (X,Y)=\Phi\left(  p\right)  ([\chi\left(  p\right)
X,\chi\left(  p\right)  Y]_{TE})$ with :

\begin{theorem}
The curvature of a principal connection is equivariant :

$\rho\left(  p,g\right)  ^{\ast}\Omega\left(  p\right)  =\rho_{p}^{\prime
}\left(  p,g\right)  \Omega(p)$
\end{theorem}

\begin{proof}
$\rho\left(  p,g\right)  ^{\ast}\Omega\left(  p\right)  (X_{p},Y_{p}%
)=\Phi\left(  (\rho\left(  p,g\right)  )\right)  ([\chi(\rho\left(
p,g\right)  \rho_{p}^{\prime}\left(  p,g\right)  X,\chi(\rho\left(
p,g\right)  \rho_{p}^{\prime}\left(  p,g\right)  Y]_{TE})$

$=\Phi(\rho\left(  p,g\right)  )\left(  [\rho\left(  p,g\right)  ^{\ast}%
\chi(p)X,\rho\left(  p,g\right)  ^{\ast}\chi(p)Y]_{TE}\right)  $

$=\Phi(\rho\left(  p,g\right)  )\rho_{p}^{\prime}\left(  p,g\right)  \left(
[\chi(p)X,\chi(p)Y]_{TE}\right)  $

$=\rho_{p}^{\prime}\left(  p,g\right)  \Phi(p)\left(  [\chi(p)X,\chi
(p)Y]_{TE}\right)  $
\end{proof}

\begin{theorem}
The \textbf{curvature form} of a principal connection with potentiel \`{A} on
a principal bundle $P\left(  V,G,\pi\right)  $ is the 2 form $\widehat{\Omega
}\in\Lambda_{2}\left(  P;T_{1}G\right)  $ such that : $\Omega=\zeta\left(
\widehat{\Omega}\right)  .$ It has the following expression in an holonomic
basis of TP and basis $\left(  \varepsilon_{i}\right)  $ of $T_{1}G$%

\begin{equation}
\widehat{\Omega}\left(  \varphi_{a}\left(  x,g_{a}\right)  \right)
=-Ad_{g_{a}^{-1}}\sum_{i}\sum_{\alpha\beta}\left(  \partial_{\alpha}\grave
{A}_{\beta}^{i}+\left[  \grave{A}_{a},\grave{A}_{\beta}\right]  _{T_{1}G}%
^{i}\right)  dx^{\alpha}\Lambda dx^{\beta}\otimes\varepsilon_{i}%
\end{equation}

\end{theorem}

\begin{proof}
The bracket $\left[  \grave{A}_{a},\grave{A}_{\beta}\right]  _{T_{1}G}^{i}%
$\ is the Lie bracket in the Lie algebra $T_{1}G$.

The curvature of a princial connection reads $:$

$\Omega=\sum_{i}\sum_{\alpha\beta}\left(  -\partial_{\alpha}\Gamma_{\beta}%
^{i}+\left[  \Gamma_{\alpha},\Gamma_{\beta}\right]  _{TG}^{i}\right)
dx^{\alpha}\Lambda dx^{\beta}\otimes\partial g_{i}\in\Lambda_{2}\left(
P;VP\right)  $

The commutator is taken on TG

$\Gamma\left(  p\right)  =\left(  R_{g}^{\prime}1\right)  \grave{A}_{a}\left(
x\right)  $ so $\Gamma$ is a right invariant vector field on TG and :

$\left[  \Gamma_{\alpha},\Gamma_{\beta}\right]  _{TG}^{i}=\left[  \left(
R_{g}^{\prime}1\right)  \grave{A}_{\alpha},\left(  R_{g}^{\prime}1\right)
\grave{A}_{\beta}\right]  _{TG}^{i}=-\left(  R_{g}^{\prime}1\right)  \left[
\grave{A}_{a},\grave{A}_{\beta}\right]  _{T_{1}G}^{i}$

$u_{1}\in T_{1}G:\zeta\left(  u_{1}\right)  \left(  \left(  \varphi_{a}\left(
x,g\right)  \right)  \right)  =\varphi_{ag}^{\prime}\left(  x,g\right)
\left(  L_{g}^{\prime}1\right)  u_{1}$

$u_{g}\in T_{g}G:\zeta\left(  \left(  L_{g^{-1}}^{\prime}g\right)
u_{g}\right)  \left(  \left(  \varphi_{a}\left(  x,g\right)  \right)  \right)
=\varphi_{ag}^{\prime}\left(  x,g\right)  u_{g}$

$\Omega\left(  p\right)  =-\sum_{\alpha\beta}dx^{\alpha}\Lambda dx^{\beta
}\otimes\zeta\left(  \left(  L_{g^{-1}}^{\prime}g\right)  \left(
R_{g}^{\prime}1\right)  \sum_{i}\left(  \partial_{\alpha}\grave{A}_{\beta}%
^{i}+\left[  \grave{A}_{a},\grave{A}_{\beta}\right]  _{T_{1}G}^{i}\right)
\varepsilon_{i}\right)  \left(  p\right)  $

$\Omega\left(  p\right)  =-\sum_{\alpha\beta}dx^{\alpha}\Lambda dx^{\beta
}\otimes\zeta\left(  Ad_{g^{-1}}\sum_{i}\left(  \partial_{\alpha}\grave
{A}_{\beta}^{i}+\left[  \grave{A}_{a},\grave{A}_{\beta}\right]  _{T_{1}G}%
^{i}\right)  \varepsilon_{i}\right)  \left(  p\right)  $ where $\left(
\varepsilon_{j}\right)  $ is a basis of $T_{1}G$
\end{proof}

At the transitions we have :

$\widehat{\Omega}\left(  \varphi_{a}\left(  x,g_{a}\right)  \right)
=Ad_{g_{a}^{-1}}\left(  \widehat{\Omega}\left(  \varphi_{a}\left(  x,1\right)
\right)  \right)  =$

$\widehat{\Omega}(\varphi_{b}(x,g_{ba}\left(  x\right)  g_{a})=Ad_{g_{a}^{-1}%
}Ad_{g_{ba}^{-1}}\left(  \widehat{\Omega}\left(  \varphi_{b}\left(
x,1\right)  \right)  \right)  $

$\widehat{\Omega}\left(  \varphi_{b}\left(  x,1\right)  \right)  =Ad_{g_{ba}%
}\left(  \widehat{\Omega}\left(  \varphi_{a}\left(  x,1\right)  \right)
\right)  $

\begin{theorem}
The \textbf{strength of a principal connection} with potentiel \`{A} on a
principal bundle $P\left(  V,G,\pi\right)  $ with atlas $\left(  O_{a}%
,\varphi_{a}\right)  _{a\in A}$ is the 2 form $%
\mathcal{F}%
\in\Lambda_{2}\left(  M;T_{1}G\right)  $ such that : $%
\mathcal{F}%
_{a}=-\mathbf{p}_{a}^{\ast}\widehat{\Omega}$ where $\mathbf{p}_{a}%
(x)=\varphi_{a}\left(  x,1\right)  .$ It has the following expression in an
holonomic basis of TM and basis $\left(  \varepsilon_{i}\right)  $ of
$T_{1}G:$%

\begin{equation}%
\mathcal{F}%
\left(  x\right)  =\sum_{i}\left(  d\grave{A}^{i}+\sum_{\alpha\beta}\left[
\grave{A}_{a},\grave{A}_{\beta}\right]  _{T_{1}G}^{i}d\xi^{\alpha}\Lambda
d\xi^{\beta}\right)  \otimes\varepsilon_{i}%
\end{equation}

At the transitions : $x\in O_{a}\cap O_{b}:%
\mathcal{F}%
_{b}\left(  x\right)  =Ad_{g_{ba}}%
\mathcal{F}%
_{a}\left(  x\right)  $
\end{theorem}

\begin{proof}
For any vector fields X,Y on M, their horizontal lifts $\chi_{L}\left(
X\right)  ,\chi_{L}\left(  Y\right)  $ is such that :

$\Omega\left(  p\right)  \left(  \chi_{L}\left(  X\right)  ,\chi_{L}\left(
Y\right)  \right)  =\left[  \chi_{L}\left(  X\right)  ,\chi_{L}\left(
Y\right)  \right]  _{TE}-\chi_{L}\left(  p\right)  \left(  \left[  X,Y\right]
_{TM}\right)  $

$\chi_{L}(\varphi_{a}(x,g))\left(  \sum_{\alpha}X_{x}^{\alpha}\partial
\xi^{\alpha}\right)  =\left(  \sum_{\alpha}X_{x}^{\alpha}\partial x^{\alpha
}-\zeta\left(  Ad_{g^{-1}}\grave{A}\left(  x\right)  X\right)  \right)
(\varphi_{a}(x,g))$

If we denote :

$%
\mathcal{F}%
\left(  x\right)  =\sum_{i}\left(  d_{M}\grave{A}^{i}+\sum_{\alpha\beta
}\left[  \grave{A}_{a},\grave{A}_{\beta}\right]  _{T_{1}G}^{i}d\xi^{\alpha
}\Lambda d\xi^{\beta}\right)  \otimes\varepsilon_{i}\in\Lambda_{2}\left(
M;T_{1}G\right)  $

then : $\widehat{\Omega}\left(  p\right)  \left(  \chi_{L}\left(  X\right)
,\chi_{L}\left(  Y\right)  \right)  =-Ad_{g^{-1}}%
\mathcal{F}%
\left(  x\right)  \left(  X,Y\right)  $

We have with $\mathbf{p}_{a}\left(  x\right)  =\varphi_{a}\left(  x,1\right)
:$

$%
\mathcal{F}%
_{a}\left(  x\right)  \left(  X,Y\right)  =-\widehat{\Omega}\left(
\mathbf{p}_{a}\left(  x\right)  \right)  \left(  \mathbf{p}_{ax}^{\prime
}\left(  x,1\right)  X,\mathbf{p}_{ax}^{\prime}\left(  x,1\right)  Y\right)  $

$%
\mathcal{F}%
_{a}=-\mathbf{p}_{a}^{\ast}\widehat{\Omega}$

At the transition :

$\widehat{\Omega}\left(  p\right)  \left(  \chi_{L}\left(  X\right)  ,\chi
_{L}\left(  Y\right)  \right)  =-Ad_{g_{a}^{-1}}%
\mathcal{F}%
_{a}\left(  x\right)  \left(  X,Y\right)  =-Ad_{g_{b}^{-1}}%
\mathcal{F}%
_{b}\left(  x\right)  \left(  X,Y\right)  $

$%
\mathcal{F}%
_{b}\left(  x\right)  =Ad_{g_{b}}Ad_{g_{a}^{-1}}%
\mathcal{F}%
_{a}\left(  x\right)  =Ad_{g_{ba}}%
\mathcal{F}%
_{a}\left(  x\right)  $
\end{proof}

$%
\mathcal{F}%
\left(  \pi\left(  p\right)  \right)  \left(  \pi^{\prime}\left(  p\right)
u_{p},\pi^{\prime}\left(  p\right)  v_{p}\right)  =%
\mathcal{F}%
\left(  x\right)  \left(  u_{x},v_{x}\right)  \Leftrightarrow%
\mathcal{F}%
=\pi_{\ast}\widehat{\Omega}$

In a change of trivialization on P :

$p=\varphi_{a}\left(  x,g_{a}\right)  =\widetilde{\varphi}_{a}\left(
x,\chi_{a}\left(  x\right)  g_{a}\right)  \Leftrightarrow\widetilde{g}%
_{a}=\chi_{a}\left(  x\right)  g_{a}$

the strength of a principal connection becomes with $g_{ba}\left(  x\right)
=\chi_{a}\left(  x\right)  $%

\begin{equation}%
\mathcal{F}%
_{a}\left(  x\right)  \rightarrow\widetilde{%
\mathcal{F}%
}_{a}\left(  x\right)  =Ad_{\chi_{a}\left(  x\right)  }%
\mathcal{F}%
_{a}\left(  x\right)
\end{equation}

\subsubsection{Covariant derivative}

\paragraph{Covariant derivative\newline}

The covariant derivative of a section S is, according to the general
definition, valued in the vertical bundle, which is here isomorphic to the Lie
algebra. So it is more convenient to define :

\begin{theorem}
The covariant derivative of a section S on the principal bundle $P\left(
M,V,\pi\right)  $\ with atlas $\left(  O_{a},\varphi_{a}\right)  _{a\in A}$ is
the map :

$\nabla:\mathfrak{X}\left(  P\right)  \rightarrow\Lambda_{1}\left(
M;T_{1}G\right)  ::S^{\ast}\Phi=\zeta\left(  \nabla S\right)  \left(  S\left(
x\right)  \right)  $

It is expressed by :%

\begin{equation}
S\left(  x\right)  =\varphi_{a}\left(  x,\sigma_{a}\left(  x\right)  \right)
:\nabla S=L_{\sigma_{a}^{-1}}^{\prime}\left(  \sigma_{a}\right)  \left(
\sigma_{a}^{\prime}\left(  x\right)  +R_{\sigma_{a}}^{\prime}\left(  1\right)
\grave{A}_{a}\left(  x\right)  \right)
\end{equation}

\end{theorem}

\begin{proof}
A section is defined by a family of maps : $\sigma_{a}:O_{a}\rightarrow G$

$S\left(  x\right)  =\varphi_{a}\left(  x,\sigma_{a}\left(  x\right)  \right)
:S^{\prime}\left(  x\right)  =\varphi_{ax}^{\prime}\left(  x,\sigma_{a}\left(
x\right)  \right)  +\varphi_{ag}^{\prime}\left(  x,\sigma_{a}\left(  x\right)
\right)  \sigma_{a}^{\prime}\left(  x\right)  =\varphi_{ax}^{\prime}\left(
x,\sigma_{a}\left(  x\right)  \right)  +\zeta\left(  L_{\sigma_{a}^{-1}%
}^{\prime}\left(  \sigma_{a}\right)  \sigma_{a}^{\prime}\left(  x\right)
\right)  \left(  S\left(  x\right)  \right)  $

For a principal connection and a section S on P, and $Y\in\mathfrak{X}\left(
M\right)  $

$S^{\ast}\Phi\left(  Y\right)  =\Phi\left(  S\left(  x\right)  \right)
\left(  S^{\prime}\left(  x\right)  Y\right)  =\zeta\left(  \left(
L_{\sigma_{a}^{-1}}^{\prime}\left(  \sigma_{a}\right)  \sigma_{a}^{\prime
}\left(  x\right)  +Ad_{\sigma_{a}^{-1}}\grave{A}_{a}\left(  x\right)
\right)  Y\right)  \left(  S\left(  x\right)  \right)  $

$\nabla_{Y}S=L_{\sigma_{a}^{-1}}^{\prime}\left(  \sigma_{a}\right)  \left(
\sigma_{a}^{\prime}\left(  x\right)  +R_{\sigma_{a}}^{\prime}\left(  1\right)
\grave{A}_{a}\left(  x\right)  \right)  Y$
\end{proof}

So with the gauge :$\mathbf{p}_{a}=\varphi_{a}\left(  x,1\right)
:\nabla\mathbf{p}_{a}=\grave{A}_{a}\left(  x\right)  $

\paragraph{Exterior covariant derivative\newline}

The exterior covariant derivative on a vector bundle E is a map: $\nabla
_{e}:\Lambda_{r}\left(  M;E\right)  \rightarrow\Lambda_{r+1}\left(
M;E\right)  .$ The exterior covariant derivative on a principal bundle is a
map : $\nabla_{e}:\Lambda_{r}\left(  P;V\right)  \rightarrow\Lambda
_{r+1}\left(  P;V\right)  $ where V is any fixed vector space.

\begin{definition}
(Kolar p.103) The \textbf{exterior covariant derivative} on a principal bundle
$P\left(  M,G,\pi\right)  $ endowed with a principal connection is the map :
$\nabla_{e}:\Lambda_{r}\left(  P;V\right)  \rightarrow\Lambda_{r+1}\left(
P;V\right)  ::$ $\nabla_{e}\varpi=\chi^{\ast}\left(  d\varpi\right)  $ where V
is any fixed vector space and $\chi^{\ast}$ is the horizontalization.
\end{definition}

\begin{theorem}
(Kolar p.103) The exterior covariant derivative $\nabla_{e}$ on a principal
bundle $P\left(  M,G,\pi\right)  $ endowed with a principal connection has the
following properties :

i) $\nabla_{e}\widehat{\Phi}=\widehat{\Omega}$

ii) Bianchi identity : $\nabla_{e}\widehat{\Omega}=0$

iii) $\forall\mu\in\Lambda_{r}\left(  P;%
\mathbb{R}
\right)  ,\varpi\in\Lambda_{s}\left(  P;V\right)  :$

$\nabla_{e}\left(  \mu\wedge\varpi\right)  =\left(  \nabla_{e}\mu\right)
\wedge\chi^{\ast}\varpi+\left(  -1\right)  ^{r}\left(  \chi^{\ast}\mu\right)
\wedge\nabla_{e}\varpi$

iv) $X\in T_{1}G:\pounds \zeta\left(  X\right)  \circ\nabla_{e}=\nabla
_{e}\circ\pounds \zeta\left(  X\right)  $

v) $\forall g\in G:\rho\left(  .,g\right)  ^{\ast}\nabla_{e}=\nabla_{e}%
\rho\left(  .,g\right)  ^{\ast}$

vi) $\nabla_{e}\circ\pi^{\ast}=d\circ\pi^{\ast}=\pi^{\ast}\circ d$

vii) $\nabla_{e}\circ\chi^{\ast}-\nabla_{e}=\chi^{\ast}i_{\Omega}$

viii) $\nabla_{e}\circ\nabla_{e}=\chi^{\ast}\circ i_{\Omega}\circ d$
\end{theorem}

\begin{theorem}
The exterior covariant derivative $\nabla_{e}$ of a r-forms $\varpi$ on P,
horizontal, equivariant and valued in the Lie algebra is:%

\begin{equation}
\nabla_{e}\varpi=d\varpi+\left[  \widehat{\Phi},\varpi\right]  _{T_{1}G}%
\end{equation}

with the connexion form $\widehat{\Phi}$ of the connection
\end{theorem}

Such a form reads :

$\varpi=\sum_{\left\{  \alpha_{1}...\alpha_{r}\right\}  }\sum_{i}%
\varpi_{\alpha_{1}...\alpha_{r}}^{i}\left(  p\right)  dx^{\alpha_{1}}%
\wedge..\wedge dx^{r}\otimes\varepsilon_{i}$

with $\sum_{i}\varpi_{\alpha_{1}...\alpha_{r}}^{i}\left(  \rho\left(
p,g\right)  \right)  \varepsilon_{i}=Ad_{g^{-1}}\sum_{i}\varpi_{\alpha
_{1}...\alpha_{r}}^{i}\left(  p\right)  \varepsilon_{i}$

the bracket is the bracket in the Lie algebra:

$\left[  \widehat{\Phi},\varpi\right]  _{T_{1}G}^{i}=\left[  \sum_{j\beta
}\widehat{\Phi}_{\beta}^{j}dx^{\beta}\otimes\varepsilon_{j},\sum_{k}%
\varpi_{\alpha_{1}...\alpha_{r}}^{k}dx^{\alpha_{1}}\wedge..\wedge
dx^{r}\otimes\varepsilon_{k}\right]  _{T_{1}G}^{i}$

$=\sum_{\beta\left\{  \alpha_{1}...\alpha_{r}\right\}  }\left[  \sum
_{j}\widehat{\Phi}_{\beta}^{j}\varepsilon_{j},\sum_{k}\varpi_{\alpha
_{1}...\alpha_{r}}^{k}\varepsilon_{k}\right]  ^{i}dx^{\beta}\Lambda
dx^{\alpha_{1}}\wedge..\wedge dx^{r}\otimes\varepsilon_{i}$

The algebra of r-forms on P, horizontal, equivariant and valued in the Lie
algebra :

$\widehat{\varpi}=\sum_{\left\{  \alpha_{1}...\alpha_{r}\right\}  }\sum
_{i}\varpi_{\alpha_{1}...\alpha_{r}}^{i}\left(  p\right)  dx^{\alpha_{1}%
}\wedge..\wedge dx^{r}\otimes\varepsilon_{i}$ on one hand, and

the algebra of r-forms on P, horizontal, equivariant and valued in $VP$ :

$\varpi=\sum_{\left\{  \alpha_{1}...\alpha_{r}\right\}  }\sum_{i}%
\varpi_{\alpha_{1}...\alpha_{r}}^{i}\left(  p\right)  dx^{\alpha_{1}}%
\wedge..\wedge dx^{r}\otimes\zeta\left(  \varepsilon_{i}\right)  \left(
p\right)  $ on the other hand,

are isomorphic so $\zeta\left(  \nabla_{e}\widehat{\varpi}\right)  =\left[
\Phi,\zeta\left(  \widehat{\Phi}\right)  \right]  $

\bigskip

\subsection{Connection on associated bundles}

\label{connection on associated bundles}

\subsubsection{Connection on general associated bundles}

\paragraph{Connection\newline}

\begin{theorem}
A principal connexion $\Phi$ with potentials $\left(  \grave{A}_{a}\right)
_{a\in A}$\ on a principal bundle $P\left(  M,G,\pi\right)  $\ with atlas
$\left(  O_{a},\varphi_{a}\right)  _{a\in A}$\ induces on any associated
bundle $E=P\left[  V,\lambda\right]  $\ with standard trivialization $\left(
O_{a},\psi_{a}\right)  _{a\in\in A}$\ a connexion at $q=\psi_{a}\left(
x,u_{a}\right)  $

$\Psi(q)\left(  v_{ax},v_{au}\right)  \sim\left(  v_{au}+\Gamma_{a}\left(
q\right)  v_{ax}\right)  \in V_{q}E$

defined by the set of Christoffel forms :%

\begin{equation}
\Gamma_{a}\left(  q\right)  =\lambda_{g}^{\prime}(1,u_{a})\grave{A}_{a}\left(
x\right)
\end{equation}

with the transition rule :%

\begin{equation}
\Gamma_{b}\left(  q\right)  =\lambda^{\prime}(g_{ba},u_{a})\left(
-Id,\Gamma_{a}\left(  q\right)  \right)
\end{equation}

\end{theorem}

\begin{proof}
i) Let $g_{ba}\left(  x\right)  \in G$\ be the transition maps of P

E is a fiber bundle $E\left(  M,V,\pi_{E}\right)  $ with atlas $\left(
O_{a},\psi_{a}\right)  _{a\in A}$ :

trivializations $\psi_{a}:O_{a}\times V\rightarrow E::\psi_{a}\left(
x,u\right)  =\Pr\left(  \left(  \varphi_{a}(x,1),u\right)  \right)  $

transitions maps : $q=\psi_{a}\left(  x,u_{a}\right)  =\psi_{b}\left(
x,u_{b}\right)  \Rightarrow u_{b}=\lambda\left(  g_{ba}\left(  x\right)
,u_{a}\right)  $

Define the maps :

$\Gamma_{a}\in\Lambda_{1}\left(  \pi_{E}^{-1}\left(  O_{a}\right)  ;TM^{\ast
}\otimes TV\right)  ::\Gamma_{a}\left(  \psi_{a}\left(  x,u_{a}\right)
\right)  =\lambda_{g}^{\prime}(1,u_{a})\grave{A}_{a}\left(  x\right)  $

\bigskip

$T_{x}M\rightarrow\overset{\grave{A}_{a}\left(  x\right)  }{\rightarrow
}\rightarrow T_{1}G\rightarrow\overset{\lambda_{g}^{\prime}(1,u_{a}%
)}{\rightarrow}\rightarrow T_{u_{a}}V$

\bigskip

At the transitions : $x\in O_{a}\cap O_{b}:q=\psi_{a}\left(  x,u_{a}\right)
=\psi_{b}\left(  x,u_{b}\right)  $

$\grave{A}_{b}\left(  x\right)  =Ad_{g_{ba}}\left(  \grave{A}_{a}\left(
x\right)  -L_{g_{ba}^{-1}}^{\prime}(g_{ba})g_{ba}^{\prime}\left(  x\right)
\right)  $

$u_{b}=\lambda\left(  g_{ba}\left(  x\right)  ,u_{a}\right)  $

$\Gamma_{b}\left(  q\right)  =\lambda_{g}^{\prime}(1,u_{b})\grave{A}%
_{b}\left(  x\right)  =\lambda_{g}^{\prime}(1,\lambda\left(  g_{ba}\left(
x\right)  ,u_{a}\right)  )Ad_{g_{ba}}\left(  \grave{A}_{a}\left(  x\right)
-L_{g_{ba}^{-1}}^{\prime}(g_{ba})g_{ba}^{\prime}\left(  x\right)  \right)  $

$=\lambda_{g}^{\prime}(g_{ba},u_{a})R_{g_{ba}}^{\prime}\left(  1\right)
Ad_{g_{ba}}\grave{A}_{a}-\lambda_{g}^{\prime}(g_{ba},u_{a})R_{g_{ba}}^{\prime
}\left(  1\right)  R_{g_{ba}^{-1}}^{\prime}(g_{ba})L_{g_{ba}}^{\prime
}(1)L_{g_{ba}^{-1}}^{\prime}(g_{ba})g_{ba}^{\prime}\left(  x\right)  $

$=\lambda_{u}^{\prime}(g_{ba},u_{a})\lambda_{g}^{\prime}(1,u_{a}%
)L_{g_{ba}^{-1}}^{\prime}(g_{ba})R_{g_{ba}}^{\prime}\left(  1\right)
Ad_{g_{ba}}\grave{A}_{a}-\lambda_{g}^{\prime}(g_{ba},u_{a})g_{ba}^{\prime
}\left(  x\right)  $

$=\lambda_{u}^{\prime}(g_{ba},u_{a})\lambda_{g}^{\prime}(1,u_{a})\grave{A}%
_{a}-\lambda_{g}^{\prime}(g_{ba},u_{a})g_{ba}^{\prime}\left(  x\right)  $

$=\lambda_{u}^{\prime}(g_{ba},u_{a})\Gamma_{a}\left(  q\right)  -\lambda
_{g}^{\prime}(g_{ba},u_{a})g_{ba}^{\prime}\left(  x\right)  $

$=\lambda^{\prime}(g_{ba},u_{a})\left(  -Id,\Gamma_{a}\left(  q\right)
\right)  $

So the maps $\Gamma_{a}$ define a connection $\Psi\left(  q\right)  $\ on E.

ii) A vector of TE reads : $v_{q}=\psi_{ax}^{\prime}\left(  x,u_{a}\right)
v_{x}+\psi_{au}^{\prime}\left(  x,u_{a}\right)  v_{au}$ and :

$\Psi(q)v_{q}=\psi_{a}^{\prime}(x,u_{a})\left(  0,v_{au}+\Gamma_{a}\left(
q\right)  v_{ax}\right)  =\psi_{au}^{\prime}(x,u_{a})v_{au}+\psi_{au}^{\prime
}(x,u_{a})\lambda_{g}^{\prime}(1,u_{a})\grave{A}_{a}\left(  x\right)  v_{ax}$

The vertical bundle is $\left\{  \left(  p_{a}\left(  x\right)  ,0\right)
\times\left(  u,v_{u}\right)  \right\}  \simeq P\times_{G}\left(  TV\right)  $

so : $\Psi(\psi_{a}\left(  x,u_{a}\right)  )v_{q}\sim\left(  v_{au}+\Gamma
_{a}\left(  q\right)  v_{ax}\right)  \in V_{q}E$
\end{proof}

It can be shown (Kolar p.107) that $\Psi\left(  p,u\right)  \left(
v_{p},v_{u}\right)  =\Pr\left(  \Phi\left(  p\right)  v_{p},v_{u}\right)  $

$\Psi(\psi_{a}\left(  x,u_{a}\right)  )v_{q}$ is a vertical vector which is
equivalent to the fundamental vector of E : $\frac{1}{2}Z\left(  \lambda
_{g}^{\prime}\left(  1,u_{a}\right)  ^{-1}v_{au}+\grave{A}_{a}v_{ax}\right)
\left(  p_{a},u_{a}\right)  $

In a change of trivialization :

$q=\psi_{a}\left(  x,u_{a}\right)  =\widetilde{\psi}_{a}\left(  x,\lambda
\left(  \chi_{a}\left(  x\right)  ,u_{a}\right)  \right)  $

$\widetilde{\Gamma}_{a}\left(  q\right)  =\lambda^{\prime}(\chi_{a}\left(
x\right)  ,u_{a})\left(  -Id,\Gamma_{a}\left(  q\right)  \right)  $

Conversely :

\begin{theorem}
(Kolar p.108) A connection $\Psi$\ on the associated bundle $P\left[
V,\lambda\right]  $ such that the map : $\zeta:T_{1}G\times P\rightarrow TP$
is injective is induced by a principal connection on P iff the Christoffel
forms $\Gamma\left(  q\right)  $ of $\Psi$\ are valued in $\lambda_{g}%
^{\prime}(1,u)T_{1}G$
\end{theorem}

\paragraph{Curvature\newline}

\begin{theorem}
The curvature $\Omega\in\Lambda_{2}\left(  E;VE\right)  $ of a connection
$\Psi$ induced on the associated bundle $E=P\left[  V,\lambda\right]  $\ with
standard trivialization $\left(  O_{a},\psi_{a}\right)  _{a\in\in A}$\ by the
connection $\Phi$\ on the principal bundle P is :

$q=\psi_{a}\left(  x,u\right)  ,u_{x},v_{x}\in T_{x}M:\Omega\left(  q\right)
\left(  u_{x},v_{x}\right)  =\psi_{a}^{\prime}\left(  \mathbf{p}_{a}\left(
x\right)  ,u\right)  \left(  0,\widehat{\Omega}\left(  q\right)  \left(
u_{x},v_{x}\right)  \right)  $ with :%

\begin{equation}
\ \widehat{\Omega}\left(  q\right)  =-\lambda_{g}^{\prime}(1,u)%
\mathcal{F}%
\end{equation}

where
$\mathcal{F}$%
\ is the strength of the connection $\Phi.$ In an holonomic basis of TM and
basis $\left(  \varepsilon_{i}\right)  $ of $T_{1}G:$

$\widehat{\Omega}\left(  q\right)  =-\sum_{i}\left(  d\grave{A}_{a}^{i}%
+\sum_{\alpha\beta}\left[  \grave{A}_{aa},\grave{A}_{a\beta}\right]  _{T_{1}%
G}^{i}d\xi^{\alpha}\Lambda d\xi^{\beta}\right)  \otimes\lambda_{g}^{\prime
}(1,u)\left(  \varepsilon_{i}\right)  $
\end{theorem}

\begin{proof}
The curvature of the induced connection is given by the Cartan formula

$\psi_{a}^{\ast}\Omega=\sum_{i}\left(  -d_{M}\Gamma^{i}+\left[  \Gamma
,\Gamma\right]  _{V}^{i}\right)  \otimes\partial u_{i}$

$\sum_{i}d_{M}\Gamma^{i}\partial u_{i}=\lambda_{g}^{\prime}(1,u)d_{M}\grave
{A}$

$\sum_{i}\left[  \Gamma,\Gamma\right]  _{V}^{i}\partial u_{i}=\sum_{i}\left[
\lambda_{g}^{\prime}(1,u)\grave{A},\lambda_{g}^{\prime}(1,u)\grave{A}\right]
_{V}^{i}\partial u_{i}=-\lambda_{g}^{\prime}(1,u)\left[  \grave{A},\grave
{A}\right]  _{T_{1}G}$

$\psi_{a}^{\ast}\Omega=-\lambda_{g}^{\prime}(1,u)\sum_{i}\left(  d_{M}%
\grave{A}+\left[  \grave{A},\grave{A}\right]  \right)  =-\lambda_{g}^{\prime
}(1,u)%
\mathcal{F}%
$

$\widehat{\Omega}\left(  q\right)  =-\sum_{i}\left(  d\grave{A}^{i}%
+\sum_{\alpha\beta}\left[  \grave{A}_{a},\grave{A}_{\beta}\right]  _{T_{1}%
G}^{i}d\xi^{\alpha}\Lambda d\xi^{\beta}\right)  \otimes\lambda_{g}^{\prime
}(1,u)\left(  \varepsilon_{i}\right)  $
\end{proof}

\paragraph{Covariant derivative\newline}

The covariant derivative $\nabla_{X}S$ of a section S on the associated bundle
$P\left[  V,\lambda\right]  $\ along a vector field X on M, induced\ by the
connection of potential \`{A}\ on the principal bundle $P\left(
M,G,\pi\right)  $\ is :

$\nabla_{X}S=\Psi\left(  S\left(  x\right)  \right)  S^{\prime}\left(
x\right)  X=\left(  \lambda_{g}^{\prime}(1,s\left(  x\right)  )\grave
{A}\left(  x\right)  X+s^{\prime}\left(  x\right)  X\right)  \in T_{s\left(
x\right)  }V$

with $S\left(  x\right)  =\left(  \mathbf{p}_{a}\left(  x\right)  ,s\left(
x\right)  \right)  $

\subsubsection{Connection on an associated vector bundle}

\paragraph{Induced linear connection\newline}

\begin{theorem}
A principal connexion $\Phi$ with potentiels $\left(  \grave{A}_{a}\right)
_{a\in A}$\ on a principal bundle $P\left(  M,G,\pi\right)  $\ with atlas
$\left(  O_{a},\varphi_{a}\right)  _{a\in A}$\ induces on any associated
vector bundle $E=P\left[  V,r\right]  $ a linear connexion $\Psi$ :

$\Psi\left(  \sum_{i}u^{i}e_{i}\left(  x\right)  \right)  \left(  \sum
_{i}v_{u}^{i}e_{i}\left(  x\right)  +\sum_{\alpha}v_{x}^{\alpha}\partial
x_{\alpha}\right)  =\sum_{i}\left(  v_{u}^{i}+\sum_{ij\alpha}\Gamma_{j\alpha
}^{i}\left(  x\right)  u^{j}v_{x}^{\alpha}\right)  \mathbf{e}_{i}\left(
x\right)  $

defined, in a holonomic basis $\left(  \left(  \mathbf{e}_{i}\left(  x\right)
\right)  _{i\in I},\left(  \partial x_{\alpha}\right)  \right)  $\ of TE, by
the Christoffel forms :%

\begin{equation}
\Gamma_{a}=r_{g}^{\prime}(1)\grave{A}_{a}\in\Lambda_{1}\left(  O_{a};%
\mathcal{L}%
\left(  V;V\right)  \right)
\end{equation}

with the transition rule :%

\begin{equation}
\Gamma_{b}\left(  x\right)  =r_{g}^{\prime}(1)Ad_{g_{ba}}\left(  \grave{A}%
_{a}\left(  x\right)  -L_{g_{ba}^{-1}}^{\prime}(g_{ba})g_{ba}^{\prime}\left(
x\right)  \right)
\end{equation}

\end{theorem}

\bigskip

\begin{proof}
i) $\widehat{\Gamma}_{a}\left(  \varphi_{a}\left(  x,1\right)  ,u_{a}\right)
=\lambda_{g}^{\prime}(1,u_{a})\grave{A}_{a}\left(  x\right)  $ reads :

$\widehat{\Gamma}_{a}\left(  \varphi_{a}\left(  x,1\right)  ,u_{a}\right)
=r_{g}^{\prime}(1)u_{a}\grave{A}_{a}\left(  x\right)  $

It is linear with respect to u :

Denote : $\Gamma_{a}\left(  x\right)  u_{a}=\widehat{\Gamma}_{a}\left(
\varphi_{a}\left(  x,1\right)  ,u_{a}\right)  $

$\Gamma_{a}\left(  x\right)  =r_{g}^{\prime}(1)\grave{A}_{a}\left(  x\right)
\in\Lambda_{1}\left(  M;%
\mathcal{L}%
\left(  V;V\right)  \right)  $

$\Gamma\left(  x\right)  =\sum_{ij\alpha}\left[  r_{g}^{\prime}(1)\grave
{A}_{a\alpha}\left(  x\right)  \right]  _{j}^{i}d\xi^{\alpha}\otimes
\mathbf{e}^{j}\left(  x\right)  \otimes\mathbf{e}_{i}\left(  x\right)  $

$\Gamma_{j\alpha}^{i}\left(  x\right)  =\left[  r_{g}^{\prime}(1)\grave
{A}_{a\alpha}\left(  x\right)  \right]  _{j}^{i}$

\bigskip

$T_{x}M\rightarrow\overset{\grave{A}_{a}\left(  x\right)  }{\rightarrow
}\rightarrow T_{1}G\rightarrow\overset{r_{g}^{\prime}\left(  1\right)
}{\rightarrow}\rightarrow%
\mathcal{L}%
\left(  V;V\right)  $

\bigskip

ii) For $x\in O_{a}\cap O_{b}:q=\left(  \varphi_{a}\left(  x,1\right)
,u_{a}\right)  =\left(  \varphi_{b}\left(  x,1\right)  ,u_{b}\right)  $

$\Gamma_{b}\left(  x\right)  =r_{g}^{\prime}(1)\grave{A}_{b}\left(  x\right)
=r_{g}^{\prime}(1)Ad_{g_{ba}}\left(  \grave{A}_{a}\left(  x\right)
-L_{g_{ba}^{-1}}^{\prime}(g_{ba})g_{ba}^{\prime}\left(  x\right)  \right)  $

iii) In a holonomic basis of E : $\mathbf{e}_{ai}\left(  x\right)  =\left[
\left(  p_{a}\left(  x\right)  ,e_{i}\right)  \right]  :$

$v_{q}\in T_{q}E::v_{q}=\sum_{i}v_{u}^{i}\mathbf{e}_{i}\left(  x\right)
+\sum_{\alpha}v_{x}^{\alpha}\partial x_{\alpha}$

So we can write :

$\Psi\left(  \sum_{i}u^{i}e_{i}\left(  x\right)  \right)  \left(  \sum
_{i}v_{u}^{i}e_{i}\left(  x\right)  +\sum_{\alpha}v_{x}^{\alpha}\partial
x_{\alpha}\right)  =\sum_{i}\left(  v_{u}^{i}+\sum_{ij\alpha}\Gamma_{j\alpha
}^{i}\left(  x\right)  u^{j}v_{x}^{\alpha}\right)  \mathbf{e}_{i}\left(
x\right)  $
\end{proof}

We have a linear connection on E, which has all the properties of a common
linear connection on vector bundles, by using the holonomic basis : $\left(
\mathbf{p}_{a}\left(  x\right)  ,e_{i}\right)  =\mathbf{e}_{ai}\left(
x\right)  . $

Remarks :

i) With the general identity : $Ad_{r\left(  g_{ba}\right)  }r^{\prime}\left(
1\right)  =r^{\prime}\left(  1\right)  Ad_{g_{ba}}$ and the adjoint map on G%
$\mathcal{L}$%
(V;V) :

$\Gamma_{b}\left(  x\right)  =Ad_{r\left(  g_{ba}\right)  }r^{\prime}\left(
1\right)  \left(  \grave{A}_{a}\left(  x\right)  -L_{g_{ba}^{-1}}^{\prime
}(g_{ba})g_{ba}^{\prime}\left(  x\right)  \right)  $

$=Ad_{r\left(  g_{ba}\right)  }\Gamma_{a}\left(  x\right)  -r^{\prime}\left(
1\right)  R_{g_{ba}^{-1}}^{\prime}(g_{ba})g_{ba}^{\prime}\left(  x\right)  $

ii) there is always a complexified $E_{%
\mathbb{C}
}=P\left[  V_{%
\mathbb{C}
},r_{%
\mathbb{C}
}\right]  $ of a real associated vector bundle with $V_{%
\mathbb{C}
}=V\oplus iV$ and

$r_{%
\mathbb{C}
}:G\rightarrow V_{%
\mathbb{C}
}::r_{%
\mathbb{C}
}\left(  g\right)  \left(  u+iv\right)  =r\left(  g\right)  u+ir\left(
g\right)  v$

so $\forall X\in T_{1}G:r_{%
\mathbb{C}
}^{\prime}\left(  1\right)  X\in G%
\mathcal{L}%
\left(  V_{%
\mathbb{C}
};V_{%
\mathbb{C}
}\right)  $ . A real connection on P has an extension on $E_{%
\mathbb{C}
}$ by : $\Gamma_{%
\mathbb{C}
}=r_{%
\mathbb{C}
}^{\prime}(1)\grave{A}_{a}\in G%
\mathcal{L}%
\left(  V_{%
\mathbb{C}
};V_{%
\mathbb{C}
}\right)  .$ However P, G and \`{A} stay the same. To define an extension to
$\left(  T_{1}G\right)  _{%
\mathbb{C}
}$ one needs an additional map : $\grave{A}_{%
\mathbb{C}
a}=\grave{A}_{a}+i\operatorname{Im}\grave{A}_{a}$

\paragraph{Curvature\newline}

\begin{theorem}
The curvature $\Omega$ of a connection $\Psi$ induced on the associated vector
bundle $P\left[  V,r\right]  $\ by the connection $\Phi$\ with potential
\`{A}\ on the principal bundle P is : \ $q=\psi_{a}\left(  x,u\right)
,u_{x},v_{x}\in T_{x}M:\Omega\left(  q\right)  \left(  u_{x},v_{x}\right)
=\psi_{a}^{\prime}\left(  q\right)  \left(  0,\widehat{\Omega}\left(
q\right)  \left(  u_{x},v_{x}\right)  \right)  $ with : \ $\widehat{\Omega
}\left(  \psi_{a}\left(  x,u\right)  \right)  =-r^{\prime}\left(  1\right)  u%
\mathcal{F}%
$ where
$\mathcal{F}$%
\ is the strength of the connection $\Phi.$ In an holonomic basis of TM and
basis $\left(  \varepsilon_{i}\right)  $ of $T_{1}G:$

$\widehat{\Omega}\left(  q\right)  =-\sum_{i}\left(  d\grave{A}_{a}^{i}%
+\sum_{\alpha\beta}\left[  \grave{A}_{aa},\grave{A}_{a\beta}\right]  _{T_{1}%
G}^{i}d\xi^{\alpha}\wedge d\xi^{\beta}\right)  \otimes r_{g}^{\prime}\left(
1\right)  u\left(  \varepsilon_{i}\right)  $

At the transitions : $x\in O_{a}\cap O_{b}:\widehat{\Omega}\left(
p_{b}\left(  x\right)  ,e_{i}\right)  =Ad_{r\left(  g_{ba}\right)  }%
\widehat{\Omega}\left(  p_{a}\left(  x\right)  ,e_{i}\right)  $
\end{theorem}

\begin{proof}
$e_{bi}\left(  x\right)  =\left(  p_{b}\left(  x\right)  ,e_{i}\right)
=r\left(  g_{ba}\left(  x\right)  \right)  e_{ai}\left(  x\right)  =r\left(
g_{ba}\left(  x\right)  \right)  \left(  p_{a}\left(  x\right)  ,e_{i}\right)
$

$%
\mathcal{F}%
_{b}=Ad_{g_{ba}}%
\mathcal{F}%
_{a}$

With the general identity : $Ad_{r\left(  g_{ba}\right)  }r^{\prime}\left(
1\right)  =r^{\prime}\left(  1\right)  Ad_{g_{ba}}$ and the adjoint map on G%
$\mathcal{L}$%
(V;V)

$\Omega\left(  p_{b}\left(  x\right)  ,e_{i}\right)  =-r^{\prime}\left(
1\right)  e_{i}%
\mathcal{F}%
_{b}=-r^{\prime}\left(  1\right)  e_{i}Ad_{g_{ba}}%
\mathcal{F}%
_{a}=-Ad_{r\left(  g_{ba}\right)  }r^{\prime}\left(  1\right)  e_{i}%
\mathcal{F}%
_{a}=Ad_{r\left(  g_{ba}\right)  }\Omega\left(  p_{a}\left(  x\right)
,e_{i}\right)  $
\end{proof}

The curvature is linear with respect to u.

The curvature form $\widetilde{\Omega}$\ of $\Psi$ : $\widetilde{\Omega}%
_{i}\left(  x\right)  =\widehat{\Omega}\left(  p_{a}\left(  x\right)
,e_{i}\right)  $

\paragraph{Exterior covariant derivative\newline}

1. There are two different exterior covariant derivatives :

i) for r-forms on M valued on a vector bundle $E\left(  M,V,\pi\right)  $ with
a linear connection:

$\widetilde{\nabla}_{e}:\Lambda_{r}\left(  M;E\right)  \rightarrow
\Lambda_{r+1}\left(  M;E\right)  ::$

$\widetilde{\nabla}_{e}\varpi=\sum_{i}\left(  d_{M}\varpi^{i}+\left(  \sum
_{j}\left(  \sum_{\alpha}\widetilde{\Gamma}_{j\alpha}^{i}d\xi^{\alpha}\right)
\wedge\varpi^{j}\right)  \right)  \otimes e_{i}\left(  x\right)  $

It is still valid on an associated vector bundle with the induced connection.

ii) for r-forms on a principal bundle P$\left(  M,G,\pi\right)  $, valued in a
fixed vector space V, with a principal connection:

$\nabla_{e}:\Lambda_{r}\left(  P;V\right)  \rightarrow\Lambda_{r+1}\left(
P;V\right)  ::$ $\nabla_{e}\varpi=\chi^{\ast}\left(  d_{P}\varpi\right)  $

There is a relation between those two concepts .

\begin{theorem}
(Kolar p.112) There is a canonical isomorphism between the space $\Lambda
_{s}\left(  M;P\left[  V,r\right]  \right)  $ of s-forms on M valued in the
associated vector bundle $P\left[  V,r\right]  $ and the space $W_{s}%
\subset\Lambda_{s}\left(  P;V\right)  $\ of horizontal, equivariant V valued
s-forms on P.
\end{theorem}

A s-form on M valued in the associated vector bundle E=$P\left[  V,r\right]  $
reads :

$\widetilde{\varpi}=\sum_{\left\{  \alpha_{1}...\alpha_{s}\right\}  }\sum
_{i}\widetilde{\varpi}_{\alpha_{1}...\alpha_{s}}^{i}\left(  x\right)
d\xi^{\alpha_{1}}\wedge..\wedge d\xi^{\alpha_{s}}\otimes\mathbf{e}_{i}\left(
x\right)  $

with a holonomic basis $\left(  d\xi^{\alpha}\right)  $ of TM*, and
$\mathbf{e}_{i}\left(  x\right)  =\psi_{a}\left(  x,e_{i}\right)  =\left(
\mathbf{p}_{a}\left(  x\right)  ,e_{i}\right)  $

A s-form on P, horizontal, equivariant, V valued reads :

$\varpi=\sum_{\left\{  \alpha_{1}...\alpha_{s}\right\}  }\sum_{i}%
\varpi_{\alpha_{1}...\alpha_{s}}^{i}\left(  p\right)  dx^{\alpha_{1}}%
\wedge..\wedge dx^{s}\otimes e_{i}$ and : $\rho\left(  .,g\right)  ^{\ast
}\varpi=r\left(  g^{-1}\right)  \varpi$

with the dual basis of $\partial x_{\alpha}=\varphi_{a}^{\prime}\left(
x,g\right)  \partial\xi_{\alpha}$\ and a basis $e_{i}$ of V

The equivariance means :

$\rho\left(  .,g\right)  ^{\ast}\varpi\left(  p\right)  \left(  v_{1}%
,..v_{s}\right)  =\varpi\left(  \rho\left(  p,g\right)  \right)  \left(
\rho_{p}^{\prime}\left(  p,g\right)  v_{1},...\rho_{p}^{\prime}\left(
p,g\right)  v_{s}\right)  =r\left(  g^{-1}\right)  \varpi\left(  p\right)
\left(  v_{1},..v_{s}\right)  $

$\Leftrightarrow\sum_{i}\varpi_{\alpha_{1}...\alpha_{s}}^{i}\left(
\rho\left(  p,g\right)  \right)  e_{i}=r\left(  g^{-1}\right)  \left(
\sum_{i}\varpi_{\alpha_{1}...\alpha_{s}}^{i}\left(  p\right)  e_{i}\right)  $

So : $\sum_{i}\varpi_{\alpha_{1}...\alpha_{s}}^{i}\left(  \varphi\left(
x,g\right)  \right)  e_{i}=r\left(  g^{-1}\right)  \left(  \sum_{i}%
\varpi_{\alpha_{1}...\alpha_{s}}^{i}\left(  \varphi\left(  x,1\right)
\right)  e_{i}\right)  $

The isomorphism : $\theta:\Lambda_{s}\left(  M;E\right)  \rightarrow W_{s}$
reads :

$v_{k}\in T_{p}P:X_{k}=\pi^{\prime}\left(  p\right)  v_{k}$

$\widetilde{\varpi}\left(  x\right)  \left(  X_{1},...,X_{s}\right)  =\left(
p_{a}\left(  x\right)  ,\varpi\left(  p_{a}\left(  x\right)  \right)  \left(
v_{1},...,v_{s}\right)  \right)  $

$\left(  p_{a}\left(  x\right)  ,\varpi\left(  \varphi_{a}\left(  x,1\right)
\right)  \left(  v_{1},...,v_{s}\right)  \right)  =\widetilde{\varpi}\left(
x\right)  \left(  \pi^{\prime}\left(  p\right)  v_{1},...,\pi^{\prime}\left(
p\right)  v_{s}\right)  $

$\theta\left(  \widetilde{\varpi}\right)  $ is called the \textbf{frame form}
of $\widetilde{\varpi}$

Then : $\theta\circ\widetilde{\nabla}_{e}=\nabla_{e}\circ\theta:\Lambda
_{r}\left(  M;E\right)  \rightarrow W_{r+1}$ where the connection on E is
induced by the connection on P.

We have similar relations between the curvatures :

The Riemann curvature on the vector bundle

$E=P\left[  V,r\right]  $ :

$R\in\Lambda_{2}\left(  M;E^{\ast}\otimes E\right)  :\widetilde{\nabla}%
_{e}\left(  \widetilde{\nabla}_{e}\widetilde{\varpi}\right)  =\sum_{ij}\left(
\sum_{\alpha\beta}R_{j\alpha\beta}^{i}d\xi^{\alpha}\wedge d\xi^{\beta}\right)
\wedge\widetilde{\varpi}^{j}\otimes\mathbf{e}_{i}\left(  x\right)  $

The curvature form on the principal bundle :

$\widehat{\Omega}\in\Lambda_{2}\left(  P;T_{1}G\right)  :\widehat{\Omega
}\left(  p\right)  =-Ad_{g^{-1}}\sum_{i}\left(  d_{M}\grave{A}^{i}+\left[
\sum_{\alpha}\grave{A}_{a},\sum_{\beta}\grave{A}_{\beta}\right]  _{T_{1}G}%
^{i}\right)  \varepsilon_{i}$

We have :

$\theta\circ R=r^{\prime}(1)\circ\widehat{\Omega}=\theta\circ\widetilde
{\nabla}_{e}\widetilde{\nabla}_{e}=\nabla_{e}\nabla_{e}\circ\theta$

\subsubsection{Metric connections}

A metric connection on a vector bundle (where it is possible to define
tensors) is a connection such that the covariant derivative of the metric is null.

\paragraph{General result\newline}

First a general result, which has a broad range of practical applications.

\begin{theorem}
If $(V,\gamma)$ is a real vector space endowed with a scalar product $\gamma$,
(V,r) an unitary representation of the group G, $P\left(  M,G,\pi\right)  $ a
real principal fiber bundle endowed with a principal connection with potential
\`{A}, then E=P[V,r] is a vector bundle endowed with a scalar product and a
linear metric connection.
\end{theorem}

\begin{proof}
i) $\gamma$ is a symmetric form, represented in a base of V by the matrix
$\left[  \gamma\right]  =\left[  \gamma\right]  ^{t}$

ii) (V,r) is a unitary representation :

$\forall u,v\in V,g\in G:\gamma\left(  r\left(  g\right)  u,r\left(  g\right)
v\right)  =\gamma\left(  u,v\right)  $

so (V,r'(1)) is a anti-symmetric representation of the Lie algebra $T_{1}G:$

$\forall X\in T_{1}G:\gamma\left(  r^{\prime}(1)Xu,r^{\prime}(1)Xv\right)  =0$

$\Leftrightarrow\left(  \left[  \gamma\right]  \left[  r_{g}^{\prime
}(1)\right]  \left[  X\right]  \right)  ^{t}+\left[  \gamma\right]  \left[
r_{g}^{\prime}(1)\right]  \left[  X\right]  =0$

iii) (V,r) is representation of G so E=P[V,r] is a vector bundle.

iv) $\gamma$ is preserved by r, so $g\left(  \left(  p_{a}\left(  x\right)
,u\right)  ,\left(  p_{a}\left(  x\right)  ,v\right)  \right)  =\gamma\left(
u,v\right)  $ is a scalar product on E

v) the principal connection on P is real so its potential \`{A} is real

vi) It induces the linear connection on E with Christoffel form :
$\Gamma\left(  x\right)  =r_{g}^{\prime}(1)\grave{A}$

vii) the connection is metric because :

$\left(  \left[  \gamma\right]  \left[  r_{g}^{\prime}(1)\right]  \left[
\grave{A}_{\alpha}\right]  \right)  ^{\ast}+\left[  \gamma\right]  \left[
r_{g}^{\prime}(1)\right]  \left[  \grave{A}_{\alpha}\right]  =0$

$\Rightarrow\left[  \grave{A}_{\alpha}\right]  ^{t}\left[  r_{g}^{\prime
}\left(  1\right)  \right]  ^{\ast}\left[  \gamma\right]  +\left[
\gamma\right]  \left[  r_{g}^{\prime}(1)\right]  \left[  \grave{A}_{\alpha
}\right]  =0$

$\Leftrightarrow\left[  \Gamma_{\alpha}\right]  ^{t}\left[  \gamma\right]
+\left[  \gamma\right]  \left[  \Gamma_{\alpha}\right]  =0$
\end{proof}

Remark : the theorem still holds for the complex case, but we need a real
structure on V with respect to which the transition maps are real, and a real connection.

\paragraph{Metric connections on the tangent bundle of a manifold\newline}

The following theorem is new. It makes a link between usual "affine
connections" (Part Differential geometry) on manifolds and connections on
fiber bundle. The demonstration gives many useful details about the definition
of both the bundle of orthogonal frames and of the induced connection, in
practical terms.

\begin{theorem}
For any real m dimensional pseudo riemannian manifold (M,g) with the signature
(r,s)\ of g, any principal connection on the principal bundle $P\left(
M,O\left(
\mathbb{R}
,r,s\right)  ,\pi\right)  $ of its orthogonal frames induces a metric, affine
connection on (M,g).\ Conversely any affine connection on (M,g) induces a
principal connection on P.
\end{theorem}

\begin{proof}
i) P is a principal bundle with atlas $\left(  O_{a},\varphi_{a}\right)
_{a\in A}$\ . The trivialization $\varphi_{a}$\ is defined by a family of maps
: $L_{a}\in C_{0}\left(  O_{a};GL\left(
\mathbb{R}
,m\right)  \right)  $ such that : $\varphi_{a}\left(  x,1\right)
=\mathbf{e}_{ai}\left(  x\right)  =\sum_{\alpha}\left[  L_{a}\right]
_{i}^{\alpha}\partial\xi_{\alpha}$ is an orthonormal basis of (M,g) and
$\varphi_{a}\left(  x,S\right)  =\rho\left(  \mathbf{e}_{ai}\left(  x\right)
,g\right)  =\sum_{\alpha}\left[  L_{a}\right]  _{j}^{\alpha}\left[  S\right]
_{i}^{j}\partial\xi_{\alpha}$ with $\left[  S\right]  \in O\left(
\mathbb{R}
,r,s\right)  .$ So in a holonomic basis of TM we have the matrices relation:
$\left[  L_{a}\left(  x\right)  \right]  ^{t}\left[  g\left(  x\right)
\right]  \left[  L_{a}\left(  x\right)  \right]  =\left[  \eta\right]  $ with
$\eta_{ij}=\pm1$. At the transitions of P : $\mathbf{e}_{bi}\left(  x\right)
=\varphi_{b}\left(  x,1\right)  =\varphi_{a}\left(  x,g_{ab}\right)
=\sum_{\alpha}\left[  L_{a}\right]  _{j}^{\alpha}\left[  g_{ab}\right]
_{i}^{j}\partial\xi_{\alpha}=\sum_{\alpha}\left[  L_{b}\right]  _{i}^{\alpha
}\partial\xi_{\alpha}$ so : $\left[  L_{b}\right]  =\left[  L_{a}\right]
\left[  g_{ab}\right]  $ with $\left[  g_{ab}\right]  \in O\left(
\mathbb{R}
,r,s\right)  .$

E is the associated vector bundle $P\left[
\mathbb{R}
^{m},\imath\right]  $ where $\left(
\mathbb{R}
^{m},\imath\right)  $ is the standard representation of $O\left(
\mathbb{R}
,r,s\right)  .$ A vector of E is a vector of TM defined in the orthonormal
basis $\mathbf{e}_{ai}\left(  x\right)  :U_{p}=\sum_{i}U_{a}^{i}%
\mathbf{e}_{ai}\left(  x\right)  =\sum_{i\alpha}U_{a}^{i}\left[  L_{a}\right]
_{i}^{\alpha}\partial\xi_{\alpha}=\sum_{\alpha}u^{\alpha}\partial\xi_{\alpha}$
with $U_{a}^{i}=\sum_{\alpha}\left[  L_{a}^{\prime}\right]  _{\alpha}%
^{i}u^{\alpha}$ and $\left[  L_{a}^{\prime}\right]  =\left[  L_{a}\right]
^{-1}.$

Any principal connexion on P induces the linear connexion on E :

$\Gamma\left(  x\right)  =\sum_{\alpha ij}\left[  \Gamma_{a\alpha}\right]
_{i}^{j}d\xi^{\alpha}\otimes\mathbf{e}_{a}^{i}\left(  x\right)  \otimes
\mathbf{e}_{aj}\left(  x\right)  $ where $\left[  \Gamma_{\alpha}\right]
=\imath^{\prime}(1)\grave{A}_{\alpha}$ is a matrix on the standard
representation $\left(
\mathbb{R}
^{m},\imath^{\prime}\left(  1\right)  \right)  $ of $o\left(
\mathbb{R}
,r,s\right)  $ so : $\left[  \Gamma_{\alpha}\right]  ^{t}\left[  \eta\right]
+\left[  \eta\right]  \left[  \Gamma_{\alpha}\right]  =0$

At the transitions : $\Gamma_{b}\left(  x\right)  =Ad_{r(g_{ba})}\Gamma
_{a}\left(  x\right)  -r_{g}^{\prime}(1)R_{g_{ba}^{-1}}^{\prime}(g_{ba}%
)g_{ba}^{\prime}\left(  x\right)  $ which reads :

$\left[  \Gamma_{b}\right]  =-\left[  \partial_{\alpha}g_{ba}\right]  \left[
g_{ab}\right]  +\left[  g_{ba}\right]  \left[  \Gamma_{a}\right]  \left[
g_{ab}\right]  =\left[  g_{ba}\right]  \left[  \partial_{\alpha}g_{ab}\right]
+\left[  g_{ba}\right]  \left[  \Gamma_{a}\right]  \left[  g_{ab}\right]  $

iii) The covariant derivative of a vector field $U=\sum_{i}U^{i}\left(
x\right)  \mathbf{e}_{i}\left(  x\right)  \in\mathfrak{X}\left(  E\right)  $ is

$\nabla U=\sum\left(  \partial_{\alpha}U^{i}+\Gamma_{a\alpha j}^{i}%
U^{j}\right)  d\xi^{\alpha}\otimes\mathbf{e}_{ai}\left(  x\right)  $

which reads in the holonomic basis of M :

$\nabla U=\left(  \sum\partial_{\alpha}\left(  \left[  L_{a}^{\prime}\right]
_{\beta}^{i}u^{\beta}\right)  +\Gamma_{a\alpha j}^{i}\left[  L_{a}^{\prime
}\right]  _{\beta}^{j}u^{\beta}\right)  d\xi^{\alpha}\otimes\left[
L_{a}\right]  _{i}^{\gamma}\partial\xi_{\gamma}$

$\nabla U=\left(  \sum\partial_{\alpha}u^{\gamma}+\widetilde{\Gamma}%
_{\alpha\beta}^{\gamma}u^{\beta}\right)  d\xi^{\alpha}\otimes\partial
\xi_{\gamma}$

with :%

\begin{equation}
\widetilde{\Gamma}_{a\alpha\beta}^{\gamma}=\left[  L_{a}\right]  _{i}^{\gamma
}\left(  \left[  \partial_{\alpha}L_{a}^{\prime}\right]  _{\beta}^{i}+\left[
\Gamma_{a\alpha}\right]  _{j}^{i}\left[  L_{a}^{\prime}\right]  _{\beta}%
^{j}\right)  \Leftrightarrow\left[  \widetilde{\Gamma}_{a\alpha}\right]
=\left[  L_{a}\right]  \left(  \left[  \partial_{\alpha}L_{a}^{\prime}\right]
+\left[  \Gamma_{a\alpha}\right]  \left[  L_{a}^{\prime}\right]  \right)
\end{equation}

As $\nabla U$ is intrinsic its defines an affine connection, at least on
$\pi_{E}^{-1}\left(  O_{a}\right)  .$

iv) At the transitions of P :

$\left[  \widetilde{\Gamma}_{b\alpha}\right]  =\left[  L_{b}\right]  \left(
\left[  \partial_{\alpha}L_{b}^{\prime}\right]  +\left[  \Gamma_{b\alpha
}\right]  \left[  L_{b}^{\prime}\right]  \right)  $

$=\left[  L_{a}\right]  \left[  g_{ab}\right]  \left(  \partial_{\alpha
}\left(  \left[  g_{ab}\right]  \left[  L_{a}^{\prime}\right]  \right)
+\left(  \left[  g_{ba}\right]  \left[  \partial_{\alpha}g_{ab}\right]
+\left[  g_{ba}\right]  \left[  \Gamma_{a}\right]  \left[  g_{ab}\right]
\right)  \left(  \left[  g_{ba}\right]  \left[  L_{a}^{\prime}\right]
\right)  \right)  $

$=\left[  L_{a}\right]  \left[  g_{ab}\right]  \left[  \partial_{\alpha}%
g_{ba}\right]  \left[  L_{a}^{\prime}\right]  +\left[  L_{a}\right]  \left[
\partial_{\alpha}L_{a}^{\prime}\right]  +\left[  L_{a}\right]  \left[
\partial_{\alpha}g_{ab}\right]  \left[  g_{ba}\right]  \left[  L_{a}^{\prime
}\right]  +\left[  L_{a}\right]  \left[  \Gamma_{a}\right]  \left[
L_{a}^{\prime}\right]  $

$=\left[  L_{a}\right]  \left(  \left[  \partial_{\alpha}L_{a}^{\prime
}\right]  +\left[  \Gamma_{a}\right]  \left[  L_{a}^{\prime}\right]  \right)
+\left[  L_{a}\right]  \left(  \left[  g_{ab}\right]  \left[  \partial
_{\alpha}g_{ba}\right]  +\left[  \partial_{\alpha}g_{ab}\right]  \left[
g_{ba}\right]  \right)  \left[  L_{a}^{\prime}\right]  $

$=\left[  \widetilde{\Gamma}_{a\alpha}\right]  +\left[  L_{a}\right]
\partial_{\alpha}\left(  \left[  g_{ab}\right]  \left[  g_{ba}\right]
\right)  \left[  L_{a}^{\prime}\right]  =\left[  \widetilde{\Gamma}_{a\alpha
}\right]  $

Thus the affine connection is defined over TM.

v) This connexion is metric with respect to g in the usual meaning for affine connection.

The condition is :

$\forall\alpha,\beta,\gamma:\partial_{\gamma}g_{\alpha\beta}=\sum_{\eta
}\left(  g_{\alpha\eta}\Gamma_{\gamma\beta}^{\eta}+g_{\beta\eta}\Gamma
_{\gamma\alpha}^{\eta}\right)  \Leftrightarrow\left[  \partial_{\alpha
}g\right]  =\left[  g\right]  \left[  \widetilde{\Gamma}_{\alpha}\right]
+\left[  \widetilde{\Gamma}_{\alpha}\right]  ^{t}\left[  g\right]  $

with the metric g defined from the orthonormal basis :

$\left[  L\right]  ^{t}\left[  g\right]  \left[  L\right]  =\left[
\eta\right]  \Leftrightarrow\left[  g\right]  =\left[  L^{\prime}\right]
^{t}\left[  \eta\right]  \left[  L^{\prime}\right]  $

$\left[  \partial_{\alpha}g\right]  =\left[  \partial_{\alpha}L^{\prime
}\right]  ^{t}\left[  \eta\right]  \left[  L^{\prime}\right]  +\left[
L^{\prime}\right]  ^{t}\left[  \eta\right]  \left[  \partial_{\alpha}%
L^{\prime}\right]  $

$\left[  g\right]  \left[  \widetilde{\Gamma}_{\alpha}\right]  +\left[
\widetilde{\Gamma}_{\alpha}\right]  ^{t}\left[  g\right]  $

$=\left[  L^{\prime}\right]  ^{t}\left[  \eta\right]  \left[  L^{\prime
}\right]  \left(  \left[  L\right]  \left(  \left[  \partial_{\alpha}%
L^{\prime}\right]  +\left[  \Gamma_{\alpha}\right]  \left[  L^{\prime}\right]
\right)  \right)  +\left(  \left[  \partial_{\alpha}L^{\prime}\right]
^{t}+\left[  L^{\prime}\right]  ^{t}\left[  \Gamma_{\alpha}\right]
^{t}\right)  \left[  L\right]  ^{t}\left[  L^{\prime}\right]  ^{t}\left[
\eta\right]  \left[  L^{\prime}\right]  $

$=\left[  L^{\prime}\right]  ^{t}\left[  \eta\right]  \left[  L^{\prime
}\right]  \left[  L\right]  \left[  \partial_{\alpha}L^{\prime}\right]
+\left[  L^{\prime}\right]  ^{t}\left[  \eta\right]  \left[  L^{\prime
}\right]  \left[  L\right]  \left[  \Gamma_{\alpha}\right]  \left[  L^{\prime
}\right]  +\left[  \partial_{\alpha}L^{\prime}\right]  ^{t}\left[
\eta\right]  \left[  L^{\prime}\right]  +\left[  L^{\prime}\right]
^{t}\left[  \Gamma_{\alpha}\right]  ^{t}\left[  \eta\right]  \left[
L^{\prime}\right]  $

$=\left[  L^{\prime}\right]  ^{t}\left[  \eta\right]  \left[  \partial
_{\alpha}L^{\prime}\right]  +\left[  \partial_{\alpha}L^{\prime}\right]
^{t}\left[  \eta\right]  \left[  L^{\prime}\right]  +\left[  L^{\prime
}\right]  ^{t}\left(  \left[  \eta\right]  \left[  \Gamma_{\alpha}\right]
+\left[  \Gamma_{\alpha}\right]  ^{t}\left[  \eta\right]  \right)  \left[
L^{\prime}\right]  $

$=\left[  L^{\prime}\right]  ^{t}\left[  \eta\right]  \left[  \partial
_{\alpha}L^{\prime}\right]  +\left[  \partial_{\alpha}L^{\prime}\right]
^{t}\left[  \eta\right]  \left[  L^{\prime}\right]  $ because $\left[
\Gamma_{\alpha}\right]  \in o(%
\mathbb{R}
,r,s)$

vi) Conversely an affine connection $\widetilde{\Gamma}_{\alpha\beta}^{\gamma
}$ defines locally the maps:

$\Gamma_{a}\left(  x\right)  =\sum_{\alpha ij}\left[  \Gamma_{a\alpha}\right]
_{i}^{j}d\xi^{\alpha}\otimes\mathbf{e}_{a}^{i}\left(  x\right)  \otimes
\mathbf{e}_{aj}\left(  x\right)  $ with $\left[  \Gamma_{a\alpha}\right]
=\left[  L_{a}^{\prime}\right]  \left(  \left[  \widetilde{\Gamma}_{\alpha
}\right]  \left[  L_{a}\right]  +\left[  \partial_{\alpha}L_{a}\right]
\right)  $

At the transitions :

$\left[  \Gamma_{b\alpha}\right]  =\left[  L_{b}^{\prime}\right]  \left(
\left[  \widetilde{\Gamma}_{\alpha}\right]  \left[  L_{b}\right]  +\left[
\partial_{\alpha}L_{b}\right]  \right)  =\left[  g_{ba}\right]  \left[
L_{a}^{\prime}\right]  \left(  \left[  \widetilde{\Gamma}_{\alpha}\right]
\left[  L_{a}\right]  \left[  g_{ab}\right]  +\left[  \partial_{\alpha}%
L_{a}\right]  \left[  g_{ab}\right]  +\left[  L_{a}\right]  \left[
\partial_{\alpha}g_{ab}\right]  \right)  $

$=\left[  g_{ba}\right]  \left[  L_{a}^{\prime}\right]  \left(  \left[
\widetilde{\Gamma}_{\alpha}\right]  \left[  L_{a}\right]  +\left[
\partial_{\alpha}L_{a}\right]  \right)  \left[  g_{ab}\right]  +\left[
g_{ba}\right]  \left[  L_{a}^{\prime}\right]  \left[  L_{a}\right]  \left[
\partial_{\alpha}g_{ab}\right]  $

$=Ad_{\left[  g_{ba}\right]  }\left[  \Gamma_{a\alpha}\right]  -\left[
\partial_{\alpha}g_{ba}\right]  \left[  g_{ab}\right]  =Ad_{\left[
g_{ba}\right]  }\left(  \left[  \Gamma_{a\alpha}\right]  -\left[
g_{ab}\right]  \left[  \partial_{\alpha}g_{ba}\right]  \right)  $

$=Ad_{r^{\prime}(1)g_{ba}}\left(  r^{\prime}(1)\grave{A}_{a\alpha}-r^{\prime
}(1)L_{g_{ba}^{-1}}^{\prime}\left(  g_{ba}\right)  \partial_{\alpha}%
g_{ba}\right)  $

$r^{\prime}(1)\grave{A}_{b\alpha}=r^{\prime}(1)Ad_{g_{ba}}\left(  \grave
{A}_{a\alpha}-L_{g_{ba}^{-1}}^{\prime}\left(  g_{ba}\right)  \partial_{\alpha
}g_{ba}\right)  $

$\grave{A}_{b}=Ad_{g_{ba}}\left(  \grave{A}_{a}-L_{g_{ba}^{-1}}^{\prime
}\left(  g_{ba}\right)  g_{ba}^{\prime}\right)  $

So it defines a principal connection on P.
\end{proof}

Remarks :

i) Whenever we have a principal bundle $P\left(  M,O\left(
\mathbb{R}
,r,s\right)  ,\pi\right)  $ of frames on TM, using the standard representation
of $O\left(
\mathbb{R}
,r,s\right)  $ we can build E, which can be assimilated to TM. E can be
endowed with a metric, by importing the metric on $%
\mathbb{R}
^{m}$ through the standard representation, which is equivalent to define g by
: $\left[  g\right]  =\left[  L^{\prime}\right]  ^{t}\left[  \eta\right]
\left[  L^{\prime}\right]  .$ (M,g) becomes a pseudo-riemannian manifold. The
bundle of frames is the geometric definition of a pseudo-riemannian manifold.
Not any manifold admit such a metric, so the topological obstructions lie also
on the level of P.

ii) With any principal connection on P we have a metric affine connexion on
TM. So such connections are not really special : using the geometric
definition of (M,g), any affine connection is metric with the induced
metric.\ The true particularity of the L\'{e}vy-Civita connection is that it
is torsion free.

\subsubsection{Connection on a Spin bundle}

The following theorem is new. With a principal bundle Sp(M,$Spin\left(
\mathbb{R}
,p,q\right)  $,$\pi_{S})$ any representation (V,r) of the Clifford algebra
Cl($%
\mathbb{R}
,r,s)$ becomes an associated vector bundle E=P[V,r], and even a fiber bundle.
So any principal connection on Sp induces a linear connection on E the usual
way. However the existence of the Clifford algebra action permits to define
another connection, which has some interesting properties and is the starting
point for the Dirac operator seen in Functional Analysis. For the details of
the demonstration see Clifford Algebras in the Algebra part.

\begin{theorem}
For any representation (V,r) of the Clifford algebra $Cl(%
\mathbb{R}
,r,s)$ and principal bundle $Sp(M,Spin\left(
\mathbb{R}
,r,s\right)  ,\pi_{S})$ the associated bundle E=Sp[V,r] is a spin bundle. Any
principal connection $\Omega$ on Sp with potential \`{A}\ induces a linear
connection with form%

\begin{equation}
\Gamma=r\left(  \upsilon\left(  \grave{A}\right)  \right)
\end{equation}

\ on E and covariant derivative $\nabla$ . Moreover, the representation $[%
\mathbb{R}
^{m},\mathbf{Ad}]$ of $Spin\left(
\mathbb{R}
,r,s\right)  $,$\pi_{S})$ leads to the associated vector bundle $F=Sp[%
\mathbb{R}
^{m},\mathbf{Ad}]$ and $\Omega$\ induces a linear connection on F with
covariant derivative $\widehat{\nabla}$\ . There is the relation :%

\begin{equation}
\forall X\in\mathfrak{X}\left(  F\right)  ,U\in\mathfrak{X}\left(  E\right)
:\nabla\left(  r\left(  X\right)  U\right)  =r\left(  \widehat{\nabla
}X\right)  U+r\left(  X\right)  \nabla U
\end{equation}

\end{theorem}

This property of $\nabla$ makes of the connection on E a "Clifford connection".

\begin{proof}
i) The ingredients are the following :

The Lie algebra $o(%
\mathbb{R}
,r,s)$\ with a basis $\left(  \overrightarrow{\kappa_{\lambda}}\right)
_{\lambda=1}^{q}$ .

$\left(
\mathbb{R}
^{m},g\right)  $ endowed with the symmetric bilinear form $\gamma$ of
signature (r,s) on $%
\mathbb{R}
^{m}$ and its basis $\left(  \varepsilon_{i}\right)  _{i=1}^{m}$

The standard representation $\left(
\mathbb{R}
^{m},\jmath\right)  $ of $SO\left(
\mathbb{R}
,r,s\right)  $ thus $\left(
\mathbb{R}
^{m},\jmath^{\prime}(1)\right)  $ is the standard representation of $o\left(
\mathbb{R}
,r,s\right)  $ and $\left[  J\right]  =\jmath^{\prime}\left(  1\right)
\overrightarrow{\kappa}$ is the matrix of $\overrightarrow{\kappa}\in o\left(
%
\mathbb{R}
,r,s\right)  $

The isomorphism : $\upsilon:o(%
\mathbb{R}
,r,s)\rightarrow T_{1}SPin(%
\mathbb{R}
,r,s)::\upsilon\left(  \overrightarrow{\kappa}\right)  =\sum_{ij}$ $\left[
\upsilon\right]  _{j}^{i}\varepsilon_{i}\cdot\varepsilon_{j}$ with $\left[
\upsilon\right]  =\frac{1}{4}\left[  J\right]  \left[  \eta\right]  $

A principal bundle $Sp(M,Spin\left(
\mathbb{R}
,r,s\right)  ,\pi_{S})$ with atlas $\left(  O_{a},\varphi_{a}\right)  _{a\in
A}$ transition maps $g_{ba}\left(  x\right)  \in Spin\left(
\mathbb{R}
,r,s\right)  $ and right action $\rho$ endowed with a connection of potential
\`{A}, $\grave{A}_{b}\left(  x\right)  =Ad_{g_{ba}}\left(  \grave{A}%
_{a}\left(  x\right)  -L_{g_{ba}^{-1}}^{\prime}(g_{ba})g_{ba}^{\prime}\left(
x\right)  \right)  $. On a Spin group the adjoint map reads:

$Ad_{s}\left(  \overrightarrow{\kappa}\right)  =\mathbf{Ad}_{s}\sigma\left(
\overrightarrow{\kappa}\right)  $ so : $\upsilon\left(  \grave{A}_{b}\left(
x\right)  \right)  =g_{ba}\cdot\upsilon\left(  \grave{A}_{a}\left(  x\right)
-g_{ab}\left(  x\right)  \cdot g_{ba}^{\prime}\left(  x\right)  \right)  \cdot
g_{ab}$

A representation (V,r) of $Cl\left(
\mathbb{R}
,r,s\right)  $, with a basis $\left(  e_{i}\right)  _{i=1}^{n}$ of V, defined
by the nxn matrices $\gamma_{i}=r\left(  \varepsilon_{i}\right)  $ and :
$\gamma_{i}\gamma_{j}+\gamma_{j}\gamma_{i}=2\eta_{ij}I,\eta_{ij}=\pm1$.

An associated vector bundle E=Sp[V,r] with atlas $\left(  O_{a},\psi
_{a}\right)  _{a\in A}$ , holonomic basis : $\mathbf{e}_{ai}\left(  x\right)
=\psi_{a}\left(  x,e_{i}\right)  ,$ $\mathbf{e}_{bi}\left(  x\right)
=r\left(  \varphi_{ba}\left(  x\right)  \right)  e_{bi}\left(  x\right)  $ and
transition maps on E are : $\psi_{ba}\left(  x\right)  =r\left(  g_{ba}\left(
x\right)  \right)  $

$\psi_{a}:O_{a}\times V\rightarrow E::\psi_{a}\left(  x,u\right)  =\left(
\varphi_{a}\left(  x,1\right)  ,u\right)  $

For $x\in O_{a}\cap O_{b}:\left(  \varphi_{a}\left(  x,1\right)
,u_{a}\right)  \sim\left(  \varphi_{b}\left(  x,1\right)  ,r\left(
g_{ba}\right)  u_{a}\right)  $

Following the diagram :

$T_{x}M\overset{\grave{A}}{\rightarrow}o(%
\mathbb{R}
,r,s)\overset{\sigma}{\rightarrow}Cl(%
\mathbb{R}
,r,s)\overset{r}{\rightarrow}G%
\mathcal{L}%
\left(  V;V\right)  $

We define the connection on E :

$\widetilde{\Gamma}_{a}\left(  \varphi_{a}\left(  x,1\right)  ,u_{a}\right)
=\left(  \varphi_{a}\left(  x,1\right)  ,r\left(  \upsilon\left(  \grave
{A}_{a}\right)  \right)  u_{a}\right)  $

The definition is consistent : for $x\in O_{a}\cap O_{b}:$

$U_{x}=\left(  \varphi_{a}\left(  x,1\right)  ,u_{a}\right)  \sim\left(
\varphi_{b}\left(  x,1\right)  ,r\left(  g_{ba}\right)  u_{a}\right)  $

$\widetilde{\Gamma}_{b}\left(  \varphi_{b}\left(  x,1\right)  ,r\left(
g_{ba}\right)  u_{a}\right)  =\left(  \varphi_{b}\left(  x,1\right)  ,r\left(
\upsilon\left(  \grave{A}_{b}\right)  \right)  r\left(  g_{ba}\right)
u_{a}\right)  $

$\sim\left(  \varphi_{a}\left(  x,1\right)  ,r\left(  g_{ab}\right)  r\left(
\upsilon\left(  \grave{A}_{b}\right)  \right)  r\left(  g_{ba}\right)
u_{a}\right)  $

$=\left(  \varphi_{a}\left(  x,1\right)  ,r\left(  g_{ab}\right)  r\left(
g_{ba}\cdot\upsilon\left(  \grave{A}_{a}\left(  x\right)  -g_{ab}\left(
x\right)  \cdot g_{ba}^{\prime}\left(  x\right)  \right)  \cdot g_{ab}\right)
r\left(  g_{ba}\right)  u_{a}\right)  $

$=\left(  \varphi_{a}\left(  x,1\right)  ,r\left(  \upsilon\left(  \grave
{A}_{a}\left(  x\right)  -g_{ab}\left(  x\right)  \cdot g_{ba}^{\prime}\left(
x\right)  \right)  \right)  u_{a}\right)  $

$=\widetilde{\Gamma}_{a}\left(  \varphi_{a}\left(  x,1\right)  ,u_{a}\right)
-\left(  \varphi_{a}\left(  x,1\right)  ,r\left(  \upsilon\left(
g_{ab}\left(  x\right)  \cdot g_{ba}^{\prime}\left(  x\right)  \right)
\right)  u_{a}\right)  $

So we have a linear connection:

$\Gamma_{a}\left(  x\right)  =r\left(  \upsilon\left(  \grave{A}_{a}\right)
\right)  =\sum_{\alpha\lambda ij}\grave{A}_{a\alpha}^{\lambda}\left[
\theta_{\lambda}\right]  _{j}^{i}d\xi^{\alpha}\otimes\mathbf{e}_{ai}\left(
x\right)  \otimes\mathbf{e}_{a}^{j}\left(  x\right)  $

with $\left[  \theta_{\lambda}\right]  =\frac{1}{4}\sum_{ij}\left(  \left[
\eta\right]  \left[  J\right]  \right)  _{l}^{k}r\left(  \varepsilon
_{k}\right)  r\left(  \varepsilon_{l}\right)  =\frac{1}{4}\sum_{kl}\left(
\left[  J_{\lambda}\right]  \left[  \eta\right]  \right)  _{l}^{k}\left(
\left[  \gamma_{k}\right]  \left[  \gamma_{l}\right]  \right)  $

The covariant derivative associated to the connection is :

$U\in\mathfrak{X}\left(  E\right)  :\nabla U=\sum_{i\alpha}\left(
\partial_{\alpha}U^{i}+\sum_{\lambda j}\grave{A}_{a\alpha}^{\lambda}\left[
\theta_{\lambda}\right]  _{j}^{i}U^{j}\right)  d\xi^{\alpha}\otimes
\mathbf{e}_{ai}\left(  x\right)  $

ii) With the representation $\left(
\mathbb{R}
^{m},\mathbf{Ad}\right)  $ of $Spin\left(
\mathbb{R}
,r,s\right)  $ we can define an associated vector bundle $F=Sp\left[
\mathbb{R}
^{m},\mathbf{Ad}\right]  $ with atlas $\left(  O_{a},\phi_{a}\right)  _{a\in
A} $ and holonomic basis : $\mathbf{\varepsilon}_{ai}\left(  x\right)
=\phi_{a}\left(  x,\varepsilon_{i}\right)  $ such that : $\mathbf{\varepsilon
}_{bi}\left(  x\right)  =\mathbf{Ad}_{g_{ba}\left(  x\right)  }\varepsilon
_{ai}\left(  x\right)  $

Because \textbf{Ad }preserves $\gamma$ the vector bundle F can be endowed with
a scalar product g. Each fiber (F(x),g(x)) has a Clifford algebra structure
Cl(F(x),g(x)) isomorphic to Cl(TM)(x) and to $Cl(%
\mathbb{R}
,r,s).$

The connection on Sp induces a linear connection on F :

$\widehat{\Gamma}\left(  x\right)  =\left(  \mathbf{Ad}\right)  ^{\prime
}|_{s=1}\left(  \upsilon\left(  \grave{A}\left(  x\right)  \right)  \right)
=\sum_{\lambda}\grave{A}_{\alpha}^{\lambda}\left(  x\right)  \left[
J_{\lambda}\right]  _{j}^{i}\mathbf{\varepsilon}_{i}\left(  x\right)
\otimes\mathbf{\varepsilon}^{j}\left(  x\right)  $

and the covariant derivative :

$X\in\mathfrak{X}\left(  F\right)  :\widehat{\nabla}X=\sum_{i\alpha}\left(
\partial_{\alpha}X^{i}+\sum_{\lambda j}\grave{A}_{\alpha}^{\lambda}\left[
J_{\lambda}\right]  _{j}^{i}X^{j}\right)  d\xi^{\alpha}\otimes
\mathbf{\varepsilon}_{i}\left(  x\right)  $

X(x) is in F(x) so in Cl(F(x),g(x)) and acts on a section of E :

$r\left(  X\right)  U=\sum_{i}r\left(  X^{i}\varepsilon_{i}\left(  x\right)
\right)  U=\sum_{i}X^{i}\left[  \gamma_{i}\right]  _{l}^{k}U^{l}\mathbf{e}%
_{k}\left(  x\right)  $

which is still a section of E. Its covariant derivative reads :

$\nabla\left(  r\left(  X\right)  U\right)  $

$=\sum\left(  \left(  \partial_{\alpha}X^{i}\right)  \left[  \gamma
_{i}\right]  _{l}^{k}U^{l}+X^{i}\left[  \gamma_{i}\right]  _{l}^{k}%
\partial_{\alpha}U^{l}+\grave{A}_{\alpha}^{\lambda}\left[  \theta_{\lambda
}\right]  _{p}^{k}X^{i}\left[  \gamma_{i}\right]  _{l}^{p}U^{l}\right)
d\xi^{\alpha}\otimes\mathbf{e}_{k}\left(  x\right)  $

On the other hand $\widehat{\nabla}X$ is a form valued in F, so we can compute :

$r\left(  \widehat{\nabla}X\right)  U=\sum\left(  \left(  \partial_{\alpha
}X^{i}+\grave{A}_{\alpha}^{\lambda}\left[  J_{\lambda}\right]  _{j}^{i}%
X^{j}\right)  \left[  \gamma_{i}\right]  _{l}^{k}\right)  U^{l}d\xi^{\alpha
}\otimes\mathbf{e}_{k}\left(  x\right)  $

and similarly :

$r\left(  X\right)  \nabla U=\sum X^{i}\left[  \gamma_{i}\right]  _{l}%
^{k}\left(  \partial_{\alpha}U^{l}+\grave{A}_{\alpha}^{\lambda}\left[
\theta_{\lambda}\right]  _{j}^{l}U^{j}\right)  d\xi^{\alpha}\otimes
\mathbf{e}_{k}\left(  x\right)  $

$\nabla\left(  r\left(  X\right)  U\right)  -r\left(  \widehat{\nabla
}X\right)  U-r\left(  X\right)  \nabla U$

$=\sum_{i\alpha}\{\left(  \partial_{\alpha}X^{i}\right)  \left[  \gamma
_{i}\right]  _{l}^{k}U^{l}+X^{i}\left[  \gamma_{i}\right]  _{l}^{k}%
\partial_{\alpha}U^{l}-\left(  \partial_{\alpha}X^{i}\right)  \left[
\gamma_{i}\right]  _{l}^{k}U^{l}-X^{i}\left[  \gamma_{i}\right]  _{l}%
^{k}\partial_{\alpha}U^{l}$

$+\sum_{\lambda j}\grave{A}_{\alpha}^{\lambda}\left[  \theta_{\lambda}\right]
_{p}^{k}X^{i}\left[  \gamma_{i}\right]  _{l}^{p}U^{l}-\grave{A}_{\alpha
}^{\lambda}\left[  J_{\lambda}\right]  _{j}^{i}X^{j}U^{l}\left[  \gamma
_{i}\right]  _{l}^{k}-X^{i}\left[  \gamma_{i}\right]  _{l}^{k}\grave
{A}_{\alpha}^{\lambda}\left[  \theta_{\lambda}\right]  _{j}^{l}U^{j}%
\}d\xi^{\alpha}\otimes\mathbf{e}_{ak}\left(  x\right)  $

$=\sum\grave{A}_{\alpha}^{\lambda}\left[  \theta_{\lambda}\right]  _{p}%
^{k}X^{i}\left[  \gamma_{i}\right]  _{l}^{p}U^{l}-\grave{A}_{\alpha}^{\lambda
}\left[  J_{\lambda}\right]  _{j}^{i}X^{j}U^{l}\left[  \gamma_{i}\right]
_{l}^{k}-X^{i}\left[  \gamma_{i}\right]  _{l}^{k}\grave{A}_{\alpha}^{\lambda
}\left[  \theta_{\lambda}\right]  _{j}^{l}U^{j}$

$=\sum_{\lambda j}\left(  \left[  \theta_{\lambda}\right]  \left[  \gamma
_{i}\right]  -\left[  \gamma_{i}\right]  \left[  \theta_{\lambda}\right]
-\sum_{p}\left[  J_{\lambda}\right]  _{i}^{p}\left[  \gamma_{p}\right]
\right)  _{l}^{k}X^{i}U^{l}\grave{A}_{\alpha}^{\lambda}$

$\left[  \theta_{\lambda}\right]  \left[  \gamma_{i}\right]  -\left[
\gamma_{i}\right]  \left[  \theta_{\lambda}\right]  =\frac{1}{4}\sum
_{pq}\left(  \left(  \left[  J_{\lambda}\right]  \left[  \eta\right]  \right)
_{q}^{p}\left[  \gamma_{p}\right]  \left[  \gamma_{q}\right]  \left[
\gamma_{i}\right]  -\left[  \gamma_{i}\right]  \left(  \left[  J_{\lambda
}\right]  \left[  \eta\right]  \right)  _{q}^{p}\left(  \left[  \gamma
_{p}\right]  \left[  \gamma_{q}\right]  \right)  \right)  $

$=\frac{1}{4}\sum_{pq}\left(  \left[  J_{\lambda}\right]  \left[  \eta\right]
\right)  _{q}^{p}\left(  \left[  \gamma_{p}\right]  \left[  \gamma_{q}\right]
\left[  \gamma_{i}\right]  -\left[  \gamma_{i}\right]  \left[  \gamma
_{p}\right]  \left[  \gamma_{q}\right]  \right)  $

$=\frac{1}{4}\sum_{pq}\left(  \left[  J_{\lambda}\right]  \left[  \eta\right]
\right)  _{q}^{p}\left(  r\left(  \varepsilon_{p}\cdot\varepsilon_{q}%
\cdot\varepsilon_{i}-\varepsilon_{i}\cdot\varepsilon_{p}\cdot\varepsilon
_{q}\right)  \right)  $

$=\frac{1}{4}\sum_{pq}\left(  \left[  J_{\lambda}\right]  \left[  \eta\right]
\right)  _{q}^{p}\left(  2r\left(  \eta_{iq}\varepsilon_{p}-\eta
_{ip}\varepsilon_{q}\right)  \right)  =\frac{1}{2}\sum_{pq}\left(  \left[
J_{\lambda}\right]  \left[  \eta\right]  \right)  _{q}^{p}\left(  \eta
_{iq}\left[  \gamma_{p}\right]  -\eta_{ip}\left[  \gamma_{q}\right]  \right)
$

$=\frac{1}{2}\left(  \sum_{p}\left(  \left[  J_{\lambda}\right]  \left[
\eta\right]  \right)  _{i}^{p}\eta_{ii}\left[  \gamma_{p}\right]  -\sum
_{q}\left(  \left[  J_{\lambda}\right]  \left[  \eta\right]  \right)  _{q}%
^{i}\eta_{ii}\left[  \gamma_{q}\right]  \right)  $

$=\frac{1}{2}\eta_{ii}\sum_{p}\left(  \left(  \left[  J_{\lambda}\right]
\left[  \eta\right]  \right)  _{i}^{p}-\left(  \left[  J_{\lambda}\right]
\left[  \eta\right]  \right)  _{p}^{i}\right)  \left[  \gamma_{p}\right]
=\eta_{ii}\sum_{p}\left(  \left[  J_{\lambda}\right]  \left[  \eta\right]
\right)  _{i}^{p}\left[  \gamma_{p}\right]  $

$=\sum_{pj}\eta_{ii}\left[  J_{\lambda}\right]  _{j}^{p}\eta_{ji}\left[
\gamma_{p}\right]  =\sum_{p}\left[  J_{\lambda}\right]  _{i}^{p}\left[
\gamma_{p}\right]  $

$\left[  \theta_{\lambda}\right]  \left[  \gamma_{i}\right]  -\left[
\gamma_{i}\right]  \left[  \theta_{\lambda}\right]  -\sum_{p}\left[
J_{\lambda}\right]  _{i}^{j}\left[  \gamma_{j}\right]  =0$
\end{proof}

Notice that the Clifford structure Cl(M) is not defined the usual way, but
starting from the principal Spin bundle by taking the Adjoint action.

\subsubsection{Chern theory}

Given a manifold M and a Lie group G, one can see that the potentials

$\sum_{i,\alpha}\grave{A}_{\alpha}^{i}\left(  x\right)  d\xi^{\alpha}%
\otimes\varepsilon_{i}$ and the forms $%
\mathcal{F}%
=\sum_{i,\alpha\beta}%
\mathcal{F}%
_{\alpha\beta}^{i}d\xi^{\alpha}\wedge d\xi^{\beta}\otimes\varepsilon_{i}$

sum all the story about the connection up to the 2nd order.\ They are forms
over M, valued in the Lie algebra, but defined locally, in the domains $O_{a}$
of the cover of a principal bundle, with transition functions depending on G.
So looking at these forms gives some insight about the topological constraints
limiting the possible matching of manifolds and Lie groups to build principal bundles.

\paragraph{Chern-Weil theorem\newline}

\begin{definition}
For any finite dimensional representation (F,r) on a field K of a Lie group G,
$I_{s}\left(  G,F,r\right)  $ is the set of symmetric s-linear maps $L_{s}\in%
\mathcal{L}%
^{s}\left(  F;K\right)  $ which are r invariant.
\end{definition}

$L_{s}\in I_{s}\left(  G,F,r\right)  :\forall\left(  X_{k}\right)  _{k=1}%
^{r}\in V,$ $\forall g\in G:$

$L_{s}\left(  r(g)X_{1},...,r\left(  g\right)  X_{s}\right)  =L_{s}\left(
X_{1},..,X_{s}\right)  $

With the product :

$\left(  L_{s}\times L_{t}\right)  \left(  X_{1},..X_{t+s}\right)  =\frac
{1}{\left(  t+s\right)  !}\sum_{\sigma\in\mathfrak{S}\left(  t+s\right)
}L_{t}\left(  X_{\sigma\left(  1\right)  },...X_{\sigma\left(  s\right)
}\right)  L_{s}\left(  X_{\sigma\left(  s+1\right)  },...X_{\sigma\left(
t+s\right)  }\right)  $

$I\left(  G,V,r\right)  =\oplus_{s=0}^{\infty}I_{r}\left(  G,V,r\right)  $ is
a real algebra on the field K.

For any finite dimensional manifold M, any finite dimensional representation
(V,r) of a group G, $L_{s}\in I_{s}\left(  G,F,r\right)  $ one defines the map :

$\widehat{L}_{s}:\Lambda_{p}\left(  M;V\right)  \rightarrow\Lambda_{sp}\left(
M;%
\mathbb{R}
\right)  $ by taking $F=\Lambda_{p}\left(  M;V\right)  $

$\forall\left(  X_{k}\right)  _{k=1}^{r}\in\mathfrak{X}\left(  TM\right)
:\widehat{L}_{s}\left(  \varpi\right)  \left(  X_{1},..,X_{rp}\right)  $

$=\frac{1}{\left(  rp\right)  !}\sum_{\sigma\in\mathfrak{S}\left(  rp\right)
}L_{r}\left(  \varpi\left(  X_{\sigma\left(  1\right)  },...X_{\sigma\left(
p\right)  }\right)  ,..,\varpi\left(  X_{\sigma\left(  \left(  s-1\right)
p+1\right)  },...X_{\sigma\left(  sp\right)  }\right)  \right)  $

In particular for any finite dimensional principal bundle $P(M,G,\pi)$ endowed
with a principal connection, the strength form $%
\mathcal{F}%
\in\Lambda_{2}\left(  M;T_{1}G\right)  $ and $\left(  T_{1}G,Ad\right)  $ is a
representation of G. So we have for any linear map $L_{r}\in I_{r}\left(
G,T_{1}G,Ad\right)  :$\ \ 

$\forall\left(  X_{k}\right)  _{k=1}^{r}\in T_{1}G:\widehat{L}_{r}\left(
\mathcal{F}%
\right)  \left(  X_{1},..,X_{2r}\right)  $

$=\frac{1}{\left(  2r\right)  !}\sum_{\sigma\in\mathfrak{S}\left(  2r\right)
}L_{r}\left(
\mathcal{F}%
\left(  X_{\sigma\left(  1\right)  },X_{\sigma\left(  2\right)  }\right)  ,..,%
\mathcal{F}%
\left(  X_{\sigma\left(  2r-1\right)  },X_{\sigma\left(  2r\right)  }\right)
\right)  $

\begin{theorem}
\textbf{Chern-Weil theorem} (Kobayashi 2 p.293, Nakahara p.422) : For any
principal bundle $P(M,G,\pi)$ and any map $L_{r}\in I_{r}\left(
G,T_{1}G,Ad\right)  :$

i) for any connection on P with strength form
$\mathcal{F}$%
\ : $d\widehat{L}_{r}\left(
\mathcal{F}%
\right)  =0$

ii) for any two principal connections with strength form $%
\mathcal{F}%
_{1},%
\mathcal{F}%
_{2}$ there is some form $\lambda\in\Lambda_{2r}\left(  M;%
\mathbb{R}
\right)  $ such that $\widehat{L}_{r}\left(
\mathcal{F}%
_{1}-%
\mathcal{F}%
_{2}\right)  =d\lambda$.

iii) the map : $\chi:L_{r}\in I_{r}\left(  G,T_{1}G,Ad\right)  \rightarrow
H^{2r}\left(  M\right)  ::\chi\left(  L_{r}\right)  =\left[  \lambda\right]  $
is linear and when extended to $\chi:I(G)\rightarrow H^{\ast}\left(  M\right)
=\oplus_{r}H^{r}\left(  M\right)  $ is a morphism of algebras.

iv) If N is a manifold and $f$ a differentiable map : $f:N\rightarrow M$ we
have the pull back of P and : $\chi_{f^{\ast}}=f^{\ast}\chi$
\end{theorem}

All the forms $\widehat{L}_{r}\left(
\mathcal{F}%
\right)  $ for any principal connection on $P(M,G,\pi)$ belong to the same
class of the cohomology space $H^{2r}\left(  M\right)  $ \ of M, which is
called the \textbf{characteristic class} of P related to the linear map
$L_{r}.$ This class does not depend on the connection, but depends both on P
and $L_{r}.$

\begin{theorem}
(Nakahara p.426) The characteristic classes of a trivial principal bundle are
trivial (meaning $\left[  0\right]  $).
\end{theorem}

\bigskip

If we have a representation (V,r) of the group G, the map $r^{\prime}%
(1):T_{1}G\rightarrow%
\mathcal{L}%
\left(  V;V\right)  $ is an isomorphism of Lie algebras. From the identity :
$Ad_{r\left(  g\right)  }r^{\prime}(1)=r^{\prime}(1)Ad_{g}$ where
$Ad_{r\left(  g\right)  }=r(g)\circ\ell\circ r(g)^{-1}$ is the conjugation
over $%
\mathcal{L}%
\left(  V;V\right)  $ we can deduce : $Ad_{g}=r^{\prime}(1)^{-1}\circ
Ad_{r(g)}\circ r^{\prime}(1).$ If we have a r linear map : $L_{r}\in%
\mathcal{L}%
^{r}\left(  V;K\right)  $ which is invariant by $Ad_{r(g)}:L\circ Ad_{r(g)}=L$
then :

$\widetilde{L}_{r}=r^{\prime}(1)^{-1}\circ L\circ r^{\prime}(1)\in%
\mathcal{L}%
^{r}\left(  T_{1}G;K\right)  $ is invariant by $Ad_{g}$ :

$\widetilde{L}_{r}=r^{\prime}(1)^{-1}\circ L\circ r^{\prime}(1)\left(
Ad_{g}\right)  =r^{\prime}(1)^{-1}\circ L\circ Ad_{r\left(  g\right)
}r^{\prime}(1)=r^{\prime}(1)^{-1}\circ L\circ r^{\prime}(1)$

So with such representations we can deduce other maps in $I_{r}\left(
G,T_{1}G,Ad\right)  $

A special case occurs when we have a real Lie group, and we want to use
complex representations.

\paragraph{Chern and Pontryagin classes\newline}

Let V be a vector space on the field K. Any symmetric s-linear maps $L_{s}\in%
\mathcal{L}%
^{s}\left(  V;K\right)  $ induces a monomial map of degree s : $Q:V\rightarrow
K::Q\left(  X\right)  =L_{s}\left(  X,X,...,X\right)  .$ Conversely by
polarization a monomial map of degree s induces a symmetric s-linear map. This
can be extended to polynomial maps and to sum of symmetric s-linear maps
valued in the field K (see Algebra - tensors). If (V,r) is a representation of
the group G, then Q is invariant by the action r of G iff the associated
linear map is invariant. So the set $I_{s}\left(  G,V,r\right)  $ is usually
defined through invariant polynomials.

Of particular interest are the polynomials \ $Q:V\rightarrow K::Q\left(
X\right)  =\det\left(  I+kX\right)  $ with $k\in K$ a fixed scalar. If (V,r)
is a representation of G, then Q is invariant by $Ad_{r(g)}:$

$Q\left(  Ad_{r(g)}X\right)  =\det\left(  I+kr(g)Xr\left(  g\right)
^{-1})\right)  =\det\left(  r(g)\left(  I+kX\right)  r(g)^{-1}\right)
=\det(r(g))\det\left(  I+kX\right)  \det r(g)^{-1}=Q(X)$

The degree of the polynomial is n=dim(V).\ They define n linear symmetric Ad
invariant maps $\widetilde{L}_{s}\in%
\mathcal{L}%
^{s}\left(  T_{1}G;K\right)  $

For any principal bundle $P(M,G,\pi)$ the previous polynomials define a sum of
maps and for each we have a characteritic class.

If $K=%
\mathbb{R}
,k=\frac{1}{2\pi},Q\left(  X\right)  =\det(I+\frac{1}{2\pi}X)$ we have the
\textbf{Pontryagin classes} denoted $p_{n}\left(  P\right)  \in\Lambda
_{2n}\left(  M;%
\mathbb{C}
\right)  \subset H^{2n}\left(  M;%
\mathbb{C}
\right)  $

If $K=%
\mathbb{C}
,k=i\frac{1}{2\pi},Q\left(  X\right)  =\det(I+i\frac{1}{2\pi}X)$ we have the
\textbf{Chern classes} denoted $c_{n}\left(  P\right)  \in\Lambda_{2n}\left(
M;%
\mathbb{R}
\right)  \subset H^{2n}\left(  M\right)  $

For 2n
$>$
dim(M) then $c_{n}\left(  P\right)  =p_{n}\left(  P\right)  =\left[  0\right]
$

\newpage

\section{BUNDLE\ FUNCTORS}

\bigskip

With fiber bundles it is possible to add some mathematical structures above
manifolds, and functors can be used for this purpose.\ However in many cases
we have some relations between the base and the fibers, involving differential
operators, such as connection or the exterior differential on the tensorial
bundle. To deal with them we need to extend the functors to jets. The category
theory reveals itself useful, as all the various differential constructs come
under a limited number of well known cases.

\bigskip

\subsection{Bundle functors}

\label{Bundle functors}

\subsubsection{Definitions}

(Kolar IV.14)

A functor is a map between categories, a bundle functor is a map between a
category of manifolds and the category of fibered manifolds. Here we use
fibered manifolds rather than fiber bundles because the morphisms are more
easily formalized : this is a pair of maps (F,f) \ F between the total spaces
and f between the bases.

\begin{notation}
$\mathfrak{M}$ is the category of manifolds (with the relevant class of
differentiability whenever necessary),

$\mathfrak{M}_{m}$ is the subcategory of m dimensional real manifolds, with
local diffeormorphisms,

$\mathfrak{M}_{\infty}$ is the category of smooth manifolds and smooth morphisms

$\mathfrak{FM}$ is the category of fibered manifolds with their morphisms,

$\mathfrak{FM}_{m}$ is the category of fibered manifolds with m dimensional
base and their local diffeormorphisms,
\end{notation}

The local diffeomorphisms in $\mathfrak{M}_{m}$\ are maps :

$f\in\hom_{\mathfrak{M}_{m}}\left(  M,N\right)  :f\in C_{1}\left(  M;N\right)
:f^{\prime}(x)\neq0$\ 

The base functor $\mathfrak{B:FM\mapsto M}$\ associates its base to a fibered
manifold and f to each morphism (F,f).

\begin{definition}
A \textbf{bundle functor} (also called natural bundle) is a functor $\digamma
$\ from the category $\mathfrak{M}_{m}$ of m dimensional real manifolds and
local diffeomorphisms to the category $\mathfrak{FM}$\ of fibered manifolds
which is :

i) Base preserving : the composition by the base functor gives the identity :

$\mathfrak{B\circ}\digamma=Id_{\mathfrak{M}}$

ii) Local : if N is a submanifold of M then $\digamma\left(  N\right)  $ is
the subfibered manifold of $\digamma\left(  M\right)  $
\end{definition}

So if N is an open submanifold of M and $\imath:N\rightarrow M$ is the
inclusion, then :

$\digamma\left(  N\right)  =\pi_{M}^{-1}\left(  N\right)  $

$\digamma\left(  \imath\right)  :\pi_{M}^{-1}\left(  N\right)  \rightarrow
\digamma\left(  M\right)  $

A bundle functor $\digamma:\mathfrak{M\mapsto FM}$ associates :

- to each manifold M a fibered manifold with base M : $M\rightarrow
\digamma_{o}\left(  M\right)  \left(  M,\pi_{M}\right)  $

- to each local diffeomorphism $f\in C_{1}\left(  M;N\right)  ,\phi^{\prime
}\left(  x\right)  \neq0$ a fibered manifold, base preserving, morphism

$\digamma_{h}\left(  f\right)  \in\hom\left(  \digamma_{o}\left(  M\right)
,\digamma_{o}\left(  N\right)  \right)  $

$\pi_{N}\circ\digamma_{h}\left(  f\right)  =\digamma_{h}\left(  f\right)
\circ\pi_{M}$

\begin{theorem}
(Kolar p.138, 204) A bundle functor is :

i) \textbf{regular} : smoothly parametrized systems of local morphisms are
transformed into smoothly parametrized systems of fibered manifold morphisms

ii) locally \textbf{of finite order} : there is r
$>$%
0 such that the operations in the functor do not involve the derivatives of
order above r
\end{theorem}

i) reads :

If $\digamma:\mathfrak{M}_{m}\mathfrak{\mapsto FM:}$ then if $f:P\times
M\rightarrow N$ is such that $f\left(  x,.\right)  \in\hom_{\mathfrak{Mm}%
}\left(  M,N\right)  $\ ,

then $\Phi:P\times\digamma\left(  M\right)  \rightarrow\digamma\left(
N\right)  $ defined by : $\Phi\left(  x,.\right)  =\digamma f\left(
x,.\right)  $ is smooth.

If $\digamma:\mathfrak{M}_{\infty}\mathfrak{\mapsto FM:}$ then if $f:P\times
M\rightarrow N$ is smooth,

then $\Phi:P\times\digamma\left(  M\right)  \rightarrow\digamma\left(
N\right)  $ defined by : $\Phi\left(  x,.\right)  =\digamma f\left(
x,.\right)  $ is smooth.

ii) reads :

$\exists r\in%
\mathbb{N}
:\forall M,N\in\mathfrak{M}_{m,r},\forall f,g\in\hom_{\mathfrak{M}}\left(
M,N\right)  ,\forall x\in M:$

$j_{x}^{r}f=j_{x}^{r}g\Rightarrow\digamma f=\digamma g$ (the equality is
checked separately at each point x inside the classes of equivalence).

As the morphisms are local diffeomorphisms $\forall x\in M,j_{x}^{r}f\in
GJ^{r}\left(  M,N\right)  $

These two results are quite powerful : they mean that there are not so many
ways to add structures above a manifold.

\begin{notation}
$\digamma^{r}$ is the set of r order bundle functors acting on $\mathfrak{M} $

$\digamma_{m}^{r}$ is the set of r order bundle functors acting on
$\mathfrak{M}_{m}$

$\digamma_{\infty}^{r}$ is the set of r order bundle functors acting on
$\mathfrak{M}_{\infty}$
\end{notation}

\paragraph{Examples of bundle functors\newline}

i) The functor which associates to any vector space V and to each manifold the
vector bundle $E\left(  M,\otimes_{s}^{r}V,\pi\right)  $

ii) The functor which associates to a manifold its tensorial bundle of r-forms

iii) the functor which associates to each manifold its tangent bundle, and to
each map its derivative, which is pointwise a linear map, and so corresponds
to the action of $GL^{1}\left(
\mathbb{R}
,m\right)  $ on $%
\mathbb{R}
^{m}.$

iii) The functor $J^{r}:\mathfrak{M\mapsto FM}$ $\ $ which gives the r jet
prolongation of a manifold$\ $

The functor $J^{r}:\mathfrak{FM}_{m}\mathfrak{\mapsto FM}$ which gives the r
jet prolongation of a fibered manifold$\ $

iv) The functor : $T_{k}^{r}:\mathfrak{M\mapsto FM}$ associates to each :

manifold M : $T_{k}^{r}(M)=J_{0}^{r}(%
\mathbb{R}
^{k},M)$

map $f\in C_{r}\left(  M;N\right)  \mapsto T_{r}^{k}f:T_{k}^{r}(M)\rightarrow
T_{k}^{r}(N)::T_{k}^{r}f\left(  j_{0}^{r}g\right)  =j_{0}^{r}\left(  f\circ
g\right)  $

This functor preserves the composition of maps. For k=r=1 we have the functor
: $f\rightarrow f^{\prime}$

v) The bi functor : $\mathfrak{M}_{m}\mathfrak{\times M\mapsto FM}$ associates
to each couple :

$\left(  M\times N\right)  \mapsto J^{r}\left(  M,N\right)  $

$\left(  f\in C_{r}\left(  M_{1},N_{1}\right)  ,g\in C_{r}\left(  M_{2}%
,N_{2}\right)  \right)  \mapsto\hom\left(  J^{r}\left(  M_{1},N_{1}\right)
,J^{r}\left(  M_{2},N_{2}\right)  \right)  ::$

$J^{r}\left(  f,g\right)  \left(  X\right)  =\left(  j_{q}^{r}g\right)  \circ
X\circ\left(  j_{p}^{r}f\right)  ^{-1}$ where $X\in J^{r}(M_{1},N_{1})$ is
such that : $q=X\left(  p\right)  $

\subsubsection{Description of r order bundle functors}

The strength of the concept of bundle functor is that it applies in the same
manner to any manifold, thus to the simple manifold $%
\mathbb{R}
^{m}.$ So when we know what happens in this simple case, we can deduce what
happens with any manifold.\ In many ways this is just the implementation of
demonstrations in differential geometry when we take some charts to come back
in $%
\mathbb{R}
^{m}.$ Functors enable us to generalize at once all these results.

\paragraph{Fundamental theorem\newline}

A bundle functor $\digamma$ acts in particular on the manifold $%
\mathbb{R}
^{m}$ endowed with the appropriate morphisms$.$The set $\digamma_{0}\left(
\mathbb{R}
^{m}\right)  =\pi_{R^{m}}^{-1}\left(  0\right)  \in\digamma\left(
\mathbb{R}
^{m}\right)  $ is just the fiber above 0 in the fibered bundle $\digamma
\left(
\mathbb{R}
^{m}\right)  $ and is called the \textbf{standard fiber} of the functor

For any two manifolds $M,N\in\mathfrak{M}_{mr},$ we have an action of
$GJ^{r}\left(  M,N\right)  $ (the invertible elements of $J^{r}\left(
M,N\right)  $) on $F\left(  M\right)  $ defined as \ follows :

$\Phi_{M,N}:GJ^{r}\left(  M,N\right)  \times_{M}\digamma\left(  M\right)
\rightarrow\digamma\left(  N\right)  ::\Phi_{M,N}\left(  j_{x}^{r}f,y\right)
=\left(  \digamma\left(  f\right)  \right)  \left(  y\right)  $

The maps $\Phi_{M,N}$\ , called the associated maps of the functor, are smooth.

For $%
\mathbb{R}
^{m}$ at x=0\ this action reads :

$\Phi_{R^{m},R^{m}}:GJ^{r}\left(
\mathbb{R}
^{m},%
\mathbb{R}
^{m}\right)  \times_{%
\mathbb{R}
^{m}}\digamma_{0}\left(
\mathbb{R}
^{m}\right)  \rightarrow\digamma_{0}\left(
\mathbb{R}
^{m}\right)  ::$

$\Phi_{R^{m},R^{m}}\left(  j_{0}^{r}f,0\right)  =\left(  \digamma\left(
f\right)  \right)  \left(  0\right)  $

But $GJ_{0}^{r}\left(
\mathbb{R}
^{m},%
\mathbb{R}
^{m}\right)  _{0}=GL^{r}\left(
\mathbb{R}
,m\right)  $ the r differential group and $\digamma_{0}\left(
\mathbb{R}
^{m}\right)  $ is the standard fiber of the functor. So we have the action :

$\ell:GL^{r}\left(
\mathbb{R}
,m\right)  \times_{%
\mathbb{R}
^{m}}\digamma_{0}\left(
\mathbb{R}
^{m}\right)  \rightarrow\digamma_{0}\left(
\mathbb{R}
^{m}\right)  $

Thus for any manifold $M\in\mathfrak{M}_{mr}$\ the bundle of r frames
$GT_{m}^{r}(M)$ is a principal bundle $GT_{m}^{r}(M)\left(  M,GL^{r}(%
\mathbb{R}
,m),\pi^{r}\right)  $

And we have the following :

\begin{theorem}
(Kolar p.140) For any bundle functor $\digamma_{m}^{r}:\mathfrak{M}%
_{m,r}\mathfrak{\mapsto FM}$ and manifold $M\in\mathfrak{M}_{mr}$ the fibered
manifold $\digamma\left(  M\right)  $ is an associated bundle \ 

$GT_{m}^{r}(M)\left[  \digamma_{0}\left(
\mathbb{R}
^{m}\right)  ,\ell\right]  $ \ to the principal bundle $GT_{m}^{r}(M)$ of r
frames of M with the standard left action $\ell$\ of $GL^{r}(%
\mathbb{R}
,m)$ on the standard fiber $\digamma_{0}\left(
\mathbb{R}
^{m}\right)  $. A fibered atlas of $\digamma\left(  M\right)  $\ is given by
the action of $\digamma$ on the charts of an atlas of M. There is a bijective
correspondance between the set $\digamma_{m}^{r}$\ of r order bundle functors
acting on m dimensional manifolds and the set of smooth left actions of the
jet group $GL^{r}\left(
\mathbb{R}
,m\right)  $ on manifolds.
\end{theorem}

That means that \textit{the result of the action of a bundle functor is always
some associated bundle}, based on the principal bundle defined on the
manifold, and a standard fiber and actions known from $\digamma\left(
\mathbb{R}
^{m}\right)  .$ Conversely a left action $\lambda:GL^{r}\left(
\mathbb{R}
,m\right)  \times V\rightarrow V$ on a manifold V defines an associated bundle
$GT_{m}^{r}(M)\left[  V,\lambda\right]  $ which can be seen as the action of a
bundle functor with $\digamma_{0}\left(
\mathbb{R}
^{m}\right)  =V,\ell=\lambda.$

\begin{theorem}
For any r order bundle functor in $\digamma_{m}^{r},$ its composition with the
functor $J^{k}:\mathfrak{FM\mapsto FM}$\ gives a bundle functor of order k+r:

$\digamma\in\digamma_{m}^{r}\Rightarrow J^{k}\circ\digamma\in\digamma
_{m}^{r+k}$
\end{theorem}

\paragraph{Vector bundle functor\newline}

\begin{definition}
A bundle functor is a \textbf{vector bundle functor} if the result is a vector bundle.
\end{definition}

\begin{theorem}
(Kolar p.141) There is a bijective correspondance between the set
$\digamma_{m}^{r}$\ of r order bundle functors acting on m dimensional
manifolds and the set of smooth representation $\left(  \overrightarrow
{V},\overrightarrow{\ell}\right)  $\ of $GL^{r}\left(
\mathbb{R}
,m\right)  .$ For any manifold M, $\digamma_{m}^{r}\left(  M\right)  $ is the
associated vector bundle $GT_{m}^{r}(M)\left[  \overrightarrow{V}%
,\overrightarrow{\ell}\right]  $
\end{theorem}

The standard fiber is the vector space $\overrightarrow{V}$\ and the left
action is given by the representation : $\overrightarrow{\ell}:GL^{r}\left(
\mathbb{R}
,m\right)  \times\overrightarrow{V}\rightarrow\overrightarrow{V}$

Example : the tensorial bundle $\otimes TM$ is the associated bundle
$GT_{m}^{1}(M)\left[  \otimes%
\mathbb{R}
^{m},\overrightarrow{\ell}\right]  $ with the standard action of GL($%
\mathbb{R}
,m)$ on $\otimes%
\mathbb{R}
^{m}$

\paragraph{Affine bundle functor\newline}

An affine r order bundle functor is such that each $\digamma\left(  M\right)
$ is an affine bundle and each $\digamma\left(  f\right)  $ is an affine
morphism. It is deduced from a vector bundle functor $\overrightarrow
{\digamma}$ defined by a smooth $\left(  \overrightarrow{V},\overrightarrow
{\ell}\right)  $ representation of $GL^{r}\left(
\mathbb{R}
,m\right)  .$The action $\ell$ of $GL^{r}\left(
\mathbb{R}
,m\right)  $\ on the standard fiber $V=\digamma_{0}R^{m}$ is given by :
$\ell\left(  g\right)  y=\ell\left(  g\right)  x+\overrightarrow{\ell}\left(
\overrightarrow{y-x}\right)  .$

Example : The bundle of principal connections on a principal fiber bundle
QP=$J^{1}E$ is an affine bundle over E, modelled on the vector bundle
$TM^{\ast}\otimes VE\rightarrow E$

\bigskip

\subsection{Natural operators}

Any structure of r-jet extension of a fiber bundle implies some rules in a
change of gauge.\ When two such structures are defined by bundle functors,
there are necessarily some relations between their objects and their
morphisms, which are given by natural transformations. The diversity of these
natural transformations is quite restricted which, conversely, lead to a
restricted diversity of bundle functors : actually the objects which can be
added onto a manifold can be deduced from the usual operators of differential geometry.

\subsubsection{Definitions}

\paragraph{Natural transformation between bundle fonctors\newline}

A \textbf{natural transformation} $\phi$\ between two functors $\digamma
_{1},\digamma_{2}:\mathfrak{M\mapsto FM}$ is a map denoted $\phi
:\digamma\hookrightarrow F_{2}$ such that the diagram commutes :

\bigskip%

\begin{tabular}
[c]{llllllll}
& $\mathfrak{M}$ &  &  & $\mathfrak{FM}$ &  &  & $\mathfrak{M}$\\
$\ulcorner$ &  & $\urcorner$ & $\ulcorner$ &  & $\ \ \ \ \ \ \ \ \ \ \urcorner
$ &  & \\
& M & $\overset{\pi_{1}}{\leftarrow}$ & $\digamma_{1o}\left(  M\right)  $ &
$\overset{\phi\left(  M\right)  }{\rightarrow}$ & $\digamma_{2o}(M)$ &
$\overset{\pi_{2}}{\rightarrow}$ & M\\
& $\downarrow$ &  & $\downarrow$ &  & $\downarrow$ &  & $\downarrow$\\
f & $\downarrow$ &  & $\downarrow\digamma_{1m}\left(  f\right)  $ &  &
$\downarrow\digamma_{2m}\left(  f\right)  $ &  & $\downarrow f$\\
& $\downarrow$ &  & $\downarrow$ &  & $\downarrow$ &  & $\downarrow$\\
& N & $\overset{\pi_{1}}{\leftarrow}$ & $\digamma_{1o}(N)$ & $\overset
{\phi\left(  N\right)  }{\rightarrow}$ & $\digamma_{2o}(N)$ & $\overset
{\pi_{2}}{\rightarrow}$ & N
\end{tabular}

\bigskip

So $\phi\left(  M\right)  \in\hom_{\mathfrak{FM}}\left(  \digamma_{1o}\left(
M\right)  ,\digamma_{2o}(M)\right)  $

Any natural transformation is comprised of base preserving morphisms, in the
meaning :

$\forall p\in\digamma_{1o}\left(  M\right)  :\pi_{2}\left(  \phi\left(
M\right)  \left(  p\right)  \right)  =\pi_{1}\left(  p\right)  $

If $\digamma$ is a vector bundle functor, then the vertical bundle $V\digamma$
is naturally equivalent to $\digamma\oplus\digamma$ (Kolar p.385)

Example : lagrangians.

A r order \textbf{lagrangian} on a fiber bundle $E\left(  M,V,\pi\right)
$\ is a m-form on M : $L\left(  \xi^{\alpha},z^{i}\left(  x\right)
,z_{\left\{  \alpha_{1}...\alpha_{s}\right\}  }^{i}\left(  x\right)  \right)
d\xi^{1}\wedge...\wedge d\xi^{m}$ where $\left(  \xi^{\alpha},z^{i}\left(
x\right)  ,z_{\left\{  \alpha_{1}...\alpha_{s}\right\}  }^{i}\left(  x\right)
\right)  $ are the coordinates of a section $Z$ of $J^{r}E.$ For a section
$S\in\mathfrak{X}\left(  E\right)  $\ the lagrangian reads $L\circ j^{r}S$

We have two functors : $J^{r}:\mathfrak{M}_{m}\mapsto\mathfrak{FM}$ and
$\Lambda_{m}:\mathfrak{M}_{m}\mapsto\mathfrak{FM}$ and $L:J^{r}\hookrightarrow
\Lambda_{r}$ is a base preserving morphism and a natural transformation.

\bigskip

\begin{theorem}
(Kolar p.142) There is a bijective correspondance between the set of all
natural transformations between functors of $\digamma_{m}^{r}$\ and the set of
smooth $GL^{r}\left(
\mathbb{R}
,m\right)  $ equivariant maps between their standard fibers.
\end{theorem}

A smooth equivariant map is a map : $\phi:\digamma_{10}R^{m}\rightarrow
\digamma_{20}R^{m}:$

$\forall g\in GL^{r}\left(
\mathbb{R}
,m\right)  ,\forall p\in\digamma_{10}R^{m}:\phi\left(  \lambda_{1}\left(
g,p\right)  \right)  =\lambda_{2}\left(  g,\phi\left(  p\right)  \right)  $

\paragraph{Local operator\newline}

\begin{definition}
Let $E_{1}\left(  M,\pi_{1}\right)  ,E_{2}\left(  M,\pi_{2}\right)  $ be two
fibered manifolds with the same base M. A \textbf{r order local operator} is a
map between their sections $D:\mathfrak{X}_{r}\left(  E_{1}\right)
\rightarrow\mathfrak{X}\left(  E_{2}\right)  $ such that

$\forall S,T\in\mathfrak{X}_{r}\left(  E_{1}\right)  ,\forall x\in
M:l=0..r:j_{x}^{l}S=j_{x}^{l}T\Rightarrow D\left(  S\right)  \left(  x\right)
=D\left(  T\right)  \left(  x\right)  $
\end{definition}

So it depends only on the r order derivatives of the sections. It can be seen
as the r jet prolongation of a morphism between fibered manifolds.

\paragraph{Naural operator\newline}

\begin{definition}
A \textbf{natural operator} between two r order bundle functors $\digamma
_{1},\digamma_{2}\in\digamma_{m}^{r}$ is :

i) a set of local operators : $D\left(  \digamma_{1}\left(  M\right)
;\digamma_{2}\left(  M\right)  \right)  :\mathfrak{X}_{\infty}\left(
\digamma_{1}\left(  M\right)  \right)  \rightarrow\mathfrak{X}_{\infty}\left(
\digamma_{2}\left(  M\right)  \right)  $

ii) a map : $\Phi:M\rightarrow D\left(  \digamma_{1}\left(  M\right)
;\digamma_{2}\left(  M\right)  \right)  $ which associates to each manifold an
operator between sections of the fibered manifolds, such that :

- $\forall S\in\mathfrak{X}_{\infty}\left(  \digamma_{1}\left(  M\right)
\right)  ,\forall f\in\hom\left(  M,N\right)  :$

$\Phi\left(  N\right)  \left(  \digamma_{1}\left(  f\right)  \circ S\circ
f^{-1}\right)  =\digamma_{2}\left(  f\right)  \circ\Phi\left(  M\right)  \circ
f^{-1}$

- for any open submanifold N of M and section S on M :

$\Phi\left(  N\right)  \left(  S|_{N}\right)  =\Phi\left(  M\right)  \left(
S\right)  |_{N}$
\end{definition}

A natural operator is an operator whose local description does not depend on
the choice of charts.

\bigskip

\begin{theorem}
There is a bijective correspondance between the set of :

r order natural operators between functors $\digamma_{1},\digamma_{2}$ and the
set of all natural transformations $J^{r}\circ\digamma_{1}\hookrightarrow
\digamma_{2}.$

k order natural operators between functors $\digamma_{1}\in\digamma_{m}%
^{r},\digamma_{2}\in\digamma_{m}^{s}$ and the set of all smooth $GL^{q}\left(
%
\mathbb{R}
,m\right)  $ equivariant maps between $T_{m}^{k}\left(  \digamma_{10}%
R^{m}\right)  $\ and $\digamma_{20}R^{m}$ where $q=\max(r+k,s)$
\end{theorem}

\bigskip

\paragraph{Examples\newline}

i) the commutator between vector fields can be seen as a bilinear natural
operator between the functors $T\oplus T$ and T, where T is the 1st order
functor : $M\mapsto TM.$ The bilinear $GL^{2}\left(
\mathbb{R}
,m\right)  $ equivariant map is the relation between their coordinates :

$\left[  X,Y\right]  ^{\alpha}=X^{\beta}\partial_{\beta}Y^{\alpha}-Y^{\beta
}X_{\beta}^{\alpha}$

And this is the only natural 1st order operator : $T\oplus T$ $\hookrightarrow
T$

ii) A r order differential operator is a local operator between two vector
bundles : $D:J^{r}E_{1}\rightarrow E_{2}$ (see Functional analysis)

\subsubsection{Theorems about natural operators}

\begin{theorem}
(Kolar p.222) All natural operators $\wedge_{k}T^{\ast}\hookrightarrow
\wedge_{k+1}T^{\ast}$ are a multiple of the exterior differential
\end{theorem}

\begin{theorem}
(Kolar p.243) The only natural operators \ $\phi:T_{1}^{r}\hookrightarrow
T_{1}^{r}$ are the maps : $X\rightarrow kX,k\in%
\mathbb{R}
$
\end{theorem}

\paragraph{Prolongation of a vector field\newline}

\begin{theorem}
Any vector field X on a manifold M induces, by a bundle functor $\digamma$, a
projectable vector field, denoted $\digamma X$ and called the
\textbf{prolongation} of X.
\end{theorem}

\begin{proof}
The flow $\Phi_{X}\left(  x,t\right)  $ \ of X is a local diffeomorphism which
has for image by $\digamma$ a fiber bundle morphism :

$\Phi_{X}\left(  .,t\right)  \in\hom\left(  M,M\right)  \longmapsto
\digamma\Phi_{X}\left(  .,t\right)  \in\hom\left(  \digamma M,\digamma
M\right)  $

$\digamma\Phi_{X}\left(  .,t\right)  =\left(  F\left(  .,t\right)  ,\Phi
_{X}\left(  .,t\right)  \right)  $

Its derivative : $\frac{\partial}{\partial t}F\left(  p.,t\right)
|_{t=0}=W\left(  p\right)  $ defines a vector field on E, which is projectable
on M as X.
\end{proof}

This operation is a natural operator : $\phi:T\hookrightarrow T\digamma
::\phi\left(  M\right)  X=\digamma X$

\paragraph{Lie derivatives\newline}

1. For any bundle functor $\digamma,$ manifold M, vector field X on M, and for
any section $S\in\mathfrak{X}\left(  \digamma M\right)  $, the Lie derivative
$\pounds _{FX}S$ exists because $\digamma X$ is a projectable vector field on
$\digamma M.$

Example : with $\digamma=$ the tensorial functors we have the Lie derivative
of a tensor field over a manifold.

$\pounds _{\digamma X}S\in V\digamma M$ the vertical bundle of $\digamma M:$
this is a section on the vertical bundle with projection S on $\digamma M$

2. For any two functors $\digamma_{1},\digamma_{2}:\mathfrak{M}_{n}%
\mathfrak{\mapsto FM}$ , the Lie derivative of a\ base preserving morphism
$f:\digamma_{1}M\rightarrow\digamma_{2}M::\pi_{2}\left(  f\left(  p\right)
\right)  =\pi_{1}\left(  p\right)  ,$ along a vector field X on a manifold M
is : $\pounds _{X}f=\pounds _{\left(  \digamma_{1}X,\digamma_{2}X\right)  }f,
$ which exists (see Fiber bundles) because $\left(  \digamma_{1}X,\digamma
_{2}X\right)  $\ are vector fields projectable on the same manifold.

3. Lie derivatives commute with linear natural operators

\begin{theorem}
(Kolar p.361) Let $\digamma_{1},\digamma_{2}:\mathfrak{M}_{n}\mathfrak{\mapsto
FM}$ \ be two functors and $D:\digamma_{1}\hookrightarrow\digamma_{2}$ a
natural operator which is linear. Then for any section S on $\digamma_{1}M,$
vector field X on M : $D\left(  \pounds _{X}S\right)  =\pounds _{X}D\left(
S\right)  $

(Kolar p.362) If D is a natural operator which is not linear we have
$VD\left(  \pounds _{X}S\right)  =\pounds _{X}DS$ where VD is the vertical
prolongation of D : $VD:V\digamma_{1}M\hookrightarrow V\digamma_{2}M$
\end{theorem}

A section Y of the vertical bundle $V\digamma_{1}$\ can always be defined as :
$Y\left(  x\right)  =\frac{d}{dt}U\left(  t,x\right)  =\frac{d}{dt}%
\varphi\left(  x,u\left(  t,x\right)  \right)  |_{t=0}$ where $U\in
\mathfrak{X}\left(  \digamma_{1}M\right)  $\ and VD(Y) is defined as :
$VD\left(  Y\right)  =\frac{\partial}{\partial t}DU\left(  t,x\right)
|_{t=0}\in V\digamma_{2}M$

\begin{theorem}
(Kolar p.365) If $E_{k},k=1..n,F$ are vector bundles over the same oriented
manifold M, D a linear natural operator $D:\oplus_{k=1}^{n}E_{k}\rightarrow F$
then : $\forall S_{k}\in\mathfrak{X}_{\infty}\left(  E_{k}\right)
,X\in\mathfrak{X}_{\infty}\left(  TM\right)  :\pounds _{X}D\left(
S_{1},..,S_{n}\right)  =\sum_{k=1}^{n}D\left(  S_{1},.,\pounds _{X}%
S_{k},.,S_{n}\right)  $

Conversely every local linear operator which satisfies this identity is the
value of a unique natural operator on $\mathfrak{M}_{n}$
\end{theorem}

\paragraph{The bundle of affine connections on a manifold\newline}

Any affine connection on a m dimensional manifold (in the usual meaning of
Differential Geometry) can be seen as a connection on the principal fiber
bundle $P^{1}\left(  M,GL^{1}\left(
\mathbb{R}
,m\right)  ,\pi\right)  $ of its first order linear frames. The set of
connections on $P^{1}$ is a quotient set $QP^{1}=J^{1}P^{1}/GL^{1}\left(
\mathbb{R}
,m\right)  $ and an affine bundle $QP^{1}M$ modelled on the vector bundle
$TM\otimes TM^{\ast}\otimes TM^{\ast}.$ The functor $QP^{1}:\mathfrak{M}%
_{m}\mapsto\mathfrak{FM}$ associates to any manifold the bundle of its affine connections.

\begin{theorem}
(Kolar p.220)

i) all natural operators : $QP^{1}\hookrightarrow QP^{1}$ are of the kind :

$\phi\left(  M\right)  =\Gamma+k_{1}S+k_{2}I\otimes\widehat{S}+k_{3}%
\widehat{S}\otimes I,k_{i}\in%
\mathbb{R}
$

with : the torsion $S=\Gamma_{\beta\gamma}^{\alpha}-\Gamma_{\gamma\beta
}^{\alpha}$ and $\widehat{S}=\Gamma_{\beta\gamma}^{\alpha}+\Gamma_{\gamma
\beta}^{\alpha}$

ii) the only natural operator acting on torsion free connections is the identity
\end{theorem}

\paragraph{Curvature like operators\newline}

The curvature $\Omega$ of a connection on a fiber bundle $E\left(
M,V,\pi\right)  $ can be seen as a 2-form on M valued in the vertical bundle :
$\Omega\left(  p\right)  \in\Lambda_{2}T_{\pi\left(  p\right)  }M^{\ast
}\otimes V_{p}E$. So, viewing the connection itself as a section of $J^{1}E,$
the map which associates to a connection its curvature is a natural operator :
$J^{1}\hookrightarrow\Lambda_{2}TB^{\ast}\otimes V$ between functors acting on
$\mathfrak{FM,}$ where B is the base functor and V the vertical functor.

\begin{theorem}
(Kolar p.231) : all natural operators $J^{1}\hookrightarrow\Lambda_{2}%
TB^{\ast}\otimes V$ are a constant multiple of the curvature operator.
\end{theorem}

\paragraph{Operators on pseudo-riemannian manifolds}

\begin{theorem}
(Kolar p.244) Any r order natural operator on pseudo-riemannian metrics with
values in a first order natural bundle factorizes through the metric itself
and the derivative, up to the order r-2, of the covariant derivative of the
curvature of the L\'{e}vy-Civita connection.
\end{theorem}

\begin{theorem}
(Kolar p.274, 276) The L\'{e}vy-Civita connection is the only conformal
natural connection on pseudo-riemannian manifolds.
\end{theorem}

With the precise meaning :

For a pseudo-riemmanian manifold (M,g), a map : $\phi\left(  g,S\right)
=\Gamma$ which defines the Christoffel form of an affine connection on M, with
S some section of a functor bundle over M, is conformal if $\forall k\in%
\mathbb{R}
:\phi\left(  k^{2}g,S\right)  =\phi\left(  g,S\right)  .$ So the only natural
operators $\phi$ which are conformal are the map which defines the
L\'{e}vy-Civita connection.

\bigskip

\subsection{Gauge functors}

Gauge functors are basically functors for building associated bundles over a
principal bundle with a fixed group. We have definitions very similar to the
previous ones.

\subsubsection{Definitions}

\paragraph{Category of principal bundles\newline}

\begin{notation}
$\mathfrak{PM}$ is the category of principal bundles
\end{notation}

It comprises :

Objects : principal bundles P$\left(  M,G,\pi\right)  $

Morphisms : $\hom\left(  P_{1},P_{2}\right)  $ are defined by a couple
$F:P_{1}\rightarrow P_{2}$ and $\chi\in\hom\left(  G_{1},G_{2}\right)  $

The principal bundles with the same group G is a subcategory $\mathfrak{PM}%
\left(  G\right)  .$ If the base manifold has dimension m we have the
subcategory $\mathfrak{PM}_{m}\left(  G\right)  $

\paragraph{Gauge functors\newline}

\begin{definition}
A \textbf{gauge functor} (or gauge natural bundle) is a functor $\digamma
:\mathfrak{PM}\left(  G\right)  \mathfrak{\mapsto FM}$ \ from the category of
principal bundles with group G to the category $\mathfrak{FM}$\ of fibered
manifolds which is :

i) Base preserving : the composition by the base functor gives the identity :
$\mathfrak{B\circ}\digamma=Id_{\mathfrak{M}}$

ii) Local : if N is a submanifold of M then $\digamma\left(  N\right)  $ is
the subfibered manifold of $\digamma\left(  M\right)  $
\end{definition}

i) Base preserving means that :

Every principal bundle $P\in\mathfrak{PM}$ is transformed in a fibered
manifold with the same base : $B\digamma\left(  P\right)  =B\left(  P\right)
$

The projections : $\pi:P\rightarrow BP=M$ form a natural transformation :
$\digamma\hookrightarrow B$

Every morphism $\left(  F:P_{1}\rightarrow P_{2},\chi=Id_{G}\right)  \in
\hom\left(  P_{1},P_{2}\right)  $ is transformed in a fibered manifold
morphism $\left(  \digamma F,f\right)  $ such that $f\left(  \pi_{1}\left(
p\right)  \right)  =\pi_{2}\left(  \digamma F\left(  p\right)  \right)  $

The morphism in G is the identity

ii) Local means that :

if N is an open submanifold of M and $\imath:N\rightarrow M$ is the inclusion,
then :

$\digamma\left(  N\right)  =\pi_{M}^{-1}\left(  N\right)  $

$\digamma\left(  i\right)  :\pi_{M}^{-1}\left(  N\right)  \rightarrow
\digamma\left(  M\right)  $

\bigskip

\begin{theorem}
(Kolar p.397) \ Any gauge functor is :

i) regular

ii) of finite order
\end{theorem}

i) Regular means that :

if $f:P\times M\rightarrow N$ is a smooth map such that $f$(x,.) is a local
diffeomorphism, then $\Phi:P\times\digamma\left(  M\right)  \rightarrow
\digamma\left(  N\right)  $ defined by : $\Phi\left(  x,.\right)  =\digamma
f\left(  x,.\right)  $ is smooth.

ii) Of finite order means that :

$\exists$r, 0$\leq r\leq\infty$ if $\forall f,g\in\hom_{\mathfrak{M}_{n}%
}\left(  M,N\right)  ,\forall p\in M:j_{p}^{r}f=j_{p}^{r}g\Rightarrow\digamma
f=\digamma g$

\paragraph{Natural gauge operators\newline}

\begin{definition}
A \textbf{natural gauge operator }between two gauge functors. $\digamma
_{1},\digamma_{2}$ is :

i) a set of local operators, maps between sections on the fibered manifolds :
$A\left(  \digamma_{1}\left(  P\right)  ;\digamma_{2}\left(  P\right)
\right)  :C_{\infty}\left(  M;\digamma_{1}\left(  P\right)  \right)
\rightarrow C_{\infty}\left(  M;\digamma_{2}\left(  P\right)  \right)  $

ii) a map : $\Phi:M\rightarrow A\left(  \digamma_{1}\left(  M\right)
;\digamma_{2}\left(  M\right)  \right)  $, such that :

- $\forall S\in C_{\infty}\left(  M;\digamma_{1}\left(  M\right)  \right)
,\forall f\in\hom\left(  M,N\right)  :$

$\Phi\left(  N\right)  \left(  \digamma_{1}\left(  f\right)  \circ S\circ
f^{-1}\right)  =\digamma_{2}\left(  f\right)  \circ\Phi\left(  M\right)  \circ
f^{-1}$

- for any open submanifold N of M and section S on M :

$\Phi\left(  N\right)  \left(  S|_{N}\right)  =\Phi\left(  M\right)  \left(
S\right)  |_{N}$
\end{definition}

Every gauge natural operator has a finite order r.

\subsubsection{Theorems}

$W^{r}P=GT_{m}^{r}(M)\times_{M}J^{r}P,W_{m}^{r}G=GL^{r}(%
\mathbb{R}
,m)\rtimes T_{m}^{r}\left(  G\right)  $ (see Jets)

\begin{theorem}
(Kolar p.398) There is a bijective correspondance between the set of r order
gauge natural operators between gauge functors $\digamma_{1},\digamma_{2}$ and
the set of all natural transformations $J^{r}\circ\digamma_{1}\hookrightarrow
\digamma_{2}.$
\end{theorem}

\begin{theorem}
(Kolar p.398) There is a canonical bijection between natural transformations
between to r order gauge functors $\digamma_{1},\digamma_{2}:\mathfrak{PM}%
_{m}\left(  G\right)  \mathfrak{\mapsto FM}$ and the $W_{m}^{r}G$ equivariant
maps : $\pi_{\digamma_{1}\left(  R^{m}\times G\right)  }^{-1}\left(  0\right)
\rightarrow\pi_{\digamma_{2}\left(  R^{m}\times G\right)  }^{-1}\left(
0\right)  $
\end{theorem}

\begin{theorem}
(Kolar p.396) For any r order gauge functor $\digamma$ , and any principal
bundle $P\in\mathfrak{PM}_{m}\left(  G\right)  $\ , $\digamma P$\ is an
associated bundle to the r jet prolongation of P: $W^{r}P=GT_{m}^{r}%
(M)\times_{M}J^{r}P,W_{m}^{r}G=GL^{r}(%
\mathbb{R}
,m)\rtimes T_{m}^{r}\left(  G\right)  $.
\end{theorem}

\begin{theorem}
(Kolar p.396) Let G be a Lie group, V a manifold and and $\lambda$ a left
action of $W_{m}^{r}G$ on V, then the action factorizes to an action of
$W_{m}^{k}G$ with $k\leq2\dim V+1.$ If m%
$>$%
1 then $k\leq\max\left(  \frac{\dim V}{m-1},\frac{\dim V}{m}+1\right)  $
\end{theorem}

\subsubsection{The Utiyama theorem}

(Kolar p.400)

Let $P(M,G,\pi)$ be a principal bundle.\ The bundle of its principal
connections is the quotient set : $QP=J^{1}P/G$ which can be assimilated to
the set of potentials : $\left\{  \grave{A}\left(  x\right)  _{\alpha}%
^{i}\right\}  .$ The adjoint bundle of P is the associated vector bundle
$E=P\left[  T_{1}G,Ad\right]  .$

$QP$ is an affine bundle over E, modelled on the vector bundle $TM^{\ast
}\otimes VE\rightarrow E$

The strength form
$\mathcal{F}$%
\ of the connection can be seen as a 2-form on M valued in the adjoint bundle
$%
\mathcal{F}%
\in\Lambda_{2}\left(  M;E\right)  .$

\ A r order lagrangian on the connection bundle is a map : $%
\mathcal{L}%
:J^{r}QP\rightarrow\mathfrak{X}\left(  \Lambda_{m}TM^{\ast}\right)  $ where m
= dim(M). This is a natural gauge operator between the functors :
$J^{r}Q\hookrightarrow\Lambda_{m}B$

The Utiyama theorem reads : all first order gauge natural lagrangian on the
connection bundle are of the form $A\circ%
\mathcal{F}%
$ where A is a zero order gauge lagrangian on the connection bundle and
$\mathcal{F}$%
\ is the strength form
$\mathcal{F}$%
\ operator.

More simply said : any first order lagrangian on the connection bundle
involves only the curvature
$\mathcal{F}$%
\ and not the potential \`{A}.

\newpage

\part{FUNCTIONAL\ ANALYSIS}

\bigskip

Functional anaysis studies functions, meaning maps which are value in a field,
which is $%
\mathbb{R}
$ or, usually, $%
\mathbb{C}
,$ as it must be complete. So the basic property of the spaces of such
functions is that they have a natural structure of topological vector space,
which can be enriched with different kinds of norms, going from locally convex
to Hilbert spaces. Using these structures they can be enlarged to "generalized
functions", or distributions.

Functional analysis deals with most of the day to day problems of applied
mathematics : differential equations, partial differential equations,
optimization and variational calculus. For this endeaour some new tools are
defined, such that Fourier transform, Fourier series and the likes. As there
are many good books and internet sites on these subjects, we will focus more
on the definitions and principles than on practical methods to solve these problems.

\bigskip

\newpage

\section{SPACES OF\ FUNCTIONS}

The basic material of functional analysis is a space of functions, meaning
maps from a topological space to a field.The field is usually $%
\mathbb{C}
$ and this is what we assume if not stated otherwise.

Spaces of functions have some basic algebraic properties, which are usefull.
But this is their topological properties which are the most relevant.
Functions can be considered with respect to their support, boundedness,
continuity, integrability. For these their domain of definition does not
matter much, because their range in the very simple space $%
\mathbb{C}
.$

When we consider differentiability complications loom. First\ the domain must
be at least a manifold M. Second the partial derivatives are no longer
functions : they are maps from M to tensorial bundles over M. It is not simple
to define norms over such spaces. The procedures to deal with such maps are
basically the same as what is required to deal with sections of a vector
bundle (indeed the first derivative f'(p) of a function over M is a vector
field in the cotangent bundle TM*). So we will consider both spaces of
functions, on any topological space, including manifolds, and spaces of
sections of a vector bundle.

In the first section we recall the main results of algebra and analysis which
will be useful, in the special view when they are implemented to functions. We
define the most classical spaces of functions of functional analysis and we
add some new results about spaces of sections on a vector bundle.\ A brief
recall of functionals lead naturally to the definition and properties of
distributions, which can be viewed as "generalised functions", with another
new result : the definition of distributions over vector bundles.

\subsection{Preliminaries}

\label{Space of functions preliminaries}

\bigskip

\subsubsection{Algebraic preliminaries}

\begin{theorem}
The space of functions : $V:E\rightarrow%
\mathbb{C}
$ is a commutative *-algebra
\end{theorem}

This is a vector space and a commutattive algebra with pointwise
multiplication : $\left(  f\cdot g\right)  \left(  x\right)  =f\left(
x\right)  g\left(  x\right)  .$ It is unital with $I:E\rightarrow%
\mathbb{C}
::I\left(  x\right)  =1,$ and a *-algebra with the involution : $\overline
{}:C\left(  E;%
\mathbb{C}
\right)  \rightarrow C\left(  E;%
\mathbb{C}
\right)  ::\overline{\left(  f\right)  }\left(  x\right)  =\overline{f}\left(
x\right)  .$

With this involution the functions $C(E;%
\mathbb{R}
)$ are the subalgebra of hermitian elements in $C(E;%
\mathbb{C}
).$

The usual algebraic definitions of ideal, commutant, self-adjoint,...functions
are fully valid. Notice that no algebraic or topological structure on E is
necessary for this purpose.

\bigskip

\begin{theorem}
The spectrum of a function in $V:E\rightarrow%
\mathbb{C}
$ is its range : Sp(f)=f(E).
\end{theorem}

\begin{proof}
The spectrum\textbf{\ }$Sp\left(  f\right)  $ of f is the subset of the
scalars $\lambda\in%
\mathbb{C}
$ such that $\left(  f-\lambda Id_{V}\right)  $ has no inverse in V$.$

$\forall y\in f\left(  E\right)  :\left(  f\left(  x\right)  -f\left(
y\right)  \right)  g\left(  x\right)  =x$ has no solution
\end{proof}

\bigskip

We have also maps from a set of functions to another : $L:C\left(  E;%
\mathbb{C}
\right)  \rightarrow C\left(  F;%
\mathbb{C}
\right)  .$ When the map is linear (on $%
\mathbb{C}
)$ it is customary to call it an \textbf{operator}. The set of operators
between two algebras of functions is itself a *-algebra.

\bigskip

\begin{definition}
The \textbf{tensorial product of two functions} $f_{1}\in C\left(  E_{1};%
\mathbb{C}
\right)  ,f_{2}\in C\left(  E_{2};%
\mathbb{C}
\right)  $ is the function :

$f_{1}\otimes f_{2}\in C\left(  E_{1}\times E_{2};%
\mathbb{C}
\right)  ::f_{1}\otimes f_{2}\left(  x_{1},x_{2}\right)  =f_{1}\left(
x_{1}\right)  f_{2}\left(  x_{2}\right)  $
\end{definition}

\begin{definition}
The \textbf{tensorial product} $V_{1}\otimes V_{2}$ of two vector spaces
$V_{1}\subset C(E_{1};%
\mathbb{C}
),V_{2}\subset C(E_{2};%
\mathbb{C}
)$ is the vector subspace of $C(E_{1}\times E_{2};%
\mathbb{C}
)$\ generated by the maps : $E_{1}\times E_{2}\rightarrow%
\mathbb{C}
::F\left(  x_{1},x_{2}\right)  =\sum_{i\in I,j\in J}a_{ij}f_{i}\left(
x_{1}\right)  f_{j}\left(  x_{2}\right)  $ where $\left(  f_{i}\right)  _{i\in
I}\in V_{1}^{I},\left(  f_{j}\right)  _{j\in J}\in V_{2}^{J},a_{ij}\in%
\mathbb{C}
$ and I,J are finite sets.
\end{definition}

\bigskip

It is customary to call \textbf{functional} the maps : $\lambda:C\left(  E;%
\mathbb{C}
\right)  \rightarrow%
\mathbb{C}
$ . The space of functionals of a vector space V of functions is the algebraic
dual of V, denoted V*. The space of continuous functionals (with the topology
on V) is its topological dual denoted V'. It is isomorphic to V iff V is
finite dimensional, so this will not be the case usually.

Notice that the convolution * of functions (see below), when defined, brings
also a structure of *-algebra, that we will call the convolution algebra.

\subsubsection{Topologies on spaces of functions}

The set of functions \ $C(E;%
\mathbb{C}
)=%
\mathbb{C}
^{E}$ is huge, and the first way to identify interesting subsets is by
considering the most basic topological properties, such that continuity. On
the $%
\mathbb{C}
$ side everything is excellent : $%
\mathbb{C}
$ is a 1 dimensional Hilbert space. So most will depend on E, and the same set
can be endowed with different topologies. It is common to have several non
equivalent topologies on a space of functions. The most usual are the
following (see Analysis-Normed vector spaces), in the pecking order of
interesting properties.

\paragraph{Weak topologies\newline}

These topologies are usually required when we look for "pointwise convergence".

1. General case:

One can define a topology without any assumption about E.

\begin{definition}
For any vector space of functions $V:E\rightarrow%
\mathbb{C}
,$ $\Lambda$\ a\ subspace of V*, the weak topology on V with respect to
$\Lambda$\ , denoted $\sigma\left(  V,\Lambda\right)  ,$ is defined by the
family of semi-norms : $\lambda\in\Lambda:p_{\lambda}\left(  f\right)
=\left\vert \lambda\left(  f\right)  \right\vert $
\end{definition}

This is the initial topology$.$ The open subsets of V are defined by the base
: $\left\{  \lambda^{-1}\left(  \varpi\right)  ,\varpi\in\Omega_{%
\mathbb{C}
}\right\}  $ where $\Omega_{%
\mathbb{C}
}$ are the open subsets of $%
\mathbb{C}
.$\ With this topology all the functionals in $\Lambda$ are continuous.

A sequence $\left(  f_{n}\right)  _{n\in%
\mathbb{N}
}\in V^{%
\mathbb{N}
}$ converges to $f\in V$ iff $\forall\lambda\in\Lambda:\lambda\left(
f_{n}\right)  \rightarrow\lambda\left(  f\right)  $

The weak topology is Hausdorff iff $\Lambda$ is separating : $\forall f,g\in
V,f\neq g,\exists\lambda\in\Lambda:\lambda\left(  f\right)  \neq\lambda\left(
g\right)  $

We have similarly :

\begin{definition}
For any vector space of functions $V:E\rightarrow%
\mathbb{C}
,$ the *weak topology on V* with respect to V\ is defined by the family of
semi-norms : $f\in V:p_{f}\left(  \lambda\right)  =\left\vert \lambda\left(
f\right)  \right\vert $
\end{definition}

This is the initial topology on V* using the evaluation maps : $\widehat
{f}:V^{\ast}\rightarrow%
\mathbb{C}
::\widehat{f}\left(  \lambda\right)  =\lambda\left(  f\right)  .$ These maps
are continuous with the *weak topology.

\bigskip

2. If V is endowed with a topology:

If V is endowed with some topology and is a vector space, it is a topological
vector space. It has a topological dual, then the weak topology on V is the
$\sigma\left(  V,V^{\prime}\right)  $ topology, with the topological dual V'
of V. This topology is Hausdorff.

Similarly the *weak topology on V* is the topology $\sigma\left(  V^{\ast
},V\right)  $ and it is also Hausdorff.

The weak and *weak topologies are not equivalent to the initial topology on V,V*.

If V is endowed with a semi-norm, the weak topology is that of a semi-normed
space, with norm : $p\left(  u\right)  =\sup_{\left\Vert \lambda\right\Vert
=1}p_{\lambda}\left(  u\right)  $ which is not equivalent to the initial norm.
This is still true for a norm.

If V is a topological vector space, the product and involution on the algebra
$C(E;%
\mathbb{C}
)$ are continuous so it is a topological algebra.

\paragraph{Fr\'{e}chet spaces\newline}

A Fr\'{e}chet space is a Hausdorff, complete, topological vector space, whose
metric comes from a countable family of seminorms $\left(  p_{i}\right)
_{i\in I}$. It is locally convex. As the space is metric we can consider
uniform convergence. But the topological dual of V is not necessarily a
Fr\'{e}chet space.

\paragraph{Compactly supported functions\newline}

The support of $f\in C\left(  E;%
\mathbb{C}
\right)  $ is the subset of E : Supp(f)=$\overline{\left\{  x\in
E:f(x)\neq0\right\}  }$ or equivalently the complement of the largest open set
where f(x) is zero. So the support is a closed subset of E.

\begin{definition}
A compactly supported function is a function in domain in a topological space
E, whose support is enclosed in some compact subset of E.
\end{definition}

\begin{theorem}
If V is a space of functions on a topological space E, then the subspace
$V_{c}$ of compactly supported functions of V is dense in V. So if V is a
Fr\'{e}chet space, then $V_{c}$ is a Fr\'{e}chet space.
\end{theorem}

\begin{proof}
Let V be a space of functions of $C\left(  E;%
\mathbb{C}
\right)  ,$ and define $V_{K}$ the subset of V of functions whose support is
the compact K, and $V_{c}$ the subset of V of functions with compact support :
$V_{c}=\cup_{K\subset E}V_{K}.$ We can define on V the final topology,
characterized by the set of opens $\Omega,$ with respect to the family of
embeddings : $\imath_{K}:V_{K}\rightarrow V$ . Then $\varpi\in\Omega$ is open
in V if $\varpi\cap V_{K}$ is open in each $V_{K}$ and we can take the sets
$\left\{  \varpi\cap V_{K},\varpi\in\Omega,K\subset E\right\}  $\ \ as base
for the topology of $V_{c}.$ and $V_{c}$\ is dense in V.

If V is a Fr\'{e}chet space then each $V_{K}$ is closed in V so is a
Fr\'{e}chet space.
\end{proof}

\begin{theorem}
If V is a space of continuous functions on a topological space E which can be
covered by countably many compacts, then the subspace $V_{c}$ of compactly
supported functions of V is a Fr\'{e}chet space, closed in V.
\end{theorem}

(it is sometimes said that E is $\sigma-$compact)

\begin{proof}
There is a countable family of compacts $\left(  \kappa_{j}\right)  _{j\in J}$
such that any compact K is contained in some $\kappa_{j}$\ . Then V$_{c}$ is a
Fr\'{e}chet space, closed in V, with the seminorms $q_{i,j}\left(  f\right)
=\sup_{j}p_{i}\left(  f|_{\kappa_{j}}\right)  $ where $p_{i}=\sup\left\vert
f|_{\kappa_{j}}\right\vert $.
\end{proof}

\paragraph{Banach spaces\newline}

A vector space of functions $V\subset C\left(  E;%
\mathbb{C}
\right)  $ is normed if we have a map : $\left\Vert {}\right\Vert
:V\rightarrow%
\mathbb{R}
_{+}$ so that :

$\forall f,g\in V,k\in%
\mathbb{C}
:$

$\left\Vert f\right\Vert \geq0;\left\Vert f\right\Vert =0\Rightarrow f=0$

$\left\Vert kf\right\Vert =\left\vert k\right\vert \left\Vert f\right\Vert $

$\left\Vert f+g\right\Vert \leq\left\Vert f\right\Vert +\left\Vert
g\right\Vert $

The topological dual of V is a normed space with the strong topology :
$\left\Vert \lambda\right\Vert =\sup_{\left\Vert f\right\Vert =1}\left\vert
\lambda\left(  f\right)  \right\vert $

If V is a *-algebra such that : $\left\Vert \overline{f}\right\Vert
=\left\Vert f\right\Vert $ and $\left\Vert f\right\Vert ^{2}=\left\Vert
\left\vert f\right\vert ^{2}\right\Vert $ then it is a normed *-algebra.

If V is normed and complete then it is a Banach space. Its topological dual is
a Banach space.

\begin{theorem}
If a vector space of functions V is a complete normed algebra, it is a Banach
algebra. Then the range of any function is a compact of $%
\mathbb{C}
.$
\end{theorem}

\begin{proof}
the spectrum of any element is just the range of f

the spectrum of any element is a compact subset of $%
\mathbb{C}
$
\end{proof}

If V is also a normed *- algebra, it is a C*-algebra.

\begin{theorem}
The norm on a C*-algebra of functions is necessarily equivalent to :
$\left\Vert f\right\Vert =\sup_{x\in E}\left\vert f\left(  x\right)
\right\vert $
\end{theorem}

\begin{proof}
the spectrum of any element is just the range of f

In a normed *-algebra the spectral radius of a normal element f is :
$r_{\lambda}\left(  f\right)  =\left\Vert f\right\Vert ,$

all the functions are normal ff*=f*f

In a C*-algebra : $r_{\lambda}\left(  f\right)  =\max_{x\in E}\left\vert
f\left(  x\right)  \right\vert =\left\Vert f\right\Vert $
\end{proof}

Notice that no property required from E.

\paragraph{Hilbert spaces\newline}

A Hilbert space H is a Banach space whose norm comes from a positive definite
hermitian form denoted $\left\langle {}\right\rangle .$ Its dual H' is a also
a Hilbert space.\ In addition to the properties of Banach spaces, Hilbert
spaces offer the existence of Hilbertian bases : any function can be written
as the series : $f=\sum_{i\in I}f_{i}e_{i}$ and of an adjoint operation :
$\overset{\ast}{}:H\rightarrow H^{\prime}::\forall\varphi\in H:\left\langle
f,\varphi\right\rangle =f^{\ast}\left(  \varphi\right)  $

\subsubsection{Vector bundles\newline}

\paragraph{Manifolds\newline}

\textit{In this part} ("Functional Analysis") we will, if not otherwise
stated, assumed that a manifold M is a \textit{finite m dimensional} real
Hausdorff class 1 manifold M. Then M has the following properties (see
Differential geometry) :

i) it has an equivalent smooth structure, so we can consider only smooth manifolds

ii) it is locally compact, paracompact, second countable

iii) it is metrizable and admits a riemannian metric

iv) it is locally connected and each connected component is separable

v) it admits an atlas with a finite number of charts

vi) every open covering $\left(  O_{i}\right)  _{i\in I}$ has a refinement
$\left(  Q_{i}\right)  _{i\in I}$ such that $Q_{i}\sqsubseteq O_{i}$ and :
$Q_{i}$ has a compact closure, $\left(  Q_{i}\right)  _{i\in I}$\ \ is locally
finite (each points of M meets only a finite number of \ $Q_{i}$), any non
empty finite intersection of $Q_{i}$\ is diffeomorphic with an open of $%
\mathbb{R}
^{m}$

So we will assume that M has a countable atlas $\left(  O_{a},\psi_{a}\right)
_{a\in A},\overline{O_{a}}$ compact.

Sometimes we will assume that M is endowed with a volume form $\varpi_{0}$\ .
A volume form induces an absolutely continuous Borel measure on M, locally
finite (finite on any compact). It can come from any non degenerate metric on
M. With the previous properties there is always a riemannian metric so such a
volume form always exist.

\paragraph{Vector bundle\newline}

\textit{In this part} ("Functional Analysis") we will, if not otherwise
stated, assumed that a vector bundle $E\left(  M,V,\pi\right)  $\ with atlas
$\left(  O_{a},\varphi_{a}\right)  _{a\in A}$, transition maps $\varphi
_{ab}\left(  x\right)  \in%
\mathcal{L}%
\left(  V;V\right)  $\ and V a Banach vector space, has the following
properties :

i) the base manifold M is a smooth finite m dimensional real Hausdorff
manifold (as above)

ii) the trivializations $\varphi_{a}\in C\left(  O_{a}\times V;E\right)  $ and
the transitions maps $\varphi_{ab}\in C\left(  O_{a}\cap O_{b};%
\mathcal{L}%
\left(  V;V\right)  \right)  $ are smooth.

As a consequence :

i) The fiber E(x) over x is canonically isomorphic to V. V can be infinite
dimensional, but\ \textit{fiberwise} there is a norm such that E(x) is a
Banach vector space.

ii) the trivilialization $\left(  O_{a},\varphi_{a}\right)  _{a\in A}$ is such
that $\overline{O_{a}}$ compact, and for any x in M there is a finite number
of $a\in A$ such that $x\in O_{a}$ and any intersection $O_{a}\cap O_{b}$ is
relatively compact.

Sometimes we will assume that E is endowed with a scalar product, meaning
there is a family of maps $\left(  g_{a}\right)  _{a\in A}$\ with domain
$O_{a} $ such that $g_{a}\left(  x\right)  $\ is a non degenerate, either
bilinear symmetric form (real case), or sesquilinear hermitian form (complex
case)\ on each fiber E(x) and on the transition : $g_{bij}\left(  x\right)
=\sum_{kl}\overline{\left[  \varphi_{ab}\left(  x\right)  \right]  }_{i}%
^{k}\left[  \varphi_{ab}\left(  x\right)  \right]  _{j}^{l}g_{akl}\left(
x\right)  .$ It is called an inner product it is definite positive.

A scalar product g on a vector space V induces a scalar product on a vector
bundle $E(M,V,\pi)$ iff the transitions maps preserve g

Notice :

i) a metric on M, riemannian or not, is of course compatible with any
tensorial bundle over TM (meaning defined through the tangent or the cotangent
bundle), and can induce a scalar product on a tensorial bundle (see Algebra -
tensorial product of maps), but the result is a bit complicated, except for
the vector bundles TM, TM*, $\Lambda_{r}TM^{\ast}$.

ii) there is no need for a scalar product on V to be induced by anything on
the base manifold. In fact, as any vector bundle can be defined as the
associated vector bundle $P\left[  V,r\right]  $\ to a principal bundle P
modelled on a group G, if (V,r) is a unitary representation of V, any metric
on V induces a metric on E. The potential topological obstructions lie
therefore in the existence of a principal bundle over the manifold M. For
instance not any manifold can support a non riemannian metric (but all can
support a riemannian one).

\paragraph{Sections of a vector bundle\newline}

A section S on a vector bundle $E\left(  M,V,\pi\right)  $\ with
trivilialization $\left(  O_{a},\varphi_{a}\right)  _{a\in A}$ is defined by a
family $\left(  \sigma_{a}\right)  _{a\in A}$ of maps $\sigma_{a}%
:O_{a}\rightarrow V$ such that : $U\left(  x\right)  =\varphi_{a}\left(
x,u_{a}\left(  x\right)  \right)  ,\forall a,b\in A,O_{a}\cap O_{b}%
\neq\varnothing,\forall x\in O_{a}\cap O_{b}:u_{b}\left(  x\right)
=\varphi_{ba}\left(  x\right)  u_{a}\left(  x\right)  $ where the transition
maps $\varphi_{ba}\left(  x\right)  \in%
\mathcal{L}%
\left(  V;V\right)  $

So we assume that the sections are defined over the same open cover as E.

The space $\mathfrak{X}\left(  E\right)  $ of sections on E is a complex
vector space of infinite dimension.

As a consequence of the previous assumptions there is, \textit{fiberwise}, a
norm for the sections of a fiber bundle and, possibly, a scalar product. But
it does not provide by itself a norm or a scalar product for $\mathfrak{X}%
\left(  E\right)  $ : as for functions we need some way to agregate the
results of each fiber. This is usually done through a volume form on M or by
taking the maximum of the norms on the fibers.

To E we can associate the dual bundle $E^{\prime}\left(  M,V^{\prime}%
,\pi\right)  $ where V' is the topological dual of V (also a Banach vector
space) and there is fiberwise the action of $\mathfrak{X}\left(  E^{\prime
}\right)  $ on $\mathfrak{X}\left(  E\right)  $

If M is a real manifold we can consider the space of complex valued functions
over M as a complex vector bundle modelled over $%
\mathbb{C}
.$ A section is then simply a function. This identification is convenient in
the definition of differential operators.

\paragraph{Sections of the r jet prolongation of a vector bundle\newline}

1. The r jet prolongation $J^{r}E$ of a vector bundle is a vector bundle

$J^{r}E\left(  E,J_{0}^{r}\left(
\mathbb{R}
^{m},V\right)  ,\pi_{0}^{r}\right)  .$ The vector space $J_{0}^{r}\left(
\mathbb{R}
^{m},V\right)  $ is a set of r components : $J_{0}^{r}\left(
\mathbb{R}
^{m},V\right)  =\left(  z_{s}\right)  _{s=1}^{r}$ which can be identified
either to multilinear symmetric maps $%
\mathcal{L}%
_{S}^{s}\left(
\mathbb{R}
^{m},V\right)  $ or to tensors $\otimes^{r}TM^{\ast}\otimes V.$

So an element of $J^{r}E\left(  x\right)  $ is identified with $\left(
z_{s}\right)  _{s=0}^{r}$ with $z_{0}\in V$

Sections $Z\in\mathfrak{X}\left(  J^{r}E\right)  $ of the r jet prolongation
$J^{r}E$ are maps :

$Z:M\rightarrow\mathfrak{X}\left(  J^{r}E\right)  ::Z\left(  x\right)
=\left(  z_{s}\left(  x\right)  \right)  _{s=.0}^{r}$

Notice that a section on $\mathfrak{X}\left(  J^{r}E\right)  $ does not need
to come from a r differentiable section in $\mathfrak{X}_{r}\left(  E\right)
$

2. Because V is a Banach space, there is a norm on the space $%
\mathcal{L}%
_{S}^{s}\left(
\mathbb{R}
^{m},V\right)  :$

$z_{s}\in%
\mathcal{L}%
_{S}^{s}\left(
\mathbb{R}
^{\dim M},V\right)  :\left\Vert z_{s}\right\Vert =\sup\left\Vert z_{s}\left(
u_{1},...,u_{s}\right)  \right\Vert _{V}$

for $\left\Vert u_{k}\right\Vert _{%
\mathbb{R}
^{\dim M}}=1,k=1...s$

So we can define, fiberwise, a norm for $Z\in J^{r}E\left(  x\right)  $ by :

either : $\left\Vert Z\right\Vert =\left(  \sum_{s=0}^{r}\left\Vert
z_{s}\right\Vert ^{2}\right)  ^{1/2}$

or : $\left\Vert Z\right\Vert =\max\left(  \left\Vert z_{s}\right\Vert
,s=0..r\right)  $

which are equivalent.

Depending on the subspace of $\mathfrak{X}\left(  J^{r}E\right)  $ we can
define the norms :

either : $\left\Vert Z\right\Vert =\sum_{s=0}^{r}\int_{M}\left\Vert
z_{s}\left(  x\right)  \right\Vert \mu$ for integrable sections, if M is
endowed with a measure $\mu$

or : $\left\Vert Z\right\Vert =\max_{x\in M}\left(  \left\Vert z_{s}\left(
x\right)  \right\Vert ,s=0..r\right)  $

\subsubsection{Miscellaneous theorems}

\begin{theorem}
(Lieb p.76) : let $f\in C\left(
\mathbb{R}
;%
\mathbb{R}
\right)  $ be a measurable function such that f(x+y)=f(x)+f(y) , then f(x)=kx
for some constant k
\end{theorem}

\paragraph{Gamma function\newline}

$\Gamma\left(  z\right)  =\int_{0}^{\infty}e^{-t}t^{z-1}dt=\left(  z-1\right)
\Gamma\left(  z-1\right)  =\left(  z-1\right)  !$

$\Gamma\left(  z\right)  \Gamma\left(  1-z\right)  =\frac{\pi}{\sin\pi z}$

$\sqrt{\pi}\Gamma\left(  2z\right)  =2^{2z-1}\Gamma\left(  z\right)
\Gamma\left(  z+\frac{1}{2}\right)  $

The area $A\left(  S^{n-1}\right)  $ of the unit sphere $S^{n-1}$ in $%
\mathbb{R}
^{n}$\ is $A\left(  S^{n-1}\right)  =\frac{2\pi^{n/2}}{\Gamma\left(  \frac
{n}{2}\right)  }$

$\int_{0}^{1}\left(  1+u\right)  ^{-x-y}u^{x-1}du=\frac{\Gamma\left(
x\right)  \Gamma\left(  y\right)  }{\Gamma\left(  x+y\right)  }$

The Lebegue volume of a ball B(0,r) in $%
\mathbb{R}
^{n}$ is $\frac{1}{n}A\left(  S^{n-1}\right)  r^{n}$

\bigskip

\subsection{Spaces of bounded or continuous maps}

\label{Space of bounded or continuous maps}

\subsubsection{Spaces of bounded or continuous functions}

Let E be a topological space, C(E,$%
\mathbb{C}
)$ the set of functions $f:E\rightarrow%
\mathbb{C}
$

\begin{notation}
$C_{c}\left(  E;%
\mathbb{C}
\right)  $ is the set of functions with compact support

$C_{0}\left(  E;%
\mathbb{C}
\right)  $ is the set of continuous functions

$C_{0b}\left(  E;%
\mathbb{C}
\right)  $ is the set of bounded continuous functions

$C_{0c}\left(  E;%
\mathbb{C}
\right)  $ is the set of continuous functions with compact support

$C_{0v}\left(  E;%
\mathbb{C}
\right)  $ is the set of continuous functions vanishing at infinity :

$\forall\varepsilon>0,\exists K$ compact in E$:\forall x\in K^{c}:\left\vert
f(x)\right\vert <\varepsilon$
\end{notation}

Let (E,d) be a metric space, $\gamma\in\left[  0,1\right]  $

\begin{notation}
C$^{\gamma}\left(  E;%
\mathbb{C}
\right)  $ is the space of order $\gamma$\ Lipschitz functions :

$\exists C>0:\forall x,y\in E:\left\vert f\left(  x\right)  -f\left(
y\right)  \right\vert \leq Cd\left(  x,y\right)  ^{\gamma}$
\end{notation}

\bigskip

\begin{theorem}
Are commutative C*-algebra with pointwise product and the norm : $\left\Vert
f\right\Vert =\max_{x\in E}\left\vert f\left(  x\right)  \right\vert $

i) $C_{b}\left(  E;%
\mathbb{C}
\right)  $

ii) $C_{0b}\left(  E;%
\mathbb{C}
\right)  $ is E is Hausdorff

iii) $C_{c}\left(  E;%
\mathbb{C}
\right)  ,C_{0v}\left(  E;%
\mathbb{C}
\right)  $ if E is Hausdorff, locally compact

iv) $C_{0}\left(  E;%
\mathbb{C}
\right)  $ is E is compact
\end{theorem}

\begin{theorem}
$C_{0c}\left(  E;%
\mathbb{C}
\right)  $\ is a commutative normed *-algebra (it is not complete) with
pointwise product and the norm : $\left\Vert f\right\Vert =\max_{x\in
E}\left\vert f\left(  x\right)  \right\vert $ if E is Hausdorff, locally
compact. Moreover $C_{0c}\left(  E;%
\mathbb{C}
\right)  $ is dense in $C_{0\nu}\left(  E;%
\mathbb{C}
\right)  $
\end{theorem}

\begin{theorem}
$C^{\gamma}\left(  E;%
\mathbb{C}
\right)  $ is a Banach space with norm : $\left\Vert f\right\Vert
=\sup\left\vert f\left(  x\right)  \right\vert +\sup_{x,y\in E}\frac
{\left\vert f\left(  x\right)  -f\left(  y\right)  \right\vert }{d\left(
x,y\right)  ^{\gamma}}$ if E is metric, locally compact
\end{theorem}

\bigskip

\begin{theorem}
Tietze extension theorem (Thill p.257) If E is a normal Hausdorff space and X
a closed subset of E, then every function in $C_{0}\left(  X;%
\mathbb{R}
\right)  $ can be extended to $C_{0}\left(  E;%
\mathbb{R}
\right)  $
\end{theorem}

\begin{theorem}
Stone-Weierstrass (Thill p.27): Let E be a locally compact Hausdorff space,
then a unital subalgebra A of $C_{0}\left(  E;%
\mathbb{R}
\right)  $ such that

$\forall x\neq y\in E,\exists f\in A:f\left(  x\right)  \neq f\left(
y\right)  $ and $\forall x\in E,\exists f\in A:f\left(  x\right)  \neq0$\ 

is dense in $C_{0}\left(  E;%
\mathbb{R}
\right)  .$
\end{theorem}

\begin{theorem}
(Thill p.258) If X is any subset of E, $C_{c}\left(  X;%
\mathbb{C}
\right)  $ is the set of the restrictions to X of the maps in $C_{c}\left(  E;%
\mathbb{C}
\right)  $
\end{theorem}

\begin{theorem}
Arzela-Ascoli (Gamelin p.82) A subset F of $C_{0}\left(  E;%
\mathbb{R}
\right)  $\ is relatively compact iff it is equicontinuous
\end{theorem}

\begin{theorem}
(Gamelin p.82) If E is a compact metric space, the inclusion : $\imath
:C^{\gamma}\left(  E;%
\mathbb{C}
\right)  \rightarrow C_{0}\left(  E;%
\mathbb{C}
\right)  $ is a compact map
\end{theorem}

\subsubsection{Spaces of bounded or continuous sections of a vector bundle}

According to our definition, a complex\ vector bundle $E\left(  M,V,\pi
\right)  $ V is a Banach vector space and there is always fiberwise a norm for
sections on the vector bundle. So most of the previous definitions can be
generalized to sections $U\in\mathfrak{X}\left(  E\right)  $\ on E by taking
the functions : $M\rightarrow%
\mathbb{R}
:\left\Vert U\left(  x\right)  \right\Vert _{E}$

\begin{definition}
Let $E\left(  M,V,\pi\right)  $ be a complex \textit{vector} bundle

$\mathfrak{X}\left(  E\right)  $ the set of sections $U:M\rightarrow E$

$\mathfrak{X}_{c}\left(  E\right)  $ the set of sections with compact support

$\mathfrak{X}_{0}\left(  E\right)  $ the set of continuous sections

$\mathfrak{X}_{0b}\left(  E\right)  $ the set of bounded continuous sections

$\mathfrak{X}_{0c}\left(  E\right)  $ the set of continuous sections with
compact support

$\mathfrak{X}_{0\nu}\left(  E\right)  $ the set of continuous sections
vanishing at infinity :

$\forall\varepsilon>0,\exists K$ compact in M$:\forall x\in K^{c}:\left\Vert
U\left(  x\right)  \right\Vert <\varepsilon$
\end{definition}

As E is necessarily Hausdorff:

\begin{theorem}
Are Banach vector spaces with the norm : $\left\Vert U\right\Vert _{E}%
=\max_{x\in M}\left\Vert U\left(  x\right)  \right\Vert $

i) $\mathfrak{X}_{0}\left(  E\right)  ,\mathfrak{X}_{0b}\left(  E\right)  $

ii) $\mathfrak{X}_{c}\left(  E\right)  ,\mathfrak{X}_{0\nu}\left(  E\right)
,\mathfrak{X}_{0c}\left(  E\right)  $ if V is finite dimensional, moreover
$\mathfrak{X}_{0c}\left(  E\right)  $ is dense in $\mathfrak{X}_{0\nu}\left(
E\right)  $
\end{theorem}

The r jet prolongation $J^{r}E$ of a vector bundle is a vector bundle, and
there is fiberwise the norm (see above) :

$\left\Vert Z\right\Vert =\max\left(  \left\Vert z_{s}\right\Vert
_{s},s=0..r\right)  $

so we have similar results for sections $\mathfrak{X}\left(  J^{r}E\right)  $

\subsubsection{Rearrangements inequalities}

They address more specifically the functions on $%
\mathbb{R}
^{m}$ which vanish at infinity. This is an abstract from Lieb p.80.

\begin{definition}
For any Lebesgue measurable set $E\subset%
\mathbb{R}
^{m}$ the \textbf{symmetric rearrangement} $E^{\times}$\ is the ball B(0,r)
centered in 0 with a volume equal to the volume of E.
\end{definition}

$\int_{E}dx=A\left(  S^{m-1}\right)  \frac{r^{m}}{m}$ where $A\left(
S^{m-1}\right)  $ is the area of the unit sphere $S^{m-1}$ in $%
\mathbb{R}
^{m}$\ : $A\left(  S^{m-1}\right)  =\frac{2\pi^{m/2}}{\Gamma\left(  \frac
{m}{2}\right)  }$

\begin{definition}
The symmetric decreasing rearrangement of any characteristic function $1_{E}$
of a measurable set E\ is the characteristic function of its symmetric
rearrangement : $1_{E^{\times}}=1_{E}$
\end{definition}

\bigskip

\begin{theorem}
Any measurable function $f\in C\left(
\mathbb{R}
^{m};%
\mathbb{C}
\right)  $ vanishes at infinity iff : $\forall t>0:\int_{\left\vert f\left(
x\right)  \right\vert >t}dx<\infty$
\end{theorem}

\bigskip

\begin{theorem}
For a measurable, vanishing at infinity function $f\in C\left(
\mathbb{R}
^{m};%
\mathbb{C}
\right)  $\ the function $f^{\times}\left(  x\right)  =\int_{0}^{\infty
}1_{\left\vert f\left(  \tau\right)  >t\right\vert }^{\times}\left(  x\right)
dt.$ has the following properties :

i) $f^{\times}\left(  x\right)  \geq0$

ii) it is radially symmetric and non increasing :

$\left\Vert x\right\Vert =\left\Vert y\right\Vert \Rightarrow f^{\times
}\left(  x\right)  =f^{\times}\left(  y\right)  $

$\left\Vert x\right\Vert \geq\left\Vert y\right\Vert \Rightarrow f^{\times
}\left(  x\right)  \geq f^{\times}\left(  y\right)  $

iii) it\ is lower semicontinuous : the sets $\left\{  x:f^{\times}\left(
x\right)  >t\right\}  $ are open

iv) $\left\vert f\right\vert ,\left\vert f^{\times}\right\vert $ are
equimeasurable :

$\left\{  x:f^{\times}\left(  x\right)  >t\right\}  =\left\{  x:\left\vert
f\left(  x\right)  \right\vert >t\right\}  ^{\times}$

$\int\left\{  x:f^{\times}\left(  x\right)  >t\right\}  dx=\int\left\{
x:\left\vert f\left(  x\right)  \right\vert >t\right\}  dx$

If $f\in L^{p}\left(
\mathbb{R}
^{m},S,dx,%
\mathbb{C}
\right)  :\left\Vert f\right\Vert _{p}=\left\Vert f^{\times}\right\Vert _{p}$

iv) If $f,g\in C_{\nu}\left(
\mathbb{R}
^{m};%
\mathbb{C}
\right)  $ and $f\leq g$ then $f^{\times}\leq g^{\times}$
\end{theorem}

\bigskip

\begin{theorem}
If f,g are positive, measurable, vanishing at infinity functions $f,g\in
C\left(
\mathbb{R}
^{m};%
\mathbb{R}
_{+}\right)  $ $f\geq0,g\geq0$ then:

$\int_{M}f\left(  x\right)  g\left(  x\right)  dx\leq\int_{M}f^{\times}\left(
x\right)  g^{\times}\left(  x\right)  dx$ possibly infinite.

If $\forall\left\Vert x\right\Vert >\left\Vert y\right\Vert \Rightarrow
f^{\times}\left(  x\right)  >f^{\times}\left(  y\right)  $ then

$\int_{M}f\left(  x\right)  g\left(  x\right)  dx=\int_{M}f^{\times}\left(
x\right)  g^{\times}\left(  x\right)  dx$ $\Leftrightarrow$ $g=g^{\times}$

if J is a non negative convex function $J:%
\mathbb{R}
\rightarrow%
\mathbb{R}
$ such that J(0)=0 then

$\int_{%
\mathbb{R}
^{m}}J\left(  f^{\times}\left(  x\right)  -g^{\times}\left(  x\right)
\right)  dx\leq\int_{%
\mathbb{R}
^{m}}J\left(  f\left(  x\right)  -g\left(  x\right)  \right)  dx$
\end{theorem}

\bigskip

\begin{theorem}
For any positive, measurable, vanishing at infinity functions $\left(
f_{n}\right)  _{n=1}^{N}\in C\left(
\mathbb{R}
^{m};%
\mathbb{R}
_{+}\right)  $, kxN matrix $A=\left[  A_{ij}\right]  $ with $k\leq N:$

$\int_{%
\mathbb{R}
^{m}}...\int_{%
\mathbb{R}
^{m}}%
{\textstyle\prod\limits_{n=1}^{N}}
f_{n}\left(  \sum_{i=1}^{k}A_{in}x_{i}\right)  dx\leq\int_{%
\mathbb{R}
^{m}}...\int_{%
\mathbb{R}
^{m}}%
{\textstyle\prod\limits_{n=1}^{N}}
f_{n}^{\times}\left(  \sum_{i=1}^{k}A_{in}x_{i}\right)  dx\leq\infty$
\end{theorem}

\bigskip

\subsection{Spaces of integrable maps}

\label{Spaces of integrable maps}

\subsubsection{$L^{p}$ spaces of functions}

\paragraph{Definition\newline}

(Lieb p.41)

Let $(E,S,\mu)$ a measured space with\ $\mu$\ a positive measure. If the
$\sigma-$algebra S is omitted then it is the Borel algebra. The Lebesgue
measure is denoted dx as usual.

Take $p\in%
\mathbb{N}
,p\neq0,$ consider the sets of measurable functions

$%
\mathcal{L}%
^{p}\left(  E,S,\mu,%
\mathbb{C}
\right)  =\left\{  f:E\rightarrow%
\mathbb{C}
:\int_{E}\left\vert f\right\vert ^{p}\mu<\infty\right\}  $

$N^{p}\left(  E,S,\mu,%
\mathbb{C}
\right)  =\left\{  f\in%
\mathcal{L}%
^{p}\left(  E,S,\mu\right)  :\int_{E}\left\vert f\right\vert ^{p}%
\mu=0\right\}  $

$L^{p}\left(  E,S,\mu,%
\mathbb{C}
\right)  =%
\mathcal{L}%
^{p}\left(  E,S,\mu,%
\mathbb{C}
\right)  /N^{p}\left(  E,S,\mu,%
\mathbb{C}
\right)  $

\begin{theorem}
$L^{p}\left(  E,S,\mu,%
\mathbb{C}
\right)  $ is a complex Banach vector space with the norm: $\left\Vert
f\right\Vert _{p}=\left(  \int_{E}\left\vert f\right\vert ^{p}\mu\right)
^{1/p}$

$L^{2}\left(  E,S,\mu,%
\mathbb{C}
\right)  $ is a complex Hilbert vector space with the scalar product :
$\left\langle f,g\right\rangle =\int_{E}\overline{f}g\mu$
\end{theorem}

Similarly the sets of bounded measurable functions :

$%
\mathcal{L}%
^{\infty}\left(  E,S,\mu,%
\mathbb{C}
\right)  =\left\{  f:E\rightarrow%
\mathbb{C}
:\exists C\in%
\mathbb{R}
:\left\vert f(x)\right\vert <C\right\}  $

$\left\Vert f\right\Vert _{\infty}=\inf\left(  C\in%
\mathbb{R}
:\mu\left(  \left\{  \left\vert f\left(  x\right)  \right\vert >C\right\}
\right)  =0\right)  $

$N^{\infty}\left(  E,S,\mu,%
\mathbb{C}
\right)  =\left\{  f\in%
\mathcal{L}%
^{\infty}\left(  E,S,\mu,%
\mathbb{C}
\right)  :\left\Vert f\right\Vert _{\infty}=0\right\}  $

$L^{\infty}\left(  E,S,\mu,%
\mathbb{C}
\right)  =%
\mathcal{L}%
^{\infty}\left(  E,S,\mu,%
\mathbb{C}
\right)  /N^{\infty}\left(  E,S,\mu,%
\mathbb{C}
\right)  $

\begin{theorem}
$L^{\infty}\left(  E,S,\mu,%
\mathbb{C}
\right)  $ is a C*-algebra (with pointwise multiplication)
\end{theorem}

Similarly one defines the spaces :

\begin{notation}
$L_{c}^{p}\left(  E,S,\mu,%
\mathbb{C}
\right)  $ is the subspace of $L^{p}\left(  E,S,\mu,%
\mathbb{C}
\right)  $ with compact support
\end{notation}

\begin{notation}
$L_{loc}^{p}\left(  E,S,\mu,%
\mathbb{C}
\right)  $ is the space of functions in C$\left(  E;%
\mathbb{C}
\right)  $ such that $\int_{K}\left\vert f\right\vert ^{p}\mu<\infty$ for any
compact K in E
\end{notation}

\paragraph{Inclusions}

\begin{theorem}
(Lieb p.43) $\forall p,q,r\in%
\mathbb{N}
:$

$1\leq p\leq\infty:$

$f\in%
\mathcal{L}%
^{p}\left(  E,S,\mu,%
\mathbb{C}
\right)  \cap%
\mathcal{L}%
^{\infty}\left(  E,S,\mu,%
\mathbb{C}
\right)  \Rightarrow f\in%
\mathcal{L}%
^{q}\left(  E,S,\mu,%
\mathbb{C}
\right)  $ for $q\geq\ p$

$1\leq p\leq r\leq q\leq\infty:$

$%
\mathcal{L}%
^{p}\left(  E,S,\mu,%
\mathbb{C}
\right)  \cap%
\mathcal{L}%
^{q}\left(  E,S,\mu,%
\mathbb{C}
\right)  \subset%
\mathcal{L}%
^{r}\left(  E,S,\mu,%
\mathbb{C}
\right)  $
\end{theorem}

Warning ! Notice that we \textit{do not have} $%
\mathcal{L}%
^{q}\left(  E,S,\mu,%
\mathbb{C}
\right)  \subset%
\mathcal{L}%
^{p}\left(  E,S,\mu,%
\mathbb{C}
\right)  $ for q
$>$
p unless $\mu$ is a finite measure

\begin{theorem}
$\lim_{p\rightarrow\infty}\left\Vert f\right\Vert _{p}=\left\Vert f\right\Vert
_{\infty}$
\end{theorem}

\subsubsection{Spaces of integrable sections of a vector bundle}

\paragraph{Definition\newline}

According to our definition, in a complex\ vector bundle $E\left(
M,V,\pi\right)  $ V is a Banach vector space, thus a normed space and there is
always pointwise a norm for sections on the vector bundle. If there is a
positive measure $\mu$ on M we can generalize the previous definitions to
sections $U\in\mathfrak{X}\left(  E\right)  $\ on E by taking the functions :
$M\rightarrow%
\mathbb{R}
:\left\Vert U\left(  x\right)  \right\Vert _{E}$

So we define on the space of sections $U\in\mathfrak{X}\left(  E\right)  $

$p\geq1:$

$%
\mathcal{L}%
^{p}\left(  M,\mu,E\right)  =\left\{  U\in\mathfrak{X}\left(  E\right)
:\int_{M}\left\Vert U\left(  x\right)  \right\Vert ^{p}\mu<\infty\right\}  $

$N^{p}\left(  M,\mu,E\right)  =\left\{  U\in%
\mathcal{L}%
^{p}\left(  M,\mu,E\right)  :\int_{M}\left\Vert U\left(  x\right)  \right\Vert
^{p}\mu=0\right\}  $

$L^{p}\left(  M,\mu,E\right)  =%
\mathcal{L}%
^{p}\left(  M,\mu,E\right)  /N^{p}\left(  M,\mu,E\right)  $

and similarly for $p=\infty$

$%
\mathcal{L}%
^{\infty}\left(  M,\mu,E\right)  =\left\{  U\in\mathfrak{X}\left(  E\right)
:\exists C\in%
\mathbb{R}
:\left\Vert U(x)\right\Vert <C\right\}  $

$\left\Vert U\right\Vert _{\infty}=\inf\left(  C\in%
\mathbb{R}
:\int_{U_{c}}\mu=0\text{ with }U_{c}=\left\{  x\in M:\left\Vert
U(x)\right\Vert >C\right\}  \right)  $

$N^{\infty}\left(  M,\mu,E\right)  =\left\{  U\in%
\mathcal{L}%
^{\infty}\left(  M,\mu,E\right)  :\left\Vert U\right\Vert _{\infty}=0\right\}
$

$L^{\infty}\left(  M,\mu,E\right)  =%
\mathcal{L}%
^{\infty}\left(  M,\mu,E\right)  /N^{\infty}\left(  M,\mu,E\right)  $

Notice:

i) the measure $\mu$ can be any measure.\ Usually it is the Lebesgue measure
induced by a volume form $\varpi_{0}.$ In this case the measure is absolutely
continuous. Any riemannian metric induces a volume form, and with our
assumptions the base manifold has always such a metric, thus has a volume form.

ii) this definition for $p=\infty$ is quite permissive (but it would be the
same in $%
\mathbb{R}
^{m})$\ with a volume form, as any submanifold with dimension
$<$
dim(M) has a null measure. For instance if the norm of a section takes very
large values on a 1 dimensional submanifold of M, it will be ignored.

\begin{theorem}
$L^{p}\left(  M,\mu,E\right)  $ is a complex Banach vector space with the
norm: $\left\Vert U\right\Vert _{p}=\left(  \int_{M}\left\Vert U\right\Vert
^{p}\varpi_{0}\right)  ^{1/p}$
\end{theorem}

If the vector bundle is endowed with an inner product g which defines the norm
on each fiber E(x), meaning that E(x) is a Hilbert space, we can define the
scalar product $\left\langle U,V\right\rangle $ of two sections U,V over the
fiber bundle with a positive measure on M :

$\left\langle U,V\right\rangle _{E}=\int_{M}g\left(  x\right)  \left(
U\left(  x\right)  ,V\left(  x\right)  \right)  \mu$

\begin{theorem}
On a vector bundle endowed with an inner product, $L^{2}\left(  M,\mu
,E\right)  $ is a complex Hilbert vector space.
\end{theorem}

\paragraph{Spaces of integrable sections of the r jet prolongation of a vector
bundle\newline}

The r jet prolongation $J^{r}E$ of a vector bundle $E$ is a vector bundle. We
have a norm on V and so fiberwise on its r-jet prolongation (see above). If
the base M is endowed with a positive measure $\mu$ we can similarly define
spaces of integrable sections of $J^{r}E$.

For each s=0...r the set of functions :

$%
\mathcal{L}%
^{p}\left(  M,\mu,%
\mathcal{L}%
_{S}^{s}\left(
\mathbb{R}
^{\dim M},V\right)  \right)  =\left\{  z_{s}\in C_{0}\left(  M;%
\mathcal{L}%
_{S}^{s}\left(
\mathbb{R}
^{\dim M},V\right)  \right)  :\int_{M}\left\Vert z_{s}\left(  x\right)
\right\Vert _{s}^{p}\mu<\infty\right\}  $ and the norm :

$\left\Vert z_{s}\right\Vert _{p}=\left(  \int_{M}\left\Vert z_{s}\left(
x\right)  \right\Vert _{s}^{p}\mu\right)  ^{1/p}$

From there we define :

$%
\mathcal{L}%
^{p}\left(  M,\mu,J^{r}E\right)  =\left\{  Z\in\mathfrak{X}\left(
J^{r}E\right)  :\sum_{s=0}^{r}\int_{M}\left\Vert z_{s}\left(  x\right)
\right\Vert _{s}^{p}\mu<\infty\right\}  $

\begin{theorem}
The space $L^{p}\left(  M,\mu,J^{r}E\right)  $ is a Banach space with the norm
: $\left\Vert Z\right\Vert _{p}=\left(  \sum_{s=0}^{r}\int_{M}\left\Vert
z\left(  x\right)  \right\Vert _{s}^{p}\varpi_{0}\right)  ^{1/p}$
\end{theorem}

\bigskip

If the vector bundle is endowed with a scalar product g which defines the norm
on each fiber E(x), meaning that E(x) is a Hilbert space, we can define the
scalar product on the fibers of $J^{r}E\left(  x\right)  .$

Z(s) reads as a tensor $V\left(  x\right)  \otimes\odot_{s}%
\mathbb{R}
^{m}$

If $V_{1},V_{2}$\ are two finite dimensional real vector space endowed with
the bilinear symmetric forms $g_{1},g_{2}$ there is a unique bilinear
symmetric form G on $V_{1}\otimes V_{2}$ such that : $\forall u_{1}%
,u_{1}^{\prime}\in V_{1},u_{2},u_{2}^{\prime}\in V_{2}:G\left(  u_{1}\otimes
u_{1}^{\prime},u_{2}\otimes u_{2}^{\prime}\right)  =g_{1}\left(  u_{1}%
,u_{1}^{\prime}\right)  g\left(  u_{2},u_{2}^{\prime}\right)  $ and G is
denoted $g_{1}\otimes g_{2}.$(see Algebra - tensorial product of maps) . So if
we define G on $V\otimes\odot_{s}%
\mathbb{R}
^{m}$ such that :

$G_{s}\left(  Z_{\alpha_{1}...\alpha_{m}}^{i}e_{i}\left(  x\right)
\otimes\varepsilon^{\alpha_{1}}..\otimes\varepsilon^{\alpha_{m}},T_{\alpha
_{1}...\alpha_{m}}^{i}e_{i}\left(  x\right)  \otimes\varepsilon^{\alpha_{1}%
}..\otimes\varepsilon^{\alpha_{m}}\right)  $

$=\sum_{i,j=1}^{n}\sum_{\alpha_{1}..\alpha_{m}}\overline{Z}_{\alpha
_{1}...\alpha_{m}}^{i}T_{\alpha_{1}...\alpha_{m}}^{j}g\left(  x\right)
\left(  e_{i}\left(  x\right)  ,e_{j}\left(  x\right)  \right)  $

This is a sesquilinear hermitian form on $V\left(  x\right)  \otimes\odot_{s}%
\mathbb{R}
^{m}$ and we have : $\left\Vert Z_{s}\left(  x\right)  \right\Vert ^{2}%
=G_{s}\left(  x\right)  \left(  Z_{s},Z_{s}\right)  $

The scalar product is extended to M by : $\left\langle Z,T\right\rangle
=\sum_{s=0}^{r}\int_{M}G_{s}\left(  x\right)  \left(  Z_{s},T_{s}\right)  \mu$

\begin{theorem}
On a vector bundle endowed with an inner product $L^{2}\left(  M,\mu
,J^{r}E\right)  $ is a complex Hilbert vector space.\bigskip
\end{theorem}

Remarks : M is not involved because the maps $%
\mathcal{L}%
_{S}^{s}\left(
\mathbb{R}
^{\dim M},V\right)  $ are defined over $%
\mathbb{R}
^{\dim M}$ and not M : there is no need for a metric on M.

Most of the following results can be obviously translated from $%
\mathcal{L}%
^{p}\left(  M,\mu,%
\mathbb{C}
\right)  $ to $%
\mathcal{L}%
^{p}\left(  M,\mu,E\right)  $ by replacing E by M.

\subsubsection{Weak and strong convergence}

The strong topology on the $L^{p}$\ spaces is the normed topology. The weak
topology is driven by the topological dual $L^{p\prime}$ (which is $L^{q}$ for
$\frac{1}{p}+\frac{1}{q}=1$ for $p<\infty$ see below): a sequence $f_{n}$
converges weakly in $L^{p}$\ if : $\forall\lambda\in L^{p^{\prime}}%
:\lambda\left(  f_{n}\right)  \rightarrow\lambda\left(  f\right)  $.

\begin{theorem}
(Lieb p.68) If O is a measurable subset of $%
\mathbb{R}
^{m},$\ $f_{n}\in L^{p}\left(  O,S,dx,%
\mathbb{C}
\right)  $ a bounded sequence, then there is a subsequence $F_{k}$\ and
\ $f\in L^{p}\left(  O,S,dx,%
\mathbb{C}
\right)  $\ such that $F_{k}$\ \ converges weakly to f.
\end{theorem}

If $1\leq p<\infty$ , or $p=\infty$ and E is $\sigma-$finite :

\begin{theorem}
(Lieb p.56) If $f\in L^{p}\left(  E,S,\mu,%
\mathbb{C}
\right)  :\forall\lambda\in L^{p}\left(  E,S,\mu,%
\mathbb{C}
\right)  \prime:\lambda\left(  f\right)  =0$ then f=0.
\end{theorem}

\begin{theorem}
(Lieb p.57) If $f_{n}\in L^{p}\left(  E,S,\mu,%
\mathbb{C}
\right)  $ converges weakly to f, then $\lim\inf_{n\rightarrow\infty
}\left\Vert f_{n}\right\Vert _{p}=\left\Vert f\right\Vert _{p}.$
\end{theorem}

\begin{theorem}
(Lieb p.57) If $f_{n}\in L^{p}\left(  E,S,\mu,%
\mathbb{C}
\right)  $ is such that $\forall\lambda\in L^{p}\left(  E,S,\mu,%
\mathbb{C}
\right)  ,$\ $\left\vert \lambda\left(  f_{n}\right)  \right\vert $ is
bounded, then $\exists C>0:\left\Vert f_{n}\right\Vert _{p}<C.$
\end{theorem}

If $1\leq p<\infty:$

\begin{theorem}
(Lieb p.57) If $f_{n}\in L^{p}\left(  E,S,\mu,%
\mathbb{C}
\right)  $ converges weakly to f then there are $c_{nj}\in\left[  0,1\right]
,\sum_{j=1}^{n}c_{nj}=1$ such that : $\varphi_{n}=\sum_{j=1}^{n}c_{nj}f_{j}$
converges strongly to f.
\end{theorem}

\subsubsection{Inequalities}%

\begin{equation}
\left(  \left\Vert f\right\Vert _{r}\right)  ^{\dfrac{1}{p}-\dfrac{1}{q}}%
\leq\left(  \left\Vert f\right\Vert _{p}\right)  ^{\dfrac{1}{r}-\dfrac{1}{p}%
}\left(  \left\Vert f\right\Vert _{q}\right)  ^{\dfrac{1}{q}-\dfrac{1}{r}}%
\end{equation}

\bigskip

\begin{theorem}
H\"{o}lder's inequality (Lieb p.45) \ For $1\leq p\leq\infty,\frac{1}{p}%
+\frac{1}{q}=1$

If $f\in%
\mathcal{L}%
^{p}\left(  E,S,\mu,%
\mathbb{C}
\right)  ,g\in%
\mathcal{L}%
^{q}\left(  E,S,\mu,%
\mathbb{C}
\right)  $ then :

i) $f\times g\in%
\mathcal{L}%
^{1}\left(  E,S,\mu,%
\mathbb{C}
\right)  $ and $\left\Vert fg\right\Vert _{1}\leq\int_{E}\left\vert
f\right\vert \left\vert g\right\vert \mu\leq\left\Vert f\right\Vert
_{p}\left\Vert g\right\Vert _{q}$

ii) $\left\Vert fg\right\Vert _{1}=\int_{E}\left\vert f\right\vert \left\vert
g\right\vert \mu$ iff $\exists\theta\in%
\mathbb{R}
,\theta=Ct:f\left(  x\right)  g\left(  x\right)  =e^{i\theta}\left\vert
f\left(  x\right)  \right\vert \left\vert g\left(  x\right)  \right\vert $
almost everywhere

iii) if $f\neq0,\int_{E}\left\vert f\right\vert \left\vert g\right\vert
\mu=\left\Vert f\right\Vert _{p}\left\Vert g\right\Vert _{q}$ iff
$\exists\lambda\in%
\mathbb{R}
,\lambda=Ct$ such that :

if $1<p<\infty:\left\vert g\left(  x\right)  \right\vert =\lambda\left\vert
f\left(  x\right)  \right\vert ^{p-1}$ almost everywhere

if $p=1:\left\vert g\left(  x\right)  \right\vert \leq\lambda$ almost
everywhere and $\left\vert g\left(  x\right)  \right\vert =\lambda$ when
f(x)$\neq0$

if $p=\infty:\left\vert f\left(  x\right)  \right\vert \leq\lambda$ almost
everywhere and $\left\vert f\left(  x\right)  \right\vert =\lambda$ when
g(x)$\neq0$
\end{theorem}

Thus : p=q=2:%

\begin{equation}
f,g\in%
\mathcal{L}%
^{2}\left(  E,S,\mu,%
\mathbb{C}
\right)  :\left\Vert fg\right\Vert _{1}\leq\left\Vert f\right\Vert
_{2}\left\Vert g\right\Vert _{2}%
\end{equation}

\begin{theorem}
The map:

$\lambda:L^{\infty}\left(  E,S,\mu,%
\mathbb{C}
\right)  \rightarrow%
\mathcal{L}%
\left(  L^{2}\left(  E,S,\mu,%
\mathbb{C}
\right)  ;L^{2}\left(  E,S,\mu,%
\mathbb{C}
\right)  \right)  ::\lambda\left(  f\right)  g=fg$

is a continuous operator $\left\Vert \lambda\left(  f\right)  \right\Vert
_{2}\leq\left\Vert f\right\Vert _{\infty}$ and an isomorphism of C*-algebra.
If $\mu$ is $\sigma-$finite then $\left\Vert \lambda\left(  f\right)
\right\Vert _{2}=\left\Vert f\right\Vert _{\infty}$
\end{theorem}

\bigskip

\begin{theorem}
Hanner's inequality (Lieb p.40) For$\ f,g\in%
\mathcal{L}%
^{p}\left(  E,S,\mu,%
\mathbb{C}
\right)  $

$\forall1\leq p\leq\infty:\left\Vert f+g\right\Vert _{p}\leq\left\Vert
f\right\Vert _{p}+\left\Vert g\right\Vert _{p}$ and if $1<p<\infty$\ the
equality holds iff $\exists\lambda\geq0:g=\lambda f$

$1\leq p\leq2:$

$\left\Vert f+g\right\Vert _{p}^{p}+\left\Vert f-g\right\Vert _{p}^{p}%
\leq\left(  \left\Vert f\right\Vert _{p}+\left\Vert g\right\Vert _{p}\right)
^{p}+\left(  \left\Vert f\right\Vert _{p}-\left\Vert g\right\Vert _{p}\right)
^{p}$

$\left(  \left\Vert f+g\right\Vert _{p}+\left\Vert f-g\right\Vert _{p}\right)
^{p}+\left\vert \left\Vert f+g\right\Vert _{p}-\left\Vert f-g\right\Vert
_{p}\right\vert ^{p}\leq2^{p}\left(  \left\Vert f\right\Vert _{p}%
^{p}+\left\Vert g\right\Vert _{p}^{p}\right)  $

$2\leq p<\infty:$

$\left\Vert f+g\right\Vert _{p}^{p}+\left\Vert f-g\right\Vert _{p}^{p}%
\geq\left(  \left\Vert f\right\Vert _{p}+\left\Vert g\right\Vert _{p}\right)
^{p}+\left(  \left\Vert f\right\Vert _{p}-\left\Vert g\right\Vert _{p}\right)
^{p}$

$\left(  \left\Vert f+g\right\Vert _{p}+\left\Vert f-g\right\Vert _{p}\right)
^{p}+\left\vert \left\Vert f+g\right\Vert _{p}-\left\Vert f-g\right\Vert
_{p}\right\vert ^{p}\geq2^{p}\left(  \left\Vert f\right\Vert _{p}%
^{p}+\left\Vert g\right\Vert _{p}^{p}\right)  $
\end{theorem}

\bigskip

\begin{theorem}
(Lieb p.51) For $1<p<\infty,f,g\in%
\mathcal{L}%
^{p}\left(  E,S,\mu,%
\mathbb{C}
\right)  ,$ the function \ : $\phi:$ $%
\mathbb{R}
\rightarrow%
\mathbb{R}
::\phi\left(  t\right)  =\int_{E}\left\vert f+tg\right\vert ^{p}\mu$ is
differentiable and

$\frac{d\phi}{dt}|_{t=0}=\frac{p}{2}\int_{E}\left\vert f\right\vert
^{p-2}\left(  \overline{f}g+f\overline{g}\right)  \mu$
\end{theorem}

\bigskip

\begin{theorem}
(Lieb p.98) There is a fixed countable family $\left(  \varphi_{i}\right)
_{i\in I},\varphi\in C\left(
\mathbb{R}
^{m};%
\mathbb{C}
\right)  $ such that for any open subset O in $%
\mathbb{R}
^{m},1\leq p\leq\infty,f\in L^{p}\left(  O,dx,%
\mathbb{C}
\right)  ,\varepsilon>0:\exists i\in I:\left\Vert f-\varphi_{i}\right\Vert
_{p}<\varepsilon$
\end{theorem}

\bigskip

\begin{theorem}
Young's inequalities (Lieb p.98): For $p,q,r>1$ such that

$\frac{1}{p}+\frac{1}{q}+\frac{1}{r}=2,$

for any \ \ $f\in L^{p}\left(
\mathbb{R}
^{m},dx,%
\mathbb{C}
\right)  ,g\in L^{q}\left(
\mathbb{R}
^{m},dx,%
\mathbb{C}
\right)  ,h\in L^{r}\left(
\mathbb{R}
^{m},dx,%
\mathbb{C}
\right)  $

$\left\vert \int_{%
\mathbb{R}
^{m}}\int_{%
\mathbb{R}
^{m}}f\left(  x\right)  g\left(  x-y\right)  h\left(  y\right)
dxdy\right\vert \leq C\left\Vert f\right\Vert _{p}\left\Vert g\right\Vert
_{q}\left\Vert h\right\Vert _{r}$

with $C=\left(  \sqrt{\frac{p^{1/p}}{p^{^{\prime}1/p^{\prime}}}}\sqrt
{\frac{q^{1/q}}{q^{^{\prime}1/q^{\prime}}}}\sqrt{\frac{r^{1/r}}{r^{^{\prime
}1/r^{\prime}}}}\right)  ^{m}$

$\frac{1}{p}+\frac{1}{p^{\prime}}=1;\frac{1}{q}+\frac{1}{q^{\prime}}%
=1;\frac{1}{r}+\frac{1}{r^{\prime}}=1$
\end{theorem}

The equality occurs iff each function is gaussian :

$f\left(  x\right)  =A\exp\left(  -p^{\prime}\left(  x-a,J\left(  x-a\right)
\right)  +ik.x\right)  $

$g\left(  x\right)  =B\exp\left(  -q^{\prime}\left(  x-b,J\left(  x-b\right)
\right)  +ik.x\right)  $

$h\left(  x\right)  =C\exp\left(  -r^{\prime}\left(  x-a,J\left(  x-a\right)
\right)  +ik.x\right)  $

with $A,B,C\in%
\mathbb{C}
,J$ a real symmetric positive definite matrix,$a=b+c\in%
\mathbb{R}
^{m}$

\bigskip

\begin{theorem}
Hardy-Litllewood-Sobolev inequality (Lieb p.106) For $p,r>1,$ $0<\lambda
<m$\ such that $\frac{1}{p}+\frac{\lambda}{m}+\frac{1}{r}=2,$

there is a constant $C\left(  p,r,\lambda\right)  $ such that :

$\forall f\in L^{p}\left(
\mathbb{R}
^{m},dx,%
\mathbb{C}
\right)  ,\forall h\in L^{r}\left(
\mathbb{R}
^{m},dx,%
\mathbb{C}
\right)  $

$\left\vert \int_{%
\mathbb{R}
^{m}}\int_{%
\mathbb{R}
^{m}}f\left(  x\right)  \left\Vert \left(  x-y\right)  \right\Vert ^{-\lambda
}h\left(  y\right)  dxdy\right\vert \leq C\left(  p,r,\lambda\right)
\left\Vert f\right\Vert _{p}\left\Vert h\right\Vert _{r}$
\end{theorem}

\subsubsection{Density theorems}

\begin{theorem}
(Neeb p.43) If E is a topological Hausdorff locally compact space and $\mu$ a
Radon measure, then the set $C_{0c}\left(  E;%
\mathbb{C}
\right)  $ is dense in $L^{2}\left(  E,\mu,%
\mathbb{C}
\right)  .$
\end{theorem}

\bigskip

\begin{theorem}
(Zuily p.14) If $O$ is an open subset of $%
\mathbb{R}
^{m},\mu$ the Lebesgue measure, then

the subset $L_{c}^{p}\left(  O,\mu,%
\mathbb{C}
\right)  $\ of $L^{p}\left(  O,\mu,%
\mathbb{C}
\right)  $ of functions with compact support is dense in $L^{p}\left(  O,\mu,%
\mathbb{C}
\right)  $ for $1\leq p<\infty$

the subset $C_{\infty c}\left(  O,\mu,%
\mathbb{C}
\right)  $\ of smooth functions with compact support is dense in $L^{p}\left(
O,\mu,%
\mathbb{C}
\right)  $ for $1\leq p<\infty$
\end{theorem}

\bigskip

\begin{theorem}
If $E$ is a Hausdorff locally compact space with its Borel algebra S, P is a
Borel, inner regular, probability, then any function $f\in L^{p}\left(  E,S,P,%
\mathbb{C}
\right)  ,1\leq p<\infty,$ is almost everywhere equal to a function in
$C_{b}\left(  E;%
\mathbb{C}
\right)  $ and there is a sequence $\left(  f_{n}\right)  \in C_{c}\left(  E;%
\mathbb{C}
\right)  ^{%
\mathbb{N}
}$ which converges to f in $L^{p}\left(  E,S,P,%
\mathbb{C}
\right)  $ almost everywhere.
\end{theorem}

\subsubsection{Integral operators}

\begin{theorem}
(Minkowski's inequality) (Lieb p.47) Let $(E,S,\mu),\left(  F,S^{\prime}%
,\nu\right)  $ $\sigma-$finite measured spaces with positive measure,

$f:E\times F\rightarrow%
\mathbb{R}
_{+}$ a measurable function,

$1\leq p<\infty:\int_{F}\left(  \int_{E}f\left(  x,y\right)  ^{p}\mu\left(
x\right)  \right)  ^{1/p}\nu\left(  y\right)  \geq\left(  \int_{E}\left(
\int_{F}f\left(  x,y\right)  \nu\left(  y\right)  \right)  ^{p}\mu\left(
x\right)  \right)  ^{1/p}$

and if one term is $<\infty$ so is the other. If the equality holds and p%
$>$%
1 then there are measurable functions $u\in C\left(  E;%
\mathbb{R}
_{+}\right)  ,v\in C\left(  F;%
\mathbb{R}
_{+}\right)  $ such that $f\left(  x,y\right)  =u\left(  x\right)  v\left(
y\right)  $ almost everywhere
\end{theorem}

\begin{theorem}
(Taylor 2 p.16) Let $(E,S,\mu)$ a measured space with\ $\mu$\ a positive
measure, $k:E\times E\rightarrow%
\mathbb{C}
$ a measurable function on E$\times$E such that : $\exists C_{1},C_{2}\in%
\mathbb{R}
:\forall x,y\in E:\int_{E}\left\vert k\left(  t,y\right)  \right\vert
\mu\left(  t\right)  \leq C_{1},\int_{E}\left\vert k\left(  x,t\right)
\right\vert \mu\left(  t\right)  \leq C_{2}$

then $\forall\infty\geq p\geq1,$ $K:L^{p}\left(  E,S,\mu,%
\mathbb{C}
\right)  \rightarrow L^{q}\left(  E,S,\mu,%
\mathbb{C}
\right)  ::K\left(  f\right)  \left(  x\right)  =\int_{E}k\left(  x,y\right)
f\left(  y\right)  \mu\left(  y\right)  $ is a linear continuous operator
called the \textbf{integral kernel} of K
\end{theorem}

K is an \textbf{integral operator} : $K\in%
\mathcal{L}%
\left(  L^{p}\left(  E,S,\mu,%
\mathbb{C}
\right)  ;L^{q}\left(  E,S,\mu,%
\mathbb{C}
\right)  \right)  $ with : $\frac{1}{p}+\frac{1}{q}=1,\left\Vert K\right\Vert
\leq C_{1}^{1/p}C_{2}^{1/q}$

The transpose $K^{t}$ of K is the operator with integral kernel $k^{t}%
(x,y)=k(y,x)$.

If p=2 the adjoint K* of K is the integral operator with kernel $k^{\ast
}(x,y)=\overline{k(y,x)}$

\begin{theorem}
(Taylor 1 p.500) Let $(E_{1},S_{1},\mu_{1}),(E_{2},S_{2},\mu_{2})\ $\ be
measured spaces with positive measures, and $T\in%
\mathcal{L}%
\left(  L^{2}(E_{1},S_{1},\mu_{1},%
\mathbb{C}
);L^{2}(E_{2},S_{2},\mu_{2},%
\mathbb{C}
)\right)  $ be a Hilbert-Schmidt operator, then there is a function $K\in
L^{2}\left(  E_{1}\times E_{2},\mu_{1}\otimes\mu_{2},%
\mathbb{C}
\right)  $ such that :

$\left\langle Tf,g\right\rangle =\int\int K\left(  x_{1},x_{2}\right)
\overline{f\left(  x_{1}\right)  }g\left(  x_{2}\right)  \mu_{1}\left(
x_{1}\right)  \mu_{2}\left(  x_{2}\right)  $

and we have $\left\Vert T\right\Vert _{HS}=\left\Vert K\right\Vert _{L^{2}}$
\end{theorem}

\begin{theorem}
(Taylor 1 p.500) Let $K_{1},K_{2}$ two Hilbert-Schmidt, integral operators on
$L^{2}\left(  E,S,\mu,%
\mathbb{C}
\right)  $ with kernels $k_{1},k_{2}.$ Then the product $K_{1}\circ K_{2}$ is
an Hilbert-Schmidt integral operator with kernel : $k\left(  x,y\right)
=\int_{E}k_{1}\left(  x,t\right)  k_{2}\left(  t,y\right)  \mu\left(
t\right)  $
\end{theorem}

\subsubsection{Convolution}

Convolution is a map on functions defined on a locally compact topological
unimodular group G (see Lie groups - integration). It is here implemented on
the abelian group $\left(
\mathbb{R}
^{m},+\right)  $ endowed with the Lebesgue measure.

\paragraph{Definition\newline}

\begin{definition}
The \textbf{convolution} is the map :

$\ast:L^{1}\left(
\mathbb{R}
^{m},dx,%
\mathbb{C}
\right)  \times L^{1}\left(
\mathbb{R}
^{m},dx,%
\mathbb{C}
\right)  \rightarrow L^{1}\left(
\mathbb{R}
^{m},dx,%
\mathbb{C}
\right)  ::$%

\begin{equation}
\left(  f\ast g\right)  \left(  x\right)  =\int_{%
\mathbb{R}
^{m}}f\left(  y\right)  g\left(  x-y\right)  dy=\int_{%
\mathbb{R}
^{m}}f\left(  x-y\right)  g\left(  x\right)  dy
\end{equation}

\end{definition}

Whenever a function is defined on an open $O\subset%
\mathbb{R}
^{m}$ it can be extended by taking $\widetilde{f}\left(  x\right)  =f\left(
x\right)  ,x\in O,\widetilde{f}\left(  x\right)  =0,x\notin O$ so that many
results still hold for functions defined in O. Convolution is well defined for
other spaces of functions.

\begin{theorem}
H\"{o}rmander : The convolution $f\ast g$ exists if $f\in C_{c}\left(
\mathbb{R}
^{m};%
\mathbb{C}
\right)  $ and $g\in L_{loc}^{1}\left(
\mathbb{R}
^{m},dx,%
\mathbb{C}
\right)  $, then f*g is continuous
\end{theorem}

\begin{theorem}
The convolution $f\ast g$ exists if $f,g\in S\left(
\mathbb{R}
^{m}\right)  $ (Schwartz functions) then $f\ast g\in S\left(
\mathbb{R}
^{m}\right)  $
\end{theorem}

\begin{theorem}
The convolution $f\ast g$ exists if $f\in L^{p}\left(
\mathbb{R}
^{m},dx,%
\mathbb{C}
\right)  ,g\in L^{q}\left(
\mathbb{R}
^{m},dx,%
\mathbb{C}
\right)  ,1\leq p,q\leq\infty$ then the convolution is a continuous bilinear
map :

$\ast\in%
\mathcal{L}%
^{2}\left(  L^{p}\left(
\mathbb{R}
^{m},dx,%
\mathbb{C}
\right)  \times L^{q}\left(
\mathbb{R}
^{m},dx,%
\mathbb{C}
\right)  ;L^{r}\left(
\mathbb{R}
^{m},dx,%
\mathbb{C}
\right)  \right)  $ with $\frac{1}{p}+\frac{1}{q}=\frac{1}{r}+1$ and
$\left\Vert f\ast g\right\Vert _{r}\leq\left\Vert f\right\Vert _{p}\left\Vert
g\right\Vert _{q}$

if $\frac{1}{p}+\frac{1}{q}=1\ $then $f\ast g\in C_{0\nu}\left(
\mathbb{R}
^{m},%
\mathbb{C}
\right)  $ \ (Lieb p.70)
\end{theorem}

This is a consequence of the Young's inequality

\begin{definition}
On the space of functions $C\left(
\mathbb{R}
^{m};%
\mathbb{C}
\right)  $\ 

the \textbf{involution} is : $f^{\ast}\left(  x\right)  =\overline{f\left(
-x\right)  }$

the \textbf{translation} is : $\tau_{a}f\left(  x\right)  =f\left(
x-a\right)  $ for $a\in%
\mathbb{R}
^{m}$
\end{definition}

So the left action is $\Lambda\left(  a\right)  f\left(  x\right)  =\tau
_{a}f\left(  x\right)  =f\left(  x-a\right)  $ and the right action is :
$P\left(  a\right)  f\left(  x\right)  =\tau_{-a}f\left(  x\right)  =f\left(
x+a\right)  $

\paragraph{Properties\newline}

\begin{theorem}
Whenever the convolution is defined on a vector space V of functions, it makes
V a commutative complex algebra without identity, and convolution commutes
with translation
\end{theorem}

$f\ast g=g\ast f$

$\left(  f\ast g\right)  \ast h=f\ast\left(  g\ast h\right)  $

$f\ast\left(  ag+bh\right)  =af\ast g+bf\ast h$

$\tau_{a}\left(  f\ast g\right)  =f\ast\tau_{a}g=\tau_{a}f\ast g$ with
$\tau_{a}:V\rightarrow V::\left(  \tau_{a}f\right)  \left(  x\right)
=f\left(  x-a\right)  $

\begin{theorem}
Whenever the convolution of f,g is defined:

Supp$\left(  f\ast g\right)  \subset Supp(f)\cap Supp\left(  g\right)  $

$\left(  f\ast g\right)  ^{\ast}=f^{\ast}\ast g^{\ast}$

If f,g are integrable, then : $\int_{%
\mathbb{R}
^{m}}\left(  f\ast g\right)  dx=\left(  \int_{%
\mathbb{R}
^{m}}f\left(  x\right)  dy\right)  \left(  \int_{%
\mathbb{R}
^{m}}g\left(  x\right)  dx\right)  $
\end{theorem}

\begin{theorem}
With convolution as internal operation and involution, $L^{1}\left(
\mathbb{R}
^{m},dx,%
\mathbb{C}
\right)  $ is a commutative Banach *-algebra and the involution, right and
left actions are isometries.

If $f,g\in L^{1}\left(
\mathbb{R}
^{m},dx,%
\mathbb{C}
\right)  $ and f or g has a derivative which is in $L^{1}\left(
\mathbb{R}
^{m},dx,%
\mathbb{C}
\right)  $ then $f\ast g$ is differentiable and :

$\frac{\partial}{\partial x_{\alpha}}\left(  f\ast g\right)  =\left(
\frac{\partial}{\partial x_{\alpha}}f\right)  \ast g=f\ast\frac{\partial
}{\partial x_{\alpha}}g$
\end{theorem}

$\forall f,g\in L^{1}\left(
\mathbb{R}
^{m},dx,%
\mathbb{C}
\right)  :\left\Vert f\ast g\right\Vert _{1}\leq\left\Vert f\right\Vert
_{1}\left\Vert g\right\Vert _{1},\left\Vert f^{\ast}\right\Vert _{1}%
=\left\Vert f\right\Vert _{1},\left\Vert \tau_{a}f\right\Vert _{1}=\left\Vert
f\right\Vert _{1}$

\paragraph{Approximation of the identity\newline}

The absence of an identity element is compensated with an approximation of the
identity defined as follows :

\begin{definition}
An \textbf{approximation of the identity} in the Banach *-algebra $\left(
L^{1}\left(
\mathbb{R}
^{m},dx,%
\mathbb{C}
\right)  ,\ast\right)  $ is a family of functions $\left(  \rho_{\varepsilon
}\right)  _{\varepsilon\in%
\mathbb{R}
}$ with $\rho_{\varepsilon}=\varepsilon^{-m}\rho\left(  \frac{x}{\varepsilon
}\right)  \in C_{\infty c}\left(
\mathbb{R}
^{m};%
\mathbb{C}
\right)  $ where $\rho\in C_{\infty c}\left(
\mathbb{R}
^{m};%
\mathbb{C}
\right)  $ such that :

Sup($\rho)\subset B\left(  0,1\right)  ,\rho\geq0,\int_{%
\mathbb{R}
^{m}}\rho\left(  x\right)  dx=1$
\end{definition}

\begin{theorem}
(Lieb p.76) The family $\rho_{\varepsilon}$ has the properties :

$\forall f\in C_{rc}\left(
\mathbb{R}
^{m};%
\mathbb{C}
\right)  ,\forall\alpha=\left(  \alpha_{1},...\alpha_{s}\right)  ,s\leq
r:D_{\left(  \alpha\right)  }\left(  \rho_{n}\ast f\right)  \rightarrow
D_{\left(  \alpha\right)  }f$ uniformly when $\varepsilon\rightarrow0$

$\forall f\in L_{c}^{p}\left(
\mathbb{R}
^{m},dx,%
\mathbb{C}
\right)  ,1\leq p<\infty,\rho_{\varepsilon}\ast f\rightarrow f$ when
$\varepsilon\rightarrow0$
\end{theorem}

These functions are used to approximate any function $f\in L^{1}\left(
\mathbb{R}
^{m},dx,%
\mathbb{C}
\right)  $ by a sequence of functions $\in C_{\infty c}\left(
\mathbb{R}
^{m};%
\mathbb{C}
\right)  $

\bigskip

\subsection{Spaces of differentiable maps}

\label{Space of differentiable maps}

The big difference is that these spaces are usually not normable. This can be
easily understood : the more regular the functions, the more difficult it is
to get the same properties for limits of sequences of these functions.
Differentiability is only defined for functions over manifolds, which can be
open subsets of $%
\mathbb{R}
^{m}.$Notice that we address only differentiability with respect to real
variables, over real manifolds. Indeed complex differentiable functions are
holomorphic, and thus C-analytic, so for most of the purposes of functional
analysis, it is this feature which is the most useful.

We start by the spaces of differentiable sections of a vector bundle, as the
case of functions follows.

\begin{notation}
$D_{\left(  \alpha\right)  }=D_{\alpha_{1}...\alpha_{s}}=\dfrac{\partial
}{\partial\xi^{\alpha_{1}}}\dfrac{\partial}{\partial\xi^{\alpha_{2}}}%
...\dfrac{\partial}{\partial\xi^{\alpha_{s}}}$ where the $\alpha_{k}%
=1...m$\ can be identical
\end{notation}

\subsubsection{Spaces of differentiable sections of a vector bundle}

\begin{theorem}
The spaces $\mathfrak{X}_{r}\left(  E\right)  $ of r $(1\leq r\leq\infty)$
continuously differentiable sections on a vector bundle E are Fr\'{e}chet spaces.
\end{theorem}

\begin{proof}
Let $E\left(  M,V,\pi\right)  $ be a complex vector bundle with a finite atlas
$\left(  O_{a},\varphi_{a}\right)  _{a=1}^{N},$ such that $\left(
F,O_{a},\psi_{a}\right)  _{a=1}^{N}$ is an atlas of the m dimensional manifold M.

$S\in\mathfrak{X}_{r}\left(  E\right)  $ is defined by a family of r
differentiable maps : $u_{a}\in C_{r}\left(  O_{a};V\right)  ::S\left(
x\right)  =\varphi_{a}\left(  x,u_{a}\left(  x\right)  \right)  $. Let be
$\Omega_{a}=\psi_{a}\left(  O_{a}\right)  \subset%
\mathbb{R}
^{m}$. $F_{a}=u_{a}\circ\psi_{a}^{-1}\in C_{r}\left(  \Omega_{a};V\right)
$\ and $D_{\alpha_{1}...\alpha_{s}}F_{a}(\xi)$ is a continuous multilinear map
$%
\mathcal{L}%
^{s}\left(
\mathbb{R}
^{m};V\right)  $ with a finite norm $\left\Vert D_{\alpha_{1}...\alpha_{s}%
}F_{a}(\xi)\right\Vert $

For each set $\Omega_{a}$ we define the sequence of sets :

$K_{p}=\left\{  \xi\in\Omega_{a},\left\Vert \xi\right\Vert \leq p\right\}
\cap\left\{  \xi\in\Omega_{a},d\left(  \xi,\Omega_{a}^{c}\right)
\geq1/p\right\}  $

It is easily shown (Zuily p.2) that : each $K_{p}$ is compact, $K_{p}%
\subset\overset{\circ}{K}_{p+1},\Omega_{a}=\cup_{p=1}^{\infty}K_{p}=\cup
_{p=2}^{\infty}\overset{\circ}{K}_{p},$ for each compact $K\subset\Omega_{a}$
there is some p such that $K\subset\Omega_{p}$

The maps $p_{n}$ :

for $1\leq r<\infty:p_{n}\left(  S\right)  =\max_{a=..n}\sum_{s=1}^{n}%
\sum_{\alpha_{1}...\alpha_{s}}\sup_{\xi\in K_{n}}\left\Vert D_{\alpha
_{1}...\alpha_{s}}F_{a}(\xi)\right\Vert $

for $r=\infty:p_{n}\left(  S\right)  =\max_{a=..n}\sum_{s=1}^{n}\sum
_{\alpha_{1}...\alpha_{n}}\sup_{\xi\in K_{n}}\left\Vert D_{\alpha_{1}%
...\alpha_{s}}F_{a}(\xi)\right\Vert $

define a countable family of semi-norms on $\mathfrak{X}_{r}\left(  E\right)
$.

With the topology induced by these semi norms a sequence $S_{n}\in
\mathfrak{X}_{r}\left(  E\right)  $ converges to $S\in\mathfrak{X}_{r}\left(
E\right)  $ iff S converges uniformly on any compact. The space of continuous,
compactly supported sections is a Banach space, so any Cauchy sequence on
$K_{p}$ converges, and converges on any compact. Thus $\mathfrak{X}_{r}\left(
E\right)  $ is complete with these semi norms.
\end{proof}

A subset A of $\mathfrak{X}_{r}\left(  E\right)  $ is bounded if : $\forall
S\in A,\forall n>1,\exists C_{n}:p_{n}\left(  S\right)  \leq C_{n}$

\bigskip

\begin{theorem}
The space $\mathfrak{X}_{rc}\left(  E\right)  $ of r differentiable, compactly
supported, sections of a vector bundle is a Fr\'{e}chet space
\end{theorem}

\begin{proof}
This is a closed subspace of $\mathfrak{X}_{r}\left(  E\right)  .$
\end{proof}

\subsubsection{Space of sections of the r jet prolongation of a vector bundle}

\begin{theorem}
The space $\mathfrak{X}\left(  J^{r}E\right)  $ of sections of the r jet
prolongation J$^{r}$E of a vector bundle is a Fr\'{e}chet space.
\end{theorem}

\begin{proof}
If E is a vector bundle on a m dimensional real manifold M and dimV=n then
$J^{r}E$\ \ is a vector bundle $J^{r}E\left(  E,J_{0}^{r}\left(
\mathbb{R}
^{m},V\right)  ,\pi_{0}^{r}\right)  $ endowed with norms fiberwise (see
above). A section $j^{r}z$ over $J^{r}E$ is a map $M\rightarrow J^{r}E$ with
coordinates :

$\left(  \xi^{\alpha},\eta^{i}\left(  \xi\right)  ,\eta_{\alpha_{1}%
..\alpha_{s}}^{i}\left(  \xi\right)  ,s=1...r,1\leq\alpha_{k}\leq\alpha
_{k+1}\leq m,i=1..n\right)  $

The previous semi-norms read :

for $1\leq s\leq r:p_{n}\left(  j^{r}z\right)  =\max_{a=..n}\sum_{s=1}^{n}%
\sum_{\alpha_{1}...\alpha_{s}}\sup_{\xi\in K_{n}}\left\Vert \eta_{\alpha
_{1}..\alpha_{s}}^{i}\left(  \xi\right)  \right\Vert $
\end{proof}

A section U on E gives rise to a section on $J^{r}E:$ $j^{r}U\left(  x\right)
=j_{x}^{r}U$ \ and with this definition : $p_{n}\left(  j^{r}U\right)
_{\mathfrak{X}\left(  J^{r}E\right)  }=p_{n}\left(  U\right)  _{\mathfrak{X}%
_{r}\left(  E\right)  }$

\subsubsection{Spaces of differentiable functions on a manifold}

\begin{definition}
$C_{r}\left(  M;%
\mathbb{C}
\right)  $ the space of r continuously differentiable functions
$f:M\rightarrow%
\mathbb{C}
$

$C_{rc}\left(  M;%
\mathbb{C}
\right)  $ the space of r continuously differentiable functions
$f:M\rightarrow%
\mathbb{C}
$ with compact support
\end{definition}

if $r=\infty$ the functions are smooth

\begin{theorem}
The spaces $C_{r}\left(  M;%
\mathbb{C}
\right)  ,C_{rc}\left(  M;%
\mathbb{C}
\right)  $ are Fr\'{e}chet spaces
\end{theorem}

\bigskip

If M is a \textit{compact} finite dimensional Hausdorff smooth manifold we
have simpler semi-norms:

\begin{theorem}
If M is a compact manifold, then :

i) the space $C_{\infty}\left(  M;%
\mathbb{C}
\right)  $ of smooth functions on M is a Fr\'{e}chet space with the family of
seminorms :

for $1\leq r<\infty:p_{r}\left(  f\right)  =\sum_{p=0}^{r}\left\vert
f^{\left(  p\right)  }\left(  p\right)  \left(  u_{1},...,u_{p}\right)
\right\vert _{\left\Vert u_{k}\right\Vert \leq1}$

ii) the space $C_{r}\left(  M;%
\mathbb{C}
\right)  $ of r differentiable functions on M is a Banach space with the norm :

$\left\Vert f\right\Vert =\sum_{p=0}^{r}\left\vert f^{\left(  p\right)
}\left(  p\right)  \left(  u_{1},...,u_{p}\right)  \right\vert _{\left\Vert
u_{k}\right\Vert \leq1}$

where the $u_{k}$ are vectors fields whose norm is measured with respect to a
riemannian metric
\end{theorem}

(which always exist with our assumptions).

Notice that we have no Banach spaces structures for differentiable functions
over a manifold if it is not compact.

\bigskip

\begin{theorem}
(Zuily p.2) A subset A of $C_{\infty}\left(  O;%
\mathbb{C}
\right)  $ is compact iff it is closed and bounded with the semi-norms
\end{theorem}

this is untrue if $r<\infty$.

\subsubsection{Spaces of compactly supported, differentiable functions on an
open of $%
\mathbb{R}
^{m}$}

There is an increasing sequence of relatively compacts open $O_{n}$ which
covers any open O. Thus with the notation above $C_{rc}\left(  O;%
\mathbb{C}
\right)  =\sqcup_{i=1}^{\infty}C_{r}\left(  O_{n};%
\mathbb{C}
\right)  .$ Each of the $C_{r}\left(  O_{n};%
\mathbb{C}
\right)  $ endowed with the seminorms is a Fr\'{e}chet space (and a Banach),
$C_{r}\left(  O_{n};%
\mathbb{C}
\right)  \subset C_{r}\left(  O_{n+1};%
\mathbb{C}
\right)  $ and the latter induces in the former the same topology. Which
entails :

\begin{theorem}
(Zuily p.10) For the space $C_{rc}\left(  O;%
\mathbb{C}
\right)  $ of compactly supported, r continuously differentiable functions on
an open subset of $%
\mathbb{R}
^{m}$

i) there is a unique topology on $C_{rc}\left(  O;%
\mathbb{C}
\right)  $ which induces on each $C_{r}\left(  O_{n};%
\mathbb{C}
\right)  $ the topology given by the seminorms

ii) A sequence $\left(  f_{n}\right)  \in C_{rc}\left(  O;%
\mathbb{C}
\right)  ^{%
\mathbb{N}
}$ converges iff : $\exists N:\forall n:Supp\left(  f_{n}\right)  \subset
O_{N},$ and $\left(  f_{n}\right)  $ converges in $C_{rc}\left(  O_{N};%
\mathbb{C}
\right)  $

iii) A linear functional on $C_{rc}\left(  O;%
\mathbb{C}
\right)  $ is continuous iff it is continuous on each $C_{rc}\left(  O_{n};%
\mathbb{C}
\right)  $

iv) A subset A of $C_{rc}\left(  O;%
\mathbb{C}
\right)  $ is bounded iff there is N such that $A\subset C_{rc}\left(  O_{N};%
\mathbb{C}
\right)  $ and A is bounded on this subspace.

v) For $0\leq r\leq\infty$ $C_{rc}\left(  O;%
\mathbb{C}
\right)  $ is dense in $C_{r}\left(  O;%
\mathbb{C}
\right)  $
\end{theorem}

So for most of the purposes it is equivalent to consider $C_{rc}\left(  O;%
\mathbb{C}
\right)  $\ or the families $C_{rc}\left(  K;%
\mathbb{C}
\right)  $\ where K is a relatively compact open in O.

\bigskip

\begin{theorem}
(Zuily p.18) Any function of $C_{rc}\left(  O_{1}\times O_{2};%
\mathbb{C}
\right)  $ is the limit of a sequence in $C_{rc}\left(  O_{1};%
\mathbb{C}
\right)  \otimes C_{rc}\left(  O_{2};%
\mathbb{C}
\right)  .$
\end{theorem}

\bigskip

Functions $f\in C_{\infty c}\left(
\mathbb{R}
^{m};%
\mathbb{C}
\right)  $ can be deduced from holomorphic functions.

\begin{theorem}
Paley-Wiener-Schwartz : For any function $f\in C_{\infty c}\left(
\mathbb{R}
^{m};%
\mathbb{C}
\right)  $ with support in the ball B(0,r), there is a holomorphic function
$F\in H\left(
\mathbb{C}
^{m};%
\mathbb{C}
\right)  $ such that :

$\forall x\in%
\mathbb{R}
^{m}:f(x)=F(x)$

$\forall p\in%
\mathbb{N}
,\exists c_{p}>0,\forall z\in%
\mathbb{C}
^{m}:\left\vert F(z)\right\vert \leq c_{p}(1+\left\vert z\right\vert
)^{-p}e^{r\left\vert \operatorname{Im}z\right\vert }$

Conversely if F is a holomorphic function $F\in H\left(
\mathbb{C}
^{m};%
\mathbb{C}
\right)  $ meeting the property above, then there is a function $f\in
C_{\infty,c}\left(
\mathbb{R}
^{m};%
\mathbb{C}
\right)  $ with support in the ball B(0,r) such that $\forall x\in%
\mathbb{R}
^{m}:f(x)=F(x)$
\end{theorem}

\subsubsection{Space of Schwartz functions}

This is a convenient space for Fourier transform. It is defined only for
functions over the whole of $%
\mathbb{R}
^{m}.$

\begin{definition}
The \textbf{Schwartz space} $S\left(
\mathbb{R}
^{m}\right)  $ of rapidly decreasing smooth functions is the subspace of
$C_{\infty}\left(
\mathbb{R}
^{m};%
\mathbb{C}
\right)  :$

$f\in C_{\infty}\left(
\mathbb{R}
^{m};%
\mathbb{C}
\right)  :\forall n\in%
\mathbb{N}
,\forall\alpha=\left(  \alpha_{1},..,\alpha_{n}\right)  :$

$\exists C_{n,\alpha}:\left\Vert D_{\alpha_{1}...\alpha_{n}}f\left(  x\right)
\right\Vert \leq C_{N,\alpha}\left\Vert x\right\Vert ^{-n}$
\end{definition}

\bigskip

\begin{theorem}
(Zuily p.108) The space $S\left(
\mathbb{R}
^{m}\right)  $ is a Fr\'{e}chet space with the seminorms: $p_{n}\left(
f\right)  =\sup_{x\in%
\mathbb{R}
^{m},\alpha,k\leq n}\left\Vert x\right\Vert ^{k}\left\Vert D_{\alpha
_{1}...\alpha_{n}}f\left(  x\right)  \right\Vert $ and a complex commutative
*-algebra with pointwise multiplication
\end{theorem}

\begin{theorem}
The product of $f\in S\left(
\mathbb{R}
^{m}\right)  $ by any polynomial, and any partial derivative of f still
belongs to $S\left(
\mathbb{R}
^{m}\right)  $

$C_{\infty c}\left(
\mathbb{R}
^{m};%
\mathbb{C}
\right)  \subset S\left(
\mathbb{R}
^{m}\right)  \subset C_{\infty}\left(
\mathbb{R}
^{m};%
\mathbb{C}
\right)  $

$C_{\infty c}\left(
\mathbb{R}
^{m};%
\mathbb{C}
\right)  $ is dense in $S\left(
\mathbb{R}
^{n}\right)  $

$\forall p:1\leq p\leq\infty:S\left(
\mathbb{R}
^{n}\right)  \subset L^{p}\left(
\mathbb{R}
^{m},dx,%
\mathbb{C}
\right)  $
\end{theorem}

\bigskip

\subsubsection{Sobolev spaces}

Sobolev spaces consider functions which are both integrable and differentiable
: there are a combination of the previous cases. To be differentiable they
must be defined at least over a manifold M, and to be integrable we shall have
some measure on M. So the basic combination is a finite m dimensional smooth
manifold endowed with a volume form $\varpi_{0}$.

Usually the domain of Sobololev spaces of functions are limited to an open of
$%
\mathbb{R}
^{m}$\ but with the previous result we can give a more extensive definition
which is useful for differential operators.

Sobolev spaces are extended (see Distributions and Fourier transform).

\paragraph{Sobolev space of sections of a vector bundle\newline}

\begin{definition}
On a vector bundle $E\left(  M,V,\pi\right)  $,with M endowed with a positive
measure $\mu,$ the Sobolev space denoted $W^{r,p}\left(  E\right)  $ is the
subspace of r differentiable sections $S\in\mathfrak{X}_{r}\left(  E\right)  $
such that their r jet prolongation: $J^{r}S\in L^{p}\left(  M,\mu
,J^{r}E\right)  $
\end{definition}

\bigskip

\begin{theorem}
$W^{r,p}\left(  E\right)  $ is a Banach space with the norm :

$\left\Vert Z\right\Vert _{p,r}=\left(  \sum_{s=1}^{r}\sum_{\alpha_{1}%
..\alpha_{s}}\int_{M}\left\Vert D_{\left(  \alpha\right)  }u\left(  x\right)
\right\Vert ^{p}\mu\right)  ^{1/p}.$ Moreover the map :

$J^{r}:W^{r,p}\left(  E\right)  \rightarrow L^{p}\left(  M,\mu,J^{r}E\right)
$ is an isometry.
\end{theorem}

\begin{proof}
$W^{r,p}\left(  E\right)  \subset L^{p}\left(  M,\varpi_{0},E\right)  $ which
is a Banach space.

Each section $S\in\mathfrak{X}_{r}\left(  E\right)  $ $S\left(  x\right)
=\varphi\left(  x,u\left(  x\right)  \right)  $\ gives a section in
$\mathfrak{X}\left(  J^{r}S\right)  $ which reads :

$M\rightarrow V\times%
{\textstyle\prod\limits_{s=1}^{r}}
\left\{
\mathcal{L}%
_{S}^{s}\left(
\mathbb{R}
^{\dim M},V\right)  \right\}  ::Z\left(  s\right)  =\left(  z_{s}\left(
x\right)  ,s=0...r\right)  $ with $z_{s}\left(  x\right)  =D_{\left(
\alpha\right)  }u\left(  x\right)  $ with, fiberwise, norms $\left\Vert
z_{s}\left(  x\right)  \right\Vert _{s}$ for each component and $%
\mathcal{L}%
^{p}\left(  M,\mu,J^{r}E\right)  =\left\{  Z\in\mathfrak{X}\left(
J^{r}E\right)  :\sum_{s=0}^{r}\int_{M}\left\Vert z_{s}\left(  x\right)
\right\Vert _{s}^{p}\mu<\infty\right\}  $

Take a sequence $S_{n}\in W^{r,p}\left(  E\right)  $ which converges to
S$\in\mathfrak{X}_{r}\left(  E\right)  $ in $\mathfrak{X}_{r}\left(  E\right)
,$ then the support of each $S_{n}$ is surely $\varpi_{0}$ integrable and by
the Lebesgue theorem $\int_{M}\left\Vert D_{\left(  \alpha\right)  }\sigma
_{n}\left(  x\right)  \right\Vert _{s}^{p}\mu<\infty.$ Thus $W^{r,p}\left(
E\right)  $ is closed in $L^{p}\left(  M,\mu,J^{r}E\right)  $ and is a Banach space.
\end{proof}

\begin{theorem}
On a vector bundle endowed with an inner product $W^{r,2}\left(  E\right)  $
is a Hilbert space, with the scalar product of $L^{2}\left(  M,\mu,E\right)  $
\end{theorem}

\begin{proof}
$W^{r,2}\left(  E\right)  $ is a vector subspace of $L^{2}\left(  M,\varpi
_{0},E\right)  $ and the map $J^{r}:W^{r,2}\left(  E\right)  \rightarrow
L^{2}\left(  M,\varpi_{0},J^{r}E\right)  $ is continuous. So $W^{r,2}\left(
E\right)  $ is closed in $L^{2}\left(  M,\varpi_{0},E\right)  $ which is a
Hilbert space
\end{proof}

\bigskip

\begin{notation}
$W_{c}^{r,p}\left(  E\right)  $ is the subspace of $W^{r,p}\left(  E\right)
$\ of sections with compact support
\end{notation}

\begin{notation}
$W_{loc}^{r,p}\left(  E\right)  $ is the subspace of $L_{loc}^{p}\left(
M,\varpi_{0},E\right)  $ comprised of r differentiable sections such that :
$\forall\alpha,\left\Vert \alpha\right\Vert \leq r,D_{\alpha}\sigma\in
L_{loc}^{p}\left(  M,\varpi_{0},E\right)  $
\end{notation}

\paragraph{Sobolev spaces of functions over a manifold\newline}

\begin{definition}
The \textbf{Sobolev space,} denoted $W^{r,p}\left(  M\right)  $ , of functions
over a manifold endowed with a positive measure $\mu$\ is the subspace of
$L^{p}\left(  M,\mu,%
\mathbb{C}
\right)  $ comprised of r differentiable functions f over M such that :
$\forall\alpha_{1},...\alpha_{s},s\leq r:D_{\left(  \alpha\right)  }f\in
L^{p}\left(  M,\mu,%
\mathbb{C}
\right)  ,1\leq p\leq\infty$\ 
\end{definition}

\bigskip

\begin{theorem}
$W^{r,p}\left(  M\right)  $ It is a Banach vector space with the norm :

$\left\Vert f\right\Vert _{W^{r,p}}=\left(  \sum_{s=1}^{r}\sum_{\alpha
_{1}..\alpha_{s}}\left\Vert D_{\left(  \alpha\right)  }f\right\Vert _{L^{p}%
}\right)  ^{1/p}.$
\end{theorem}

\begin{theorem}
If there is an inner product in E, $W^{r,2}\left(  M\right)  =H^{r}\left(
M\right)  $ is a Hilbert space with the scalar product : $\left\langle
\varphi,\psi\right\rangle =\sum_{s=1}^{r}\sum_{\alpha_{1}..\alpha_{s}%
}\left\langle D_{\left(  \alpha\right)  }\varphi,D_{\alpha}\psi\right\rangle
_{L^{2}}$
\end{theorem}

\bigskip

\begin{notation}
$H^{r}\left(  M\right)  $ is the usual notation for $W^{r,2}\left(  M\right)
$
\end{notation}

\paragraph{Sobolev spaces of functions over $%
\mathbb{R}
^{m}$\newline}

\begin{theorem}
For any open subset O of $%
\mathbb{R}
^{m}:$

i)$\ \forall r>r^{\prime}:H^{r}\left(  O\right)  \subset H^{r^{\prime}}\left(
O\right)  $ and if O is bounded then the injection $\imath:H_{c}^{r}\left(
O\right)  \rightarrow H_{c}^{r^{\prime}}\left(  O\right)  $ is compact

ii) $C_{\infty}\left(  O;%
\mathbb{C}
\right)  $ is dense in $W_{loc}^{1,1}\left(  O\right)  $ and $H^{1}\left(
O\right)  $
\end{theorem}

\begin{theorem}
$C_{\infty c}\left(
\mathbb{R}
^{m};%
\mathbb{C}
\right)  $ is dense in $H^{r}\left(
\mathbb{R}
^{m}\right)  $
\end{theorem}

but this is no true for $O\subset%
\mathbb{R}
^{m}$

\begin{theorem}
(Lieb p.177) For any functions $u,v\in H^{1}\left(
\mathbb{R}
^{m}\right)  $ :

$\forall k=1..m:\int_{%
\mathbb{R}
^{m}}u\left(  \partial_{k}v\right)  dx=-\int_{%
\mathbb{R}
^{m}}v\left(  \partial_{k}u\right)  dx$
\end{theorem}

If v is real and $\Delta v\in L_{loc}^{1}\left(
\mathbb{R}
^{m},dx,%
\mathbb{R}
\right)  $ then

$\int_{%
\mathbb{R}
^{m}}u\left(  \Delta v\right)  dx=-\int_{%
\mathbb{R}
^{m}}\sum_{k=1}^{m}\left(  \partial_{k}v\right)  \left(  \partial_{k}u\right)
dx$

\begin{theorem}
(Lieb p.179) For any real valued functions $f,g\in H^{1}\left(
\mathbb{R}
^{m}\right)  $ :

$\int_{%
\mathbb{R}
^{m}}\sum_{\alpha}\left(  \partial_{\alpha}\sqrt{f^{2}+g^{2}}\right)
^{2}dx\leq\int_{%
\mathbb{R}
^{m}}\sum_{\alpha}\left(  \left(  \partial_{\alpha}f\right)  ^{2}+\left(
\partial_{\alpha}g\right)  ^{2}\right)  dx$

and if $g\geq0$ then the equality holds iff $f=cg$ almost everywhere for a
constant c
\end{theorem}

\begin{theorem}
(Zuilly p.148) : For any class r manifold with boundary M in $%
\mathbb{R}
^{m}$ :

i) $C_{\infty c}\left(  M;%
\mathbb{C}
\right)  $ is dense in $H^{r}\left(  \overset{\circ}{M}\right)  $

ii) There is a map : $P\in%
\mathcal{L}%
\left(  H^{r}\left(  \overset{\circ}{M}\right)  ;H^{r}\left(
\mathbb{R}
^{m}\right)  \right)  $ such that $P(f\left(  x\right)  )=f\left(  x\right)  $
on $\overset{\circ}{M}$

iii) If $k>\frac{m}{2}+l:H^{k}\left(  \overset{\circ}{M}\right)  \subset
C_{l}\left(  M;%
\mathbb{C}
\right)  $ so $\cup_{k\in%
\mathbb{N}
}H^{k}\left(  \overset{\circ}{M}\right)  \subset C_{\infty}\left(  M;%
\mathbb{C}
\right)  $

iv) $\forall r>r^{\prime}:H^{r}\left(  \overset{\circ}{M}\right)  \subset
H^{r^{\prime}}\left(  \overset{\circ}{M}\right)  $ and if M is bounded then
the injection $\imath:H^{r}\left(  \overset{\circ}{M}\right)  \rightarrow
H^{r^{\prime}}\left(  \overset{\circ}{M}\right)  $ is compact
\end{theorem}

\newpage

\section{DISTRIBUTIONS}

\subsection{Spaces of functionals}

\label{Space of functionals}

\subsubsection{Reminder of elements of normed algebras}

see the section Normed Algebras in the Analysis part.

Let V be a topological *-subalgebra of $C\left(  E;%
\mathbb{C}
\right)  ,$ with the involution $f\rightarrow\overline{f}$ and pointwise multiplication.

The subset of functions of V valued in $%
\mathbb{R}
$ is a subalgebra of hermitian elements, denoted $V_{R},$ which is also the
real subspace of V with the natural real structure on V : $f=\operatorname{Re}%
f+i\operatorname{Im}f.$ The space $V_{+}$\ of positive elements of V is the
subspace of $V_{R}$ of maps in $C\left(  E:%
\mathbb{R}
_{+}\right)  .$ If V is in $C\left(  E;%
\mathbb{R}
\right)  $ then $V=V_{R}$ and is hermitian.

A linear functional is an element of the algebraic dual $V^{\ast}=L(V;%
\mathbb{C}
)$ so this is a complex linear map. With the real structures on V and $%
\mathbb{C}
,$\ any functional reads : $\lambda\left(  \operatorname{Re}%
f+i\operatorname{Im}f\right)  =\lambda_{R}\left(  \operatorname{Re}f\right)
+\lambda_{I}\left(  \operatorname{Im}f\right)  +i\left(  -\lambda_{I}\left(
\operatorname{Re}f\right)  +\lambda_{R}\left(  \operatorname{Im}f\right)
\right)  $ with two real functionals $\lambda_{R},\lambda_{I}\in C\left(
V_{R};%
\mathbb{R}
\right)  .$\ In the language of algebras, $\lambda$ is hermitian if
$\lambda\left(  \overline{f}\right)  =\overline{\lambda\left(  f\right)  }$
that is $\lambda_{I}=0.$ Then $\forall f\in V_{R}:\lambda\left(  f\right)  \in%
\mathbb{R}
.$

In the language of normed algebras a linear functional $\lambda$\ is weakly
continuous if $\forall f\in V_{R}$ the map $g\in V\rightarrow\lambda\left(
\left\vert g\right\vert ^{2}f\right)  $ is continuous.\ As the map : $f,g\in
V\rightarrow\left\vert g\right\vert ^{2}f$ is continuous on V (which is a
*-topological algebra) then here weakly continuous = continuous on V$_{R}$

\subsubsection{Positive linear functionals}

\begin{theorem}
A linear functional $\lambda:V\rightarrow%
\mathbb{C}
$ on a space of functions V is positive iff $\lambda\left(  \overline
{f}\right)  =\overline{\lambda\left(  f\right)  }$ and $\lambda\left(
f\right)  \geq0$ when $f\geq0$
\end{theorem}

\begin{proof}
Indeed a linear functional $\lambda$\ is positive\ (in the meaning viewed in
Algebras) if $\lambda\left(  \left\vert f\right\vert ^{2}\right)  \geq0$ and
any positive function has a square root.
\end{proof}

The variation of a positive linear functional is :

$v\left(  \lambda\right)  =\inf_{f\in V}\left\{  \gamma:\left\vert
\lambda\left(  f\right)  \right\vert ^{2}\leq\gamma\lambda\left(  \left\vert
f\right\vert ^{2}\right)  \right\}  $.

If it is finite then $\left\vert \lambda\left(  f\right)  \right\vert ^{2}\leq
v\left(  \lambda\right)  \lambda\left(  \left\vert f\right\vert ^{2}\right)  $

A positive linear functional $\lambda$, continuous on $V_{R},$ is a
state\textbf{\ }if $v\left(  \lambda\right)  =1$ , a quasi-state\textbf{\ }if
$v\left(  \lambda\right)  \leq1.$ The set of states and of quasi-states are convex.

If V is a normed *-algebra :

i) \ a quasi-state is continuous on V with norm $\left\Vert \lambda\right\Vert
\leq1$

\begin{proof}
It is $\sigma-$contractive, so $\left\vert \lambda\left(  f\right)
\right\vert \leq r_{\lambda}\left(  f\right)  =\left\Vert f\right\Vert $
because f is normal
\end{proof}

ii) the variation of a positive linear functional is $v\left(  \lambda\right)
=\lambda\left(  I\right)  $ where I is the identity element if V is unital.

iii) if V has a state the set of states has an extreme point (a pure state).

If V is a Banach *-algebra :

i) a positive linear functional $\lambda$\ is continuous on $V_{R}.$ If
$v\left(  \lambda\right)  <\infty$ it is continuous on V, if $v\left(
\lambda\right)  =1$ it is a state.

ii) a state (resp. a pure state) $\lambda$ on a closed *-subalgebra can be
extended to a state (resp. a pure state) on V

If V is a C*-algebra : a positive linear functional is continuous and
$v\left(  \lambda\right)  =\left\Vert \lambda\right\Vert ,$ it is a state iff
$\left\Vert \lambda\right\Vert =\lambda\left(  I\right)  =1$

As a consequence:

\begin{theorem}
A positive functional is continuous on the following spaces :

i) $C_{b}\left(  E;%
\mathbb{C}
\right)  $ of bounded functions if E is topological

ii) $C_{0b}\left(  E;%
\mathbb{C}
\right)  $ of bounded continuous functions if E Hausdorff

iii) $C_{c}\left(  E;%
\mathbb{C}
\right)  $ of functions with compact support, $C_{0v}\left(  E;%
\mathbb{C}
\right)  $ of continuous functions vanishing at infinity, if E is Hausdorff,
locally compact

iv) $C_{0}\left(  E;%
\mathbb{C}
\right)  $ of continuous functions if E is compact
\end{theorem}

\begin{theorem}
If E is a locally compact, separable, metric space, then a positive functional
$\lambda\in L\left(  C_{0c}\left(  E;%
\mathbb{R}
\right)  ;%
\mathbb{R}
\right)  $ can be uniquely extended to a functional in $%
\mathcal{L}%
\left(  C_{0\nu}\left(  E;%
\mathbb{R}
\right)  ;%
\mathbb{R}
\right)  $
\end{theorem}

\subsubsection{Functional defined as integral}

\paragraph{Function defined on a compact space\newline}

\begin{theorem}
(Taylor 1 p.484) If E is a compact metric space, $C\left(  E;%
\mathbb{C}
\right)  ^{\prime}$ is isometrically isomorphic to the space of complex
measures on E endowed with the total variation norm.
\end{theorem}

\paragraph{L$^{p}$ spaces\newline}

\begin{theorem}
(Lieb p.61) For any measured space with a positive measure $(E,S,\mu)$ ,
$1\leq p\leq\infty,\frac{1}{p}+\frac{1}{q}=1$

The map : $L_{p}:$ $L^{q}\left(  E,S,\mu,%
\mathbb{C}
\right)  \rightarrow L^{p}\left(  E,S,\mu,%
\mathbb{C}
\right)  ^{\ast}::L_{p}\left(  f\right)  \left(  \varphi\right)  =\int
_{E}f\varphi\mu$ is continuous\ and an isometry : $\left\Vert L_{p}\left(
f\right)  \right\Vert _{p}=\left\Vert f\right\Vert _{q}$ so $L_{p}\left(
f\right)  \in L^{p}\left(  E,S,\mu,%
\mathbb{C}
\right)  ^{\prime}$

i) If $1<p<\infty,$ or if p=1 and the measure $\mu$ is $\sigma$-finite
(meaning E is the countable union of subsets of finite measure), then $L_{p}$
is bijective in the topological dual $L^{p}\left(  E,S,\mu,%
\mathbb{C}
\right)  ^{\prime}$ of $L^{p}\left(  E,S,\mu,%
\mathbb{C}
\right)  $ , which is isomorphic to $L^{q}\left(  E,S,\mu,%
\mathbb{C}
\right)  $

ii) If $p=\infty$ : elements of the dual $L^{\infty}\left(  E,S,\mu,%
\mathbb{C}
\right)  ^{\prime}$ can be identified with bounded signed finitely additive
measures on S that are absolutely continuous with respect to $\mu$.
\end{theorem}

So :

if f is real then $L_{p}\left(  f\right)  $ is hermitian, and positive if
$f\geq0$ almost everywhere for $1<p<\infty$

Conversely:

\begin{theorem}
(Doob p.153) For any measured space with a $\sigma-$finite measure
$(E,S,\mu),$ and any continuous linear functional on $L^{1}\left(  E,S,\mu,%
\mathbb{C}
\right)  $ there is a function f, unique up to a set of null measure, bounded
and integrable, such that : $\ell\left(  \varphi\right)  =\int_{E}f\varphi
\mu.$ Moreover : $\left\Vert \ell\right\Vert =\left\Vert f\right\Vert
_{\infty}$ and $\ell$ is positive iff $f\geq0$ almost everywhere.
\end{theorem}

\paragraph{Radon measures\newline}

A Radon measure $\mu$ is a Borel (defined on the Borel $\sigma-$algebra S of
open subsets), locally finite, regular, signed measure on a topological
Hausdorff locally compact space $\left(  E,\Omega\right)  $. So :

$\forall X\in S:\mu\left(  X\right)  =\inf\left(  \mu\left(  Y\right)
,X\sqsubseteq Y,Y\in\Omega\right)  $

$\forall X\in S,\mu\left(  X\right)  <\infty:\mu\left(  X\right)  =\sup\left(
\mu\left(  K\right)  ,K\sqsubseteq X,K\text{ compact}\right)  $

A measure is locally compact if it is finite on any compact

\begin{theorem}
(Doob p.127-135 and Neeb p.42) Let $E$ be a Hausdorff, locally compact,
topological space, and S its Borel $\sigma-$algebra.

i) For any positive, locally finite Borel measure $\mu$ the map :
$\lambda_{\mu}\left(  f\right)  =\int_{E}f\mu$ is a positive linear functional
on $C_{0c}\left(  E;%
\mathbb{C}
\right)  $ (this is a Radon integral). Moreover it is continuous on
$C_{0}\left(  K;%
\mathbb{C}
\right)  $ for any compact K of E.

ii) Conversely for any positive linear functional $\lambda$ on $C_{0c}\left(
E;%
\mathbb{C}
\right)  $ there is a unique Radon measure, called the Radon measure
associated to $\lambda$ ,such that $\lambda=\lambda_{\mu}$

iii) For any linear functional on $C_{0v}\left(  E;%
\mathbb{C}
\right)  $ there is a unique complex measure $\mu$ on (E,S) such that
$\left\vert \mu\right\vert $ is regular and $\forall f\in C_{0v}\left(  E;%
\mathbb{C}
\right)  :\lambda\left(  f\right)  =\int_{E}f\mu$

iv) For any continuous positive linear functional $\lambda$\ on $C_{c}\left(
E;%
\mathbb{C}
\right)  $ with norm 1 (a state) there is a Borel, inner regular, probability
P such that : $\forall f\in C_{c}\left(  E;%
\mathbb{C}
\right)  :\lambda\left(  f\right)  =\int_{E}fP.$ So we have $P(E)=1,P\left(
\varpi\right)  \geq0$
\end{theorem}

\begin{theorem}
Riesz Representation theorem (Thill p.254): For every positive linear
functional $\ell$ on the space $C_{c}\left(  E;%
\mathbb{C}
\right)  $ where E is a locally compact Hausdorff space, bounded with norm 1,
there exists a unique inner regular Borel measure $\mu$ such that :

$\forall\varphi\in C_{c}\left(  E;%
\mathbb{C}
\right)  :\ell\left(  \varphi\right)  =\int_{E}\varphi\mu$

On a compact K of E, $\mu\left(  K\right)  =\inf\left\{  \ell\left(
\varphi\right)  :\varphi\in C_{c}\left(  E;%
\mathbb{C}
\right)  ,1_{K}\leq\varphi\leq1_{E}\right\}  $
\end{theorem}

\subsubsection{Multiplicative linear functionals}

For an algebra A of functions a multiplicative linear functional is an element
$\lambda$\ of the algebraic dual A* such that $\lambda\left(  fg\right)
=\lambda\left(  f\right)  \lambda\left(  g\right)  $ and $\lambda
\neq0\Rightarrow\lambda\left(  I\right)  =1.$ It is necessarily continuous
with norm $\left\Vert \lambda\right\Vert \leq1$\ if A is a Banach *-algebra.

If E is a locally compact Hausdorff space, then the set of multiplicative
linear functionals $\Delta\left(  C_{0v}\left(  E;%
\mathbb{C}
\right)  \right)  $ is homeomorphic to $E:$

For $x\in E$ fixed $:\delta_{x}:C_{0}\left(  E;%
\mathbb{C}
\right)  \rightarrow%
\mathbb{C}
$ with norm $\left\Vert \lambda\right\Vert \leq1::\delta_{x}\left(  f\right)
=f\left(  x\right)  $

So the only multiplicative linear functionals are the Dirac distributions.

\bigskip

\subsection{Distributions on functions}

\label{Distributions on fn functions}

Distributions, also called generalized functions, are a bright example of the
implementation of duality (due to L.Schwartz). The idea is to associate to a
given space of functions its topological dual, meaning the space of linear
continuous functionals. The smaller the space of functions, the larger the
space of functionals. We can extend to the functionals many of the operations
on functions, such as derivative, and thus enlarge the scope of these
operations, which is convenient in many calculii, but also give a more unified
understanding of important topics in differential equations. But the facility
which is offered by the use of distributions is misleading. Everything goes
fairly well when the functions are defined over $%
\mathbb{R}
^{m}$\ , but this is another story when they are defined over manifolds.

\subsubsection{ Definition}

\begin{definition}
A \textbf{distribution} is a continuous linear functional on a Fr\'{e}chet
space V of functions, called the space of \textbf{test functions}.
\end{definition}

\begin{notation}
V' is the space of distributions over the space of test functions V
\end{notation}

There are common notations for the most used spaces of distributions but, in
order to avoid the introduction of another symbol, I find it simpler to keep
this standard and easily understood notation, which underlined the true nature
of the set.

Of course if V is a Banach space the definition applies, because a Banach
vector space is a Fr\'{e}chet space. When V is a Hilbert space, V' is a
Hilbert space, and when V is a Banach space, V' is a Banach space, and in both
cases we have powerful tools to deal with most of the problems. But the spaces
of differentiable maps are only Fr\'{e}chet spaces and it is not surprising
that the most usual spaces of tests functions are space of differentiable functions.

\paragraph{Usual spaces of distributions on $%
\mathbb{R}
^{m}$\newline}

Let O be an open subset of $%
\mathbb{R}
^{m}.$ We have the following spaces of distributions:

$C_{\infty c}\left(  O;%
\mathbb{C}
\right)  ^{\prime}:$ usually denoted $\mathfrak{D}\left(  O\right)  $

$C_{\infty}\left(  O;%
\mathbb{C}
\right)  ^{\prime}:$ usually denoted $\mathfrak{E}^{\prime}\left(  O\right)  $

$C_{rc}\left(  O;%
\mathbb{C}
\right)  ^{\prime}$

$S\left(
\mathbb{R}
^{m}\right)  ^{\prime}$ called the space of \textbf{tempered distributions. }

$S\in S\left(
\mathbb{R}
^{m}\right)  ^{\prime}\Leftrightarrow S\in L\left(  S\left(
\mathbb{R}
^{m}\right)  ;%
\mathbb{C}
\right)  ,\exists p,q\in%
\mathbb{N}
,\exists C\in%
\mathbb{R}
:\forall\varphi\in S\left(
\mathbb{R}
^{m}\right)  :$

$\left\vert S\left(  \varphi\right)  \right\vert \leq C\sum_{k\leq p,,l\leq
q}\sum_{\left(  \alpha_{1}..\alpha_{k}\right)  \left(  \beta_{1}...\beta
_{l}\right)  }\sup_{x\in%
\mathbb{R}
^{m}}\left\vert x_{\alpha_{1}}...x_{\alpha_{k}}D_{\beta_{1}...\beta_{l}%
}\varphi\left(  x\right)  \right\vert $

We have the following inclusions : the larger the space of tests functions,
the smaller the space of distributions.

\begin{theorem}
$\forall r\geq1\in%
\mathbb{N}
:C_{\infty}\left(  O;%
\mathbb{C}
\right)  ^{\prime}\subset C_{rc}\left(  O;%
\mathbb{C}
\right)  ^{\prime}\subset C_{\infty c}\left(  O;%
\mathbb{C}
\right)  ^{\prime}$

$C_{\infty}\left(  O;%
\mathbb{C}
\right)  ^{\prime}\subset S\left(
\mathbb{R}
^{m}\right)  ^{\prime}\subset C_{\infty c}\left(
\mathbb{R}
^{m};%
\mathbb{C}
\right)  ^{\prime}$
\end{theorem}

\begin{theorem}
(Lieb p.150) Let O be an open in $%
\mathbb{R}
^{m},$ if $\left(  S_{k}\right)  _{k=1}^{n},S_{k}\in C_{\infty c}\left(  O;%
\mathbb{C}
\right)  ^{\prime}$ and $S\in C_{\infty c}\left(  O;%
\mathbb{C}
\right)  ^{\prime}$ such that : $\forall\varphi\in\cap_{k=1}^{n}\ker
S_{k}:S\left(  \varphi\right)  =0$ then there are $\ c_{k}\in%
\mathbb{C}
:$ $S=\sum_{k=1}^{n}c_{k}S_{k}$
\end{theorem}

\paragraph{Usual spaces of distributions on a manifold\newline}

For any real finite dimensional manifold M, we have the following spaces of
distributions :

$C_{\infty c}\left(  M;%
\mathbb{C}
\right)  ^{\prime}:$ usually denoted $\mathfrak{D}\left(  M\right)  $

$C_{\infty}\left(  M;%
\mathbb{C}
\right)  ^{\prime}:$ usually denoted $\mathfrak{E}^{\prime}\left(  M\right)  $

$C_{rc}\left(  M;%
\mathbb{C}
\right)  ^{\prime}$

\subsubsection{Topology}

As a Fr\'{e}chet space V is endowed with a countable family $(p_{i})_{i\in I}
$ of semi-norms, which induces a metric for which it is a complete locally
convex Hausdorff space. The strong topology on V implies that a functional
$S:V\rightarrow%
\mathbb{C}
$ is continuous iff for any bounded subset W of V, that is a subset such that :

$\forall i\in I,\exists D_{W_{i}}\in%
\mathbb{R}
,\forall f\in W:p_{i}\left(  f\right)  \leq D_{Wi}$ ,

we have : $\exists C_{W}\in%
\mathbb{R}
:\forall f\in W,\forall i\in I:\left\vert S\left(  f\right)  \right\vert \leq
C_{W}p_{i}\left(  f\right)  $

Equivalently a linear functional (it belongs to the algebraic dual V*) S is
continuous if for any sequence $\left(  \varphi_{n}\right)  _{n\in%
\mathbb{N}
}\in V^{%
\mathbb{N}
}$ :

$\forall i\in%
\mathbb{N}
:p_{i}\left(  \varphi_{n}\right)  \rightarrow0\Rightarrow$ $S\left(
\varphi_{n}\right)  \rightarrow0$.

The most common cases are when $V=\cup_{K\subset E}V_{K}$ where $V_{K}$ are
functions with compact support in $K\subset E$ and E is a topological space
which is the countable union of compacts subsets. Then a continuous functional
can be equivalently defined as a linear functional whose restriction on
$V_{K},\ $where K is any compact, is continuous.

The topological dual of a Fr\'{e}chet space V is usually not a Fr\'{e}chet
space. So the natural topology on V' is the *weak topology : a sequence
$\left(  S_{n}\right)  ,S_{n}\in V^{\prime}$ converges to S in V' if :
$\forall\varphi\in V:S_{n}\left(  \varphi\right)  \rightarrow S\left(
\varphi\right)  $

Notice that S must be defined and belong to V' prior to checking the
convergence, the simple convergence of $S_{n}\left(  \varphi\right)  $ is
usually not sufficient to guarantee that there is S in V' such that $\lim
S_{n}\left(  \varphi\right)  =S\left(  \varphi\right)  $. However :

\begin{theorem}
(Zuily p.57) If a sequence $\left(  S_{n}\right)  _{n\in%
\mathbb{N}
}\in\left(  C_{\infty c}\left(  O;%
\mathbb{C}
\right)  ^{\prime}\right)  ^{%
\mathbb{N}
}$ is such that $\forall\varphi\in C_{\infty c}\left(  O;%
\mathbb{C}
\right)  :S_{n}\left(  \varphi\right)  $ converges, then there is $S\in
C_{\infty c}\left(  O;%
\mathbb{C}
\right)  ^{\prime}$ such that $S_{n}\left(  \varphi\right)  \rightarrow
S\left(  \varphi\right)  .$
\end{theorem}

\subsubsection{Identification of functions with distributions\newline}

One of the most important feature of distributions is that functions can be
"assimilated" to distributions, meaning that there is a map T between some
space of functions W and the space of distributions V', W being larger than V.

There are many theorems which show that, in most of the cases, a distribution
is necessarily an integral for some measure. This question is usually treated
rather lightly.\ Indeed it is quite simple when the functions are defined over
$%
\mathbb{R}
^{m}$ but more complicated when they are defined over a manifold. So it needs attention.

Warning ! there is not always a function associated to a distribution.

\paragraph{General case\newline}

As a direct consequences of the theorems on functionals and integrals :

\begin{theorem}
For any measured space (E,S,$\mu)$ with\ a positive measure $\mu$\ and
Fr\'{e}chet vector subspace $V\subset L^{p}\left(  E,S,\mu,%
\mathbb{C}
\right)  $ with $1\leq p\leq\infty$ and $\mu$ is $\sigma-$finite if
p=1,$\frac{1}{p}+\frac{1}{q}=1$, the map : $T\left(  f\right)  :L^{q}\left(
E,S,\mu,%
\mathbb{C}
\right)  \rightarrow V^{\prime}::T\left(  f\right)  \left(  \varphi\right)
=\int_{E}f\varphi\mu$ is linear, continuous, injective and an isometry, so it
is a distribution in V'
\end{theorem}

i) T is continuous : if the sequence $\left(  f_{n}\right)  _{n\in%
\mathbb{N}
}$ converges to f in $L^{p}\left(  E,S,\mu,%
\mathbb{C}
\right)  $ then $T\left(  f_{n}\right)  \rightarrow T\left(  f\right)  $ in
$V^{\prime}$

ii) it is an isometry : $\left\Vert T\left(  f\right)  \right\Vert _{L^{p}%
}=\left\Vert f\right\Vert _{L^{q}}$

iii) Two functions f,g give the same distribution iff they are equal almost
everywhere on E.

iv) T is surjective for $V=L^{p}\left(  E,S,\mu,%
\mathbb{C}
\right)  $ : to any distribution $S\in L^{p}\left(  E,S,\mu,%
\mathbb{C}
\right)  ^{\prime}$ one can associate a function $f\in L^{q}\left(  E,S,\mu,%
\mathbb{C}
\right)  $ such that $T\left(  f\right)  =S$ but, because V' is larger than
$L^{p}\left(  E,S,\mu,%
\mathbb{C}
\right)  ^{\prime}$ , it is not surjective on V'.

This theorem is general, but when the tests functions are defined on a
manifold we have two very different cases.

\paragraph{Functions defined in $%
\mathbb{R}
^{m}$\newline}

\begin{theorem}
For any Fr\'{e}chet vector subspace $V\subset L^{1}\left(  O,dx,%
\mathbb{C}
\right)  $ where O is an open subset of $%
\mathbb{R}
^{m},$ for any $f\in L^{\infty}\left(  O,\mu,%
\mathbb{C}
\right)  $ the map : $T\left(  f\right)  :V\rightarrow%
\mathbb{C}
::T\left(  f\right)  \left(  \varphi\right)  =\int_{O}f\varphi dx^{1}%
\wedge...\wedge dx^{m}$ is a distribution in V', and the map : $T:L^{\infty
}\left(  O,dx,%
\mathbb{C}
\right)  \rightarrow V^{\prime}$ is linear, continuous and injective
\end{theorem}

This is just the application of the previous theorem : $f\varphi dx^{1}%
\wedge...\wedge dx^{m}\simeq f\varphi dx^{1}\otimes...\otimes dx^{m}$ defines
a measure which is locally finite.

More precisely we have the following associations (but they are not bijective !)

$C_{\infty c}\left(  O;%
\mathbb{C}
\right)  ^{\prime}\leftrightarrow W=L_{loc}^{p}\left(  O,dx,%
\mathbb{C}
\right)  ,1\leq p\leq\infty$

$C_{\infty}\left(  O;%
\mathbb{C}
\right)  ^{\prime}\leftrightarrow W=L_{c}^{p}\left(  O,dx,%
\mathbb{C}
\right)  ,1\leq p\leq\infty$ if the function is compactly supported so is the distribution

$S\left(
\mathbb{R}
^{m}\right)  ^{\prime}\leftrightarrow W=L^{p}\left(
\mathbb{R}
^{m},dx,%
\mathbb{C}
\right)  ,1\leq p\leq\infty$ (Zuily p.113)

$S\left(
\mathbb{R}
^{m}\right)  ^{\prime}\leftrightarrow$ $W=\{f\in C\left(
\mathbb{R}
^{m};%
\mathbb{C}
\right)  $ measurable, $\left\vert f(x)\right\vert \leq P\left(  x\right)  $
where P is a polynomial \} (Zuily p.113)

In an all cases : $f\in W:T\left(  f\right)  \left(  \varphi\right)  =\int
f\varphi dx$ and the map : $T:W\rightarrow V^{\prime}$ is injective and
continuous (but not surjective). So :

$T\left(  f\right)  =T\left(  g\right)  \Leftrightarrow f=g$ almost everywhere

Moreover :

$T\left(  C_{\infty c}\left(  O;%
\mathbb{C}
\right)  \right)  $ is dense in $C_{\infty c}\left(  O;%
\mathbb{C}
\right)  ^{\prime}$ and $\left(  C_{\infty}\left(  O;%
\mathbb{C}
\right)  \right)  ^{\prime}$

$T\left(  S\left(
\mathbb{R}
^{m}\right)  \right)  $ is dense in $S\left(
\mathbb{R}
^{m}\right)  ^{\prime}$

In all the cases above the measure is absolutely continuous, because it is a
multiple of the Lebesgue measure.\ For $C_{\infty c}\left(  O;%
\mathbb{C}
\right)  $ we have more:

\begin{theorem}
(Lieb p.161) Let O be an open in $%
\mathbb{R}
^{m}$ and $S\in C_{\infty c}\left(  O;%
\mathbb{C}
\right)  ^{\prime}$ such that : $\forall\varphi\geq0:S\left(  \varphi\right)
\geq0$ then there is a unique positive Radon measure $\mu$ on O such that :
$S\left(  \varphi\right)  =\mu\left(  \varphi\right)  =\int_{O}\varphi\mu.$
Conversely any Radon measure defines a positive distribution on O.
\end{theorem}

So on $C_{\infty c}\left(  O;%
\mathbb{C}
\right)  $\ distributions are essentially Radon measures (which are not
necessarily absolutely continuous).\ 

\paragraph{Functions over a manifold\newline}

If $V\subset C\left(  M;%
\mathbb{C}
\right)  $ and M is some m dimensional manifold, other than an open in $%
\mathbb{R}
^{m}$\ the quantity $f\left(  x\right)  d\xi^{1}\wedge...\wedge d\xi^{m}$ is
not a m form and does not define a measure on M. The first solution is the
simple implementation of the previous theorem, by using a given m-form on M.

\begin{theorem}
For an oriented Hausdorff class 1 real manifold M, any continuous m
form\ $\varpi$ on M and its induced Lebesgue measure $\mu,$ any Fr\'{e}chet
vector space of functions $V\subset L^{p}\left(  M,\mu,%
\mathbb{C}
\right)  $ with $1\leq p\leq\infty$ , any function $f\in L^{q}\left(  O,\mu,%
\mathbb{C}
\right)  $ with $\frac{1}{p}+\frac{1}{q}=1,$\ the map : $T\left(  f\right)
:V\rightarrow%
\mathbb{C}
::T\left(  f\right)  \left(  \varphi\right)  =\int_{M}\varphi f\varpi$ is a
distribution in V', and the map : $T:L^{\infty}\left(  O,dx,%
\mathbb{C}
\right)  \rightarrow V^{\prime}$ is linear, continuous and injective.
\end{theorem}

\begin{proof}
any continuous m form defines a Radon measure $\mu$, and M is $\sigma-$finite

this measure $\mu$ can be decomposed into two positive measure $\mu_{+}%
,\mu_{-}$ which are still Radon measure (they are locally compact because
$\mu\left(  K\right)  <\infty\Rightarrow\mu_{+}\left(  K\right)  ,\mu
_{-}\left(  K\right)  <\infty)$

From there it suffices to apply the \ theorem above.
\end{proof}

As $\mu$ is locally finite the theorem holds for any space V of bounded
functions with compact support. In particular $C_{0c}\left(  M;%
\mathbb{C}
\right)  $ is dense in $L^{p}\left(  M,\mu,%
\mathbb{C}
\right)  $ for each $\infty\geq p\geq1$.

However as we will see below this distribution T(f) is not differentiable. So
there is another solution, where the distribution is defined by a r-form
itself : we have $T\left(  \varpi\right)  $ and not T(f):

\bigskip

\begin{theorem}
For an oriented Hausdorff class 1 real manifold M, any Fr\'{e}chet vector
space of functions $V\subset C_{0c}\left(  M,%
\mathbb{C}
\right)  $\ the map : $T:\Lambda_{m}\left(  M;%
\mathbb{C}
\right)  \times V\rightarrow%
\mathbb{C}
::T\left(  \varpi\right)  \left(  \varphi\right)  =\int_{M}\varphi\varpi$
defines a distribution $T\left(  \varpi\right)  $ in V'.
\end{theorem}

\begin{proof}
The Lebesgue measure $\mu$ induced by $\varpi$\ is locally finite, so
$\varphi$ is integrable if it is bounded with compact support.

$\mu$ can be decomposed in 4 real positive Radon measures : $\mu=\left(
\mu_{r+}-\mu_{r-}\right)  +i\left(  \mu_{I+}-\mu_{I-}\right)  $ and for each
$T$ is continuous on $C_{0}\left(  K;%
\mathbb{C}
\right)  $ for any compact K of M so it is continuous on $C_{0c}\left(  M,%
\mathbb{C}
\right)  $
\end{proof}

\bigskip

Warning ! It is clear that $f\neq T\left(  f\right)  $, even\ If is common to
use the same symbol for the function and the associated distribution.\ By
experience this is more confusing than helpful. So we will stick to :

\begin{notation}
T(f) is the distribution associated to a function f or a form by one of the
maps T above, usually obvious in the context.
\end{notation}

And conversely, if for a distribution S there is a function f such that
S=T(f), we say that S is induced by f.

\subsubsection{Support of a distribution\newline}

\begin{definition}
Let V a Fr\'{e}chet space in $C(E;%
\mathbb{C}
).$ The \textbf{support} of a distribution S in V' is defined as the subset of
E, complementary of the largest open O in E such as : $\forall\varphi\in
V,Supp(\varphi)\subset O\Rightarrow S\left(  \varphi\right)  =0.$
\end{definition}

\begin{notation}
(V')$_{c}$ is the set of distributions of V' with compact support
\end{notation}

\begin{definition}
For a Fr\'{e}chet space $V\subset C\left(  E;%
\mathbb{C}
\right)  $ of tests functions, and a subset $\Omega$ of E, the
\textbf{restriction} $S|_{\Omega}$ of a distribution $S\in V^{\prime}$ is the
restriction of S to the subspace of functions : $V\cap C\left(  \Omega;%
\mathbb{C}
\right)  $
\end{definition}

Notice that, contrary to the usual rule for functions, the map : $V^{\prime
}\rightarrow\left(  V\cap C\left(  \Omega;%
\mathbb{C}
\right)  \right)  ^{\prime}$ is neither surjective or injective. This can be
understood by the fact that $V\cap C\left(  \Omega;%
\mathbb{C}
\right)  $ is a smaller space, so the space of its functionals is larger.

\begin{definition}
If for a Fr\'{e}chet space $V\subset C\left(  E;%
\mathbb{C}
\right)  $ of tests functions there is a map : $T:W\rightarrow V^{\prime}$ for
some subspace of functions on E, the \textbf{singular support} of a
distribution S is the subset \ SSup(S) of E, complementary of the largest open
O in E such as : $\exists f\in W\cap C\left(  O;%
\mathbb{C}
\right)  :T\left(  f\right)  =S|_{O}.$
\end{definition}

So S cannot be represented by a function which has its support in SSup(S).

Then :

$SSup(S)=\varnothing\Rightarrow\exists f\in W:T\left(  f\right)  =S$

$SSup(S)\subset Sup(S)$

If $S\in\left(  C_{c\infty}\left(  O;%
\mathbb{C}
\right)  \right)  ^{\prime},f\in C_{\infty}\left(  O;%
\mathbb{C}
\right)  :$

$SSup\left(  fS\right)  =SSup(S)\cap Supp\left(  f\right)  ,SSup\left(
S+T(f)\right)  =SSup(S)$

\begin{theorem}
(Zuily p.120) The set $\left(  C_{\infty c}\left(  O;%
\mathbb{C}
\right)  \right)  _{c}^{\prime}$ of distributions with compact support can be
identified with the set of distributions on the smooth functions : $\left(
C_{\infty}\left(  O;%
\mathbb{C}
\right)  \right)  ^{\prime}\equiv\left(  C_{\infty c}\left(  O;%
\mathbb{C}
\right)  \right)  _{c}^{\prime}$and $\left(  C_{\infty c}\left(
\mathbb{R}
^{m};%
\mathbb{C}
\right)  \right)  _{c}^{\prime}$ is dense in $S\left(
\mathbb{R}
^{m}\right)  ^{\prime}$
\end{theorem}

\begin{theorem}
For any family of distributions $\left(  S_{i}\right)  _{i\in I},$ $S_{i}\in
C_{\infty c}\left(  O_{i};%
\mathbb{C}
\right)  ^{\prime}$\ where $\left(  O_{i}\right)  _{i\in I}$ is an open cover
of O in $%
\mathbb{R}
^{m}$ , such that : $\forall i,j\in I,S_{i}|_{O_{i}\cap O_{j}}=S_{j}%
|_{O_{i}\cap O_{j}}$ there is a unique distribution $S\in C_{\infty c}\left(
O;%
\mathbb{C}
\right)  ^{\prime}$ such that $S|_{O_{i}}=S_{i}$
\end{theorem}

\subsubsection{Product of a function and a distribution}

\begin{definition}
The product of a distribution $S\in V^{\prime}$ where $V\subset C\left(  E;%
\mathbb{C}
\right)  $\ and a function $f\in C\left(  E;%
\mathbb{C}
\right)  $\ is the distribution : $\forall\varphi\in V:\left(  fS\right)
\left(  \varphi\right)  =S\left(  f\varphi\right)  ,$ defined whenever
$f\varphi\in V.$
\end{definition}

The operation $f\varphi$ is the pointwise multiplication : $\left(
f\varphi\right)  \left(  x\right)  =f\left(  x\right)  \varphi\left(
x\right)  $

The product is well defined for :

$S\in C_{rc}\left(  O;%
\mathbb{C}
\right)  ^{\prime},1\leq r\leq\infty$ and $f\in C_{\infty}\left(  O;%
\mathbb{C}
\right)  $

$S\in S\left(
\mathbb{R}
^{m}\right)  ^{\prime}$ and $f$ any polynomial in $%
\mathbb{R}
^{m}$

When the product is well defined : $Supp(fS)\subset Supp(f)\cap Supp(S)$

\subsubsection{Derivative of a distribution}

This is the other important property of distributions, and the main reason for
their use : distributions are smooth. So, using the identification of
functions to distributions, it leads to the concept of derivative "in the
meaning of distributions". However caution is required on two points. First
the "distributional derivative", when really useful (that is when the function
is not itself differentiable) is not a function, and the habit of using the
same symbol for the function and its associated distribution leads quickly to
confusion. Second, the distributional derivative is simple only when the
function is defined over $%
\mathbb{R}
^{m}$. Over a manifold this is a bit more complicated and the issue is linked
to the concept of distribution on a vector bundle seen in the next subsection.

\paragraph{Definition\newline}

\begin{definition}
The r derivative $D_{\alpha_{1}..\alpha_{r}}S$ of a distribution $S\in
V^{\prime}$ on a Fr\'{e}chet space of r differentiable functions on a class r
m dimensional manifold M, is the distribution :%

\begin{equation}
\forall\varphi\in V:\left(  D_{\alpha_{1}..\alpha_{r}}S\right)  \left(
\varphi\right)  =\left(  -1\right)  ^{r}S\left(  D_{\alpha_{1}..\alpha_{r}%
}\varphi\right)
\end{equation}

\end{definition}

Notice that the test functions $\varphi$ must be differentiable.

By construction, the derivative of a distribution is well defined if :
$\forall\varphi\in V:D_{\alpha_{1}..\alpha_{r}}\varphi\in V$ which is the case
for all the common spaces of tests functions. In particular :

If $S\in C_{rc}\left(  O;%
\mathbb{C}
\right)  ^{\prime}$: then $\forall s\leq r:\exists D_{\alpha_{1}..\alpha_{s}%
}S\in C_{rc}\left(  O;%
\mathbb{C}
\right)  ^{\prime}$ and $D_{\alpha_{1}..\alpha_{r}}S\in C_{r+1,c}\left(  O;%
\mathbb{C}
\right)  ^{\prime}$

if $S\in S\left(
\mathbb{R}
^{m}\right)  ^{\prime}$: then $\forall\left(  \alpha_{1}...\alpha_{r}\right)
:\exists D_{\alpha_{1}..\alpha_{r}}S\in S\left(
\mathbb{R}
^{m}\right)  ^{\prime}$

As a consequence if V is a space of r differentiable functions, any
distribution is r differentiable.

\paragraph{Fundamental theorems\newline}

\begin{theorem}
Let V be a Fr\'{e}chet space of r differentiable functions $V\subset
C_{rc}\left(  O,%
\mathbb{C}
\right)  $ on an open subset O of $%
\mathbb{R}
^{m}$. If the distribution S is induced by the integral of a r differentiable
function $f\in C_{rc}\left(  O;%
\mathbb{C}
\right)  $ then we have :%

\begin{equation}
D_{\alpha_{1}..\alpha_{r}}\left(  T\left(  f\right)  \right)  =T\left(
D_{\alpha_{1}..\alpha_{r}}f\right)
\end{equation}

\end{theorem}

meaning that the derivative of the distribution is the distribution induced by
the derivative of the function.

\begin{proof}
The map T reads : $T:W\rightarrow V^{\prime}::T\left(  f\right)  \left(
\varphi\right)  =\int_{O}f\varphi d\xi^{1}\wedge...\wedge d\xi^{m}$

We start with r=1 and $D_{\alpha}=\partial_{\alpha}$\ with $\alpha\in1...m$
and denote: $d\xi=d\xi^{1}\wedge...\wedge d\xi^{m}$

$i_{\partial_{\alpha}}d\xi=\left(  -1\right)  ^{\alpha-1}d\xi^{1}%
\wedge...\wedge\left(  \widehat{d\xi^{\alpha}}\right)  \wedge..\wedge d\xi
^{m}$

$d\left(  i_{\partial_{\alpha}}\varpi\right)  =0$

$\left(  d\varphi\right)  \wedge i_{\partial_{\alpha}}d\xi=\sum_{\beta}\left(
-1\right)  ^{\alpha-1}\left(  \partial_{\beta}\varphi\right)  d\xi^{\beta
}\wedge d\xi^{1}\wedge...\wedge\left(  \widehat{d\xi^{\alpha}}\right)
\wedge..\wedge d\xi^{m}=\left(  \partial_{\beta}\varphi\right)  dx $

$d\left(  f\varphi i_{\partial_{\alpha}}d\xi\right)  =d\left(  f\varphi
\right)  \wedge i_{\partial_{\alpha}}d\xi-\left(  f\varphi\right)  \wedge
d\left(  i_{\partial_{\alpha}}d\xi\right)  =\varphi\left(  df\right)  \wedge
i_{\partial_{\alpha}}d\xi+fd\left(  \varphi\right)  \wedge i_{\partial
_{\alpha}}d\xi=\left(  f\partial_{\alpha}\varphi\right)  d\xi+\left(
\varphi\partial_{\alpha}f\right)  d\xi$

Let N be a manifold with boundary in O. The Stockes theorem gives :

$\int_{N}d\left(  f\varphi i_{\partial_{\alpha}}d\xi\right)  =\int_{\partial
N}f\varphi i_{\partial_{\alpha}}d\xi=\int_{N}\left(  f\partial_{\alpha}%
\varphi\right)  d\xi+\int_{N}\left(  \varphi\partial_{\alpha}f\right)  d\xi$

N can always be taken such that $Supp\left(  \varphi\right)  \subset
\overset{\circ}{N}$ and then :

$\int_{N}\left(  f\partial_{\alpha}\varphi\right)  d\xi+\int_{N}\left(
\varphi\partial_{\alpha}f\right)  d\xi=\int_{O}\left(  f\partial_{\alpha
}\varphi\right)  d\xi+\int_{O}\left(  \varphi\partial_{\alpha}f\right)  d\xi$

$\int_{\partial N}f\varphi i_{\partial_{\alpha}}d\xi=0$ because $Supp\left(
\varphi\right)  \subset\overset{\circ}{N}$

So : $\int_{O}\left(  f\partial_{\alpha}\varphi\right)  d\xi=-\int_{O}\left(
\varphi\partial_{\alpha}f\right)  d\xi\Leftrightarrow T\left(  f\right)
\left(  \partial_{\alpha}\varphi\right)  =-T\left(  \partial_{\alpha}f\right)
\left(  \varphi\right)  =-\partial_{\alpha}T\left(  f\right)  \left(
\varphi\right)  $

$T\left(  \partial_{\alpha}f\right)  \left(  \varphi\right)  =\partial
_{\alpha}T\left(  f\right)  \left(  \varphi\right)  $

By recursion over r we get the result
\end{proof}

The result still holds if $O=%
\mathbb{R}
^{m}$

\bigskip

For a manifold the result is different. If T is defined by : $T\left(
f\right)  \left(  \varphi\right)  =\int_{M}f\varphi\varpi$ where
$\varpi=\varpi_{0}d\xi$ the formula does not hold any more. But we have seen
that it is possible to define a distribution $T\left(  \varpi\right)  \left(
\varphi\right)  =\int_{M}\varphi\varpi$ associated to a m-form, and it is differentiable.

\begin{theorem}
Let $V\subset C_{rc}\left(  M,%
\mathbb{C}
\right)  $ be a Fr\'{e}chet space of r differentiable functions on an oriented
Hausdorff class r real manifold M, If the distribution S is induced by a r
differentiable m form $\varpi=\varpi_{0}d\xi^{1}\wedge...\wedge d\xi^{m}$ then
we have :%

\begin{equation}
\left(  D_{\alpha_{1}...\alpha_{r}}T\left(  \varpi\right)  \right)  \left(
\varphi\right)  =\left(  -1\right)  ^{r}T\left(  \left(  D_{\alpha
_{1}...\alpha_{r}}\varpi_{0}\right)  d\xi^{1}\wedge...\wedge d\xi^{m}\right)
\left(  \varphi\right)
\end{equation}

\end{theorem}

By r differentiable m form we mean that the function $\varpi_{0}$ is r
differentiable on each of its domain ($\varpi_{0}$ changes according to the
rule : $\varpi_{b0}=\det\left[  J_{ba}\right]  \varpi_{a0})$

\begin{proof}
The map T reads : $T:W\rightarrow V^{\prime}::T\left(  \varpi\right)  \left(
\varphi\right)  =\int_{M}\varphi\varpi$

We start with r=1 and $D_{\alpha}=\partial_{\alpha}$\ with $\alpha\in1...m$
and denote: $d\xi^{1}\wedge...\wedge d\xi^{m}=d\xi,\varpi=\varpi_{0}d\xi$

$i_{\partial_{\alpha}}\varpi=\left(  -1\right)  ^{\alpha-1}\varpi_{0}d\xi
^{1}\wedge...\wedge\left(  \widehat{d\xi^{\alpha}}\right)  \wedge..\wedge
d\xi^{m}$

$d\left(  i_{\partial_{\alpha}}\varpi\right)  =\sum_{\beta}\left(  -1\right)
^{\alpha-1}\left(  \partial_{\beta}\varpi_{0}\right)  d\xi^{\beta}\wedge
d\xi^{1}\wedge...\wedge\left(  \widehat{d\xi^{\alpha}}\right)  \wedge..\wedge
d\xi^{m}=\left(  \partial_{\alpha}\varpi_{0}\right)  d\xi$

$\left(  d\varphi\right)  \wedge i_{\partial_{\alpha}}\varpi=\sum_{\beta
}\left(  -1\right)  ^{\alpha-1}\left(  \partial_{\beta}\varphi\right)
\varpi_{0}d\xi^{\beta}\wedge d\xi^{1}\wedge...\wedge\left(  \widehat
{d\xi^{\alpha}}\right)  \wedge..\wedge d\xi^{m}=\left(  \partial_{\beta
}\varphi\right)  \varpi_{0}d\xi=\left(  \partial_{\beta}\varphi\right)
\varpi$

$d\left(  \varphi i_{\partial_{\alpha}}\varpi\right)  =d\varphi\wedge
i_{\partial_{\alpha}}\varpi-\varphi d\left(  i_{\partial_{\alpha}}%
\varpi\right)  =\left(  \partial_{\alpha}\varphi\right)  \varpi-\varphi\left(
\partial_{\alpha}\varpi_{0}\right)  d\xi$

Let N be a manifold with boundary in M. The Stockes theorem gives :

$\int_{N}d\left(  \varphi i_{\partial_{\alpha}}\varpi\right)  =\int_{\partial
N}\varphi i_{\partial_{\alpha}}\varpi=\int_{N}\left(  \partial_{\alpha}%
\varphi\right)  \varpi-\int_{N}\varphi\left(  \partial_{\alpha}\varpi
_{0}\right)  d\xi$

N can always be taken such that $Supp\left(  \varphi\right)  \subset
\overset{\circ}{N}$ and then :$\int_{\partial N}\varphi i_{\partial_{\alpha}%
}\varpi=0$

$\int_{M}\left(  \partial_{\alpha}\varphi\right)  \varpi=\int_{M}%
\varphi\left(  \partial_{\alpha}\varpi_{0}\right)  d\xi$

$\left(  \partial_{\alpha}T\left(  \varpi\right)  \right)  \left(
\varphi\right)  =-T\left(  \partial_{\alpha}\varpi_{0}d\xi\right)  \left(
\varphi\right)  $

By recursion over r we get :

$\left(  D_{\alpha_{1}...\alpha_{r}}T\left(  \varpi\right)  \right)  \left(
\varphi\right)  =\left(  -1\right)  ^{r}T\left(  D_{\alpha_{1}...\alpha_{r}%
}\varpi_{0}d\xi\right)  \left(  \varphi\right)  $
\end{proof}

This is why the introduction of m form is useful. However the factor $\left(
-1\right)  ^{r}$ is not satisfactory. It comes from the very specific case of
functions over $%
\mathbb{R}
^{m}$, and in the next subsection we have a better solution.

\paragraph{Derivative "in the meaning of distributions"\newline}

For a function f defined in $%
\mathbb{R}
^{m},$ which is not differentiable, but is such that there is some
distribution T(f) which is differentiable, its derivative "in the sense of
distributions" (or distributional derivative) is the derivative of the
distribution $D_{\alpha}T\left(  f\right)  $ sometimes denoted $\left\{
D_{\alpha}f\right\}  $ and more often simply $D_{\alpha}f$ . However the
distributional derivative of f is represented by a function iff f itself is differentiable:

\begin{theorem}
(Zuily p.53) Let $S\in C_{\infty c}\left(  O;%
\mathbb{C}
\right)  ^{\prime}$ with O an open in $%
\mathbb{R}
^{m}$\ .\ The following are equivalent :

i) $\exists f\in C_{r}\left(  O;%
\mathbb{C}
\right)  :S=T(f)$

ii) $\forall\alpha_{1}..\alpha_{s}=1...m,s=0...r:\exists g\in C_{0}\left(  O;%
\mathbb{C}
\right)  :\partial_{\alpha_{1}...\alpha_{s}}S=T\left(  g\right)  $
\end{theorem}

So, if we can extend extend "differentiability" to many functions which
otherwise are not differentiable, we must keep in mind that usually $\left\{
D_{\alpha}f\right\}  $\ is not a function, and is defined with respect to some
map T and some measure.

Expressions like "a function f such that its distributional derivatives belong
to $L^{p}$" are common. They must be interpreted as "f is such that
$D_{\alpha}\left(  T\left(  f\right)  \right)  =T\left(  g\right)  $ with
$g\in L^{p}"$ . But if the distributional derivative of f is represented by a
function, it means that f is differentiable, and so it would be simpler and
clearer to say "f such that $D_{\alpha}f\in L^{p}$"\ .

Because of all these problems we will always stick to the notation T(f) to
denote the distribution induced by a function.

In the most usual cases we have the following theorems :\ 

\begin{theorem}
If $f\in L_{loc}^{1}\left(
\mathbb{R}
^{m},dx,%
\mathbb{C}
\right)  $ is locally integrable :

$\forall\left(  \alpha_{1}...\alpha_{r}\right)  :\exists D_{\alpha_{1}%
..\alpha_{r}}\left(  T\left(  f\right)  \right)  \in C_{\infty c}\left(
\mathbb{R}
^{m};%
\mathbb{C}
\right)  ^{\prime}$
\end{theorem}

\begin{theorem}
(Lieb p.175) Let O be an open in $%
\mathbb{R}
^{m}.$ If $f\in H^{1}\left(  O\right)  ,g\in C_{\infty}\left(  O;%
\mathbb{C}
\right)  $ and has bounded derivatives, then :

$f\times g\in H^{1}\left(  O\right)  $ and $\partial_{\alpha}T\left(
fg\right)  \in C_{\infty c}\left(  O;%
\mathbb{C}
\right)  ^{\prime}$
\end{theorem}

\begin{theorem}
(Zuily p.39) $S\in C_{\infty c}\left(
\mathbb{R}
;%
\mathbb{C}
\right)  ^{\prime},a\in C_{\infty}\left(
\mathbb{R}
;%
\mathbb{C}
\right)  ,f\in C_{0}\left(
\mathbb{R}
;%
\mathbb{C}
\right)  $

$\frac{dS}{dx}+aS=T\left(  f\right)  \Leftrightarrow\exists g\in C_{1}\left(
\mathbb{R}
;%
\mathbb{C}
\right)  :S=T\left(  g\right)  $
\end{theorem}

\paragraph{Properties of the derivative of a distribution\newline}

\begin{theorem}
Support : For a distribution S, whenever the derivative $D_{\alpha_{1}%
..\alpha_{r}}S$ exist : $Supp\left(  D_{\alpha_{1}..\alpha_{r}}S\right)
\subset Supp\left(  S\right)  $
\end{theorem}

\begin{theorem}
Leibnitz rule: For a distribution S and a function f, whenever the product and
the derivative exist :%

\begin{equation}
\partial_{\alpha}\left(  fS\right)  =f\partial_{\alpha}S+\left(
\partial_{\alpha}f\right)  S
\end{equation}

\end{theorem}

Notice that for f this is the usual derivative.

\begin{theorem}
Chain rule (Lieb p.152): Let $O$ open in $%
\mathbb{R}
^{m},y=\left(  y_{k}\right)  _{k=1}^{n},y_{k}\in W_{loc}^{1,p}\left(
O\right)  ,F\in C_{1}\left(
\mathbb{R}
^{n};%
\mathbb{C}
\right)  $ with bounded derivative,\ then :

$\frac{\partial}{\partial x_{j}}T\left(  F\circ y\right)  =\sum_{k}%
\frac{\partial F}{\partial y_{k}}\frac{\partial}{\partial x_{j}}T\left(
y_{k}\right)  $
\end{theorem}

\begin{theorem}
Convergence : If the sequence of distributions $\left(  S_{n}\right)  _{n\in%
\mathbb{N}
}\in V^{%
\mathbb{N}
}$ converges to S in V', and have derivatives in V', then $D_{\alpha}%
S_{n}\rightarrow D_{\alpha}S$
\end{theorem}

\begin{theorem}
Local structure of distributions (Zuily p.76): For any distribution S in
$C_{\infty c}\left(  O;%
\mathbb{C}
\right)  ^{\prime}$ with O an open in $%
\mathbb{R}
^{m},$ and any compact K in O, there is a finite family $\left(  f_{i}\right)
_{i\in I}\in C_{0}\left(  O;%
\mathbb{C}
\right)  ^{I}:S|_{K}=\sum_{i\in I,\alpha_{1}...\alpha_{r}}D_{\alpha
_{1}...\alpha_{r}}T\left(  f_{i}\right)  $
\end{theorem}

As a consequence a distribution whose derivatives are null is constant :

$S\in C_{\infty c}\left(
\mathbb{R}
^{m};%
\mathbb{C}
\right)  ^{\prime},\alpha=1..m:\frac{\partial}{\partial x^{\alpha}%
}S=0\Leftrightarrow S=Cst$

\begin{theorem}
(Lieb p.145) Let O be an open in $%
\mathbb{R}
^{m},S\in C_{\infty c}\left(  O;%
\mathbb{C}
\right)  ^{\prime},\varphi\in C_{\infty c}\left(  O;%
\mathbb{C}
\right)  ,y\in%
\mathbb{R}
^{m}$ such that $\forall t\in\left[  0,1\right]  :ty\in O$ then :

$S\left(  \psi\right)  -S\left(  \varphi\right)  =\int_{0}^{1}\sum_{k=1}%
^{m}y_{k}\partial_{k}S\left(  \varphi\left(  ty\right)  \right)  dt$

with $\psi:\left\{  ty,t\in\left[  0,1\right]  \right\}  \rightarrow%
\mathbb{C}
::\psi\left(  x\right)  =\varphi\left(  ty\right)  $

If $f\in W_{loc}^{1,1}\left(
\mathbb{R}
^{m}\right)  $ then : $\forall\varphi\in C_{\infty c}\left(  O;%
\mathbb{C}
\right)  :$

$\int_{O}\left(  f\left(  x+y\right)  -f\left(  x\right)  \right)
\varphi\left(  x\right)  dx=\int_{O}\left(  \int_{0}^{1}\sum_{k=1}^{m}%
y_{k}\partial_{k}f\left(  x+ty\right)  dt\right)  \varphi\left(  x\right)  dx$
\end{theorem}

\begin{theorem}
Jumps formula (Zuily p.40) Let $f\in C\left(
\mathbb{R}
;%
\mathbb{R}
\right)  $ be a function continuous except at the isolated points $a_{i}$
where it is semi-continuous : $\exists\lim_{\epsilon\rightarrow0}f\left(
a_{i}\pm\epsilon\right)  $. Then the derivative of f reads : $\frac{d}%
{dx}T\left(  f\right)  =f^{\prime}+\sum_{i}\sigma_{i}\delta_{a_{i}}$ with
$\sigma_{i}=\lim_{\epsilon\rightarrow0}f\left(  a_{i}+\epsilon\right)
-\lim_{\epsilon\rightarrow0}f\left(  a_{i}-\epsilon\right)  $ and f' the usual
derivative where it is well defined.
\end{theorem}

\subsubsection{Heaviside and Dirac functions}

\paragraph{Definitions\newline}

\begin{definition}
For any Fr\'{e}chet space V\ in C(E;$%
\mathbb{C}
)$ the \textbf{Dirac's distribution} is : $\delta_{a}:V\rightarrow%
\mathbb{C}
::\delta_{a}\left(  \varphi\right)  =\varphi\left(  a\right)  $ where $a\in E.
$ Its derivatives are : $D_{\alpha_{1}...\alpha_{r}}\delta_{a}\left(
\varphi\right)  =\left(  -1\right)  ^{r}D_{\alpha_{1}...\alpha_{r}}%
\varphi|_{x=a}$
\end{definition}

Warning ! If $E\neq%
\mathbb{R}
^{m}$ usually $\delta_{0}$ is meaningless

\begin{definition}
The \textbf{Heaviside function} on $%
\mathbb{R}
$ is the function given by $H\left(  x\right)  =0,x\leq0,H(x)=1,x>0$
\end{definition}

\paragraph{General properties\newline}

The Heaviside function is locally integrable and the distribution T(H) is
given in $C_{\infty c}\left(
\mathbb{R}
;%
\mathbb{C}
\right)  $ by : T(H)$\left(  \varphi\right)  =\int_{0}^{\infty}\varphi dx.$ It
is easy to see that : $\frac{d}{dx}T\left(  H\right)  =\delta_{0}.$ And we can
define the Dirac \textit{function} : $\delta:%
\mathbb{R}
\rightarrow%
\mathbb{R}
::\delta\left(  0\right)  =1,x\neq0:\delta\left(  x\right)  =0.$ So $\left\{
\frac{d}{dx}H\right\}  =T\left(  \delta\right)  =\delta_{0}.$ But there is no
function such that $T(F)=\frac{d^{2}}{dx^{2}}T\left(  H\right)  $

\bigskip

\begin{theorem}
Any distribution S in $C_{\infty}\left(  O;%
\mathbb{C}
\right)  ^{\prime}$ which has a support limited to a unique point a has the
form :

$S\left(  \varphi\right)  =c_{0}\delta_{a}\left(  \varphi\right)
+\sum_{\alpha}c_{\alpha}D_{\alpha}\varphi|_{a}$ with $c_{0},c_{\alpha}\in%
\mathbb{C}
$
\end{theorem}

If $S\in C_{\infty c}\left(  O;%
\mathbb{C}
\right)  ^{\prime},xS=0\Rightarrow S=\delta_{0}$

If $f\in L^{1}\left(
\mathbb{R}
^{m},dx,%
\mathbb{C}
\right)  ,\varepsilon>0:\lim_{\varepsilon\rightarrow0}\varepsilon^{-m}f\left(
\varepsilon^{-1}x\right)  =\delta_{0}\int_{%
\mathbb{R}
^{m}}f\left(  x\right)  dx$

\paragraph{Laplacian\newline}

On $C_{\infty}\left(
\mathbb{R}
^{m};%
\mathbb{R}
\right)  $ the laplacian is the differential operator : $\Delta=\sum
_{\alpha=1}^{m}\frac{\partial^{2}}{\left(  \partial x^{\alpha}\right)  ^{2}}$

The function : $f\left(  x\right)  =\left(  \sum_{i}\left(  x^{i}%
-a^{i}\right)  ^{2}\right)  ^{1-\frac{m}{2}}=\frac{1}{r^{m-2}}$ is such that

For $m\geq3:\Delta\left(  T\left(  f\right)  \right)  =\left(  2-m\right)
A\left(  S_{m-1}\right)  \delta_{a}$ where $A\left(  S_{m-1}\right)  $ is the
Lebesgue surface of the unit sphere in $%
\mathbb{R}
^{m}.$

For m=3 : $\Delta T\left(  \left(  \sum_{i}\left(  x^{i}-a^{i}\right)
^{2}\right)  ^{-1/2}\right)  =-4\pi\delta_{a}$

For m=2 : $\Delta T\left(  \ln\left(  \sum_{i}\left(  x^{i}-a^{i}\right)
^{2}\right)  ^{1/2}\right)  =2\pi\delta_{a}$

\bigskip

\begin{theorem}
(Taylor 1 p.210) If $S\in S\left(
\mathbb{R}
^{m}\right)  ^{\prime}$ is such that $\Delta S=0$ then S=T$\left(  f\right)  $
with f a polynomial in $%
\mathbb{R}
^{m}$
\end{theorem}

\subsubsection{Tensorial product of distributions}

\begin{theorem}
(Zuily p.18,64) For any open subsets $O_{1},O_{2}\subset%
\mathbb{R}
^{m}:$

i) any function in $C_{\infty c}\left(  O_{1}\times O_{2};%
\mathbb{C}
\right)  $ is the limit of a sequence of functions of $C_{\infty c}\left(
O_{1};%
\mathbb{C}
\right)  \otimes C_{\infty c}\left(  O_{2};%
\mathbb{C}
\right)  $

ii) for any distributions $S_{1}\in C_{\infty c}\left(  O_{1};%
\mathbb{C}
\right)  ^{\prime},S_{2}\in C_{\infty c}\left(  O_{2};%
\mathbb{C}
\right)  ^{\prime}$ there is a unique distribution $S_{1}\otimes S_{2}\in
C_{\infty c}\left(  O_{1}\times O_{2};%
\mathbb{C}
\right)  ^{\prime}$\ such that :

$\forall\varphi_{1}\in C_{\infty c}\left(  O_{1};%
\mathbb{C}
\right)  ,\varphi_{2}\in C_{\infty c}\left(  O_{2};%
\mathbb{C}
\right)  :S_{1}\otimes S_{2}\left(  \varphi_{1}\otimes\varphi_{2}\right)
=S_{1}\left(  \varphi_{1}\right)  S_{2}\left(  \varphi_{2}\right)  $

iii) $\forall\varphi\in C_{\infty c}\left(  O_{1}\times O_{2};%
\mathbb{C}
\right)  :S_{1}\otimes S_{2}\left(  \varphi\right)  =S_{1}\left(  S_{2}\left(
\varphi\left(  x_{1},.\right)  \right)  \right)  =S_{2}\left(  S_{1}\left(
\varphi\left(  .,x_{2}\right)  \right)  \right)  $

iv) If $S_{1}=T\left(  f_{1}\right)  ,S_{2}=T\left(  f_{2}\right)  $ then
$S_{1}\otimes S_{2}=T\left(  f_{1}\right)  \otimes T\left(  f_{2}\right)
=T\left(  f_{1}\otimes f_{2}\right)  $

v) $\frac{\partial}{\partial x_{1}^{\alpha}}\left(  S_{1}\otimes S_{2}\right)
=\left(  \frac{\partial}{\partial x_{1}^{\alpha}}S_{1}\right)  \otimes S_{2}$
\end{theorem}

\begin{theorem}
Schwartz kernel theorem (Taylor 1 p.296) : let M,N be compact finite
dimensional real manifolds. $L:C_{\infty}\left(  M;%
\mathbb{C}
\right)  \rightarrow C_{\infty c}\left(  N;%
\mathbb{C}
\right)  ^{\prime}$ a continuous linear map, B the bilinear map :
$B:C_{\infty}\left(  M;%
\mathbb{C}
\right)  \times C_{\infty}\left(  N;%
\mathbb{C}
\right)  \rightarrow%
\mathbb{C}
::B\left(  \varphi,\psi\right)  =L\left(  \varphi\right)  \left(  \psi\right)
$\ separately continuous in each factor. Then there is a distribution : $S\in
C_{\infty c}\left(  M\times N;%
\mathbb{C}
\right)  ^{\prime}$ such that :

$\forall\varphi\in C_{\infty}\left(  M;%
\mathbb{C}
\right)  ,\psi\in C_{\infty}\left(  N;%
\mathbb{C}
\right)  :S\left(  \varphi\otimes\psi\right)  =B\left(  \varphi,\psi\right)  $
\end{theorem}

\subsubsection{Convolution of distributions}

\paragraph{Definition\newline}

Convolution of distributions are defined so as we get back the convolution of
functions when a distribution is induced by a function :%

\begin{equation}
T\left(  f\right)  \ast T\left(  g\right)  =T\left(  f\ast g\right)
\end{equation}

It is defined only for functions on $%
\mathbb{R}
^{m}.$

\begin{definition}
The \textbf{convolution of the distributions} $S_{1},S_{2}\in C_{rc}\left(
\mathbb{R}
^{m};%
\mathbb{C}
\right)  ^{\prime}$ is the distribution $S_{1}\ast S_{2}\in C_{rc}\left(
\mathbb{R}
^{m};%
\mathbb{C}
\right)  ^{\prime}:$

$\forall\varphi\in C_{\infty c}\left(
\mathbb{R}
^{m};%
\mathbb{C}
\right)  ::\left(  S_{1}\ast S_{2}\right)  \left(  \varphi\right)  =\left(
S_{1}\left(  x_{1}\right)  \otimes S_{2}\left(  x_{2}\right)  \right)  \left(
\varphi\left(  x_{1}+x_{2}\right)  \right)  $%

\begin{equation}
\left(  S_{1}\ast S_{2}\right)  \left(  \varphi\right)  =S_{1}\left(
S_{2}\left(  \varphi\left(  x_{1}+x_{2}\right)  \right)  \right)
=S_{2}\left(  S_{1}\left(  \varphi\left(  x_{1}+x_{2}\right)  \right)
\right)
\end{equation}

It is well defined when at least one of the distributions has a compact
support.\ If both have compact support then $S_{1}\ast S_{2}$\ has a compact support.
\end{definition}

The condition still holds for the product of more than two distributions : all
but at most one must have compact support.

If $S_{1},S_{2}\in C_{\infty c}\left(
\mathbb{R}
^{m};%
\mathbb{C}
\right)  ^{\prime}$ and their support meets the condition:

$\forall R>0,\exists\rho:x_{1}\in SupS_{1},x_{2}\in SupS_{2},\left\Vert
x_{1}+x_{2}\right\Vert \leq R\Rightarrow\left\Vert x_{1}\right\Vert \leq
\rho,\left\Vert x_{2}\right\Vert \leq\rho$

then the convolution $S_{1}\ast S_{2}$ is well defined, even if the
distributions are not compacly supported.

If the domain of the functions is some relatively compact open O in $%
\mathbb{R}
^{m}$ we can always consider them as compactly supported functions defined in
the whole of $%
\mathbb{R}
^{m}$\ by : $x\notin O:\varphi\left(  x\right)  =0$

\paragraph{Properties\newline}

\begin{theorem}
Convolution of distributions, when defined, is associative, commutative. With
convolution $C_{rc}\left(
\mathbb{R}
^{m};%
\mathbb{C}
\right)  _{c}^{\prime}$ is an unital commutative algebra with unit element the
Dirac distribution $\delta_{0}$
\end{theorem}

\begin{theorem}
Over $C_{rc}\left(
\mathbb{R}
^{m};%
\mathbb{C}
\right)  _{c}^{\prime}$ the derivative of the convolution product is :%

\begin{equation}
D_{\alpha_{1}...\alpha_{r}}\left(  S\ast U\right)  =\left(  D_{\alpha
_{1}...\alpha_{r}}S\right)  \ast U=S\ast D_{\alpha_{1}...\alpha_{r}}U
\end{equation}

\end{theorem}

\begin{theorem}
If the sequence $\left(  S_{n}\right)  _{\in%
\mathbb{N}
}\in\left(  C_{rc}\left(
\mathbb{R}
^{m};%
\mathbb{C}
\right)  _{c}^{\prime}\right)  ^{%
\mathbb{N}
}$ converges to S then $\forall U\in C_{rc}\left(
\mathbb{R}
^{m};%
\mathbb{C}
\right)  ^{\prime}:S_{n}\ast U\rightarrow S\ast U$ in $C_{rc}\left(
\mathbb{R}
^{m};%
\mathbb{C}
\right)  _{c}^{\prime}$
\end{theorem}

\paragraph{Convolution of distributions induced by a function\newline}

\begin{theorem}
(Zuily p.69) The convolution of a distribution and a distribution induced by a
function gives a distribution induced by a function. If both distribution and
function are compactly supported then the product is a compactly supported function.
\end{theorem}

\begin{equation}
S\ast T\left(  f\right)  =T\left(  S_{y}\left(  f\left(  x-y\right)  \right)
\right)
\end{equation}

and $S_{y}\left(  f\left(  x-y\right)  \right)  \in C_{r}\left(
\mathbb{R}
^{m};%
\mathbb{C}
\right)  $

With :

$S\in C_{rc}\left(
\mathbb{R}
^{m};%
\mathbb{C}
\right)  ^{\prime},f\in C_{rc}\left(
\mathbb{R}
^{m};%
\mathbb{C}
\right)  $ \ 

or $S\in C_{rc}\left(
\mathbb{R}
^{m};%
\mathbb{C}
\right)  _{c}^{\prime},f\in C_{r}\left(
\mathbb{R}
^{m};%
\mathbb{C}
\right)  $

In particular : $\delta_{a}\ast T\left(  f\right)  =T\left(  \delta_{a}\left(
t\right)  f\left(  x-t\right)  \right)  =T\left(  f\left(  x-a\right)
\right)  $

$Supp(S\ast T(f))\subset Supp(S)+Supp(T(f))$

$SSup(S\ast U)\subset SSup(S)+SSup(U)$

\subsubsection{Pull back of a distribution}

\paragraph{Definition\newline}

\begin{theorem}
(Zuily p.82) If $O_{1}$ is an open in $%
\mathbb{R}
^{m},O_{2}$ an open in $%
\mathbb{R}
^{n},$ $F:O_{1}\rightarrow O_{2}$ a smooth submersion, there is a map
$F^{\ast}:C_{\infty c}\left(  O_{2};%
\mathbb{C}
\right)  ^{\prime}\rightarrow C_{\infty c}\left(  O_{1};%
\mathbb{C}
\right)  ^{\prime}$ such that : $\forall f_{2}\in C_{0}\left(  O_{2};%
\mathbb{C}
\right)  ,$ $F^{\ast}T\left(  f_{2}\right)  $ is the unique\ functional
$F^{\ast}T\left(  f_{2}\right)  \in C_{\infty c}\left(  O_{1};%
\mathbb{C}
\right)  ^{\prime}$ such that
\begin{equation}
F^{\ast}T\left(  f_{2}\right)  \left(  \varphi_{1}\right)  =T\left(
f_{2}\circ F\right)  \left(  \varphi_{1}\right)  =T\left(  F^{\ast}%
f_{2}\right)  \left(  \varphi_{1}\right)
\end{equation}

\end{theorem}

F is a submersion = F is differentiable and rank $F^{\prime}(p)=n\leq m$.

The definition is chosen so that the pull back of the distribution given by
$f_{2}$ is the distribution given by the pull back of $f_{2}.$ So this is the
assimilation with functions which leads the way. But the map F* is valid for
any distribution.

\paragraph{Properties\newline}

\begin{theorem}
(Zuily p.82) The pull back of distributions has the following properties :

i) $Supp\left(  F^{\ast}S_{2}\right)  \subset F^{-1}\left(  Supp\left(
S_{2}\right)  \right)  $

ii) $F^{\ast}S_{2}$ is a positive distribution if S$_{2}$ is positive

iii) $\frac{\partial}{\partial x_{1}^{\alpha}}F^{\ast}S_{2}=\sum_{\beta=1}%
^{n}\frac{\partial F_{\beta}}{\partial x_{1}^{\alpha}}F^{\ast}\left(
\frac{\partial S_{2}}{\partial x_{2}^{\beta}}\right)  $

iv) $g_{2}\in C_{\infty}\left(  O_{2};%
\mathbb{C}
\right)  :F^{\ast}\left(  g_{2}S_{2}\right)  =\left(  g_{2}\circ F\right)
F^{\ast}S_{2}=\left(  F^{\ast}g_{2}\right)  \times F^{\ast}S_{2}$

v) if $G:O_{2}\rightarrow O_{3}$ is a submersion, then $\left(  G\circ
F\right)  ^{\ast}S_{3}=F^{\ast}\left(  G^{\ast}S_{3}\right)  $

vi) If the sequence $\left(  S_{n}\right)  _{\in%
\mathbb{N}
}\in\left(  C_{\infty c}\left(  O_{2};%
\mathbb{C}
\right)  ^{\prime}\right)  ^{%
\mathbb{N}
}$ converges to S then $F^{\ast}S_{n}\rightarrow F^{\ast}S$
\end{theorem}

\paragraph{Pull back by a diffeomorphism\newline}

\begin{theorem}
If $O_{1},O_{2}$ are open subsets in $%
\mathbb{R}
^{m},$ $F:O_{1}\rightarrow O_{2}$ a smooth diffeomorphism, $V_{1}\subset
C\left(  O_{1};%
\mathbb{C}
\right)  ,V_{2}\subset C\left(  O_{2};%
\mathbb{C}
\right)  $ two Fr\'{e}chet spaces, the pull back on $V_{1}^{\prime}$ of a
distribution $S_{2}\in V_{2}^{\prime}$ by a smooth diffeomorphism is the map :%

\begin{equation}
F^{\ast}S_{2}\left(  \varphi_{1}\right)  =\left\vert \det F^{\prime
}\right\vert ^{-1}S_{2}\left(  \varphi_{1}\circ F^{-1}\right)
\end{equation}

\end{theorem}

$\forall\varphi_{1}\in C_{\infty c}\left(  O_{1};%
\mathbb{C}
\right)  ,f_{2}\in C_{0}\left(  O_{2};%
\mathbb{C}
\right)  :$

$F^{\ast}T\left(  f_{2}\right)  \left(  \varphi_{1}\right)  =\int_{O_{1}}%
f_{2}\left(  F\left(  x_{1}\right)  \right)  \varphi_{1}\left(  x_{1}\right)
dx_{1}$

$=\int_{O_{2}}f_{2}\left(  x_{2}\right)  \varphi_{1}\left(  F^{-1}\left(
x_{2}\right)  \right)  \left\vert \det F^{\prime}\left(  x_{2}\right)
\right\vert ^{-1}dx_{2}$

$\Leftrightarrow$ $F^{\ast}T\left(  f_{2}\right)  \left(  \varphi_{1}\right)
=T\left(  f_{2}\right)  \left(  \varphi_{1}\left(  F^{-1}\left(  x_{2}\right)
\right)  \left\vert \det F^{\prime}\left(  x_{2}\right)  \right\vert
^{-1}\right)  $

Notice that we have the absolute value of $\det F^{\prime}\left(
x_{2}\right)  ^{-1}:$ this is the same formula as for the change of variable
in the Lebesgue integral.

We have all the properties listed above. Moreover, if $S\in S\left(
\mathbb{R}
^{m}\right)  ^{\prime}$ then $F^{\ast}S\in S\left(
\mathbb{R}
^{m}\right)  ^{\prime}$

\paragraph{Applications\newline}

For distributions $S\in C_{\infty c}\left(
\mathbb{R}
^{m};%
\mathbb{C}
\right)  ^{\prime}$

1. Translation :

$\tau_{a}(x)=x-a$

$\tau_{a}^{\ast}S\left(  \varphi\right)  =S_{y}\left(  \varphi\left(
y+a\right)  \right)  =\int_{O_{2}}S_{y}(y)\varphi\left(  y+a\right)  dy$

$\delta_{a}=\tau_{a}^{\ast}\delta_{0}$

2. Similitude : $,$

$k\neq0\in R:\lambda_{k}(x)=kx$

$\lambda_{k}^{\ast}S\left(  \varphi\right)  =\frac{1}{\left\vert k\right\vert
}S_{y}\left(  \varphi\left(  \frac{y}{k}\right)  \right)  =\frac{1}{\left\vert
k\right\vert }\int_{O_{2}}S_{y}(y)\varphi\left(  \frac{y}{k}\right)  dy$

3. Reflexion :

$R(x)=-x$

$R^{\ast}S\left(  \varphi\right)  =S_{y}\left(  \varphi\left(  -y\right)
\right)  =\int_{O_{2}}S_{y}(y)\varphi\left(  -y\right)  dy$

\subsubsection{Miscellaneous operations with distributions\newline}

\paragraph{Homogeneous distribution\newline}

\begin{definition}
$S\in C_{\infty c}\left(
\mathbb{R}
^{m};%
\mathbb{C}
\right)  ^{\prime}$ is \textbf{homogeneous} of degree r if

$\forall k>0$ : $S\left(  \varphi\left(  kx\right)  \right)  =k^{-m-r}S\left(
\varphi(x)\right)  $
\end{definition}

The definition coincides with the usual is $S=T\left(  f\right)  $

S is homogeneous of degree r iff $\sum_{\alpha=1}^{m}x^{\alpha}\dfrac{\partial
S}{\partial x^{\alpha}}=rS$

\paragraph{Distribution independant with respect to a variable\newline}

Let us define the translation along the canonical vector $e_{i}$ of $%
\mathbb{R}
^{m}$ as :

$h\in%
\mathbb{R}
,\tau_{i}\left(  h\right)  :%
\mathbb{R}
^{m}\rightarrow%
\mathbb{R}
^{m}::\tau_{i}\left(  h\right)  \left(  \sum_{\alpha=1}^{m}x^{\alpha}%
e_{\alpha}\right)  =\sum_{\alpha=1}^{m}x^{\alpha}e_{\alpha}-he_{i}$

A distribution $S\in C_{\infty c}\left(
\mathbb{R}
^{m};%
\mathbb{C}
\right)  $ is said to be independant with respect to the variable i if
$\forall h:\tau_{i}\left(  h\right)  ^{\ast}S=S\Leftrightarrow\frac{\partial
S}{\partial x^{i}}=0$

\paragraph{Principal value\newline}

1. Principal value of a function :

let $f:%
\mathbb{R}
\rightarrow%
\mathbb{C}
$ be some function such that :

either $\int_{-\infty}^{+\infty}f\left(  x\right)  dx$ is not defined but
$\exists\lim_{X\rightarrow\infty}\int_{-X}^{+X}f(x)dx$

or $\int_{a}^{b}f\left(  x\right)  dx$ is not defined, but $\exists
\lim_{\varepsilon\rightarrow0}\left(  \int_{a}^{c-\varepsilon}f(x)dx+\int
_{c+\varepsilon}^{b}f(x)dx\right)  $

then we define the principal value integral : $pv\int f(x)dx$ as the limit

2. The \textbf{principal value} distribution is defined as $pv\left(  \frac
{1}{x}\right)  \in C_{\infty c}\left(
\mathbb{R}
;%
\mathbb{C}
\right)  ^{\prime}::$

$pv\left(  \frac{1}{x}\right)  \left(  \varphi\right)  =\lim_{\varepsilon
\rightarrow0}\left(  \int_{-\infty}^{-\varepsilon}\frac{f(x)}{x}%
dx+\int_{+\varepsilon}^{\infty}\frac{f(x)}{x}dx\right)  $

If $S\in C_{\infty c}\left(
\mathbb{R}
;%
\mathbb{C}
\right)  ^{\prime}:xS=1$ then $S=pv\left(  \frac{1}{x}\right)  +C\delta_{0}$
and we have : $\frac{d}{dx}T\left(  \ln\left\vert x\right\vert \right)
=pv\left(  \frac{1}{x}\right)  $

3. It can be generalized as follows (Taylor 1 p.241):

Take $f\in C_{\infty}\left(  S^{m-1};%
\mathbb{C}
\right)  $ , where $S^{m-1}$ is the unit sphere of $%
\mathbb{R}
^{m}$ , such that : $\int_{S^{m-1}}f\varpi_{0}=0$ with the Lebesgue volume
form on $S^{m-1}.$ Define : $f_{n}\left(  x\right)  =\left\Vert x\right\Vert
^{-n}f\left(  \frac{x}{\left\Vert x\right\Vert }\right)  ,x\neq0,n\geq-m$

then : $T\left(  f_{n}\right)  :T\left(  f_{n}\right)  \left(  \varphi\right)
=\int_{S^{m-1}}f_{n}\varphi\varpi_{0}$ is a homogeneous distribution in
S'$\left(
\mathbb{R}
^{m}\right)  $ called the principal value of $f_{n}:T\left(  f_{n}\right)
=pv\left(  f_{n}\right)  $

\subsubsection{Distributions depending on a parameter}

\paragraph{Definition\newline}

1. Let V be Fr\'{e}chet space of functions on some manifold M, J an open
interval in $%
\mathbb{R}
$\ \ and a map : $S:J\rightarrow F^{\prime}::S\left(  t\right)  $ is a family
of distributions acting on V.

2. The result of the action on V is a function $S\left(  t\right)  \left(
\varphi\right)  $ on J. So we can consider the map : $\widetilde{S}:J\times
V\rightarrow%
\mathbb{C}
::S\left(  t\right)  \left(  \varphi\right)  $ and the existence of a partial
derivative $\frac{\partial}{\partial t}$ ot this map with respect to the
parameter t. We say that S is of class r if : $\forall\varphi\in
V:\widetilde{S}\left(  .\right)  \left(  \varphi\right)  \in C_{r}\left(  J;%
\mathbb{C}
\right)  $ that we denote : $S\in C_{r}\left(  J;V^{\prime}\right)  $

3. We can go further and consider families of functions depending on a
parameter : $u\in C\left(  J;V\right)  $ so that $u\left(  t\right)  \in V.$

If we have on one hand a family of functions $u\in C\left(  J;V\right)  $ and
on the other hand a family of distributions $S\in C\left(  J;V^{\prime
}\right)  $ we can consider quantities such that : $S\left(  t\right)  u(t)\in%
\mathbb{C}
$ and the derivative with respect to t of the scalar map : $J\rightarrow%
\mathbb{C}
::S\left(  t\right)  u(t).$ For this we need a partial derivative of u with
respect to t, which assumes that the space V is a normed vector space.

If $V\subset C\left(  E;%
\mathbb{C}
\right)  $ we can also consider $u\in C\left(  E\times J;%
\mathbb{C}
\right)  $ and distributions $S\in C\left(  E\times J;%
\mathbb{C}
\right)  ^{\prime}$ if we have some way to agregate the quantities $S\left(
t\right)  u(t)\in%
\mathbb{C}
.$

\paragraph{Family of distributions in $C_{\infty c}\left(  O;%
\mathbb{C}
\right)  ^{\prime}$\newline}

\begin{theorem}
(Zuily p.61) Let O be an open of $%
\mathbb{R}
^{m},$ J be an open interval in $%
\mathbb{R}
$, if $S\in C_{r}\left(  J;C_{\infty c}\left(  O;%
\mathbb{C}
\right)  ^{\prime}\right)  $ then :

$\forall s:0\leq s\leq r:\exists S_{s}\in C_{r-s}\left(  J;C_{\infty c}\left(
O;%
\mathbb{C}
\right)  ^{\prime}\right)  :$

$\forall\varphi\in C_{\infty c}\left(  O;%
\mathbb{C}
\right)  :\left(  \frac{d}{dt}\right)  ^{s}\left(  S\left(  t\right)  \left(
\varphi\right)  \right)  =S_{s}\left(  t\right)  \left(  \varphi\right)  $
\end{theorem}

Notice that the distribution $S_{s}$ is not the derivative of a distribution
on $J\times%
\mathbb{R}
^{m},$ even if is common to denote $S_{s}\left(  t\right)  =\left(
\frac{\partial}{\partial t}\right)  ^{s}S$

The theorem still holds for Su if : $S\in C\left(  J;C_{\infty c}\left(  O;%
\mathbb{C}
\right)  _{c}^{\prime}\right)  $ and $u\in C\left(  J;C_{\infty}\left(  O;%
\mathbb{C}
\right)  \right)  $

We have, at least in the most simple cases, the chain rule :

\begin{theorem}
(Zuily p.61) Let O be an open of $%
\mathbb{R}
^{m},$ J be an open interval in $%
\mathbb{R}
$, $S\in C_{1}\left(  J;C_{\infty c}\left(  O;%
\mathbb{C}
\right)  ^{\prime}\right)  ,u\in C_{1}\left(  J;C_{\infty c}\left(  O;%
\mathbb{C}
\right)  \right)  $ then :

$\frac{d}{dt}\left(  S\left(  t\right)  \left(  u\left(  t\right)  \right)
\right)  =\frac{\partial S}{\partial t}\left(  u\left(  t\right)  \right)
+S\left(  t\right)  \left(  \frac{\partial u}{\partial t}\right)  $
\end{theorem}

$\frac{\partial S}{\partial t}$ is the usual derivative of the map
$S:J\rightarrow C_{\infty c}\left(  O;%
\mathbb{C}
\right)  ^{\prime}$

The theorem still holds if : $S\in C\left(  J;C_{\infty c}\left(  O;%
\mathbb{C}
\right)  _{c}^{\prime}\right)  $ and $u\in C\left(  J;C_{\infty}\left(  O;%
\mathbb{C}
\right)  \right)  $

\bigskip

Conversely we can consider the integral of the scalar map : $J\rightarrow%
\mathbb{C}
::S\left(  t\right)  u(t)$ with respect to t

\begin{theorem}
(Zuily p.62) Let O be an open of $%
\mathbb{R}
^{m},$ J be an open interval in $%
\mathbb{R}
$, $S\in C_{0}\left(  J;C_{\infty c}\left(  O;%
\mathbb{C}
\right)  ^{\prime}\right)  $,$u\in C_{c\infty}\left(  J\times O;%
\mathbb{C}
\right)  $ then the map :

$\widehat{S}:C_{\infty c}\left(  J\times O;%
\mathbb{C}
\right)  \rightarrow%
\mathbb{C}
::\widehat{S}\left(  u\right)  =\int_{J}\left(  S\left(  t\right)  u\left(
t,.\right)  \right)  dt$

is a distribution $\widehat{S}:C_{\infty c}\left(  J\times O;%
\mathbb{C}
\right)  ^{\prime}$
\end{theorem}

The theorem still holds if : $S\in C_{0}\left(  J;C_{\infty c}\left(  O;%
\mathbb{C}
\right)  _{c}^{\prime}\right)  $ and $u\in C_{\infty}\left(  J\times O;%
\mathbb{C}
\right)  $

\begin{theorem}
(Zuily p.77) Let J be an open interval in $%
\mathbb{R}
$, $S\in C_{r}\left(  J;C_{\infty c}\left(
\mathbb{R}
^{m};%
\mathbb{C}
\right)  _{c}^{\prime}\right)  $,$U\in C_{\infty c}\left(
\mathbb{R}
^{m};%
\mathbb{C}
\right)  ^{\prime}$ then the convolution : $S\left(  t\right)  \ast U\in
C_{r}\left(  J;C_{\infty c}\left(
\mathbb{R}
^{m};%
\mathbb{C}
\right)  ^{\prime}\right)  $ and

$\forall s,0\leq s\leq r:\left(  \frac{\partial}{\partial t}\right)
^{s}\left(  S\left(  t\right)  \ast U\right)  =\left(  \left(  \frac{\partial
}{\partial t}\right)  ^{s}S\right)  \ast U$
\end{theorem}

\begin{theorem}
(Zuily p.77) Let J be an open interval in $%
\mathbb{R}
$, $S\in C_{r}\left(  J;C_{\infty c}\left(
\mathbb{R}
^{m};%
\mathbb{C}
\right)  _{c}^{\prime}\right)  ,\varphi\in C_{\infty}\left(
\mathbb{R}
^{m};%
\mathbb{C}
\right)  $ then :

$S\left(  t\right)  \ast T\left(  \varphi\right)  =T\left(  S\left(  t\right)
_{y}\left(  \varphi\left(  x-y\right)  \right)  \right)  $ with $S\left(
t\right)  _{y}\left(  \varphi\left(  x-y\right)  \right)  \in C_{r}\left(
J\times%
\mathbb{R}
^{m};%
\mathbb{C}
\right)  $
\end{theorem}

\subsubsection{Distributions and Sobolev spaces}

This is the first extension of Sobolev spaces, to distributions.

\paragraph{Dual of Sobolev's spaces on $%
\mathbb{R}
^{m}$\newline}

\begin{definition}
The Sobolev space, denoted $H^{-r}\left(  O\right)  ,r\geq1$\ is the
topological dual of the closure $\overline{C_{\infty c}\left(  O;%
\mathbb{C}
\right)  }$ of $C_{\infty c}\left(  O;%
\mathbb{C}
\right)  $ in $H_{c}^{r}\left(  O\right)  $ where O is an open subset of $%
\mathbb{R}
^{m}$
\end{definition}

\begin{theorem}
(Zuily p.87) $H^{-r}\left(  O\right)  $ is a vector subspace of $C_{\infty
c}\left(  O;%
\mathbb{C}
\right)  ^{\prime}$ which can be identified with :

the topological dual $\left(  H_{c}^{r}\left(  O\right)  \right)  ^{\prime}$
of $H_{c}^{r}\left(  O\right)  $

the space of distributions :

$\left\{  S\in C_{\infty c}\left(  M;%
\mathbb{C}
\right)  ^{\prime};S=\sum_{s=0}^{r}\sum_{\alpha_{1}...\alpha_{s}}D_{\alpha
_{1}...\alpha_{s}}T\left(  f_{\alpha_{1}...\alpha_{s}}\right)  ,f_{\alpha
_{1}...\alpha_{s}}\in L^{2}\left(  O,dx,%
\mathbb{C}
\right)  \right\}  $

This is a Hilbert space with the norm : $\left\Vert S\right\Vert _{H^{-r}%
}=\inf_{f_{\left(  \alpha\right)  }}\left(  \sum_{s=0}^{r}\sum_{\left(
\alpha\right)  }\left\Vert f_{\left(  \alpha\right)  }\right\Vert _{L^{2}%
}\right)  ^{1/2}$
\end{theorem}

\begin{theorem}
(Zuily p.88) $\forall\varphi\in L^{2}\left(  O,dx,%
\mathbb{C}
\right)  ,\psi\in H_{c}^{r}\left(  O\right)  :$

$\left\vert \int_{M}\overline{\varphi}\psi\mu\right\vert \leq\left\Vert
\varphi\right\Vert _{H^{-r}}\left\Vert \psi\right\Vert _{H^{r}}$
\end{theorem}

There is a generalization for compact manifolds (Taylor 1 p.282) .

\paragraph{Sobolev's inequalities\newline}

This is a collection of inequalities which show that many properties of
functions are dictated by their first order derivative.

\begin{theorem}
(Zuily p.89) Poincar\'{e} inequality : For any subset O open in $%
\mathbb{R}
^{m}$ with diameter $d<\infty$ : $\forall\varphi\in H_{c}^{1}\left(  O\right)
:\left\Vert \varphi\right\Vert _{L^{2}}\leq2d\sum_{\alpha=1}^{m}\left\Vert
\frac{\partial\varphi}{\partial x^{\alpha}}\right\Vert _{L^{2}}$
\end{theorem}

\begin{theorem}
(Zuily p.89) If O is bounded in $%
\mathbb{R}
^{m}$ the quantity $\sum_{\alpha_{1}...\alpha_{r}}\left\Vert D_{\alpha
_{1}...\alpha_{r}}\varphi\right\Vert _{L^{2}}$ is a norm on $H_{c}^{r}\left(
O\right)  $ equivalent to $\left\Vert {}\right\Vert _{H^{r}\left(  O\right)
}$
\end{theorem}

\bigskip

For the following theorems we define the spaces of functions on $%
\mathbb{R}
^{m}:$

$D^{1}\left(
\mathbb{R}
^{m}\right)  =\left\{  f\in L_{loc}^{1}\left(
\mathbb{R}
^{m},dx,%
\mathbb{C}
\right)  \cap C_{\nu}\left(
\mathbb{R}
^{m};%
\mathbb{C}
\right)  ,\partial_{\alpha}T\left(  f\right)  \in T\left(  L^{2}\left(
\mathbb{R}
^{m},dx,%
\mathbb{C}
\right)  \right)  \right\}  $

$D^{1/2}\left(
\mathbb{R}
^{m}\right)  =\left\{  f\in L_{loc}^{1}\left(
\mathbb{R}
^{m},dx,%
\mathbb{C}
\right)  \cap C_{\nu}\left(
\mathbb{R}
^{m};%
\mathbb{C}
\right)  ,\int_{%
\mathbb{R}
^{2m}}\frac{\left\vert f\left(  x\right)  -f\left(  y\right)  \right\vert
^{2}}{\left\Vert x-y\right\Vert ^{m+1}}dxdy<\infty\right\}  $

\begin{theorem}
(Lieb p.204) For m%
$>$%
2 If $f\in D^{1}\left(
\mathbb{R}
^{m}\right)  $ then $f\in L^{q}\left(
\mathbb{R}
^{m},dx,%
\mathbb{C}
\right)  $ with $q=\frac{2m}{m-2}$ and :

$\left\Vert f^{\prime}\right\Vert _{2}^{2}\geq C_{m}\left\Vert f\right\Vert
_{q}^{2}$ with $C_{m}=\frac{m\left(  m-2\right)  }{4}A\left(  S^{m}\right)
^{2/m}=\frac{m\left(  m-2\right)  }{4}2^{2/m}m^{1+\frac{1}{m}}\Gamma\left(
\frac{m+1}{2}\right)  ^{-2/m}$

The equality holds iff f is a multiple of $\left(  \mu^{2}+\left\Vert
x-a\right\Vert ^{2}\right)  ^{-\frac{m-2}{2}}$ ,$\mu>0,a\in%
\mathbb{R}
^{m}$
\end{theorem}

\begin{theorem}
(Lieb p.206) For m%
$>$%
1 If $f\in D^{1/2}\left(
\mathbb{R}
^{m}\right)  $ then $f\in L^{q}\left(
\mathbb{R}
^{m},dx,%
\mathbb{C}
\right)  $ with $q=\frac{2m}{m-2}$ and :

$\frac{\Gamma\left(  \frac{m+1}{2}\right)  }{2\pi^{\frac{m+1}{2}}}\int_{%
\mathbb{R}
^{2m}}\frac{\left\vert f\left(  x\right)  -f\left(  y\right)  \right\vert
^{2}}{\left\Vert x-y\right\Vert ^{m+1}}dxdy\geq\frac{m-1}{2}\left(
C_{m}\right)  ^{1/m}\left\Vert f\right\Vert _{q}^{2}$ with

$C_{m}=\frac{m-1}{2}2^{1/m}\pi^{\frac{m+1}{2m}}\Gamma\left(  \frac{m+1}%
{2}\right)  ^{-1/m}$

The equality holds iff f is a multiple of $\left(  \mu^{2}+\left\Vert
x-a\right\Vert ^{2}\right)  ^{-\frac{m-1}{2}}$ ,$\mu>0,a\in%
\mathbb{R}
^{m}$
\end{theorem}

The inequality reads :

$\int_{%
\mathbb{R}
^{2m}}\frac{\left\vert f\left(  x\right)  -f\left(  y\right)  \right\vert
^{2}}{\left\Vert x-y\right\Vert ^{m+1}}dxdy\geq C_{m}^{\prime}\left\Vert
f\right\Vert _{q}^{2}$

with $C_{m}^{\prime}=\left(  \frac{m-1}{2}\right)  ^{1+\frac{1}{m}}\left(
2\pi^{\frac{m+1}{2}}\Gamma\left(  \frac{m+1}{2}\right)  ^{-1}\right)
^{1+\frac{1}{m^{2}}}$

\begin{theorem}
(Lieb p.210) Let $\left(  f_{n}\right)  $ a sequence of functions $f_{n}\in
D^{1}\left(
\mathbb{R}
^{m}\right)  ,m>2$ such that $f_{n}^{\prime}\rightarrow g$ weakly in
$L^{2}\left(
\mathbb{R}
^{m},dx,%
\mathbb{C}
\right)  $ then :

i) $g=f^{\prime}$ for some unique function $f\in D^{1}\left(
\mathbb{R}
^{m}\right)  .$

ii) there is a subsequence which converges to f for almost every $x\in%
\mathbb{R}
^{m}$

iii) If A is a set of finite measure in $%
\mathbb{R}
^{m}$ then $1_{A}f_{n}\rightarrow1_{A}f$ strongly in $L^{p}\left(
\mathbb{R}
^{m},dx,%
\mathbb{C}
\right)  $ for $p<\frac{2m}{m-2}$
\end{theorem}

\begin{theorem}
Nash's inequality (Lieb p.222) For every function

$f\in H^{1}\left(
\mathbb{R}
^{m}\right)  \cap L^{1}\left(
\mathbb{R}
^{m},dx,%
\mathbb{C}
\right)  :$

$\left\Vert f\right\Vert _{2}^{1+\frac{2}{m}}\leq C_{m}\left\Vert f^{\prime
}\right\Vert _{2}\left\Vert f\right\Vert _{1}^{2/m}$ where

$C_{m}^{2}=2m^{-1+\frac{2}{m}}\left(  1+\frac{m}{2}\right)  ^{1+\frac{2}{m}%
}\lambda\left(  m\right)  ^{-1}A\left(  S^{m-1}\right)  ^{-2/m}$ and
$\lambda\left(  m\right)  $ is a constant which depends only on m
\end{theorem}

\begin{theorem}
Logarithmic Sobolev inequality (Lieb p.225) Let $f\in H^{1}\left(
\mathbb{R}
^{m}\right)  $,$a>0$ then :

$\frac{a^{2}}{m}\int_{%
\mathbb{R}
^{m}}\left\Vert f^{\prime}\right\Vert ^{2}dx\geq\int_{%
\mathbb{R}
^{m}}\left\vert f\right\vert ^{2}\ln\left(  \frac{\left\vert f\right\vert
^{2}}{\left\Vert f\right\Vert _{2}^{2}}\right)  dx+m\left(  1+\ln a\right)
\left\Vert f\right\Vert _{2}^{2}$

The equality holds iff f is, up to translation, a multiple of $\exp\left(
-\pi\frac{\left\Vert x\right\Vert ^{2}}{2a^{2}}\right)  $
\end{theorem}

\begin{theorem}
(Lieb p.229) Let $(E,S,\mu)$ be a $\sigma-$finite\ measured space with\ $\mu
$\ a positive measure, U(t) a one parameter semi group which is also symmetric
and a contraction on both $L^{p}\left(  E,S,\mu,%
\mathbb{C}
\right)  ,p=1,2,$ $\ $with infinitesimal generator S whose domain is D(S),
$\gamma\in\left[  0,1\right]  $ a fixed scalar.\ Then the following are equivalent:

i) $\exists C_{1}\left(  \gamma\right)  >0:\forall f\in L^{1}\left(  E,S,\mu,%
\mathbb{C}
\right)  :\left\Vert U\left(  t\right)  f\right\Vert _{\infty}\leq
C_{1}t^{-\frac{\gamma}{1-\gamma}}\left\Vert f\right\Vert _{1}$

ii) $\exists C_{2}\left(  \gamma\right)  >0:\forall f\in L^{1}\left(  E,S,\mu,%
\mathbb{C}
\right)  \cap D\left(  S\right)  :\left\Vert f\right\Vert _{2}^{2}\leq
C_{2}\left(  \left\langle f,Sf\right\rangle \right)  ^{\gamma}\left\Vert
f\right\Vert _{1}^{2\left(  1-\gamma\right)  }$
\end{theorem}

U(t) must satisfy :

$U(0)f=f,U(t+s)=U(t)\circ U(s),\left\Vert U\left(  t\right)  f-U\left(
s\right)  f\right\Vert _{p}\rightarrow_{t\rightarrow s}0$

$\left\Vert U(t)f\right\Vert _{p}\leq\left\Vert f\right\Vert _{p},\left\langle
f,U\left(  t\right)  g\right\rangle =\left\langle U(t)f,g\right\rangle $

\bigskip

\subsection{Extension of distributions}

\label{Extension of distributions}

\subsubsection{Colombeau algebras}

\paragraph{Colombeau algebras\newline}

It would be nice if we had distributions $S\left(  \varphi\times\psi\right)
=S\left(  \varphi\right)  S\left(  \psi\right)  $ meaning S is a
multiplicative linear functional in V', so $S\in\Delta\left(  V\right)  $. But
if E is a locally compact Hausdorff space, then $\Delta\left(  C_{0v}\left(
E;%
\mathbb{C}
\right)  \right)  $ is homeomorphic to E and the only multiplicative
functionals are the Dirac distributions : $\delta_{a}\left(  \varphi
\psi\right)  =\delta_{a}\left(  \varphi\right)  \delta_{a}\left(  \psi\right)
.$ This is the basis of a celebrated "no go" theorem by L.Schwartz stating
that there is no possibility to define the product of distributions. However
there are ways around this issue, by defining quotient spaces of functions and
to build "Colombeau algebras" of distributions.\ The solutions are very
technical. See Nigsch on this topic.They usually involved distributions acting
on m forms, of which we give an overview below as it is a kind of introduction
to the following subsection.

\paragraph{Distributions acting on m forms\newline}

1.The general idea is to define distributions acting, not on functions, but on
m forms on a m dimensional manifold.\ To do this the procedure is quite
similar to the one used to define Lebesgue integral of forms. It is used, and
well defined, with $V=C_{\infty c}\left(  M;%
\mathbb{C}
\right)  $.

2. Let M be a m dimensional, Hausdorff, smooth, real, orientable manifold,
with atlas $\left(  O_{i},\varphi_{i}\right)  _{i\in I},\varphi_{i}\left(
O_{i}\right)  =U_{i}\in%
\mathbb{R}
^{m}$\ 

A m form $\varpi$\ on M has for component in the holonomic basis $\varpi
_{i}\in C\left(  O_{i};%
\mathbb{R}
\right)  $ with the rule in a transition between charts :$\varphi_{ij}%
:O_{i}\rightarrow O_{j}::\varpi_{j}=\det\left[  \varphi_{ij}^{\prime}\right]
^{-1}\varpi_{i}$

The push forward of $\varpi$ by each chart gives a m form on $%
\mathbb{R}
^{m}$ whose components in each domain $U_{i}$ is equal to $\varpi_{i}:$

$\varphi_{i\ast}\varpi\left(  \xi\right)  =\varpi_{i}\left(  \varphi_{i}%
^{-1}\left(  \xi\right)  \right)  e^{1}\wedge...\wedge e^{m}$ \ in the
canonical basis $\left(  e^{k}\right)  _{k=1}^{m}$ of $%
\mathbb{R}
^{m\ast}$

If each of the functions $\varpi_{i}$ is smooth and compactly supported :
$\varpi_{i}\in C_{\infty c}\left(  O_{i};%
\mathbb{R}
\right)  $ we will say that $\varpi$ is smooth and compactly supported and
denote the space of such forms : $\mathfrak{X}_{\infty c}\left(  \wedge
_{m}TM^{\ast}\right)  .$ Obviously the definition does not depend of the
choice of an atlas.

Let $S_{i}\in C_{\infty c}\left(  U_{i};%
\mathbb{R}
\right)  ^{\prime}$ then $S_{i}\left(  \varpi_{i}\right)  $ is well defined.

Given the manifold M, we can associate to $\mathfrak{X}_{\infty c}\left(
\wedge_{m}TM^{\ast}\right)  $ and any atlas, families of functions $\left(
\varpi_{i}\right)  _{i\in I},\varpi_{i}\in C_{\infty c}\left(  U_{i};%
\mathbb{R}
\right)  $ meeting the transition properties. As M is orientable, there is an
atlas such that for the transition maps $\det\left[  \varphi_{ij}^{\prime
}\right]  >0$

Given a family $\left(  S_{i}\right)  _{i\in I}\in\left(  C_{\infty c}\left(
U_{i};%
\mathbb{R}
\right)  ^{\prime}\right)  ^{I}$ \ when it acts on a family $\left(
\varpi_{i}\right)  _{i\in I},$ representative of a m form, we have at the
intersection $S_{i}|_{U_{i}\cap U_{j}}=S_{j}|_{U_{i}\cap U_{j}}$ , so there is
a unique distribution $\widehat{S}\in C_{\infty c}\left(  U;%
\mathbb{R}
\right)  ^{\prime},U=\cup_{i\in I}U_{i}$ such that $\widehat{S}|_{U_{i}}%
=S_{i}.$

We define the pull back of $\widehat{S}$ on M by the charts and we have a
distribution S on M with the property :

$\forall\psi\in C_{\infty c}\left(  O_{i};%
\mathbb{R}
\right)  :\left(  \varphi_{i}^{-1}\right)  ^{\ast}S\left(  \psi\right)
=\widehat{S}|_{U_{i}}\left(  \psi\right)  $

and we say that $S\in\mathfrak{X}_{\infty c}\left(  \wedge_{m}TM^{\ast
}\right)  ^{\prime}$ and denote $S\left(  \varpi\right)  =\widehat{S}\left(
\varphi_{\ast}\varpi\right)  $

On the practical side the distribution of a m form is just the sum of the
value of distributions on each domain, applied to the component of the form,
with an adequate choice of the domains.

3. The pointwise product of a smooth function f and a m form $\varpi\in\left(
\wedge_{m}TM^{\ast}\right)  _{\infty c}$ is still a form $f\varpi
\in\mathfrak{X}_{\infty c}\left(  \wedge_{m}TM^{\ast}\right)  $ so we can
define the product of such a function with a distribution. Or if $\varpi
\in\mathfrak{X}_{\infty}\left(  \wedge_{m}TM^{\ast}\right)  $ is a smooth m
form, and $\varphi\in C_{\infty c}\left(  M;%
\mathbb{R}
\right)  $ then $\varphi\varpi\in\mathfrak{X}_{\infty c}\left(  \wedge
_{m}TM^{\ast}\right)  .$

4. The derivative takes a different approach. If X a is smooth vector field on
M, then the Lie derivative of a distribution $S\in\left(  \mathfrak{X}_{\infty
c}\left(  \wedge_{m}TM^{\ast}\right)  \right)  ^{\prime}$ is defined as :

$\forall X\in\mathfrak{X}_{\infty}\left(  TM\right)  ,\varpi\in\mathfrak{X}%
_{\infty c}\left(  \wedge_{m}TM^{\ast}\right)  :\left(  \pounds _{X}S\right)
\left(  \varpi\right)  =-S\left(  \pounds _{X}\varpi\right)  =-S\left(
d\left(  i_{X}\varpi\right)  \right)  $

If there is a smooth volume form $\varpi_{0}$ on M, and $\varphi\in C_{\infty
c}\left(  M;%
\mathbb{R}
\right)  $ then

$\left(  \pounds _{X}S\right)  \left(  \varphi\varpi_{0}\right)  =-S\left(
\left(  divX\right)  \varphi\varpi_{0}\right)  =-\left(  divX\right)  S\left(
\varphi\varpi_{0}\right)  $

If M is manifold with boundary we have some distribution :

$S\left(  d\left(  i_{X}\varpi_{0}\right)  \right)  =S_{\partial M}\left(
i_{X}\varpi_{0}\right)  =-\left(  \pounds _{X}S\right)  \left(  \varpi
_{0}\right)  $

\subsubsection{Distributions on vector bundles}

In fact the construct above can be generalized to any vector bundle in a more
convenient way, but we take the converse of the previous idea : distributions
act on sections of vector bundle, and they can be assimilated to m forms
valued on the dual vector bundle.

\paragraph{Definition\newline}

If $E\left(  M,V,\pi\right)  $ is a complex vector bundle, then the spaces of
sections $\mathfrak{X}_{r}\left(  E\right)  ,$ $\mathfrak{X}\left(
J^{r}E\right)  ,\mathfrak{X}_{rc}\left(  E\right)  ,$ $\mathfrak{X}_{c}\left(
J^{r}E\right)  $ are Fr\'{e}chet spaces. Thus we can consider continuous
linear functionals on them and extend the scope of distributions.

The space of test functions is here $\mathfrak{X}_{\infty,c}\left(  E\right)
.$ We assume that the base manifold M is m dimensional and V is n dimensional.

\begin{definition}
A distribution on the vector bundle E is a linear, continuous functional on
$\mathfrak{X}_{\infty,c}\left(  E\right)  $
\end{definition}

\begin{notation}
$\mathfrak{X}_{\infty,c}\left(  E\right)  ^{\prime}$ is the space of
distributions on E
\end{notation}

Several tools for scalar distributions can be extended to vector bundle distributions.

\paragraph{Product of a distribution and a section\newline}

\begin{definition}
The product of the distribution $S\in\mathfrak{X}_{\infty,c}\left(  E\right)
^{\prime}$ by a function $f\in C_{\infty}\left(  M;%
\mathbb{C}
\right)  $ is the distribution $fS\in\mathfrak{X}_{\infty,c}\left(  E\right)
^{\prime}::\left(  fS\right)  \left(  X\right)  =S\left(  fX\right)  $
\end{definition}

The product of a section $X\in\mathfrak{X}_{\infty,c}\left(  E\right)  $ by a
function $f\in C_{\infty}\left(  M;%
\mathbb{C}
\right)  $ still belongs to $\mathfrak{X}_{\infty,c}\left(  E\right)  $

\paragraph{Assimilation to forms on the dual bundle\newline}

\begin{theorem}
To each continuous m form $\lambda\in\Lambda_{m}\left(  M;E^{\prime}\right)  $
valued in the topological bundle E' can be associated a distribution in
$\mathfrak{X}_{\infty,c}\left(  E\right)  ^{\prime}$
\end{theorem}

\begin{proof}
The dual E' of $E\left(  M,V,\pi\right)  $ is the vector bundle $E^{\prime
}\left(  M,V^{\prime},\pi\right)  $ where V' is the topological dual of V (it
is a Banach space). A m form $\lambda\in\Lambda_{m}\left(  M;E^{\prime
}\right)  $ reads in a holonomic basis $\left(  e_{a}^{i}\left(  x\right)
\right)  _{i\in I}$ of E' and $\left(  d\xi^{\alpha}\right)  _{\alpha=1}^{m}$
of TM* :

$\lambda=\sum\lambda_{i}\left(  x\right)  e_{a}^{i}\left(  x\right)  \otimes
d\xi^{1}\wedge...\wedge d\xi^{m}$

It acts fiberwise on a section $X\in\mathfrak{X}_{\infty,c}\left(  E\right)  $
by :

$\sum_{i\in I}X^{i}\left(  x\right)  \lambda_{i}\left(  x\right)  d\xi
^{1}\wedge...\wedge d\xi^{m}$

With an atlas $\left(  O_{a},\psi_{a}\right)  _{a\in A}$ of M and $\left(
O_{a},\varphi_{a}\right)  _{a\in A}$ of E, at the transitions (both on M and
E) (see Vector bundles) :

$\lambda_{b}^{i}=\sum_{j}\det\left[  \psi_{ba}^{\prime}\right]  \left[
\varphi_{ab}\right]  _{i}^{j}\lambda_{a}^{j}$

$X_{b}^{i}=\sum_{j\in I}\left[  \varphi_{ba}\right]  _{j}^{i}X_{a}^{j}$

$\mu_{b}=\sum_{i\in I}X_{b}^{i}\left(  x\right)  \lambda_{bi}\left(  x\right)
=\det\left[  \psi_{ba}^{\prime}\right]  \sum_{i\in I}\left[  \varphi
_{ba}\right]  _{i}^{j}\lambda_{a}^{j}\left[  \varphi_{ab}\right]  _{k}%
^{i}X_{a}^{k} $

$=\det\left[  \psi_{ba}^{\prime}\right]  \sum_{i\in I}\lambda_{a}^{i}X_{a}%
^{i}$

So $:\mu=\sum_{i\in I}X^{i}\left(  x\right)  \lambda_{i}\left(  x\right)
d\xi^{1}\wedge...\wedge d\xi^{m}$ is a m form on M and the actions reads :

$\mathfrak{X}_{0}\left(  E^{\prime}\right)  \times\mathfrak{X}_{\infty
,c}\left(  E\right)  \rightarrow\Lambda_{m}\left(  M;%
\mathbb{C}
\right)  ::\mu=\sum_{i\in I}X^{i}\left(  x\right)  \lambda_{i}\left(
x\right)  d\xi^{1}\wedge...\wedge d\xi^{m}$

$\mu$ defines a Radon measure on M, locally finite. And because X is compactly
supported and bounded and $\lambda$ continuous the integral of the form on M
is finite.
\end{proof}

We will denote T the map :%

\begin{equation}
T:\Lambda_{0,m}\left(  M;E^{\prime}\right)  \rightarrow\mathfrak{X}_{\infty
,c}\left(  E\right)  ^{\prime}::T\left(  \lambda\right)  \left(  X\right)
=\int_{M}\lambda\left(  X\right)
\end{equation}

Notice that the map is not surjective.\ Indeed the interest of the concept of
distribution on vector bundle is that \textit{this not a local operator but a
global operator}, which acts on sections. The m forms of $\Lambda_{m}\left(
M;E^{\prime}\right)  $\ can be seen as the local distributions.

\paragraph{Pull back, push forward of a distribution\newline}

\begin{definition}
For two smooth complex finite dimensional vector bundles $E_{1}\left(
M,V_{1},\pi_{1}\right)  ,E_{2}\left(  M,V_{2},\pi_{2}\right)  $ on the same
real manifold and a base preserving morphism $L:E_{1}\rightarrow E_{2}$ the
pull back of a distribution is the map :%

\begin{equation}
L^{\ast}:\mathfrak{X}_{\infty,c}\left(  E_{2}\right)  ^{\prime}\rightarrow
\mathfrak{X}_{\infty,c}\left(  E_{1}\right)  ^{\prime}::L^{\ast}S_{2}\left(
X_{1}\right)  =S_{2}\left(  LX_{1}\right)
\end{equation}

\end{definition}

\begin{theorem}
In the same conditions, there is a map :

$L^{t\ast}:\Lambda_{m}\left(  M;E_{2}^{\prime}\right)  \rightarrow\Lambda
_{m}\left(  M;E_{1}^{\prime}\right)  $ such that : $L^{\ast}T\left(
\lambda_{2}\right)  =T\left(  L^{t\ast}\left(  \lambda_{2}\right)  \right)  $
\end{theorem}

\begin{proof}
The transpose of $L:E_{1}\rightarrow E_{2}$ is $L^{t}:E_{2}^{\prime
}\rightarrow E_{1}^{\prime}.$ This is a base preserving morphism such that :
$\forall X_{1}\in\mathfrak{X}\left(  E_{1}\right)  ,Y_{2}\in\mathfrak{X}%
\left(  E_{1}^{\prime}\right)  $:$L^{t}\left(  Y_{2}\right)  \left(
X_{1}\right)  =Y_{2}\left(  LX_{1}\right)  $

It acts on $\Lambda_{m}\left(  M;E_{2}^{\prime}\right)  :$

$L^{t\ast}:\Lambda_{m}\left(  M;E_{2}^{\prime}\right)  \rightarrow\Lambda
_{m}\left(  M;E_{1}^{\prime}\right)  ::L^{t\ast}\left(  \lambda_{2}\right)
\left(  x\right)  =L^{t}\left(  x\right)  \left(  \lambda_{2}\right)  \otimes
d\xi^{1}\wedge...\wedge d\xi^{m}$

If $S_{2}=T\left(  \lambda_{2}\right)  $ then $L^{\ast}S_{2}\left(
X_{1}\right)  =S_{2}\left(  LX_{1}\right)  =T\left(  \lambda_{2}\right)
\left(  LX_{1}\right)  =\int_{M}\lambda_{2}\left(  LX_{1}\right)  =\int
_{M}L^{t}\left(  \lambda_{2}\right)  \left(  X_{1}\right)  =T\left(  L^{t\ast
}\left(  \lambda_{2}\right)  \right)  \left(  X_{1}\right)  $
\end{proof}

Similarly :

\begin{definition}
For two smooth complex finite dimensional vector bundles $E_{1}\left(
M,V_{1},\pi_{1}\right)  ,E_{2}\left(  M,V_{2},\pi_{2}\right)  $ on the same
real manifold and a base preserving morphism $L:E_{2}\rightarrow E_{1}$ the
push forward of a distribution is the map :%

\begin{equation}
L_{\ast}:\mathfrak{X}_{\infty,c}\left(  E_{1}\right)  ^{\prime}\rightarrow
\mathfrak{X}_{\infty,c}\left(  E_{2}\right)  ^{\prime}::L_{\ast}S_{1}\left(
X_{2}\right)  =S_{1}\left(  LX_{2}\right)
\end{equation}

\end{definition}

\paragraph{Support of a distribution\newline}

The support of a section X is defined as the complementary of the largest open
O in M such as : $\forall x\in O:X\left(  x\right)  =0.$

\begin{definition}
The support of a distribution S in $\mathfrak{X}_{\infty,c}\left(  E\right)
^{\prime}$ is the subset of M, complementary of the largest open O in M such
that :

$\forall X\in\mathfrak{X}_{\infty,c}\left(  E\right)  ,Supp(X)\subset
O\Rightarrow S\left(  X\right)  =0.$
\end{definition}

\paragraph{r jet prolongation of a vectorial distribution\newline}

Rather than the derivative of a distribution it is more logical to look for
its r jet prolongation.\ But this needs some adjustments.

\begin{definition}
The action of a distribution $S\in\mathfrak{X}_{\infty,c}\left(  E\right)
^{\prime}$ on a section $Z\in\mathfrak{X}_{\infty,c}\left(  J^{r}E\right)  $
is defined as the map :

$S:\mathfrak{X}_{\infty,c}\left(  J^{r}E\right)  \rightarrow\left(
\mathbb{C}
,%
\mathbb{C}
^{sm},s=1..r\right)  ::$%

\begin{equation}
S\left(  Z_{\alpha_{1}...\alpha_{s}}\left(  x\right)  ,s=0..r,\alpha
_{j}=1...m\right)  =\left(  S\left(  Z_{\alpha_{1}...\alpha_{s}}\right)
,s=0..r,\alpha_{j}=1...m\right)
\end{equation}

\end{definition}

The r jet prolongation of the vector bundle $E\left(  M,V,\pi\right)  $ is the
vector bundle $J^{r}E\left(  M,J_{0}^{r}\left(
\mathbb{R}
^{m},V\right)  _{0},\pi_{0}^{r}\right)  .$A section $Z\in\mathfrak{X}%
_{\infty,c}\left(  J^{r}E\right)  $ can be seen as the set $\left(
Z_{\alpha_{1}...\alpha_{s}}\left(  x\right)  ,s=0..r,\alpha_{j}=1...m\right)
$ where $Z_{\alpha_{1}...\alpha_{s}}\left(  x\right)  \in E\left(  x\right)  $

The result is not a scalar, but a set of scalars : $\left(  k_{\alpha
_{1}...\alpha_{s}},s=0..r,\alpha_{j}=1...m\right)  $

\bigskip

\begin{definition}
The r jet prolongation of a distribution $S\in\mathfrak{X}_{\infty,c}\left(
E\right)  ^{\prime}$ is the map denoted $J^{r}S$ such that%

\begin{equation}
\forall X\in\mathfrak{X}_{\infty,c}\left(  E\right)  :J^{r}S\left(  X\right)
=S\left(  J^{r}X\right)
\end{equation}

\end{definition}

$J^{r}E$ is a vector bundle, and the space of its sections, smooth and
compactly supported $\mathfrak{X}_{\infty,c}\left(  J^{r}E\right)  ,$ is well
defined, as the space of its distributions. If $X\in\mathfrak{X}_{\infty
,c}\left(  E\right)  $ then $J^{r}X\in\mathfrak{X}_{\infty,c}\left(
J^{r}E\right)  :$

$J^{r}X=\left(  X,\sum_{i=1}^{n}\left(  D_{\alpha_{1}..\alpha_{s}}%
X^{i}\right)  \mathbf{e}_{i}\left(  x\right)  ,i=1..n,s=1...r,\alpha
_{j}=1...m\right)  $

So the action of $S\in\mathfrak{X}_{\infty,c}\left(  E\right)  ^{\prime}$ is :

$S\left(  J^{r}X\right)  =\left(  S\left(  D_{\alpha_{1}..\alpha_{s}}X\right)
,s=0...r,\alpha_{j}=1...m\right)  $

and this leads to the definition of $J^{r}S:$

$J^{r}S:\mathfrak{X}_{\infty,c}\left(  E\right)  \rightarrow\left(
\mathbb{C}
,%
\mathbb{C}
^{sm},s=1..r\right)  ::J^{r}S\left(  X\right)  =\left(  S\left(  D_{\alpha
_{1}..\alpha_{s}}X\right)  ,s=0...r,\alpha_{j}=1...m\right)  $

\bigskip

\begin{definition}
The derivative of a distribution $S\in\mathfrak{X}_{\infty,c}\left(  E\right)
^{\prime}$ with respect to $\left(  \xi^{\alpha_{1}},...,\xi^{\alpha_{s}%
}\right)  $ on M is the distribution :%

\begin{equation}
\left(  D_{\alpha_{1}..\alpha_{s}}S\right)  \in\mathfrak{X}_{\infty,c}\left(
E\right)  ^{\prime}:\forall X\in\mathfrak{X}_{\infty,c}\left(  E\right)
:\left(  D_{\alpha_{1}..\alpha_{s}}S\right)  \left(  X\right)  =S\left(
D_{\alpha_{1}..\alpha_{s}}X\right)
\end{equation}

\end{definition}

The map : $\left(  D_{\alpha_{1}..\alpha_{s}}S\right)  :\mathfrak{X}%
_{\infty,c}\left(  E\right)  \rightarrow%
\mathbb{C}
::\left(  D_{\alpha_{1}..\alpha_{s}}S\right)  \left(  X\right)  =S\left(
D_{\alpha_{1}..\alpha_{s}}X\right)  $ is valued in $%
\mathbb{C}
$\ , is linear and continuous, so it is a distribution.\ 

\begin{theorem}
The r jet prolongation of a distribution $S\in\mathfrak{X}_{\infty,c}\left(
E\right)  ^{\prime}$ is the set of distributions :%

\begin{equation}
J^{r}S=\left(  \left(  D_{\alpha_{1}..\alpha_{s}}S\right)  ,s=0..r,\alpha
_{j}=1...m\right)
\end{equation}

\end{theorem}

The space $\left\{  J^{r}S,S\in\mathfrak{X}_{\infty,c}\left(  E\right)
^{\prime}\right\}  $ is a vector subspace of $\left(  \mathfrak{X}_{\infty
,c}\left(  E\right)  ^{\prime}\right)  ^{N}$ where $N=\sum_{s=0}^{r}\frac
{m!}{s!\left(  m-s\right)  !}$

\bigskip

\begin{theorem}
If the distribution $S\in\mathfrak{X}_{\infty,c}\left(  E\right)  ^{\prime}$
is induced by the m form $\lambda\in\Lambda_{m}\left(  M;E^{\prime}\right)  $
then $:$%

\begin{equation}
J^{r}\left(  T\left(  \lambda\right)  \right)  =T\left(  J^{r}\lambda\right)
\text{ with }J^{r}\lambda\in\Lambda_{m}\left(  M;J^{r}E^{\prime}\right)
\end{equation}

\end{theorem}

\begin{proof}
The space $\Lambda_{m}\left(  M;E^{\prime}\right)  \simeq E^{\prime}%
\otimes\left(  TM^{\ast}\right)  ^{m}$ and its r jet prolongation is :
$J^{r}E^{\prime}\otimes\left(  TM^{\ast}\right)  ^{m}\simeq\Lambda_{m}\left(
M;J^{r}E^{\prime}\right)  $

$\left(  D_{\alpha_{1}..\alpha_{s}}T\left(  \lambda\right)  \right)  \left(
X\right)  =T\left(  \lambda\right)  \left(  D_{\alpha_{1}..\alpha_{s}%
}X\right)  =\int_{M}\lambda\left(  D_{\alpha_{1}..\alpha_{s}}X\right)  $

The demonstration is similar to the one for derivatives of distributions on a manifold.

We start with r=1 and $D_{\alpha}=\partial_{\alpha}$\ with $\alpha\in1...m$
and denote: $d\xi=d\xi^{1}\wedge...\wedge d\xi^{m}$

$i_{\partial_{\alpha}}\left(  \lambda_{i}d\xi\right)  =\left(  -1\right)
^{\alpha-1}\lambda_{i}d\xi^{1}\wedge...\wedge\left(  \widehat{d\xi^{\alpha}%
}\right)  \wedge..\wedge d\xi^{m}$

$d\left(  i_{\partial_{\alpha}}\left(  \lambda_{i}d\xi\right)  \right)
=\sum_{\beta}\left(  -1\right)  ^{\alpha-1}\left(  \partial_{\beta}\lambda
_{i}\right)  d\xi^{\beta}\wedge d\xi^{1}\wedge...\wedge\left(  \widehat
{d\xi^{\alpha}}\right)  \wedge..\wedge d\xi^{m}$

$=\left(  \partial_{\alpha}\lambda_{i}\right)  d\xi$

$\left(  dX^{i}\right)  \wedge i_{\partial_{\alpha}}\left(  \lambda_{i}%
d\xi\right)  =\sum_{\beta}\left(  -1\right)  ^{\alpha-1}\left(  \partial
_{\beta}X^{i}\right)  \lambda_{i}d\xi^{\beta}\wedge d\xi^{1}\wedge
...\wedge\left(  \widehat{d\xi^{\alpha}}\right)  \wedge..\wedge d\xi^{m}$

$=\left(  \partial_{\beta}X^{i}\right)  \lambda_{i}d\xi$

$d\left(  X^{i}i_{\partial_{\alpha}}\left(  \lambda_{i}d\xi\right)  \right)
=dX^{i}\wedge i_{\partial_{\alpha}}\left(  \lambda_{i}d\xi\right)
-X^{i}d\left(  i_{\partial_{\alpha}}\left(  \lambda_{i}d\xi\right)  \right)  $

$=\left(  \partial_{\alpha}X^{i}\right)  \left(  \lambda_{i}d\xi\right)
-X^{i}\left(  \partial_{\alpha}\lambda_{i}\right)  d\xi$

$d\left(  \sum_{i}X^{i}i_{\partial_{\alpha}}\lambda_{i}\right)  =\sum
_{i}\left(  \left(  \partial_{\alpha}X^{i}\right)  \left(  \lambda_{i}%
d\xi\right)  -X^{i}\left(  \partial_{\alpha}\lambda_{i}\right)  d\xi\right)  $

Let N be a compact manifold with boundary in M. The Stockes theorem gives :

$\int_{N}d\left(  \sum_{i}X^{i}i_{\partial_{\alpha}}\lambda_{i}\right)
=\int_{\partial N}\sum_{i}X^{i}i_{\partial_{\alpha}}\lambda_{i}=\int_{N}%
\sum_{i}\left(  \left(  \partial_{\alpha}X^{i}\right)  \left(  \lambda_{i}%
d\xi\right)  -X^{i}\left(  \partial_{\alpha}\lambda_{i}\right)  d\xi\right)  $

Because X has a compact support we can always choose N such that $Supp\left(
X\right)  \subset\overset{\circ}{N}$ and then :

$\int_{\partial N}\sum_{i}X^{i}i_{\partial_{\alpha}}\lambda_{i}=0$

$\int_{N}\sum_{i}\left(  \left(  \partial_{\alpha}X^{i}\right)  \left(
\lambda_{i}d\xi\right)  -X^{i}\left(  \partial_{\alpha}\lambda_{i}\right)
d\xi\right)  =\int_{M}\sum_{i}\left(  \left(  \partial_{\alpha}X^{i}\right)
\lambda_{i}d\xi-X^{i}\left(  \partial_{\alpha}\lambda_{i}\right)  d\xi\right)
=0$

$T\left(  \partial_{\alpha}\lambda d\xi\right)  \left(  X\right)  =T\left(
\lambda\right)  \left(  \partial_{\alpha}X\right)  =\partial_{\alpha}\left(
T\left(  \lambda\right)  \right)  \left(  X\right)  $

By recursion over r we get :

$T\left(  D_{\alpha_{1}...\alpha_{r}}\lambda\right)  \left(  X\right)
=\left(  D_{\alpha_{1}...\alpha_{r}}T\left(  \lambda\right)  \right)  \left(
X\right)  $

$J^{r}T\left(  \lambda\right)  =\left(  \left(  D_{\alpha_{1}..\alpha_{s}%
}T\left(  \lambda\right)  \right)  ,s=0..r,\alpha_{j}=1...m\right)  $

$=\left(  T\left(  D_{\alpha_{1}...\alpha_{r}}\lambda\right)  ,s=0..r,\alpha
_{j}=1...m\right)  =T\left(  J^{r}\lambda\right)  $
\end{proof}

Notice that we have gotten rid off the $(-1)^{r}$ in the process, which was
motivated only for the consistency for functions in $%
\mathbb{R}
^{m}.$

\newpage

\section{THE\ FOURIER\ TRANSFORM}

\label{Fourier transform}

\subsection{General definition}

The Fourier transform is defined for functions on an abelian, locally compact,
topological group (G,+). G has a Haar measure $\mu$, which is both left and
right invariant and is a Radon measure, the spaces of integrable complex
valued functions on G : $L^{p}\left(  G,\mu,%
\mathbb{C}
\right)  ,1\leq p\leq\infty$ are well defined (see Functional analysis) and
are Banach vector spaces, L%
${{}^2}$
is a Hilbert space. The Pontryagin dual of G is the set : $\widehat
{G}=\left\{  \chi:G\rightarrow T\right\}  $ where $T=\left\{  z\in%
\mathbb{C}
,\left\vert z\right\vert =1\right\}  .$ G is isomorphic to its bidual by the
Gelf'and transform (see Lie groups) :

$\widehat{}:G\rightarrow\widehat{\left(  \widehat{G}\right)  }::\widehat
{g}\left(  \chi\right)  =\chi\left(  g\right)  \in T$

\begin{theorem}
(Neeb p.150-163) If G is a locally compact, abelian, topological group,
endowed with a Haar measure $\mu$ , $\widehat{G}$ its Pontryagin dual, then
the \textbf{Fourier transform} is the map :%

\begin{equation}
\symbol{94}:L^{1}\left(  G,\mu,%
\mathbb{C}
\right)  \rightarrow C_{0\nu}\left(  \widehat{G};%
\mathbb{C}
\right)  :\widehat{\varphi}(\chi)=\int_{G}\varphi(g)\overline{\chi\left(
g\right)  }\mu\left(  g\right)
\end{equation}

$\widehat{\varphi}$ is well defined, continuous and vanishes at infinity :
$\widehat{\varphi}\in C_{0\nu}\left(  \widehat{G};%
\mathbb{C}
\right)  $

Conversely, for each Haar measure $\mu,$ there is a unique Haar measure $\nu$
on $\widehat{G}$\ such that :%

\begin{equation}
\forall\varphi\in L^{1}\left(  G,\mu,%
\mathbb{C}
\right)  :\varphi\left(  g\right)  =\int_{\widehat{G}}\widehat{\varphi}\left(
\chi\right)  \chi\left(  g\right)  \nu\left(  \chi\right)
\end{equation}

for almost all g. If $\varphi$ is continuous then the identity holds for all g
in G.

ii) If $\varphi\in C_{0c}\left(  G;%
\mathbb{C}
\right)  $ the Fourier transform is well defined and :

$\widehat{\varphi}\in L^{2}\left(  \widehat{G},\nu,%
\mathbb{C}
\right)  $ and $\int_{G}\left\vert \varphi\left(  g\right)  \right\vert
^{2}\mu\left(  g\right)  =\int_{\widehat{G}}\left\vert \widehat{f}\left(
\chi\right)  \right\vert ^{2}\nu\left(  \chi\right)  \Leftrightarrow\left\Vert
f\right\Vert _{2}=\left\Vert \widehat{f}\right\Vert _{2}$

iii) There is a unique extension
$\mathcal{F}$%
\ of the Fourier transform to a continuous linear map between the Hilbert
spaces :

$%
\mathcal{F}%
:L^{2}\left(  G,\mu,%
\mathbb{C}
\right)  \rightarrow L^{2}\left(  \widehat{G},\nu,%
\mathbb{C}
\right)  $

There is a continuous adjoint map $%
\mathcal{F}%
^{\ast}:L^{2}\left(  \widehat{G},\nu,%
\mathbb{C}
\right)  \rightarrow L^{2}\left(  G,\mu,%
\mathbb{C}
\right)  $ such that :

$\int_{G}\left(  \overline{\varphi_{1}\left(  g\right)  }\right)  \left(
\left(
\mathcal{F}%
^{\ast}\varphi_{2}\right)  \left(  g\right)  \right)  \mu\left(  g\right)
=\int_{\widehat{G}}\left(  \overline{%
\mathcal{F}%
\left(  \varphi_{1}\right)  \left(  \chi\right)  }\right)  \left(  \varphi
_{2}\left(  \chi\right)  \right)  \nu\left(  \chi\right)  $

If $\varphi\in L^{1}\left(  G,\mu,%
\mathbb{C}
\right)  \cap L^{2}\left(  G,\mu,%
\mathbb{C}
\right)  $ then $%
\mathcal{F}%
^{\ast}=%
\mathcal{F}%
^{-1}:$ $%
\mathcal{F}%
^{\ast}\left(  h\right)  \left(  g\right)  =\int_{\widehat{G}}h(\chi
)\chi\left(  g\right)  \nu\left(  \chi\right)  $

iv) the map : $F:\widehat{G}\rightarrow L^{1}\left(  G,S,\mu,%
\mathbb{C}
\right)  ^{\prime}::F\left(  \chi\right)  \left(  \varphi\right)
=\widehat{\varphi}\left(  \chi\right)  $ is bijective
\end{theorem}

\begin{theorem}
$L^{1}\left(  G,S,\mu,%
\mathbb{C}
\right)  $ is a commutative C*-algebra with convolution product as internal operation.

The only linear multiplicative functionals $\lambda$\ on the algebra$\ \left(
L^{1}\left(  G,\mu,%
\mathbb{C}
\right)  ,\ast\right)  :\lambda\left(  f\ast g\right)  =\lambda\left(
f\right)  \lambda\left(  f\right)  $ are the functionals induced by\ the
Fourier transform : $\lambda\left(  \chi\right)  :L^{1}\left(  G,\mu,%
\mathbb{C}
\right)  \rightarrow%
\mathbb{C}
::\lambda\left(  \chi\right)  \left(  f\right)  =\widehat{f}\left(
\chi\right)  $

$\forall f,g\in L^{1}\left(  G,\mu,%
\mathbb{C}
\right)  :\widehat{f\ast g}=\widehat{f}\times\widehat{g}$
\end{theorem}

\bigskip

Abelian Lie groups are isomorphic to the products of vector spaces and tori.

\subsubsection{Compact abelian Groups}

\begin{theorem}
(Neeb p.79) For any compact abelian group G there is a Haar measure $\mu$ such
that $\mu\left(  G\right)  =1$ and a family $\left(  \chi_{n}\right)  _{n\in%
\mathbb{N}
}$ of functions which constitute a Hermitian basis of $L^{2}\left(  G,%
\mathbb{C}
,\mu\right)  $ so that every function $\varphi\in L^{2}\left(  G,%
\mathbb{C}
,\mu\right)  $ is represented by the \textbf{Fourier} \textbf{Series} :%

\begin{equation}
\varphi=\sum_{n\in%
\mathbb{N}
}\left(  \int_{G}\overline{\chi_{n}(g)}\varphi\left(  g\right)  \mu\right)
\chi_{n}%
\end{equation}

\end{theorem}

\begin{proof}
If G is compact (it is isomorphic to a torus) then it has a finite Haar
measure $\mu$, which can be normalized to $1=\int_{G}\mu.$ And any continuous
unitary representation is completely reducible in the direct sum of orthogonal
one dimensional irreducible representations.

$\widehat{G}\subset C_{0b}\left(  G;%
\mathbb{C}
\right)  $ so $\forall p:1\leq p\leq\infty:$ $\widehat{G}\subset L^{p}\left(
G,%
\mathbb{C}
,\mu\right)  $

$\mu$ is finite, $L^{2}\left(  G,\mu,%
\mathbb{C}
\right)  \subset L^{1}\left(  G,\mu,%
\mathbb{C}
\right)  $ so that the Fourier transform extends to $L^{2}\left(  G,\mu,%
\mathbb{C}
\right)  $ and the Fourier transform is a unitary operator on L%
${{}^2}$%
.

$\widehat{G}$ is an orthonormal subset of $L^{2}\left(  G,%
\mathbb{C}
,\mu\right)  $ with the inner product

$\left\langle \chi_{1},\chi_{2}\right\rangle =\int_{G}\overline{\chi_{1}%
(g)}\chi_{2}\left(  g\right)  \mu=\delta_{\chi_{1},\chi_{2}}$

$\widehat{G}$ is discrete so : $\widehat{G}=\left\{  \chi_{n},n\in%
\mathbb{N}
\right\}  $\ and we can take $\chi_{n}$\ as a Hilbert basis of $L^{2}\left(
G,%
\mathbb{C}
,\mu\right)  $

The Fourier transform reads : $\widehat{\varphi}_{n}=\int_{G}\overline
{\chi_{n}\left(  g\right)  }\varphi\left(  g\right)  \mu$

$\varphi\in L^{2}\left(  G,S,%
\mathbb{C}
,\mu\right)  :\varphi=\sum_{n\in%
\mathbb{N}
}\left\langle \chi_{n},\varphi\right\rangle \chi_{n}=\sum_{n\in%
\mathbb{N}
}\left(  \int_{G}\overline{\chi_{n}(g)}\varphi\left(  g\right)  \mu\right)
\chi_{n}=\sum_{n\in%
\mathbb{N}
}\widehat{\varphi}_{n}\chi_{n}$

which is the representation of $\varphi$\ by a Fourier series.
\end{proof}

For any unitary representation (H,f) of G, then the spectral measure is :

$P:S=\sigma\left(
\mathbb{N}
\right)  \rightarrow%
\mathcal{L}%
\left(  H;H\right)  ::P_{n}=\int_{G}\overline{\chi_{n}\left(  g\right)
}f(g)\mu$

\subsubsection{Non compact abelian groups}

Then G is isomorphic to a finite dimensional vector space E.

\begin{theorem}
The Fourier transform reads:
\end{theorem}

\begin{equation}
\symbol{94}:L^{1}\left(  E,\mu,%
\mathbb{C}
\right)  \rightarrow C_{0\nu}\left(  E^{\ast};%
\mathbb{C}
\right)  :\widehat{\varphi}(\lambda)=\int_{E}\varphi(x)\exp\left(
-i\lambda\left(  x\right)  \right)  \mu\left(  x\right)
\end{equation}

The Fourier transform has an extension
$\mathcal{F}$%
\ on $L^{2}\left(  E,\mu,%
\mathbb{C}
\right)  $\ which is a unitary operator, but
$\mathcal{F}$%
\ is usually not given by any explicit formula.

\bigskip

\subsection{Fourier series}

\label{Fourier series}

Fourier series are defined for \textit{periodic functions} defined on $%
\mathbb{R}
^{m}$ or on bounded boxes in $%
\mathbb{R}
^{m}$. This is a special adaptation of the compact case above.

\subsubsection{Periodic functions}

A periodic function $f:%
\mathbb{R}
^{m}\rightarrow%
\mathbb{C}
$ with period $A\in%
\mathbb{R}
^{m}$ is a function such that : $\forall x\in%
\mathbb{R}
^{m}:f\left(  x+A\right)  =f\left(  x\right)  .$

If we have a space V of functions defined on a bounded box B in $%
\mathbb{R}
^{m}:B=\left\{  \xi^{\alpha}\in\lbrack a_{\alpha},b_{\alpha}[,\alpha
=1..m\right\}  $ one can define the periodic functions on $%
\mathbb{R}
^{m}:\forall f\in V,y\in B,Z\in%
\mathbb{Z}
^{m}:F\left(  y+ZA\right)  =f\left(  y\right)  $ with $A=\left\{  b_{\alpha
}-a_{\alpha},\alpha=1...m\right\}  .$

For A fixed in $%
\mathbb{R}
^{m}$\ the set : $%
\mathbb{Z}
A=\left\{  zA,z\in%
\mathbb{Z}
\right\}  $ is a closed subgroup so the set $G\mathfrak{=}%
\mathbb{R}
^{m}/\mathfrak{%
\mathbb{Z}
}^{m}A$ is a commutative compact group. The classes of equivalence are :
$x\sim y\Leftrightarrow x-y=ZA,Z\in%
\mathbb{Z}
^{m}.$

So we can define periodic functions as functions $\varphi:G\mathfrak{=}%
\mathbb{R}
^{m}/\mathfrak{%
\mathbb{Z}
}^{m}A\rightarrow%
\mathbb{C}
$.

With the change of variable : $x=\frac{a}{2\pi}\theta$ it is customary and
convenient to write :

$T^{m}=\left\{  \left(  \theta_{k}\right)  _{k=1}^{m},0\leq\theta_{k}\leq
2\pi\right\}  $

and we are left with functions $f\in C\left(  T^{m};%
\mathbb{C}
\right)  $ with the same period $\left(  2\pi,...,2\pi\right)  $ defined on
$T^{m}$

In the following, we have to deal with two kinds of spaces :

i) The space of periodic functions $f\in C\left(  T^{m};%
\mathbb{C}
\right)  :$ the usual spaces of functions, with domain in $T^{m},$ valued in $%
\mathbb{C}
.$ The Haar measure is proportional to the Lebesgue measure $d\theta=\left(
\otimes d\xi\right)  ^{m}$ meaning that we integrate on $\xi\in\left[
0,2\pi\right]  $ for each occurence of a variable.

ii) The space $C\left(
\mathbb{Z}
^{m};%
\mathbb{C}
\right)  $ of functions with domain in $%
\mathbb{Z}
^{m}$ \ and values in $%
\mathbb{C}
$ : they are defined for any combinations of m signed integers. The
appropriate measure (denoted $\nu$ above) is the Dirac measure $\delta
_{z}=\left(  \delta_{z}\right)  _{z\in%
\mathbb{Z}
}$\ and the equivalent of the integral in $L^{p}\left(
\mathbb{Z}
^{m},\delta_{z},%
\mathbb{C}
\right)  $ is a series from $z=-\infty$ to $z=+\infty$

$\left\langle z,\theta\right\rangle =\sum_{k=1}^{m}z_{k}\theta_{k}$

\subsubsection{Fourier series}

See Taylor 1 p.177

\begin{theorem}
The \textbf{Fourier transform} \textbf{of periodic functions} is the map :%

\begin{equation}
\symbol{94}:L^{1}\left(  T^{m},d\theta,%
\mathbb{C}
\right)  \rightarrow C_{0\nu}\left(
\mathbb{Z}
^{m};%
\mathbb{C}
\right)  :\widehat{f}(z)=\left(  2\pi\right)  ^{-m}\int_{T^{m}}f\left(
\theta\right)  \exp\left(  -i\left\langle z,\theta\right\rangle \right)
d\theta
\end{equation}

The \textbf{Fourier coefficients} $\widehat{f}(z)$ vanish when $z\rightarrow
\pm\infty:$

$\forall Z\in%
\mathbb{Z}
^{m}:\sup_{Z\in%
\mathbb{Z}
^{m}}\left(  1+\sum_{k=1}^{m}z_{k}^{2}\right)  ^{Z}\left\vert \widehat
{f}(z)\right\vert <\infty$
\end{theorem}

\begin{theorem}
The space $L^{2}\left(
\mathbb{Z}
^{m},\delta_{z},%
\mathbb{C}
\right)  $ is a Hilbert space with the scalar product $:\left\langle
\varphi,\psi\right\rangle =\sum_{z\in%
\mathbb{Z}
^{m}}\overline{\varphi\left(  z\right)  }\psi\left(  z\right)  $ and the
Hilbert basis : $\left(  \exp iz\right)  _{z\in%
\mathbb{Z}
^{m}}k=1...m$

The space $L^{2}\left(  T^{m},d\theta,%
\mathbb{C}
\right)  $ is a Hilbert space with the scalar product $:\left\langle
f,g\right\rangle =\left(  2\pi\right)  ^{-m}\int_{T^{m}}\overline{f\left(
\theta\right)  }g\left(  \theta\right)  d\theta$ and the Hilbert basis :
$\left(  \exp iz\theta\right)  _{z\in%
\mathbb{Z}
^{m}}$
\end{theorem}

We have the identity :

$\left\langle \exp iz_{1}\theta,\exp iz_{2}\theta\right\rangle =\left(
2\pi\right)  ^{-m}\int_{T^{m}}e^{-i\left\langle z_{1},\theta\right\rangle
}e^{i\left\langle z_{2},\theta\right\rangle }d\theta=\delta_{z_{1},z_{2}}$

\begin{theorem}
The Fourier transform is a continuous operator between the Hilbert spaces :

$%
\mathcal{F}%
:L^{2}\left(  T^{m},d\theta,%
\mathbb{C}
\right)  \rightarrow L^{2}\left(
\mathbb{Z}
^{m},\delta_{z},%
\mathbb{C}
\right)  ::%
\mathcal{F}%
\left(  f\right)  \left(  z\right)  =\left(  2\pi\right)  ^{-m}\int_{T^{m}%
}f\left(  \theta\right)  \exp\left(  -i\left\langle z,\theta\right\rangle
\right)  d\theta$

and its inverse is the adjoint map :

$%
\mathcal{F}%
^{\ast}:L^{2}\left(
\mathbb{Z}
^{m},\delta_{z},%
\mathbb{C}
\right)  \rightarrow L^{2}\left(  T^{m},d\theta,%
\mathbb{C}
\right)  ::%
\mathcal{F}%
^{\ast}\left(  \varphi\right)  \left(  \theta\right)  =\sum_{z\in%
\mathbb{Z}
^{m}}\varphi\left(  z\right)  \exp i\left\langle z,\theta\right\rangle $

It is an isometry : $\left\langle
\mathcal{F}%
\left(  f\right)  ,\varphi\right\rangle _{L^{2}\left(
\mathbb{Z}
^{m},\nu,%
\mathbb{C}
\right)  }=\left\langle f,%
\mathcal{F}%
^{\ast}\varphi\right\rangle _{L^{2}\left(  T^{m},\mu,%
\mathbb{C}
\right)  }$

Parseval identity : $\left\langle
\mathcal{F}%
\left(  f\right)  ,%
\mathcal{F}%
\left(  h\right)  \right\rangle _{L^{2}\left(
\mathbb{Z}
^{m}\delta_{z},%
\mathbb{C}
\right)  }=\left\langle f,h\right\rangle _{L^{2}\left(  T^{m},d\theta,%
\mathbb{C}
\right)  }$

Plancherel identity : $\left\Vert
\mathcal{F}%
\left(  f\right)  \right\Vert _{L^{2}\left(
\mathbb{Z}
^{m},\delta_{z},%
\mathbb{C}
\right)  }=\left\Vert f\right\Vert _{L^{2}\left(  T^{m},d\theta,%
\mathbb{C}
\right)  }$
\end{theorem}

\begin{theorem}
(Taylor 1 p.183) Any periodic function $f\in L^{2}\left(  T^{m},d\theta,%
\mathbb{C}
\right)  $ can be written as the series :%

\begin{equation}
f\left(  \theta\right)  =\sum_{z\in%
\mathbb{Z}
^{m}}%
\mathcal{F}%
\left(  f\right)  \left(  z\right)  \exp i\left\langle z,\theta\right\rangle
\end{equation}

The series is still absolutely convergent if $f\in C_{r}\left(  T^{m};%
\mathbb{C}
\right)  $ and r
$>$
m/2 or if $f\in C_{\infty}\left(  T^{m};%
\mathbb{C}
\right)  $
\end{theorem}

\subsubsection{Operations with Fourier series}

\begin{theorem}
If f is r differentiable : $\widehat{D_{\alpha_{1}...\alpha_{r}}f}\left(
z\right)  =\left(  i\right)  ^{r}\left(  z_{\alpha_{1}}...z_{\alpha_{r}%
}\right)  \widehat{f}\left(  z\right)  $
\end{theorem}

\begin{theorem}
For $f,g\in L^{1}\left(  T^{m},d\theta,%
\mathbb{C}
\right)  :$

$\widehat{f}\left(  z\right)  \times\widehat{g}\left(  z\right)
=\widehat{\left(  f\ast g\right)  }\left(  z\right)  $ with $\left(  f\ast
g\right)  \left(  \theta\right)  =\left(  2\pi\right)  ^{-m}\int_{T^{m}%
}f\left(  \zeta\right)  g\left(  \zeta-\theta\right)  d\zeta$

$\left(  \widehat{f\times g}\right)  \left(  z\right)  =\sum_{\zeta\in%
\mathbb{Z}
^{m}}\widehat{f}\left(  z-\zeta\right)  \widehat{g}\left(  \zeta\right)  $
\end{theorem}

\begin{theorem}
Abel summability result (Taylor 1 p.180) : For any $f\in L^{1}\left(
T^{m},d\theta,%
\mathbb{C}
\right)  $ the series

$J_{r}f\left(  \theta\right)  =\sum_{z\in%
\mathbb{Z}
^{m}}\widehat{f}\left(  z\right)  r^{\left\Vert z\right\Vert }e^{i\left\langle
z,\theta\right\rangle }$ with $\left\Vert z\right\Vert =\sum_{k=1}%
^{m}\left\vert z_{k}\right\vert $

converges to f when $r\rightarrow1$

The convergence is uniform on $T^{m}$ if f is continuous or if $f\in
L^{p}\left(  T^{m},d\theta,%
\mathbb{C}
\right)  ,1\leq p\leq\infty$
\end{theorem}

If m=1 $J_{r}f\left(  \theta\right)  $ can be written with $z=r\sin\theta$\ :

$J_{r}f\left(  \theta\right)  =PI\left(  f\right)  \left(  z\right)
=\frac{1-\left\vert z\right\vert ^{2}}{2\pi}\int_{\left\vert \zeta\right\vert
=1}\frac{f\left(  \zeta\right)  }{\left\vert \zeta-z\right\vert ^{2}}d\zeta$

and is called the Poisson integral.

This is the unique solution to the Dirichlet problem :

$u\in C\left(
\mathbb{R}
^{2};%
\mathbb{C}
\right)  :\Delta u=0$ in $\left\vert x^{2}+y^{2}\right\vert <1,$ $u=f$ on
$\left\vert x^{2}+y^{2}\right\vert =1$

\bigskip

\subsection{Fourier integrals}

\label{Fourier integral}

The Fourier integral is defined for functions on $%
\mathbb{R}
^{m}$ or an open subset of $%
\mathbb{R}
^{m}.$

\subsubsection{Definition}

The Pontryagin dual of $G=\left(
\mathbb{R}
^{m},+\right)  $ is $\widehat{G}=\left\{  \exp i\theta\left(  g\right)  ,t\in%
\mathbb{R}
^{m\ast}\right\}  $ so $\chi\left(  g\right)  =\exp i\sum_{k=1}^{m}t_{i}%
x_{i}=\exp i\left\langle t,x\right\rangle $

The Haar measure is proportional to the Lebesgue measure, which gives several
common definitions of the Fourier transform, depending upon the location of a
$2\pi$ factor. The chosen solution gives the same formula for the inverse (up
to a sign). See Wikipedia "Fourier transform" for the formulas with other
conventions of scaling.

\bigskip

\begin{theorem}
(Lieb p.125) The \textbf{Fourier transform of functions} is the map :%

\begin{equation}
\symbol{94}:L^{1}\left(
\mathbb{R}
^{m},dx,%
\mathbb{C}
\right)  \rightarrow C_{0\nu}\left(
\mathbb{R}
^{m};%
\mathbb{C}
\right)  :\widehat{f}(t)=\left(  2\pi\right)  ^{-m/2}\int_{%
\mathbb{R}
^{m}}f(x)e^{-i\left\langle t,x\right\rangle }dx
\end{equation}

with $\left\langle x,t\right\rangle =\sum_{k=1}^{m}x_{k}t_{k}$

then $\widehat{f}\in L^{\infty}\left(
\mathbb{R}
^{m},dx,%
\mathbb{C}
\right)  $ and $\left\Vert \widehat{f}\right\Vert _{\infty}\leq\left\Vert
f\right\Vert _{1}$

Moreover :

If $f\in L^{2}\left(
\mathbb{R}
^{m},dx,%
\mathbb{C}
\right)  \cap L^{1}\left(
\mathbb{R}
^{m},dx,%
\mathbb{C}
\right)  $

then $\widehat{f}\in L^{2}\left(
\mathbb{R}
^{m},dx,%
\mathbb{C}
\right)  $ and $\left\Vert \widehat{f}\right\Vert _{2}=\left\Vert f\right\Vert
_{1}$

If $f\in S\left(
\mathbb{R}
^{m}\right)  $ then $\widehat{f}\in S\left(
\mathbb{R}
^{m}\right)  $

For $\widehat{f}\in L^{1}\left(
\mathbb{R}
^{m},dx,%
\mathbb{C}
\right)  $\ there is an inverse given by :%

\begin{equation}%
\mathcal{F}%
^{-1}\left(  \widehat{f}\right)  (x)=\left(  2\pi\right)  ^{-m/2}\int_{%
\mathbb{R}
^{m}}\widehat{f}(t)e^{i\left\langle t,x\right\rangle }dt
\end{equation}

for almost all t. If f is continuous then the identity holds for all x in $%
\mathbb{R}
^{m}$
\end{theorem}

The Fourier transform is a bijective, bicontinuous, map on $S\left(
\mathbb{R}
^{m}\right)  .$ As $S\left(
\mathbb{R}
^{m}\right)  \subset L^{p}\left(
\mathbb{R}
^{m},dx,%
\mathbb{C}
\right)  $ for any p, it is also an unitary map (see below).

Warning ! the map is usually not surjective, there is no simple
characterization of the image

\begin{theorem}
(Lieb p.128,130) There is a unique extension
$\mathcal{F}$%
\ of the Fourier transform to a continuous operator $%
\mathcal{F}%
\in%
\mathcal{L}%
\left(  L^{2}\left(
\mathbb{R}
^{m},dx,%
\mathbb{C}
\right)  ;L^{2}\left(
\mathbb{R}
^{m},dx,%
\mathbb{C}
\right)  \right)  .$

Moreover
$\mathcal{F}$%
\ is an unitary isometry :

Its inverse is the adjoint map : $%
\mathcal{F}%
^{\ast}=%
\mathcal{F}%
^{-1}$ with $%
\mathcal{F}%
^{-1}\left(  \widehat{f}\right)  \left(  x\right)  =\left(
\mathcal{F}%
\widehat{f}\right)  \left(  -x\right)  .$

Parserval formula : $\left\langle
\mathcal{F}%
\left(  f\right)  ,g\right\rangle _{L^{2}}=\left\langle f,%
\mathcal{F}%
^{\ast}\left(  g\right)  \right\rangle _{L^{2}}\Leftrightarrow\int_{%
\mathbb{R}
^{m}}\overline{%
\mathcal{F}%
\left(  f\right)  \left(  t\right)  }g\left(  t\right)  dt=\int_{%
\mathbb{R}
^{m}}\overline{f\left(  x\right)  }\left(
\mathcal{F}%
^{\ast}g\right)  \left(  x\right)  dx$

Plancherel identity: $\left\Vert
\mathcal{F}%
\left(  f\right)  \right\Vert _{L^{2}\left(
\mathbb{R}
^{m},dx,%
\mathbb{C}
\right)  }=\left\Vert f\right\Vert _{L^{2}\left(
\mathbb{R}
^{m},dx,%
\mathbb{C}
\right)  }$
\end{theorem}

However
$\mathcal{F}$%
\ is not given by any explicit formula but by an approximation with a sequence
in $L^{1}\cap L^{2}$ which is dense in L%
${{}^2}$%
. Of course whenever f belongs also to $L^{1}\left(
\mathbb{R}
^{m},dx,%
\mathbb{C}
\right)  $ then $%
\mathcal{F}%
\equiv\symbol{94}$

Remark : If O is an open subset of $%
\mathbb{R}
^{m}$ then any function $f\in L^{1}\left(  O,dx,%
\mathbb{C}
\right)  $ can be extended to a function $\widetilde{f}\in L^{1}\left(
\mathbb{R}
^{m},dx,%
\mathbb{C}
\right)  $ with $\widetilde{f}\left(  x\right)  =f(x),x\in O,\widetilde
{f}\left(  x\right)  =0,x\neq O$ (no continuity condition is required).

\subsubsection{Operations with Fourier integrals}

\begin{theorem}
Derivatives : whenever the Fourier transform is defined :%

\begin{equation}%
\mathcal{F}%
\left(  D_{\alpha_{1}...\alpha_{r}}f\right)  \left(  t\right)  =i^{r}\left(
t_{\alpha_{1}}..t_{\alpha_{r}}\right)
\mathcal{F}%
\left(  f\right)  \left(  t\right)
\end{equation}

\begin{equation}%
\mathcal{F}%
(x^{\alpha_{1}}x^{\alpha_{2}}...x^{\alpha_{r}}f)=i^{r}D_{\alpha_{1}%
...\alpha_{r}}\left(
\mathcal{F}%
\left(  f\right)  \right)
\end{equation}

\end{theorem}

with, of course, the usual notation for the derivative : $D_{\left(
\alpha\right)  }=D_{\alpha_{1}...\alpha_{s}}=\dfrac{\partial}{\partial
\xi^{\alpha_{1}}}\dfrac{\partial}{\partial\xi^{\alpha_{2}}}...\dfrac{\partial
}{\partial\xi^{\alpha_{s}}}$

\begin{theorem}
(Lieb p.181) If $f\in L^{2}\left(
\mathbb{R}
^{m},dx,%
\mathbb{C}
\right)  $ then $\widehat{f}\in H^{1}\left(
\mathbb{R}
^{m}\right)  $ iff the function : $g:%
\mathbb{R}
^{m}\rightarrow%
\mathbb{R}
^{m}::g\left(  t\right)  =\left\Vert t\right\Vert \widehat{f}\left(  t\right)
$ is in $L^{2}\left(
\mathbb{R}
^{m},dx,%
\mathbb{C}
\right)  $ . And then 

$%
\mathcal{F}%
\left(  D_{\alpha_{1}...\alpha_{r}}f\right)  \left(  t\right)  =\left(
i\right)  ^{r}\left(  t_{\alpha_{1}}..t_{\alpha_{r}}\right)
\mathcal{F}%
\left(  f\right)  $ 

$\left\Vert f\right\Vert _{2}=\int_{%
\mathbb{R}
^{m}}\left\vert \widehat{f}\left(  t\right)  \right\vert ^{2}\left(
1+\left\Vert t\right\Vert ^{2}\right)  dt$
\end{theorem}

\begin{theorem}
Translation : For $f\in L^{1}\left(
\mathbb{R}
^{m},dx,%
\mathbb{C}
\right)  ,a\in%
\mathbb{R}
^{m},$

$\tau_{a}\left(  x\right)  =x-a,\tau_{a}^{\ast}f=f\circ\tau_{a}$

$%
\mathcal{F}%
\left(  \tau_{a}^{\ast}f\right)  \left(  t\right)  =e^{-i\left\langle
a,t\right\rangle }%
\mathcal{F}%
\left(  f\right)  \left(  t\right)  $

$%
\mathcal{F}%
\left(  e^{i\left\langle a,x\right\rangle }f\right)  =\tau_{a}^{\ast}%
\mathcal{F}%
\left(  f\right)  $
\end{theorem}

\begin{theorem}
Scaling : For $f\in L^{1}\left(
\mathbb{R}
^{m},dx,%
\mathbb{C}
\right)  ,a\neq0\in%
\mathbb{R}
,$

$\lambda_{a}\left(  x\right)  =ax,\lambda_{a}^{\ast}f=f\circ\lambda_{a}$

$%
\mathcal{F}%
\left(  \lambda_{a}^{\ast}f\right)  =\frac{1}{\left\vert a\right\vert }%
\lambda_{1/a}^{\ast}%
\mathcal{F}%
\left(  f\right)  \Leftrightarrow%
\mathcal{F}%
\left(  \lambda_{a}^{\ast}f\right)  \left(  t\right)  =\frac{1}{\left\vert
a\right\vert }%
\mathcal{F}%
\left(  f\right)  \left(  \frac{t}{a}\right)  $
\end{theorem}

\begin{theorem}
For $f\in L^{1}\left(
\mathbb{R}
^{m},dx,%
\mathbb{C}
\right)  ,$ $L\in GL\left(
\mathbb{R}
^{m};%
\mathbb{R}
^{m}\right)  $

$y=\left[  A\right]  x$ with $\det A\neq0:L_{\ast}\left(
\mathcal{F}%
\left(  f\right)  \right)  =\left\vert \det A\right\vert
\mathcal{F}%
\left(  \left(  L^{t}\right)  ^{\ast}f\right)  $
\end{theorem}

\begin{proof}
$L_{\ast}\left(
\mathcal{F}%
\left(  f\right)  \right)  =\left(  L^{-1}\right)  ^{\ast}\left(
\mathcal{F}%
\left(  f\right)  \right)  =\left(
\mathcal{F}%
\left(  f\right)  \right)  \circ L^{-1}=\left(  2\pi\right)  ^{-m/2}\int_{%
\mathbb{R}
^{m}}e^{-i\left\langle A^{-1}t,x\right\rangle }f\left(  x\right)  dx$

$=\left(  2\pi\right)  ^{-m/2}\int_{%
\mathbb{R}
^{m}}e^{-i\left\langle t,\left(  A^{t}\right)  ^{-1}x\right\rangle }f\left(
x\right)  dx=\left(  2\pi\right)  ^{-m/2}\left\vert \det A\right\vert \int_{%
\mathbb{R}
^{m}}e^{-i\left\langle t,y\right\rangle }f\left(  A^{t}y\right)  dy=\left\vert
\det A\right\vert
\mathcal{F}%
\left(  f\circ L^{t}\right)  $
\end{proof}

\begin{theorem}
Conjugation : For $f\in L^{1}\left(
\mathbb{R}
^{m},dx,%
\mathbb{C}
\right)  :$%

\begin{equation}
\overline{%
\mathcal{F}%
\left(  f\right)  \left(  t\right)  }=%
\mathcal{F}%
\left(  f\right)  \left(  -t\right)
\end{equation}

\end{theorem}

\begin{theorem}
Convolution (Lieb p.132) If $f\in L^{p}\left(
\mathbb{R}
^{m},dx,%
\mathbb{C}
\right)  ,g\in L^{q}\left(
\mathbb{R}
^{m},dx,%
\mathbb{C}
\right)  $ with $1+\frac{1}{r}=\frac{1}{p}+\frac{1}{q},1\leq p,q,r\leq2$ then :%

\begin{equation}%
\mathcal{F}%
\left(  f\ast g\right)  =\left(  2\pi\right)  ^{m/2}%
\mathcal{F}%
\left(  f\right)
\mathcal{F}%
\left(  g\right)
\end{equation}

\begin{equation}%
\mathcal{F}%
\left(  fg\right)  =\left(  2\pi\right)  ^{-m/2}%
\mathcal{F}%
\left(  f\right)  \ast%
\mathcal{F}%
\left(  g\right)
\end{equation}

\end{theorem}

\paragraph{Some usual Fourier transforms for m=1:\newline}

$%
\mathcal{F}%
\left(  He^{-ax}\right)  =\frac{1}{\sqrt{2\pi}}\frac{1}{a+it}$

$%
\mathcal{F}%
\left(  e^{-ax^{2}}\right)  =\frac{1}{\sqrt{2a}}e^{-\frac{t^{2}}{4a}}$

$%
\mathcal{F}%
\left(  e^{-a\left\vert x\right\vert }\right)  =\sqrt{\frac{2}{\pi}}\frac
{a}{a^{2}+t^{2}}$

\paragraph{Partial Fourier transform\newline}

\begin{theorem}
(Zuily p.121) Let $X=\left(  x,y\right)  \in%
\mathbb{R}
^{n}\times%
\mathbb{R}
^{p}=%
\mathbb{R}
^{m}$

The partial Fourier transform of $f\in L^{1}\left(
\mathbb{R}
^{n+p};dx,%
\mathbb{C}
\right)  $ on the first n components is the function :

$\widehat{f}(t,y)=\left(  2\pi\right)  ^{-n/2}\int_{%
\mathbb{R}
^{n}}f(x,y)e^{-i\left\langle t,x\right\rangle }dx\in C_{0\nu}\left(
\mathbb{R}
^{m};%
\mathbb{C}
\right)  $

It is still a bijective map on $S\left(
\mathbb{R}
^{m}\right)  $ and the inverse is :

$f(x,y)=\left(  2\pi\right)  ^{-n/2}\int_{%
\mathbb{R}
^{n}}\widehat{f}(t,y)e^{i\left\langle t,x\right\rangle }dt$
\end{theorem}

All the previous properties hold whenever we consider the Fourier transform on
the first n components.

\paragraph{Fourier transforms of radial functions\newline}

A radial function in $%
\mathbb{R}
^{m}$ is a function F(x)=f(r) where r=$\left\Vert x\right\Vert $

$\widehat{F}\left(  t\right)  =\left(  2\pi\right)  ^{-m/2}\int_{0}^{\infty
}f\left(  r\right)  \phi\left(  r\left\Vert t\right\Vert \right)  r^{n-1}dr$

$=\left\Vert t\right\Vert ^{1-\frac{m}{2}}\int_{0}^{\infty}f\left(  r\right)
J_{\frac{m}{2}-1}\left(  r\left\Vert t\right\Vert \right)  r^{\frac{m}{2}}dr$

with

$\phi\left(  u\right)  =\int_{S^{m-1}}e^{iut}d\sigma_{S}=A_{m-2}\int_{-1}%
^{1}e^{irs}\left(  1-s^{2}\right)  ^{\left(  m-3\right)  /2}ds=\left(
2\pi\right)  ^{m/2}u^{1-m/2}J_{\frac{m}{2}-1}\left(  u\right)  $

with the Bessel function defined for $\operatorname{Re}\nu>-1/2:$

$J_{\nu}\left(  z\right)  =\left(  \Gamma\left(  \frac{1}{2}\right)
\Gamma\left(  \nu+\frac{1}{2}\right)  \right)  ^{-1}\left(  \frac{z}%
{2}\right)  ^{\nu}\int_{-1}^{1}\left(  1-t^{2}\right)  ^{\nu-\frac{1}{2}%
}e^{izt}dt$

which is solution of the ODE :

$\left(  \frac{d^{2}}{dr^{2}}+\frac{1}{r}\frac{d}{dr}+\left(  1-\frac{\nu^{2}%
}{r^{2}}\right)  \right)  J_{\nu}\left(  r\right)  =0$

\paragraph{Paley-Wiener theorem}

\begin{theorem}
(Zuily p.123) For any function $f\in C_{\infty c}\left(
\mathbb{R}
^{m};%
\mathbb{C}
\right)  $ with support Supp(f) = $\left\{  x\in%
\mathbb{R}
^{m}:\left\Vert x\right\Vert \leq r\right\}  $ there is a holomorphic function
F on $%
\mathbb{C}
^{m}$ such that :

$\forall x\in%
\mathbb{R}
^{m}:F\left(  x\right)  =\widehat{f}\left(  x\right)  $

(1) $\forall n\in%
\mathbb{N}
,\exists C_{n}\in%
\mathbb{R}
:\forall z\in%
\mathbb{C}
^{m}:\left\vert F\left(  z\right)  \right\vert \leq C_{n}\left(  1+\left\Vert
z\right\Vert \right)  ^{-n}e^{r\left\vert \operatorname{Im}z\right\vert }$

Conversely for any holomorphic function F on $%
\mathbb{C}
^{m}$ which meets the condition (1) there is a function $f\in C_{\infty
c}\left(
\mathbb{R}
^{m};%
\mathbb{C}
\right)  $ such that Supp(f)=$\left\{  x\in%
\mathbb{R}
^{m}:\left\Vert x\right\Vert \leq r\right\}  $ and $\forall x\in%
\mathbb{R}
^{m}:$ $\widehat{f}\left(  x\right)  =F\left(  x\right)  $
\end{theorem}

This theorem shows in particular that the Fourier transform of functions with
compact support is never compactly supported (except if it is null). This is a
general feature of Fourier transform : $\widehat{f}$ is always "more spread
out" than f. And conversely
$\mathcal{F}$%
* is "more concentrated" than f.

\paragraph{Asymptotic analysis\newline}

\begin{theorem}
(Zuily p.127) For any functions $\varphi\in C_{\infty}\left(
\mathbb{R}
^{m};%
\mathbb{R}
\right)  ,f\in C_{\infty c}\left(
\mathbb{R}
^{m};%
\mathbb{C}
\right)  $ the asymptotic value of $I\left(  t\right)  =\int_{%
\mathbb{R}
^{m}}e^{it\varphi\left(  x\right)  }f\left(  x\right)  dx$ when $t\rightarrow
+\infty$ is :

i) If $\forall x\in Supp(f),\varphi^{\prime}\left(  x\right)  \neq0$ then
$\forall n\in%
\mathbb{N}
,\exists C_{n}\in%
\mathbb{R}
:\forall t\geq1:\left\vert I\left(  t\right)  \right\vert \leq C_{n}t^{-n}$

ii) If $\varphi$\ \ has a unique critical point $a\in Supp(f)$\ and it is not
degenerate ($\det\varphi"\left(  a\right)  \neq0)$ then :

$\forall n\in%
\mathbb{N}
,\exists\left(  a_{k}\right)  _{k=0}^{n},a_{k}\in%
\mathbb{C}
,\exists C_{n}>0,r_{n}\in C\left(
\mathbb{R}
;%
\mathbb{R}
\right)  :$

$I(t)=r_{n}\left(  t\right)  +e^{it\varphi\left(  a\right)  }\sum_{k=0}%
^{n}a_{k}t^{-\frac{\pi}{2}-k}$

$\left\vert r_{n}\left(  t\right)  \right\vert \leq C_{n}t^{-\frac{\pi}{2}-n}$

$a_{0}=\frac{\left(  2\pi\right)  ^{m/2}}{\sqrt{\left\vert \det\varphi"\left(
a\right)  \right\vert }}e^{i\frac{\pi}{4}\epsilon}f\left(  a\right)  $ with
$\epsilon=$sign $\det\varphi"\left(  a\right)  $
\end{theorem}

\bigskip

\subsection{Fourier transform of distributions}

\label{Fourier transform of distributions}

For an abelian topological group G endowed with a Haar measure $\mu$ the
Fourier transform is well defined as a continuous linear map : $\symbol{94}%
:L^{1}\left(  G,\mu,%
\mathbb{C}
\right)  \rightarrow C_{0\nu}\left(  G;%
\mathbb{C}
\right)  $ and its has an extension
$\mathcal{F}$%
\ to $L^{2}\left(  G,\mu,%
\mathbb{C}
\right)  $ as well.

So if V is some subspace of $L^{1}\left(  G,\mu,%
\mathbb{C}
\right)  $ or $L^{2}\left(  G,\mu,%
\mathbb{C}
\right)  $\ we can define the Fourier transform on the space of distributions
V' as :

$%
\mathcal{F}%
:V^{\prime}\rightarrow V^{\prime}::%
\mathcal{F}%
\left(  S\right)  \left(  \varphi\right)  =S\left(
\mathcal{F}%
\left(  \varphi\right)  \right)  $ whenever $%
\mathcal{F}%
\left(  \varphi\right)  \in V$

If G is a compact group $%
\mathcal{F}%
\left(  \varphi\right)  $ is a function on $\widehat{G}$ which is a discrete
group and cannot belong to V (except if G is itself a finite group). So the
procedure will work only if G is isomorphic to a finite dimensional vector
space E, because $\widehat{G}$ is then isomorphic to E.

\subsubsection{Definition}

The general rules are :%

\begin{equation}%
\mathcal{F}%
\left(  S\right)  \left(  \varphi\right)  =S\left(
\mathcal{F}%
\left(  \varphi\right)  \right)  \text{ whenever }%
\mathcal{F}%
\left(  \varphi\right)  \in V
\end{equation}

\begin{equation}%
\mathcal{F}%
\left(  T\left(  f\right)  \right)  =T\left(
\mathcal{F}%
\left(  f\right)  \right)  \text{ whenever }S=T(f)\in V^{\prime}%
\end{equation}

Here T is one of the usual maps associating functions on $%
\mathbb{R}
^{m}$ to distributions.%

$\mathcal{F}$%
,%
$\mathcal{F}$%
* on functions are given by the usual formulas.

\bigskip

\begin{theorem}
Whenever the Fourier transform is well defined for a function f, and there is
an associated distribution S=T(f), the Fourier transform of S is the
distribution :%

\begin{equation}%
\mathcal{F}%
\left(  T\left(  f\right)  \right)  =T\left(
\mathcal{F}%
\left(  f\right)  \right)  =\left(  2\pi\right)  ^{-m/2}T_{t}\left(
S_{x}\left(  e^{-i\left\langle x,t\right\rangle }\right)  \right)
\end{equation}

\end{theorem}

\begin{proof}
For $S=T(f),f\in L^{1}\left(
\mathbb{R}
^{m},dx,%
\mathbb{C}
\right)  $

$S\left(
\mathcal{F}%
\left(  \varphi\right)  \right)  =T(f)\left(
\mathcal{F}%
\left(  \varphi\right)  \right)  )=\left(  2\pi\right)  ^{-m/2}\int_{%
\mathbb{R}
^{m}}f\left(  x\right)  \left(  \int_{%
\mathbb{R}
^{m}}e^{-i\left\langle x,t\right\rangle }\varphi\left(  t\right)  dt\right)
dx $

$=\int_{%
\mathbb{R}
^{m}}\varphi\left(  t\right)  \left(  \left(  2\pi\right)  ^{-m/2}\int_{%
\mathbb{R}
^{m}}e^{-i\left\langle x,t\right\rangle }f\left(  x\right)  dx\right)
dt=T\left(  \widehat{f}\right)  \left(  \varphi\right)  $

$=\left(  2\pi\right)  ^{-m/2}\int_{%
\mathbb{R}
^{m}}\varphi\left(  t\right)  \left(  T_{x}(f)\left(  e^{-i\left\langle
x,t\right\rangle }\right)  \right)  dt$

$=\left(  2\pi\right)  ^{-m/2}T_{t}\left(  S_{x}\left(  e^{-i\left\langle
x,t\right\rangle }\right)  \right)  \left(  \varphi\right)  =%
\mathcal{F}%
\left(  S\right)  \left(  \varphi\right)  $
\end{proof}

Which implies that it suffices to compute $S_{x}\left(  e^{-i\left\langle
x,t\right\rangle }\right)  $ to get the Fourier transform of the distribution.

\begin{theorem}
The map :

$F:L^{2}\left(
\mathbb{R}
^{m},dx,%
\mathbb{C}
\right)  \rightarrow L^{2}\left(
\mathbb{R}
^{m},dx,%
\mathbb{C}
\right)  ::F\left(  f\right)  \left(  x\right)  =\left(  2\pi\right)
^{-m/2}T_{t}(f)\left(  e^{-i\left\langle x,t\right\rangle }\right)  $

is a bijective isometry.
\end{theorem}

\begin{theorem}
(Zuily p.114) The Fourier transform
$\mathcal{F}$%
\ and its inverse
$\mathcal{F}$%
* are continuous bijective linear operators on the space $S\left(
\mathbb{R}
^{m}\right)  ^{\prime}$ of tempered distributions

$%
\mathcal{F}%
:S\left(
\mathbb{R}
^{m}\right)  ^{\prime}\rightarrow S\left(
\mathbb{R}
^{m}\right)  ^{\prime}::%
\mathcal{F}%
\left(  S\right)  \left(  \varphi\right)  =S\left(
\mathcal{F}%
\left(  \varphi\right)  \right)  $

$%
\mathcal{F}%
^{\ast}:S\left(
\mathbb{R}
^{m}\right)  ^{\prime}\rightarrow S\left(
\mathbb{R}
^{m}\right)  ^{\prime}::%
\mathcal{F}%
^{\ast}\left(  S\right)  \left(  \varphi\right)  =S\left(
\mathcal{F}%
^{\ast}\left(  \varphi\right)  \right)  $
\end{theorem}

\begin{theorem}
(Zuily p.117) The Fourier transform
$\mathcal{F}$%
\ and its inverse
$\mathcal{F}$%
* are continuous linear operators on the space $C_{rc}\left(
\mathbb{R}
^{m};%
\mathbb{C}
\right)  _{c}^{\prime}$ of distributions with compact support.

$%
\mathcal{F}%
:C_{rc}\left(
\mathbb{R}
^{m};%
\mathbb{C}
\right)  _{c}^{\prime}\rightarrow C_{rc}\left(
\mathbb{R}
^{m};%
\mathbb{C}
\right)  _{c}^{\prime}::%
\mathcal{F}%
\left(  S\right)  \left(  \varphi\right)  =S\left(
\mathcal{F}%
\left(  \varphi\right)  \right)  $

$%
\mathcal{F}%
\ast:C_{rc}\left(
\mathbb{R}
^{m};%
\mathbb{C}
\right)  _{c}^{\prime}\rightarrow C_{rc}\left(
\mathbb{R}
^{m};%
\mathbb{C}
\right)  _{c}^{\prime}::%
\mathcal{F}%
^{\ast}\left(  S\right)  \left(  \varphi\right)  =S\left(
\mathcal{F}%
^{\ast}\left(  \varphi\right)  \right)  $

Moreover : if $S\in C_{\infty c}\left(
\mathbb{R}
^{m};%
\mathbb{C}
\right)  _{c}^{\prime}\equiv C_{\infty}\left(
\mathbb{R}
^{m};%
\mathbb{C}
\right)  ^{\prime}$ then $S_{x}\left(  e^{-i\left\langle x,t\right\rangle
}\right)  \in C_{\infty}\left(
\mathbb{R}
^{m};%
\mathbb{C}
\right)  $ and can be extended to a holomorphic function in $%
\mathbb{C}
^{m}$
\end{theorem}

\begin{theorem}
Paley-Wiener-Schwartz (Zuily p.126): For any distribution $S\in\left(
C_{rc}\left(
\mathbb{R}
^{m};%
\mathbb{C}
\right)  \right)  _{c}^{\prime}$ with support Supp(S)=$\left\{  x\in%
\mathbb{R}
^{m}:\left\Vert x\right\Vert \leq\rho\right\}  $ there is a holomorphic
function F on $%
\mathbb{C}
^{m}$ such that :

$\forall\varphi\in C_{rc}\left(
\mathbb{R}
^{m};%
\mathbb{C}
\right)  :T\left(  F\right)  \left(  \varphi\right)  =\left(
\mathcal{F}%
\left(  S\right)  \right)  \left(  \varphi\right)  $

(1) $\exists C\in%
\mathbb{R}
:\forall z\in%
\mathbb{C}
^{m}:\left\vert F\left(  z\right)  \right\vert \leq C_{n}\left(  1+\left\Vert
z\right\Vert \right)  ^{-r}e^{\rho\left\vert \operatorname{Im}z\right\vert }$

Conversely for any holomorphic function F on $%
\mathbb{C}
^{m}$ which meets the condition (1) there is a distribution $S\in\left(
C_{rc}\left(
\mathbb{R}
^{m};%
\mathbb{C}
\right)  \right)  _{c}^{\prime}$ with support Supp(S)=$\left\{  x\in%
\mathbb{R}
^{m}:\left\Vert x\right\Vert \leq\rho\right\}  $ such that $\forall\varphi\in
C_{rc}\left(
\mathbb{R}
^{m};%
\mathbb{C}
\right)  :T\left(  F\right)  \left(  \varphi\right)  =\left(
\mathcal{F}%
\left(  S\right)  \right)  \left(  \varphi\right)  $
\end{theorem}

So $%
\mathcal{F}%
\left(  S\right)  =T\left(  F\right)  $ and as $S\in S\left(
\mathbb{R}
^{m}\right)  ^{\prime}:$

$S=%
\mathcal{F}%
^{\ast}T\left(  F\right)  =T\left(
\mathcal{F}%
^{\ast}F\right)  =\left(  2\pi\right)  ^{-m/2}T_{t}\left(  S_{x}\left(
e^{i\left\langle x,t\right\rangle }\right)  \right)  \Rightarrow F\left(
t\right)  =S_{x}\left(  e^{i\left\langle x,t\right\rangle }\right)  $

\subsubsection{Properties}

\begin{theorem}
Derivative : Whenever the Fourier transform and the derivative are defined :%

\begin{equation}%
\mathcal{F}%
\left(  D_{\alpha_{1}...\alpha_{r}}S\right)  =i^{r}\left(  t_{\alpha_{1}%
}...t_{\alpha_{r}}\right)
\mathcal{F}%
\left(  S\right)
\end{equation}

\begin{equation}%
\mathcal{F}%
(x^{\alpha_{1}}...x^{\alpha_{r}}S)=i^{r}D_{\alpha_{1}...\alpha_{r}}\left(
\mathcal{F}%
\left(  S\right)  \right)
\end{equation}

\end{theorem}

\begin{theorem}
(Zuily p.115) Tensorial product : For $S_{k}\in S\left(
\mathbb{R}
^{m}\right)  ^{\prime},k=1..n:$%

\begin{equation}%
\mathcal{F}%
\left(  S_{1}\otimes S_{2}...\otimes S_{n}\right)  =%
\mathcal{F}%
\left(  S_{1}\right)  \otimes...\otimes%
\mathcal{F}%
\left(  S_{n}\right)
\end{equation}

\end{theorem}

\begin{theorem}
Pull-back : For $S\in S\left(
\mathbb{R}
^{m}\right)  ^{\prime}$ , $L\in GL\left(
\mathbb{R}
^{m};%
\mathbb{R}
^{m}\right)  ::y=\left[  A\right]  x$ with detA$\neq0:$%

\begin{equation}%
\mathcal{F}%
\left(  L^{\ast}S\right)  =S\left(
\mathcal{F}%
\left(  L^{t}\right)  ^{\ast}\right)
\end{equation}

\end{theorem}

\begin{proof}
$L^{\ast}S_{y}\left(  \varphi\right)  =\left\vert \det A\right\vert
^{-1}S\left(  \left(  L^{-1}\right)  ^{\ast}\varphi\right)  $

$%
\mathcal{F}%
\left(  L^{\ast}S_{y}\right)  \left(  \varphi\right)  =L^{\ast}S_{y}\left(
\mathcal{F}%
\left(  \varphi\right)  \right)  =\left\vert \det A\right\vert ^{-1}%
S_{y}\left(  \left(  L^{-1}\right)  ^{\ast}\left(
\mathcal{F}%
\left(  \varphi\right)  \right)  \right)  $

$=\left\vert \det A\right\vert ^{-1}S_{y}\left(  \left\vert \det A\right\vert
\mathcal{F}%
\left(  \varphi\circ L^{t}\right)  \right)  =S_{y}\left(
\mathcal{F}%
\left(  \left(  L^{t}\right)  ^{\ast}\varphi\right)  \right)  $
\end{proof}

\begin{theorem}
Homogeneous distribution : If S is homogeneous of order n in $S\left(
\mathbb{R}
^{m}\right)  $ then
$\mathcal{F}$%
(S) is homogenous of order -n-m
\end{theorem}

\begin{theorem}
Convolution :.For $S\in C_{\infty c}\left(
\mathbb{R}
^{m};%
\mathbb{C}
\right)  _{c}^{\prime},U\in S\left(
\mathbb{R}
^{m}\right)  ^{\prime}:$%

\begin{equation}%
\mathcal{F}%
\left(  U\ast S\right)  =S_{t}\left(  e^{-i\left\langle x,t\right\rangle
}\right)
\mathcal{F}%
\left(  U\right)
\end{equation}

\end{theorem}

which must be understood as the product of the function of x : $S_{t}\left(
e^{-i\left\langle x,t\right\rangle }\right)  $ by the distribution $%
\mathcal{F}%
\left(  U\right)  ,$ which is usually written, : $%
\mathcal{F}%
\left(  U\ast S\right)  =\left(  2\pi\right)  ^{m/2}%
\mathcal{F}%
\left(  S\right)  \times%
\mathcal{F}%
\left(  U\right)  $ , incorrectly, because the product of distributions is not defined.

\paragraph{Examples of Fourier transforms of distributions:\newline}

$%
\mathcal{F}%
\left(  \delta_{0}\right)  =%
\mathcal{F}%
^{\ast}\left(  \delta_{0}\right)  =\left(  2\pi\right)  ^{-m/2}T\left(
1\right)  ;$

$%
\mathcal{F}%
\left(  T\left(  1\right)  \right)  =%
\mathcal{F}%
^{\ast}\left(  T\left(  1\right)  \right)  =\left(  2\pi\right)  ^{m/2}%
\delta_{0};$

$S=T\left(  e^{i\epsilon a\left\Vert x\right\Vert ^{2}}\right)  ,a>0,\epsilon
=\pm1:%
\mathcal{F}%
\left(  S\right)  =\left(  \frac{1}{2a}\right)  ^{m/2}e^{mi\epsilon\frac{\pi
}{4}}T\left(  e^{-i\epsilon\frac{\left\Vert x\right\Vert ^{2}}{a}}\right)  $

$S=vp\frac{1}{x}\in S\left(
\mathbb{R}
\right)  ^{\prime}:%
\mathcal{F}%
\left(  S\right)  =T\left(  i\sqrt{2\pi}\left(  2-H\right)  \right)  $

with the Heaviside function H

(Zuily p.127) Let be $\sigma_{r}$\ the Lebesgue measure on the sphere of
radius r in $%
\mathbb{R}
^{m}.\sigma_{r}\in\left(  C_{\infty c}\left(
\mathbb{R}
^{m};%
\mathbb{C}
\right)  \right)  _{c}^{\prime}$

For m=3 \ $%
\mathcal{F}%
\left(  \sigma_{r}\right)  =\sqrt{\frac{2}{\pi}}rT\left(  \frac{\sin
r\left\Vert x\right\Vert }{\left\Vert x\right\Vert }\right)  $

Conversely for any m%
$>$%
0 : supp$%
\mathcal{F}%
^{\ast}\left(  T\left(  \frac{\sin r\left\Vert x\right\Vert }{\left\Vert
x\right\Vert }\right)  \right)  \subset\left\{  \left\Vert x\right\Vert \leq
r\right\}  $

\subsubsection{Extension of Sobolev spaces}

We have seen the Sobolev spaces $H^{r}\left(  O\right)  $ for functions with
$r\in%
\mathbb{N}
$, extended to r%
$<$%
0 as their dual (with distributions).\ Now we extend to $r\in%
\mathbb{R}
.$

\paragraph{Definition\newline}

\begin{definition}
For $m\in%
\mathbb{N}
,$ $s\in%
\mathbb{R}
,$ the Sobolev Space denoted $H^{s}\left(
\mathbb{R}
^{m}\right)  $ is the subspace of distributions $S\in S\left(
\mathbb{R}
^{m}\right)  ^{\prime}$ induced by a function f : S=T(f) such that $f\in
C\left(
\mathbb{R}
^{m};%
\mathbb{C}
\right)  $ with $\left(  1+\left\Vert x\right\Vert ^{2}\right)  ^{s/2}%
\mathcal{F}%
\left(  f\right)  \in L^{2}\left(
\mathbb{R}
^{m};dx,%
\mathbb{C}
\right)  $
\end{definition}

$\forall s\in%
\mathbb{N}
$ , $H^{s}\left(
\mathbb{R}
^{m}\right)  $ coincides with $T\left(  H^{s}\left(
\mathbb{R}
^{m}\right)  \right)  $\ where $H^{s}\left(
\mathbb{R}
^{m}\right)  $\ is\ the usual Sobolev space

\begin{theorem}
Inclusions (Taylor 1 p.270, Zuily p.133,136): We have the following inclusions :

i) $\forall s_{1}\geq s_{2}:H^{s_{1}}\left(
\mathbb{R}
^{m}\right)  \subset H^{s_{2}}\left(
\mathbb{R}
^{m}\right)  $ and the embedding is continuous

ii) $\delta_{0}\subset H^{s}\left(
\mathbb{R}
^{m}\right)  $ iff $s<-m/2$

iii) $\forall s<m/2:T\left(  L^{1}\left(
\mathbb{R}
^{m};dx,%
\mathbb{C}
\right)  \right)  \subset H^{s}\left(
\mathbb{R}
^{m}\right)  $

iv) $\forall s\in%
\mathbb{R}
:T\left(  S\left(
\mathbb{R}
^{m}\right)  \right)  \subset H^{s}\left(
\mathbb{R}
^{m}\right)  $ and is dense in $H^{s}\left(
\mathbb{R}
^{m}\right)  $

v) $\forall s\in%
\mathbb{R}
:T\left(  C_{\infty c}\left(
\mathbb{R}
^{m};%
\mathbb{C}
\right)  \right)  \subset H^{s}\left(
\mathbb{R}
^{m}\right)  $ and is dense in $H^{s}\left(
\mathbb{R}
^{m}\right)  $

vi) $\forall s>m/2+r,r\in%
\mathbb{N}
:H^{s}\left(
\mathbb{R}
^{m}\right)  \subset T\left(  C_{r}\left(
\mathbb{R}
^{m};%
\mathbb{C}
\right)  \right)  $

vi) $\forall s>m/2:H^{s}\left(
\mathbb{R}
^{m}\right)  \subset T\left(  C_{0b}\left(
\mathbb{R}
^{m};%
\mathbb{C}
\right)  \right)  $

vii) $\forall s>m/2+\gamma,0<\gamma<1:H^{s}\left(
\mathbb{R}
^{m}\right)  \subset T\left(  C^{\gamma}\left(
\mathbb{R}
^{m};%
\mathbb{C}
\right)  \right)  $ (Lipschitz functions)

vi) $C_{\infty c}\left(
\mathbb{R}
^{m};%
\mathbb{C}
\right)  _{c}^{\prime}\subset\cup_{s\in%
\mathbb{R}
}H^{s}\left(
\mathbb{R}
^{m}\right)  $
\end{theorem}

\paragraph{Properties\newline}

\begin{theorem}
(Taylor 1 p.270, Zuily p.133,137) $H^{s}\left(
\mathbb{R}
^{m}\right)  $ is a Hilbert space, its dual is a Hilbert space which is
anti-isomorphic to $H^{-s}\left(
\mathbb{R}
^{m}\right)  $ by :

$\tau:H^{-s}\left(
\mathbb{R}
^{m}\right)  \rightarrow\left(  H^{s}\left(
\mathbb{R}
^{m}\right)  \right)  ^{\prime}::\tau\left(  U\right)  =%
\mathcal{F}%
\left(  \psi\left(  -x\right)  \right)  $ where $U=T\left(  \psi\right)  $
\end{theorem}

The scalar product on $H^{s}\left(
\mathbb{R}
^{m}\right)  $ is :

$S,U\in H^{s}\left(
\mathbb{R}
^{m}\right)  ,S=T\left(  \varphi\right)  ,U=T\left(  \psi\right)  :$

$\left\langle S,U\right\rangle =\left\langle \left(  1+\left\Vert x\right\Vert
^{2}\right)  ^{s/2}%
\mathcal{F}%
\left(  \varphi\right)  ,\left(  1+\left\Vert x\right\Vert ^{2}\right)  ^{s/2}%
\mathcal{F}%
\left(  \psi\right)  \right\rangle _{L%
{{}^2}%
}$

$=\int_{%
\mathbb{R}
^{m}}\left(  1+\left\Vert x\right\Vert ^{2}\right)  ^{s}\overline{%
\mathcal{F}%
\left(  \varphi\right)  }%
\mathcal{F}%
\left(  \psi\right)  dx$

\begin{theorem}
(Zuily p.135) Product with a function :

$\forall s\in%
\mathbb{R}
:\forall\varphi\in S\left(
\mathbb{R}
^{m}\right)  ,S\in H^{s}\left(
\mathbb{R}
^{m}\right)  :\varphi S\in H^{s}\left(
\mathbb{R}
^{m}\right)  $
\end{theorem}

\begin{theorem}
(Zuily p.135) Derivatives : $\forall s\in%
\mathbb{R}
,\forall S\in H^{s}\left(
\mathbb{R}
^{m}\right)  ,\forall\alpha_{1},...\alpha_{r}:D_{\alpha_{1}...\alpha_{r}}S\in
H^{s-r}\left(
\mathbb{R}
^{m}\right)  ,\left\Vert D_{\alpha_{1}...\alpha_{r}}S\right\Vert _{H^{s-r}%
}\leq\left\Vert S\right\Vert _{H^{s}}$
\end{theorem}

\begin{theorem}
Pull back : For $L\in G%
\mathcal{L}%
\left(
\mathbb{R}
^{m};%
\mathbb{R}
^{m}\right)  $ the pull back $L^{\ast}S$\ of $S\in S\left(
\mathbb{R}
^{m}\right)  ^{\prime}$ is such that $L^{\ast}S\in S\left(
\mathbb{R}
^{m}\right)  ^{\prime}$ and $\forall s\in%
\mathbb{R}
:L^{\ast}:H^{s}\left(
\mathbb{R}
^{m}\right)  \rightarrow H^{s}\left(
\mathbb{R}
^{m}\right)  $
\end{theorem}

\begin{theorem}
(Zuily p.141) Trace :The map :

$Tr_{m}:S\left(
\mathbb{R}
^{m}\right)  \rightarrow S\left(
\mathbb{R}
^{m-1}\right)  ::Tr_{m}\left(  \varphi\right)  \left(  x_{1},..x_{m-1}\right)
=\varphi\left(  x_{1},..x_{m-1},0\right)  $

is continuous and $\forall s>\frac{1}{2}$ there is a unique extension to a
continuous linear surjective map :

$Tr_{m}:H^{s}\left(
\mathbb{R}
^{m}\right)  \rightarrow H^{s-\frac{1}{2}}\left(
\mathbb{R}
^{m-1}\right)  $
\end{theorem}

\begin{theorem}
(Zuily p.139) If K is a compact of $%
\mathbb{R}
^{m},H^{s}\left(  K\right)  $ the subset of $H^{s}\left(
\mathbb{R}
^{m}\right)  $ such that $S=T(f)$\ with Supp(f) in K, then $\forall s^{\prime
}>s$ the map $H^{s}\left(  K\right)  \rightarrow H^{s^{\prime}}\left(
\mathbb{R}
^{m}\right)  $ is compact.
\end{theorem}

\paragraph{Sobolev spaces on manifolds\newline}

(Taylor 1 p.282) The space $H^{s}\left(  M\right)  $\ of Sobolev distributions
on a m dimensional manifold M\ is defined as the subset of distributions $S\in
C_{\infty c}\left(  M;%
\mathbb{C}
\right)  ^{\prime}$ such that, for any chart $\left(  O_{i},\psi_{i}\right)
_{i\in I}$ of M, $\psi_{i}\left(  O_{i}\right)  =U_{i}$ where $U_{i}$ is
identified with its embedding in $%
\mathbb{R}
^{m}$ and any $\varphi\in C_{\infty c}\left(  O_{i};%
\mathbb{C}
\right)  :\left(  \psi_{i}^{-1}\right)  ^{\ast}\varphi S\in H^{s}\left(
U_{i}\right)  $.

If M is compact then we have a Sobolev space $H^{s}\left(  M\right)  $ with
some of the properties of $H^{s}\left(
\mathbb{R}
^{m}\right)  .$

The same construct can be implemented to define Sobolev spaces on compact
manifolds with boundary.

\newpage

\section{DIFFERENTIAL\ OPERATORS}

\label{Differential operators}

Differential operators are the key element of any differential equation,
meaning a map relating an unknown function and its partial derivative to some
known function and subject to some initial conditions. And indeed solving a
differential equation can often be summarized to finding the inverse of a
differential operator.

When dealing with differential operators, meaning with maps involving both a
function and its derivative, a common hassle comes from finding a simple
definition of the domain of the operator : we want to keep the fact that the
domain of y and y' are somewhat related. The simplest solution is to use the
jet formalism. It is not only practical, notably when we want to study
differential operators on fiber bundles, but it gives a deeper insight of the
meaning of locality, thus making a better understanding of the difference
between differential operators and pseudo differential operators.

\subsection{Definitions}

\begin{definition}
A \textbf{r order differential operator} is a fibered manifold, base
preserving, morphism $D:J^{r}E_{1}\rightarrow E_{2}$ between two smooth
complex finite dimensional vector bundles $E_{1}\left(  M,V_{1},\pi
_{1}\right)  ,E_{2}\left(  M,V_{2},\pi_{2}\right)  $\ on the same real manifold.
\end{definition}

\begin{definition}
A \textbf{r order scalar differential operator} on a space F of r
differentiable complex functions on a manifold M is a map : $D:J^{r}%
F\rightarrow C\left(  M;%
\mathbb{C}
\right)  $ where $J^{r}F$ is the set of r-jet prolongations of the maps in F.
\end{definition}

\subsubsection{Locality}

$J^{r}E_{1},E_{2}$\ are two fibered manifolds \textit{with the same base M},
so $\forall x\in M:\pi_{2}\circ D=\pi_{1}.$ That we will write more plainly :
D(x) maps fiberwise $Z\left(  x\right)  \in J^{r}E_{1}\left(  x\right)  $ to
$Y\left(  x\right)  \in E_{2}\left(  x\right)  $

So a differential operator is \textit{local}, in the meaning that it can be
fully computed from data related at one point and these data involve not only
the value of the section at x, but also the values at x of its partial
derivatives up to the r order. This is important in numerical analysis,
because generally one can find some algorithm to compute a function from the
value of its partial derivative. This is fundamental here as we will see that
almost all the properties of D use locality. We will see that pseudo
differential operators are not local.

\subsubsection{Sections on $J^{r}E_{1}$ and on $E_{1}$}

D maps sections of $J^{r}E_{1}$ to sections of $E_{2}:$%

\begin{equation}
D:\mathfrak{X}\left(  J^{r}E_{1}\right)  \rightarrow\mathfrak{X}\left(
E_{2}\right)  ::D\left(  Z\right)  \left(  x\right)  =D\left(  x\right)
\left(  Z\left(  x\right)  \right)
\end{equation}

D, by itself, does not involve any differentiation.

$J^{r}E_{1}$ is a vector bundle $J^{r}E_{1}\left(  M,J_{0}^{r}\left(
\mathbb{R}
^{m},V_{1}\right)  _{0},\pi^{r}\right)  $.\ A section Z on $J^{r}E_{1}$ is a
map $M\rightarrow J^{r}E_{1}$ and reads :

$Z=\left(  z,z_{\alpha_{1}...\alpha_{s}},1\leq\alpha_{k}\leq m,s=1...r\right)
$

with $z\in E_{1},$ $z_{\alpha_{1}...\alpha_{s}}\in V_{1}$ symmetric in all
lower indices, dimM=m

D reads in a holonomic basis of $E_{2}$ :

$D\left(  x\right)  \left(  Z\right)  =\sum_{i=1}^{n_{2}}D^{i}\left(
x,z,z_{\alpha},z_{\alpha\beta},....z_{\alpha_{1}...\alpha_{r}}\right)
e_{2i}\left(  x\right)  $

Any class r section on $E_{1}$ gives rise to a section on $J^{r}E_{1}$ thus we
have a map :%

\begin{equation}
\widehat{D}:\mathfrak{X}_{r}\left(  E_{1}\right)  \rightarrow\mathfrak{X}%
\left(  E_{2}\right)  ::\widehat{D}\left(  U\right)  =D\left(  J^{r}U\right)
\end{equation}

and this is $\widehat{D}=D\circ J^{r}$ which involves the derivatives.

If $X\in\mathfrak{X}_{r}\left(  E_{1}\right)  ,X=\sum_{i=1}^{n_{1}}%
X^{i}\left(  x\right)  e_{1i}\left(  x\right)  $ then $z_{\alpha_{1}%
...\alpha_{s}}^{i}\left(  x\right)  =D_{\alpha_{1}...\alpha_{s}}X^{i}\left(
x\right)  $ with

$D_{\left(  \alpha\right)  }=D_{\alpha_{1}...\alpha_{s}}=\dfrac{\partial
}{\partial\xi^{\alpha_{1}}}\dfrac{\partial}{\partial\xi^{\alpha_{2}}}%
...\dfrac{\partial}{\partial\xi^{\alpha_{s}}}$ where the $\alpha_{k}%
=1...m$\ can be identical and$\left(  \xi^{\alpha}\right)  _{\alpha=1}^{m}$
are the coordinates in a chart of M.

\subsubsection{Topology}

$\mathfrak{X}_{r}\left(  E_{1}\right)  ,\mathfrak{X}\left(  J^{r}E_{1}\right)
$ are Fr\'{e}chet spaces, $\mathfrak{X}\left(  E_{2}\right)  $ can be endowed
with one of the topology see previously, $\mathfrak{X}_{0}\left(
E_{2}\right)  $ is a Banach space. The operator will be assumed to be
continuous with the chosen topology. Indeed the domain of a differential
operator is quite large (and can be easily extended to distributions), so
usually more interesting properties will be found for the restriction of the
operator to some subspaces of $\mathfrak{X}\left(  J^{r}E_{1}\right)  $ or
$\mathfrak{X}_{r}\left(  E_{1}\right)  .$

\subsubsection{Manifold with boundary}

If M is a manifold with boundary in N, then $\overset{\circ}{M}$ is an open
subset and a submanifold of N.\ The restriction of a differential operator D
from $\mathfrak{X}\left(  J^{r}E_{1}\right)  $ to the sections of
$\mathfrak{X}\left(  J^{r}E_{1}\right)  $ with support in $\overset{\circ}{M}$
is well defined. It can de defined on the boundary $\partial M$ if D is
continuous on N.

\subsubsection{Composition of differential operators}

Differential operators can be composed as follows :

$D_{1}:\mathfrak{X}\left(  J^{r_{1}}E_{1}\right)  \rightarrow\mathfrak{X}%
\left(  E_{2}\right)  $

$D_{2}:\mathfrak{X}\left(  J^{r_{2}}E_{2}\right)  \rightarrow\mathfrak{X}%
\left(  E_{3}\right)  $

$D_{2}\circ J^{r_{2}}\circ D_{1}=\widehat{D}_{2}\circ D_{1}:\mathfrak{X}%
\left(  J^{r_{1}}E_{1}\right)  \rightarrow\mathfrak{X}\left(  E_{3}\right)  $

It implies that $D_{1}\left(  \mathfrak{X}\left(  E_{2}\right)  \right)
\subset\mathfrak{X}_{r_{2}}\left(  E_{2}\right)  .$ The composed operator is
in fact of order $r_{1}+r_{2}$ : $\widehat{D}_{3}:\mathfrak{X}_{r_{1}+r_{2}%
}\left(  E_{1}\right)  \rightarrow\mathfrak{X}\left(  E_{3}\right)
::\widehat{D}_{3}=\widehat{D}_{2}\circ\widehat{D}_{1}$

\subsubsection{Parametrix}

\begin{definition}
A \textbf{parametrix} for a differential operator : $D:J^{r}E\rightarrow E$ on
the vector bundle E is a map : $Q:\mathfrak{X}\left(  E\right)  \rightarrow
\mathfrak{X}\left(  E\right)  $ such that $Q\circ\widehat{D}-Id$ and
$\widehat{D}\circ Q-Id$ are compact maps $\mathfrak{X}\left(  E\right)
\rightarrow\mathfrak{X}\left(  E\right)  $
\end{definition}

Compact maps "shrink" a set. A parametrix can be seen as a proxy for an
inverse map : $Q\approx\widehat{D}^{-1}.$ The definition is relative to the
topology on $\mathfrak{X}\left(  E\right)  .$ Usually Q is not a differential
operator but a pseudo-differential operator, and it is not unique.

\subsubsection{Operators depending on a parameter}

Quite often one meets family of differential operators depending on a scalar
parameter t (usually the time) : $D\left(  t\right)  :J^{r}E_{1}\rightarrow
E_{2}$ where t belongs to some open T of $%
\mathbb{R}
.$

One can extend a manifold M to the product TxM, and similarly a vector bundle
$E\left(  M,V,\pi\right)  $ with atlas $\left(  O_{a},\varphi_{a}\right)
_{a\in A}$ to a vector bundle $E_{T}\left(  T\times M,V,\pi_{T}\right)  $ with
atlas $\left(  T\times O_{a},\psi_{a}\right)  _{a\in A}:\psi_{a}\left(
\left(  t,x\right)  ,u\right)  =\left(  \varphi_{a}\left(  x,u\right)
,t\right)  $ and transitions maps : $\psi_{ba}\left(  \left(  t,x\right)
\right)  =\varphi_{ba}\left(  x\right)  $. The projection is : $\pi_{T}\left(
\left(  \varphi_{a}\left(  x,u\right)  ,t\right)  \right)  =\left(
t,x\right)  .$ All elements of the fiber at (t,x) of $E_{T}$ share the same
time t so the vector space structure on the fiber at t is simply that of V.
$E_{T}$ is still a vector bundle with fibers isomorphic to V. A section
$X\in\mathfrak{X}\left(  E_{T}\right)  $ is then identical to a map : $%
\mathbb{R}
\rightarrow\mathfrak{X}\left(  E\right)  .$

The r jet prolongation of $E_{T}$ is a vector bundle $J^{r}E_{T}\left(
\mathbb{R}
\times M,J_{0}^{r}\left(
\mathbb{R}
^{m+1},V\right)  _{0},\pi_{T}\right)  .$ We need to enlarge the derivatives to
account for t. A section $X\in\mathfrak{X}\left(  E_{T}\right)  $ has a r jet
prolongation and for any two sections the values of the derivatives are taken
at the same time.

A differential operator between two vector bundles depending on a parameter is
then a base preserving map : $D:J^{r}E_{T1}\rightarrow E_{T2}$ so $D\left(
t,x\right)  :J^{r}E_{T1}\left(  t,x\right)  \rightarrow E_{T2}\left(
t,x\right)  .$ It can be seen as a family of operators D(t), but the locality
condition imposes that the time is the same both in $Z_{1}\left(  t\right)  $
and $\ Z_{2}\left(  t\right)  =D\left(  t\right)  \left(  Z_{1}\left(
t\right)  \right)  $

\bigskip

\subsection{Linear differential operators}

\label{Linear differential operators}

\subsubsection{Definitions}

Vector bundles are locally vector spaces, so the most "natural" differential
operator is linear.

\begin{definition}
A r order \textbf{linear differential operator} is a linear, base preserving
morphism $D:\mathfrak{X}\left(  J^{r}E_{1}\right)  \rightarrow\mathfrak{X}%
\left(  E_{2}\right)  $ between two smooth complex finite dimensional vector
bundles $E_{1}\left(  M,V_{1},\pi_{1}\right)  ,E_{2}\left(  M,V_{2},\pi
_{2}\right)  $\ on the same real manifold.
\end{definition}

\paragraph{Components expression\newline}

$\forall x\in M:D\left(  x\right)  \in%
\mathcal{L}%
\left(  J^{r}E_{1}\left(  x\right)  ;E_{2}\left(  x\right)  \right)  $

It reads in a holonomic basis $\left(  e_{2,i}\left(  x\right)  \right)
_{i=1}^{n}$\ of E$_{2}$%

\begin{equation}
D(J^{r}z\left(  x\right)  )=\sum_{s=0}^{r}\sum_{\alpha_{1}...\alpha_{s}=1}%
^{m}\sum_{i=1}^{n_{2}}\sum_{j=1}^{n_{1}}A\left(  x\right)  _{j}^{i,\alpha
_{1}...\alpha_{s}}z_{\alpha_{1}...\alpha_{s}}^{j}\left(  x\right)
e_{2,i}\left(  x\right)
\end{equation}

where $A\left(  x\right)  _{j}^{i,\alpha_{1}...\alpha_{s}},z_{\alpha
_{1}...\alpha_{s}}^{j}\left(  x\right)  \in%
\mathbb{C}
$ and $z_{\alpha_{1}...\alpha_{s}}^{j}$ is symmetric in all lower indices.

Or equivalently :%

\begin{equation}
D(J^{r}z\left(  x\right)  )=\sum_{s=0}^{r}\sum_{\alpha_{1}...\alpha_{s}=1}%
^{m}A\left(  x\right)  ^{\alpha_{1}...\alpha_{s}}Z_{\alpha_{1}...\alpha_{s}%
}\left(  x\right)
\end{equation}

where

$Z_{\alpha_{1}...\alpha_{s}}\left(  x\right)  =\sum_{j=1}^{n_{1}}z_{\alpha
_{1}...\alpha_{s}}^{i}\left(  x\right)  e_{1,i}\left(  x\right)  \in
E_{1}\left(  x\right)  ,$

$A\left(  x\right)  ^{\alpha_{1}...\alpha_{s}}\in%
\mathcal{L}%
\left(  E_{1}\left(  x\right)  ;E_{2}\left(  x\right)  \right)  \sim%
\mathcal{L}%
\left(  V_{1};V_{2}\right)  $

x appear only through the maps $A\left(  x\right)  ^{\alpha_{1}...\alpha_{s}%
},D$ is linear in all the components.

A scalar linear differential operator reads :

$D(J^{r}z\left(  x\right)  )=\sum_{s=0}^{r}\sum_{\alpha_{1}...\alpha_{s}%
=1}^{m}A\left(  x\right)  ^{\alpha_{1}...\alpha_{s}}z_{\alpha_{1}...\alpha
_{s}}$

where $A\left(  x\right)  ^{\alpha_{1}...\alpha_{s}},z_{\alpha_{1}%
...\alpha_{s}}\in%
\mathbb{C}
$

\paragraph{Quasi-linear operator\newline}

A differential operator is said to be \textbf{quasi-linear} if it reads :

$D(J^{r}z\left(  x\right)  )=\sum_{s=0}^{r}\sum_{\alpha_{1}...\alpha_{s}%
=1}^{m}A\left(  x,z\right)  ^{\alpha_{1}...\alpha_{s}}Z_{\alpha_{1}%
...\alpha_{s}}\left(  x\right)  $ where the coefficients depend on the value
of z only, and not $z_{\alpha_{1}...\alpha_{s}}^{i}$ (so they depend on X and
not its derivatives).

\paragraph{Domain and range of D\newline}

We assume that the maps : $A^{\alpha_{1}...\alpha_{s}}:M\rightarrow%
\mathcal{L}%
\left(  V_{1};V_{2}\right)  $ are smooth.

So on $\mathfrak{X}_{0}\left(  J^{r}E_{1}\right)  $ we have $D\left(
\mathfrak{X}_{0}\left(  J^{r}E_{1}\right)  \right)  \subset\mathfrak{X}%
_{0}\left(  E_{2}\right)  $ and $\widehat{D}\left(  \mathfrak{X}_{r}\left(
E_{1}\right)  \right)  \subset\mathfrak{X}_{0}\left(  E_{2}\right)  .$

If there is a Radon measure $\mu$\ on M, and if the coefficients $A\left(
x\right)  _{j}^{i,\alpha_{1}...\alpha_{s}}$ are bounded on M (notably if M is
compact) then we have :

$D\left(  L^{p}\left(  M,\mu,J^{r}E_{1}\right)  \right)  \subset L^{p}\left(
M,\mu,E_{2}\right)  $ and $\widehat{D}\left(  W^{r,p}\left(  E_{1}\right)
\right)  \subset L^{p}\left(  M,\mu,E_{2}\right)  $

Any linear differential operator can also be seen as acting on sections of
$E_{1}:$

$D^{\#}\left(  x\right)  :E_{1}\left(  x\right)  \rightarrow J^{r}E_{2}\left(
x\right)  ::D^{\#}\left(  x\right)  \left(  X\right)  =\left(  A\left(
x\right)  ^{\alpha_{1}...\alpha_{s}}X,s=0..r,1\leq\alpha_{k}\leq m\right)  $
so $Z^{\alpha_{1}...\alpha_{s}}=A\left(  x\right)  ^{\alpha_{1}...\alpha_{s}%
}X$

\subsubsection{Differential operators and distributions}

\paragraph{Scalar operators\newline}

Let $V\subset C_{r}\left(  M;%
\mathbb{C}
\right)  $ be a Fr\'{e}chet space such that $\forall\varphi\in V,D_{\alpha
_{1}...\alpha_{r}}\varphi\in V$ with the associated distributions V'. Then :

$\forall S\in V^{\prime},s\leq r,\exists D_{\alpha_{1}...\alpha_{s}%
}S::D_{\alpha_{1}...\alpha_{s}}S\left(  \varphi\right)  =\left(  -1\right)
^{s}S\left(  D_{\alpha_{1}...\alpha_{s}}\varphi\right)  $

and let us denote : $J^{r}V^{\prime}=\left\{  D_{\alpha_{1}...\alpha_{s}%
}S,\alpha_{k}=1...m,s=0...r,S\in V^{\prime}\right\}  $

A linear differential operator on V' is a map : $D^{\prime}:J^{r}V^{\prime
}\rightarrow V^{\prime}$ such that there is a linear differential operator D
on V with :%

\begin{equation}
\forall\varphi\in V::D^{\prime}S\left(  \varphi\right)  =S\left(
DJ^{r}\varphi\right)
\end{equation}

$DJ^{r}\varphi=\sum_{s=0}^{r}\sum_{\alpha_{1}...\alpha_{s}=1}^{m}A\left(
x\right)  ^{\alpha_{1}...\alpha_{s}}D_{\alpha_{1}...\alpha_{s}}\varphi$ with
$A\left(  x\right)  ^{\alpha_{1}...\alpha_{s}}\in C\left(  M;%
\mathbb{C}
\right)  $

$S\left(  DJ^{r}\varphi\right)  =\sum_{s=0}^{r}\sum_{\alpha_{1}...\alpha
_{s}=1}^{m}A\left(  x\right)  ^{\alpha_{1}...\alpha_{s}}S\left(  D_{\alpha
_{1}...\alpha_{s}}\varphi\right)  $

$=\sum_{s=0}^{r}\left(  -1\right)  ^{s}\sum_{\alpha_{1}...\alpha_{s}=1}%
^{m}A\left(  x\right)  ^{\alpha_{1}...\alpha_{s}}D_{\alpha_{1}...\alpha_{s}%
}S\left(  \varphi\right)  $

So to any linear scalar differential operator D on V is associated the linear
differential operator on V' :

$D^{\prime}S=\sum_{s=0}^{r}\left(  -1\right)  ^{s}\sum_{\alpha_{1}%
...\alpha_{s}=1}^{m}A\left(  x\right)  ^{\alpha_{1}...\alpha_{s}}D_{\alpha
_{1}...\alpha_{s}}S$

If M is an open O of $%
\mathbb{R}
^{m}$ and V' defined by a map $T:W\rightarrow V^{\prime}$ then :

$D^{\prime}T\left(  f\right)  =\sum_{s=0}^{r}\left(  -1\right)  ^{s}%
\sum_{\alpha_{1}...\alpha_{s}=1}^{m}A\left(  x\right)  ^{\alpha_{1}%
...\alpha_{s}}D_{\alpha_{1}...\alpha_{s}}T\left(  f\right)  $

$=\sum_{s=0}^{r}\left(  -1\right)  ^{s}\sum_{\alpha_{1}...\alpha_{s}=1}%
^{m}A\left(  x\right)  ^{\alpha_{1}...\alpha_{s}}T\left(  D_{\alpha
_{1}...\alpha_{s}}\left(  f\right)  \right)  $

$D^{\prime}T\left(  f\right)  \left(  \varphi\right)  =\sum_{s=0}^{r}\left(
-1\right)  ^{s}\sum_{\alpha_{1}...\alpha_{s}=1}^{m}A\left(  x\right)
^{\alpha_{1}...\alpha_{s}}\int_{O}\varphi\left(  y\right)  \left(
D_{\alpha_{1}...\alpha_{s}}\left(  f\right)  \right)  d\xi^{1}\wedge...\wedge
d\xi^{m}$

If V' is given by a set of r-differentiable m-forms $\lambda\in\Lambda
_{m}\left(  M;%
\mathbb{C}
\right)  $ then :

$D^{\prime}T\left(  \lambda_{0}d\xi^{1}\wedge...\wedge d\xi^{m}\right)  $

$=\sum_{s=0}^{r}\left(  -1\right)  ^{s}\sum_{\alpha_{1}...\alpha_{s}=1}%
^{m}A\left(  x\right)  ^{\alpha_{1}...\alpha_{s}}D_{\alpha_{1}...\alpha_{s}%
}T\left(  \lambda_{0}d\xi^{1}\wedge...\wedge d\xi^{m}\right)  $

$=\sum_{s=0}^{r}\sum_{\alpha_{1}...\alpha_{s}=1}^{m}A\left(  x\right)
^{\alpha_{1}...\alpha_{s}}T\left(  D_{\alpha_{1}...\alpha_{s}}\left(
\lambda_{0}\right)  d\xi^{1}\wedge...\wedge d\xi^{m}\right)  $

$D^{\prime}T\left(  f\right)  \left(  \varphi\right)  =\sum_{s=0}^{r}%
\sum_{\alpha_{1}...\alpha_{s}=1}^{m}A\left(  x\right)  ^{\alpha_{1}%
...\alpha_{s}}\int_{M}\varphi\left(  y\right)  \left(  D_{\alpha_{1}%
...\alpha_{s}}\left(  \lambda_{0}\right)  d\xi^{1}\wedge...\wedge d\xi
^{m}\right)  $

This works only for linear operators, because the product of distributions is
not defined. This can be extended to linear differential operators on vector
bundles, using the theory of distributions on vector bundles introduced previously.

\paragraph{Differential operators on vector bundles\newline}

A distribution on a vector bundle E is a functional : $S:\mathfrak{X}%
_{\infty,c}\left(  E\right)  \rightarrow%
\mathbb{C}
$

The action of a distribution $S\in\mathfrak{X}_{\infty,c}\left(  E\right)
^{\prime}$ on a section $Z\in\mathfrak{X}_{\infty,c}\left(  J^{r}E\right)  $
is the map :

$S:\mathfrak{X}_{\infty,c}\left(  J^{r}E\right)  \rightarrow\left(
\mathbb{C}
,%
\mathbb{C}
^{sm},s=1..r\right)  ::$

$S\left(  Z_{\alpha_{1}...\alpha_{s}}\left(  x\right)  ,s=0..r,\alpha
_{j}=1...m\right)  =\left(  S\left(  Z_{\alpha_{1}...\alpha_{s}}\right)
,s=0..r,\alpha_{j}=1...m\right)  $

The derivative of $S\in\mathfrak{X}_{\infty,c}\left(  E\right)  ^{\prime}$
with respect to $\left(  \xi^{\alpha_{1}},...,\xi^{\alpha_{s}}\right)  $ on M
is the distribution :

$\left(  D_{\alpha_{1}..\alpha_{s}}S\right)  \in\mathfrak{X}_{\infty,c}\left(
E\right)  ^{\prime}$: $\forall X\in\mathfrak{X}_{\infty,c}\left(  E\right)
:\left(  D_{\alpha_{1}..\alpha_{s}}S\right)  \left(  X\right)  =S\left(
D_{\alpha_{1}..\alpha_{s}}X\right)  $

The r jet prolongation of $S\in\mathfrak{X}_{\infty,c}\left(  E\right)
^{\prime}$ is the map $J^{r}S$ such that $\forall X\in\mathfrak{X}_{\infty
,c}\left(  E\right)  :J^{r}S\left(  X\right)  =S\left(  J^{r}X\right)  $

$J^{r}S:\mathfrak{X}_{\infty,c}\left(  E\right)  \rightarrow\left(
\mathbb{C}
,%
\mathbb{C}
^{sm},s=1..r\right)  ::J^{r}S\left(  X\right)  =\left(  S\left(  D_{\alpha
_{1}..\alpha_{s}}X\right)  ,s=0...r,\alpha_{j}=1...m\right)  $

If the distribution $S\in\mathfrak{X}_{\infty,c}\left(  E\right)  ^{\prime}$
is induced by the m form $\lambda\in\Lambda_{m}\left(  M;E^{\prime}\right)  $
then its r jet prolongation $:J^{r}\left(  T\left(  \lambda\right)  \right)
=T\left(  J^{r}\lambda\right)  $ with $J^{r}\lambda\in\Lambda_{m}\left(
M;J^{r}E^{\prime}\right)  $

\bigskip

\begin{definition}
Let $E_{1}\left(  M,V_{1},\pi_{1}\right)  ,E_{2}\left(  M,V_{2},\pi
_{2}\right)  $\ be two smooth complex finite dimensional vector bundles on the
same real manifold$.$ A differential operator for the distributions
$\mathfrak{X}_{\infty c}\left(  E_{2}\right)  ^{\prime}$ is a map :
$D^{\prime}:J^{r}\mathfrak{X}_{\infty c}\left(  E_{2}\right)  ^{\prime
}\rightarrow\mathfrak{X}_{\infty c}\left(  E_{1}\right)  ^{\prime}$ such that
:
\begin{equation}
\forall S_{2}\in\mathfrak{X}_{\infty c}\left(  E_{2}\right)  ^{\prime}%
,X_{1}\in\mathfrak{X}_{\infty c}\left(  E_{1}\right)  :D^{\prime}J^{r}%
S_{2}\left(  X_{1}\right)  =S_{2}\left(  DJ^{r}\left(  X_{1}\right)  \right)
\end{equation}

for a differential operator $D:\mathfrak{X}\left(  J^{r}E_{1}\right)
\rightarrow\mathfrak{X}\left(  E_{2}\right)  $
\end{definition}

\bigskip

\begin{theorem}
To any linear differential operator D between the smooth finite dimensional
vector bundles on the same real manifold $E_{1}\left(  M,V_{1},\pi_{1}\right)
,E_{2}\left(  M,V_{2},\pi_{2}\right)  $\ is associated a differential operator
for the distributions : $D^{\prime}:J^{r}\mathfrak{X}_{\infty c}\left(
E_{2}\right)  ^{\prime}\rightarrow\mathfrak{X}_{\infty c}\left(  E_{1}\right)
^{\prime}$ which reads :%

\begin{equation}
D(Z)=\sum_{s=0}^{r}\sum_{\alpha_{1}...\alpha_{s}=1}^{m}A\left(  x\right)
^{\alpha_{1}...\alpha_{s}}z_{\alpha_{1}...\alpha_{s}}\rightarrow D^{\prime
}=\sum_{s=0}^{r}\sum_{\alpha_{1}...\alpha_{s}=1}^{m}\left(  A\left(  x\right)
^{\alpha_{1}...\alpha_{s}}\right)  ^{t}D_{\alpha_{1}...\alpha_{s}}%
\end{equation}

\end{theorem}

\begin{proof}
The component expression of D is :

$D(Z)=\sum_{s=0}^{r}\sum_{\alpha_{1}...\alpha_{s}=1}^{m}A\left(  x\right)
^{\alpha_{1}...\alpha_{s}}z_{\alpha_{1}...\alpha_{s}}$ where $A\left(
x\right)  ^{\alpha_{1}...\alpha_{s}}\in%
\mathcal{L}%
\left(  E_{1}\left(  x\right)  ;E_{2}\left(  x\right)  \right)  $

Define :

$D^{\prime}J^{r}S_{2}\left(  X_{1}\right)  =\sum_{s=0}^{r}\sum_{\alpha
_{1}...\alpha_{s}=1}^{m}\left(  A^{\alpha_{1}...\alpha_{s}}\right)  ^{\ast
}S_{2}\left(  D_{\alpha_{1}...\alpha_{s}}X_{1}\right)  $

with the pull-back $\left(  A^{\alpha_{1}...\alpha_{s}}\right)  ^{\ast}S_{2}$
of $S_{2}::$

$\left(  A^{\alpha_{1}...\alpha_{s}}\right)  ^{\ast}S_{2}\left(  D_{\alpha
_{1}...\alpha_{s}}X_{1}\right)  =S_{2}\left(  A\left(  x\right)  ^{\alpha
_{1}...\alpha_{s}}\left(  D_{\alpha_{1}...\alpha_{s}}X_{1}\right)  \right)  $

$D^{\prime}J^{r}S_{2}\left(  X_{1}\right)  =\sum_{s=0}^{r}\sum_{\alpha
_{1}...\alpha_{s}=1}^{m}S_{2}\left(  A\left(  x\right)  ^{\alpha_{1}%
...\alpha_{s}}\left(  D_{\alpha_{1}...\alpha_{s}}X_{1}\right)  \right)  $

$=S_{2}\left(  \sum_{s=0}^{r}\sum_{\alpha_{1}...\alpha_{s}=1}^{m}A\left(
x\right)  ^{\alpha_{1}...\alpha_{s}}\left(  D_{\alpha_{1}...\alpha_{s}}%
X_{1}\right)  \right)  =S_{2}\left(  DJ^{r}\left(  X_{1}\right)  \right)  $

$A^{\alpha_{1}...\alpha_{s}}\in%
\mathcal{L}%
\left(  E_{1}\left(  x\right)  ;E_{2}\left(  x\right)  \right)  \Rightarrow
\left(  A^{\alpha_{1}...\alpha_{s}}\right)  ^{\ast}=\left(  A^{\alpha
_{1}...\alpha_{s}}\right)  ^{t}\in%
\mathcal{L}%
\left(  E_{2}^{\prime}\left(  x\right)  ;E_{1}^{\prime}\left(  x\right)
\right)  $

So : $D^{\prime}J^{r}S_{2}=\sum_{s=0}^{r}\sum_{\alpha_{1}...\alpha_{s}=1}%
^{m}\left(  A^{\alpha_{1}...\alpha_{s}}\right)  ^{t}S_{2}\circ D_{\alpha
_{1}...\alpha_{s}}$

$=\sum_{s=0}^{r}\sum_{\alpha_{1}...\alpha_{s}=1}^{m}\left(  A^{\alpha
_{1}...\alpha_{s}}\right)  ^{t}D_{\alpha_{1}...\alpha_{s}}S_{2}$
\end{proof}

If $E_{1}=E_{2}$ and D is such that : $A\left(  x\right)  ^{\alpha
_{1}...\alpha_{s}}=a\left(  x\right)  ^{\alpha_{1}...\alpha_{s}}I$ with
$a\left(  x\right)  ^{\alpha_{1}...\alpha_{s}}$ a function and I the identity
map then

$S_{2}\left(  A\left(  x\right)  ^{\alpha_{1}...\alpha_{s}}\left(
D_{\alpha_{1}...\alpha_{s}}X_{1}\left(  x\right)  \right)  \right)  =a\left(
x\right)  ^{\alpha_{1}...\alpha_{s}}S_{2}\left(  D_{\alpha_{1}...\alpha_{s}%
}X_{1}\left(  x\right)  \right)  $

$=a\left(  x\right)  ^{\alpha_{1}...\alpha_{s}}D_{\alpha_{1}...\alpha_{s}%
}S_{2}\left(  X_{1}\left(  x\right)  \right)  $

$D^{\prime}J^{r}S_{2}=\sum_{s=0}^{r}\sum_{\alpha_{1}...\alpha_{s}=1}%
^{ms}a\left(  x\right)  ^{\alpha_{1}...\alpha_{s}}D_{\alpha_{1}...\alpha_{s}%
}S_{2}$

so D' reads with the same coefficients as D.

\bigskip

For the distributions $S\in\mathfrak{X}_{\infty,c}\left(  E_{2}\right)
^{\prime}$ induced by m forms $\lambda_{2}\in\Lambda_{m}\left(  M;E_{2}%
^{\prime}\right)  $ then :

$S_{2}=T\left(  \lambda_{2}\right)  \Rightarrow D^{\prime}J^{r}T\left(
\lambda_{2}\right)  \left(  X_{1}\right)  =T\left(  \lambda_{2}\right)
\left(  DJ^{r}\left(  X_{1}\right)  \right)  =D^{\prime}T\left(  J^{r}%
\lambda_{2}\right)  \left(  X_{1}\right)  $

$D^{\prime}J^{r}T\left(  \lambda_{2}\right)  =\sum_{s=0}^{r}\sum_{\alpha
_{1}...\alpha_{s}=1}^{m}\left(  A^{\alpha_{1}...\alpha_{s}}\right)
^{t}T\left(  D_{\alpha_{1}...\alpha_{s}}\lambda_{2}\right)  $

$=\sum_{s=0}^{r}\sum_{\alpha_{1}...\alpha_{s}=1}^{m}T\left(  \left(
A^{\alpha_{1}...\alpha_{s}}\right)  ^{t}D_{\alpha_{1}...\alpha_{s}}\lambda
_{2}\right)  $

$=T\left(  \sum_{s=0}^{r}\sum_{\alpha_{1}...\alpha_{s}=1}^{m}\left(
A^{\alpha_{1}...\alpha_{s}}\right)  ^{t}D_{\alpha_{1}...\alpha_{s}}\lambda
_{2}\right)  $

that we can write :%

\begin{equation}
D^{\prime}J^{r}T\left(  \lambda_{2}\right)  =T\left(  D^{t}\lambda_{2}\right)
\text{ with }D^{t}=\sum_{s=0}^{r}\sum_{\alpha_{1}...\alpha_{s}=1}^{m}\left(
A^{\alpha_{1}...\alpha_{s}}\right)  ^{t}D_{\alpha_{1}...\alpha_{s}}%
\end{equation}

Example : $E_{1}=E_{2},r=1$

$D\left(  J^{1}X\right)  =\sum_{ij}\left(  \left[  A\right]  _{j}^{i}%
X^{j}+\left[  B^{\alpha}\right]  _{j}^{i}\partial_{\alpha}X^{j}\right)
e_{i}\left(  x\right)  $

$D^{\prime}T\left(  J^{1}\lambda\right)  =T\left(  \sum_{ij}\left(  \left[
A\right]  _{j}^{i}\lambda_{i}+\left[  B^{\alpha}\right]  _{j}^{i}%
\partial_{\alpha}\lambda_{i}\right)  e^{j}\left(  x\right)  \right)  $

\subsubsection{Adjoint of a differential operator}

\paragraph{Definition\newline}

The definition of the adjoint of a differential operator follows the general
definition of the adjoint of a linear map with respect to a scalar product.

\begin{definition}
A map $\widehat{D}^{\ast}:\mathfrak{X}\left(  E_{2}\right)  \rightarrow
\mathfrak{X}\left(  E_{1}\right)  $\ is the adjoint of a linear differential
operator D between the smooth finite dimensional vector bundles on the same
real manifold $E_{1}\left(  M,V_{1},\pi_{1}\right)  ,E_{2}\left(  M,V_{2}%
,\pi_{2}\right)  $\ endowed with scalar products $G_{1},G_{2}$ on sections if :%

\begin{equation}
\forall X_{1}\in\mathfrak{X}\left(  E_{1}\right)  ,X_{2}\in\mathfrak{X}\left(
E_{2}\right)  :G_{2}\left(  \widehat{D}\left(  X_{1}\right)  ,X_{2}\right)
=G_{1}\left(  X_{1},\widehat{D}^{\ast}X_{2}\right)
\end{equation}

\end{definition}

\begin{definition}
A linear differential operator on a vector bundle is said to be \textbf{self
adjoint} if $D=D^{\ast}$
\end{definition}

An adjoint map is not necessarily a differential operator. If D has bounded
smooth coefficients, the map : $\widehat{D}=D\circ J^{r}:W^{2,r}\left(
E\right)  \rightarrow W^{2,r}\left(  E\right)  $ is continuous on the Hilbert
space $W^{2,r}\left(  E\right)  $ , so $\widehat{D}$ has an adjoint
$\widehat{D}^{\ast}\in%
\mathcal{L}%
\left(  W^{2,r}\left(  E\right)  ;W^{2,r}\left(  E\right)  \right)  $. However
this operator is not necessarily local. Indeed if D is scalar, it is also a
pseudo-differential operator, and as such its adjoint is a pseudo-differential
operator, whose expression is complicated (see next section).

\paragraph{Condition of existence\newline}

In the general conditions of the definition :

i) The scalar products induce antilinear morphisms :

$\Theta_{1}:\mathfrak{X}\left(  E_{1}\right)  \rightarrow\mathfrak{X}\left(
E_{1}\right)  ^{\prime}::\Theta_{1}\left(  X_{1}\right)  \left(  Y_{1}\right)
=G_{1}\left(  X_{1},Y_{1}\right)  $

$\Theta_{2}:\mathfrak{X}\left(  E_{2}\right)  \rightarrow\mathfrak{X}\left(
E_{2}\right)  ^{\prime}::\Theta_{2}\left(  X_{2}\right)  \left(  Y_{2}\right)
=G_{2}\left(  X_{2},Y_{2}\right)  $

They are injective, but not surjective, because the vector spaces are infinite dimensional.

ii) To the operator : $\widehat{D}:\mathfrak{X}\left(  E_{1}\right)
\rightarrow\mathfrak{X}\left(  E_{2}\right)  $ the associated operator on
distributions (which always exists) reads:

$\widehat{D}^{\prime}:\mathfrak{X}\left(  E_{2}\right)  ^{\prime}%
\rightarrow\mathfrak{X}\left(  E_{1}\right)  ^{\prime}::\widehat{D}^{\prime
}S_{2}\left(  X_{1}\right)  =S_{2}\left(  \widehat{D}\left(  X_{1}\right)
\right)  $

iii) Assume that, at least on some vector subspace F of $\mathfrak{X}\left(
E_{2}\right)  $ there is $X_{1}\in\mathfrak{X}\left(  E_{1}\right)
::\Theta_{1}\left(  X_{1}\right)  =$ $\widehat{D}^{\prime}\circ\Theta
_{2}\left(  X_{2}\right)  $ then the operator : $\widehat{D}^{\ast}=\Theta
_{1}^{-1}\circ\widehat{D}^{\prime}\circ\Theta_{2}$ is such that :

$G_{1}\left(  X_{1},\widehat{D}^{\ast}X_{2}\right)  =G_{1}\left(  X_{1}%
,\Theta_{1}^{-1}\circ\widehat{D}^{\prime}\circ\Theta_{2}\left(  X_{2}\right)
\right)  $

$=\overline{G_{1}\left(  \Theta_{1}^{-1}\circ\widehat{D}^{\prime}\circ
\Theta_{2}\left(  X_{2}\right)  ,X_{1}\right)  }=\overline{\widehat{D}%
^{\prime}\circ\Theta_{2}\left(  X_{2}\right)  \left(  X_{1}\right)  }$

$=\overline{\Theta_{2}\left(  X_{2}\right)  \left(  \widehat{D}\left(
X_{1}\right)  \right)  }=\overline{G_{2}\left(  X_{2},\widehat{D}\left(
X_{1}\right)  \right)  }=G_{2}\left(  \widehat{D}\left(  X_{1}\right)
,X_{2}\right)  $

$\widehat{D}^{\ast}=\Theta_{1}^{-1}\circ\widehat{D}^{\prime}\circ\Theta_{2}$
is $%
\mathbb{C}
$-linear, this is an adjoint map and a differential operator. As the adjoint,
when it exists, is unique, then $\widehat{D}^{\ast}$ is the adjoint of D on F.
So whenever we have an inverse $\Theta_{1}^{-1}$ we can define an adjoint.
This leads to the following theorem, which encompasses the most usual cases.

\paragraph{Fundamental theorem\newline}

\begin{theorem}
A linear differential operator D between the smooth finite dimensional vector
bundles on the same real manifold $E_{1}\left(  M,V_{1},\pi_{1}\right)
,E_{2}\left(  M,V_{2},\pi_{2}\right)  $\ endowed with scalar products
$G_{1},G_{2}$ on sections defined by scalar products $g_{1},g_{2}$ on the
fibers and a volume form $\varpi_{0}$ on M has an adjoint which is a linear
differential operator with same order as D.
\end{theorem}

The procedure above can be implemented because $\Theta_{1}$ is inversible.

\begin{proof}
Fiberwise the scalar products induce antilinear isomorphisms :

$\tau:E\left(  x\right)  \rightarrow E^{\prime}\left(  x\right)  ::\tau\left(
X\right)  \left(  x\right)  =\sum_{ij}g_{ij}\left(  x\right)  \overline{X}%
^{i}\left(  x\right)  e^{j}\left(  x\right)  $

$\Rightarrow\tau\left(  X\right)  \left(  Y\right)  =\sum_{ij}g_{ij}%
\overline{X}^{i}Y^{j}=g\left(  X,Y\right)  $

$\tau^{-1}:E^{\prime}\left(  x\right)  \rightarrow E\left(  x\right)
::\tau\left(  \lambda\right)  \left(  x\right)  =\sum_{ij}\overline{g}%
^{ki}\left(  x\right)  \overline{\lambda}_{k}\left(  x\right)  e_{i}\left(
x\right)  $

$\Rightarrow g\left(  \tau\left(  \lambda\right)  ,Y\right)  =\sum_{ij}%
g_{ij}g^{ki}\lambda_{k}Y^{j}=\lambda\left(  Y\right)  $

with $\overline{g}^{ki}=g^{ik}$ for a hermitian form.

The scalar products for sections are : $G\left(  X,Y\right)  =\int_{M}g\left(
x\right)  \left(  X\left(  x\right)  ,Y\left(  x\right)  \right)  \varpi_{0}$
and we have the maps :

$T:\Lambda_{m}\left(  M;E^{\prime}\right)  \rightarrow\mathfrak{X}\left(
E\right)  ^{\prime}::T\left(  \lambda\right)  \left(  Y\right)  =\int
_{M}\lambda\left(  x\right)  \left(  Y\left(  x\right)  \right)  \varpi_{0}$

$=\int_{M}g\left(  x\right)  \left(  X\left(  x\right)  ,Y\left(  x\right)
\right)  \varpi_{0}$

$\Theta:\mathfrak{X}\left(  E\right)  \rightarrow\mathfrak{X}\left(  E\right)
^{\prime}::\Theta\left(  X\right)  =T\left(  \tau\left(  X\right)
\otimes\varpi_{0}\right)  $

If $D(Z)=\sum_{s=0}^{r}\sum_{\alpha_{1}...\alpha_{s}=1}^{m}A\left(  x\right)
^{\alpha_{1}...\alpha_{s}}z_{\alpha_{1}...\alpha_{s}}$ where $A\left(
x\right)  ^{\alpha_{1}...\alpha_{s}}\in%
\mathcal{L}%
\left(  E_{1}\left(  x\right)  ;E_{2}\left(  x\right)  \right)  $

$\widehat{D}^{\prime}\circ\Theta_{2}\left(  X_{2}\right)  =D^{\prime}%
J^{r}T\left(  \tau_{2}\left(  X_{2}\right)  \otimes\varpi_{0}\right)
=T\left(  D^{t}\tau_{2}\left(  X_{2}\right)  \otimes\varpi_{0}\right)  $

$D^{t}\tau_{2}\left(  X_{2}\right)  =\sum_{s=0}^{r}\sum_{\alpha_{1}%
...\alpha_{s}=1}^{m}\left(  A^{\alpha_{1}...\alpha_{s}}\right)  ^{t}\left(
D_{\alpha_{1}...\alpha_{s}}\left(  \sum_{ij}g_{2ij}\left(  x\right)
\overline{X_{2}}^{i}\left(  x\right)  \right)  e_{2}^{j}\left(  x\right)
\right)  $

$\left(  A^{\alpha_{1}...\alpha_{s}}\right)  ^{t}\left(  D_{\alpha
_{1}...\alpha_{s}}\left(  \sum_{ij}g_{2ij}\left(  x\right)  \overline{X_{2}%
}^{i}\left(  x\right)  \right)  e_{2}^{j}\left(  x\right)  \right)  $

$=\sum_{j}\left[  A^{\alpha_{1}...\alpha_{s}}\right]  _{j}^{k}D_{\alpha
_{1}...\alpha_{s}}\left(  \sum_{kl}g_{2lk}\left(  x\right)  \overline{X_{2}%
}^{l}\left(  x\right)  \right)  e_{1}^{j}\left(  x\right)  $

$=\sum_{kj}\left[  A^{\alpha_{1}...\alpha_{s}}\right]  _{j}^{k}\Upsilon
_{k\alpha_{1}...\alpha_{s}}\left(  J^{s}\overline{X_{2}}\right)  e_{1}%
^{j}\left(  x\right)  $

where $\Upsilon_{k\alpha_{1}...\alpha_{s}}$ is a s order linear differential
operator on $\overline{X_{2}}$

$D^{t}\tau_{2}\left(  X_{2}\right)  =\sum_{s=0}^{r}\sum_{\alpha_{1}%
...\alpha_{s}=1}^{m}\sum_{jk}\left[  A^{\alpha_{1}...\alpha_{s}}\right]
_{j}^{k}\Upsilon_{k\alpha_{1}...\alpha_{s}}\left(  J^{s}\overline{X_{2}%
}\right)  e_{1}^{j}\left(  x\right)  $

$\widehat{D}^{\ast}\left(  X_{2}\right)  =\sum_{ijk}g_{1}^{ik}\sum_{s=0}%
^{r}\sum_{\alpha_{1}...\alpha_{s}=1}^{m}\overline{\left[  A^{\alpha
_{1}...\alpha_{s}}\right]  }_{k}^{j}\overline{\Upsilon_{j\alpha_{1}%
...\alpha_{s}}\left(  J^{s}\overline{X_{2}}\right)  }e_{1i}\left(  x\right)  $

So we have a r order linear differential operator on $X_{2}.$
\end{proof}

\bigskip

Example : $E_{1}=E_{2},r=1$

$D\left(  J^{1}X\right)  =\sum_{ij}\left(  \left[  A\right]  _{j}^{i}%
X^{j}+\left[  B^{\alpha}\right]  _{j}^{i}\partial_{\alpha}X^{j}\right)
e_{i}\left(  x\right)  $

$D^{\prime}T\left(  J^{1}\lambda\right)  =T\left(  \sum_{ij}\left(  \left[
A\right]  _{j}^{i}\lambda_{i}+\left[  B^{\alpha}\right]  _{j}^{i}%
\partial_{\alpha}\lambda_{i}\right)  e^{j}\left(  x\right)  \right)  $

$\tau\left(  X\right)  \left(  x\right)  =\sum_{pq}g_{pq}\left(  x\right)
\overline{X}^{p}\left(  x\right)  e^{q}\left(  x\right)  $

$D^{\prime}T\left(  J^{1}\tau\left(  X\right)  \right)  =T\left(  \sum\left(
\left[  A\right]  _{j}^{q}g_{pq}\overline{X}^{p}+\left[  B^{\alpha}\right]
_{j}^{q}\left(  \left(  \partial_{\alpha}g_{pq}\right)  \overline{X}%
^{p}+g_{pq}\partial_{\alpha}\overline{X}^{p}\right)  \right)  e^{j}\left(
x\right)  \right)  $

$\widehat{D}^{\ast}\left(  X\right)  =\sum g^{ij}\left(  \overline{\left[
A\right]  }_{j}^{q}\overline{g}_{pq}X^{p}+\overline{\left[  B^{\alpha}\right]
}_{j}^{q}\left(  \left(  \partial_{\alpha}\overline{g}_{pq}\right)
X^{p}+\overline{g}_{pq}\partial_{\alpha}X^{p}\right)  \right)  e_{i}\left(
x\right)  $

$\widehat{D}^{\ast}\left(  X\right)  =\sum\left(  \left(  \left[
g^{-1}\right]  \left[  A\right]  ^{\ast}\left[  g\right]  +\left[
g^{-1}\right]  \left[  B^{\alpha}\right]  ^{\ast}\left[  \partial_{\alpha
}g\right]  \right)  X+\left(  \left[  g^{-1}\right]  \left[  B^{\alpha
}\right]  ^{\ast}\left[  g\right]  \right)  \partial_{\alpha}X\right)
^{i}e_{i}\left(  x\right)  $

\bigskip

Notice that the only requirement on M is a volume form, which can come from a
non degenerate metric, but not necessarily. And this volume form is not
further involved. If the metric is induced by a scalar product on V then
$\left[  \partial_{\alpha}g\right]  =0.$

\subsubsection{Symbol of a linear differential operator}

\paragraph{Definition\newline}

\begin{definition}
The \textbf{symbol} of a linear r order differential operator $D:\mathfrak{X}%
\left(  J^{r}E_{1}\right)  \rightarrow\mathfrak{X}\left(  E_{2}\right)  $ is
the r order symmetric polynomial map :

$P\left(  x\right)  :%
\mathbb{R}
^{m\ast}\rightarrow E_{2}\left(  x\right)  \otimes E_{1}\left(  x\right)
^{\ast}::$%

\begin{equation}
P\left(  x\right)  \left(  u\right)  =\sum_{i=1}^{n_{1}}\sum_{j=1}^{n_{2}}%
\sum_{s=0}^{r}\sum_{\alpha_{1}...\alpha_{s}}\left[  A\left(  x\right)
^{\alpha_{1}...\alpha_{s}}\right]  _{i}^{j}u_{\alpha_{1}}..u_{\alpha_{s}%
}e_{2j}\left(  x\right)  \otimes e_{1}^{i}\left(  x\right)
\end{equation}

with $e_{1}^{i}\left(  x\right)  ,e_{2,j}\left(  x\right)  $ are holonomic
bases of $E_{1}\left(  x\right)  ^{\ast},E_{2}\left(  x\right)  $ and
$u=\left(  u_{1},...u_{m}\right)  \in%
\mathbb{R}
^{m\ast}$

The r order part of $P\left(  x\right)  \left(  u\right)  $ is the
\textbf{principal symbol} of D : $\sigma_{D}\left(  x,u\right)  \in%
\mathcal{L}%
\left(  E_{1}\left(  x\right)  ;E_{2}\left(  x\right)  \right)  $
\end{definition}%

\begin{equation}
\sigma_{D}\left(  x,u\right)  =\sum_{i=1}^{n_{1}}\sum_{j=1}^{n_{2}}%
\sum_{\alpha_{1}...\alpha_{r}}A\left(  x\right)  _{i}^{j,\alpha_{1}%
...\alpha_{r}}u_{\alpha_{1}}..u_{\alpha_{r}}e_{2j}\left(  x\right)  \otimes
e_{1}^{i}\left(  x\right)
\end{equation}

Explanation : A linear operator is a map : $D\left(  x\right)  \in%
\mathcal{L}%
\left(  J^{r}E_{1}\left(  x\right)  ;E_{2}\left(  x\right)  \right)
=J^{r}E_{1}\left(  x\right)  ^{\ast}\otimes E_{2}\left(  x\right)  .$
$J^{r}E_{1}\left(  x\right)  $ can be identified to $E_{1}\left(  x\right)
\otimes\sum_{s=0}^{r}\odot^{s}%
\mathbb{R}
^{m}$ so we can see D as a tensor $D\in\sum_{s=0}^{r}\odot^{s}%
\mathbb{R}
^{m}\otimes E_{1}{}^{\ast}\otimes E_{2}$ which acts on vectors of $%
\mathbb{R}
^{m\ast}.$

Conversely, given a r order symmetric polynomial it defines uniquely a linear
differential operator.

Formally, it sums up to replace $\frac{\partial}{\partial\xi^{\alpha}}$ by
$u_{\alpha}.$

Notice that : $X\in\mathfrak{X}\left(  E_{1}\right)  :P\left(  x\right)
\left(  u\right)  \left(  X\right)  =D\left(  Z\right)  $

with $Z=\sum_{i=1}^{n_{1}}\sum_{s=0}^{r}\sum_{\alpha_{1}...\alpha_{s}%
}u_{\alpha_{1}}..u_{\alpha_{s}}X^{i}\left(  x\right)  e_{1i}\left(  x\right)
$

\paragraph{Composition of operators\newline}

\begin{theorem}
If $D_{1},D_{2}$ are two order r differential on vector bundles on the same
manifold, the principal symbol of their compose $D_{1}\circ D_{2}$ is a 2r
order symmetric polynomial map given by : $\left[  \sigma_{D_{1}\circ D_{2}%
}\left(  x,u\right)  \right]  =\left[  \sigma_{D_{1}}\left(  x,u\right)
\right]  \left[  \sigma_{D_{2}}\left(  x,u\right)  \right]  $
\end{theorem}

it is not true for the other components of the symbol

\begin{proof}
$\sigma_{D_{1}\circ D_{2}}\left(  x,u\right)  $

$=\sum_{i,j,k=1}^{n}\sum_{\alpha_{1}...\alpha_{r}}\sum_{\beta_{1}...\beta_{r}%
}A_{1}\left(  x\right)  _{k}^{j,\alpha_{1}...\alpha_{r}}A_{2}\left(  x\right)
_{i}^{k,\beta_{1}...\beta_{r}}u_{\alpha_{1}}..u_{\alpha_{r}}u_{\beta_{1}%
}..u_{\beta_{r}}e_{3j}\left(  x\right)  \otimes e_{1}^{i}\left(  x\right)  $

In matrix notation:

$\left[  \sigma_{D_{1}\circ D_{2}}\left(  x,u\right)  \right]  =\left(
\sum_{\alpha_{1}...\alpha_{r}}u_{\alpha_{1}}..u_{\alpha_{r}}\left[
A_{1}\left(  x\right)  \right]  ^{\alpha_{1}...\alpha_{r}}\right)  \left(
\sum_{\beta_{1}...\beta_{r}}u_{\beta_{1}}..u_{\beta_{r}}\left[  A_{2}\left(
x\right)  \right]  ^{\beta_{1}...\beta_{r}}\right)  $
\end{proof}

\paragraph{Image of a differential operator\newline}

A constant map $L:\mathfrak{X}\left(  E_{1}\right)  \rightarrow\mathfrak{X}%
\left(  E_{2}\right)  $ between two vector bundles $E_{1}\left(  M,V_{1}%
,\pi_{1}\right)  ,$ $E_{2}\left(  M,V_{2},\pi_{2}\right)  $ on the same
manifold is extended to a map

$\widehat{L}:\mathfrak{X}\left(  J^{r}E_{1}\right)  \rightarrow\mathfrak{X}%
\left(  J^{r}E_{2}\right)  ::\widehat{L}=J^{r}\circ L$

$\widehat{L}\left(  Z_{\alpha_{1}...\alpha_{s}}^{i}e_{1i}^{\alpha_{1}%
..\alpha_{s}}\right)  =\sum_{j}\left[  L\right]  _{j}^{i}Z_{\alpha
_{1}...\alpha_{s}}^{j}e_{2i}^{\alpha_{1}..\alpha_{s}}$

A differential operator $D:J^{r}E_{1}\rightarrow E_{1}$ gives a differential
operator :

$D_{2}:J^{r}E_{1}\rightarrow E_{2}::D_{2}=D\circ J^{r}\circ L=\widehat{D}%
\circ\widehat{L}$

The symbol of $D_{2}$ is : $\sigma_{D_{2}}\left(  x,u\right)  =\sigma
_{D}\left(  x,u\right)  \circ L\in%
\mathcal{L}%
\left(  E_{1}\left(  x\right)  ;E_{2}\left(  x\right)  \right)  $

$P\left(  x\right)  \left(  u\right)  $

$=\sum_{i=1}^{n_{1}}\sum_{j=1}^{n_{2}}\sum_{s=0}^{r}\sum_{\alpha_{1}%
...\alpha_{s}}\left[  A\left(  x\right)  ^{\alpha_{1}...\alpha_{s}}\right]
_{k}^{j}\left[  L\right]  _{i}^{k}u_{\alpha_{1}}..u_{\alpha_{s}}e_{2j}\left(
x\right)  \otimes e_{1}^{i}\left(  x\right)  $

\paragraph{Adjoint\newline}

\begin{theorem}
If we have scalar products $\left\langle {}\right\rangle $ on two vector
bundles $E_{1}\left(  M,V_{1},\pi_{1}\right)  ,$ $E_{2}\left(  M,V_{2},\pi
_{2}\right)  ,$ and two linear r order differential operators $D_{1}%
:\mathfrak{X}\left(  J^{r}E_{1}\right)  \rightarrow\mathfrak{X}\left(
E_{2}\right)  ,D_{2}:\mathfrak{X}\left(  J^{r}E_{2}\right)  \rightarrow
\mathfrak{X}\left(  E_{1}\right)  $ such that $\ D_{2}$ is the adjoint of
$D_{1}$ then : $P_{D_{1}}\left(  x,u\right)  =P_{D_{2}}\left(  x,u\right)
^{\ast}$ the adjoint with respect to the scalar products.
\end{theorem}

\begin{proof}
We have : $\forall X_{1}\in\mathfrak{X}\left(  J^{r}E_{1}\right)  ,X_{2}%
\in\mathfrak{X}\left(  J^{r}E_{2}\right)  :$

$\left\langle D_{1}J^{r}X_{1},X_{2}\right\rangle _{E_{2}}=\left\langle
X_{1},D_{2}J^{r}X_{2}\right\rangle _{E_{1}}$

take : $X_{1}\left(  x\right)  =X_{1}\exp\left(  \sum_{\alpha=1}^{m}u_{\alpha
}\xi^{\alpha}\right)  $ with X some fixed vector in V$_{1}$.

$D_{\alpha_{1}..\alpha_{s}}X_{1}\left(  x\right)  =u_{\alpha_{1}}%
..u_{\alpha_{s}}X_{1}\exp\left(  \sum_{\alpha=1}^{m}u_{\alpha}\xi^{\alpha
}\right)  $

And similarly for $X_{2}$

$D_{1}J^{r}X_{1}=P_{D_{1}}\left(  x,u\right)  X_{1}\exp\left(  \sum_{\alpha
=1}^{m}u_{\alpha}\xi^{\alpha}\right)  ,$

$D_{1}J^{r}X_{2}=P_{D_{2}}\left(  x,u\right)  X_{2}\exp\left(  \sum_{\alpha
=1}^{m}u_{\alpha}\xi^{\alpha}\right)  $

$\forall X_{1}\in V_{1},X_{2}\in X_{2}:\left\langle P_{D_{1}}\left(
x,u\right)  X_{1},X_{2}\right\rangle =\left\langle X_{1},P_{D_{2}}\left(
x,u\right)  X_{2}\right\rangle $
\end{proof}

\paragraph{Elliptic operator\newline}

\begin{definition}
For a linear r order differential operator $D:\mathfrak{X}\left(
J^{r}E\right)  \rightarrow\mathfrak{X}\left(  E\right)  $ with principal
symbol $\sigma_{D}\left(  x,u\right)  $ over a vector bundle $E\left(
M,V,\pi\right)  $\ :

The \textbf{characteristic set} of D is the set : $Char(D)\subset M\times%
\mathbb{R}
^{m}$ where $\sigma_{D}\left(  x,u\right)  $ is an isomorphism.

D\ is said to be \textbf{elliptic (}or weakly elliptic\textbf{)} if its
principal symbol $\sigma_{D}\left(  x,u\right)  $ is an isomorphism whenever
$u\neq0.$

If V is a Hilbert space with inner product $\left\langle {}\right\rangle $ :

D is said to be \textbf{strongly} (or uniformly) \textbf{elliptic} if:

$\exists C\in%
\mathbb{R}
:\forall x\in M,\forall u,v\in T_{x}M^{\ast},\left\langle v,v\right\rangle
=1:\left\langle \sigma_{D}\left(  x,u\right)  v,v\right\rangle \geq
C\left\langle v,v\right\rangle $

D is said to be \textbf{semi elliptic} if:

$\forall x\in M,\forall u,v\in T_{x}M^{\ast},\left\langle v,v\right\rangle
=1:\left\langle \sigma_{D}\left(  x,u\right)  v,v\right\rangle \geq0$
\end{definition}

\begin{theorem}
A necessary condition for an operator to be strongly elliptic is dim(M) even.
\end{theorem}

\begin{theorem}
The composition of two weakly elliptic operators is still a weakly operator.
\end{theorem}

\begin{theorem}
A weakly elliptic operator on a vector bundle with compact base has a (non
unique) parametrix.
\end{theorem}

As a consequence the kernel of such an operator is finite dimensional in the
space of sections and is a Fredholm operator.

\begin{theorem}
If is D weakly elliptic, then D*D, DD* are weakly elliptic.
\end{theorem}

\begin{proof}
by contradiction : let be $u\neq0,X\neq0:\sigma_{D^{\ast}\circ D}\left(
x,u\right)  X=0$

$\sigma_{D^{\ast}\circ D}\left(  x,u\right)  =\sigma_{D^{\ast}}\left(
x,u\right)  \sigma_{D}\left(  x,u\right)  =\sigma_{D}\left(  x,u\right)
^{\ast}\sigma_{D}\left(  x,u\right)  $
\end{proof}

\begin{definition}
A linear differential operator D' on a space of distributions is said to be
\textbf{hypoelliptic} if the singular support of D'(S) is included in the
singular support of S.
\end{definition}

So whenever $f\in W$ then $\exists g\in W:T\left(  g\right)  =D\left(
T\left(  f\right)  \right)  .$ If D is strongly elliptic then the associated
operator D' is hypoelliptic. The laplacian and the heat kernel are hypoelliptic.

Remark : contrary at what we could expect a hyperbolic operator is not a
differential operator such that the principal symbol is degenerate (see PDE).

\paragraph{Index of an operator\newline}

There is a general definition of the index of a linear map (see Banach
spaces).\ For differential operators we have :

\begin{definition}
A linear differential operator $D:F_{1}\rightarrow\mathfrak{X}\left(
E_{2}\right)  $ between two vector bundle $E_{1},E_{2}$ on the same base, with
$F_{1}\subset\mathfrak{X}\left(  J^{r}E_{1}\right)  $ is a Fredholm operator
if $\ker D$ and $\mathfrak{X}\left(  E_{2}\right)  /D(F_{1})$ are finite
dimensional. The index (also called the analytical index) of D is then :
Index(D)=$\dim\ker D-\dim\mathfrak{X}\left(  E_{2}\right)  /D(F_{1})$
\end{definition}

Notice that the condition applies to the full vector space of sections (and
not fiberwise).

\begin{theorem}
A weakly elliptic operator on a vector bundle with compact base is Freholm
\end{theorem}

This is the starting point for a set of theorems such that the Atiyah-Singer
index theorem. For a differential operator $D:J^{r}E_{1}\rightarrow E_{2}$
between vector bundles on the same base manifold M, one can define a
topological index (this is quite complicated) which is an integer deeply
linked to topological invariants of M. The most general theorem is the
following :

\begin{theorem}
Teleman index theorem : For any abstract elliptic operator on a closed,
oriented, topological manifold, its analytical index equals its topological index.
\end{theorem}

It means that the index of a differential operator depends deeply of the base
manifold. We have more specifically the followings :

\begin{theorem}
(Taylor 2 p.264) If D is a linear elliptic first order differential operator
between two vector bundles $E_{1},E_{2}$ on the same \textit{compact}
manifold, then $D:W^{2,r+1}\left(  E_{1}\right)  \rightarrow W^{2,r}\left(
E_{2}\right)  $ is a Fredholm operator, $\ker D$ is independant of r, and
$D^{t}:W^{2,-r}\left(  E_{1}\right)  \rightarrow W^{2,-r-1}\left(
E_{2}\right)  $ has the same properties. If $D_{s}$ is a family of such
operators, continuously dependant of the parameter s, then the index of
$D_{s}$ does not depend on s.
\end{theorem}

\begin{theorem}
(Taylor 2 p.266) If D is a linear elliptic first order differential operator
on a vector bundle with base a compact manifold with odd dimension then Index(D)=0
\end{theorem}

On any oriented manifold M the exterior algebra can be split between the forms
of even E or odd F order. The sum $D=d+\delta$\ of the exterior differential
and the codifferential exchanges the forms between E and F. If M is compact
the topological index of D is the Euler characteristic of the Hodge cohomology
of M, and the analytical index is the Euler class of the manifold. The index
formula for this operator yields the Chern-Gauss-Bonnet theorem.

\subsubsection{Linear differential operators and Fourier transform}

\begin{theorem}
For a linear scalar differential operator D over $%
\mathbb{R}
^{m}$\ : $D:C_{r}\left(
\mathbb{R}
^{m};%
\mathbb{C}
\right)  \rightarrow C_{0}\left(
\mathbb{R}
^{m};%
\mathbb{C}
\right)  $ and for $f\in S\left(
\mathbb{R}
^{m}\right)  :$%

\begin{equation}
Df=\left(  2\pi\right)  ^{-m/2}\int_{%
\mathbb{R}
^{m}}P(x,it)\widehat{f}(t)e^{i\left\langle t,x\right\rangle }dt
\end{equation}

where P is the symbol of D
\end{theorem}

As $\widehat{f}\in S\left(
\mathbb{R}
^{m}\right)  $ and $P(x,it)$ is a polynomial in t, then $P(x,it)\widehat
{f}(t)\in S\left(
\mathbb{R}
^{m}\right)  $ and $Df=%
\mathcal{F}%
_{t}^{\ast}\left(  P(x,it)\widehat{f}(t)\right)  $

As $\widehat{f}\in S\left(
\mathbb{R}
^{m}\right)  $ and the induced distribution $T\left(  \widehat{f}\right)  \in
S\left(
\mathbb{R}
^{m}\right)  ^{\prime}$ we have : $Df=\left(  2\pi\right)  ^{-m/2}T\left(
\widehat{f}\right)  _{t}\left(  P(x,it)e^{i\left\langle t,x\right\rangle
}\right)  $

\begin{proof}
The Fourier transform is a map : $%
\mathcal{F}%
:S\left(
\mathbb{R}
^{m}\right)  \rightarrow S\left(
\mathbb{R}
^{m}\right)  $ and

$D_{\alpha_{1}...\alpha_{s}}f=%
\mathcal{F}%
^{\ast}\left(  i^{r}\left(  t_{\alpha_{1}}..t_{\alpha_{s}}\right)  \widehat
{f}\right)  $

So : $Df=\sum_{s=0}^{r}\sum_{\alpha_{1}...\alpha_{s}=1}^{m}A\left(  x\right)
^{\alpha_{1}...\alpha_{s}}D_{\alpha_{1}...\alpha_{s}}f\left(  x\right)  $

$=\left(  2\pi\right)  ^{-m/2}\sum_{s=0}^{r}\sum_{\alpha_{1}...\alpha_{s}%
=1}^{m}i^{s}A\left(  x\right)  ^{\alpha_{1}...\alpha_{s}}\int_{%
\mathbb{R}
^{m}}t_{\alpha_{1}}...t_{\alpha_{s}}\widehat{f}(t)e^{i\left\langle
t,x\right\rangle }dt$

$=\left(  2\pi\right)  ^{-m/2}\int_{%
\mathbb{R}
^{m}}\left(  \sum_{s=0}^{r}\sum_{\alpha_{1}...\alpha_{s}=1}^{m}A\left(
x\right)  ^{\alpha_{1}...\alpha_{s}}\left(  it_{\alpha_{1}}\right)  ...\left(
it_{\alpha_{s}}\right)  \right)  \widehat{f}(t)e^{i\left\langle
t,x\right\rangle }dt$

$=\left(  2\pi\right)  ^{-m/2}\int_{%
\mathbb{R}
^{m}}P(x,it)\widehat{f}(t)e^{i\left\langle t,x\right\rangle }dt$
\end{proof}

\bigskip

\begin{theorem}
For a linear scalar differential operator D over $%
\mathbb{R}
^{m}$\ and for $f\in S\left(
\mathbb{R}
^{m}\right)  ,A\left(  x\right)  ^{\alpha_{1}...\alpha_{s}}\in L^{1}\left(
\mathbb{R}
^{m},dx,%
\mathbb{C}
\right)  :$%

\begin{equation}
Df=\left(  2\pi\right)  ^{-m}\int_{%
\mathbb{R}
^{m}}\sum_{s=0}^{r}\sum_{\alpha_{1}...\alpha_{s}=1}^{m}\left(  \widehat
{A}^{\alpha_{1}...\alpha_{s}}\ast\left(  \left(  it_{\alpha_{1}}\right)
...\left(  it_{\alpha_{s}}\right)  \widehat{f}\right)  \right)
e^{i\left\langle t,x\right\rangle }dt
\end{equation}

\end{theorem}

\begin{proof}
$A\left(  x\right)  ^{\alpha_{1}...\alpha_{s}}D_{\alpha_{1}...\alpha_{s}%
}f\left(  x\right)  $

$=%
\mathcal{F}%
^{\ast}\left(  \left(  2\pi\right)  ^{-m/2}%
\mathcal{F}%
\left(  A\left(  x\right)  ^{\alpha_{1}...\alpha_{s}}\right)  \ast%
\mathcal{F}%
\left(  D_{\alpha_{1}...\alpha_{s}}f\left(  x\right)  \right)  \right)  $

$=\left(  2\pi\right)  ^{-m/2}%
\mathcal{F}%
^{\ast}\left(  \widehat{A}\left(  x\right)  ^{\alpha_{1}...\alpha_{s}}%
\ast\left(  \left(  it_{\alpha_{1}}\right)  ...\left(  it_{\alpha_{s}}\right)
\widehat{f}\left(  x\right)  \right)  \right)  $

$=\left(  2\pi\right)  ^{-m}\int_{%
\mathbb{R}
^{m}}\widehat{A}\left(  x\right)  ^{\alpha_{1}...\alpha_{s}}\ast\left(
\left(  it_{\alpha_{1}}\right)  ...\left(  it_{\alpha_{s}}\right)  \widehat
{f}\left(  x\right)  \right)  e^{i\left\langle t,x\right\rangle }dt$
\end{proof}

\bigskip

\begin{theorem}
For a linear differential operator D' on the space of tempered distributions
$S\left(
\mathbb{R}
^{m}\right)  ^{\prime}$ $:$%

\begin{equation}
D^{\prime}S=%
\mathcal{F}%
^{\ast}\left(  P(x,it)\widehat{S}\right)
\end{equation}

where P is the symbol of D'
\end{theorem}

\begin{proof}
The Fourier transform is a map : $%
\mathcal{F}%
:S\left(
\mathbb{R}
^{m}\right)  ^{\prime}\rightarrow S\left(
\mathbb{R}
^{m}\right)  ^{\prime}$ and $D_{\alpha_{1}...\alpha_{s}}S=%
\mathcal{F}%
^{\ast}\left(  i^{r}\left(  t_{\alpha_{1}}..t_{\alpha_{s}}\right)  \widehat
{S}\right)  $

So : $D^{\prime}S=\sum_{s=0}^{r}\sum_{\alpha_{1}...\alpha_{s}=1}^{m}A\left(
x\right)  ^{\alpha_{1}...\alpha_{s}}D_{\alpha_{1}...\alpha_{s}}S$

$=\sum_{s=0}^{r}\sum_{\alpha_{1}...\alpha_{s}=1}^{m}A\left(  x\right)
^{\alpha_{1}...\alpha_{s}}%
\mathcal{F}%
^{\ast}\left(  it_{\alpha_{1}}...it_{\alpha_{s}}\widehat{S}\right)  $

$=%
\mathcal{F}%
^{\ast}\left(  \sum_{s=0}^{r}\sum_{\alpha_{1}...\alpha_{s}=1}^{m}A\left(
x\right)  ^{\alpha_{1}...\alpha_{s}}\left(  it_{\alpha_{1}}\right)  ...\left(
it_{\alpha_{s}}\right)  \widehat{S}\right)  $

$=%
\mathcal{F}%
^{\ast}\left(  P(x,it)\widehat{S}\right)  $
\end{proof}

Whenever $S=T(f),f\in L^{1}\left(
\mathbb{R}
^{m},dx,%
\mathbb{C}
\right)  :\widehat{S}=T\left(  \widehat{f}\right)  =\left(  2\pi\right)
^{-m/2}T_{t}\left(  S_{x}\left(  e^{-i\left\langle x,t\right\rangle }\right)
\right)  $ and we get back the same formula as 1 above.

\subsubsection{Fundamental solution}

Fundamental solutions and Green functions are ubiquitous tools in differential
equations, which come in many flavors. We give here a general definition,
which is often adjusted with respect to the problem in review. Fundamental
solutions for classic operators are given in the present section and in the
section on PDE.

\paragraph{Definition\newline}

\begin{definition}
A \textbf{fundamental solution} at a point\ $x\in M$ of a scalar linear
differential operator D' on distributions V' on a space of functions $V\subset
C_{r}\left(  M;%
\mathbb{C}
\right)  $ over a manifold M is a distribution $U_{x}\in V^{\prime}$ such that
: $D^{\prime}U_{x}=\delta_{x}$
\end{definition}

$\forall\varphi\in V::D^{\prime}U_{x}\left(  \varphi\right)  =\varphi\left(
x\right)  =\sum_{s=0}^{r}\sum_{\alpha_{1}...\alpha_{s}=1}^{m}A\left(
x\right)  ^{\alpha_{1}...\alpha_{s}}U_{x}\left(  D_{\alpha_{1}...\alpha_{s}%
}\varphi\right)  =\sum_{s=0}^{r}\left(  -1\right)  ^{s}\sum_{\alpha
_{1}...\alpha_{s}=1}^{m}A\left(  x\right)  ^{\alpha_{1}...\alpha_{s}}%
D_{\alpha_{1}...\alpha_{s}}U_{x}\left(  \varphi\right)  $

where D is the differential operator on V :

$DJ^{r}\varphi=\sum_{s=0}^{r}\sum_{\alpha_{1}...\alpha_{s}=1}^{m}A\left(
x\right)  ^{\alpha_{1}...\alpha_{s}}D_{\alpha_{1}...\alpha_{s}}\varphi$ with
$A\left(  x\right)  ^{\alpha_{1}...\alpha_{s}}\in C\left(  M;%
\mathbb{C}
\right)  $

So to any linear scalar differential operator D on V is associated the linear
differential operator on V' :

$D^{\prime}S=\sum_{s=0}^{r}\left(  -1\right)  ^{s}\sum_{\alpha_{1}%
...\alpha_{s}=1}^{m}A\left(  x\right)  ^{\alpha_{1}...\alpha_{s}}D_{\alpha
_{1}...\alpha_{s}}S$

If V' is given by a set of r-differentiable m-forms $\lambda\in\Lambda
_{m}\left(  M;%
\mathbb{C}
\right)  $ then a fundamental solution is $\lambda\left(  x,y\right)  d\xi
^{1}\wedge...\wedge d\xi^{m}:$

$\sum_{s=0}^{r}\sum_{\alpha_{1}...\alpha_{s}=1}^{m}A\left(  x\right)
^{\alpha_{1}...\alpha_{s}}\int_{M}\varphi\left(  y\right)  \left(
D_{\alpha_{1}...\alpha_{s}}\left(  \lambda\left(  x,y\right)  \right)
d\xi^{1}\wedge...\wedge d\xi^{m}\right)  =\varphi\left(  x\right)  $

\bigskip

If M is an open O of $%
\mathbb{R}
^{m}$ and V' defined by a map $T:W\rightarrow V^{\prime}$ then a fundamental
solution is $T\left(  G\left(  x,.\right)  \right)  $ with a function G(x,y)
called a \textbf{Green' function :}

$D^{\prime}T\left(  G\left(  x,y\right)  \right)  \left(  \varphi\right)
=\varphi\left(  x\right)  $

$=\sum_{s=0}^{r}\left(  -1\right)  ^{s}\sum_{\alpha_{1}...\alpha_{s}=1}%
^{m}A\left(  x\right)  ^{\alpha_{1}...\alpha_{s}}\int_{O}\varphi\left(
y\right)  \left(  D_{\alpha_{1}...\alpha_{s}}\left(  G\left(  x,y\right)
\right)  \right)  d\xi^{1}\wedge...\wedge d\xi^{m}$

Example :

$D=\frac{\partial^{2}}{\partial x^{2}}$ then $U=xH$ with H the Heaviside function.

$\frac{\partial}{\partial x}\left(  xH\right)  =H+x\delta_{0}$

$\frac{\partial^{2}}{\partial x^{2}}\left(  xH\right)  =2\delta_{0}%
+x\delta_{0}^{^{\prime}}$

$\frac{\partial^{2}}{\partial x^{2}}\left(  xH\right)  \left(  \varphi\right)
=2\varphi\left(  0\right)  -\delta_{0}\left(  \frac{\partial}{\partial
x}\left(  x\varphi\right)  \right)  =2\varphi\left(  0\right)  -\delta
_{0}\left(  \varphi+x\varphi^{\prime}\right)  =\varphi\left(  0\right)
=\delta_{0}\left(  \varphi\right)  $

If $D=\sum_{i=1}^{n}\frac{\partial^{2}}{\partial x_{i}^{2}}$ then
$U=\sum_{i=1}^{n}x_{i}H\left(  x_{i}\right)  $

\bigskip

Notice that :

i) the result is always a Dirac distribution, so we\ need a differential
operator acting on distributions

ii) a Dirac distribution is characterized by a support in one point. On a
manifold it can be any point.

iii) $\delta_{x}$ must be in V'

\paragraph{Fundamental solution of PDE\newline}

Its interest, and its name, come from the following:.

\begin{theorem}
If $U_{0}$ is a fundamental solution at 0 for a scalar linear differential
operator D' on $C_{\infty c}\left(  O;%
\mathbb{C}
\right)  ^{\prime}$\ with an open O of $%
\mathbb{R}
^{m}$ , then for any $S\in$ $\left(  C_{\infty c}\left(  O;%
\mathbb{C}
\right)  ^{\prime}\right)  _{c}$\ , $U_{0}\ast S$ is a solution of D'X=S
\end{theorem}

\begin{proof}
$D^{\prime}\left(  U_{0}\ast S\right)  =D^{\prime}\left(  U_{0}\right)  \ast
S=\delta_{0}\ast S=S$
\end{proof}

\begin{theorem}
If V is a Fr\'{e}chet space of complex functions in an open O of $%
\mathbb{R}
^{m},$\ D a linear differential operator on V and U(y) a fundamental solution
at y of the associated differential operator D', then for any compactly
supported function f, $u=U\left(  y\right)  _{t}\left(  f\left(  x+y-t\right)
\right)  $ is a solution of Du = f
\end{theorem}

\begin{proof}
As we are in $%
\mathbb{R}
^{m}$\ we can use convolution. The convolution of $U(y)$ and T(f) is well
defined, and :

$D^{\prime}\left(  U\left(  y\right)  \ast T\left(  f\right)  \ast\delta
_{-y}\right)  =\left(  D^{\prime}U\left(  y\right)  \right)  \ast T\left(
f\right)  \ast\delta_{-y}=\delta_{y}\ast\delta_{-y}\ast T\left(  f\right)
=\delta_{0}\ast T\left(  f\right)  =T(f)$

$U\left(  y\right)  \ast T\left(  f\right)  \ast\delta_{-y}=U\left(  y\right)
\ast T_{x}\left(  f\left(  x+y\right)  \right)  =T\left(  U\left(  y\right)
_{t}\left(  f\left(  x+y-t\right)  \right)  \right)  =T\left(  u\right)  $

and $u\in C_{\infty}\left(  O;%
\mathbb{C}
\right)  $ so $D^{\prime}T\left(  u\right)  =T\left(  Du\right)  $ and

$T\left(  f\right)  =T\left(  Du\right)  \Rightarrow Du=f$
\end{proof}

If there is a Green's function then \ : $u\left(  x\right)  =\int_{O}G\left(
x,y\right)  f\left(  y\right)  dy$ is a solution of Du = f.

\paragraph{Operators depending on a parameter\newline}

Let V be a Fr\'{e}chet space of complex functions on a manifold M, J some
interval in $%
\mathbb{R}
.$ We consider functions in V depending on a parameter t. We have a family
D(t) of linear scalar differential operators $D\left(  t\right)  :V\rightarrow
V$ , depending on the same parameter $t\in J.$

We can see this as a special case of the above, with $J\times M$ as manifold.
A fundamental solution at (t,y) with y some fixed point in M is a family
U(t,y) of distributions acting on $C_{\infty c}\left(  J\times M;%
\mathbb{C}
\right)  ^{\prime}$ and such that : $D^{\prime}\left(  t\right)  U\left(
t,y\right)  =\delta_{\left(  t,y\right)  }.$

\paragraph{Fourier transforms and fundamental solutions\newline}

The Fourier transform gives a way to fundamental solutions for scalar linear
differential operators.

Let D' be a linear differential operator on the space of tempered
distributions $S\left(
\mathbb{R}
^{m}\right)  ^{\prime}$ then $:D^{\prime}S=%
\mathcal{F}%
^{\ast}\left(  P(x,it)\widehat{S}\right)  $ where P is the symbol of D'. If U
is a fundamental solution of D' at 0 then $P(x,it)\widehat{U}=%
\mathcal{F}%
\left(  \delta_{0}\right)  =\left(  2\pi\right)  ^{-m/2}T\left(  1\right)  $

$P(x,it)\widehat{U}=%
\mathcal{F}%
\left(  \delta_{0}\right)  =\left(  2\pi\right)  ^{-m/2}T\left(  1\right)
=\left(  \sum_{s=0}^{r}\sum_{\alpha_{1}...\alpha_{s}=1}^{m}A^{\alpha
_{1}...\alpha_{s}}(-i)^{s}t_{\alpha_{1}}...t_{\alpha_{s}}\right)  \widehat{U}$

which gives usually a way to compute U.

\paragraph{Parametrix\newline}

A parametrix is a proxy" for a fundamental solution.

Let V be a Fr\'{e}chet space of complex functions, V' the associated space of
distributions, W a space of functions such that : $T:W\rightarrow V^{\prime}$,
a scalar linear differential operator D' on V'. A \textbf{parametrix} for D'
is a distribution $U\in V^{\prime}:D^{\prime}\left(  U\right)  =\delta
_{y}+T(u)$\ with $u\in W.$

\subsubsection{Connection and differential operators}

\paragraph{Covariant derivative as a differential operator\newline}

A covariant derivative induced by a linear connection on a vector bundle
$E\left(  M,V,\pi\right)  $ is a map acting on sections of E.

$\nabla:\mathfrak{X}\left(  E\right)  \rightarrow\Lambda_{1}\left(
M;E\right)  :\nabla S=\sum_{i\alpha}\left(  \partial_{\alpha}X^{i}%
+\Gamma_{\alpha j}^{i}\left(  x\right)  X^{j}\right)  e_{i}\left(  x\right)
\otimes dx^{\alpha}$

We can define the tensorial product of two vector bundles, so we have
$TM^{\ast}\otimes E$ and we can consider the covariant derivative as a linear
differential operator :

$\widehat{\nabla}:\mathfrak{X}\left(  J^{1}E\right)  \rightarrow
\mathfrak{X}\left(  E\otimes TM^{\ast}\right)  :\widehat{\nabla}%
Z=\sum_{i\alpha}\left(  Z_{\alpha}^{i}+\Gamma_{\alpha j}^{i}\left(  x\right)
Z^{j}\right)  e_{i}\left(  x\right)  \otimes dx^{\alpha}$

and for a section of E, that is a vector field : $\widehat{\nabla}\circ
J^{1}=\nabla$

Notice that $E\otimes TM^{\ast}$ involves the cotangent bundle, but
$J^{1}E=J^{1}E\left(  E,%
\mathbb{R}
^{m\ast}\otimes V,\pi\right)  $ does not :

$X\in E\left(  x\right)  \otimes T_{x}M^{\ast}:X=\sum_{i\alpha}X_{\alpha}%
^{i}e_{i}\left(  x\right)  \otimes dx^{\alpha}$

$Z\in J^{1}E\left(  x\right)  :Z=\left(  \left(  \sum_{i}Z^{i}e_{i}\left(
x\right)  ,\sum_{i\alpha}Z_{\alpha}^{i}e_{i}^{\alpha}\left(  x\right)
\right)  \right)  $

We can define higher order connections in the same way as r linear
differential operators acting on E :

$\nabla^{r}:\mathfrak{X}\left(  J^{r}E\right)  \rightarrow\Lambda_{1}\left(
M;E\right)  ::$

$\nabla^{r}Z=\sum_{s=0}^{r}\sum_{\beta_{1}..\beta_{s}}\Gamma_{\alpha j}%
^{\beta_{1..}\beta_{s}i}\left(  x\right)  Z_{\beta_{1}..\beta_{s}}^{j}%
e_{i}\left(  x\right)  \otimes dx^{\alpha}$

$\nabla^{r}X=\sum_{s=0}^{r}\sum_{\beta_{1}..\beta_{s}}\Gamma_{\alpha j}%
^{\beta_{1..}\beta_{s}i}\left(  x\right)  \left(  \partial_{\beta_{1}%
..\beta_{s}}^{j}X^{j}\right)  e_{i}\left(  x\right)  \otimes dx^{\alpha}$

\paragraph{Exterior covariant derivative\newline}

1. The curvature of the connection is a linear map, but not a differential
operator :

$\widehat{\Omega}:\mathfrak{X}\left(  E\right)  \rightarrow\Lambda_{2}\left(
M;E\right)  ::\widehat{\Omega}\left(  x\right)  \left(  X\left(  x\right)
\right)  $

$=\sum_{\alpha\beta}\sum_{j\in I}\left(  -\partial_{\alpha}\Gamma_{j\beta}%
^{i}\left(  x\right)  +\sum_{k\in I}\Gamma_{j\alpha}^{k}\left(  x\right)
\Gamma_{k\beta}^{i}\left(  x\right)  \right)  X^{j}\left(  x\right)
dx^{\alpha}\wedge dx^{\beta}\otimes e_{i}\left(  x\right)  $

2. The exterior covariant derivative is the map :

$\nabla_{e}:\Lambda_{r}\left(  M;E\right)  \rightarrow\Lambda_{r+1}\left(
M;E\right)  ::\nabla_{e}\varpi=\sum_{i}\left(  d\varpi^{i}+\sum_{j}%
\Gamma_{\alpha j}^{i}d\xi^{\alpha}\wedge\varpi^{j}\right)  e_{i}\left(
x\right)  $

This is a first order linear differential operator : $d\varpi^{i}$ is the
ordinary exterior differential on M.

For r=0 we have just $\nabla_{e}=\widehat{\nabla}$

For r=1 the differential operator reads :

$\widehat{\nabla}_{e}:J^{1}\left(  E\otimes TM^{\ast}\right)  \rightarrow
E\otimes\Lambda_{2}TM^{\ast}::$

$\widehat{\nabla}_{e}\left(  Z_{\alpha}^{i}e_{i}\left(  x\right)  \otimes
dx^{\alpha},Z_{\alpha,\beta}^{i}e_{i}^{\beta}\left(  x\right)  \otimes
dx^{\alpha}\right)  =\sum_{i\alpha\beta}\left(  Z_{\alpha,\beta}^{i}%
+\Gamma_{\alpha j}^{i}Z_{\beta}^{j}\right)  e_{i}\left(  x\right)  \otimes
dx^{\alpha}\wedge dx^{\beta}$

and we have : $\nabla_{e}\left(  \nabla X\right)  =-\widehat{\Omega}\left(
X\right)  $ which reads : $\widehat{\nabla}_{e}\circ J^{1}\circ\widehat
{\nabla}\left(  X\right)  =-\widehat{\Omega}\left(  X\right)  $

$\widehat{\nabla}_{e}J^{1}\left(  \nabla_{\beta}X\otimes dx^{\beta}\right)
=\widehat{\nabla}_{e}\left(  \left(  \nabla_{\beta}X^{i}\right)  e_{i}\left(
x\right)  \otimes dx^{\beta},\partial_{\alpha}\nabla_{\beta}X^{i}e_{i}\left(
x\right)  \otimes dx^{\alpha}\otimes dx^{\beta}\right)  $

$=\sum_{i\alpha\beta}\left(  \partial_{\alpha}\nabla_{\beta}X^{i}%
+\Gamma_{\alpha j}^{i}\left(  \nabla_{\beta}X^{i}\right)  \right)
e_{i}\left(  x\right)  \otimes dx^{\alpha}\wedge dx^{\beta}=-\widehat{\Omega
}\left(  X\right)  $

3. If we apply two times the exterior covariant derivative we get :

$\nabla_{e}\left(  \nabla_{e}\varpi\right)  =\sum_{ij}\left(  \sum
_{\alpha\beta}R_{j\alpha\beta}^{i}d\xi^{\alpha}\wedge d\xi^{\beta}\right)
\wedge\varpi^{j}\otimes e_{i}\left(  x\right)  $

Where $R=\sum_{\alpha\beta}\sum_{ij}R_{j\alpha\beta}^{i}d\xi^{\alpha}\wedge
d\xi^{\beta}\otimes e^{j}\left(  x\right)  \otimes e_{i}\left(  x\right)  $
and $R_{j\alpha\beta}^{i}=\partial_{\alpha}\Gamma_{j\beta}^{i}+\sum_{k}%
\Gamma_{k\alpha}^{i}\Gamma_{j\beta}^{k}$

$R\in\Lambda_{2}\left(  M;E^{\ast}\otimes E\right)  $ has been called the
Riemannian curvature. This is not a differential operator (the derivatives of
the section are not involved) but a linear map.

\paragraph{Adjoint\newline}

1. The covariant derivative along a vector field W on M is a differential
operator on the same vector bundle :

$\widehat{\nabla}_{W}:\mathfrak{X}\left(  J^{1}E\right)  \rightarrow
\mathfrak{X}\left(  E\right)  :::\widehat{\nabla}_{W}Z=\sum_{i\alpha}\left(
Z_{\alpha}^{i}+\Gamma_{\alpha j}^{i}\left(  x\right)  Z^{j}\right)  W^{\alpha
}e_{i}\left(  x\right)  $ with W some fixed vector field in TM.

2. If there is a scalar product g on E and a volume form on M\ the adjoint map
of $\widehat{\nabla}_{W}$\ is defined as seen above (Adjoint).

with $\left[  A\right]  =\sum_{\alpha}\left[  \Gamma_{\alpha}\right]
W^{\alpha},\left[  B_{\alpha}\right]  =W^{\alpha}\left[  I\right]  $

$\widehat{\nabla}_{W}^{\ast}\left(  X\right)  =\sum_{i,\alpha}\left(  \left(
\left[  g^{-1}\right]  \left[  \Gamma_{\alpha}\right]  ^{\ast}\left[
g\right]  +\left[  g^{-1}\right]  \left[  \partial_{\alpha}g\right]  \right)
X+\partial_{\alpha}X\right)  ^{i}\overline{W}^{\alpha}e_{i}\left(  x\right)  $

W is real so : $\nabla^{\ast}:\mathfrak{X}\left(  E\right)  \rightarrow
\mathfrak{\Lambda}_{1}\left(  M;E\right)  $

$\nabla^{\ast}\left(  X\right)  =\sum_{i,\alpha}\left(  \left[  g^{-1}\right]
\left(  \left(  \left[  \Gamma_{\alpha}\right]  ^{\ast}+\left[  \partial
_{\alpha}g\right]  \left[  g^{-1}\right]  \right)  X+\partial_{\alpha
}X\right)  \left[  g\right]  \right)  ^{i}e_{i}\left(  x\right)  \otimes
d\xi^{\alpha}$

is such that :

$\forall W\in\mathfrak{X(}TM),\forall Z\in J^{1}E\left(  x\right)  ,Y\in
E\left(  x\right)  :\left\langle \widehat{\nabla}_{W}Z,Y\right\rangle
_{E\left(  x\right)  }=\left\langle Z,\widehat{\nabla}_{W}^{\ast
}Y\right\rangle _{J^{1}E\left(  x\right)  }$

3. The symbol of $\widehat{\nabla}$ and $\widehat{\nabla}^{\ast}$ are:

$P\left(  x\right)  \left(  u\right)  =\sum_{\alpha}\Gamma_{\alpha j}^{i}%
e_{i}\left(  x\right)  \otimes dx^{\alpha}\otimes e^{j}\left(  x\right)
+u_{\alpha}e_{i}\left(  x\right)  \otimes dx^{\alpha}\otimes e^{i}\left(
x\right)  $

$P^{\ast}\left(  x\right)  \left(  u\right)  =\sum_{\alpha}\left(  \left[
g^{-1}\right]  \left[  \Gamma_{\alpha}\right]  ^{\ast}\left[  g\right]
+\left[  g^{-1}\right]  \left[  \partial_{\alpha}g\right]  \right)  _{j}%
^{i}e_{i}\left(  x\right)  \otimes dx^{\alpha}\otimes e^{j}\left(  x\right)
+u_{\alpha}e_{i}\left(  x\right)  \otimes dx^{\alpha}\otimes e^{i}\left(
x\right)  $

so : $\sigma_{\nabla}\left(  x,u\right)  =\sigma_{\nabla^{\ast}}\left(
x,u\right)  =\sum_{\alpha}\sum_{i=1}^{n}u_{\alpha}e_{i}\left(  x\right)
\otimes dx^{\alpha}\otimes e^{i}\left(  x\right)  $

\subsubsection{Dirac operators}

\paragraph{Definition of a Dirac operator\newline}

\begin{definition}
A first order linear differential operator D on a vector bundle endowed with a
scalar product g is said to be a \textbf{Dirac operator} if it is weakly
elliptic and the principal symbol of $D^{\ast}\circ D$ is scalar
\end{definition}

Meaning that $\sigma_{D^{\ast}\circ D}\left(  x,u\right)  =\gamma\left(
x,u\right)  Id\in%
\mathcal{L}%
\left(  E\left(  x\right)  ;E\left(  x\right)  \right)  $

D* is the adjoint of D with respect to g

\paragraph{Properties of a Dirac operator\newline}

\begin{theorem}
A Dirac operator D on a vector bundle E with base M induces a riemannian
metric on TM* :

$G^{\ast}\left(  x\right)  \left(  u,v\right)  =\frac{1}{2}\left(
\gamma\left(  x,u+v\right)  -\gamma\left(  x,u\right)  -\gamma\left(
x,v\right)  \right)  $
\end{theorem}

\begin{proof}
$\sigma_{D^{\ast}\circ D}\left(  x,u\right)  =\sigma_{D^{\ast}}\left(
x,u\right)  \sigma_{D}\left(  x,u\right)  =\sigma_{D}\left(  x,u\right)
^{\ast}\sigma_{D}\left(  x,u\right)  =\gamma\left(  x,u\right)  Id$

As D is elliptic, $D^{\ast}\circ D$ is elliptic and :$\gamma\left(
x,u\right)  \neq0$ whenever $u\neq0$

By polarization $G^{\ast}\left(  x\right)  \left(  u,v\right)  =\frac{1}%
{2}\left(  \gamma\left(  x,u+v\right)  -\gamma\left(  x,u\right)
-\gamma\left(  x,v\right)  \right)  $ defines a bilinear symmetric form
$G^{\ast}$ on $\otimes^{2}TM^{\ast}$ which is definite positive and induces a
riemannian metric on $TM^{\ast}$ and so a riemannian metric G on TM.
\end{proof}

Notice that G is not directly related to g on V (which can have a dimension
different from M). E can be a complex vector bundle.

With $D(J^{1}Z\left(  x\right)  )=B\left(  x\right)  Z\left(  x\right)
+\sum_{\alpha=1}^{m}A\left(  x\right)  ^{\alpha}Z_{\alpha}\left(  x\right)  $

$G^{\ast}\left(  x\right)  \left(  u,v\right)  I_{m\times m}=\frac{1}{2}%
\sum_{\alpha,\beta}\left[  A\left(  x\right)  ^{\ast}\right]  ^{\alpha}\left[
A\left(  x\right)  \right]  ^{\beta}\left(  u_{\alpha}v_{\beta}+u_{\beta
}v_{\alpha}\right)  $

$=\sum_{\alpha,\beta}\left[  A\left(  x\right)  ^{\ast}\right]  ^{\alpha
}\left[  A\left(  x\right)  \right]  ^{\beta}u_{\alpha}v_{\beta}=\left[
\sigma_{D}\left(  x,u\right)  \right]  ^{\ast}\left[  \sigma_{D}\left(
x,v\right)  \right]  $

\begin{theorem}
If the Dirac operator D on the vector bundle $E\left(  M,V,\pi\right)  $
endowed with a scalar product g is self adjoint there is an algebra morphism
$\Upsilon:Cl\left(  TM^{\ast},G^{\ast}\right)  \rightarrow%
\mathcal{L}%
\left(  E;E\right)  ::\Upsilon\left(  x,u\cdot v\right)  =\sigma_{D}\left(
x,u\right)  \circ\sigma_{D}\left(  x,v\right)  $
\end{theorem}

\begin{proof}
i) Associated to the vector space (TM*(x),G*(x)) there is a Clifford algebra Cl(TM*(x),G*(x))

ii) G*(x) is a Riemmanian metric, so all the Clifford algebras
Cl(TM*(x),G*(x))\ are isomorphic and we have a Clifford algebra Cl(TM*,G*)
over M, which is isomorphic to Cl(TM,G)

iii) $%
\mathcal{L}%
\left(  E\left(  x\right)  ;E\left(  x\right)  \right)  $ is a complex algebra
with composition law

iv) the map :

$L:T_{x}M^{\ast}\rightarrow%
\mathcal{L}%
\left(  E\left(  x\right)  ;E\left(  x\right)  \right)  ::L\left(  u\right)
=\sigma_{D}\left(  x,u\right)  $ is such that :

$L\left(  u\right)  \circ L\left(  v\right)  +L\left(  v\right)  \circ
L\left(  u\right)  =\sigma_{D}\left(  x,u\right)  \circ\sigma_{D}\left(
x,v\right)  +\sigma_{D}\left(  x,v\right)  \circ\sigma_{D}\left(  x,u\right)
=2G^{\ast}\left(  x\right)  \left(  u,v\right)  I_{m\times m}$

so, following the universal property of Clifford algebra, there exists a
unique algebra morphism :

$\Upsilon_{x}:Cl\left(  T_{x}M^{\ast},G^{\ast}\left(  x\right)  \right)
\rightarrow%
\mathcal{L}%
\left(  E\left(  x\right)  ;E\left(  x\right)  \right)  $ such that
$L=\Upsilon_{x}\circ\imath$ where \i\ is the canonical map : $\imath
:T_{x}M^{\ast}\rightarrow Cl\left(  T_{x}M^{\ast},G^{\ast}\left(  x\right)
\right)  $

and $Cl\left(  T_{x}M^{\ast},G^{\ast}\left(  x\right)  \right)  $ is the
Clifford algebra over $T_{x}M^{\ast}$ endowed with the bilinear symmetric form
$G^{\ast}\left(  x\right)  .$

v) The Clifford product of vectors translates as : $\Upsilon\left(  x,u\cdot
v\right)  =\sigma_{D}\left(  x,u\right)  \circ\sigma_{D}\left(  x,v\right)  $
\end{proof}

\paragraph{Dirac operators on a spin bundle\newline}

Conversely, it we have a connection on a spin bundle E we can build a
differential operator on $\mathfrak{X}\left(  E\right)  $\ which is a Dirac
like operator. This procedure is important in physics so it is useful to
detail all the steps.

A reminder of a theorem (see Fiber bundle - connections) : for any
representation (V,r) of the Clifford algebra Cl($%
\mathbb{R}
,r,s)$ and principal bundle $Sp(M,Spin\left(
\mathbb{R}
,r,s\right)  ,\pi_{S})$ the associated bundle E=Sp[V,r] is a spin bundle. Any
principal connection $\Omega$ on Sp with potential \`{A}\ induces a linear
connection with form $\Gamma=r\left(  \upsilon\left(  \grave{A}\right)
\right)  $\ on E and covariant derivative $\nabla$ . Moreover, the
representation $[%
\mathbb{R}
^{m},\mathbf{Ad}$] of $Spin\left(
\mathbb{R}
,r,s\right)  $,$\pi_{S})$ leads to the associated vector bundle $F=Sp[%
\mathbb{R}
^{m},\mathbf{Ad}]$ and $\Omega$\ induces a linear connection on F with
covariant derivative $\widehat{\nabla}$\ . There is the relation :

$\forall X\in\mathfrak{X}\left(  F\right)  ,U\in\mathfrak{X}\left(  E\right)
:\nabla\left(  r\left(  X\right)  U\right)  =r\left(  \widehat{\nabla
}X\right)  U+r\left(  X\right)  \nabla U$

The ingredients are :

The Lie algebra $o(%
\mathbb{R}
,r,s)$\ with a basis $\left(  \overrightarrow{\kappa_{\lambda}}\right)
_{\lambda=1}^{q}$ and r+s = m

$\left(
\mathbb{R}
^{m},\eta\right)  $ endowed with the symmetric bilinear form $\eta$ of
signature (r,s) on $%
\mathbb{R}
^{m}$ and its orthonormal basis $\left(  \varepsilon_{\alpha}\right)
_{\alpha=1}^{m}$

$\upsilon$ is the isomorphism : $o(%
\mathbb{R}
,r,s)\rightarrow T_{1}SPin(%
\mathbb{R}
,r,s)::\upsilon\left(  \overrightarrow{\kappa}\right)  =\sum_{\alpha\beta}$
$\left[  \upsilon\right]  _{\beta}^{\alpha}\varepsilon_{\alpha}\cdot
\varepsilon_{\beta}$ with $\left[  \upsilon\right]  =\frac{1}{4}\left[
J\right]  \left[  \eta\right]  $ where $\left[  J\right]  $ is the mxm matrix
of $\overrightarrow{\kappa}$ in the standard representation of $o(%
\mathbb{R}
,r,s)$

The representation (V,r) of $Cl\left(
\mathbb{R}
,r,s\right)  $, with a basis $\left(  e_{i}\right)  _{i=1}^{n}$ of V, is
defined by the nxn matrices $\gamma_{\alpha}=r\left(  \varepsilon_{\alpha
}\right)  $ and : $\gamma_{i}\gamma_{j}+\gamma_{j}\gamma_{i}=2\eta_{ij}%
I,\eta_{ij}=\pm1$.

The linear connection on E is :

$\Gamma\left(  x\right)  =r\left(  \upsilon\left(  \grave{A}_{a}\right)
\right)  =\sum_{\alpha\lambda ij}\grave{A}_{\alpha}^{\lambda}\left[
\theta_{\lambda}\right]  _{j}^{i}d\xi^{\alpha}\otimes e_{i}\left(  x\right)
\otimes e^{j}\left(  x\right)  $

with $\left[  \theta_{\lambda}\right]  =\frac{1}{4}\sum_{kl}\left(  \left[
J_{\lambda}\right]  \left[  \eta\right]  \right)  _{l}^{k}\left(  \left[
\gamma_{k}\right]  \left[  \gamma_{l}\right]  \right)  $

and its covariant derivative :

$U\in\mathfrak{X}\left(  E\right)  :\nabla U=\sum_{i\alpha}\left(
\partial_{\alpha}U^{i}+\sum_{\lambda j}\grave{A}_{\alpha}^{\lambda}\left[
\theta_{\lambda}\right]  _{j}^{i}U^{j}\right)  d\xi^{\alpha}\otimes
e_{i}\left(  x\right)  $

The connection on Sp induces a linear connection on F :

$\widehat{\Gamma}\left(  x\right)  =\left(  \mathbf{Ad}\right)  ^{\prime
}|_{s=1}\left(  \upsilon\left(  \grave{A}\left(  x\right)  \right)  \right)
=\sum_{\lambda}\grave{A}_{\alpha}^{\lambda}\left(  x\right)  \left[
J_{\lambda}\right]  _{j}^{i}\varepsilon_{i}\left(  x\right)  \otimes
\varepsilon^{j}\left(  x\right)  $

and its covariant derivative :

$X\in\mathfrak{X}\left(  F\right)  :\widehat{\nabla}X=\sum_{i\alpha}\left(
\partial_{\alpha}X^{i}+\sum_{\lambda j}\grave{A}_{\alpha}^{\lambda}\left[
J_{\lambda}\right]  _{j}^{i}X^{j}\right)  d\xi^{\alpha}\otimes\varepsilon
_{i}\left(  x\right)  $

\bigskip

\begin{definition}
The Dirac operator is the first order differential operator : $\widehat
{D}:J^{1}E\rightarrow E::\widehat{D}Z=\sum_{\alpha}\left[  \gamma^{\alpha
}\right]  \left(  Z_{\alpha}+\left[  \Gamma_{\alpha}\right]  Z\right)  $
\end{definition}

\begin{proof}
i) The Clifford algebra of the dual $Cl\left(
\mathbb{R}
^{m\ast},\eta\right)  $ is isomorphic to $Cl(%
\mathbb{R}
,r,s).$ (V*,r*) is a representation of Cl$\left(
\mathbb{R}
^{m\ast},\eta\right)  .$ It is convenient to take as generators : $r^{\ast
}\left(  \varepsilon^{i}\right)  =\gamma^{i}=\eta_{ii}\gamma_{i}%
\Leftrightarrow\left[  \gamma^{i}\right]  =-\left[  \gamma_{i}\right]  ^{-1}$

ii) Because \textbf{Ad }preserves $\eta$ the vector bundle F is endowed with a
scalar product g for which the holonomic basis $\varepsilon_{i}\left(
x\right)  $ is orthonormal : $g\left(  x\right)  \left(  \varepsilon_{a\alpha
}\left(  x\right)  ,\varepsilon_{a\beta}\left(  x\right)  \right)
=\eta\left(  \varepsilon_{\alpha},\varepsilon_{\beta}\right)  =\eta
_{\alpha\beta}$

iii) F is m dimensional, and the holonomic basis of F can be expressed in
components of the holonomic basis of M : $\varepsilon_{i}\left(  x\right)
=\left[  L\left(  x\right)  \right]  _{i}^{\beta}\partial_{\beta}\xi.$

The scalar product g on F is expressed as :

$g\left(  x\right)  _{\alpha\beta}=g\left(  x\right)  \left(  \partial
_{\alpha}\xi,\partial_{\beta}\xi\right)  =\sum\left[  L\left(  x\right)
^{-1}\right]  _{\alpha}^{i}\left[  L\left(  x\right)  ^{-1}\right]  _{\beta
}^{j}g\left(  x\right)  \left(  \varepsilon_{i}\left(  x\right)
,\varepsilon_{j}\left(  x\right)  \right)  =\sum_{ij}\left[  L\left(
x\right)  ^{-1}\right]  _{\alpha}^{i}\left[  L\left(  x\right)  ^{-1}\right]
_{\beta}^{j}\eta_{ij}$

and the associated scalar product on TM* is

$g^{\ast}\left(  x\right)  ^{\alpha\beta}=g^{\ast}\left(  x\right)  \left(
d\xi^{\alpha},d\xi^{\beta}\right)  =\sum_{ij}\left[  L\left(  x\right)
\right]  _{i}^{\alpha}\left[  L\left(  x\right)  \right]  _{j}^{\beta}%
\eta^{ij}$

iv) So we can define the action :

$R^{\ast}\left(  x\right)  :T_{x}M^{\ast}\times E\left(  x\right)  \rightarrow
E\left(  x\right)  ::$

$R^{\ast}\left(  d\xi^{\alpha}\right)  \left(  e_{i}\left(  x\right)  \right)
=r^{\ast}\left(  \sum_{i}\left[  L\left(  x\right)  \right]  _{k}^{\alpha
}\varepsilon^{k}\right)  \left(  e_{i}\right)  =\sum_{kj}\left[  L\left(
x\right)  \right]  _{k}^{\alpha}\left[  \gamma^{k}\right]  _{i}^{j}%
e_{j}\left(  x\right)  $

or with $\left[  \gamma^{\alpha}\right]  =\sum_{k}\left[  L\left(  x\right)
\right]  _{k}^{\alpha}\left[  \gamma^{k}\right]  :R^{\ast}\left(  d\xi
^{\alpha}\right)  \left(  e_{i}\left(  x\right)  \right)  =\sum_{j}\left[
\gamma^{\alpha}\right]  _{i}^{j}e_{j}\left(  x\right)  $

v) The Dirac operator is defined as :

$DU=\sum_{i\alpha}\left(  \nabla_{\alpha}U^{i}\right)  R^{\ast}\left(
d\xi^{\alpha}\right)  \left(  e_{i}\left(  x\right)  \right)  =\sum_{ij\alpha
}\left[  \gamma^{\alpha}\right]  _{i}^{j}\left[  \nabla_{\alpha}U^{i}\right]
e_{j}\left(  x\right)  $

Written as a usual differential operator :

$\widehat{D}:J^{1}E\rightarrow E::\widehat{D}Z=\sum_{ij\alpha}\left[
\gamma^{\alpha}\right]  _{j}^{i}\left(  Z_{\alpha}^{j}+\Gamma_{\alpha k}%
^{j}Z^{k}\right)  e_{i}\left(  x\right)  =\sum_{\alpha}\left[  \gamma^{\alpha
}\right]  \left(  Z_{\alpha}+\left[  \Gamma_{\alpha}\right]  Z\right)  $
\end{proof}

\begin{theorem}
D is a first order weakly elliptic differential operator and the principal
symbol of D%
${{}^2}$
is scalar
\end{theorem}

\begin{proof}
i) The symbol of D is :

$P\left(  x,u\right)  =\sum_{\alpha,i,j}\left(  \left[  \gamma^{\alpha
}\right]  \left[  \Gamma_{\alpha}\right]  +\left[  \gamma^{\alpha}\right]
u_{\alpha}\right)  _{i}^{j}e_{j}\left(  x\right)  \otimes e^{i}\left(
x\right)  $

and its principal symbol :

$\sigma_{D}\left(  x,u\right)  =\sum_{\alpha}u_{\alpha}\left[  \gamma^{\alpha
}\right]  =\sum_{\alpha}u_{\alpha}R^{\ast}\left(  d\xi^{\alpha}\right)
=R^{\ast}\left(  u\right)  $.

As $r$ is an algebra morphism $R^{\ast}\left(  u\right)  =0\Rightarrow u=0.$
Thus D is weakly elliptic, and DD*,D*D are weakly elliptic.

ii) As we have in the Clifford algebra :

$u,v\in T_{x}M^{\ast}:u\cdot v+v\cdot u=2g^{\ast}\left(  x\right)  \left(
u,v\right)  $

and $R^{\ast}$\ is an algebra morphism :

$R^{\ast}\left(  u\cdot v+v\cdot u\right)  =R^{\ast}\left(  u\right)  \circ
R^{\ast}\left(  v\right)  +R^{\ast}\left(  v\right)  \circ R^{\ast}\left(
u\right)  =2g^{\ast}\left(  u,v\right)  Id$

$\sigma_{D}\left(  x,u\right)  \circ\sigma_{D}\left(  x,v\right)  +\sigma
_{D}\left(  x,v\right)  \circ\sigma_{D}\left(  x,u\right)  =2g^{\ast}\left(
u,v\right)  Id$

$\sigma_{D\circ D}\left(  x,u\right)  =\sigma_{D}\left(  x,u\right)
\circ\sigma_{D}\left(  x,u\right)  =g^{\ast}\left(  u,u\right)  Id$
\end{proof}

So D%
${{}^2}$
is a scalar operator : it is sometimes said that D is the "square root" of the
operator with symbol $g^{\ast}\left(  u,v\right)  Id$

\begin{theorem}
The operator D is self adjoint with respect to the scalar product G on E iff :

$\sum_{\alpha}\left(  \left[  \gamma^{\alpha}\right]  \left[  \Gamma_{\alpha
}\right]  +\left[  \Gamma_{\alpha}\right]  \left[  \gamma^{\alpha}\right]
\right)  =\sum_{\alpha}\left[  \gamma^{\alpha}\right]  \left[  G^{-1}\right]
\left[  \partial_{\alpha}G\right]  $

If the scalar product G is induced by a scalar product on V then the condition
reads :

$\sum_{\alpha}\left(  \left[  \gamma^{\alpha}\right]  \left[  \Gamma_{\alpha
}\right]  +\left[  \Gamma_{\alpha}\right]  \left[  \gamma^{\alpha}\right]
\right)  =0$
\end{theorem}

\begin{proof}
i) The volume form on M is $\varpi_{0}\sqrt{\left\vert \det g\right\vert }%
d\xi^{1}\wedge...\wedge d\xi^{m}$. If we have a scalar product fiberwise on
E(x) with matrix $\left[  G\left(  x\right)  \right]  $ then, applying the
method presented above (see Adjoint), the adjoint of D is with $\left[
A\right]  =\sum_{\alpha}\left[  \gamma^{\alpha}\right]  \left[  \Gamma
_{\alpha}\right]  ,\left[  B^{\alpha}\right]  =\left[  \gamma^{\alpha}\right]
:$

$D^{\ast}\left(  X\right)  =\sum(\left(  \left[  G^{-1}\right]  \sum_{\alpha
}\left[  \Gamma_{\alpha}\right]  ^{\ast}\left[  \gamma^{\alpha}\right]
^{\ast}\left[  G\right]  +\left[  G^{-1}\right]  \left[  \gamma^{\alpha
}\right]  ^{\ast}\left[  \partial_{\alpha}G\right]  \right)  X$

$+\left(  \left[  G^{-1}\right]  \left[  \gamma^{\alpha}\right]  ^{\ast
}\left[  G\right]  \right)  \partial_{\alpha}X^{i}e_{i}\left(  x\right)  )$

The operator D is self adjoint $D=D^{\ast}$ iff both :

$\forall\alpha:\left[  \gamma^{\alpha}\right]  =\left[  G^{-1}\right]  \left[
\gamma^{\alpha}\right]  ^{\ast}\left[  G\right]  $

$\sum_{\alpha}\left[  \gamma^{\alpha}\right]  \left[  \Gamma_{\alpha}\right]
=\sum_{\alpha}\left[  G^{-1}\right]  \left[  \Gamma_{\alpha}\right]  ^{\ast
}\left[  \gamma^{\alpha}\right]  ^{\ast}\left[  G\right]  +\left[
G^{-1}\right]  \left[  \gamma^{\alpha}\right]  ^{\ast}\left[  \partial
_{\alpha}G\right]  $

ii) As $\left[  L\right]  $ is real the first condition reads :

$\forall\alpha:\sum_{j}\left[  L\left(  x\right)  \right]  _{j}^{\alpha
}\left[  \gamma^{j}\right]  ^{\ast}\left[  G\left(  x\right)  \right]
=\left[  G\left(  x\right)  \right]  \sum_{j}\left[  L\left(  x\right)
\right]  _{j}^{\alpha}\left[  \gamma^{j}\right]  $

By multiplication with $\left[  L\left(  x\right)  ^{-1}\right]  _{\alpha}%
^{i}$ and summing we get :

$\left[  \gamma^{i}\right]  ^{\ast}=\left[  G\left(  x\right)  \right]
\left[  \gamma^{i}\right]  \left[  G\left(  x\right)  \right]  ^{-1}$

Moreover :

$\partial_{\alpha}\left(  \left[  G\left(  x\right)  \right]  \left[
\gamma^{i}\right]  \left[  G\left(  x\right)  \right]  ^{-1}\right)  =0$

$=\left(  \partial_{\alpha}\left[  G\left(  x\right)  \right]  \right)
\left[  \gamma^{i}\right]  \left[  G\left(  x\right)  \right]  ^{-1}-\left[
G\left(  x\right)  \right]  \left[  \gamma^{i}\right]  \left[  G\left(
x\right)  \right]  ^{-1}\left(  \partial_{\alpha}\left[  G\left(  x\right)
\right]  \right)  \left[  G\left(  x\right)  \right]  ^{-1}$

$\left[  G\left(  x\right)  \right]  ^{-1}\left(  \partial_{\alpha}\left[
G\left(  x\right)  \right]  \right)  \left[  \gamma^{i}\right]  =\left[
\gamma^{i}\right]  \left[  G\left(  x\right)  \right]  ^{-1}\left(
\partial_{\alpha}\left[  G\left(  x\right)  \right]  \right)  $

iii) With $\left[  \gamma^{\alpha}\right]  ^{\ast}=\left[  G\right]  \left[
\gamma^{\alpha}\right]  \left[  G^{-1}\right]  $ the second condition gives :

$\sum_{\alpha}\left[  \gamma^{\alpha}\right]  \left[  \Gamma_{\alpha}\right]
=\sum_{\alpha}\left[  G^{-1}\right]  \left[  \Gamma_{\alpha}\right]  ^{\ast
}\left[  G\right]  \left[  \gamma^{\alpha}\right]  +\left[  \gamma^{\alpha
}\right]  \left[  G^{-1}\right]  \left[  \partial_{\alpha}G\right]  $

$=\left(  \sum_{\alpha}\left[  G^{-1}\right]  \left[  \Gamma_{\alpha}\right]
^{\ast}\left[  G\right]  +\left[  G^{-1}\right]  \left[  \partial_{\alpha
}G\right]  \right)  \left[  \gamma^{\alpha}\right]  $

iv) $\left[  \Gamma_{\alpha}\right]  =\sum_{\lambda}\grave{A}_{\alpha
}^{\lambda}\left[  \theta_{\lambda}\right]  $

$\left[  \theta_{\lambda}\right]  =\frac{1}{4}\sum_{kl}\left(  \left[
J_{\lambda}\right]  \left[  \eta\right]  \right)  _{l}^{k}\left[  \gamma
_{k}\right]  \left[  \gamma_{l}\right]  =\frac{1}{4}\sum_{kl}\left[
J_{\lambda}\right]  _{l}^{k}\left[  \gamma_{k}\right]  \left[  \gamma
^{l}\right]  $

$\left[  \Gamma_{\alpha}\right]  =\frac{1}{4}\sum_{kl}\sum_{\lambda}\grave
{A}_{\alpha}^{\lambda}\left[  J_{\lambda}\right]  _{l}^{k}\left[  \gamma
_{k}\right]  \left[  \gamma^{l}\right]  =\frac{1}{4}\sum_{kl}\left[
\widehat{\Gamma}_{\alpha}\right]  _{l}^{k}\left[  \gamma_{k}\right]  \left[
\gamma^{l}\right]  $

\`{A} and J are real so :

$\left[  \Gamma_{\alpha}\right]  ^{\ast}=\frac{1}{4}\sum_{kl}\left[
\widehat{\Gamma}_{\alpha}\right]  _{l}^{k}\left[  \gamma^{l}\right]  ^{\ast
}\left[  \gamma_{k}\right]  ^{\ast}=\frac{1}{4}\left[  G\left(  x\right)
\right]  \left(  \sum_{kl}\left[  \widehat{\Gamma}_{\alpha}\right]  _{l}%
^{k}\left[  \gamma^{l}\right]  \left[  \gamma_{k}\right]  \right)  \left[
G\left(  x\right)  \right]  ^{-1}$

Using the relation $\left[  \widehat{\Gamma}_{\alpha}\right]  ^{t}\left[
\eta\right]  +\left[  \eta\right]  \left[  \widehat{\Gamma}_{\alpha}\right]
=0$ (see Metric connections on fiber bundles):

$\sum_{kl}\left[  \widehat{\Gamma}_{\alpha}\right]  _{l}^{k}\left[  \gamma
^{l}\right]  \left[  \gamma_{k}\right]  =-\sum_{kl}\left[  \widehat{\Gamma
}_{\alpha}\right]  _{k}^{l}\eta_{kk}\eta_{ll}\left[  \gamma^{l}\right]
\left[  \gamma_{k}\right]  =-\sum_{kl}\left[  \widehat{\Gamma}_{\alpha
}\right]  _{k}^{l}\left[  \gamma_{l}\right]  \left[  \gamma^{k}\right]
=-\sum_{kl}\left[  \widehat{\Gamma}_{\alpha}\right]  _{l}^{k}\left[
\gamma_{k}\right]  \left[  \gamma^{l}\right]  =-\left[  \Gamma_{\alpha
}\right]  $

$\left[  \Gamma_{\alpha}\right]  ^{\ast}=-\left[  G\left(  x\right)  \right]
\left[  \Gamma_{\alpha}\right]  \left[  G\left(  x\right)  \right]  ^{-1}$

v) The second condition reads:

$\sum_{\alpha}\left[  \gamma^{\alpha}\right]  \left[  \Gamma_{\alpha}\right]
=\left[  G^{-1}\right]  \sum_{\alpha}\left(  -\left[  G\right]  \left[
\Gamma_{\alpha}\right]  \left[  G\right]  ^{-1}\left[  G\right]  +\left[
\partial_{\alpha}G\right]  \right)  \left[  \gamma^{\alpha}\right]  $

$=\sum_{\alpha}\left(  -\left[  \Gamma_{\alpha}\right]  +\left[
G^{-1}\right]  \left[  \partial_{\alpha}G\right]  \right)  \left[
\gamma^{\alpha}\right]  $

$\sum_{\alpha}\left[  \gamma^{\alpha}\right]  \left[  \Gamma_{\alpha}\right]
+\left[  \Gamma_{\alpha}\right]  \left[  \gamma^{\alpha}\right]  =\sum
_{\alpha}\left[  G^{-1}\right]  \left[  \partial_{\alpha}G\right]  \left[
\gamma^{\alpha}\right]  =\sum_{\alpha}\left[  \gamma^{\alpha}\right]  \left[
G^{-1}\right]  \left[  \partial_{\alpha}G\right]  $

vi ) If G is induced by a scalar product on V then $\left[  \partial_{\alpha
}G\right]  =0$
\end{proof}

\bigskip

\subsection{Laplacian}

\label{Laplacian (Funct.analysis)}

The laplacian comes in mathematics with many flavors and definitions (we have
the connection laplacian, the Hodge laplacian, the Lichnerowicz laplacian, the
Bochner laplacian,...). We will follow the more general path, requiring the
minimum assumptions for its definition. So we start from differential geometry
and the exterior differential (defined on any smooth manifold) and
codifferential (defined on any manifold endowed with a metric) acting on forms
(see Differential geometry). From there we define the laplacian $\Delta$ as an
operator acting on forms over a manifold. As a special case it is an operator
acting on functions on a manifold, and furthermore as an operator on functions
in $%
\mathbb{R}
^{m}.$

In this subsection :

M is a pseudo riemannian manifold : real, smooth, m dimensional, endowed with
a non degenerate scalar product g (when a definite positive metric is required
this will be specified as usual) which induces a volume form $\varpi_{0}$. We
will consider the vector bundles of r forms complex valued $\Lambda_{r}\left(
M;%
\mathbb{C}
\right)  $ and $\Lambda\left(  M;%
\mathbb{C}
\right)  =\oplus_{r=0}^{m}\Lambda_{r}\left(  M;%
\mathbb{C}
\right)  $

The metric g can be extended fiberwise to a hermitian map $G_{r}$\ on
$\Lambda_{r}\left(  M;%
\mathbb{C}
\right)  $ and to an inner product on the space of sections of the vector
bundle : $\lambda,\mu\in\Lambda_{r}\left(  M;%
\mathbb{C}
\right)  :\left\langle \lambda,\mu\right\rangle _{r}=\int_{M}G_{r}\left(
x\right)  \left(  \lambda\left(  x\right)  ,\mu\left(  x\right)  \right)
\varpi_{0}$ which is well defined for $\lambda,\mu\in L^{2}\left(
M,\varpi_{0},\Lambda_{r}\left(  M;%
\mathbb{C}
\right)  \right)  .$

We denote $\epsilon=(-1)^{p}$ where p is the number of - in the signature of
g. So $\epsilon=1$ for a riemannian manifold.

Most of the applications are in differential equations and involve initial
value conditions, so we will also consider the case where M is a manifold with
boundary, embedded in a real, smooth, m dimensional manifold (it is useful to
see the precise definition of a manifold with boundary).

\subsubsection{Laplacian acting on forms}

\paragraph{Hodge dual\newline}

The Hodge dual $\ast\lambda_{r}$ of $\lambda\in L^{2}\left(  M,\varpi
_{0},\Lambda_{r}\left(  M;%
\mathbb{C}
\right)  \right)  $ is m-r-form, denoted $\ast\lambda,$ such that :

$\forall\mu\in L^{2}\left(  M,\varpi_{0},\Lambda_{r}\left(  M;%
\mathbb{C}
\right)  \right)  :\ast\lambda_{r}\wedge\mu=G_{r}\left(  \lambda,\mu\right)
\varpi_{0}$

with $G_{r}\left(  \lambda,\mu\right)  $ the scalar product of r forms.

This is an anti-isomorphism $\ast:L^{2}\left(  M,\varpi_{0},\Lambda_{r}\left(
M;%
\mathbb{C}
\right)  \right)  \rightarrow L^{2}\left(  M,\varpi_{0},\Lambda_{m-r}\left(
M;%
\mathbb{C}
\right)  \right)  $ and its inverse is :

$\ast^{-1}\lambda_{r}=\epsilon(-1)^{r\left(  m-r\right)  }\ast\lambda
_{r}\Leftrightarrow\ast\ast\lambda_{r}=\epsilon(-1)^{r\left(  m-r\right)
}\lambda_{r}$

\paragraph{Codifferential\newline}

1. The exterior differential d is a first order differential operator on
$\Lambda\left(  M;%
\mathbb{C}
\right)  :$

$d:\Lambda\left(  M;%
\mathbb{C}
\right)  \rightarrow\Lambda_{r+1}\left(  M;%
\mathbb{C}
\right)  $

With the jet symbolism :

$\widehat{d}:J^{1}\left(  \Lambda_{r}\left(  M;%
\mathbb{C}
\right)  \right)  \rightarrow\Lambda_{r+1}\left(  M;%
\mathbb{C}
\right)  ::$

$\widehat{d}\left(  Z_{r}\right)  =\sum_{\left\{  \alpha_{1}...\alpha
_{r}\right\}  }\sum_{\beta}Z_{\alpha_{1}...\alpha_{r}}^{\beta}d\xi^{\beta
}\wedge d\xi^{\alpha_{1}}\wedge...d\xi^{\alpha_{r}}$

The symbol of $\widehat{d}$ is :

$P\left(  x,d\right)  \left(  u\right)  \left(  \mu_{r}\right)  =\left(
\sum_{\beta}u_{\beta}d\xi^{\beta}\right)  \wedge\left(  \sum_{\left\{
\alpha_{1}...\alpha_{r}\right\}  }\mu_{r}d\xi^{\alpha_{1}}\wedge
...d\xi^{\alpha_{r}}\right)  =u\wedge\mu$

2. The codifferential is the operator :

$\delta:\mathfrak{X}_{1}\left(  \Lambda_{r+1}\left(  M;%
\mathbb{C}
\right)  \right)  \rightarrow\mathfrak{X}_{0}\left(  \Lambda_{r}\left(  M;%
\mathbb{C}
\right)  \right)  ::$

$\delta\lambda_{r}=\epsilon(-1)^{r(m-r)+r}\ast d\ast\lambda_{r}=\left(
-1\right)  ^{r}\ast d\ast^{-1}\lambda_{r}$

With the jet symbolism :

$\widehat{\delta}:J^{1}\left(  \Lambda_{r+1}\left(  M;%
\mathbb{C}
\right)  \right)  \rightarrow\Lambda_{r}\left(  M;%
\mathbb{C}
\right)  ::\widehat{\delta}Z_{r}=\epsilon(-1)^{r(m-r)+r}\ast\widehat{d}\circ
J^{1}\left(  \ast Z_{r+1}\right)  $

The symbol of $\widehat{\delta}:P\left(  x,\delta\right)  \left(  u\right)
\mu_{r+1}=-i_{u^{\ast}}\mu_{r+1}$ with $u^{\ast}=\sum g^{\alpha\beta}%
\overline{u}_{\beta}\partial x_{\alpha}$

\bigskip

\begin{proof}
As $\left\langle d\lambda,\mu\right\rangle =\left\langle \lambda,\delta
\mu\right\rangle $ (see below) we have :

$\left\langle P\left(  x,d\right)  \left(  u\right)  \lambda_{r},\mu
_{r+1}\right\rangle _{r+1}=\left\langle \lambda_{r},P\left(  x,\delta\right)
\left(  u\right)  \mu_{r+1}\right\rangle _{r}=\left\langle u\wedge\lambda
_{r},\mu_{r+1}\right\rangle _{r+1}$

$\sum_{\left\{  \alpha_{1}...\alpha_{r+1}\right\}  }\left(  -1\right)
^{k-1}\overline{u}_{\alpha_{k}}\overline{\lambda}_{\alpha_{1}...\widehat
{\alpha_{k}}...\alpha_{r+1}}\mu^{\alpha_{1}...\alpha_{r+1}}$

$=\sum_{\left\{  \alpha_{1}...\alpha_{r}\right\}  }\overline{\lambda}%
_{\alpha_{1}.....\alpha_{r}}\left[  P\left(  x,\delta\right)  \left(
u\right)  \mu_{r+1}\right]  ^{\alpha_{1}..\alpha_{r}}$

$\left(  -1\right)  ^{k-1}\overline{u}_{\alpha_{k}}\mu^{\alpha_{1}%
...\alpha_{r+1}}=\left[  P\left(  x,\delta\right)  \left(  u\right)  \mu
_{r+1}\right]  ^{\alpha_{1}..\alpha_{r}}$

$\left[  P\left(  x,\delta\right)  \left(  u\right)  \mu_{r+1}\right]
_{\alpha_{1}...\alpha_{r}}=\sum_{k}\left(  -1\right)  ^{k-1}\sum_{\alpha_{k}%
}\overline{u}^{\alpha_{k}}\mu_{\alpha_{1}...\alpha_{r+1}}$
\end{proof}

\bigskip

As $\sigma_{\widehat{\delta}}\left(  x,u\right)  \lambda_{r}=\epsilon
(-1)^{r(m-r)+r}\ast\sigma_{\widehat{d}}\left(  x,u\right)  \ast\lambda
_{r}=\epsilon(-1)^{r(m-r)+r}\ast\left(  u\wedge\left(  \ast\lambda_{r}\right)
\right)  $ we have :

$\ast\left(  u\wedge\left(  \ast\lambda_{r}\right)  \right)  =-\epsilon
(-1)^{r(m-r)+r}i_{u^{\ast}}\lambda_{r}$

3. Properties :

$d^{2}=\delta^{2}=0$

For $f\in C\left(  M;%
\mathbb{R}
\right)  :\delta f=0$

For $\mu_{r}\in\Lambda_{r}TM^{\ast}:$

$\ast\delta\mu_{r}=\left(  -1\right)  ^{m-r-1}d\ast u_{r}$

$\delta\ast\mu_{r}=\left(  -1\right)  ^{r+1}\ast d\mu_{r}$

For r=1 : $\delta\left(  \sum_{i}\lambda_{\alpha}dx^{\alpha}\right)
=(-1)^{m}\frac{1}{\sqrt{\left\vert \det g\right\vert }}\sum_{\alpha,\beta
=1}^{m}\partial_{\alpha}\left(  g^{\alpha\beta}\lambda_{\beta}\sqrt{\left\vert
\det g\right\vert }\right)  $

4. The codifferential is the adjoint of the exterior differential :

$d,\delta$ are defined on $\Lambda\left(  M;%
\mathbb{C}
\right)  ,$ with value in $\Lambda\left(  M;%
\mathbb{C}
\right)  .$ The scalar product is well defined if $dX,\delta X\in L^{2}\left(
M,\varpi_{0},\Lambda_{r}\left(  M;%
\mathbb{C}
\right)  \right)  $

So we can consider the scalar products on $W^{1,2}\left(  \Lambda_{r}\left(
M;%
\mathbb{C}
\right)  \right)  :$\ 

$\lambda\in W^{1,2}\left(  \Lambda_{r}\left(  M;%
\mathbb{C}
\right)  \right)  ,\mu\in W^{1,2}\left(  \Lambda_{r+1}\left(  M;%
\mathbb{C}
\right)  \right)  :$

$\left\langle \delta\mu,\lambda\right\rangle =\int_{M}G_{r}\left(  \delta
\mu,\lambda\right)  \varpi_{0},\left\langle \mu,d\lambda\right\rangle
=\int_{M}G_{r+1}\left(  \mu,d\lambda\right)  \varpi_{0}$

We have the identity for any manifold M :%

\begin{equation}
\left\langle \delta\mu,\lambda\right\rangle -\left\langle \mu,d\lambda
\right\rangle =(-1)^{r-m}\int_{M}d\left(  \ast\mu\wedge\lambda\right)
\end{equation}

\begin{theorem}
If M is a manifold with boundary :%

\begin{equation}
\left\langle \delta\mu,\lambda\right\rangle -\left\langle \mu,d\lambda
\right\rangle =(-1)^{r-m}\int_{\partial M}\ast\mu\wedge\lambda
\end{equation}

\end{theorem}

\begin{proof}
$d\left(  \ast\mu\wedge\lambda\right)  =\left(  d\ast\mu\right)  \wedge
\lambda+\left(  -1\right)  ^{m-r-1}\ast\mu\wedge d\lambda$

$=(-1)^{r-m}\ast\delta\mu\wedge\lambda+\left(  -1\right)  ^{m-r-1}\ast
\mu\wedge d\lambda$

with $\ast\delta\mu=(-1)^{m-r}d\ast\mu$

$d\left(  \ast\mu\wedge\lambda\right)  =(-1)^{r-m}\left(  \ast\delta\mu
\wedge\lambda-\ast\mu\wedge d\lambda\right)  $

$=(-1)^{r-m}\left(  G_{r}\left(  \delta\mu,\lambda\right)  -G_{r+1}\left(
\mu,d\lambda\right)  \right)  \varpi_{0}$

$d\left(  \ast\mu\wedge\lambda\right)  \in\Lambda_{m}\left(  M;%
\mathbb{C}
\right)  $

$(-1)^{r-m}\int_{M}\left(  G_{r}\left(  \delta\mu,\lambda\right)
-G_{r+1}\left(  \mu,d\lambda\right)  \right)  \varpi_{0}=\left\langle
\delta\mu,\lambda\right\rangle -\left\langle \mu,d\lambda\right\rangle $

$=\int_{M}d\left(  \ast\mu\wedge\lambda\right)  =\int_{\partial M}\ast
\mu\wedge\lambda$
\end{proof}

\begin{theorem}
The codifferential is the adjoint of the exterior derivative with respect to
the interior product on $W^{1,2}\left(  \Lambda\left(  M;%
\mathbb{C}
\right)  \right)  $
\end{theorem}

in the following meaning :%

\begin{equation}
\left\langle d\lambda,\mu\right\rangle =\left\langle \lambda,\delta
\mu\right\rangle
\end{equation}

Notice that this identity involves $\widehat{d}\circ J^{1}=d,\widehat{\delta
}\circ J^{1}=\delta$ so we have $d^{\ast}=\delta$

\begin{proof}
Starting from : $\left\langle \delta\mu,\lambda\right\rangle -\left\langle
\mu,d\lambda\right\rangle =\int_{M}\left(  G_{r}\left(  \delta\mu
,\lambda\right)  -G_{r+1}\left(  \mu,d\lambda\right)  \right)  \varpi
_{0}=(-1)^{r-m}\int_{M}d\left(  \ast\mu\wedge\lambda\right)  $

With the general assumptions about the manifold M there is a cover of M such
that its closure gives an increasing, countable, sequence of compacts. The
space $W_{c}^{1,2}\left(  \Lambda_{r}\left(  M;%
\mathbb{C}
\right)  \right)  $ of r forms on M, continuous with compact support is dense
in $W^{1,2}\left(  \Lambda_{r}\left(  M;%
\mathbb{C}
\right)  \right)  ,$ which is a Banach space.\ So any form can be estimated by
a convergent sequence of compactly supported forms, and for them we can find a
manifold with boundary N in M such that :

$\left\langle \delta\mu,\lambda\right\rangle -\left\langle \mu,d\lambda
\right\rangle =\int_{N}\left(  G_{r}\left(  \delta\mu,\lambda\right)
-G_{r+1}\left(  \mu,d\lambda\right)  \right)  \varpi_{0}$

$=(-1)^{r-m}\int_{\partial N}\ast\mu\wedge\lambda=0$

So the relation above extends to $W^{1,2}\left(  \Lambda_{r}\left(  M;%
\mathbb{C}
\right)  \right)  .$
\end{proof}

This theorem is true only if M is without boundary but, with our precise
definition, a manifold with boundary is not a manifold.

\begin{theorem}
The operator $d+\delta$ is self adjoint on $W^{1,2}\left(  \Lambda_{r}\left(
M;%
\mathbb{C}
\right)  \right)  $
\end{theorem}

\paragraph{Laplacian\newline}

On the pseudo Riemannian manifold M the Laplace-de Rahm (also called Hodge
laplacian) operator is :%

\begin{equation}
\Delta:\mathfrak{X}_{2}\left(  \Lambda_{r}\left(  M;%
\mathbb{C}
\right)  \right)  \rightarrow\mathfrak{X}_{0}\left(  \Lambda_{r}\left(  M;%
\mathbb{C}
\right)  \right)  ::\Delta=-\left(  \delta d+d\delta\right)  =-\left(
d+\delta\right)  ^{2}%
\end{equation}

Remark : one finds also the definition $\Delta=\left(  \delta d+d\delta
\right)  .$

We have :

$\left\langle \Delta\lambda,\mu\right\rangle =-\left\langle \left(  \delta
d+d\delta\right)  \lambda,\mu\right\rangle =-\left\langle \delta d\lambda
,\mu\right\rangle -\left\langle \left(  d\delta\right)  \lambda,\mu
\right\rangle =-\left\langle d\lambda,d\mu\right\rangle -\left\langle
\delta\lambda,\delta\mu\right\rangle $

\begin{theorem}
(Taylor 1 p.163) The principal symbol of $\Delta$ is scalar :%

\begin{equation}
\sigma_{\Delta}\left(  x,u\right)  =-\left(  \sum_{\alpha\beta}g^{\alpha\beta
}u_{\alpha}u_{\beta}\right)  Id
\end{equation}

\end{theorem}

It follows that the laplacian (or $-\Delta)$ is an elliptic operator iff the
metric g is definite positive.

\begin{theorem}
The laplacian $\Delta$ is a self adjoint operator with respect to $G_{r}$ on
$W^{2,2}\left(  \Lambda_{r}\left(  M;%
\mathbb{C}
\right)  \right)  :$
\end{theorem}

\begin{equation}
\forall\lambda,\mu\in\Lambda_{r}\left(  M;%
\mathbb{C}
\right)  :\left\langle \Delta\lambda,\mu\right\rangle =\left\langle
\lambda,\Delta\mu\right\rangle
\end{equation}

\begin{proof}
$\left\langle \Delta\lambda,\mu\right\rangle =-\left\langle \left(  \delta
d+d\delta\right)  \lambda,\mu\right\rangle =-\left\langle \delta d\lambda
,\mu\right\rangle -\left\langle \left(  d\delta\right)  \lambda,\mu
\right\rangle =-\left\langle d\lambda,d\mu\right\rangle -\left\langle
\delta\lambda,\delta\mu\right\rangle =-\left\langle \lambda,\delta
d\mu\right\rangle -\left\langle \lambda,d\delta\mu\right\rangle =\left\langle
\lambda,\Delta\mu\right\rangle $
\end{proof}

\begin{theorem}
If the metric g sur M is definite positive then the laplacian on
$W^{2,2}\left(  \Lambda_{r}\left(  M;%
\mathbb{C}
\right)  \right)  $ is such that :

i) its spectrum is a locally compact subset of $%
\mathbb{R}
$

ii) its eigen values are real, and constitute either a finite set or a
sequence converging to 0

iii) it is a closed operator : if $\mu_{n}\rightarrow\mu$ in $W^{1,2}\left(
\Lambda_{r}\left(  M;%
\mathbb{C}
\right)  \right)  $ then $\Delta\mu_{n}\rightarrow\Delta\mu$
\end{theorem}

\begin{proof}
If $X\in W^{2,2}\left(  \Lambda_{r}\left(  M;%
\mathbb{C}
\right)  \right)  \Leftrightarrow X\in\mathfrak{X}_{2}\left(  \Lambda
_{r}\left(  M;%
\mathbb{C}
\right)  \right)  ,J^{2}X\in L^{2}\left(  M,\varpi_{0},\Lambda_{r}\left(  M;%
\mathbb{C}
\right)  \right)  $ then $\delta dX,d\delta X\in L^{2}\left(  M,\varpi
_{0},\Lambda_{r}\left(  M;%
\mathbb{C}
\right)  \right)  $ and $\Delta X\in L^{2}\left(  M,\varpi_{0},\Lambda
_{r}\left(  M;%
\mathbb{C}
\right)  \right)  .$ So the laplacian is a map : $\Delta:$ $W^{2,2}\left(
\Lambda_{r}\left(  M;%
\mathbb{C}
\right)  \right)  \rightarrow L^{2}\left(  M,\varpi_{0},\Lambda_{r}\left(  M;%
\mathbb{C}
\right)  \right)  $

If the metric g is definite positive, then $W^{2,2}\left(  \Lambda_{r}\left(
M;%
\mathbb{C}
\right)  \right)  \subset L^{2}\left(  M,\varpi_{0},\Lambda_{r}\left(  M;%
\mathbb{C}
\right)  \right)  $\ are Hilbert spaces and as $\Delta$\ is self adjoint the
properties are a consequence of general theorems on operators on C*-algebras.
\end{proof}

\paragraph{Harmonic forms\newline}

\begin{definition}
On the pseudo Riemannian manifold M a r-form is said to be \textbf{harmonic}
if $\Delta\mu=0$
\end{definition}

Then for $\mu\in\Lambda_{r}\left(  M;%
\mathbb{C}
\right)  $: $\Delta\mu=0\Leftrightarrow d\delta\mu=\delta d\mu=0,$

\begin{theorem}
(Taylor 1 p.354) On a Riemannian compact manifold M the set of harmonic r-form
is finite dimensional, isomorphic to the space of order r cohomology
$H^{r}\left(  M\right)  $
\end{theorem}

The space $H^{r}\left(  M\right)  $ is defined in the Differential Geometry
part (see cohomology). It is independant of the metric g, so the set of
harmonic forms is the same whatever the riemannian metric.

\begin{theorem}
(Taylor 1 p.354) On a Riemannian compact manifold M

$\forall\mu\in W^{k,2}\left(  \Lambda_{r}\left(  M;%
\mathbb{C}
\right)  \right)  :\mu=d\delta G\mu+\delta dG\mu+P_{r}\mu$

where $G:W^{k,2}\left(  \Lambda_{r}\left(  M;%
\mathbb{C}
\right)  \right)  \rightarrow W^{k+2,2}\left(  \Lambda_{r}\left(  M;%
\mathbb{C}
\right)  \right)  $ and $P_{r}$ is the orthogonal projection of $L^{2}\left(
M,\Lambda_{r}\left(  M;%
\mathbb{C}
\right)  ,\varpi_{0}\right)  $ onto the space of r harmonic forms.\ The three
terms are mutually orthogonal in $L^{2}\left(  M,\Lambda_{r}\left(  M;%
\mathbb{C}
\right)  ,\varpi_{0}\right)  $
\end{theorem}

This is called the \textbf{Hodge decomposition}.\ There are many results on
this subject, mainly when M is a manifold with boundary in $%
\mathbb{R}
^{m}$ (see Axelsson).

\paragraph{Inverse of the laplacian\newline}

We have a stronger result if the metric is riemannian and M compact :

\begin{theorem}
(Taylor 1 p.353) On a smooth Riemannian compact manifold M the operator :

$\Delta:W^{1,2}\left(  \Lambda_{r}\left(  M;%
\mathbb{C}
\right)  \right)  \rightarrow W^{-1,2}\left(  \Lambda_{r}\left(  M;%
\mathbb{C}
\right)  \right)  $

is such that : $\exists C_{0},C_{1}\geq0:-\left\langle \mu,\Delta
\mu\right\rangle \geq C_{0}\left\Vert \mu\right\Vert _{W^{1,2}}^{2}%
-C_{1}\left\Vert \mu\right\Vert _{W^{-1,2}}^{2}$

The map : $-\Delta+C_{1}:H^{1}\left(  \Lambda_{r}\left(  M;%
\mathbb{C}
\right)  \right)  \rightarrow H^{-1}\left(  \Lambda_{r}\left(  M;%
\mathbb{C}
\right)  \right)  $ is bijective and its inverse is a self adjoint compact
operator on $L^{2}\left(  M,\Lambda_{r}\left(  M;%
\mathbb{C}
\right)  ,\varpi_{0}\right)  $
\end{theorem}

The space $H^{-1}\left(  \Lambda_{r}\left(  M;%
\mathbb{C}
\right)  \right)  =W^{-1,2}\left(  \Lambda_{r}\left(  M;%
\mathbb{C}
\right)  \right)  $ is defined as the dual of $\overline{\mathfrak{X}_{\infty
c}\left(  \Lambda_{r}TM^{\ast}\otimes%
\mathbb{C}
\right)  }$ in $W^{1,2}\left(  \Lambda_{r}\left(  M;%
\mathbb{C}
\right)  \right)  $

\subsubsection{Scalar Laplacian}

\paragraph{Coordinates expressions}

\begin{theorem}
On a smooth m dimensional real pseudo riemannian manifold (M,g) :%

\begin{equation}
f\in C_{2}\left(  M;%
\mathbb{C}
\right)  :\Delta f=(-1)^{m+1}div\left(  gradf\right)
\end{equation}

\end{theorem}

\begin{theorem}
$\Delta f=(-1)^{m+1}\sum_{\alpha,\beta=1}^{m}g^{\alpha\beta}\partial
_{\alpha\beta}^{2}f+\left(  \partial_{\beta}f\right)  \left(  \partial
_{\alpha}g^{\alpha\beta}-\frac{1}{2}\sum_{\lambda\mu}g_{\lambda\mu}%
\partial_{\alpha}g^{\mu\lambda}\right)  $
\end{theorem}

So that if $g^{\alpha\beta}=\eta^{\alpha\beta}=Cte:\Delta f=\sum_{\alpha
,\beta=1}^{m}\eta^{\alpha\beta}\partial_{\alpha\beta}^{2}f$ and in euclidean
space we have the usual formula : $\Delta f=(-1)^{m+1}\sum_{\alpha=1}^{m}%
\frac{\partial^{2}f}{\partial x_{\alpha}^{2}}$

The principal symbol of $\Delta$\ is $\sigma_{\Delta}\left(  x,u\right)
=(-1)^{m+1}\sum_{\alpha,\beta=1}^{m}g^{\alpha\beta}u_{\alpha}u_{\beta}$

\begin{proof}
$\delta f=0\Rightarrow\Delta f=-\delta df$

$\Delta f=-\delta\left(  \sum_{\alpha}\partial_{\alpha}fdx^{\alpha}\right)
=-(-1)^{m}\frac{1}{\sqrt{\left\vert \det g\right\vert }}\sum_{\alpha,\beta
=1}^{m}\partial_{\alpha}\left(  g^{\alpha\beta}\partial_{\beta}f\sqrt
{\left\vert \det g\right\vert }\right)  $

$\Delta f=(-1)^{m+1}\sum_{\alpha,\beta=1}^{m}g^{\alpha\beta}\partial
_{\alpha\beta}^{2}f+\left(  \partial_{\beta}f\right)  \left(  \partial
_{\alpha}g^{\alpha\beta}+g^{\alpha\beta}\frac{\partial_{\alpha}\sqrt
{\left\vert \det g\right\vert }}{\sqrt{\left\vert \det g\right\vert }}\right)
$

$\frac{\partial_{\alpha}\sqrt{\epsilon\det g}}{\sqrt{\epsilon\det g}}=\frac
{1}{2}\frac{\epsilon\partial_{\alpha}\det g}{\epsilon\det g}=\frac{1}{2}%
\frac{1}{\det g}\left(  \det g\right)  Tr\left(  \left[  \partial_{\alpha
}g\right]  \left[  g^{-1}\right]  \right)  =\frac{1}{2}\sum_{\lambda\mu}%
g^{\mu\lambda}\partial_{\alpha}g_{\lambda\mu}=-\frac{1}{2}\sum_{\lambda\mu
}g_{\lambda\mu}\partial_{\alpha}g^{\mu\lambda}$
\end{proof}

The last term can be expressed with the L\'{e}vy Civita connection :

\begin{theorem}
$\Delta f=(-1)^{m+1}\sum_{\alpha,\beta=1}^{m}g^{\alpha\beta}\left(
\partial_{\alpha\beta}^{2}f-\sum_{\gamma}\Gamma_{\alpha\beta}^{\gamma}%
\partial_{\gamma}f\right)  $
\end{theorem}

\begin{proof}
$\sum_{\gamma}\Gamma_{\gamma\alpha}^{\gamma}=\frac{1}{2}\frac{\partial
_{\alpha}\left\vert \det g\right\vert }{\left\vert \det g\right\vert }%
=\frac{\partial_{\alpha}\left(  \sqrt{\left\vert \det g\right\vert }\right)
}{\sqrt{\left\vert \det g\right\vert }}$

$\sum_{\alpha}\partial_{\alpha}g^{\alpha\beta}+g^{\alpha\beta}\sum_{\gamma
}\Gamma_{\gamma\alpha}^{\gamma}=\sum_{\alpha\gamma}-g^{\beta\alpha}%
\Gamma_{\gamma\alpha}^{\gamma}-g^{\alpha\gamma}\Gamma_{\alpha\gamma}^{\beta
}+g^{\alpha\beta}\Gamma_{\gamma\alpha}^{\gamma}=-\sum_{\alpha\gamma}%
g^{\alpha\gamma}\Gamma_{\alpha\gamma}^{\beta}$

$\Delta f=(-1)^{m+1}\sum_{\alpha,\beta=1}^{m}g^{\alpha\beta}\partial
_{\alpha\beta}^{2}f-\left(  \partial_{\beta}f\right)  \sum_{\gamma}\left(
g^{\alpha\gamma}\Gamma_{\alpha\gamma}^{\beta}\right)  $

$\Delta f=(-1)^{m+1}\sum_{\alpha,\beta=1}^{m}g^{\alpha\beta}\left(
\partial_{\alpha\beta}^{2}f-\sum_{\gamma}\Gamma_{\alpha\beta}^{\gamma}%
\partial_{\gamma}f\right)  $
\end{proof}

If we write : $p=gradf$ then $\Delta f=(-1)^{m+1}divp$

\begin{proof}
$p=gradf\Leftrightarrow p^{\alpha}=g^{\alpha\beta}\partial_{\beta
}f\Leftrightarrow\partial_{\alpha}f=g_{\alpha\beta}p^{\beta}$

$\Delta f=(-1)^{m+1}\sum_{\alpha,\beta\gamma=1}^{m}g^{\alpha\beta}\left(
p^{\gamma}\partial_{\beta}g_{\alpha\gamma}+g_{\alpha\gamma}\partial_{\beta
}p^{\gamma}-\Gamma_{\alpha\beta}^{\gamma}g_{\gamma\eta}p^{\eta}\right)  $

$=(-1)^{m+1}\sum_{\alpha,\beta\gamma=1}^{m}\left(  p^{\gamma}g^{\alpha\beta
}\left(  \sum_{\eta}g_{\gamma\eta}\Gamma_{\alpha\beta}^{\eta}+g_{\alpha\eta
}\Gamma_{\beta\gamma}^{\eta}\right)  +\partial_{\beta}p^{\beta}-\Gamma
_{\alpha\beta}^{\gamma}g^{\alpha\beta}g_{\gamma\eta}p^{\eta}\right)  $

$=(-1)^{m+1}\sum_{\alpha,\beta\gamma\eta=1}^{m}\left(  g^{\alpha\beta}\left(
p^{\gamma}g_{\gamma\eta}\Gamma_{\alpha\beta}^{\eta}-\Gamma_{\alpha\beta}%
^{\eta}g_{\gamma\eta}p^{\gamma}\right)  +p^{\gamma}\Gamma_{\beta\gamma}%
^{\beta}+\partial_{\beta}p^{\beta}\right)  $

$=(-1)^{m+1}\sum_{\alpha=1}^{m}\left(  \partial_{\alpha}p^{\alpha}+\sum
_{\beta}p^{\alpha}\Gamma_{\beta\alpha}^{\beta}\right)  =\sum_{\alpha=1}%
^{m}\nabla_{\alpha}p^{\alpha}$
\end{proof}

\bigskip

Warning ! When dealing with the scalar laplacian, meaning acting on functions
over a manifold, usually one drops the constant $(-1)^{m+1}.$ So the last
expression gives the alternate definition : $\Delta f=div\left(  grad\left(
f\right)  \right)  .$ We will follow this convention.

The riemannian Laplacian in $%
\mathbb{R}
^{m}$ with spherical coordinates has the following expression:

$\Delta=\left(  \frac{\partial^{2}}{\partial r^{2}}+\frac{m-1}{r}%
\frac{\partial}{\partial r}+\frac{1}{r^{2}}\Delta_{S^{n-1}}\right)  $

\paragraph{Wave operator\newline}

As can be seen from the principal symbol $\Delta$ is elliptic iff the metric g
on M is riemannian, which is the usual case. When the metric has a signature
(p + ,q -), as for Lorentz manifolds, we have a \textbf{d'Alambertian} denoted
$\square$. If q=1 usually there is a folliation of M in space like
hypersurfaces $S_{t}$, p dimensional manifolds endowed with a riemannian
metric, over which one considers a purely riemannian laplacian $\Delta_{x}$\ .
So $\square$\ is split in $\Delta_{x}$\ and a "time component" which can be
treated as $\frac{\partial^{2}}{\partial t^{2}}$\ , and the functions are then
$\varphi\left(  t,x\right)  \in C\left(
\mathbb{R}
;C\left(  S_{t};%
\mathbb{C}
\right)  \right)  $ .This is the "wave operator" which is seen in the PDE sections.

\paragraph{Domain of the laplacian\newline}

As many theorems about the laplacian use distributions it is necessary to
understand how we get from on side of the question (functions) to the other (distributions).

1. The scalar riemannian laplacian acts on functions. If $f\in C_{r}\left(  M;%
\mathbb{C}
\right)  $ then $\Delta f$ is still a function if $r\geq2.$ Thus $\Delta f$ is
in L%
${{}^2}$
if $f\in W^{r,2}\left(  M\right)  =H^{r}\left(  M\right)  $ with r
$>$
1.

2. The associated operator acting on distributions $\mu\in C_{\infty c}\left(
M;%
\mathbb{C}
\right)  ^{\prime}$ is $\Delta^{\prime}\mu\left(  \varphi\right)  =\mu\left(
\Delta\varphi\right)  $ .\ It has same coefficients as $\Delta$

$\Delta^{\prime}\mu=\sum_{\alpha,\beta=1}^{m}g^{\alpha\beta}\partial
_{\alpha\beta}^{2}\mu+\left(  \partial_{\alpha}g^{\alpha\beta}-\frac{1}{2}%
\sum_{\lambda\mu}g_{\lambda\mu}\partial_{\alpha}g^{\mu\lambda}\right)  \left(
\partial_{\beta}\mu\right)  $

3. Distributions acting on a space of functions can be defined through m
forms, with all the useful properties (notably the derivative of the
distribution is the derivative of the function). As noticed above the operator
D' on distributions can be seen as D acting on the unique component of the
m-form : $D^{\prime}J^{r}T\left(  \left(  \lambda_{0}d\xi^{1}\wedge...\wedge
d\xi^{m}\right)  \right)  =T\left(  \left(  D\lambda_{0}\right)  d\xi
^{1}\wedge...\wedge d\xi^{m}\right)  $ so it is usual to "forget" the
$d\xi^{1}\wedge...\wedge d\xi^{m}$ part.

If $\ \lambda_{0}\in L_{loc}^{p}\left(  O,dx,%
\mathbb{C}
\right)  ,1\leq p\leq\infty$ then $T(\lambda_{0})\in C_{\infty c}\left(  O;%
\mathbb{C}
\right)  ^{\prime}$ so we can consider $\Delta^{\prime}T\left(  \lambda
_{0}\right)  =T\left(  \lambda_{0}\right)  \Delta.$

The Sobolev spaces have been extended to distributions : $H^{-r}\left(
M\right)  $ is a vector subspace of $C_{\infty c}\left(  M;%
\mathbb{C}
\right)  ^{\prime}$ identified to the distributions : $H^{-r}\left(  M\right)
=\mu\in C_{\infty c}\left(  M;%
\mathbb{C}
\right)  ^{\prime};\mu=\left\{  \sum_{\left\Vert \alpha\right\Vert \leq
r}D_{\alpha}T\left(  f_{\alpha}\right)  ,f_{\alpha}\in L^{2}\left(
M,\varpi_{0},%
\mathbb{C}
\right)  \right\}  $. The spaces $H^{k}\left(  M\right)  $, $k\in%
\mathbb{Z}
$ are Hilbert spaces.

If $f\in H^{r}\left(  M\right)  $ then $\Delta^{\prime}T\left(  f\right)  \in
H^{r-2}\left(  M\right)  $ . It is a distribution induced by a function if r
$>$
1 and then $\Delta^{\prime}T\left(  f\right)  =T\left(  \Delta f\right)  $

4. $\Delta$ is defined in $H^{2}\left(  M\right)  \subset L^{2}\left(
M,\varpi_{0},%
\mathbb{C}
\right)  ,$ which are Hilbert spaces. It has an adjoint $\Delta^{\ast}$ on
$L^{2}\left(  M,\varpi_{0},%
\mathbb{C}
\right)  $ defined as :

$\Delta^{\ast}\in L\left(  D\left(  \Delta^{\ast}\right)  ;L^{2}\left(
M,\varpi_{0},%
\mathbb{C}
\right)  \right)  ::\forall u\in H^{2}\left(  M\right)  ,v\in D\left(
\Delta^{\ast}\right)  :\left\langle \Delta u,v\right\rangle =\left\langle
u,\Delta^{\ast}v\right\rangle $ and $H^{2}\left(  M\right)  \subset D\left(
\Delta^{\ast}\right)  $ so $\Delta$ is symmetric. We can look for extending
the domain beyond $H^{2}\left(  M\right)  $ in $L^{2}\left(  M,\varpi_{0},%
\mathbb{C}
\right)  $\ (see Hilbert spaces). If the extension is self adjoint and unique
then $\Delta$ is said to be essentially self adjoint. We have the following :

\begin{theorem}
(Gregor'yan p.6) On a riemannian connected smooth manifold M $\Delta$ has a
unique self adjoint extension in $L^{2}\left(  M,\varpi_{0},%
\mathbb{C}
\right)  $ to the domain $\left\{  f\in\overline{C_{\infty c}\left(  M;%
\mathbb{C}
\right)  }:\Delta^{\prime}T\left(  f\right)  \in T\left(  L^{2}\left(
M,\varpi_{0},%
\mathbb{C}
\right)  \right)  \right\}  $ where the closure is taken in $L^{2}\left(
M,\varpi_{0},%
\mathbb{C}
\right)  .$ If M is geodesically complete then $\Delta$ is essentially self
adjoint on $C_{\infty c}\left(  M;%
\mathbb{C}
\right)  $
\end{theorem}

\begin{theorem}
(Taylor 2 p.82-84) On a riemannian compact manifold with boundary M $\Delta$
has a self adjoint extension in $L^{2}\left(  \overset{\circ}{M},%
\mathbb{C}
,\varpi_{0}\right)  $\ to the domain $\left\{  f\in H_{c}^{1}\left(
\overset{\circ}{M}\right)  :\Delta^{\prime}T\left(  f\right)  \in T\left(
L^{2}\left(  \overset{\circ}{M},%
\mathbb{C}
,\varpi_{0}\right)  \right)  \right\}  $ . If M is smooth $\Delta$ is
essentially self adjoint on the domains

$f\in C_{\infty}\left(  M;%
\mathbb{C}
\right)  :f=0$ on $\partial M$ and on $f\in C_{\infty}\left(  M;%
\mathbb{C}
\right)  :\frac{\partial f}{\partial n}=0$ on $\partial M$ (n is the normal to
the boundary)
\end{theorem}

\begin{theorem}
(Zuily p.165) On a compact manifold with smooth boundary M in $%
\mathbb{R}
^{m}$\ :

$\left\{  f\in H_{c}^{1}\left(  \overset{\circ}{M}\right)  :\Delta^{\prime
}T\left(  f\right)  \in T\left(  L^{2}\left(  \overset{\circ}{M},%
\mathbb{C}
,\varpi_{0}\right)  \right)  \right\}  \equiv H_{c}^{1}\left(  \overset{\circ
}{M}\right)  \cap H^{2}\left(  \overset{\circ}{M}\right)  $
\end{theorem}

\paragraph{Green's identity\newline}

\begin{theorem}
On a pseudo riemannian manifold, for $f,g\in W^{2,2}\left(  M\right)  $ :%

\begin{equation}
\left\langle df,dg\right\rangle =-\left\langle f,\Delta g\right\rangle
=-\left\langle \Delta f,g\right\rangle
\end{equation}

\end{theorem}

\begin{proof}
$\left\langle df,dg\right\rangle =\left\langle f,\delta dg\right\rangle $ and
$\delta g=0\Rightarrow\Delta g=-\delta dg$

$\left\langle df,dg\right\rangle =\overline{\left\langle dg,df\right\rangle
}=-\overline{\left\langle g,\Delta f\right\rangle }=-\left\langle \Delta
f,g\right\rangle $
\end{proof}

As a special case : $\left\langle df,df\right\rangle =-\left\langle f,\Delta
f\right\rangle =\left\Vert df\right\Vert ^{2}$

As a consequence :

\begin{theorem}
On a pseudo riemannian manifold with boundary M, for $f,g\in W^{2,1}\left(
M\right)  $ :%

\begin{equation}
\left\langle \Delta f,g\right\rangle -\left\langle f,\Delta g\right\rangle
=\int_{\partial M}\left(  \frac{\partial f}{\partial n}g-f\frac{\partial
g}{\partial n}\right)  \varpi_{1}%
\end{equation}

\end{theorem}

where $\frac{\partial f}{\partial n}=f^{\prime}(p)n$ and n is a unitary normal
and $\varpi_{1}$ the volume form on $\partial M$ induced by $\varpi_{0}$

\paragraph{Spectrum of the laplacian\newline}

The spectrum of $\Delta$ is the set of complex numbers $\lambda$\ such that
\ $\Delta-\lambda I$ has no bounded inverse. So it depends on the space of
functions on which the laplacian is considered : regularity of the functions,
their domain and on other conditions which can be imposed, such as "the
Dirichlet condition". The eigen values are isolated points in the spectrum, so
if the spectrum is discrete it coincides with the eigenvalues. The eigen value
0 is a special case : the functions such that $\Delta f=0$ are harmonic and
cannot be bounded (see below).\ So if they are compactly supported they must
be null and thus cannot be eigenvectors and 0 is not an eigenvalue.

One key feature of the laplacian is that the spectrum is different if the
domain is compact or not. In particular the laplacian has eigen values iff the
domain is relatively compact. The eigen functions are an essential tool in
many PDE.

\bigskip

\begin{theorem}
(Gregor'yan p.7) In any non empty relatively compact open subset O of a
riemannian smooth manifold M the spectrum of $-\Delta$ on

$\left\{  f\in\overline{C_{\infty c}\left(  O;%
\mathbb{C}
\right)  }:\Delta^{\prime}T\left(  f\right)  \in T\left(  L^{2}\left(
O,\varpi_{0},%
\mathbb{C}
\right)  \right)  \right\}  $ is discrete and consists of an increasing
sequence $\left(  \lambda_{n}\right)  _{n=1}^{\infty}$ with $\lambda_{n}\geq0$
and $\lambda_{n}\rightarrow_{n\rightarrow\infty}\infty$ .

If $M\backslash\overline{O}$ is non empty then $\lambda_{1}>0.$
\end{theorem}

If the eigenvalues are counted with their multplicity we have the Weyl's
formula: $\lambda_{n}\sim C_{m}\left(  \frac{n}{Vol\left(  O\right)  }\right)
^{2/m}$ with : $C_{m}=\left(  2\pi\right)  ^{2}\left(  \frac{m\Gamma\left(
\frac{m}{2}\right)  }{2\pi^{m/2}}\right)  ^{m/2}$ and Vol(O)=$\int_{O}%
\varpi_{0}$

\begin{theorem}
(Taylor 1 p.304-316) On a riemannian compact manifold with boundary M :

i) The spectrum of -$\Delta$ on $\left\{  f\in H_{c}^{1}\left(  \overset
{\circ}{M}\right)  :\Delta^{\prime}T\left(  f\right)  \in T\left(
L^{2}\left(  \overset{\circ}{M},\varpi_{0},%
\mathbb{C}
\right)  \right)  \right\}  $ is discrete and consists of an increasing
sequence $\left(  \lambda_{n}\right)  _{n=1}^{\infty}$ with $\lambda_{n}\geq0$
and $\lambda_{n}\rightarrow_{n\rightarrow\infty}\infty$ .If the boundary is
smooth, then $\lambda_{1}>0$

ii) The eigenvectors $e_{n}$ of -$\Delta$\ \ belong to $C_{\infty}\left(  M;%
\mathbb{C}
\right)  $ and constitute a countable Hilbertian basis of $L^{2}\left(
\overset{\circ}{M},%
\mathbb{C}
,\varpi_{0}\right)  .$ If the boundary is smooth, then $e_{1}\in H_{c}%
^{1}\left(  \overset{\circ}{M}\right)  $, $e_{1}$\ is nowhere vanishing in the
interior of M, and if $e_{1}$=0 on $\partial M$ then the corresponding
eigenspace is unidimensional.
\end{theorem}

Notice that by definition M is closed and includes the boundary.

\begin{theorem}
(Zuily p.164) On a \textit{bounded} open subset O of $%
\mathbb{R}
^{m}$ the eigenvalues of $-\Delta$ are an increasing sequence $\left(
\lambda_{n}\right)  _{n=1}^{\infty}$ with $\lambda_{n}>0$ and $\lambda
_{n}\rightarrow_{n\rightarrow\infty}\infty$ . The eigenvectors $\left(
e_{n}\right)  _{n=1}^{\infty}$ $\in H_{c}^{1}\left(  O\right)  ,\left\Vert
e_{n}\right\Vert _{H^{1}\left(  O\right)  }=\lambda_{n}$ and can be chosen to
constitute an orthonomal basis of $L^{2}\left(  O,dx,%
\mathbb{C}
\right)  $
\end{theorem}

When O is a sphere centered in 0 then the $e_{n}$ are the spherical harmonics
: polynomial functions which are harmonic on the sphere (see Representation theory).

If the domain is an open non bounded of $%
\mathbb{R}
^{m}$ the spectrum of $-\Delta$\ is $\left[  0,\infty\right]  $ and the
laplacian\ has no eigen value$.$

\paragraph{Fundamental solution for the laplacian\newline}

On a riemannian manifold a fundamental solution of the operator -$\Delta$ is
given by a Green's function through the heat kernel function p(t,x,y) which is
itself given through the eigenvectors of -$\Delta$\ (see Heat kernel)

\begin{theorem}
(Gregor'yan p.45) On a riemannian manifold (M,g) if the Green's function
$G\left(  x,y\right)  =\int_{0}^{\infty}p\left(  t,x,y\right)  dt$ is such
that $\forall x\neq y\in M:G\left(  x,y\right)  <\infty$ then a fundamental
solution of -$\Delta$ is given by T(G) : $-\Delta_{x}TG\left(  x,y\right)
=\delta_{y}$

It has the following properties :

i) $\forall x,y\in M:G\left(  x,y\right)  \geq0,$

ii) $\forall x:G(x,.)\in L_{loc}^{1}\left(  M,\varpi_{0},%
\mathbb{R}
\right)  $ is harmonic and smooth for $x\neq y$

iii) it is the minimal non negative fundamental solution of -$\Delta$ on M.

The condition $\forall x\neq y\in M:G\left(  x,y\right)  <\infty$ is met (one
says that G is finite) if :

i) M is a non empty relatively compact open subset of a riemannian manifold N
such that $N\backslash\overline{M}$ is non empty

ii) or if the smallest eigen value\ $\lambda_{\min}$ of -$\Delta$ is $>0$
\end{theorem}

On $%
\mathbb{R}
^{m}$ the fundamental solution of $\Delta U=\delta_{y}$ is : $U\left(
x\right)  =T_{y}\left(  G\left(  x,y\right)  \right)  $ where G(x,y) is the
function :

$m\geq3:G\left(  x,y\right)  =\frac{1}{\left(  2-m\right)  A\left(
S_{m-1}\right)  }\left\Vert x-y\right\Vert ^{2-m}$ where $A\left(
S_{m-1}\right)  =\frac{2\pi^{m/2}}{\Gamma\left(  \frac{m}{2}\right)  }$ is the
Lebesgue surface of the unit sphere in $%
\mathbb{R}
^{m}.$

$m=3:G\left(  x,y\right)  =-\frac{1}{4\pi}\frac{1}{\left\Vert x-y\right\Vert }
$

$m=2:G\left(  x,y\right)  =-\frac{1}{2\pi}\ln\left\Vert x-y\right\Vert $

\paragraph{Inverse of the laplacian\newline}

The inverse of the laplacian exists if the domain is bounded.\ It is given by
the Green's function.

\begin{theorem}
(Taylor 1 p.304-316) On a riemannian compact manifold with boundary M :

$\Delta:H_{c}^{1}\left(  \overset{\circ}{M}\right)  \rightarrow H^{-1}\left(
\overset{\circ}{M}\right)  $ is a bijective map

The inverse of\ $\Delta$\ is then a compact, self adjoint differential
operator on $L^{2}\left(  \overset{\circ}{M},%
\mathbb{C}
,\varpi_{0}\right)  $
\end{theorem}

\begin{theorem}
(Gregor'yan p.45) On a riemannian manifold (M,g), if the smallest eigen value
$\lambda_{1}$ of $-\Delta$ on M is
$>$
0 then the operator : $\left(  Gf\right)  \left(  x\right)  =\int_{M}G\left(
x,y\right)  f\left(  y\right)  \varpi_{0}\left(  y\right)  $ is the inverse
operator of $-\Delta$ in $L^{2}\left(  M,\varpi_{0},%
\mathbb{R}
\right)  $
\end{theorem}

The condition is met if O is a non empty relatively compact open subset of a
riemannian manifold (M,g) such that $M\backslash\overline{O}$ is non empty.
Then $\forall f\in L^{2}\left(  O,\varpi_{0},%
\mathbb{R}
\right)  ,\varphi\left(  x\right)  =\int_{O}G\left(  x,y\right)  f\left(
y\right)  \varpi_{0}\left(  y\right)  $\ is the unique solution of
$-\Delta\varphi=f$

\begin{theorem}
(Zuily p.163) On a \textit{bounded} open subset O of $%
\mathbb{R}
^{m}:$

$-\Delta:\left\{  f\in H_{c}^{1}\left(  O\right)  :\Delta^{\prime}T\left(
f\right)  \in T\left(  L^{2}\left(  O,\varpi_{0},%
\mathbb{C}
\right)  \right)  \right\}  \rightarrow T\left(  L^{2}\left(  O,\varpi_{0},%
\mathbb{C}
\right)  \right)  $ is an isomorphism

Its inverse is an operator compact, positive and self adjoint.\ Its spectrum
is comprised of 0 and a sequence \ of positive numbers which are eigen values
and converges to 0.
\end{theorem}

\paragraph{Maximum principle\newline}

\begin{theorem}
(Taylor 1 p.309) , For a first order linear real differential operator D with
smooth coefficients on $C_{1}\left(  M;%
\mathbb{R}
\right)  ,$ with M a connected riemannian compact manifold with boundary :

i) $-\Delta+D:H_{c}^{1}\left(  \overset{\circ}{M}\right)  \rightarrow
H^{-1}\left(  \overset{\circ}{M}\right)  $ is a Freholm operator of index
zero, thus it is surjective iff it is injective.

ii) If $\partial M$\ is smooth, for $f\in C_{1}\left(  \overline{M};%
\mathbb{R}
\right)  \cap C_{2}\left(  \overset{\circ}{M};%
\mathbb{R}
\right)  $ such that $\left(  D+\Delta\right)  \left(  f\right)  \geq0$ on
$\overset{\circ}{M}$, and $y\in\partial M:\forall x\in\overset{\circ}%
{M}:f\left(  y\right)  \geq f\left(  x\right)  $ then $f^{\prime}(y)n>0$ where
n is an outward pointing normal to $\partial M.$

iii) If $\partial M$\ is smooth, for $f\in C_{0}\left(  \overline{M};%
\mathbb{R}
\right)  \cap C_{2}\left(  \overset{\circ}{M};%
\mathbb{R}
\right)  $ such that $\left(  D+\Delta\right)  \left(  f\right)  \geq0$ on
$\overset{\circ}{M}$, then either f is constant or $\forall x\in\overset
{\circ}{M}:f\left(  x\right)  <\sup_{y\in\partial M}f\left(  y\right)  $
\end{theorem}

\begin{theorem}
(Taylor 1\ p.312) For a first order scalar linear differential operator D on
$C_{1}\left(  O;%
\mathbb{R}
\right)  $ with smooth coefficients, with O an open, bounded open subset of $%
\mathbb{R}
^{m}$ with boundary $\partial O=\overset{\cdot}{O}$\ :

If $f\in C_{0}\left(  \overline{O};%
\mathbb{R}
\right)  \cap C_{2}\left(  O;%
\mathbb{R}
\right)  $ and $\left(  D+\Delta\right)  \left(  f\right)  \geq0$ on O then
$\sup_{x\in O}f\left(  x\right)  =\sup_{y\in\partial O}f\left(  y\right)  $

Furthermore if $\left(  D+\Delta\right)  \left(  f\right)  =0$ on $\partial O$
then $\sup_{x\in O}\left\vert f\left(  x\right)  \right\vert =\sup
_{y\in\partial M}\left\vert f\left(  y\right)  \right\vert $
\end{theorem}

\paragraph{Harmonic functions in $%
\mathbb{R}
^{m}$\newline}

\begin{definition}
A function $f\in C_{2}\left(  O;%
\mathbb{C}
\right)  $ where O is an open subset of $%
\mathbb{R}
^{m}$\ is said to be \textbf{harmonic} if : $\Delta f=\sum_{j=1}^{m}%
\frac{\partial^{2}f}{\left(  \partial x^{i}\right)  ^{2}}=0$
\end{definition}

\begin{theorem}
(Lieb p.258) If $S\in C_{\infty c}\left(  O;%
\mathbb{R}
\right)  $ , where O is an open subset of $%
\mathbb{R}
^{m},$\ is such that $\Delta S=0,$ there is a harmonic function $f:S=T(f)$
\end{theorem}

Harmonic functions have very special properties : they are smooth, defined
uniquely by their value on a hypersurface, have no extremum except on the boundary.

\begin{theorem}
A harmonic function is indefinitely R-differentiable
\end{theorem}

\begin{theorem}
(Taylor 1 p.210) A harmonic function $f\in C_{2}\left(
\mathbb{R}
^{m};%
\mathbb{C}
\right)  $ which is bounded is constant
\end{theorem}

\begin{theorem}
(Taylor 1 p.189) On a smooth manifold with boundary M in $%
\mathbb{R}
^{m},$ if $u,v\in C_{\infty}\left(  \overline{M};%
\mathbb{R}
\right)  $ are such that $\Delta u=\Delta v=0$ in $\overset{\circ}{M}$ and
$u=v=f$ on $\partial M$ then $u=v$ on all of M.
\end{theorem}

The result \ still holds if M is bounded and u,v continuous on $\partial M.$

\begin{theorem}
A harmonic function $f\in C_{2}\left(  O;%
\mathbb{R}
\right)  $ on an open connected subset O of $%
\mathbb{R}
^{m}$ has no interior minimum of maximum unless it is constant. In particular
if O is bounded and f continuous on the border $\partial O$ of O then
$\sup_{x\in O}f\left(  x\right)  =\sup_{y\in\partial O}f\left(  y\right)  $
.If f is complex valued : $\sup_{x\in O}\left\vert f\left(  x\right)
\right\vert =\sup_{y\in\partial O}\left\vert f\left(  y\right)  \right\vert $
\end{theorem}

The value of harmonic functions equals their averadge on balls :

\begin{theorem}
(Taylor 1 p.190) Let B(0,r) the open ball in $%
\mathbb{R}
^{m},f\in C_{2}\left(  B\left(  0,r\right)  ;%
\mathbb{R}
\right)  \cap C_{0}\left(  B\left(  0,r\right)  ;%
\mathbb{R}
\right)  ,\Delta f=0$ then $\ f\left(  0\right)  =\frac{1}{A\left(  B\left(
0,r\right)  \right)  }\int_{\partial B\left(  0,r\right)  }f\left(  x\right)
dx$
\end{theorem}

Harmonic radial functions : In $%
\mathbb{R}
^{m}$ a harmonic function which depends only of $r=\sqrt{x_{1}^{2}%
+..+x_{m}^{2}}$ must satisfy the ODE :

$f(x)=g(r):\Delta f=\frac{d^{2}g}{dr^{2}}+\frac{n-1}{r}\frac{dg}{dr}=0$

and the solutions are :

$m\neq2:g=Ar^{2-m}+B$

$m=2:g=-A\ln r+B$

Because of the singularity in 0 the laplacian is :

$m\neq2:\Delta^{\prime}T\left(  f\right)  =T\left(  \Delta f\right)
-(m-2)2\pi^{m/2}/\Gamma(m/2)\delta_{0}$

$m=2:\Delta^{\prime}T\left(  f\right)  =T\left(  \Delta f\right)  -2\pi
\delta_{0}$

\paragraph{Harmonic functions in $%
\mathbb{R}
^{2}$\newline}

\begin{theorem}
A harmonic function in an open O of $%
\mathbb{R}
^{2}$ is smooth and real analytic.
\end{theorem}

\begin{theorem}
If $f:\Omega\rightarrow%
\mathbb{C}
$ \ is holomorphic on an open $\Omega$ in $%
\mathbb{C}
$, then the function $\widehat{f}(x,y)=f(x+iy)$ is harmonic : $\frac
{\partial^{2}\widehat{f}}{\partial x^{2}}+\frac{\partial^{2}\widehat{f}%
}{\partial y^{2}}=0.$ Its real and imaginary components are also harmonic.
\end{theorem}

This is the direct consequence of the Cauchy relations. Notice that this
result does not hold for $O\subset%
\mathbb{R}
^{2n},n>1$

\begin{theorem}
(Schwartz III p.279) A harmonic function $P\in C_{2}\left(  O;%
\mathbb{R}
\right)  $ on a simply connected open O in $%
\mathbb{R}
^{2}$:

i) is in an infinitely many ways the real part of an holomorphic function :
$f\in H\left(  \widehat{O};%
\mathbb{C}
\right)  $ where $\widehat{O}=\left\{  x+iy,x,y\in\Omega\right\}  .$ and f is
defined up to an imaginary constant.

ii) has no local extremum or is constant

iii) if B(a,R) is the largest disc centered in a, contained in $O,\gamma
=\partial B$ then

$f\left(  a\right)  =\frac{1}{2\pi R}%
{\textstyle\oint\limits_{\gamma}}
f=\frac{1}{2\pi R^{2}}\int\int_{B}f(x,y)dxdy$
\end{theorem}

\bigskip

\subsection{The heat kernel}

\label{Heat kernel}

The heat kernel comes from the process of heat dissipation inside a medium
which follows a differential equation like $\frac{\partial u}{\partial
t}-\Delta_{x}u=0.$ This operator has many fascinating properties. In many ways
it links the volume of a region of a manifold with its area, and so is a
characteristic of a riemannian manifold.

\subsubsection{Heat operator}

\paragraph{Heat operator\newline}

\begin{definition}
The \textbf{heat operator} on a riemannian manifold (M,g) with metric g is the
scalar linear differential operator $D\left(  t,x\right)  =\frac{\partial
u}{\partial t}-\Delta_{x}u$ \ acting on functions on $(%
\mathbb{R}
_{+}\times M,g)$.
\end{definition}

According to our general definition a fundamental solution of the operator
D(t,x) at a point (t,y)$\in%
\mathbb{R}
_{+}\times M$\ is a distribution U(t,y) depending on the parameters (t,y) such
that : $D^{\prime}U\left(  t,y\right)  =\delta_{t,y}.$ To account for the
semi-continuity in t=0 one imposes :

$\forall\varphi\in C_{\infty c}\left(
\mathbb{R}
_{+}\times M;%
\mathbb{R}
\right)  :D^{\prime}U\left(  t,y\right)  \varphi\left(  t,x\right)
\rightarrow_{t\rightarrow0_{+}}\varphi\left(  t,y\right)  $

A fundamental solution is said to be \textbf{regular} if : $U\left(
t,y\right)  =T\left(  p\left(  t,x,y\right)  \right)  $ is induced by a smooth
map :

$p:M\rightarrow C_{\infty}\left(
\mathbb{R}
_{+};%
\mathbb{R}
\right)  ::u(t,x)=p(t,x,y)$ such that :

$p\left(  t,x,y\right)  \geq0,\forall t\geq0:$ $\int_{M}u(t,x)\varpi_{0}\leq1$
where $\varpi_{0}$\ is the volume form induced by g.

A fundamental regular solution can then be extended by setting $u\left(
t,x\right)  =0$ for $t\leq0$. Then u(t,x) is smooth on $%
\mathbb{R}
\times M\backslash\left(  0,y\right)  $ and satisfies $\frac{\partial
u}{\partial t}-\Delta_{x}u=\delta_{\left(  0,y\right)  }$ for $y\in M$

\paragraph{Heat kernel operator\newline}

Fundamental regular solutions of the heat operator are given by the heat
kernel operator. See One parameter group in the Banach Spaces section.

\begin{definition}
The \textbf{heat kernel} of a riemannian manifold (M,g) with metric g is the
semi-group of operators $\left(  P\left(  t\right)  =e^{t\Delta}\right)
_{t\geq0}$ where $\Delta$ is the laplacian on M acting on functions on M
\end{definition}

The domain of P(t) is well defined as :

$\left\{  f\in\overline{C_{\infty c}\left(  M;%
\mathbb{C}
\right)  }:\Delta^{\prime}T\left(  f\right)  \in T\left(  L^{2}\left(
M,\varpi_{0},%
\mathbb{C}
\right)  \right)  \right\}  $

where the closure is taken in $L^{2}\left(  M,\varpi_{0},%
\mathbb{C}
\right)  $ . It can be enlarged :

\begin{theorem}
(Taylor 3 p.38) On a riemannian compact manifold with boundary M, the heat
kernel $\left(  P\left(  t\right)  =e^{t\Delta}\right)  _{t\geq0}$ is a
strongly continuous semi-group on the Banach space : $\left\{  f\in
C_{1}\left(  M;%
\mathbb{C}
\right)  ,f=0\text{ on }\partial M\right\}  $
\end{theorem}

\begin{theorem}
(Taylor 3 p.35) On a riemannian compact manifold with boundary M :
$e^{z\Delta}$ is a holomorphic semi-group on $L^{p}\left(  M,\varpi_{0},%
\mathbb{C}
\right)  $ for $1\leq p\leq\infty$
\end{theorem}

\begin{theorem}
(Gregor'yan p.10) For any function $f\in L^{2}\left(  M,\varpi_{0},%
\mathbb{C}
\right)  $ on a riemannian smooth manifold (without boundary) (M,g), the function

$u(t,x)=P(t,x)f(x)\in C_{\infty}\left(
\mathbb{R}
_{+}\times M;%
\mathbb{C}
\right)  $ and satisfies the heat equation : $\frac{\partial u}{\partial
t}=\Delta_{x}u$ with the conditions \ : $u(t,.)\rightarrow f$ in $L^{2}\left(
M,\varpi_{0},%
\mathbb{C}
\right)  $ when $t\rightarrow0_{+}$ and $\inf(f)\leq u\left(  t,x\right)
\leq\sup\left(  f\right)  $
\end{theorem}

\paragraph{Heat kernel function\newline}

The heat kernel operator, as a fundamental solution of a heat operator, has an
associated Green's function.

\begin{theorem}
(Gregor'yan p.12) For any riemannian manifold (M,g) there is a unique function
p, called the \textbf{heat kernel function}

$p\left(  t,x,y\right)  \in C_{\infty}\left(
\mathbb{R}
_{+}\times M\times M;%
\mathbb{C}
\right)  $ such that \ 

$\forall f\in L^{2}\left(  M,\varpi_{0},%
\mathbb{C}
\right)  ,\forall t\geq0:\left(  P\left(  t\right)  f\right)  \left(
x\right)  =\int_{M}p\left(  t,x,y\right)  f\left(  y\right)  \varpi_{0}\left(
y\right)  .$
\end{theorem}

p(t) is the integral kernel of the heat operator $P(t)=e^{t\Delta_{x}}$
(whence its name) and $U\left(  t,y\right)  =T\left(  p\left(  t,x,y\right)
\right)  $ is a regular fundamental solution on the heat operator at y.

As a regular fundamental solution of the heat operator:

\begin{theorem}
$\forall f\in L^{2}\left(  M,\varpi_{0},%
\mathbb{C}
\right)  ,u\left(  t,x\right)  =\int_{M}p\left(  t,x,y\right)  f\left(
y\right)  \varpi_{0}\left(  y\right)  $ is solution of the PDE :

$\frac{\partial u}{\partial t}=\Delta_{x}u$ for t
$>$
0

$\lim_{t\rightarrow0_{+}}u\left(  t,x\right)  =f\left(  x\right)  $

and $u\in C_{\infty}(%
\mathbb{R}
_{+}\times M;%
\mathbb{R}
)$
\end{theorem}

\begin{theorem}
The heat kernel function has the following properties :

i) $p(t,x,y)=p(t,y,x)$

ii) $p(t+s,x,y)=\int_{M}p(t,x,z)p(s,z,y)\varpi_{0}\left(  z\right)  $

iii) $p(t,x,.)\in L^{2}\left(  M,\varpi_{0},%
\mathbb{C}
\right)  $

iv) As $p\geq0$ and $\int_{M}p\left(  t,x,y\right)  \varpi_{0}\left(
x\right)  \leq1$ the domain of the operator P(t) can be extended to any
positive or bounded measurable function f on M by : $\left(  P\left(
t\right)  f\right)  \left(  x\right)  =\int_{M}p\left(  t,x,y\right)  f\left(
y\right)  \varpi_{0}\left(  y\right)  $

v) Moreover $G\left(  x,y\right)  =\int_{0}^{\infty}p\left(  t,x,y\right)  dt$
is the Green's function of $-\Delta$ (see above).
\end{theorem}

Warning ! because the heat operator is not defined for t%
$<$%
0 we cannot have p(t,x,y) for t%
$<$%
0

The heat kernel function depends on the domain in M

\begin{theorem}
If O, O' are open subsets of M, then $O\subset O^{\prime}\Rightarrow
p_{O}\left(  t,x,y\right)  \leq p_{O^{\prime}}\left(  t,x,y\right)  $ . If
$\left(  O_{n}\right)  _{n\in%
\mathbb{N}
}$ is a sequence $O_{n}\rightarrow M$ then $p_{O_{n}}\left(  t,x,y\right)
\rightarrow p_{M}\left(  t,x,y\right)  $
\end{theorem}

With this meaning p is the minimal fundamental positive solution of the heat
equation at y

\begin{theorem}
If (M$_{1},g_{1}),$(M$_{2},g_{2})$ are two riemannian manifolds, $\left(
M_{1}\times M_{2},g_{1}\otimes g_{2}\right)  $ is a riemannian manifold, the
laplacians $\Delta_{1},\Delta_{2}$ commute, $\Delta_{M_{1}\times M_{2}}%
=\Delta_{1}+\Delta_{2},$ $P_{M_{1}\times M_{2}}\left(  t\right)  =P_{1}\left(
t\right)  P_{2}\left(  t\right)  ,p_{M_{1}\times M_{2}}\left(  t,\left(
x_{1},x_{2}\right)  ,\left(  y_{1},y_{2}\right)  \right)  =p_{1}\left(
t,x_{1},y_{1}\right)  p_{2}\left(  t,x_{2},y_{2}\right)  $
\end{theorem}

\begin{theorem}
Spectral resolution : if $dE\left(  \lambda\right)  $ is the spectral
resolution of $-\Delta$ then P(t)=$\int_{0}^{\infty}e^{-\lambda t}dE\left(
\lambda\right)  $
\end{theorem}

If M is some open, \textit{relatively compact}, in a manifold N, then the
spectrum of $-\Delta$ is an increasing countable sequence $\left(  \lambda
_{n}\right)  _{n=1}^{\infty},\lambda_{n}>0$ with eigenvectors $e_{n}\in
L^{2}\left(  M,\varpi_{0},%
\mathbb{C}
\right)  :$ $-\Delta e_{n}=\lambda_{n}e_{n}$\ \ forming an orthonormal basis.
Thus for t$\geq0,x,y\in M:$

$p(t,x,y)=\sum_{n=1}^{\infty}e^{-\lambda_{n}t}e_{n}\left(  x\right)
e_{n}\left(  y\right)  $

$\lim_{t\rightarrow\infty}\frac{\ln p\left(  t,x,y\right)  }{t}=-\lambda_{1} $

$\int_{M}p\left(  t,x,x\right)  \varpi_{0}\left(  x\right)  =\sum
_{n=1}^{\infty}e^{-\lambda_{n}t}$

\paragraph{Heat kernel, volume and distance\newline}

The heat kernel function is linked to the relation between volume and distance
in (M,g).

1. In a neighborhood of y we have :

$p(t,x,y)$ $\sim\frac{1}{\left(  4\pi t\right)  ^{m/2}}\left(  \exp
\frac{-d^{2}\left(  x,y\right)  }{4t}\right)  u(x,y)$ when $t\rightarrow0$

where m=dimM, d(x,y) is the geodesic distance between x,y and u(x,y) a smooth function.

2. If $f:M\rightarrow M$ is an isometry on (M,g), meaning a diffeomorphism
preserving g, then it preserves the heat kernel function :

$p(t,f\left(  x\right)  ,f\left(  y\right)  )=p(t,x,y)$

3. For two Lebesgue measurable subsets A,B of M :

$\int_{A}\int_{B}p(t,x,y)\varpi_{0}\left(  x\right)  \varpi_{0}\left(
y\right)  \leq\sqrt{\varpi_{0}\left(  A\right)  \varpi_{0}\left(  B\right)
}\left(  \exp\frac{-d^{2}\left(  A,B\right)  }{4t}\right)  $

4. The heat kernel function in $%
\mathbb{R}
^{m}$ is : $p(t,x,y)=\frac{1}{\left(  4\pi t\right)  ^{m/2}}\left(  \exp
\frac{-\left\Vert x-y\right\Vert ^{2}}{4t}\right)  $

\subsubsection{Brownian motion}

\paragraph{Manifold stochastically complete \newline}

\begin{definition}
A riemannian manifold (M,g) is said to be stochastically complete if $\forall
y\in M,t\geq0:\int_{M}p(t,x,y)\varpi_{0}\left(  x\right)  =1$
\end{definition}

then any fundamental solution of the heat equation at y is equal to the heat
kernel and is unique.

\begin{theorem}
(Gregor'yan p.15) The following conditions are equivalent :

i) (M,g) is stochastically complete

ii) any non negative bounded solution of $\Delta f-f=0$ on (M,g) is null

iii) the bounded Cauchy problem on (M,g) has a unique solution
\end{theorem}

A compact manifold is stochastically complete, $%
\mathbb{R}
^{m}$ is stochastically complete

A geodesic ball centered in y in M is the set of points B(y,r)=$\left\{  p\in
M,d\left(  p,y\right)  <r\right\}  .$ A manifold is geodesically complete iff
the geodesic balls are relatively compact.

If a manifold M is geodesically complete and $\exists y\in M:\int_{0}^{\infty
}\frac{r}{\log Vol\left(  B\left(  y,r\right)  \right)  }dr=\infty$ then M is
stochastically complete

\paragraph{Brownian motion\newline}

Beacause it links a time parameter and two points x,y of M, the heat kernel
function is the tool of choice to define brownian motions, meaning random
paths on a manifold. The principle is the following (see Gregor'yan for more).

1. Let (M,g) be a riemannian manifold, either compact or compactified with an
additional point $\left\{  \infty\right\}  .$

Define a path in M as a map : $\varpi:\left[  0,\infty\right]  \rightarrow M$
such that if $\exists t_{0}:\varpi\left(  t_{0}\right)  =\infty$ then $\forall
t>t_{0}:\varpi\left(  t\right)  =\infty$

Denote $\Omega$ the set of all such paths and $\Omega_{x}$\ the set of paths
starting at x..

Define on $\Omega_{x}$ the $\sigma-$algebra comprised of the paths such that :

$\exists N,\left\{  t_{1},...,t_{N}\right\}  ,\varpi\left(  0\right)
=x,\varpi\left(  t_{i}\right)  \in A_{i}$ where $A_{i}$ is a mesurable subset
of M. Thus we have measurables sets $\left(  \Omega_{x},S_{x}\right)  $

2. Define the random variable : $X_{t}=\varpi\left(  t\right)  $ and the
stochastic process with the transition probabilities :

$P_{t}\left(  x,A\right)  =\int_{A}p\left(  t,x,y\right)  \varpi_{0}\left(
y\right)  $

$P_{t}\left(  x,\infty\right)  =1-P_{t}\left(  x,M\right)  $

$P_{t}\left(  \infty,A\right)  =0$

The heat semi group relates the transition probabilities by : $P_{t}\left(
x,A\right)  =P\left(  t\right)  1_{A}\left(  x\right)  $ whith the
characteristic function $1_{A}$\ of A.

So the probability that a path ends at $\infty$\ is null if M is
stochastically complete (which happens if it is compact).

The probability that a path is such that : $\varpi\left(  t_{i}\right)  \in
A_{i},\left\{  t_{1},...,t_{N}\right\}  $ is :

$P_{x}\left(  \varpi\left(  t_{i}\right)  \in A_{i}\right)  $

$=\int_{A_{1}}...\int_{A_{N}}P_{t_{1}}\left(  x,y_{1}\right)  P_{t_{2}-t_{1}%
}\left(  y_{1},y_{2}\right)  ...P_{t_{N}-t_{N-1}}\left(  y_{N-1},y_{N}\right)
\varpi_{0}\left(  y_{1}\right)  ...\varpi_{0}\left(  y_{N}\right)  $

So : $P_{x}\left(  \varpi\left(  t\right)  \in A\right)  =P_{t}\left(
x,A\right)  =P\left(  t\right)  1_{A}\left(  x\right)  $

We have a stochastic process which meets the conditions of the Kolmogoroff
extension (see Measure).

Notice that this is one stochastic process among many others : it is
characterized by transition probabilities linked to the heat kernel function.

3. Then for any bounded function f on M we have :

$\forall t\geq0,\forall x\in M:\left(  P\left(  t\right)  f\right)  \left(
x\right)  =E_{x}\left(  f\left(  X_{t}\right)  \right)  =\int_{\Omega_{x}%
}f\left(  \varpi\left(  t\right)  \right)  P_{x}\left(  \varpi\left(
t\right)  \right)  $

\bigskip

\subsection{Pseudo differential operators}

\label{Pseudodifferential operator}

A linear differential operator D on the space of complex functions over $%
\mathbb{R}
^{m}$\ can be written :

if $f\in S\left(
\mathbb{R}
^{m}\right)  :Df=\left(  2\pi\right)  ^{-m/2}\int_{%
\mathbb{R}
^{m}}P(x,it)\widehat{f}(t)e^{i\left\langle t,x\right\rangle }dt$

where P is the symbol of D.

In some way we replace the linear operator D by an integral, as in the
spectral theory. The interest of this specification is that all the practical
content of D is summarized in the symbol. As seen above this is convenient to
find fundamental solutions of partial differential equations. It happens that
this approach can be generalized, whence the following definition.

\subsubsection{Definition}

\begin{definition}
A \textbf{pseudo differential operator} is a map, denoted P(x,D), on a space F
of complex functions on $%
\mathbb{R}
^{m}$ : $P(x,D):F\rightarrow F$ such that there is a function $P\in C_{\infty
}\left(
\mathbb{R}
^{m}\times%
\mathbb{R}
^{m};%
\mathbb{C}
\right)  $ called the symbol of the operator :

$\forall\varphi\in F:$%

\begin{equation}
P\left(  x,D\right)  \varphi=\left(  2\pi\right)  ^{-m/2}\int_{y\in%
\mathbb{R}
^{m}}\int_{t\in%
\mathbb{R}
^{m}}e^{i\left\langle t,x-y\right\rangle }P\left(  x,t\right)  \varphi\left(
y\right)  dydt=\int_{t\in%
\mathbb{R}
^{m}}e^{i\left\langle t,x\right\rangle }P\left(  x,t\right)  \widehat{\varphi
}\left(  t\right)  dt
\end{equation}

\end{definition}

Using the same function P(x,t) we can define similarly a pseudo differential
operator acting on distributions.

\begin{definition}
A pseudo differential operator, denoted P(x,D'), acting on a space F' of
distributions is a map : $P(x,D\prime):F^{\prime}\rightarrow F^{\prime}$ such
that there is a function $P\in C_{\infty}\left(
\mathbb{R}
^{m}\times%
\mathbb{R}
^{m};%
\mathbb{C}
\right)  $ called the symbol of the operator :%

\begin{equation}
\forall S\in F^{\prime},\forall\varphi\in F:P\left(  x,D^{\prime}\right)
\left(  S\right)  \varphi=S_{t}\left(  P\left(  x,t\right)  e^{i\left\langle
t,x\right\rangle }\widehat{\varphi}\left(  t\right)  \right)
\end{equation}

\end{definition}

These definitions assume that the Fourier transform of a test function
$\varphi$ is well defined.\ This point is seen below.

\paragraph{H\"{o}rmander classes\newline}

A pseudo differential operator is fully defined through its symbol : the
function P(x,t). They are classified according to the space of symbols P. The
usual classes are the \textbf{H\"{o}rmander classes} denoted $D_{\rho b}^{r}:$

r$\in%
\mathbb{R}
$ is the order of the class$,\rho,b\in\left[  0,1\right]  $ are parameters
(usually $\rho=1,b=0)$ such that :

$\forall\left(  \alpha\right)  =\left(  \alpha_{1},..\alpha_{k}\right)
,\left(  \beta\right)  =\left(  \beta_{1},..,\beta_{l}\right)  :\exists
C_{\alpha\beta}\in%
\mathbb{R}
:$

$\left\vert D_{\alpha_{1}..\alpha_{k}}\left(  x\right)  D_{\beta_{1}%
..\beta_{l}}\left(  t\right)  P\left(  x,t\right)  \right\vert \leq
C_{\alpha\beta}\left(  \left(  1+\left\Vert t\right\Vert ^{2}\right)
^{1/2}\right)  ^{r-\rho k+bl}$

and one usually indifferently says that $P(x,D)\in D_{\rho b}^{r}$ or
$P(x,t)\in D_{\rho b}^{r}$

The concept of principal symbol is the following : if there are smooth
functions $P_{k}\left(  x,t\right)  $ homogeneous of degree k in t, and :

$\forall n\in%
\mathbb{N}
:P\left(  x,t\right)  -\sum_{k=0}^{n}P_{k}\left(  x,t\right)  \in
D_{1,0}^{r-n}$ one says that $P\left(  x,t\right)  \in D^{r}$ and
$P_{r}\left(  x,t\right)  $ is the principal symbol of P(x,D).

\paragraph{Comments\newline}

From the computation of the Fourier transform for linear differential
operators we see that the definition of pseudo differential operators is
close. But it is important to notice the differences.

i) Linear differential operators of order r with smooth bounded coefficients
(acting on functions or distributions) are pseudo-differential operators of
order r : their symbols is a map polynomial in $t.$ Indeed the H\"{o}rmander
classes are equivalently defined with the symbols of linear differential
operators. But the converse is not true.

ii) Pseudo differential operators are not local : to compute the value
$P\left(  x,D\right)  \varphi$\ at x we need to know the whole of $\varphi$ to
compute the Fourier transform, while for a differential operator it requires
only the values of the function and their derivatives at x. So a pseudo
differential operator is not a differential operator, with the previous
definition (whence the name pseudo).

On the other hand to compute a pseudo differential operator we need only to
know the function, while for a differential operator we need to know its r
first derivatives. A pseudo differential operator is a map acting on sections
of F, a differential operator is a map acting (locally) on sections of
$J^{r}F.$

iii) Pseudo differential operators are linear with respect to functions or
distributions, but they are not necessarily linear with respect to sections of
the J$^{r}$ extensions.

iv) and of course pseudo differential operators are scalar : they are defined
for functions or distributions on $%
\mathbb{R}
^{m}$, not sections of fiber bundles.

So pseudo differential operators can be seen as a "proxy" for linear scalar
differential operators. Their interest lies in the fact that often one can
reduce a problem in analysis of pseudo-differential operators to a sequence of
algebraic problems involving their symbols, and this is the essence of
microlocal analysis.

\subsubsection{General properties}

\paragraph{Linearity\newline}

Pseudo-differential operators are linear endomorphisms on the spaces of
functions or distributions (but not their jets extensions) :

\begin{theorem}
(Taylor 2 p.3) The pseudo differential operators in the class $D_{\rho b}^{r}
$ are such that :

$P(x,D):S\left(
\mathbb{R}
^{m}\right)  \rightarrow S\left(
\mathbb{R}
^{m}\right)  $

$P(x,D^{\prime}):S\left(
\mathbb{R}
^{m}\right)  ^{\prime}\rightarrow S\left(
\mathbb{R}
^{m}\right)  ^{\prime}$ if b
$<$
1

and they are linear continuous operators on these vector spaces.
\end{theorem}

So the same function P(x,t) can indifferently define a pseudo differential
operator acting on functions or distributions, with the Fr\'{e}chet space
$S\left(
\mathbb{R}
^{m}\right)  $ and its dual $S\left(
\mathbb{R}
^{m}\right)  ^{\prime}.$

\begin{theorem}
(Taylor 2 p.17) If $P(x,D)\in D_{\rho b}^{0}$ and $b<\rho,s\in%
\mathbb{R}
$ then :

$P(x,D):L^{2}\left(
\mathbb{R}
^{m}dx,%
\mathbb{C}
\right)  \rightarrow L^{2}\left(
\mathbb{R}
^{m}dx,%
\mathbb{C}
\right)  $

$P(x,D):H^{s}\left(
\mathbb{R}
^{m}\right)  \rightarrow H^{s-r}\left(
\mathbb{R}
^{m}\right)  $

(Taylor 3 p.18) If $P(x,D)\in D_{1b}^{0}$ and $b\in\left[  0,1\right]  ,$ then :

$P(x,D):L^{p}\left(
\mathbb{R}
^{m}dx,%
\mathbb{C}
\right)  \rightarrow L^{p}\left(
\mathbb{R}
^{m}dx,%
\mathbb{C}
\right)  $ for $1<p<\infty$
\end{theorem}

\paragraph{Composition\newline}

Pseudo-differential are linear endomorphisms, so they can be composed and we
have :

\begin{theorem}
(Taylor 2 p.11) The composition $P_{1}\circ P_{2}$ of two pseudo-differential
operators is again a pseudo-differential operator $Q\left(  x,D\right)  $\ :
\end{theorem}

If $P_{1}\in D_{\rho_{1}b_{1}}^{r_{1}},P_{2}\in D_{\rho_{2}b_{2}}^{r_{2}%
}:b_{2}\leq\min\left(  \rho_{1},\rho_{2}\right)  =\rho,\delta=\max\left(
\delta_{1},\delta_{2}\right)  $

$Q\left(  x,D\right)  \in D_{\rho b}^{r_{1}+r_{2}}$

$Q\left(  x,t\right)  \sim\sum_{\left\Vert \alpha\right\Vert \geq0}%
\frac{i^{\left\Vert \alpha\right\Vert }}{\alpha!}D_{\alpha}\left(
t_{1}\right)  P\left(  x,t_{1}\right)  D_{\alpha}\left(  x\right)
P_{2}\left(  x,t\right)  $ when $x,t\rightarrow\infty$

The commutator of two pseudo differential operators is defined as :

$\left[  P_{1},P_{2}\right]  =P_{1}\circ P_{2}-P_{2}\circ P_{1}$ and we have :

$\left[  P_{1},P_{2}\right]  \in D_{\rho b}^{r_{1}+r_{2}-\rho-b}$

\paragraph{Adjoint of a pseudo differential operator\newline}

$\forall p:1\leq p\leq\infty:S\left(
\mathbb{R}
^{m}\right)  \subset L^{p}\left(
\mathbb{R}
^{m},dx,%
\mathbb{C}
\right)  $ so $S\left(
\mathbb{R}
^{m}\right)  \subset L^{2}\left(
\mathbb{R}
^{m},dx,%
\mathbb{C}
\right)  $ with the usual inner product : $\left\langle \varphi,\psi
\right\rangle =\int_{%
\mathbb{R}
^{m}}\overline{\varphi}\psi dx$

A pseudo differential operator $P(x,D):S\left(
\mathbb{R}
^{m}\right)  \rightarrow S\left(
\mathbb{R}
^{m}\right)  $ is a continuous operator on the Hilbert space $S\left(
\mathbb{R}
^{m}\right)  $ and has an adjoint : $P(x,D)^{\ast}:S\left(
\mathbb{R}
^{m}\right)  \rightarrow S\left(
\mathbb{R}
^{m}\right)  $ such that : $\left\langle P\left(  x,D\right)  \varphi
,\psi\right\rangle =\left\langle \varphi,P(x,D)^{\ast}\psi\right\rangle $

\begin{theorem}
(Taylor 2 p.11)The adjoint of a pseudo differential operator is also a
pseudo-differential operator :
\end{theorem}

$P\left(  x,D^{\ast}\right)  \varphi=\int_{t\in%
\mathbb{R}
^{m}}e^{i\left\langle t,x\right\rangle }P\left(  x,t\right)  ^{\ast}%
\widehat{\varphi}\left(  t\right)  dt$

If $P(x,D)\in D_{\rho b}^{r}$ then $P(x,D)^{\ast}\in D_{\rho b}^{r}$ with :

$P\left(  x,t\right)  ^{\ast}\sim\sum_{k\geq0}\frac{i^{k}}{\beta_{1}%
!\alpha_{2}!..\beta_{m}!}i^{k}D_{\alpha_{1}..\alpha_{k}}\left(  t\right)
D_{\alpha_{1}..\alpha_{k}}\left(  x\right)  P\left(  x,t\right)  $ when
$x,t\rightarrow\infty$ where $\beta_{p}$ is the number of occurences of the
index p in $\left(  \alpha_{1},...,\alpha_{k}\right)  $

\paragraph{Transpose of a pseudo differential operator:\newline}

\begin{theorem}
The transpose of a pseudo differential operator $P(x,D):S\left(
\mathbb{R}
^{m}\right)  \rightarrow S\left(
\mathbb{R}
^{m}\right)  $ is a pseudo operator $P(x,D)^{t}:S\left(
\mathbb{R}
^{m}\right)  ^{\prime}\rightarrow S\left(
\mathbb{R}
^{m}\right)  ^{\prime}$ with symbol $P\left(  x,t\right)  ^{t}=P\left(
t,x\right)  $
\end{theorem}

\begin{proof}
$P\left(  x,t\right)  ^{t}$ is such that :

$\forall S\in S\left(
\mathbb{R}
^{m}\right)  ^{\prime},\forall\varphi\in S\left(
\mathbb{R}
^{m}\right)  :$

$P(x,D)^{t}\left(  S\right)  \left(  \varphi\right)  =S\left(  P(x,D)\varphi
\right)  =S_{t}\left(  P\left(  x,t\right)  ^{t}e^{i\left\langle
t,x\right\rangle }\widehat{\varphi}\left(  t\right)  \right)  $

$=S_{t}\left(  \left(  2\pi\right)  ^{m/2}%
\mathcal{F}%
_{x}^{\ast}\left(  P\left(  x,t\right)  ^{t}\widehat{\varphi}\left(  t\right)
\right)  \right)  =\left(  2\pi\right)  ^{m/2}%
\mathcal{F}%
_{x}^{\ast}\left(  S_{t}\left(  P\left(  x,t\right)  ^{t}\widehat{\varphi
}\left(  t\right)  \right)  \right)  $

$S\left(  P(x,D)\varphi\right)  =S_{x}\left(  \int_{t\in%
\mathbb{R}
^{m}}e^{i\left\langle t,x\right\rangle }P\left(  x,t\right)  \widehat{\varphi
}\left(  t\right)  dt\right)  $

$=S_{x}\left(  \left(  2\pi\right)  ^{m/2}%
\mathcal{F}%
_{t}^{\ast}\left(  P\left(  x,t\right)  \widehat{\varphi}\left(  t\right)
\right)  \right)  =\left(  2\pi\right)  ^{m/2}%
\mathcal{F}%
_{t}^{\ast}\left(  S_{x}\left(  P\left(  x,t\right)  \widehat{\varphi}\left(
t\right)  \right)  \right)  $

with $%
\mathcal{F}%
^{\ast}:S\left(
\mathbb{R}
^{m}\right)  ^{\prime}\rightarrow S\left(
\mathbb{R}
^{m}\right)  ^{\prime}::%
\mathcal{F}%
^{\ast}\left(  S\right)  \left(  \varphi\right)  =S\left(
\mathcal{F}%
^{\ast}\left(  \varphi\right)  \right)  $
\end{proof}

\subsubsection{Schwartz kernel}

For $\varphi,\psi\in S\left(
\mathbb{R}
^{m}\right)  $ the bilinear functional :

$K\left(  \varphi\otimes\psi\right)  =\int_{%
\mathbb{R}
^{m}}\varphi\left(  x\right)  \left(  P\left(  x,D\right)  \psi\right)  dx$

$=\int\int\int_{%
\mathbb{R}
^{m}\times%
\mathbb{R}
^{m}\times%
\mathbb{R}
^{m}}e^{i\left\langle t,x-y\right\rangle }\varphi\left(  x\right)  \psi\left(
y\right)  P\left(  x,t\right)  dxdydt=\int\int_{%
\mathbb{R}
^{m}\times%
\mathbb{R}
^{m}}k\left(  x,y\right)  \varphi\left(  x\right)  \psi\left(  y\right)  dxdy$

with $k\left(  x,y\right)  =\int_{%
\mathbb{R}
^{m}}P\left(  x,t\right)  e^{i\left\langle t,x-y\right\rangle }dt$ and
$\varphi\otimes\psi\left(  x,y\right)  =\varphi\left(  x\right)  \psi\left(
y\right)  $

can be extended to the space $C_{\infty}\left(
\mathbb{R}
^{m}\times%
\mathbb{R}
^{m};%
\mathbb{C}
\right)  $ and we have the following :

\begin{theorem}
(Taylor 2 p.5) For any pseudo differential operator $P(x,D)\in D_{\rho b}^{r}
$ on $S\left(
\mathbb{R}
^{m}\right)  $ there is a distribution $K\in\left(  C_{\infty c}\left(
\mathbb{R}
^{m}\times%
\mathbb{R}
^{m};%
\mathbb{C}
\right)  \right)  ^{\prime}$ called the \textbf{Schwartz kernel} of P(x,D)
such that : $\forall\varphi,\psi\in S\left(
\mathbb{R}
^{m}\right)  :K\left(  \varphi\otimes\psi\right)  =\int_{%
\mathbb{R}
^{m}}\varphi\left(  x\right)  \left(  P\left(  x,D\right)  \psi\right)  dx$
\end{theorem}

K is induced by the function $k\left(  x,y\right)  =\int_{%
\mathbb{R}
^{m}}P\left(  x,t\right)  e^{i\left\langle t,x-y\right\rangle }dt$

If $\rho>0$ then $k\left(  x,y\right)  $ belongs to $C_{\infty}\left(
\mathbb{R}
^{m}\times%
\mathbb{R}
^{m};%
\mathbb{C}
\right)  $ for $x\neq y$

$\exists C\in%
\mathbb{R}
:\left\Vert \beta\right\Vert >-m-r:\left\vert D_{\beta}\left(  x,y\right)
k\right\vert \leq C\left\Vert x-y\right\Vert ^{-m-r-\left\Vert \beta
\right\Vert }$

The Schwartz kernel is characteristic of P(x,D) and many of the theorems about
pseudo differential operators are based upon its properties.

\subsubsection{Support of a pseudo-differential operator}

\paragraph{Elliptic pseudo differential operators\newline}

\begin{definition}
(Taylor 2 p.14) A pseudo differential operator $P(x,D)\in D_{\rho b}^{r}%
,\rho<b$ on $S\left(
\mathbb{R}
^{m}\right)  $ is said to be \textbf{elliptic} if

$\exists c\in%
\mathbb{R}
:\forall\left\Vert t\right\Vert >c:\left\vert P\left(  x,t\right)
^{-1}\right\vert \leq c\left\Vert t\right\Vert ^{-r}$
\end{definition}

Then if

$H_{c}:%
\mathbb{R}
^{m}\rightarrow%
\mathbb{R}
::\left\Vert t\right\Vert \leq c:H_{c}\left(  t\right)  =0,\left\Vert
t\right\Vert >c:H_{c}\left(  t\right)  =1$

we have $H_{c}\left(  t\right)  P\left(  x,t\right)  ^{-1}=Q\left(
x,t\right)  $ and the pseudo differential operator $Q(x,D)\in D_{\rho b}^{-r}$
is such that :

$Q\left(  x,D\right)  \circ P\left(  x,D\right)  =Id+R_{1}\left(  x,D\right)
$

$P\left(  x,D\right)  \circ Q\left(  x,D\right)  =Id-R_{2}\left(  x,D\right)
$

with $R_{1}\left(  x,D\right)  ,R_{2}\left(  x,D\right)  \in D_{\rho
b}^{-\left(  \rho-b\right)  }$

so Q(x,D) is a proxy for a left and right inverse of P(x,D) : this is called a
two sided \textbf{parametrix}.

Moreover we have for the singular supports :

$\forall S\in S\left(
\mathbb{R}
^{m}\right)  ^{\prime}:SSup\left(  P(x,D)S\right)  =SSup(S)$

which entails that an elliptic pseudo differerential operator does not add any
critical point to a distribution (domains where the distribution cannot be
identified with a function). Such an operator is said to be microlocal. It can
be more precise with the following.

\paragraph{Characteristic set\newline}

(Taylor 2 p.20):

The \textbf{characteristic set} of a pseudo differential operator $P(x,D)\in
D^{r}$ with principal symbol $P_{r}\left(  x,t\right)  $ is the set :

$Char(P(x,D))=\left\{  \left(  x,t\right)  \in%
\mathbb{R}
^{m}\times%
\mathbb{R}
^{m},\left(  x,t\right)  \neq\left(  0,0\right)  :P_{r}\left(  x,t\right)
\neq0\right\}  $

The \textbf{wave front set} of a distribution $S\in H^{-\infty}\left(
\mathbb{R}
^{m}\right)  =\cup_{s}H^{-s}\left(
\mathbb{R}
^{m}\right)  $ is the set :

$WF\left(  S\right)  =\cap_{P}\left\{  Char\left(  P(x,D\right)  ,P(x,D)\in
S^{0}:P(x,D)S\in T\left(  C_{\infty}\left(
\mathbb{R}
^{m};%
\mathbb{C}
\right)  \right)  \right\}  $

then $\Pr_{1}\left(  WF\left(  S\right)  \right)  =SSup(S)$ with the projection

$\Pr_{1}:%
\mathbb{R}
^{m}\times%
\mathbb{R}
^{m}\rightarrow%
\mathbb{R}
^{m}::\Pr_{1}\left(  x,t\right)  =x$

\paragraph{Essential support\newline}

(Taylor 2 p.20)

A pseudo differential operator $P(x,D)\in D_{\rho b}^{r}$ is said to be of
order $-\infty$\ at $\left(  x_{0},t_{0}\right)  \in%
\mathbb{R}
^{m}\times%
\mathbb{R}
^{m}$\ if :

$\forall\left(  \alpha\right)  =\left(  \alpha_{1},..\alpha_{k}\right)
,\left(  \beta\right)  =\left(  \beta_{1},..,\beta_{l}\right)  ,\forall N\in%
\mathbb{N}
:$

$\exists C\in%
\mathbb{R}
,\forall\xi>0:\left\vert D_{\alpha_{1}..\alpha_{k}}\left(  x\right)
D_{\beta_{1}..\beta_{l}}\left(  t\right)  P\left(  x,t\right)  |_{\left(
x_{0},\xi t_{0}\right)  }\right\vert \leq C\left\Vert \xi t_{0}\right\Vert
^{-N}$

A pseudo differential operator $P(x,D)\in D_{\rho b}^{r}$ is said to be of
order $-\infty$\ in an open subset $U\subset%
\mathbb{R}
^{m}\times%
\mathbb{R}
^{m}$ if it is of order $-\infty$\ at any point $\left(  x_{0},t_{0}\right)
\in U$

The \textbf{essential support} $ESup(P(x,D))$ of a pseudo differential
operator $P(x,D)\in D_{\rho b}^{r}$ is the smallest closed subset of $%
\mathbb{R}
^{m}\times%
\mathbb{R}
^{m}$ on the complement of which P(x,D) is of order $-\infty$

For the compose of two pseudo-differential goperators $P_{1},P_{2}$ :

$ESup\left(  P_{1}\circ P_{2}\right)  \subset ESup\left(  P_{1}\right)  \cap
ESup\left(  P_{2}\right)  $

If $S\in H^{-\infty}\left(
\mathbb{R}
^{m}\right)  ,P(x,D)\in D_{\rho b}^{r},\rho>0,b<1$ then $WF(P(x,D)S)\subset
WF(S)\cap ESup\left(  P\left(  x,D\right)  \right)  $

If $S\in H^{-\infty}\left(
\mathbb{R}
^{m}\right)  ,P(x,D)\in D_{\rho b}^{r},b<\rho$ is elliptic, then P(x,D) has
the microlocal regularity property : WF(P(x,D)S)=WF(S).

For any general solution U of the scalar hyperbolic equation : $\frac{\partial
S}{\partial t}=iP(x,D)S$ on $S\left(
\mathbb{R}
^{m}\right)  ^{\prime}$ with $P(x,D)\in D^{1}$ with real principal symbol :
WF(U)=C(t)WF(S) where C(t) is a smooth map.

\newpage

\section{DIFFERENTIAL\ EQUATIONS}

Differential equations are usually met as equations whose unknown variable is
a map $f:E\rightarrow F$ such that $\forall x\in E:L\left(  x,f\left(
x\right)  ,f^{\prime}\left(  x\right)  ,...,f^{\left(  r\right)  }\left(
x\right)  \right)  =0$ subject to conditions such as $\forall x\in A:M\left(
x,f\left(  x\right)  ,f^{\prime}\left(  x\right)  ,...,f^{\left(  r\right)
}\left(  x\right)  \right)  =0$ for some subset A of E.

Differential equations raise several questions : existence, unicity and
regularity of a solution, and eventually finding an explicit solution, which
is not often possible. The problem is "well posed" when, for a given problem,
there is a unique solution, depending continuously on the parameters.

Ordinary differential equations (ODE) are equations for which the map f
depends on a unique real variable. For them there are general theorems which
answer well to the first two questions, and many ingenious more or less
explicit solutions.

Partial differential equations (PDE) are equations for which the map f depends
of more than one variable, so x is in some subset of $%
\mathbb{R}
^{m},m>1.$ There are no longer such general theorems. For linear PDE there are
many results, and some specific equations of paramount importance in physics
will be reviewed with more details. Non linear PDE are a much more difficult
subject, and we will limit ourselves to some general results.

Anyway the purpose is not here to give a list of solutions : they can be found
at some specialized internet sites such that :

http://eqworld.ipmnet.ru/en/solutions/ode.htm, 

and in the exhaustive handbooks of A.Polyanin and V.Zaitsev.

\newpage

\subsection{Ordinary Differential equations (ODE)}

\label{Differentiela equations ordinary}

\subsubsection{Definitions}

As $%
\mathbb{R}
$\ is simply connected, any vector bundle over $%
\mathbb{R}
$ is trivial and a r order ordinary differential equation is an evolution
equation :

$D:J^{r}X\rightarrow V_{2}$ is a differential operator

the unknown function $X\in C\left(  I;V_{1}\right)  $

I is some interval of $%
\mathbb{R}
,$ $V_{1}$ a m dimensional complex vector space, $V_{2}$ is a n dimensional
complex vector space.

The Cauchy conditions are : $X\left(  a\right)  =X_{0},X^{\prime}%
(a)=X_{1},...X^{\left(  r\right)  }\left(  a\right)  =X_{r}$ for some value
$x=a\in I\subset%
\mathbb{R}
$

An ODE of order r can always be replaced by an equivalent ODE of order 1 :

Define : $Y_{k}=X^{\left(  k\right)  },k=0...r$ ,$Y\in W=\left(  V_{1}\right)
^{r+1}$

Replace $J^{r}X=J^{1}Y\in C\left(  I;W\right)  $

Define the differential operator :

$G:J^{1}W\rightarrow V_{2}::G\left(  x,j_{x}^{1}Y\right)  =D\left(
x,j_{x}^{r}X\right)  $

with initial conditions : $Y\left(  a\right)  =\left(  X_{0},...,X_{r}\right)
$

Using the implicit map theorem (see Differential geometry, derivatives). a
first order ODE can then be put in the form :$\frac{dX}{dx}=L\left(
x,X\left(  x\right)  \right)  $

\subsubsection{Fundamental theorems}

\paragraph{Existence and unicity\newline}

Due to the importance of the Cauchy problem, we give several theorems about
existence and unicity.

\bigskip

\begin{theorem}
(Schwartz 2 p.351) The 1st order ODE : $\frac{dX}{dx}=L\left(  x,X\left(
x\right)  \right)  $ with :

i) $X:I\rightarrow O$, I an interval in $%
\mathbb{R}
$\ , O an open subset of an affine Banach space E

ii) $L:I\times O\rightarrow E$ a continuous map, globally Lipschitz with
respect to the second variable :

$\exists k\geq0:\forall\left(  x,y_{1}\right)  ,\left(  x,y_{2}\right)  \in
I\times O:\left\Vert L\left(  x,y_{1}\right)  -L\left(  x,y_{2}\right)
\right\Vert _{E}\leq k\left\Vert y_{1}-y_{2}\right\Vert _{E}$

iii) the Cauchy condition : $x_{0}\in I,y_{0}\in O:X\left(  x_{0}\right)
=y_{0}$

has a unique solution and : $\left\Vert X\left(  x\right)  -y_{0}\right\Vert
\leq e^{k\left\vert x-x_{0}\right\vert }\int_{\left\vert x_{0},x\right\vert
}\left\Vert L\left(  \xi,y_{0}\right)  \right\Vert d\xi$
\end{theorem}

The problem is equivalent to the following : find X such that :

$X\left(  x\right)  =y_{0}+\int_{x_{0}}^{x}L\left(  \xi,f\left(  \xi\right)
\right)  d\xi$

and the solution is found by the Picard iteration method :

$X_{n+1}\left(  x\right)  =y_{0}+\int_{x_{0}}^{x}L\left(  \xi,X_{n}\left(
\xi\right)  \right)  d\xi::X_{n}\rightarrow X$.

Moreover the series : $X_{0}+\sum_{k=1}^{n}\left(  X_{k}-X_{k-1}\right)
\rightarrow X$ is absolutely convergent on any compact of I.

if L is not Lipshitz :

i) If E is finite dimensional there is still a solution (Cauchy-Peano theorem
- Taylor 1 p.110), but it is not necessarily unique.\ 

ii) If E is infinite dimensional then the existence itself is not assured.

\bigskip

2. In the previous theorem L is globally Lipschitz.\ This condition can be
weakened as follows :

\begin{theorem}
(Schwartz 2 p.364) The 1st order ODE : $\frac{dX}{dx}=L\left(  x,X\left(
x\right)  \right)  $ with

i) $X:I\rightarrow O$, I an interval in $%
\mathbb{R}
$\ , O an open subset of a finite dimensional affine Banach space E

ii) $L:I\times O\rightarrow\overrightarrow{E}$ a continuous map, locally
Lipschitz with respect to the second variable:

$\forall a\in I,\forall y\in O,\exists n\left(  a\right)  \subset I,n\left(
y\right)  \subset O:\exists k\geq0:$

$\forall\left(  x,y_{1}\right)  ,\left(  x,y_{2}\right)  \in n\left(
a\right)  \times n\left(  y\right)  :\left\Vert L\left(  x,y_{1}\right)
-L\left(  x,y_{2}\right)  \right\Vert _{E}\leq k\left\Vert y_{1}%
-y_{2}\right\Vert _{E}$

iii) the Cauchy condition $x_{0}\in I,y_{0}\in O:X\left(  x_{0}\right)
=y_{0}$

has a unique solution in a maximal interval $\left\vert a,b\right\vert
:\ I^{\prime}\subset\left\vert a,b\right\vert \subset I$, and for any compact
C in O, $\exists\varepsilon>0:b-x<\varepsilon\Rightarrow$ $X\left(  x\right)
\notin C$
\end{theorem}

(meaning that X(x) tends to the border of O, possibly infinite).

The solution is still given by the integrals $X_{n}\left(  x\right)
=y_{0}+\int_{x_{0}}^{x}L\left(  \xi,X_{n-1}\left(  \xi\right)  \right)  d\xi$
and the series : $X_{0}+\sum_{k=1}^{n}\left(  X_{k}-X_{k-1}\right)  $

\bigskip

3. The previous theorem gives only the existence and unicity of local
solutions. We can have more.

\begin{theorem}
(Schwartz 2 p.370) The 1st order ODE : $\frac{dX}{dx}=L\left(  x,X\left(
x\right)  \right)  $ with :

i) $X:I\rightarrow E$, I an interval in $%
\mathbb{R}
$\ , E an affine Banach space

ii) $L:I\times E\rightarrow\overrightarrow{E}$ a continuous map such that :

$\exists\lambda,\mu\geq0,A\in E:\forall\left(  x,y\right)  \in I\times
E:\left\Vert L\left(  x,y\right)  \right\Vert \leq\lambda\left\Vert
y-A\right\Vert +\mu$

$\forall\rho>0$ L is Lipschitz with respect to the second variable on $I\times
B\left(  0,\rho\right)  $

iii) the Cauchy condition $x_{0}\in I,y_{0}\in O:X\left(  x_{0}\right)
=y_{0}$

has a unique solution defined on I
\end{theorem}

We have the following if $E=%
\mathbb{R}
^{m}$ :

\begin{theorem}
(Taylor 1 p.111) The 1st order ODE : $\frac{dx}{dx}=L\left(  x,X\left(
x\right)  \right)  $ with

i) $X:I\rightarrow O$, I an interval in $%
\mathbb{R}
$\ , O an open subset of $%
\mathbb{R}
^{m}$

ii) $L:I\times O\rightarrow%
\mathbb{R}
^{m}$ a continuous map such that :

$\forall\left(  x,y_{1}\right)  ,\left(  x,y_{2}\right)  \in I\times
O:\left\Vert L\left(  x,y_{1}\right)  -L\left(  x,y_{2}\right)  \right\Vert
_{E}\leq\lambda\left(  \left\Vert y_{1}-y_{2}\right\Vert \right)  $ where :
$\lambda\in C_{0b}\left(
\mathbb{R}
_{+};%
\mathbb{R}
_{+}\right)  $ is such that $\int\frac{ds}{\lambda\left(  s\right)  }=\infty$

iii) the Cauchy condition $x_{0}\in I,y_{0}\in O:X\left(  x_{0}\right)
=y_{0}$

has a unique solution defined on I
\end{theorem}

\paragraph{Majoration of solutions\newline}

\begin{theorem}
(Schwartz 2 p.370) Any solution g of the scalar first order ODE : $\frac
{dX}{dx}=L(x,X(x)),$ with

i) $X:\left[  a,b\right]  \rightarrow%
\mathbb{R}
_{+}$,

ii) $L:\left[  a,b\right]  \times%
\mathbb{R}
_{+}\rightarrow%
\mathbb{R}
_{+}$ continuous,

iii) $F\in C_{1}\left(  \left[  a,b\right]  ;E\right)  ,$ E affine normed
space such that :

$\exists A\in E:\forall x\in\left[  a,b\right]  :\left\Vert F^{\prime
}(x)\right\Vert <L\left(  x,\left\Vert F(x)-A\right\Vert \right)  $

$y_{0}=\left\Vert F\left(  a\right)  -A\right\Vert $

iv) the Cauchy condition $X\left(  a\right)  =y_{0}$

is such that : $\forall x>a:\left\Vert F\left(  x\right)  -A\right\Vert <g(x)$
\end{theorem}

\paragraph{Differentiability of solutions\newline}

\begin{theorem}
(Schwartz 2 p.377) Any solution of the 1st order ODE : $\frac{dX}{dx}=L\left(
x,X\left(  x\right)  \right)  $ with

i) $X:I\rightarrow O$, J an interval in $%
\mathbb{R}
$\ , O an open subset of an affine Banach space E

ii) $L:I\times O\rightarrow E$ a class r map,

iii) the Cauchy condition $x_{0}\in I,y_{0}\in O:X\left(  x_{0}\right)
=y_{0}$

on $I\times O$ is a class r+1 map on O
\end{theorem}

\bigskip

If E is a finite m dimensional vector space, and if in the neighborhood
$n\left(  x_{0}\right)  $\ of the Cauchy conditions $\left(  x_{0}%
,y_{0}\right)  \in I\times O$ the Cauchy problem has a unique solution, then
there are exactly m independant conservations laws, locally defined on
$n(x_{0}).$

Generally there are no conservations laws globally defined on the whole of O.

\paragraph{Solution depending on a parameter\newline}

1. Existence and continuity

\begin{theorem}
(Schwartz 2 p.353) Let be the 1st order ODE : $\frac{\partial X}{\partial
x}=L\left(  x,\lambda,X\left(  x,\lambda\right)  \right)  $ with :

i) $X:I\times\Lambda\rightarrow O$, I an interval in $%
\mathbb{R}
$\ , O an open subset of an affine Banach space E, $\Lambda$ a topological space

ii) $L:I\times\Lambda\times O\rightarrow E$ a continuous map, globally
Lipschitz with respect to the second variable :

$\exists k\geq0:\forall x\in I,y_{1},y_{2}\in O,\lambda\in\Lambda:\left\Vert
L\left(  x,\lambda,y_{1}\right)  -L\left(  x,\lambda,y_{2}\right)  \right\Vert
_{E}\leq k\left\Vert y_{1}-y_{2}\right\Vert _{E}$

iii) the Cauchy condition : $X\left(  x_{0}\left(  \lambda\right)
,\lambda\right)  =y_{0}\left(  \lambda\right)  $ where

$y_{0}:\Lambda\rightarrow O,x_{0}:\Lambda\rightarrow J$ are continuous maps

Then for any $\lambda_{0}\in\Lambda$\ the Cauchy problem :

$\frac{dX}{dx}=L\left(  x,\lambda_{0},X\left(  x,\lambda_{0}\right)  \right)
,f\left(  x_{0}\left(  \lambda_{0}\right)  ,\lambda_{0}\right)  =y_{0}\left(
\lambda_{0}\right)  $

has a unique solution $X\left(  x,\lambda_{0}\right)  $ and $X\left(
x,\lambda_{0}\right)  \rightarrow X\left(  x,\lambda_{1}\right)  $ uniformly
on any compact of I when $\lambda_{0}\rightarrow\lambda_{1}$
\end{theorem}

2. We have a theorem with weaker Lipschitz conditions :

\begin{theorem}
(Schwartz 2 p.374) Let be the 1st order ODE : $\frac{\partial X}{\partial
x}=L\left(  x,\lambda,X\left(  x,\lambda\right)  \right)  $ with :

i) $X:I\times\Lambda\rightarrow O$, I an interval in $%
\mathbb{R}
$\ , O an open subset of an affine Banach space E, $\Lambda$ a topological space

ii) $L:I\times\Lambda\times O\rightarrow\overrightarrow{E}$ a continuous map,
locally Lipschitz with respect to the second variable :

$\forall\left(  a,y,\mu\right)  \in I\times O\times\Lambda:\exists n\left(
a\right)  \subset I,n\left(  p\right)  \subset O,n\left(  \mu\right)
\in\Lambda,\exists k\geq0:$

$\forall\left(  x,y_{1},\mu\right)  ,\left(  x,y_{2},\nu\right)  \in n\left(
a\right)  \times n\left(  y\right)  \times n\left(  \mu\right)  :$

$\left\Vert L\left(  x,\lambda,y_{1}\right)  -L\left(  x,\lambda,y_{2}\right)
\right\Vert _{E}\leq k\left\Vert y_{1}-y_{2}\right\Vert _{E}$

iii) the Cauchy condition : $X\left(  x_{0}\left(  \lambda\right)
,\lambda\right)  =y_{0}\left(  \lambda\right)  $

where $y_{0}:\Lambda\rightarrow O,x_{0}:\Lambda\rightarrow I$ are continuous maps

Then for any $\lambda_{0}\in\Lambda$\ , there is an interval $I_{0}\left(
\lambda_{0}\right)  \subset I$\ such that :

i) for any compact $K\subset I_{0}\left(  \lambda_{0}\right)  ,$\ there are a
neighborhood $n\left(  X\left(  K,\lambda_{0}\right)  \right)  $ of $X\left(
K,\lambda_{0}\right)  ,$ a neighborhood n$\left(  \lambda_{0}\right)  $ of
$\lambda_{0}$

such that the Cauchy problem has a unique solution on K$\times$n$\left(
\lambda_{0}\right)  $ valued in $n\left(  X\left(  K,\lambda_{0}\right)
\right)  $

ii) $X\left(  .,\lambda\right)  \rightarrow X\left(  .,\lambda_{0}\right)  $
when $\lambda\rightarrow\lambda_{0}$ uniformly on K

iii) f is continuous in $I_{0}\left(  \lambda_{0}\right)  \times\Lambda$
\end{theorem}

3. Differentiability of the solution with respect to the parameter :

\begin{theorem}
(Schwartz 2 p.401) Let be the 1st order ODE : $\frac{\partial X}{\partial
x}=L\left(  x,\lambda,X\left(  x,\lambda\right)  \right)  $ with :

i) $X:I\times\Lambda\rightarrow O$, I an interval in $%
\mathbb{R}
$\ , O an open subset of an affine Banach space E, $\Lambda$ a topological space

ii) $L:I\times\Lambda\times O\rightarrow\overrightarrow{E}$ a continuous map,
with a continuous partial derivative $\frac{\partial L}{\partial y}%
:I\times\Lambda\times O\rightarrow%
\mathcal{L}%
\left(  \overrightarrow{E};\overrightarrow{E}\right)  $

iii) the Cauchy condition : $X\left(  x_{0}\left(  \lambda\right)
,\lambda\right)  =y_{0}\left(  \lambda\right)  $ where $y_{0}:\Lambda
\rightarrow O,x_{0}:\Lambda\rightarrow I$ are continuous maps

If for $\lambda_{0}\in\Lambda$ the ODE has a solution $X_{0}$ defined on I\ then:

i) there is a neighborhood n$\left(  \lambda_{0}\right)  $ such that the ODE
has a unique solution $X\left(  x,\lambda\right)  $ for $\left(
x,\lambda\right)  \in I\times n\left(  \lambda_{0}\right)  $

ii) the map $X:n\left(  \lambda\right)  \rightarrow C_{b}\left(
I;\overrightarrow{E}\right)  $ is continuous

iii) if $\Lambda$ is an open subset of an affine normed space F, $L\in
C_{r}\left(  I\times\Lambda\times O;\overrightarrow{E}\right)  $,$r\geq1$ then
the solution $X\in C_{r+1}\left(  I\times\Lambda;\overrightarrow{E}\right)  $

iv) Moreover if $x_{0}\in C_{1}\left(  \Lambda;I\right)  ,y_{0}\in
C_{1}\left(  \Lambda;O\right)  $\ the derivative

$\varphi\left(  x\right)  =\frac{\partial X}{\partial\lambda}\left(
x,\lambda\right)  \in C\left(  I;%
\mathcal{L}%
\left(  \overrightarrow{F};\overrightarrow{E}\right)  \right)  $ is solution
of the ODE :

$\frac{d\varphi}{dx}=\frac{\partial L}{\partial y}\left(  x_{0}\left(
\lambda\right)  ,\lambda_{0},X_{0}\left(  x\right)  \right)  \circ
\varphi\left(  x\right)  +\frac{\partial L}{\partial\lambda}\left(
x_{0}\left(  \lambda\right)  ,\lambda_{0},X_{0}\left(  x\right)  \right)  $

with the Cauchy conditions :

$\varphi\left(  x_{0}\left(  \lambda_{0}\right)  \right)  =\frac{dy_{0}%
}{d\lambda}\left(  \lambda_{0}\right)  -L\left(  x_{0}\left(  \lambda
_{0}\right)  ,\lambda_{0},X_{0}\left(  x_{0}\left(  \lambda_{0}\right)
\right)  \right)  \frac{dx_{0}}{d\lambda}\left(  \lambda_{0}\right)  $
\end{theorem}

6. Differentiability of solutions with respect to the initial conditions

\begin{theorem}
(Taylor 1.p.28) If there is a solution $Y\left(  x,x_{0}\right)  $ of the 1st
order ODE : $\frac{dX}{dx}=L\left(  X\left(  x\right)  \right)  $ with:

i) $X:I\rightarrow O$, I an interval in $%
\mathbb{R}
$\ , O an open convex subset of a Banach vector space E

ii) $L:O\rightarrow E$ a class 1 map

iii) the Cauchy condition $x_{0}\in I,y_{0}\in O:X\left(  x_{0}\right)
=y_{0}$

over I, whenever $x_{0}\in I,$ then

i) Y is continuously differentiable with respect to $x_{0}$

ii) the partial derivative : $\varphi\left(  x,x_{0}\right)  =\frac{\partial
Y}{\partial x_{0}}$ is solution of the ODE :

$\frac{\partial\varphi}{\partial x}\left(  x,x_{0}\right)  =\frac{\partial
}{\partial y}L\left(  f\left(  x,x_{0}\right)  \right)  \varphi\left(
x,x_{0}\right)  ,\varphi\left(  x_{0},x_{0}\right)  =y_{0}$

iii) If $L\in C_{r}\left(  I;O\right)  $ then $Y\left(  x,x_{0}\right)  $ is
of class r in $x_{0}$

iv) If E is finite dimensional and L real analytic, then $Y\left(
x,x_{0}\right)  $ is real analytic in $x_{0}$
\end{theorem}

\paragraph{ODE on manifolds\newline}

So far we require only initial conditions from the solution. One can extend
the problem to the case where X is a path on a manifold.

\begin{theorem}
(Schwartz p.380) A class 1 vector field V on a a real class r
$>$%
1 m dimensional manifold M with charts $\left(  O_{i},\varphi_{i}\right)
_{i\in I}$ is said to be locally Lipschitz if for any p in M there is a
neighborhood and an atlas of M such that the maps giving the components of V
in a holonomic basis $v_{i}:\varphi_{i}\left(  O_{i}\right)  \rightarrow%
\mathbb{R}
^{m}$ are Lipschitz.

The problem, find :

$c:I\rightarrow M$ where I is some interval of $%
\mathbb{R}
$\ which comprises 0

$c^{\prime}(t)=V\left(  c\left(  t\right)  \right)  $

$V(0)=p$ where p is some fixed point in M

defines a system of 1st order ODE, expressed in charts of M and the components
$\left(  v^{\alpha}\right)  _{\alpha=1}^{m}$ of V.

If V is locally Lipschitz, then for any p in M, there is a maximal interval J
such that there is a unique solution of these equations.
\end{theorem}

\subsubsection{Linear ODE}

(Schwartz 2 p.387)

\paragraph{General theorems\newline}

\begin{definition}
The first order ODE : $\frac{dX}{dx}=L(x)X(x)$ with:

i) $X:I\rightarrow E$, I an interval in $%
\mathbb{R}
$\ , E a Banach vector space

ii) $\forall x\in I:L(x)\in%
\mathcal{L}%
\left(  E;E\right)  $

iii) the Cauchy conditions : $x_{0}\in I,y_{0}\in E:X\left(  x_{0}\right)
=y_{0}$

is a linear 1st order ODE
\end{definition}

\begin{theorem}
For any Cauchy conditions a first order linear ODE has a unique solution
defined on I

If $\forall x\in I:\left\Vert L\left(  x\right)  \right\Vert \leq k$ then
$\forall x\in J:$ $\left\Vert X\left(  x\right)  \right\Vert \leq\left\Vert
y_{0}\right\Vert e^{k\left\vert x\right\vert }$

If $L\in C_{r}\left(  I;%
\mathcal{L}%
\left(  E;E\right)  \right)  $ then the solution is a class r map : $X\in
C_{r}\left(  I;E\right)  $

The set V of solutions of the Cauchy problem, when $\left(  x_{0}%
,y_{0}\right)  $ varies on $I\times E$, is a vector subspace of $C_{r}\left(
I;E\right)  $ and the map : $F:V\rightarrow E::F\left(  X\right)  =X\left(
x_{0}\right)  $ is a linear map.
\end{theorem}

If E is finite m dimensional then V is m dimensional.

If m=1 the solution is given by : $X\left(  x\right)  =y_{0}\exp\int_{x_{0}%
}^{x}A\left(  \xi\right)  d\xi$

\paragraph{Resolvent\newline}

\begin{theorem}
For a first order linear ODE there is a unique map : $R:E\times E\rightarrow%
\mathcal{L}%
\left(  E;E\right)  $ called evolution operator (or resolvent) characterized
by :%

\begin{equation}
\forall X\in V,\forall x_{1},x_{2}\in E:X\left(  x_{2}\right)  =R\left(
x_{2},x_{1}\right)  X\left(  x_{1}\right)
\end{equation}

\end{theorem}

R has the following properties :

$\forall x_{1},x_{2},x_{3}\in E:$

$R\left(  x_{3},x_{1}\right)  =R\left(  x_{3},x_{2}\right)  \circ R\left(
x_{2},x_{1}\right)  $

$R\left(  x_{1},x_{2}\right)  =R\left(  x_{2},x_{1}\right)  ^{-1}$

$R\left(  x_{1},x_{1}\right)  =Id$

R is the unique solution of the ODE :

$\forall\lambda\in E:\frac{\partial R}{\partial x}\left(  x,\lambda\right)
=L\left(  x\right)  R\left(  x,\lambda\right)  $

$R\left(  x,x\right)  =Id$

If $L\in C_{r}\left(  I;%
\mathcal{L}%
\left(  E;E\right)  \right)  $ then the resolvant is a class r map : $R\in
C_{r}\left(  I\times I;%
\mathcal{L}%
\left(  E;E\right)  \right)  $

\paragraph{Affine equation\newline}

\begin{definition}
An affine 1st order ODE (or inhomogeneous linear ODE) is
\begin{equation}
\frac{dX}{dx}=L(x)X(x)+M\left(  x\right)
\end{equation}

with :

i) $X:I\rightarrow E$, I an interval in $%
\mathbb{R}
$\ , E a Banach vector space

ii) $\forall x\in I:L(x)\in%
\mathcal{L}%
\left(  E;E\right)  $

iii) $M:I\rightarrow E$ is a continuous map

iv) Cauchy conditions : $x_{0}\in I,y_{0}\in E:X\left(  x_{0}\right)  =y_{0}$
\end{definition}

The homogeneous linear ODE associated is given by $\frac{dX}{dx}=L(x)X(x)$

If g is a solution of the affine ODE, then any solution of the affine ODE is
given by : $X=g+\varphi$ where $\varphi$ is the general solution of :
$\frac{d\varphi}{dx}=L(x)\varphi(x)$

For any Cauchy conditions $\left(  x_{0},y_{0}\right)  $ the affine ODE has a
unique solution given by :%

\begin{equation}
X\left(  x\right)  =R\left(  x,x_{0}\right)  y_{0}+\int_{x_{0}}^{x}R\left(
x,\xi\right)  M\left(  \xi\right)  d\xi
\end{equation}

where R is the resolvent of the associated homogeneous equation.

If $E=%
\mathbb{R}
$ the solution reads :

$X\left(  x\right)  =e^{\int_{x_{0}}^{x}A\left(  \xi\right)  d\xi}\left(
y_{0}+\int_{x_{0}}^{x}M\left(  \xi\right)  e^{-\int_{x_{0}}^{\xi}A\left(
\eta\right)  d\eta}d\xi\right)  $

$X\left(  x\right)  =y_{0}e^{\int_{x_{0}}^{x}A\left(  \xi\right)  d\xi}%
+\int_{x_{0}}^{x}M\left(  \xi\right)  e^{\int_{\xi}^{x}A\left(  \eta\right)
d\eta}d\xi$

If L is a constant operator $L\in%
\mathcal{L}%
\left(  E;E\right)  $ and M a constant vector\ then the solution of the affine
equation is :

$X\left(  x\right)  =e^{(x-x_{0})A}\left(  y_{0}+\int_{x_{0}}^{x}%
Me^{-(\xi-x_{0})A}d\xi\right)  $

$=e^{(x-x_{0})A}y_{0}+\int_{x_{0}}^{x}Me^{(x-\xi)A}d\xi$

$R(x_{1},x_{2})=\exp[(x_{1}-x_{2})A]$

\paragraph{Time ordered integral\newline}

For the linear 1st order ODE $\frac{dX}{dx}=L(x)X(x)$ with:

i) $X:I\rightarrow%
\mathbb{R}
^{m}$, I an interval in $%
\mathbb{R}
$\ 

ii) $L\in C_{0}\left(  J;%
\mathcal{L}%
\left(
\mathbb{R}
^{m};%
\mathbb{R}
^{m}\right)  \right)  $

iii) Cauchy conditions : $x_{0}\in I,y_{0}\in%
\mathbb{R}
^{m}:X\left(  x_{0}\right)  =y_{0}$

the map L and the resolvent R are m$\times$m matrices.

For $a,b\in I$ , $n\in%
\mathbb{N}
$ , and a partition of $\left[  a,b\right]  :$

$\left[  a=t_{0},t_{1},...,t_{n}=b\right]  ,\Delta_{k}=t_{k}-t_{k-1}$ ,

if X is Riemann integrable and $X\left(  x\right)  \geq0$ then :

$R\left(  b,a\right)  =\lim_{n\rightarrow\infty}%
{\textstyle\prod\limits_{k=0}^{k=n}}
\exp\left(  \Delta_{k}L\left(  x_{k}\right)  \right)  =\lim_{n\rightarrow
\infty}%
{\textstyle\prod\limits_{k=0}^{k=n}}
\left(  I+L\left(  x_{k}\right)  \Delta_{k}\right)  $

$=\lim_{n\rightarrow\infty}%
{\textstyle\prod\limits_{k=0}^{k=n-1}}
\int_{x_{k}}^{x_{k+1}}\left(  \exp L\left(  \xi\right)  \right)  d\xi$

So :

$X\left(  x\right)  =R\left(  x,x_{0}\right)  X\left(  x_{0}\right)  =\left(
\lim_{n\rightarrow\infty}%
{\textstyle\prod\limits_{k=0}^{k=n}}
\exp\left(  \Delta_{k}L\left(  x_{k}\right)  \right)  \right)  y_{0}$

\bigskip

\subsection{Partial differential equations (PDE)}

\label{PDE}

\subsubsection{General definitions and results}

\paragraph{Definition of a PDE\newline}

The most general and, let us say, the "modern" way to define a differential
equation is through the jet formalism.

\bigskip

\begin{definition}
A differential equation of order r is a closed subbundle F of the r jet
extension $J^{r}E$ of a fibered manifold E.
\end{definition}

\bigskip

If the fibered manifold E is $E(N,\pi)$ then :

- the base of F is a submanifold M of N and $\pi_{F}^{r}=\pi_{E}^{r}|_{M}.$

- the fibers of F over M are themselves defined usually through a family of
differential operators with conditions such as: $D_{k}\left(  X\right)  =0.$

- a solution of the PDE is a section $X\in\mathfrak{X}_{r}\left(  E\right)  $
such that

$\forall x\in M:J^{r}X\left(  x\right)  \in F$

- the set of solutions is a the subset \ $S=\left(  J^{r}\right)  ^{-1}\left(
F\right)  \subset\mathfrak{X}_{r}\left(  E\right)  $

If S is a vector space we have a homogeneous PDE. Then the "superposition
principle" applies : any linear combination (with fixed coefficients) of
solutions is still a solution.

If S is an affine space then the underlying vector space gives "general
solutions" and any solution of the PDE is obtained as the sum of a particular
solution and any general solution.

Differential equations can be differentiated, and give equations of higher
order.\ If the s jet extension $J^{s}F$ of the subbundle F is a differential
equation (meaning a closed subbundle of $J^{r+s}E)$ the equation is said to be
regular. A necessary condition for a differential equation to have a solution
is that the maps : $J^{s}F\rightarrow F$ are onto, and then, if it is regular,
there is a bijective correspondance between the solution of the r+s order
problem and the r order problem.

Conversely an \textbf{integral} of a differential equation is a section
$Y\in\mathfrak{X}_{k}\left(  E\right)  ,k<r$ such that $J^{r-k}Y\in F.$ In
physics it appears quite often as a conservation law : the quantity Y is
preserved inside the problem. Indeed if 0 belongs to F then Y=Constant brings
a solution. It can be used to lower the order of a differential equation : F
is replaced by a a subbundle G of $J^{k}E$ defined through the introduction of
additional parameters related to Y and the problem becomes of order k :
$J^{k}X\in G.$

The most common and studied differential equations are of the kinds :

\paragraph{Dirichlet problems\newline}

$D:J^{r}E_{1}\rightarrow E_{2}$ is a r order differential operator between two
smooth complex finite dimensional vector bundles $E_{1}\left(  N,V_{1},\pi
_{1}\right)  ,E_{2}\left(  N,V_{2},\pi_{2}\right)  $\ on the same real
manifold N.

M is a manifold with boundary (so M itself is closed) in N, which defines two
subbundles in $E_{1},E_{2}$ with base $\overset{\circ}{M},$ denoted
$M_{1},M_{2}$\ and their r jet prolongations.

A solution of the PDE is a section $X\in\mathfrak{X}\left(  E_{1}\right)
$\ such that :

i) for $x\in\overset{\circ}{M}$ $:D\left(  J^{r}X\right)  =Y_{0}$ where
$Y_{0}$ is a given section on $M_{2}$\ meaning : X is a r differentiable map
for $x\in\overset{\circ}{M}$ and $\forall x\in\overset{\circ}{M}:D\left(
x\right)  \left(  j_{x}^{r}X\right)  =Y_{0}\left(  x\right)  $

ii) for $x\in\partial M:X\left(  x\right)  =Y_{1}$ where $Y_{1}$ is a given
section on $M_{1}$

So if $Y_{0},Y_{1}=0$ the problem is homogeneous.

\paragraph{Neumann problems\newline}

$D:J^{r}E\rightarrow E$ is a r order scalar differential operator on complex
functions over a manifold N : $E=C_{r}\left(  M;%
\mathbb{C}
\right)  $.

M is a manifold with smooth riemannian boundary ($\partial M$ is a
hypersurface) in N, which defines the subbundle with base $\overset{\circ}{M}$

A solution of the PDE is a section $X\in\mathfrak{X}\left(  E\right)  $\ such
that :

i) for $x\in\overset{\circ}{M}$ $:D\left(  J^{r}X\right)  =Y_{0}$ where
$Y_{0}$ is a given section on $\overset{\circ}{M}$\ meaning : X is a r
differentiable map for $x\in\overset{\circ}{M}$ and $\forall x\in
\overset{\circ}{M}:D\left(  x\right)  \left(  j_{x}^{r}X\right)  =Y_{0}\left(
x\right)  $

ii) for $x\in\partial M:X^{\prime}\left(  x\right)  n=0$ where n is the
outward oriented normal to $\partial M$

So if $Y_{0}=0$ the problem is homogeneous.

\paragraph{Evolution equations\newline}

N is a m dimensional real manifold.

D is a r order scalar differential operator acting on complex functions on $%
\mathbb{R}
\times N$ (or $%
\mathbb{R}
_{+}\times N)$ seen as maps $u\in C\left(
\mathbb{R}
;C\left(  N;%
\mathbb{C}
\right)  \right)  $

So there is a family of operators D(t) acting on functions u(t,x) for t fixed

There are three kinds of PDE :

1. Cauchy problem :

The problem is to find $u\in C\left(
\mathbb{R}
;C\left(  N;%
\mathbb{C}
\right)  \right)  $ such that :

i) $\forall t,\forall x\in N:D\left(  t\right)  \left(  x\right)  \left(
J_{x}^{r}u\right)  =f\left(  t,x\right)  $ where f is a given function on $%
\mathbb{R}
\times N$

ii) with the initial conditions, called Cauchy conditions :

u(t,x) is continuous for t=0 (or $t\rightarrow0_{+})$ and

$\forall x\in N:\frac{\partial^{s}}{\partial t^{s}}u\left(  0,x\right)
=g_{s}\left(  x\right)  ,s=0...r-1$

2. Dirichlet problem :

M is a manifold with boundary (so M itself is closed) in N

The problem is to find $u\in C\left(
\mathbb{R}
;C\left(  \overset{\circ}{M};%
\mathbb{C}
\right)  \right)  $ such that :

i) $\forall t,\forall x\in\overset{\circ}{M}:D\left(  t\right)  \left(
x\right)  \left(  J_{x}^{r}u\right)  =f\left(  t,x\right)  $ where f is a
given function on $%
\mathbb{R}
\times\overset{\circ}{M}$

ii) with the initial conditions, called Cauchy conditions :

u(t,x) is continuous for t=0 (or $t\rightarrow0_{+})$ and

$\forall x\in\overset{\circ}{M}:\frac{\partial^{s}}{\partial t^{s}}u\left(
0,x\right)  =g_{s}\left(  x\right)  ,s=0...r-1$

iii) and the Dirichlet condition :

$\forall t,\forall x\in\partial M:u\left(  t,x\right)  =h\left(  t,x\right)  $
where h is a given function on $\partial M$

3. Neumann problem :

M is a manifold with smooth riemannian boundary ($\partial M$ is a
hypersurface) in N

The problem is to find $u\in C\left(
\mathbb{R}
;C\left(  \overset{\circ}{M};%
\mathbb{C}
\right)  \right)  $ such that :

i) $\forall t,\forall x\in\overset{\circ}{M}:D\left(  t\right)  \left(
x\right)  \left(  J_{x}^{r}u\right)  =f\left(  t,x\right)  $ where f is a
given function on $%
\mathbb{R}
\times\overset{\circ}{M}$

ii) with the initial conditions, called Cauchy conditions :

u(t,x) is continuous for t=0 (or $t\rightarrow0_{+})$ and

$\forall x\in\overset{\circ}{M}:\frac{\partial^{s}}{\partial t^{s}}u\left(
0,x\right)  =g_{s}\left(  x\right)  ,s=0...r-1$

iii) and the Neumann condition :

$\forall t,\forall x\in\partial M:\frac{\partial}{\partial x}u\left(
t,x\right)  n=0$ where n is the outward oriented normal to $\partial M$

\paragraph{Box boundary\newline}

The PDE is to find X such that :

$DX=0$ in $\overset{\circ}{M}$ where $D:J^{r}E_{1}\rightarrow E_{2}$ is a r
order differential operator

$X=U$ on $\partial M$

and the domain M is a rectangular box of $%
\mathbb{R}
^{m}:$ $M=\left\{  a_{\alpha}\leq x^{\alpha}\leq b_{\alpha},\alpha
=1...m\right\}  ,A=\sum_{\alpha}a^{\alpha}\varepsilon_{\alpha}$

X can always be replaced by Y defined in $%
\mathbb{R}
^{m}$\ and periodic :

$\forall Z\in Z^{m},x\in M:Y\left(  x+ZA\right)  =X\left(  x\right)  $ with
$A=\sum_{\alpha}\left(  b_{\alpha}-a_{\alpha}\right)  \varepsilon_{\alpha}$

and the ODE becomes:

Find Y such that : $DY=0$ in $%
\mathbb{R}
^{m}$ and use series Fourier on the components of Y.

\subsubsection{Linear PDE}

\paragraph{General theorems\newline}

\begin{theorem}
Cauchy-Kowalesky theorem (Taylor 1 p.433) : The PDE :

Find $u\in C\left(
\mathbb{R}
^{m+1};%
\mathbb{C}
\right)  $ such that :

$Du=f$

$u\left(  t_{0},x\right)  =g_{0}\left(  x\right)  ,...\frac{\partial^{s}%
}{\partial t^{s}}u\left(  t_{0},x\right)  =g_{s}\left(  x\right)  ,s=0..r-1$

where D is a scalar linear differential operator on $C\left(
\mathbb{R}
^{m+1};%
\mathbb{C}
\right)  $ :

$D\left(  \varphi\right)  =\frac{\partial^{r}}{\partial t^{r}}\varphi
+\sum_{s=0}^{r-1}A^{\alpha_{1}...\alpha_{s}}\left(  t,x\right)  \frac
{\partial^{s}}{\partial\xi^{\alpha_{1}}..\partial\xi^{\alpha_{s}}}\varphi$

If $A^{\alpha_{1}...\alpha_{s}}$ are real analytic in a neighborhood $n_{0} $
of $\left(  t_{0},x_{0}\right)  $ and $g_{s}$ are real analytic in a
neighborhood of $\left(  x_{0}\right)  ,$then in $n_{0}$\ there is a unique
solution u(t,x)
\end{theorem}

\begin{theorem}
(Taylor 1 p.248) The PDE : find $u\in C\left(
\mathbb{R}
^{m};%
\mathbb{C}
\right)  :Du=f$ in $B_{R}=\left\{  \left\Vert x\right\Vert <R\right\}  $ where
D is a scalar linear r order differential operator D on $%
\mathbb{R}
^{m}$ with constant coefficients, and $f\in C_{\infty c}\left(
\mathbb{R}
^{m};%
\mathbb{C}
\right)  ^{\prime},$\ has always a solution.
\end{theorem}

\paragraph{Fundamental solution\newline}

see Linear Differential operator

\paragraph{General linear elliptic boundary problems\newline}

(Taylor 1 p.380-395) The PDE is : find $X\in\mathfrak{X}_{r}\left(
E_{1}\right)  $ such that :

$DX=Y_{0}$ on $\overset{\circ}{M}$

$D_{j}X=Y_{j},j=1...N$ on $\partial M$

where :

$E_{1},E_{2}$ are vector bundles on the same smooth compact manifold M with boundary

$F_{j}$ are vector bundles on $\partial M$

$D:\mathfrak{X}_{r}\left(  E_{1}\right)  \rightarrow\mathfrak{X}\left(
E_{2}\right)  $ is a weakly elliptic r order linear differential operator

$D_{j}:\mathfrak{X}_{r}\left(  E_{1}\right)  \rightarrow\mathfrak{X}\left(
F_{j}\right)  ,j=1..N$ are $r_{j}$ order linear differential operators

The problem is said to be regular if for any solution :

$\exists s\in%
\mathbb{R}
,\exists C\in%
\mathbb{R}
:$

$\left\Vert X\right\Vert _{H^{s+r}\left(  E_{1}\right)  }^{2}\leq C\left(
\left\Vert DX\right\Vert _{H^{r}\left(  E_{2}\right)  }^{2}+\sum_{j}\left\Vert
D_{j}X\right\Vert _{H^{r+s-r_{j}-1/2}\left(  F_{j}\right)  }^{2}+\left\Vert
X\right\Vert _{H^{r+s-1}\left(  E_{1}\right)  }^{2}\right)  $

If the problem is regular elliptic, then for $k\in%
\mathbb{N}
$ the map :

$\phi:H^{r+k}\left(  E_{1}\right)  \rightarrow H^{k}\left(  E_{1}\right)
\oplus\oplus_{j=1}^{N}H^{r+k-r_{j}-1/2}\left(  F_{j}\right)  $

defined by :

$\phi\left(  X\right)  =DX$ on $\overset{\circ}{M}$ and $\phi\left(  X\right)
=D_{j}X=Y_{j},j=1...N$ on $\partial M$

is Fredholm. $\phi$ has a finite dimensional kernel, closed range, and its
range has finite codimension. It has a right Fredholm inverse.

So the problem has a solution, which is given by the inverse of $\phi.$

As a special case we have the following

\begin{theorem}
(Taylor 1 p.393) M is a riemannian manifold with boundary, n is the unitary
normal outward oriented to $\partial M$.

The PDE find $u\in\Lambda_{r}\left(  M;%
\mathbb{C}
\right)  $ such that $\ \Delta u=f$ on $\overset{0}{M}$ and

Problem 1 : $n\wedge u=g_{0},n\wedge\delta u=g_{1}$ on $\partial M$

Problem 2 : $i_{n}u=g_{0},i_{n}du=g_{1}$ on $\partial M$

are both regular and have a solution.
\end{theorem}

\paragraph{Hyperbolic PDE\newline}

A PDE is said to be \textbf{hyperbolic} at a point u if it is an evolution
equation with Cauchy conditions such that in a neighborhood of p, there is a
unique solution for any initial conditions.

\begin{theorem}
(Taylor 1 p.435) The PDE is to find $u\in C\left(  M;%
\mathbb{C}
\right)  $ such that :

$Du=f$ on M

$u(x)=g_{0}$ on $S_{0},$

$Yu=g_{t}$ on $S_{t}$

where

M is a m+1 dimensional manifold endowed with a Lorentz metric of signature
(m-,1+), folliated by compact space like hypersurfaces $S_{t}.$

D is the scalar operator $D=\square+P$ with a\ first order differential
operator P

Y is a vector field transverse to the $S_{t}$

If $f\in H^{k-1}\left(  M\right)  ,g_{0}\in H^{k}\left(  S_{0}\right)
,g_{t}\in H^{k-1}\left(  S_{t}\right)  ,k\in%
\mathbb{N}
,k>0$ then there is a unique solution $u\in H^{k}\left(  M\right)  $ which
belongs to $H^{1}\left(  \Omega\right)  ,\Omega$ being the region swept by
$S_{t},t\in\left[  0,T\right]  $
\end{theorem}

\subsubsection{Poisson like equations}

They are PDE with the scalar laplacian on a riemannian manifold as operator.
Fundamental solutions for the laplacian are given through the Green's function
denoted here G, itself built from eigen vectors of $-\Delta$ (see Differential operators)

\paragraph{Poisson equation\newline}

The problem is to find a function u on a manifold M such that : $-\Delta u=f$
where f is a given function. If f=0 then it is called the \textbf{Laplace}
\textbf{equation} (and the problem is then to find harmonic functions).

This equation is used in physics whenever a force field is defined by a
potential depending on the distance from the sources, such that the electric
or the gravitational field (if the charges move or change then the wave
operator is required).

\bigskip

1. Existence of solutions :

\begin{theorem}
(Gregor'yan p.45) If O is a relatively compact open in a riemannian manifold M
such that $M\backslash\overline{O}\neq\varnothing$\ then the Green function G
on M is finite and : $\forall f\in L^{2}\left(  M,\varpi_{0},%
\mathbb{R}
\right)  ,u\left(  x\right)  =\int_{M}G\left(  x,y\right)  f\left(  y\right)
\varpi_{0}\left(  y\right)  $ is the unique solution of $-\Delta u=f$
\end{theorem}

The solutions are not necessarily continuous or differentiable (in the usual
sense of functions). Several cases arise (see Lieb p.262). However :

\begin{theorem}
(Lieb p.159) If O is an open subset of $%
\mathbb{R}
^{m}$\ and $f\in L_{loc}^{1}\left(  O,dx,%
\mathbb{C}
\right)  $ then $u\left(  x\right)  =\int_{O}G\left(  x,y\right)  f\left(
y\right)  dy$ is such that $-\Delta u=f$ and $u\in L_{loc}^{1}\left(  O,dx,%
\mathbb{C}
\right)  .$ Moreover for almost every x : $\partial_{\alpha}u=\int_{O}%
\partial_{\alpha}G\left(  x,y\right)  f\left(  y\right)  dy$ where the
derivative is in the sense of distribution if needed. \ If $f\in L_{c}%
^{p}\left(  O,dx,%
\mathbb{C}
\right)  $ with p
$>$
m then u is differentiable. The solution is given up to a harmonic function :
$\Delta u=0$
\end{theorem}

\begin{theorem}
(Taylor 1 p.210) If $S\in S\left(
\mathbb{R}
^{m}\right)  ^{\prime}$ is such that $\Delta S=0$ then $S=T\left(  f\right)  $
with f a polynomial in $%
\mathbb{R}
^{m}$
\end{theorem}

\bigskip

2. Newtons's theorem : in short it states that a spherically symmetric
distribution of charges can be replaced by a single charge at its center.

\begin{theorem}
If $\mu_{+},\mu_{-}$ are positive Borel measure on $%
\mathbb{R}
^{m},\mu=\mu_{+}-\mu_{-},\nu=\mu_{+}+\mu_{-}$ such that $\int_{%
\mathbb{R}
^{m}}\phi_{m}\left(  y\right)  \nu\left(  y\right)  <\infty$ then $V\left(
x\right)  =\int_{%
\mathbb{R}
^{m}}G\left(  x,y\right)  \mu\left(  y\right)  \in L_{loc}^{1}\left(
\mathbb{R}
^{m},dx,%
\mathbb{R}
\right)  $

If $\mu$ is spherically symmetric (meaning that $\mu\left(  A\right)
=\mu\left(  \rho\left(  A\right)  \right)  $ for any rotation $\rho)$ then :
$\left\vert V\left(  x\right)  \right\vert \leq\left\vert G\left(  0,x\right)
\right\vert \int_{%
\mathbb{R}
^{m}}\nu\left(  y\right)  $

If for a closed ball B(0,r) centered in 0 and with radius r $\forall A\subset%
\mathbb{R}
^{m}$ : $A\cap B\left(  0,r\right)  =\varnothing\Rightarrow\mu\left(
A\right)  =0$ then : $\forall x:\left\Vert x\right\Vert >r:V\left(  x\right)
=G\left(  0,x\right)  \int_{%
\mathbb{R}
^{m}}\nu\left(  y\right)  $
\end{theorem}

The functions $\phi_{m}$ are :

m%
$>$%
2 : $\phi_{m}\left(  y\right)  =\left(  1+\left\Vert y\right\Vert \right)
^{2-m}$

m=2 :$\phi_{m}\left(  y\right)  =\ln\left(  1+\left\Vert y\right\Vert \right)
$

m=1 : $\phi_{m}\left(  y\right)  =\left\Vert y\right\Vert $

\paragraph{Dirichlet problem\newline}

\begin{theorem}
(Taylor 1 p.308) The PDE : find $u\in C\left(  M;%
\mathbb{C}
\right)  $\ such that :

$\Delta u=0$ on $\overset{\circ}{M}$

$u=f$ on $\partial M$

where

M is a riemannian compact manifold with boundary.

$f\in C_{\infty}\left(  \partial M;%
\mathbb{C}
\right)  $\ 

has a unique solution u=PI(f) and the map PI has a unique extension :

$PI:H^{s}\left(  \partial M\right)  \rightarrow H^{s+\frac{1}{2}}\left(
\overset{\circ}{M}\right)  $.
\end{theorem}

This problem is equivalent to the following :

find $v:\Delta v=-\Delta F$ on $\overset{\circ}{M},v=0$ on $\partial M$ where
$F\in C_{\infty}\left(  M;%
\mathbb{C}
\right)  $ is any function such that : $F=f$ on $\partial M$

Then there is a unique solution v with compact support in $\overset{\circ}{M}$
(and null on the boundary) given by the inverse of $\Delta$ and u=F + v$. $

If M is the unit ball in $%
\mathbb{R}
^{m}$ with boundary $S^{m-1}$ the map

$PI:H^{s}\left(  S^{m-1}\right)  \rightarrow H^{s+\frac{1}{2}}\left(
B\right)  $ for $s\geq1/2$\ \ is :

$PI(f)\left(  x\right)  =u\left(  x\right)  =\frac{1-\left\Vert x\right\Vert
^{2}}{A\left(  S^{m-1}\right)  }\int_{S^{m-1}}\frac{f\left(  y\right)
}{\left\Vert x-y\right\Vert ^{m}}d\sigma\left(  y\right)  $ with $A\left(
S^{m-1}\right)  =\frac{2\pi^{m/2}}{\Gamma\left(  \frac{m}{2}\right)  }$

\paragraph{Neumann problem\newline}

\begin{theorem}
(Taylor 1 p.350) The PDE : find $u\in C\left(  M;%
\mathbb{C}
\right)  $ such that

$\Delta u=f$ on $\overset{\circ}{M},$

$\frac{\partial u}{\partial n}=0$ on $\partial M$

where

M is a riemannian compact manifold with smooth boundary,

$f\in L^{2}\left(  M,\varpi_{0},%
\mathbb{C}
\right)  $

has a solution $u\in H^{2}\left(  \overset{\circ}{M}\right)  $ iff $\int
_{M}f\varpi_{0}=0$ .Then the solution is unique up to an additive constant and
belongs to $H^{r+2}\left(  \overset{\circ}{M}\right)  $ if $f\in H^{r}\left(
\overset{\circ}{M}\right)  ,r\geq0$
\end{theorem}

\subsubsection{Equations with the operator $-\Delta+P$}

They are equations with the scalar operator $D=-\Delta+P$ where P is a first
order differential operator, which can be a constant scalar, on a riemannian manifold.

The PDE : find $u:-\Delta u=\lambda u$ in $\overset{\circ}{M}$, $u=g$ on
$\partial M$ where $\lambda$\ is a constant scalar comes to find eigenvectors
e such that $e=g$ on $\partial M.$ There are solutions only if $\lambda$ is
one of the eigenvalues (which depend only on M). Then the eigenvectors are
continuous on M and we have the condition : $e_{n}\left(  x\right)
=g_{0}\left(  x\right)  $ on $\partial M.$

\begin{theorem}
(Taylor 1 p.304) The differential operator :

$D=-\Delta+P::H_{c}^{1}\left(  \overset{\circ}{M}\right)  \rightarrow
H^{-1}\left(  \overset{\circ}{M}\right)  $

on a smooth compact manifold M with boundary in a riemannian manifold N,

with a smooth first order differential operator P with smooth coefficients

is Fredholm of index 0.\ It is surjective iff it is injective.

A solution of the PDE :

find \ $u\in H_{c}^{1}\left(  \overset{\circ}{M}\right)  $ such that Du=f on
$\overset{\circ}{M}$ with $f\in H^{k-1}\left(  \overset{\circ}{M}\right)
,k\in%
\mathbb{N}
,$

belongs to $H^{k+1}\left(  \overset{\circ}{M}\right)  $
\end{theorem}

\begin{theorem}
(Zuily p.93) The differential operator : $D=-\Delta+\lambda$ where
$\lambda\geq0$ is a fixed scalar,\ is an isomorphism $H_{c}^{1}\left(
O\right)  \rightarrow H^{-1}\left(  O\right)  $ on an open subset O of $%
\mathbb{R}
^{m},$ it is bounded if $\lambda=0.$
\end{theorem}

\begin{theorem}
(Zuily p.149) The differential operator : $D=-\Delta+\lambda$ where
$\lambda\geq0$ is a fixed scalar,

is, $\forall k\in%
\mathbb{N}
,$ an isomorphism $H^{k+2}(\overset{\circ}{M})\cap H_{c}^{1}(\overset{\circ
}{M})\rightarrow H^{k}(\overset{\circ}{M}),$

and an isomorphism $\left(  \cap_{k\in%
\mathbb{N}
}H^{k}(\overset{\circ}{M})\right)  \cap H_{c}^{1}(\overset{\circ}%
{M})\rightarrow\cap_{k\in%
\mathbb{N}
}H^{k}(\overset{\circ}{M})$

where M is a smooth manifold with boundary of $%
\mathbb{R}
^{m},$ compact if $\lambda=0$
\end{theorem}

\subsubsection{Helmoltz equation}

Also called "scattering problem". The differential operator is $\left(
-\Delta+k^{2}\right)  $ where k is a real scalar

\paragraph{Green's function\newline}

\begin{theorem}
(Lieb p.166) In $%
\mathbb{R}
^{m}$ the fundamental solution of $\left(  -\Delta+k^{2}\right)  U\left(
y\right)  =\delta_{y}$ is $U\left(  y\right)  =T\left(  G\left(  x,y\right)
\right)  $ where the Green's function G is given for $m\geq1,k>0$ by:
$G\left(  x,y\right)  =\int_{0}^{\infty}\left(  4\pi\zeta\right)  ^{-m/2}%
\exp\left(  -\frac{\left\Vert x-y\right\Vert ^{2}}{4\zeta}-k^{2}\zeta\right)
d\zeta$
\end{theorem}

G is called the "Yukawa potential"

G is symmetric decreasing,
$>$
0 for $x\neq y$

$\int_{%
\mathbb{R}
^{m}}\int_{0}^{\infty}\left(  4\pi\zeta\right)  ^{-m/2}\exp\left(
-\frac{\left\Vert \xi\right\Vert ^{2}}{4\zeta}-k^{2}\zeta\right)  d\zeta
d\xi=k^{-2}$

when $\xi\rightarrow0:\int_{0}^{\infty}\left(  4\pi\zeta\right)  ^{-m/2}%
\exp\left(  -\frac{\left\Vert \xi\right\Vert ^{2}}{4\zeta}-k^{2}\zeta\right)
d\zeta\rightarrow1/2k$ for m=1, and $\sim\frac{1}{\left(  2-m\right)  A\left(
S_{m-1}\right)  }\left\Vert \xi\right\Vert ^{2-m}$ for m%
$>$%
1

when $\xi\rightarrow\infty:-\ln\left(  \int_{0}^{\infty}\left(  4\pi
\zeta\right)  ^{-m/2}\exp\left(  -\frac{\left\Vert \xi\right\Vert ^{2}}%
{4\zeta}-k^{2}\zeta\right)  d\zeta\right)  \sim k\left\Vert \xi\right\Vert $

The Fourier transform of $\int_{0}^{\infty}\left(  4\pi\zeta\right)
^{-m/2}\exp\left(  -\frac{\left\Vert \xi\right\Vert ^{2}}{4\zeta}-k^{2}%
\zeta\right)  d\zeta$ is :

$\left(  2\pi\right)  ^{-m/2}\left(  \left\Vert t\right\Vert ^{2}%
+k^{2}\right)  ^{-1}$

\paragraph{General problem}

\begin{theorem}
If $f\in L^{p}\left(
\mathbb{R}
^{m};dx,%
\mathbb{C}
\right)  ,1\leq p\leq\infty,$ then $u\left(  x\right)  =\int_{%
\mathbb{R}
^{m}}G\left(  x,y\right)  f\left(  y\right)  dy$ is the unique solution of
$\left(  -\Delta+k^{2}\right)  u=f$ such that $u\left(  x\right)  \in
L^{r}\left(
\mathbb{R}
^{m};dx,%
\mathbb{C}
\right)  $ for some r.
\end{theorem}

\begin{theorem}
(Lieb p.257) If $f\in L^{p}\left(
\mathbb{R}
^{m};dx,%
\mathbb{C}
\right)  ,1\leq p\leq\infty$ is such that : $\left(  -\Delta+k^{2}\right)
T\left(  f\right)  =0$ then f=0
\end{theorem}

\paragraph{Dirichlet problem\newline}

This is the "scattering problem" proper.

\begin{theorem}
(Taylor 2 p.147) The PDE is to find a function $u\in C\left(
\mathbb{R}
^{3};%
\mathbb{C}
\right)  $ such that :

$-\left(  \Delta+k^{2}\right)  u=0$ in O with a scalar k
$>$
0

$u=f$ on $\partial K$

$\left\Vert ru(x)\right\Vert <C,r\left(  \frac{\partial u}{\partial
r}-iku\right)  \rightarrow0$ when $r=\left\Vert x\right\Vert \rightarrow
\infty$

where K is a compact connected smooth manifold with boundary in $%
\mathbb{R}
^{3}$ with complement the open O

i) if f=0 then the only solution is u=0

ii) if $f\in H^{s}\left(  \partial K\right)  $ there is a unique solution u in
H$_{loc}^{s+\frac{1}{2}}\left(  O\right)  $
\end{theorem}

\subsubsection{Wave equation}

In physics, the mathematical model for a force field depending on the distance
to the sources is no longer the Poisson equation when the sources move or the
charges change, but the wave equation, to account for the propagation of the field.

\paragraph{Wave operator\newline}

On a manifold endowed with a non degenerate metric g of signature (+1,-p) and
a folliation in space like hypersurfaces $S_{t}$, p dimensional manifolds
endowed with a riemannian metric, the \textbf{wave operator} is the
d'Alembertian : $\square u=\dfrac{\partial^{2}u}{\partial t^{2}}-\Delta_{x}$
acting on families of functions $u\in C\left(
\mathbb{R}
;C\left(  S_{t};%
\mathbb{C}
\right)  \right)  $. So $\square$ splits in $\Delta_{x}$\ and a "time
component" which can be treated as $\frac{\partial^{2}}{\partial t^{2}}$\ ,
and the functions are then $\varphi\left(  t,x\right)  \in C\left(
\mathbb{R}
;C\left(  S_{t};%
\mathbb{C}
\right)  \right)  .$ The operator is the same for functions or distributions :
$\square^{\prime}=\square$

\paragraph{Fundamental solution of the wave operator in $%
\mathbb{R}
^{m}$\newline}

The \textbf{wave operator} is the d'Alembertian : $\square u=\dfrac
{\partial^{2}u}{\partial t^{2}}-\sum_{\alpha=1}^{m}\frac{\partial^{2}%
u}{\partial x_{\alpha}^{2}}$ acting on families of functions $u\in C\left(
\mathbb{R}
;C\left(
\mathbb{R}
^{m};%
\mathbb{C}
\right)  \right)  $ or distributions $u\in C\left(
\mathbb{R}
;C\left(
\mathbb{R}
^{m};%
\mathbb{C}
\right)  ^{\prime}\right)  $ . The operator is symmetric with respect to the
inversion of t : $t\rightarrow-t.$

\begin{theorem}
The fundamental solution of the wave operator $\square u=\dfrac{\partial^{2}%
u}{\partial t^{2}}-\sum_{\alpha=1}^{m}\frac{\partial^{2}u}{\partial x_{\alpha
}^{2}}$\ acting on families of functions $u\in C\left(
\mathbb{R}
;C\left(
\mathbb{R}
^{m};%
\mathbb{C}
\right)  \right)  $ is the distribution : $U\in C\left(
\mathbb{R}
;S^{\prime}\left(
\mathbb{R}
^{m}\right)  \right)  :$

$U\left(  \varphi\left(  t,x\right)  \right)  =\left(  2\pi\right)
^{-m/2}\int_{0}^{\infty}\left(  \int_{R^{m}}e^{i\xi x}\frac{\sin\left(
t\left\Vert \xi\right\Vert \right)  }{\left\Vert \xi\right\Vert }%
\varphi\left(  t,x\right)  d\xi\right)  dt$
\end{theorem}

\begin{proof}
It is obtained from a family of distribution through Fourier transform

$%
\mathcal{F}%
_{x}\square U\left(  t\right)  =%
\mathcal{F}%
_{x}\dfrac{\partial^{2}}{\partial t^{2}}U-\sum_{k}%
\mathcal{F}%
_{x}\left(  \frac{\partial}{\partial x_{k}}\right)  ^{2}U=\left(
\dfrac{\partial^{2}}{\partial t^{2}}%
\mathcal{F}%
_{x}U\right)  -\left(  -i\right)  ^{2}\sum_{k}\left(  \xi_{k}\right)  ^{2}%
\mathcal{F}%
_{x}\left(  U\right)  $

$=\left(  \dfrac{\partial^{2}}{\partial t^{2}}+\sum_{k}\left(  \xi_{k}\right)
^{2}\right)  \left(
\mathcal{F}%
_{x}U\right)  $

If $\square U=\delta_{0}\left(  t,x\right)  =\delta_{0}\left(  t\right)
\otimes\delta_{0}\left(  x\right)  $

then $%
\mathcal{F}%
_{x}U=%
\mathcal{F}%
_{x}\left(  \delta_{0}\left(  t\right)  \otimes\delta_{0}\left(  x\right)
\right)  =%
\mathcal{F}%
_{x}\left(  \delta_{0}\left(  t\right)  \right)  \otimes%
\mathcal{F}%
_{x}\left(  \delta_{0}\left(  x\right)  \right)  =\delta_{0}\left(  t\right)
\otimes\left[  1\right]  _{\xi}$

and $%
\mathcal{F}%
_{x}\left(  \delta_{0}\left(  t\right)  \right)  =\delta_{0}\left(  t\right)
$ because $:%
\mathcal{F}%
_{x}\left(  \delta_{0}\left(  t\right)  \right)  \left(  \varphi\left(
t,x\right)  \right)  =\delta_{0}\left(  t\right)  \left(
\mathcal{F}%
_{x}\varphi\left(  t,x\right)  \right)  =%
\mathcal{F}%
_{x}\left(  \varphi\left(  0,x\right)  \right)  $

Thus we have the equation : $\left(  \dfrac{\partial^{2}}{\partial t^{2}%
}+\left\Vert \xi\right\Vert ^{2}\right)  \left(
\mathcal{F}%
_{x}U\right)  =\delta_{0}\left(  t\right)  \otimes\left[  1\right]  _{\xi}$

$%
\mathcal{F}%
_{x}U=u\left(  t,\xi\right)  $ . For $t\neq0$ the solutions of the ODE
$\left(  \dfrac{\partial^{2}}{\partial t^{2}}+\left\Vert \xi\right\Vert
^{2}\right)  u\left(  t,\xi\right)  =0$ are :

$u\left(  t,\xi\right)  =a\left(  \left\Vert \xi\right\Vert \right)  \cos
t\left\Vert \xi\right\Vert +b\left(  \left\Vert \xi\right\Vert \right)  \sin
t\left\Vert \xi\right\Vert $

So for $t\in%
\mathbb{R}
$ we take :

$u\left(  t,\xi\right)  =H(t)\left(  a\left(  \left\Vert \xi\right\Vert
\right)  \cos t\left\Vert \xi\right\Vert +b\left(  \left\Vert \xi\right\Vert
\right)  \sin t\left\Vert \xi\right\Vert \right)  $ with the Heavyside
function H(t)=1 for t$\geq0$

and we get :

$\left(  \dfrac{\partial^{2}}{\partial t^{2}}+\left\Vert \xi\right\Vert
^{2}\right)  u\left(  t,\xi\right)  =\delta_{0}\left(  t\right)
\otimes\left\Vert \xi\right\Vert b\left(  \left\Vert \xi\right\Vert \right)
+\frac{d}{dt}\delta_{0}\left(  t\right)  \otimes a\left(  \left\Vert
\xi\right\Vert \right)  =\delta_{0}\left(  t\right)  \otimes\left[  1\right]
_{\xi}$

$\Rightarrow$%
$\mathcal{F}$%
$_{x}U=H\left(  t\right)  \frac{\sin\left(  t\left\Vert \xi\right\Vert
\right)  }{\left\Vert \xi\right\Vert }$

$U\left(  t\right)  =$%
$\mathcal{F}$%
$_{x}^{\ast}\left(  H\left(  t\right)  \frac{\sin\left(  t\left\Vert
\xi\right\Vert \right)  }{\left\Vert \xi\right\Vert }\right)  =\left(
2\pi\right)  ^{-m/2}\int_{R^{m}}e^{i\xi x}H\left(  t\right)  \frac{\sin\left(
t\left\Vert \xi\right\Vert \right)  }{\left\Vert \xi\right\Vert }d\xi$
\end{proof}

The fundamental solution has the following properties (Zuily) :

U(t,x) = 0 for $\left\Vert x\right\Vert >t$ which is interpreted as
propagation at speed 1

If $m>2$ and m odd then $SuppU(t,.)\subset\left\{  \left\Vert x\right\Vert
=t\right\}  $ so we have a "ligth cone"

If m=3 then U can be expressed through the the Lebesgue measure $\sigma$ on
the unique sphere $S^{2}$

$U:U\left(  \varphi\right)  =\int_{0}^{\infty}\left(  \frac{t}{4\pi}%
\int_{S^{2}}\psi_{t}\left(  s\right)  \sigma\right)  dt$ where $\psi
_{t}\left(  s\right)  =\varphi\left(  tx\right)  |_{\left\Vert x\right\Vert
=1}$

$V\left(  t\right)  \left(  \varphi\right)  =\frac{t}{4\pi}\int_{S^{2}}%
\psi_{t}\left(  s\right)  \sigma,V\in C_{\infty}\left(
\mathbb{R}
;C_{\infty}\left(
\mathbb{R}
^{3};%
\mathbb{C}
\right)  ^{\prime}\right)  ,$

$V(0)=0,\frac{dV}{dt}=\delta_{0},\frac{d^{2}V}{dt^{2}}=0$

\bigskip

(Taylor 1 p.222) As $-\Delta$ is a positive operator in the Hilbert space
$H^{1}\left(
\mathbb{R}
^{m}\right)  $ it has a square root $\sqrt{-\Delta}$ with the following
properties :

$U\left(  t,x\right)  =\left(  \sqrt{-\Delta}\right)  ^{-1}\circ\sin
\sqrt{-\Delta}\circ\delta\left(  x\right)  $

$\frac{\partial U}{\partial t}=\cos t\sqrt{-\Delta}\circ\delta\left(
x\right)  $

If $f\in S\left(
\mathbb{R}
^{m}\right)  :$ with $r=\left\Vert x\right\Vert $

$f\left(  \sqrt{-\Delta}\right)  \delta\left(  x\right)  =\frac{1}{\sqrt{2\pi
}}\left[  -\frac{1}{2\pi r}\frac{\partial}{\partial r}\right]  ^{k}\widehat
{f}\left(  t\right)  $ if $m=2k+1$

$f\left(  \sqrt{-\Delta}\right)  \delta\left(  x\right)  =\frac{1}{\sqrt{\pi}%
}\int_{t}^{\infty}\left[  -\frac{1}{2\pi s}\frac{\partial}{\partial s}%
\widehat{f}\left(  s\right)  \right]  ^{k}\frac{s}{\sqrt{s^{-}-r^{2}}}ds$ if
$m=2k$

\paragraph{Cauchy problem on a manifold\newline}

\begin{theorem}
(Taylor 1 p.423) The PDE : find $u\in C\left(
\mathbb{R}
;C\left(  M;%
\mathbb{C}
\right)  \right)  $ such that :

$\square u=\dfrac{\partial^{2}u}{\partial t^{2}}-\Delta_{x}u=0$

u(0,x)=f(x)

$\frac{\partial u}{\partial t}(0,x)=g\left(  x\right)  $

where M is a geodesically complete riemannian manifold (without boundary).

$f\in H_{c}^{1}\left(  M\right)  ,g\in L_{c}^{2}\left(  M,\varpi_{0},%
\mathbb{C}
\right)  $ have non disjointed support

has a unique solution u, $u\in C\left(
\mathbb{R}
;H^{1}\left(  M\right)  \right)  \cap C\left(
\mathbb{R}
;L^{2}\left(  M,\varpi_{0},%
\mathbb{C}
\right)  \right)  $ and has a compact support in M for all t.
\end{theorem}

\paragraph{Cauchy problem in $%
\mathbb{R}
^{m}$\newline}

\begin{theorem}
(Zuily p.170) The PDE : find $u\in C\left(  I;C\left(
\mathbb{R}
^{m};%
\mathbb{C}
\right)  \right)  :$

$\square u=\dfrac{\partial^{2}u}{\partial t^{2}}-\sum_{\alpha=1}^{m}%
\frac{\partial^{2}u}{\partial x_{\alpha}^{2}}=0$ in $I\times O$

$u\left(  t_{0},x\right)  =f\left(  x\right)  ,$

$\frac{\partial u}{\partial t}\left(  t_{0},x\right)  =g\left(  x\right)  ,$

where

O is a bounded open subset of $%
\mathbb{R}
^{m},$

I an interval in $%
\mathbb{R}
$ with t$_{0}\in I$

$f\in H_{c}^{1}\left(  O\right)  ,g\in L^{2}\left(  O,dx,%
\mathbb{C}
\right)  $

has a unique solution : $u\left(  t,x\right)  =U\left(  t\right)  \ast
g\left(  x\right)  +\frac{dU}{dt}\ast f\left(  x\right)  $ and it belongs to
$C\left(  I;H_{c}^{1}\left(  O\right)  \right)  $\ .\ 

$u(t,x)=\sum_{k=1}^{\infty}\{\left\langle e_{k},f\right\rangle \cos
(\sqrt{\lambda_{k}}\left(  t-t_{0}\right)  )+\dfrac{\left\langle
e_{k},g\right\rangle }{\sqrt{\lambda_{k}}}\sin(\sqrt{\lambda_{k}}\left(
t-t_{0}\right)  )\}e_{k}\left(  x\right)  $

where $\left(  \lambda_{n},e_{n}\right)  _{n\in%
\mathbb{N}
}$ are eigen values and eigen vectors of the -laplacian -$\Delta$, the $e_{n}$
being chosen to be a Hilbertian basis of $L^{2}\left(  O,dx,%
\mathbb{C}
\right)  $
\end{theorem}

\begin{theorem}
(Taylor 1 p.220) The PDE : find $u\in C\left(
\mathbb{R}
;C\left(
\mathbb{R}
^{m};%
\mathbb{C}
\right)  ^{\prime}\right)  :$

$\square u=\dfrac{\partial^{2}u}{\partial t^{2}}-\sum_{\alpha=1}^{m}%
\frac{\partial^{2}u}{\partial x_{\alpha}^{2}}=0$ in $%
\mathbb{R}
\times%
\mathbb{R}
^{m}$

$u\left(  0,x\right)  =f\left(  x\right)  $

$\frac{\partial u}{\partial t}\left(  0,x\right)  =g\left(  x\right)  $

where $f,g\in S\left(
\mathbb{R}
^{m}\right)  ^{\prime}$

has a unique solution $u\in C_{\infty}\left(
\mathbb{R}
;S\left(
\mathbb{R}
^{m}\right)  ^{\prime}\right)  $\ .\ It reads : $u\left(  t,x\right)
=U\left(  t\right)  \ast g\left(  x\right)  +\frac{dU}{dt}\ast f\left(
x\right)  $
\end{theorem}

\paragraph{Cauchy problem in $%
\mathbb{R}
^{4}$\newline}

(Zuily)

1. Homogeneous problem:

\begin{theorem}
The PDE: find $u\in C\left(
\mathbb{R}
;C\left(
\mathbb{R}
^{m};%
\mathbb{C}
\right)  ^{\prime}\right)  $ such that :

$\square u=\dfrac{\partial^{2}u}{\partial t^{2}}-\sum_{\alpha=1}^{3}%
\frac{\partial^{2}u}{\partial x_{\alpha}^{2}}=0$ in $%
\mathbb{R}
_{+}\times%
\mathbb{R}
^{3}$

$u\left(  0,x\right)  =f\left(  x\right)  $

$\frac{\partial u}{\partial t}\left(  0,x\right)  =g\left(  x\right)  $

has a unique solution : $u\left(  t,x\right)  =U\left(  t\right)  \ast
g\left(  x\right)  +\frac{dU}{dt}\ast f\left(  x\right)  $ which reads :

$u\left(  t,x\right)  =\frac{1}{4\pi t}\int_{\left\Vert y\right\Vert
=t}g\left(  x-y\right)  \sigma_{ty}+\frac{\partial}{\partial t}\left(
\frac{1}{4\pi t}\int_{\left\Vert y\right\Vert =t}f\left(  x-y\right)
\sigma_{ty}\right)  $
\end{theorem}

2. Properties of the solution :

If $f,g\in C_{\infty}\left(
\mathbb{R}
^{3};%
\mathbb{C}
\right)  $ then $u(t,x)\in C_{\infty}\left(
\mathbb{R}
^{4};%
\mathbb{C}
\right)  $

When $t\rightarrow\infty$ \ $u(t,x)\rightarrow0$ and $\exists M>0:$
$t\geq1:\left\vert u(t,x)\right\vert \leq\dfrac{M}{t}$

The quantity (the "energy") is constant :

$W(t)=\int\{\left(  \dfrac{\partial u}{\partial t}\right)  ^{2}+\sum_{i=1}%
^{3}\left(  \dfrac{\partial u}{\partial x_{i}}\right)  ^{2}\}dx=Ct$

Propagation at the speed 1 :

If $\left\Vert x\right\Vert >R\Rightarrow f\left(  x\right)  =g\left(
x\right)  =0$ then :

$u(t,x)=0$ for $\left\Vert x\right\Vert \leq R+t$ or $\left\{  t>R,\left\Vert
x\right\Vert \leq t-R\right\}  $

The value of u(t,x) in $\left(  t_{0},x_{0}\right)  $ depends only on the
values of f and g on the hypersuface $\left\Vert x-x%
{{}^\circ}%
\right\Vert =t%
{{}^\circ}%
$

Plane waves :

if $f(x)=-k.x=-\sum_{l=1}^{3}k_{l}x^{l}$ with k fixed, $g\left(  x\right)  =1$
then : $u\left(  t,x\right)  =t-k.x$

\bigskip

3. Inhomogeneous problem :

\begin{theorem}
The PDE : find $u\in C\left(
\mathbb{R}
_{+};C\left(
\mathbb{R}
^{3};%
\mathbb{C}
\right)  \right)  $ such that :

$\square u=\dfrac{\partial^{2}u}{\partial t^{2}}-\sum_{\alpha=1}^{3}%
\frac{\partial^{2}u}{\partial x_{\alpha}^{2}}=F$ in $%
\mathbb{R}
_{+}\times%
\mathbb{R}
^{3}$

$u\left(  0,x\right)  =f\left(  x\right)  $

$\frac{\partial u}{\partial t}\left(  0,x\right)  =g\left(  x\right)  $

where $F\in C_{\infty}\left(
\mathbb{R}
_{+};C\left(
\mathbb{R}
^{3};%
\mathbb{C}
\right)  \right)  $

has a unique solution : $u\left(  t,x\right)  =v=u+U\ast F$ ,where U is the
solution of the homogeneous problem, which reads :

$t\geq0:u(t,x)=$

$\frac{1}{4\pi t}\int_{\left\Vert y\right\Vert =t}g\left(  x-y\right)
\sigma_{ty}+\frac{\partial}{\partial t}\left(  \frac{1}{4\pi t}\int
_{\left\Vert y\right\Vert =t}f\left(  x-y\right)  \sigma_{ty}\right)
+\int_{t=0}^{\infty}\frac{1}{4\pi s}\left(  \int_{\left\Vert x\right\Vert
=s}F\left(  t-s,x-y\right)  \sigma_{yt}\right)  ds$
\end{theorem}

\subsubsection{Schr\"{o}dinger operator}

It is of course linked to the celebrated Schr\"{o}dinger equation of quantum mechanics.

\paragraph{Definition\newline}

This is the scalar operator $D=\frac{\partial}{\partial t}-i\Delta_{x}$ acting
on functions $f\in V\subset C\left(  J\times M;%
\mathbb{C}
\right)  $ where J is some interval in $%
\mathbb{R}
$\ and M a manifold upon which the laplacian is defined.\ Usually $M=%
\mathbb{R}
^{m}$ and $V=S\left(
\mathbb{R}
\times%
\mathbb{R}
^{m}\right)  .$

If $V=C\left(  J;F\right)  $ , where F is some Fr\'{e}chet space of functions
on M, f can be seen as a map : $f:J\rightarrow F.$ If there is a family of
distributions : $S:J\rightarrow F^{\prime}$ then $\widetilde{S}\left(
f\right)  =\int_{J}S\left(  t\right)  _{x}\left(  f\left(  t,x\right)
\right)  dt$ defines a distribution $\widetilde{S}\in V^{\prime}$

This is the basis of the search for fundamental solutions.

\paragraph{Cauchy problem in $%
\mathbb{R}
^{m}$\newline}

From Zuily p.152

\begin{theorem}
If $S\in S\left(
\mathbb{R}
^{m}\right)  ^{\prime}$ there is a unique family of distributions $u\in
C_{\infty}\left(
\mathbb{R}
;S\left(
\mathbb{R}
^{m}\right)  ^{\prime}\right)  $ such that :

$D\widetilde{u}=0$

$u(0)=S$

where $\widetilde{u}:$ $\forall\varphi\in S\left(
\mathbb{R}
\times%
\mathbb{R}
^{m}\right)  :\widetilde{u}\left(  \varphi\right)  =\int_{%
\mathbb{R}
}u(t)\left(  \varphi\left(  t,.\right)  \right)  dt$

which is given by :

$\widetilde{u}\left(  \varphi\right)  =\int_{%
\mathbb{R}
}\left(  \int_{%
\mathbb{R}
^{m}}%
\mathcal{F}%
_{\xi}^{\ast}\left(  e^{-it\left\Vert \xi\right\Vert ^{2}}\widehat{S}\left(
\varphi\right)  \right)  dx\right)  dt=\int_{%
\mathbb{R}
}\int_{%
\mathbb{R}
^{m}}\phi\left(  t,x\right)  \varphi\left(  t,x\right)  dxdt$
\end{theorem}

Which needs some explanations...

Let $\varphi\in S\left(
\mathbb{R}
\times%
\mathbb{R}
^{m}\right)  $ .The Fourier transform $\widehat{S}$ of S is a distribution
$\widehat{S}\in S\left(
\mathbb{R}
^{m}\right)  ^{\prime}$ such that :

$\widehat{S}\left(  \varphi\right)  =S\left(
\mathcal{F}%
_{y}\left(  \varphi\right)  \right)  =S_{\zeta}\left(  \left(  2\pi\right)
^{-m/2}\int_{%
\mathbb{R}
^{m}}e^{-i\left\langle y,\zeta\right\rangle }\varphi\left(  t,y\right)
dy\right)  $ (S acts on the $\zeta$ function)

This is a function of t, so $e^{-it\left\Vert \xi\right\Vert ^{2}}\widehat
{S}\left(  \varphi\right)  $ is a function of t and $\xi\in%
\mathbb{R}
^{m}$

$e^{-it\left\Vert \xi\right\Vert ^{2}}\widehat{S}\left(  \varphi\right)
=e^{-it\left\Vert \xi\right\Vert ^{2}}S_{\zeta}\left(  \left(  2\pi\right)
^{-m/2}\int_{%
\mathbb{R}
^{m}}e^{-i\left\langle y,\zeta\right\rangle }\varphi\left(  t,y\right)
dy\right)  $

$=S_{\zeta}\left(  \left(  2\pi\right)  ^{-m/2}\int_{%
\mathbb{R}
^{m}}e^{-it\left\Vert \xi^{2}\right\Vert -i\left\langle y.,\zeta\right\rangle
}\varphi\left(  t,y\right)  dy\right)  $

Its inverse Fourier transform $%
\mathcal{F}%
_{\xi}^{\ast}\left(  e^{-it\left\Vert \xi\right\Vert ^{2}}\widehat{S}\left(
\varphi\right)  \right)  $ is a function of t and $x\in%
\mathbb{R}
^{m}$\ (through the interchange of $\xi,x)$

$%
\mathcal{F}%
_{\xi}^{\ast}\left(  e^{-it\left\Vert \xi\right\Vert ^{2}}\widehat{S}\left(
\varphi\right)  \right)  =\left(  2\pi\right)  ^{-m/2}\int_{%
\mathbb{R}
^{m}}e^{i\left\langle \xi,x\right\rangle }S_{\zeta}\left(  \left(
2\pi\right)  ^{-m/2}\int_{%
\mathbb{R}
^{m}}e^{-it\left\Vert \xi^{2}\right\Vert -i\left\langle y.,\zeta\right\rangle
}\varphi\left(  t,y\right)  dy\right)  d\xi$

$=S_{\zeta}\left(  \left(  2\pi\right)  ^{-m}\int_{%
\mathbb{R}
^{m}}\int_{%
\mathbb{R}
^{m}}e^{-it\left\Vert \xi^{2}\right\Vert +i\left\langle x,\xi\right\rangle
-i\left\langle y.,\zeta\right\rangle }\varphi\left(  t,y\right)
dyd\xi\right)  $

and the integral with respect both to x and t gives :

$\widetilde{u}\left(  \varphi\right)  =\int_{%
\mathbb{R}
}\left(  \int_{%
\mathbb{R}
^{m}}S_{\zeta}\left(  \left(  2\pi\right)  ^{-m}\int_{%
\mathbb{R}
^{m}}\int_{%
\mathbb{R}
^{m}}e^{-it\left\Vert \xi^{2}\right\Vert +i\left\langle x,\xi\right\rangle
-i\left\langle y.,\zeta\right\rangle }\varphi\left(  t,y\right)
dyd\xi\right)  dx\right)  dt$

By exchange of x and y :

$\widetilde{u}\left(  \varphi\right)  =\int_{%
\mathbb{R}
}\int_{%
\mathbb{R}
^{m}}S_{\zeta}\left(  \left(  2\pi\right)  ^{-m}\int_{%
\mathbb{R}
^{m}}\int_{%
\mathbb{R}
^{m}}e^{-it\left\Vert \xi^{2}\right\Vert +i\left\langle y,\xi\right\rangle
-i\left\langle x.,\zeta\right\rangle }dyd\xi\right)  \varphi\left(
t,x\right)  dxdt$

$\widetilde{u}\left(  \varphi\right)  =\int_{%
\mathbb{R}
}\int_{%
\mathbb{R}
^{m}}\phi\left(  t,x\right)  \varphi\left(  t,x\right)  dxdt$

where $\phi\left(  t,x\right)  =\left(  2\pi\right)  ^{-m}S_{\zeta}\left(
e^{-i\left\langle x.,\zeta\right\rangle }\int_{%
\mathbb{R}
^{m}}\int_{%
\mathbb{R}
^{m}}e^{-it\left\Vert \xi^{2}\right\Vert +i\left\langle y,\xi\right\rangle
}d\xi dy\right)  $

\bigskip

If S=T(g) with $g\in S\left(
\mathbb{R}
^{m}\right)  $ then :

$\phi\left(  t,x\right)  =\left(  2\pi\right)  ^{-m}\int_{%
\mathbb{R}
^{m}}e^{-it\left\Vert \xi^{2}\right\Vert +i\left\langle y,\xi\right\rangle
+i\left\langle x.,\xi\right\rangle }\widehat{g}\left(  \xi\right)  d\xi\in
C_{\infty}\left(
\mathbb{R}
;S\left(
\mathbb{R}
^{m}\right)  \right)  $

If $S\in H^{s}\left(
\mathbb{R}
^{m}\right)  ,s\in%
\mathbb{R}
$ then $\phi\left(  t,x\right)  \in C_{0}\left(
\mathbb{R}
;H^{s}\left(
\mathbb{R}
^{m}\right)  \right)  $

If $S\in H^{k}\left(
\mathbb{R}
^{m}\right)  $,$k\in%
\mathbb{N}
$ then $\left\Vert \phi\left(  t,.\right)  \right\Vert _{H^{k}}=\left\Vert
g\right\Vert _{H^{k}};\frac{\partial^{r}\phi}{\partial x_{\alpha_{1}%
}..\partial x_{\alpha_{r}}}\in C_{0}\left(
\mathbb{R}
;H^{s-2k}\left(
\mathbb{R}
^{m}\right)  \right)  ;$

If S=T(g),$g\in L^{2}\left(
\mathbb{R}
^{m},dx,%
\mathbb{C}
\right)  ,\forall r>0:\left\Vert x\right\Vert ^{r}g\in L^{2}\left(
\mathbb{R}
^{m},dx,%
\mathbb{C}
\right)  $ then

$\forall t\neq0:\phi\left(  t,.\right)  \in C_{\infty}\left(
\mathbb{R}
^{m};%
\mathbb{C}
\right)  $

If S=T(g),$g\in L^{1}\left(
\mathbb{R}
^{m},dx,%
\mathbb{C}
\right)  ,$ then $\forall t\neq0:\left\Vert \phi\left(  t,.\right)
\right\Vert _{\infty}\leq\left(  4\pi\left\vert t\right\vert \right)
^{-m/2}\left\Vert g\right\Vert _{1}$

\subsubsection{Heat equation}

The operator is the heat operator : $D=\frac{\partial}{\partial t}-\Delta_{x}$
on $C\left(
\mathbb{R}
_{+};C\left(  M;%
\mathbb{C}
\right)  \right)  $

\paragraph{Cauchy problem\newline}

M is a riemannian manifold, the problem is to find $u\in C\left(
\mathbb{R}
_{+};C\left(  M;%
\mathbb{C}
\right)  \right)  $ such that $\frac{\partial u}{\partial t}-\Delta_{x}u=0$ on
M and u(0,x)=g(x) where g is a given function.

\begin{theorem}
(Gregor'yan p.10) The PDE : find $u\in C_{\infty}\left(
\mathbb{R}
_{+}\times M;%
\mathbb{C}
\right)  $ such that :

$\frac{\partial u}{\partial t}=\Delta_{x}u$

$u(t,.)\rightarrow g$ in $L^{2}\left(  M,\varpi_{0},%
\mathbb{C}
\right)  $ when $t\rightarrow0_{+}$

where

M is a riemannian smooth manifold (without boundary)

$g\in L^{2}\left(  M,\varpi_{0},%
\mathbb{C}
\right)  $

has a unique solution

If M is an open, \textit{relatively compact}, in a manifold N, then $u\in
C_{0}\left(
\mathbb{R}
_{+};H^{1}\left(  M\right)  \right)  $ and is given by $u\left(  t,x\right)
=\sum_{n=1}^{\infty}\left\langle e_{n},\widehat{g}\right\rangle e^{-\lambda
_{n}t}e_{n}\left(  x\right)  $ where $\left(  e_{n},\lambda_{n}\right)  $ are
the eigen vectors and eigen values of $-\Delta,$ the $e_{n}$ being chosen to
be a Hilbertian basis of $L^{2}\left(  M,dx,%
\mathbb{C}
\right)  .$
\end{theorem}

\paragraph{Dirichlet problem\newline}

\begin{theorem}
(Taylor 1 p.416) The PDE : find $u\in C\left(
\mathbb{R}
_{+};C\left(  N;%
\mathbb{C}
\right)  \right)  $ such that :

$\frac{\partial u}{\partial t}-\Delta_{x}u=0$ on $\overset{\circ}{N}$

$u\left(  0,x\right)  =g\left(  x\right)  $

$u\left(  t,x\right)  =0$ if $x\in\partial N$

where

M is a compact smooth manifold with boundary in a riemannian manifold N

$g\in L^{2}\left(  N,dx,%
\mathbb{C}
\right)  $

has a unique solution, and it belongs to $C\left(
\mathbb{R}
_{+};H^{-s}\left(  \overset{\circ}{M}\right)  \right)  \cap C_{1}\left(
\mathbb{R}
_{+};H^{-s+2}\left(  \overset{\circ}{M}\right)  \right)  .$ It reads :
$u\left(  t,x\right)  =\sum_{n=1}^{\infty}\left\langle e_{n},\widehat
{g}\right\rangle e^{-\lambda_{n}t}e_{n}\left(  x\right)  $ where $\left(
\lambda_{n},e_{n}\right)  _{n\in%
\mathbb{N}
}$ are eigen values and eigen vectors of -$\Delta$ ,- the laplacian, on
$\overset{\circ}{N},$ the $e_{n}$ being chosen to be a Hilbertian basis of
$L^{2}\left(  \overset{\circ}{M},dx,%
\mathbb{C}
\right)  $
\end{theorem}

\subsubsection{Non linear partial differential equations}

There are few general results, and the diversity of non linear PDE is such
that it would be impossible to give even a hint of the subject. So we will
limit ourselves to some basic definitions.

\paragraph{Cauchy-Kowalesky theorem\newline}

There is an extension of the theorem to non linear PDE.

\begin{theorem}
(Taylor 3 p.445) The scalar PDE : find $u\in C\left(
\mathbb{R}
\times O;%
\mathbb{C}
\right)  $

$\frac{\partial^{r}u}{\partial t^{r}}=D\left(  t,x,u,\frac{\partial^{s}%
u}{\partial x_{\alpha_{1}}...\partial x_{\alpha_{s}}},\frac{\partial^{s+k}%
u}{\partial x_{\alpha_{1}}...\partial x_{\alpha_{s}}\partial t^{k}}\right)  $
where s=1...r,k=1..r-1,$\alpha_{j}=1...m$

$\frac{\partial^{k}u}{\partial t^{k}}\left(  0,x\right)  =g_{k}\left(
x\right)  ,k=0,...r-1$

where O is an open in $%
\mathbb{R}
^{m}:$

If $D,g_{k}$ are real analytic for $x_{0}\in O,$ then there is a unique real
analytic solution in a neighborhood of $\left(  0,x_{0}\right)  $
\end{theorem}

\paragraph{Linearization of a PDE\newline}

The main piece of a PDE is a usually differential operator : $D:J^{r}%
E_{1}\rightarrow E_{2}$. If D is regular, meaning at least differentiable, in
a neighborhood $n\left(  Z_{0}\right)  $ of a point $Z_{0}\in J^{r}%
E_{1}\left(  x_{0}\right)  ,x_{0}$ fixed in M, it can be Taylor expanded with
respect to the r-jet Z. The resulting linear operator has a principal symbol,
and if it is elliptic D is said to be locally elliptic.

\paragraph{Locally elliptic PDE\newline}

Elleiptic PDE usually give smooth solutions.

If the scalar PDE : find $u\in C\left(  O;%
\mathbb{C}
\right)  $ such that :$D\left(  u\right)  =g$ in O, where O is an open in $%
\mathbb{R}
^{m}$ has a solution $u_{0}\in C_{\infty}\left(  O;%
\mathbb{C}
\right)  $ at $x_{0}$ and if the scalar r order differential operator is
elliptic in $u_{0}$\ then, for any s, there are functions\ $u\in C_{s}\left(
O;%
\mathbb{C}
\right)  $ which are solutions in a neighborhood n($x_{0}).$ Moreover is g is
smooth then $u\left(  x\right)  -u_{0}\left(  x\right)  =o\left(  \left\Vert
x-x_{0}\right\Vert ^{r+1}\right)  $

\paragraph{Quasi linear symmetric hyperbolic PDE\newline}

Hyperbolic PDE are the paradigm of "well posed" problem : they give unique
solution, continuously depending on the initial values. One of their
characteristic is that a variation of the initial value propagated as a wave.

\begin{theorem}
(Taylor 3 p.414) A quasi linear symmetric PDE is a PDE : find $u\in C\left(
J\times M;%
\mathbb{C}
\right)  $ such that :

$B\left(  t,x,u\right)  \frac{\partial u}{\partial t}=\sum_{\alpha=1}%
^{m}A^{\alpha}\left(  t,x,u\right)  \frac{\partial u}{\partial x^{\alpha}%
}+g\left(  t,x,u\right)  $

$u\left(  0,x\right)  =g\left(  x\right)  $

where

$E\left(  M,V,\pi\right)  $ is a vector bundle on a m dimensional manifold M,

$J=\left[  0,T\right]  \subset%
\mathbb{R}
$

$B\left(  t,x,u\right)  ,A^{\alpha}\left(  t,x,u\right)  \in%
\mathcal{L}%
\left(  V;V\right)  $ and the $A^{\alpha}$\ are represented by self adjoint
matrices $\left[  A^{\alpha}\right]  =\left[  A^{\alpha}\right]  ^{\ast}$

Then if $g\in H^{n}\left(  E\right)  $ with $n>1+\frac{m}{2}$ the PDE has a
unique solution u in a neighborhood of t=0 and $u\in C\left(  J;H^{n}\left(
E\right)  \right)  $
\end{theorem}

The theorem can be extended to a larger class of semi linear non symmetric
equations. There are similar results for quasi linear second order PDE.

\paragraph{Method of the characteristics\newline}

This method is usually, painfully and confusely, explained in many books about
PDE. In fact it is quite simple with the r jet formalism. A PDE of order r is
a closed subbundle F of the r jet extension $J^{r}E$ of a fibered manifold E.
A subbundle has, for base, a submanifold of M. Clearly a solution
$X:M\rightarrow\mathfrak{X}\left(  E\right)  $ is such that there is some
relationship between the coordinates $\xi_{\alpha}$ of a point x in M and the
value of $X^{i}\left(  x\right)  .$ This is all the more true when one adds
the value of the derivatives, linked in the PDE, with possibly other coercives
conditions. When one looks at the solutions of classical PDE one can see they
are some deformed maps of these constraints (notably the shape of a boundary).
So one can guess that solutions are more or less organized around some
subbundles of E, whose base is itself a submanifold of M. In each of these
subbundles the value of $X^{i}\left(  x\right)  $ , or of some quantity
computed from it, are preserved. So the charcateristics" method sums up to
find "primitives", meaning sections $Y\in\mathfrak{X}_{k}\left(  E\right)
,k<r$ such that $J^{r-k}Y\in F.$

One postulates that the solutions X belong to some subbundle, defined through
a limited number of parameters (thus these parameters define a submanifold of
M). So each derivative can be differentiated with respect to these parameters.
By reintroducing these relations in the PDE one gets a larger system (there
are more equations) but usually at a lower order. From there one can reduce
the problem to ODE.

The method does not give any guarantee about the existence or unicity of
solutions, but usually gives a good insight of particular solutions. Indeed
they give primitives, and quantities which are preserved along the
transformation of the system. For instance in hyperbolic PDE they can describe
the front waves or the potential discontinuities (when two waves collide).

The classical example is for first order PDE.

Let be the scalar PDE in $%
\mathbb{R}
^{m}:$ find $u\in C_{1}\left(
\mathbb{R}
^{m};%
\mathbb{C}
\right)  $ such that $D\left(  x,u,p\right)  =0$ where $p=\left(
\frac{\partial u}{\partial x_{\alpha}}\right)  _{\alpha=1}^{m}.$ In the r jet
formalism the 3 variables x,u,p can take any value at any point x. $J^{r}E$ is
here the 1 jet extension which can be assimilated with $%
\mathbb{R}
^{m}\otimes%
\mathbb{R}
\otimes%
\mathbb{R}
^{m\ast}$

One looks for a curve in $J^{1}E$ , that is a map : $%
\mathbb{R}
\rightarrow J^{1}E::Z\left(  s\right)  =\left(  x\left(  s\right)  ,u\left(
s\right)  ,p\left(  s\right)  \right)  $ with the scalar parameter s, which is
representative of a solution. It must meet some conditions. So one adds all
the available relations coming from the PDE and the defining identities :
$\frac{du}{ds}=p\frac{dx}{ds}$ and $d\left(  \sum_{\alpha}p_{\alpha}%
dx^{\alpha}\right)  =0$ . Put all together one gets the Lagrange-Charpit
equations for the characteristics :

$\frac{\overset{\cdot}{x_{\alpha}}}{D_{p_{\alpha}}^{\prime}}=\frac
{\overset{\cdot}{p_{\alpha}}}{D_{x_{\alpha}}^{\prime}+D_{u}^{\prime}p_{\alpha
}}=\frac{\overset{\cdot}{u}}{\sum_{\alpha}p_{\alpha}D_{p_{\alpha}}^{\prime}}$
with $\overset{\cdot}{x_{\alpha}}=\frac{dx_{\alpha}}{ds},\overset{\cdot
}{p_{\alpha}}=\frac{dp_{\alpha}}{ds},\overset{\cdot}{u}=\frac{du}{ds}$

which are ODE in s.

A solution u(x(s)) is usually specific, but if one has some quantities which
are preserved when s varies, and initial conditions, these solutions are of
physical significance.

\newpage

\section{VARIATIONAL\ CALCULUS}

\label{Variational Calculus}

The general purpose of variational calculus is to find functions such that the
value of some functional is extremum. This problem could take many different
forms.\ We start with functional derivatives, which is an extension of the
classical method to find the extremum of a function.\ The general and
traditional approach of variational calculus is through lagrangians. It gives
the classical Euler-Lagrange equations, and is the starting point for more
elaborate studies on the "variational complex", which is part of topological algebra.

We start by a reminder of definitions and notations which will be used all
over this section.

\subsubsection{Notations}

1. Let M be a m dimensional real manifold with coordinates in a chart
$\psi\left(  x\right)  =\left(  \xi^{\alpha}\right)  _{\alpha=1}^{m}.$ A
holonomic basis with this chart is $\partial\xi_{\alpha}$ and its dual
$d\xi^{\alpha}.$ The space of antisymmetric p covariant tensors on $TM^{\ast}$
is a vector bundle, and the set of its sections : $\mu:M\rightarrow\Lambda
_{p}TM^{\ast}::\mu=\sum_{\left\{  \alpha_{1}...\alpha_{p}\right\}  }%
\mu_{\alpha_{1}...\alpha_{p}}\left(  x\right)  d\xi^{\alpha_{1}}\wedge..\wedge
d\xi^{\alpha_{p}}$ is denoted as usual $\Lambda_{p}\left(  M;%
\mathbb{R}
\right)  \equiv\mathfrak{X}\left(  \Lambda_{p}TM^{\ast}\right)  .$ We will use
more often the second notation to emphasize the fact that this is a map from M
to $\Lambda_{p}TM^{\ast}.$

2. Let $E\left(  M,V,\pi\right)  $ be a fiber bundle on M, $\varphi$ be some
trivialization of E, so an element p of E is : $p=\varphi\left(  x,u\right)  $
for $u\in V.$ A section $X\in\mathfrak{X}\left(  E\right)  $ reads :
X(x)=$\varphi\left(  x,\sigma\left(  x\right)  \right)  $

Let $\phi$ be some chart of V, so a point u in V has the coordinates
$\phi\left(  u\right)  =\left(  \eta^{i}\right)  _{i=1}^{n}$

The tangent vector space $T_{p}E$ at p has the basis denoted :

$\partial x_{\alpha}=\varphi_{x}^{\prime}\left(  x,u\right)  \partial
\xi_{\alpha},\partial u_{i}=\varphi_{u}^{\prime}\left(  x,u\right)
\partial\eta_{i}$ for $\alpha=1...m,i=1...n$ with dual $\left(  dx^{\alpha
},du^{i}\right)  $

3. $J^{r}E$ is a fiber bundle $J^{r}E\left(  M,J_{0}^{r}\left(
\mathbb{R}
^{m},V\right)  _{0},\pi^{r}\right)  .$ A point Z in $J^{r}E$ reads :
$Z=\left(  z,z_{\alpha_{1}...\alpha_{s}}:1\leq\alpha_{k}\leq m,s=1..r\right)
.$ It is projected on a point x in M and has for coordinates :

$\Phi\left(  Z\right)  =\zeta=\left(  \xi^{\alpha},\eta^{i},\eta_{\alpha
_{1}...\alpha_{s}}^{i}:1\leq\alpha_{1}\leq\alpha_{2}...\leq\alpha_{s}\leq
m,i=1...n,s=1..r\right)  $

A section $Z\in\mathfrak{X}\left(  J^{r}E\right)  $ reads : $Z=\left(
z,z_{\alpha_{1}...\alpha_{s}}:1\leq\alpha_{k}\leq m,s=1..r\right)  $ where
each component $z_{\alpha_{1}...\alpha_{s}}$ can be seen as independant
section of $\mathfrak{X}\left(  E\right)  .$

A section X of E induces a section of $J^{r}E:J^{r}X\left(  x\right)  $ has
for coordinates : $\Phi\left(  J^{r}X\right)  =\left(  \xi^{\alpha},\sigma
^{i}\left(  x\right)  ,\frac{\partial^{s}\sigma^{i}}{\partial\xi^{\alpha_{1}%
}..\partial\xi^{\alpha_{s}}}|_{x}\right)  $

The projections are denoted:

$\pi^{r}:J^{r}E\rightarrow M:\pi^{r}\left(  j_{x}^{r}X\right)  =x$

$\pi_{0}^{r}:J^{r}E\rightarrow E:\pi_{0}\left(  j_{x}^{r}X\right)  =X\left(
x\right)  $

$\pi_{s}^{r}:J^{r}E\rightarrow J^{s}E:\pi_{s}^{r}\left(  j_{x}^{r}X\right)
=j_{x}^{s}X$

4. As a manifold $J^{r}E$ has a tangent bundle with holonomic basis :

$\left(  \partial x_{\alpha},\partial u_{i}^{\alpha_{1}...\alpha_{s}}%
,1\leq\alpha_{1}\leq\alpha_{2}...\leq\alpha_{s}\leq m,i=1...n,s=0...r\right)
$ and a\ cotangent bundle with basis $\left(  dx^{\alpha},du_{\alpha
_{1}...\alpha_{s}}^{i},1\leq\alpha_{1}\leq\alpha_{2}...\leq\alpha_{s}\leq
m,i=1...n,s=0...r\right)  .$

The vertical bundle is generated by the vectors $\partial u_{i}^{\alpha
_{1}...\alpha_{s}}.$

A vector on the tangent space $T_{Z}J^{r}E$ of $J^{r}E$ reads :

$W_{Z}=w_{p}+\sum_{s=1}^{r}\sum_{\alpha_{1\leq}..\leq\alpha_{s}}w_{\alpha
_{1}...\alpha_{\alpha_{s}}}^{i}\partial u_{i}^{\alpha_{1}...\alpha_{\alpha
_{s}}}$

with $w_{p}=w^{\alpha}\partial x_{\alpha}+w^{i}\partial u_{i}\in T_{p}E$

The projections give :

$\pi^{r\prime}\left(  Z\right)  :T_{Z}J^{r}E\rightarrow T_{\pi^{r}\left(
Z\right)  }M:\pi^{r\prime}\left(  Z\right)  W_{Z}=w^{\alpha}\partial
\xi_{\alpha}$

$\pi_{0}^{r\prime}\left(  Z\right)  :T_{Z}J^{r}E\rightarrow T_{\pi_{0}%
^{r}\left(  Z\right)  }E:\pi^{r\prime}\left(  Z\right)  W_{Z}=w_{p}$

The space of antisymmetric p covariant tensors on $TJ^{r}E^{\ast}$ is a vector
bundle, and the set of its sections : $\varpi:J^{r}E\rightarrow\Lambda
_{p}TJ^{r}E^{\ast}$ is denoted $\mathfrak{X}\left(  \Lambda_{p}TJ^{r}E^{\ast
}\right)  \equiv\Lambda_{p}\left(  J^{r}E\right)  .$

A form on $J^{r}E$ is $\pi^{r}$ horizontal if it is null for any vertical
vertical vector. It reads :

$\varpi=\sum_{\left\{  \alpha_{1}...\alpha_{p}\right\}  }\varpi_{\alpha
_{1}...\alpha_{p}}\left(  Z\right)  dx^{\alpha_{1}}\wedge..\wedge
dx^{\alpha_{p}}$ where $Z\in J^{r}E$

The set of p horizontal forms on $J^{r}E$ is denoted $\mathfrak{X}^{H}\left(
\Lambda_{p}TJ^{r}E^{\ast}\right)  $

5. A projectable vector field $W\in\mathfrak{X}\left(  TE\right)  $ on E is
such that :

$\exists Y\in\mathfrak{X}\left(  TM\right)  :\forall p\in E:\pi^{\prime
}\left(  p\right)  W\left(  p\right)  =Y\left(  \pi\left(  p\right)  \right)
$

Its components read : $W=W^{\alpha}\partial x_{\alpha}+W^{i}\partial u_{i}$
where $W^{\alpha}$ does not depend on u and $W^{\alpha}\equiv0$ if it is vertical.

It defines, at least locally, with any section X on E a one parameter group of
fiber preserving morphisms on E through its flow :

$U\left(  t\right)  :\mathfrak{X}\left(  E\right)  \rightarrow\mathfrak{X}%
\left(  E\right)  ::U\left(  t\right)  X\left(  x\right)  =\Phi_{W}\left(
X(\Phi_{Y}\left(  x,-t\right)  ),t\right)  $

This one parameter group of morphism on E has a prolongation as a one
parameter group of fiber preserving morphisms on $J^{r}E$ :.

$J^{r}U\left(  t\right)  :\mathfrak{X}\left(  J^{r}E\right)  \rightarrow
\mathfrak{X}\left(  J^{r}E\right)  ::J^{r}U\left(  t\right)  Z\left(
x\right)  =j_{\Phi_{Y}\left(  x,t\right)  }^{r}\left(  \Phi_{W}\left(
X(\Phi_{Y}\left(  x,-t\right)  ),t\right)  \right)  $

with any section X on E such that $Z\left(  x\right)  =j^{r}X\left(  x\right)
.$

A projectable vector field W on E has a prolongation as a projectable vector
field $J^{r}W$ on $J^{r}E$ defined through the derivative of the one parameter
group : $J^{r}W\left(  j^{r}X\left(  x\right)  \right)  =\frac{\partial
}{\partial t}J^{r}\Phi_{W}\left(  j^{r}X\left(  x\right)  ,t\right)  |_{t=0}$.

By construction the r jet prolongation of the one parameter group of morphism
induced by W is the one parameter group of morphism induced by the r jet
prolongation of W: $J^{r}\Phi_{W}=\Phi_{J^{r}W}$. So the Lie derivative
$\pounds _{J^{r}W}Z$ of a section Z of $J^{r}E$ is a section of the vertical
bundle : $VJ^{r}E.$ And if W is vertical the map : $\pounds _{J^{r}%
W}Z:\mathfrak{X}\left(  J^{r}E\right)  \rightarrow VJ^{r}E$ is fiber
preserving and $\pounds _{J^{r}W}Z=J^{r}W\left(  Z\right)  $

\bigskip

\subsection{Functional derivatives}

At first functional derivatives are the implementation of the general theory
of derivatives to functions whose variables are themselves functions. The
theory of distributions gives a rigorous framework to a method which is
enticing because it is simple and intuitive.

The theory of derivatives holds whenever the maps are defined over affine
normed spaces, even if it is infinite dimensional. So, for any map :
$\ell:E\rightarrow%
\mathbb{R}
$ where E is some normed space of functions, one can define derivatives, and
if E is a product $E_{1}\times E_{2}..\times E_{p}$ partial derivatives, of
first and higher order. A derivative such that $\frac{d\ell}{dX}$ , where X is
a function, is a linear map : $\frac{d\ell}{dX}:E\rightarrow%
\mathbb{R}
$ meaning a distribution if it is continuous. So, in many ways, one can see
distributions as the "linearisation" of functions of functions. All the
classic results apply to this case and there is no need to tell more on the subject.

It is a bit more complicated when the map : $\ \ell:\mathfrak{X}\left(
J^{r}E\right)  \rightarrow%
\mathbb{R}
$ is over the space $\mathfrak{X}\left(  J^{r}E\right)  $\ of sections of the
r jet prolongation $J^{r}E$ of a vector bundle E. $\mathfrak{X}\left(
J^{r}E\right)  $ is an infinite dimensional complex vector space.This is not a
normed space but only a Fr\'{e}chet space. So we must look for a normed vector
subspace $F_{r}$ of \ $\mathfrak{X}\left(  J^{r}E\right)  .$ Its choice
depends upon the problem.

\subsubsection{Definition of the functional derivative}

\begin{definition}
A functional $L=\ell\circ J^{r}$ with $\ell:J^{r}E\rightarrow%
\mathbb{R}
$ , where E is a vector bundle $E\left(  M,V,\pi\right)  ,$ defined on a
normed vector subspace $F\subset\mathfrak{X}\left(  E\right)  $ of the
sections of E, has a \textbf{functional derivative} $\frac{\delta L}{\delta
X}$ at $X_{0}$ if there is a distribution $\frac{\delta L}{\delta X}\in
F^{\prime}$ such that :%

\begin{equation}
\forall\delta X\in F:\lim_{\left\Vert \delta X\right\Vert _{F}\rightarrow
0}\left\vert L\left(  X_{0}+\delta X\right)  -L\left(  X_{0}\right)
-\frac{\delta L}{\delta X}\left(  \delta X\right)  \right\vert =0
\end{equation}

\end{definition}

1. To understand the problem, let us assume that $\forall X\in F,J^{r}X\in
F_{r}\subset\mathfrak{X}\left(  J^{r}E\right)  $ a normed \ vector space. The
functional $\ell$ reads :

$L\left(  X\right)  =\ell\left(  X,D_{\alpha_{1}...\alpha_{s}}%
X,s=1...r\right)  $ and if $\ell$ is differentiable at $J^{r}X_{0}$ in $F_{r}$
then :

$L\left(  X_{0}+\delta X\right)  -L\left(  X_{0}\right)  $

$=\sum_{s=0}^{r}\sum_{\alpha_{1}...\alpha_{s}}\frac{\partial\ell}{\partial
D_{\alpha_{1}...\alpha_{s}}X}\left(  J^{r}X_{0}\right)  \left(  D_{\alpha
_{1}...\alpha_{s}}\delta X\right)  +\varepsilon\left\Vert J^{r}\delta
X\right\Vert $

The value of $J^{r}\delta X$ is computed as the derivatives of $\delta
X:D_{\alpha_{1}...\alpha_{s}}\delta X\left(  x\right)  .$ The quantities
$\frac{\partial\ell}{\partial D_{\alpha_{1}...\alpha_{s}}X}$ are the first
order partial derivative of $\ell$ with respect to each of the $D_{\alpha
_{1}...\alpha_{s}}X:$ they are linear maps : $\pi_{s-1}^{s}\left(  J^{r}\delta
X\right)  \rightarrow%
\mathbb{R}
.$ Notice that all the quantities depend on x, thus the relation above shall
hold $\forall x\in M.$

As we can see a priori there is no reason why $D_{\alpha_{1}...\alpha_{s}%
}\delta X$ should be related to $\delta X$ alone.

2. Notice that the order r is involved in the definition of $\ell$ and we do
not consider the derivative of the map : $\frac{\delta L}{\delta
X}:F\rightarrow F^{\prime}$

3. If $\ell$ is a continuous linear map then it has a functional derivative :
$\ell\left(  Z_{0}+\delta Z\right)  -\ell\left(  Z_{0}\right)  =\ell\left(
\delta Z\right)  $ and $\frac{d\ell}{dZ}=\ell$

4. A special interest of functional derivatives is that they can be
implemented with composite maps.

If $F\left(  N,W,\pi_{F}\right)  $ is another vector bundle, and
$Y:F\rightarrow E$ is a differentiable map between the manifolds F,E, if L is
a functional on E, and $X\in\mathfrak{X}_{r}\left(  E\right)  $ has a
functional derivative on E, then $\widehat{L}=L\circ X$ has a functional
derivative with respect to Y given by $\frac{\delta\widehat{L}}{\delta
Y}=\frac{dX}{dY}\frac{\delta L}{\delta X}$ which is the product of the
distribution $\frac{\delta L}{\delta X}$ by the function $\frac{dX}{dY}:$

$X\left(  Y+\delta Y\right)  -X\left(  Y\right)  =\frac{\partial X}{\partial
Y}\delta Y+\varepsilon\left\Vert \delta Y\right\Vert $

$\widehat{L}\left(  Y+\delta Y\right)  -\widehat{L}\left(  Y\right)  =L\left(
X\left(  Y+\delta Y\right)  \right)  -L\left(  X\left(  Y\right)  \right)  $

$=L\left(  X\left(  Y\right)  +\frac{\partial X}{\partial Y}\delta
f+\varepsilon\left\Vert \delta Y\right\Vert \right)  -L\left(  X\left(
f\right)  \right)  $

$=\frac{\delta L}{\delta X}|_{X(Y)}\left(  \frac{\partial X}{\partial Y}\delta
Y+\varepsilon\left\Vert \delta Y\right\Vert \right)  =\frac{\delta L}{\delta
X}|_{X(Y)}\frac{\partial X}{\partial Y}\delta Y+\varepsilon^{\prime}\left\Vert
\delta Y\right\Vert $

\subsubsection{Functional defined as an integral}

\begin{theorem}
A functional $L:\mathfrak{X}_{r}\left(  E\right)  \rightarrow%
\mathbb{R}
::L\left(  X\right)  =\int_{M}\lambda\left(  J^{r}X\right)  $ where
$\lambda\in C_{r+1}\left(  J^{r}E;\Lambda_{m}TM^{\ast}\right)  $ has the
functional derivative :%

\begin{equation}
\frac{\delta L}{\delta X}=\sum_{s=0}^{r}\sum_{\alpha_{1}...\alpha_{s}=1}%
^{m}\left(  -1\right)  ^{s}D_{\alpha_{1}...\alpha_{s}}T\left(  \frac
{\partial\lambda_{a}}{\partial z_{\alpha_{1}...\alpha_{s}}}\right)
\end{equation}

over the space of tests maps : $\delta X\in\mathfrak{X}_{2r,c}\left(
E\right)  $
\end{theorem}

\begin{proof}
M is a class r m dimensional real manifold with atlas $\left(  O_{a},\psi
_{a}\right)  _{a\in A}.$

$\lambda$ reads : $\lambda=\lambda_{a}\left(  z,z_{\alpha_{1}...\alpha_{s}%
},s=1..r,1\leq\alpha_{k}\leq m\right)  d\xi^{1}\wedge...\wedge d\xi^{m}$ where
$\lambda_{a}\in C_{r+1}\left(  J^{r}E;%
\mathbb{R}
\right)  $ changes at the transition of opens $O_{a}$ according to the usual rules.

The definition : $\ell:\mathfrak{X}\left(  J^{r}E\right)  \rightarrow%
\mathbb{R}
::\ell\left(  Z\right)  =\int_{M}\lambda\left(  Z\left(  x\right)  \right)  $
makes sense because, for any section $Z\in\mathfrak{X}\left(  J^{r}E\right)  $

$Z=\left(  z,z_{\alpha_{1}...\alpha_{s}},s=1..r,1\leq\alpha_{k}\leq m\right)
\in\mathfrak{X}\left(  J^{r}E\right)  $ has a value for each $x\in M$.

$J^{r}E$ is a fiber bundle $J^{r}E\left(  M,J_{0}^{r}\left(
\mathbb{R}
^{m},V\right)  _{0},\pi^{r}\right)  .$ So $z_{\alpha_{1}...\alpha_{s}}\in
C\left(  M;V\right)  $

With the first order derivative of $\lambda_{a}$\ with respect to Z, in the
neighborhood of $Z_{0}\in\mathfrak{X}\left(  J^{r}E\right)  $ and for $\delta
Z\in J^{r}E$

$\lambda_{a}\left(  Z_{0}+\delta Z\right)  =\lambda\left(  Z_{0}\right)
+\sum_{s=0}^{r}\sum_{\alpha_{1}...\alpha_{s}=1}^{m}\frac{\partial\lambda_{a}%
}{\partial z_{\alpha_{1}...\alpha_{s}}}\left(  Z_{0}\right)  \left(  \delta
z_{\alpha_{1}...\alpha_{s}}\right)  +\varepsilon\left(  \delta Z\right)
\left\Vert \delta Z\right\Vert $

For a given $Z_{0}\in\mathfrak{X}\left(  J^{r}E\right)  ,$ and $\alpha
_{1},...,\alpha_{s}$ fixed, s
$>$
0, $\frac{\partial\lambda_{a}}{\partial z_{\alpha_{1}...\alpha_{s}}}\left(
Z_{0}\right)  $ is a 1-form on V, acting on $z_{\alpha_{1}...\alpha_{s}},$
valued in $%
\mathbb{R}
.$

The r-form $\varpi_{\alpha_{1}...\alpha_{s}}=\frac{\partial\lambda_{a}%
}{\partial z_{\alpha_{1}...\alpha_{s}}}\left(  Z_{0}\left(  x\right)  \right)
d\xi^{1}\wedge...\wedge d\xi^{m}\in\Lambda_{m}\left(  M;V^{\prime}\right)  $
defines a distribution :

$T\left(  \varpi_{\alpha_{1}...\alpha_{s}}\right)  :C_{rc}\left(
J^{r}E;V\right)  \rightarrow%
\mathbb{R}
::\varphi\rightarrow\int_{M}\frac{\partial\lambda_{a}}{\partial z_{\alpha
_{1}...\alpha_{s}}}\left(  Z_{0}\left(  x\right)  \right)  \left(
\varphi\right)  d\xi^{1}\wedge...\wedge d\xi^{m}$

By definition of the derivative of a distribution $T\left(  \varpi_{\alpha
_{1}...\alpha_{s}}\right)  \left(  D_{\alpha_{1}...\alpha_{s}}\varphi\right)
=\left(  -1\right)  ^{s}D_{\alpha_{1}...\alpha_{s}}T\left(  \varpi_{\alpha
_{1}...\alpha_{s}}\right)  \left(  \varphi\right)  $

If we take $\delta Z=J^{r}\delta X,\delta X\in\mathfrak{X}_{2r,c}\left(
E\right)  $ then :

$T\left(  \varpi_{\alpha_{1}...\alpha_{s}}\right)  \left(  D_{\alpha
_{1}...\alpha_{s}}\delta X\right)  =\left(  -1\right)  ^{s}D_{\alpha
_{1}...\alpha_{s}}T\left(  \varpi_{\alpha_{1}...\alpha_{s}}\right)  \left(
\delta X\right)  $

L reads :

$L\left(  X_{0}+\delta X\right)  $

$=L\left(  X_{0}\right)  +\sum_{s=0}^{r}\sum_{\alpha_{1}...\alpha_{s}=1}%
^{m}\left(  -1\right)  ^{s}D_{\alpha_{1}...\alpha_{s}}T\left(  \varpi
_{\alpha_{1}...\alpha_{s}}\right)  \left(  \delta X\right)  +\int
_{M}\varepsilon\left(  \delta X\right)  \left\Vert \delta X\right\Vert
d\xi^{1}\wedge...\wedge d\xi^{m}$

So L has the functional derivative :

$\frac{\delta L}{\delta X}=\sum_{s=0}^{r}\sum_{\alpha_{1}...\alpha_{s}=1}%
^{m}\left(  -1\right)  ^{s}D_{\alpha_{1}...\alpha_{s}}T\left(  \frac
{\partial\lambda_{a}}{\partial z_{\alpha_{1}...\alpha_{s}}}\right)  \left(
\delta X\right)  $
\end{proof}

$D_{\alpha_{1}...\alpha_{s}}T\left(  \frac{\partial\lambda_{a}}{\partial
z_{\alpha_{1}...\alpha_{s}}}\right)  $\ is computed by the total differential :

$D_{\alpha_{1}...\alpha_{s}}T\left(  \frac{\partial\lambda_{a}}{\partial
z_{\alpha_{1}...\alpha_{s}}}\right)  \left(  \delta X\right)  =d_{\alpha_{1}%
}d_{\alpha_{2}}...d_{\alpha_{s}}\left(  \frac{\partial\lambda\left(  x\right)
}{\partial z_{\alpha_{1}...\alpha_{s}}}\left(  \delta X\left(  x\right)
\right)  \right)  $

\bigskip

\subsection{Lagrangian}

This is the general framework of variational calculus. It can be implemented
on any fibered manifold E. The general problem is to find a section of E for
which the integral on the base manifold M: $\ell\left(  Z\right)  =\int_{M}%
\mathcal{L}%
\left(  Z\right)  $ is extremum.
$\mathcal{L}$%
\ is a m form depending on a section of the r jet prolongation of E called the lagrangian.

\subsubsection{Lagrangian form}

\begin{definition}
A \textbf{r order lagrangian} is a base preserving map $%
\mathcal{L}%
:J^{r}E\rightarrow\Lambda_{m}\left(  M;%
\mathbb{R}
\right)  $ where $E\left(  M,V,\pi\right)  $ is a fiber bundle on a real m
dimensional manifold.
\end{definition}

Comments :

i) Base preserving means that the operator is local : if $Z\in J^{r}E,\pi
^{r}\left(  Z\right)  =x\in M$ and $L\left(  x\right)  :J^{r}E\left(
x\right)  \rightarrow\Lambda_{m}T_{x}M^{\ast}$

ii) L\ is a differential operator as defined previously and all the results
and definitions hold. The results which are presented here are still valid
when E is a fibered manifold.

iii) Notice that
$\mathcal{L}$%
\ is a m form, the same dimension as M, and it is real valued.

iv) This definition is consistent with the one given by geometers (see Kolar
p.388). Some authors define the lagrangian as a horizontal form on $J^{r}E$ ,
which is more convenient for studies on the variational complex. With the
present definition we preserve, in the usual case of vector bundle and in all
cases with the lagrangian function, all the apparatus of differential
operators already available.

v) $J^{r}:E\mapsto J^{r}E$ and $\Lambda_{m}:M\mapsto\Lambda_{m}\left(  M;%
\mathbb{R}
\right)  $ are two bundle functors and $%
\mathcal{L}%
:J^{r}\hookrightarrow\Lambda_{m}$ is a natural operator.

With the bases and coordinates above,
$\mathcal{L}$%
\ reads :

$\lambda\left(  \xi^{\alpha},\eta^{i},\eta_{\alpha_{1}...\alpha_{s}}%
^{i}\right)  d\xi^{1}\wedge..\wedge d\xi^{m}$

that we denote $%
\mathcal{L}%
\left(  Z\right)  =\lambda\left(  Z\right)  d\xi^{1}\wedge...\wedge d\xi^{m}$

For a section $X\in\mathfrak{X}\left(  E\right)  $ one can also write :

$%
\mathcal{L}%
\left(  J^{r}X\left(  x\right)  \right)  =%
\mathcal{L}%
\circ J^{r}\left(  X\left(  x\right)  \right)  =\widehat{%
\mathcal{L}%
}\left(  X\left(  x\right)  \right)  =J^{r}X^{\ast}%
\mathcal{L}%
\left(  x\right)  $

with the pull back of the map
$\mathcal{L}$%
\ by the map : $J^{r}X:M\rightarrow J^{r}E$. But this is not the pull back of
a m form defined on $J^{r}E.$ So we have :

$%
\mathcal{L}%
\left(  J^{r}X\left(  x\right)  \right)  =J^{r}X^{\ast}\lambda\left(
x\right)  d\xi^{1}\wedge...\wedge d\xi^{m}$

\subsubsection{Scalar lagrangian}

The definition above is a geometric, intrinsic definition. In a change of
charts on M the value of $\lambda(Z)$ changes as any other m form. In a change
of holonomic basis :

$d\xi^{\alpha}\rightarrow d\widetilde{\xi}^{\alpha}:%
\mathcal{L}%
\left(  Z\right)  =\lambda\left(  Z\left(  x\right)  \right)  d\xi^{1}%
\wedge...\wedge d\xi^{m}=\widetilde{\lambda}\left(  Z\left(  x\right)
\right)  d\widetilde{\xi}^{1}\wedge...\wedge d\widetilde{\xi}^{m}$

with $\widetilde{\lambda}=\lambda\det\left[  J^{-1}\right]  $ where J is the
jacobian $J=\left[  F^{\prime}(x)\right]  \simeq\left[  \frac{\partial
\widetilde{\xi}^{\alpha}}{\partial\xi^{\beta}}\right]  $

If M is endowed with a volume form $\varpi_{0}$ then
$\mathcal{L}$%
(Z) can be written L(Z)$\varpi_{0}$. Notice that any volume form (meaning that
is never null) is suitable, and we do not need a riemannian structure for this
purpose. For any pseudo-riemannian metric we have $\varpi_{0}=\sqrt{\left\vert
\det g\right\vert }d\xi^{1}\wedge...\wedge d\xi^{m}$ and L(z)=$\lambda\left(
Z\left(  x\right)  \right)  /\sqrt{\left\vert \det g\right\vert } $

In a change of chart on M the component of $\varpi_{0}$ changes according to
the rule above, thus L(Z) does not change : this is a function.

Thus $L:J^{r}E\rightarrow C\left(  M;%
\mathbb{R}
\right)  $ is a r order scalar differential operator that we will call the
\textbf{scalar lagrangian} associated to
$\mathcal{L}$%
.

\subsubsection{Covariance}

1. The scalar lagrangian is a function, so it shall be invariant by a change
of chart on the manifold M.

If we proceed to the change of charts :

$\xi^{\alpha}\rightarrow\widetilde{\xi}^{\alpha}:\widetilde{\xi}=F\left(
\xi\right)  $ with the jacobian : $\left[  \frac{\partial\widetilde{\xi
}^{\alpha}}{\partial\xi^{\beta}}\right]  =K$ and its inverse J.

the coordinate $\eta^{i}$\ on the fiber bundle will not change, if E is not
some tensorial bundle, that we assume.

The coordinates $\eta_{\alpha_{1}...\alpha_{s}}^{i}$ become $\widetilde{\eta
}_{\alpha_{1}...\alpha_{s}}^{i}$ with $\widetilde{\eta}_{\beta\alpha
_{1}...\alpha_{s}}^{i}=\sum_{\gamma}\frac{\partial\xi^{\gamma}}{\partial
\widetilde{\xi}^{\beta}}d_{\gamma}\eta_{\alpha_{1}...\alpha_{s}}^{i}$ where
$d_{\gamma}=\frac{\partial}{\partial\xi^{\gamma}}+\sum_{s=1}^{r}\sum
_{\beta_{1\leq}...\leq\beta_{s}}y_{\gamma\beta_{1}...\beta_{s}}^{i}%
\frac{\partial}{\partial y_{\beta_{1}...\beta_{s}}^{i}}$ is the total differential.

2. The first consequence is that the coordinates $\xi$ of points of M cannot
appear explicitely in the Lagrangian.

3. The second consequence is that there shall be some relations between the
partial derivatives of L. They can be found as follows with the simple case
r=1 as example :

With obvious notations : $L\left(  \eta^{i},\eta_{\alpha}^{i}\right)
=\widetilde{L}\left(  \widetilde{\eta}^{i},\widetilde{\eta}_{\alpha}%
^{i}\right)  =\widetilde{L}\left(  \eta^{i},J_{\alpha}^{\beta}\eta_{\beta}%
^{i}\right)  $

The derivative with respect to $J_{\mu}^{\lambda},\lambda,\mu$ fixed is :

$\sum_{\alpha\beta i}\frac{\partial\widetilde{L}}{\partial\widetilde{\eta
}_{\alpha}^{i}}\eta_{\beta}^{i}\delta_{\lambda}^{\beta}\delta_{\alpha}^{\mu
}=0=\sum_{i}\frac{\partial\widetilde{L}}{\partial\widetilde{\eta}_{\mu}^{i}%
}\eta_{\lambda}^{i}$

The derivative with respect to $\eta_{\lambda}^{j},\lambda,j$ fixed is :

$\sum_{\alpha\beta i}\frac{\partial\widetilde{L}}{\partial\widetilde{\eta
}_{\alpha}^{i}}J_{\alpha}^{\beta}\delta_{\lambda}^{\beta}\delta_{j}^{i}%
=\frac{\partial L}{\partial\eta_{\lambda}^{j}}=\sum_{\alpha}\frac
{\partial\widetilde{L}}{\partial\widetilde{\eta}_{\alpha}^{j}}J_{\alpha
}^{\lambda}$

For : $J_{\mu}^{\lambda}=\delta_{\mu}^{\lambda}$ it comes :

$\forall\lambda:\sum_{i}\frac{\partial\widetilde{L}}{\partial\widetilde{\eta
}_{\lambda}^{i}}\eta_{\lambda}^{i}=0$

$\forall\lambda:\frac{\partial L}{\partial\eta_{\lambda}^{j}}=\frac
{\partial\widetilde{L}}{\partial\widetilde{\eta}_{\lambda}^{j}}$

So there is the identity: $\forall\alpha:\sum_{i}\frac{\partial L}%
{\partial\eta_{\alpha}^{i}}\eta_{\alpha}^{i}=0$

4. The third consequence is that some partial derivatives can be considered as
components of tensors. In the example above :

$\frac{\partial L}{\partial\eta^{i}}$ do not change, and are functions on M

$\sum\frac{\partial L}{\partial\eta_{\lambda}^{j}}\partial\xi^{\alpha}%
=\sum\frac{\partial\widetilde{L}}{\partial\widetilde{\eta}_{\alpha}^{j}%
}J_{\alpha}^{\lambda}\partial\xi^{\alpha}=\sum\frac{\partial\widetilde{L}%
}{\partial\widetilde{\eta}_{\alpha}^{j}}\partial\widetilde{\xi}^{\alpha}$ so
$\forall i:$ $\left(  \frac{\partial L}{\partial\eta_{\lambda}^{j}}\right)
_{\alpha=1}^{m}$ are the components of a vector field.

This remark comes handy in many calculations on Lagrangians.

5. The Lagrangian can also be equivariant in some gauge transformations. This
is studied below with Noether currents.

\bigskip

\subsection{Euler-Lagrange equations}

Given a Lagrangian $%
\mathcal{L}%
:J^{r}E\rightarrow\Lambda_{m}\left(  M;%
\mathbb{R}
\right)  $ the problem is to find a section $X\in\mathfrak{X}\left(  E\right)
$ such that the integral $\ell\left(  Z\right)  =\int_{M}%
\mathcal{L}%
\left(  Z\right)  $ is extremum.

\subsubsection{Vectorial bundle}

\paragraph{General result\newline}

If E is a vectorial bundle we can use the functional derivative : indeed the
usual way to find an extremum of a function is to look where its derivative is
null. We use the precise conditions stated in the previous subsection.

1. If E$\left(  M,V,\pi\right)  $ is a class 2r vector bundle, $\lambda$\ a
r+1 differentiable map then the map : $\ell\left(  J^{r}X\right)  =\int_{M}%
\mathcal{L}%
\left(  J^{r}X\right)  $ has a functional derivative in any point $X_{0}%
\in\mathfrak{X}_{2r,c}\left(  E\right)  $, given by :

$\frac{\delta\widehat{\ell}}{\delta X}\left(  X_{0}\right)  =T\left(
\sum_{s=0}^{r}\sum_{\alpha_{1}...\alpha_{s}=1}^{m}\left(  -1\right)
^{s}d_{\alpha_{1}}d_{\alpha_{2}}...d_{\alpha_{s}}\frac{\partial\lambda
}{\partial z_{\alpha_{1}...\alpha_{s}}}\left(  J^{2r}X_{0}\right)  \right)  $

So :

$\forall\delta X\in\mathfrak{X}_{2r,c}\left(  E\right)  :$

$\lim_{\left\Vert \delta X\right\Vert _{W^{r,p}\left(  E\right)  }%
\rightarrow0}\left\vert L\left(  X_{0}+\delta X\right)  -L\left(
X_{0}\right)  -\frac{\delta\widehat{\ell}}{\delta X}\left(  X_{0}\right)
\delta X\right\vert =0$

2. The condition for a local extremum is clearly : $\frac{\delta\widehat{\ell
}}{\delta X}\left(  X_{0}\right)  =0$ which gives the Euler-Lagrange equations :%

\begin{equation}
i=1...n:\sum_{s=0}^{r}\sum_{\alpha_{1}...\alpha_{s}=1}^{m}\left(  -1\right)
^{s}d_{\alpha_{1}}d_{\alpha_{2}}...d_{\alpha_{s}}\frac{\partial\lambda
}{\partial\eta_{\alpha_{1}...\alpha_{s}}^{i}}\left(  J^{2r}X_{0}\right)  =0
\end{equation}

We have a system of n partial differential equations of order 2r

For r=1 :%

\begin{equation}
\frac{\partial L}{\partial\eta^{i}}-\sum_{\alpha=1}^{n}\frac{d}{d\xi^{\alpha}%
}\frac{\partial L}{\partial\eta_{\alpha}^{i}}=0;i=1...n
\end{equation}

For r=2 : $\frac{\partial L}{\partial\eta^{i}}-\frac{d}{d\xi^{\alpha}}\left(
\frac{\partial L}{\partial\eta_{\alpha}^{i}}-\sum_{\beta}\frac{d}{d\xi^{\beta
}}\frac{\partial L}{\partial\eta_{\alpha\beta}^{i}}\right)  =0;i=1...n,\alpha
=1...m$

The derivatives are evaluated for $X\left(  \xi\right)  :$ they are total
derivatives $\frac{d}{d\xi^{\alpha}}$

$\alpha=1...m:d_{\alpha}\lambda\left(  \xi^{\alpha},\eta^{i},\eta_{\alpha
_{1}...\alpha_{s}}^{i}\right)  =\frac{\partial\lambda}{\partial\xi^{\alpha}%
}+\sum_{s=1}^{r}\sum_{\beta_{1}...\beta_{s}}\frac{\partial\lambda}%
{\partial\eta_{\alpha\beta_{1}...\beta_{s}}^{i}}\eta_{\alpha\beta_{1}%
...\beta_{s}}^{i}$

3. We could equivalently consider $\lambda$ compactly supported, or M a
compact manifold with boundary and $\lambda$ continuous on the boundary.. Then
the test functions are $\mathfrak{X}_{2r}\left(  E\right)  .$

\paragraph{First order Lagrangian\newline}

The \textbf{stress energy tensor} (there are many definitions) is the quantity :

$\sum_{\alpha\beta}T_{\beta}^{\alpha}\partial_{\alpha}x\otimes dx^{\beta}%
=\sum_{\alpha\beta}\left(  \sum_{i=1}^{n}\frac{\partial\lambda}{\partial
\eta_{\alpha}^{i}}\eta_{\beta}^{i}-\delta_{\beta}^{\alpha}\lambda\right)
\partial_{\alpha}x\otimes dx^{\beta}\in TM\otimes TM^{\ast}$

It is easy to check the identity :

$\sum_{\alpha=1}^{m}\frac{d}{d\xi^{\alpha}}T_{\beta}^{\alpha}=-\frac
{\partial\lambda}{\partial\xi^{\beta}}+\sum_{i=1}^{p}\mathfrak{E}_{i}\left(
\lambda\right)  \eta_{\beta}^{i}$

where $\mathfrak{E}_{i}\left(  \lambda\right)  =\left(  \frac{\partial
L}{\partial Z^{i}}-\sum_{\alpha=1}^{m}\frac{d}{d\xi^{\alpha}}\left(
\frac{\partial L}{\partial Z_{\alpha}^{i}}\right)  \right)  d\xi^{1}%
\wedge...\wedge d\xi^{m}\in\Lambda_{m}TM$

and the \textbf{Euler-Lagrange form} is :

$\mathfrak{E}\left(  \lambda\right)  =\sum_{i}\mathfrak{E}_{i}\left(
\lambda\right)  du^{i}\otimes d\xi^{1}\wedge...\wedge d\xi^{m}\in E^{\ast
}\otimes\Lambda_{m}TM$

So if $\lambda$ does not depend on $\xi$ (as it should because of the
covariance) there is a primitive : $\sum_{\alpha=1}^{m}\frac{d}{d\xi^{\alpha}%
}T_{\beta}^{\alpha}=0$

\subsubsection{First order lagrangian on $%
\mathbb{R}
$}

This is indeed the simplest but one of the most usual case. The theorems below
use a method similar to the functional derivative. We give them because they
use more precise, and sometimes more general conditions.

\paragraph{Euler Lagrange equations\newline}

(Schwartz 2 p.303)

\begin{problem}
Find, in $C_{1}\left(  \left[  a,b\right]  ;F\right)  ,$ a map $X:I\rightarrow
F$ such that :

$\ell\left(  X\right)  =\int_{a}^{b}\lambda\left(  t,X\left(  t\right)
,X^{\prime}\left(  t\right)  \right)  dt$ is extremum.

$I=\left[  a,b\right]  $ is a closed interval in $%
\mathbb{R}
,$ F is a normed affine space, U is an open subset of F$\times$F

$\lambda:I\times U\rightarrow%
\mathbb{R}
$ is a continuous function
\end{problem}

We have the following results :

i) the set of solutions is an open subset O of $C_{1}\left(  I;F\right)  $

ii) the function : $\ell:C_{1}\left(  I;E\right)  \rightarrow%
\mathbb{R}
$ is continuously differentiable in O, and its derivative in $X_{0}$ is :

$\frac{\delta\ell}{\delta X}|_{X=X_{0}}\delta X=\int_{a}^{b}\left(
\lambda^{\prime}(t,X_{0}(t),X_{0}^{\prime}(t))(0,\delta X(t),\left(  \delta
X\right)  ^{\prime}(t))\right)  dt$

$=\int_{a}^{b}[\frac{\partial\lambda}{\partial X}\delta X+\frac{\partial
\lambda}{\partial X^{\prime}}\left(  \delta X\right)  ^{\prime}]dt$

iii) If $\lambda$ is a class 2 map, $X_{0}$ a class 2 map then :

$\frac{\delta\ell}{\delta X}|_{X=X_{0}}\delta X=[\frac{\partial\lambda
}{\partial X^{\prime}}(t,X_{0}(t),X_{0}^{\prime}(t))\delta X(t)]_{a}^{b}%
+\int_{a}^{b}[\frac{\partial\lambda}{\partial X}-\dfrac{d}{dt}\frac{\partial
L}{\partial X^{\prime}}]\delta Xdt$

and if X is an extremum of $\ell$ with the conditions $X\left(  a\right)
=\alpha,X\left(  b\right)  =\beta$\ \ then it is a solution of the ODE :

$\dfrac{\partial\lambda}{\partial X}=\dfrac{d}{dt}(\dfrac{\partial\lambda
}{\partial X^{\prime}});X\left(  a\right)  =\alpha,X\left(  b\right)  =\beta$

If $\lambda$ does not depend on t then for any solution the quantity :
$\lambda-\dfrac{\partial\lambda}{\partial X^{\prime}}X^{\prime}=Ct$

iv) If X is a solution, it is a solution of the variational problem on any
interval in I:

$\forall\left[  a_{1},b_{1}\right]  \subset\left[  a,b\right]  ,\int_{a_{1}%
}^{b_{1}}\lambda(t,X,X^{\prime})dt$ is extremum for X

\paragraph{Variational problem with constraints\newline}

\begin{problem}
(Schwartz 2 p.323) Find, in $C_{1}\left(  \left[  a,b\right]  ;F\right)  ,$ a
map $X:I\rightarrow F$ such that :

$\ell\left(  X\right)  =\int_{a}^{b}\lambda\left(  t,X\left(  t\right)
,X^{\prime}\left(  t\right)  \right)  dt$ is extremum

and for $k=1..n:J_{k}=\int_{a}^{b}M_{k}\left(  t,X\left(  t\right)
,X^{\prime}\left(  t\right)  \right)  dt$ where $J_{k}$ is a given scalar

F is a normed affine space, U is an open subset of FxF
\end{problem}

Then a solution must satisfy the ODE :

$\dfrac{\partial\lambda}{\partial X}-\dfrac{d}{dt}(\dfrac{\partial\lambda
}{\partial X^{\prime}})=\sum_{k=1}^{n}c_{k}[\dfrac{\partial M_{k}}{\partial
X}-\dfrac{d}{dt}\dfrac{\partial M_{k}}{\partial X^{\prime}}]$ with some
scalars $c_{k}$

\begin{problem}
(Schwartz 2 p.330) Find , in $C_{1}\left(  \left[  a,b\right]  ;F\right)  ,$ a
map $X:I\rightarrow F$ and the 2 scalars a,b such that :

$\ell\left(  X\right)  =\int_{a}^{b}\lambda\left(  t,X\left(  t\right)
,X^{\prime}\left(  t\right)  \right)  dt$ is extremum

F is a normed affine space, U is an open subset of FxF, $\lambda:I\times
U\rightarrow%
\mathbb{R}
$ is a class 2 function
\end{problem}

Then the derivative of $\ell$ with respect to X,a,b at $(X_{0},a_{0},b_{0})$
is :

$\frac{\delta\ell}{\delta X}\left(  \delta X,\delta\alpha,\delta\beta\right)
$

$=\int_{a_{0}}^{b_{0}}[\frac{\partial\lambda}{\partial X}\delta X-\dfrac
{d}{dt}\frac{\partial\lambda}{\partial X^{\prime}}]\left(  t,X_{0}\left(
t\right)  ,X_{0}^{\prime}\left(  t\right)  \right)  \delta Xdt$

$+\left(  \frac{\partial\lambda}{\partial X^{\prime}}\left(  b,X_{0}\left(
b\right)  ,X_{0}^{\prime}\left(  b\right)  \right)  \delta X\left(  b\right)
+\lambda\left(  b,X_{0}\left(  b\right)  ,X_{0}^{\prime}\left(  b\right)
\right)  \right)  $

$-\left(  \frac{\partial\lambda}{\partial X^{\prime}}\left(  a,X_{0}\left(
a\right)  ,X_{0}^{\prime}\left(  b_{0}\right)  \right)  \delta X\left(
a\right)  +\lambda\left(  a,X_{0}\left(  a\right)  ,X_{0}^{\prime}\left(
a\right)  \right)  \right)  $

With the notation $X\left(  a\right)  =\alpha,X\left(  b\right)  =\beta$ this
formula reads :

$\delta X\left(  b\right)  =\delta\beta-X^{\prime}\left(  b_{0}\right)  \delta
b,\delta X\left(  a\right)  =\delta\alpha-X^{\prime}\left(  a_{0}\right)
\delta a$

\paragraph{Hamiltoninan formulation\newline}

This is the classical formulation of the variational problem in physics, when
one variable (the time t) is singled out.

\begin{problem}
(Schwartz 2 p.337) Find , in $C_{1}\left(  \left[  a,b\right]  ;F\right)  ,$ a
map $X:I\rightarrow F$ such that : $\ell\left(  X\right)  =\int_{a}^{b}%
\lambda\left(  t,X\left(  t\right)  ,X^{\prime}\left(  t\right)  \right)  dt$
is extremum

$I=\left[  a,b\right]  $ is a closed interval in $%
\mathbb{R}
,$ F is a finite n dimensional vector space, U is an open subset of F$\times$F
, $\lambda:I\times U\rightarrow%
\mathbb{R}
$ is a continuous function
\end{problem}

$\lambda$ is denoted here (with the usual 1 jet notation) : $L\left(
t,y^{1},...y^{n},z^{1},...z^{m}\right)  $ where for a section $X\left(
t\right)  :z^{i}=\frac{dy^{i}}{dt}$

By the change of variables :

$y^{i}$ replaced by $q_{i}$

$z^{i}$ replaced by $p_{i}=\frac{\partial L}{\partial z^{1}},i=1...n$ which is
called the momentum conjugate to $q_{i}$

one gets the function, called Hamiltonian :%

\begin{equation}
H:I\times F\times F^{\ast}\rightarrow%
\mathbb{R}
::H\left(  t,q_{1},...,q_{n},p_{1},...p_{n}\right)  =\sum_{i=1}^{n}p_{i}%
z^{i}-L
\end{equation}

Then :

$dH=\sum_{i=1}^{n}z^{i}dp_{i}-\sum_{i=1}^{n}\frac{\partial\lambda}{\partial
y^{i}}dq_{i}-\frac{\partial\lambda}{\partial t}$

$\frac{\partial H}{\partial t}=-\frac{\partial\lambda}{\partial t}%
,\frac{\partial H}{\partial q_{i}}=-\frac{\partial\lambda}{\partial y^{i}%
},\frac{\partial H}{\partial p_{i}}=z^{i}$

and the Euler-Lagrange equations read :%

\begin{equation}
\frac{dq_{i}}{dt}=\frac{\partial H}{\partial p_{i}},\frac{dp_{i}}{dt}%
=-\frac{\partial H}{\partial q_{i}}%
\end{equation}

which is the Hamiltonian formulation.

If $\lambda$ does not depend on t then the quantity $H=Ct$ for the solutions.

\subsubsection{Variational problem on a fibered manifold}

Given a Lagrangian $%
\mathcal{L}%
:J^{r}E\rightarrow\Lambda_{m}\left(  M;%
\mathbb{R}
\right)  $ on any fibered manifold the problem is to find a section
$X\in\mathfrak{X}\left(  E\right)  $ such that the integral $\ell\left(
Z\right)  =\int_{M}%
\mathcal{L}%
\left(  Z\right)  $ is extremum. The first step is to define the extremum of a function.

\paragraph{Stationary solutions\newline}

If E is not a vector bundle the general method is to use the Lie derivative.

The - clever - idea is to "deform" a section X according to some rules that
can be parametrized, and to look for a stationary state of deformation, such
that in a neighborhood of some $X_{0}$ any other transformation does not
change the value of $\ell(J^{r}X).$ The focus is on one parameter groups of
diffeomorphisms, generated by a projectable vector field W on E.

It defines a base preserving map :

$\mathfrak{X}\left(  J^{r}E\right)  \rightarrow\mathfrak{X}\left(
J^{r}E\right)  ::J^{r}U\left(  t\right)  Z\left(  x\right)  =\Phi_{J^{r}%
W}\left(  Z\left(  x\right)  ,t\right)  $

One considers the changes in the value of $\lambda\left(  Z\right)  $ when Z
is replaced by $J^{r}U\left(  t\right)  Z:\lambda\left(  Z\right)
\rightarrow\lambda\left(  J^{r}U\left(  t\right)  Z\right)  $ and the Lie
derivative of $\lambda$\ along W is : $\pounds _{W}\lambda=\frac{d}{dt}%
\lambda\left(  J^{r}U\left(  t\right)  Z\right)  |_{t=0}$\ . Assuming that
$\lambda$ is differentiable with respect to Z at $Z_{0}$ : $\pounds _{W}%
\lambda=\lambda^{\prime}\left(  Z_{0}\right)  \frac{d}{dt}J^{r}U\left(
t\right)  Z|_{t=0}=\lambda^{\prime}\left(  Z_{0}\right)  \pounds _{J^{r}W}Z.$
The derivative $\lambda^{\prime}\left(  Z_{0}\right)  $ is a continuous linear
map from the tangent bundle $TJ^{r}E$ to $%
\mathbb{R}
$ and $\pounds _{J^{r}W}Z$ is a section of the vertical bundle $VJ^{r}E.$ So
$\pounds _{W}\lambda$ is well defined and is a scalar.%

\begin{equation}
\pounds _{W}%
\mathcal{L}%
\left(  X\right)  =\frac{\partial}{\partial t}%
\mathcal{L}%
\left(  J^{r}U\left(  t\right)  J^{r}X\right)  |_{t=0}=\pounds _{W}\lambda
d\xi^{1}\wedge...\wedge d\xi^{m}\
\end{equation}

is called the \textbf{variational derivative} of the Lagrangian
$\mathcal{L}$%
\ . This is the Lie derivative of the natural operator (Kolar p.387) : $%
\mathcal{L}%
:J^{r}\hookrightarrow\Lambda_{m}$ :

$\pounds _{W}%
\mathcal{L}%
\left(  p\right)  =\pounds _{\left(  J^{r}W,\Lambda_{m}TM^{\ast}Y\right)  }%
\mathcal{L}%
\left(  p\right)  =\frac{\partial}{\partial t}\Phi_{TM^{\ast}Y}\left(
\mathcal{L}%
\left(  \Phi_{J^{r}W}\left(  p,-t\right)  \right)  ,t\right)  |_{t=0}%
:J^{r}Y\rightarrow\Lambda_{m}TM^{\ast}$

A section $X_{0}\in\mathfrak{X}\left(  E\right)  $\ is a \textbf{stationary
solution} of the variational problem if $\int_{M}\pounds _{W}%
\mathcal{L}%
\left(  X\right)  =0$ for any projectable vector field.

The problem is that there is no simple formulation of $\pounds _{J^{r}W}Z.$

Indeed : $\pounds _{J^{r}W}Z=-\frac{\partial\Phi_{J^{r}W}}{\partial Z}\left(
Z,0\right)  \frac{dZ}{dx}Y+J^{r}W\left(  Z\right)  \in V_{p}E$ where Y is the
projection of W on TM (see Fiber bundles).

For $Z=J^{r}X,X\in\mathfrak{X}\left(  E\right)  $

$\Phi_{W}\left(  J^{r}X\left(  x\right)  ,t\right)  =j_{\Phi_{Y}\left(
x,t\right)  }^{r}\left(  \Phi_{W}\left(  X(\Phi_{Y}\left(  x,-t\right)
),t\right)  \right)  $

$=\left(  D_{\alpha_{1}..\alpha_{s}}\Phi_{W}\left(  X(\Phi_{Y}\left(
x,-t\right)  ),t\right)  ,s=0..r,1\leq\alpha_{k}\leq m\right)  $

$\frac{\partial\Phi_{J^{r}W}}{\partial Z}\left(  J^{r}X,0\right)  =\left(
D_{\alpha_{1}..\alpha_{s}}\Phi_{W}\left(  X,0\right)  ,s=0..r,1\leq\alpha
_{k}\leq m\right)  $

$\pounds _{J^{r}W}J^{r}X$

$=J^{r}W\left(  J^{r}X\right)  -\sum_{s=0}^{r}\sum_{\alpha_{1}...\alpha_{s}%
=1}^{m}\left(  D_{\alpha_{1}..\alpha_{s}}\Phi_{W}\left(  X,0\right)  \right)
\left(  \sum_{\beta}Y^{\beta}D_{\beta\alpha_{1}..\alpha_{s}}X\right)  $

The solution, which is technical, is to replace this expression by another
one, which is "equivalent", and gives someting nicer when one passes to the integral.

\paragraph{The Euler-Lagrange form\newline}

\begin{theorem}
(Kolar p.388) For every r order lagrangian $%
\mathcal{L}%
:J^{r}E\rightarrow\Lambda_{m}\left(  M;%
\mathbb{R}
\right)  $ there is a morphism $K\left(
\mathcal{L}%
\right)  :J^{2r-1}E\rightarrow VJ^{r-1}E^{\ast}\otimes\Lambda_{m-1}TM^{\ast}$
and a unique morphism : $\mathfrak{E}\left(
\mathcal{L}%
\right)  :J^{2r}E\rightarrow VE^{\ast}\otimes\Lambda_{m}TM^{\ast}$ such that
for any vertical vector field W on E and section X on E: $\pounds _{W}%
\mathcal{L}%
=\mathfrak{D}\left(  K\left(
\mathcal{L}%
\right)  \left(  J^{r-1}W\right)  \right)  +\mathfrak{E}\left(
\mathcal{L}%
\right)  \left(  W\right)  $
\end{theorem}

$\mathfrak{D}$ is the total differential (see below).

The morphism K(%
$\mathcal{L}$%
), called a \textbf{Lepage equivalent} to
$\mathcal{L}$%
,\ reads:

$K\left(
\mathcal{L}%
\right)  =\sum_{s=0}^{r-1}\sum_{\beta=1}^{m}\sum_{i=1}^{n}\sum_{\alpha_{1}%
\leq..\leq\alpha_{s}}K_{i}^{\beta\alpha_{1}..\alpha_{s}}d\eta_{\alpha
_{1}..\alpha_{s}}^{i}\otimes d\xi^{1}\wedge.\widehat{\wedge d\xi^{\beta}%
}.\wedge d\xi^{m}$

It is not uniquely determined with respect to
$\mathcal{L}$%
. The mostly used is the \textbf{Poincar\'{e}-Cartan equivalent}
$\Theta\left(  \lambda\right)  $ defined by the relations (Krupka 2002):

$\Theta\left(  \lambda\right)  =\left(  \lambda+\sum_{\alpha=1}^{m}\sum
_{s=0}^{r-1}K_{i}^{\alpha\beta_{1}...\beta_{s}}y_{\beta_{1}..\beta_{s}}%
^{i}\right)  \wedge d\xi^{1}\wedge..\wedge d\xi^{m}$

$K_{i}^{\beta_{1}...\beta r+1}=0$

$K_{i}^{\beta_{1}...\beta s}=\frac{\partial\lambda}{\partial\eta_{\beta
_{1}..\beta._{s}}^{i}}-\sum_{\gamma}d_{\gamma}K_{i}^{\gamma\beta_{1}...\beta
s},s=1..r$

$y_{\beta_{1}..\beta_{s}}^{i}=d\eta_{\beta_{1}..\beta_{s}}^{i}-\sum_{\gamma
}\eta_{\beta_{1}..\beta_{s}\gamma}^{i}d\xi^{\gamma}$

$y_{\beta}=\left(  -1\right)  ^{\beta-1}d\xi^{1}\wedge.\widehat{\wedge
d\xi^{\beta}}.\wedge d\xi^{m}$

It has the property that : $\lambda=h\left(  \Theta\left(  \lambda\right)
\right)  $ where h is the horizontalization (see below) and $h\left(
\Theta\left(  J^{r+1}X\left(  x\right)  \right)  \right)  =\Theta\left(
J^{r}X\left(  x\right)  \right)  $

For r = 1 : $\Theta\left(  \lambda\right)  =\lambda+\sum_{\alpha\beta=1}%
^{m}\sum_{i}\frac{\partial\lambda}{\partial\eta_{\alpha}^{i}}y_{\beta}%
^{i}\wedge y_{\beta}$

$=\lambda+\sum_{\alpha\beta=1}^{m}\sum_{i}\frac{\partial\lambda}{\partial
\eta_{\alpha}^{i}}\left(  d\eta^{i}\wedge\left(  -1\right)  ^{\beta-1}d\xi
^{1}\wedge.\widehat{\wedge d\xi^{\beta}}.\wedge d\xi^{m}-\eta_{\beta}^{i}%
d\xi^{1}\wedge..\wedge d\xi^{m}\right)  $

The \textbf{Euler-Lagrange form} $\mathfrak{E}\left(
\mathcal{L}%
\right)  $ is :

$\mathfrak{E}\left(
\mathcal{L}%
\right)  =\sum_{i=1}^{n}\mathfrak{E}\left(
\mathcal{L}%
\right)  _{i}du^{i}\wedge d\xi^{1}\wedge..\wedge d\xi^{m}$

$\mathfrak{E}\left(
\mathcal{L}%
\right)  _{i}=\sum_{s=0}^{r}\left(  -1\right)  ^{s}\sum_{\alpha_{1}\leq
..\leq\alpha_{s}}d_{\alpha_{1}}d_{\alpha_{2}}..d_{\alpha_{s}}\frac
{\partial\lambda}{\partial\eta_{\alpha_{1}..\alpha_{s}}^{i}.}$ where
$d_{\alpha}$ is the total differentiation

$d_{\alpha}f=\frac{\partial f}{\partial\xi^{\alpha}}+\sum_{s=1}^{r}\sum
_{\beta_{1}\leq...\leq\beta_{s}}\frac{\partial f}{\partial\eta_{\alpha
\beta_{1}...\beta_{s}}^{i}}\eta_{\alpha\beta_{1}...\beta_{s}}^{i}$

$\mathfrak{E}\left(
\mathcal{L}%
\right)  $ is a linear natural operator which commutes with the Lie derivative :

\begin{theorem}
(Kolar p.390) For any projectable vector field W on E :

$\pounds _{W}\mathfrak{E}\left(
\mathcal{L}%
\right)  =\mathfrak{E}\left(  \pounds _{W}%
\mathcal{L}%
\right)  $
\end{theorem}

\paragraph{Solutions\newline}

The solutions are deduced from the previous theorem.

The formula above reads :

$J^{r-1}W$ is a vertical vector field in $VJ^{r-1}E\ $\ :

$J^{r-1}W=W^{i}\partial u_{i}+\sum_{s=1}^{r-1}\sum_{\alpha_{1\leq}%
...\leq\alpha_{s}}\left(  d_{\alpha_{1}}d_{\alpha_{2}}...d_{\alpha_{s}}%
W^{i}\right)  \partial u_{i}^{\alpha_{1}...\alpha_{s}}$

so $F=K\left(
\mathcal{L}%
\right)  \left(  J^{r-1}W\right)  =\sum_{\alpha=1}^{m}F_{\alpha}d\xi^{1}%
\wedge.\widehat{\wedge d\xi^{\beta}}..\wedge d\xi^{m}\in\Lambda_{m-1}TM^{\ast
}$

$\mathfrak{D}$ is the total differential :

$\mathfrak{D}F=\sum_{\alpha,\beta=1}^{m}\left(  d_{\alpha}F_{\beta}\right)
d\xi^{\alpha}\wedge d\xi^{1}\wedge.\widehat{\wedge d\xi^{\beta}}.\wedge
d\xi^{m}$

$=\sum_{\alpha,\beta=1}^{m}\left(  -1\right)  ^{\alpha-1}\left(  d_{\alpha
}F_{\alpha}\right)  d\xi^{1}\wedge...\wedge d\xi^{m}$

with : $d_{\alpha}F=\frac{\partial F}{\partial\xi^{\alpha}}+\sum_{s=1}^{r}%
\sum_{\beta_{1}\leq..\leq\beta_{s}}\frac{\partial F}{\partial\eta_{\beta
_{1}...\beta_{s}}^{i}}\eta_{\alpha\beta_{1}...\beta_{s}}^{i}$

$\mathfrak{E}\left(
\mathcal{L}%
\right)  \left(  W\right)  =\sum_{i=1}^{n}\mathfrak{E}\left(
\mathcal{L}%
\right)  _{i}W^{i}d\xi^{1}\wedge...\wedge d\xi^{m}$

$=\sum_{i=1}^{n}\sum_{s=0}^{r}\left(  -1\right)  ^{s}\sum_{\alpha_{1}%
\leq..\leq\alpha_{s}}W^{i}\left(  d_{\alpha_{1}}d_{\alpha_{2}}..d_{\alpha_{s}%
}\frac{\partial\lambda}{\partial\eta_{\alpha_{1}..\alpha_{s}}^{i}.}\right)
d\xi^{1}\wedge...\wedge d\xi^{m}$

The first term $\mathfrak{D}\left(  K\left(
\mathcal{L}%
\right)  \left(  J^{r-1}W\right)  \right)  $ is the differential of a m-1 form :

$\mathfrak{D}\left(  K\left(
\mathcal{L}%
\right)  \left(  J^{r-1}W\right)  \right)  =d\mu$

The second reads : $\mathfrak{E}\left(
\mathcal{L}%
\right)  \left(  W\right)  =i_{W}\mathfrak{E}\left(
\mathcal{L}%
\right)  $

So : $\ell\left(  X\right)  =\int_{M}\left(  d\mu+i_{W}\mathfrak{E}\left(
\mathcal{L}%
\right)  \right)  $

Any open subset O of M, relatively compact, gives a manifold with boundary and
with the Stokes theorem :

$\ell\left(  X\right)  =\int_{\partial O}\mu+\int_{M}i_{W}\mathfrak{E}\left(
\mathcal{L}%
\right)  $

It shall hold for any projectable vector field W. For W with compact support
in O the first integral vanishes. We are left with : $\ell\left(  X\right)
=\int_{M}i_{W}\mathfrak{E}\left(
\mathcal{L}%
\right)  $ which is linearly dependant of $W^{i}.$ So we must have :
$J^{2r}X^{\ast}\mathfrak{E}\left(
\mathcal{L}%
\right)  =0$

\begin{theorem}
A section $X\in\mathfrak{X}_{2r}\left(  E\right)  $ is a stationary solution
of $\ell\left(  J^{r}X\right)  $\ only if $J^{2r}X^{\ast}\mathfrak{E}\left(
\mathcal{L}%
\right)  =0$
\end{theorem}

Notice that there is no guarantee that the solution is a maximum or a minimum,
and it may exist "better solutions" which do not come from a one parameter
group of morphisms.

The Euler Lagrange equations read :

$\sum_{k=0}^{r}\left(  -1\right)  ^{k}d_{\alpha_{1}}d_{\alpha_{2}}%
...d_{\alpha_{k}}\frac{\partial L}{\partial\eta_{\alpha_{1}..\alpha_{k}.}^{i}%
}\left(  j^{2r}X\right)  =0;i=1...n$

So they are exactly the same as the equations that we have found for a
vectorial bundle with the functional derivatives.

\subsubsection{Noether currents}

\paragraph{Principle\newline}

The symmetries in the model provided by a lagrangian are of physical great
importance. They can resort to different categories : they can be physical
(ex: spherical or cylindrical symmetry) so depending on the problem (and the
model should account for them), or be gauge symmetries. They can be modelled
by an automorphism on E : $G:E\rightarrow E$ such that : $\forall
X\in\mathfrak{X}\left(  E\right)  :%
\mathcal{L}%
\left(  J^{r}\left(  G\left(  X\right)  \right)  \right)  =%
\mathcal{L}%
\left(  J^{r}X\right)  .$ Similarly they keep inchanged the scalar lagrangian.
Usually they are studied by one parameter groups of automorphisms,
parametrized by their infinitesimal generator, which is a vector field, but
not necessarily a vector field on E.

\paragraph{Classical symmetries\newline}

\begin{definition}
A vector field $W_{r}\in\mathfrak{X}\left(  TJ^{r}E\right)  $ is said to be
the generator of a (classical, exact) symmetry of
$\mathcal{L}$%
\ if : $\forall Z\in J^{r}E:\pounds _{W_{r}}\pi^{r\ast}%
\mathcal{L}%
\left(  Z\right)  =0$
\end{definition}

Some explanations...

i) $\pi^{r\ast}%
\mathcal{L}%
\left(  Z\right)  $\ is the pull back of a lagrangian
$\mathcal{L}$%
\ by the projection : $\pi^{r}:J^{r}E\rightarrow M$ .\ This is a horizontal m
form\ on $J^{r}E$ , such that :

$\forall W^{k}\in\mathfrak{X}\left(  TJ^{r}E\right)  ,k=1...m:$

$\pi^{r\ast}%
\mathcal{L}%
\left(  Z\right)  \left(  W^{1},..,W^{m}\right)  =%
\mathcal{L}%
\left(  \pi^{r}\left(  Z\right)  \right)  \left(  \pi^{r}\left(  Z\right)
^{\prime}W^{1},...\pi^{r}\left(  Z\right)  ^{\prime}W^{m}\right)  $

ii) $\pounds _{W_{r}}$\ is the Lie derivative (in the usual meaning on the
manifold $J^{r}E)$ defined by the flow $\Phi_{W_{r}}$ of a vector field
$W_{r}\in\mathfrak{X}\left(  TJ^{r}E\right)  $ :

$\Phi_{W_{r}}\ :\frac{\partial}{\partial t}\pi^{r\ast}%
\mathcal{L}%
\left(  Z\right)  \Phi_{W_{r}}\left(  X\left(  \Phi_{W_{r}}\left(
Z,-t\right)  \right)  ,t\right)  =\pounds _{W_{r}}\pi^{r\ast}%
\mathcal{L}%
\left(  Z\right)  $

$W_{r}$ is not necessarily projectable on TM.

So the set of such vectors $W_{r}$ provides a large range of symmetries of the
model. It has the structure of a vector space and of a Lie algebra with the
commutator (Giachetta p.70).

\begin{theorem}
First Noether theorem : If $W_{r}$ is a classical symmetry for
$\mathcal{L}$%
\ then $\mathfrak{D}\left(  K\left(
\mathcal{L}%
\right)  \left(  J^{r-1}W_{r}\right)  \right)  =0$ for the solutions
\end{theorem}

It is common to say that a property is satisfied "on the shell" when it is
satisfied for the solutions of the Euler-Lagrange equations. As a consequence
the quantity called a Noether current

$\mathfrak{I=}K\left(
\mathcal{L}%
\right)  \left(  J^{r-1}W_{r}\right)  =\sum_{\alpha}\mathfrak{I}_{\alpha}%
d\xi^{1}\wedge.\widehat{\wedge d\xi^{\alpha}}.\wedge d\xi^{m}$ is conserved on
the shell.

\begin{theorem}
Noether-Bessel-Hagen theorem (Kolar p.389): A projectable vector field W on E
is a generalized infinitesimal automorphism of the r order lagrangian $%
\mathcal{L}%
:J^{r}E\rightarrow\Lambda_{m}\left(  M;%
\mathbb{R}
\right)  $ iff $\mathfrak{E}\left(  \pounds _{W}%
\mathcal{L}%
\right)  =0$
\end{theorem}

\paragraph{Gauge symmetries\newline}

1. Gauge symmetries arise if there is a principal bundle $P\left(  M,G,\pi
_{P}\right)  .$ Then a change of trivialization on P induces a change in the
components of a connection (the potential \`{A}\ and the strength of the
field
$\mathcal{F}$%
) and of sections on associated bundles. When the change of trivialization is
initiated by a one parameter group defined by a section $\kappa$ on the
adjoint bundle $P\left[  T_{1}G,Ad\right]  $ then the new components are
parametrized by t and $\kappa:\widetilde{X}\left(  t\right)  =F\left(
X,t\right)  $ and the transformation extends to the r-jet prolongation (see
Morphisms on Jets)$.$ A lagrangian $L:J^{r}E\rightarrow%
\mathbb{R}
$ is invariant by a change of trivialization. Then we must have : $L\left(
J^{r}\widetilde{X}\left(  t\right)  \right)  =L\left(  J^{r}X\right)  $ for
any t and change of gauge $\kappa.$ By differentiating with respect to t at
t=0 we get :

$\forall\kappa:\sum\frac{\partial L}{\partial u_{\alpha_{1}...\alpha_{s}}^{i}%
}\frac{d}{dt}\frac{\partial\widetilde{u}^{i}}{\partial\xi^{\alpha_{1}%
}...\partial\xi^{\alpha_{s}}}=0$

that is a set of identities between $J^{r}X$ and the partial derivatives
$\frac{\partial L}{\partial u_{\alpha_{1}...\alpha_{s}}^{i}}$ of L.

An example of this method can be found in Dutailly (2014).

\bigskip

\subsection{The variationnal bicomplex}

The spaces of p forms on r prolongation of fibered manifolds can be endowed
with algebraic structures which are actively studied in the frame work of
variational calculus as well as algebraic topology. We give just a brief
introduction on the subject. We use the same notations as above.

\subsubsection{Infinite jet prolongation}

If M, V and the trivializations are smooth a r jet can be extended to
$r=\infty.$ The infinite jet prolongation $J^{\infty}E$ is a Fr\'{e}chet space
and not a true manifold.

The projections yield a direct sequence of forms on the jet prolongations :

$\mathfrak{X}\left(  T_{p}M^{\ast}\right)  \overset{\pi^{\ast}}{\rightarrow
}\mathfrak{X}\left(  \Lambda_{p}TE^{\ast}\right)  \overset{\pi_{0}^{1\ast}%
}{\rightarrow}\mathfrak{X}\left(  \Lambda_{p}TJ^{1}E^{\ast}\right)
\overset{\pi_{1}^{2\ast}}{\rightarrow}\mathfrak{X}\left(  \Lambda_{p}%
TJ^{1}E^{\ast}\right)  ...$

with the pull back of forms :

$\varpi_{p}\in\mathfrak{X}\left(  \Lambda_{p}TJ^{r-1}E^{\ast}\right)
,W_{r}\in\mathfrak{X}\left(  TJ^{r}E\right)  :\pi_{r-1}^{r\ast}\varpi
_{p}\left(  Z_{r}\right)  W_{r}$

$=\varpi_{p}\left(  \pi_{r-1}^{r}\left(  Z_{r}\right)  \right)  \pi_{r-1}%
^{r}\left(  Z_{r}\right)  ^{\prime}W_{r}$

In particular a p form $\lambda$ on M defines a p horizontal form $\pi^{r\ast
}\varpi_{p}$ on $J^{r}E$

It exists a direct limit, and p forms can be defined on $J^{\infty}E$ \ with
the same operations (exterior differential and exterior product).

Any closed form $\varpi$\ on $J^{\infty}E$ can be decomposed as : $\varpi
=\mu+d\lambda$ where $\mu$ is closed on E.

\subsubsection{Contact forms}

A key point, in order to understand what follows, is to keep in mind that, as
usual in jet formalism, a point of $J^{r}E$ does not come necessarily from a
section on E, which assumes that the components are related. Indeed the
prolongation $J^{r}X$ of a section X of E imposes the relations between the
coordinates : $z_{\alpha_{1}..\alpha_{s}}^{i}=\frac{\partial^{s}\sigma^{i}%
}{\partial\xi^{\alpha_{1}}..\partial\xi^{\alpha_{s}}}|_{x}$ and the image of M
by the map $J^{r}X$ is a subset of $J^{r}E.$ So a p form on $J^{r}E$ takes
specific values $\varpi\left(  Z\right)  $ when it is evaluated at the r jet
prolongation $Z=J^{r}X\left(  x\right)  $ of a section of E and there is a map
: $M\rightarrow\Lambda_{p}TJ^{r}E^{\ast}$::$\varpi\left(  J^{r}X\left(
x\right)  \right)  =J^{r}X^{\ast}\varpi\left(  x\right)  $ with the usual
notation for the pull back of a map. Notice that this is not the pullback to M
of the p form on $J^{r}E.$

\paragraph{Contact forms\newline}

To fully appreciate this remark there are non zero\ forms on $J^{r}E$ which
are identically null when evaluated by the prolongation of a section. This can
be understood : in accounting for the relations between coordinates in
$du_{\alpha_{1}...\alpha_{s}}^{i}$ the result can be 0. This leads to the
following definitions :

\begin{definition}
(Vitolo) A p form on $J^{r}E:\varpi\in\Lambda_{p}J^{r}E$ is said to be
\textbf{contact} if \ $\forall X\in\mathfrak{X}_{r}\left(  E\right)
:J^{r}X^{\ast}\varpi=0.$ It is said to be a k-contact form if it generated by
k powers (with the exterior product) of contact forms.
\end{definition}

Any p\ form with p
$>$
m is contact. If $\varpi$ is contact then $d\varpi$ is contact. A 0 contact
form is an ordinary form.

The space of 1-contact p-forms is generated by the forms :

$\varpi_{\beta_{1}..\beta_{s}}^{i}=d\eta_{\beta_{1}..\beta_{s}}^{i}%
-\sum_{\gamma}\eta_{\beta_{1}..\beta_{s}\gamma}^{i}dx^{\gamma},s=0..r-1$ and
$d\varpi_{\beta_{1}..\beta_{r-1}}^{i}$

The value of a contact form vanishes in any computation using sections on E.
So they can be discarded, or conversely, added if it is useful.

\paragraph{Horizontalization\newline}

\begin{theorem}
(Vitolo) There is a map, called (p,q) horizontalisation, such that:
$h_{(k,p-k)}:\mathfrak{X}\left(  T_{p}J^{r}E^{\ast}\right)  ^{k}%
\rightarrow\mathfrak{X}\left(  T_{k}J^{r+1}E^{\ast}\right)  ^{k}%
\wedge\mathfrak{X}^{H}\left(  T_{p-k}J^{r+1}E^{\ast}\right)  $ where
$\mathfrak{X}\left(  T_{p}J^{r+1}E^{\ast}\right)  ^{k}$ is the set of k
contacts q forms.\ 
\end{theorem}

The most important is the map with k=0 usually denoted simply h. It is a
morphism of exterior algebras (Krupka 2000).

So any p form can be projected on a horizontal p form : the result is :
$h_{\left(  0,p\right)  }\left(  \varpi\right)  =\Omega_{\alpha_{1}%
...\alpha_{p}}dx^{\alpha_{1}}\wedge...\wedge dx^{\alpha_{p}}$ where \ $\Omega
$\ is a combination of the components of $\varpi.$ Horizontal forms give :
$h_{\left(  0,p\right)  }\left(  \varpi\right)  =\varpi$

The most important properties of horizontalization are :

$\forall\varpi\in\mathfrak{X}\left(  T_{p}J^{r}E^{\ast}\right)  ,X\in
\mathfrak{X}\left(  E\right)  :J^{r}X^{\ast}\varpi=J^{r+1}X^{\ast}h_{\left(
0,p\right)  }\left(  \varpi\right)  $

$\ker h_{(k,p-k)}=\mathfrak{X}\left(  T_{p}J^{r}E^{\ast}\right)  ^{k+1}$

\subsubsection{Variational bicomplex}

(Vitolo)

1. The de Rahm complex (see cohomologie) is defined by the sequence
$0\rightarrow\Lambda_{0}TM^{\ast}\overset{d}{\rightarrow}\Lambda_{1}TM^{\ast
}\overset{d}{\rightarrow}\Lambda_{2}TM^{\ast}\overset{d}{\rightarrow}....$
There is something similar, but with two dimensions (so it is called a
bicomplex) and two differential operators.

The variational bicomplex uses the maps :

$i_{H}:\Lambda_{p}J^{r}E\rightarrow\Lambda_{p}J^{r+1}E::i_{H}=i_{D^{r+1}}%
\circ\left(  \pi_{r}^{r+1}\right)  ^{\ast}$

$i_{V}:\Lambda_{p}J^{r}E\rightarrow\Lambda_{p}J^{r+1}E::i_{V}=i_{\varpi^{r+1}%
}\circ\left(  \pi_{r}^{r+1}\right)  ^{\ast}$

$d_{H}:\Lambda_{p}J^{r}E\rightarrow\Lambda_{p}J^{r+1}E::d_{H}=i_{H}\circ
d-d\circ i_{H}$

$d_{V}:\Lambda_{p}J^{r}E\rightarrow\Lambda_{p}J^{r+1}E::d_{V}=i_{V}\circ
d-d\circ i_{V}$

with :

$D^{r+1}=\sum_{\gamma}dx^{\gamma}\otimes\left(  \partial\xi_{\gamma}%
+\sum_{s=1}^{r}\sum_{\alpha_{1}\leq...\leq\alpha_{s}}\eta_{\gamma\alpha
_{1}...\alpha_{s}}^{i}\partial u_{\alpha_{1}...\alpha_{s}}^{i}\right)  $

$\varpi^{r+1}=\sum_{\gamma}\left(  \partial u_{\alpha_{1}...\alpha_{s}}%
^{i}-\eta_{\gamma\alpha_{1}...\alpha_{s}}^{i}dx^{\gamma}\right)
\otimes\partial u_{\alpha_{1}...\alpha_{s}}^{i}$

They are fully defined through their action on functions :

$d_{H}f=\sum_{\gamma}\left(  \frac{\partial f}{\partial\xi^{\gamma}}%
+\sum_{s=1}^{r}\sum_{\alpha_{1}\leq...\leq\alpha_{s}}\eta_{\gamma\alpha
_{1}...\alpha_{s}}^{i}\frac{\partial f}{\partial\eta_{\alpha_{1}...\alpha_{s}%
}^{i}}\right)  dx^{\gamma}$

$d_{H}dx^{\alpha}=0,d_{H}\left(  du_{\alpha_{1}...\alpha_{s}}^{i}\right)
=-\sum_{\beta}du_{\beta\alpha_{1}...\alpha_{s}}^{i}\wedge dx^{\beta},$

$d_{V}f=\sum_{\alpha_{1}..\alpha_{s}}\frac{\partial f}{\partial\eta
_{\alpha_{1}...\alpha_{s}}^{i}}\varpi_{\alpha_{1}...\alpha_{s}}^{i}$

$d_{V}dx^{\alpha}=0,d_{V}\left(  du_{\alpha_{1}...\alpha_{s}}^{i}\right)
=\sum_{\beta}du_{\beta\alpha_{1}...\alpha_{s}}^{i}\wedge dx^{\beta}$

with the properties :

$d_{H}^{2}=d_{V}^{2}=0$

$d_{H}\circ d_{V}+d_{V}\circ d_{H}=0$

$d_{H}+d_{V}=\left(  \pi_{r}^{r+1}\right)  ^{\ast}d$

$\left(  J^{r+1}X\right)  ^{\ast}\circ d_{V}=0$

$d\circ\left(  J^{r}X\right)  ^{\ast}=\left(  J^{r+1}X\right)  ^{\ast}\circ
d_{H}$

$d_{H}=\mathfrak{D}$ the total external differentiation used previously.

On the algebra $\Lambda J^{\infty}E$ these relations simplify a bit :
$d_{H}+d_{V}=d$ and the Lie derivatives : \pounds $_{d_{\alpha}}=d_{\alpha
}\circ d+d\circ d_{\alpha}$ are such that :

$\pounds _{d_{\alpha}}\left(  \varpi\wedge\mu\right)  =\left(
\pounds _{d_{\alpha}}\varpi\right)  \wedge\mu+\varpi\wedge\pounds _{d_{\alpha
}}\mu$

2. On this algebra the space of p forms can be decomposed as follows :

$\mathfrak{X}\left(  T_{p}J^{r}E^{\ast}\right)  =\oplus_{k}\mathfrak{X}\left(
T_{k}J^{r}E^{\ast}\right)  ^{k}\wedge h_{(0,p-k)}\left(  \mathfrak{X}\left(
T_{p-k}J^{r}E^{\ast}\right)  ^{0}\right)  $

The first part are contact forms (they will vanish with sections on E), the
second part are horizontal forms.

Then with the spaces :

$F_{0}^{0q}=h_{(0,q)}\left(  \mathfrak{X}\left(  \Lambda_{q}TJ^{r}E^{\ast
}\right)  ^{0}\right)  ,F_{0}^{pq}$

$=\mathfrak{X}\left(  T_{p}J^{r}E^{\ast}\right)  ^{p}\wedge h_{(0,q)}\left(
\mathfrak{X}\left(  T_{q}J^{r}E^{\ast}\right)  ^{0}\right)  ,F_{1}^{pn}%
=F_{0}^{pn}/d_{H}\left(  F_{0}^{pn-1}\right)  $

one can define a bidirectional sequence of spaces of forms, similar to the de
Rahm cohomology, called the variational bicomplex, through which one navigates
with the maps $d_{H},d_{V}.$

The variational sequence is :

$0\rightarrow%
\mathbb{R}
\rightarrow F_{0}^{00}\overset{d_{H}}{\rightarrow}...F_{0}^{0n-1}%
\overset{d_{H}}{\rightarrow}F0+_{0}^{0n}\rightarrow...$

3. The variational bicomplex is extensively\ used to study symmetries. In
particular :

\begin{theorem}
(Vitolo p.42) A lagrangian $%
\mathcal{L}%
:J^{r}E\rightarrow\wedge_{m}TM^{\ast}$ defines by pull back a horizontal m
form $\varpi=\left(  \pi^{r}\right)  ^{\ast}%
\mathcal{L}%
$ on $J^{r}E.$ If $d_{V}\varpi=0$ then $\exists\mu\in h^{0m-1}\left(
\nu\right)  ,\nu\in T_{m-1}J^{r-1}E^{\ast}$ such that : $\varpi=d_{H}\mu.$
\end{theorem}

\begin{theorem}
First variational formula (Vitolo p.42) For $W\in\mathfrak{X}\left(
TE\right)  ,\varpi\in F_{1}^{pm}$ the identity :$\pounds _{W}\varpi=i_{W}%
d_{V}\varpi+d_{V}\left(  i_{W}\varpi\right)  $ holds
\end{theorem}

It can be seen as a generalization of the classic identity : $\pounds _{W}%
\varpi=i_{W}d\varpi+d\left(  i_{W}\varpi\right)  $

\newpage

\part{\textbf{BIBLIOGRAPHY}}

\begin{center}
\bigskip
\end{center}

\label{Bibliography}

David Applebaum \textit{Universal Malliavin calculus in Fock and L\'{e}vy-Ito
spaces} arXiv:0808-2593v1 19 Aug 2008

Stephane Attal, Ion Nechita \textit{Discrete approximation on the free Fock
space} arXiv:0810.4070v1 22 oct 2008

A.Axelsson, A.McIntosh \textit{Hodge decomposition on weakly Lipschtiz domain
}Internet paper (1991)

C.Berge \textit{Espaces topologiques} Dunond (1965)

N.Bourbaki \textit{General topology} Addison-Wesley (1968)

O.Bratteli, D.W.Robinson \textit{Operators algebras and quantum mechanics 1}
2nd edition Springer (1987)

J.L.Doob \textit{Measure theory} Springer (1994)

J.J.Duistermaat, J.A.C.Kolk \textit{Lie groups} Springer (1999) sep

J.C.Dutailly \textit{Particles and Fields} hal-00933043 v1 19 janvier 2014

W.Fulton, J.Harris \textit{Representation theory} Springer (1991)

Th.W.Gamelin, R.E.Greene \textit{Introduction to topology }2nd edition Dover (1999)

G.Giachetta, L.Mangiarotti,,G.Sardanashvily \textit{Advanced classical field
theory} World Scientific (2009)

A.Gregor'yan \textit{Heat kernel on weighted manifolds and applications} (2006)

P de la Harpe \textit{The Clifford algebra and the spinor group of a Hilbert
space} Compositio Mathematica tome 25 N$%
{{}^\circ}%
3$ (1972)

H.Hofer,E.Zehnder \textit{Symplectic invariants and hamiltonian dynamics}
Birkha\"{u}ser (1994)

D.Husemoller \textit{Fiber bundle} 3d Edition Springer (1994)

M.Kashiwara, P.Shapira \textit{Categories and sheaves} Springer (2006)

S.C.Kleene \textit{Mathematical logic} Wiley (1967)

A.W.Knapp \textit{Lie groups : beyond an introduction (2nd edition)}
Birkh\"{a}user (2005)

A.W.Knapp \textit{Representation theory of semi-simple Lie groups }Princeton
landmarks (1986)

E.H.Lieb, M.Loss \textit{Analysis} 2nd edition American Mathematical Society (2001)

I.Kol\`{a}r,P.Michor,J.Slovak \textit{Natural operations in differential
geometry} Springer (1993)

Y.Kosmann-Schwarzbach \textit{Groupes et sym\'{e}tries} Editions de l'Ecole
Polytechnique (2006)

D.Krupka \textit{On the local structure of the Euler-Lagrange mapping of the
calculus of variations }Mathematical Institute Opava (2002)

D.Krupka \textit{Some geometric aspects of variational problems in fibered
manifolds }Mathematical Institute Opava (2001)

D.Krupka, M.Krupka \textit{Jets and contacts elements} Proceedings of the
seminar on differential geometry Opava (2000)

O.Krupkova, G.E.Prince \textit{Second order differential equations in jets
bundles and the inverse problem of the calcul of variations} Handbook of
global analysis Elsevier (2007)

J.Lafontaine \textit{Introduction aux vari\'{e}t\'{e}s diff\'{e}rentiables}
EDP (1996)

S.M.Lane \textit{Categories for the working mathematician 2nd edition}
Springer (1969)

A.D.Lewis \textit{Affine connections and distributions} Reports on
mathematical physics 42 (1/2) 125-164 (1998)

D.Lovelock,H.Rund \textit{Tensors, differential forms and variational
principles} Dover (1989)

P.Malliavin \textit{G\'{e}om\'{e}trie diff\'{e}rentielle intrins\`{e}que}
Hermann (1972)

J.Munkres \textit{Topology} (2nd ed) Pearson (2003)

M.Nakahara \textit{Geometry, topology and physics} 2nd edition IoP (2003)

K.H.Neeb \textit{An introduction to unitary representations of Lie groups}
Internet Paper (june 2 2010)

E.Nigsch \textit{Colombeau generalized functions on Manifolds} Technische
Universit\"{a}t Wien

A.Polyanin, V.Zaitsev \textit{Handbook of exact solutions for ordinary
differential equations} Chapman \& Hall (2003)

A.Polyanin, V.Zaitsev \textit{Handbook of linear partial differential
equations} for engineers Chapman \& Hall (2003)

A.Polyanin, V.Zaitsev \textit{Handbook of non linear partial differential
equations} Chapman \& Hall (2004)

L.Schwartz \textit{Analyse} 4 tomes Hermann (1997)

Jerzy B.Stochel \textit{Representation of generalized creation and
annhiliation operators in Fock space} Universitatis Iagellonicae Acta
Mathematica (1997)

G.Svetlichny \textit{Preparation for gauge theories} arXiv:math-ph 9902027 v3
12 march 1999

M.E.Taylor \textit{Partial differential equations }3 volumes Springer (1999)

M.Thill \textit{Introduction to normed *-algebras and their representations
7th edition} arXiv 1011-1558v1 [math.0A] 6 nov.2010

M.Thill \textit{Representation of hermitian commutative *-algebras by
unbounded operators} arXiv 0908-3267v2 [math.0A] 20 nov.2009

R.Vitolo \textit{Variational sequences} (http:%
$\backslash$%
$\backslash$%
poincare.unile.it) (2007)

R.M.Wald \textit{General relativity} University of Chicago Press (1984)

A.Wilansky \textit{Topology for analysis} Dover (2008)

C.Zuily \textit{El\'{e}ments de distributions et d'\'{e}quations aux
d\'{e}riv\'{e}es partielles} Dunod (2002)

\end{document}